\documentclass[twoside,titlepage]{book} 
\usepackage{latexsym}
\usepackage{amssymb}
\usepackage{amsmath}
\usepackage{graphicx}
\usepackage{enumitem}
\usepackage{bbold}
\usepackage{slashed}
\usepackage{hyperref}
\usepackage{xcolor}
\usepackage{apacite}
\usepackage{tikz-cd}
\usepackage{yfonts}
\usepackage{mathrsfs}

\def\be{\begin{equation}}
\def\ee{\end{equation}}
\newcommand\bra[1]{{\langle {#1}|}}
\newcommand\ket[1]{{|{#1}\rangle}}
\def\Tr{{\rm Tr}}
\def\dd{\mbox{d}}

\def\Om{\Omega}
\def\om{\omega}
\def\bra{\langle}
\def\ket{\rangle}
\def\a{\alpha}
\def\b{\beta}
\def\g{\gamma}
\def\d{\delta}
\def\D{\Delta}

\def\g{\gamma}
\def\G{\Gamma}
\def\e{\epsilon}
\def\ve{\varepsilon}

\def\f{\phi}
\def\F{\Phi}

\def\vf{\varphi}
\def\k{\kappa}
\def\l{\lambda}
\def\L{\Lambda}
\def\m{\mu}
\def\n{\nu}
\def\s{\sigma}
\def\Si{\Sigma}

\def\r{\rho}
\def\t{\tau}
\def\th{\theta}

\def\pa{\partial}
\newcommand{\ti}[1]{\tilde{#1}}

\newcommand{\sm}[1]{\mbox{\scriptsize #1}}
\newcommand{\tn}[1]{\mbox{\tiny #1}}
\renewcommand{\@}[1]{\sqrt{#1}}
\renewcommand{\le}[1]{\label{#1}\end{eqnarray}}
\newcommand{\bea}{\begin{eqnarray}}
\newcommand{\eea}{\end{eqnarray}}
\newcommand{\eq}[1]{(\ref{#1})}
\def\nn{\nonumber\\}

\def\na{\nabla}
\def\half{{1\over2}\,}

\begin{document}
\rm\large
\thispagestyle{empty}
\titlepage
\vspace*{15pt}
\begin{center}
{\scshape\huge
\setlength\baselineskip{30pt}
The Philosophy and}\\ 
\vspace{.3cm}
{\scshape\huge
\setlength\baselineskip{30pt} Physics of Duality
\par}
\vspace{5cm}
\begin{center}
{\Large\bf Sebastian De Haro~\,and~\,Jeremy Butterfield}

\end{center}
\vspace{2cm}
\today
\end{center}
\vspace{4cm}

\noindent This is the book manuscript as submitted for production in October 2024. The book is published Open Access by Oxford University Press and can be freely downloaded here as a PDF: \\\href{https://global.oup.com/academic/product/the-philosophy-and-physics-of-duality-9780198846338}{https://global.oup.com/academic/product/the-philosophy-and-physics-of-duality-9780198846338}. \\ When citing this work, please refer to the published version.

\newpage
\normalsize
\pagenumbering{roman}
\thispagestyle{plain}

{\large{\it To the memory of my father, who sparked my interests} (SDH)}\\
\\

{\large{\it To Mari} (JNB)}

\tableofcontents
\newpage
\pagenumbering{arabic}
\frenchspacing

\markboth{}{}

\section*{Preface}\addcontentsline{toc}{section}{\it Preface}

For several decades, dualities have been centre-stage in modern physics: both in empirically well-confirmed theories, e.g.~within statistical mechanics and quantum field theory; and in emerging theories, such as string theory. So the time is ripe for a philosophical assessment of dualities: which we undertake in this book to provide. We leave to Chapter 1 the task of a fuller description. Here, we will just describe the origin of the book, and make our acknowledgements.
 
This book arose from our collaboration which began in 2014; and from the resulting papers, both joint and sole-authored. (There is also joint work with Nicholas Teh, Daniel Mayerson, Jeroen van Dongen and Manus Visser; and work by SDH with Dennis Dieks, Jeroen van Dongen, Henk de Regt and Elena Castellani.) In the writing of the book, 
JNB drafted parts of: Chapter 2, Sections 1.1, 1.3, 3.1, 3.2, 3.3, 4.1, 4.4 and 13.1; and SDH wrote the other Sections and Chapters, which JNB also commented on and helped to edit.\\

\smallskip

{\it Acknowledgements.} It is a pleasure to thank many people for the help they have given us. Our gratitude goes, first of all, for their comments on the manuscript, to: Enrico Cinti, Richard Dawid, Hans Halvorson, Klaas Landsman, Jill North, James Read, Dean Rickles, Johan van Benthem, Jasper van Wezel and James Weatherall. We also thank the members of the History and Philosophy of Physics group of the University of Amsterdam for extensive comments on the manuscript. We thank Amal Salman for her skilful assistance with the manuscript and drawing the figures.

For discussions, at various stages of this project, about the materials presented in this book, we thank: Thomas Barrett, Silvester Borsboom, Elena Castellani, Hasok Chang, Valeriya Chasova, Henk de Regt, Neil Dewar, Dennis Dieks, Nick Huggett, Baptiste Le Bihan, Keizo Matsubara, F.~A.~Muller, Oliver Pooley, Kostas Skenderis, Sonja Smets, Andrew Strominger, Nicholas Teh, David Tong, Cumrun Vafa, Benno van den Berg, Jeroen van Dongen, Sanne Vergouwen, Erik Verlinde and Manus Visser. We also thank audiences at many graduate classes, seminars and conferences where this material was presented, especially at Urbino, Viterbo, Geneva, Munich, Amsterdam and Oxford.

We thank our institutions for their support over the years of this project, the University of Amsterdam (SDH) and Trinity College, Cambridge (JNB). Additionally, SDH thanks Trinity College, Cambridge and the Department of History and Philosophy of Science of the University of Cambridge for their support during the first years of this project. We also thank the Black Hole Initiative, Harvard University, for supporting long visits in 2018-2019. SDH also thanks the Tsinghua Sanya International Mathematics Forum and the Galilei Institute for Advanced Studies for supporting his visits in 2019.

We also thank the editorial and production staff at OUP for their help; especially Peter Momtchiloff, April Peake and Tara Werger. We are also grateful for the funding for the open access publication of this book, that we have received from NWO, under project number 36.201.106 (SDH); and from Trinity College, Cambridge (JNB).

SDH's research was funded by the Dutch National Science Agenda (Nationale Wetenschapsagenda, NWA) by NWO, under project numbers NWA.1418.22.029, Small Innovative NWA projects; and NWA.1436.20.002, The quantum/nano-revolution.\\

Sections \ref{salientstipul}, \ref{dynlsymm}, \ref{sptthies} and \ref{expleGalil} reuse materials from De Haro and Butterfield (2021:~pp.~2986-2990, 2996-3000), under a Creative Commons CC BY licence by Springer Nature (29 May 2019). 

Figure \ref{SV} is taken from De Haro et al.~(2020:~pp.~95), under Elsevier License Number 5440740026222 (2 December 2022).

Some text suggestions have been taken from www.bing.com.

\chapter{Introduction}\label{Intro}
\markboth{\small{\textup{Introduction}}}{\textup{\small{Introduction}}}

\section{Dualities in physics and philosophy}\label{dpp}

This book is about {\it dualities} in physics. In short, a duality is an equivalence between two physical theories, that is often surprising because the theories are disparate. In recent decades, dualities have become a large and rich topic, in many fields of physics. In statistical physics, condensed matter physics, quantum field theory and string theory, a great deal of research has been devoted to finding, understanding and then exploiting, various dualities. This is especially true in string theory, where for more than twenty years, various dualities have been a (arguably: the) primary focus of research.

Familiar examples of dualities include position-momentum duality in quantum mechanics and electric-magnetic duality in the Maxwell theory. Less familiar dualities in quantum field theory relate bosons and fermions, and ordinary particles to spatially extended (soliton) solutions of non-linear equations of motion. In statistical mechanics, Ising model duality relates high- and low-temperature spin lattices (usually called `Kramers-Wannier duality').

Thus as we will see, dualities involve not just the formal equivalence between theories, but also the discovery of new, interesting solutions, with rich physics. For example, electric-magnetic duality in Yang-Mills theory relates electric point particles to extended magnetic monopoles. In string theory, dualities relate theories on different spaces: gravitation is thought to be equivalent, through gauge-gravity duality, to a quantum field theory on the boundary of the space. And since dualities relate such disparate physical systems, they are useful tools for theory construction.

The idea of an equivalence between two theories, especially a surprising one, is a natural topic for philosophy of science---concerned as it is with giving an account of theories, and relations between them such as (various sorts of) equivalence. Thus one might well ask: how can two different, indeed disparate, theories be called `equivalent'? Besides, physics' use of dualities as a heuristic for developing theories, or for finding new ones, is also a natural topic for philosophy. 

But the philosophical analysis of duality in physics is an enterprise that has hardly begun. Although there has been a recent stream of research articles in the philosophy of physics literature about dualities, there is still no book-length philosophical account. Hence this book undertakes two main tasks: (i) to explain and illustrate dualities in physics by presenting our own account of them (and we do this in conversation with the philosophical literature about dualities, from which we gather the threads); and (ii) to relate our account to other themes in philosophy of physics and in philosophy of science, so as to build a convincing account of dualities and their scientific and conceptual roles. 

The book has three Parts. In Part I, from Chapter \ref{Thies} to Chapter \ref{Simple} we present our main views about what dualities {\it are} (Chapters \ref{Thies} and \ref{Schema}), and then give some simple examples (Chapter \ref{Simple}). These Chapters, and this Introduction, are written with our two intended readerships, philosophers and physicists, equally in mind. In particular, we will introduce the jargons of the two disciplines. In Part II, from Chapter \ref{Advan} to Chapter \ref{HABHM}, we present more advanced examples of dualities. We start with examples in a low number of spatial dimensions (i.e.~one or two rather than three); then we present electric-magnetic duality and dualities in condensed matter theory and quantum field theories (Chapters \ref{EMDuality} and \ref{EMYM}); and then we discuss examples in string theory (Chapters \ref{String} to \ref{HABHM}). Thus from Chapter \ref{Simple} to Chapter \ref{HABHM}, the technicality of our examples increases; so that there is a learning curve, especially for readers who are philosophers. In Part III (Chapters \ref{Theor} and \ref{Heuri}), philosophy returns to the foreground. Here, we discuss dualities' implications for philosophical issues. Chapter \ref{Concl} concludes. 

So: what is a duality? We said it is a surprising equivalence between two physical theories. It is also often said that a duality is: a {\it symmetry} between two apparently different theories. So what, one asks, is a symmetry in physics? To explain this, we will in the next Section introduce some jargon from physics, which will be used throughout the book (Section \ref{thsq}). With this jargon in hand, we then give a first statement of our account of duality, namely as a {\it symmetry} or {\it isomorphism} between theories (Section \ref{giantS}). (The discussion in these two Sections is informal and elementary, and meant as an invitation: more rigorous discussions follow in Chapter \ref{Thies}, and in Part III.)

\subsection{Theories, states and quantities: some jargon from physics}\label{thsq}

Let us first introduce the idea of a {\bf theory}, in physics. At its simplest, a physical theory describes a certain kind of object by ascribing to it numerically measurable properties like position, or momentum (i.e.~mass times velocity) or energy. In the jargon of physics, the objects are called {\bf systems}; their properties like position etc.~are called {\bf quantities} (also: {\it magnitudes}, but we will not use this word); and the amounts or degrees of such properties that are ascribed are called {\bf values} (almost always real numbers). In the jargon of philosophy, the quantities are {\it determinables}, and each of their values is a {\it determinate}. This is analogous to standard philosophical examples such as colour being a determinable, of which scarlet is a determinate. Then the {\bf state} of a system, according to a physical theory that describes it, is the list, or conjunction, stating what are the system's values for the various quantities that apply to it.\footnote{Classical and quantum states of course have important differences: for example, a quantum state does not stipulate values for all the quantities of the theory---hence the phrase `for the various quantities that apply to it'. But, in this Section, our discussion will be as simple as possible, and we will not need to emphasise these differences.}

Agreed: various questions already arise, such as: should the quantities include extrinsic properties of the system, i.e.~properties concerning how the system described by the theory relates to the rest of the world? Could the values of a quantity be a complex, rather than a real, number? And what about probabilities? We will come to these and similar questions later, but for the moment it is enough to keep things very simple: in particular, considering only quantities for intrinsic properties, with real-number values, and ignoring probabilities.

 The state of course changes over time, as the values of the various quantities go up or down. A physical theory gives descriptions of these changes. In the paradigm cases (to which we confine ourselves in this Section), the theory provides an equation stating exactly how the state (the values of all the system's quantities) changes over time. This is the system's {\bf equation of motion}. Typically, it is a {\it differential equation} that fixes the rate of change of some chosen quantity (or quantities) of interest, as a function of the values of that quantity, and usually also other quantities, at some initial time. Given those other values, and thereby the rate of change of the chosen quantity, one then solves the equation so as to find the value of the chosen quantity at later times. In short, one predicts the future values of that quantity. 
 
The archetypal case is, of course, Newtonian mechanics. Imagine that a small solid object, say a sphere, is our system of interest (in Newtonian mechanics, usually called a {\it body}). If we know the forces that are now, and that will later, be exerted on the sphere (say by other bodies, e.g.~gravitational forces or electric forces), and we also know the sphere's present position and momentum: then the equation of motion for its position can be solved. That is: the position at later times (and so also the momentum) can be calculated. 
 
Agreed: even in this archetypal case, qualifications are needed. The two main ones concern: (i) whether there is an explicit solution; (ii) what to say about collisions. As to (i): We only mean that the equation can be solved, the later positions calculated, {\it in principle}. One must admit that the equation may be very complicated to solve, and it may be impossible to write down an explicit formula for the position at later times (even a very long formula) in terms of familiar functions, such as using the operations of addition, multiplication, taking a square root, or using trigonometric functions like sine and cosine. As to (ii): Of course, when two bodies collide, what happens is very complicated: they usually distort each other, or even break up, so that describing what happens will soon outstrip the resources of Newtonian mechanics---for example, because the collision generates heat. 

But again, we can for the moment set these qualifications aside. So we envisage in general that each theory is of limited scope in that it describes a selected set of objects, i.e.~systems, in an appropriate {\bf regime}. For example, in Newtonian mechanics, we might require that the bodies to be described do not collide. And for systems in this regime, the theory ascribes to each system a selected set of quantities. For example, fluid mechanics, as usually understood, treats a fluid as a continuum, and also ignores thermal notions, like heat and temperature. So it will not describe rigid bodies, nor ascribe temperature. In philosophy, the jargon corresponding to `regime' is {\bf domain of application} (or for short: `domain');\footnote{But there is a difference. Physics' word `regime' connotes an abstract class of theoretically possible phenomena or behaviours, e.g.~motions without collisions, fluid flow without temperature, while `domain of application' connotes a concrete part of the actual empirical world, which the theory describes, maybe accurately, but never fully, since there are bound to be features that the theory is not concerned with.}

Thus a theory envisages a set of possible states for each of its systems (the objects to which the theory applies): a {\bf state-space}. A {\it possible history} (or just: {\it history}) of such a system, according to the theory, is a sequence of states: a sequence of elements of the state-space. We can think of it as a {\it curve} or {\it trajectory} through the state-space. Note: although the word `history' connotes the past, we envisage a curve passing through a given state as including both directions in time: as including states occurring (possessed by the system) later than the given state, as well as states earlier than it. 

In the paradigm cases just considered, in which the theory gives an equation of motion, not all such curves are possible for the system: only those that are solutions to the equation of motion. More jargon: for a given system, the framework consisting of its states and the state-space (and thereby all the mathematically definable curves through the state-space), with the quantities defined on it, is called {\bf kinematics}. When we add an equation of motion---say, because we know the forces that are now, and that will later, be exerted on the system---we cut down the set of curves being considered to those that obey the equation of motion. Then we are doing {\bf dynamics}, and the curves that obey the equation of motion are called the {\bf dynamically possible trajectories/histories/curves}. 

Here, we can also give a first statement of {\it determinism}. There are various precise formulations, but the general idea is of course that the state at one time determines the state at other times. The most common precise formulation is that any state in the state-space determines the sequence of states for all future, and indeed all past, times. In terms of curves in the state-space: through any point (element) of the state-space, there is a unique curve to the future, and indeed to the past. 

Again, Newtonian mechanics is the archetypal case. Provided we are not asking for an explicit formula (cf.~(i) above), and also are willing to set aside collisions (cf.~(ii) above), Newtonian mechanics provides us with a very neat picture of a body's possible motions. Namely: given the forces that are exerted on the body, not just at the given time (`now') but throughout the past and the future, any state in the state-space has passing through it a unique history (curve through the state-space).

To sum up: this characterisation of theories in terms of states, quantities, and dynamics is standard in physics, both classical and quantum. But in closing, we stress again that there are qualifications. For example, some theories provide as their dynamics, not an equation of motion of certain quantities, but only a principle stating the system's tendency to maximise its entropy. So the cases considered above are the simplest or central cases. But we will occasionally consider alternative `shapes' that a theory can have, i.e.~different ways in which a theory can be formulated. 

\subsection{Duality as a giant symmetry; and as an isomorphism}\label{giantS}

So much by way of introducing states, quantities and associated notions like state-space and determinism. Now we can give the general idea of a {\bf symmetry}. 

It is a function, a map, on states that preserves the values of quantities. That is: a symmetry associates to each state in a certain set ${\cal S}$ of states---usually ${\cal S}$ is the theory's entire state-space---another state in the state-space that has the same values for quantities. Here, the equality of values will in general not hold for {\it all} the quantities that apply to the system in question, but only for all the quantities in some large but salient subset of quantities. 

Various questions immediately arise. The over-arching one is: what is the physical significance, and the scientific importance, of symmetries as thus defined? Another one, obviously related to the first, is: what determines the large but salient subset of quantities whose values are to match? Another is: what about other (perhaps rival) definitions of symmetries, such as {\it dynamical symmetries}? (These are maps on states that preserve dynamical possibility, in the sense that if one considers how the map sends curves through the state-space to other curves, it has the feature that any dynamically possible history is mapped into another dynamically possible history.) We will of course discuss these questions in the sequel (especially Sections \ref{Symm} and \ref{dualsym}). But for now, we only need the general idea of a symmetry as a map on states that preserves the values of (salient) quantities. For that is enough for us to give our first statement of what a duality is.\footnote{There is a substantial philosophical literature about symmetries. Compare e.g.~Brading and Castellani (2003). Also (shorter) Brading et al.~(2017). The philosophical literature about dualities is much smaller and recent: in Chapter \ref{Schema} we will of course relate our account to this literature.}\\

We said above that a duality is a `symmetry between two apparently very different theories'. Now we can see better what this comes to.\footnote{We use the word `apparent' to mean just that: `as it appears'. In our usage, `apparent' does not imply, or connote, `merely apparent', `seemingly so, but in fact not so'. For us, `apparent' leaves the matter open: `maybe merely apparent, but maybe not'.} 
So we are presented with two theories, each with their states, and so state-space, and with their set of quantities: states and quantities that they ascribe to the systems they describe. The theories are `apparently different'. That is: the system they describe, and-or the states and quantities they ascribe to them, appear to be different. 

\begin{figure}
\begin{center}
\includegraphics[height=2.5cm]{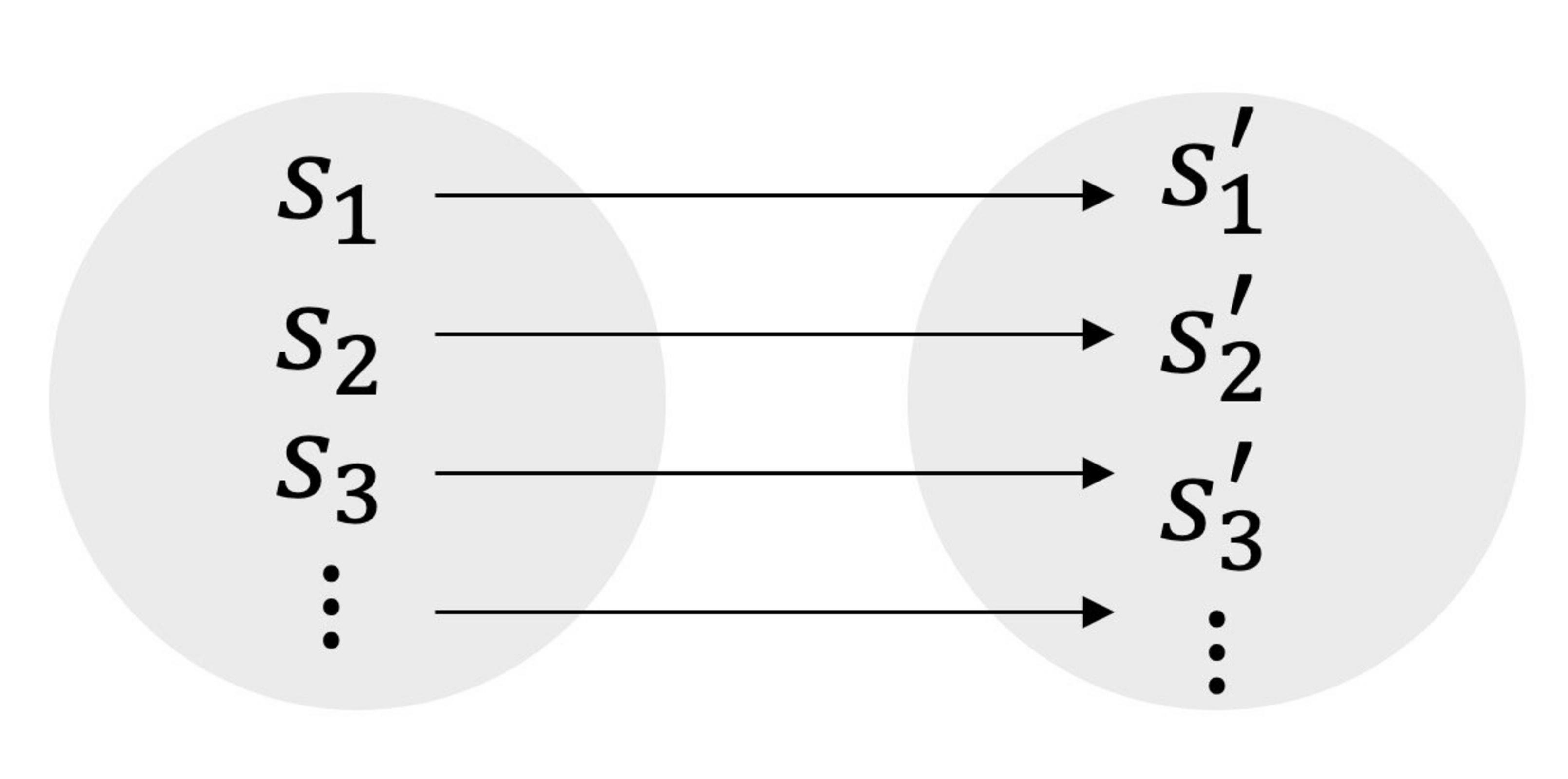}
\caption{\small On the left, a theory with the set of states $s_1,s_2,s_3,\ldots$ and, on the right, a theory with the set of states $s'_1,s'_2,s'_3,\ldots$. These states might describe a cold system (on the left) and a hot system (on the right). The duality maps the states of one theory to the states of the other theory, such that the values of their respective quantities match.}
\label{t1tot2}
\end{center}
\end{figure}

It is natural to visualize this in terms of one theory being on our left, and the other on our right: see Figure \ref{t1tot2}. For example, we are presented on our left with a theory that ascribes to a particle appropriate values of position (but not of momentum), and on our right with a theory that ascribes to a particle appropriate values of momentum (but not of position). Another example: on our left, a theory that describes a cold system; and on our right, a theory that describes a hot system. 

So the idea of duality is: despite these apparent differences, there is a map from the states and quantities of one theory to the states and quantities of the other {\it that makes the values match} and preserves the dynamics. So a duality is like a `giant symmetry'. While a symmetry maps a state of the system into another state of the same system (with equal values of appropriate quantities): in a duality, an entire theory---state-space and quantities---is mapped into another theory, with the values of quantities and the dynamics being appropriately mapped from one to the other.

We can summarise so far by saying that a duality is an {\bf isomorphism} between two theories. Indeed, this one-line summary is common usage among physicists, and mathematicians discussing dualities in physics. (In pure mathematics, `duality' often means an isomorphism that is self-inverse, i.e.~an isomorphism that applied twice yields the identity map.\footnote{For a discussion of duality in mathematics, and especially in category theory, see Corfield (2017).} 
We will occasionally touch on this contrast; but it will not be central.) We end this Section with more details about duality as isomorphism.\\

An isomorphism is of course, by definition, a bijection (a one-one and onto map) that preserves structure. But one must always ask: what structure? Obviously, the structure must concern the features that match between the two theories, and the isomorphism will map those features of the one to matching features of the other. Since we said above that a duality maps states and quantities so as to match the values of physical quantities in states, it is clear that for us, these common features include assignments of values to quantities.\footnote{If the sets of states and quantities come defined with a structure, then this defining structure is also preserved: see Section \ref{intext} (A), and Section \ref{dualsym} for the preservation of symmetries.}

\begin{figure}
\begin{center}
\includegraphics[height=2.5cm]{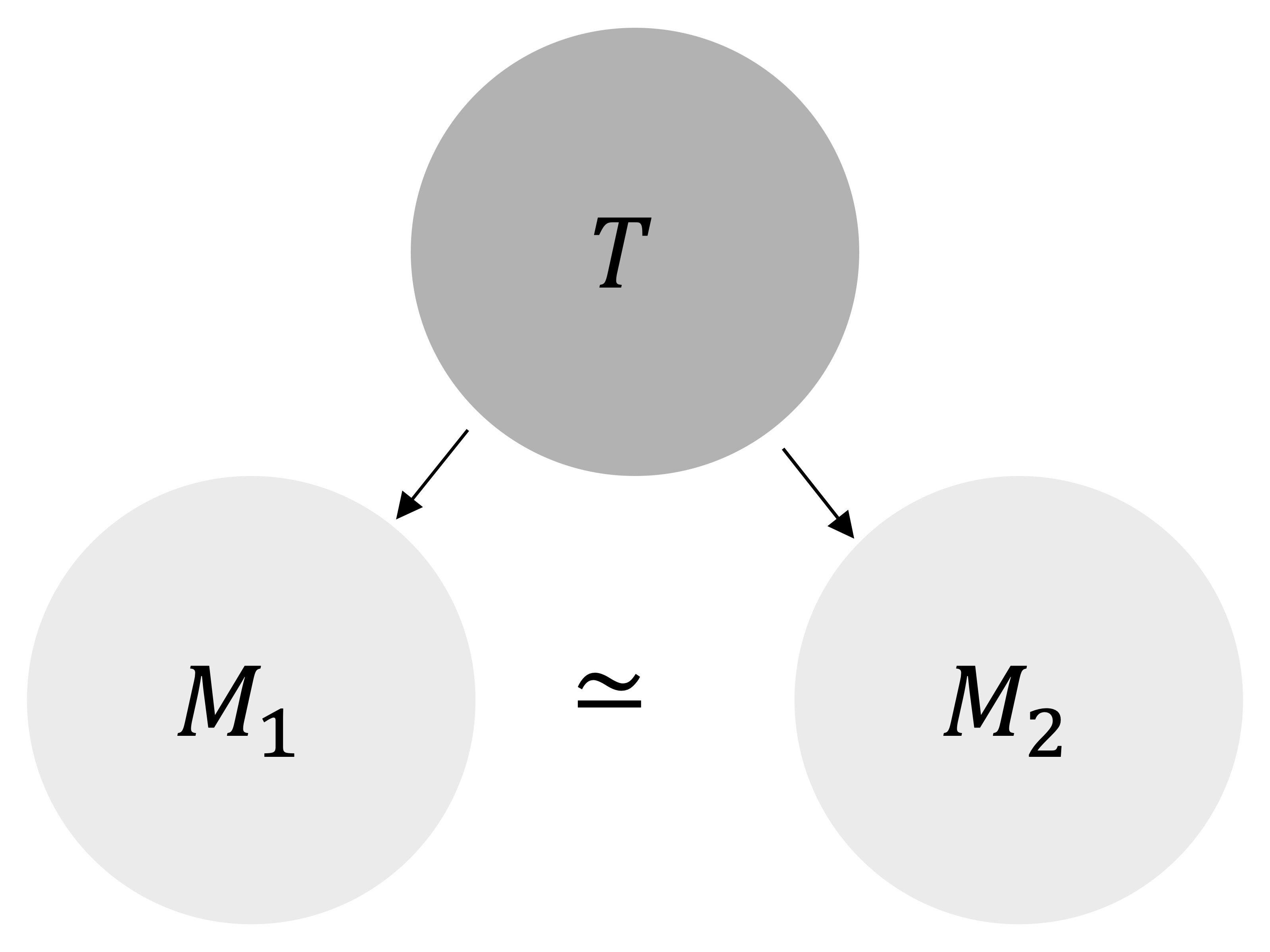}
\caption{\small A common core theory, $T$, with two dual (i.e.~isomorphic) representations, $M_1$ and $M_2$.}
\label{tm1m2}
\end{center}
\end{figure}

So far, we have said about the idea of a `physical theory' only that it postulates states (and so a state-space), quantities, and so values of quantities, and (at least usually) a dynamics (equations of motion, or other dynamical laws). We shall say more about theories in Chapter \ref{Thies}. But this will indeed be our main conception of them: a state-space, a set of quantities, and a dynamics. 

And since a duality is an isomorphism between two theories, and dual theories have other features that make them apparently different, we think of the dual theories as formulations, or versions, or realizations, of a {\bf common core theory} (see Figure \ref{tm1m2}). In fact, they are almost always {\bf representations} of the common core theory, in the sense of representation theory.\footnote{That is; they are homomorphic copies of the common core theory. (`Homomorphism' means a structure-preserving map that need not be one-one, nor onto.)} For the common core theory will consist of a state-space, a set of quantities and a dynamics, and so it will be precisely enough defined that one can sensibly talk of representations (homomorphic copies) of it. We will see this in the examples of Chapters \ref{Simple} to \ref{HABHM}.\footnote{We stress that here we mean `representation' in the mathematical, not the philosophical, sense. We will discuss the philosophical sense in Section \ref{itm}.} 

Two representations, i.e.~homomorphic copies, of a given mathematical object are of course in general {\it not} isomorphic to each other---with respect to the structure for which they are homomorphic copies of the given object. But we say: {\it a duality is a matter of two representations of a common core theory, being isomorphic with respect to the structure of that core.}\\

In short: our account---we will call it our {\bf Schema} for duality---says that a duality between two physical theories requires that:

(a): the two theories share a {\it common core}; the common core is itself a theory, the {\it common core theory}; and 

(b): the two given theories are {\it isomorphic formulations}---almost always representations in the sense of representation theory---of the common core theory.

Here, a word about what we mean by `account' is called for. Given a concept of philosophical interest---for us, duality---philosophers differ about what their task is. The tradition of {\it conceptual analysis}, which stretches from Plato's dialogues (asking `What is courage?, `What is knowledge? What is justice?') to analytic philosophy, says: the task is to analyse the concept, as it is, in terms that are already well-understood (and in that sense are not problematic, in the way the given concept is assumed to be). Since to present any doctrine whatsoever one must use language, this means, in linguistic terms: to provide a definition of (the word for) the concept, that is faithful to it. 

But there is also a more `liberal' tradition, to which we adhere. It allows that the given concept may be vague and-or in some way defective; and if so, it is legitimate, perhaps even mandatory, for the philosophical account of the concept to make it precise and-or to alter it so as to mend the defects. A word commonly used for this endeavour is {\it explication}, as against analysis. `Explication' also connotes (unlike `analysis') non-uniqueness. For there could be judgment calls about how best to make precise, or mend, the given concept: judgments about which there need be no objectively correct answer. As we said, this is what we will mean by giving an `account' of duality. (`Explication' is Carnap's word; but we will not need further details of his notion of explication.)\footnote{See Carnap (1945:~p.~513; 1947:~pp.~7-8; 1950:~Chapter 1).\label{explication}}

Accordingly, we propose our Schema for duality in an undogmatic spirit. We accept that there will be rough edges in matching it to physicists' usage of the word `duality'; and that maybe it could be improved. For there may be more mathematically precise conceptions of what a duality is than the one that we will defend. But that said, we are sanguine. We will see that the Schema fares very well. For it will prove apt to describe both simple and advanced physics examples, to cast light on relevant philosophical questions about dualities, and to give undoubtedly reasonable judgments of theoretical equivalence.

\section{Philosophy and physics: friends or foes?}\label{friendf}

Our motivation for writing with a readership of both philosophers and physicists in mind, and for thus spending six Chapters of Part II detailing the {\it physics} of dualities, deserves a brief comment. For there are three ways in which our overall project requires us to delve (in Part II) into advanced physics examples.

The first is the obvious pedagogic way in which acquaintance with examples of dualities in quantum field theory and string theory is essential for philosophers with an interest in dualities and, more generally, current condensed matter physics and high-energy physics. For example, just as a book on the philosophy of quantum mechanics would be incomplete without covering several relevant versions of Bell's theorem, so would a book on dualities be incomplete without covering examples of dualities. Indeed, the full appreciation of the philosophical argumentation in Part III requires some acquaintance with advanced examples of dualities, on which Part III builds. The Prospectus in the next Section will clarify how philosophers can best read Part II.

Second, dualities have important interpretative and heuristic uses: they are central tools for theory construction on which physicists have focussed for several decades, and can answer questions that are not currently answerable using other tools, such as perturbative quantum field theory or axiomatic quantum field theory.\footnote{See, for example, the debate between Fraser (2009, 2011) and Wallace (2011) about the relative merits of algebraic quantum field theory and conventional (i.e.~Lagrangian-based, usually perturbative) quantum field theory. Although we will not enter into this debate, Section \ref{intext} will illustrate our Schema with the interpretation of algebraic quantum field theory; and Chapter \ref{EMYM} will illustrate how the non-perturbative duality methods give results that are not reached using either the perturbative or the algebraic approach, and thus that dualities give philosophers yet another toolkit, beyond the familiar ones.} 
Thus apart from answering questions that are {\it mathematically} intractable in perturbative physics, they also help us to construct the {\it interpretation} of dual theories. For example, dualities help establish the ontology of a theory, e.g.~the number and nature of the states of a quantum field theory. 

Ultimately, the proof of the pudding is in the eating: that is, in how Part III will pick up the themes from Part II and, if succesful, illustrate their relevance for a philosophical account of theoretical equivalence and other philosophical questions.

Third, about the relation between physics and philosophy: dualities open up new questions in the philosophy and foundations of physics: about the relativity of the notions of `elementary' and `composite' particles, about whether the dimensionality of spacetime is part of the ontology of a spacetime theory, and about the possibility that `a final theory of all of physics' is not a single unified theory but rather more like a manifold covered by locally overlapping patches. Thus to study dualities is to enter a vast world of contemporary physics ripe for philosophical analysis. One reason why philosophers sometimes delay their engagement with contemporary, or still ongoing, physics is their preference for theories that have attained a level of maturity, and in particular of clarity and rigour, that makes them suitable for philosophical analysis and interpretation. We in part sympathise with this view. But it is also true that several philosophers have convincingly argued against such delays.\footnote{See the philosophical motivations to engage with various aspects of quantum gravity in van Dongen and De Haro (2004:~pp.~509-510), Dawid (2006:~p.~319), Rickles (2008:~pp.~268-284), Huggett and W\"uthrich (2013:~p.~284), Butterfield and Bouatta (2015:~p.~64), and Dieks et al.~(2015:~pp.~203-204). See also Butterfield and Isham's (2001:~pp.~39-40) three ways to pursue the philosophy of quantum gravity.\label{motivQG}} 
After all, philosophy has always been, and still is, strongly motivated by recent advances in physics---and there is accordingly a growing line of philosophical work on various aspects of dualities. Furthermore, Part III will argue that dualities bear on general questions in philosophy of science: such as theoretical equivalence, the formulation of scientific theories, and scientific realism (i.e.~the idea that well-confirmed scientific theories are true, or approximately true, and correctly describe the nature of both observable and unobservable entities: and that this view accounts for the success and progress of science).

Another reason for delaying the engagement with the still ongoing physics of quantum gravity is the lack of direct empirical confirmation of quantum gravity effects, and so the lack of empirical evidence for theories like string theory. In reply to this: first, as Chapters \ref{Simple} to \ref{EMYM} will show, the interest in dualities is not restricted to the regime of quantum gravity or string theory, since dualities also cast light on empirically well-established pieces of physics, such as spins on a lattice, solitons, superconductivity and the Higgs mechanism. Furthermore, there are various independent philosophical motivations for studying quantum gravity theories, which have in fact been an object of interest to philosophers of physics for several decades.\footnote{For a discussion of this independent motivation, see footnote \ref{motivQG} and, in addition, Crowther (2021) and Crowther and De Haro (2022). For an early collection of articles on quantum gravity, see Callender and Huggett (2001). For a collection of papers on string theory, see the special issue De Haro et al.~(2013). For a collection of papers on dualities, see the special issue Castellani and Rickles (2017).}
Thus the study of dualities is an important part not only of the study of established physical theories, but also of the emerging field of the philosophy of quantum gravity. 

For readers with a more specific philosophy of physics interest, dualities are relevant for two additional reasons. First, they sometimes give exact answers to important questions in physics and in the foundations of physics that are not answerable using other methods, thereby making many aspects of quantum field theory and string theory amenable to philosophical scrutiny. For example, Kramers-Wannier (1941) duality was instrumental to Onsager's (1944) exact solution of the Ising model (Section \ref{dualpf0}). Dualities were also instrumental to the study of soliton solutions in quantum field theory (Chapter \ref{Advan}), to the formulation of the 't Hooft-Mandelstam mechanism of confinement, to the understanding of the spectrum of supersymmetric quantum field theories (Section \ref{N=4SYM}), to Seiberg and Witten's (1994a, 1994b) solution of ${\cal N}=2$ supersymmetric Yang-Mills theory, and to Strominger and Vafa's (1996) microscopic calculation of black hole states in string theory (Section \ref{ocs}). All in all, there is here a broad and deep physics literature awaiting philosophical analysis: yet another reason for our engaging with it in Part II.

Second, we wish to argue that the ideas from non-perturbative quantum field theory and string theory, which are used to study dualities, ought to become part of philosophers' toolkits. For, just as ideas from mathematics and formal logic, and axiomatic quantum field theory and general relativity, were embraced by philosophers as useful tools for the analysis of (physical) theories, so we argue that these ideas, developed by theoretical and mathematical physicists over the past decades, are of interest to philosophers of modern physics.

While we do not claim that we will treat all physics topics at the level of rigour required for technical philosophy of physics research---which would distract us from our main philosophical aims in Part III, and would require a different type of book (or books!)---we aim to treat them at an accessible, intermediate level of rigour that is suitable for our philosophical project.\footnote{For two philosophical papers that make a start on rigorous analysis of the geometry of gauge-gravity dualities and bosonization, see De Haro (2017b) and De Haro and Butterfield (2018), respectively. See the discussion of the balance of rigour and scientific importance, in the next Section. } 
Thus in discussing the physics of dualities, we aim not to provide a fully critical philosophy of physics assessment of the whole literature, but an exposition of the physics that is as clear as possible in a way that is relevant to philosophy: mostly with minor adaptations of the physics ``as we found it'', although in appropriate places our exposition is more explicit than the ones we found in the literature. 

\section{Prospectus}\label{prospectus}

As announced in Section \ref{dpp}, the aims of this book are:

(i) to present an account of duality in physics (and the weakening of duality that, in Chapter \ref{Schema}, we will dub `quasi-duality');

(ii) to relate it to other themes in philosophy of physics (especially symmetry and the interpretation of physical theories);

(iii) to illustrate it with various examples; and 

(iv) to relate it to some broader themes in the philosophy of science.\\
Broadly speaking, in the Chapters to follow, we will address these four aims in order. 

At the heart of our account, (i), is a well-nigh formal definition of a duality as an isomorphism of two physical theories, each presented as an appropriate mathematical structure (which is of course also physically interpreted): an isomorphism which will ignore some features of the individual theories. These ignored features are specific to each theory, i.e.~different between the two theories. And they may well be prominent in how we first formulate the theory, and how we think about it. It is in this way that the duality---the isomorphism once these features are set aside---can be very surprising; and scientifically and heuristically valuable. 

We call our account of duality, which we present in Chapters \ref{Thies} and \ref{Schema}, a {\it Schema} for duality: its core proposal being the formal definition of a duality as an isomorphism, given in Section \ref{isomdef}. In the following six Chapters, \ref{Simple} to \ref{STII}, we illustrate various aspects of this Schema: first, with simple examples, some of which are not called `dualities', though they are of course recognized as a kind of isomorphism (Chapter \ref{Simple}); and then with advanced examples, from statistical physics through to string theory (Chapters \ref{Advan} to \ref{HABHM}: this last Chapter also contains a discussion of the hole argument). 

In choosing these examples, we have tried to balance (a) simplicity and rigour (so that Chapter \ref{Simple} contains such well-known examples as position-momentum duality) with (b) current scientific importance (so that Chapters \ref{String} to \ref{HABHM} contain advanced examples from string theory). Our attempt at a balance between these desiderata (a) and (b) is shown in particular by the examples of Chapters \ref{Advan} and \ref{EMYM} (in statistical mechanics and quantum field theory), which are both rigorous and currently active areas of research. 

The Schema also leads to philosophical themes, such as the conditions for two theories to be equivalent (called: {\it theoretical equivalence}), {\it under-determination}, {\it emergence}, `flat' vs.~`structured' {\it conceptions of theories} and the {\it heuristic} use of dualities for constructing new theories, and the aims of understanding and explanation. These themes are the subject of Chapters \ref{Realism} to \ref{Understand}. Of course, since debates continue about these themes, this discussion is not a matter of just unproblematically illustrating the Schema in relation to accepted definitions of e.g.~emergence or understanding. Rather, it is a matter of testing the Schema against the debates. This amounts to: showing that the Schema meshes well with sensible, albeit perhaps controvertible, positions in the debates.

\subsection{Three ways to read this book}

\begin{figure}
\begin{center}
\includegraphics[height=17cm]{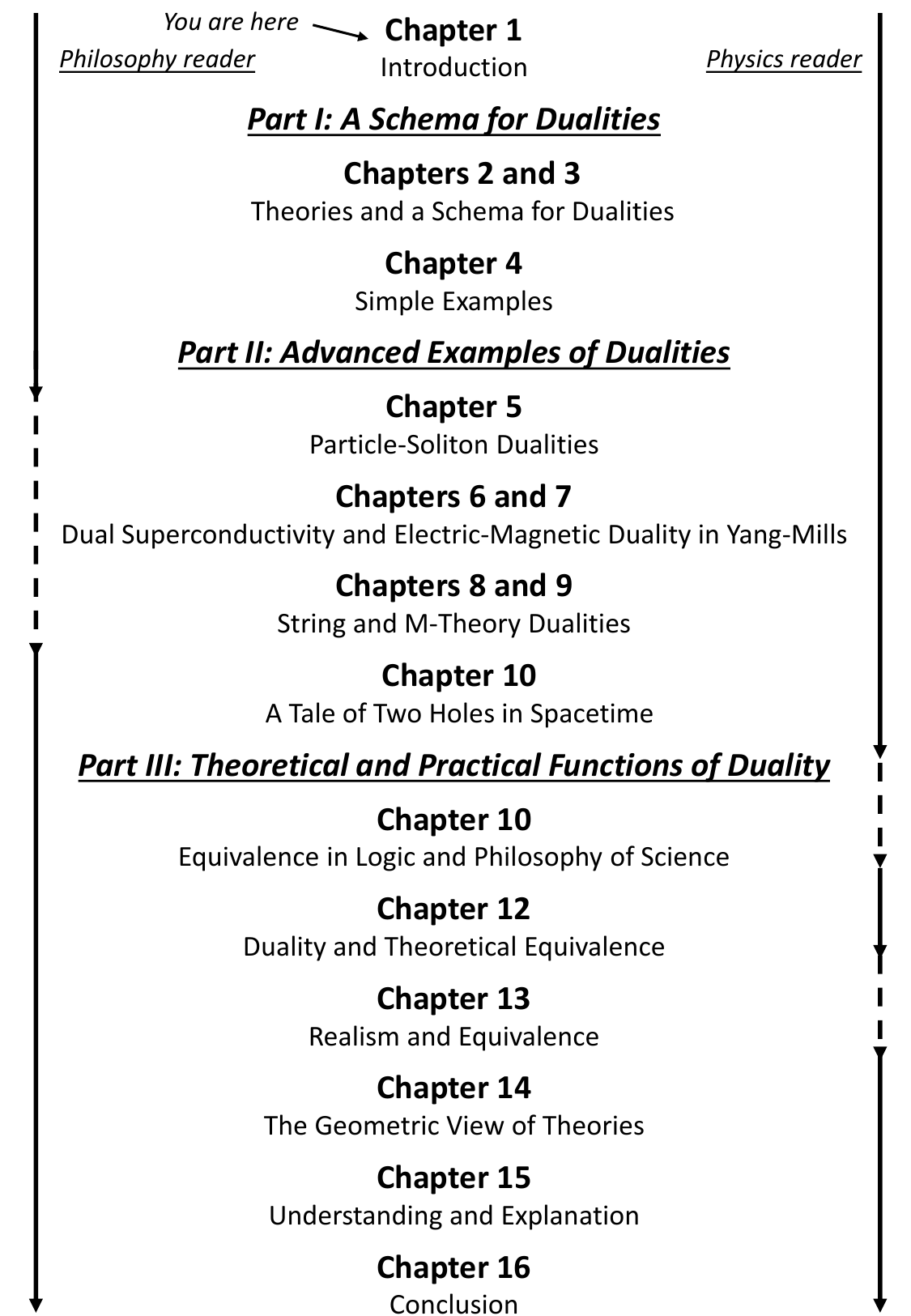}
\caption{\small Plan of the book and its three Parts. The arrows on the left propose a reading sequence for philosophy readers, and on the right for physics readers. We expect philosophy of physics readers to be in a superposition of two ``timelines''. Solid lines indicate recommended Chapters. Broken lines indicate Chapters that readers may wish to skim through, depending on their interests. See the main text for a more detailed exposition of the content of the Chapters.}
\label{bookplan}
\end{center}
\end{figure}

This book is written with a readership of both philosophers and physicists in mind: this is indicated in Figure \ref{bookplan}, which is our proposal for how more philosophy-inclined and more physics-inclined readers may best benefit from reading this book. All readers are urged to read Part I, with our Schema for dualities illustrated by simple examples. 

The remaining two Parts can best be read as follows:\\
\\
{\it Philosophers} may wish to skim through Part II: e.g.~they could read, in each Chapter, the Conclusion, and also the Sections on `illustrating the Schema': and the first few Sections of Chapter \ref{Advan} on particle-soliton dualities, Chapter \ref{EMDuality} on superconductivity and confinement, Section \ref{holeA} on the hole argument, and skim through the rest of the Chapters of Part II. Then they may return to the Chapters in Part III, which discuss the bearing of dualities on various topics in contemporary philosophy of science such as theoretical equivalence, scientific realism, emergence, understanding and explanation.\\
\\
{\it Philosophers of physics} are encouraged to read Parts II and III. Part II is written at an introductory graduate-physics level. Chapter \ref{Advan} reports basic and well-established examples of dualities between bosons and fermions that are also mathematically rigorous (and bosonization will be one of our ``running examples'', in Part III). Chapter \ref{EMDuality} discusses much physics that will be familiar to philosophers of physics (viz.~superconductivity and the Higgs mechanism), and that is here reviewed from the point of view of dualities---thus leading in Part III, to related philosophical ideas such as theory succession, analogies, and understanding. Chapter \ref{EMYM} on electric-magnetic duality in quantum field theory contains material that may be new to some philosophers of physics (namely, non-abelian magnetic monopoles and supersymmetry) but that is here presented in a self-contained and hopefully accessible way. In fact, the materials in this Chapter constitute much of the ``meat'' about dualities and other developments in high-energy physics of the past few decades---and which, as we discussed in Section \ref{friendf}, are ripe for philosophical analysis. Because electric-magnetic duality in quantum field theory is one of the exemplars illustrating how physicists think about dualities, Part III builds on these materials, especially to develop a `geometric view of theories' about the structure of physical theories. Thus the Seiberg-Witten theory will be another ``running example'', of quasi-dualities. Chapters \ref{String} and \ref{STII} are self-contained and include introductions to string theory and to how dualities suggest the existence of a mysterious new theory, M-theory. This then closes a circle of highly influential ideas for theory construction in string theory, which are discussed further in Part III. Chapter \ref{HABHM} discusses the role of dualities in two arguments regarding holes in spacetime: the hole argument, and black hole microstate counting.\\
\\
{\it Physicists} with an interest in learning about dualities, and about the main historically motivating examples, may wish to read all of Part II. The emphasis will of course lie on the concepts and on how they connect to the philosophical issues, rather than on the technical details---for which we will cue in to the relevant literature. 

Some physics readers might wish to skim through Chapters \ref{Theor} and \ref{Realism} in Part III. Thus Chapter \ref{Theor} first discusses formal equivalence in logic and in logic-oriented philosophy of science, and might be skipped by readers with little interest in formal philosophy: however, this introductory Chapter fulfils the important function (like Chapter \ref{Advan} does in Part II for dualities in physics) of discussing rigorous work about equivalence, and of collecting important philosophical themes that are further developed in Part III: namely, the idea of {\it translation}, the requirements for the interpretation of physical theories, and the relation between dualities and traditional conceptions of `theory' and `model' in philosophy of science. Chapter \ref{physeq} discusses how dualities bear on the interpretation of physical theories and illustrates this for bosonization. Chapter \ref{Realism} relates dualities to one of the oldest philosophical questions: namely, realism. As such, physics readers may wish to read its first Section and skim through later Sections. 

\subsection{Chapter-by-chapter prospectus}

With these preliminaries in hand, we can now give a Chapter-by-Chapter prospectus that emphasises the content of the book; as follows.\\ 

Part I is a basic introduction to our Schema for dualities, and some simple examples. Chapter \ref{Thies} describes our conception of {\it physical theories}, of {\it interpretation} of theories, and of {\it symmetries}. These are topics on which so much has been written, in the philosophy of science in general and the philosophy of physics in particular, that some early orientation to our own views, and our jargon, is in order. Thus one topic that philosophers will naturally expect to be addressed is how our discussion relates to {\it realism}, as understood in philosophy. But about this, our main point will be that most of our claims in this book do {\it not} depend on espousing realism, nor even the more specific doctrine within philosophy of science called {\it scientific realism}. We are in fact scientific realists. But our discussion does not depend on this; and accordingly we will postpone details about realism to Chapter \ref{Realism}. However, we will need in Chapter \ref{Thies} to spell out our commitments about the interpretation, i.e.~the {\it semantics}, of physical theories. Symmetry is also included in this Chapter since, as we said in Section \ref{thsq} a duality is like a `giant symmetry'. 

Chapter \ref{Schema} opens with the Schema's core proposal: i.e.~{\it duality as isomorphism}. It also presents the main relations between the Schema and (a) interpretation of theories and (b) symmetries of theories. Chapter \ref{Schema} also includes a brief discussion of the various roles and types of dualities. The main contrast is between what we call the `theoretical' and `practical' functions of dualities. This contrast looks ahead to Chapters \ref{Theor}-\ref{Realism} and \ref{Heuri}-\ref{Understand} respectively. The Chapter ends by comparing our account of duality with the views of other authors.

Chapter \ref{Simple} presents simple examples of the Schema. Namely: (i) position-momentum duality in elementary quantum mechanics; (ii) elementary classical electric-magnetic duality; and (iii) Kramers-Wannier duality in statistical mechanics, i.e.~the duality between high-temperature and low-temperature versions of the Ising model of spins on a lattice. This Chapter also contrasts our conception of duality with the notions of `wave-particle duality' and `complementarity' in quantum theory, which in general are {\it not} cases of duality according to our Schema. By doing so, the Chapter looks ahead to Part III, where we will also ``go beyond'' the Schema.\\ 

In Part II, we present some advanced examples of the Schema, drawn from statistical mechanics, quantum field theory, and string theory. Chapter \ref{Advan} discusses examples with low spatial dimension, i.e.~spatial dimension 1 or 2, where dualities relate particles and solitons. It begins with boson-fermion dualities i.e.~bosonization, in quantum field theory; and particle-vortex duality. It ends with elementary (classical and quantum) electric-magnetic duality of electromagnetism in four dimensions (i.e.~the Maxwell theory)---which is developed for Yang-Mills theories in Chapter \ref{EMYM}.

Chapter \ref{EMDuality} discusses the analogies and approximate electric-magnetic dualities between type II superconductors and the Higgs mechanism. It first reviews superconductivity and discusses colour electric flux tubes as approximate (analogue) electric duals of magnetic Abrikosov vortices in type II superconductors. Then it discusses the proposal by 't Hooft and Mandelstam to explain quark confinement as the (analogue) magnetic dual of a condensate of electric Cooper pairs: namely, a condensate of magnetic monopoles in which pairs of quarks and anti-quarks are held together by a linear electric potential that satisfies Wilson's area law. This proposal illustrates the heuristic fruitfulness of approximate dualities combined with analogies, i.e.~between the experimentally confirmed existence of Cooper pairs within type II superconductors, and monopole condensation as a proposed explanation of colour charge confinement.

Chapter \ref{EMYM} opens with an exposition of 't Hooft-Polyakov monopole solutions in Yang-Mills-Higgs theory and goes on to formulate the Montonen-Olive electric-magnetic duality conjecture. Although the conjecture stumbles for the familiar Yang-Mills-Higgs theories, it is realized in supersymmetric Yang-Mills theory, to which the Chapter gives a gentle and self-contained introduction. The last two Sections expound the ground-breaking work by Seiberg and Witten, who solved the low-energy regime of supersymmetric Yang-Mills theory, thus paving the way for the duality revolution in string theory, and the M-theory conjectures, from 1995 onwards.

Chapters \ref{String} and \ref{STII} discuss those developments: of all the Chapters, these are of course the ones that draw on the most technically demanding physics. So Chapter \ref{String} first gives a brief overview of the general idea of the dualities between the five string theories and M-theory. The Chapter then gives a brief introduction to string theory, and to the simplest string duality: namely, T-duality, between strings on circles of large and small radii. Then we introduce D-branes, which are higher-dimensional analogues of strings whose existence is argued to be required by T-duality.

Chapter \ref{STII} introduces the eleven-dimensional M-theory perspective on string duality, and discusses the relations between the eleven-dimensional objects (supermembranes) and ten-dimensional objects (superstrings and D-branes). We then discuss in order two main dualities that play an important role in the search for a non-perturbative formulation of string theory. These are: 

(i) S-duality, i.e.~a self-duality of string theory that is analogous to electric-magnetic duality; and

(ii) gauge-gravity duality, between theories of gravity and gauge theories without gravity.

This last duality is the springboard for discussing, in Chapter \ref{HABHM}, one of the most striking successes of string theory. Namely: its providing, in certain cases, an understanding of black hole entropy as a statistical mechanical entropy of strings. Chapter \ref{HABHM} also discusses the hole argument, especially in spacetimes with a cosmological constant, and classifies the relevant classes of diffeomorphisms from the perspectives of both the gravity theory, and the dual theory on the boundary.\\

Part III (Chapters \ref{Theor} to \ref{Understand}) develops the philosophical issues that arise from the examples in Part II. We adopt the contrast introduced at the end of Chapter \ref{Schema}: between what we call the `theoretical' and `practical' functions of dualities. Chapters \ref{Theor} to \ref{Realism} are about the theoretical functions and Chapters \ref{Heuri}-\ref{Understand} about the practical functions. Thus the first philosophical theme we mentioned at the start of this Section---the {\it equivalence of theories}---falls under the theoretical functions of duality. 

Chapter \ref{Theor} first discusses equivalence in logic and in formal philosophy of science, emphasising two influential proposals for theoretical equivalence, by philosophers Glymour (1970) and Quine (1975). It then introduces and defends the model-theoretic criterion of {\it isomorphism} of models, and compares dualities to this mainstream semantic criterion. The Chapter ends with a discussion of interpretative requirements that good criteria of theoretical equivalence in science ought to satisfy. 

Chapter \ref{physeq} is a central philosophical chapter, which discusses the bearing of dualities on discussions of {\it theoretical equivalence} in philosophy of science, and gives our main dualities-based proposal for theoretical equivalence. This proposal satisfies the requirements, from Chapter \ref{Theor}, for the ``meshing'' of formal criteria and interpretation. 

Chapter \ref{Realism} introduces realism in philosophy and the more specific view of {\it scientific realism}. It discusses, in the light of dualities, one of the main arguments against scientific realism, namely the {\it empirical under-determination} argument (we call it: `in-principle' under-determination). The question is whether dualities are new ammunition for the anti-realist, in the sense of giving scientifically sound examples of empirically equivalent but theoretically inequivalent theories. It also advocates a `cautious' approach to scientific realism, motivated by dualities.

Finally, Chapters \ref{Heuri} and \ref{Understand} discuss the practical functions of dualities, that ``go beyond'' our Schema. Chapter \ref{Heuri} takes the main practical function to be using a duality to articulate a ``deeper'' theory suggested by---but ``better than''---the two dual theories. This is vividly illustrated in string theory (especially by the M-theory programme), where in the last twenty-five years, physicists have used dualities for exactly this purpose. This also suggests what we shall dub the {\it geometric view of theories}, which ties in to recent discussions about the formulation and nature of scientific theories that often contrast `flat' and `structured' views of theories. The practical use of dualities also bears on the second philosophical theme mentioned at the beginning of this Section---{\it emergence}. For physicists discussing these dualities, especially gauge-gravity duality, often say that one of the dual theories is emergent from the other. (Namely, the gravity theory is emergent from the gravity-free theory.) This raises issues about the type of emergence involved, as well as other issues, such as {\it fundamentality}, and the dizzying idea of a `deeper' theory that does not use a spacetime framework---which are discussed in Chapter \ref{Understand}.

Thus Chapter \ref{Understand} opens with a review of the notions of {\it understanding and explanation} in (recent) philosophy of science, and discusses critically whether, and if so how, theories that purport to have no spacetime at the fundamental level can provide explanations and understanding. 

The book ends with a short conclusion, Chapter \ref{Concl}, by way of narrating what has, and what has not, been achieved: including listing questions that remain open. 

\newpage
\thispagestyle{empty}
$ $
\begin{center}
\vskip7cm
{\bf\Huge Part I. A Schema for Dualities}
\end{center}
\addcontentsline{toc}{chapter}{Part I. A Schema for Dualities}

\newpage
\thispagestyle{plain}

\chapter{Theories, Interpretations and Symmetries}\label{Thies}
\markboth{\small{\textup{Theories and Symmetries}}}{\textup{\small{Theories and Symmetries}}}

The main aim of this Chapter is to present our philosophical views about physical theories and their interpretation. We discuss physical theories in Section \ref{thsmodels}. This will expand on what Section \ref{thsq} said: namely that a physical theory typically postulates a space of states, a set of quantities and a dynamics. But we also need to discuss the notion of a {\it model}, since there are various conflicting usages of the words `theory' and `model'. In Section \ref{ThisB}, we present our preferred usages, which are prompted by our account of duality. In Section \ref{itm}, we turn to the interpretation of theories. By `interpretation' we here mean {\it semantics}, the systematic presentation of the meanings of words. `Interpretation' also has of course a much wider meaning, as understanding. But for reasons given in Section \ref{itm}, we can (until Chapter \ref{Realism}) focus on semantics. The Chapter ends with a discussion of symmetries (Section \ref{Symm}). This is needed because, as we said in Chapter \ref{Intro}, a duality is like a `giant symmetry'.

\section{Theories and models}\label{thsmodels}

 In this Section we will present our philosophical views about physical theories, and our usage of the terms `theory' and `model'. Our views are mainstream. But we will need to be careful about the usages of these terms. It is not just that in the literature they have various (sometimes conflicting) usages. Also, our account of duality will prompt a special usage. 

For we said in Section \ref{thsq} that on our account, two dual theories share a {\it common core}, which is typically itself a theory, in the sense of specifying a state-space, quantities and a dynamics. Recall that we called this the `common core theory'; and that dual theories are isomorphic formulations (almost always {\it representations}, in the sense of representation theory) of the common core theory. So to explain that account, we first need to explain our use of the term `theory'. Besides, we will co-opt the term {\it model} for these formulations of the common core. So we also need to discuss the word `model'. We address both `theory' and `model' in Sections \ref{thmscph} and \ref{Ourthm}. The first reviews the usual views and usages; the second presents our views and usages, as prompted by our account of duality.

This will lead to our distinguishing within (what we call!) a model, two parts. The first part we will call the {\it model root} (in most cases: a {\it model triple}), and we will call the second part the {\it specific structure}. This is done in Section \ref{modelrootss}. Finally, in this Section, we will introduce notation for the values of physical quantities on states and for dynamics (Section \ref{valuesQ}). 

\subsection{`Theory' and `model' in science and philosophy}\label{thmscph} 

In science and philosophy, indeed in everyday language, a `theory' is a connected body of doctrine, true or false, about some topic. `Theory' also connotes generality and organization, though perhaps not completeness. The theory purports to give the general, rather than the specific, facts about the topic; and to present them in an organized, though perhaps not complete, way. Here, there is a broad contrast with the word `model'. A model also represents some topic: but `model' connotes that the representation is both specific and omits (usually deliberately) some facts. 

But there are some differences in usage, between science, especially physics, and philosophy, that we should signal. But rest assured: the broad contrast that a theory is more general, and a model more specific, will be preserved. 

In physics, the word `theory' is often used for bodies of doctrine that are of very large scope (i.e.~have a very large or wide topic) and are formulated in detail; and it is especially used for well-established and successful theories. For example: the kinetic theory of gases, Newtonian gravitation theory, general relativity theory etc. The word is also often used---both in classical and quantum physics---for more specific bodies of doctrine that are defined by the forces that (are known or postulated to) act; and so also defined by the ensuing equations of motion. These forces are often encoded in a Lagrangian or Hamiltonian, subject to an action principle, from which the set of equations of motion is derived. In philosophy, on the other hand, the word `theory' can be used casually for mere scraps of doctrine of very limited scope. For example, a philosopher might say that what we know about the layout of my home town is `my theory' of its street-plan. Actually, this can be a problem when doing philosophy of science, and it is a problem underlying some of these discussions, and here... we do well to adopt more specific notions of `theory' in philosophy of science. Luckily, there is a long and venerable tradition of doing so.\footnote{For a recent account of the history and the discussions about the usage of the words `theory' and `model', see Frigg (2023).}

In physics, a body of doctrine that is of more limited scope, or is formulated in less detail, or is less established or successful, is often called a `model'. As such, models are often seen as ``mediators'' between theory (general, and abstract) and phenomena (concrete, and tangible).\footnote{See, for example, Morgan and Morrison (1999) and Morrison (2015).} There are countless examples at all levels: from school level, such as the planetary model of the solar system and the simple harmonic oscillator as a model of a spring, to advanced, such as Bohr's model of the atom and the liquid-drop model of the atomic nucleus. The word `model' also connotes contingency, with the parameters (such as the amplitude and frequency of the spring's oscillations) being set by experiment. The features listed---limited scope (specificity), the lack of detail, being less established or successful, contingency---all go along with the model being tentative and, often, even known to be partly false. 

The fact that despite these limitations, models are useful for description, prediction and explanation, prompts the philosophical literature to study in detail how such approximations and idealizations work. Of course, `model' is often used in this sort of way outside physics. In everyday language, any representation of a comparatively specific topic or phenomenon (which again, may lack detail or be tentative or even be known to be partly false) is called a model of it. And philosophy goes along with this general usage.\footnote{Agreed, `model' does not always have these connotations of `modesty' (of scope, or detail, or success): neither in science nor in everyday language. Thus in physics, we talk of `the standard model of particle physics', and `the standard model of cosmology'; and in linguistics, of a grammarian's model of a language. But for the most part, these connotations of `modesty' hold good.} 

So far, so straightforward. These usages are very much as one would expect. But the situation has been complicated by overlapping developments in twentieth-century logic and analytic philosophy: especially in philosophy of science, as influenced by logic. 

At first, these developments engendered a more precise conception of theory, viz.~as {\it a set of sentences}. This conception was adopted both by logic and by early twentieth-century philosophy of science; though with the difference that in logic, the language of the sentences is always one of logic's formal languages with their associated vocabularies, while in philosophy of science, it need not be---it can be a natural language, e.g.~English augmented as necessary with technical words e.g.~from mathematics. Nowadays, logic continues to use `theory' like this: maintaining its in-house restriction to formal languages. 

These developments also led logic (and more generally, formal semantics) to use `model' for a precise notion, which was also tied to a formal language. Namely, a model (also called an {\it interpretation}) is an assignment of referents (also known as: semantic values) to the expressions and formulas of the language. Such an assignment almost always involves: a set $D$ of objects, called the {\bf domain of quantification} or just {\bf domain}; an assignment to each of the language's singular terms (e.g.~individual constants) of an element of $D$; an assignment to each of the language's one-place predicates of a subset of $D$, thought of as the predicate's set of instances; an assignment to the two-place (binary) predicates of a subset of $D^2$, thought of as the pairs that instantiate the binary relation; and so on. The result is that one can formally define what it is for a sentence of the language, whether simple or compound, to be true according to such an assignment. If so, we say that the assignment (interpretation, model) {\it is a model of} (or: {\it models}) the sentence. Similarly for a set of sentences, or a theory: we say that such an assignment is a model of a theory if it makes every sentence in the set true.\footnote{To distinguish these two kinds of models, the former kind (i.e.~the planetary model, Bohr's model of the atom, etc.) are often called `iconic' or `representational' models, while the latter (i.e.~the ones in logic) are often called `mathematical' or `logical' models. Although these two model kinds are independent, they are not mutually exclusive, and a given model can belong to both, as models in physics often do: they are not just replicas or visual aids that represent a target system, but come equipped with a mathematical structure that serves to apply a theory to a target system. For a discussion, see Frigg (2023:~pp.~46-48). For Suppes (1962:~p.~252), the difference between an experimental model and a model in logic is of degree (of formalization) rather than kind: `the notion of model used in any serious statistical treatment of a theory and its relation to experiment does not differ in any essential way from the logical notion of model'. Thus he goes on to describe a hierarchy of models in the empirical sciences (p.~259): see also Suppe (2000:~p.~105; 1974:~pp.~98-99).}

Again: this seems, so far, straightforward; and to reflect the idea implicit above that, while a theory is general, a model is specific. For the specificity of the domain $D$, and of which of its elements are assigned as referents and instances of singular terms and predicates, means that a model represents one way that the topic of a theory could be, while making the theory true (i.e.~true according to the model). We will return to this in Chapter \ref{Theor}.

But in the late twentieth-century some philosophers of science criticized the conception of a scientific theory as a set of sentences---dubbed {\bf the syntactic conception of theories}---and suggested a replacement: which they called the {\bf the semantic} or {\bf structural conception of theories}. And their suggested replacement's central notion was named with that over-used word ... wait for it! . . . `model'. So their proposal was that a theory is a set, not of sentences, but of (what they called) models. And for them, a model is like logic's notion which we just summarized---but with the important difference that for them, a model is not tied to a formal language. As we have just seen: for a logician, any model has `written on its face' its associated formal language: a change in the formal language {\it ipso facto} changes the model. But that is {\it not} so for these philosophers' models. For those philosophers' whole idea was to get away from what they considered the strictures of language. It is thus typical for them to present what they call a `model' as just a set $D$ together with some selected elements of it, some selected subsets of it, some selected subsets of $D^2$, etc. In short: like a model {\it \`a la} logic, but stripped of the associated vocabulary whose semantic values these selected items are. 

So nowadays, the situation is a bit complicated, and the terminology can be confusing. The {\it locus classicus} reporting both the syntactic conception of theories (also called `the received view': see Putnam, 1962:~p.~240) and the main critiques of it, is Suppe (1974). Among the major critiques of the received view are the lack of axiomatization in many instances of actual science that is assumed by the syntactic conception (pp.~63-65), and the untenability of two key distinctions closely linked with the syntactic conception: namely, the observational-theoretical distinction, and the analytic-synthetic distinction (Suppe, 1974:~pp.~66-72; Putnam, 1962:~p.~240). As we reported above, the syntactic conception was also criticized because of its alleged language-dependence (van Fraassen, 1980:~p.~44; 1989:~p.~221).\footnote{For a recent summary of this discussion and response to the criticisms, see Lutz (2017:~pp.~320-330; 2013:~p.~80-91). Suppe (2000:~pp.~102-104) is a personal recollection of how Hempel came to abandon the received view, and of what, in Suppe's view, was and was not wrong about it.}

Early proponents of the semantic conception of theories are Beth (1960), Suppes (1962) and van Fraassen (1970). Beth (1960:~pp.~172-175) used the semantic conception to formalize both classical and quantum mechanics, in a way that makes rigorous the relation between the laws of the theory and the world (even though his paper does not use the word `model'). Van Fraassen (1970:~pp.~328-329) further developed Beth's work and gave a more general semantic construal of a physical theory (to which the Schema's characterization in the next Section is close). Van Fraassen's critique of the syntactic conception aimed to stay close to scientific practice: `it does not present anything like a faithful picture of actual foundational work in physics'. And he acknowledged that `[t]here are natural interrelations between the two approaches, [that] would make implausible any claim of philosophical superiority for either approach' (p.~326).

The next Section introduces our usage of `theories' and `models'.

\section{Theories and models in this book}\label{ThisB}

Section \ref{thsmodels} has set the stage. This Section now introduces our players: our views about physical theories and models, and especially our usages of the confusing terms, `theory' and `model'---as prompted by our account of duality, which we call the Schema. We state the main points in Section \ref{Ourthm}. This will lead to our distinguishing within (what we call!) a model, two parts. The first part we will call the {\it model root} (in most cases, it will be: a {\it model triple}), and we will call the second part the {\it specific structure}. This is done in Section \ref{modelrootss}. Then we introduce notation for the values of physical quantities on states and for dynamics (Section \ref{valuesQ}). 

\subsection{Our usage of `theory' and `model'}\label{Ourthm}

We will begin with a specific decision about how to use `theory' and `model', that is prompted by our account of duality. (Recall the statement in italics at the end of Section \ref{giantS}.) In short: that account prompts us to shift the usages of each of these words ``one level up''. This decision will lead to more details about, first, models; and then theories. \\

Recall from Section \ref{giantS} that on our account, a duality involves two ``levels'': the {\it common core} ``above'' and the two dual theories ``below''. The dual theories are formulations, or versions, or realizations, of the common core: almost always {\it representations} of it, in the sense of mathematical representation theory. Besides, we will see that in most cases of dualities, the common core is more general and-or more abstract, and the dual theories more specific and-or more concrete. So this contrast matches the contrast described in Section \ref{thmscph}. For there, we said that a theory is general and-or abstract: it ``stands above'', and ``is common to'', its more specific and-or concrete models. 

So in discussing dualities, the question arises: should we allocate the word `theory' to the common core, or to each of the two dual theories? In other words: what jargon is best for keeping clear the two levels: that of the two dual theories, and the level `above', of the common core? 

We take our cue from the fact that although both the words, `theory' and `model', have (unfortunately) many different usages, these usages match in two major respects---that we saw throughout Section \ref{thsmodels}. Namely: a theory is more general than a model; and a theory corresponds to more than one model. So there is a strong connotation that a theory should be `one level up' from a model. 

Accordingly, we will henceforth dub the common core as {\bf the common core theory}. And we will call the two dual theories that instantiate the common core theory {\bf models}. So our proposed jargon is that for dualities, the uses of the words `model' and `theory' go ``one level up''. So what was called a `theory' (i.e.~one of two dual theories) is now a `model'. 

As a simple example to illustrate our usage, consider position-momentum duality in elementary quantum mechanics. This will be Chapter \ref{Simple}'s first simple example (Section \ref{pmd}). For the moment, we just point out that the position and momentum representations might have been discovered, in a counterfactual history, as different ``theories'': for the Schr\"odinger equation takes very different forms in the position and momentum bases, as do the operators and wave-functions. Then, one can imagine, the discovery of the Fourier transformation between them would reveal that they are in fact two representations of a common core theory: namely, quantum mechanics formulated in the basis-free language of Hilbert space. Thus what in the counterfactual history are at first thought to be two distinct ``theories'', are seen to be just two models, i.e.~representations, of a common core theory, viz.~what we in fact call elementary quantum mechanics.

This decision prompts a few further comments about: first, `model', about which we want to disavow some tempting connotations; and then `theory', about which we will emphasise a theory being a triple of states, quantities and dynamics.

We stress that the rest of this Section and the next two Sections (\ref{modelrootss} and \ref{valuesQ}) will {\it not} use the idea of duality, nor depend on our account of duality as an isomorphism of models (in the sense of `model' just adopted). All this material sorts out ideas and jargon about the two levels: but isomorphism at the lower level will only be a topic in Chapter \ref{Schema}. \\
 
\noindent {\bf `Model'}:--- \\
So beware! Our use of `model' rejects three common connotations of the word: connotations that in the course of Sections \ref{thsq} and \ref{thsmodels} it may well have gathered in the reader's mind. Namely, the connotations that a model is: 

(i): a specific solution (at a single time: or for all times, i.e.~a possible history) for the physical system concerned, whereas the `theory' encompasses all solutions---and in many cases, for a whole class of systems; 

(ii): an approximation, in particular an approximate solution (maybe about a specific regime or system the theory is concerned with); whereas the `theory' deals with exact solutions; 

(iii): a part (in particular, an empirical or observable part) of the physical world---a hunk of reality!---that gives an interpretation (or part of an interpretation) of the theory; which is itself {\it not} part of the physical world, but a representation of it (in the philosophical, not mathematical sense)---and so stands in need of interpretation.\\
We will of course be concerned from time to time with each of the ideas (i) to (iii): in particular, Section \ref{itm} will focus on (iii). But they are not part of our definition of `model'. 

So we stress that our use of `model' rejects all three connotations. As we just said: for us, a model is a specific realization---one might say `version'---of a theory. So a model adds details---we shall say: `specific structure'---to its theory. But these details are {\it not} a matter of specifying: (i) a solution or history of the system; or (ii) approximation(s); or (iii) interpretation(s). Rather, the extra details are extra mathematical structure: just like a representation of a group or an algebra has extra details or structure, beyond that of the group or algebra of which it is a representation. Section \ref{modelrootss} will develop this thought, with some examples and notation.\footnote{{\label{modelusage}}{Agreed: (i) to (iii) mean that for our notion, the word `model' has disadvantages. But note that other words also have disadvantages. For example: `formulation' connotes that any two formulations of a theory are `notational variants', i.e.~fully equivalent: they make exactly the same claims about the world. But that is far from true for our notion (and this matches the connotations of `model'). For us, two models of a common core theory are in general not isomorphic, and so also not equivalent; and so typically, it is surprising to find two isomorphic models, i.e.~to find a duality. Besides, as we shall stress: even if they are isomorphic, they might not make the same claims about the world. Other examples of the struggle to choose the best word: `realization', `instance' and `instantiation' connote being part of the physical world, as in `the mechanism/hardware which realizes some specific function/software', or `the object is an instance/instantiation of the predicate'---which is the misleading connotation (iii) above. We should also recall the usage of `model' in theoretical physics (cf.~Section \ref{thmscph}). It is, roughly, between: (a) our usage, and (b) (ii) and (iii) above. Think for example of the `massive Thirring model' or the `sine-Gordon model'.}} 

To sum up: a model is for us a specific realization (or version or formulation) of a theory. That is: it `models' (verb!) another theory ``above'', which in general also has other such models. Almost always, it is a representation of the theory above, in the sense of mathematical representation theory---`representation' being another word with confusingly diverse uses. Recall the example of position vs.~momentum representations.

And as we have announced: the main case for us of ``the theory above'' will be the {\it common core} theory, of two dual theories (our `models'). It is sometimes best to think of the common core theory as uninterpreted, in which case we will call it a {\bf bare theory}.\footnote{Thus we endorse the widespread view that the formulation of a scientific theory can be fruitfully conceptualised as a `two-step' procedure. The first step determines the theory's formal structure (what we call the `bare theory'), and the second adds an interpretation through a referential semantics (which we will discuss in Section \ref{itm}). Although these two steps can be synchronic, and as we will also discuss they are related, it is nevertheless useful in foundational discussions to distinguish them. For a summary of this `standard account', see e.g.~Ruetsche (2002:~pp.~349-350).\label{Ruetsche2002}} 
But it may also be interpreted: cf.~the above example of elementary quantum mechanics. So we say the two given theories are {\it models} of the common core theory; and their being {\it isomorphic} models (isomorphic as regards the structure of the common core theory) is what makes them duals. (We will discuss the interpretation of theories and models in Section \ref{itm}.)

But as we stressed above: dualities are not our only cases of ``theory above and model, i.e.~realization, below''. For that general idea obviously applies even if no pair of models is isomorphic, as regards the structure of the theory above. This gives another rationale for using the term {\it bare theory}, in addition to `common core theory'. That is: even if no duality, no pair of isomorphic models, is at issue, we will sometimes consider a theory and one of its realizations, i.e.~models in our sense. In such a case, we will often, for clarity, call the uninterpreted theory `the bare theory'. For although, as just mentioned, it may be interpreted, yet it lacks the extra details, the specific structure, that its model has. We will give details in Section \ref{modelrootss}. \\

\noindent {\bf `Theory'}:---\\
As to our usage of `theory', and our ``putting the theory above the model'', the first thing to say is that it is a mainstream usage: i.e.~typical of the literature, especially the literature on the semantic conception of theories as applied to physics. 

Apart from this, there are three comments to make. The first is, for us, minor; it rebuts the allegation that the notion of theory is unimportant for understanding science. The second and third comments are more positive, and more important to our account. For they develop the idea (already expounded in Section \ref{thsq}) that a theory comprises a space (i.e.~set) of states for the system or systems concerned, a set of quantities and a dynamics. And the third introduces notation that we will use throughout the sequel.\\

\noindent (1) {\it Defending the notion of theory}:---\\
Agreed: recent philosophy of science has emphasised many aspects of scientific endeavour that hardly invoke the notion of {\it scientific theory}, central though this notion was for discussions by both the positivists and their successors. For example, aspects such as experiment (calibration of instruments etc.), causation (mechanistic explanation etc.), and the social dimensions of knowledge (testimony etc.) have recently been discussed with a strong emphasis on models (in a more usual sense than ours!), rather than theories.

We agree that these aspects of scientific endeavour are important for our philosophical understanding of science.\footnote{See De Haro and De Regt (2018:~pp.~635-636), and also Section \ref{pragmatic}.} But even if these aspects did not need the notion of theory, still the notion may well be useful for other aspects. Indeed, we believe it is indispensable for discussion of symmetries and dualities in physics.\footnote{We also believe it useful, even indispensable, in other discussions. One main one is understanding renormalization---a topic for which, again, there has been scepticism about its usefulness: e.g.~Kaiser (2005:~pp.~377-387). For a defence of the notion, cf.~Butterfield (2014:~Section IV.1). In fact, the notions of theory and model are pervasive in Parts II and III of this book.} 

Furthermore, notice de Regt's (2017:~p.~98) observation that the `claim of theory-independence is not as radical as it appears to be at first sight ... What they oppose is a strong ``theory-driven view'', on which models simply follow from theories by procedures of de-idealization ... In sum, while models are ``autonomous agents'' that do not follow deductively from theory, they are almost never completely theory-free'.\footnote{See also Su\'arez and Cartwright (2008:~p.~66) and Morrison (2007:~p.~198). For an extensive discussion of this understanding of models, see Frigg (2023:~pp.~363-383).} 

A different sort of critique of scientific theories has been voiced by French (2020), who defends eliminativism about theories, and proposes that the content of theories is made true by scientific practices, rather than by the existence of some entities that we call `scientific theories'. However, note that French's eliminativism is not a prohibition against the use of scientific theories or models, which are `useful fictions' that summarize our scientific practices: but rather, against the illusion that scientific theories or models have a special ontological status. 

We agree with Halvorson (2021:~p.~611) that the question about the ontological status of scientific theories cannot be separated from other perennial philosophical questions about e.g.~the nature of propositions and of mathematical objects.\footnote{While we are {\it not} Platonists about scientific theories, our views about scientific theories are closely related to our views about laws of nature, which can be summarized as a `best systems' view.}
Theories are less than Platonic abstract objects, but also more than useful summaries of scientific practices: they are also summaries of objective states of affairs, made true by the way the world is, independently of us. Luckily for us, these questions are only tangentially relevant to our project, and we can set them aside until Chapter \ref{Realism} (see also Section \ref{srpost}).\footnote{See also Dewar's (2021) review of French (2020).}\\

\noindent (2) {\it Adopting an intermediate notion of theory and of model}:---\\
There is an obvious contrast between Section \ref{thsq}'s idea of a physical theory with a single state-space that is intended to describe a single type of system, and the idea of a theory that treats many types of system. Of course, the contrast depends on how finely one wishes to define `type of system'. But as noted in Section \ref{thmscph}, it is usual to think of Newtonian mechanics, or Hamiltonian mechanics, or elementary quantum mechanics, as a theory; and each of these theories postulates many different state-spaces, so as to treat diverse systems. 

To take a very simple example: consider the kinematics of $N$ point-particles in Euclidean space $\mathbb{R}^3$. Even if, for simplicity, we take the state of the point-particles to be just their positions, i.e.~we ignore the momentum information, the instantaneous state of the system is a point in $\mathbb{R}^{3N}$: in an obvious notation, $\langle x_1, y_1, z_1, x_2, y_2, z_2,\ldots , x_N, y_N, z_N \rangle$. Similarly, if the state included---as specifying a dynamics would require---the momenta of the point-particles: the state would be a point in $\mathbb{R}^{6N}$. So as the number $N$ varies, the dimension of the state-space varies. And most authors would {\it not} say that each specific value of $N$ corresponds to a theory. Rather, the `theory' is what is in common between the various state-spaces, equipped with suitably related quantities and suitably related dynamics. (Here, `suitably related' is made precise in the obvious way: the quantities for the $N$ particle case are to include positions and momenta for all $N$ particles, and the dynamics is given by forces between particles, e.g.~gravitational attraction between all pairs.) 

Bearing in mind the discussion of `model' above, and especially its connotation (i) (`a specific solution'), we see that Section \ref{thsq}'s idea of a physical theory with a single state-space is an {\it intermediate} idea of a theory. That is: this idea gathers together a class of states (and so of solutions through time: trajectories through state-space). So it ``is more general than'', ``stands above'', individual states and solutions. But it is not as general as Section \ref{thmscph}'s very common usage of `theory', which ``generalizes over $N$'', so that for example Newtonian mechanics is a theory.

Despite this very common usage, it is often useful to think of a physical theory as having a single state-space, with quantities and dynamics. And it will be especially convenient for our account of duality. So we henceforth adopt this intermediate idea as our primary meaning for `theory'. Our occasional use of other meanings for `theory' will almost always be clear from the context; and when needed, we will declare them.

Besides, following our decision at the start of this Section that for dualities, the words `model' and `theory' should shift one level up: we similarly adopt this idea of a single state-space as our primary meaning for `model'.\\

\noindent (3): {\it Notation for states, quantities and dynamics}\\
We introduce some notation for this intermediate idea of a bare theory or model, which will be developed in Section \ref{modelrootss}; and then make two further comments about it. This notation flows directly from (2)'s adoption of single state-spaces, for both theories and models. 

A {\bf bare theory or model} {\it in our senses} is a {\it triple}, consisting of: a space of states, $\cal S$, for the systems it is concerned with; a set of physical quantities, $\cal Q$, which take values on the states; and a dynamics $\cal D$ which describes how the values of quantities change over time. (Symmetries and dualities will of course concern the preservation or matching of these values. Also, once a bare theory is interpreted, the {\it measurable quantities} are the {\it values} of various physical quantities: we will turn to intepretation in Section \ref{itm}.) 

So we write a bare theory, or a model in our sense, schematically as: 
 \be\label{introtriple}
 \langle {\cal S}, {\cal Q}, {\cal D} \rangle. 
\ee
This applies to both classical and quantum physics. In any specific example---whether bare theory or model, classical or quantum---each of $\cal S$, $\cal Q$ and $\cal D$ will of course have further structure (see Section \ref{intext} (A)). In giving some details about this, we will until the end of this Section just say `theory', to avoid clumsily repeating `theory or model'.\\
\\
{\it Example: position and momentum representations in quantum mechanics.} Take elementary quantum mechanics in one dimension as our common core theory: 

(i)~~The {\it state-space} is the Hilbert space over $\mathbb{C}$ of denumerable dimension (which is unique up to isomorphism). It is endowed, in the usual way, with a semi-positive inner product $\bra\psi|\f\ket$; (more details about this elementary example are in Section \ref{pmd}, which, as we will discuss, is also a {\it duality}). 

(ii)~~The {\it quantities} ${\cal Q}$ are the operators position and momentum, $X$ and $P$, which satisfy the algebra: $[X,P]=i\hbar$. (We also consider functions of these operators, respecting the self-adjointness.) We set aside the mathematical subleties about domains for unbounded operators, since they will not affect our discussion at all (here or in later Chapters). For details, see e.g.~Prugovecki (1981:~p.~191ff), Jauch (1968:~pp.~40-42), and Ruetsche (2011:~pp.~37-43).

(iii)~~The {\it dynamics} is the Schr\"odinger time evolution: thus it singles out, from the set of operators ${\cal Q}$, a distinguished self-adjoint operator $H(P,X)$ as the theory's Hamiltonian.

This common core theory has two obvious models, in each of which the Hilbert space is represented as $L^2(\mathbb{R})$ with respect to the Lebesgue measure over the real line; but which differ by, roughly speaking, choosing a basis of position or of momentum.\footnote{There are of course countless bases that one can choose for the same Hilbert space. One can e.g.~choose a basis of eigenstates of any self-adjoint operator of ${\cal Q}$.} 

In the position representation, the states of the Hilbert space are represented by wave-functions $\psi(x)$ that are functions of position. The quantities of ${\cal Q}$ are operators (which we indicate by hats) built from $\hat X$ and $\hat P$, which are defined to act on wave-functions as follows:
\bea\label{QP}
\hat X\psi(x)&:=&x\,\psi(x)\nn
\hat P\psi(x)&:=&-i\hbar\,{\dd\over\dd x}\,\psi(x)\,.
\eea
The Hamiltonian that gives the dynamics is the operator $H(\hat X,\hat P)$. This gives us a model model $M_x$ that is a representation of the common core theory $T$ in the position representation. (Elements of the common core theory are mapped to the model by a representation map, i.e.~$h_{\cal S}^x:\psi\mapsto\psi(x)$ and $h_{\cal Q}^x:X\mapsto\hat X$. The maps are homomorphisms---here, isomorphisms---respecting the inner product.)

The second model is the momentum representation, $M_p$. This model represents the states of the Hilbert space by wave-functions that are functions of the wave number (i.e.~the momentum), i.e.~$\ti\psi(p)$. The quantities of ${\cal Q}$ are represented as follows:
\bea\label{FQP}
\hat X\ti\psi(p)&:=&i\hbar\,{\dd\over\dd p}\,\ti\psi(p)\nn
\hat P\,\ti\psi(p)&:=&p\,\ti\psi(p)\,.
\eea
The two models are dual in that the Fourier transformation maps the states in position representation to states in momentum representation, and vice versa, and since the Fourier transformation is unitary, it thereby preserves the inner product, $\bra \psi|\f\ket$. The Fourier transformation also induces the transformation of the operators in position representation into momentum representation and vice versa (and likewise for the Schr\"odinger equation), i.e.~it transforms Eq.~\eq{QP} and Eq.~\eq{FQP} into each other. (Again, this model is a representation of the common core theory in the sense of there being a homomorphism that maps the theory into the model, $h_{\cal S}^p:\psi\mapsto\ti\psi(p)$ and $h_{\cal Q}^p:X\mapsto\hat X$.) Section \ref{pmd} will discuss that the Fourier transformation is a {\it duality} between the two models.\\
\\
Thus more generally, the main points about our formulation of theories and models are:---

(i): For a quantum theory, $\cal S$ will usually be a Hilbert space, or a set of density matrices. For a classical theory, it will usually be a manifold, or a set of probability distributions. 

(ii): For almost any theory, $\cal Q$ will be an algebra over the real or complex numbers, allowing quantities to be added and multiplied. 

(iii): For almost any theory, the dynamics $\cal D$ can be understood as the states changing over time while the quantities remain fixed over time. This was how we spoke in Section \ref{thsq}: the changes of state simply encode the changing values. In quantum theories, this is called `the Schr\"{o}dinger picture'. But for almost any theory, one can describe the dynamics equivalently (though perhaps less intuitively) with states staying fixed, and quantities changing so as to encode the changes of values. In quantum theories, this is called `the Heisenberg picture'. Though this Schr\"{o}dinger/Heisenberg jargon is used only for quantum theories, the same choice about how to describe the dynamics applies equally to classical theories. We will return to this at the end of Section \ref{valuesQ}. Finally, to accommodate Euclidean theories in which, roughly speaking, time is treated formally like space, we should really say that $\cal D$ describes how the values of quantities change over time {\it and-or over space}. Thus in general, for almost any theory in physics, the dynamics is an equation that distinguishes, on the set of kinematically possible states ${\cal S}$ and quantities ${\cal Q}$, those that have a specific time and-or space dependence: usually determined by a Hamiltonian or a Lagrangian function (see the discussion of kinematics and dynamics in Section \ref{thsq}). We will illustrate how spacetime theories like general relativity can be cast as a triple, in Section \ref{sptthies}.\\ 

Two clarifications about this treatment of a bare theory (or a model in our sense) as a triple: the first formal, the second interpretative. 

First, note that we said: `it is {\it often} useful to think' of a physical theory as a triple'; and `{\it almost any}' in (ii) and (iii). For we agree that not every theory is presented, or best thought of, in this way. Theories in statistical and quantum physics are often formulated in terms of partition functions and-or path integrals with sources, and related concepts like sets of correlation functions, rather than in terms of states and quantities. And in field theories, the dynamics is often presented as field equations holding at each spacetime point---and so not naturally thought of in terms of the Schr\"{o}dinger or Heisenberg pictures with their ``background time''. But for most of this book (including the examples in Part II) we can think of theories (and models, in our sense) as triples of states, quantities, and dynamics. Besides, all our morals will carry over to these other ways of formulating theories.\footnote{Of course, one can draw connections between formulations with states, quantities and dynamics---our triple conception---and other types of formulation, like partition functions, path integrals and field equations. For example, a partition function with a source that couples to an operator in the Lagrangian of a quantum field theory is standardly used to calculate, by taking functional derivatives, the correlation functions of that operator in the vacuum state. See for example Duncan (2012:~pp.~91-95, 245-268), Zinn-Justin (2002:~pp.~146-166) and De Haro et al.~(2017:~pp.~75-76).\label{nontriple}} 

Second: We admit that of course, a physical theory is almost never presented to us as a tidy triple of state-space, quantities and dynamics. Almost always, the theory appears to us messier than that: more vaguely defined and-or more complicated. The triple needs to be extracted from that ``mess''. Indeed, there are two points here. 

(i): The complicated appearance is of course in part due to those aspects such as experiment mentioned in (1) above. But this complexity, and the need to allow for such aspects (and to assess them philosophically), does not militate against extracting a triple as a concept useful for e.g.~understanding symmetries. 

(ii): We make no claim that there is always, or even typically, a unique best definition of the triple. So presenting a theory as a triple usually involves: (a) choices about exactly what to take as the state-space etc.; and even (b) judgment about interpretative and perhaps controversial matters. We will see some examples of this variety, in later Chapters and we will see that these choices may affect verdicts of interest, e.g.~whether there is an isomorphism, or a duality (see the notion of `augmentation', in Section \ref{aa}).
 
This conception of a bare theory or model as a triple leads in to the next Sections. They introduce more jargon and notation which we will use for most of this book.\\

\subsection{`Model root' and `specific structure'---distinguished}\label{modelrootss}

As announced under {\it `Model'} in Section \ref{Ourthm}, we now give some jargon, notation and examples, about the relation between a theory ``above'', a bare theory, and its models. As explained there: (i) we here favour `bare theory' over `common core theory' since in this Section and the next, our topic is the two levels, not duality i.e.~isomorphism at the bottom level; and (ii) the bare theory may be interpreted.\footnote{We take the point, from Read and M\o ller-Nielsen (2020:~pp.~284-286), that in what is called a `duality' not every common core need be so rich as to lead to a theory (in the present case, not every common core needs to allow being written as a triple). But we will restrict our attention to cases in which the common core {\it is} as rich as a theory, which we submit are the most interesting cases---and indeed they seem to be the most common cases in the literature on dualities. Hence we take duals to be isomorphs whose common core is a bare theory. See also De Haro (2021:~pp.~138-139).\label{richcc}} 

The role of the jargon and notation will be to distinguish the parts or aspects of models that are ``shadows'' of corresponding parts or aspects of the bare theory, from those that are not. We will call the former the {\it model root}; and the latter will be the model's {\it specific structure}. This jargon and notation will be used in the definition of duality in Section \ref{isomdef}, and in the examples in Chapters \ref{Simple} to \ref{HABHM}.\\

The jargon and notation is clearest for the common case, which we will focus on for most of this book. Namely, the case where the realization of the bare theory is a (mathematical) representation, and the bare theory is a triple comprising a state-space $\cal S$, a set of quantities $\cal Q$ and a dynamics $\cal D$ (cf.~Section \ref{Ourthm} and footnote \ref{nontriple}).

This is common in physics, where theories are presented as mathematical structures (here, as triples), and the models of the theory, if they are to be recognized as `models of a given theory', should respect that structure. The notion of `respecting structure' immediately brings us, in the context of physical theories, which are usually formulated using {\it set theory} (indeed, all of our examples in Part II are defined using set theory), to the notion of the (mathematical) representation of structure. 

The reasons for this are not only mathematical, but also interpretative. For the interpretations of theories and of models often depend on structural considerations (see e.g.~North (2021:~pp.~4, 6, 60)), and so to develop an interpretation of a model that is compatible with the interpretation of the theory of which it is a model, one should require that the model is a representation in the mathematical sense. (For example, Chapter \ref{physeq} will discuss dual models that have the same interpretation as their common core theory, and also the logico-semantic relations between a theory and its models: all of which requires mathematical representations.)

In this common case: by the very definition of `representation', the model gives a {\it homomorphic copy} of the bare theory. (This is of course irrespective of there being a duality, i.e.~another model to which the given one is isomorphic.) That is: the bare theory and the model are, {\it both of them}, triples; and there are appropriate structure-preserving maps from the states, quantities and dynamics of the bare theory to the model's homomorphic copy. To be precise: there is a pair of structure-preserving maps---from states in the bare theory to states in the model, and from quantities in the bare theory to quantities in the model. And there is a meshing condition on the model's dynamics that makes it implement that of the bare theory. The details are as follows: though we can mostly take them in our stride, and just say `homomorphic copy'.\footnote{So the relation of representation between a bare theory and a model of it will involve not just one map as in, say, group representation theory (the homomorphism from the abstract group to e.g.~a set of matrices), but at least two maps. These two maps are related to each other, because states and quantities are dual (in mathematicians' sense!) to one another: see De Haro and Butterfield (2018:~pp.~344-345).\label{dualMo}}

With the notation of Section \ref{Ourthm}, we write the bare theory as a triple $T = \bra {\cal S}, {\cal Q}, {\cal D} \ket$; and similarly its model $M = \bra {\cal S}_M, {\cal Q}_M, {\cal D}_M \ket$. Here, the subscripts signal that the state-space etc.~are different from that of the bare theory. If we think of the dynamics in the Schr\"{o}dinger picture as a map on the state-space (cf.~the end of Section \ref{Ourthm}), and write the homomorphism from ${\cal S}$ to ${\cal S}_M$ as $h$, then the meshing condition on the model's dynamics will be the commuting diagram in Figure \ref{meshdynamicsbasic}. \\

\begin{figure}
\begin{center}
\bea
\begin{array}{ccc}{\cal S}&\xrightarrow{\makebox[.6cm]{$\cal{D}$}}&{\cal S}\\
~~\Big\downarrow {\sm{$h$}}&&~~\Big\downarrow {\sm{$h$}} \\
{\cal S}_M&\xrightarrow{\makebox[.6cm]{${\cal {D}}_M$}}&{\cal S}_M\nonumber
\end{array}\nonumber
\eea
\caption{\small The model's (Schr\"{o}dinger) dynamics implements that of the bare theory.}
\label{meshdynamicsbasic}
\end{center}
\end{figure}

We now turn to this Section's main distinction. (As we said: it is important irrespective of whether there is a duality.) Namely, the distinction between:

(i): the parts and aspects of the model which together express its realizing the bare theory;

(ii): the parts and aspects which do not express the realization. \\
We will call (i) the {\bf model root}. And in the common case where the bare theory, and so also the model, is a triple of states, quantities and dynamics, we will call (i) the {\bf model triple}. 

And we will call (ii) the model's {\bf specific structure}. We can think of it as the `ingredients' or `building blocks' by which the representation of the bare theory, i.e.~(i), is built. Of course, ingredients are present in the cooked dish, and building blocks are present in the built house. Similarly here: often, an item of specific structure is in the model root, though (by definition) it is not part of the representation of the bare theory.\footnote{In Section \ref{intext}, we will see that this distinction implies a correlative distinction between two kinds of interpretation. In short: an {\bf internal} interpretation is one that only interprets the model root; while an {\bf external} interpretation also interprets (some or all of) the specific structure. Thus the distinction between specific structure (`building blocks'), and model root (`what gets built') is physically significant, in that it constrains interpretation. It is formal in that it can be stated without giving an interpretation: but it has consequences for interpretation. Details in Section \ref{itm}.}

Group representations provide obvious---and countless---examples of these ideas, while letting us avoid the complication of having to consider two basic sets, states and quantities, and also values and dynamics. So let the bare theory be a very ``toy'' theory: an abstract group $G$. Then there are two broad kinds of extra detail or structure whereby a model, i.e.~a representation, differs from $G$. The first adds detail or structure to $G$, and the second subtracts it:--- 

(A): The `concreteness' of a specific mathematical object: such as $\mbox{GL}(n,\mathbb{C})$, the general linear group over $\mathbb{C}^n$, or any subgroup of it---any of which is a `concrete', not abstract, group. (Agreed, the concrete vs.~abstract contrast is flexible; but this will not matter for anything that follows.) 

(B): Since representation requires homomorphism, not isomorphism (i.e.~it allows non-injectivity and non-surjectivity), two representations of $G$ can be non-isomorphic to $G$ as groups; and of course also, non-isomorphic to each other.

Of course, these kinds (A) and (B) of `extra detail' usually occur together. Just think of how every abstract group can be represented by the trivial one element subgroup of $\mbox{GL}(n,\mathbb{C})$, the $n \times n$ identity matrix. \\

We stress that even within a given model, the distinction between root and specific structure is not `God-given'. It is relative to how exactly we define the bare theory, and thereby the homomorphism from it to the model. And for a physical theory, as usually presented to us informally and even vaguely, there need be no best or most natural way to make this exact definition. Recall the end of Section \ref{Ourthm}: how exactly to present a theory as a triple of states, quantities and dynamics is a matter of choice and even judgment. 

We will also see this flexibility in connection with dualities, in Chapter \ref{Schema}. For in a duality, the isomorphism of models is an isomorphism with respect to the structure of the common core theory---which is the structure that the models represent. So the isomorphism that is the duality is an isomorphism of model roots. And in the common case of triples of states, quantities and dynamics: it is an isomorphism of model triples.\footnote{A clarification: This is not to say that the specific structure (the `building blocks') is always `invisible' to the other side of the duality, i.e.~that no part of the specific structure is mapped across by a duality to the other model. Often, some part of the specific structure is mapped across. That is unsurprising. For the model root is built from specific structure: so one expects that in order to map the model roots, one into the other, the duality must map at least part of the specific structure, one into the other. In our earlier example of quantum mechanics, the Fourier transformation maps a function of position, i.e.~$\psi(x)$, to a function of momentum, i.e.~$\ti\psi(p)$.\label{notinvisible}} So the flexibility about the definition of a bare theory, and so about what a model must represent, leads to flexibility about exactly what the duality mapping is, i.e.~what is the isomorphism between model roots. \\

It is worth having a notation distinguishing between the model root (which is usually a model triple of states etc.) and the specific structure. So given a model $M$, we now write $m$ for its model root. This is usually a model triple: which we write as $\bra{\cal S}_M,{\cal Q}_M,{\cal D}_M \ket$---as at the beginning of this Section. Again, the subscripts indicate that the state-space etc.~are different from those of the bare theory. We also write $\bar M$ for the specific structure. 

So we write a model $M$ of a bare theory $T$ as
\be\label{eqmodel}
 M =:\bra m ; \bar M\ket = \bra{\cal S}_M,{\cal Q}_M,{\cal D}_M ; \bar M\ket\,,
\ee
 where: (i) the colon to the right of the = signals, as usual, that the adjacent expression, i.e.~to the right of =:, is being defined; (ii) the semi-colons in the defined angle-brackets signal the contrast between root and specific structure, and the last equation just expresses the usual case of the root being a triple. 
 
But {\it beware}: one should not think of $M$ as just the ordered pair of independently given $m$ and $\bar M$. For $m$ is built by using the specific structure $\bar M$, and so it is not given independently of $\bar M$. Again, the obvious example of group representations illustrates. We should not think of a matrix representation of, say, the symmetric group $S_N$ as containing a copy of $S_N$ `beside' its specific structure of a vector space $V$ and $N!$ linear maps on $V$; (or maybe less than $N!$ maps---recall that a representation need only be a homomorphism). Rather, $V$ and the (upto) $N!$ linear maps realize---are used to build a `concrete copy' of---$S_N$. Similarly, as we shall see, in physics examples. Typically, the specific structure $\bar M$ consists of a set of fields, endowed with a set of symmetries, a dynamics for the fields, and a set of states of the fields. (So here, fields, and the quantities constructed from them, play the role of states and quantities in our conception of theories as triples, though of course not all fields are observable: in quantum theories, self-adjoint.) These fields etc.~are used to build a `concrete copy' (maybe a `coarse-grained' copy) of the bare theory's structure.

Rather, $m$ encodes $M$'s representing the bare theory $T$. So one might well write $T_M$ instead of $m$, since having a subscript $M$ on the right hand side of the equation $M = \bra T_M; \bar M \ket$ signals that $M$ is not an ordered pair of two independently given items. In other words: the notation $T_M$ emphasises that the model triple: (i) is a representation of $T$, (ii) is built from $M$'s specific structure viz.~$\bar M$, and (iii) is itself a theory (hence the letter `T'). In short, $T_M$ is not `given before' $M$ itself: rather, $T_M$ realizes $T$ by making use of $\bar M$. 

In our example of simple quantum mechanics from the previous Section, the model $M_x$ in the position representation contains wave-functions $\psi(x)$ and quantities, such as $\hat X$, whose definition depends on the specific structure, i.e.~on the coordinate variable $x$: see Eq.~\eq{QP} (and likewise for the momentum representation: see Eq.~\eq{FQP}). 

Notice that the choice of `the part of the structure that is common to all of the theory's models' goes into the definition of a model, and singles out the core physics that is described by the theory, because it is represented in all its models. Thus the distinction between model root and specific structure constrains what kinds of systems the model is able to describe.\\

Here is an illustration of this notation in use. As announced: we say that a duality is an isomorphism of model roots, with respect to the structure of the bare theory. So if $M_1, M_2$ are models of a bare theory $T$, their being duals means: $m_1 \cong m_2$. And in the usual case of model triples, i.e.~$m_i = \bra{\cal S}_{M{_i}},{\cal Q}_{M{_i}},{\cal D}_{M{_i}} \ket\,,~i = 1,2$: this isomorphism of roots will be a matter of two appropriately meshing isomorphisms, one between the state-spaces ${\cal S}_{M{_i}}$ and one between the quantity algebras ${\cal Q}_{M{_i}}$. Details in Chapter \ref{Schema}.

\subsection{Values of quantities on states; dynamics}\label{valuesQ}

We now introduce notation for: the value that a quantity takes on a state; and for dynamics, i.e.~how the value of a quantity changes over time. So suppose we are given a set of states ${\cal S}$, a set of quantities ${\cal Q}$ and a dynamics ${\cal D}$: $\bra {\cal {S}}, {\cal {Q}}, {\cal {D}} \ket$. In this Section, it will not matter whether this triple is a bare theory ``above'', or one of its realizations, i.e.~a model, ``below''.\footnote{Two other aspects, which were also discussed in Section \ref{Ourthm}, will not affect this Section. Namely: (i) in any specific case, ${\cal S}$, ${\cal Q}$ and ${\cal D}$ will each have a lot of structure beyond being sets---but we will not need these details in what follows; (ii) not all theories are presented, or best thought of, as such triples---but what we say will carry over to other formulations using e.g.~partition functions; cf.~footnote \ref{nontriple}.} 

Then we write the value of quantity $Q$ in state $s$ as
\be\label{pairing}
\bra Q , s\ket\,.
\ee
It is these values that are to be preserved, or suitably matched, by the duality, i.e.~by the isomorphism of model triples: cf.~Eq.~\eq{obv1}. Besides, as one would expect from the idea that a duality is a `giant symmetry', it is these values that are preserved by a symmetry map (which we will discuss in Section \ref{Symm}).
 
For classical physics, one naturally represents (that word again!) a quantity as a real-valued function on states: $Q: s \mapsto Q(s)$. Given such a function representing the quantity, $\langle Q , s \rangle := Q(s) \in \mathbb{R}$ 
 is the system's possessed or intrinsic value, in state $s$, of the quantity $Q$. Similarly for quantum physics: one naturally represents quantities as linear operators on a Hilbert space of states, so that $\langle Q , s \rangle := \langle s |{\hat Q} | s \rangle \in \mathbb{R}$ is the system's Born-rule expectation value of the quantity. But we shall see that in fact, a quantum duality often preserves off-diagonal matrix elements $\langle s_1 |{\hat Q} | s_2 \rangle \in \mathbb{C}$. 
 
We turn to dynamics. Here, and for all the examples in Part II, we will assume the dynamics is deterministic, in the sense explained at the end of Section \ref{thsq}. We now state this more precisely, building on the discussion after Eq.~\eq{introtriple} (Section \ref{Ourthm}). 

We recall that there are two main ways of describing the changes over time of the values of quantities. Adopting the quantum terminology, we call them the {\it Schr\"{o}dinger picture} and the {\it Heisenberg picture}, respectively. But the ideas apply in exactly the same way in classical physics. The former is more intuitive; and we already adopted it in Section \ref{thsq}'s description of the history of a system as a curve in state-space, i.e.~a sequence of states as a function of time.

Thus in the Schr\"{o}dinger picture, the system's state changes over time, while the mathematical representation of a quantity such as energy is given once and for all: it is time-independent. The changing state represents how the values change over time. This is certainly an intuitive way to describe change. But under determinism, there is a mathematically equivalent Heisenberg picture, in which all the time-dependence is assigned to the quantities, not the state. The state is given once and for all; but the mathematical representation of a quantity such as energy is time-{\it dependent}. So in this picture, the change of values over time is represented by having each of the quantities change.

 For simplicity, we will adopt the Schr\"{o}dinger picture. Then we can formulate determinism as: the state at one time determines the state at other times. More precisely: any state in the state-space determines the sequence of states for all future, and indeed all past, times. In terms of curves in the state-space: through any point (element) of the state-space, there is a unique curve to the future, and indeed to the past. So determinism has a straightforward geometric expression, in terms of state-space being filled by a family of non-intersecting curves.\footnote{Notice that here we reap the benefit of having adopted our {\it intermediate idea} of a theory, with a single state-space; (cf.~Section \ref{thmscph}). For on the more common use of `theory', where a theory encompasses many different state-spaces each with their quantities and dynamics (e.g.~`Newtonian mechanics'), a statement of determinism must be more prolix: namely a conjunction with, for each state-space, a conjunct given by our straightforward geometric expression.} 

All this can be put in the jargon of group actions. On the Schr\"{o}dinger picture, the dynamics is an action $D_S$ of the real line $\mathbb{R}$ representing time on the state-space ${\cal S}$. A possible history of the system is an orbit of the action, i.e.~a curve in ${\cal S}$. On the equivalent Heisenberg picture, the dynamics is an action $D_H$ of $\mathbb{R}$ representing time on the set of quantities ${\cal Q}$. (So a possible history {\it for a single quantity}, e.g.~a possible course of values for energy, is an orbit of the action, i.e.~a curve in ${\cal Q}$.) The two pictures are related by, in an obvious notation:
\bea\label{dyns1}
D_S: \mathbb{R} \times {{\cal S}} \ni (t,s) &\mapsto &D_S(t,s) =: s(t) \in {{\cal S}} \; ~{\mbox{iff}} \nn
D_H: \mathbb{R} \times {{\cal Q}} \ni (t,Q) &\mapsto& D_H(t,Q) =: Q(t) \in {{\cal Q}} \,,
\eea 
where for all $s \in {{\cal S}}$ considered as the initial state, and all quantities $Q \in {{\cal Q}}$, the values of physical quantities at the later time $t$ agree in the two pictures:
\be\label{dyns2}
\langle Q, s(t) \rangle = \langle Q(t), s \rangle\,.
\ee 

\section{Interpretations of theories and models}\label{itm}

We turn to the interpretation of physical theories. A large subject! In this Section, we will just state our main views and jargon about the subject. The jargon will be used in subsequent Chapters; and the views (which are mainstream) will be illustrated in Chapter \ref{Schema} and further developed in Part III. 

But we stress that almost all of this Section will {\it not} use the idea of duality, nor our Schema: i.e.~our account of duality as an isomorphism of models (in our sense of `model'). Nor will this Section much need our distinction of two levels, with theory above and model below. For the material concerns how any theory, or model in our sense, relates to the real world: a subject on which one can hope to settle one's views independently of physical theory providing examples of dualities, and so before one settles on some account of dualities. We think this hope is fulfilled: regardless of dualities, we advocate the views of this Section. But this does not mean this Section is `surplus to requirements': that it gives no support to our account of duality. For our views here will be deployed in later Chapters. So the status of this Section---as preparation and support for our account, but not invoking it---is like Section \ref{ThisB}: as we announced at the start of Section \ref{Ourthm}.\footnote{We also agree that the hope, though natural, is defeasible. One might get bad news: that one's views about interpretation in general conflict with some dualities in physics; or more likely, with some interpretative or philosophical construal or aspect of them that one wants to endorse. But as we said at the end of Section \ref{giantS}, we are sanguine: we maintain that our account of duality, including our views about interpretation and our Schema, fares well.}

Broadly speaking, this Section will proceed from the general and more contentious, to the specific and less contentious. Thus we first endorse referential semantics and relate this to scientific realism (respectively, Sections \ref{refsr} and \ref{srpost}). Then in Section \ref{ints}, we endorse intensional semantics (which is a species of referential semantics).
Finally, in Section \ref{intext} we relate our endorsement of intensional semantics to our distinction of levels, with theory above and model below. This will prompt a new jargon, namely {\it internal interpretations} and {\it external interpretations}. 

\subsection{Referential semantics}\label{refsr}

We endorse the endeavour of providing theories (and so: models in our sense) with a referential semantics, of the kind familiar in modern logic and semantics. Such a semantics assigns references in the empirical world to appropriate elements of (bare) theories; with the reference assigned to a compound being a function of the references assigned to its constituents. At the end of Section \ref{thmscph}, we summarized how this proceeds, when explaining how the word `model' is used in logic. Indeed, as we mentioned: a model in this sense is often called an `interpretation' or an `assignment of values', with the values, i.e.~referents, of singular terms being elements of the domain of quantification $D$.

Of course, providing such a semantics is in no way special to scientific theories. Linguists and logicians provide such semantics for all sorts of languages, including (large fragments of) natural languages. And since the endeavour is so widely accepted as unproblematic, indeed mandatory, one might ask why we bother to announce our endorsement: why make a fuss?

As we see it, there are two broad reasons to pause over the issue. The first is very general; the second will relate to the debate over scientific realism.\\

First: as we have noted, some philosophers entirely reject this endeavour as wrong-headed or unilluminating, for one reason or another. For example, they say that: (i) a scientific theory is itself non-linguistic, i.e.~not a set of sentences; or (ii) the non-linguistic aspects of science are all-important; or (iii) the very notion of scientific theory is not useful. We have already briefly set our face against (ii) and (iii). Recall, as to (ii) and (iii): item (1) of `Theory' in Section \ref{Ourthm} (we will return to (iii) in Section \ref{UaS}). Here we will add some detail about (i) and (ii). 

(i): We endorse the recent literature in denying that the distinction between the syntactic and semantic conceptions of theory is so firm or useful.\footnote{See Glymour (2013:~pp.~287-288), van Fraassen (2014:~pp.~279-281), Lutz (2013:~p.~93; 2017:~pp.~341, 345-347), and Frigg (2023:~p.~167).\label{syn=sem}}
Once we recall from the end of Section \ref{thmscph} that logic's notion of model is tied to a formal language, it will be clear that to logicians, this denial will not be a surprise. After all, they will say, a theory as a set of sentences in a formal language determines its set of models, i.e.~the interpretations of that language in which it is true; and a set of interpretations (of a common language) determines a theory, viz.~the sentences of that language true in all of them. But, as this literature argues, even for a notion of model that is not tied to a formal language, the distinction is weak. One main reason for this is that both the scientist advocating a theory, and the philosopher advocating the semantic conception of a theory as set, not of sentences, but of models (sets with selected elements, subsets etc.), need to use language. They cannot formulate a theory---even if it is a non-linguistic entity, e.g.~a set of models---without using some language.\footnote{Indeed, they cannot convey a theory to the mind of the hearer without using some kind of language. This is of course not to deny that there are important aspects of science that need not necessarily involve language: but we cannot have scientific theories without using some language.} 
(This reason is in play in the articles cited: see footnote \ref{syn=sem}.)

We will return to this in Chapters \ref{Theor} and \ref{physeq}, where we will discuss the distinction between the model-theoretic semantics that is adopted by logicians and the physical semantics: both of which are used by physical theories, and we will develop interpretative principles for the latter. We will argue that the distinction of two levels, introduced by dualities, with theory above and model below, suggests a structured view of theories (and also a more specific view, that we will dub the {\it geometric view of theories}).

(ii): We see no conflict between providing a referential semantics and accepting the importance of non-linguistic aspects of science, including such aspects of scientific theories. Here, we recall the irenic moral of Lewis' (1975) seminal paper about the philosophy of language. Lewis argues that two views of the topic and aim of philosophy of language, that are usually opposed to each other, should instead be considered complementary. He labels the two views `Thesis' and `Antithesis'; and the task of his paper, and his moral, is to reconcile them in a `Synthesis'. The Thesis begins by saying that {\it a language} is an assignment `of meanings to certain sequences of types of sound or of marks ...' (p.~3). So the Thesis is advocacy of a referential semantics; (indeed an intensional semantics---cf.~Section \ref{intext}).\footnote{For more on intensional semantics, see Carnap (1947:~pp.~177-182) and Lewis (1980). For an excellent introduction to intensional semantics in connection with sentence structure, see Heim and Kratzer (1988:~Chapters 2 and 12). For an overview of theories of meaning, see Speaks (2019).} 
The Antithesis begins by saying that `{\it language} is a social phenomenon wherein people utter strings of vocal sounds ... and wherein people respond by thought or action to the sounds they observe to have been produced' (ibid). So the Antithesis is advocacy of an account of language emphasising people's propositional attitudes (intentions, beliefs, desires etc.): both as what is communicated by language, and as what underpins that communication. Thus Lewis' Synthesis is his account of what it is for {\it a} language $L$, {\it \`a la} the Thesis, to be {\it the} language used by a human population, {\it \`a la} the Antithesis. His main idea is that this is a matter of the population having conventions (in Lewis' (1969) sense of regularities of behaviour sustained by common knowledge) of speaking truthfully in $L$ and trusting to what they hear spoken in $L$. Thus he knits the Thesis and Antithesis together in a detailed way, using his theory of conventions; (and this yields replies to various objections). He ends by saying: `According to the proposal we have presented, the philosophy of language is a single subject. The thesis and antithesis have been the property of different schools; but in fact they are complementary essential ingredients in any adequate account either of languages or of language' (p.~35).

To which we say: `Hear, hear!'. That is: we claim a similar reconciliation between our advocating a referential semantics for scientific theories, and various lines of philosophical work that downplay, or even do not mention, reference or theories. For example: much philosophical writing about experiment (calibration of instruments etc.) emphasizes non-linguistic skills, practices and norms; and much philosophical writing about the social dimensions of knowledge emphasizes the functioning of scientific communities and institutions in e.g.~the maintenance of norms and accreditation. Such emphases are entirely appropriate, say we.\footnote{We will discuss the pragmatic aspects of explanation and of understanding in Chapter \ref{Understand}, and the `use' of scientific theories will be centre stage in Section \ref{pragmatic}. See, for example, De Haro and De Regt (2018:~pp.~633-637).} 
Study of experiment should of course emphasize non-linguistic skills; and so on. But such emphases in no way militate against developing a referential semantics for scientific language, and so for scientific theories. 

In short, we think Lewis' synthesis gives a valuable, because irenic, perspective on this work's relation to referential semantics.\footnote{More controversially: we think Lewis' synthesis gives a valuable, because deflating, perspective on the burgeoning literature about scientific representation; viz.~along the lines Callender and Cohen's claim that we should analyse representation in science in terms drawn from philosophy of language and mind (2006, especially Section 3).}

So much by way of endorsing referential semantics in general, even outside science. In Sections \ref{ints}, we will discuss a more specific programme within it, of intensional semantics. But first we will discuss whether, and if so how, the question of scientific realism bears on referential semantics. 

\subsection{Varieties of scientific realism; postponed}\label{srpost}

The second broad reason for discussing our commitment to referential semantics is its relation to controversial matters, especially the debate about scientific realism. For of course, as soon as one starts sketching a referential semantics for a scientific theory, many questions arise. As we see matters, these fall into three main groups, which we can label with corresponding kinds of `realism'. We will sketch these groups, and then focus on scientific realism: though that also will be largely postponed to Chapter \ref{Realism}.

First, there are questions about whether one's referential semantics commits one to some variety of `semantic realism': i.e.~some doctrine to the effect that reference, truth and meaning are objective in some controvertible sense. Second, there are questions about whether it commits one to some variety of `metaphysical realism': i.e.~some doctrine to the effect that even if a scientific theory, or perhaps even the conjunction of all our theories, was epistemically ideal, i.e.~satisfied all our desiderata of being well-confirmed, explanatory, simple, yet it could be false.\footnote{This is Putnam's (1978:~p.~126) critique of metaphysical realism, and its associated account of reference, which Lewis (1984:~p.~224) calls `global descriptivism'. In reply, Lewis (1984:~pp.~227-228) argues that some interpretations of our theories are more natural or eligible than others, and this is often called `reference magnetism'.} 
Third, there are questions about whether it commits one to some variety of scientific realism: i.e.~some doctrine to the effect that the well-confirmed theories of mature sciences---in particular, for us: well-confirmed contemporary physical theories---are approximately true. 

Needless to say, the questions in these groups are closely related. One obvious example is the question whether Quine's (1968:~pp.~185-212; 1960:~pp.~24-26, 68-79) advocacy of indeterminacy of reference are right. If so, a denial of semantic realism could imply a denial of metaphysical realism. For if there is indeed no fact of the matter about whether `gavagai' in the native's mouth refers to an enduring rabbit or to a time-slice of a rabbit (or to a perduring rabbit), then {\it ex hypothesi} some versions of semantic realism are false. Namely, versions that take objectivity of reference to require determinacy about endurance vs.~perdurance. And if one also takes metaphysical realism to require such determinacy, then it also is false.

But in this book, we will leave aside most of these questions. This is not just a matter of limiting our to-do list: of saying `sufficient unto the day is the work thereof'. Also, our account of duality, and the reasons for (or against!) it, will not depend on the answers to these questions. For semantic realism and metaphysical realism, such independence is hardly surprising. (We will return to scientific realism shortly.) For one can usually assess claims in philosophy of physics without having to answer general (perennial!) questions about the objectivity of reference, or whether an epistemically ideal theory could be false. 

But here, we should address two objections one might make about referential semantics' use of mathematical functions that map a term, as an argument of the function, to its referent as the value. In Section \ref{intext} we will endorse this idea, with what we will call {\it interpretation maps}. These will be functions, and so will have, by definition, a set as their co-domain (and as their range). This is standard practice in referential semantics, throughout logic and linguistics as well as philosophy.

But there are two possible objections to this.\footnote{Here, we are indebted to Klaas Landsman and Benno van den Berg for discussion.}
We think there is a definitive reply to the first. But we will admit that there can be no definitive reply to the second, which raises deep, perhaps perennial, controversies about how language relates to the world. 

The first objection takes its cue from a widespread practice in set theory, considered as a branch of pure mathematics. Namely, to work only with sets all of whose elements are themselves sets.\footnote{A bit more precisely: to work only with {\it pure} sets, i.e.~sets that are ``built up'' only from other sets, perhaps ``bottoming out'' only in the empty set. This formulation is not wholly precise since it does not address the questions whether (a) to require {\it ur}-elements (usually: just the empty set), and-or (b) to have a set-class distinction. But it is precise enough for us.} 
If so, then since the ``concrete''---in particular, contingently-existing---objects that we refer to (in everyday talk and thought, just as much as in technical science) are surely not themselves sets, they also cannot even be elements of sets. So they cannot be in the co-domain or range of the interpretation maps that we propose. And so the enterprise of referential semantics---to understand the relation between language and ``concrete'' objects, precisely, by invoking the idea of functions---stumbles. 

To this, our reply is that we see no reason for the enterprise of semantics (or more generally of interpreting our language, or our theories) to follow this set-theoretic practice. Agreed: a philosopher of pure mathematics might want the truth of pure mathematics, as formulated in set theory, to not depend on the existence of contingently-existing objects (or even one such).\footnote{Recall Russell's notorious struggles with the axiom of infinity, which---in set-theoretic terms---required the bottom level to contain an infinity of non-sets.}
But that viewpoint in no way prevents someone, who is doing something other than pure mathematics, from using a set theory that {\it does} ``bottom out'' in contingently-existing objects (usually called `individuals' or `atoms' in the philosophy of set theory). In short: once set theory is accepted as a framework, nothing prevents us (or any referential semanticist) from using impure set theory---and so from specifying contingently-existing objects as referents, i.e.~as elements of the range of an interpretation map. 

The second objection is mistier. It begins by recalling Frege's requirement that the notion of an object requires each object to have a criterion of identity (whether or not the object is a set); and that logic, as it later developed, has endorsed this requirement: witness Quine's slogan `no entity without identity'---though of course controversy continues about what exactly the requirement amounts to. Besides, in any set theory (pure or impure) a set is individuated by its elements. For the axiom of extensionality says that sets $x, y$ are identical, $x=y$, iff each element of $x$ is an element of $y$ and vice versa: which surely requires us to understand the elements as objects subject to a criterion of identity. Thus the objection is that by taking a referent to be an element of a set, we (or any referential semanticist) assume that the world consists of objects subject to criteria of identity (of course: perhaps a different criterion for different kinds of object). But this assumption is contentious. To put it metaphorically: what does the world, or Nature, ``know or care'' about our human division of the world in to objects, or kinds of objects, with various criteria of identity?

To this, there can of course be no definitive reply. We are in deep waters. What can we know, or warrantably believe, about the intrinsic nature of the world, independent of its being articulated by humans' cognitive apparatus? Must it consist of objects, enjoying properties and relations to one another? And if so, how are those objects, properties and relations related to the various kinds of objects etc.~articulated by us humans, either in everyday talk and thought or in technical science? We must leave such questions to the likes of a Kant (or in our own day, a Putnam (e.g.~1978) or a Lewis (e.g.~1984)). We will take it that, whatever their answers, there is some stable sense in which the world consists of objects (and properties and relations) with criteria of identity precise enough to be elements of sets. That is all our referential semantics will need.

These objections, and our answers to them, were prompted by discussion of semantic realism and metaphysical realism. But a similar independence also holds as regards scientific realism: accepting referential semantics for scientific theories is independent, or at least largely independent, of the debate over scientific realism. 

The reason for this is that nowadays, both scientific realists and their opponents agree on the broad shape of the semantics of physical theories: and that agreed shape will be enough for our account of duality. For the two sides of the debate agree that all the statements of a theory, both the observational statements and the non-observational statements, should be given a referential semantics. The arch (anti-realist) empiricist, van Fraassen (1980:~pp.~14, 43, 57), explicitly accepts a literal construal of the theoretical claims of---i.e.~a referential semantics for---scientific theories (1980:~p.~14): as do other influential positions that reject realism, such as Fine's `natural ontological attitude' (1984:~pp.~96-99, 1986:~p.~171). What scientific realists and their opponents {\it disagree} about is epistemology. That is, they disagree about which statements of our various theories we have a warrant, i.e.~justification, for believing. In this disagreement, the scientific realists are more `cognitively optimistic'. For example, they say that for well-confirmed scientific theories, we should believe even the non-observational statements to be true, or at least approximately true (in some sense of `approximately true'). 

We are in fact scientific realists. But again, our account of duality does not depend on this. Indeed, not only does our Schema and its illustrations in various physics examples not so depend. Also: our views about topics like theoretical equivalence and the heuristic value of dualities are largely independent of scientific realism. So we will postpone details about realism to Chapter \ref{Realism}, where we will discuss the epistemic warrant for believing an interpretation of a physical theory, and a principle for selecting interpretations that we shall dub a {\it cautious scientific realism}. All we will need until then is our commitment to referential semantics: in particular, to intensional semantics as in the next Subsection.

\subsection{Intensional semantics}\label{ints}

So far, our sketch of referential semantics (in Section \ref{refsr}) said simply that words were assigned semantic values in an interpretation---for singular terms, objects in the domain; and for predicates, subsets of it---so that whole sentences were determined as true or false relative to the interpretation. To the newcomer, this framework has the striking feature---one is bound to think: defect---that it allows the semantics to be utterly faithless to the existing meanings of words. 

For there is no requirement that an interpretation (so-called!) should be thus faithful. So an interpretation can make a sentence true, while none of the elements of its domain $D$ are, according to the received understanding of the sentence, part of its topic. And even if some such elements are, the sentence can be made true by assigning predicates to subsets of $D$ regardless of their meanings. Suppose Frances and George are both in $D$, and Frances but not George is in fact tall. Still there is nothing to be said against an interpretation that assigns to the name `Frances' Frances, and to the name `George' George---so far, so good---but then assigns `is tall' a subset of $D$ that lacks Frances, and contains George. This interpretation makes `George is tall' true. No wonder, one might say, that this discipline is called `{\em formal} semantics'! 

In the face of this, it is natural to propose that one should focus on some (single?) interpretation that assigns semantic values faithfully to words' meanings, and so makes true, according to the official definition, just the sentences that are in fact true. But this proposal is still liable to obscure the relations between meanings and contingent facts. 

For suppose Frances is, as a matter of contingent fact, the tallest philosopher. This proposal will then make both `Frances is tall' and `the tallest philosopher is tall' true. Again, so far so good; since they are in fact true. But what about sentences describing {\em how things might have been}? For example: `Frances might not have been tall'. If our only conception of meaning is the semantic value (i.e.~for singular terms like `Frances': referent) assigned by this favoured faithful interpretation, then `Frances might not have been tall' and `the tallest philosopher might not have been tall' must be treated alike. In particular, they must be assigned the same truth-value. But the former is surely true, and the latter either false or ambiguous.

The answer is to accept that there are two notions of meaning, along the lines of Frege's distinction (1892) [1948] between reference (in our jargon: referent) and {\bf sense}; (his words were `Bedeutung' and `Sinn'). Although `Frances' and `the tallest philosopher' have the same referent, they have a different sense. So in different circumstances, one and the same sense can attach to different referents. Thus in our example, one naturally says: `Had Frances's life gone differently, Frances might not have been the tallest philosopher'. The semantic treatment now proposed is: relative to some non-actual alternative course of history, in particular of Frances's life, `the tallest philosopher' is assigned a referent different from Frances. This accords with Frege's idea of sense. For he calls it the expression's `mode of presentation' of the referent (1892) [1948:~p.~210]: (so `sense' is close to `linguistic meaning') Thus in the actual course of history, it is Frances who gets presented by `the tallest philosopher'; but in an alternative course of history, it is someone else. 

So overall, we can think of the sense of a singular term as the assignment to each possible course of history of the referent therein of the term. (We should of course allow that in some courses of history, there is no referent: the assignment is partial (see Lewis, 1970:~p.~25).) Similarly for predicates like `... is a philosopher'. Referential semantics says that its reference (semantic value) is its set of instances, a subset of the interpretation's domain. But now we also say that the predicate has a sense: the assignment to each possible course of history of the predicate's set of instances therein.

Thus we have arrived at the framework---widely used in modern philosophy, logic and semantics---of {\bf possible worlds}. Although this notion has various historical antecedents (especially in Leibniz), it was Carnap who saw how it provided a framework that naturally implemented Frege's ideas. Carnap's word for reference was {\bf extension}; and for sense {\bf intension}. Hence the label: {\bf intensional semantics}. After Carnap, the framework was developed by many philosophers, logicians and linguists, including especially Kripke, Lewis and Montague (e.g.~Kripke (1972:~pp.~59, 106, 135), Lewis (1970:~pp.~19, 22-26; 1973; 1986), Montague (1973:~pp.~229-237)). We can summarize this discussion, and how the semantics works, with a passage from Lewis (1975:~p.~18):

\begin{quote}\small
Referential semantics tried to answer the question [what are the meanings of sub-sentential items]. It was a near miss, failing because contingent facts got mixed up with the meanings. The cure, discovered by Carnap, is to do referential semantics not just in our actual world, but in every possible world. A meaning [sense/intension] for a name can be a function from worlds to possible individuals; for a common noun [predicate] a function from worlds to sets [of possible individuals]; for a sentence, a function from worlds to truth-values (or more simply, the set of worlds where that function takes the value truth). 
\end{quote}

Although this framework was originally developed for formal languages in logic and for (fragments of) natural languages, it is readily adapted to physical theories with their talk of systems, states and quantities. The idea will be that: systems are objects, i.e.~referents of singular terms; quantities are relations of objects to numbers (relative to a system of units); a state is a value-assignment to all the quantities that apply to a given object; and a possible world is the total histories of the states of all the objects in the world.

We should notice two main choices to be made in adapting the framework to physical theories. We will not have to make either of these choices. But it will be important to have formulated them, since doing so will enable us to focus on the simpler cases while knowing how to read the discussion back to more complicated cases (which some of the advanced examples in Part III will use). 

The first choice arises from the contrast between our intermediate notion of theory (and of model) as each using a single state-space, and the more usual notion comprising {\em many} state-spaces for the many examples of the theory's kind of system. Thus recall from (2) of {\em Theory} in Section \ref{Ourthm} that the theory, Newtonian mechanics, is usually understood to comprise state-spaces for any number of point-particles (and perhaps also: for any number of rigid bodies, and-or deformable bodies); although of course any use of the theory, theoretical or experimental, will focus on one such state-space appropriate to the problem at hand. So an intensional semantics for Newtonian mechanics so understood will postulate a vast number of possible worlds. Even if one considers only point-particles: for each positive integer $N$, there will be a possible world for each of the (no doubt: continuously many) possible total histories of the $N$ particles. 

On the other hand, an intensional semantics for our notion of theory (or model) will be simpler. For since there is just one state-space at issue, a possible world will be an appropriate curve through the state-space; (here, as usual, we invoke the Schr\"odinger picture of time-evolution). Besides, the discussion can often be further simplified by abstracting away from the contrast between instantaneous states and possible worlds (histories).

The second choice arises from the contrast between the syntactic and semantic conceptions of theories, that we sketched in Section \ref{thmscph}. The syntactic conception obviously fits intensional semantics well. After all, one gives semantics for pieces of syntax: a linguistic expression is mapped onto its extension---an object, set of objects or truth-value, for singular terms, predicates and sentences, respectively---by its intension. 

On the other hand, the semantic conception seems to have already applied such maps---as the label `semantic conception' indeed suggests. That is: one naturally expects that the codomains, and their elements, of the intensions will be the sort of item that {\em is} a model, or is a constituent of a model, of the theory, according to the semantic conception. Broadly speaking, that is right. For as we already remarked in Section \ref{thmscph}: a model, i.e.~an interpretation {\em \`a la} referential semantics, often has specific features additional to those making a given sentence or theory true in the model---so that a model represents one way that the topic of a theory could be, while making the theory true according to the model. We can now see that remark as prescient: it says that a model of a theory is a prototype (dare we say: toy-model!) for a possible world. 

But here we should also remember that Section \ref{thmscph} warned against making the distinction between the syntactic and semantic conceptions too sharply. For a model of the kind the semantic conception advocates is just a set that is equipped with some structures, viz.~by having certain elements, or subsets, or subsets of its Cartesian powers singled out; (with the singling out usually signalled by putting all these items in an ordered $n$-tuple). And that is... just a set. The set and its structures (and-or the $n$-tuple) needs to get associated with some interpreted language if it or they are to represent anything. So after all, there is a need for interpretation (genuine interpretation!) of models {\em \`a la} the semantic conception. And it can be provided by intensional semantics, for example: this element of the set is mapped by the intension for the singular term `the particle of mass $m$', relative to a possible world with one such particle, to that particle in the world. (As we mentioned, most intensions will be partial functions: many possible worlds as input, i.e.~argument, will yield no output i.e.~value. So unsurprisingly, the literatures on the semantic conception of theories and on intensional semantics overlap (see e.g.~van Fraassen (1970:~pp.~328, 335)).

For the purposes of this book, the upshot about these two choices is, fortunately, simple. For each choice, we can adopt the second option. We can apply intensional semantics to our intermediate notion of theory (and of model in our sense) as a triple of state-space, quantities and dynamics, $\bra {\cal S}, {\cal Q}, {\cal D} \ket$. The core idea will be partial functions defined on either ${\cal S}$ or ${\cal Q}$, that behave like intensions (Fregean senses) and so deliver as values, extensions (worldly referents). But we will not need details of how to do this, since a distinction between meanings as intensions and as extensions will only be made in Section \ref{emergence} when we discuss emergence: details can be found in De Haro and Butterfield (2018:~pp.~322-325 and (applied to dualities) pp.~341-344). 

\subsection{Interpretations as internal and external; and as partial maps}\label{intext}

So far within Section \ref{itm}, our discussion of interpretation has set aside the two levels, theory ``above'' and model ``below'', introduced in Section \ref{Ourthm}. But we now return to that contrast. For it led in Section \ref{modelrootss} to the distinction between two aspects of a model: the model root which is the representation of the theory above (in most cases: a model triple) and specific structure. We now note that this gives a correlative distinction between two kinds of interpretation of a model. (So this distinction does not apply to the theory ``above''. But it will be important to our account of dualities since according to us, the two duals are indeed models of a common core theory ``above''.) Thus we say that an {\bf internal interpretation} is one that only interprets the model root: while an {\bf external interpretation} also interprets (some or all of) the specific structure.\footnote{For more discussion, cf.~De Haro (2020a:~Section 1.1.2; 2021:~Section 2.2.3), De Haro and Butterfield (2018: Section 3.2.2). The contrast between external and internal interpretations was introduced in Dieks et al.~(2015:~p.~209), and further articulated in De Haro (2017a:~p.~116; 2020:~p.~263), as a contrast between interpretations of dual models that are anchored on something external to the duality relation, i.e.~external to the common core (e.g.~a theory of measurement, or some other external piece of physics), and those that are not so anchored. Our current definition is the formal correlate of this, in that external interpretations also map the specific structure (which is external to the duality relation), while internal interpretations only map structure that is common to the duals: thus also providing an interpretation of the common core theory, as we will discuss in Chapter \ref{Theor}. Sklar (1982:~pp.~93-94) deploys a contrast in how `meanings are grasped', similar to the former version of the external vs.~external contrast, to distinguish non-interesting (or not surprising) forms of equivalence from substantive ones.\label{II}}\\
\\
{\it Interpretations as partial, structure-preserving, maps}. Since we have formulated theories and models in terms of sets of states and quantities (i.e.~Eqs.~\eq{introtriple} and \eq{eqmodel}), it is natural to formulate interpretations in terms of interpretation maps, i.e.~functions (in general, partial functions) mapping items in the theory or model to items in the world.\footnote{For a definition of a partial relation or partial function, see Frigg (2023:~p.~205).} 
This is analogous to the assignments of referents to the expressions of the language, which is standard in the syntactic conception of theories, and we already discussed in Section \ref{thmscph}, now applied to theories formulated in the language of mathematical physics.\footnote{The distinction between a formalism and its interpretation is also well-known in physics (see footnote \ref{Ruetsche2002}): Muller (1997b:~pp.~242-243) ascribes this introduction to Heisenberg. This distinction indeed already appears in Heisenberg's (1925b:~p.~879), which bears the title `Quantum-theoretical re-interpretation of kinematic and mechanical relations'. See also Heisenberg (1926:~pp.~992-994; 1929:~p.~493).}

Thus in general, an interpretation of a bare theory, $T$, is a (partial) function, $i$, preserving appropriate structure, from the theory to a domain of application in the world, $D$, i.e.~a map $i: T \rightarrow D$. If the theory is presented as a triple of states, quantities, and dynamics, then $i$ denotes a triple of functions, one for each component. In general, we expect that the interpretation maps preserve appropriate structure (see point (A) below), and so that they are partial homomorphisms.

Likewise, given a model $M$ of a bare theory $T$, presented in terms of its model root and it specific structure, i.e.~$M=\bra m;\bar M\ket$, an interpretation of this model is also a map, preserving appropriate structure, between the model and a domain of application. To distinguish the model's map from the theory's interpretation map, we temporarily use a different letter for it, viz.~$j:M\rightarrow D'$. The domain of application $D'$ is in general different from the theory's domain of application $D$, since a theory can have models with various (disjoint) domains of application.\footnote{Since a model is for us almost always a representation (see Section \ref{modelrootss}), i.e.~a homomorphic copy, of the theory $T$, it would appear natural to require that the interpretation and representation maps form a commuting diagram. Thus, given: (i) the interpretation maps for the theory and the model, $i:T\rightarrow D$ and $j:M\rightarrow D'$, respectively; (ii) the model's representation map (i.e.~a homomorphism, see Figure \ref{meshdynamicsbasic}), $h:T\rightarrow M$, and a homomorphism between the domains, $k:D\rightarrow D'$: one might require that they commute, i.e.~$j\circ h=k\circ i$. In particular, if this condition is satisfied, then the theory can be interpreted through the model's domain, $D'$, i.e.~the model $M$ gives an interpretation map for the theory into the domain $D'$: namely, $j\circ h: T\rightarrow D'$. This then reproduces one of the standard connotations of the word `model' that we rejected in point (iii) under the title {\it `Model'} in Section \ref{Ourthm}. But we still in general reject this condition: for the interpretation maps $i$ and $j$ are only {\it partial}, i.e.~may not map some elements, and so in general this diagram does not commute. Thus we will resist the temptation of imposing such (stronger) requirements for the maps, in general. However, in Section \ref{semanticeq}, when we consider dualities between {\it several} models, and the condition of theoretical equivalence, we {\it will} consider conditions of this type.}

This gives a natural realization of the distinction introduced above, between internal and external interpretations, in terms of this map: namely, an internal interpretation maps all of and only the model root, $m$ to items in the world, while an external interpretation also maps (some of the) the specific structure, $\bar M$, to items in the world.\\
\\
{\it Requirements on interpretation maps for scientific theories.} In order for formal maps thus defined to give suitable interpretations of theories and models, we will further clarify two points. By `suitable interpretation', we here mean `suitable for the scientific aims of description and prediction'. (In general, science also has other aims: especially explanation and understanding. But these other aims require additional conditions, and we can postpone this topic until Chapter \ref{Understand}).\footnote{For example, Ruetsche (2011:~p.~5) has noted that the interpretations given for the aims of description and prediction may differ from those given for the aims of explanation and understanding: `It is often a theory {\it under an interpretation} that predicts, explains, and promotes understanding ... there may not be a single interpretation under which a given theory accomplishes all those things.'} 

This `suitability for description and prediction' is an epistemic issue that we will discuss in more detail in Chapters \ref{Realism} to \ref{Understand}. Here, we make make two comments---the first is structural, the second is empirical---where the second, more general, comment guides the first, more specific one: \\
\\
(A):~{\it Preserving the structure of theories and models.} The first comment is about the structure that it is desirable for interpretations to preserve, for the aims of description and prediction, which we mentioned in our phrase `preserving appropriate structure'. This requirement, that an interpretation map ought to preserve structure, guides the construction of interpretations {\it as much as} it guides the construction of bare theories and models. In general, different types of interpretations will preserve different (amounts of) structure (and which structure it is appropriate to preserve is of course an empirical matter: see point (B)). 

Thus there are two types of structures that should, in this sense, be preserved, because they are part of the defining structure of a theory (or model):

(1):~{\it Defining structure of the triples.} An interpretation should respect the way a theory (or model) is defined as a triple. Thus if we denote, by $i_{\cal S}:{\cal S}\rightarrow D$, the function that assigns to the state $s\in{\cal S}$ a particular state of affairs in the world, and by $i_{\cal Q}:{\cal Q}\rightarrow D$ the function that assigns to the quantity $Q\in{\cal Q}$ a quantity in the world, then both of these maps should respect the operations with which ${\cal S}$ and ${\cal Q}$ are endowed. 

For example, if the state space ${\cal S}$ is a linear vector space, we require an appropriate (i.e.~systematic) relation between the state of affairs in the world that corresponds to a linear combination of states, and the states of affairs that correspond to the individual states, i.e.~a systematic relation between $i_{\cal S}(s_1+s_2)$, $i_{\cal S}(s_1)$ and $i_{\cal S}(s_2)$ that reflects the linearity of the state space: and likewise for the algebra of quantities in ${\cal Q}$. 

Also, we require that there is a map $i_v$, mapping the {\it values of quantities} (see Eq.~\eq{pairing}) to elements in $D$, i.e.~$i_v(\bra Q,s\ket)$, that is related in an appropriate way to $i_{\cal Q}(Q)$ and $i_{\cal S}(s)$. 

\begin{figure}
\begin{center}
\includegraphics[height=3.5cm]{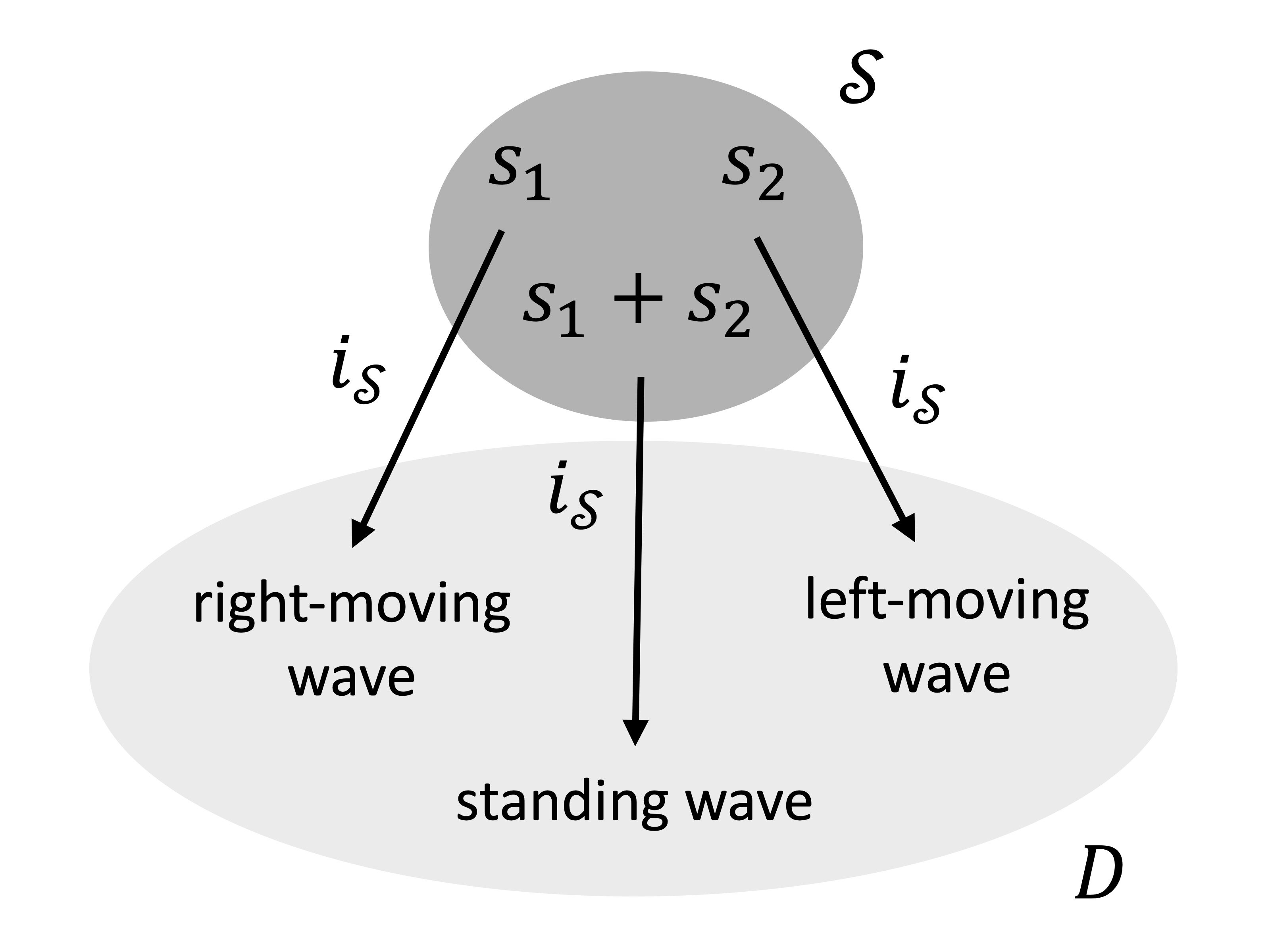}
\caption{\small The interpretation map $i_{\cal S}: {\cal S}\rightarrow D$ respects the additive structure of the state space. The state $s_1$ is mapped to a right-moving wave in the domain of application, and the state $s_2$ is mapped to a left-moving wave. Then $s_1+s_2$ is mapped to a standing wave.}
\label{standing-w}
\end{center}
\end{figure}

Figure \ref{standing-w} shows three states in ${\cal S}$, namely $s_1$, $s_2$, and $s_1+s_2$. The state $s_1$ represents a right-moving wave, and the state $s_2$ represents a left-moving wave. Then it is required that $i_{\cal S}(s_1+s_2)$ is an item in $D$ with an appropriate relation to left- and right-moving waves: namely, it is a {\it standing wave}.

(2):~{\it Symmetries.} We also require that an interpretation preserves the theory's {\it stipulated symmetries}, i.e.~those symmetries that are part of the definition of a theory (for details, see Section \ref{salientstipul}).

Thus our triples, i.e.~our theories, are by definition structured: they are not `flat' collections of models, and so they are not subject to some of the usual criticisms of the semantic view. We will return in more detail to this topic in Part III: Chapter \ref{Theor} will discuss the relation between the Schema's formulation of theories and the usual syntactic and semantic conceptions, as well as some criticisms of the semantic conception: we will see that the Schema's view of theories is structured, and we will develop this structure further in our geometric view of theories, in Chapter \ref{Heuri}.\\
\\
(B):~{\it Empirical confirmation.} Since scientific theories aim to describe the empirical world and make predictions about it, we of course require that the interpretation map is in agreement with the available empirical evidence and confirmation of the theory in its domain of application. But we do not need to restrict attention to already well-confirmed or successful theories such as classical mechanics or general relativity, and we will also consider theories, such as string theory, that are being developed and have not yet been tested beyond the domains of the old theories---so long as they can make a serious claim to a domain of reality! Such standards for a good theory are of course set by our relation to the empirical world, and can of course change (i.e.~they become stricter in time). 

Thus even though one aims for theories that are as rigorously defined as possible, the degree of precision of the interpretation maps is set by the degree of precision that is relevant to the empirical testability of the theory in question.\\
\\
{\bf An illustration in algebraic quantum field theory.} There is a natural relation between: (i) the Schema's notions that we have so far introduced, viz.~models as representations of a bare theory, model roots vs.~specific structure, and internal vs.~external interpretations, and (ii) the philosophical discussions of algebraic quantum theory in the literature (in particular, we will here follow Ruetsche (2002, 2011)).\footnote{For an excellent introduction to von Neumann algebras that emphasises projections, see Chapter 6 of Redei (1998). Probabilistic interpretations of algebraic quantum field theory and quantum statistical mechanics are discussed in Ruetsche and Earman (2011). Issues of relativistic causality and locality are discussed in Earman and Valente (2014). For a general introduction to algebraic quantum field theory, see Halvorson (2007).} 
Although we will not discuss algebraic quantum field theory in this book, highlighting this connection serves to illustrate the generality of the Schema, regardless of dualities.

We first briefly introduce two interpretations of quantum theories that Ruetsche (2002:~p.~361f.; 2011:~p.~132f.) and Arageorgis (1995:~pp.~132, 140) dub `algebraic imperialism' and `Hilbert space conservatism', together with some of their drawbacks: then we will sketch the relation to the Schema.

Consider a quantum theory, defined by an abstract C$^*$-algebra ${\cal A}$ (for example, the Weyl algebra), and equipped with a set of states, understood as linear maps from the algebra to the complex numbers, i.e.~$\om:{\cal A}\rightarrow\mathbb{C}$, that are positive-definite and with norm 1. The value $\om(A)$, for $A$ a self-adjoint member of ${\cal A}$, is to be thought of as the expectation value of the observable represented by $A$. According to the interpretation that Ruetsche (2002, 2011) and Arageorgis (1995) {\bf algebraic imperialism}, the state-space of this quantum theory is the space of algebraic states on the abstract algebra, and two such quantum theories are physically equivalent iff their associated C$^*$-algebras are isomorphic.

This algebraic imperialist interpretation is exceedingly austere. Thus to see whether it can be improved, consider representing the algebra on a {\it concrete Hilbert space}, i.e.~through a representation map $\pi:{\cal A}\rightarrow{\cal B}({\cal H})$, where ${\cal B}({\cal H})$ is the set of bounded linear operators on the Hilbert space, which is complete with respect to the Hilbert space operator norm. This space is also a C$^*$-algebra. This can be done by the GNS construction, whereby a vector $\psi\in{\cal H}$, often called a `vacuum state', represents $\om$, and the expectation values of operators are as follows: $\forall A\in{\cal A}, \om(A)=\bra\psi|\pi(A)|\psi\ket$.\footnote{This slightly simplified discussion follows Ruetsche (2002). For a discussion of Hilbert space representations that uses von Neumann algebras, see Ruetsche (2011:~Chapter 6). In that framework, the relevant set of observable quantities are the self-adjoint members of the von Neumann algebra affiliated with a GNS representation of a state of the abstract algebra.}

Having represented the abstract algebra on a concrete Hilbert space, the theory admits an interpretation, which Ruetsche (2002, 2011) dubs {\bf Hilbert space conservatism}, that is closer to that of ordinary quantum mechanics. For the Hilbert space conservative, it is the states represented in Hilbert space, rather than the abstract algebraic states, that count as the theory's state-space. This Hilbert space comes with {\it specific structure}: notably, since the representation map is in general not surjective, its image is a proper subset of the set of bounded operators, i.e.~$\mbox{Im}(\pi)\subset{\cal B}({\cal H})$. This means that there are self-adjoint operators on the Hilbert space that do not have a correlate in the abstract algebra ${\cal A}$. Ruetsche suggests calling these {\it parochial observables}: a jargon that has caught on in the philosophical literature.

Parochial observables are a challenge for algebraic imperialism, because some of them are operators that we normally take to be physically significant: for example, the number operator, the position and momentum operators, and the stress-energy tensor, are parochial observables. Thus the algebraic imperialist does not have direct access to these familiar operators, since they do not have a correlate in the imperialist's preferred algebra.

But, in quantum field theory, there is a challenge for Hilbert space conservatism, because of the existence of unitarily inequivalent representations of the algebra.\footnote{In elementary quantum mechanics, the Stone-von Neumann theorem secures that any two irreducible representations of the canonical commutation relations of the position and momentum operators (in Weyl form, so that the operators are bounded) are unitarily equivalent. However, the assumptions of the Stone-von Neumann theorem, especially the assumption of finitely many degrees of freedom, do not hold in quantum field theory, and there are unitarily inequivalent representations. For a discussion, see e.g.~Ruetsche (2011:~pp.~41, 46, 59-65).}
The states that can be represented as density matrices through the GNS construction are called a {\it folium}. Unitarily inequivalent GNS constructions, based on different states $\om$ and $\om'$, thus involve different (unitarily inequivalent) folia. Thus for the Hilbert space conservative, the physically possible states lie in a single folium. This means that some parochial observables will be missing in the conservative's preferred folium and, more generally, that there is no clear general criterion by which the Hilbert space conservative can choose a preferred Hilbert space.\footnote{In these two paragraphs, we say `challenge', rather than e.g.~`objection', so as to indicate that it may be surmountable. See for example Feintzeig (2018).} 

It is clear how to reformulate these ideas in the Schema's terminology. First, the different Hilbert space representations are {\it models}. These models are representations of a {\it bare theory}, which is the abstract algebra ${\cal A}$, together with its set of states. This bare theory is the {\it common core} that different Hilbert space representations have in common.\footnote{Recall, from Section \ref{Ourthm} (just before `Theory'), that, even if no duality is at issue, the Schema is still a useful characterisation of theories, models and interpretations. Thus even if no duality is at issue, we consider an uninterpreted, bare theory, and its models as representations of it: and we still distinguish, within a model, model root from specific structure, with the model root of each model being the image of a representation map. Thus we can still speak of the bare theory as the {\it common core} of these models, i.e.~the common structure that all the models represent, even if this structure is not isomorphic to its models, and the models are not duals. Indeed, Sections \ref{abstraction} and \ref{lsr} will discuss that the interpretation of a common core is more abstract, and thus logically weaker, than that of its models: and that it can be logically weaker than the model roots.} 
The {\it specific structure} of each model includes some parochial observables that are, in general, not part of the bare theory, and also not part of the other models. Parochial observables do not have a correlate in the common core, and so they are always part of the specific structure. The {\it model root} is the image of the representation map, i.e.~$\mbox{Im}(\pi)$, which is, roughly speaking, the ``part'' of the Hilbert space that has a correlate in the abstract algebra.

Algebraic imperialism thus corresponds to an {\it internal interpretation} of such a quantum theory: it only interprets the common core theory, and it interprets the models only in so far as they have correlates in the bare theory (i.e.~such an interpretation maps only the model root and not the specific structure). Hilbert space conservatism corresponds to an {\it external interpretation}, because it only interprets the features of a particular model, including its specific structure, and not the specific structure of other models, which is incompatible with the specific structure of the model whose interpretation one takes to be external.

We expect that a quantum-theoretic example of unitary equivalence (as against the cases of unitary inequivalence discussed above) would be an example of the Schema. Indeed, Section \ref{pmd} will discuss position-momentum duality in quantum mechanics as being both a case of unitary equivalence, and a duality according to the Schema. And Part II will give further examples such as bosonization (see Section \ref{bosoniz}).

We cannot add further detail about algebraic quantum field theory. We will discuss interpretative issues further in Part III. But we submit that the discussions of interpretation, theoretical equivalence, ontology, epistemic warrant and emergence, that the literature has reported for the Schema and for dualities more generally, is relevant for the discussion of algebraic imperialism and Hilbert space conservatism (see, in particular, Chapters \ref{physeq} and \ref{Realism}). 

Furthermore, we suggest that the geometric view of theories that we will discuss in Chapter \ref{Heuri}, and which is in part motivated by the phenomenon of symmetry breaking in the Ising model (Section \ref{dualpf0}) and in the Seiberg Witten theory (Section \ref{effD}), is bound to cast light more generally on the interpretation of (algebraic) quantum field theories.

\section{Symmetries}\label{Symm}

On the topic of symmetries, we have so far said only that (i) a symmetry is a map on states that preserves the values of quantities (not all quantities but a large and salient subset of them), and (ii) a duality is like a `giant symmetry' in that it is a map between theories that preserves values (Section \ref{giantS}). In this Section, we will expand on (i), but we will leave the development of (ii) to the next Chapter.

There are two broad ways in which we need to expand on the idea (i). First: in relation to existing discussions of symmetry; and so regardless of the two-levels contrast we emphasised in Section \ref{Ourthm}, of the theory above and the model (in our sense) below that represents the theory. For those discussions ignore the contrast, because it seems irrelevant to them. That is hardly surprising, since the contrast is part of the stage-setting needed for our account of duality. But second: in fact, there are things to say about (i) in relation to the contrast, even before one uses the contrast to address the topic of dualities.

So we will proceed in that order. We will first discuss symmetries without regard to the contrast. But we will use Section \ref{Ourthm}'s notation of triples $\bra {\cal S}, {\cal Q}, {\cal D} \ket$ of states, quantities and dynamics: which, as we have stressed, applies equally well at the two levels, the theory above and its representation below. And we will of course also use Section \ref{valuesQ}'s notation for values and for dynamics. Thus we will discuss in order: symmetries as maps on states (Section \ref{dual}); the factors that influence what counts as the salient set of quantities whose values are to be preserved (Section \ref{salientstipul}); the idea of dynamical symmetries (Section \ref{dynlsymm}); how our treatment compares with the (different) way that symmetries are usually defined for spacetime theories (Section \ref{sptthies}).

Then we will bring in the two-levels contrast. There will be no need to repeat the general claims of the previous Sections in the context of this contrast: the situation will be clear. But with both the previous discussion and the contrast in hand, we can relate a simple and historically important example in detail to our notions: namely Galilean transformations. We do this in Section \ref{expleGalil}. It will also illustrate Section \ref{intext}'s distinction between internal and external interpretations. \\

A final preliminary: in this Section we will not engage with two debates that have been prominent in the recent philosophical literature about symmetries. They both concern whether symmetry is always a sign of `surplus structure', `redundancy' or `gauge': roughly, whether a pair of symmetry-related states (solutions of a theory) should be interpreted as `the same'. So these are debates about interpretation; and since a duality is a `giant symmetry', they have obvious analogies with the corresponding over-arching question whether two duals make the same claims about the world. So it will be clearest if we postpone discussion of these debates until after we have stated our Schema for duality; so for our views, we refer to Sections \ref{dualsym} and \ref{semanticeq}. 

However, since this Section {\it does} discuss some interpretative issues, it will be helpful to briefly note here what these debates are, and our broad view about them. So in short:---

(1): Should two symmetry-related solutions of a theory: (a) be interpreted {\it ab initio} as representing the same physical state of affairs?; or (b) be taken merely to motivate searching for a common ontology that secures such an interpretation? This debate is articulated by M\o ller-Nielsen (2017) and, in relation to dualities, Read and M\o ller-Nielsen (2018). These authors defend option (b). Broadly speaking, we agree with them, about symmetries as well as dualities: (cf.~De Haro (2020a:~pp.~267-270), Butterfield (2021:~p.~47) and De Haro (2019b:~pp.~5158-5162)).

(2): Given a theory whose solution-space is partitioned by a group of symmetries---i.e.~solutions in the same cell are symmetry-related---should we: (a) try to write down a `quotiented' (also known as: `reduced') theory whose solutions correspond to the cells of the partition; or (b) resist quotienting the given theory, but take its symmetry-related solutions to be isomorphic? This debate is articulated, with (a) and (b) labelled `reduction' and `sophistication' respectively, by Dewar (2015: Sections 4 to 6; 2017); see also Caulton (2015:~pp.~156-157) and Weatherall (2016a). Dewar defends sophistication. Broadly speaking, we are sympathetic: reduction is not to be undertaken lightly (De Haro and Butterfield, 2021:~p.~2976).

\subsection{The idea of symmetry}\label{dual}

We begin with the notion of symmetry announced in Section \ref{giantS}: as a map $a$ on states, $a: {\cal S} \rightarrow {\cal S}$, that preserves the values of a salient set of quantities: usually a large set, though not necessarily all the quantities. The map $a$ must also respect the structure of ${\cal S}$, e.g.~topological or differential structure. (Thus `$a$' is for `automorphism'.) But this requirement will be in the background in the sequel: the emphasis will be on the state $s$ and the image-state $a(s)$ having the same values for quantities in the salient set. 

\begin{figure}
\begin{center}
\includegraphics[height=2.9cm]{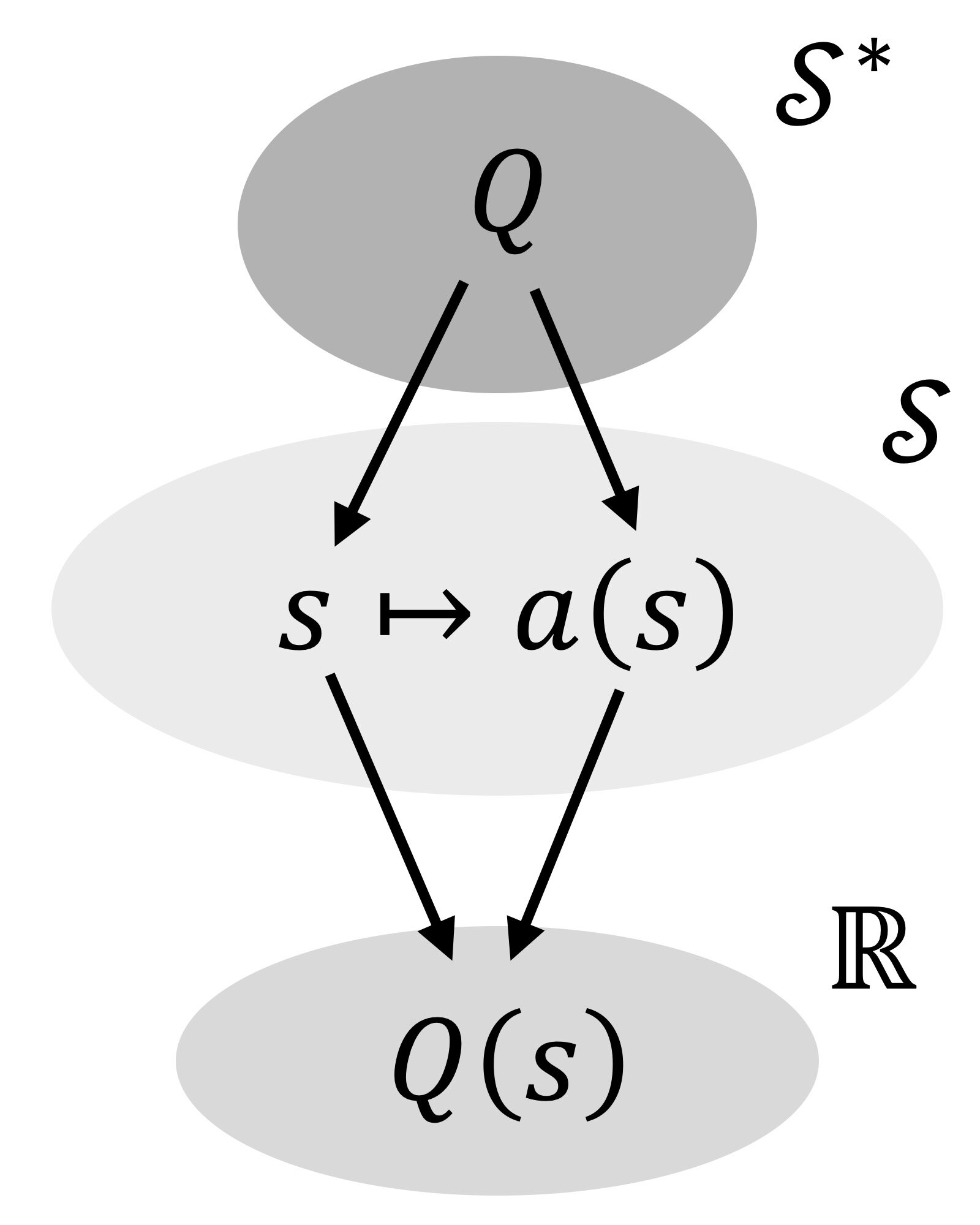}~~~~~~~~~~~~~~~~~~~~~
\includegraphics[height=3cm]{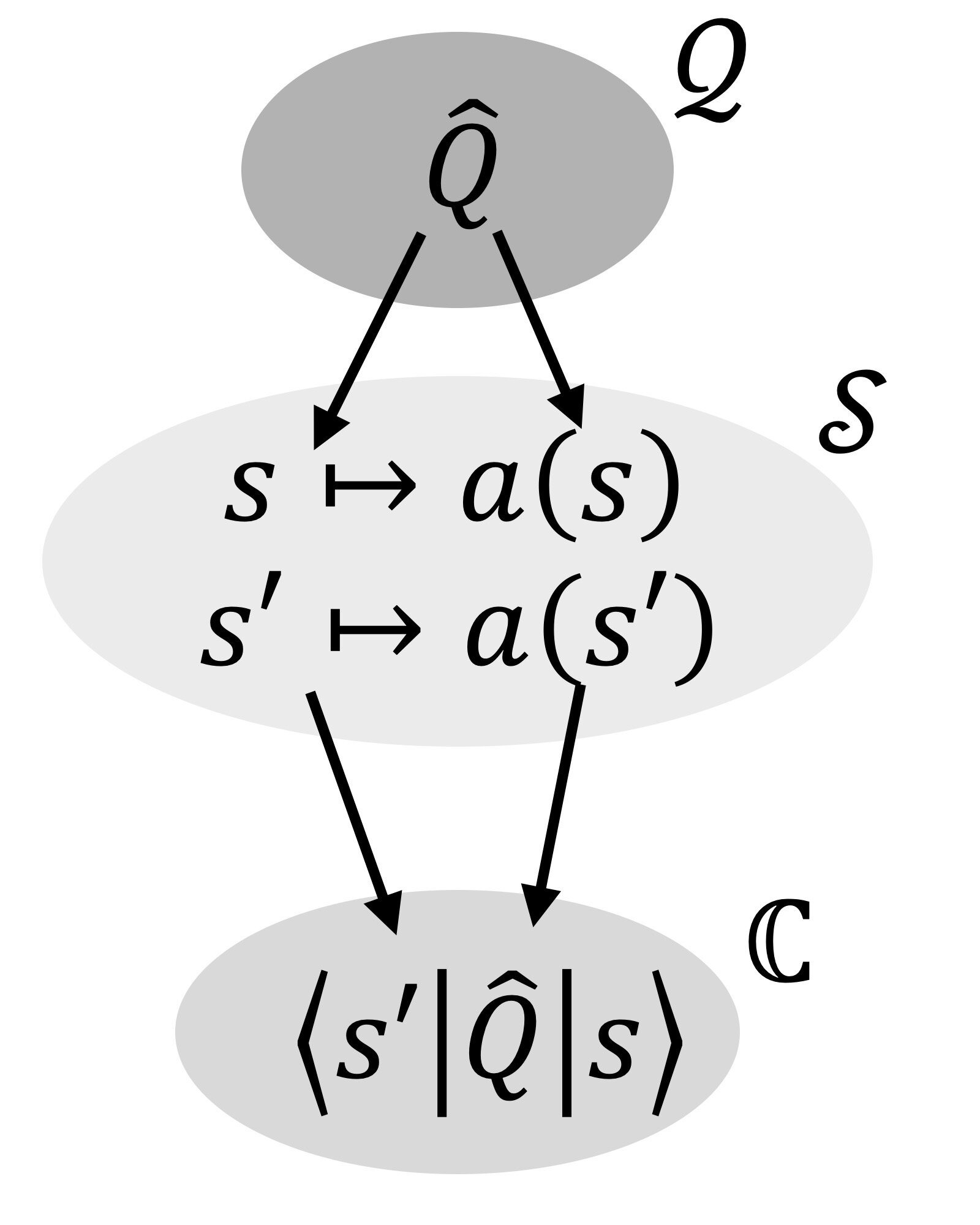}
\caption{\small Preservation of values: (a) Left: classical possessed or quantum expectation value. (b) Right: matrix elements, for quantum theories.}
\label{sQs}
\end{center}
\end{figure}

Here, values are understood as in Section \ref{valuesQ}: $\langle Q , s \rangle$ is a classical possessed value or a quantum expectation value (or more generally, a matrix element, $\langle s' |\,{\hat Q}\, | s \rangle \in \mathbb{C}$: see below). The equality of values, for a symmetry $a$, is then
\be\label{symbasic}
\langle Q , a(s) \rangle = \langle Q , s \rangle\,.
\ee 
More generally: we take quantum symmetries to also preserve off-diagonal matrix elements:
\bea\label{qsyms}
\forall\,s,s'\in{\cal S}~,~~~~\bra a(s') |\,\hat Q\, | a(s) \ket=\bra s'|\,\hat Q\,|\,s\ket\,,
\eea
for a salient subset of operators in ${\cal Q}$, usually including the Hamiltonian. (And this condition can be weakened, to hold only for a salient subset of states in ${\cal S}$.) The preservation of the values of quantities on states is illustrated in Figure \ref{sQs}.\\
\\
{\it Example: rotations}. As an elementary example, consider states that are the relative position vectors between two particles in space, i.e.~${\bf x}={\bf x}_1-{\bf x}_2$, and take the symmetries to be the group of relative rotations between the two particles, i.e.~${\bf x}\mapsto {\bf x}'=R\,\cdot\,{\bf x}$, where $R\in \mbox{SO}(3)$ is a $3\times 3$ rotation matrix. The distance between the two particles is a quantity that is preserved by such rotations, i.e.~$|{\bf x}'|=|{\bf x}|$.

This notion of symmetry is very simple. But suitably adapted and augmented, it will be sufficient for our purposes. We will see that the notion is adaptable, including to dynamics and to spacetime theories.\footnote{De Haro and Butterfield (2021:~p.~2995) show that symmetries can be seen as maps on {\it quantities}, rather than on states: namely, dual (in mathematicians' sense not ours!) maps.\label{symq}}

\subsection{Salient quantities and states: stipulated symmetries}\label{salientstipul}

We said that a symmetry preserves values for `a salient, usually large, set of quantities/states'. This general formulation deliberately uses the vague word `salient', since it varies from case to case which quantities/states it is noteworthy to preserve the values of. But it is worth noticing three sorts of consideration that often mould the choice of quantities/states, i.e.~which quantities/states count as salient. The first, which we label {\it stipulated symmetries}, is part of the defining structure that we discussed in Section \ref{intext} (A); the second and third will get a Subsection of their own (Sections \ref{dynlsymm} and \ref{sptthies}).

Comment (A) in Section \ref{intext} already mentioned that a theory is not always given to us with prescribed sets of states and of quantities, so that the set of symmetries is thereby fixed, once some precise meaning of `salient' is fixed. For often in physics, we begin our theorizing by adhering to some symmetry principles. That is: we define a theory's sets of states and quantities in order that they carry a representation of a given abstract symmetry group: spacetime symmetry groups such as the Poincar\'{e} group being a standard example. In such a case, we will say the symmetries are {\it stipulated}---they are part of the definition of the theory. Usually, we think of these symmetries as maps on states: unitary representations of a spacetime symmetry group on a Hilbert space being a standard quantum example. 

In general, then, a theory $T$ that is formulated as a triple, $\bra{\cal S},{\cal Q},{\cal D}\ket$, is said to have a {\bf stipulated symmetry} if it is formulated as having an automorphism of the state-space, $a:{\cal S}\rightarrow{\cal S}$, that preserves some salient subset of the quantities. The stipulated symmetry thus comes with a specification of which quantities count as salient, so that their values are ``worth'' preserving. This specification is encoded in the definition of the triple and its stipulated symmetry, and it of course bears on the theory's interpretation---it moulds the kinds of interpretations that the theory can be given. 

But note that stipulating a symmetry does {\it not} imply that every state has its value preserved for {\it every} quantity definable on the state-space. (And {\it mutatis mutandis} if we conceive a symmetry as a map on quantities; cf.~footnote \ref{symq}.) For example, one might stipulate rotational symmetry: more precisely, that in a quantum theory the Hamiltonian is rotationally invariant, so that the Hilbert space carries representations of SO(3). But this still allows using quantities $Q$ whose expectation values on some states are not invariant under rotations. Thus even with a stipulated symmetry, there is a question of selecting the salient quantities. This point will recur in Section \ref{dynlsymm}.

\subsection{Dynamical symmetries}\label{dynlsymm}

We have presented symmetries as maps that preserve the salient quantities' values (and respect the structure of $\cal S$ or $\cal Q$: cf.~Section \ref{dual}). But we have not mentioned time, i.e.~the fact that values change over time. It is indeed very usual to define a symmetry as a map that `preserves the dynamics'. Taking a symmetry, as usual, as a map on states, this means, roughly: if a sequence of states is possible according to the dynamics, so is the sequence of image-states. 
 
We can make this precise by using the framework for dynamics, in both Schr\"{o}dinger and Heisenberg pictures, given in Section \ref{valuesQ}, Eqs.~\eq{dyns1} and \eq{dyns2}. We will favour the former. (As noted there, a full discussion would also address the treatment of time in theories not best thought of as triples, e.g.~theories formulated using partition functions; again, cf.~footnote \ref{nontriple} at the end of Section \ref{Ourthm}.)

On Schr\"{o}dinger dynamics, a dynamically possible total history of the system is a curve through the state-space ${\cal S}$ parameterized by time $t$: with each point $s(t)$ defining the values $\langle Q , s(t) \rangle$ of the various quantities $Q$ at $t$. Then we can define a dynamical symmetry as a map $a$ on $\cal S$ that (a) respects $\cal S$'s structure and (b) maps any dynamically possible history (curve through state-space) to another such history. That is: if $s(t)$ is a dynamically possible history, the sequence $a(s(t))$ of states is also dynamically possible.\footnote{Preserving the dynamics in this sense is of course a commutation i.e.~equivariance condition. For if we write $s(t) = D_{t, t_0}(s(t_0))$ with $D_{t, t_0}$ representing the dynamics (cf.~eq \ref{dyns1}), then preservation of the dynamics is: $a(s(t)) \equiv a(D_{t, t_0}(s(t_0))) = D_{t, t_0}(a(s(t_0)))$.\label{dynsequivariance}} 

On Heisenberg dynamics, the definition of a dynamical symmetry is (we think!) more complicated, because the representation of a dynamically possible history is more complicated. A history is given by a fixed $s \in {\cal S}$ and a family of curves through ${\cal Q}$, all parameterized by time $t$: with $Q_1, Q_2$ on a common curve representing the same physical quantity, e.g.~energy, at two times $t_1, t_2$. So for a single history, there are as many curves through $Q$ in the family as there are physical quantities pertaining to the system. So a dynamical symmetry must be a map whose domain is, not $\cal Q$, but the set of all such families of curves (or all such families that are indeed dynamically possible, once some $s \in {\cal S}$ is chosen). So the map will have to suitably respect, not so much $\cal Q$'s structure, but the structure $\cal Q$ induces on this set of families of curves. And for the map to be a dynamical symmetry, it must leave invariant the dynamically possible families (allowing, no doubt, for a change of state $s \in {\cal S}$). But for most of this book, we can focus on symmetries as maps on states; and so we will not here further consider the Heisenberg picture. 

Given this definition of dynamical symmetry in Schr\"{o}dinger picture, as a map on ${\cal S}$ that commutes with the dynamics (footnote \ref{dynsequivariance}), the obvious first question is: how is this related to our initial idea of symmetry as a map on ${\cal S}$ that preserves quantities' values, regardless of time? 

{\it A priori}, they seem very different. Indeed the notion of dynamical symmetry seems weaker in that it requires the transform $a(s(t))$ of each dynamically possible history $s(t)$ `only' to be itself dynamically possible---it need not have any distinctive relation to $s(t)$, e.g.~by being in some sense a `replica' of $s(t)$. But in fact the notions are drawn together by dynamical symmetry's requirement that the map on histories be induced by a map on states. That is: writing a history as a set of points in ${\cal S}$ for brevity, $\{ s(t) \}$: a dynamical symmetry requires that the map on histories $\{ s(t) \} \mapsto \{ s'(t) \}$ be induced by a map on instantaneous states, i.e.~$s(t) \mapsto a(s(t))$. This turns out to be a strong requirement, thanks to the `sensitivity' of dynamics to the values of many quantities. That is: it turns out to force $a$ to preserve the values of many quantities---leading us back to our initial idea of symmetry. But note that this implication is not {\it a priori}: it depends on what dynamical evolution, in typical theories, in fact depends on. 

In the elementary example of rotations from the previous Section, the two particles can have a dynamics, so that their relative position vector is time-dependent, i.e.~${\bf x}(t)={\bf x}_1(t)-{\bf x}_2(t)$. Then the group of rotations still preserves the time-dependent distance, i.e.~$\forall t~|{\bf x}'(t)|=|{\bf x}(t)|$, as well as their relative speed, i.e.~$|{\bf v}|=|\dot{\bf x}(t)|$.

The point here is well illustrated by spacetime symmetries, as mentioned in Section \ref{salientstipul}. Take for example, spatial translation or spatial rotation in Euclidean space $\mathbb{R}^3$; and consider any of Newtonian mechanics, relativistic mechanics, quantum mechanics, or indeed their `cousin' field theories. Each of these is of course a framework for theorising: not a specific theory, with specific particle and-or field contents, and their dynamics (equations of motion). But it turns out that most such specific theories that have been empirically successful,\footnote{`Successful' within limits, of course: for example, relative velocities small compared with $c$ for Newtonian mechanics to be successful, and typical actions large compared with $h$ for Newtonian or relativistic mechanics to be successful.} once set in Euclidean space $\mathbb{R}^3$, do have spatial translations and spatial rotations as dynamical symmetries. And since a dynamical symmetry is to be induced by a single map $a$ on instantaneous states, e.g.~by spatial translation of 1 mile due East applied to every state, the transform $a(s(t))$ of each dynamically possible history $s(t)$ will indeed be a `replica' of $s(t)$, e.g.~spatially translated by 1 mile. Besides, the dynamics is `sensitive' to the values of many quantities, in the sense that a dynamical symmetry must {\it not} alter them: so its map $a$ on instantaneous states is indeed a symmetry {\it \`{a} la} our initial idea. Again, spatial translations and spatial rotations give a standard illustration; as follows. As we said, most of the empirically successful specific theories written in any of the above frameworks, and set in Euclidean space $\mathbb{R}^3$, have these as dynamical symmetries. But they ``don't allow squeezing''! That is, a dynamical symmetry $a$ must preserve all the relative distances, and relative velocities, between the constituents of the system: if $a$ is, or includes, a spatial translations or rotation, it must be a rigid one, of the system as a whole. It must preserve the values of many quantities---the relative ones. 

This discussion of a dynamical symmetry leads in to Section \ref{sptthies}. Also, the cautionary note at the end of Section \ref{salientstipul} above applies again. That is: stipulating that a symmetry be dynamical---indeed, stipulating more specifically: both a symmetry and a dynamics it respects---does not imply that every state has its value preserved for every quantity definable on the state-space. Again, the example of $\mbox{SO}(3)$ in quantum theory suffices. One might stipulate that $\mbox{SO}(3)$ be a symmetry, and that the Hamiltonian be $\mbox{SO}(3)$-invariant (i.e.~in obvious notation: $[H, U_R] = 0$ for all $R \in \mbox{SO}(3)$). This does not imply that one can only use quantities $Q$ that are rotationally invariant (i.e.~$[Q, U_R] = 0$). 

\subsection{Spacetime theories and their symmetries}\label{sptthies}

In philosophical and foundational discussions of spacetime theories, it is usual to define symmetries in an apparently different way from ours.

Besides, it is usual to define such theories, not as a triple $\bra {\cal S}, {\cal Q}, {\cal D} \ket$ as we have done, but as a set of, so to speak, possible universes. That is: as a set of $n$-tuples, consisting of a spacetime manifold $M$, equipped with both chrono-geometric structure (encoded in a metric field, a connection etc.) and matter fields (encoded in tensor and spinor fields obeying equations of motion). Each such $n$-tuple represents a (total, 4-dimensional) solution of the theory: a `possible universe'.\footnote{So these $n$-tuples are usually called `dynamically possible models' of the theory: but we will resist yet another use of the over-worked word `model'!} A symmetry is then usually defined along the following lines. (We will give the example of Galilean transformations in Section \ref{expleGalil}.) It is a bijection of the manifold that:

(a)~~respects its topological and differential structure (technically: is a diffeomorphism); and whose induced maps on tensor fields, connections etc.:

(b)~~fix the chrono-geometric structure (i.e.~maps the metric field, connection etc.~into themselves) and also 

(c)~~send the matter fields into another solution of the equations of motion---another sequence of values over time that is dynamically allowed/possible.

So we need to link our construal of a theory as a triple, and of a symmetry as a map on a state-space, to these ideas. The main link is of course that while a physical theory usually has as its subject-matter some limited kind of system, for which we think of the instantaneous state (values of quantities) changing over time, a spacetime theory takes the universe-throughout-all-time as its subject-matter. So in our construal, a dynamically possible total history of the system is, on Schr\"{o}dinger picture dynamics, as discussed in Section \ref{dynlsymm}: a curve through ${\cal S}$ parameterized by time $t$, with each point $s(t)$ defining the values $\langle Q , s(t) \rangle$ of the various quantities $Q$ at $t$.\footnote{As in Section \ref{dynlsymm} , one can have a corresponding discussion using the Heisenberg picture. But here, we set this aside.} But in a spacetime theory, a dynamically possible total history of the system is just an $n$-tuple. 

One natural way to link to our construal is to make a space-vs.-time split within the spacetime theory's manifold. An alternative way to link to the usual covariant formulations of spacetime theories takes a model of $T$ to have as its state-space the set of all admissible spacetimes that are solutions of the theory's dynamical equations (e.g.~in general relativity, the Einstein field equations). But adopting either of these ways, one faces issues, if a state is a whole spacetime, about how to ``get inside a spacetime'', so as to distinguish a quantity's differing values at different spacetime points. Thus one widely adopted approach is to associate quantities not with points, but with extended regions of different types, i.e.~one gets `quasi-local quantities': see Penrose (1988) and Brown and York (1993); for reviews, see Wang (2015) and Szabados (2009). There are also other issues about which much could be said, such as: rigorously defining quasi-local quantities; fixed fields; the definition of diffeomorphism invariance and its interplay with boundary conditions. (See for example Pooley (2017:~p.~117) and De Haro (2017b, 2023).) That is: we take the spacetime theory to have:

(1)~~a state-space $\cal S$ of the instantaneous states of a notional 3-manifold $\Sigma$, which we take as a fiducial spacelike slice of the spacetime manifold; 

(2)~~a set of quantities $\cal Q$ defined on $\Sigma$ (local densities of matter fields etc.); and 

(3)~~a dynamics $\cal D$ determining the evolution of the instantaneous state of $\Sigma$. \\
Then a dynamically possible total history is a foliation of spacetime, whose leaves are time-evolutes of $\Sigma$, equipped with fields. In other words: it is a curve through $\cal S$ parameterized by a time $t$ labelling the leaves of the foliation. 

Combining (1)-(3) with our discussion of dynamical symmetries in Section \ref{dynlsymm}, we can now see that our initial simple idea of a symmetry, as a map $a$ on the system's state-space $\cal S$ that respects its structure and preserves the values of a salient set of quantities, is, after all, similar to the usual definition of symmetry for spacetime theories, (a)-(c) above. For making the space-vs.-time split, (1)-(3), renders this usual definition of symmetry, (a)-(c), like Section \ref{dynlsymm}'s definition of dynamical symmetry. Indeed, most of the familiar (and empirically successful) theories set in the framework of Newtonian mechanics, or relativistic mechanics or quantum mechanics (or their `cousin' field theories) mentioned in Section \ref{dynlsymm} can be written down as spacetime theories, i.e.~as postulating a spacetime manifold equipped with chrono-geometric structure and matter fields: with the different notions of symmetry being linked by using a space-vs.-time split. 

There are of course several issues here, about which much more could be said. Among them are:

(i)~~the interplay of the structures (topological, differential, metrical etc.) of physical space, spacetime, and state-space; 

(ii)~~the justification for singling out, in the spacetime definition of symmetry, the chrono-geometric structure of the manifold as having to be fixed (requirement (b)), while the matter fields need not be (especially in the context of general relativity); and 

(iii)~~the justification for making the space-vs.-time split, especially in the context of relativity theory (especially general relativity: see Section \ref{sptthies}, just before points (1) to (3)). \\
And of course, such issues are mutually related. For example, (i) and (ii): one might argue that the requirement (b), to fix the chrono-geometric structure, reflects the requirement in our initial idea of symmetry, that $a$ must respect the structure of the state-space $\cal S$, and the fact that $\cal S$'s structure is largely determined by the chrono-geometric structure of spacetime. However, for all of this book, we will not need to explore these issues: sufficient unto the day is the trouble thereof.

\subsection{Galilean transformations}\label{expleGalil}

Newtonian mechanics---in particular, Galilean transformations---provides a simple illustration of several important notions developed in this Chapter: not just symmetries as in this Section, but also Section \ref{modelrootss}'s notions of model root and specific structure, and Section \ref{intext}'s distinction between internal and external interpretations. So it is good way to sum up the material so far.

The idea is as follows. The bare theory $T$ is Newtonian mechanics, of say $N$ gravitating point-particles, set in a {\it Galilean} (also known as: neo-Newtonian) spacetime: i.e.~in a spacetime manifold that is `globally like $\mathbb{R}^4$', with Euclidean geometry in its instantaneous time-slices, and a flat 4-dimensional connection, but without any preferred absolute rest. This bare theory is modelled (in our sense: i.e.~realized, represented) by formulations of Newtonian mechanics of $N$ gravitating particles, set in a {\it Newtonian} spacetime, i.e.~in a spacetime that is `globally like $\mathbb{R}^4$' but that {\it does} have an absolute rest.\footnote{See Geroch (1978:~Section 3) and Malament (2012:~Chapter 4.1).} 

Famously (notoriously!), the difference in such formulations' specifications of absolute rest is not experimentally detectable, since specifications that are each boosted with respect to the other specify the same flat 4-dimensional connection, and a boost maps a solution of the equations of motion to another solution. Or as it is usually put, in more physical terms: no experiment in Newtonian mechanics can distinguish one specification of rest from the others, because the theory is invariant under boosts (`has boosts as a symmetry'). Hence, of course, the debate between Newton and Leibniz, as articulated in the Leibniz-Clarke correspondence, and with its long legacy down to the present day.\footnote{See e.g.~Earman (1989:~Chapter 1) and Huggett (1999:~pp.~143-168). For the original correspondence, see Ariew (2000:~p.~19).}

 So in this example, it is natural to say that the specific structure of each model includes its specification of absolute rest. Using this specification, the model defines a flat 4-dimensional connection---viz.~the same connection as is defined by the other models---and thereby builds a homomorphic copy of $T$. 
 
We can make this example simpler and precise, and yet still a worthwhile illustration, by taking the bare theory $T$ to be just the abstract Galilean group Gal(3). This is a 10-dimensional Lie group, whose generators are three spatial rotations, four (space and time) translations, and three boosts. That is: its generators are usually thus described, by way of justifying their commutation relations. But of course the abstract group can be defined by the commutation relations, free of a physical interpretation. 
 
Gal(3) is usually presented in its fundamental representation. Namely, as a concrete group of transformations on (bijections of) $\mathbb{R}^4$, written in terms of the standard coordinates $({\bf x},t) \in \mathbb{R}^4$, with $g \in$ Gal(3) represented as the function
\bea\label{galilean}
g({\bf x},t):=(R\cdot {\bf x}+{\bf v}_0\,t+{\bf r}_0,t+t_0)\,,
\eea
where $R$ is a $3\times3$ spatial rotation matrix, ${\bf v}_0$ is the velocity boost, ${\bf r}_0$ is the spatial translation vector, and $t_0$ is the time translation. This fundamental representation can also be expressed in a coordinate-free way as an action on the affine space of $\mathbb{R}^4$, i.e.~on Euclidean 4-space. But we will not need the details of affine spaces (cf.~e.g.~Auslander and MacKenzie (1963:~Chapter 1)).

Now to this we can adjoin a choice of an inertial coordinate system as specific structure, and we can take this to give the model's specification of absolute rest.\footnote{We say `give'---meaning `determine'---rather than `be', simply because a coordinate system includes choices of spatial and temporal origins and units, and of an orientation of spatial axes, as well as the absolute rest, i.e.~the timelike congruence of inertial worldlines.\label{coordcong}} The standard coordinate system defined by the components of $\mathbb{R}^4$ itself is then just one choice among many, determining one specification of absolute rest among many. Natural though we find it for writing down the fundamental action of Gal(3), as we did in Eq~\eq{galilean}, the action can of course be written down in any inertial coordinate system. And any such system can be taken to give a model's specification of absolute rest. 

Thus each Galilean boost maps an adjoined choice of absolute rest, represented mathematically by an inertial coordinate system, into another such choice, i.e.~another inertial coordinate system.\footnote{Of course, Galilean transformations that are not boosts keep fixed the choice of absolute rest: cf.~footnote \ref{coordcong}.}

So much by way of how Galilean transformations illustrate Section \ref{modelrootss}'s notions of model root and specific structure. We end with two comments, (A) and (B), connecting this example to (i) spacetime symmetries and (ii) internal vs.~external interpretations (Sections \ref{sptthies} \ref{intext} respectively). For simplicity and brevity, we will restrict both comments to the simple ``vacuum'' scenario which we have concentrated on: i.e.~$\mathbb{R}^4$ as a description of either neo-Newtonian or Newtonian spacetime, without regard to the $N$ gravitating point-particles we mentioned at the start of this Section. Thus recall that we concentrated on taking the bare theory $T$ to be just the abstract Galilean group Gal(3), and considered its action on $\mathbb{R}^4$. But this is only for simplicity: these two comments carry over to the non-vacuum scenario, where there are particles.\footnote{It is just that it would take too long to spell out the non-vacuum scenario. To glimpse why, we briefly note some of the issues one confronts. Obviously, one must consider the particles' state-space: which one would build from their configuration space (the ``$q$s''), by adding either velocities (``$\dot q$s'': defining velocity phase space, in the Lagrangian framework) or canonical momenta (``$p$s'': defining phase space, in the Hamiltonian framework). At first sight, the $N$ particles' configuration space is ``just'' $\mathbb{R}^{3N}$. But there are subtleties to be dealt with. Indeed: not only the topics mentioned above, of passing to the affine space so as to ``rub out'' the preferred origin, and whether to have an absolute rest; but also whether to excise collision points, i.e.~whether to forbid point-particles to be in the very same place. Assuming these subtleties are dealt with, and the Lagrangian or Hamiltonian state-space is constructed, one would then consider the action of the Euclidean group on this state-space, lifted from its action on $\mathbb{R}^3$. Again, there are subtleties about this lifted action; and to treat boosts and so represent the Galilean group, one needs to ``add a time axis'', defining what is often called `extended (velocity) phase space'. For a philosophical introduction to all these subtleties, cf.~e.g.~Butterfield (2007: Section 2.3). \label{subtleties}} \\

(A): {\it Agreement with usual verdicts about symmetries}:--- The first comment looks back to Section \ref{sptthies}'s discussion of symmetries in a spacetime theory, i.e.~a theory that postulates a spacetime with certain chrono-geometric structures like metrics and connection. (Such theories of course also postulate matter and radiation, particles and fields, in the spacetime; but as just announced, we are setting that aside.) 

We saw there that in a spacetime theory, a symmetry is usually defined as a map on the spacetime that (once its domain of definition is extended in the natural way to include chrono-geometric structures and matter fields): (i) fixes, i.e.~does not alter, the chrono-geometric structures, and (ii) maps a matter-field solution of the equations of motion to another solution. (We also saw how this relates to our more basic and general notion of symmetry.) Accordingly, boosts are a symmetry of neo-Newtonian spacetime: for a boost preserves the chrono-geometric structures, i.e.~the spatial and temporal metrics and the flat affine connection, and maps solutions to solutions. But boosts are not a symmetry of Newtonian spacetime (i.e.~a spacetime that is globally like $\mathbb{R}^4$, but that has a specification of absolute rest). For a boost does not fix a specification of rest.\footnote{On the other hand: spatial rotations and spatiotemporal translations are symmetries of both spacetimes.} These points are, in effect, the modern mathematical expression of the famous (notorious!) point with which this Section began: that no mechanical experiment can discern which is the putatively correct standard of rest. 

Our discussion above, and our notions of model root etc., accords with this. Mapping one model's specification of absolute rest into another's is not the same as fixing the referent of the given specification, i.e.~not the same as a symmetry in spacetime theories' usual sense. Thus this example illustrates how a bare theory can have a symmetry, viz.~boosts, that (some or even all) its models lack.
 
(B): {\it Internal and external interpretations}:--- The example also illustrates Section \ref{intext}'s distinction between internal and external interpretations. For it allows us to formulate the disagreement between Newton (Clarke) and Leibniz---the question whether absolute rest is physically real---in terms of the internal vs.~external contrast. 

For recall that an internal interpretation interprets only the model root, but not the specific structure. More precisely, we define this as meaning that specific structure which is ``in'' a model root as a building block, does not get interpreted. Therefore models that are isomorphic, i.e.~have isomorphic roots, and so differ only in their specific structure, must receive the same interpretation. Thus in our example: an internal interpretation of a model simply does not interpret the specification of rest (as given by an inertial coordinate system considered as rest). In short: the specification of rest is not part of what is physical. Thus an internal interpretation articulates Leibniz' relationist views.

On the other hand, an external interpretation (by definition) does interpret (some or all of) the specific structure. Thus an external interpretation can take any isomorphic model roots with different specifications of rest to have distinct interpretations. This kind of external interpretation thus articulates the Newton-Clarke view: in short, that giving all material bodies the same boost makes a physical difference.

\section{Conclusion}

This Chapter has introduced philosophers' jargon for `theory' and `model', and motivated our own choice. While our ``intermediate'' choice of jargon is motivated by dualities, it is independent of it: as we discussed, our usage aims to remain faithful to the relation between a theory and its models: namely, to models as formulations or realizations of a theory---almost always representations i.e.~homomorphic copies of it---thus distinguishing two levels: the theory ``above'' and the models ``below''. Thus, we argued, since in coming Chapters dualities will prompt us to shift our talk of theories ``one level up'', we also shift our talk of models ``one level up''. Note that all of this is regardless of interpretation, i.e.~it regards what we have called `bare theories' and `bare models'.

We distinguished syntactic and semantic conceptions of theories, and adopted a semantic conception, i.e.~of theories (and models) as set-theoretic structures. More specifically, they are structured triples of states, quantities, and dynamics. In a model, we distinguished the model root from the specific structure, i.e.~the image of the representation map of a bare theory, from the ingredients out of which the representation is built. Chapter \ref{Theor} will discuss how this conception of theories and models, and in particular the specific structure, relates to the syntactic conception. 

Also our views on interpretation are mainstream: namely, referential semantics allows us to formulate interpretations as structure-preserving partial maps. In turn, this enables the distinction between internal and external interpretations, which will allow, in the next Chapter, the formulation of our conception of physical equivalence. 

In the last Section, we discussed the idea of symmetry, as an automorphism that preseves a salient subset of quantities, and we distinguished stipulated symmetries, dynamical symmetries, and spacetime symmetries. Chapter \ref{Schema} will discuss the analogy between this conception of symmetry and our conception of duality as a ``giant symmetry'', and other symmetries that it is natural to consider given our distinction between theories and models. 

This Chapter has emphasized the logico-semantic aspects of theories and models, rather than the epistemic ones, which for the most part we postpone until the second part of Part III, namely Chapters \ref{Realism} to \ref{Understand}. 

However, Section \ref{intext} already indicated one way in which theories and models depend on epistemic issues: through the structures, especially symmetries, that interpretation maps ought to preserve. Other epistemic issues will surface when we introduce dualities in Chapter \ref{Schema}. 

\chapter{Duality as Isomorphism}\label{Schema}
\markboth{\small{\textup{Duality as Isomorphism}}}{\textup{\small{Duality as Isomorphism}}}

\begin{quote}
{\it A science can never determine its subject-matter except up to an isomorphic representation} (Hermann Weyl, 1934).
\end{quote}

This Chapter presents and discusses the core proposal of our account, our {\it Schema}, for duality. Namely, that duality is an isomorphism of two theories---which we call `models'---with respect to a structure of a common core, a bare theory, that they both instantiate (usually by being representations of it in the mathematical sense). The definition of this isomorphism is given in Section \ref{isomdef}. There follow discussions of how it is related to (i) the interpretation of theories, and (ii) symmetry (Sections \ref{dualint} and \ref{dualsym} respectively). 

To avoid these three Sections being too abstract, we will use four ``running'' examples. Namely: (i) position-momentum duality in elementary quantum mechanics (already introduced in Section \ref{Ourthm}); (ii) elementary classical electric-magnetic duality; (iii) Kramers-Wannier duality between the low- and high-temperature Ising models on a square lattice; and (iv) gauge-gravity duality. As we will see, each of them illustrates the Schema. (They will get a much fuller treatment later, in Chapter \ref{Simple}.)

Then in Section \ref{featurerole}, we discuss some scientifically important features that dualities often have: such as making a hard problem easy, or at least easier. We also give a brief prospective discussion of the various roles of dualities. Among roles, the main contrast is between what we call the `theoretical' and `practical' functions of dualities: this contrast looks ahead to Chapters \ref{Theor} and \ref{Heuri} respectively. 

Then, in Section \ref{comparison}, we compare our account of dualities with the views of other authors. Our terminology, and indeed our views, are sometimes in common with other authors. For example, the word `common core' is used by Rickles and others.\\ 
\\
So to summarize: the main features of our Schema will be that:

(1): we distinguish uninterpreted theories, which we call {\it bare theories}, from interpreted theories;

(2): we emphasize that, wholly independently of issues of interpretation, a bare theory can have many realizations, which we call {\it models};

(3): we take duality to be an isomorphism between two models of a single bare theory.\\
Of these three features, it is (2) and (3) that are the distinctive ones. For several authors also define duality in terms of uninterpreted theories. This has the advantage of making verdicts of duality not beholden to semantic issues, and so less vague or even controversial. And it allows cases of duality without any sort of physical or semantic equivalence---which certainly occur, e.g.~Kramers-Wannier duality between the high and low temperature regimes of the statistical mechanics of a lattice. But features (2) and (3) make duality an equivalence (formally: an isomorphism) between items that are not only uninterpreted, but also {\it more specific} than an uninterpreted theory: viz.~realizations---which we will call `models'---of a (single) bare theory. A prototypical example is: taking a bare theory to be an abstract algebra of quantities (maybe also equipped with a dynamics, viz.~as a 1-parameter group of automorphisms, and a set of abstract states, i.e.~rules for evaluating (i.e.~assinging values to) the quantities): a model or realization is given by a representation (in the mathematical sense) of the algebra, together with a realization of the rules for evaluating the quantities, for the representation in question, i.e.~a set of maps to the relevant field, of complex or real numbers. Indeed, we saw in Section \ref{intext}, how algebraic quantum field theory gives such examples.

\begin{figure}
\begin{center}
\includegraphics[height=3cm]{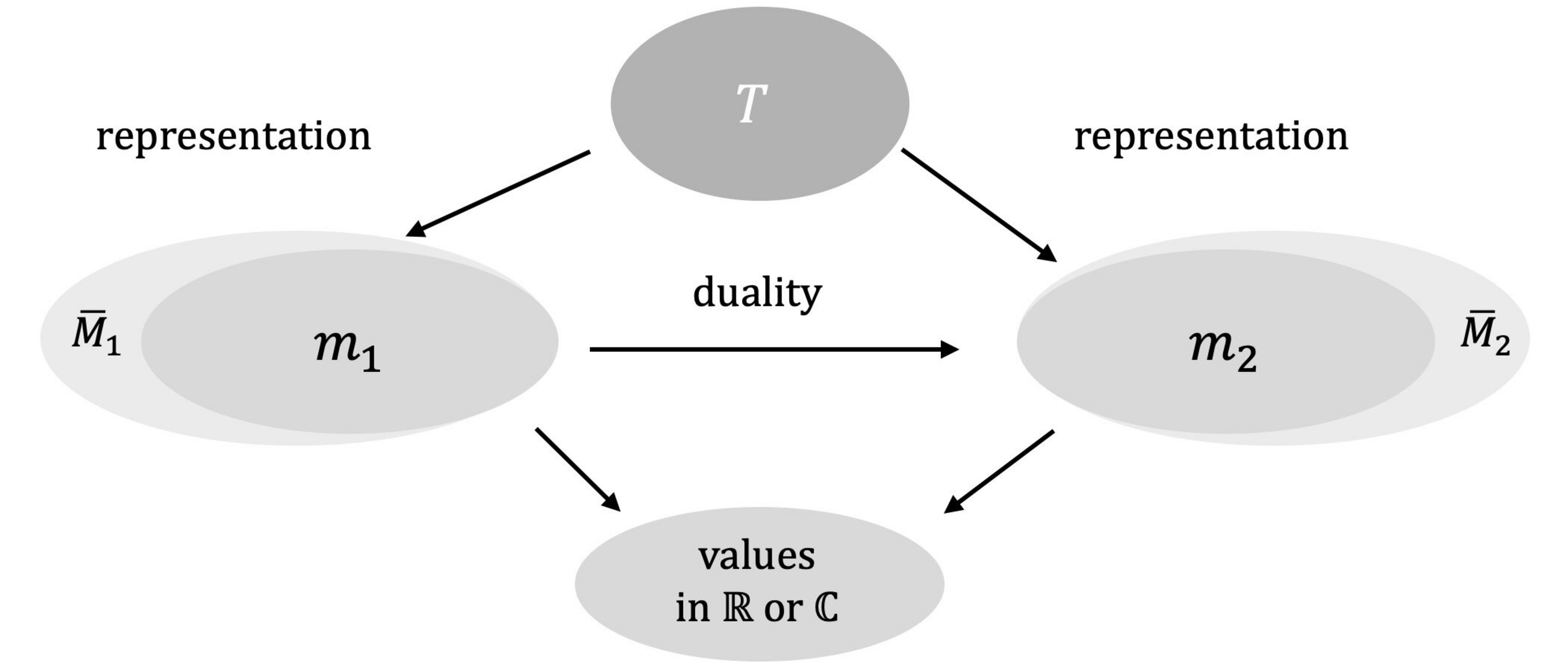}
\caption{\small Relation between the bare theory ``above'' and its two duals, as representations, ``below''. The duality maps one model into the other. The isomorphism preserves the values of quantities.}
\label{Trep}
\end{center}
\end{figure}

\section{The isomorphism defined}\label{isomdef}

The definition of the isomorphism is as follows;\footnote{See De Haro and Butterfield (2018:~p.~338), De Haro (2020a:~p.~264), and Butterfield (2021:~p.~50).} 
the jargon and notation is as introduced in Section \ref{thsmodels}. Namely: we speak of quantities taking values on states. We write the value of a quantity $Q \in {\cal Q}$ on a state $s \in {\cal S}$, as $\langle Q , s \rangle$. We take a bare theory schematically as a triple of the state space, the set of quantities and the dynamics, $ T=\langle {\cal S}, {\cal Q}, {\cal D} \rangle$. We can think of the dynamics ${\cal D}$ either in the Schr\"{o}dinger picture as a map on states, or in the Heisenberg picture as a map on quantities. And we think of the bare theory as being ``at a higher level'', with the models ``below'' being representations of it, with model roots $m_1$ and $m_2$ (see also Figure \ref{tm1m2}).

Thus we define:--- A {\bf duality} between $m_1 = \bra {\cal S}_{M_1}, {\cal Q}_{M_1}, {\cal D}_{M_1}\ket$ and $m_2 = \bra {\cal S}_{M_2}, {\cal Q}_{M_2}, {\cal D}_{M_2}\ket$ requires:

an {\it isomorphism between the state-spaces} (almost always: Hilbert spaces, or for classical theories, manifolds): 
\be
d_{\cal S}: {{\cal S}_{M_1}} \rightarrow {{\cal S}_{M_2}} \;\; {\mbox{using $d$ for `duality'}} \; ;
\ee 

and an {\it isomorphism between the sets} (almost always: algebras) of quantities\\
\be
d_{\cal Q}: {{\cal Q}_{M_1}} \rightarrow {{\cal Q}_{M_2}} \;\; {\mbox{using $d$ for `duality'}} \; ;
\ee 

such that: (i) the {\it values of quantities match}: 
\be\label{obv1}
\forall s_1 \in {{\cal S}_{M_1}}\,\forall Q_1 \in {{\cal Q}_{M_1}}\,,~~~\langle Q_1, s_1 \rangle_1 = \langle d_{\cal Q}(Q_1), d_{\cal S}(s_1) \rangle_2\,;
\ee
and: (ii) $d_{\cal S}$ is {\it equivariant for the two triples' dynamics}, $D_{S:1}, D_{S:2}$, in the Schr\"{o}dinger picture; and $d_{\cal Q}$ is equivariant for the two triples' dynamics, $D_{H:1}, D_{H:2}$, in the Heisenberg picture. (And likewise, in quantum theories, for more general matrix elements $\bra Q;s_1,s_2\ket:=\bra s_1|\hat Q|s_2\ket$.) 

Eq.~\eq{obv1} appears to favour $m_1$ over $m_2$; but in fact does not, thanks to the maps $d$ being bijections.

The relation between a theory and its duals, and the preservation of values, are shown in Figure \ref{Trep}. The equivariance condition is expressed in terms of the commuting diagrams of Figure \ref{obv2}.

\begin{figure}
\begin{center}
\bea
\begin{array}{ccc}{\cal S}_{M_1}&\xrightarrow{\makebox[.6cm]{$\sm{$d_{\cal S}$}$}}&{\cal S}_{M_2}\\
~~\Big\downarrow {\sm{$D_{S:1}$}}&&~~\Big\downarrow {\sm{$D_{S:2}$}}\\
{\cal S}_{M_1}&\xrightarrow{\makebox[.6cm]{\sm{$d_{\cal S}$}}}&{\cal S}_{M_2}
\end{array}~~~~~~~~~~~~
\begin{array}{ccc}{\cal Q}_{M_1}&\xrightarrow{\makebox[.6cm]{$\sm{$d_{\cal Q}$}$}}&{\cal Q}_{M_2}\\
~~\Big\downarrow {\sm{$D_{H:1}$}}&&~~\Big\downarrow {\sm{$D_{H:2}$}}\\
{\cal Q}_{M_1}&\xrightarrow{\makebox[.6cm]{\sm{$d_{\cal Q}$}}}&{\cal Q}_{M_2}
\end{array}\nonumber
\eea
\caption{\small Equivariance of duality and dynamics, for states and quantities.}
\label{obv2}
\end{center}
\end{figure}

Here as an illustration is a sketch of how this definition of duality applies to two of the four running examples which we announced in this Chapter's preamble.\\

{\it Electric-magnetic duality}: this relates two models by mapping the electric charges of one model to the magnetic charges of the other. Furthermore, it does so by mapping a small electric charge to a large magnetic charge. Nevertheless, the common structure is the same in the two models, i.e.~the quantum theory is invariant under the replacement of one gauge group by its dual. 

{\it Gauge-gravity duality}: in this case, the models differ in the dimensions they assign to spacetime ($D$ in the gravity model, $D-1$ in the gauge model), in their field content and classical equations of motion, and in much more. The duality maps the states and quantities of a theory of gravitation in $D$ dimensions (i.e.~the states that encode the geometry and the matter fields, and the quantities constructed from the fields) to the states and quantities of a conformal field theory. The correspondence (in its semi-classical approximation) relates bulk fields to operators of the boundary theory, where the mass of the bulk field is matched by the conformal dimension of the boundary operator. Under this relation, the partition function of the bulk theory (and other quantities derived from it) matches the correlation functions of the conformal field theory for the corresponding operators. In this example, the common core consists only in a class of asymptotic operators and a conformal class of $(D-1)$-dimensional metrics. Of course, it is very surprising to learn that a gauge theory model in $D-1$ dimensions, and a model of quantum gravity in $D$ dimensions, despite their very disparate guises, nevertheless have the same common core, and represent the same theory.\footnote{See De Haro~(2020a) for a discussion in the context of this Schema.}\\
\\
{\bf The reason for requiring that the common core is a triple.} By definition, any isomorphism preserves stucture: but one must specify the structure it preserves. Our requirement, that models share a common core theory that is a triple, is the correct isomorphism criterion {\it for models that are formulated as triples}. 

Furthermore, since the common core, like the models, is a triple of states, quantities, and dynamics, this secures that models share a sufficient amount of structure: which both avoids that dualities can be trivial (i.e.~that any old partial isomorphism between models could be called a `duality') and secures that the common core can itself be physically significant, i.e.~that it can be a {\it theory}. 

As we discussed earlier (see our second clarification at the end of `Theory' in Section \ref{Ourthm}), the formulation of a model as a triple, as well as the distinction between the model root and the specific structure, is constrained by initial interpretative choices (e.g.~about the states that are included in the state-space, and the quantities that are deemed physical), and so it is not purely formal.\footnote{For a discussion, see De Haro (2020a:~p.~260). North (2021:~p.~175) makes the same point: `Some interpretive assumptions will go into choosing a formalism to begin with'.}
(Indeed, this distinguishes a theory in mathematical physics from a theory in pure mathematics.) As we have also stressed, it is useful to make this conceptual distinction between the bare theory and the interpretation, i.e.~the referential and intensional semantics discussed in Section \ref{itm}, which does not need to reflect how a theory or model was arrived at historically. 

But this means that our definition of duality as an isomorphism that respects a common core that is a bare theory, i.e.~a triple of states, quantities, and dynamics, is {\it strong}. For such a triple is regarded as capable of physical interpretation. This condition is required for the structure that is common to duals to be physically significant (so that dualities are scientifically interesting!). (Chapter \ref{physeq} will discuss how a common core thus defined can inherit an interpretation from its models, so that it can indeed itself be a theory: and it will also discuss the common core's desired logical strength. Also, we anticipate occasionally breaking our own rule for purposes of either pedagogy or rigour, i.e.~considering a very simple structure, like a group, and calling that a `theory', even if it does not qualify as a physical theory. But for the most part, our examples {\it will} be physical theories.)\\
\\
{\bf Dualities for other formulations of model roots.} As we discussed in Section \ref{modelrootss}, model roots do not always come presented as model triples: rather than using the language of states, quantities, and dynamics, models are sometimes presented: 

(i)~~in statistical mechanics: in terms of a Hamiltonian and a partition function that is a sum of Boltzmann weights (and also other quantities that can be derived from the partition function); 

(ii)~~in quantum field theory: in terms of a Lagrangian and a path integral (and also other quantities that can be derived from the path integral). 

While there are relations between e.g.~the path integral and Hilbert space formalisms,\footnote{See De Haro et al.~(2017:~p.~75).}
different formalisms have of course different rationales and merits; and there is no advantage in attempting to formulate every example in terms of triples. Thus we allow the following generalization of our notion of duality, which does not require the models to be presented as triples: duality is an {\it isomorphism of model roots} (as introduced in Section \ref{modelrootss}) such that all the quantities match. Thus, for a model formulated using a partition function, we require that the partition function and all the correlation functions of the variables summed over in the partition function have the same values in both duals, up to a constant factor (i.e.~a factor independent of the variables in the partition function, as in Eq.~\eq{Zmod} below). These correlation functions can be obtained by taking functional derivatives of the partition function coupled to a source that is set to zero after taking the derivative (this gives the {\it moments} of the variables that are in the partition function). We will use this formulation for Kramers-Wannier duality, in Section \ref{dualpf0} (see, in particular, Eq.~\eq{Isingb}).

In quantum field theories formulated using a path integral, the definition is analogous: we require the equality of the partition function, and of all its derivatives, up to constants (we will use this formulation in Section \ref{quantumD}). Alternatively, we will work with the models' Lagrangians, and require that they agree (we will do this in Sections \ref{pvdah} and \ref{cdbm}).\\
\\
{\bf Generality and strength of the proposal:} we should briefly discuss whether our proposal has the desired generality and strength. Specifically, there are two questions, or possible worries, that we will address in turn: 

(i)~~Is our characterisation of a duality, as an isomorphism of state-spaces and (algebras of) quantities, equivariant for the dynamics, {\it too general or abstract, and thus liable to being devoid of content}?\footnote{We thank James Weatherall for a discussion of point (i) below.} 

(ii)~~More specifically: should an account of duality make a clear distinction between dualities and other relations, especially {\it symmetries}, that in some cases may also be formulated as isomorphism? And, if so, {\it does our account sufficiently make this distinction?}

{\it About (i)}: Although we of course agree that the notion of isomorphism is empty without mention of what structure is to be preserved: this is a trivial point, and certainly we do not believe that we have commmitted this error. For we have specified the kinds of state-spaces and sets of quantities that dualities are supposed to map (for quantum theories: Hilbert spaces, with their operator algebras), and the types of structures that they are supposed to preserve (for quantum theories, correlation functions, as well as symmetries: see Section \ref{dualsym}). In Parts II and III, we will see that our conceptions of theory and of duality fruitfully cover the examples.

We should also distinguish our project---of specifying notions of theory and of duality of the right logical strength---from a riskier and more controversial project: of characterising, in full generality, what kind of {\it mathematical object} a scientific theory (any scientific theory) is. Two comments about this: First, dualities {\it do} contribute to our understanding of the kinds of mathematical objects that scientific theories can be. For Chapter \ref{Heuri} will propose that many examples of physical theories are best understood as geometric objects: specifically, differentiable manifolds and, more generally, algebraic varieties. But second, we are in principle suspicious of the essentialist claim that {\it any} scientific theory should be captured by some single kind of precisely defined mathematical object. For, while we are sympathetic to this attempt, we think that the jury is still out. For example, algebraic varieties may not capture everything that scientists are justified in calling a `theory'. It may be that we require {\it more than one kind} of mathematical object to cover all cases. In any case, our conception is sufficiently clear and precise to cast light on theories in mathematical physics: and we welcome further work towards making the conceptions of theories and models, and the isomorphisms between them, more mathematically rigorous.

{\it About (ii)}: Although we do think that there is a distinction between dualities and symmetries,\footnote{For example, not all symmetries preserve all the kinematically possible states, nor the whole set of quantities, but only a salient subset of them needs to be preserved: see e.g.~Caulton (2015:~p.~159).} 
we deny that this distinction should in all cases be committed to entirely different structures for symmetries and for dualities. Rather, we argue that the evidence from the examples in physics points to a distinction of levels, i.e.~a hierarchy: or, better, a network with different levels. For example, we deny that `a duality is a gauge symmetry', because choosing duals is not like choosing gauges in a gauge theory. But this denial does not contradict the fact that a more general conception of what we call theories and models, and the isomorphisms between them, can subsume both dualities and symmetries (with at least some types of symmetries often seen as self-dualities, i.e.~{\it automorphisms} of a single model, rather than isomorphisms between distinct models: cf.~Section \ref{std}).

This process, of collecting models into theories, and then finding dualities between theories (thus rebranding theories as models, i.e.~representations of a yet more general theory), can be repeated more than once, thus generating several levels of structure. Indeed, this seems to be one of the main overall lessons of dualities: that what we thought were unrelated theories, are in fact models of yet more general structures. There is no problem of recursivity here, but only a question of hierarchy, i.e.~of how many levels we require for a given theory, and what kinds of structures are required to incorporate both the higher and the lower levels.

Note that this ``unified'' view of dualities and symmetries is pervasive in the physics literature, where `duality symmetry' is a common phrase, rather than an oxymoron. In the physics literature, dualities for quantum theories are usually seen as unitary transformations on the Hilbert space and its operator algebra. But a unitary transformation of this kind is a symmetry: which vindicates our statement that the isomorphisms, by which we define a duality, can be used to describe both dualities and symmetries, at different levels in the structure.

Having said that, we {\it will} distinguish, in Section \ref{featurerole}, different themes and special types of dualities. These themes and types can be used to identify particularly salient types of dualities: for example, {\it quantum dualities} form a salient type, where the quantum models are duals, but their corresponding classical limits are {\it not} duals. (We will discuss this type in Sections \ref{theoreq} and \ref{lsrr}.)

Perhaps surprisingly, our {\it syntactic} reformulation of a duality in Chapters \ref{Theor} and \ref{physeq} (as against the semantic formulation on which we have focussed so far) will allow us to be more precise about what dualities, especially quantum dualities, are. Namely, we will argue that the specific structure of duals adds both vocabulary (i.e.~new items in the signature, i.e.~the non-logical vocabulary, of the theory) and new statements that use that vocabulary, without contradicting any of the theory's statements. For quantum dualities, the new statements are in effect bridge laws connecting the quantum theory with one of its classical limits. This vindicates another general statement that we defended earlier: that the formulation of scientific theories involves both syntactic and semantic aspects. 

\section{Duality and theoretical equivalence}\label{dualint}

This Section does two main jobs. The first is straightforward; the second philosophical and so inevitably controversial. First, we discuss how the duality maps $d_{\cal S}, d_{\cal Q}$ in the Schema mesh with interpreting the dual models: in particular, with the interpretation maps, for intensions and extensions, given by intensional semantics (cf.~Sections \ref{refsr} and \ref{ints}). 

Second: we address the question under what general conditions two dual models, once interpreted, make the very same claims about the world. We shall call this `making the very same claims' {\it physical equivalence}.\footnote{`Physical equivalence' is the phrase that most philosophers of dualities (and many physicists) have used to refer to `making the very same claims about the world'. In the philosophy of science literature, the usual phrase is `theoretical equivalence'. Although both literatures are ultimately concerned with a common set of questions, their preoccupations have been different. Thus throughout most of this Chapter, we will temporarily follow the usage of philosophers of physics, and then revert to talk about `theoretical equivalence' in Part III, where we will explicitly relate the discussion of dualities to the relevant discussions in logic and in philosophy of science.
\label{usePE}} 
It is clear that, because our Schema for duality is formal---i.e.~demands nothing interpretative of two dual models---models can be dual yet not physically equivalent. (We also see this in some of our four ``running'' elementary examples.) So we ask: what further conditions are necessary, or even perhaps sufficient, for physical equivalence? There will be a full discussion of this in the first half of Chapter \ref{physeq}. But this Section gives a preliminary discussion that will be enough for us to reach reasonable verdicts about physical equivalence for the various examples in Chapters \ref{Simple} to \ref{HABHM}. 

So far, the discussion of interpretation has concerned a {\it single} theory or model. Since duality is about relations between models, there is, at first sight, little to say about duality and interpretation. That is: interpretation should simply proceed independently on the two sides of the duality---for example, we just require the interpretation-symmetry commuting diagram on both sides of the duality. Indeed: we said already in the preamble of this Chapter that in some cases of duality, the two sides were clearly not---nor intended to be---physically or semantically equivalent: e.g.~the high and low temperature regimes in Kramers-Wannier duality. And our definition of duality as formal (viz.~an isomorphism of model triples) certainly allows this idea of `distinct but isomorphic sectors of reality'---namely as the codomains of the interpretation maps on the two sides of the duality, as in Figure \ref{phineq}.

\begin{figure}
\begin{center}
\bea
\begin{array}{ccc}{\cal S}_{M_1}&\xrightarrow{\makebox[.6cm]{$\sm{$d_{\cal S}$}$}}&{\cal S}_{M_2}\\
~~\Big\downarrow {\sm{$i_{{\cal S}_1}$}}&&~~\Big\downarrow {\sm{$i_{{\cal S}_2}$}}\\
D_{{\cal S}_{M_1}}&\not=&D_{{\cal S}_{M_2}}
\end{array}~~~~~~~~~~~~
\begin{array}{ccc}{\cal Q}_{M_1}&\xrightarrow{\makebox[.6cm]{$\sm{$d_{\cal Q}$}$}}&{\cal Q}_{M_2}\\
~~\Big\downarrow {\sm{$i_{{\cal Q}_1}$}}&&~~\Big\downarrow {\sm{$i_{{\cal Q}_2}$}}\\
D_{{\cal Q}_{M_1}}&\not=&D_{{\cal Q}_{M_2}}
\end{array}\nonumber
\eea
\caption{\small Two dual models are mapped to `different sectors of reality', for states and quantities. The state spaces of the two models, ${\cal S}_{M_1}$ and ${\cal S}_{M_2}$, are mapped by the interpretation maps, $i_{{\cal S}_1}$ and $i_{{\cal S}_2}$, to different domains of application, i.e.~$D_{{\cal S}_{M_1}}$ and $D_{{\cal S}_{M_2}}$: and likewise for quantities.}
\label{phineq}
\end{center}
\end{figure}
 
This verdict---`there is little to say'---is true, so far as it goes. And of course, it does not forbid the other sort of case: where the two sides of the duality {\it are} physically equivalent, i.e.~do describe `the same sector of reality'. In the Schema, this would be modelled by the interpretation maps on the two sides having the same images in their codomain---so as to give a {\it triangular}, rather than {\it square}, commuting diagram. For states and for quantities being mapped by interpretation maps, $i_1$ and $i_2$, the two sides of a duality describing `the same sector of reality' amounts to the diagram in Figure \ref{interp1} (as against the square diagram in Figure \ref{phineq}).\footnote{We have not introduced notation for what one might call the `realm of intension', or `meanings', and for what one might call the `realm of extension', or the `world', i.e.~for the Fregean distinction between `sense' and `reference'. And to avoid a lot of extra notation, which would be cumbersone, and without compensating advantages, we will not introduce this. In short: these diagrams can be drawn for both extensions and intensions (see De Haro and Butterfield, 2018:~pp.~333-335).}

\begin{figure}
\begin{center}
\bea
\begin{array}{ccc}{\cal S}_{M_1}\!\!&\xrightarrow{\makebox[.6cm]{$\sm{$d_{\cal S}$}$}}&\!{\cal S}_{M_2}\\
~~{\mbox{$i_{{\cal S}_1}$}}\searrow&&\swarrow \mbox{$i_{{\cal S}_2}$}~~\\
&D_{\cal S}&\end{array}~~~~~~~~~
\begin{array}{ccc}{\cal Q}_{M_1}\!\!&\xrightarrow{\makebox[.6cm]{$\sm{$d_{\cal Q}$}$}}&\!{\cal Q}_{M_2}\\
~~{\mbox{$i_{{\cal Q}_1}$}}\searrow&&\swarrow \mbox{$i_{{\cal Q}_2}$}~~\\
&D_{\cal Q}&\end{array}\nonumber
\eea
\caption{\small Two dual models describe `the same sector of reality', for states and quantities. The state spaces of the two models, ${\cal S}_{M_1}$ and ${\cal S}_{M_2}$, are mapped by the interpretation maps, $i_{{\cal S}_1}$ and $i_{{\cal S}_2}$, to the same domain of application, i.e.~$D_{\cal S}$: and likewise for quantities.}
\label{interp1}
\end{center}
\end{figure}

So far, so straightforward. Our argument so far says that we can have cases of dualities between models that can be either physically equivalent or inequivalent, depending on whether they map to different or to the same domains of application, and that the Schema models this using interpretation maps that either commute or do not commute with the duality map, thus leading to a triangle or a square diagram. This straightforward argument prompts the question of what determines whether the two sides of a duality are physically equivalent, i.e.~describe the same `sector of reality'. 

\subsection{What determines theoretical equivalence?}\label{wde}

As we mentioned in the preamble, this Section's aim is to give a preliminary discussion of the conditions that determine theoretical i.e.~physical equivalence, which will be enough for us to reach reasonable verdicts for the examples in Chapters \ref{Simple} to \ref{HABHM}. We will first sketch examples of physical equivalence, and physical inequivalence, and then use our notions of internal and external interpretations to give some natural sufficient conditions for two dual models to be physically equivalent.\footnote{Philosophers and physicists tend to have different intuitions about physical equivalence: while physicists tend to think that thorough-going dualities {\it do} relate physically equivalent models (e.g.~for dualities that map all the operators on the Hilbert space that are deemed physical), philosophers often want an account of how the ontologies of dual models can be the same, before they accept a verdict of theoretical equivalence.} 

We have seen from the elementary examples that both options are certainly possible. Position-momentum duality is an example of a duality between physically equivalent models: see our discussion of a `counterfactual history' in Section \ref{Ourthm}.

As an example of physically inequivalent models, we take the Ising model on a two-dimensional lattice at {\it high temperature}, which is dual to the Ising model at {\it low temperature}. Its Hamiltonian is:
\bea\label{Hisotropic}
H_{\sm{2D Ising}}=-J\sum_{(ij)}s_i\,s_j\,,
\eea
with spin value $s_i=\pm1$ on each lattice site $i$. The partition function $Z$, i.e.~the sum over the Boltzmann weights of the whole lattice, satisfies the following property, called {\it Kramers-Wannier duality} (details in Section \ref{dualpf0}):
\bea\label{Zmod}
\frac{Z_{\sm{2D Ising}}(\b)}{\sinh^{N/2}2\b}=\frac{Z_{\sm{2D Ising}}(\ti\b)}{\sinh^{N/2}2\ti\b}\,,
\eea
where the {\it inverse temperatures} are defined by: $\b:=J/kT$, and $\ti\b:=\ti J/k\ti T$ in the dual model. The temperatures are related by:
\bea\label{Itemp}
\sinh2\b\,\sinh2\ti\b=1\,,
\eea
so that, assuming that the couplings are fixed, the high-temperature regime of one model (i.e.~$T$ large) is indeed mapped to the low-temperature regime of the dual (i.e.~$\ti T$ small). 

In this case, there is no temptation to say that the models are physically equivalent, because their temperatures differ: namely, measuring the temperature using a thermometer (or coupling the Ising model to some other physical system) will break the duality and establish the physical inequivalence.

More precisely, the value of the temperature of each model, i.e.~$T$ and $\ti T$, is an obvious candidate for non-matching {\it specific structure}, whose interpretation renders dual models physically {\it inequivalent}: high on one side, and low on the other. Given dual lattices realized in a lab, one with high temperature and the other with low temperature, an example of `other facts, external to the model triples', that determine the distinct ranges, $i({\bar M}_1)$ and $i({\bar M}_2)$ (in the notation of Eq.~\eq{eqmodel}), is the fixed standard used to measure the temperature, which is incorporated in e.g.~the thermometer readings. In short: these additional facts about the standard for the temperature determine different ranges for the interpretation maps.\\

An elementary reason why, in general, duals can be physically inequivalent is the fact that, regardless of dualities, a given model usually admits multiple interpretations: for example, depending on how the symbols of the heat equation are interpreted, this equation can be interpreted as an equation for the spreading of heat in a region of space, or as a diffusion equation for e.g.~the motion of Brownian particles on the surface of a fluid. 

Also, elementary quantum mechanics famously admits mutually incompatible interpretations, at least some of which rely on the same set of formulas, i.e.~what Muller (2014:~p.~14) has dubbed `minimal quantum mechanics'. Thus it is not the case that, if models are isomorphic (or even equal), their interpretations are automatically the same:\footnote{On the other hand, in the case of internal interpretations discussed below, this multiplicity of interpretations for a given model does not exclude the possibility that dual models can be physically equivalent. We will return to the general topic of `duality and under-determination' in the second half of Chapter \ref{Realism}.\label{multipleI}} 
formalisms do not always have a unique semantics, not even up to isomorphism (we will return to this, in the context of simple logics, in the beginning of Chapter \ref{Theor}). 

Thus in the paper in which Schr\"odinger (1926b:~p.~58) argues for the equivalence of matrix and wave mechanics (see Section \ref{wpd}), he distinguishes between {\it mathematical equivalence} and {\it physical equivalence}. First, the mathematical formalism of one dual might have specific structure that the other dual lacks (but this is not the case for quantum mechanics: there is an isomorphism between matrix and wave mechanics that involves all of the structure).\footnote{In this passage, Schr\"odinger states clearly that, if there were such differences between duals, he would prefer the simpler model: or, in our language, he would favour the common core. Muller (1997a:~p.~35) argues that Schr\"odinger's claims were not successful, in that he did not prove full equivalence in 1926. This was only done by von Neumann in (1932).} 
Second, one formalism may be extendable to {\it new physical situations}, while the other may not be: and he gives the example of two expressions for the electrostatic energy, only one of which remains useful in electrodynamics. 

The distinction, from Section \ref{intext}, between internal and external interpretations allows us to give a sufficient condition of theoretical i.e.~physical equivalence. Namely, even in cases where dual models have isomorphic model triples that are mapped by their interpretation maps to a common subdomain of application, the ranges of the interpretation maps will differ on their specific structure, which is different for the two models. Thus external interpretations (which do map the specific structure) in general differ, i.e.~we have a square diagram as in Figure \ref{phineq}. In physics jargon, the specific structure will often include other pieces of physics to which the system described by each model triple is coupled---with different pieces of physics on the two sides of the duality. This coupling `breaks the symmetry' between the two sides, so that the two model triples are about distinct, albeit isomorphic, subject matters (`sectors of reality'). In our jargon: such a coupling provides an `external interpretation' of the model triple.

But regardless of this specific structure, we can also construct pairs of internal interpretations that form a triangle diagram as in Figure \ref{interp1}, i.e.~a commuting diagram where the interpretation maps satisfy: $i_1=i_2\,\circ\,d$, with the same domain of application.

Because the duality $d$ maps only the common core and not the specific structure, such interpretation maps do not map the specific structure onto the domain of application.\footnote{This holds symmetrically for the two interpretation maps, because $i_2=i_1\,\circ\,d^{-1}$, so that both $i_1$ and $i_2$ are internal interpretations.} 
Also, because of the commutativity condition, they map duality-related states and quantities onto the same items in the domain of application, so that the models are physically equivalent, namely: their formal structures are isomorphic, and also their interpretations are the same. Thus to secure physical equivalence it suffices that we require internal interpretations that form a commuting triangle, as in Figure \ref{interp1}.\footnote{In cases where dual models have multiple {\it internal} interpretations (see footnote \ref{multipleI}), these come in pairs, and so the members of such pairs are physically equivalent.} 
The common structure that is mapped by the duality is then capable of being a physical theory. This is secured by its being a triple with the models' structure of states, quantities, and dynamics: and this is why it was important for us to require, in Section \ref{isomdef}, that the common core of two duals is a bare theory. We will develop this idea further in Chapter \ref{physeq}.

Let us summarize the semantic statement of the equivalence as follows: first, given a pair of dual models, $M_1$ and $M_2$, the invariant content under the duality, contained in the model triples, is the content common to the two models (i.e.~what, since Section \ref{giantS}, we have called the `common core theory'). Second, an internal interpretation for either model gives an internal interpretation for the other. Namely, since the duality map, $d:M_1\rightarrow M_2$, is an isomorphism we can form, from an internal interpretation for e.g.~the first model, i.e.~$i_1:M_1\rightarrow D$, an interpretation map for the second: $i_2:=i_1\circ d^{-1}:M_2\rightarrow D$ (and vice versa). Thus the interpretation and duality maps form a commuting diagram, as in Figure \ref{interp1}, and $i_2$ thus defined is an {\it internal} interpretation for $M_2$. 

If the common core theory is isomorphic to the two model triples (which we assume in the rest of this Section),\footnote{Indeed, since our discussion does not depend on this, we will assume it in all of Parts I and II. We will return to the more general case, of representations that are not isomorphic to the common core, in Chapter \ref{physeq} when we discuss recovering the common core theory from a set of dual models.} 
this argument extends to the common core theory, so that it also gets an internal interpretation. Thus, such thorough-going dualities can be taken to give {\it theoretical i.e.~physical equivalence} between apparently very different models.

In other words, given a theory with two dual models, i.e.~given $T$, $M_1$, and $M_2$, an internal interpretation of any one of them gives an internal interpretation for the others. Therefore, we propose that, under an internal interpretation thus understood, two models are essentially the same, i.e.~are theoretically equivalent representations of the same common core theory, as follows:

(i):~Isomorphism is an appropriate formal standard of equivalence for models of the kind discussed here, i.e.~sets with structures defined on them. Thus the formalisms of the two models ``say the same thing'', i.e.~they contain equivalent sets of states, quantities, and dynamics.

(ii):~Their ontologies are also the same, since the domains of application of the interpretation maps are the same.

In other words, such dual models describe exactly the same physics, and any difference between them is, {\it under such an interpretation,} a mere difference of formulation and not a difference in the physics described. In other words, although obviously there are differences in how the isomorphic triples are formulated, the internal interpretation adopted does not map these differences to anything in the world---the internal interpretation does not assign any physical relevance to these differences, because they lack a counterpart in the world. 

\subsection{Epistemic considerations postponed}\label{epistemicc}

There is a question whether verdicts of equivalence and inequivalence, such as the ones we discussed in the previous Section, are justified. For example, one might worry that, although the standard of temperature is external to the Ising model, this might be due to the fact that the Ising model is a simplified description of a lattice of spins. Perhaps it is possible to develop a more sophisticated version of the Ising model that determines its own temperature standard `from within', so that temperature measurements can be included into the model root. If so, then the domain and range of the duality map could perhaps be extended such that it maps the standard of temperature of one model to the standard of the other.

While this does {\it not} seem to happen for the Ising model, where the most natural extensions, rather than making the models equivalent, add to the inequivalence, something like this {\it does} happen for T-duality in string theory, where a large circle is mapped to a small circle, and yet these circles cannot be distinguished by string theory, because they are ``defined in the same way'', on both sides of the duality: we will discuss this duality in Chapter \ref{String}. 

This `possibility to augment or extend dual models', so that initially inequivalent models end up agreeing, is of course not an objection to the semantic distinction between physical equivalence and physical inequivalence given by the commuting and non-commuting diagrams in Figures \ref{interp1} and \ref{phineq}. Namely, a semantic verdict of physical (in)equivalence does not need to consider extending or augmenting the dual models, because this extension amounts to changing the models, while verdicts of physical (in)equivalence hold for models that are {\it given}, i.e.~fixed for interpretative purposes.

We ask you whether the weather is the same today as it was yesterday, and you answer that it is. You mean, as we do, the weather {\it here}: and, given that the weather here {\it was} in fact the same, you {\it are} right. If we then say that today's weather is different from yesterday's somewhere else on earth, we change our subject, and we answer a different question.

The question whether or not it is appropriate to extend a model is an {\it epistemic} question regarding the (empirical, and perhaps also theoretical) justification for considering a particular model as `given', or for extending it, given the way our world is, or could be. Thus because of this broad distinction between semantics and epistemology (which also surfaced in our brief---and postponed---discussion of scientific realism in Section \ref{srpost}), we will be able, for the same reasons, to postpone this epistemic discussion to Chapter \ref{Realism} (as we also said, in Section \ref{intext}, in connection with the aims of scientific theories). 

Epistemic considerations like the ones above have led a number of authors to consider dual models that are putative models of the whole universe, i.e.~putative cosmologies.\footnote{See Rickles (2013:~p.~318), Dieks et al.~(2015:~p.~209), De Haro, Mayerson and Butterfield~(2016:~p.~1416), Huggett (2017:~p.~86), Rickles (2017:~p.~64), Matsubara (2013:~p.~484), Butterfield (2021:~p.~46).} 
For relative to a model that purports to describe the whole universe, including all the forces in it, there are no physical facts beyond those about the system they describe (viz.~the universe), and so such dual models can be taken to be physically equivalent. Thus for dualities between such models, there can be no coupling to other pieces of physics that gives an analogue of our `external measurement of the temperature'. (Gauge-gravity duality provides, of course, a putative example of such a duality between theories of the universe: see Chapter \ref{STII}.)

\section{Duality and symmetry}\label{dualsym}

This Section describes how a duality relation, understood according to our Schema, meshes with the symmetries of the duals. The main points are straightforward. But there are interesting subtleties. We shall emphasise:

(i): The differences between (a) symmetries of the bare theory (the common core of the two duals) that each dual is required to represent (which we call `stipulated symmetries'); (b) symmetries of the bare theory that each dual is {\it not} required to represent, though it may do so (which we call `accidental symmetries'); (c) symmetries of a dual, i.e.~of a model, that are not shared by all other models, i.e.~by all other representations of the bare theory, or by the bare theory (which we call `proper symmetries of models').

(ii): The analogy between duality and symmetry, and in particular the comparison of duality with gauge symmetry. Here our main point will be a contrast. For `gauge' connotes descriptive redundancy, i.e.~that states related by a gauge symmetry represent the same physical scenario. But as stressed in Section \ref{dualint}, two dual theories need not be physically equivalent, i.e.~need not be making the very same claims about the world.\\ 
\\
(i):~{\it Duality and symmetries}:--- Our considerations about how dualities and symmetries mesh first follow the relation between a bare theory and its dual models. Given this relation, there are three (mutually exclusive) types of symmetries that we can have:\footnote{De Haro and Butterfield (2021:~p.~3007) considered a fourth type of symmetry (`improper symmetry of the models'), so that the four types jointly cover all the possible cases of symmetries of theories and models. We will not discuss this case here, since it will play no role in our examples in Part II.}

(a):~{\bf Stipulated symmetries}: almost always in (especially relativistic) physics, theories are required to reflect certain symmetries (e.g.~Poincar\'e invariance) that are included in the definition of the theory (see the discussion in Section \ref{intext} (2)). For example, in quantum field theory, the Lagrangian is taken to be Poincar\'e invariant, so that the fields transform according to various spin and mass representations of this symmetry.\footnote{See Weinberg (1995:~Sections 5.2-5.6).}
They also often carry non-abelian charges, with associated gauge symmetries (as we will discuss in Chapter \ref{EMYM}). Since `models', in our main sense, are mathematical representations, i.e.~`homomorphic copies', of a bare theory, they necessarily carry representations of stipulated symmetries, which are the defining symmetries of the bare theory.\footnote{These representations need not be {\it faithful}, i.e.~the representation map need not have trivial kernel.} Thus for a stipulated symmetry $a$ on the theory's state space, ${\cal S}$, each model triple is required to have a symmetry $a_M$ on its state space, ${\cal S}_M$, that is a `shadow' of the theory's symmetry, i.e.~such that the diagram in Figure \ref{shadowS} commutes.

\begin{figure}
\begin{center}
\bea
\begin{array}{ccc}{\cal S}&\xrightarrow{\makebox[.6cm]{$a$}}&{\cal S}\\
~\Big\downarrow {\sm{$h$}}&&~\Big\downarrow {\sm{$h$}} \\
{\cal S}_M&\xrightarrow{\makebox[.6cm]{$a_M$}}&{\cal S}_M\nonumber
\end{array}\nonumber
\eea
\caption{\small The symmetry and representation maps commute for stipulated symmetries.}
\label{shadowS}
\end{center}
\end{figure}

The unitary symmetry of quantum mechanics is a well-known example. This symmetry acts on both states and self-adjoint operators, in such a way that all the matrix elements are preserved, and also the dynamics (i.e.~in the Schr\"odinger picture, the Schr\"odinger equation) is preserved. But also all the models, i.e.~all the representations of the Hilbert space concerned (e.g.~in wave mechanics, the position vs.~momentum representations) carry a representation of this symmetry, such that their matrix elements and dynamics are preserved. 

Notice that gauge symmetries are often {\it not} stipulated. For example, in gauge-gravity duality, the non-abelian gauge symmetry of the quantum field theory on the boundary does not get mapped to a corresponding symmetry of the gravity model (see Section \ref{ggd}). The situation for the diffeomorphisms of the gravity model is more subtle: as we will discuss in Section \ref{holeA}, there are `hole' diffeomorphisms that go to the identity at the boundary, and are isometries of the boundary metric. These diffeomorphisms do not map across the duality, i.e.~they are part of the specific structure (i.e.~they are proper symmetries of the gravity model, in the sense of (c) below). But other diffeomorphisms act as conformal transformations on the boundary metric, and {\it do} map to the conformal symmetries of the boundary model.

(b):~{\bf Accidental symmetries}: the requirement that the models have a shadow of a symmetry, so that the diagram in Figure \ref{shadowS} commutes, suggests another sort of case, where this requirement is not satisfied, i.e.~the diagram fails to commute. While we will not focus on this case here, it is easy to see the reason for this failure. Namely, if the bare theory has a symmetry that is not part of the theory's definition (and in that sense it is `accidental', i.e.~it is not stipulated as a symmetry), then the representation maps to the theory's models cannot be required to respect this symmetry, and so in general there will be models where this symmetry is not realized.\footnote{For a simple example, cf.~De Haro and Butterfield (2021:~pp.~3003-3005, 3011).}

For us, the most interesting case, because often surprising, will be the following:

(c):~{\bf Proper symmetries of models}: since models have specific structure, it is clearly possible for them to have symmetries that do not have a counterpart in the bare theory (and perhaps also not in the other models). Thus any symmetries of the specific structure $\bar M$ of a model $M=\bra m;\bar M\ket$ will be of this type. 

For us, this is the most interesting case, because it can be scientifically very surprising that models with very different symmetry groups are nevertheless isomorphic, while all the quantities are preserved. For example, in gauge-gravity dualities, the quantum field theory on the boundary has a gauge symmetry that does not have a counterpart in the gravity theory. And vice versa: the diffeomorphism symmetry of the gravity theory in the bulk is not (fully) reflected by the symmetries of the boundary theory, so that `hole diffeomorphisms' do not have a counterpart in the boundary theory (see Section \ref{holecosm}).\footnote{See De Haro (2017b:~p.~1468). For more on proper symmetries, see De Haro and Butterfield (2021:~pp.~3005-3006). In the physics literature, some examples of proper symmetries of models are provided in Horowitz and Polchinski (2009:~p.~178) and Polchinski (2017:~p.~9).}
These extra symmetries are not symmetries of the common core (neither stipulated nor accidental), and so they do not participate in the isomorphism.\\
\\
(ii):~{\it Analogy with symmetry}:--- The analogy is (as is often remarked) that `a duality is like a symmetry, but at the level of a theory or model'. That is, the analogy is: while a symmetry carries a state to another state that is `the same' or `matches it', a duality carries a model to another model that is `the same' or `matches it'. We will endorse this analogy. So the interesting questions, for both sides of the analogy, will concern the different ways to make precise `the same' or `matches'. We give details in Chapters \ref{Theor} and \ref{physeq}. But the questions about making precise `the same'/`matches' can be introduced as follows. 

In our notation from Section \ref{dual}, a symmetry $a$ carries a state $s$ in a state space ${\cal S}$ to another state $a(s)$: thanks to $a$ being a symmetry, the two states $s$ and $a(s)$ assign the same values to all the quantities (i.e.~magnitudes) in some salient, usually large, set of quantities. The question then arises: do $s$ and $a(s)$ represent the very same physical state of affairs, or scenario---or in philosophers' jargon: the same possible world? (Recall our discussion in the preamble to Section \ref{Symm}.)

The answer, in full generality, is of course: `No'. That is: not always. But for a large enough set of quantities being preserved; and in particular for a theory that is a `toy cosmology' (i.e.~a theory whose system of interest is a cosmos, with no external environment, so that there are no relational quantities whose values are {\it not} preserved by $a$): there is a tradition of answering `Yes'. 

Debate then ensues about:

(i):~what are the general conditions for the `Yes' answer being correct? and 

(ii)~what does the `Yes' answer imply about the propriety of---perhaps even the requirement of---moving to a reduced formalism, i.e.~one in which states are taken as the orbits, in the given formalism, of the action of the symmetry $a$?\footnote{A bit more precisely: states would be taken as the union of the orbits for all the symmetries for which the `Yes' answer is true. See the overview of the debate about (i) and (ii) in the preamble to Section \ref{Symm}.}

So, turning to our topic of dualities: we endorse this analogy. For a model $M_1$ is mapped by duality to a model $M_2$ which is `the same', or `matches it', formally, in the sense that each state of the first model, $s\in M_1$, is taken into a state $d_{\cal S}(s_1)\in M_2$ of the other (and vice versa), in such a way that the values of all of the models' quantities under the map match (recall our theme (i) from Section \ref{wde}). 

As for the question whether $s$ and $d_{\cal S}(s)$ represent the very same physical states of affairs, the general answer is also `No'. Thus Section \ref{dualint} discussed that the `Yes' answer requires a commuting diagram as in Figure \ref{interp1}, i.e.~that the duality map and the interpretation maps commute, i.e.~$\forall s\in{\cal S}_{M_1}$, $i_{{\cal S}_1}(s)=i_{{\cal S}_2}(d_{\cal S}(s))$. And this condition is not in general satisfied, but it can be secured if the interpretation maps are internal, so that only the common structure is mapped (Chapter \ref{physeq} will discuss in more detail our proposal for necessary and sufficient conditions for the `Yes' answer). As for (ii), i.e.~about the propriety of moving to a reduced formalism, the way the Schema is set up---with a common core theory `above' and the dual models, which are representations of this theory, `below' (see Figure \ref{tm1m2})---is our initial answer. (Chapters \ref{Theor}, \ref{physeq} and \ref{Heuri} will add detail about the desired logical strength of the common core theory, and about the desirability of moving to a successor theory.)

\section{Themes, roles and types}\label{featurerole}

This Section has three main aims: first, to list some scientifically important features that dualities often have, and that will be running themes in the examples in the next Chapter and in Part II; second, to discuss the various roles of dualities; and third, to discuss two special types of dualities. 

\subsection{Themes associated with dualities}\label{themesd}

Our examples in Part II will illustrate five themes that make dualities both surprising and useful:

(1):~{\bf Hard-easy}: making a hard problem easy, or at least easier. Often, this is because dualities relate weak interactions to strong interactions, i.e.~weak couplings to strong couplings (see e.g.~Eq.~\eq{betag} in Chapter \ref{Advan}). This then allows one to turn a strong-coupling problem into a weak-coupling one, thereby making it easier. This hard-easy contrast is also prompted by the next one, with which it is often combined:

(2):~{\bf Elementary-composite}: mapping an elementary solution of the field equations to a composite solution, or a topologically trivial (perhaps localized) solution to a topologically non-trivial (extended) one, or a fundamental field (i.e.~linear in the creation and annihilation operators of the vacuum) to a composite one (typically, an infinite series of creation and annihilation operators). This contrast is typically associated with a change of Noether and topological conserved currents and charges under a duality. And the literature often connects this to the contrast between `fundamental' and `emergent' (see Chapter \ref{Heuri}).

(3):~{\bf Unification}: dualities unify apparently very different theories. The unifying role of string dualities is part of the motivation for the (still incomplete) M-theory programme (Chapters \ref{String} and \ref{STII}). But in some cases, unification prompts physicists to {\it look beyond} a set of duals, to find a successor theory (Chapter \ref{Heuri} will discuss this idea). This also suggests the contrast between exact and effective dualities (i.e.~our fifth theme below).

(4):~{\bf Symmetry-breaking}: connecting different vacua or phases of systems. Dualities often connect an ordered state of a system to a disordered one, or a vacuum with a symmetry to a vacuum with a different symmetry, where the symmetry breaking is characterised by a distinguished parameter. Namely, this parameter takes values in a space whose regions characterise the phases and-or vacua of a theory, each of which corresponds to the domain of application of a different (dual, or effectively dual) model.\footnote{This discussion anticipates the geometric view of theories which will be developed in Chapter \ref{Heuri} .} 
Also, in quantum field theories the space of values of this parameter often comes equipped with structures (e.g.~a non-trivial topology, a metric, etc.) that relate (effectively) dual models. This leads in to the following distinction:

(5):~{\bf Exact-effective}: even if there is no duality, but only something similar to it, still the idea of duality is very useful, and it guides the search for important relations between the models and the theory (a common core theory, or a successor theory).

An {\bf effective duality} is a map between models or theories that approximates a duality in an appropriate regime of parameters (usually, at low energies, or at large distance scales): so that, if the parameters are taken to a limit, there could be or not be a duality. (Although having an idealisation is not a necessary condition for an effective duality: in our examples in Part II, the approximation will always be an idealisation, so that there is some physical system that the duals refer to, under the appropriate interpretation map.) 

More generally, {\bf quasi-duality} will be our umbrella term for `almost a duality': this includes effective dualities, isomorphisms with a domain that is not the whole state-space or not the whole set of quantities, and bijections that fall short of being a full isomorphism (i.e.~they respect some, but not all, structure).

\subsection{Roles of dualities}

There are two main types of function or role that a duality can have: we call them `theoretical' and `practical'. They will be developed in detail in Chapters \ref{Theor}-\ref{Realism} and \ref{Heuri}-\ref{Understand} respectively. But broadly speaking, the theoretical functions of a duality are closely tied to satisfying the Schema: i.e.~a bare theory being represented by two isomorphic models. One main theoretical function of a duality, pursued in Chapters \ref{Theor} and \ref{physeq}, is the conditions for two theories to be equivalent. Another, developed in Chapter \ref{Realism}, is the implications of duality for scientific realism. On the other hand, the practical functions of a duality ``look beyond'' the satisfaction of the Schema. For example, one main practical function of a duality is to guess a theory ``beyond'' the bare theory: of which the given duals would {\it not} be isomorphic models, nor perhaps models (precise representations) at all---they would only be approximations, and their resulting relation is typically an effective duality (see theme (5) above). This is pursued in Chapter \ref{Heuri}.\footnote{For a discussion of the practical function of dualities, and recent authors who have explored it, see De Haro (2019a) and Section \ref{uds}.}

\subsection{Special types of dualities}\label{std}

Our fifth theme above identified a significant contrast between dualities and quasi-dualities that we will return to in Parts II and III. And, while further work should articulate different types of dualities and effective dualities, our examples will single out two specific types of dualities as having particular significance:

(A):~{\bf Quantum duality:} this is a duality that is realized only at the quantum level, i.e.~the corresponding classical models are not duals.

(B):~{\bf Self-duality:} a duality within models of the same type, at different values of the parameters (or for different states), as opposed to dualities that relate very different models. The reason for the name `self-duality' is that physicists often call models that depend on a free parameter (for example, in models of $N$ Newtonian point-particles, the parameter $N$) `the same model' (i.e.~they ``generalize over $N$''), while our `intermediate' notion of theory and model (see Section \ref{Ourthm} (2)) regards models at different values of the parameter as being different models.

Some string theorists have used a restricted definition of `duality', which corresponds to our `quantum duality', i.e.~(A) above. For example, Sen (1999:~p.~1636) writes:\footnote{Also Polchinski (2017:~p.~7) endorses this restricted conception: `The interpretation of a duality is then that we have a single quantum system that has two classical limits'.}
\begin{quote}\small
Duality in its most general form is a statement of equivalence between two or more `apparently different' quantum theories. Here by apparently different theories we refer to theories whose corresponding classical theories are genuinely different, i.e.~there is no change of variables which can relate these classical theories. Two such theories will be called dual to each other if they are identical as quantum theories, i.e.~if there is a unitary transformation relating the Hilbert spaces of the two theories under which all correlation functions in one theory are mapped exactly to the corresponding correlation functions in the other theory. Thus a dual pair of theories represent two theories which are identical as quantum theories, but yet their classical limits are genuinely different.
\end{quote}

While the second part of the quote agrees with our own definition of duality for quantum models, the first part is inconsistent with our general definition of duality, because a duality can obtain between classical i.e.~non-quantized models: one does not need to require, as Sen does, that the classical limits of dual models are formally inequivalent. Dual models may in fact be wholly classical, as in e.g.~Ising model duality and electric-magnetic duality.

Thus, while we reject `quantum duality' as a {\it definition} of duality,\footnote{In fact, the requirement that `there is no change of variables which can relate these classical theories' seems to exclude some commonly accepted {\it string} dualities from being dualities in Sen's sense. For example, for T-duality there is a relatively straightforward change of variables that takes us from the classical Type IIA to the Type IIB string theory. This is another reason to reject Sen's characterisation as the definition of a duality, for being too strict.} 
because inconsistent with the extant examples of dualities between classical theories, it is an important {\it special case} of a duality. Gauge-gravity dualities are important examples (see Section \ref{ggd}). 

Indeed, we will argue that the reason why dualities are often so {\it surprising} is because they relate quantum models that have {\it very different (semi-classical) limits}. Thus, for a quantum duality, a single quantum theory can give, in two very different regimes of validity, models that behave semi-classically and that are very different from each other.

If two dual quantum mechanical models are regarded as different versions of a single theory, i.e.~interpreted as equivalent, then quantum dualities imply the interesting possibility that a single theory can have more than one classical limit: namely, one for each dual (alternatively, that inequivalent classical models give the same quantum theory). This also seems to be the best explanation for the existence of gauge-gravity dualities: as weak and strong-coupling limits of a single theory.\footnote{In quantum field theories, this can be explained using the idea of a {\it master field}: namely, a configuration of the field that dominates the path integral in an appropriate limit (see Witten (1980:~Section IV) and Coleman (1985:~p.~392)). In this limit, the master field's contribution to the path integral is the dominant one, and the theory becomes essentially classical, with the master field satisfying a classical equation of motion. In this limit, the relevant operators have zero standard deviation from their classical values, as well as factorization properties characteristic of c-numbers.} 
One reason for this behaviour is that the classical limit, $\hbar\rightarrow0$, is not unique. As we will see for electric-magnetic duality, where the electric charge satisfies the Dirac quantization condition (see Eq.~\eq{Diracq}), there are two ways to take the limit $\hbar\rightarrow0$: either taking the electric charge to zero (keeping the magnetic charge fixed), or taking the magnetic charge to zero (keeping the electric charge fixed). 

\section{Comparison with other philosophical work on dualities}\label{comparison}

This Section summarizes key aspects of dualities that have been studied in a recent stream of research articles in the philosophical literature, and which Part III will discuss in more detail, using the Schema. First, Section \ref{substd} distinguishes four issues on which the discussion has focussed: the definition of duality, symmetries, roles of duality, and theoretical equivalence. This last topic is so big that it deserves a Subsection of its own: thus Section \ref{contcons} focusses on theoretical equivalence and the associated question of under-determination. Section \ref{uds} mentions a few topics that are still under-developed, such as the heuristic uses of duality and the distinction between elementary and composite objects (theme (2) in Section \ref{featurerole}), as well as topics that are specific to particular examples of dualities.

\subsection{Four issues}\label{substd}

Philosophical discussions of dualities have focussed on at least four substantive issues, as follows:

(i)~~{\it The definition of duality}: The main difference is between authors who define dualities formally,\footnote{This includes Teh (2013), Huggett (2017), Castellani (2017), Rickles (2017), Huggett and W\"uthrich (2023), and ourselves.} 
and those who, in addition, require duals to be empirically equivalent.\footnote{This includes Read and M\o ller-Nielsen (2020:~p.~263) and Matsubara (2013:~pp.~477-479).} 

There are two ways in which this difference influences an account of duality. The first way is in terms of scope: if duality is defined formally, then there are more cases, e.g.~Kramers-Wannier is a case of duality, which agrees with how physicists use the word. 

The second way is more substantive: for requiring that duals are empirically equivalent makes the notion of duality dependent on a conception of `empirical equivalence' (which we will discuss in Section \ref{eronos}). This can have consequences for other issues, such as under-determination (see Section \ref{srr}). 

Dawid (2007:~pp.~14, 22) appears to sometimes use the word `duality' in our sense of `effective duality' (see theme (5) in Section \ref{featurerole}). This presumably follows the usage of some string theorists, who use `duality' for both dualities and quasi-dualities. (However, it {\it is} standard in string theory to distinguish between a duality and an effective duality.)

(ii)~~{\it Symmetries and dualities}: The literature has emphasised the analogy between duality and symmetry, i.e.~Section \ref{giantS}'s theme of duality as a ``giant'' symmetry, and the similar interpretative questions that arise about whether symmetry-related states (and quantities) are the same or different (see point (ii) in Section \ref{dualsym}, and also the preamble to Section \ref{Symm}). Especially, Read and M\o ller-Nielsen's (2020) analysis\footnote{Read (2016:~p.~224) considers the analogy between dualities and the hole argument (and their disanalogies), where the worry, rather than being about indeterminism as in the hole argument, is about under-determination. We will discuss this analogy in Chapters \ref{HABHM} and \ref{Realism}.} 
of the interpretative options for dualities follows closely the corresponding discussion for symmetries. Thus their interpretational vs.~motivational contrast (see Section \ref{contcons}) is an extension of the one that applies to the interpretation of symmetries. 

Commentators have emphasised that dual models can have different sets of symmetries, i.e.~what we call `proper symmetries of models' (see (i)-(c) in Section \ref{dualsym}): especially Polchinski (2017:~p.~9) and Horowitz and Polchinski (2009:~p.~178).\footnote{Although Teh (2013) does not explicitly mention the differences in symmetries between dual models, it is clear from his exposition of AdS-CFT that he has such differences in mind, i.e.~that he has proper symmetries of models in mind. This also appears to underpin his statement that dualities are not cases of isomorphism between models, and that categorical equivalence and definitional equivalence are better criteria to characterize dualities (2013:~pp.~302, 310). While we will return to the latter criteria in Chapter \ref{Theor}, our difference with Teh lies in what we take to be the defining structure of dual models---with respect to which the isomorphism is defined. Teh considers {\it all} of the theory's structure, while our proper symmetries of models are by definition not stipulated by the bare theory, and so they {\it cannot} in general enter in the isomorphism between triples (see Section \ref{dualsym}).}

Finally, Rickles (2017:~p.~66; 2011:~p.~55) has the view that dualities are (perhaps generalized cases of) ``gauge symmetries'', where `gauge' here means that there are `different [mathematical] realisations of the same physical state'. Hence dualities help one identify unphysical degrees of freedom. We will discuss this in Section \ref{contcons}.

(iii)~~{\it Themes and roles}: Philosophers of dualities have addressed some of the themes associated with dualities, as well as the theoretical and practical roles of dualities (see Section \ref{featurerole}). 

Theme (2) of elementary vs.~composite solutions has been stressed by Castellani (2017), who discusses whether one should take solutions to be elementary or composite in an epistemic or in an ontological sense. Castellani also stresses the fact that dualities relate Noether charges to topological conserved charges---a topic that we will indeed develop in Chapter \ref{Advan}.

According to Castellani, dualities relate physically equivalent situations, and so one should not take the fact that an elementary solution (e.g.~a free particle) is mapped by a duality into a composite solution (i.e.~a soliton) as an ontological statement. Rather, it is a matter of the model's descriptions. This surely ties into the idea that models (and their solutions) have specific structure, and that on an internal interpretation such structure is not mapped to anything in the world. But, on an external interpretation, this view would not be vindicated: what is elementary and what is composite would be different in the two models, so that the models disagree. We will return to this in Chapters \ref{EMYM} and \ref{Heuri}.

Dawid (2006~p.~310, 2007) has developed Section \ref{featurerole}'s theme (3) of unification along the lines of his thesis, discussed earlier, that string theory is a structurally unique theory: dualities are evidence for the idea that the structure of string theory, including its parameters, is unique. For example, the symmetries of string theory determine the dimension of spacetime ($D=10$, and $D=11$ in the case of M-theory) and the matter content of the theory, so that the low-energy Lagrangian is fixed by the symmetries alone, and contains a single parameter: namely, the string length.\footnote{However, there are a large number of internal manifolds on which one can compactify the theory from ten to four dimensions, each type of compactification giving rise to a different vacuum, with a different set of massless and massive fields---this is known as the `landscape' of string theory, and by conservative estimates it contains $10^{500}$ possible vacua. According to Dawid (2006:~p.~312), `[t]he compactification of the six extra dimensions is a matter of the theory's dynamics, and its outcome constitutes the ground state of the theory'. He acknowledges that `string theory would require some physical vacuum selection mechanism that reduces the number of physically possible ground states'. This situation might be improved by swampland considerations, which are a set of conjectured criteria that aim to distinguish consistent quantum theories of gravity from inconsistent ones, and may provide some guidance for finding a vacuum selection mechanism. For an accessible introduction, and an overview of the conjectures, see Palti (2019) and van Riet and Zoccarato (2023).}

(iv)~~{\it Theoretical equivalence and the interpretation of dualities}: Arguably {\it the} main substantive issue that philosophers of duality have focussed on is the interpretation of dualities. This includes the question whether or not dual models ought to be taken as theoretically equivalent (often, in the context of dualities, also called `physically equivalent': see footnote \ref{usePE}), and questions about the criteria that lead to such verdicts. Thus several authors give {\it taxonomies} of possible interpretations of dualities.\footnote{See Matsubara (2013:~p.~484), Read (2016:~pp.~223-224), Huggett (2017:~pp.~83-84), and Huggett and W\"uthrich (2023:~Section 7.2.1).} 
Also, there are the related questions of scientific realism and under-determination of theory by data. These are big topics that we will discuss in a separate Section \ref{contcons} (indeed, we will be concerned with these topics in much of Part III of this book).

Note that the issues (i), (ii), and (iv), on which the literature has focussed, i.e.~the definition of duality, its relation to symmetry, and the question of theoretical equivalence, belong to the theoretical function of dualities, i.e.~duality as isomorphism of models with respect to a common core. 

But some authors have also discussed the issue (iii), which belongs to the practical functions of duality, i.e.~how the idea of a duality functions in going beyond the common core---especially, how a duality can function in theory succession, so that discovering a duality suggests the existence of a successor theory that will not itself instantiate the duality other than in an effective sense: see e.g.~Read (2016:~p.~228). We will return to these topics in our examples in Part II, and in Chapters \ref{Heuri} and \ref{Understand}. 

\subsection{Theoretical equivalence}\label{contcons}

A common feature of the taxonomies of interpretations of dualities is that the first main interpretative option is a contrast between duals that are theoretically equivalent (i.e.~describe the same domain of application, the same set of worlds, the same physical situations) and those that are inequivalent. The taxonomies then differ in their further interpretative options, i.e.~in the ways in which duals can describe the same, or different, domains of application. 

It is interesting to follow the subsequent interpretative options, which often agree with our own views, a bit further: for the (often apparently small) differences between them are illuminating. 

One important difference is about duals that are taken to be {\it theoretically inequivalent}: while Huggett (2017:~p.~83) and Huggett and W\"uthrich (2023:~Section 7.2.2) initially consider this as a live option, they ultimately reject it for string dualities, because string theories are purported to be total theories about the physical world, without an external point of view (and thus this is akin to adopting an internal interpretation: cf.~Section \ref{itm}).\footnote{Huggett discusses a second interpretative option that arises for duals that describe the same possible world: namely, the question is whether duals assert the same things about the world or not, i.e.~whether the interpretation of the terms is the same (as in the case of an internal interpretation that interprets only the common core theory, see Section \ref{giantS}) or different. In the latter case, duals only appear to be incompatible because they are written in different languages and assign different meanings to the same words (e.g.~one theory says that the radius or a circle is $R$, and the other is $1/R$). However, this second interpretation (duals agree, despite their being written in different languages) appears to be available only in very special cases like T-duality (and only the simplest case, i.e.~T-duality between bosonic string theories, since T-duality between Type IIA and Type IIB string theories is not merely a matter of a different language: see Section \ref{T-d}). See Huggett (2017:~pp.~84-85) and Huggett and W\"uthrich (2023:~Section 2.2).} 
By contrast, Read and Matsubara take these cases of theoretically inequivalent duals much more seriously, and further distinguish duals that are irreconcilable and are worrying cases of under-determination, from duals that are reconcilable, and can e.g.~be embedded into a deeper theory (i.e.~what in Chapter \ref{Heuri} we will call a `successor theory', see also Section \ref{contcons}), and thus are not worrying cases of under-determination.

In these taxonomies, something like the distinction between external and internal interpretations has been defended by several philosophers of dualities. For example, Rickles' (2017:~p.~64, especially footnote 14) defence of the theoretical equivalence of duals assumes that an internal interpretation is being adopted, as does Huggett's (2017:~p.~86). Also Read (2016:~p.~228) draws an analogy between adopting an internal interpretation and viewing duals as equivalent gauges.\footnote{However, there is a disanalogy: in the case of local gauge symmetries, a natural move is to adopt a fibre bundle formalism that makes the gauge equivalence explicit (namely, in terms of a principal bundle, on which the gauge field is a connection), where a choice of gauge corresponds to a choice of local trivialization of the total space of the bundle. In the case of dualities, which are isomorphic representations of a common core, there seems to be, in most examples so far, no immediate temptation to move to a `reduced' formalism: we simply keep the explicit representations, see Section \ref{isomdef}. There are (at least) three reasons for this. First, because the common core theory may be strictly more abstract than the dual models, and so the dual models may contain more information than the common core (recall, from Section \ref{intext}, the idea of parochial observables in the example of algebraic quantum field theory; see also our discussion in Section \ref{abstraction}). Second, the common core theory may have non-dual as well as dual models: and so, in addition to specifying the common core, one needs to specify the relevant class of models. Third, it is not always true that the common core theory `has fewer variables than the duals', since formulating a common core can adopt the use of Lagrange multiplier fields (see e.g.~Section \ref{MEMD}). More generally, in our `structured view of theories' in Chapters \ref{physeq} and \ref{Heuri}, the representation relations between a theory and its models have interesting logico-semantic consequences, so that we do not wish, and there is no need, to excise the models: see Chapter \ref{physeq}. In Dewar's (2019:~p.~492) jargon, a common core theory can be either a `reduced' or an `internally sophisticated' theory. However, note that what Dewar calls `semantics' only counts as what Chapter \ref{Theor} will call a `model-theoretic semantics'. In addition to the model-theoretic semantic, Chapter \ref{Theor} will also consider a `physical semantics'. Thus we endorse the critical evaluation of sophistication by Martens and Read (2021).}

Regarding the interpretion of dual models, Read and M\o ller-Nielsen (2020:~pp.~266, 276) distinguish `motivationalist' and `interpretationalist' attitudes towards dualities (see our previous short discussions of this issue in Section \ref{dualsym}-(ii), the preamble to Section \ref{Symm} and Section \ref{substd}). In the presence of a duality, the interpretationalist sees themselves warranted to take the dual models to be theoretically equivalent, even absent an explication of the duals' common ontology. By contrast, the motivationalist sees themselves warranted to take dual models to be theoretically equivalent only once such a common ontology is provided.\footnote{Read and M\o ller-Nielsen's formulation of motivationalism is ambiguous, since it is unclear whether (i) the interpretationalist takes duals as theoretically equivalent solely on the basis of their being duals, or (ii) some additional requirements---short of the motivationalists' ontological requirements---can, or even must, be satisfied before inferring theoretical equivalence. For example, the interpretationalist might require that duals are physically equivalent only if they model the whole universe, and are interpreted internally (this is e.g.~Huggett's interpretationalist position). We take Read and M\o ller-Nielsen to be saying that it is (ii), i.e.~saying that an interpretationalist can make a verdict of theoretical equivalence, while lacking an explication of the common ontology of the models (as the motivationalist would require) {\it provided additional requirements are satisfied}. For otherwise, i.e.~if the interpretationalist sees themselves as {\it automatically} warranted to accept theoretical equivalence from the fact of duality alone, regardless of any further analysis, then the contrast between the interpretationalist and the motivationalist does not cover all possible positions, and none of the philosophers of dualities that we have discussed would qualify as interpretationalists, since {\it all} of them require additional conditions.} 
We endorse motivationalism, because, as we will discuss in Chapters \ref{physeq} and \ref{Realism}, theoretical equivalence has an interpretative component for which one needs to provide epistemic justification (see our discussion in Section \ref{epistemicc}).

One issue that philosophers of dualities disagree about is whether dualities give genuine examples of under-determination of theory by the empirical data, i.e.~of systems of the world that predict the same observations, but whose theoretical descriptions otherwise disagree. The question is, roughly, whether dual models are in {\it competition} as putative descriptions of the world, or whether they are compatible, and perhaps even fully equivalent, descriptions.\footnote{The challenge of under-determination will get a fuller treatment in Section \ref{srr}.} While philosophers agree that dualities, at least {\it prima facie}, give cases of under-determination, they disagree whether such under-determination creates a problem for scientific realism.

Matsubara (2013), Huggett and W\"uthrich (2013), and Read (2016) all consider a number of competing interpretations of dualities, at least some of which lead to a problem of under-determination. If duals are interpreted as describing different physical situations, yet have the same empirical content, then they are clear cases of under-determination. Furthermore, these authors take this under-determination to be problematic for scientific realism. Matsubara (2013:~p.~484) puts the anti-realist worry thus: `the world may in reality be more like one dual description than the other, but we have no empirical way of knowing this'. Le Bihan and Read (2018:~p.~5) put it as follows: `under-determination gives rise to a sceptical challenge: absent a means of determining which of a class of worlds is the actual world, {\it we have at hand no determinate picture of what the world is really like}'.\footnote{For example, Huggett and W\"uthrich (2013:~p.~281) argue that some of the properties that we usually assign to spacetimes and to strings are not `physically determinate in the fundamental picture of the world in string theory'. For example, in T-duality `there is no physical fact of the matter how long strings are, or how many times they are wrapped around space' (for more on T-duality, see Section \ref{T-d}). We postpone a more detailed discussion of Le Bihan and Read (2018) to Chapter \ref{Realism}.}

One answer to the under-determination challenge to scientific realism is to endorse a form of structural realism, i.e.~the thesis that we should be realists about the {\it structure} that scientific theories assign to the world (as against e.g.~objects and their properties)\footnote{For an overview of structural realism, see Ladyman (1998, 2014) and Frigg and Votsis (2011). More details are in Worrall (1989) and Ladyman et al.~(2007). For a recent discussion of the interaction between structural realism and the syntactic and semantic conceptions of theories (as a contrast between a `language-first' and a `maths-first' approach)), see Wallace (2022c).} 
Matsubara (2013:~p.~484) indeed proposes to take this route.\footnote{Depending on the interpretation that is given to the duality (i.e.~duals as describing different possible worlds, or the same world), Matsubara considers epistemic and ontic versions of structural realism. However, he also warns that `structure' is a vague notion, and expresses doubts that it can be explicated in a satisfactory manner (p.~487). Note that according to our Schema, structural realism would amount to realism about something specific: namely, the triples of states, quantities and dynamics (Eq.~\eq{introtriple}), defined up to isomorphism by duality maps, as in Section \ref{isomdef}. Our own scientific realism differs from this in that the referential semantics from Section \ref{refsr} does not assume that the domain of application is a mathematical structure. Of course, we will have to explain whether, and if so how, such scientific realism is viable: which we will do in Chapter \ref{Realism}. This structure-realist route is also discussed by Rickles (2011).}
Namely, on a structural realist view one does not need to worry about two models describing the world in different ways (e.g.~in terms of a large or a small radius), since all that matters is the structure that they ascribe to the world: and since these structures are isomorphic, there is no real incompatibility.\footnote{For a critique, see Section \ref{lsps}.}

Dawid (2006, 2007) addresses under-determination with a stronger structuralist claim, in three steps: 

(1)~While most other commentators consider several interpretative options, only some of which may present a problem of under-determination, Dawid is sanguine that `string dualities ... deliver the lethal blow to ontological realism' (2007:~p.~28). The reason for this claim is that string objects do not have a real existence in an external world, but only in a {\it mathematical framework} (p.~27). 

(2)~Dawid puts forward a notion of the {\it structural uniqueness} that is being discovered in string theory, and that dualities point to (see Chapters \ref{String} to \ref{HABHM}). This structure unifies the various previously known string theories. On such a view, dual descriptions give identical physics (Dawid, 2006:~p.~317). Thus the structural uniqueness thesis dissolves the under-determination, since the differences between duals are merely {\it apparent}.\footnote{More precisely, Dawid (2013:~Chapter 6) distinguishes `local' and `global' limitations to the possible alternatives that, combined, make string theory an ultimately unique theory. The {\it global} limitations are requirements that any theory of quantum gravity must satisfy (such as reproducing quantum field theory and general relativity in appropriate limits, gauge symmetry, and renormalizability) and that point to string theory's {\it structural uniqueness}, as the only viable theory that satisfies these requirements. He also discusses three arguments that reduce the alternatives in a {\it local} way: a no-alternatives argument, an argument from unexpected explanatory coherence, and a meta-inductive argument from the success of other theories in the research programme. See the discussion in Chapter \ref{Realism}.} 

(3)~The structural uniqueness thesis, seen as a claim over the {\it final theory} of physics, introduces `a new conception of a scientific process that is characterized by intratheoretical progress instead of theory succession' (Dawid, 2006:~p.~319). Dawid (2013) has worked out this picture further in his book {\it String Theory and the Scientific Method}. (We will return to this issue, which falls within what we call the `practical function' of dualities, in Chapters \ref{Realism} and \ref{Heuri}.)

Castellani (2017), by contrast, mitigates the threat of under-determination by downplaying the ontological consequences of string dualities (see also Rickles (2011:~p.~66)). She argues against the `literal reading' of duality-related string models: for example, the `elementary-composite' contrast from Section \ref{featurerole} should be taken as an {\it epistemic}, rather than as an ontological, contrast. The apparent ontological differences between dual models are {\it artefacts} of the particular approximations or limits taken. Thus they are artefacts of our models, rather than reflecting genuine differences in the ontology. (We will discuss some of the reasons for this view at the end of Chapter \ref{Heuri}.)

Huggett (2017) and Rickles (2011, 2017) argue differently against the under-determination worry. Both take dualities to be best interpreted as cases of theoretical equivalence, so that the under-determination problem does not arise. (In short: the `common core theory', from Chapter \ref{Thies}, gives a shared ontology to dual models.) Thus, even rejecting a deflationary ontological view \`a la Castellani, and reading the ontology of dual pairs literally, the incompatibility turns out to be only apparent. Furthermore, Rickles (2017:~pp.~63-64) states that dualities signal a representational redundancy, and that they point to a more fundamental description of the underlying theory, in which this representational redundancy is eliminated.\\ 

These differences in views about under-determination aside, philosophers of dualities agree about at least three basic points: the first is about scientific theories in general, and the other two about the interpretation of dualities.

First, all recent philosophers of dualities seem to agree about the importance of distinguishing (i) the formalism of a theory, from (ii) its interpretation, and in their endorsing referential semantics (see Section \ref{itm}, and Section \ref{lsps}). Second, the major substantive issue discussed by various authors has been the question of theoretical equivalence. Thus combining the first and second points, it is agreed that one should distinguish two parts, or sides, to the question of theoretical equivalence: (i) whether two theories or models are formally equivalent, and (ii) whether they have the same interpretation. Third, all recent authors agree that the answer given to the question `are duals theoretically equivalent?' does not have an {\it automatic} `yes' or `no' answer, and that a definite answer requires non-trivial philosophical assumptions, or decisions, about e.g.~how best to interpret dual models: to which, as we discussed in Section \ref{contcons}, different authors then give different answers. (Chapters \ref{Theor} to \ref{Realism} will investigate this issue from a more systematic perspective.)

\subsection{Under-developed and specific topics}\label{uds}

Given the recent focus of philosophers on dualities {\it in string theory}, the literature has naturally gravitated towards the set of questions that we discussed in Sections \ref{substd} and \ref{contcons}, especially (i)-(iv), most of which are indeed about the theoretical function: or, when the practical function of dualities has been addressed, it has often been in the context of theory succession and how to break possible cases of under-determination.

But the consideration of other dualities, such as the simpler particle-soliton dualities in quantum field theory (in Part II), will stress other important practical aspects of dualities: e.g.~to discover and study new physical mechanisms (cf.~Chapter \ref{EMDuality}), and the geometric view of theories (Chapter \ref{Heuri}). 

Also, with the exception of Castellani (2017), the elementary-composite distinction has hardly been discussed in the recent wave of philosophy of physics papers,\footnote{Our own work on this distinction is De Haro and Butterfield (2018) and Castellani and De Haro (2020).}
as has the theme of symmetry breaking (see Chapters \ref{Advan} and \ref{EMDuality}.) Likewise for the distinction between dualities and quasi-dualities, which we will examine in Chapter \ref{Heuri}.\\

Besides the general issues that apply to many (or even all) examples of dualities, philosophers have also examined questions that are specific to particular dualities: especially, about the nature of spacetime, both in its classical and quantum aspects. For example, the hole argument and the associated positions of relationism and sophisticated substantivalism are relevant for dualities that involve classical spacetime theories, especially gauge-gravity dualities, and have been discussed by Read (2016) and De Haro, Teh, and Butterfield (2016, 2017).\footnote{Read (2016:~pp.~219-222) uses the hole argument as a springboard to elaborate a taxonomy of interpretations of dualities. Based on a comparison between dualities and gauge symmetries in general relativity, Rickles (2017) also mentions the hole argument in the context of dualities.}
This is also the case for the emergence of the spacetime of one dual from the other, which has been discussed mostly in the context of gauge-gravity dualities\footnote{See e.g.~Rickles (2013), Teh (2013), and Dieks et al.~(2015).}
and T-duality (Huggett and W\"uthrich, 2023:~Section 2.4). We will investigate these issues in Chapters \ref{String} to \ref{HABHM}, and \ref{Heuri}.

\section{Conclusion}

This Chapter defined duality as an isomorphism of models with respect to the structure of a common core, itself a bare theory, that the models instantiate (usually, they are mathematical representations of it). Once dual models are interpreted, the duality and the interpretation maps can form either a diagram that does not commute, i.e.~a square diagram, or a triangle diagram (see Figures \ref{phineq} and \ref{interp1}). When the diagram commutes, the two models describe the same sector of reality, for states and for quantities, so that the models are physically equivalent. 

Section \ref{wde} gave a preliminary discussion of the conditions under which duals are physically equivalent. We stressed that equivalence is not automatic, and argued that an internal interpretation is a sufficient condition for equivalence, i.e.~an interpretation that maps only the common core of the models and not their specific structure. Questions of course remain: about necessary conditions for equivalence, and about the epistemic justification for verdicts of equivalence. Chapters \ref{Theor} and \ref{physeq} will address them (a preliminary discussion of the literature is in Section \ref{comparison}).

Section \ref{dualsym} mentioned three types of symmetries that, wholly independent of dualities, can appear given the relations between a bare theory and its models. It went on to endorse the analogy between duality and symmetry: the main similarities are about the questions of physical equivalence and whether, presented with dualities or symmetries that are cases of physical equivalence, one should develop a reduced formalism that has the equivalence ``written on its face''.

The themes associated with dualities (Section \ref{featurerole}) include features that make dualities useful (e.g.~making a hard problem easy) or surprising (e.g.~the equivalence between elementary and composite quantities), and themes like unification that are important for theory development.

In Section \ref{comparison}, we discussed the theoretical equivalence of duals as the major theme that the literature has addressed, with various extant taxonomies that philosophers have developed: and all hands agree that {\it theoretical equivalence is not automatic}. We will return to this more systematically in Chapters \ref{Theor} and \ref{physeq}, where we will place this discussion in the wider context of logic and philosophy of science. 

\chapter{Simple Examples}\label{Simple}
\markboth{\small{\textup{Simple Examples}}}{\textup{\small{Simple Examples}}}

Chapter \ref{Schema} adverted intermittently to three elementary examples of our Schema. Namely:\\
(i)~~position-momentum duality in elementary quantum mechanics;\\
(ii)~~elementary classical electric-magnetic duality; and\\
(iii)~~Kramers-Wannier duality between low- and high-temperature Ising models on a lattice.
 
In this Chapter, these dualities will each get a more thorough treatment, in Sections \ref{pmd}, \ref{EMduality}, and \ref{dualpf0}, respectively. Section \ref{wpd} will discuss wave-particle duality, and its relation to {\it complementarity}, from a historical and philosophical perspective. In Section \ref{IMD}, Kramers-Wannier duality will be generalized to spin-dislocation duality. Wave-particle duality, electric-magnetic duality, and spin-dislocation duality are preludes to more advanced dualities in Part II.

The examples in this and the next Chapter show that the phenomenon of duality can be illustrated and proven in theories that are well under control. Thus this Chapter and the next one set the stage for discussing examples of dualities that are conjectures: namely, the examples of duality in quantum field theory (Chapters \ref{EMDuality} and \ref{EMYM}) and string theory (Chapters \ref{String} to \ref{HABHM}). In this sense, Chapters \ref{Simple} and \ref{Advan} are complementary to Chapters \ref{EMDuality} to \ref{HABHM}. 

{\bf How to read this Chapter:} although this Chapter uses only undergraduate physics and mathematics and so should be widely accessible, readers who are wholly unfamiliar with electrodynamics and statistical mechanics may wish to read Sections \ref{pmd} and \ref{wpd} on position-momentum duality and wave-particle duality, and then skim through Sections \ref{EMduality}-\ref{dualpf0}. Section \ref{summary4} concludes.

\section{Position-momentum duality in quantum mechanics}\label{pmd}

This Section gives our first and most familiar example of a duality, in elementary quantum mechanics: position-momentum duality. This duality is both elementary (it ends up being a Fourier transformation between the position and momentum representations of quantum mechanics) and of scientific importance. Furthermore, it is an archetypal duality in the Schema's sense, because the position and momentum formulations are distinct {\it representations} (namely, different choices of basis) of the Hilbert space and of the algebra of operators, these two choices being related by a unitary transformation. Furthermore, as we will discuss, dualities in quantum field theory are usually also unitary transformations.

In presenting this (as indeed: any other) example, one must judge how mathematically rigorous and-or abstract a presentation to adopt. We will take a {\it via media}.\footnote{We mostly follow Jordan (1969:~Sections 14-18). More rigorous but brief discussions include Jauch (1968:~Section 4.7), Prugovecki (1981:~Sections 7.1-7.2), Gustafson and Sigal (2006:~Section 25.14), and Takhtajan (2008:~Section 2.2).}
For example, we will assume a notion of Hilbert space, and thereby introduce $l^2$ and $L^2$ spaces; but we will only gesture at the need for Lebesgue integrals.\footnote{To keep this Section short, we will {\it not} present the spectral theorem for quantities with a continuous spectrum.} 

\subsection{Introduction: Hilbert spaces}\label{iHs}

A {\it Hilbert space} $\cal H$ is a vector space (for us, always over $\mathbb{C}$) with a positive-definite sequilinear inner product---which we write as $\bra\m\, \psi|\l\, \phi\ket = \m^*\l\, \bra\psi|\phi\ket$, so that it is antilinear in the first argument---that is complete in the norm induced by the inner product. That is: every Cauchy sequence converges to a vector in the space. $\cal H$ is {\it separable} iff it has a countable (finite or denumerable) basis. We shall in general assume that all Hilbert spaces we discuss are separable. 

Two examples that will be important in our discussion of matrix and wave mechanics later on, as follows:

(1)~$l^2 : = \{ (x_1, x_2,...) \; | \; x_n \in \mathbb{C} , \sum_n |x_n|^2 < \infty \}$ has an orthonormal basis $(1,0,0,...), (0,1,0,0,...), ...$ =: $\{\phi_n \}$. So each vector is $\sum_n \, x_n \phi_n$. In general: each vector $\psi$ has a unique expression in terms of any orthonormal basis $\{\psi_n \}$: $\psi = \sum_n \, \bra\psi_n|\psi\ket\, \psi_n $. 

(2)~~$L^2$ spaces, i.e.~square-integrable functions on the real line. We write (taking $[\cdot]$ to form equivalence classes of functions, and understanding the equivalence relation to be: almost everywhere equality):\footnote{Treating {\it spaces of functions} needs care, for two main reasons. With wave mechanics in mind, one {\it wants to} say: $\int \dd x\, \psi^*(x)\,\phi(x)$ is an inner product. But in Riemann (i.e.~elementary) integration theory, one does not get an inner product with the desired properties. There are two problems: (i) A semi-positive definite inner product requires that $\bra\psi|\psi\ket=0$ iff $\psi=0$. But there are many non-zero functions with zero norm. (ii) There are Cauchy sequences that do not converge. But both problems are solved by adopting Lebesgue integration. We will not give details of this and the associated measure theory. For present purposes, the benefits of adopting Lebesque integration can be summed up, in terms of our two problems, as follows. As to (i): We define an equivalence relation between functions on, say $[0,1]$, $\psi: [0,1] \rightarrow \mathbb{C}$: $f \sim f'$ iff $f$ and $f'$ are equal {\it almost everywhere}, meaning `equal everywhere except on a set of (Lebesque) measure 0'. Then the equivalence classes $[f]$ themselves form a vector space, in a natural way. That is: the inner product we intuitively want to have, viz.~$\int\dd x \, f^*(x)\, g(x)$, is well-defined on the equivalence classes, i.e.~when we read $f^*$ and $g$ in the integrand as referring to equivalence classes. For the integral's value is independent of the representatives $f, g$ that are chosen. Thus returning to (i): the equivalence class of the zero-function, $[0]$ is the unique vector with norm zero. As to (ii): This inner product space whose elements are equivalence classes (under: almost everywhere equality) of Lebesque-integrable functions $f$ with finite square integral on, say $[0,1]$, i.e.~$\int^1_0 \, | f |^2 \, dx < \infty$, is {\em complete}. That is: it is a Hilbert space. Similarly for square-integrable functions on the whole real line.}
\bea\label{L2}
L^2([0,1]) &:=& \{ [\psi] \; | \; \psi: [0,1] \rightarrow \mathbb{C}, \int^1_0\dd x\, | \psi |^2 < \infty \} \,; \\ \nonumber
L^2(\mathbb{R}) &:=& \{ [\psi] \; | \; \psi: \mathbb{R} \rightarrow \mathbb{C}, \int_{\mathbb{R}} \dd x\, | \psi |^2 < \infty \}\,.
\eea 
These two $L^2$ spaces are both separable.\footnote{For example, the functions $\{1, \surd 2 \cos 2 \pi kx, \surd 2 \sin 2 \pi kx, ... \}$, with $k = 1,2,3,...$, are orthonormal in $L^2([0,1])$; and the theory of Fourier series teaches us that they are an orthonormal basis: every Lebesgue-square-integrable function on $[0,1]$ is a limit of linear combinations of these trigonometric functions.}

We stress that any two Hilbert spaces over $\mathbb{C}$ of equal dimension are isomorphic as Hilbert spaces: ``just map one orthonormal basis onto another''.\footnote{See Prugove\v{c}ki (1981:~p.~41).} 
This applies equally to the infinite-dimensional cases. So any infinite-dimensional separable Hilbert space, e.g.~$L^2([0,1])$, is isomorphic to $l^2$. In Section \ref{dualism}, we will recall that this is the formal core of the often-cited equivalence between Schr\"odinger's wave mechanics and Heisenberg's matrix mechanics (though we will also note conceptual and historical subtleties). The wider point hereabouts concerns how expositions in physics texts often say that two pieces of formalism involve ``different Hilbert spaces''. Namely: it should be borne in mind that in almost all such expositions, the Hilbert spaces concerned are of the same dimension, and so isomorphic---and so some more fine-grained (discriminating) criterion for when to say two Hilbert spaces are ``the same'' must be meant. We will return to this later: the main such criterion will of course be unitary equivalence (already mentioned in Section \ref{dualsym}), which requires a bijection of quantities, i.e.~linear operators, on the Hilbert spaces---not just their being isomorphic.

\subsection{Fourier transformation as duality}\label{Fourier}

We begin as usual with Schr\"odinger's position representation. So we consider $L^2(\mathbb{R}^3) \ni \psi({\bf x}) \equiv \psi(x_1, x_2, x_3)$, where we think of ${\bf x}$ as position in Euclidean space; and the inner product is $\bra\phi|\psi\ket = \int_{\mathbb{R}^3} \dd^3x~\phi^*({\bf x})\,\psi({\bf x})$. In the position representation, for $l=1,2,3$, we define the position operator $X_l$ as follows (we will also use the vector notation ${\bf X} := (X_1, X_2,X_3)$):
\bea\label{position}
(X_l \,\psi) ({\bf x}) = x_l \,\psi({\bf x})\,;~~\mbox{or}~~({\bf X}\,\psi)({\bf x})={\bf x}\,\psi({\bf x})\,,
\eea
i.e.~it is a diagonal operator in this representation. For $m=1,2,3$, we define the operator $P_m$ by:
\bea\label{momentum}
 (P_m \psi)({\bf x}) := -i\hbar \frac{\partial }{\partial x_m} \psi({\bf x}) \; ;\; \mbox{ or}~~({\bf P} \psi)({\bf x}) := -i\hbar({\bf {\nabla}} \psi)({\bf x}) \,.
\eea
Both of these operators are self-adjoint, i.e.~$X^\dagger_l=X_l$ and $P^{\dagger}_m = P_m$.\footnote{If an operator $A$ has a {\it dense domain}, i.e.~a domain whose topological closure is all of $\cal H$, then we can define the {\it adjoint} of $A$, $A^{\dagger}$. Namely: dom($A^{\dagger}$) := $\{\psi \in {\cal H} \, | \, \mbox{there is a vector } \tilde{\psi} \mbox{ such that } \forall \phi \in \mbox{dom}(A): (\phi, \tilde{\psi}) = (A\phi,\psi) \, .\}$. Then we define $A^{\dagger}$ by the following assignment: $A^{\dagger}: \psi \in \mbox{dom}(A^{\dagger}) \mapsto \tilde{\psi}$. This defines $A^{\dagger}(\psi)$ uniquely (because dom($A$) is dense); and $A^{\dagger}$ is linear, and dom($A^{\dagger}$) is closed under linear combination. We say that a linear operator $A$ is {\it symmetric} iff: $A$ has a dense domain, and $(\phi, A \psi) = (A\phi, \psi)$ for all $\phi, \psi$ in the domain of $A$. Then for such an operator: for all $\psi$ in the domain of $A$, $A^{\dagger}(\psi)$ is defined, and $A^{\dagger}(\psi) = A(\psi)$. That is: for a symmetric operator, $A^{\dagger}$ is an {\it extension} of $A$. If in fact the domains are the same, i.e.~we have $A^{\dagger} = A$, then we say that $A$ is {\it self-adjoint} or {\it Hermitian}. In the case at hand, the operators $P_m$ have the symmetric property $(\phi, P_m \psi) = (P_m \phi, \psi)$ (i.e.~integration by parts), and they have dense domain, so that $P^{\dagger}_m$ is defined, and one can show that $P^\dagger_m=P_m$; similarly of course, $X_l$.}

We will not linger on the interpretation of the operators $P_m$ as representing momentum. We just note that one is led to it by deep analogies with Hamilton-Jacobi theory and with Hamiltonian mechanics' treatment of Poisson brackets: analogies which were of course in the minds of the theory's inventors, especially Schr\"odinger and Dirac.\footnote{See Dirac (1925:~pp.~314-317; 1927a:~pp.~631-632; 1927b:~pp.~244, 251), Born, Heisenberg and Jordan (1925:~pp.~328-330), and Schr\"odinger (1926a:~p.~27). For an account focussing on Pascual Jordan's implementation of canonical transformations in quantum mecanics, see Duncan and Janssen (2009).} 

The analogy with the Poisson brackets is illustrated by the canonical commutation relations between the position and momentum operators, i.e.~between Eqs.~\eq{position} and \eq{momentum}:
\bea\label{xpcomm}
[X_l,P_m]=i\hbar\, \d_{lm}\,.
\eea

The main point for us is that the spectral representation of the momentum operators is given by {\it Fourier transforms} of the spectral representation of the position operators, discussed above; as in the following:\\
\\
{\bf Theorem:} For any $\psi({\bf x}) \in L^2(\mathbb{R}^3)$, the sequence of vectors $\chi_n$, $n \in {\mathbb Z}^+$, defined by:
\be
\chi_n({\bf p}) := (2 \pi)^{-\frac{3}{2}} \int^n_{-n} \dd x_1 \int^n_{-n} \dd x_2 \int^n_{-n} \dd x_3 \;e^{-i {\bf p\cdot x}} \, \psi({\bf x})\,,
\ee 
converges to a limit vector ${\cal F}\psi$ (${\cal F}$ for `Fourier') such that $|| {\cal F}\psi ||^2 = || \psi ||^2$. ${\cal F}\psi$ is the {\bf Fourier transform} of $\psi$. Besides, the sequence of vectors
\be
\psi_n({\bf x}) := (2 \pi)^{-\frac{3}{2}} \int^n_{-n} \dd p_1\int^n_{-n} \dd p_2 \; \int^n_{-n} \dd p_3~e^{i{\bf p \cdot x}} \, ({\cal F}\psi)({\bf p}) 
\ee 
converges to $\psi$. Taking the limit $n\rightarrow\infty$, and reinserting $\hbar$ by a rescaling of ${\bf p}$ after we take the limit, we get:\footnote{The rescaling of ${\bf p}$ by $\hbar$ amounts to a {\it choice of physical units} for ${\bf p}$ and ${\bf x}$.}
\bea
({\cal F}\psi)({\bf p}) &=& (2 \pi\hbar)^{-\frac{3}{2}} \int_{\mathbb{R}^3}\dd^3x ~e^{-{i\over\hbar}\, {\bf p \cdot x}} \, \psi({\bf x})\label{Fpsi} \\
\psi({\bf x})&=& (2 \pi\hbar)^{-\frac{3}{2}} \int_{\mathbb{R}^3} \dd^3p~e^{{i\over\hbar}\, {\bf p \cdot x}} \, ({\cal F} \psi)({\bf p}) \,. \label{InvFrr}
\eea
{\bf Plancherel's theorem:} since ${\cal F}$ preserves norm and has an inverse, it is {\it unitary}, and so preserves inner products:\footnote{See Gustafson and Sigal (2020:~p.~406) and Jauch (1968:~p.~64).} 
\bea\label{innerpr}
\bra{\cal F}\f|{\cal F}\psi\ket=\bra\f|\psi\ket\,.
\eea
As we will discuss in the next Section, this is a central result for position-momentum duality's illustrating the Schema.

We write the inverse of ${\cal F}$ as:
\be\label{Finverse}
({\cal F}^{-1} \phi) ({\bf x}) = (2 \pi\hbar)^{-\frac{3}{2}} \int_{\mathbb{R}^3}\dd^3p~ e^{{i\over\hbar}\, {\bf p \cdot x}} \, \phi({\bf p}) \,\Leftrightarrow\,({\cal F}^{-1} \phi) ({\bf x}) = ({\cal F} \phi)( - {\bf x}) \, . 
\ee
\\
{\bf The momentum representation}. The inverse of the Fourier transformation, Eq.~\eq{Finverse}, allows us to give expressions for $P_m$ and $X_l$ in the momentum representation (recall that, at the beginning of this Section, these operators were defined in the position representation). Eq.~\eq{InvFrr} implies that, in the position representation:
\bea
(P_m \psi)({\bf x}) = (2 \pi\hbar)^{-\frac{3}{2}} \int_{\mathbb{R}^3}\dd^3p~e^{{i\over\hbar}\, {\bf p \cdot x}}\,p_m \, ({\cal F} \psi)({\bf p}) \,.
\eea

A vector $\psi$ is in the domain of $P_m$ iff $p_m({\cal F}\psi)({\bf p})$ is square-integrable, in which case (cf.~integration by parts with boundary term vanishing):\footnote{See also Landsman (2017:~p.~181).}
\bea\label{beta}
({\cal F}P_m \psi)({\bf p}) = p_m\, ({\cal F} \psi)({\bf p}) \, .
\eea
We can rewrite this as (where the universal quantification on the left allows us to `peel off the wave function' on the right, and $p_m$ is the argument of the wave-function $\psi({\bf p})$):
\bea\label{FPF}
\forall\psi\forall{\bf p}\,,~({\cal F}P_m{\cal F}^{-1}\,{\cal F}\psi)({\bf p})=p_m\,({\cal F}\psi)({\bf p})~~\Rightarrow~~{\cal F}P_m{\cal F}^{-1}=p_m\,.
\eea
Thus the momentum operator {\it in the momentum representation}, i.e.~${\cal F}^{-1}P_m{\cal F}$, is diagonal.

Likewise, we can write the position operator $X_l$ in the momentum representation. Writing Eq.~\eq{Fpsi} for the wave-function $X_l\,\psi$ (i.e.~using the same square-integrability and integration by parts arguments), and using Eq.~\eq{position} on the right-hand side, we derive:
\bea\label{QlFpsi}
({\cal F}X_l\psi)({\bf p})&=&(2\pi\hbar)^{-{3\over 2}}\int_{\mathbb{R}^3}\dd^3x~e^{-{i\over\hbar}\,{\bf p}\cdot{\bf x}}\, x_m\,\psi({\bf x})=i\hbar\,{\pa\over\pa p_l}\,{\cal F}\psi({\bf p})\,\nn
\Rightarrow~~{\cal F}X_l{\cal F}^{-1}&=&i\hbar\,{\pa\over\pa p_l}\,,
\eea
where $p_l$ is the argument of the wave-function that the operator acts on.

We can similarly now connect with Dirac notation for momentum-space. We do not need to exhibit the details. But for example:
\bea\label{P3first=last}
{\bf P} = \int_{\mathbb{R}^3} \dd^3p~ {\bf p\, |p \rangle \langle p}| \, .
\eea 

This discussion of the Fourier transform as a unitary map is a template for our later discussions of {\it unitary equivalence}:\footnote{Unitary equivalence will for example recur in Section \ref{bosoniz}, in the example of bosonization in two-dimensional quantum field theory. For a quantum system with infinitely many degrees of freedom, i.e.~a quantum field or a quantum statistical mechanical system in the limit of infinitely many components (e.g.~an infinite lattice), one may need---in order to describe the various possible physical behaviours of the system---unitarily {\it inequivalent} algebra of operators. But in a certain precise sense: a finite system does not need unitarily inequivalent algebras. See also, at the end of Section \ref{intext}, the brief discussion of unitarily inequivalent representations in the algebraic approach to quantum theory.} 
i.e.~the idea that a single unitary operator $U$ ``carries'' each operator $A$ in an algebra of operators ${\cal A} \ni A$ that represents the system to the corresponding element of another algebra: namely to $ U A\, U^{-1} \in {\cal A}' := U{\cal A}\,U^{-1}$. This is clear in the last two equations above: Eq.~\eq{FPF} and \eq{QlFpsi}. A single unitary operator ${\cal F}$ carries the operators defined in the position representation to the corresponding operators in the momentum representation (and this argument generalizes to functions of these operators). In short, we have a unitary equivalence between the two abelian i.e.~commutative (and maximal) von Neumann algebras of functions of $Q$, and of $P$, respectively. This will be the basis of how, in the next Subsection, the Fourier transformation between position and momentum illustrates the Schema for duality.\footnote{As we said at the end of Section \ref{iHs}, unitary equivalence amounts to more than Hilbert space isomorphism, since the structure of an algebra of operators is being preserved. So again, as we mentioned there: saying that unitary equivalence means there is ``just one Hilbert space'' in question, or that unitary inequivalence is a matter of ``different Hilbert spaces'', is rough speaking.}

\subsection{Illustrating the Schema}\label{illusSc}

Here, we lay out how position-momentum duality illustrates our Schema: in particular, the isomorphism from Section \ref{isomdef}. The idea here is that duality is a passive transformation between expansions of the state. 

To illustrate the Schema with a single spinless non-relativistic particle in three spatial dimensions, one naturally takes the bare theory to be the Hilbert space $L^2(\mathbb{R}^3)$, equipped with its algebra of linear operators.\footnote{For simplicity, take bounded operators. Unbounded operators like $Q$ and $P$ are then treated in terms of their spectral projectors.} 
Then the idea will of course be: one dual, the ``left dual'' $M_1$, is ``a matter of'' position, while the other, the ``right dual'' $M_2$, is ``a matter of'' momentum. We write the duals as $M_x$ and $M_p$, respectively.

It is usual to think of specifying the left or right dual as just `a choice of basis'. More precisely, so as to respect the fact that position and momentum are continuous quantities, and so do not have eigenvectors: as `choices of a spectral family of projectors'. 

The unitary Fourier transformation is then a passive transformation sending an arbitrary physical state, expressed in position representation, to the momentum representation of the same physical state: and so preserving all quantities' expectation values. 

A bit more precisely: our state-space ${\cal S}$ is the (abstract) infinite-dimensional separable Hilbert space ${\cal H}$ over $\mathbb{C}$. The set of quantities ${\cal Q}$ is the set of position and momentum operators, $\hat X_l$ and $\hat P_m$, satisfying the canonical commutation relation, Eq.~\eq{xpcomm} (and other quantities that are self-adjoint functions of these operators). The hats here indicate that these are operators on the Hilbert space, for which we have not yet chosen a representation in terms of position or momentum. The dynamics is determined by a choice of a Hamiltonian $H(\hat P,\hat X)$, so that (in the Schr\"odinger picture) the wave-function satisfies the Schr\"odinger equation.

\begin{figure}
\begin{center}
\includegraphics[height=2.2cm]{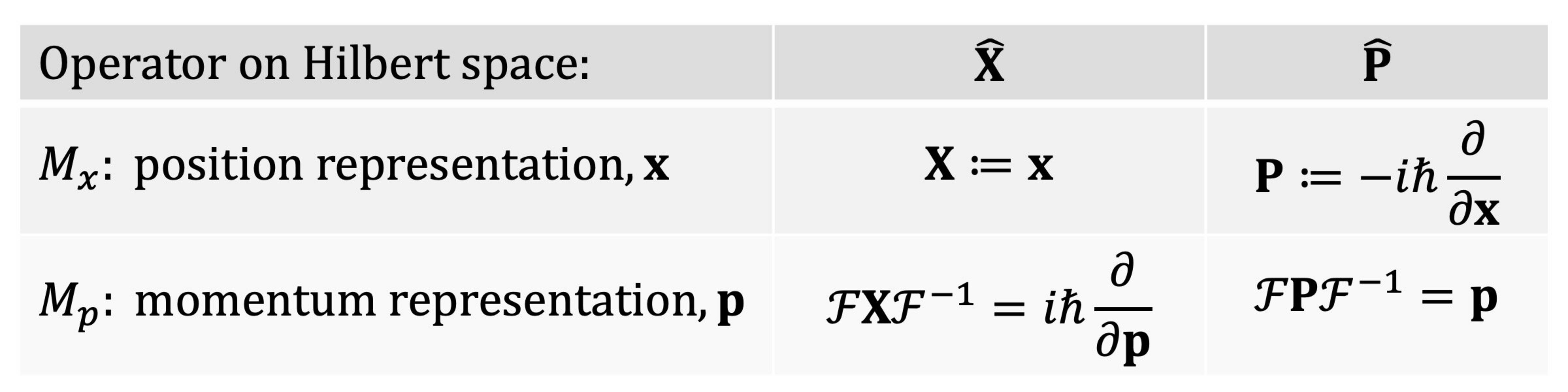}
\caption{\small Position and momentum operators, and their representations. Top row: representation-independent operators, ${\bf\hat X}$ and ${\bf\hat P}$. Middle row: model in the position representation, $M_x$. Bottom row: model in the momentum representation.}
\label{xp-rep}
\end{center}
\end{figure}

The two models, $M_x$ and $M_p$, are the position and momentum representations of this Hilbert space, that are obtained by the corresponding choice of a spectral family of projectors (thus our discussion here is a generalization of the one-dimensional example given in item (3) of `Theory', in Section \ref{Ourthm}). The state-spaces of these models are both $L^2(\mathbb{R}^3)$, in either representation. This is summarized in Figure \ref{xp-rep}.

Thus $M_x$ has as its state-space ${\cal S}_x$ the space of square-integrable wave-functions $\psi({\bf x})$ (i.e.~the three-dimensional version of the state-space Eq.~\eq{L2}). The quantities ${\cal Q}_x$ are the operators $X_l$ and $P_l$ in the position representation, i.e.~given in Eqs.~\eq{position} and \eq{momentum} (and functions of these operators). The dynamics is given by the Hamiltonian $H(P,X)$, written in the position representation.

Likewise for $M_p$, which has as its state-space ${\cal S}_p$ the space of square-integrable functions of the wave-number, $\ti\psi({\bf p})$, where the position and momentum operators are represented as in Eqs.~\eq{FPF} and \eq{QlFpsi}.

(i)~~The duality map $d_{\cal S}$ from Section \ref{isomdef} for {\it states} is the Fourier transformation, ${\cal F}:L^2(\mathbb{R}^3)\rightarrow L^2(\mathbb{R}^3)$, i.e.~$d_{\cal S}={\cal F}$. It assigns: $d_{\cal S}:\psi({\bf x})\mapsto\ti\psi({\bf p})=({\cal F}\psi)({\bf p})$, as in Eq.~\eq{Fpsi}. 

(ii)~~The duality map $d_{\cal Q}$ for {\it quantities} maps an operator $Q$ in the position representation to the corresponding operator, $d_{\cal Q}(Q)={\cal F}Q{\cal F}^{-1}$, in the momentum representation. For the operator $X_l$ (with value $x_l$ on a wave-function) in the position representation (see Eq.~\eq{position}), this gives: $d_{\cal Q}(X_l)=i\hbar\,{\pa\over\pa p_l}$, and for the operator $P_m$ in the position representation, i.e.~Eq.~\eq{momentum}, we get: $d_{\cal Q}(P_m)=p_m$ (the last equality gives the value on a wave-function in the momentum representation). 

(iii)~~The equivariance with respect to the dynamics is the fact that the dynamics of $M_x$ is the Schr\"odinger equation with the Hamiltonian $H(P,X)$ and wave-function $\psi({\bf x})$ in the position representation, and the dynamics of $M_p$ is the appropriately transformed Schr\"odinger equation, i.e.~with Hamiltonian $d_{\cal Q}(H)={\cal F}H{\cal F}^{-1}$ and wave-function $\ti\psi({\bf p})$ in the momentum representation. 

(iv)~~The preservation of any expectation value between the `left' and the `right' amounts, essentially, to an inner product being the same when evaluated in two different orthobases, related by a unitary map: see Eq.~\eq{innerpr}. Thus for any quantity $Q\in{\cal Q}$, and for any state $\psi$, we have that their Fourier transforms are $\ti Q={\cal F}Q{\cal F}^{-1}$ and $\ti\psi={\cal F}\psi$, respectively. Then we have (relating the notation in Eq.~\eq{obv1} to the inner product in quantum mechanics):
\bea
\bra Q,\psi\ket_x&:=&\bra\psi|\,Q\,|\psi\ket_x=\bra{\cal F}^{-1}\ti\psi\,|\,{\cal F}^{-1}\ti Q\,{\cal F}\,|\,{\cal F}^{-1}\ti\psi\ket_x\\
&=&\bra\ti\psi|\,\ti Q\,|\ti\psi\ket_p=:\bra\ti Q,\ti\psi\ket_p
=\bra d_{\cal Q}(Q),d_{\cal S}(\psi)\ket_p\,,\nonumber
\eea
where the subscripts $x$ and $p$ indicate the inner products in the corresponding representations. Thus we reproduce Eq.~\eq{obv1}. (This generalizes, in the obvious way, to matrix elements between different states.)

\section{Wave-particle duality}\label{wpd}

Perhaps the best-known, and oldest, use of the word `duality' in modern physics is for `wave-particle duality', which was seen as one of the central foundational insights of quantum mechanics. 

However, one should beware that wave-particle duality is not a single duality, but rather a collection of different principles with varying degrees of (im)precision. Thus this Section takes up three main tasks:

(1)~~Distinguish the different uses of `wave-particle duality'. 

(2)~~Clarify the relation of these various uses to the conception of duality used in our Schema. 

(3)~~Distinguish `duality' from another much-used notion: {\it complementarity}.

Section \ref{dualism} will take up tasks (1) and (2): about (1), we will distinguish three different meanings of `wave-particle duality', roughly speaking as it was used: (i) in the old quantum theory; (ii) in the new quantum theory, i.e.~quantum mechanics; (iii) in early (quantum) field theory. 

About (2), our answer will be that, of these three meanings, it is only (ii) and (iii), and not (i), that are something like the meaning in our Schema. Namely, the equivalence between matrix and wave mechanics turns on the equivalence between $L^2$ and $l^2$ Hilbert spaces, as completed by von Neumann (1932).

Sections \ref{complement} and \ref{adcom} take up (3), i.e.~they will contrast duality and complementarity, and argue that these are different concepts. This brief review of the debates surrounding wave-particle duality and complementarity will be a prelude to ideas currently being used in quantum field theory and black hole physics, as well as to debates about scientific understanding (discussed in Chapter \ref{Understand}).\\
\\
{\bf On the history of the word `duality' in quantum theory.} To the best of our knowledge, a history of the term `duality' in physics is yet to be written. And its usage in modern physics, in the sense of our Schema, is likely to be the result of converging lines of work, originating in quantum theory and field theory (perhaps with influences from the usage of `duality' in mathematics). 

Although we do not intend to give definitive answers to these historical questions, there are two preliminary points that we can make in connection with the use of `duality' in quantum theory and in electrodynamics. For it is clear that Heisenberg, Dirac, Pauli, and Jordan (not Bohr) were influential figures in both establishing something like the modern usage of the term `duality', and in aligning it (by equivocation, we would now want to say) with the usage in the old quantum theory. For already in his 1930 Chicago lectures, Heisenberg (1984) used the word `duality' in both senses (i) and (ii) that we discuss below: (i) in terms of the dual interpretation of light as a wave-like interference or as a corpuscular phenomenon in the old quantum theory (ibid, p.~121); and (ii) as a mathematical equivalence between matrix and wave mechanics. His use of the word `duality' in these two senses in a single well-known set of lectures was a conscious choice (ibid, p.~164): 

\begin{quote}\small
The problem of the quantum theory centers on the fact that the particle picture and the wave picture are merely two different aspects of one and the same physical reality. Although this is a problem of purely physical nature it is satisfying to find a counterpart to this duality in the mathematical apparatus of the theory. The analogy consists in the fact that one and the same set of mathematical equations can be interpreted at will in terms of either picture.
\end{quote}
Thus we will argue that, although he {\it did} distinguish the two meanings (i) and (ii): by claiming that there is an analogy between them, he was contributing to establishing {\it a new meaning for an old word}.

Second, Dirac (1931:~p.~71) contributed to establishing the notion of duality, in the different context of {\it electrodynamics}, through his discovery of magnetic monopoles (we will discuss this in Section \ref{EMduality}): `The theoretical reciprocity between electricity and magnetism is perfect'.\footnote{In this paper, Dirac did not use the word `duality', but rather `reciprocity'.} 

Our distinction between the different usages of `duality', and the contrast with `complementarity', should cast light on how the word came to have its current meaning in physics.

\subsection{`Dualism' and matrix vs.~wave mechanics}\label{dualism}

With these preliminaries in hand, we move on to our first task, of distinguishing the three different meanings of `wave-particle duality' in early quantum mechanics: (i) duality of wave and particle {\it properties of radiation and matter}, often called wave-particle `dualism'; (ii) duality of {\it matrix mechanics and wave mechanics}; (iii) duality of waves and particles for {\it non-relativistic fields}.

As we will discuss, of these three putative dualities, the equivalence between matrix mechanics and wave mechanics, i.e.~(ii), taken in the sense of von Neumann (1932:~p.~29), agrees best with the modern sense of `duality' as an isomorphism between state spaces and quantities that is equivariant for the dynamics (see Section \ref{dualism}-(ii)).\\
\\
(i)~~{\it Wave-particle dualism}. In the early days of quantum theory, talk of wave-particle `dualism', or {\bf dualism between waves and corpuscles}, arose from the need to ascribe to both matter and radiation, in interaction, both {\it wave} and {\it particle} properties.\footnote{For a recent discussion, see Duncan and Janssen (2023:~pp.~117-126). Among the authors that we are going to discuss in some detail, see Heisenberg (1925a:~p.~841; 1926:~p.~994), Bohr (1985:~p.~8), Jordan and Klein (1927:~p.~751), de Broglie (1930:~pp.~9, 10, 145; 1960:~pp.~89-91, 287, 289) [but he also uses `duality': see de Broglie, 1930:~pp.~3, 6, 134)]. As one can see from his collected works, Bohr very infrequently used the word, in either its `dualism' or `duality' versions: between 1926-1932 it appears only twice (Bohr, 1985:~pp.~27, 108) and slightly more frequently (in nine different places) between 1933-1958 (Bohr, 1996). We will contrast complementarity with duality in Section \ref{complement}. This contrasts with his prolific use of {\it complementarity}, about 200 times in each volume of his collected works between 1926-1958. For example, while in the {\it Nature} write-up of his famous Como lecture there is no talk of `duality', `complementarity' is used abundantly and in different phrases: `complementary features of the description', `complementarity theory' (Bohr, 1928:~p.~580), `complementary pictures of the phenomena', `complementary nature of the description' (ibid, p.~581), `complementary character of the description' (ibid, pp.~581, 583, 584).\label{fcomp}}
This dualism arose between 1905 and 1925 in the efforts to explain the experimental results in the old quantum theory\footnote{The application of the dualism in these experiments built on Planck's earlier idea of energy quantisation, which he used for the description of black body radiation in 1900. Planck's quantum hypothesis was that the electrically charged harmonic oscillators in a black body cavity are harmonic oscillators with an individual energy $E=h\nu$, where $\nu$ is their frequency of oscillation and $h$ is Planck's constant (with the dimension of action, i.e.~energy $\times$ time, i.e.~$\sm{J}\cdot\sm{s}$), which Bohr called `Planck's quantum of action'. Planck's key assumption was Boltzmann's entropy formula $S=k\log W$ for the oscillators, where $W$ is the number of ways to distribute the total energy among the $N$ oscillators in equal parts. For more details, see Duncan and Janssen (2023:~pp.~71-77), Kuhn (1978~pp.~100-110)
Jammer (1989:~pp.~8-20), and Murdoch (1987:~pp.~1-3).}---and given the apparently contradictory features of `wave' and `particle', was of course seen as a problem. We now give some of the details of this story.

In Einstein's 1905 explanation of the photo-electric effect, electrons were ejected from a metal by the energy transferred by light quanta, later called photons, incident on the metal (ultraviolet, X-rays, and $\g$-rays). Since the energy of the ejected electrons increases linearly with the frequency of the incident radiation, the photon explanation is {\it incompatible} with the Maxwell theory, which describes light as an excitation of the classical electromagnetic field, i.e.~a wave, and predicts that the electron's energy is proportional to the incident radiation's {\it intensity}. By contrast, Einstein assumed that light consists of {\bf discrete energy quanta}, i.e.~photons, whose individual energies are given by Planck's formula:\footnote{In 1916, Einstein used Planck's law, as well as Bohr's idea of the stationary states of the atom, to calculate the rate of emission and absorption of radiation by an atom.}
\bea\label{Planckf}
E=h\nu\,.
\eea

Seventeen years later, in 1922, the corpuscular nature of light was further confirmed by the {\it Compton effect}, in which a high-energy photon collides with an electron, thereby deflecting the electron's trajectory (thus transferring momentum to it) and reducing the photon's energy (and thus its wavelength). In the Compton effect, even though the photon is characterized by a wavelike property (namely, its wavelength), the collision is accurately described as a collision of two {\it pointlike particles} with definite energy and momentum, and the deflection angle of the photon can be calculated using the localized energy and momentum conservation laws for point particles.\footnote{See Stuewer (1975:~pp.~217-273).}

Furthermore, de Broglie's 1923 doctoral dissertation extended the idea of wave-particle duality from radiation to {\it matter} (in particular, electrons). He proposed that, just like light, bits of matter have a wave-like nature that is characterized by the frequency, $\nu$ in Planck's formula Eq.~\eq{Planckf}, and the following wavelength:
\bea\label{deB}
\l={h\over p}~\Rightarrow~ p=\hbar k\,,
\eea
where the {\it reduced Planck constant} $\hbar$, and the wave number $k$, are defined by: $\hbar:=h/2\pi$ and $k:=2\pi/\l$. De Broglie's wave hypothesis was subsequently confirmed in wave diffraction experiments in the following four years.\footnote{For a brief history of de Broglie's work between 1922 and 1925, see Klein (1964:~pp.~31-33). For a history of the experimental confirmation of de Broglie's hypothesis between 1925 and 1927, see (ibid, pp.~38-39). De Broglie's pilot wave theory incorporates waves and particles into a single description, where the phase of the wave guides the motion of the particle through a guidance equation that relates the velocity of the particle to the gradient of the phase. In de Broglie's (1930:~p.~80; 1960:~pp.~91, 99) `double solution' theory, only the phase and not the amplitude of the wave is physical. But the double solution does not entail a {\it duality} between waves and particles, since there is only an approximate formal relationship between the (linear) wave equation and the (non-linear) equation of motion of the particle. Furthermore, the particle features only appear in the non-linear regime. For de Broglie's own recounting of the Solvay conference, including his change of views and his views on Bohm's papers, see de Broglie (1930:~pp.~1-6; 1960:~pp.~90-92). For a general dilscussion, see Cushing (1984:~pp.~118-121, 126-128).}\\
\\
(ii)~~{\it Duality between matrix mechanics and wave mechanics}. The duality between matrix and wave mechanics was only completed by von Neumann's (1932:~p.~29) use of a theorem of Fischer and Riesz that establishes the isomorphism between the spaces of functions defined on a discrete configuration space and on a continuous configuration space (i.e.~between sequences which are functions on the discrete configuration space, $l^2$, and wave functions on the continuous configuration space, $L^2$: see (1) and (2) in Sections \ref{iHs}, especially Eq.~\eq{L2}). Indeed, all complex infinite-dimensional separable Hilbert spaces are isomorphic to $l^2$.\footnote{See Prugove\v{c}ki (1981:~p.~41). For a discussion of the status of the equivalence between matrix and wave mechanics in 1926, see Muller (1997a-b). For further discussion of the relevant isomorphism, see De Haro (2021:~p.~5151).}

With the new quantum theory developed between 1925 and 1927, the notion of wave-particle duality was applied, but also transformed, within the new framework of matrix mechanics and wave mechanics (recall, from the preamble of this Section, that Heisenberg gave ``a new meaning to an old word''). In this framework, it was often understood as the contrast between discontinuity and continuity.

Heisenberg saw matrix mechanics as giving a {\it discontinuous} description of nature: a discontinuity that he took to be demanded by Planck's quantum postulate. This discontinuity was not put in by assuming a `quantum state' (as in Bohr's stationary states of the hydrogen atom), but rather as `quantum numbers which are indeed no more than distinguishing indices' (Born and Jordan, 1925:~pp.~300-301). The whole theory followed from his commutation relations, Eq.~\eq{xpcomm},\footnote{These were first written down explicitly by Born and Jordan (1925:~p.~282).} 
which reproduced both discrete and continuous energy spectra. 

Heisenberg's claim of the theoretical equivalence of wave and particle models rested on Dirac's (1927a:~pp.~627-628, 635) {\it transformation theory}, i.e.~the idea that any quantum system can be described by different sets of dynamical variables, and that there is a unitary transformation that relates them.\footnote{But we note that Schr\"odinger had already stated the equivalence, of heuristic terms, in one of his great 1926 papers (1926b).} 
By using a canonical transformation that diagonalizes the Hamiltonian, Dirac argued that matrix mechanics and wave mechanics are special cases of his more general formalism, and are related by a unitary transformation.\footnote{He also showed that the matrix elements of any physical quantity are equal to the inner products of the corresponding wave functions. The transformation introduced by Dirac included his delta function, which von Neumann (1932:~pp.~ix; 28) criticized as `Dirac's mathematical fiction', and which allowed the transformation between discrete and continuous sets of indices (see Section \ref{dualism}-(ii)).}

We cannot here dwell on Heisenberg's changing views on wave vs.~particle duality during this period:\footnote{This interesting story is told in Beller (1999:~pp.~18-26, 68-77) and Camilleri (2009:~pp.~67-72), both of whom also contrast Heisenberg and Bohr. In 1926, Heisenberg already held a view similar to his `mature' view (see (iii) below), for he wrote in a letter to Dirac: `I see the real progress made by Schr\"odinger's theory in this: that the same mathematical equation can be interpreted as point-mechanics in a non-classical kinematics {\it and} as a wave theory according to Schr\"odinger' (Mehra and Rechenberg (2000:~p.~202)). At the same time, he was critical of Schr\"odinger's attempts to interpret his own theory as a theory of waves.\label{lettD}}
but in his polemic with Schr\"odinger, Heisenberg adopted the view that quantum mechanics---especially in his preferred matrix mechanics formulation---is a {\bf corpuscular theory}.\footnote{See Camilleri (2009:~pp.~69, 72) and Beller (1999:~pp.~71, 77).}

Having said that, Heisenberg's view that matrix mechanics is a theory of particles is disputable, not just in the light of the correspondence between matrix and wave mechanics, but for two other reasons: First: matrix mechanics was itself built using {\it wave} concepts (see Beller, 1999:~pp.~19, 24-26, 71-75). Second, consider the general conception of {\bf particle}: having discrete properties is surely too weak a condition for something to be a particle, since, in addition, localization in space (at a point, or over a region of space) is usually also required.\footnote{For a discussion, see Teller (1995), especially Chapters 2 to 4, Duncan (2012:~pp.~98-99, 111-114), and Kuhlmann (2010:~pp.~84-85).}

Schr\"odinger (1926a:~p.~27), on the other hand, delighted in his {\it undulatory} representation of the mechanics of atoms, which he regarded as {\it physical}, and which gave him {\it understanding} (p.~30).\footnote{For the competition between Heisenberg and Schr\"odinger, see Beller (1999:~pp.~17, 26, 29, 32-39). For a discussion of {\it Anschaulichkeit}, see (ibid, pp.~67-70) and de Regt (2017:~pp.~226-251). We will also discuss the requirement of {\it Anschaulichkeit} for understanding, at the end of Section \ref{adcom}. We will give a philosophical discussion of understanding in Chapter \ref{Understand}.}
Using a wave theory, he hoped to `eliminate the irrationalities of quantum jumping' (Beller, 1999:~p.~74). (We will discuss these philosophical points in Section \ref{adcom}.)

But, just as Heisenberg's matrix mechanics is not a pure theory of particles, neither is Schr\"odinger's mechanics a theory of waves in the ordinary sense. At best, these are {\it probability} waves in an {\it abstract space}: the $3N$-dimensional configuration space of $N$ (spinless) particles. Thus, except for a single spinless particle ($N=1$), the wave-function $\psi$ {\it cannot} be interpreted as a wave in three-dimensional space. Nor does it give a straightforward transition from micro- to macro-mechanics, as Schr\"odinger originally hoped. Beller (1999:~p.~39) aptly summarizes the outcome of the debates between Heisenberg (and the Copenhagen cabal) and Schr\"odinger as follows: `Schr\"odinger's methods proved indispensable. His philosophy [i.e.~his wave interpretation] did not'.\\
\\
Heisenberg's more mature interpretation of wave-particle duality, in his quotation given in the preamble of this Section, is relevant here, because it is closer to the interpretation of duality we advocate in this book: more specifically, the idea that the two duals ``point to'' a common core theory (see Section \ref{isomdef}). Namely, the ``new meaning'' of `duality' was that (up to a unitary transformation) a single set of mathematical equations can be interpreted using a classical picture that involves either particles or waves. And, although the classical theories are different, their corresponding quantum theories are the same. Thus this is analogous to our idea, in Section \ref{std}, of {\it quantum duality}, i.e.~of classical models that are not duals but whose corresponding quantum models {\it are} duals.\footnote{The reason for saying `analogous' here lies in our reservations, mentioned above, about identifying Heisenberg's matrix mechanics with `a classical corpuscular model', and Schr\"odinger's mechanics of waves with `a classical undular model', even in special classical limits.}

In this mature and `definitive' version of quantum theory (already foreshadowed in his 1926 letter to Dirac: see footnote \ref{lettD}), Heisenberg interpreted wave-particle duality as the {\it formal and physical equivalence} of the matrix and wave formulations of quantum mechanics.
Thus, in effect, Heisenberg advocated that the models are theoretically equivalent, i.e.~they describe the `same sector of reality', as in an internal interpretation of a duality (see Section \ref{dualint}).

On Heisenberg's view, the resolution of the conundrum of duality, i.e.~the incompatibility of the wave and particle pictures, was that `waves' and `corpuscles' are interpretations according to {\it incompatible classical models} whose corresponding {\it quantum mechanical models} are fully equivalent. For Heisenberg, it was the development of quantum field theory that made this a viable interpretation.\\
\\
(iii)~~{\it Enter quantum fields: revisiting wave-particle duality.} From 1927 onwards, the quantum theory of fields saw a rapid development that, by contrast to the wave-particle `dualism' of the old quantum theory (cf.~(i)), included a reasonably precise version of wave-particle `duality', in our sense of `duality'.

Unless complemented by a notion of locality, the reservations in Section \ref{dualism}-(ii) also apply to the sense in which `particle' is here used. In modern quantum field theory, the {\bf one-particle states} are identified with irreducible representations of the symmetry group (usually, the Poincar\'e group). The states of particles that are sent in to collide in an accelerator, and the states of the particles that come out, are taken to be direct products of one-particle states, with their spacetime dependence (see e.g.~Weinberg (1995:~p.~107)). The spacetime dependence can be approximated by free particle states that are arbitraily close to interacting states, provided the interaction potential is sufficiently short-ranged (Duncan, 2012:~p.~99). Beyond these idealizations, the applicability of the notion of `particle' in quantum field theory, for both rigorous and heuristic reasons, is limited: and so, `particle' needs to be understood differently from ordinary quantum mechanics. Among the rigorous reasons are Malament's (1996) theorem, which amounts to saying that, in relativistic quantum mechanics with a fixed number of particles, particles cannot be localized;\footnote{See also Hegerfeldt (1998a-b), where, for either relativistic or non-relativistic theories, localization and positivity of the Hamiltonian are shown to allow super-luminal propagation. For a rebuttal of objections to these theorems, see Halvorson and Clifton (2001).} 
as well as the Unruh effect (but see Arageorgis et al., 2002). Among the heuristic reasons is the presence of vacuum fluctuations. But the notion of a field is of course also not the same as in classical field theory.\footnote{For a summary of the issues that arise for particles in quantum field theory in curved spacetimes, see Wald (1994:~pp.~3-6). Among recent philosophical discussions, see Sebens (2022:~p.~380) and Kuhlmann (2010:~pp.~99-103, 117).}

Starting with Dirac's (1927b) seminal work on quantum electrodynamics, i.e.~the quantum theory of the interactions between atoms and radiation, Jordan, Klein and Wigner\footnote{See Dirac (1927, 1958), Jordan (1927), Jordan and Klein (1927) and Jordan and Wigner (1928). These works of course build on earlier work, especially on the {\it Dreim\"annerarbeit} by Born, Heisenberg and Jordan (1925), who generalized matrix mechanics to an arbitrary number of (pointlike) harmonic oscillators, and related this to the quantized electromagnetic field. See also Heisenberg's (1984:~pp.~161-165) proof of equivalence of the quantum theory of particles and of waves. For a detailed discussion of Jordan's work, see Duncan and Janssen (2008). Our focus here is not the early history of quantum field theory {\it per se}, but rather its significance for wave-particle duality. A beautiful brief historical introduction to quantum field theory is in Weinberg (1995:~pp.~1-38): see also Duncan (2012:~pp.~1-56). Detailed historical accounts are Darrigol (1986), Schweber (1994:~Chapters 1 and 2), and Cao (1997:~Chapter 7).}
developed quantum theories for systems of many particles that made a version of wave-particle duality explicit. The idea was to quantize the wave-equation describing {\it matter} \`a la de Broglie: namely, the (non-relativistic) interactions between an atom and its surrounding radiation, thereby obtaining a discrete spectrum of states that could be reformulated in terms of an interacting gas of quantum oscillators, i.e.~point particles, satisfying either Bose or Fermi statistics. As it turned out, the corpuscular property of having a discrete number of particles was described by a number operator canonically conjugate to the wave's phase.\footnote{See also Schweber (1994:~p.~27). This procedure in non-relativistic quantum field theory is called `second quantization' (Dirac, 1958:~pp.~230, 292; Schweber, 1961:~pp.~130-137): the idea was that the Schr\"odinger equation for a wave gets quantized, so that the wave becomes a field. In modern quantum field theory, there is a single quantization procedure: namely, one quantizes the field's equation of motion. See e.g.~Peskin and Schroeder (1995:~p.~19). Weinberg (1995:~p.~28) recommends that `It would be a good thing if the misleading expression `second quantization' were permanently retired'. Also, in modern quantum field theory the number of particles is {\it not a constant}: see Weinberg (1995:~p.~3) and Malament (1996:~p.~6).\label{retire}}

Heisenberg, along with Jordan, Klein and Wigner, pointed to a key difference between their new wave theories of matter, i.e.~quantum field theories, and Schr\"odinger's wave theory: unlike Schr\"odinger's theory, which described a wave on a $3N$-dimensional configuration space of $N$ particles, quantum field theories were defined on an ordinary three-dimensional space (four-dimensional in the case of relativistic quantum field theories). Their amplitudes were not probability amplitudes \`a la Born, but amplitudes that count numbers of particles.\footnote{For a discussion of the interpretation of Schr\"odinger's wave-function, see Camilleri (2006:~pp.~302-304). For Heisenberg, Jordan, Klein, and Wigner's view that the quantized waves lived in ordinary 3-space, see (ibid, pp.~306-308).} 

These developments strengthened Heisenberg's view that particle-wave duality gives a complete {\it equivalence} between pictures, i.e.~the differences being a matter of (imperfect) interpretation of equivalent sets of equations.\footnote{Schr\"odinger (1926b:~pp.~58-59), in the paper in which he argues for the equivalence of matrix and wave mechanics, was careful to make a distinction between formal or mathematical equivalence and physical equivalence. Mathematical equivalence is isomorphism; physical equivalence in addition requires that two models have the same domains of application, even in future domains (thus he anticipated our {\it unextendability} condition: see Section \ref{ejpe}).}
Any process could be described in terms of particles or in terms of waves; one description might, for practical purposes such as e.g.~visualization, be more convenient in a particular situation, but the two pictures are in principle equivalent. To express this, he used the words `symmetry' and `equivalence'.\footnote{See Heisenberg (1984:~p.~164), i.e.~Chapter 11 of his Chicago lectures. For a discussion, see Camilleri (2006:~pp.~299). This was also a crucial point in the disagreement between Heisenberg and Bohr: see Beller (1999:~pp.~226-227) and Camilleri (2006:~p.~310).}

\subsection{Complementarity}\label{complement}

The previous Section discussed Heisenberg's judgment that wave-particle duality, in sense (ii), is a case of {\it equivalence}: and so, that it is a duality in our sense in this book. This contrasts with Bohr, who did not hold an equivalence, but rather a {\it complementarity}, thesis. For Bohr, complementarity was the answer to the {\it paradoxes} of wave-particle duality. 

Bohr was a notoriously obscure thinker. But in recent years commentators have made progress extracting a reasonably coherent picture of two of his views about complementarity that are of interest to our topic, especially in his 1927 Como lecture and in his 1935 reply to EPR.\footnote{See Faye (2019) and Murdoch (1987). Beller (1999) is a dialogical view that complements, and to a certain extent corrects, previous work. Other interesting accounts include Scheibe (1973), Jammer (1989), and Held (1994).}
Although Bohr used the word `complementarity' in multiple ways (see footnote \ref{fcomp}), we will focus on its two main meanings (i.e.~(A) and (B) below).\footnote{As we mentioned in footnote \ref{fcomp}, Bohr rarely used the word `duality' in writing. Commentators have distinguished two main senses in which he used it, both of them along the lines discussed immediately below: (i) Formal, i.e.~the sense contained in the two main {\it equations} of the old quantum theory by Planck, Einstein, and de Broglie, i.e.~Eqs.~\eq{Planckf} and \eq{deB}. In this formal sense, the quantities $(E,p)$ are typical of particles, while $(\nu,\l)$ are typical of waves. (ii) Empirical, i.e.~the ways in which experiments are explained: the Compton effect requires a particle model of the collision between photons and electrons, while intereference effects require a wave model. See Murdoch (1987:~p.~62).}

In physics, two concepts are {\bf complementary} if they:\footnote{Hilgevoord and Uffink (2016:~Section 3.1) also mention `complementary phenomena'. However, we restrict our discussion to the epistemic aspects of complementarity, i.e.~complementary concepts, descriptions, and properties ascribed to objects or physical systems.}

(a) have different meanings: indeed, they are incompatible, i.e.~mutually exclusive; and

(b) are individually insufficient, but jointly sufficient, for a complete description of a physical system.

In Bohr's view, complementarity resolved the paradoxes of wave-particle duality, because (by (b)) the particle and wave descriptions, taken jointly, give a full description of a system.

But one must distinguish {\it two types of complementarity} that Bohr used (both of which he interpreted as being examples of (a) and (b)). The first aimed to resolve the traditional problem of `wave-particle dualism' (see Section \ref{dualism}). But the second type was novel:\\
\\
(A)~~{\it Wave-particle complementarity}: this is the use of the wave and particle {\it pictures}, or {\it models}, that correspond to the classical notions of either waves or particles, i.e.~wave or particle properties are ascribed to an object, thereby satisfying the conditions (a) and (b) above.\\
\\
(B)~~{\it Kinematic-dynamic complementarity}: this is the complementarity between spatiotemporal or `kinematic' descriptions, and momentum-energy or `dynamic' descriptions of an object. The dynamic description is associated with the conservation laws of energy and momentum, and Bohr dubbed this description `causal'. 

Bohr saw Heisenberg's uncertainty principle as an important consequence, or formal expression, of kinematic-dynamic complementarity, because a description in terms of spacetime variables $q$ and $t$ is incompatible with one in terms of energy and momentum variables $p$ and $E$, but both are required for the description of different types of experiments: e.g.~an experiment where position is measured, or an experiment where momentum is measured, and these experiments are themselves mutually exclusive.\\

There are two important contrasts between the above two complementarities, (A) and (B), that determine the kinds of phenomena to which they apply: 

First, wave-particle complementarity can apply to a {\it single experiment}, while kinematic-dynamic complementarity applies to {\it different experiments}. For example, in Compton scattering (discussed in Section \ref{dualism}), the {\it Compton shift} is measured, i.e.~the increase in the photon's {\it wavelength} (thus assigning a wave property to the photon), while the conservation laws that apply to point particles are used to describe the {\it collision} between the photon and the electron (thus using a particle model). Thus the wave and the particle complementary descriptions are used in the same experiment. This is not the case for kinematic-dynamic complementarity, as in Bohr's well-known statement that a measurement of the position of a particle requires a fixed experimental set-up, while a measurement of the momentum requires a movable one.

Second, the wave and particle concepts describe properties that are {\it classically incompatible}, i.e.~a given object is either a wave or a point particle but not both, while the kinematic and dynamic concepts are {\it classically compatible}, since particles have both position and momentum.\footnote{Commentators have warned against a uniform linking of these two types of complementarities. For example, Bohr in some cases associates kinematic descriptions, i.e.~an assignment of $(q,t)$, with a particle model; and dynamic descriptions, i.e.~$(p,E)$, with a wave model: namely, for the stationary states of the atom. In other cases, the kinematic description is the wave model and the dynamic description is the particle model: namely, in the Compton effect: see Murdoch (1987:~p.~66) and Beller (1999:~p.~118).}

Commentators have noted that Bohr's mature, {\it post-EPR} view on complementarity is the kinematic-dynamic one, understood in terms of {\it incompatible observables}, i.e.~measurable quantities like $q$ and $p$ cannot be said to have simultaneously well-defined values, and the types of experiments required to measure them are mutually exclusive. For example, Bub and Clifton (1996:~p.~216) articulate Bohr's mature view of complementarity as saying that an observable has a determinate value only in the context of a specific, classically describable, experiment: and in general a pair of such experiments are mutually exclusive.\footnote{Halvorson and Clifton (1999:~p.~2472) articulate Bohr's view of {\it measurement} as `selecting for beable status the maximal ... subalgebra determined by the ``privileged'' pointer observable $R$ of the measuring system and the pure entangled state ... of the composite measured/measuring system'. Thus the act of measurement in an EPR-type of scenario is not made determinate via a physical disturbance, but by constructing `the maximal set of observables that, together with the pointer observable $R$, can have simultaneously determinate values'. They also argue (2002:~p.~3) against interpreting Bohr in positivistic terms.}

The next Section will discuss the analogies between duality and complementarity, and how these notions relate to the advanced examples of Part II.

\subsection{Duality and complementarity: same or different?}\label{adcom}

Since contemporary debates in quantum field theory and quantum gravity crucially use the notions of `duality' and `complementarity', it is important to clarify the differences between these two concepts: and this is readily done in the case of quantum mechanics. As we have already mentioned, the difference between duality and complementarity is best explicated by contrasting the differences between Heisenberg's and Bohr's views on wave-particle duality. We here spell out their similarities and differences, especially in terms of Section \ref{complement}'s aspects, (a) and (b), of complementarity, and we will then compare these conclusions with the themes and roles of duality. At the end of the Section, we will return to the topic of {\it Anschaulichkeit}.

Heisenberg denied that the wave and particle models satisfy the condition (b) of complementarity of descriptions (see Section \ref{complement}):\footnote{He might well agree with a version of Bohr's condition (a), i.e.~he would agree that, even though the wave and particle models are quantum mechanically the same, waves and particles are interpretations that are practically, and classically, different and incompatible.}
rather, his view was that each dual model gives a {\it complete description} of the physical system. While one dual may be practically more useful than the other for some purposes, the two duals give equivalent descriptions of the same system. (Heisenberg's pragmatic admission regarding duals---that, for some purposes, one dual may be more useful than the other---is of course like our theme (1), in Section \ref{themesd}, of {\it hard-easy}.)

Thus the completeness of one of the two descriptions, i.e.~the denial of complementarity's aspect (b), is the most significant difference between duality and complementarity. Duals are not intended to complement each other in the sense that they each give incomplete descriptions: they each give complete descriptions, whereas complementaries give descriptions that cannot be equivalent and are not equally good: one description is better in some cases, and the other is better in others. Both complementaries are required for a full description of a system. 

Bohr's condition (b) that complementary descriptions are individually insufficient, but jointly sufficient, to describe a physical system, is analogous to the use of {\it quasi dualities}, especially our constrast (5) in Section \ref{themesd} between dualities and effective dualities, and the related contrast between the theoretical and the heuristic function of dualities. Chapter \ref{EMYM} will discuss effective dualities in quantum field theory: in such cases, a physical system cannot be fully described using a single model, but requires several models for a full description of the `theory space' (i.e.~the theory's moduli space: see Figure \ref{SWduality}). And Chapter \ref{String} will discuss the M-theory programme in these terms: namely, the idea that different perturbative string theories describe different limits of M-theory, or different `patches' or `corners' of the whole `theory space' (see Figure \ref{Mthfig}). Chapter \ref{Heuri} will dub this the `geometric view' of theories, by analogy with how a non-trivial $n$-dimensional spacetime manifold is described not by a single map to $\mathbb{R}^n$ but by a collection of `patches' on the manifold, each of them mapping to a subset of $\mathbb{R}^n$, with transition functions defined on the overlaps.

Bohr and Heisenberg agreed that the two classical pictures of a system, e.g.~the kinematic or the dynamic one (i.e.~(B) in Section \ref{complement}), are practically useful in different situations. For example, when the wave-function of a quantum system or object is sharply localized in position space, the classical ascription of a definite position to that object is useful because it gives a good approximation to that system's state, while if the wave-function is localized in momentum space, the classical ascription of a definite momentum to that object is more useful.\footnote{Bohr would probably not formulate this statement in terms of an object's wave-function, since he was first and foremost concerned with the details of experimental situations. But this translation into the language of wave-functions is admissible and closer to the Schema.}
These two classical pictures are valid for quantities in specific states, and for specific approximations. This is analogous to the idea, in Section \ref{std} (A), of a {\it quantum duality}, i.e.~a duality that obtains at the quantum level. Thus in such cases, we can have classical models that are {\it not} duals and that {\it emerge} as e.g.~different limits of a given quantum theory. In Chapters \ref{STII} and \ref{Heuri}, we will argue that this is the situation in gauge-gravity dualities, where evidence points to the existence of a single quantum theory underlying two duals, of which one limit (one dual) is best described as a semi-classical gravity model in $d+1$ dimensions (i.e.~the `bulk' model), and the other limit (dual) as a perturbative quantum field theory in $d$ dimensions (i.e.~the `boundary' theory). As we will discuss in Chapter \ref{STII}, these two limits are mutually exclusive, in that when one model is weakly coupled, the other is strongly coupled, and vice versa (thus again illustrating our theme (1) of {\it hard-easy}, from Section \ref{themesd}).

A final remark: the quantum field theories and string theories in Part II will not give support to a precise version of wave-particle duality:\footnote{Section \ref{dualism}-(iii) already pointed to difficulties of interpreting quantum field theories in terms of particles or in terms of fields.} 
but there is an analogous case of dualities that exchange particles and solitons, as in our theme (2) in Section \ref{featurerole} of {\it elementary-composite}.\footnote{The general feature of particle-soliton (quasi-)dualities is that a state with discrete quantum numbers that are the conserved charges of a Noether current (and thus reflect a symmetry) is mapped to a state where the quantum numbers are topological charges, obtained from configurations of a field that are extended over the whole space: see Chapter \ref{Advan}.}\\

A major {\it philosophical} topic of debate between the matrix and wave theorists was which of the two models gives understanding: the issue was, as they dubbed it, the model's {\bf Anschaulichkeit}, i.e.~visualizability, including its connotation of `intelligibility'. 

De Regt (2017:~pp.~226-251) gives a detailed historical and philosophical account of this debate, and in fact his theory of scientific understanding is inspired by Heisenberg's construal of {\it Anschaulichkeit}.\footnote{Beller (1999:~pp.~67-79) discusses the significance of this debate for the emergence of Heisenberg's (1927) uncertainty paper.}
As de Regt (2017:~p.~231) notices, already the old quantum theory resisted ordinary visualizability in space and time, and in particular the `quantum jumps' of Bohr's model of the hydrogen atom were regarded as unvisualizable. More generally, Bohr (1928:~p.~580) often mentioned the `irrationality' of the quantum postulate. 

As we mentioned in Section \ref{dualism} (i), Schr\"odinger regarded his own undulatory theory as physical and intelligible because it did not require, as the matrix theory appeared to do, abandoning the idea of the location and path of an electron in space. For Schr\"odinger (1926a:~p.~27), {\it Anschaulichkeit}, as visualizability in space and time, was a necessary condition for intelligibility (see also De Regt: 2017:~p.~239). Thus Schr\"odinger's (1926b:~p.~46) sneer at Heisenberg's matrix mechanics is well-known: `I ... was discouraged, if not repelled, by what appeared to me as very difficult methods of transcendental algebra, and by the want of perspicuity ({\it Anschaulichkeit})'.

De Regt (2017:~p.~242) argues that Schr\"odinger (1926b:~p.~59) predicted that his wave mechanics would be more useful, because it could be used in an intuitive way in other situations: for example, in extending quantum mechanics to the Maxwell theory. And de Regt argues that this was in some sense confirmed in the late 1920s, when wave mechanics was indeed widely used: `wave mechanics completely overshadowed matrix mechanics in the late 1920s'. 

Pauli and Heisenberg argued that visualization in space and time was not essential for intelligibility. Heisenberg's (1927:~p.~478) uncertainty paper was titled `\"Uber den anschaulichen Inhalt der quantentheoretischen Kinematik und Mechanik', translated in his collective works as `The Perceptible Content of the Quantum Theoretical Kinematics and Mechanics'. As de Regt (2017:~p.~244) argues, this article involved an attempted {\it re-definition} of Schr\"odinger's notion of {\it Anschaulichkeit}: and thus, a second example of a ``new meaning for an old word''!

The question of visualizability in space and time becomes even more critical in quantum gravity, where it is sometimes claimed that spacetime emerges from an underlying non-spatiotemporal structure. We will return to this in Chapter \ref{Understand}.

\section{Classical electric-magnetic duality}\label{EMduality}

This Section discusses the electric-magnetic duality of the classical Maxwell theory,\footnote{For introductions, see Olive (1997), Figueroa-O'Farril (1998), and Griffiths (1999:~pp.~536-537).}
i.e.~the invariance of the set of Maxwell's equations under a suitable exchange of the electric and magnetic fields. 

This duality is interesting for a number of reasons. First, unlike the dualities in the previous two Sections, electric-magnetic duality is not a quantum duality, since the classical models are already duals. Indeed, although, as we mentioned in Section \ref{wpd}, Dirac (1931) completed electric-magnetic duality by his introduction of magnetic monopoles (which we will also discuss), and this involved considering the quantum theory, electric-magnetic duality already exists in the simplest case, of the Maxwell equations in vacuum. 

Furthermore, this duality will be a blueprint for more advanced dualities in Part II: namely, in Yang-Mills theory, and in string theory, where it is related to important physical phenomena such as colour confinement and the Higgs mechanism. 

Section \ref{EM-duality} first introduces this elementary duality, and Section \ref{EmdS} shows how it illustrates the Schema. Section \ref{FHodge} reformulates the duality using the Faraday tensor and Hodge duality, which we will use extensively in Part II, and Section \ref{illH} then shows how this formulation illustrates the Schema. Section \ref{Dqc} then discusses one of the implications of electric-magnetic duality: namely, Dirac's monopole solution, which is suggested by the idea of electric-magnetic duality. 

\subsection{Duality and the Maxwell equations}\label{EM-duality}

At its simplest, electric-magnetic duality is the invariance of the set of four Maxwell's equations in vacuum:\footnote{In this Section we use SI units, and keep factors of $c$ and $\e_0$ in the formulas. For other unit conventions used in the literature, see Griffiths (1999:~pp.~558-561) and Jackson (1962:~pp.~611-621).}
\bea\label{4Max}
\nabla\cdot{\bf E}=0\,,&&\na\times{\bf E}+~~{\pa{\bf B}\over\pa t}~=0\nn
\nabla\cdot{\bf B}=0\,,&&\na\times{\bf B}-{1\over c^2}{\pa{\bf E}\over\pa t}=0\,,
\eea
under the following exchange of the electric and magnetic fields, i.e.~our putative duality map:
\bea\label{EMd}
{\bf E}/c&\mapsto&{\bf B}\nn
{\bf B}&\mapsto&-{\bf E}/c\,.
\eea
By `invariance' of the equations, we mean that the {\it set} of Maxwell's equations is invariant, i.e.~the map exchanges the equations for the divergences of the fields, and also the equations for the curls of the fields. Thus, under the map Eq.~\eq{EMd}, Gauss' law in vacuum is exchanged with the condition that there are no monopoles (i.e.~the magnetic field is divergenceless), and Faraday's law and Amp\`ere's laws exchange roles: the set of equations is mapped onto itself by the putative duality map, so that the dynamics of the Maxwell theory remains unmodified (recall that, in defining a duality as a formal map, we are dealing with bare models, so that we have not yet interpreted the equations as being either `electric' or `magnetic'). 

In this formulation, the state space ${\cal S}$ is a pair of square-integrable vector fields ${\bf E}:\mathbb{R}^3\rightarrow\mathbb{R}^3$ and ${\bf B}:\mathbb{R}^3\rightarrow\mathbb{R}^3$. The square-integrability condition secures that the total electromagnetic energy $E$ in a volume $V$ is finite:
\bea\label{emE}
E={1\over2}\int_V\dd^3x\left(\e_0\,{\bf E}^2+{1\over\m_0}\,{\bf B}^2\right)={1\over2\m_0}\int_V\dd^3x\left({{\bf E}^2\over c^2} +{\bf B}^2\right),
\eea
where the permeability $\e_0$ and the permittivity, $\m_0$, of the vacuum are related by $\e_0\m_0=1/c^2$. Section \ref{EmdS} will discuss how electric-magnetic duality thus defined illustrates the Schema from Section \ref{isomdef}.\\
\\
{\bf Duality groups: SO(2) and U(1)}. There is a generalisation of electric-magnetic duality that will be important in Part II. The set of Maxwell's equations have a group of invariances that is larger than the discrete $\mathbb{Z}_4$-group generated by the duality map, Eq.~\eq{EMd}. Indeed, the equations are invariant under arbitrary SO(2) mixings of the electric and magnetic fields:
\bea\label{so2}
\left(\begin{array}{c}{\bf E}'/c\\{\bf B'}\end{array}\right)=\left(\begin{array}{cc}\cos\th&\sin\th\\-\sin\th&\cos\th\end{array}\right)\left(\begin{array}{c}{\bf E}/c\\{\bf B}\end{array}\right).
\eea
One easily checks that also the energy and the Poynting vector are invariant under this map. Thus for each value of $\th$, we have again a duality that maps states to states, quantities to quantities, and is equivariant for the dynamics. The set of all such dualities is thus parametrized by the group SO(2). A set of dualities that form a (discrete, or continuous) group is called a {\bf duality group}.

A useful reformulation of the SO(2) duality group is in terms of U(1), which is isomorphic to SO(2). To this end, we rewrite Maxwell's equations in terms of a complex vector, defined as follows:\footnote{See Silberstein (1907a:~p.~581; 1907b:~p.~783).}
\bea\label{calE}
{\cal E}:={\bf E}/c+i\,{\bf B}\,.
\eea
In terms of this vector, the Maxwell equations are written in compact form:
\bea
\nabla\cdot{\cal E}&=&0\nn
\nabla\times{\cal E}&=&{i\over c}\,{\pa{\cal E}\over\pa t}\,,
\eea
which implies that ${\cal E}$ satisfies the massless Klein-Gordon equation, so that its solutions are linear combinations of plane waves whose four-momenta are null vectors. This implies that each plane wave has null energy-momentum, i.e.~its energy and momentum are related by $E^2 = {\bf p}^2 c^2$. Therefore, each plane wave travels at the speed of light in any inertial frame of reference. 

\begin{figure}
\begin{center}
\includegraphics[height=4cm]{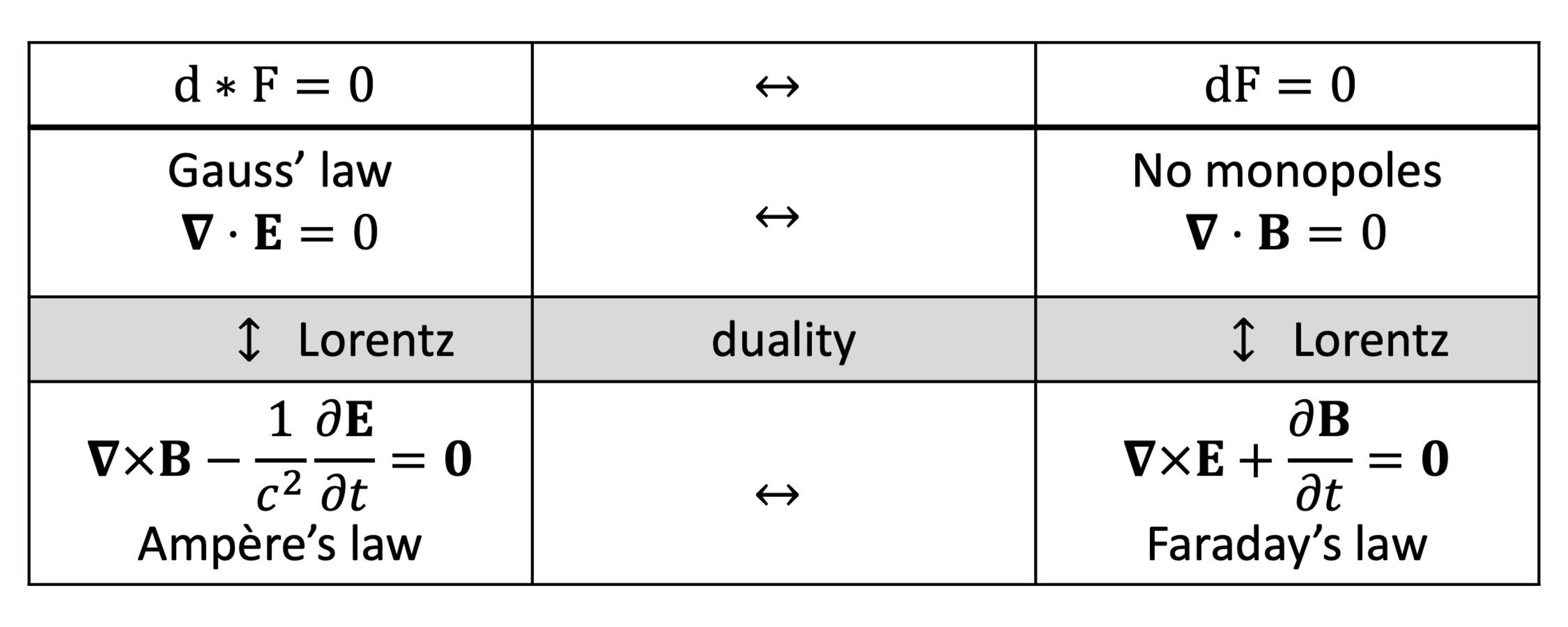}
\caption{\small Maxwell's equations and how they are mapped under Lorentz transformations (vertical arrows) and duality (horizontal arrows). Lorentz transformations generate linear transformations of equations along the vertical direction; duality transformations exchange equations along the horizontal direction. The analogy is complete if the duality group SO(2), which generates linear combinations of dual equations, acts along the horizontal direction.}
\label{Maxfig}
\end{center}
\end{figure}

By virtue of the isomorphism between the groups SO(2) and U(1), the duality group SO(2) in Eq.~\eq{so2} acts as the U(1) group on the complex variable ${\cal E}$:
\bea\label{eith}
{\cal E}'=e^{-i\th}{\cal E}\,.
\eea
We recover the original duality, i.e.~Eq.~\eq{EMd1}, by setting $\th={\pi\over2}$, so that the action of the duality in Eq.~\eq{EMd} on the fields ${\bf E}$ and ${\bf B}$ induces the following action on ${\cal E}$: ${\cal E}\mapsto-i\,{\cal E}$.\\
\\
{\bf Interpretation: an analogy with the Lorentz transformations}. The idea that duality exchanges electric and magnetic fields is not as unusual as one might think. For after all, Lorentz transformations already do that: they map electric and magnetic fields into one another under changes of coordinates. Thus for example, a Lorentz transformation maps Amp\`ere's law into a linear combination of Amp\`ere's law and Gauss' law, as the vertical arrow in the left column of Figure \ref{Maxfig} shows. And Faraday's law is mapped into a linear combination of Faraday's law and the condition that there are no monopoles, i.e.~the right column. Duality makes an analogous exchange along the horizontal direction.\footnote{This kind of mixing between electric and magnetic forces is of course the effect discussed in the opening paragraph of Einstein's (1905) paper on special relativity. The effect of the Lorentz transformations is indeed striking if we introduce sources into Maxwell's equations. Thus consider e.g.~the force between an infinitely extended line current with zero net electric field (in the rest frame of the wire) and a test point charge at a fixed distance from the wire, moving parallel to it. In the wire's frame of reference, there is no net electric field coming from the wire: but, since the point charge moves, there is a non-zero {\it magnetic Lorentz force} on the charge. (The current's magnetic field is given, by Amp\`ere's law with a line current, in terms of the current and the distance to the test charge.) On the other hand, in the rest frame of the test charge, the charge's velocity is zero and there is no magnetic Lorentz force: but because of length contraction effects, the wire carries a net line charge, and there is a non-zero electrostatic field from the wire that produces an {\it electric force} on the charge: namely, the force of a cylindrical electric field on a point charge, as described by Gauss' law with an electric line charge. Furthermore, one can of course prove that the {\it electric force} in the charge's rest frame transforms into the {\it magnetic Lorentz force} in the wire's rest frame---thus generalizing the mixing between Amp\`ere's law and Gauss' law to the case with non-zero charges and currents. Thus imagine the relativistic world of the system just described, where magnetic forces had not yet been discovered (for in the charge's rest frame there are no magnetic forces): in such a world, the magnetic Lorentz force could be discovered by going to a frame in which the point charge is moving. See also Griffiths (1999:~pp.~522-525).\label{magnfi}}

The aim here is not to advocate the existence of a common ontology underlying the two duals, which would require anticipating the discussions in Parts II and III. Rather, the aim is to point to the analogy with the familiar Lorentz transformations, where the relation between electric and magnetic fields is widely accepted. Thus footnote \ref{magnfi} illustrates how, in the case of Lorentz transformations, one uses electric charges and currents to establish such equivalence relations.\footnote{For philosophical discussions of electric-magnetic duality, see Dieks et al.~(2015:~pp.~209-210) and Weatherall (2020:~pp.~1176-1179).}

\subsection{Illustrating the Schema for electric-magnetic duality}\label{EmdS}

The previous Section introduced the basic case of electric-magnetic duality, Eq.~\eq{EMd}, and a generalization to a duality group. In this Section, we discuss how this illustrates the Schema for dualities from Section \ref{isomdef}. There are three things to discuss: 

(i)~~the isomorphism between the state-spaces of the two models; 

(ii)~~the isomorphism between their sets of quantities; and

(iii)~~the equivariance for the two triples' dynamics.

About (i): as we discussed in the previous Section, the state-space of the Maxwell theory is the space of square-integrable pairs of vectors, i.e.~$({\bf E}/c,{\bf B}):\mathbb{R}^3\rightarrow\mathbb{R}^6$. Since the duality exchanges the two factors in $\mathbb{R}^6$, the two state-spaces, ${\cal S}_{M_1}$ and ${\cal S}_{M_2}$, carry different but isomorphic representations of the Lorentz group: this amounts to saying that the difference between the two state-spaces is in ``which variable is electric and which variable is magnetic''. Namely, electric and magnetic fields transform differently under Lorentz transformations, and the two state-spaces carry different but isomorphic representations of the Lorentz group.\footnote{Under Lorentz transformations, the electric and magnetic fields along the directions perpendicular to the relative motion transform with different relative signs. Alternatively, we say that electric fields are polar vectors, and so they are parity-odd: while magnetic fields are axial vectors, and so they are parity-even, viz.~they are curls of polar vectors. This means that the state spaces ${\cal S}_{M_1}$ and ${\cal S}_{M_2}$ carry different but isomorphic representations of the Lorentz group, and in particular they carry different but isomorphic representations, i.e.~differing by a sign, of the {\it improper Lorentz transformations} of parity inversion and time reversal.\label{imL}} 

The duality map $d_{\cal S}$ between the state spaces ${\cal S}_{M_1}$ and ${\cal S}_{M_2}$ thus defined in given in the obvious way:
\bea\label{EMd1}
d_{\cal S}:~({\bf E}/c,{\bf B})&\mapsto&({\bf E}'/c,{\bf B}'):=\,({\bf B},-{\bf E}/c)\,.
\eea

About (ii): there are two main quantities to consider, and for whose values we need to check the matching: (1) the energy, i.e.~Eq.~\eq{emE}, and (2) the Poynting vector, which quantifies the energy flow through the boundary of a volume. (We will discuss other quantities when we discuss the Faraday tensor formulation in the next Section.) As for (1): one easily checks that the energy density has the same value in the two models, i.e.~${\bf E}^2/c^2+{\bf B}^2={\bf E}'{}^2/c^2+{\bf B}'{}^2$. (2) The Poynting vector is ${\bf S}:={1\over\m_0}\,{\bf E}\times{\bf B}$, and one easily checks that it also has the same value, i.e.~${\bf S}={\bf S'}$. (And this invariance also obtains for the generalization of the duality map, Eq.~\eq{EMd1}, to the SO(2) duality group in Eq.~\eq{so2}.)

More formally, using our notation for the values of quantities from Section \ref{isomdef}: (1) the energy, Eq.~\eq{emE}, is the value of the Hamitonian $H:{\cal S}\rightarrow\mathbb{R}_{\geq0}$, which is a map from states to energy values, i.e.~positive real numbers. Thus using the notation of Eq.~\eq{obv1}, i.e.~$E_i=\bra H_i,s_i\ket_i$, where $i=1$ labels the model on the left, and $i=2$ the model on the right, the states are pairs of electric and magnetic vectors, $s=({\bf E}/c,{\bf B})$, for either model, as in (i). Then we have: 
\bea
E_1=\bra H_1,s_1\ket_1=\bra d_{\cal Q}(H_1),d_{\cal S}(s_1)\ket_2=\bra H_2,s_2\ket_2=E_2\,,
\eea
which is the Schema's requirement that the values of quantities are preserved.

(2) The Poynting vector is likewise the value of a map ${\cal P}$ from the state space to the vector space $\mathbb{R}^3$: namely, ${\cal P}$ assigns, to a state $s=({\bf E}/c,{\bf B})$, a Poynting vector ${\bf S}$, as follows: ${\cal P}:{\cal S}\rightarrow\mathbb{R}^3$, where ${\bf S}={\cal P}(s)=\bra{\cal P},s\ket$, using our bracket notation in Eq.~\eq{pairing}. Then we have again:
\bea
{\bf S}_1=\bra{\cal P}_1,s_1\ket_1=\bra d_{\cal Q}({\cal P}_1),d_{\cal S}(s_1)\ket_2=\bra{\cal P}_2,s_2\ket_2={\bf S}_2\,,
\eea
where the values of the Poynting vector are as above, i.e.~${\bf S}_i={\bf E}_i\times{\bf B}_i/\m_0$.

About (iii): the dynamics of the Maxwell theory is the set of four Maxwell equations, which by definition remain invariant under electric-magnetic duality: thus the dynamics of one model is correctly mapped onto the dynamics of the other model, and the duality is equivariant for the dynamics.

The electric-magnetic duality map does not square to the identity, but rather to minus the identity: $d_{\cal S}^2=-\mbox{id}$. This is because, applying the duality map twice, the electric and magnetic fields return to minus themselves: $({\bf E}''/c,{\bf B}'')=-({\bf E}/c,{\bf B})$. Also, after one includes charges, this amounts to changing the signs of both electric and magnetic charge sources, i.e.~it is charge conjugation (see the next Section). 

\subsection{Faraday tensor formulation and Hodge duality}\label{FHodge}

Formulating the Maxwell theory in terms of the Faraday tensor allows us to reformulate electric-magnetic duality in the more modern language of Hodge duality, which makes Lorentz invariance manifest. 

The electric and magnetic fields together have six independent components. The smallest representation of the Lorentz group that one can put these six degrees of freedom into is an antisymmetric two-tensor, or matrix.\footnote{Our guiding principle here is Wigner's (1939:~p.~53) principle that the one-particle states are the representation spaces of irreducible representations of the Poincar\'e group. See the discussions in Weinberg (1995:~pp.~62-66), Gilmore (2008:~p.~275), Georgi (1999:~pp.~138-143), and Maggiore (2005:~pp.~23-24). In the case of the Maxwell equations in vacuum, the one-particle states are the classical photon states of the field (see footnote \ref{photon}). (The Maxwell equations in vacuum are linear i.e.~free field equations, and so the usual difficulties of defining in and out states in interacting quantum field theories do not arise: see Section \ref{dualism}-(iii).) Earlier treatments of irreducible representations as ``basic building blocks'' of solution spaces of the Schr\"odinger or the field equation, including in non-relativistic quantum mechanics, include Wigner (1931:~pp.~73, 79, 85, 129) and Bargmann and Wigner (1948:~p.~216). Note that a {\it symmetric} two-tensor also has 6 independent components, but such a representation is not irreducible, and so the physics it describes is very different from what we need here. For the connection between the `photon' description, of a massless state of spin 1, and the two-index antisymmetric tensor field that satisfies the Maxwell equations, see Gilmore (2008:~p.~275), who discusses all the unitary irreducible representations of the Poincar\'e group (ibid, pp.~266-274). Also see de Wit and Smith (1986:~pp.~137-141), Jeevanjee (2011:~p.~203), Aitchison and Hey (2013:~pp.~197-199) and Greiner and Reinhardt (1996:~p.~157).} 
Thus we introduce the Faraday tensor $F^{\m\n}$. We choose its components, relative to a given Minkowskian system of coordinates, in the usual way:\footnote{For an introduction to the tensor calculus used in the rest of this Section, see Jeevanjee (2011).}
\bea\label{EBfields}
F^{0i}&=&E^i/c\nn
F^{ij}&=&\e^{ijk}\,B_k\,,~~i,j,k=1,2,3\,,
\eea
where $\e^{ijk}$ is the Levi-Civita antisymmetric symbol. 

We will also require the {\bf Hodge dual} of the Faraday tensor, denoted by $G^{\m\n}$:\footnote{For more details, see e.g.~Greiner and Reinhardt (1996:~p.~142) and Hamilton (2017:~pp.~406-410).} 
\bea\label{Gdual}
G^{\m\n}:=\half\e^{\m\n\l\s}\,F_{\l\s}\,,
\eea
where $\e^{\m\n\s\l}$ is the totally antisymmetric pseudo-tensor with components $\e^{0ijk}=\e^{ijk}$. With these definitions, $F^{\m\n}$ is the unique Lorentz-covariant quantity that is linear in the electric and magnetic fields, while $G^{\m\n}$ is an antisymmetric pseudo-2-tensor. This is because, while $G^{\m\n}$ is covariant under the restricted Lorentz group, $\mbox{SO}^+(1,3)$ (also called the proper, orthochronous Lorentz group), i.e.~the set of Lorentz transformations that preserve both orientation and time direction, it changes sign under changes of orientation and under time reversal.\footnote{Under a change of coordinates $\L$, the transformation rule of a pseudo-tensor $\e$ includes a factor of $\det\L$. As a differential form, the Hodge dual is the orthogonal complement of the Faraday two-form in the space of two-forms, so that $F\wedge*F=\mbox{vol}_g\times\bra F,F\ket_g$, where $\mbox{vol}_g=\sqrt{|g|}\,\dd x^1\wedge\cdots\wedge\dd x^4$ is the volume form of the four-manifold endowed with metric $g$, and $\bra\cdot,\cdot\ket_g$ is the inner product induced by the metric on the space of two-forms (and the orthogonality condition is $\bra F,*F\ket_g=0$). See e.g.~Nakahara (2003:~p.~292).}

In what follows, it will be useful to use form notation, i.e.~a notation that suppresses the explicit indices, so that $F$ is a 2-form whose components in a Lorentzian system of coordinates are $F^{\m\n}$, and $G^{\m\n}$ are the components of the Hodge dual of $F$, written as: $G=*F$. An important property of the Hodge star is that it satisfies: $*^2=-\mbox{id}$, as can be checked from Eq.~\eq{Gdual}.\footnote{In covariant notation, the equations of motion and Bianchi identity, respectively, with sources added take the form: $\pa_\n F^{\m\n}=\m_0J^\m_{\tn e}$ and $\pa_\n G^{\m\n}=\m_0J^\m_{\tn m}$, where $J^\m_{\tn e}=(c\r_{\tn e},{\bf J}_{\tn e})$ and $J^\m_{\tn m}=(c\r_{\tn m},{\bf J}_{\tn m})$. We will return to the form notation including the sources in Chapter \ref{String}.\label{Jdual}}

In terms of $F$ and its dual, the source-free Maxwell's equations, Eq.~\eq{4Max}, take the form:\footnote{See Hamilton (2017:~p.~419).}
\bea\label{Max2}
\mbox{Bianchi identity:}~~~~~~~\dd F&=&0~\Leftrightarrow~ \dd*G=0\nn
\mbox{Equation of motion:}~~~~\dd*F&=&0~\Leftrightarrow~ ~\dd G=0\,.
\eea
Although {\it both} equations, written in terms of the electric and magnetic fields Eq.~\eq{EBfields}, summarise the Maxwell equations Eq.~\eq{4Max}, we followed the common usage of dubbing the first equation the `Bianchi identity', and the second the `equation of motion'. The reason for this terminology is that the Faraday tensor is given mathematically in terms of a four-vector potential $A$, namely the gauge field (see Section \ref{Dqc}):\footnote{More precisely, the Faraday tensor gives the components of the curvature associated with the connection form, $A$, on the principal U(1) bundle. Since the gauge group is the Abelian group U(1), the curvature is an exact two-form.}
$F=\dd A$. Given the Faraday tensor in this form, the Bianchi identity then follows as a mathematical identity. The equation of motion in Eq.~\eq{Max2} is then the equation of motion derived from the following Maxwell action, varied with respect to the four-vector potential $A$:
\bea\label{FAA}
S_{\tn{Maxwell}}[A]&=&-{1\over4e^2}\int\dd^4x~F_{\m\n}[A]\,F^{\m\n}[A]\nn
F_{\m\n}[A]&=&\pa_\m A_\n-\pa_\n A_\m\,.
\eea
Here, $e$ is an arbitrary coupling constant (often reabsorbed into the definition of $A$), which at this stage need not be identified with the electron charge. Varying this action with respect to the four-vector potential gives the equation of motion $\pa_\n F^{\m\n}=0$.\footnote{The Faraday tensor is of course invariant under gauge transformations of the four-vector potential, $A_\m'=A_\m+\pa_\m\l$, for an arbitrary real function $\l:\mathbb{R}^4\rightarrow\mathbb{R}$.} 

The main implication of the reformulation of the Mawell equations in terms of the Faraday tensor is that {\it Hodge duality is electric-magnetic duality}, for two-forms: namely, Hodge duality, $F\mapsto*F$, written out in components, gives the duality map in Eq.~\eq{EMd}. And also, Hodge duality exchanges the Bianchi identity and the equation of motion in Eq.~\eq{EBfields}, which written out in components gives the duality of the Maxwell equations, in Figure \ref{Maxfig}. 

Before we return to this in the next Section, we give an important reformulation of Hodge duality in terms of complex two-forms, which will allow us to arrive at a simple result for Hodge duality that generalizes Section \ref{Mth}'s description of electric-magnetic duality in terms of a complex vector $\cal{E}$ as in Eq.~\eq{eith}.\\
\\
{\bf Complex two-forms.} In form notation, the state-space is the vector space of pairs of normalizable two-forms $(F,G)$, analogous to the pairs of electric and magnetic vectors, in Eq.~\eq{EMd1}.\footnote{This space is endowed with a complex structure: recall that a {\bf complex structure}, $J$, on a real vector space $V$, is an automorphism $J:V\rightarrow V$ that squares to minus the identity, i.e.~$J^2=-\mbox{id}_V$. Thus $J$ is analogous to $i=\sqrt{-1}$, and allows one to define multiplication by complex numbers on the vector space, as follows: $(a+ib)\,v:=av+b\,J(v)$, where $a,b\in\mathbb{R}$ and $v\in V$. In the present case, the complex structure can be represented by a matrix:
$*\left(\begin{array}{c}F\\G\end{array}\right)=\left(\begin{array}{cc}0&1\\-1&0\end{array}\right)\left(\begin{array}{c}F\\G\end{array}\right),$
and the matrix squares to minus the identity. This is a clockwise rotation about $\pi/2$. The Hodge star, and thus the action of the duality map on the states, can be diagonalized by considering complex two-forms, as in the main text.}
Thus by analogy with the complex vector in Eq.~\eq{calE}, two-forms can also be conveniently written in terms of a {\it complex two-form}:
\bea\label{calF}
\mathscr{F}:=F+i\,G\,,
\eea
whose Hodge dual is proportional to the two-form itself:
\bea\label{HcalF}
*\,\mathscr{F}=-i\,\mathscr{F}\,.
\eea
Thus the convenience of the complex form notation is that the Hodge dual acts on the two-form as $-\sqrt{-1}$, and so we have ``diagonalized the action'' of the Hodge duality, as in Eq.~\eq{eith}, of which this is the Lorentz-covariant version. Indeed, the $0i$-component of the complex two-form is the (spatial component of the) complex vector we defined above, in Eq.~\eq{calE}: namely, ${\mathscr{F}}^{0i}={\cal E}^i$. 

The complex two-form and the action of Hodge duality on it as multiplication by $-i$ is an important result that we will use later.\footnote{The complex conjugate of $\mathscr{F}$, viz.~$\bar{\mathscr{F}}$, transforms with $i$: $*\,\bar{\mathscr{F}}=i\,\bar{\mathscr{F}}$.}

\subsection{Illustrating the Schema for Hodge duality}\label{illH}

This Section illustrates our Schema from Section \ref{isomdef} for electric-magnetic duality in terms of the Faraday tensor and its dual, i.e.~(i) the isomorphism between the state-spaces; (ii) the isomorphism between the quantities; and (iii) the equivariance for the dynamics (cf.~Section \ref{EmdS}). 

About (i): in this formulation, the state-space is the space of pairs of square-integrable two-forms $(F,*F)$ (or, alternatively, the set of square-integrable complex two-forms $\mathscr{F}$, defined in Eq.~\eq{calF}). Since, as before, the Faraday tensor and its Hodge dual give different but isomorphic representations of the Lorentz group (i.e.~including transformations that reverse the orientation or the time orientation of the spacetime), each model has a state-space that is a representation of the Lorentz group that is isomorphic to that of the other model (see the discussion in Section \ref{EmdS}-(i), and especially footnote \ref{imL}). The duality map on the states then follows from the application of the Hodge dual, which we discussed in the previous Section:
\bea\label{EMd2}
d_{\cal S}:{\cal S}_{M_1}\rightarrow {\cal S}_{M_2}\,,~~(F,G)=(F,*F)~\mapsto~(F',G'):=\,(G,-F)\,.
\eea
Because the Hodge star satisfies $*^2=-\mbox{id}$, the duality map squares to minus the identity, as it should. Thus the duality map on the states, Eq.~\eq{EMd2}, is the Hodge star itself applied to pairs of two forms: $d_{\cal S}=*$. 

Since we already discussed that Hodge duality in effect maps the equation of motion of one model into the Bianchi identity of the other model in Eq.~\eq{Max2}, there is no need to say more about equivariance for the dynamics, i.e.~(iii). So we now focus on quantities, i.e.~on (ii).

About (ii): we first show, for an important quantity (namely, the stress-energy tensor), that the duality map $d_{\cal Q}$ is well-defined: and that, as required by Eq.~\eq{obv1}, the values of this quantity match betwen duals. Then we sketch a similar argument for all physical quantities, i.e.~those that respect the structure of the state space.

In the Maxwell theory, the main quantity of physical interest is the symmetric electromagnetic stress-energy tensor. This is a {\it rank two,} symmetric tensor, whose components in a Minkowskian system of coordinates are given as follows:
\bea\label{Tmunu}
T_{\m\n}=-{1\over\m_0}\,F_{\m\l}F^\l{}_\n-{1\over4\m_0}\,\eta_{\m\n}F_{\l\s}\,F^{\l\s}\,.
\eea
Indeed, relative to a choice of coordinates, the electromagnetic stress-energy tensor contains the {\it energy density}, which is integrated over in Eq.~\eq{emE}; the {\it Poynting vector} (see the discussion after Eq.~\eq{EMd1}); and the Maxwell {\it stress-tensor}, which describes the electromagnetic stress, i.e.~force per unit area including both pressure and shear, acting on a surface. The energy density is the 00-component, $T_{00}$; the components of the Poynting vector are the mixed components, $T_{0i}$; and the spatial components, $T_{ij}$, give the Maxwell stress-tensor.

As in Section \ref{isomdef}, we regard the stress-energy tensor as the value of a quantity ${\cal T}\in{\cal Q}$ that maps states to tensors (or, to simplify our notation: to the {\it values} of the stress-energy tensor),\footnote{The stress-energy tensor is itself of course a map from the tangent space to $S^2(\mathbb{R}^4)$. But as we mentioned, we only need to focus on its values.} 
i.e.~${\cal T}:{\cal S}\rightarrow S^2(\mathbb{R}^4)$, where $S^2$ is the symmetric product of $\mathbb{R}^4$ with itself, which is the space where the stress-energy tensor takes values. Then we have: $T=\bra{\cal T},s\ket$, where $T$ is the value of the stress-energy tensor, whose components in a Minkowskian system of coordinates we reported in Eq.~\eq{Tmunu}. 

It is somewhat tedious, but straightforward, to check that the values of $T$ in any Minkowskian system of coordinates are invariant under the duality map Eq.~\eq{EMd1}, extended to a map on quantities. Namely, the values of the stress-energy tensor and its dual are equal under Eq.~\eq{EMd1}: $T_{\m\n1}=T_{\m\n2}$. Thus for a Minkowskian system of coordinates, our requirement of preservation of quantities, i.e.~Eq.~\eq{obv1}, is safisfied: namely, $T_1=\bra{\cal T}_1,s_1\ket_1=\bra d_{\cal Q}({\cal T}_1),d_{\cal S}(s_1)\ket_2=\bra{\cal T}_2,s_2\ket_2=T_2$.

The above argument can be generalized to other quantities in ${\cal Q}$: namely, the values of quantities match between the two models (sometimes up to a state-independent sign) for physical quantities (see our comment on overall factors of quantities that are independent of the state variables, i.e.~up to overall constants, at the end of Section \ref{isomdef}). 

This general argument rests on two main premisses that we endorse: first (i) that physical quantities respect the structure of the state-space; second (ii) that the set of quantities ${\cal Q}$ is ``sufficiently large''. As follows: (i) The values of the physical quantities respect the structure of the state-space: in particular, we only consider quantities that are Lorentz covariant. This implies that we can write the quantities in ${\cal Q}$ as Lorentz-covariant functions of the Faraday tensor and its Hodge dual.\footnote{In the classical theory that we are discussing here, there is no gain in considering holonomies of the gauge field around smooth and orientable loops, and the Faraday tensor suffices to construct all the physical quantities.}
(ii) As we already discussed, many quantities, like the energy density ${\bf E}^2/c^2+{\bf B}^2$, have values that are invariant under the duality. But other quantities are not thus invariant: for example, ${\bf E}^2/c^2$ is mapped to ${\bf B}^2$, and vice versa. Thus requiring that ${\cal Q}$ is ``sufficiently large'' means that it contains both members of {\it all such pairs} of quantities (and such pairs of quantities {\it are} usually considered in physics as both belonging to ${\cal Q}$). In other words, functions of $F$ are exchanged with functions of $*F$, and both are included in ${\cal Q}$. Such pairs are exchanged between the duals, and so our duality map $d_{\cal Q}$ is well-defined, and the requirement in Eq.~\eq{obv1} that the values of the quantities match is satisfied.

One might worry that this argument depends on our already knowing that there is a duality, i.e.~on using the duality of the state-spaces to construct the sets of quantities, ${\cal Q}_i$. But there is nothing wrong with this, since (as we discussed at the end of Section \ref{Ourthm})) sets of quantities in physics are not God-given, but are constructed by using physical and structural principles such as symmetries---with duality also being one such principle. Section \ref{abstraction} will argue more generally (as we already briefly discussed in Section \ref{epistemicc}, and as we will discuss in other examples in the next Chapter) that augmentation is a legitimate procedure to formulate a bare theory: namely, if one has models that are almost duals, duality can be used to augment these models to make them duals (a procedure similar to a definitional extension). 

\subsection{A quantum duality: Dirac's monopole}\label{Dqc}

In the previous Sections, we discussed electric-magnetic duality in vacuum, i.e.~without sources. It is clear that introducing charged particles, e.g.~electrons, into the Maxwell equations in the usual way, i.e.~by adding an electric charge density $\r_{\sm e}$ on the right-hand side of Gauss' law in Eq.~\eq{4Max}, and introducing a current ${\bf J}_{\sm e}$ on the right-hand side of Amp\`ere's law, breaks the duality. 

An important question for the physical significance of electric-magnetic duality is therefore whether it can be extended to Maxwell's equations with matter. This requires the introduction of both a magnetic charge density, $\r_{\sm m}$, and a magnetic current, ${\bf J}_{\sm m}$. In turn, this requires the introduction of magnetic monopoles, represented by magnetic charge density $\r_{\sm m}$, as the duals of pointlike electric charges. 

There are two possible objections that need to be addressed. We will state and address these objections and then expound the details.\\
\\
{\bf Two objections:}\\
\\
(i) Magnetic monopoles have never been observed in nature. Rather, all the observed magnetic field lines are closed, i.e.~return to their sources, e.g.~for magnetic dipoles with zero total magnetic charge. And so, we will need to motivate why one might want to introduce objects into the theory that have never been observed.\\
\\
(ii) In the Maxwell theory, the classical degrees of freedom of a wave of light (i.e.~a photon)\footnote{Section \ref{wpd} discusses the history of wave-particle duality, and in that context `photon' of course has the connotation of a quantum of light that cannot be described by the Maxwell theory. But in classical and quantum field theory we do not need to hold on to this terminology: there are no disadvantages to our using the term `photon' in the context of the Maxwell theory in flat space, where the photon's key properties are its masslessness and its two transverse physical polarizations.\label{photon}}
are encoded in the scalar and vector potentials, respectively $V$ and ${\bf A}$, which are the starting point of most generalizations of the theory (including its quantum version). However, since the magnetic field and the vector potential are related by ${\bf B}=\nabla\times{\bf A}$, the divergence of the magnetic field is identically zero, i.e.~$\nabla\cdot{\bf B}=0$, which is {\it incompatible} with having a non-zero magnetic charge density. Thus it appears that, according to this widespread reformulation of the Maxwell theory using the vector potential, magnetic monopoles {\it cannot exist}.\\

In reply to (i), one can make two points. First, one must admit that part of the motivation for introducing monopoles is indeed theoretical. For the purposes of this Chapter, monopoles allow us to generalise the basic electric-magnetic duality in vacuum, thereby getting a better understanding of how duality works: in a rich, yet still simple, example. With respect to the theoretical uses of monopoles, Dirac monopoles are important for understanding 't Hooft-Polyakov monopoles, which play an important heuristic role in our current understanding of physical phenomena such as the confinement of quarks (which we will introduce in Chapter \ref{EMYM}). 

Second, Dirac's quantization condition, to be derived below (cf.~\eq{Diracq}), is the only known mechanism that predicts the experimentally observed quantization of electric charges in nature. This is, by itself, a good reason to examine the mechanism in detail, and its more sophisticated analogues in Part II, regardless of whether one thinks that there is good motivation to believe in the existence of monopoles---a question that ultimately can only be answered experimentally. Furthermore, there may also be good reason, which goes back to Dirac, why magnetic monopoles have not been observed: namely, their magnetic charges are large, and the force between opposite magnetic charges is much larger than the Coulomb force between opposite electric charges, so that magnetic monopoles would tend to recombine into magnetically neutral particles.\\

The second point, (ii), rules out the existence of vector potentials ${\bf A}$ that are everywhere smooth (even if we exclude the origin, where the monopole is sitting). Indeed, if the vector potential is smooth, then the divergence of the magnetic field is identically zero, and the Maxwell equations, Eq.~\eq{4Max}, do not admit a source of magnetic charge. Thus Dirac's (1931:~p.~68; 1948:~p.~819) method to introduce monopoles was to allow for a singular vector potential, i.e.~one that is not only singular at the location of the monopole, but along an entire semi-infinite line from the origin to spatial infinity: the {\it Dirac string.} If Dirac's quantization condition is satisfied, then (as we will see below) the magnetic flux along the Dirac string does not give rise to any {\it observable} phases of the type that appear in the Aharonov-Bohm (1959:~pp.~486-487) effect for solenoids.\footnote{See e.g.~Weinberg (1984:~pp.~4-5) and Preskill (1984:~p.~467).}

Notice that the singularity of the Dirac string is quite different from the singularity of the monopole at a single point. The latter is regarded as a physical singularity that results from an {\it idealisation}, i.e.~the point-like singularity models a physical particle whose internal structure is unknown in the current theory. By contrast, the singularity of the {\it string} is not supposed to be physical: there is, in this theory, nothing of physical significance there (in particular, no energy or momentum), the more so because the location of the string depends on a choice of gauge.

Agreed, there are two reasons one may worry about this argument. First, the admissibility of the Dirac string---and, with it, the admissibility of the monopole attached to it---depends on the singularity's lack of {\it observability.} But if the worry is that the Dirac string is {\it singular,} why would its being unobservable make it admissible? Thus the worry is that this depends on a verificationist position that many philosophers find questionable, since a theory's class of physically meaningful statements outstrips its class of observable statements. Second, the physical parameter that correlates with the observability of the Dirac string (namely, the Aharonov-Bohm {\it phase factor,} which we will discuss below) is ill-defined if the path of an electron near the monopole crosses the singular Dirac string. Thus Wu and Yang (1975:~p.~3847) say that `this difficulty {\it must} be resolved in order to use a ... phase factor as a fundamental concept to describe electromagnetism'. 

This last argument appears to contradict the claim that the singularity is not observable: but note that, by contrast with the previous argument, Wu and Yang's argument for avoiding singularities is {\it not} motivated by the observable effects of singularities. To derive his monopole-cum-string in such a way that the string is not physical (i.e.~it carries no energy or momentum), Dirac (1948:~p.~822) had to `impose the condition that a {\it string must never pass through a charged particle'.} But subsequent theoretical developments increasingly modified the Dirac string, thus giving it a {\it physical} interpretation, in other theories, as a {\it dual string} that binds together a pair of monopoles, and thus carries energy and momentum (i.e.~a precursor of string theory: see Section \ref{analo2} and Nambu (1974:~p.~4264)).

Fortunately for us, both worries can be bypassed by introducing monopoles in a way that does not use a singular field. In this light, the appearance of a singularity stems from the use of poor mathematical resources. The idea is to allow for {\it locally (but not globally) defined potentials,} i.e.~roughly, potentials that are defined by different functions in different regions of space. When comparing potentials between different regions, a local gauge transformation is in general required. This implies that the gauge potential is not a function defined over the whole space, but only defined locally. The mathematical formalism to describe such gauge potentials is the formalism of fibre bundles, where the gauge potential is a connection on a principal U(1) bundle that is a larger space consisting of copies of the U(1) group attached to each point of the base space.\footnote{Fibre bundles were first introduced into the physics of electromagnetism by Wu and Yang (1975:~pp.~3845, 3851). However, their use of `global' and `local' differs from the modern one that we adopt here. For introductions to this formalism, see Hamilton (2017:~pp.~207-212, 261-266) and Nakahara (2003:~p.~366, 374-383).} Although we will not need to pursue this formalism here, the formulas we will give below for the Dirac monopole can be easily recast in fibre-bundle language.\footnote{There are many elementary introductions to Dirac monopoles, along the lines that we will discuss below. See for example Harvey (1996:~pp.~10-11), Figueroa-O'Farrill (1998:~pp.~14-15), Naber (2011:~pp.~1-7), and Nakahara (2003:~pp.~61-63). Dirac's original papers are (1931, 1948). The fibre-bundle description of the Dirac monopole is in Nakahara (2003:~p.~401).}\\
\\
{\bf Exposition.} We begin by showing how electric-magnetic duality generalises when we include both electric and magnetic sources:\footnote{The factors of $c$ in the new source terms have been conveniently chosen so that the magnetic sources have the same units as the electric ones, and the duality map takes the form in Eqs.~\eq{rJ}-\eq{qig}.}
\bea\label{elmagn}
\nabla\cdot{\bf E}={1\over\e_0}\,\r_{\sm e}\,,&&\na\times{\bf E}+~~{\pa{\bf B}\over\pa t}~=-\m_0c~{\bf J}_{\sm m}\nn
\nabla\cdot{\bf B}={1\over\e_0 c}\,\r_{\sm m}\,,&&\na\times{\bf B}-{1\over c^2}{\pa{\bf E}\over\pa t}=\m_0\,{\bf J}_{\sm e}\,.
\eea
With the complex field defined in Eq.~\eq{calE}, and defining also complex charge and current densities:
\bea\label{rJ}
\r&:=&\r_{\sm e}+i\,\r_{\sm m}\nn
{\bf J}&:=&{\bf J}_{\sm e}+i\,{\bf J}_{\sm m}\,,
\eea
one easily sees that the Maxwell equations with sources are invariant under the generalised duality maps Eq.~\eq{so2} and Eq.~\eq{eith}, provided the matter fields transform in the following way:
\bea\label{rpr}
\r'&=&e^{-i\th}~\r\nn
{\bf J}'&=&e^{-i\th}~{\bf J}\,.
\eea
In particular, if we consider pointlike sources with electric charge $q$ and magnetic charge $g$, they transform as follows:
\bea\label{qig}
q+ig~\mapsto~ e^{-i\th}\,(q+ig)\,.
\eea

Consider a point-like source of the magnetic field, in analogy with the electric field of a point particle:
\bea\label{Bpoint}
{\bf B}={g\over4\pi\e_0\, r^3}\,{\bf r}\,.
\eea
Here, $g$ is the magnetic charge. It can be defined as the total magnetic flux passing through any closed two-surface $S$ that contains the origin, ${\bf r}=0$, where the charge is located:
\bea\label{gcharge}
g=\e_0\int_S\dd{\bf S}\cdot{\bf B}\,.
\eea
Since $S$ is closed, this would, by Stokes' theorem, be zero if the magnetic field were given in terms of a vector potential that is defined over the entire surface, ${\bf B}=\na\times{\bf A}$. However, as we noted earlier, the vector potential is not defined over the whole surface $S$.\footnote{Because of the combination $g/\e_0$ that appears in the two equations above, Eqs.\eq{Bpoint} and \eq{gcharge}, the literature on magnetic monopoles usually absorbs $\e_0$ into $g$, so that $\e_0$ in effect disappears. Also, units are usually chosen where $c=1$. We will follow these convenient conventions in the rest of this Subsection.\label{convention}}
To describe the field of the monopole everywhere in the space, we have to choose at least {\it two} vector potentials, each of them associated with a region of space.\footnote{Any configuration of a {\it single} vector potential ${\bf A}$ that reproduces the monopole magnetic field Eq.~\eq{Bpoint} will have the Dirac string singularity that we mentioned earlier (however, the location of the string depends on the choice of the vector potential).} 

It is simplest to divide the sphere $S$ that appears in the magnetic charge, Eq.~\eq{gcharge}, into two parts, e.g.~a northern and a southern hemisphere, and define two different vector potentials, i.e.~different vector potentials, on the two hemispheres. On the northern hemisphere, take the following vector potential:

\bea\label{AN}
{\bf A}_{\tn N}={g\over4\pi r}{1-\cos\th\over\sin\th}\,\hat{\bf e}_\f\,,
\eea
with the obvious label `N' for northern hemisphere, and (below) `S' for southern hemisphere. We use spherical coordinates $(r,\th,\f)$, where $r$ is the radius of the sphere $S$. The above vector potential is defined everywhere on the sphere, {\it except} at $\th=\pi$, i.e.~at the south pole, which lies outside the region covered by this potential: this would have been the location of the Dirac string for this potential, if there had been one. (And, in spite of its appearance, the limit of the vector potential is smooth, and goes to zero, at the north pole, i.e.~for $\th\rightarrow0$.)

\begin{figure}
\begin{center}
\includegraphics[height=3.5cm]{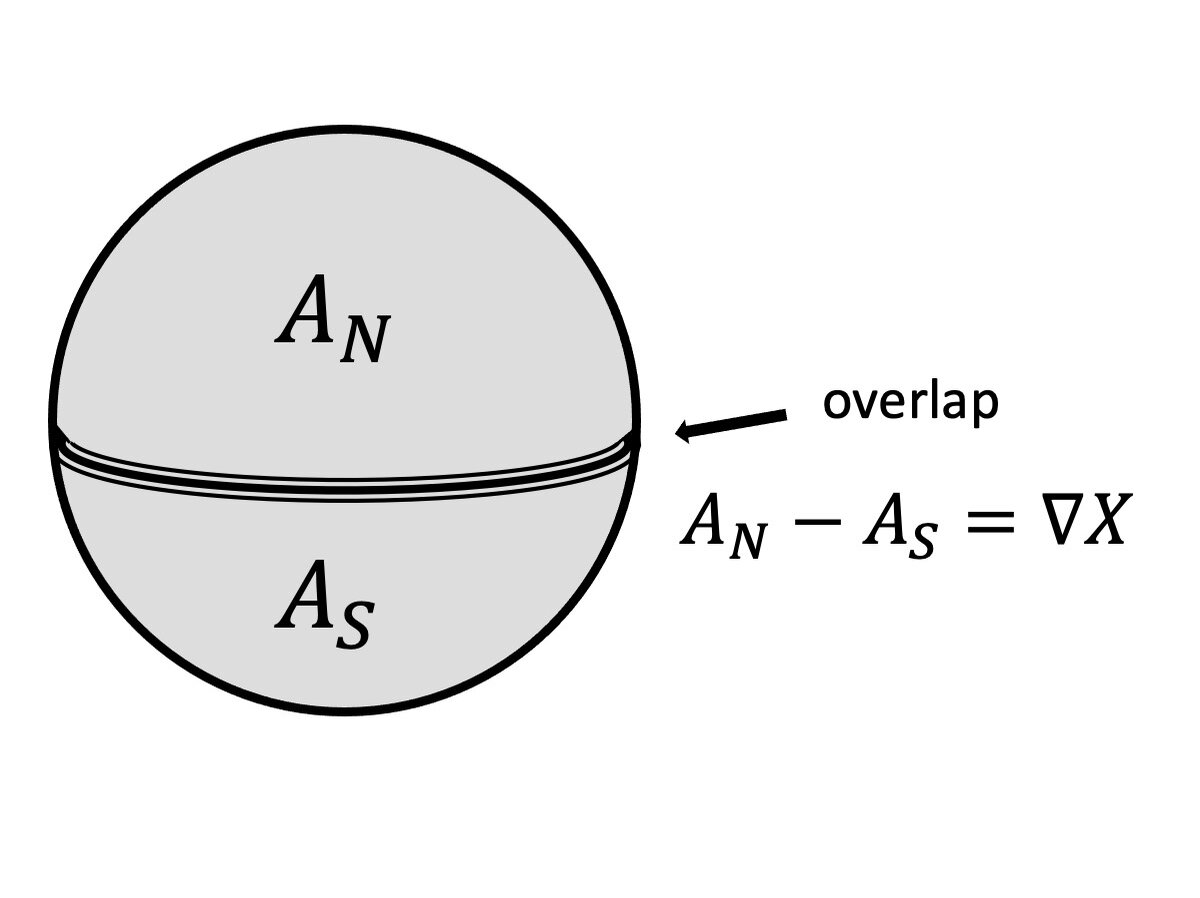}
\caption{\small Two gauge potentials, covering the northern and southern regions of a two-sphere. Where they overlap, they differ by the gradient of $\chi$.}
\label{Two-sphere}
\end{center}
\end{figure}

On the southern hemisphere, we define a second vector potential:
\bea\label{AS}
{\bf A}_{\tn S}=-{g\over4\pi r}{1+\cos\th\over\sin\th}\,\hat{\bf e}_\f\,,
\eea
which again is well-defined everywhere except at the north pole, where $\th=0$, and lies outside the region covered by this potential: this would have been the location of the Dirac string for this potential, if there had been one. But by choosing two different functions on the two regions, we have eliminated the Dirac string. Both potentials satisfy $\na\times{\bf A}_{\tn{N,S}}={\bf B}$, where ${\bf B}$ is the magnetic field of the monopole, Eq.~\eq{Bpoint}, and so both describe the monopole's magnetic field: one on the northern hemisphere, the other on the southern hemisphere. See Figure \ref{Two-sphere}.

Note that the above vector potentials are well-defined not only on a (hemisphere of a) two-sphere. ${\bf A}_{\tn N}$ is well-defined everywhere except on the half-line $\th=\pi$ (i.e.~the half-line starting at the origin and extending along the negative $z$-axis), and ${\bf A}_{\tn S}$ is well-defined everywhere except on the half-line $\th=0$ (i.e.~the half-line starting at the origin and extending along the positive $z$-axis). Where the two vector potentials overlap, their difference satisfies: $\nabla\times({\bf A}_{\tn N}-{\bf A}_{\tn S})=0$; hence this difference can be written as a total divergence, ${\bf A}_{\tn N}-{\bf A}_{\tn S}=\nabla\chi$. Since ${\bf A}_{\tn N}-{\bf A}_{\tn S}$ is defined everywhere except on the would-be Dirac string (i.e.~the $z$-axis), the space where $\chi$ is defined is not simply connected, and so the function $\chi$ need not be single-valued. We find: $A_{\tn N}-A_{\tn S}={g\over2\pi r\sin\th}=\na\chi$, so that (up to an irrelevant integration constant) $\chi={g\over2\pi}\,\f$, where $\f\in[0,2\pi)$.

Using Stokes' theorem, one can check that Eq.~\eq{gcharge} is indeed satisfied for these vector potentials: $g=\oint_C\dd{\bf\ell}\cdot\nabla\chi=\chi|^{\f=2\pi}_{\f=0}$, where $C$ is the equator. And thus, the {\it pair} of local vector potentials Eqs.~\eq{AN}-\eq{AS} correctly defines a monopole solution with magnetic field Eq.~\eq{Bpoint}, and no Dirac string singularity.\\
\\
{\bf Dirac's quantization argument}\\
\\
Dirac's argument for the quantization of charges combines the above analysis with the framework of quantum mechanics. Thus Dirac considers the wave-function of an electrically charged particle, of charge $q$ and mass $m$, in the field of the putative magnetic monopole charge. The Schr\"odinger equation in the presence of a (not necessarily everywhere-defined) vector potential ${\bf A}$ takes the following form:
\bea\label{SA}
i\hbar\,{\pa\psi\over\pa t}=-{\hbar^2\over2m}\left(\nabla-{i\over\hbar}\,q{\bf A}\right)^2\psi+V\psi\,.
\eea
This equation is invariant under gauge transformations ${\bf A}\mapsto{\bf A}+\na\chi$, provided the wave-function also transforms by a corresponding phase, which may depend on space but not on time:
\bea
\psi~\mapsto~ e^{{i\over\hbar}q\chi}\,\psi\,.
\eea
In other words, in quantum mechanics, and in the presence of an electromagnetic potential, the wave-function also transforms non-trivially from one patch to another.

Back to the monopole. In this case, we have that the wave-functions on the northern and southern regions are related by:
\bea
\psi_{\tn N}=e^{{i\over\hbar}{qg\f\over2\pi}}\,\psi_{\tn S}\,,
\eea
where $\f$ is the angular coordinate describing the equatorial circle. This formula tells us how to match the two wave-functions: we require that both wave-functions are single-valued when we go around the circle by an angle of $2\pi$, i.e.~$\psi_{\tn{N,S}}(r,\th,\f)=\psi_{\tn{N,S}}(r,\th,\f+2\pi)$. This implies that $e^{iqg/\hbar}=1$, and so there must be an integer $n$ such that:\footnote{The conventions used here are as in footnote \ref{convention}. To restore SI units and usual conventions for the magnetic charge, one puts back a factor of $c$ on the right-hand side and rescales the magnetic charge by $\e_0$ so that $e$ and $g$ have the same units. One can also check the conventions by comparing with the fine structure constant in SI units, $\a=e^2/4\pi\e_0\hbar c$.}
\bea\label{Diracq}
qg=2\pi n\hbar\,.
\eea
This is the {\bf Dirac quantization condition}. Although it does not predict the value of the electric charge, since the values of $g$ and $n$ are unknown, it does lead to an astonishing conclusion. Namely, that if there is a monopole with magnetic charge $g$ somewhere in the universe, then {\it any} electrically charged particle that is described by the Schr\"odinger equation, Eq.~\eq{SA}, has an electric charge that is an integer multiple of $2\pi\hbar/g$.\footnote{For a discussion of cases with several charges, and also of dyons, i.e.~particles with both electric and magnetic charge, see Goddard and Olive (1978:~pp.~1367, 1373).}

This again illustrates the idea, from Section \ref{featurerole}, of {\it hard-easy}. Since the charges $e$ and $g$ measure, respectively, the strength of the electric and magnetic forces of matter particles, a {\it small} magnetic charge means a {\it large} electric charge, and vice versa.\footnote{Dirac's argument is in the context of quantum mechanics. As we will discuss in Chapter \ref{EMYM}, renormalization effects in quantum field theory can spoil the condition Eq.~\eq{Diracq}, and so one needs to understand whether, and if so how, this relation continues to hold when quantum corrections are taken into account.}

Notice also that Eq.~\eq{Diracq} is precisely the condition that there is {\it no} Aharonov-Bohm (1959:~pp.~486-487) effect for the monopole:\footnote{The Aharonov-Bohm effect was first verified experimentally by Chambers (1960:~p.~3).} namely, with this condition the flux along the Dirac string does not give observable effects in e.g.~an experiment whereby electrons are made to interfere with the Dirac string. For example, consider an interferometer where a biprism splits a beam of electrons in the horizontal plane, such that the two arms enclose the location of a Dirac string that is placed just behind the biprism and runs vertically. According to Aharonov and Bohm (1959:~p.~486), the only observable effect\footnote{This point, i.e.~that only the {\it phase factor} $\exp(-{iq\over\hbar}\oint\dd{\bf \ell}\cdot{\bf A})$, rather than the loop integral $-{iq\over\hbar}\oint\dd{\bf \ell}\cdot{\bf A}$ itself, is physically significant, is developed by Wu and Yang (1975:~pp.~3848-3849), especially in their Theorem 1.} of the presence of the solenoid (in this case, a Dirac string) is in the interference between the two electron beams, which is given by the relative phase of their wave-functions, given by: $\exp(-{iq\over\hbar}\oint_C\dd{\bf \ell}\cdot{\bf A})=\exp(-iqg/\hbar)=\exp(-2\pi in)=1$. Here, $C$ is a closed curve, topologically a circle, between the source and the plane of observation, containing the two arms of the interferometer, and enclosing the Dirac string. Everywhere in the horizontal plane that lies outside the region enclosed by the curve $C$, the Faraday tensor is zero. Notice that the measurable phase does not depend on the length or precise shape of the path.\footnote{This is proven by Wu and Yang (1975:~pp.~3847-3848), by using the group properties of the phase factors.} 
That this phase is equal to one is equivalent to the Dirac condition, Eq.~\eq{Diracq}. So (unlike the case of the usual Aharonov-Bohm effect with a solenoid) there is no interference iff the Dirac condition is satisfied, so that the presence of the Dirac string has no observable effect.\footnote{For a brief discussion of `observability', see our reply to point (ii) in Section \ref{Dqc}. For more details about electron interference experiments with monopoles, see Weinberg (1984:~pp.~4-5) and Preskill (1984:~p.~467).} 

\section{Kramers-Wannier duality in statistical mechanics}\label{dualpf0}

We now expound, in this Section, {\it Kramers-Wannier duality}. It relates a two-dimensional Ising model on a square lattice at low temperature, to a duplicate Ising model (i.e.~on a two-dimensional square lattice) at high temperature. Thus in effect, a cold, ordered ferromagnetic lattice is related to a hot, disordered one. A bit more precisely: the main idea is that an expression for the partition function---i.e.~the function which encodes much of the physics of statistical mechanical systems---that is valid at high temperatures is identical to an expression for the partition function on an isomorphic lattice (called the {\it dual lattice}) that is valid at low temperatures. 

This example is important---both for our exposition and scientifically---for four main reasons; as follows.

(1): It is simple. Indeed, within statistical physics, it is the simplest duality that is nevertheless scientifically important. Of course, this is due in part to the choice of model. In statistical mechanics, lattice models have long been heuristically invaluable for understanding real systems.\footnote{For the history of the Ising model, cf.~Brush (1967).}

Besides, the Kramers-Wannier duality itself is also simple in that its dual lattice is isomorphic to the original one; while in general, for dualities of lattice models, this is not so. (The dual lattice being isomorphic to the original is, in this simple case, a case of self-duality: see Section \ref{std}.\footnote{However, we will refine this statement at the end of Section \ref{IMD}.})

(2): Despite its simplicity, Kramers-Wannier duality has the scientifically crucial feature of mapping between disparate regimes, so that problems that are too hard to solve in one regime can be more easily addressed---and with luck, solved---by being formulated in the other regime. As we mentioned, the duality maps the physics of a hot, disordered lattice to the physics of a cold, ordered lattice. In Section \ref{featurerole}, we called this feature {\it hard-easy}. 

Thus solving for the thermodynamic limit of the Ising model will give us a first example of what Section \ref{mvd} will call the `geometric view of theories'.

(3): It is the prototype for many other dualities, not only in statistical mechanics but also in field theory. It is also the prototype for the interplay between dualities and phase transitions, which we will study in Part II. In particular, the ideas involved in proving it can be adapted to many other systems. So we shall here, and in Appendix 4.A, give some details about these ideas and thus pave the way for Chapter 5. We say `also in field theory' , since various important ideas and results in statistical mechanics map over to, or have analogues in, field theory (especially Euclidean field theory). So roughly speaking, the notions of high vs.~low temperature in statistical mechanics map to large vs.~small coupling constants in field theory. (So in field theory, a duality's {\it hard-easy} feature from Section \ref{themesd} will usually be a matter of mapping a hard problem with a large coupling constant to an easier one with a small coupling constant.) Also, the idea of spontaneous magnetisation as an {\it order parameter} for a macroscopic phase of matter, and its associated {\it symmetry-breaking} feature, is a central idea in both statistical mechanics and quantum field theories.

(4): We note also that the duality was historically important. For in their paper (1941), Kramers and Wannier used it to identify the critical point of the two-dimensional Ising model, i.e.~the value of the critical temperature $T_{\sm c}$ for the ferromagnetic phase transition, i.e.~the temperature at which some thermodynamic quantities (in particular, the specific heat) diverge. This was some two years before Onsager solved the model completely, i.e.~calculated the partition function, and its thermodynamic limit, exactly (for zero applied magnetic field).

So to sum up these four reasons: It would be hard to over-state the importance of this duality for the study of lattice models in statistical mechanics: not just because of the centrality of the Ising model, but also because the ideas of Kramers-Wannier duality can be generalised in various ways.

\subsection{Duality of the partition function and order parameters}\label{dualpf}

In this Section, we will follow the exposition and notation of Savit (1980:~pp.~455-457), and will confine ourselves to the duality for the partition function. Section \ref{IMD} will give the details of the duality for other quantities such as spin correlation functions.\footnote{Savit (1980:~p.~457f.) gives details of how these ideas generalise to other statistical mechanical systems. Other expositions of Kramers-Wannier duality include: Thompson (1972: Section 6-2), Baxter (1982:~pp.~73-78), Lavis and Bell (1999: Sections 8.2-8.3, 206-211) and Kadanoff (2000:~pp.~359-369). Of course the details vary. For example: Thompson works in terms of power series expansions for the partition function; Baxter exhibits the duality only for an infinite lattice (i.e.~the thermodynamic limit); but his treatment, and Kadanoff's, contain a generalisation to an anisotropic lattice, i.e.~where the coupling $J$ is different along the $x$- and $y$-axes of the lattice.}
That Section will discuss the important notion of a `disorder parameter', which will recur in later Chapters.

Consider a two-dimensional square lattice, where the nodes, or lattice sites, are labelled by a pair of integers, $n_x$ and $n_y$, which we will collectively denote by $i=(n_x,n_y)$. On each lattice site is a variable, $s_i$, that can take on values $\pm1$. So the variable $s$ is a toy-model of a local degree of freedom---about the simplest one can imagine. It is called a `spin', since (as hinted in (1) of the Preamble), the collective behaviour of $s$ across all sites, in this and similar models, turns out to give good models of magnetism. The {\bf Ising model} Hamiltonian associated with this lattice is:
\bea\label{isingH}
H_{\sm{2D Ising}}=-\sum_{(ij)}J_{ij}\,s_i\,s_j\,,
\eea
where $(ij)$ denotes a sum over all nearest-neighbour pairs. If the couplings, which give the strength of the lattice bonds, are isotropic, i.e.~$J_{ij}=J$, the Hamiltonian reduces to Eq.~\eq{Hisotropic}, although we will not always assume this. We introduce the notation: $\b_{ij}:=J_{ij}/k_{\tn B}T$ and $\b:=J/k_{\tn B}T$, where $k_{\tn B}$ is Boltzmann's constant, and $T$ is the temperature (so that the $\b$'s are called `inverse temperatures'---though note that they also depend on the couplings $J$). 

The {\bf ferromagnetic} case is $J>0$. In a ferromagnet, the tendency is for neighbouring spins to point in the same direction: for with $J>0$, the energy is reduced when neighbouring values of $s_i$ are equal. In this way, a macroscopic magnetisation appears, which can be taken as the ferromagnet's {\it order parameter.} We will here focus on the ferromagnetic case; we will return to non-ferromagnetic cases in Section \ref{IMD}.

We now recall that the fundamental idea of (Gibbsian) statistical mechanics is to assign to each state of the system, a probabilistic weight (`Boltzmann factor'), which is an exponential of minus the energy of the state divided by the temperature. These weights are not normalized: that is, their sum $Z$ over all possible states is not unity. But the functional form of $Z$, the partition function, contains a great deal of information. For one can obtain various thermodynamic quantities as functions of $Z$ and its derivatives. 

More precisely, we say that a state of our system, the lattice, is a specification, $s=\{s_i\}$, of the spin values at all the lattice sites. The Boltzmann factor for each state $s$ is then: $\exp (-H(s)/ k_{\tn B}T)$, where $H(s)$ is the energy of the state $s$. So the partition function $Z$ of the lattice, equipped with the Hamiltonian in Eq.~\eq{isingH}, is given by:
\bea\label{ising0}
Z_{\sm{2D Ising}}(\{\b_{ij}\})=\sum_{\{s\}}\exp\left(\sum_{(ij)}\b_{ij}\,s_i\,s_j\right) =\sum_{\{s\}}\prod_{(ij)}\exp\left(\b_{ij}\, s_i\,s_j\right),
\eea
where we sum over all the possible states. If the lattice system is isotropic, so that $\b_{ij}=\b$, this simplifies to:
\bea\label{ising}
Z_{\sm{2D Ising}}(\b)=\sum_{\{s\}}\exp\left(\b\sum_{(ij)}s_i\,s_j\right) =\sum_{\{s\}}\prod_{(ij)}\exp\left(\b s_i\,s_j\right).
\eea

In this Section, our goal is to show that this partition function is invariant (up to a numerical factor, specified below) under an appropriate inversion of the temperature, and evaluation on another two-dimensional square lattice, called the dual lattice (see below). A full proof is long. In Appendix 4.A, we sketch the main steps: which, as mentioned, are a useful orientation for examples in Section \ref{IMD}. 

In short, the {\bf dual lattice} is obtained by placing, in the centre of each plaquette (tile) of the given lattice, a site (vertex), and then connecting nearest-neighbour sites by a link. This new lattice is again a two-dimensional square lattice, just displaced from the original by half a lattice-spacing in each of the $x$ and $y$ directions. And we also associate a new (dual) spin variable $\s_i = \pm 1$ with each site of the dual lattice. 

With these preliminaries, using result (A) in Appendix 4.A (especially Eq.~\eq{Savit2.9a}), the above partition function can be rewritten in terms of spin variables defined on the dual lattice, as follows:
\bea\label{Savit2.9b}
Z_{\sm{2D Ising}}(\b)&=& \frac{1}{2}\,(\sinh 2\ti\b)^{-N} \, \sum_{\{\s\}} \exp\left(\ti\b \sum_{(ij)} \s_i\s_j\right),
\eea
where the $\sigma_i$'s are spin variables on dual lattice sites, and the sum in the exponent is over nearest-neighbour pairs in the dual lattice. Here, we have defined a {\bf dual inverse temperature} $\ti \b$ by:
\bea\label{dualb}
\ti \b:=-\frac{1}{2}\,\ln\tanh\b\,.
\eea
The result on the right-hand side of Eq.~\eq{Savit2.9b} is again the partition function of the Ising model on a square lattice, i.e.~Eq.~\eq{ising}, but at inverse temperature $\ti\b$. In other words, the partition function has the same form as the original one. (The spins are now labelled by dual variables $\s_i$ rather than $s_i$, but this is not important, since they are dummy variables, and we sum over their values.) Thus we conclude that:
\bea\label{dualKW}
Z_{\sm{2D Ising}}(\b)&=&\frac{1}{2}\, \sinh^{-N}(2\ti\b)~\sum_{\{\s\}}\prod_{(ij)} \exp(\ti\b\,\s_i\,\s_j)\nn
&=&\frac{1}{2}\,\sinh^{-N}(2\ti\b)~Z_{\sm{2D Ising}}(\ti\b)\,.
\eea

This is the central {\it duality formula} for the Ising model, written (by using Eq.~\eq{dualb}) in more symmetric form in Eq.~\eq{Zmod}. In words: the partition function for a system at inverse temperature $\b$ is equal to, apart from an overall spin-independent factor, the partition function for a system of Ising spins at an inverse temperature $\ti\b$. \\

We note that $\ti\b$ is a monotonically decreasing function of $\b$, satisfying $\ti\b(0)=\infty$ and $\ti\b(\infty)=0$. So the high temperature regime for the original lattice, i.e.~Eq.~\eq{ising}, is mapped to the low temperature regime for the dual lattice, and vice versa. As we announced in (2) at the beginning of this Section: this is scientifically important. It illustrates the general idea that a duality enables us to address, and even solve, problems about a regime where the problem is hard (usually because the parameter in a perturbative expansion gets too large), by solving a ``dual problem'' which is easy, or anyway easier (usually because the parameter is smaller). Thus for the Ising model, we can trust a power series in $\b$ when $\b$ is small, i.e.~at high temperature; but not at low temperature. But thanks to this duality, we can compute low temperature properties of the Ising model simply by performing a high-temperature expansion in $\ti\b$, i.e.~using Eq.~\eq{dualKW}.

As we also mentioned in (1) at the beginning of this Section: this is sometimes called a `self-duality', since the duality map leads to an isomorphic lattice, and to the very same partition function, apart from an overall factor. (The lattice and the dual lattice being isomorphic makes matters simpler, not least in notation: this no longer holds for other lattices.)\footnote{Thus the hexagonal lattice is dual to the triangular lattice: see Baxter (1982:~pp.~78-80) and Fisher and Ferdinand (1967:~p.~170).}
This is illustrated by the symmetric form of the duality:
\bea\label{ZZbar}
\frac{Z_{\sm{2D Ising}}(\b)}{\sinh^{N/2}2\b}=\frac{Z_{\sm{2D Ising}}(\ti\b)}{\sinh^{N/2}2\ti\b}\,,\nn
\sinh2\b\,\sinh2\ti\b=1\,.~~~~
\eea
{\bf The thermodynamic limit, and order parameters.} Finally, we turn to give some details about (4) at the beginning of this Section: how the duality allows one to identify, in the thermodynamic limit, the critical point of the model. This illustrates how dualities can be heuristically valuable; (and as we reported in (4), this argument was given by Kramers and Wannier themselves). The idea is to assume that, in the limit of an infinite number of lattice sites, i.e.~$N\rightarrow\infty$, the partition function of the given lattice (equivalently: the partition function of the dual lattice!) has only one singularity. Then it is straightforward to show (see below) that this singularity must occur when the two parameters $\b$ and $\ti \b$ satisfy a simple algebraic relation.
 
Thus in the limit of an infinite number of lattice sites, i.e.~$N\rightarrow\infty$, we can define an expression for the free energy (which, away from any critical point, has a finite value): $F:=-\lim_{N\rightarrow\infty}\frac{1}{N}\,\ln{Z_{\sm{2D Ising}}}$, and satisfies:
\bea\label{sinhb}
F(\b)=-\sqrt{\frac{\sinh2\ti\b}{\sinh2\b}}+F(\ti\b)\,.
\eea
Assuming that there is a single point $\b_{\sm c}$ where the free energy is divergent (see Kramers and Wannier (1941:~p.~261)), we see from the above that this must be at the self-dual value $\b=\ti\b$. Then Eq.~\eq{ZZbar} gives $\sinh2\b_{\sm c}=1$ and $\b_{\sm c}=0.440687$. 

Thus at $\b=\b_{\sm c}$, the system undergoes a phase transition at finite temperature. Notice that, because $\b=J/k_{\tn B}T$, it is---in this model---not just the temperature that is critical, but the ratio of the temperature to the coupling, although we will still sometimes talk about the `critical temperature' $T=T_{\sm c}$.

There is another way to characterise and understand the phase transition at $T=T_{\sm c}$ that we have derived, in terms of the {\bf magnetisation}, i.e.~the average of the spins, i.e.~$M=1/N\sum_{i=1}^N\bra s_i\ket$. The magnetisation can be calculated by the usual method of `coupling to an external source', i.e.~coupling the system to an external magnetic field $B_i$. The Hamiltonian is: $H=-\b\sum_{(ij)}s_is_j-\sum_iB_is_i$. Taking the derivative of the free energy $F$, and then setting the field to zero, gives the average of the $i$th spin:
\bea\label{Isingb}
\bra s_i\ket=\lim_{B_i\rightarrow0}\,{\pa\over\pa B_i}\, \ln\sum_{\{s\}}\exp\left(\b\sum_{(ij)}s_i\,s_j+\sum_i\,B_i\,s_i\right),
\eea
and hence the magnetisation $M$ is an {\bf order parameter} in the sense that, in the limit $N\rightarrow\infty$, it is non-zero at low temperatures (i.e.~$\b>\b_{\sm c}$), i.e.~when the spins are aligned over large macroscopic regions, but it is zero above the critical temperature (i.e.~$\b<\b_{\sm c}$). As a function of the reduced temperature, $t:=(T-T_{\sm c})/T_{\sm c}$, the magnetisation, near the critical temperature, is: 
\bea\label{Mt}
M(t):=\lim_{N\rightarrow\infty}{1\over N}\sum_{i=1}^N~\bra s_i\ket=\left\{
\begin{array}{cc}0~,&t>0\\
m\,(-t)^\b~,&~~t\rightarrow0^-
\end{array}\right. ,
\eea
and $m$ and $\b$ are numbers, independent of it. The exponential $\b$ is known as a {\bf critical exponent}. In mean field theory, these coefficients are easy to calculate: $m=3$ and $\b=1/2$. However, the correct value for the two-dimensional Ising model (e.g.~from Onsager's solution) is $\b=1/8$.\footnote{See Baxter (1982:~p.~44). The exact result for the spontaneous magnetisation is in Yang (1952:~p.~815).}
(The fact that for some models, the critical exponents as calculated by mean field theory are wrong is a ``symptom'' of the need for a more subtle treatment of the coarse-graining involved in macroscopic variables like magnetisation: a treatment that is given by renormalisation group methods. This will be important in various places later on: especially in Sections \ref{cmds} and \ref{emergence}.) 

Thus the magnetisation is a continuous but non-differentiable function of the temperature. And, in the thermodynamic limit, the specific heat, which is obtained by taking the second derivative of the free energy with respect to the inverse temperature, diverges at the critical temperature (logarithmically in $t$: see Onsager (1944:~p.~117)). At the critical temperature, there is a second-order phase transition, with no latent heat.

\subsection{Spin-dislocation duality and disorder parameters}\label{IMD}

This Section extends Kramers-Wannier duality, which the previous Section established for the partition function, to two-point correlation functions. We shall see that the dual of a two-point correlation function between the sites $i$ and $j$, on a homogeneous lattice, is the partition function on a dual lattice with a topological defect, or dislocation, between $i$ and $j$. This extension of the duality will be a useful introduction to the vortex and monopole-like solutions of Sections \ref{PVD} and \ref{mmYM}. 

The relation between the $s_i$'s and the dual $\s_i$'s is not simple, but it is given indirectly by the solution of the delta-function constraint in Appendix 4.A, i.e.~Eq.~\eq{kmi}. To get more insight into this relation, we will study the two-point functions of the spins, which are our real interest because of how they look in the dual picture.

Thus we begin with the two-point function. It will be easiest to begin with the dual spins and to calculate their two-point function, $\bra\s_i\,\s_j\ket$ (we will then transform this back to the original spin variables $s_i$):
\bea\label{sigmaij}
\bra\s_i\,\s_j\ket_{\ti\b}:={1\over Z_{\sm{2D Ising}}(\ti\b)}~\sum_{\{\s\}}\s_i\,\s_j\,\exp\,(\ti\b\sum_{(kl)}\s_k\,\s_l)\,,
\eea
calculated at the dual inverse temperature $\ti\b$, which is related to the original temperature through Eq.~\eq{dualb}. 

\begin{figure}
\begin{center}
\includegraphics[height=3.5cm]{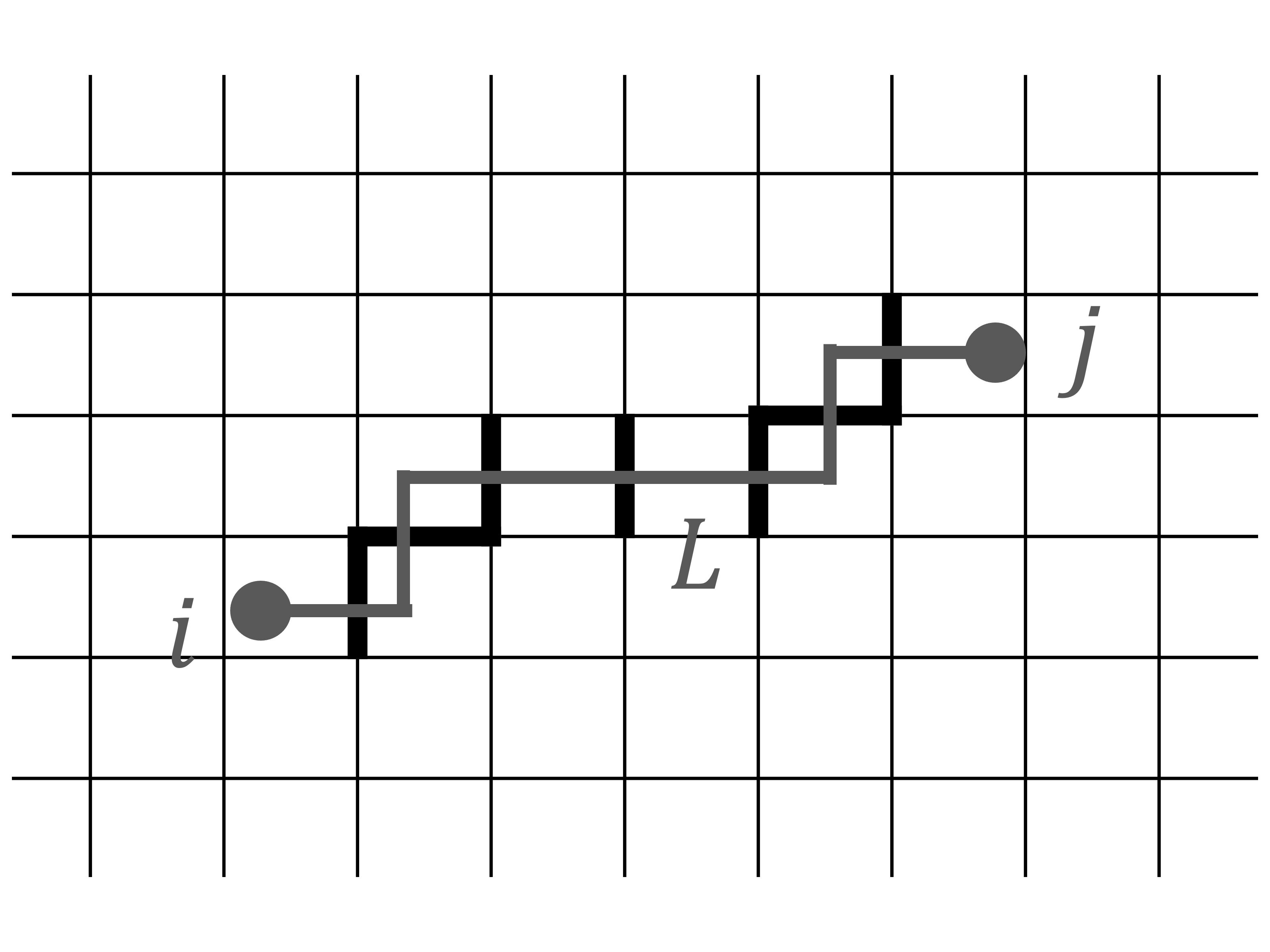}
\caption{\small The given lattice, with dual spins $\s_i$ and $\s_j$ on the sites $i$ and $j$ of the dual lattice (the dual lattice is not drawn), with a line $L$ between the two spins. The two-point function $\bra\s_i\s_j\ket_{\ti\b}$ of the dual lattice is given by the partition function of the given lattice, with anti-ferromagnetic couplings on the (thinckened) links that cross the line $L$.}
\label{HT2}
\end{center}
\end{figure}

Using the result (B) in Appendix 4.A, the two-point function of the {\it dual} Ising model, Eq.~\eq{sigmaij}, can be rewritten in terms of the original lattice, as follows:
\bea\label{spinglass}
\bra\s_i\,\s_j\ket_{\ti\b}&=&{1\over Z_{\sm{2D Ising}}(\b)}\sum_{\{s\}}\exp\left(\sum_{(kl)}\b_{kl}\,s_k\,s_l\right),\\
\b_{kl}&=&+\b\,,~\forall\,(k,l)\not\perp L_{ij}\nn
\b_{kl}&=&-\b\,,~\forall\,(k,l)\perp L_{ij}\,,\nonumber
\eea
where $(k,l)\perp L_{ij}$ means that the link $(k,l)$ of the original lattice intersects the line $L_{ij}$, between the sites $i$ and $j$ of the dual lattice, perpendicularly (see Figure \ref{HT2}). Also, $\b$ is related to the dual inverse temperature $\ti\b$ as in Eq.~\eq{dualb}. 

Thus we have expressed the dual two-point function Eq.~\eq{sigmaij} in terms of the original spin variables $s_i$.

Recall the definition $\b_{kl}:= J_{kl}/k_{\tn B}T$, where (from Eq.~\eq{spinglass}) the coupling is negative on the links that are traversed by the path $L$ connecting $k$ and $l$, i.e.~$J_{kl}<0$. Thus the above result for the two-point function means that the lattice on which the dual Ising model is defined is not a pure ferromagnet, i.e.~does not only have ferromagnetic bonds, but has {\bf anti-ferromagnetic} bonds (the signs differ, but all the links have the same strength of coupling). 

In this way, we see that the dual picture is quite different from the original one, and the correlation functions between the two pictures are not related in a simple way. We will dub such configurations, i.e.~lines or loops of anti-ferromagnetic couplings against a background of ferromagnetic couplings, {\bf dislocations}. 

As we mentioned, the result only depends on the topology of the line $L$, i.e.~its starting and end-point, and not on its detailed geometry (and crossings are also allowed). 

The quantity Eq.~\eq{spinglass} is best interpreted as the free energy of a {\it spin glass}, normalised by the ferromagnetic free energy. A spin glass is a dilute magnetic alloy (e.g.~a dilute solution of Mn in Cu), and its properties can be modelled theoretically by an Ising model with a random distribution of ferromagnetic and anti-ferromagnetic couplings $J_{ij}$ (see Edwards and Anderson, 1975). Just as the ferromagnetic bonds $J_{ij}>0$ favour configurations where the spins on the sites $i$ and $j$ are aligned, anti-ferromagnetic bonds favour configurations where the spins point in opposite directions. This leads to the phenomenon of {\bf frustration}, where the signs of the couplings give conflicting forces on the spins. In our square lattice, one easily checks that this happens when there are an odd number of anti-ferromagnetic couplings on a given square; in turn, frustated squares in the Ising model correspond to squares with a spin at their centre, in the dual model. 

In sum: {\it the dual of the two-point function of a ferromagnet is a partition function of a ferromagnet with dislocations}, with inverse temperatures related by Eq.~\eq{dualb}. Since the correlation functions of locally defined variables (spins) are dual to topological defects, or dislocations, this is a {\bf spin-dislocation duality}.

\begin{figure}
\begin{center}
\includegraphics[height=2.5cm]{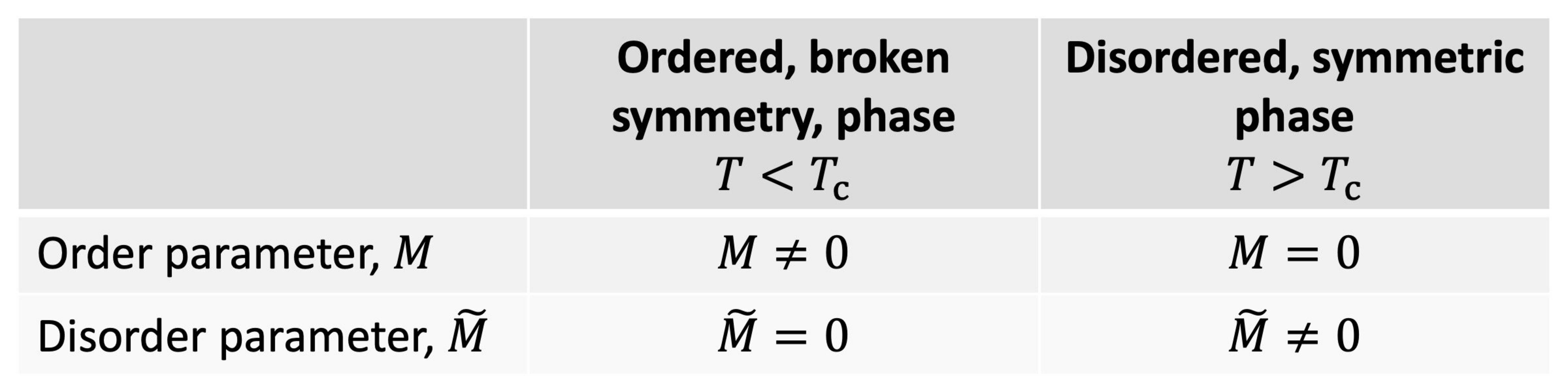}
\caption{\small The spontaneous magnetisation, $M$, is an order parameter, whose value indicates the ordered phase. The disorder parameter, $\ti M=\bra\mu\ket$, indicates the disordered phase.}
\label{DisorderP}
\end{center}
\end{figure}

The duality relation Eq.~\eq{spinglass} generalises to correlation functions with an arbitrary number $n$ of spins; these can be evaluated by the free energy of the corresponding dual Ising model with a defect. The spins are then located at the centre of each dual square, or frustrated plaquette.\footnote{See e.g.~Savit (1980:~p.~459). Kadanoff and Ceva (1971:~pp.~3922, 3926-3927, 3932-3938) consider both higher-point functions, and mixed correlation functions of order and disorder parameters.}\\
\\
{\bf Disorder parameters.} There is an alternative formulation of this duality in terms of a new disorder parameter $\mu({\bf r})$, defined in the thermodynamic limit, such that its two-point function has a defect between the points ${\bf r}$ and ${\bf r}'$.\footnote{These variables are defined in Kadanoff and Ceva (1971:~p.~3919). The literature also uses the phrases `disorder variable' and `disorder operator'. However, we will retain the more usual `disorder parameter', which indicates a specific type of variable that distinguishes different phases of a material.} 
The effect of adding a pair of disorder parameters (also called `disorder operators') in a correlation function is equivalent to inserting a line of anti-ferromagnetic bonds along the path on the dual lattice, i.e.~adding a defect. The correlation function $\bra\m({\bf r})\m({\bf r}')\ket_{\ti\b}$ is evaluated, in the dual Ising model, on points ${\bf r}$ and ${\bf r}'$ that correspond, in the continuum limit, to the lattice sites $i$ and $j$ of the given lattice. Just like the dual spins $\s_i$'s, this disorder parameter has the property that, in the thermodynamic limit, $\bra\mu({\bf r})\ket=0$ in the ordered {\bf ferromagnetic phase} $T<T_{\sm c}$ (i.e.~the broken symmetry phase, where spins are aligned along the same direction, and so the macroscopic rotational symmetry is broken), and $\bra\mu({\bf r})\ket\not=0$ in the disordered {\bf paramagnetic phase} at $T>T_{\sm c}$ (i.e.~the phase where the spins are randomly aligned and not correlated which has a macroscopic rotational symmetry: this symmetry can be broken by applying an external magnetic field, hence the word `paramagnetic'): see Figure \ref{DisorderP}.

We can study the spatial variation of the disorder variables from their two-point function, given as follows:
\bea\label{Mtilde}
\bra\m({\bf r})\,\m({\bf r}')\ket_{\ti\b}=\left\{\begin{array}{cc}{c_1\,e^{-\k|{\bf r}-{\bf r}'|}\over\sqrt{|{\bf r}-{\bf r}'|}}\,,&T<T_{\sm c}\,\\{c_2\over|{\bf r}-{\bf r}'|^{1/4}}\,,&T=T_{\sm c}\,\\|\bra\m({\bf r})\ket|^2+{\cal O}(e^{-\k|{\bf r}-{\bf r}'|})\,,&T>T_{\sm c}\end{array}\right.
\eea
where $\bra\m({\bf r})\ket=c_4\,t^{1/8}$ is location-independent and has the same temperature-dependence near the critical point as the magnetisation Eq.~\eq{Mt}, but in the dual regime of $T>T_{\sm c}$ rather than $T<T_{\sm c}$. $1/\k$ is the {\bf correlation length} of the material\footnote{See Fradkin (2017:~p.~429), Baxter (1982:~pp.~115-120), Fisher and Ferdinand (1967:~p.~170) and Onsager (1944:~p.~141).}

At $T<T_{\sm c}$, the two-point function decays exponentially with the distance, which indicates that the correlations extend over regions whose typical radius is given by the correlation length, and decay exponentially to zero outside. 

Above the critical temperature, the two-point function is position-independent and factorizes. By the {\bf factorization of the two-point function}, we mean: 
\bea\label{factorize2}
\bra\m({\bf r})\,\m({\bf r}')\ket=\bra\m({\bf r})\ket\,\bra\m({\bf r}')\ket\,,
\eea
so that the disorder parameter in effect behaves as a deterministic, rather than as a statistical, variable: in effect, there is a single macroscopic configuration of the spins that produces the macroscopic disorder, i.e.~the symmetric phase, and gives the disorder parameter its non-zero, position-independent, value. 

When the two-point function factorizes, the correlation function, $C({\bf r},{\bf r}'):=\bra\m({\bf r})\,\m({\bf r}')\ket-\bra\m({\bf r})\ket\,\bra\m({\bf r})\ket$ is zero (here, up to terms that vanish exponentially). Thus in the disordered phase, the correlation function is zero outside of a radius of the order of the correlation length. Since the disorder parameter has no local variations, there are no long-range correlations. In Chapters \ref{EMDuality} and \ref{EMYM}, we will discuss the analogous behaviour in quantum field theory.

\subsection{Illustrating the Schema for Kramers-Wannier duality}\label{illKW}

Kramers-Wannier duality is our first example of a duality where the model roots are {\it not} formulated as triples of state-space, set of quantities, and dynamics (as we already discussed at the end of Section \ref{isomdef}): but rather in terms of a Hamiltonian and a partition function that is the sum of the Boltzmann weights of all the states; as well as various other quantities that can be derived from it, such as the spontaneous magnetisation and the higher-point functions. 

As we discussed in our derivation of the duality relation for the partition function, Eq.~\eq{dualKW}, the Hamiltonian of the dual has the same form as the original ferromagnetic Hamiltonian, i.e.~Eq.~\eq{isingH}: except that it is evaluated at the dual value of the inverse temperature, i.e.~Eq.~\eq{Itemp}.

The values of the quantities in the set ${\cal Q}$ obtained by taking derivatives of the partition function match, even though the matching values are for different quantities. For example, the dual of the two-point function (i.e.~Eq.~\eq{sigmaij}) is the partition function at the dual inverse temperature with a {\it topological defect} (i.e.~Eq.~\eq{spinglass}: this also generalizes to higher-point functions. Thus dual models look very different: in one model, we calculate correlation functions, while in the dual we calculate the partition function for various choices of the defect on the lattice.

In the formulation of the duality at the end of Section \ref{IMD}, in the thermodynamic limit, order variables in the given lattice are replaced by disorder variables in the dual lattice, and these two sets of variables have opposite behaviour, as in Figure \ref{DisorderP}. 

\section{Conclusion}\label{summary4}

This Chapter has illustrated the Schema for dualities in three main examples from the history of physics. In elementary quantum mechanics and its Fourier transformation, duality is unitary equivalence of Hilbert spaces, together with their algebras of physical quantities. Electric-magnetic duality, besides anticipating themes in quantum field theory and string theory that we will elaborate on in Part II, shows the heuristic power of duality, through Dirac's prediction of the quantization of the electric charge. Kramers-Wannier duality shows how the Schema works in a formalism of partition functions and correlation functions. 

These examples illustrated the themes, roles, and types of dualities from Section \ref{featurerole}. As we saw, the {\it hard-easy} theme is ubiquitous: in one way or another, it is illustrated by all of our examples. The theme of {\it elementary-composite} is illustrated by the duality for the two-point function of the Ising model, where a ``local'' correlation function between two spins at definite lattice sites in a uniform lattice is dual to the correlation function of dual spins in an Ising model with an extended defect, where the elementary spins of one model are not the elementary spins of the other model. This relation between local and extended objects is a prelude to the idea of particle-soliton duality in the next Chapter. Kramers-Wannier duality also illustrates {\it symmetry-breaking}, which will be one of our themes in Parts II and III, and will motivate our `geometric view of theories'.

Although {\it wave-particle duality} is perhaps the most familiar example of duality in physics, it is often taken in only one of three relevant senses, namely the `dualism' of the old quantum theory: while it is the senses (ii) and (iii) that are closest to the modern sense of `duality'. 

Nevertheless, our study of wave-particle duality shows that the founding fathers of quantum mechanics anticipated many of the aspects of dualities that are also relevant in recent discussions. This statement of course depends on the way in which we have argued that `duality', in its senses (ii) and (iii) in quantum theory, resembles `duality' in its current usages. Thus Heisenberg anticipated the idea of a common core theory with an internal interpretation, and Schr\"odinger anticipated unextendability as a necessary condition for a warranted claim of theoretical equivalence (see Chapter \ref{Realism}). Both Heisenberg and, especially, Schr\"odinger made a clear distinction between formal and interpretative aspects of theoretical equivalence. Heisenberg's view of wave-particle duality meshes with the idea of {\it quantum duality}, because the quantum models are equivalent but their corresponding classical models are incompatible. 

The contrast between duality and Bohr's complementarity is particularly interesting and relevant to current physics. As we argued, the main difference is that duality drops the aspect of complementarity that we called (b): that each of the two complementary descriptions is by itself incomplete (cf.~Section \ref{complement}). That is: duals are taken to {\it each} give a complete description of a system: complementarity is closer to the idea of {\it effective duality}, where duals describe a system under different conditions or aspects.

\section*{Appendix 4.A. Dual partition function and dislocations}
\addcontentsline{toc}{section}{Appendix 4.A. Dual partition function and dislocations}

In this Appendix, we give two results that were used in the main text of Sections \ref{dualpf} and \ref{IMD}, respectively: (A) a proof of the dual partition function, Eq.~\eq{Savit2.9a}; and (B) a sketch of a proof of the partition function with dislocations, Eq.~\eq{spinglass}.\\
\\
{\bf (A)~~Proof of the dual partition function, Eq.~\eq{Savit2.9b}}. We substitute into the partition function of the Ising model, Eq.~\eq{ising}, the following identity, which is immediate, using the fact that $s_i=\pm1$:
\bea\label{expandC}
e^{\b\,s_i\,s_j}=\cosh\b+s_i\,s_j\,\sinh\b=\sum_{n=0}^1C_n(\b)\,(s_i\,s_j)^n\,,
\eea
where $C_0(\b)=\cosh\b$, $C_1(\b)=\sinh\b$. Putting this into Eq.~\eq{ising}, we get:
\bea
Z_{\sm{2D Ising}}(\b)=\sum_{\{s\}}\prod_{(ij)}\sum_{n=0}^1C_n(\b)\,s_i^n\,s_j^n\,.
\eea
The product over $(ij)$ is a product over all pairs that are nearest neighbours, i.e.~over {\it links} on the lattice, and on each such pair $(ij)$ there is a label $n$ that is being summed over. 

\begin{figure}
\begin{center}
\includegraphics[height=3cm]{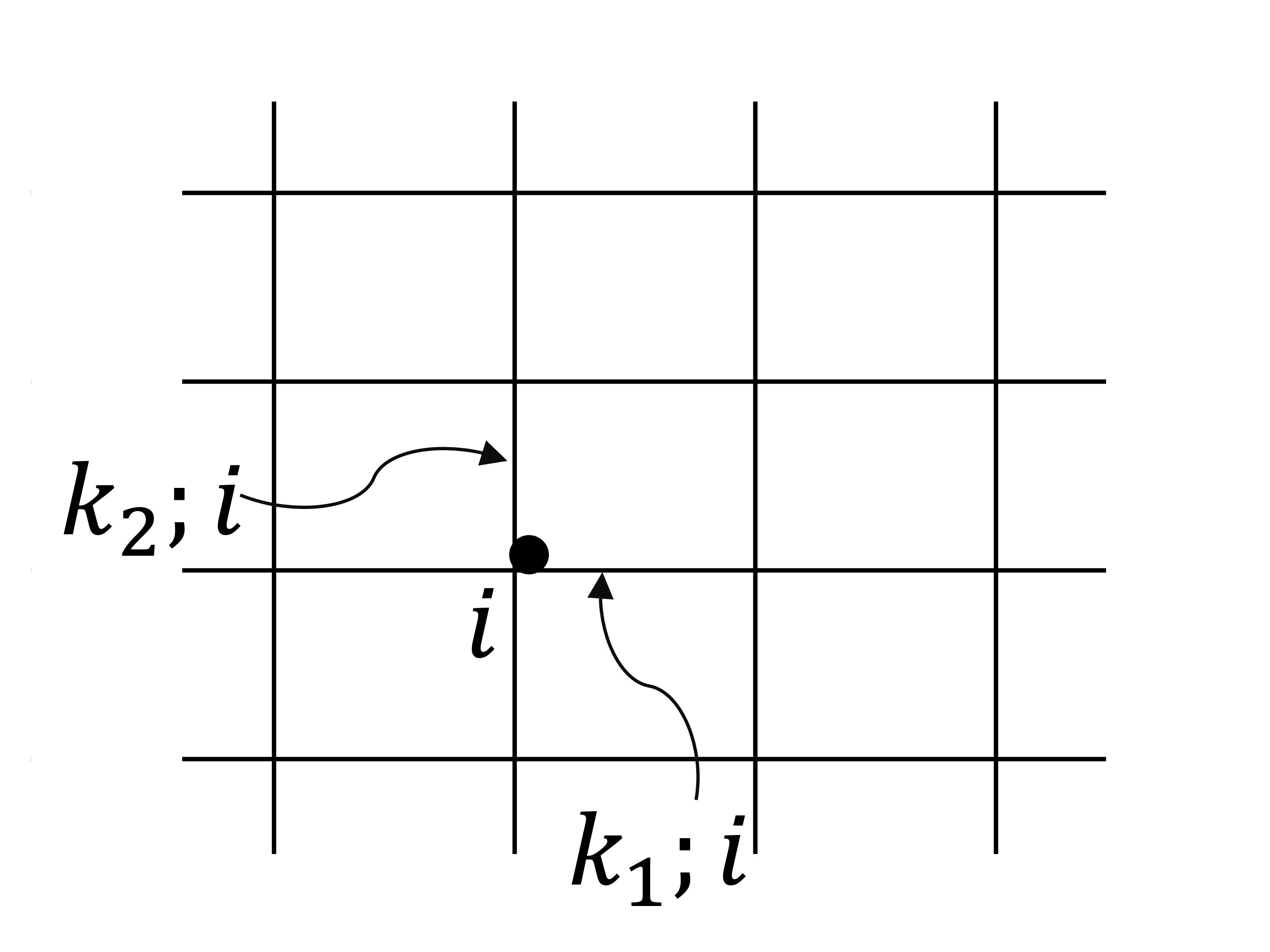}~~~
\caption{\small The variables $k_{\m;i}$ used in the duality transformation are associated with horizontal and vertical links.}
\label{HT}
\end{center}
\end{figure}

We want to exchange the products over pairs $(ij)$ and the summation over $n$, and to group together the factors $s_i$ associated with a given site $i$. To this end, we introduce different summation variables $k$, each one associated to a link. Since there are two kinds of links, namely along the $x$- and $y$-directions, we replace the $n$s by link variables $k_{\m i}$, where $i$ indicates the lattice site, and $\m$ denotes the direction, i.e.~$\m=1$ for a link along the $x$-direction, and $\m=2$ for the $y$-direction. Cf.~Figure \ref{HT}. (We associate any link with the site at its lower or left end.) 

Using these new variables, we can rewrite $Z$ as
\bea\label{Zkmi}
Z_{\sm{2D Ising}}(\b)&=&\sum_{\{s\}}\sum_{\{k_{\m i}\}}\prod_l C_{k_{\m i}}(\b)\prod_i(s_i)^{\sum_ik_{\m i}} \nn
&=&\sum_{\{k_{\m i}\}}\prod_l C_{k_{\m i}}(\b)\,\prod_i \sum_{\{s_i=\pm1\}}(s_i)^{\sum_ik_{\m i}}~;
\eea
where:

(i) the product over $l$ (links) is a product of all the $C_{k_{\m i}}(\b)$ (one for each link);

(ii) the product over $i$ is a product over all sites; and 

(iii) $\sum_ik_{\m i}$ means the sum of the four $k_{\m}$'s for the four links that impinge on a given site $i$ of the lattice.\\

Using $s_i^{\sm{even}\,\#}=1$, $\sum_{s_i=\pm1}s_i^2=2$, and $\sum_{s_i=\pm1}s_i=0$, we find that $\sum_{s=\pm1}s_i^n=2\,\d_n$, where $\d_n$ is the Kronecker delta of $n$ modulo 2, i.e.~$\d_n=1$ if $n$ is even, and $\d_n=0$ if $n$ is odd. Following Savit (1980:~p.~456), we denote this as $\d_2(n)$. So the result is:
\bea\label{Zdelta}
Z_{\sm{2D Ising}}(\b)=\sum_{\{k_{\m i}\}}\prod_l C_{k_{\m i}}(\b)\,\prod_i 2\,\d_2\left(\sum_ik_{\m i}\right).
\eea

The next step, which is crucial in proving the duality, is to find a representation for the $k_{\m i}$ which automatically satisfies $\sum_ik_{\m i}=0~\mbox{mod}~2$. This is done by using the {\it dual lattice}, which we introduced in Section \ref{dualpf}, which has dual spin variables $\s_i = \pm 1$. In this way, we associate, with each link $k_{\m i}$ of the original lattice, a unique pair of spins $\s_i$ of the dual lattice: namely, those that lie at the ends of the dual lattice link which intersects the lattice link marked by $k_{\m i}$. This implies that we can write each $k_{\m i}$ in the form: 
\bea\label{kmi}
k_{\m i}=\frac{1}{2}\left(1-\s_i\,\s_{i-\hat \n}\right),~~\m\not=\n\,.
\eea
This expresion relates the variables of the two lattices, the given lattice and the dual one: and so, it will be the starting point of the relation between the Ising model and its dual. The notation is as follows:

(a): $\s_i$ is the spin on the dual lattice at the centre of the plaquette we have labelled by $s_i$ (in its bottom-left corner); and, explaining the $\hat \n$ notation;

(b): $\s_{i-\hat \n}$ is the spin located in the centre of the other plaquette adjacent to $k_{\m i}$, i.e.~$\s_i$ is translated in the (negative) direction $\n$, which is perpendicular to $\m$. (So in Eq.~\eq{kmi}, the site label on $k$ refers to the original lattice, while the site label on $\s$ refers to the corresponding site of the dual lattice.) 

If we label the four dual sites that surround a given site of the original lattice by 1 through 4, we have:
\bea\label{Savit2.6}
\sum_{\m} k_{\m i} = 2 - \frac{1}{2}\,(\s_1\s_2 + \s_2\s_3 + \s_3\s_4 + \s_4\s_1) \,,
\eea
which is even for any set of $\{ \s_i = \pm1\}$. (For more details about the relation between the $s_i$ and $\s_i$, see Savit (1980:~p.~456-457).) 

Applying Eq.~\eq{kmi} to Eq.~\eq{Zdelta}, we have
\bea\label{Savit2.7}
Z_{\sm{2D Ising}}(\b)=\frac{1}{2}\, 2^N \sum_{\{\s\}}\prod_{l_d} C_{(1 - \s_i\s_j)/2}(\b).
\eea
where: (i) $N$ is the number of lattice sites, (ii) the product over $l_d$ is a product over links of the dual lattice, and (iii) the factor $\frac{1}{2}$ results from the fact that our now summing over $\{\s\}$ rather than $\{k\}$ counts each configuration of $\{k\}$ twice (since any $\{\s_i\} \mapsto \{-\s_i\}$ gives the same $\{k\}$).

Using $k=0$ or 1, and Eq.~\eq{kmi}, the coefficients $C_{k_\m i}(\b)$ can be readily written in terms of the $\{\s\}$ variables as: $C_{k_{\m i}}(\b)=\sqrt{\cosh\b\sinh\b}~\exp\left(-\s_i\,\s_j\,\frac{1}{2}\ln\tanh\b\right)$. We insert this in to Eq.~\eq{Savit2.7}; so that the partition function gives the following result:
\bea\label{Savit2.9a}
Z_{\sm{2D Ising}}(\b)&=& {1\over2}\,(2 \cosh \b \sinh\b)^N\sum_{\{\s\}} \exp\left(- \frac{1}{2} \ln \tanh \b \sum_{(ij)} \s_i\s_j\right),
\eea
where the $\sigma_i$'s are the spin variables on the dual lattice sites, and the sum in the exponent is over nearest-neighbour pairs in the dual lattice. 

Defining the inverse temperature as in Eq.~\eq{dualb}, the above partition function becomes Eq.~\eq{Savit2.9b}. $\Box$\\
\\
{\bf (B)~~Sketch of proof of the relation between the {\it dual} two-point function, Eq.~\eq{sigmaij}, and the partition function with dislocations, Eq.~\eq{spinglass}.} The steps are as in Section \ref{dualpf}: we expand the exponential in the dual two-point function Eq.~\eq{sigmaij} as in Eq.~\eq{expandC}, and find an expression analogous to our previous result, Eq.~\eq{Zdelta}. The difference with the previous calculation is that Eq.~\eq{sigmaij} now contains additional factors of $\s_i$ and $\s_j$, and so the powers in $(s_i)^{\sum_ik_{\m i}}$ in Eq.~\eq{Zkmi} are shifted by one unit for the sites $i$ and $j$, i.e.~we get $(s_i)^{1+\sum_ik_{\m i}}$, and likewise for $s_j$. Recall that, in moving from Eq.~\eq{Zkmi} to Eq.~\eq{Zdelta}, the effect of these powers of the spins was to produce delta functions whose arguments are defined modulo 2, i.e.~$\d_2(\sum_ik_{\m i})$. The difference is thus that the arguments are now shifted by one, e.g.~$\d_2(1+\sum_ik_{\m i})$. Since this shift by one only occurs for the sites $i$ and $j$, the delta functions over all the other sites are unshifted. Thus, in effect, we get three types of delta functions in Eq.~\eq{Zdelta}: one for $i$, one for $j$, and one for all other sites. We denote the powers of $\sigma_i$ by $k^*_{\m i}$, i.e.~the {\it dual link variables,} since the $\s_i$'s are located on sites of the dual lattice. The result is proportional to:
\bea\label{dpr}
\sum_{\{k^*_{\m l}\}}\prod_lC_{k^*_{\m l}}(\b)\,\d_2(1+\sum_ik^*_{\m i})\,\d_2(1+\sum_jk^*_{\m j})\prod_m{}'\,\d_2(\sum_mk^*_{\m m})\,,
\eea
where the dash denotes the product over all dual sites {\it except} $i$ and $j$, since those get a delta function whose argument is shifted by one. 

Next, we need to solve the delta function equations, in the same way we solved Eq.~\eq{kmi}, in terms of the dual variables. The solution will be similar to Eq.~\eq{kmi}, but the details are different. Namely, the sum of factors should add up to an even number in the third delta function, i.e.~$\d_2(\sum_mk^*_{\m m})$, but they should add up to an {\it odd} number for the dual sites $i$ and $j$ (and also, we have replaced the original spins, $s_i$, for the dual ones, $\s_i$). This is because the arguments of their delta functions are shifted by one. In other words, $1+\sum_ik^*_{\m i}$ should be even, and therefore $\sum_ik^*_{\m i}$ should be odd. 

All the solutions of the type Eq.~\eq{kmi} gave even sums; to get an odd sum, we will need to shift some signs around, but not all. It turns out that the most economic way to do it is as follows. 

We draw a line, $L$, along the lattice, connecting the dual lattice sites, or sites, $i$ and $j$, as in Figure \ref{HT2}. The idea is that we keep a solution like Eq.~\eq{kmi} for all the dual links that are not on this line, but change the sign for the dual links on the line: so that, as we will see, the corresponding bonding in the {\it given or original lattice} will come out to be anti-ferromagnetic: accordingly, we also find the set of {\it original spins}, $s_i$. The following change of variables solves for the delta functions, i.e.~they give rise to sums which are zero modulo 2 when inserted into the delta functions in Eq.~\eq{dpr}:
\bea\label{kss}
k^*_{\m i}&=&\half(1-s_i\,s_{i+\hat\n})\,,~~~k^*_{\m i}\not\in L\nn
k^*_{\m i}&=&\half(1+s_i\,s_{i+\hat\n})\,,~~~k^*_{\m i}\in L\,.
\eea
Thus for the sites that are not on the line $L$ between $i$ and $j$, the solution is unchanged (compare the first equation in Eq.~\eq{kss} with Eq.~\eq{kmi}); but for the sites on the line we change the sign, so that we will get an odd number.

The line $L$ is obviously not unique (hence our phrase `the most economic way to do it'), but it turns out that the result is independent of $L$'s precise trajectory: it only depends on the starting point $i$ and end-point $j$.\footnote{This can be relatively easily shown with the techniques we illustrate below. This path-independence is an expression of a $\mathbb{Z}_2$ gauge symmetry: see Savit (1980:~p~458).}

\begin{figure}
\begin{center}
\includegraphics[height=2cm]{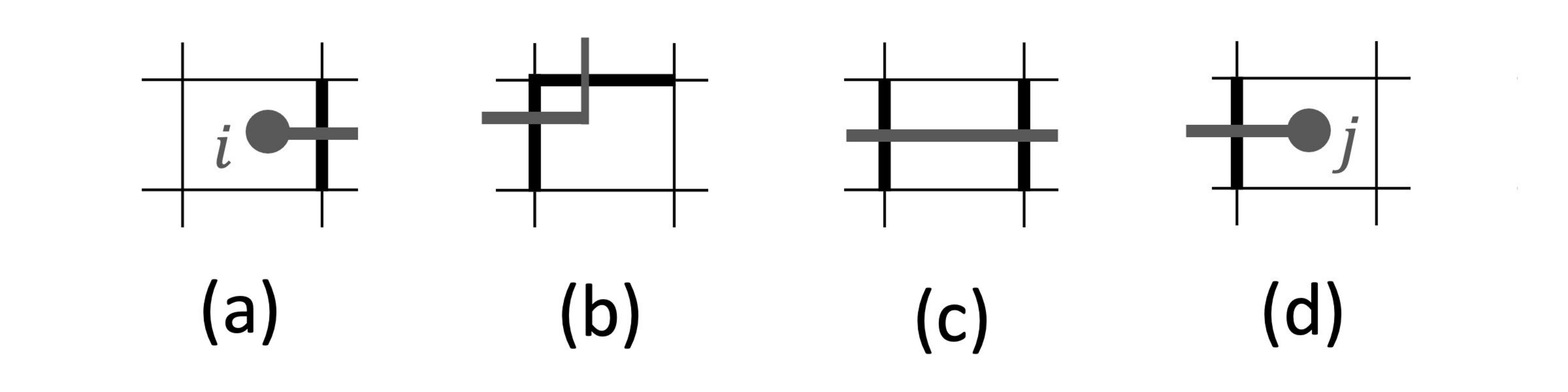}
\caption{\small Four examples of the calculation of the sum of lattice links, $\sum_\m k_{\m i}^*$, for a single square, using Eq.~\eq{kss}. The result is a number that is: (a) odd, (b) even, (c) even, (d) odd.}
\label{HT3}
\end{center}
\end{figure}

Let us check that the above solution, Eq.~\eq{kss}, gives the correct, i.e.~even resp.~odd numbers, in the arguments of the delta functions. To this end, we calculate the sums $\sum_\m k^*_{\m i}$ for each square, as in Figure \ref{HT3}.

For example, consider the first square, Figure \ref{HT3} (a). The dual lattice links are denoted by $k^*_{\m i}$. Label the four spins on this square, clockwise and starting from the upper-right corner, by $s_1,\ldots,s_4$. Their corresponding links are $k^*_{1i},\ldots,k^*_{4i}$ (for simplicity, we will drop the subscript $i$) i.e.~$k^*_1$ is the dual link that intersects the link between $s_1$ and $s_2$, $k_2^*$ is the dual link that intersects the link between $s_2$ and $s_3$, etc. Thus the dual link $k^*_1$ is the only one that lies on the grey line $L$. Thus, according to Eq.~\eq{kss}, $k^*_1$ gets a plus sign, and the other dual links get a minus:
\bea\label{k1k4}
k_1^*=\half(1+s_1s_2)\,,&~~~~&k_3^*=\half(1-s_3s_4)\nn
k_2^*=\half(1-s_2s_3)\,,&~~~~&k_4^*=\half(1-s_4s_1)\,.
\eea
Adding this up, we get:
\bea\label{sumk1}
\sum_\m k^*_{\m i}=2+\half(s_1s_2-s_2s_3-s_3s_4-s_4s_1).
\eea
Calculating this for arbitrary combinations of values of the spins, $\{s_j\}=\pm1$, one easily sees that this number is always odd. The fact that it is odd is due to the presence of a single plus sign (i.e.~an odd number of plus signs) in Eq.~\eq{k1k4}, and that is related to the fact that the site $i$ is the endpoint of the line $L$.

Contrast the above result with the next square, Figure \ref{HT3} (b), where the line $L$ goes in and out of the square: this square has two dual links that lie along $L$ (i.e.~$k_3$ and $k_4$), and the number of plus signs is now even:
\bea
k_1=\half(1-s_1s_2)\,,&~~~~&k_3=\half(1+s_3s_4)\nn
k_2=\half(1-s_2s_3)\,,&~~~~&k_4=\half(1+s_4s_1)\,,
\eea
so that adding them up we get:
\bea\label{sumk2}
\sum_\m k^*_{\m,(b)}=2+\half(-s_1s_2-s_2s_3+s_3s_4+s_4s_1),
\eea
and one easily checks that this is always even. Likewise, we get, for Figure \ref{HT3} (c) and (d), respectively:
\bea\label{sumk3}
\sum_\m k^*_{\m(c)}&=&2+\half(s_1s_2-s_2s_3+s_3s_4-s_4s_1)\nn
\sum_\m k^*_{\m(d)}&=&2+\half(-s_1s_2-s_2s_3+s_3s_4-s_4s_1)\,.
\eea
Again, Figure \ref{HT3} (c) has an even number of dual links that lie on the line $L$, and so the first sum in Eq.~\eq{sumk3} has an even number of plus signs, and so it is always an even number. On the other hand, Figure \ref{HT3} (d) contains the end-point of the line $L$ at the site $j$, and so it has an odd number of dual links lying on $L$. The corresponding sum in Eq.~\eq{sumk3} also has an odd number of plus signs, and so it is always an odd number. 

Having discussed the mechanics of the solution Eq.~\eq{kss} for the delta functions in Eq.~\eq{dpr}, we can now insert this solution back into Eqs.~\eq{sigmaij}-\eq{dpr}. This gives us an expression that is similar to the partition function itself, but in terms of the original spins $s_i$ and $s_j$, and with signs that depend on the signs of the spins that appear in sums like Eqs.~\eq{sumk1}, \eq{sumk2}, and \eq{sumk3}. Clearly, these signs depend on whether the corresponding dual link lies on the line $L$. In other words, it depends on whether the corresponding original link, denoted by thick black lines in Figure \ref{HT3}, crosses the grey line $L$. Thus the links that cross $L$ get the opposite sign from the ones that do not cross $L$: the latter correspond to ferromagnetic couplings, where the former correspond to anti-ferromagnetic couplings. The result is the partition function with dislocations, i.e.~Eq.~\eq{spinglass}.

\newpage
\thispagestyle{empty}
$ $
\begin{center}
\vskip7cm
{\bf\Huge Part II. Advanced Examples of Dualities}
\end{center}
\addcontentsline{toc}{chapter}{Part II. Advanced Examples of Dualities}

\newpage
\thispagestyle{plain}

\chapter{Particle-Soliton Dualities}\label{Advan}
\markboth{\small{\textup{Particle-Dislocation Dualities}}}{\textup{
\small{Particle-Dislocation Dualities}}}

The main aim of this Chapter is to introduce an important new class of examples of dualities that exchange {\it particles} and {\it solitons.} There are three primary justifications for the significance of these examples in contemporary theoretical physics. First, the duality between bosons and fermions (Section \ref{SGT}) is a highly non-trivial example of a duality for {\it interacting} models in quantum field theory, which will inform our philosophical discussions in Part III. Second, the influence of soliton solutions of field theories on the developments in theoretical physics in the past four decades can hardly be overestimated. For the physics of these solitons, and their associated dualities, is part and parcel of physicists' current understanding of key physical phenomena such as quark confinement and the Higgs mechanism (see Section \ref{PVD}). Furthermore, these solitons, and their dualities, are central for discussing analogies between condensed matter physics and high-energy physics (e.g.~the analogy between the Meissner-Ochsenfeld effect in superconductivity and quark confinement) that played a historically important role. (Confinement, superconductivity and the Higgs mechanism will be further developed in Chapter \ref{EMDuality}.) Third, the example of electric-magnetic duality (Section \ref{MEMD}) is paradigmatic of many of the dualities studied in recent decades in physics. Indeed, it is the paradigm case of almost any discussion of duality in string theory. (Accordingly, we will further develop it in Chapter \ref{EMYM}.) 

So far as we are aware, these models have received little attention in the philosophical literature. Indeed, while recent philosophical discussions of classical and quantum field theories often emphasise the important theoretical and structural aspects of these theories---gauge symmetries, renormalisation, reduction, explanation, etc.---the theories' solutions are equally important: and there is a wealth of soliton solutions, vortices, monopoles, etc., whose behaviour is worth studying in more detail than we can do here.\footnote{Hartmann (1999, 2001) discusses the MIT bag model of the 1970s, which predates several developments we will discuss in this and the next Chapter.} 

In short, a {\bf soliton} is a solution of the non-linear equations of motion of a (field) theory that has finite energy and is spatially extended, i.e.~localized within a finite region. The solution is topologically stable: it cannot simply decay into point particles. Furthermore, solitons can enter into scattering processes with other solitons and with particles, and are stable under these interactions. Their topological stability is expressed by a topological charge, which is discrete but does {\it not} correspond, as in Noether's theorem, to a symmetry of the Lagrangian.

Three subsequent Sections will present examples in order of increasing number of spatial dimensions: Section \ref{SGT} works in two spacetime dimensions (one space and one time), where the solitons are {\it kink} solutions. Section \ref{PVD} works in three spacetime dimensions (two space and one time), where the solitons are {\it vortices}. Section \ref{MEMD} works in four spacetime dimensions, where the solitons are {\it monopoles} (but these exist only in the non-linear theory, which will be discussed in Chapter \ref{EMDuality}). 

To keep the Chapter short, we can only begin scratching the surface of what is, essentially, virgin territory for philosophers of physics. Thus the emphasis will be on: (i) presenting the relevant solutions and the associated physical phenomena; (ii) discussing how the examples illustrate the features of dualities from Section \ref{featurerole}; and (iii) paving the way for more advanced examples. Indeed, the conclusion will make substantive points that will be further developed in Parts II and III.

\section{Dualities between bosons and fermions}\label{SGT}

In this Section, we will study dualities between bosons and fermions in two dimensions (i.e.~one dimension of space and one dimension of time), also called `bosonization'. Section \ref{soliton2d} discusses the duality between a scalar field model with a non-trivial potential, and a massive fermion model with a quartic interaction. This example exhibits how particle and solitonic solutions are exchanged by duality.\footnote{Since the model is interacting, we will not deal with arbitrary solutions of the equations of motion, but focus on the static ones, whose behaviour is already very rich.} 
Section \ref{bosoniz} will then discuss the simpler case of boson-fermion duality for the corresponding {\it free}, massless models. Because of the simplicity of these two models, we can in this case analyse the full state spaces, thereby illustrating the Schema.

\subsection{Sine-Gordon-Thirring duality}\label{soliton2d}

As we mentioned in Chapter \ref{Schema}, dualities are isomorphisms between theories: often described through the theories' {\it solutions.} Thus the soliton solutions of the sine-Gordon model, with their corresponding states and quantities, are dual to the particle solutions of the Thirring model, and their states and quantities. This relation is really a full equivalence between quantum theories, a version of which will be presented in Section \ref{bosoniz}.\footnote{The Thirring-sine-Gordon model duality is quite important in condensed matter systems: see Giamarchi (2003) and Atland and Simons (2010). The duality was discovered in a series of papers in the 1970s: see Dell'Antonio et al.~(1972), Coleman (1975), Mandelstam (1975b), and Schroer and Truong (1977).}

The existence of solitons appears to be a consequence of two main features of classical field theories:\footnote{Of these two features, (i) is a prerequisite for (ii), because any solution $\f({\bf x},t)$ of a linear equation is topologically trivial, in the sense that it is homotopic to the trivial solution, $\f({\bf x},t)=0$---and homotopy is one of the main topological notions that characterise solitons: a topic that we cannot enter into here. See Manton and Sutcliffe (2004:~pp.~75-82).}

(i) The {\it non-linearity} of the dynamics (i.e.~the potential being of order higher than quadratic) gives lump-like solutions, distinct from the solutions of the linear theory.

(ii) The existence of a {\it topological current} makes the soliton stable against perturbations: so that it can interact with other solitons, collide, etc., thus showing its own particle-like behaviour. The {\it charge} corresponding to the topological current normally takes integer values, which characterise the possible {\it states} of the soliton, and depend on its behaviour at infinity.\footnote{This will also emphasise a recurrent theme, which we believe has not received the attention that it merits in the philosophical literature, that the possible {\it states} of a dynamical system are determined by its possible {\it boundary conditions}.}

Below, (A) discusses these two features in the {\bf bosonic sine-Gordon model}, and (B) discusses them in the {\bf fermionic Thirring model}; as follows.\\
\\
(A)~~We begin our description of solitons with the simplest example: a two-dimensional {\bf scalar field model} with a potential $V(\f)$, whose action is:
\bea\label{sineGact}
S&=&\int\dd^2x\left(\half\pa_\m\f\,\pa^\m\f-V(\f)\right).
\eea
For the potential $V(\f)$, we consider two possible choices. First the quartic, {\bf double-well, potential}:
\bea\label{VMex}
V_{\sm{dw}}(\f)={1\over4}\,\l\left(\f^2-{m^2\over\l}\right)^2\,.
\eea
Second, the potential of the {\bf sine-Gordon model}:
\bea\label{sGp}
V_{\tn{SG}}(\f)=-{\m^2\over\b^2} \left(\cos(\b\f)-1\right)={2\m^2\over\b^2}\,\sin^2{\b\f\over2}\,.
\eea
The first form, in terms of the cosine, is the most widespread; in the second form, we used the sine addition formula, and this sinusoidal expression will be useful later. The parameters $\l$, $m$, $\m$, and $\b$ are arbitrary positive real constants. 

In classical field theory, one often studies the solutions through a perturbative expansion, i.e.~a Taylor series around the point $\f=0$, where the system is non-interacting. One easily checks, by expanding the potential $V(\f)$ for the two choices above, that for small $\f$, quadratic terms in $\f$ appear. Their coefficients are: $-\half m^2$ for the double-well potential (the negative sign thus denotes an unstable local maximum at $\f=0$), and $+\half\m^2$ for the sine-Gordon potential, which is the usual coefficient of the mass term. In other words, $m$ plays the role of a ``tachyonic mass'', and $\m$ is the mass of the particle excitations around $\f=0$.

However, we are here interested in the full vacua of the potential, and in particular in its absolute minima.\footnote{In the case of the double-well potential, $V_{\sm{dw}}(\f)$, the solution $\f=0$ is a maximum, rather than a minimum, of the potential: and so, it is not a stable solution, for any small perturbation will draw the solution away from it.}
To find solutions that correspond to single solitons, i.e.~non-trivial solutions of the scalar field equations with finite energy that from a distance look like particles, one can begin by considering static solutions. (Since the equations of motion are Lorentz invariant, it is easy to then boost the solutions to finite speed; however, there is no claim here of this being the most general solution, which in fact it is not: only of its being a physically significant solution). 

The energy associated with a given configuration of the field is:
\bea\label{Ekink}
E=\int_{-\infty}^\infty\dd x\left[{1\over2}\left({\pa\f\over\pa t}\right)^2+{1\over2}\left({\pa\f\over\pa x}\right)^2+V(\f)\right].
\eea
For static (i.e.~time-independent) solutions, the first term is zero, and the second and third terms are related to each other through the equations of motion, so that the energy only depends on the (integral of the) potential. For the sine-Gordon potential (Eq.~\eq{sGp}), we can evaluate the integral in terms of the mass and the coupling constant, $\b$, as follows:\footnote{The details are straightforward: (i) For a static solution, $\pa\f/\pa t=0$, the equation of motion of the scalar field takes the following form: ${1\over2}\left({\sm d\f\over\sm d x}\right)^2=V(\f)\Rightarrow{\sm d\f\over\sm dx}=\sqrt{2V(\f)}$. This result is obtained as follows: the equation of motion for a static solution is: ${\sm d^2\f\over\sm d x^2}={\pa V\over\pa\f}$. Multiplying both sides by ${\sm d\f\over\sm dx}$, we get: $\half{\sm d\over\sm d x}\left({\sm d\f\over\sm dx}\right)^2={\sm d V\over\sm dx}$. Integrating both sides with respect to $x$, we get the desired formula. (ii) Substituting the result of (i) into the energy, Eq.~\eq{Ekink}, we find: $E=\int_{-\infty}^\infty\dd x\left({\dd\f\over\dd x}\right)^2=\int_{-\infty}^\infty\dd x\,{\dd\f\over\dd x}{\dd\f\over\dd x}=\int_0^{2\pi/\b}\dd\f\,{\dd\f\over\dd x}$. Using again the (static) equation of motion from (i), ${\sm d\f\over\sm d x}=\sqrt{2V(\f)}$, we find: $E=\int_0^{2\pi/\b}\dd\f\,\sqrt{2V(\f)}={2\m\over\b}\int_0^{2\pi/\b}\dd\f\,\sin{\b\f\over2}$. The result of this elementary integral is the value reported in Eq.~\eq{Eresult}.}
\bea\label{Eresult}
E=\int_0^{2\pi/\b}\dd\f\,\sqrt{2V(\f)}={8\m\over\b^2}\,,
\eea
which is indeed finite.\footnote{The result is the same if the integral runs between any two minima of the potential that the kink solution interpolates between: see Eq.~\eq{staticSG}.}
(Our choice of boundary conditions for the field at infinity is in the choice of range of the above integral. We made the natural choice: we integrate between the first two zeroes of the potential, as in our kink solution Eq.~\eq{dwSG} below.)

This can be done for other potentials too. Requiring that the energy is again finite, the field must thus approach one of the zeroes of the potential, $V(\f)=0$, at infinity (analogously to the ranges of integration, from $0$ to $2\pi/\b$, in Eq.~\eq{Eresult}). Thus for the $\f^4$ potential, it approaches $\f\rightarrow\pm{m\over\sqrt{\l}}$ as $x\rightarrow\pm\infty$ (and either combination of signs is possible). We will see that a ``kink''-solution arises when the field approaches $\f\overset{x\rightarrow\infty}{\rightarrow}{m\over\sqrt{\l}}$, and $\f\overset{x\rightarrow-\infty}{\rightarrow}-{m\over\sqrt{\l}}$, while the opposite case is called an `anti-kink'.

We will now show that the fact that the potential has a discrete set of minima renders the topological charge, when normalised by an appropriate overall constant, an integer number. First we define the topological current.\\
\\
For any field model with a scalar $\f$ in two spacetime dimensions, we can define the following {\it conserved current}:
\bea\label{curr1d}
J_\f^\m:=\a~\e^{\m\n}\pa_\n\f\,.
\eea
This current is {\it not} a Noether current, since it is not derived from the Lagrangian and there is no symmetry of the sine-Gordon model associated with it. It is conserved identically, i.e.~independent of the equations of motion: $\pa_\m J^\m=\e^{\m\n}\pa_\m\pa_\n\f=0$. Furthermore, $\a$ can be chosen such that the associated {\it charge} is an integer:
\bea\label{solitonQ}
Q:=\int_{-\infty}^\infty \dd x~J^0=\a\int_{-\infty}^\infty\dd x~{\pa\f\over\pa x}=\a\left(\f(\infty)-\f(-\infty)\right).
\eea
Choosing $\a={\l\over2m}$ for the $\f^4$ model, and $\a={\b\over2\pi}$ for the sine-Gordon model, one indeed finds that kink solutions have $Q=1$, and anti-kink solutions have $Q=-1$ (solutions that are not kinks have $Q=0$). 

To find the values of the charge, we only need to know the values of the field at infinity, i.e.~the minima of the potential. Thus, as we expect from a topological charge, the local physics is not needed.

We can exhibit the explicit solutions of the equations that show this behaviour. The Euler-Lagrange equations of motion can be, in both cases, integrated to give the following static solutions:
\bea\label{dwSG}
\f_{\sm{dw}}(x)&=&\pm{m\over\sqrt{\l}}\,\tanh\left({m\over\sqrt{2}}(x-x_0)\right)\\
\f_{\tn{SG}}(x)&=&{2\pi n\over\b}\pm {4\over\b}\,\arctan\left(e^{\pm{\m\over\b}(x-x_0)}\right)~,~~n\in\mathbb{Z}\,.\label{staticSG}
\eea
The solution of the sine-Gordon model\footnote{We pick the two signs (in front of the artctan and in the exponential) to be plus. If the first sign is taken minus and the second plus, the solution goes to $2\pi/\b(n-1)$ at $x\rightarrow\infty$. Picking the minus sign inside the exponential simply exchanges the two asymptotic regions.}
goes to ${2\pi\over\b}\,n$ as $x\rightarrow-\infty$, and to ${2\pi\over\b}\,(n+1)$ as $x\rightarrow\infty$. Thus the sine-Gordon model has an infinite set of degenerate asymptotic minima, namely $\f=2\pi n/\b$ for any $n\in\mathbb{Z}$, and the above solution interpolates smoothly between $2\pi n/\b$ on the left, and ${2\pi\over\b}\,(n+1)$ on the right. (The solution of the $\phi^4$-model goes to $\pm{m\over\sqrt{\l}}$ as $x\rightarrow\pm\infty$, as required.) Thus, for both solutions, the charges are $Q=\pm1$.\footnote{Solutions with several kinks, moving with different velocities, colliding and scattering, also exist. See Rajaraman (1975:~pp.~274-277), Manton and Sutcliffe (2004:~pp.~119-123).}\\
\\
(B)~~The sine-Gordon model is dual to the {\bf massive fermionic Thirring model} (1958:~pp.~93-98), with action:
\bea\label{Thirrac}
S_{\tn{mThirr}}=\int\dd^2x\left(i\bar\psi\g^\m\pa_\m\psi-m\bar\psi\psi-\half g\,\bar\psi\g^\m\psi\bar\psi\g_\m\psi\right),
\eea
where $\psi$ is a Dirac spinor (i.e.~its two components are {\it complex}). This model is renormalisable, and exactly solvable, for $g\hbar>-\pi$. It has a symmetry under the following global U(1) {\it change of phase}: $\psi\mapsto\psi'=e^{i\a}\psi$. The Noether current, associated with this symmetry, is the following vector current:\footnote{Thus the quartic term in the action can be written as the square of this vector current.}
\bea\label{ThirrJ}
J^\m_{\tn{Thirr}}=\bar\psi\g^\m\psi\,.
\eea
The associated charge is defined as before, and (in the quantum theory) it is an integer times $\hbar$.

The duality between the Thirring and sine-Gordon models will be discussed in detail in the next section. The relation between the fermionic field $\psi$ and the bosonic field $\f$ (both of them treated as quantum fields) is non-local, and thereby also exchanges (the operator versions of) their equations of motion, and their currents, $J^\m_\f$ and $J^\m_{\tn{Thirr}}/\hbar$.\footnote{From the next Section onwards, we will set $\hbar=1$.} Thus the duality also interchanges their charges, i.e.~it exchanges one set of integers with another. Since, for static solutions, the charges label the states, the duality also exchanges the states, as expected. 

The coupling constants of the two models are related by the duality as follows:
\bea\label{betag}
{\beta^2\hbar\over4\pi}={1\over1+g\hbar/\pi}\,.
\eea 
We see that the region where the sine-Gordon model is weakly coupled ($\beta\rightarrow0$) corresponds to a strongly coupled Thirring model, i.e.~$g\rightarrow\infty$. Also, a free fermion with zero coupling ($g=0$) corresponds to $\b^2\hbar=4\pi$, i.e.~a coupling in the sine-Gordon model of order one (infinite coupling in the sine-Gordon model corresponds to $g\hbar=-\pi$). This illustrates our general theme, from Section \ref{featurerole}, of {\it hard-easy}: for example, before Mandelstam (1975b:~3027) discovered the full duality map, Coleman (1975:~2093), using perturbative field theory methods, was only able to prove the finiteness of the sine-Gordon model at $\beta=0$ (where the model is a free massive Bose field theory) and at $\beta^2\hbar=4\pi$, where it is a free massive Fermi field model.

The above relation between the couplings is independent of the classical mass parameters, $m$ and $\m$. As we will see in the next Section, the existence of a duality map between massless bosons and fermions can be used to extend this map to the massive case.

Perhaps the most striking aspect of the duality is that, although it exchanges two fully quantum systems, it in effect allows us to describe the quantum behaviour of the massive Thirring fermion, with its usual fermionic-like excitations, in terms of a quantum soliton that is a superposition of coherent states of the $\phi$-field (Coleman, 1975:~p.~2095-2096): `the particles which are fundamental in one description are composite in the other: in the Thirring model, the fermion is fundamental and the boson a fermion-antifermion bound state; in the sine-Gordon equation, the boson is fundamental and the fermion a coherent bound state'. This is our theme {\it elementary-composite} from Section \ref{featurerole}. 

\subsection{Bosonization}\label{bosoniz}

As we mentioned in the preamble at the beginning of this Section, the basic {\it bosonization} case is a special case of the previous one: namely, the duality, in two dimensions, between:

(1)~~the free, massless {\bf bosonic scalar field}; and

(2)~~the free, massless {\bf Dirac} (i.e.~complex) {\bf fermion field}.

Since the masses and the potentials are set to zero, the actions of these models are just the kinetic terms of Eqs.~\eq{sineGact} and \eq{Thirrac}. Because the actions are so simple, we can exhibit the duality map, and the common core theory, in general.\footnote{For more details, see De Haro and Butterfield (2018:~Sections 4 and 5).} 
We begin by: 

(a) giving the most general solution of the equations of motion, 

(b) analysing the symmetries of the two models, and

(c) giving the associated Noether currents associated with the symmetries.

Then, in a final common step, we work out the duality map, which we argue is a unitary transformation of the Hilbert space of the model triples.

For free theories in two dimensions, it is useful to work with complex coordinates parametrizing the complex plane, $\mathbb{C}\simeq\mathbb{R}^2$, i.e.~$z:=x^0+ix^1$, $\bar z:=x^0-ix^1$. \\
\\
(A)~~We begin with the {\bf bosonic model} in three steps:

(a)~~Its {\it equation of motion} is the massless Klein-Gordon equation for the scalar field written in complex coordinates, i.e.~$\F(z,\bar z)$, and takes the form: $\partial\bar\partial\Phi(z,\bar z)=0$, where $\pa:=\pa/\pa z$ and $\bar\pa:=\pa/\pa\bar z$. The general solution of this equation is the sum of a holomorphic and an anti-holomorphic function (i.e.~functions of $z$ and $\bar z$, respectively):\footnote{Classically, one may indeed require the solutions to be holomorphic and anti-holomorphic functions. Quantum mechanically, the fields have isolated singularities, and so we will allow them to be meromorphic and anti-meromorphic functions. For a discussion, see De Haro and Butterfield (2018:~p.~349).}
$\F(z,\bar z)=\f(z)+\bar\f(\bar z)$. Since $\F$ is a real field, $\bar\f$ is the complex conjugate of $\f$. 

(b)~~{\it Symmetries}: the classical action and the equations of motion are both invariant under two types of symmetries: (i) {\it conformal transformations}, i.e.~scale transformations with a variable scale factor, such that the angles are preserved. (ii) {\it Affine current algebra transformations}: these are translations of the field $\F$ by an arbitrary holomorphic function $\varphi(z)$, i.e.~$\F(z,\bar z)\mapsto\F(z,\bar z)+\varphi(z)$ (and likewise for anti-holomorphic translations). 

(c)~~The {\it conserved currents} obtained from these symmetries through Noether's second theorem: (i) the conserved currents for the conformal transformations are the components of the stress-energy tensor (again with holomorphic and anti-holomorphic components; and, anticipating the model's quantization that we will discuss below, we consider the quantum field $\hat\f$ corresponding to $\f$, and normal-order the product):
\bea\label{TTbar}
T(z)=-{1\over2}\,:\pa\hat\f\,\pa\hat\f:=-\half :J(z)J(z):\,,
\eea
where $J(z)$ and its conjugate $\bar J(\bar z)$ are defined to be:\footnote{The conserved currents for affine current algebra transformations, $\bar J(z)$ and $\bar J(\bar z)$, are the complex versions of the topological conserved current of the sine-Gordon model, Eq.~\eq{curr1d}, which did {\it not} follow from Noether's theorem. The reason for this difference is that only the free, massless, and not the sine-Gordon, model is symmetric under affine current algebra transformations. Indeed, the sine-Gordon model, with potential Eq.~\eq{sGp}, only has the {\it discrete} symmetry $\f\mapsto\f+2\pi n/\b$, which gives a family of degenerate minima, and so Noether's theorem does not apply: but a topological current still exists.}
(ii) the conserved currents for the affine current algebra transformations (which we will simply call `affine currents'):
\bea\label{affJ}
J(z):=\pa\hat\f(z)~~~~\mbox{and}~~~~\bar J(\bar z):=\bar\pa\hat{\bar\f}(\bar z)\,.
\eea

Upon quantization, the components of the stress-energy tensor, which are quadratic in the field, are defined by normal ordering, as in Eq.~\eq{TTbar}. Also, using the fact that the holomorphic and anti-holomorphic parts of the field can be written as a Laurent series in $z$ and $\bar z$ (i.e.~a generalization of the Taylor series that includes negative powers of $z$ and $\bar z$), one finds that the coefficients of this series satisfy a closed algebra. 

This algebra is called the {\it enveloping Virasoro algebra}, and in general it depends on two arbitary coefficients, $c$ and $k$. In our case, these coefficients are both fixed to one, $c=k=1$. $c$ is called the {\bf central charge} of the algebra, and it comes from an anomalous term (called the `Virasoro anomaly': i.e.~it appears as a numerical constant added to an angular momentum-type algebra, also called a `central' element because it commutes with all the generators, and is required for the quantum consistency of the algebra).\footnote{In general, the central charge is given by $c=k\,{\sm{dim}}\,G/(k+C)$, where $G$ is the Lie group associated with the enveloping Virasoro algebra, and $C$ is the eigenvalue of the Casimir operator of the adjoint representation.} 
$k$ is another kind of central extension, which in general makes the Kac-Moody algebra non-abelian, but here it is abelian because $k=1$ (a Kac-moody algebra is an infinite-dimensional generalization of a Lie algebra).\footnote{The enveloping Virasoro algebra combines two algebras: the algebra of the stress-tensor coefficients (i.e.~the Virasoro algebra) and the algebra of the affine conserved currents (an abelian Kac-Moody algebra). For details, see De Haro and Butterfield (2018:~pp.~351-354). For more on affine Lie algebras, see Di Francesco et al.~(1997:~pp.~557-559) and Kac (1990). A simple example of a non-abelian affine current algebra, namely the affine Kac-Moody SU(2) algebra at level $k=1$, involving a free boson on a circle of self-dual radius, is in Ginsparg (1988:~pp.~131-132).} 

This algebra is the central result we need, because, together with (i) the relation between the two currents, Eq.~\eq{TTbar} (appropriately normally ordered) and (ii) the Laurent expansions, this algebra contains all the information about the quantum model. Namely, {\it any} solution of the equations of motion can be thus quantized, and satisfies these relations. 

The {\it state space} ${\cal S}_{\tn B}$ of the bosonic model is the vector space on which the algebra acts, endowed with an appropriate semi-positive norm.\footnote{More specifically, the irreducible representations are constructed from a left-Verma module, as we discuss below: cf.~footnote \ref{verma}.} 
The {\it quantities}, ${\cal Q}_{\tn B}$, and the dynamics, ${\cal D}_{\tn B}$, are constructed from the operators that satisfy the algebra (as we also discuss below).\\
\\
(B)~~The {\bf fermionic model} can be worked out using the same three steps: 

(a)~~We decompose the Dirac spinor into chiral (i.e.~Weyl) fermions, $\psi$ and $\ti\psi$, which are left- and right-chiral, respectively. Upon solving the {\it equations of motion}, these spinors are holomorphic and anti-holomorphic, respectively. 

(b)~~The {\it symmetries} are of two kinds: (i) like the bosonic model, the fermionic model is invariant under {\it conformal transformations}. (ii) Furthermore, the action is invariant under {\it left-holomorphic-chiral (and right-holomorphic-chiral) transformations} of the spinors, i.e.~$\psi\mapsto\psi'=e^{i\a(z)}\psi$, where $\a$ is an arbitrary holomorphic function. Note that this transformation is {\it different} from the affine current algebra transformation of the bosonic model, since the fields are also different. Thus the symmetries (i) are the same as the bosonic ones, but the transformations (ii) are different.

(c)~~{\it Conserved currents}: the currents associated with the fermionic symmetries, (i) and (ii), are different from the currents of the bosonic model, since they are built from the fermionic fields. Nevertheless, the functional relation between the two currents, i.e.~the quadratic relation Eq.~\eq{TTbar}, is the same in the two models (this relation is called the {\it Sugawara construction}).\footnote{The significance of the Sugawara construction is that it simplifies the theory, because the primary fields (i.e.~fields whose transformation rules under conformal transformations have well-defined `conformal weights': see De Haro and Butterfield (2018:~pp.~357, 373) and Ginsparg (1988:~p.~13)) that are associated with both currents then are the same. See Frishman and Sonnenschein (2010:~p.~52). The (holomorphic and anti-holomorphic) conformal weights $h$ and $\bar h$ of a field in a conformal field theory are related to the spin $s$ and scaling dimension $\Delta$ of the field as follows: $s=h-\bar h$ and $\Delta=h+\bar h$.\label{primaries}}
The fermionic current is given by $J_{\tn F}(z)=\,:\hat\psi^\dagger\hat\psi:$, and thus $T_{\tn F}(z)=-\half:J(z)J(z):$. 

The {\it algebra} that one derives for the Laurent expansion of the currents has the {\it same} form as the bosonic one: it is the enveloping Virasoro algebra with $c=k=1$. This means that the state-space has the same form as in the bosonic model, since it is constructed as the vector space of semi-positive norm, on which the algebra acts. In other words, the two state spaces are different representations of the same algebra---one proceeds from fermionic, and the other from bosonic, variables.\\
\\
{\bf Duality as unitary equivalence of representations of algebras.} In both models, the Hilbert space is obtained as the representation space (more precisely, a module) of the enveloping Virasoro algebra, in two different representations. In more detail, the vacuum state is a highest-weight state (i.e.~the state annihilated by the affine current operators, which act as ladder operators---this can be understood as a regularity condition of the state at the origin, $z=0$).\footnote{The origin of coordinates in the complex plane, $z=0$, has the significance that, in appropriate coordinates, it corresponds to the infinite past, and so it is a boundary condition for asymptotic states. See De Haro and Butterfield (2018:~p.~351) and Ginsparg (1988:~p.~16).}
The Hilbert space is then constructed from the irreducible representations of the algebra, analogous to the way SU(2) angular momentum representations are constructed by acting with ladder operators on the vacuum. Starting from the top state i.e.~the vacuum, one obtains the lower states by the application of the field operators.\footnote{The Verma module is the set of (linearly independent) states obtained from the highest state and acting with the lowering operators in all inequivalent ways. See Frishman and Sonnenschein (2010:~p.~24), Ginsparg (1988:~p.~48), Di Francesco et al.~(1997:~p.~158), and Polchinski (1998a:~p.~386).\label{verma}}
The states have quantum numbers that keep track of how many operators were applied. In this way, the whole Hilbert space can be constructed in either representation, i.e.~the bosonic or the fermionic one.

There is a uniqueness result for the enveloping Virasoro algebra: namely, the irreducible unitary representations of the enveloping algebra (constructed in the way just outlined) are unique up to unitary equivalence.\footnote{See Frishman and Sonnenschein (2010:~p.~133), and also the detailed discussion in De Haro and Butterfield (2018:~p.~359).} 
This means that the bosonic and fermionic Hilbert spaces thus constructed, ${\cal H}_{\tn B}$ and ${\cal H}_{\tn F}$ respectively, as endowed with (an algebra of) quantities (i.e.~self-adjoint operators), are {\it unitarily equivalent}. In other words, the duality map $d_{\cal S}$ is a unitary transformation. The duality map acts on the currents in the obvious way: it exchanges bosonic and fermionic currents, as follows: 
\bea\label{currco}
J_{\tn B}(z)&\leftrightarrow&J_{\tn F}(z)\nn
T_{\tn B}(z)&\leftrightarrow&T_{\tn F}(z)\,.
\eea
One can work out how the duality map acts on the field operators themselves. The top formula gives the relation between a bosonic field and a {\it pair} of fermionic fields, as follows: $\pa\hat\f(z)\,\leftrightarrow\,\,:\hat\psi^\dagger(z)\hat\psi(z):$. The inverse of this duality map maps a single fermionic field to a composite operator made of bosons. This is readily worked out, and the result is the following normally ordered exponential:
\bea\label{expf}
\hat\psi(z)\,\,\leftrightarrow~:e^{i\hat\f(z)}:\,,
\eea
and likewise for the anti-holomorphic components. 

Note that, although $\hat\f(z)$ is a bosonic operator, its exponential has {\it fermionic statistics}.\footnote{This can be shown by calculating the anti-commutator of two fermions and using the canonical commutation relations for the boson and its canonically conjugate momentum. To evaluate the comutator of the exponentials, one uses the formula $e^Ae^B=e^{[A,B]}e^Be^A$, i.e.~the special case of the Baker-Campbell-Hausdorff formula that obtains when $[A,B]$ commutes with both $A$ and $B$.} 
This change of statistics is a quantum mechanical effect with no classical analogue.

Let us summarise how this illustrates duality as defined in Section \ref{isomdef}: duality is here shown to be an isomorphism of Hilbert spaces together with algebras acting on them: namely, as a {\it unitary equivalence of representations} (or modules). Therefore, all the numerical values of the correlation functions, $\bra s|Q|s'\ket$, agree between the bosonic and the fermionic models. Duality thus defined is also equivariant for the dynamics, because the Hamiltonians (i.e.~the $00$-components of the stress-energy tensor) also match under the unitary equivalence. Thus there is a triple of states, quantities, and dynamics that is a common core theory, of which the two models are representations.

The common core theory thus defined has a manifest conformal symmetry: namely, the theory's spacetime symmetry, which is represented in the same way in the two models, viz.~by their stress-energy tensors, which are mapped onto each other by the duality map. Thus conformal symmetry is a {\it stipulated symmetry}, in the sense of Section \ref{salientstipul}.\footnote{Both models also represent the affine current symmetry algebra, although in very different ways. This symmetry is represented as {\it translations} of the field in the bosonic model, but as a {\it chiral symmetry}, in the fermionic model. Yet their corresponding currents satisfy the same algebra, and so these symmetries act in the same way on the states and on the quantities.}
This symmetry restricts the class of admissible quantities, and it does that in the same way for the two models. \\
\\
{\bf Generalizations of bosonization.} In the massive, interactive case (discussed in Section \ref{soliton2d}), the exponential map relating the bosonic and the fermionic fields, i.e.~Eq.~\eq{expf}, includes the parameter $\b$ and the mass. The algebra underlying the model triple can then still be written as the Virasoro enveloping algebra with $c=1$, but the level now depends on the coupling: $k=\b^2/4\pi=1/(1+g/\pi)$ (setting $\hbar=1$), so that in the limit of zero fermionic coupling, we indeed reproduce the algebra at level 1. 

Bosonization can be generalized to cases with non-abelian symmetry groups: there are also cases of bosonization between fermions with non-abelian gauge groups and Wess-Zumino-Witten models (see Witten (1984) and De Haro and Butterfield (2018:~pp.~371-372)). Bosonization can also be generalized to higher dimensions, where it is not a duality, but an {\it effective duality}. We will return to this when we discuss successor theories and the heuristic function of dualities, in Section \ref{bcc}. \\
\\
{\bf Significance of bosonization.} Bosonization illustrates the Schema for dualities in a non-trivial example of a quantum field theory that exhibits a duality exactly. 

Furthermore, the exponential map in Eq.~\eq{expf} illustrates Section \ref{featurerole}'s theme (1) of {\it hard-easy}, because a calculation that is easy in terms of the fermion $\hat\psi$ (e.g.~the expansion of the fermion $\hat\psi$ in creation and annihilation operators) can be difficult to do in terms of the boson $\hat\f$, since it requires an infinite number of terms in the Taylor series of the exponential $e^{i\hat\f(z)}$ and vice versa.\footnote{Another way to say this is that an eigenstate of the fermionic field $\hat\psi(z)$ corresponds, under duality, to a {\it coherent} state of the boson, i.e.~to an eigenstate of the exponential operator in Eq.~\eq{expf}. Importantly, these kinds of operators appear naturally in string theory as vertex operators that create incoming and outgoing strings (see Section \ref{stringth}). Namely, $e^{ikX}$, where $k_\m$ is the momentum and $X^\m$ is the embedding coordinate of the string worldsheet into the spacetime, is an operator that represents the emission or absorption of a string state of momentum $k_\m$ and zero spin (i.e.~a scalar, which happens to be a tachyon). See Del Giudice et al.~(1972:~p.~381), Green et al.~(1987:~p.~35) and Polchinski (1998a:~p.~65).} 
On the other hand, calculations that only use the common core quantities, i.e.~the conserved currents, are done in the same way in both models. 

A very surprising fact about the fermionization relation Eq.~\eq{expf} is its implication that, if we allow quantities that are defined by complex exponentials, then {\it the bosonic Hilbert space has fermionic states!} This is not something that one would easily consider in a quantum field theory, but these states are naturally defined using the duality map.\footnote{Agreed, one might naturally consider fermionic states constructed out of bosons if one were interested in conformal field theory or string theory. But note that these fermionic operators were constructed by Mandelstam (1975b:~p.~3027) in his search for creation and annihilation operators of quantum sine-Gordon solitons, i.e.~so as to extend Coleman's bosonization duality, which was previously restricted to the relation between the boson and the fermion {\it bilinear}. In order words, the construction of these fermionic operators out of bosons was motivated by duality. The algebra of analogous vertex operators in string theory was known: see e.g.~Del Giudice et al.~(1972:~p.~381).}
This is also our {\it hard-easy} theme: namely, such states are difficult to define in the bosonic model, but easy to define in the dual.

Thus imagine that one's original bosonic model did not include operators with fermionic statistics, such as the exponential operator in Eq.~\eq{expf}, in the its set of quantities. In that case, the fermionic and the bosonic models are {\it not} duals: but they {\it are} duals if augmented with the appropriate operators. Thus given two models $M_{\tn B}$ and $M_{\tn F}$ thus defined that are not duals, the discovery of a (partial) duality map between them, such as Eq.~\eq{expf}, may lead one to {\bf augment} one of the models---in this case, the bosonic model---by adding to it the operators and states required for the duality map to exist, thus in effect obtaining a larger, richer model, $M_{\tn B}'$ (as we already discussed in Section \ref{epistemicc}). 

There is here an analogy with the idea in logic, which we will discuss in Section \ref{sse}, of {\it definitional extension}. In Section \ref{abstraction}, we will also discuss this as part of the procedure of augmentation, by which one can obtain a common core theory from dual models. 

\section{Particle-vortex duality}\label{PVD}

In this Section, we explore solitons in two spatial dimensions (i.e.~$d=2$) that generalise the kinks discussed in Section \ref{SGT}. These solutions describe {\it magnetic vortices}, while their duals describe electrically charged particles. We follow the usage of calling a `point particle' a configuration of the field with the usual (perturbative) properties. By `the usual' properties, we mean that there exist linear(ized) solutions whose currents couple electrically to the electromagnetic field, i.e.~through a term $J^\m A_\m$ in the action, so that they contribute {\it electric} charge on the right-hand side of the Maxwell equations, Eq.~\eq{elmagn}. Such solutions satisfy the minimal requirements for point particles from Section \ref{wpd} (see especially Section \ref{dualism}-(ii)), in particular localization. (Having discussed wave-particle duality in Section \ref{wpd}, there is no danger of confusion here, because, as footnote \ref{retire} in Section \ref{wpd} also stressed, there is but a single quantization step in quantum field theory.)

Adding charge is required for static solutions, and this will suggest a method to construct the duals. For unlike solitons in $d=1$ (where $d$ is the spatial dimension), a simple scalar field model in $d=2$ and $d=3$ does not admit solutions describing a single vortex or monopole with total finite energy (see Eq.~\eq{Ekink}). However, when coupled to a vector field, the energy can be rendered finite. Thus we will begin with the solutions of the scalar field model with infinite energy, and then study finite-energy solutions where the scalar field is coupled to a vector field.\footnote{Here we treat the case $d=2$, and $d=3$ in Chapter \ref{EMDuality}. For a detailed treatment of vortex solutions and their properties, see Weinberg (2012:~pp.~40-41, 45-47), and also Manton and Sutcliffe (2004:~chapter 7). Particle-vortex duality is discussed in Zee (2003: chapter VI.3).\label{finiteE}}

\subsection{Vortex solutions of classical field theory}\label{simplev}

We take as our model the three-dimensional (i.e.~$D=d+1=2+1$) version of the scalar field model with the double-well potential from Section \ref{SGT}, but now for a scalar field $\f$ that can take {\it complex} values (anticipating that the main characteristic of the vortex solutions is the jump in the phase of the scalar field around a circle enclosing the ``core'' of the vortex, located e.g.~at $\f=0$). This is usually called the {\bf Mexican hat} potential, of which the double-well potential is a slice. The Lagrangian density is:
\bea\label{Lphi}
{\cal L}=\half\pa_\m\f^*\pa^\m\f-{\l\over4}\left(|\f|^2-v^2\right)^2,
\eea
where $\l$ and $v$ are positive constants (compared to Eq.~\eq{VMex}, we have defined: $v^2:=m^2/\l$). This model has a global U(1) gauge symmetry, namely the symmetry under constant changes of phase: $\f\mapsto e^{i\a}\,\f$. 

As before, we are interested in the global (stable) minimum of the potential, where the modulus of the field takes the value: $|\f|=v$. Because the field is complex, we can parametrise it as follows:
\bea\label{parametrisef}
\f(x)=\r(x)~e^{i\chi(x)}\,,
\eea
with $\r$ and $\chi$ being real functions, and $\r(x)\geq0$. In these new variables, at the minimum of the potential we have $\r=v$, while $\chi$ can take any value. This means that there is a one-parameter family, parametrised by $\chi$, of solutions at the minimum. The variable $\chi$ has the interpretation as a massless Goldstone boson for the global U(1) gauge symmetry, which is spontaneously broken by the solution.\footnote{In addition, the linear excitations of the field around the solutions Eq.~\eq{parametrisef} contain a real scalar, with mass $2\l\,v^2$.}

The solutions of the Euler-Lagrange equation of motion of the scalar field fall into two broad classes. First, there are the topologically trivial `particle' solutions (see the preamble of Section \ref{PVD}), where $\chi$ is single-valued and $\r$ is smooth. We will denote the phase of these solutions by $\chi_{\tn{reg}}$, for `regular'.

Second, there are solutions where $\chi$ is not single-valued, but jumps by an integer when we go around a circle containing the core of the vortex in its interior, i.e.~a circle around the vortex. We will denote these solutions by $\chi_{\tn{vor}}$. Denote the circle at infinity by $C$; then, for any smooth configuration $\f$, we can define the following line integral:
\bea\label{vorticity}
n_C={1\over2\pi}\oint_C\dd{\bf l}\cdot\nabla(\mbox{arg}\,\f)={1\over2\pi}\oint_C\dd{\bf l}\cdot\na\chi\,.
\eea
This number is called the {\it vorticity} of the solution, and it is well-defined as long as $\f$ is non-zero everywhere on $C$. $n_C$ counts the number of rotations of the phase of the field, $\chi$, as we go around the circle $C$, and so it is an integer. Since it is an integer, it does not vary under continuous deformations of the contour $C$, as long as the contour does not cross a point $P$ where the field is zero, $\f=0$. At such a point, the vorticity is not well-defined, and it changes value if the contour $C$ crosses such a point. A configuration of $\f$ with non-zero $n_C$ is called a {\bf vortex solution}, or simply a vortex. 

Clearly, for this particular solution, $n_C$ is analogous to the topological charge, $Q$, defined for solitons in one dimension of space, in Eq.~\eq{solitonQ}. In cylindrical coordinates $(r,\th)$, the asymptotic form of a single vortex solution is: $\f=\r\,e^{i\chi_{\tn{vor}}}\overset{r\rightarrow\infty}{\rightarrow} v\,e^{i\th}$, and an anti-vortex is: $\f\overset{r\rightarrow\infty}{\rightarrow} v\,e^{-i\th}$. Their vorticity is, respectively, $+1$ and $-1$. 

We can now see why the energy of single static vortex solutions is unbounded (see the preamble of Section \ref{PVD}). The total energy is:\footnote{The unboundedness comes from the gradient $(\na\f)^2$ in the energy expression Eq.~\eq{Ekink}. For an arbitrary dimension $d$, this term decays, at large distances from the origin, as $r^{-2}$. Including the volume factor, the corresponding term in the integral has the following radial dependence: $\int^\infty\dd r~r^{d-3}\sim r^{d-2}$, which goes to zero in one space dimension ($d=1$), but is logarithmically divergent in two space dimensions ($d=2$), and linearly divergent in three dimensions ($d=3$).}
\bea\label{Venergy}
E=\int_0^{2\pi}\dd\th\int_a^R\dd r~r\left[\half(\na\r)^2+\half\r^2\,(\na\chi_{\tn{vor}})^2+{\l\over4}(\r^2-v^2)^2\right].
\eea
We have regularised this integral, by assuming the vortex's core to be of size $a$, so that $r$ runs from $a$ to $R$, for $R\gg a$. For large $r$, the first and third term decay fast enough that their contribution to the integral from the large-$r$ region is finite, but the second term is divergent, because it goes to zero as: $\half r\r^2(\nabla\chi_{\tn{vor}})^2\overset{r\rightarrow\infty}= \half rv^2{1\over r^2}\left({\pa\chi_{\tn{vor}}\over\pa\th}\right)^2\overset{r\rightarrow\infty}{=} {v^2\over2r}$, where in the last step we used the fact that, asymptotically for a {\it single} vortex, $\chi_{\tn{vor}}\overset{r\rightarrow\infty}{\rightarrow}\pm\th$. Upon integration, this gives a logarithmically divergent, large-scale contribution to the energy, $E_{\tn{div}}=\pi v^2\ln(R/a)$. Thus, for $R\rightarrow\infty$, the vortex solution has infinite total energy, which is surely an undesirable property: there exists no `soliton' solution, in the sense of the preamble of this Chapter. 

In spite of this, we can use the solution with finite $R$ to study heuristically the dynamics of vortex pairs, and their dissociation, as a function of the temperature, as we will do below.\\
\\
{\bf Intermezzo: heuristics of vortex-anti-vortex pairs and a phase transition}\\
\\
Even though the single-vortex solutions do not have well-defined energy, we will show, later in this Section, that their energy is rendered finite by adding an electromagnetic field. Thus, since the unboundedness can be dealt with, we can consider in more detail the physics of the vortices themselves, since they illustrate the important idea of a transition from a disordered phase at high temperatures (consisting of dissociated vortices) to a {\bf topologically ordered phase} at low temperatures, in which the vortices are bound in vortex-anti-vortex pairs.

The heuristic idea of a phase transition towards a vacuum state that is a {\bf condensate of solitons in pairs} has been highly influential in all of physics, not least as a possible explanation of the confinement of quarks: which, as we will see, is analogous to the confinement of vortices in pairs.\footnote{See Nielsen and Olesen (1973), Nambu (1974), 't Hooft (1975), and Mandelstam (1975a, 1976). We will discuss these ideas in Chapter \ref{EMDuality}.} 
It also leads to the idea that, in addition to point particles, solitons must also be used to study the vacua of a field theory. Thus it is worth pausing to calculate the critical temperature, based on the regularized model in Eq.~\eq{Venergy} for the vortices. 

Note that this regularized energy can be used to show that vortex-anti-vortex pairs have {\it finite} energy, as follows.

An {\it anti-vortex} is defined as a vortex with negative vorticity, i.e.~$n_C=-1$, for a single anti-vortex. Therefore, its contribution to the phase $\chi_{\tn{vor}}$ at infinity cancels that of the vortex, and the total vortex number is zero. The dominant contribution to the energy of a vortex-anti-vortex solution can likewise be estimated to be of the order $v^2\ln(R'/a)$, where $R'$ is the {\it separation} between the vortex and the anti-vortex, and, as before, $a$ is the vortex size. 

Thus vortex-anti-vortex pairs have finite energy: they attract each other through the Coulomb force (which grows logarithmically in two space dimensions), and a {\it gas} of multiple vortices and anti-vortices would appear as a gas of charged particles. This also means that, in this model, separating a vortex-anti-vortex pair, e.g.~by carrying the vortex out to infinity, costs infinite energy: and that in trying to do so, it is energetically more favourable to create a new vortex-anti-vortex pair out of the vacuum. Thus, in effect, any vortex is bound to an anti-vortex in a pair at a finite distance from each other, and it is only pairs that can be carried off to infinity. 

To calculate the temperature of the phase transition between a phase with and a phase without free vortices, we combine this qualitative picture of vortices as approximate solutions of the equations of motion of the Lagrangian Eq.~\eq{Lphi}, with logarithmically growing energies, with statistical mechanical considerations about a system of large, but finite, size $R$, of many such vortices, in equilibrium at temperature $T$. This description is due to Kosterlitz and Thouless (1972, 1973);\footnote{See also Berezinskii (1971, 1972).} 
and we note in this connection that Thouless and Kosterlitz, together with Haldane, were awarded the 2016 Nobel Prize.

For a system of finite size $R$, the single-vortex solutions have finite energy Eq.~\eq{Venergy}, but it is the {\it free energy} that determines, at various values of the temperature, whether free vortices (i.e.~vortices that can travel to the boundary of the region) are favoured, or suppressed in favour of vortex-anti-vortex pairs. A state with vortex-anti-vortex pairs and no free vortices is a state with high long-range order, i.e.~an order that spans through the entire system. In this case, the order is topological, because vortices are topological solutions, carefully ordered so that no free vortices are left. It is a {\it metastable state,} i.e.~roughly, a state that is not the ground state (its energy is clearly non-zero),\footnote{Note that, at large distances, the solutions we are considering are at the minimum of the Mexican-hat potential in Eq.~\eq{Lphi}, i.e.~$V(\f)=0$, and thus the third term in the energy Eq.~\eq{Venergy} is asymptotically zero. Nevertheless, the non-trivial phase $\chi_{\tn{vor}}$ gives a non-zero contribution in the second term, so that these are indeed not ground states, despite their minimizing the potential asymptotically.} 
but is a local minimum and is stable against {\it small} perturbations. Thus the vortices are bound in pairs, or in clusters of zero total vorticity, below a critical temperature, as in a Bose condensate: while they can dissociate above this temperature.

By contrast, a state where vortex-anti-vortex pairs may dissociate is disordered. The question is what is the order parameter that can distinguish these two phases. Traditionally in statistical mechanics without solitons, order parameters are given by two-point correlation functions of the fields involved. Near the critical temperature of the phase transitions, these correlation functions exhibit singularities, which are characterised by their {\it critical exponents} (see Section \ref{dualpf}).

However, Kosterlitz and Thouless (1972:~p.~L124; 1973:~p.~1181-1182) noticed that the two-point correlation functions for vortex solutions are zero at low temperatures, and so cannot be order parameters. The underlying reason is that vortices are topological solutions, determined by the phase of the field, $\chi_{\tn{vor}}$, far away from the centre, rather than the familiar fluctuations of the modulus of the field, $\r$ (cf.~Eq.~\eq{parametrisef}). Therefore, they proposed to define the long-range order of a two-dimensional system with vortex-type solutions (such as a superfluid, a solid or a spin model---the latter could sustain vortices of the kind discussed here)\footnote{They discuss these various models in Kosterlitz and Thouless (1973:~pp.~1184-1200).} 
based on the overall properties of the system. More specifically, Kosterlitz and Thouless (1973) take, depending on the model they consider, different quantities as the order parameter. For example: the mean square separation of the vortices in a pair (p.~1184) and the density of dissociated pairs (p.~1189). These quantities are seen to diverge at the critical temperature. They called this {\bf topological long-range order.} 

The idea is analogous to the behaviour of a liquid near the freezing point: its local structure is similar to that of a solid, but there is some concentration of dislocations, which breaks the order. For large systems of this type at low temperatures, dislocations are energetically disfavoured and cannot occur because, as we saw, the energy grows logarithmically with the size of the system. However, pairs of dislocations with different signs {\it can} occur, even at low temperature. If the system has a  temperature, even a low one, such pairs of dislocations and anti-dislocations will occur because of the thermal excitation. Above a certain critical temperature, such dislocation pairs can dissociate. 

For the vortex solution considered above, the critical temperature can be found as follows. The Helmholtz free energy associated with a single vortex is estimated to be: $F_{\tn{div}}=E_{\tn{div}}-T\,S_{\tn{div}}\simeq(\pi v^2-T)\ln(R/a)$, where $S_{\tn{div}}$ is the entropy, and the subscript indicates that we are only considering the leading (in $R$, for large $R$) contribution to the free energy. At low temperatures, the energy term dominates over the entropy term: and so, if the system is large, the probability of a single dislocation (i.e.~a vortex) appearing is small. The probability (i.e.~the relative contribution of a single vortex to the partition function and to other average quantities) is suppressed\footnote{The overall normalisations are calculated in the various models in the appropriate order parameters, see e.g.~Kosterlitz and Thouless (1973:~p.~1184).} by a factor of $e^{-F_{\tn{div}}/T}\simeq(R/a)^{1-\pi v^2/T}\simeq(R/a)^{-\pi v^2/T}$, where in the last approximation we used $T\ll \pi v^2$ in addition to $R\gg a$. Thus the probability of the free vortices is suppressed by the negative exponential. However, in this regime of low temperature, bound states of vortices and anti-vortices can exist, along with ordinary smooth excitations of the field, i.e.~the topologically non-trivial excitations mentioned above Eq.~\eq{vorticity}. 

As we increase the temperature, i.e.~for $T>\pi v^2$, the entropy term in the free energy takes over, and the probability is proportional to $e^{-F_{\tn{div}}/T}\simeq(R/a)^{1-\pi v^2/T}\simeq R/a$ at high temperatures, $T\gg\pi v^2$. In this regime, free vortices will appear spontaneously because they are energetically favoured. The critical temperature at which this happens is $T_{\sm c}=\pi v^2$.\footnote{This estimate gives only an upper bound on the critical temperature, because it is the temperature at which a pair of vortices dissociates, but it ignores the effect of other pairs in the system: `These other pairs will relax in the field of the first pair, thereby renormalizing the rigidity modulus etc.~downwards and consequently reducing the critical temperature' (Kosterlitz and Thouless (1973:~p.~1183). The corrections to this critical temperature are considered on p.~1187.
} Thus, for $T>T_{\sm c}$ we {\it can} have free vortices, whereas below $T=T_{\sm c}$ the vortex-anti-vortex pairs condense. This is the celebrated {\bf Berezinskii-Kosterlitz-Thouless phase transition}. \\
\\
{\bf Vortices with finite energy}\\
\\
We saw that, in the simple scalar field model Eq.~\eq{Lphi}, single vortices have infinite energy. To get vortex solutions with finite energy, we will couple the model scalar field $\f$ to a gauge field $A_\m$, thereby making the global U(1) gauge symmetry of the scalar into a local U(1) gauge symmetry shared by $A_\m$ and $\f$. The Lagrangian density, again in $2+1$ dimensions, is:
\bea\label{Lvortex}
{\cal L}&=&-{1\over4}\,F_{\m\n}\,F^{\m\n}+\half (D_\m\f)^*(D^\m\f)-{\l\over4}\left(|\f|^2-v^2\right)^2\nn
D_\m\f&=&\pa_\m\f+ie\,A_\m\,\f\,,
\eea
where $e$ is the electric charge. For obvious reasons, this Lagrangian is called the {\bf Abelian Higgs model.} It is invariant under the combined local transformations $\f(x)\mapsto e^{ie\L(x)}\f(x)$, $A_\m(x)\mapsto A_\m(x)-\pa_\m \L(x)$, for an arbitrary gauge function $\L:\mathbb{R}^3\rightarrow\mathbb{R}$. 

As before, the potential of the scalar field is minimized at $\r=v$. This minimum does not enjoy the U(1) symmetry, and so this is a `spontaneously broken model'.\footnote{Elitzur's theorem in lattice gauge theory implies that local gauge symmetries cannot be broken. As such, it does not forbid our use of the phrase `spontaneous symmetry breaking' in the sense that we will discuss in Chapter \ref{EMDuality}; (indeed, a quotient of the local gauge group {\it can} be broken). For an introduction to this topic, see Beekman et al.~(2019:~Chapter 7). We thank Silvester Borsboom and Jasper van Wezel for discussions of this point.} 
The excitations around this minimum are a massive scalar (with the same mass as before) and a massive vector, with mass $m_A=ev$. 

The vorticity of the scalar field is defined just as before, i.e.~through Eq.~\eq{vorticity}, and it is invariant under local gauge transformations, because these are single-valued. 

Let us see how the energy of the vortex is rendered finite. Recall that the origin of the divergence was the fact that the term $\nabla\chi$ in Eq.~\eq{Venergy} did not go fast enough to zero at infinity. But the relevant quantity is now the covariant derivative, which contains the contribution of the gauge field:
\bea\label{Dphi}
{\bf D}\f=e^{i\chi_{\tn{vor}}}\left(\nabla\r+i\r\,(\nabla\chi_{\tn{vor}}-e\,{\bf A})\right),
\eea
The gauge field will cancel the leading $1/r$ behaviour of $\nabla\chi$ at infinity, if it is given (asymptotically) by ${\bf A}=1/e\,\nabla\chi$. With this leading behaviour, we also get that the vorticity is given by the magnetic flux through the surface enclosed by the curve $C$:
\bea
n_C={e\over2\pi}\int_S\dd^2x\,B\,,
\eea
where $S$ is a surface bounded by the curve $C$. Taking $C$ to be the circle at infinity, the total magnetic flux $\F$ is related to the vorticity by:
\bea
\F={2\pi n\over e}\,.
\eea
This physical realization of the vortex is much more satisfying: the vortex is a topological configuration or defect with a {\it magnetic flux}, whose value is inversely proportional to the electric charge.

The above result led to speculations, which are verified in supersymmetric field theories, that the confinement of electric charges (quarks) could be seen as the electric-magnetic dual of the confinement of vortices that takes place below the critical temperature. More precisely, since confinement takes place in three not two spatial dimensions, it is the three-dimensional analogue of vortices that should play a role: namely, magnetic monopoles. Thus the question is whether confinement can be understood as the electric-magnetic dual effect of the condensation of magnetic monopoles. We will return to this in Chapters \ref{EMDuality} and \ref{EMYM}.

But first, in Section \ref{pvdah}, we study the {\it duality between particles and vortices} in two spatial dimensions. We will show that duality exchanges the vorticity with the U(1) Noether current, and how this can be used to describe a phase transition to a topologically ordered phase. Then Section \ref{MEMD} will prepare the ground by discussing electric-magnetic duality.

\subsection{Particle-vortex duality for the Abelian Higgs model}\label{pvdah}

The $d=1$ sine-Gordon-Thirring model duality exchanges the topological current of the bosonic model with the Noether current of the fermionic model. In $d=2$, this suggests that, by analogy, we seek a {\it dual model}, in which the topological vortex, which contributes magnetic flux, is exchanged with a point particle, which couples to the electromagnetic field as an electric current.\footnote{For the conception of a `point particle' here used, see the discussion in the preamble of Section \ref{PVD}.}
It turns out that there is a transformation that achieves this: we will find it by considering the path integral for the action Eq.~\eq{Lvortex}.\footnote{For a discussion of particle-vortex duality, see e.g.~Zee (2003: Chapter VI.3), which we follow with some modifications. Our approach to duality here follows the Lagrangian and path integral method, rather than the state space description.}

But since the Lagrangian now has both a scalar field and a gauge field, the definition of the duality requires an additional idea: to dualize the gauge field using {\it Hodge duality}, because requiring that the vortex current couples electrically to the (Hodge) dual gauge field will automatically give us the expression for the (topological) vortex current.

Our treatment will use two main (related) simplifications, which are standard in the literature: 

(1) Since the main effect of the vorticity comes from the massless scalar $\chi$ (cf.~Eq.~\eq{vorticity}), we will substitute the Ansatz $\f(x)=v\,e^{i\chi(x)}$ into the action Eq.~\eq{Lvortex}. While this Ansatz obscures the fact that the vortex core is located at the zeroes of the scalar field, $\f=0$, it gives an adequate description of the asymptotic region, where we measure the magnetic flux. Indeed, this suffices to describe the infrared physics, i.e.~low energies and-or weak coupling. Thus, in effect, we can neglect the interactions between the massive scalar $\r$ and the Goldstone boson $\chi$, and set $\r$ to be the value that minimizes the potential, $\r=v$.\footnote{A fuller treatment would keep the oscillations of the field $\r$ around its asymptotic value $v$. For a rigorous treatment, see Peskin (1978).}
(Thus our discussion here is complementary to that in the previous Section, which concerned the local dynamics of vortex-anti-vortex pairs.)

(2) We will drop the kinetic term for the gauge field $A_\m$, i.e.~the first (Maxwell) term in the action Eq.~\eq{Lvortex}. Thus we treat the electromagnetic field as a source and not as a dynamical field, which is justified because we will restrict attention to the vortex, i.e.~the topological non-trivial solution of the scalar field. As we discussed at the end of the previous Section, $A_\m$'s main role is to render the vortex's energy finite, and to measure its magnetic flux at infinity. This depends on the boundary condition for $A_\m$ at infinity, rather than on its full dynamics.

Following (2), the action and its corresponding partition function, Eq.~\eq{Lvortex}, simplify to:
\bea\label{coupA}
S_{\tn{AHiggs}}[\phi,A]&=&\int\dd^3x\left[\half(D_\m\f)^*(D^\m\f) -{\l\over4}\left(|\f|^2-v^2\right)^2\right]\nn
Z_{\tn{AHiggs}}[A]&=&\int{\cal D}\f~\exp\left(iS_{\tn{AHiggs}}[\f,A]\right).
\eea
The partition function is thus a functional of the (fixed) gauge field $A$.\footnote{The Abelian Higgs action is the (continuum limit of the) XY-action, where `XY' refers to the classical XY-model, which is a statistical-mechanical model for the interaction between 2-dimensional spins, i.e.~a generalisation of the Ising model, and a special case of the $\mbox{O}(n)$ vector model where $n=2$ (the Ising model being the case $n=1$). Notice that the quadratic term in the potential, $a\,|\f|^2$, has a negative coefficient, i.e.~$a=-\l\,v^2/2<0$, and so we expect there to be a phase transition at $T=T_{\sm c}$, where the quadratic coefficient is proportional to $T-T_{\sm c}$: see Section \ref{topphases}, and also Peskin (1978).}

The above action has a conserved Noether current for the U(1) symmetry (for later use, we temporarily consider solutions with zero gauge field, i.e.~$A=0$), given as follows:\footnote{For a philosophical discussion of Noether's theorems, see Read and Teh (2022).}
\bea\label{JNoether}
J_\m^{\tn{Noether}}={ie\over2}\left(\f\,\pa_\m\f^*-\f^*\pa_\m\f\right),
\eea
and it is conserved in virtue of the Klein-Gordon equation that holds for the scalar field when the gauge field is zero (and taking $|\f|=v$).

We now implement our first simplification, (1), above, i.e.~we write $\f=v\,e^{i\chi}$. (In the path integral, this amounts to doing a saddle-point approximation around vortex solutions, valid at low energies). The Noether current Eq.~\eq{JNoether} then simplifies:
\bea\label{Npachi}
J_\m^{\tn{Noether}}=ev^2\,\pa_\m\chi\,,
\eea
as does the action:
\bea\label{Llin}
S[\chi,A]={v^2\over2}\int\dd^3x\,(\pa_\m\chi-eA_\m)^2\,.
\eea
One might think that the first term can be cancelled by a gauge transformation of the gauge field $A$. However, this is not possible, because $A$ is a fixed field, while we integrate over all the values of $\chi$ in the path integral Eq.~\eq{coupA}. 

It is also clear that the (fixed) photon $A_\m$ has acquired a mass, through the Higgs mechanism.

In our next step, we wish to ``integrate out'' the regular part of the field $\chi$ in the path integral Eq.~\eq{coupA}; (this is jargon for: `we wish to perform the path integral for the regular part of the field $\chi$'). However, before we do this, it is useful to linearise the action, by introducing an auxiliary variable $\xi_\m$ (the components of a one-form), i.e.~a variable that appears quadratically in the new action, such that it reproduces the original action Eq.~\eq{Llin} when the Gaussian integral over $\xi_\m$ is performed:
\bea\label{linear}
S[\chi,\xi,A]=\int\dd^3x\left(-{1\over2v^2}\,\xi_\m\xi^\m+\xi^\m(\pa_\m\chi-eA_\m)\right).
\eea
Since this model is equivalent to the original model Eq.~\eq{coupA} for the solution $\f=v\,e^{i\chi}$, we expect from Section \ref{simplev} that the solutions will include vortices, located at the points where $\f=0$. For such solutions, going in a closed curve around these zeroes, $\chi$ changes by a factor of $2\pi$.

However, the field also has topologically trivial configurations, where $\chi$ is regular. The path integral in Eq.~\eq{coupA} sums over all of these configurations. Thus we split $\chi$ into two pieces: the vortex solutions, which we will denote by $\chi_{\tn{vor}}$, and the regular solutions, i.e.~$\chi=\chi_{\tn{vor}}+\chi_{\tn{reg}}$. Since we are interested in the physics of the vortices, and the action is linear in $\chi$, we can integrate out the regular part. Partial integration of the $\chi$-term gives (up to a boundary term, whose details we will here neglect) $-\chi_{\sm{reg}}\,\pa_\m\xi^\m$, which upon integration of $\chi_{\sm{reg}}$ results in the constraint $\pa_\m\xi^\m=0$. 

In three dimensions, this equation can in general be solved in terms of a new ({\it dual}!) gauge field $a_\m$, as follows:
\bea\label{curla}
\xi^\m=\e^{\m\n\l}\pa_\n a_\l\,.
\eea
The one-form $\xi$ is the {\bf Hodge dual} of the Faraday two-form of the one-form $a$, in three dimensions (i.e.~the curvature, $\xi^\m=*F^\m[a]$).\footnote{In four dimensions, the Hodge dual of a two-form is another two-form, but in three dimensions it is a one-form. For the idea of Hodge duality, see Section \ref{FHodge}, especially Eq.~\eq{Gdual}.}
Plugging this back into the action, we get:
\bea\label{LF}
S[\chi,a,A]=\int\dd^3x\left(-{1\over4v^2}\,F_{\m\n}^2[a]+\e^{\m\n\l}\pa_\n\,a_\l(\pa_\m\chi_{\tn{vor}}-eA_\m)\right).
\eea
Here, $F_{\m\n}[a]$ is the Faraday tensor for the dual field, $a_\m$: a new dynamical field with a U(1) gauge symmetry. There are two ways to read the second term:

First, it is a {\bf magnetic coupling} between the dual gauge field $a$ and the conserved Noether current, Eq.~\eq{Npachi}. It is called `magnetic' because it does not couple the gauge field and the Noether current, i.e.~$a^\m J_\m^{\tn{Noether}}$, but rather the Hodge dual of the curvature two-form and the current, i.e.~$*F^\m[a] J_\m^{\tn{Noether}}$.

Second, we can require that it is an {\it electric} coupling between the gauge field $a$ and a conserved {\bf topological vortex current}, $J^\m_{\tn{vortex}}$. We find the required expression by partial integration:
\bea\label{vortexj}
J^\l_{\tn{vortex}}:={1\over2\pi}\,\e^{\l\m\n}\pa_\m\pa_\n\chi_{\tn{vor}}\,.
\eea
In terms of it, the action is: 
\bea\label{jvortex}
S[a,J,A]=\int\dd^3x\left(-{1\over4v^2}\,F_{\m\n}^2[a]+2\pi\,a_\m\,J^\m_{\tn{vortex}}-eA_\m\e^{\m\n\l}\,\pa_\n\,a_\l\right).
\eea
The topological vortex current {\it is} the Noether current for the U(1) gauge symmetry of the dual gauge field $a$, to which it is electrically coupled.\footnote{For a discussion of Noether's second theorem and the interpretation of the conserved current, see De Haro (2022:~pp.~206-207; 214-219).}
Notice that the original Noether current and the (dual) topological vortex current satisfy the following relation:
\bea\label{jjdual}
J^\l_{\tn{vortex}}={1\over2\pi ev^2}\,\e^{\l\m\n}\,\pa_\m J^{\tn{Noether}}_\n\,.
\eea
Despite the antisymmetry of the $\e$-tensor and the symmetry of the partial derivatives, this expression is non-zero, because the vortex solution $\chi_{\tn{vor}}$ is not single-valued. In fact, integrating the time component of this current over the plane returns the vorticity, Eq.~\eq{vorticity}, which is an integer. Thus the vorticity is the Noether charge of the local U(1) gauge symmetry of the {\it dual} vector field $a_\m$. This charge is obtained from the current $J^\m_{\tn{vortex}}$ that couples to the dual gauge field. 

Thus we have achieved our goal: to couple the vortices in the action as {\it particles,} i.e.~through an electric current---one that couples to the dual gauge potential $a$.\footnote{There is also an electromagnetic current for the {\it original field} $A$: $J^\m_{\tn{EM}}=e\,\e^{\m\n\l}\,\pa_\n\,a_\l$.}
This also illustrates our theme from Section \ref{SGT}, that {\it the topological charge is the Noether charge of the dual model}. 

\section{Electric-magnetic duality in four dimensions}\label{MEMD}

This Section discusses the {\it classical and quantum} electric-magnetic duality of the Maxwell theory.\footnote{The generalizations discussed in this Section can be found in many places. We draw especially from Witten (1995b) and Seiberg and Witten (1994a:~pp.~28-30). A version of it has recently been used by Armoni (2023:~p.~2) to give a magnetic dual of the three-dimensional Maxwell--Chern-Simons model.}
Compared to the electric-magnetic duality of the Maxwell theory discussed in Section \ref{EMduality}, this Section adds the {\it derivation of a Lagrangian} for the common core theory, which is then used to define the quantum version of the duality, and is a stepping stone, to the dualities for interacting quantum field theories in Chapter \ref{EMYM}. 

Section \ref{cdbm} first discusses the classical case, and introduces a {\it common core theory} from which the electric and magnetic models are derived. Section \ref{quantumD} first introduces an innovation into the Lagrangian action of the Maxwell theory: a $\th$-term that will be important when we discuss supersymmetric theories in Chapter \ref{EMYM}. It then discusses the quantum duality, and finds that, up to an overall numerical factor, the partition function of the Maxwell theory is invariant under the duality map.\footnote{In jargon, although we will not need these details: it is in general a modular form.}

\subsection{Classical duality between models}\label{cdbm}

This Section reformulates electric-magnetic duality (see Section \ref{EMduality}) by introducing a common core theory ``above'', of which which the Lagrangian actions of the electric and magnetic models are representations ``below''. 

The idea can be summarised as follows. We begin with the theory ``above'' defined by the action $S[{\cal F},B]$ depending on two fields: an antisymmetric two-tensor ${\cal F}$ and a four-vector field $B$ (a gauge potential with no dynamics).\footnote{The gauge potential $B$ is an auxiliary vector field, because its equation of motion imposes a constraint on ${\cal F}$ and does not have dynamics of its own; but this need not concern us here.} There are two alternative ways to treat this action, labelled (A) and (B) below, which give rise to two models of the common core theory. The action of each model is obtained from the action of the common core theory by substituting a specific value for the field ${\cal F}$ (substituting one value yields an electric model, and substituting its `dual value' yields the magnetic dual model). Since the models are obtained by defining the common core theory's field ${\cal F}$ using specific structure, i.e.~through an equation that specifies ${\cal F}$ in terms of another field, the models are, as in our Schema, representations of the common core theory. 

The two treatments of the action are as follows:

(A) Vary the action $S[{\cal F},B]$ with respect to the four-vector field $B$. This gives a constraint for the antisymmetric tensor ${\cal F}$, namely the condition $\dd{\cal F}=0$, which implies the existence of a four-vector potential $A$ such that the antisymmetric tensor ${\cal F}$ is the Faraday tensor for it, i.e.~${\cal F}=F[A]=\dd A$ (see Section \ref{EmdS}, especially after Eq.~\eq{Max2}). Substituting this into the action $S[{\cal F},B]$, we get a new action that is precisely the {\it Maxwell action} for the field $A$, i.e.~Eq.~\eq{FAA}, with coupling constant $e$. We dub this the `A-model'.

(B) Vary the action $S[{\cal F},B]$ with respect to the antisymmetric tensor ${\cal F}$ instead. This gives an equation that relates the Hodge dual $*{\cal F}$ to the Faraday tensor of $B$ (see Eq.~\eq{eF} below). Substituting this in $S[{\cal F},B]$, the action is again of the Maxwell form Eq.~\eq{FAA}, with the gauge potential $B$ and a coupling constant $e'$ that is $e$'s reciprocal: $e'=4\pi/e$. We dub this the `B-model'.

The derivation is as follows: we begin from the following action (the integrals are over $\mathbb{R}^4$): 
\bea
S[{\cal F},B;e]&=&\int\dd^4x\left(-{1\over4e^2}\,{\cal F}^2-{1\over8\pi}*F[B]\,{\cal F}\right),\label{SFBnew}\\
&=&\int\dd^4x\left(-{1\over4e^2}\,{\cal F}^2-{1\over4\pi}\, B_\m\,\pa_\n*{\cal F}^{\m\n}\right).\label{SFB}
\eea
where the step from the first to the second line uses partial integration.\footnote{The steps are as follows, where the boundary term of the partial integration is zero: $\int*F[B]\,{\cal F}=\int F[B]*{\cal F}=\int(\pa_\m B_\n-\pa_\n B_\m)*{\cal F}^{\m\n}=-2\int\pa_\n B_\m*{\cal F}^{\m\n} =2\int B_\m\,\pa_\n*{\cal F}^{\m\n}$.\label{FBF}}
The notation is as follows:\footnote{See Hamilton (2017:~p.~414).} 
${\cal F}^2:={\cal F}_{\m\n}{\cal F}^{\m\n}$, and $*F$ is the Hodge dual of $F$ (defined in Eq.~\eq{Gdual}; and likewise for ${\cal F}$). Thus the first term takes the same form as the Maxwell action (Eq.~\eq{FAA}), with the important difference that ${\cal F}$ is {\it not} exact, i.e.~it is not the Faraday tensor (the curvature) of a gauge potential.\\
\\
(A)~~We vary the action with respect to $B$ and set the result to zero (we use the second form, Eq.~\eq{SFB}). This gives the following condition: 
\bea\label{*F}
\pa_\n*{\cal F}^{\m\n}=0~~~\mbox{i.e.}~~~\dd{\cal F}=0\,,
\eea
which is the Bianchi identity for ${\cal F}$ (see Eq.~\eq{Max2}), and implies the existence of a four-vector potential $A$, such that: 
\bea\label{eomA}
{\cal F}=F[A]=\dd A\,.
\eea
Substituting this into the action Eq.~\eq{SFB}, the result is the Maxwell action, Eq.~\eq{FAA}, for $A$.\footnote{The Maxwell action comes entirely from the first term in Eq.~\eq{SFB}, after the substitution ${\cal F}=\dd A$, because the second term in Eq.~\eq{SFB} is zero due to the Bianchi identity Eq.~\eq{*F}.}

The equation of motion of the Maxwell action of the A-model is the usual one: in terms of the original two-form, it is $\dd*{\cal F}=0$. Thus in the A-model, ${\cal F}$ is the Faraday tensor for the gauge potential $A$, and it satisfies the Bianchi identity $\dd{\cal F}=0$ and the equation of motion $\dd*{\cal F}=0$.\\
\\
(B)~~We perform the variation of the action with respect to ${\cal F}$ (in its first form, Eq.~\eq{SFBnew}). This gives the condition:
\bea\label{FFB}
{\cal F}=-{e^2\over4\pi}*F[B]\,.
\eea
As we will see in a moment, this is significant, because it means that the antisymmetric tensor ${\cal F}$, properly normalized, is the Hodge dual of $F[B]$ (rather than the Faraday tensor of $A$, as in (A)). Substituting this into the action Eq.~\eq{SFBnew}, and using $*^2=-1$, we get back the Maxwell action:
\bea\label{dMax}
S_{\tn{Maxwell}}[B;e']=-{1\over4e'{}^2}\int\dd^4x~F^2[B]\,,
\eea
where we defined a new coupling constant $e'$, so that the action has the same form as the Maxwell action Eq.~\eq{FAA}, for the Faraday tensor $F[B]$. The coupling constants $e$ and $e'$ are related by the following strong vs.~weak coupling relation:
\bea\label{dadb}
e'={4\pi\over e}\,.
\eea
This is Dirac's quantization condition Eq.~\eq{Diracq} with $n=2$ (in units where $\hbar=1$). However, one should be cautious, because there are no monopoles here, and the precise numerical coefficient can be changed by changing the numerical coefficient of the second term in the action Eq.~\eq{SFBnew} (i.e.~by a rescaling of $B$; our choice of normalization anticipates the understanding of the next Section.) The relation between the charges Eq.~\eq{dadb} allows us to rewrite Eq.~\eq{FFB} in the following symmetric form:
\bea\label{eF}
{1\over e}\,{\cal F}=-{1\over e'}*F[B]\,,
\eea
an important relation that will allow us to relate the quantities of the two models. Note that, in virtue of the Bianchi identity for $B$, ${\cal F}$ satisfies $\dd*{\cal F}=0$ (which looks like the Maxwell equations, but is here derived as a Bianchi identity).

The equation of motion of the Maxwell action of the B-model is $\dd*F[B]=0$, which written in terms of ${\cal F}$ is $\dd{\cal F}=0$, and looks like the Bianchi identity for ${\cal F}$, but here follows as the equation of motion for $B$.

\begin{figure}
\begin{center}
\includegraphics[height=5cm]{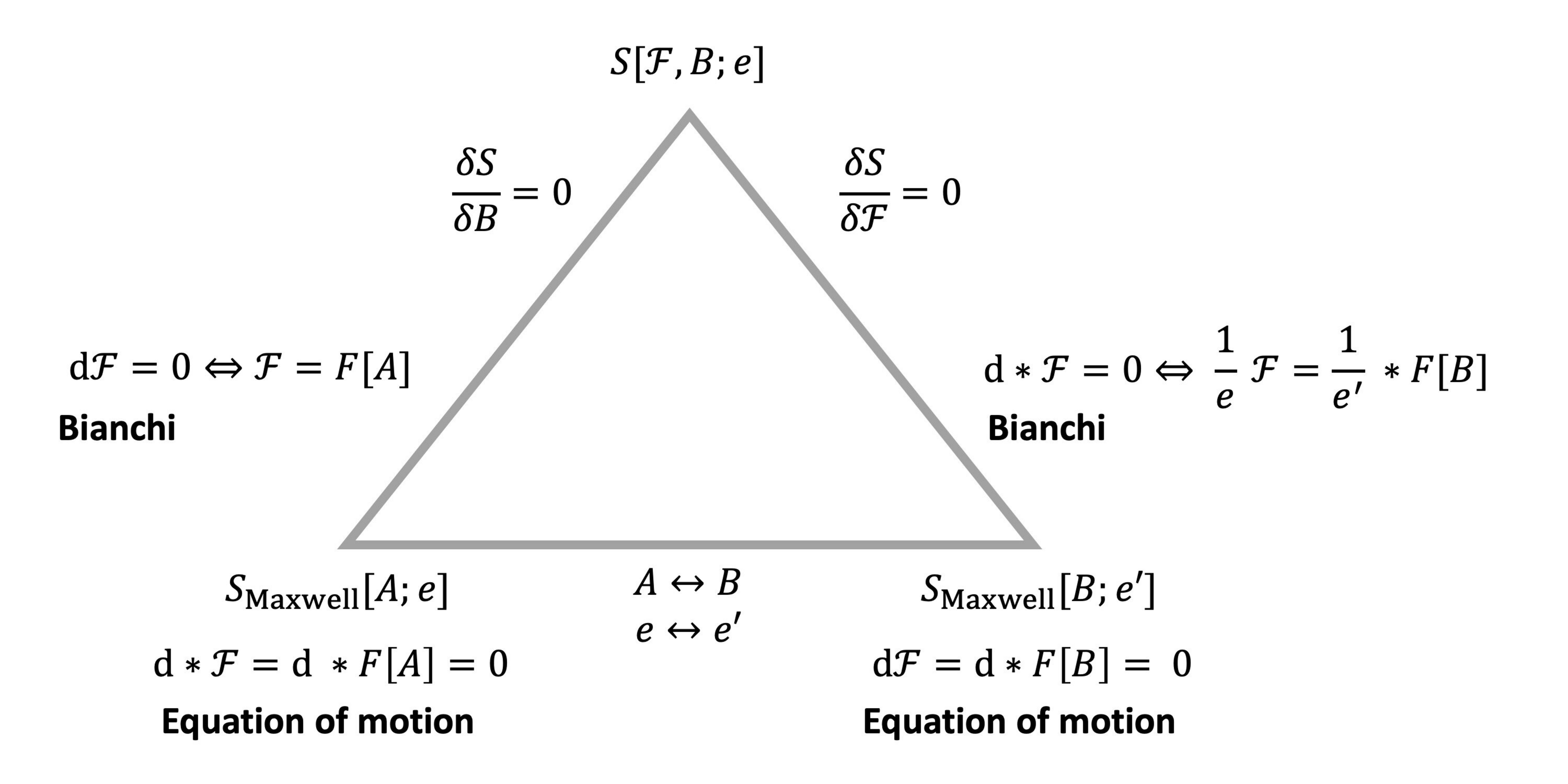}
\caption{\small Relation between the common core theory above and its two dual models below (A on the left, B on the right), which represent it in two different ways. Compare this with Figure \ref{tm1m2}.}
\label{Triality}
\end{center}
\end{figure}

The relations are summarized in Figure \ref{Triality}. The two models, with actions $S_{\tn{Maxwell}}[A;e]$ and $S_{\tn{Maxwell}}[B;e']$ (where the notation indicates the dependence on the coupling constant), are related through the ``theory above'', i.e.~Eq.~\eq{SFBnew}. Each of them is a representation of the theory above, in terms of different variables. (The two representations reflect the two orders in which the equations of motion are derived.) The models are also related by a duality that maps the states of the two models as $d_{\cal S}(A)=B$.\footnote{This duality map straightforwardly relates the curvatures of the two models. To find the direct relationship between the gauge potentials, one must solve the equation: ${1\over e}\,F[A]={1\over e'}*F[B]$, which gives a non-local expression. This non-locality is already familiar from e.g.~Kramers-Wannier duality and particle-vortex duality.\label{locald}} 

Upon deriving the equations of motion, the two models satisfy the same pair of equations, $\dd{\cal F}=0$ and $\dd*{\cal F}=0$, written in terms of either $A$ or $B$. In the A-model, these are respectively the Bianchi identity and the equation of motion for $A$. In the B-model, these roles are interchanged: $\dd{\cal F}=0$ is the equation of motion, and $\dd*{\cal F}=0$ is the Bianchi identity for $B$ (and this agrees with the discussion in Section \ref{EM-duality}, especially Figure \ref{Maxfig}). Yet the equations of the two models have the same content. One can check, along the lines of Section \ref{EmdS}, that the quantities of the two models are also the same.\footnote{In showing this, one uses the proper normalization of the quantities as in the action, i.e.~each Faraday tensor is normalized with a factor of $1/e$ ($1/e'$ in the dual model). (The electromagnetic stress-energy tensor in Eq.~\eq{Tmunu} contained the permittivity of the vacuum, here replaced by the coupling constant $e^2$ in the action.) The relation Eq.~\eq{eF} then secures that the quantities of the two models are the same.}

If $e$ is interpreted as the {\it electric charge} of the common core theory, e.g.~as given by an external source in Gauss' law (see Eq.~\eq{elmagn}), then the A-model is {\it electric}, while the B-model is {\it magnetic}. Relative to a source with charge $e'$, the B-model is {\it electric}, i.e.~it couples to that source as to an electric point-particle, and the A-model and the common core are {\it magnetic}, i.e.~they couple to it as to a magnetic monopole. (Recall, from particle-vortex duality, that an electric field couples to the source of a point-particle, while a magnetic field couples to the source of a soliton: in this case, a magnetic monopole.)\\
\\
{\bf Illustrating the Schema: the common core theory and its models.} As we already discussed, the theory $T$ ``above'' is a common core theory for the duals, in the Schema's sense. Namely, the duals are obtained by giving the two-form ${\cal F}$ in terms of a four-vector potential: $A$ or its dual, $d_{\cal S}(A)=B$. The thus defined models are representations of $T$, in the sense of representation theory: the axioms that are satisfied by the common core theory $T$ are also satisfied by either model, on the specific structures defined for ${\cal F}$: namely, as $A$ or as $B$. Thus specified in either model, the field ${\cal F}$ satisfies additional properties, i.e.~axioms, that are not provided by the common core theory, and are thus part of the specific structure: viz.~the two-form ${\cal F}$ is exact in the A-model, and its Hodge dual is exact in the B-model. 

One perhaps prima facie surprising feature of the common core theory $T$ is that it is ``more encompassing'' than the duals, because it has two independent fields rather than one: indeed, intuitively, one way to construct this common core theory from the duals is by constructing an action that includes the fields of both duals, and reproduces their equations of motion. 

That $T$ can have additional fields of its own is to be expected: additional fields can always be introduced as Lagrange multipliers (i.e.~a field that, like $B$ in Eq.~\eq{SFB}, only enters the action linearly, so that its variation enforces the constraint Eq.~\eq{*F}), and so no classical Lagrangian theory has a unique presentation. The important point is that the duals are {\it representations} of $T$, i.e.~homomorphisms that map $T$ to its duals. $T$'s ``job'' is to enable a simple proof of the duality.\footnote{The duality is a {\it symmetry} of $T[{\cal F},B;e]$, i.e.~the theory with action Eq.~\eq{SFBnew}. Substituting the saddle point of the Lagrange multiplier $B$ gives the electric model with Maxwell action Eq.~\eq{FAA}. And $T[{\cal F},A;e']$, with a different name for the gauge field and inverse coupling, by substituting the saddle point of $A$, gives the magnetic model with dual action, Eq.~\eq{dMax}. In $T$ thus formulated, duality is a matter of renaming $A$ and $B$ and taking the coupling to its inverse, rather than solving the non-local equation $\dd A={e\over e'}\,*\dd B$ (see footnote \ref{locald}). Thus $T$ ``localizes'' the duality map. (This also happens in the quantum case: see Section \ref{quantumD}).}

This deserves three further comments: about the common core theory, about its relation to the models, and about a successor theory.

First, from the point of view of a bare theory, it is no surprise that $T$ has new fields relative to its models, to which it is dynamically equivalent. These new fields are Lagrange multipliers that do not introduce new physical degrees of freedom. For, as we discussed in item (3) of `Theory' in Section \ref{Ourthm}, the state space ${\cal S}$ of a bare theory is the set of kinematically possible states: and this is larger than its subset of dynamically possible states, which is specified by the dynamics ${\cal D}$. (Section \ref{abstraction} will discuss that, more generally: (i) a common core theory need not be unique; (ii) to get a common core theory, models are sometimes {\it augmented}.)

Allowing $T$ to be a common core theory, as in the Schema, seems to undermine the view that dualities are like gauge symmetries and that we should always move towards what Section \ref{dualsym} called a {\it reduced formalism} for the common core. Instead, our common core theory $T$ uses an augmented formalism.

Second, although our A- and B-models are duals, $T$ is not isomorphic to them, because its set of kinematically possible states is larger (only the subset of $T$'s dynamically possible states is isomorphic to the corresponding subsets of states of each model.) And this feature is admitted by the Schema. 

Section \ref{ica} will argue that a common core theory that is isomorphic to its models is preferable from a logico-semantic point of view.\footnote{Notice that, as discussed, this is a small difference in this case, because the states that the common core theory has in addition to its models, are non-dynamical.}
But for the purpose of finding a successor theory, i.e.~a theory that is more comprehensive than its duals, an augmented theory is usually preferred, because it can have more states, and can describe more situations. This leads in to our last comment:

Third, a useful way to think about the second point is to understand the duality as originating in the equivalence between the {\it two saddle points} of the common core theory for the two-form ${\cal F}$: one of which yields the A-model, and the other yields the B-model. Because $B$ is a Lagrange multiplier, the two saddle points are one and the same, and the corresponding models are isomorphic. 

But if $B$ were a dynamical field, one would expect these two saddle points to yield inequivalent physics, and one could have a quasi-duality. Section \ref{mvd} will argue that quasi-duals can often be understood in this way: as different saddle points or points of expansion of a given successor theory, to which the quasi-dual models give approximations.

\subsection{A Quantum Duality}\label{quantumD}

The above electric-magnetic duality between classical models carries over to their quantum versions, as we will now show. Furthermore, the quantum case has a number of features that make it worth discussing---thus anticipating electric-magnetic duality in supersymmetric theories and string theory.\footnote{Electric-magnetic duality for Abelian gauge theories was discussed, using lattice regularization, in Cardy and Rabinovici (1982) and Shapere and Wilczek (1989). Electric-magnetic duality of the non-Abelian model was discussed by 't Hooft (1979:~p.~149). For the duality of the Abelian model on a curved manifold, see Witten (1995b).}

We will first generalize the classical models by including a {\bf vacuum angle} $\th$, or $\th$-term, in the Maxwell action:\footnote{To reduce our use of form notation, we write $\dd^4x\,F[A]*F[A]$ rather than its proper version in form notation, i.e.~$F[A]\wedge F[A]$ (which amounts to $\dd^4x\,{\bf E}\cdot{\bf B}$). The star indicates that the first Faraday tensor is contracted with the Hodge dual of the second, i.e.~$F[A]*F[A]:={1\over2}\,\e^{\m\n\l\s}F_{\m\n}[A]\,F_{\l\s}[A]$. For this use (and, admittedly, slight abuse) of notation, see Harvey (1996:~p.~25).}
\bea\label{Swtheta}
S[A;e,\th]=\int\dd^4x\left(-{1\over4e^2}\,F[A]^2-{\th\over32\pi^2}\,F[A]*F[A]\right).
\eea
The new term is a total derivative and therefore it does not modify the classical equations of motion (Nakahara, 2003:~p.~407): however, as we will see, such a term is important upon quantization.\footnote{So far as we are aware, the vacuum angle $\th$ was first introduced into gauge theories by Jackiw and Rebbi (1976:~p.~174) and Callan et al.~(1976:~p.~335), to describe the vacuum, as a topological quantum number associated with a time history of the gauge field. They argued that $0\leq\th\leq2\pi$ parametrizes vacuum ground states of independent and in general physically inequivalent sectors, between which the field can tunnel. This term violates the $CP$ invariance of the Maxwell theory, and was subsequently used by Witten (1979:~p.~283) to calculate the electric charge of a magnetic monopole.}

The new term gives a natural complexification of the coupling constant $e$. For as we will see, electric-magnetic duality takes a simple form in terms of a complex coupling defined as follows:
\bea\label{deftau}
\t={\th\over2\pi}+{4\pi i\over e^2}\,.
\eea
Indeed, the action can be rewritten in terms of this complexified coupling and the complex two-forms, $\mathscr{F}$ and $\bar{\mathscr{F}}$, introduced earlier in Eq.~\eq{calF}. The action can then be written as follows:
\bea\label{SAeth}
S[A;e,\th]=-{1\over32\pi}\int\dd^4x~\mbox{Im}\left(\t{\mathscr{F}}^2\right).
\eea
We will show that, under duality, $\t$ maps as follows:
\bea\label{taumintau}
\t\mapsto \tau'=-{1\over\t}\,,
\eea
and $\mathscr{F}$ transforms into its complex conjugate, as in Eq.~\eq{HcalF}. This will generalise the duality between the A- and the B-models (defined analogously to the previous Section) to the case that the action has a $\th$-term.

The real and imaginary parts of the transformation rule for $\tau$, Eq.~\eq{taumintau}, give the transformation rules for the coupling $e$ and the theta-angle $\th$:
\bea\label{eth}
{1\over e'{}^2}={1\over e^2}\,{1\over{\th^2\over4\pi^2}+{16\pi^2\over e^4}}~~~~\mbox{and}~~~~\th'=-{\th\over{\th^2\over4\pi^2}+{16\pi^2\over e^4}}\,.
\eea
Note that, if $\th=0$, this reproduces our earlier result for the dual coupling, Eq.~\eq{dadb}. The above Maxwell theory will define our quantum A-model (for gauge field $A$ and couplings $e,\th$) and our quantum B-model (for gauge field $B$ and dual couplings $e',\th'$).

The steps taking the common core theory to the two models are analogous to the classical case, but at the level of the path integral. We begin with the following definition of the common core theory ``above'':
\bea\label{ZFB}
Z_T=\int{\cal D}{\cal F}\,{\cal D}B~\exp\left[i\int\dd^4x\left(-{1\over 4e^2}\,{\cal F}^2-{\th\over32\pi^2}\,{\cal F}*{\cal F}-{1\over8\pi}\,F[B]*{\cal F}\right)\right].
\eea
The rationale for the last term will become clear in a moment: it is the quantum analogue of the last term in Eq.~\eq{SFBnew}. It gives a Dirac delta function that in effect sets $\pa_\n*{\cal F}^{\m\n}=0$, which is the Bianchi identity, and implies that there exists a gauge field $A$ such that ${\cal F}$ is its Faraday tensor, i.e.~as in Eq.~\eq{eomA} (see Seiberg and Witten (1994a:~p.~29); Alvarez-Gaum\'e and Hassan (1997:~p.~206)). It is the term of the action that, after exponentiation in the path integral, gives rise to a Dirac delta function that imposes the Bianchi identity as a constraint. 

As in (A) and (B) of the previous Section, to get two quantum models we evaluate this expression in two different ways: namely, we evaluate the path integrals in different orders:

(qA) We first do the path integral over $B$ (in jargon: we `integrate out $B$'). 

(qB) We first do the path integral over ${\cal F}$ instead. 

In both cases, the result is a path integral over a single field. As we will see, the resulting path integrals will be each others' duals, in the sense of exchanging the integration variables $A$ and $B$, and taking the dual couplings.\\
\\
(qA)~~To integrate out $B$, we use the form Eq.~\eq{SFB} of the $B$-${\cal F}$ term in the action. There is only one $B$-dependent term under the $B$-integral sign, and the result of this integral is a product of Dirac delta functions:\footnote{This path integral is analogous to the integral representation of the Dirac delta function of a single real variable $x$, $\d(p)={1\over2\pi}\int_{-\infty}^\infty\dd x~e^{ipx}$. We have one such integral for each spacetime point.}
\bea\label{intB}
\int{\cal D}B~e^{-{i\over4\pi}\int{\sm d}^4x\,B_\m\,\pa_\n*{\cal F}^{\m\n}}~\rightarrow~\prod_{\m,x}8\pi^2\,\d\left(\pa_\n*{\cal F}^{\m\n}\right).
\eea
As indicated by the arrow, this expression can be rendered well-defined by using a lattice regularisation of the spacetime and by renormalisation of the multiplicative factors of $8\pi^2$. This lattice regularisation is here only an intermediate step, because in the next step below we will revert to the continuum limit. All these expressions can be defined on the lattice.\footnote{For introductions, see Aitchison and Hey (2013:~pp.~158-161).}

The Dirac delta function in Eq.~\eq{intB} sets $\pa_\n*{\cal F}^{\m\n}=0$, which is the Bianchi identity, and implies that there exists a gauge field $A$ such that ${\cal F}$ is its Faraday tensor, i.e.~as in Eq.~\eq{eomA}.

Thus an integration measure over the gauge field $A$ can be defined as follows: 
\bea\label{intDA}
\int{\cal D}A:=\int{\cal D}F\,\prod_x\d(\pa_\n*F^{\m\n})\,,
\eea
where the Dirac delta function correctly picks out the configurations of the two-form $F$ for which it is a Faraday tensor for the gauge field, $A$. Substituting Eq.~\eq{intB} into the common core theory Eq.~\eq{ZFB} and using Eq.~\eq{intDA}, we are left with a path integral over $A$, which is our {\it quantum A-model}:\footnote{As in the classical case, the $F[B]*{\cal F}$ term in Eq.~\eq{ZFB} is zero because of footnote \ref{FBF} and \eq{intB}.}
\bea\label{Maxwth}
Z_A=\int{\cal D}A~\exp\left[i\int\dd^4x\left(-{1\over 4e^2}\,F[A]^2-{\th\over32\pi^2}\,F[A]*F[A]\right)\right].
\eea
This is the path integral of the Maxwell model for $A$, with the $\th$-term added. This shows that the common core theory ``above'' simplifies to the quantum A-model ``below'' thus defined: it is a representation of the common core theory obtained by integrating out the field $B$.\\
\\
(qB)~~We integrate out ${\cal F}$ instead of $B$.\footnote{This is possible because ${\cal F}$ does not satisfy the Bianchi identity, and we integrate over all its values.} 
We rewrite the common core theory's partition function (Eq.~\eq{ZFB}) in a way that facilitates this:
\bea
Z_T=\int{\cal D}B\,{\cal D}{\cal F}~\exp\left[i\int\dd^4x\left({\cal F}_{\m\n}\,a^{\m\n}{}_{\l\s}\,{\cal F}^{\l\s} +b_{\m\n}\,{\cal F}^{\m\n}\right)\right],
\eea
where we have switched the order of the integrals, and used $F(B)*{\cal F}=*F(B)\,{\cal F}$. The new tensors $a$ and $b$ are ${\cal F}$-independent, and are defined as follows:
\bea\label{Ab}
a^{\m\n}_{\l\s}&:=&-{1\over8e^2}\left(\d^\m_\l\d^\n_\s-\d^\m_\s\d^\n_\l\right)-{\th\over64\pi^2}\,\e^{\m\n}{}_{\l\s} \nn
b_{\m\n}&:=&-{1\over8\pi}*F_{\m\n}[B]\,.
\eea
The tensor $a$ is antisymmetric in its upper and in its lower indices, as is $b$ with respect to its indices. 

To derive our final result, we use the following (regularised) path integral:\footnote{This path integral is analogous to the following integral over a set of $n$ variables $x_i$: $\int_{-\infty}^\infty\left(\prod_{i=1}^n\dd x_i\right)e^{ix\,\cdot\, a\,\cdot\, x+i b \,\cdot\, x}=\sqrt{(i\pi)^n\over\det(a)}~e^{-{{i\over4}b\,\cdot\, a^{-1}\,\cdot\, b}}$, where the dots indicate vector multiplication of $n$ variables. This integral can be defined by using a Gaussian regulator.\label{ndeta}}
\bea\label{DFFAF}
\int{\cal D}{\cal F}~e^{i\int{\sm d}^4x\,\left({\cal F}\,\cdot\, a\,\cdot\,{\cal F} +b\,\cdot\,{\cal F}\right)}=C~e^{{i\over256\pi^2}\int{\sm d}^4x~*F[B]\,\cdot\, a^{-1}\,\cdot\, *\,F[B]}\,,
\eea
where we used the definition Eq.~\eq{Ab} of $b$, and $C$ is a field-independent number. Note that, unlike the A-model, this multiplicative number is coupling-dependent. Thus by renormalizing this expression we lose sight of the overall coupling-dependent numerical factor multiplying the path integral (we will discuss the significance of this below).\footnote{We will absorb the coupling-dependent but field-independent numerical factor multiplying the exponential in Eq.~\eq{DFFAF} into the renormalisation of the path integral. In appoaches that attempt to fix the overall coupling-dependence of the partition function so that it transforms as a modular form, an overall factor of the coupling constant is included in the partition function so as to make it independent of the (lattice) regularization used. In effect, these approaches multiply the partition function by a factor of $\sqrt{4\pi}/e=({\sm{Im}}\t)^{1/2}$ for each of the $n$ integration components of ${\cal F}$, thus rescaling ${\cal F}$ as in in Eq.~\eq{eF}. This then cancels out the factor of $\sqrt{(i\pi)^n/\det(a)}$ that results from the regularised path integral (see footnote \ref{ndeta}). For a detailed discussion of this coupling-dependence on a four-dimensional Euclidean manifold, see Witten (1995b:~pp.~390-391) and Verlinde (1995:~pp.~213-214).\label{renormaliseZ}} 

Using this expression gives our final result for the partition function of the quantum B-model:\footnote{We use the following expression for the inverse of the matrix $a$: $(a^{-1})^{\m\n}{}_{\l\s}={1\over{\th^2\over256\pi^4}+{1\over4e^4}}\left(-{1\over2e^2}\left(\d^\m_\l\d^\n_\s-\d^\m_\s\d^\n_\l\right)+{\th\over16\pi^2}\,\e^{\m\n}{}_{\l\s}\right)$.}
\bea\label{Maxwth2}
Z_B=\int{\cal D}B~\exp\left[i\int\dd^4x\left(-{1\over 4e'{}^2}\,F[B]^2-{\th'\over32\pi^2}\,F[B]*F[B]\right)\right],
\eea
where $e'$ and $\th'$ are the dual couplings, defined exactly as in Eq.~\eq{eth}. 

Comparing this with Eq.~\eq{Maxwth}, we find that, up to an overall coupling-dependent factor, the partition function of the quantum B-model takes the same form as that of the quantum A-model, except that it depends on the {\it dual coupling constants}. For in the quantum A-model, we integrate over $A$ (cf.~Eq.~\eq{Maxwth}), while in the quantum B-model we integrate over $B$ (cf.~Eq.~\eq{Maxwth2}), and these are merely names for dummy integration variables. Thus starting in the B-model: by renaming the integration variable of the path integral, we recover, up to an overall factor, the partition function of the A-model, with the dual coupling constants. (Cf.~Eq.~\eq{Zmod}, where the partition function and its dual matched up to an overall coupling-dependent factor.)

Since the partition functions of the two models are the same except for an overall factor and the dual couplings, this partition function must be an invariant function of $\tau$ under the duality map, Eq.~\eq{taumintau}. Such functions have been calculated by Witten (1995b:~p.~387-388), Vafa and Witten (1994:~pp.~46-50), and Verlinde (1995:~pp.~216-218), and they are otherwise well-known in conformal field theory: for simple examples, see Di Francesco et al.~(1997:~pp.~340-344).\footnote{More precisely, under appropriate renormalisation of the partition function (see footnote \ref{renormaliseZ}), on a Euclidean four-manifold, the partition function is a modular function with weights that are given by the Euler characteristic and signature of the four-manifold: see Witten (1995b:~p.~385). For a discussion of the duality properties of lattice gauge theories, see Cardy (1982), Cardy and Rabinovici (1982) and Shapere and Wilczek (1989).}

\section{Conclusion}

Solitons make two major contributions to our understanding of the physics of dualities. First, the examples of sine-Gordon-Thirring model duality and particle-vortex duality are characterised by their exchanging {\it topological and Noether currents}, and their associated charges. (This is surprising, and illustrates our theme of {\it hard-easy} well beyond the exchange of weak and strong coupling regimes. It is reminiscent of wave-particle duality, in Section \ref{wpd}.) The dualities exchange {\it topologically trivial} and {\it topologically non-trivial} configurations of the fields: trivial configurations are possibly localised, while non-trivial configurations are non-local and depend on the boundary conditions at infinity.

Second, these solitons, when used in statistical physics, describe {\it ordered and disordered phases} of a system. In two dimensions, the transition from a disordered phase at high temperature to a topologically ordered phase at low temperature can be understood through the condensation of vortices into pairs. Furthermore, duality can be used to study vortices in the disordered phase by dualizing them into point particles that couple to a dual electromagnetic field. (This is reminiscent of Kramers-Wannier duality in Section \ref{IMD}, where a non-local duality map allowed us to treat dislocations as spins on a dual lattice.) 

The import of our examples for the ontology of quantum field theories is that they prioritise the study of soliton solutions, and their non-trivial vacua, as a prerequisite for discussions of the ontology of quantum field theories. (We will develop this theme more fully in upcoming Chapters). Quantum field theories, even if they have a single ground state energy, in general have multiple local minima: and studying the states of the theory requires the study of these other vacua through the use of solitons. For example, vacua where solitons are condensed into pairs are not accessible to usual AQFT and perturbative methods: they require a non-linear and non-perturbative approach.\footnote{For a development of this point of view, see Vergouwen (2022:~Ch.~3) and Vergouwen and De Haro (2024).}

Philosophical studies of quantum field theory have often focussed on its structural aspects: Hilbert spaces and their algebras, symmetries, renormalization, etc. Solitons and their dualities show that this focus needs to be expanded by a study of exact solutions of the non-linear equations of motion. By exhibiting the non-perturbative information that the study of solitons can give, Chapter \ref{EMYM} will give some pay-offs of this expanded focus. (This conclusion may not surprise philosophers of spacetime, who have long studied solutions of the Einstein field equations: see Chapter \ref{HABHM}.)

Solitons also bear on our Schema. First, through the procedure of {\it augmentation} that we discussed for bosonization, where two partially isomorphic models are rendered dual through augmentation by either states or quantities. (As we will see in Chapter \ref{String}, this played a role in the discovery of D-branes in string theory, when in 1995 Witten realized that non-perturbative solitonic states predicted by string dualities should be added to the spectrum of string theory.)

Second, we saw several examples of a {\it common core theory} behind two duals, especially for electric-magnetic duality: the common core is not sparing in its number of variables, since indeed it uses more variables that the models do. The common core theory can thus be more comprehensive than the models, which are representations of the common core theory with fewer kinematically possible states but the same amount of dynamically possible states as the common core theory. This agrees with our Schema. As we saw, the role of the common core is to {\it localize the derivation of the duality map}: while a direct duality map between the gauge fields of the classical A- and B-models is non-local, the representation maps follow from variables in the common core theory ``above'' by standard methods. Thus ``going one level up'' gives a simple way to study the duality.

\chapter{Quark Confinement as Dual Superconductivity}\label{EMDuality}
\markboth{\small{\textup{Quark Confinement as Dual Superconductivity}}}{\textup{
\small{Quark Confinement as Dual Superconductivity}}}

This is the first of two Chapters on {\it electric-magnetic duality}: it expounds the {\it heuristic power} of duality, as witnessed by developments in condensed matter theory and particle physics in the late 1970s and early 80s. (Chapter \ref{EMYM} will give mathematically more advanced illustrations of electric-magnetic duality in quantum field theory, from the 1990s onwards.)

Previously, Chapters \ref{Simple} and \ref{Advan} introduced examples of {\it particle-soliton duality}. If the soliton is charged, this is usually associated with {\it electric-magnetic duality}. From the statistical physics of solitons, especially vortices, we found examples of phase transitions in low dimensions. 

This Chapter brings together these influential ideas around the physics of symmetry breaking and colour confinement. Its main aim is to demonstrate how the idea of duality enables the formulation of one of the main candidate explanations of colour confinement in particle physics: namely, the condensation of magnetic monopoles that screen colour electric charge, which is confined to the interior of electric colour tubes. In short, the proposed mechanism, also known as the {\it 't Hooft-Mandelstam mechanism of confinement}, is {\it dual superconductivity}.\footnote{We here use `mechanism' in the physicists' sense, which, in contrast to its usual diachronic-causal sense, admits broader synchronic, mereological, and modal aspects. The screening of electric colour charge will be explained as screening by the non-zero vacuum expectation value of a magnetically charged field (as in the Meissner effect), realized microscopically by the condensation of magnetic monopoles into pairs (like Cooper pairs in a superconductor). Comparing this state to a state with either no monopoles, or free monopoles, does not require that we regard charge confinement as a diachronic process. The synchronic and modal aspects of course do not exclude the realization of confinement as a real-time process in the {\it deconfinement phase transition}, that is believed to take place in QCD, as the temperature is lowered below the critical temperature: from a quark-gluon plasma to a confining phase. For a discussion, see Shuryak (2021:~pp.~68).\label{mechanism}} 

Besides duality, two further ideas catalysed this proposal: 

(i) The discovery of the 't Hooft-Polyakov magnetic monopole solution: a soliton solution of the Yang-Mills-Higgs equations that, from a distance, looks like a Dirac monopole, but has a non-singular core. (Electric-magnetic duality of course also plays a major role in {\it this} idea.) 

(ii) The discovery, as part of a series of two-way analogies between condensed matter physics and quantum field theory, of the Higgs mechanism. 

The detail of the 't Hooft-Polyakov magnetic monopole solution, i.e.~(i), will be discussed in Chapter \ref{EMYM}. This Chapter first expounds the Higgs mechanism, i.e.~(ii), and its associated physics: superconductivity, phase transitions, and spontaneous symmetry breaking. It then goes on to give the magnetic quasi-dual of the Higgs mechanism as a proposed explanation of quark confinement.

{\bf Confinement} is an {\it outstanding open problem} in quantum chromodynamics (QCD): indeed, it is a Millennium Prize Problem to prove that (quantum) Yang-Mills theory, even without introducing quarks, has a {\it mass gap}---a problem that, as we will see, is closely related to confinement. Confinement is the phenomenon that quarks, and other particles with colour charge, are never detected in isolation: they always appear in colourless combinations (i.e.~colour singlets of the local colour gauge group, SU(3)).\footnote{We are here describing `colour confinement'. The word `confinement' is sometimes applied more restrictively, to the confinement of quarks: it is then called `quark confinement'. For a discussion of different kinds of confinement, see Greensite (2020:~p.~21).\label{quarkC}} Thus quarks are confined in hadrons (there are three quarks in baryons such as the proton and the neutron, and a quark and an anti-quark in mesons such as the pion). Likewise, gluons, which transmit the strong nuclear force, are not detected in isolation, but can form colourless bound states, or `glueballs'.\footnote{For recent reviews of the experimental evidence, see Abazov et a.~(2021), Ochs (2013:~p.~14) and Crede and Meyer (2009:~p.~106).}

Several explanations of confinement have been proposed.\footnote{As Bali (2000:~p.~18) notes, these various explanations do not exclude each other. For a discussion of some of these mechanisms, see Greensite (2020).}
 't Hooft (1975) and Mandelstam (1976) suggested that the condensation of magnetic monopoles in the vacuum plays a key role.\footnote{Polyakov (1977) subsequently showed that monopole condensation is indeed present as the mechanism of confinement in a version of three-dimensional quantum electrodynamics (QED), i.e.~in two dimensions of space and one of time: namely, a variant of the Georgi-Glashow electroweak model. Subsequent work by Banks, Myerson and Kogut (1977), Osterwalder and Seiler (1978), and Fradkin and Shenker (1979), showed that in this three-dimensional model the quarks are {\it always confined.} Monte Carlo simulations by DeGrand and Toussaint (1980) explicity identified the behaviour of the monopoles that contributes to the confinement (viz.~their screening of an external magnetic field, as in a ``magnetic superconductor'') and, for four dimensions, they found evidence of the transition from the confining phase to the Coulomb phase, due to the ``unbinding'' of closed loops of monopole strings. For a review of recent lattice simulation results, see Bali (2000). Shuryak (2021) and Greensite (2020) are good overviews of recent work.} 
Their argument can be seen as based on the analogy with the condensation of electrons in Cooper pairs in a {\it superconductor,} which leads to the celebrated Meissner-Ochsenfeld effect (1933) and the screening of the magnetic charge. Thus quark confinement could be analogous to a {\it magnetic superconductor} that screens colour electric charge (cf.~footnote \ref{mechanism}). Our aim in this Chapter is to understand the logical and heuristic role of {\it electric-magnetic duality} in this proposal.\\

Recent philosophical work, in large part in response to Earman (2003, 2004a, 2004b), who set the agenda at an early stage, has addressed several aspects of the Higgs mechanism and spontaneous symmetry breaking.\footnote{Smeenk (2006) is a reply to Earman's worries about the gauge-dependence of the Higgs mechanism and the consequences of Elitzur's theorem. The discussion was re-opened by Lyre (2008) and St\"oltzner (2012), with rejoinders by Struyve (2011), Friederich (2013), Karaca (2013), and Rivat (2014). A recent overview is Berghofer et al.~(2023).\label{SBphil}} 
There is also a broader philosophical discussion of the empirical significance of global and local gauge symmetries.\footnote{See Greaves and Wallace (2014), Brading and Brown (2004) and Gomes (2021). For a recent summary of this discussion, see Berghofer et al.~(2023). For recent work on the significance of Noether's two theorems, see Read and Teh (2022). See also footnote \ref{empsym} in Section \ref{ncsr}.}

By contrast, the 't Hooft-Mandelstam mechanism of confinement (and indeed, quark confinement in general), like the physics discussed in the previous Chapter, is virtually uncharted territory for philosophers.\footnote{In the philosophical literature, we are aware only of a passing mention by Polchinski (2017:~p.~11). Hartmann (1999:~pp.~330-340, 2001:~pp.~289-292) discusses some aspects of quark confinement, especially in the light of the MIT bag-model of confinement, which predates the developments we will discuss in this Chapter. See also Healey (2007:~p.~94).} 

Agreed: quark confinement is a hard problem not yet solved in physics. But, in addition to our argument in this book, that philosophy ought to not narrow its scope to ``solved'' problems or well-understood physics,\footnote{See Huggett and W\"uthrich (2013:~p.~284), and our defence of the `engagement with still ongoing physics' in Section \ref{friendf}.} 
this situation is unsatisfactory, not just because of the topic of quark confinement, but also because of the wide influence of the 't Hooft-Mandelstam argument on contemporary quantum field theory. 

Even physics texts do not always make this wide influence explicit.\footnote{Two commendable recent textbooks that discuss this mechanism in detail are Greensite (2020) and Shuryak (2021).} 
For, although many discuss magnetic monopoles, few canvass the logical, historical, and heuristic structure of the 't Hooft-Mandelstam argument and the role of duality, which---along with monopoles and other solitons---has contributed to the study of exact solutions, phases, and non-perturbative properties, of quantum field theories: not to mention further developments in string theory. 

\begin{figure}
\begin{center}
\includegraphics[height=5cm]{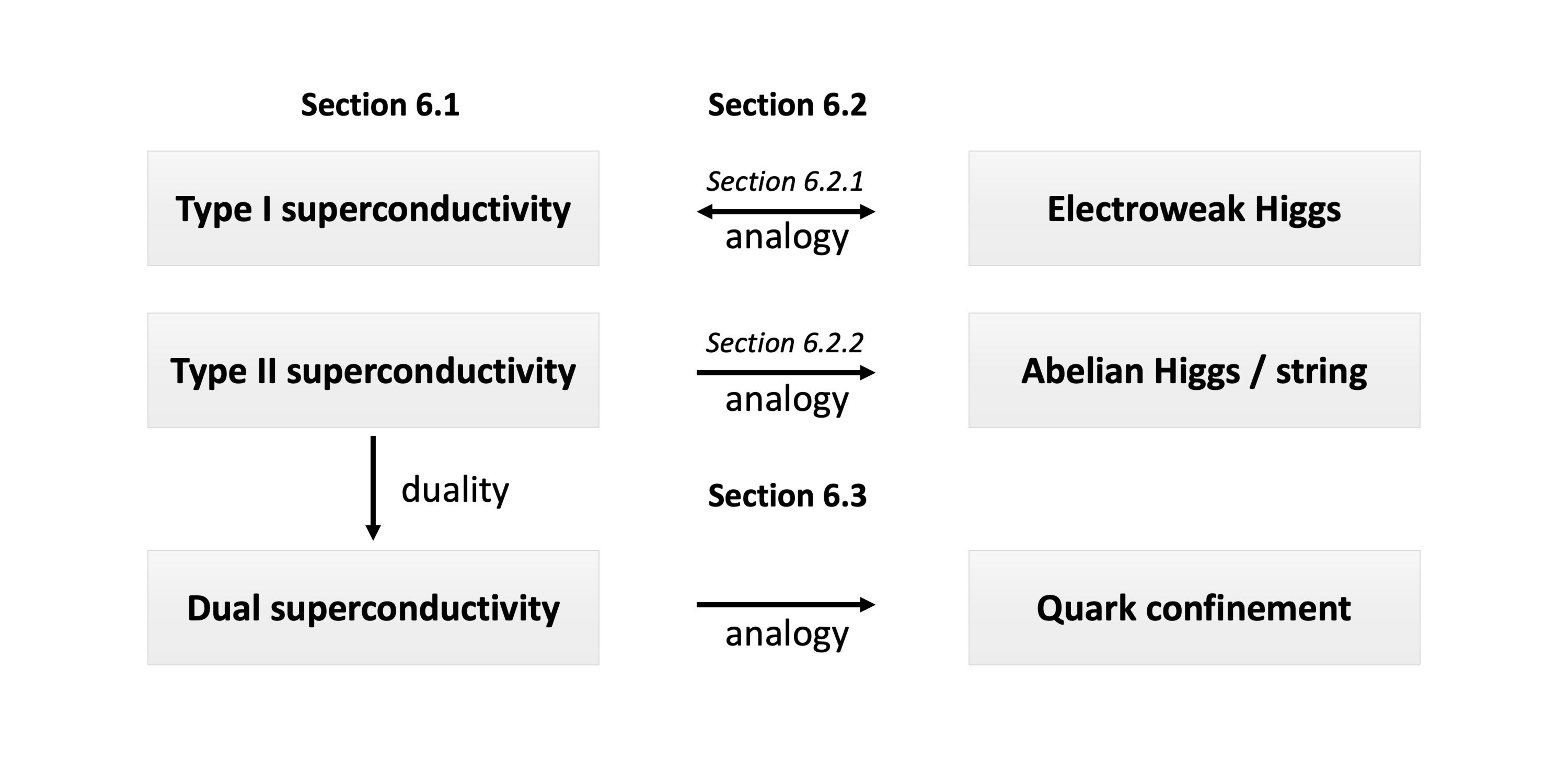}
\caption{\small Logical and heuristic structure of 't Hooft and Mandelstam's proposed mechanism of {\it quark confinement as dual conductivity}.}
\label{dualSuperC}
\end{center}
\end{figure}

The objective of this Chapter is to start charting this territory: Figure \ref{dualSuperC} summarises the logical and heuristic structure of the argument, and reflects the structure of this Chapter. The argument is a three-step analogy between superconductivity and particle physics, in three consecutive steps:

(1)~~The Higgs mechanism as analogous to superconductivity. 

(2)~~Vortex and string solutions as relativistic analogues of vortices in the {\it mixed phase} of superconductivity (i.e.~`type II' superconductivity).

(3)~~Quark confinement as the analogue of {\it dual} superconductivity (i.e.~a magnetic dual of superconductivity).

Section \ref{topphases} introduces topological phase transitions and superconductivity (in its two types, type I and type II), i.e.~the two squares on the top-left of Figure \ref{dualSuperC}. The two analogies between superconductivity and the Higgs mechanism are developed in Section \ref{analogies} (i.e.~the top two rows in the Figure). Section \ref{cmds} motivates 't Hooft and Mandelstam's proposal, that quark confinement is a relativistic analogue of dual superconductivity (i.e.~the third row), and discusses the phases of gauge theories more generally. 

\section{On the topological phases of matter}\label{topphases}

Previous Chapters studied soliton solutions and their duals, thereby focussing primarily on {\it single solitons}; (Sections \ref{IMD} and \ref{PVD} also considered states of many solitons). This Section expounds the {\it statistical physics} of solitons, i.e.~it considers {\it ensembles} of them: Section \ref{ssp} first gathers the threads from previous Chapters about soliton-induced phase transitions. Section \ref{LBCS} introduces basic aspects of superconductivity. Section \ref{GLtheory} then gives the Ginzburg-Landau phenomenological theory of phase transitions, on which later Sections will build. 

\subsection{Solitons and duality in statistical physics}\label{ssp}

This Section characterises phase transitions that are driven by the {\it condensation of pairs of solitons} at a critical value of some parameter, usually the temperature. The archetypal such transitions are the ferromagnetic and Berezinskii-Kosterlitz-Thouless transitions from disordered high temperature to low-temperature phases with topological order (see Chapters \ref{Simple} and \ref{Advan}).\footnote{Not all phase transitions are of course explained by the condensation of solitons, and not all that do, instantiate all the properties that we will mention in this Section. However, the properties that we here list will recur in later Sections.}

In Section \ref{PVD}, a phase transition between a disordered phase at high temperatures and a phase with a topological order below a critical temperature was described in terms of the condensation of vortices and anti-vortices into metastable {\it pairs} that minimize the free energy. The transition is characterised by an {\it order parameter} that has a zero value in the disordered phase, and non-zero in the ordered phase. In the case of the ferromagnetic phase transition, this parameter is the average magnetisation. That the phase is ordered can usually also be understood as the divergence of a correlation length near the critical temperature, so that there are long-range correlations in the material. 

The Ginzburg-Landau model of continuous phase transitions near the critical point gives a powerful {\it phenomenological} description of phase transitions that captures the main features of superconductivity, without requiring the microscopic detail (see Section \ref{GLtheory}).\footnote{The Ginzburg-Landau model can also model first-order phase transitions, e.g.~in type I superconductors (see below), although it was not originally presented in this way.} 
The order parameter appears as a field in the free energy, with phenomenological coefficients that depend on external parameters such as temperature, pressure, magnetic field, etc. In the simplest cases, the potential function is taken to be quartic in the order parameter, and depending on the coefficients it may have a single minimum or several (see Figure \ref{Phi4potential} below). In a superconductor, as the temperature is lowered below the critical temperature, the phase transition is from the normal state, where the potential has a single minimum and the order parameter has zero value, i.e.~$\psi=0$, to the superconducting phase, where it is non-zero and the energy has a lower minimum than the normal phase. In the microscopic BCS (for `Bardeen, Cooper, and Schrieffer', 1957) theory of superconductivity, the value of the order parameter $\psi=\bra\psi_\uparrow\psi_\downarrow\ket$ is interpreted as the expectation value of a condensate of electrons in a Cooper pair, with one electron spin up and the other down. Since this order parameter takes complex values, the ground state below the critical temperature has a U(1) degeneracy for the different values of the complex phase of the order parameter that do not change the energy, and this phase is interpreted as the sum of the phases of the electron fields in the Cooper pairs.

The above can also be understood as the {\it spontaneous breaking of the symmetry} of the electron fields. The normal state, with $\psi=0$, has a global U(1) chiral symmetry. This chiral symmetry is spontaneously broken by the order parameter's taking a {\it non-zero value} at the degenerate minimum.\\
\\
{\bf Spontaneous symmetry breaking:} a widely used definition of spontaneously broken symmetry is that the model has a {\it global symmetry} that is broken by a choice of vacuum state. A more precise and convenient way to define it is as follows: consider a system described by a Lagrangian (usually out of a smooth family of Lagrangians) with a symmetry under a gauge group. If a critical point $\f_0$ of the Lagrangian does not enjoy that symmetry, but only a proper subgroup of it (viz.~the stabilizer of $\f_0$), then we say that the model has the capacity for spontaneous symmetry breaking, and the linearized model about $\f_0$ is the {\it spontaneously broken} model.\footnote{We owe this definition to Nicholas Teh.}

In the examples of ferromagnetism and vortices, {\it particle-soliton duality} maps soliton configurations to particle configurations (also interchanging high- and low-temperatures, and coupling regimes), and so the duality can be used to study the solitonic phase.\footnote{As mentioned in Section \ref{dualpf}, the duality can be used to give a numerical value of the critical temperature. In Section \ref{PVD}, the duality was used to describe vortices as electrically charged particles, i.e.~as particles that couple to a magnetic gauge field $a$ through an {\it electric coupling}, i.e.~through the normal coupling between conserved currents and gauge fields, $a_\m J^\m$.}

Phase transitions towards topologically ordered phases realise these features more generally in interesting ways.\footnote{As one sees from Figure \ref{dualSuperC}, duality does not play a role in the phenomenology of the Higgs mechanism of the electroweak theory, but only in the attempts to explain the strong nuclear force using dual superconductivity (see Section \ref{cmds}).} 
As we will see in Section \ref{cmds}, electric-magnetic duality is a key guiding principle in attempts to understand colour confinement. As a preparation, the next Section first recalls the features of type I conductivity (i.e.~the top-left square of Figure \ref{dualSuperC}).

\subsection{London and BCS models of type I superconductivity}\label{LBCS}

Three years after he first liquefied helium in 1908 in Leiden, Kamerlingh-Onnes found that, if a metal such as mercury (later also tin and lead) was immersed in liquid helium and cooled to about 4K, the {\it electrical resistance} fell abruptly to zero.\footnote{For some of the history of the first experiments and further references, see Tinkham (1996:~pp.~1-6), Saint-James et al.~(1969:~p.~3).} 

Until 1933, it was not clearly understood that a superconductor differs from a perfect conductor in how it reacts to an external magnetic field. The phenomenological theories that we use today were worked out in the wake of the Meissner-Ochsenfeld (1933) effect (`Meissner effect', for short): when a superconducting material is placed under an external magnetic field and cooled down to reach its superconducting state, the magnetic field is {\it expelled} from the material: the magnetic field is zero in the interior, and the superconductor is a {\it perfect diamagnet}. This is unlike the perfect conductor, which locks in the magnetic field if the latter is applied before the material is cooled down.\footnote{If the magnetic field is applied {\it after} a perfect conductor is cooled down, then the perfect conductor also repels the magnetic field. For a clear account of this difference, see London (1950:~pp.~13-14, 27). See also Saint-James et al.~(1969:~p.~4) and Tinkham (1996:~pp.~2-3).}

When an external magnetic field, $H_0$, is applied, the superconductor forms a `supercurrent' on its surface, which creates a magnetic field in the interior of the material that cancels the external magnetic field. Thus, in effect, the external magnetic field is ``expelled'' from the material. In the {\bf London model}, this expulsion is described by the {\it London equation,} which relates the supercurrent to the (curl of) the external magnetic field, and replaces Ohm's law for conductors.\footnote{The brothers F.~and H.~London (1935) proposed two equations for the electric and magnetic fields of a superconductor. In the main text, we are concerned with the equation for the magnetic field. It is worth noting that London and London (1935:~pp.~86-87) already wrote down the equations for the currents and charge density that follow from a relativistic generalisation of the Schr\"odinger equation, and which provide the heuristics for the Ginzburg-Landau model. London and London rederived some of their main results from these equations.} 
Combining the London equation with Amp\`ere's law, one finds that, for a longitudinal applied external magnetic field (i.e.~with no components normal to the surface of the material), the magnetic field $H$ in the interior of the superconductor decays exponentially with the distance to the surface:\footnote{Recall, from elementary electrodynamics, that the magnetic field in a material is defined by: ${\bf H}:={\bf B}-4\pi{\bf M}$, where ${\bf M}$ is the magnetisation of the material (in the Gaussian-cgs units used in this Section, which are standard in the literature on superconductivity: for the conversion to SI units, see Tinkham (1996:~p.~434), and also Griffiths (1999:~pp.~558-561) and Jackson (1962:~pp.~611-621)). As we will see below, the magnetic field ${\bf H}$ is here associated with the superconducting current ${\bf J}$, which contains the contributions of both the charge carriers in the superconductor (the Cooper pairs, as we will see) {\it and} the vector potential, ${\bf A}$. The literature often uses ${\bf h}({\bf r})$, rather than ${\bf H}$ or ${\bf B}$, for the magnetic field inside the material, reserving the upper-case letter for the macroscopic field, i.e.~the average of the local field ${\bf h}({\bf r})$ over (a region of) the material (see e.g.~Tinkham (1996:~pp.~4, 435) and London (1950:~p.~30)). Since, in this Chapter, we never explicitly require the average, we use the upper-case ${\bf H}$.}
\bea\label{Londoneq}
H(z)=H_0~e^{-z/\l_{\tn L}},~~~~\l_{\tn L}:=\sqrt{mc^2\over4\pi n_se^2}\,.
\eea
The magnitude of the applied external magnetic field, $H_0$, is the value at the surface of the material, and $z$ is the direction normal to the surface, $z=0$, i.e.~the surface is the $x-y$ plane. $\l_{\tn L}$ is the London length, also called the {\it penetration depth,} and $n_s$ is the density of superconducting electrons, i.e.~the number of superconducting electrons per unit volume, which depends on state parameters such as the temperature. Thus London's straightforward application of the Maxwell equations gives a solution where the magnetic field in the interior quickly approaches zero at distances larger than the penetration depth. 

The London (1948:~p.~564) equation predicts that superconductors are not only {\it ideal conductors} (i.e.~they can sustain a non-zero steady macroscopic current even if the electric field in the material is zero), but also {\it ideal diamagnets}, i.e.~under an external magnetic field the magnetisation is non-zero, such that the total macroscopic\footnote{`Macroscopic' here means `averaged over a region', since the local magnetic field need not be zero, especially at the boundary of the superconductor.} magnetic field in the interior of the material is zero.\footnote{London's energy conservation equation also predicts that, in the interior of the superconductor, the power density (i.e.~the total energy density in the interior of the supercondcutor per unit time) is a total divergence, so that in the interior there is {\it no energy dissipation through heat}: the production of heat is localised on the surface of the superconductor. See London and London (1935:~p.~78).\label{Ldiss}}

If the temperature increases, the density of superconducting electrons $n_s$ decreases, so that the penetration depth, $\l_{\tn L}(T)$, increases---signalling that the magnetic field can penetrate deeper into the material, thereby losing its superconductivity. At the critical temperature, superconductivity is lost, the material is in its normal state, the density of the superconducting electrons is zero, the penetration length diverges, and the magnetic field is not screened.

A second key parameter for superconductors is the {\bf coherence length,} $\xi$, which was first introduced by Pippard (1953:~p.~553): it is the {\it characteristic quantum length} of the electrons within an energy of order $kT_{\sm c}$ of the Fermi energy, where $T_{\sm c}$ is the critical temperature ($\xi$ is proportional to $1/\sqrt{T_{\sm c}}$: we will derive a formula for it, in the Ginzburg-Landau model, in Eq.~\eq{corrL} below).\footnote{The coherence length was originally introduced by Pippard (1953:~pp.~552-553) as an additional length scale characterising, roughly, the size of the region of space over which the electromagnetic potential contributes to the current. This was introduced by analogy with the ``mean free path'' in the metal, which, roughly, characterises the dimension of the region of space over which the electric field is averaged to find its contribution to the current (and thus generalising Ohm's law).} For type I superconductors, such as tin and aluminium, the coherence length is much larger than the penetration depth, i.e.~$\xi\gg\l_{\tn L}(0)$. We will also be interested in `type II' superconductors, where $\xi\ll\l_{\tn L}(0)$.\\

The quantum nature of superconductivity is best explained by the {\bf BCS model of superconductivity}.\footnote{See Bardeen et al.~(1957:~p.~1175). However, we will not require its mathematical details.} 
At low temperatures, the interactions between electrons and phonons\footnote{Phonons are collective excitations of the periodically arranged atoms.} 
cause electrons with opposite spins to pair up into {\bf Cooper pairs}\footnote{Thus called after Cooper (1956:~p.~1189).} 
(here, the electrons are treated as forming a gas of electrons in their ground state at the Fermi energy). This attractive force between pairs of electrons with opposite spin, and whose energies are close to each other, results from the exchange of phonons, and dominates their (screened) Coulomb repulsions.\footnote{This happens when the energy difference between the electron states involved is less than the phonon energy: see Bardeen et al.~(1957:~pp.~1175-1176).} 
Cooper pairs are bosons that have twice the electron charge, and below $T_{\sm c}$ they condense, i.e.~there are enough of them to form a macroscopic superconducting quantum state of the material, where charged particles can travel without resistance.\footnote{Therefore, this state is also called a {\it superfluid} or Bose-Einstein condensate of the Cooper pairs.}
The Cooper pairs are the superconducting charge carriers that had been anticipated by the London model of superconductivity.\footnote{First attempts towards a ``molecular theory of superconductivity'' are in London and London (1948:~p.~572), and in more detail in London (1950:~pp.~147-153).} 
In the BCS model, the characteristic or coherence length, $\xi$, expresses the uncertainty in the electron's wave-function, which is given by the typical size of the Cooper pair, i.e.~the distance between the two electrons. 

A central feature of the BCS model is that the Cooper pairs are relatively stable, with a minimum energy of order $kT_{\sm c}$ required to break them. Thus there is an energy gap of order $kT_{\sm c}$ between: (i) the ground state, in which the electrons in an appropriate energy range\footnote{These are the electrons within an energy shell of order $kT_{\sm c}$ around the Fermi surface.} condense, and (ii) the first excited state, in which a Cooper pair has broken up into separate electrons. 

Thus the BCS state is a `cooperative many-particle state' (Bardeen et al., 1957:~p.~1176), whose energy is lower than the energy of the `normal state', i.e.~the state in which the electrons are not bound in pairs (recall the discussion of the quartic potential in the previous Section). 

This {\it energy gap} is key to explaining the phenomenology of superconductivity (as it is also, and we will discuss later, in explaining confinement!). Notably, it is of crucial importance in explaining both the lack of energy dissipation (cf.~footnote \ref{Ldiss}) and the drop of the electrical resistance when the magnetic flux is expelled from the superconductor. 

\subsection{The Ginzburg-Landau model of superconductivity}\label{GLtheory}

The London and BCS theories of superconductivity just discussed are at opposite ends of the spectrum: the London theory is a macroscopic electromagnetic theory, while the BCS model is a microscopic field theory. 

As we anticipated in Section \ref{ssp}, the Ginzburg-Landau (1950) theory is a powerful intermediate, coarse-grained, approach that captures the main features of superconductivity, without requiring the detail of the BCS model. It is a {\it phenomenological theory} based on the general theory of {\bf continuous phase transitions}, i.e.~where, across the transition parameter, the value of the first derivative of an appropriate thermodynamical potential is continuous, while the second- or higher-order derivatives are discontinuous.\footnote{The terminology of `continuous phase transition' was introduced by Fisher (1967:~p.~617), who contrasted it with Ehrenfest's 1933 terminology of `phase transitions of the second kind', which required the {\it discontinuity of the second derivative}. The theory of continuous phase transitions is expounded in Landau and Lifshitz (1980:~pp.~451-455). It was announced in Landau (1936:~pp.~841-842) and developed in (1937:~pp.~195-203). For an overview of various approaches to phase transitions, including Ehrenfest's classification and later applications and extensions, see Jaeger (1998). The first edition of Landau and Lifschitz, published in 1938, already contains the relevant treatment of a phase transition of the second kind, also called a {\it Curie point} transition (see pp.~205-208).} 
By an `appropriate thermodynamical potential', we mean the thermodynamic functions of the state of the system, such as entropy and free energy. These are continuous through the transition, so that (unlike in phase transitions of the first kind) no heat is absorbed or transmitted.\footnote{For a proof, see Landau and Lifshitz (1980:~p.~454). For the singularities of the specific heat, see idem (pp.~483-493).} 

As we discussed in Section \ref{ssp}, Ginzburg and Landau characterised a continuous phase transition by an order parameter $\psi$ that takes a non-zero value in the {\it asymmetric phase} (usually at low temperatures) and a zero value in the {\it symmetric phase} (usually at high temperatures).\footnote{Landau's (1937:~p.~196) theory of phase transitions was initially formulated wholly in terms of symmetries: specifically, in group-theoretic terms. The idea is that the free energy is invariant under the symmetries of the system. This restricts the form of the free energy and the kind of phase transition.}
While the values of external parameters like the pressure and the temperature can be specified arbitrarily, the value of the order parameter is determined by the condition of thermal equilibrium, i.e.~by minimizing the free energy.\footnote{The equilibrium condition can of course be formulated in terms of either the Gibbs free energy, where the pressure and temperature are held fixed at the minimum, or the Helmholtz free energy, where the temperature and the volume are held fixed.} 

The assumption of continuity states that, near the critical point $T_{\sm c}$, the order parameter varies smoothly and takes small values, and is a continuous function of space, $\psi({\bf r})$. Thus the free energy can be expanded as a Taylor series in powers of $\psi({\bf r})$, including spatial derivatives. The coefficients of this series in effect determine the existence of solutions where the order parameter is different from zero, which indicates the breaking of the symmetry.\footnote{For a continuous phase transition, one usually requires that only even powers can appear in the Taylor series. As we will discuss, the sign of the coefficient of the quadratic term determines the continuous phase transition.} 

For the superconductor, the modulus squared of the order parameter represents the density of superconducting electrons, $n_s$, which appears in the penetration depth, Eq.~\eq{Londoneq}.\footnote{An important improvement of the Ginzburg-Landau model over the London theory is that this density, $|\psi({\bf r})|^2$, is now a {\it local} density. Ginzburg and Landau (1950:~p.~548) took $\psi$ to represent `some ``effective'' wave-function of the ``superconducting electrons''{}'. As the authors recognise, $\psi$ satisfies a non-linear equation rather than the Schr\"odinger equation, and so this is only a heuristic analogous reasoning that helps theory construction. As we reported above, there was a precedent, in London and London (1935:~p.~86) (which Ginzburg and Landau do not cite, although they do cite other work by F.~London), for considering $\psi$ as satisfying an equation that looks formally like the Schr\"odinger equation. We will see below that the BCS model casts light on the appropriate interpretation of $\psi$.} 
Since the superconducting electrons are coupled to an external magnetic field, the complex-valued field $\psi({\bf r})$ is coupled to the vector potential in the usual gauge-invariant way, i.e.~replacing the operator $-i\hbar\nabla$ by $-i\hbar\nabla-q{\bf A}$, where $q$ is the total electric charge of a Cooper pair, $q=2e$.\footnote{This value is fixed by the BCS model, where the order parameter is interpreted as characterising the local condensation of Cooper {\it pairs} of electrons.} 
Thus, summing kinetic and potential energies, the free energy is (assuming $\b>0$ for positivity):\footnote{This expression is defined up to an additive, temperature-dependent, function that is independent of $\psi$ and ${\bf A}$. We suppress it, because it does not affect the variation of the free energy, which we are here interested in. Also, odd powers of $\psi$ are excluded because the free energy must be real, and odd powers of $|\psi|$ are not analytic at $\psi=0$.
Keeping terms up to quartic in $\psi$ is a good approximation so long as the temperature is close to the critical temperature, where $|\psi|$ is small. However, the non-linear behaviour coming from the quartic term is essential for the correct description of superconductivity, and is a second major improvement over the London theory. In view of this, it is required that $\b(T)$ is positive, because otherwise no stable solutions exist, and the approximation made by keeping only the first few terms of the Taylor series would be inadequate.}
\bea\label{GLfreeE}
F={\hbar^2\over2m}\left|\left(\nabla-{i q\over\hbar c}\,{\bf A}\right)\psi\right|^2+\a(T)\,|\psi|^2+{\b(T)\over2}\,|\psi|^4\,.
\eea
The equilibrium configuration is obtained by varying the total free energy with respect to both $\psi$ and ${\bf A}$. (Below we will focus on the equation for the order parameter, $\psi$, thus in effect considering the vector potential as given.)\footnote{Under the analogy with the time-independent Sch\"odinger equation mentioned before, the first term of this equation corresponds to the kinetic energy operator (i.e.~it is the gauge-invariant momentum operator squared), and the other terms correspond to potential energy terms. More precisely, Ginzburg and Landau interpret this equation as an eigenvalue equation for the eigenfunction $\psi$. For example, for small $\psi$ and zero vector potential ${\bf A}$, the eigenvalue is $-\a$. For the superconducting half-space discussed above, and in the type II case, Ginzburg and Landau (1950:~p.~556) show that the equation reduces to the harmonic oscillator equation.} 

\begin{figure}
\begin{center}
\includegraphics{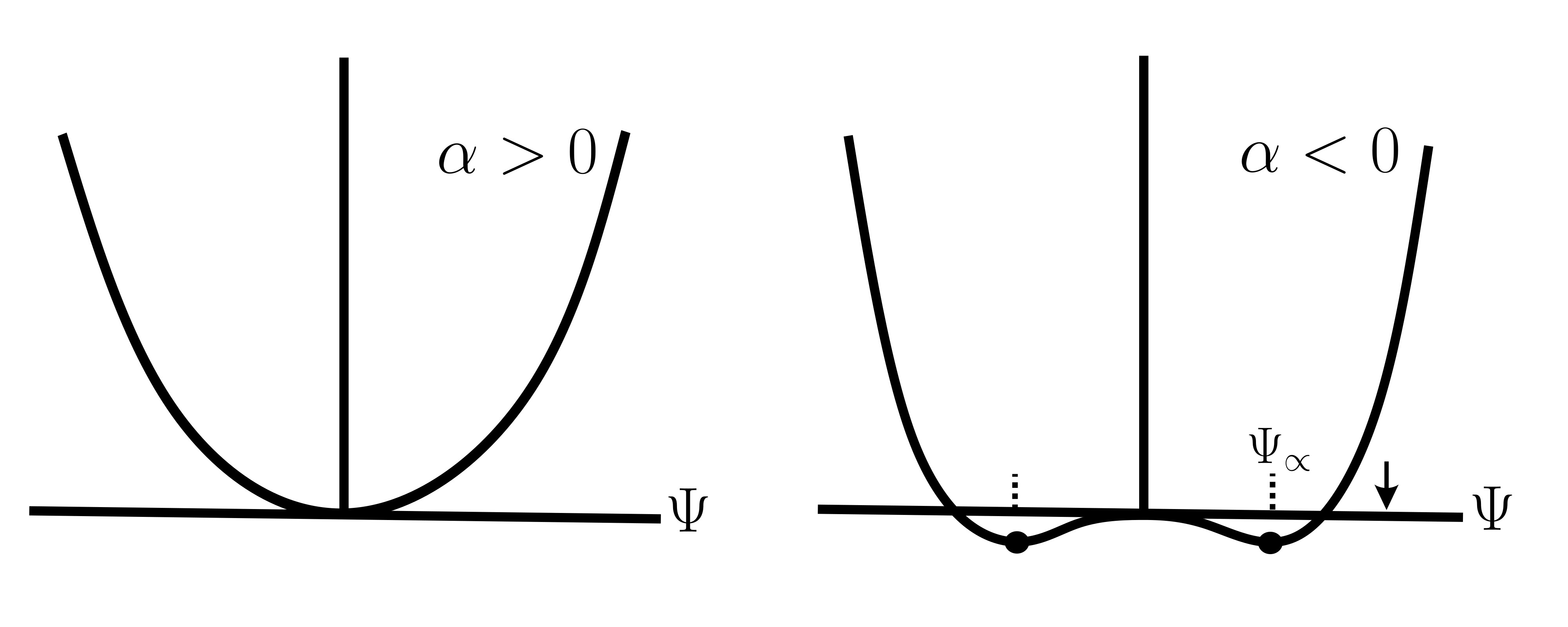}
\caption{\small Phases of the Ginzburg-Landau potential: (i) $\a>0$ (left), and (ii) $\a<0$ (right).}
\label{Phi4potential}
\end{center}
\end{figure}

Before we proceed to analyse the phase transition, let us comment on Gor'kov's (1959:~p.~1366) proof that the Ginzburg-Landau model reduces to the microscopic BCS model (for a philosophical discussion of reduction of theories, see Chapter \ref{Heuri}).\footnote{Cyrot (1973:~pp.~105-106) discusses three simplifying assumptions that allow the derivation of the Ginzburg-Landau equations from the BCS model: namely, assuming the temperature is near the critical temperature, assuming that the system is at equilibrium, and assuming time-independence. He also discusses, in Section 6, the degree to which the Ginzburg-Landau model can be extended to apply when these simplying assumptions do not hold.} 
To do this reduction, Gor'kov identified the order parameter, $\psi$, with an appropriate two-point function of the fermionic field of the electrons, which is also proportional to the {\it energy gap}: namely, the binding energy between the electrons in a Cooper pair, which is also the difference between the energy of the minimum in the normal state and in the superconducting state, i.e.~the amount of energy required to dissociate a Cooper pair.\footnote{See Gor'kov (1958:~pp.~506-507).} 
Thus a zero minimum value of $\psi$ indicates the normal phase, while a non-zero value indicates the Bose condensation of Cooper pairs in the superconducting phase.\\
\\
{\bf Phase transition and the superconducting state}\\
\\
The superconducting behaviour depends critically on the sign of the coefficient of the quadratic term, $\a(T)$. Figure \ref{Phi4potential} shows the potential terms of the free energy for different values of $\a$: 

(i)~~For $\a>0$, the minimum of the free energy is at $|\psi|=0$: and since the order parameter takes zero value, this describes the {\it normal, i.e.~non-superconducting, state} of the superconductor. We assume that this happens at high temperatures, $T>T_{\sm c}$.

(ii)~~For $\a<0$, the potential in Figure \ref{Phi4potential} has a degenerate minimum for the following value of the order parameter:\footnote{The minimum value Eq.~\eq{psimin} only involves the modulus of $\psi$, and so the phase is not fixed by the minimum, and there is a U(1) degeneracy, corresponding to the possible values that the phase can have. Also, note that we have here calculated the minimum of the potential, rather than the total free energy. Strictly speaking, we look for the absolute minimum of the free energy: but, this being a non-linear problem, it cannot be found analytically. Thus the approach taken here is to take a perfect Meissner effect as our idealisation, i.e.~in effect to consider solutions that are spatially homogeneous. Expanding around such solutions, one obtains solutions that {\it do} minimize the free energy, including its spatial dependence. The idealisation can then indeed be seen to correspond to a perfect Meissner effect, where surface effects are neglected (see footnote \ref{linearT}). See also the discussion in Saint-James et al.~(1969:~p.~26).}
\bea\label{psimin}
|\psi|_{\tn{min}}=-{\a\over\b}>0\,.
\eea
We assume that this happens at low temperatures, $T<T_{\sm c}$. Thus the coefficient $\a(T)$ changes from being positive, at $T>T_{\sm c}$, to being negative at $T<T_{\sm c}$. Assuming that $\a$ has a well-defined Taylor series near $T_{\sm c}$, to leading order in $T-T_{\sm c}$ it must therefore be of the form:
\bea\label{alphaT}
\a(T)=a\,(T-T_{\sm c})\,,
\eea
for some temperature-independent and positive $a>0$. The boundedness of the potential requires that $\b(T)$ is non-zero and positive throughout the transition: thus to leading order, it can be approximated by its value at the critical temperature, i.e.~$\b(T)=\b(T_{\sm c}):=b$. Thus to leading order in $T-T_{\sm c}$, the order parameter is given by:
\bea\label{psi2T}
|\psi|^2={a\over b}\,(T_{\sm c}-T)\,.
\eea

This temperature dependence can be used to confirm that the penetration depth {\it rises} with the temperature as it approaches the critical point and the normal phase. For the penetration depth can be calculated by reproducing the London equation (which relates the electric current and the magnetic field) from the equilibrium equation obtained from the variation of the free energy, Eq.~\eq{GLfreeE}. The result is:\footnote{For a derivation, see e.g.~Saint-James et al.~(1969:~p.~27).}
$\lambda(T)=\sqrt{mc^2\over16\pi q{}^2|\psi|^2}$. This indeed agrees with the London expression, under the already-discussed identification (up to a numerical constant) of the square of the modulus, $|\psi|^2$, with the local electron density of superconducting electrons, i.e.~$n_s$. Substituting Eq.~\eq{psi2T}, the temperature dependence of the penetration depth is: $\l\sim1/\sqrt{T_{\sm c}-T}$. Thus at the critical temperature, the penetration depth is infinite: there are no superconducting electrons, the Meissner effect disappears, and the material is in its normal state.\footnote{For more details, see Tinkham (1996:~pp.~113-114).}

Recall that the coherence length was defined as the characteristic quantum length of the superconducting electrons. We can estimate it by noting that, in the absence of an external magnetic field (i.e.~when ${\bf A}=0$), the free energy, Eq.~\eq{GLfreeE}, has a single (temperature-dependent) length scale, defined as follows: 
\bea\label{corrL}
\xi(T):=\sqrt{\hbar^2\over2m|\a(T)|}\sim{1\over\sqrt{T_{\sm c}-T}}\,.
\eea
While this is not Pippard's coherence length, it does approximate it at low temperatures. $\xi(T)$ is the length over which the size of the condensate, $\psi$, will vary as a function of the temperature.\footnote{As mentioned above, it can be shown that the characteristic length scale of the fluctuations of the order parameter, $\psi(T)$, away from its minimum value in Eq.~\eq{psimin}, is given by $\xi(T)$. Namely, a short calculation in Tinkham (1996:~p.~119) shows that (for ${\bf A}$=0, where the order parameter $\psi$ can be taken to be real), the order parameter can be solved to linear order around its equilibrium value, and is given as follows: $\psi=-{\a\over\b}\,(1+e^{-\sqrt{2}z/\xi(T)})$. This linearisation is valid at distances $z$ that are sufficiently larger than the characteristic length, $\xi$. Note that this fluctuation away from the constant value of the order parameter describes a {\it surface effect} that is a departure from a perfect Meissner effect (i.e.~a constant condensate of Cooper pairs throughout the material). For the fluctuation depends on the distance $z$ to the surface of the superconductor: it is zero far away from the surface, but it dominates the order parameter at the surface (and so, this approximation is not meant to be reliable near the surface).\label{linearT}}

The two length scales in the problem, namely the penetration depth, $\l$, and the coherence length, $\xi$, have the same temperature-dependence near the critical temperature. Thus, to leading order, the temperature dependence disappears from the following dimensionless ratio:
\bea\label{kappa}
\k:={\l(T)\over\xi(T)}\,.
\eea
As we mentioned earlier, this parameter determines the difference between ordinary, i.e.~type I, superconductors, and type II superconductors: which we will shortly discuss below, because it will lead, in later Subsections, to our two analogies in quantum field theory: with the Higgs mechanism, and with quark confinement (see Figure \ref{dualSuperC}).\\

So far we have discussed the modulus of the order parameter: but its {\it phase} generates the supercurrent on the surface of the superconductor, which adjusts itself so that the external magnetic field is cancelled, yielding the Meissner effect. 

A key property of the {\it supercurrent} is that it is quantised on a superconducting ring i.e.~an annulus: its topological stability is given by the winding of the order parameter around a closed loop, so that it does not decay over time. Thus for $\psi$ to be single-valued, its phase must be periodic, i.e.~takes values between 0 and $2\pi n$, where $n\in\mathbb{Z}$. The magnetic field is quantised as follows:\footnote{We here give the argument for the quantization of the magnetic flux. This argument was first given by London (1950:~p.~152) and was subsequently confirmed experimentally. London suggested that the flux trapped by a superconducting ring is quantized; the argument is of course also implicit in Abrikosov (1957:~pp.~1178-1179). The derivation is also given in Saint-James et al.~(1969:~pp.~45-62) and Tinkham (1996:~pp.~161-167). First, the following current can be derived from the free energy, Eq.~\eq{GLfreeE}, with the value $q=2e$ (this in effect generalises Eq.~\eq{JNoether} to three dimensions): ${\bf J}=-{ie\hbar\over m}\left(\psi^*\na\psi-\psi\na\psi^*\right)-{4e^2\over mc}\,|\psi|^2{\bf A}$. Writing $\psi({\bf r})=\r({\bf r})\,e^{i\chi({\bf r})}$, we find that this current can be written in the following simple form: ${\bf J}={2e\over m}\,\r^2\left(\hbar\,\na\chi-{2e\over c}\,{\bf A}\right)$, which can be used to solve for the vector potential: ${\bf A}=-{mc\over4e^2\r^2}\,{\bf J}+{\hbar c\over2e}\,\na\chi$. Recall that the phase of the order parameter, $\chi$, is periodic, with period $2\pi n$, for $n\in\mathbb{Z}$. This means that the integral of its divergence around a closed loop $\g$ inside the superconductor, i.e.~$\oint_\g{\dd\bf l}\cdot\na\chi$, need not be zero, but can take the value $2\pi n$, for any integer $n\in\mathbb{Z}$. Thus if we integrate the two sides of the equation around a closed loop $\g$, choosing $\g$ to not intersect any regions where the current ${\bf J}$ is non-zero, the contribution of the current to the circle integral is zero, and we obtain a quantization condition for the magnetic flux. The condition that $\g$ does not intersect any regions where the current ${\bf J}$ is non-zero can be achieved by choosing $\g$ to only pass through regions where the current is zero, in particular to not cross the core of any vortex in a type II superconductor, nor be close to the boundary. This sufficient condition is physically clear, but it is not necessary that ${\bf J}=0$ on the closed loop. The necessary condition is that the closed loop $\g$ intersect a region with non-vanishing current in the direction perpendicular to the current (without crossing the core of the vortex, where $\r=0$, and the denominator of the above vector potential would be ill-defined).\label{mfq}}
\bea\label{fquant}
\F={2\pi n\hbar c\over2e}\,,~~n\in\mathbb{Z}.
\eea
This, in effect, generalises Section \ref{PVD}'s quantisation condition for two-dimensional vortices to any three-dimensional superconductor.\footnote{The quantization condition is a general consequence of the form of the current, and so it holds for both type I and type II superconductors. In a type I superconductor, $n$ is non-zero only if the geometry is multiply connected, as e.g.~in a ring. In type II superconductors, the quantisation condition can be understood in terms of the formation of vortices with quantised magnetic field in their core. For a discussion of the differences, see Saint-James et al.~(1969:~p.~45).} 
The flux is thus quantized in units of the elementary flux quantum ${hc/2e}$, called the {\it fluxoid quantum}.\\
\\
{\bf Two types of superconductors, and Abrikosov vortices}\\
\\
The major difference in the behaviour of superconductors (see Figure \ref{dualSuperC}) is the dependence of the Meissner effect on the external magnetic field, $H_0$:\\
\\
{\bf (Type I)}:~~the superconducting state is destroyed abruptly, so that the Meissner effect disappears, when the applied magnetic field exceeds a critical value, i.e.~at $H_0>H_{\sm c}$. Thus at $H_0>H_{\sm c}$, the external magnetic field penetrates the material, so that the total average magnetic field $\bar H$ in the interior, which is zero in the superconducting state, suddenly rises.\footnote{Most pure metals, like aluminium, lead, and mercury, are type I superconductors. Almost all impure and compound superconductors are of type II.}\\
{\bf (Type II)}:~~there is an intermediate {\bf Shubnikov phase} or {\it mixed state}, in a range $(H_{\sm c1},H_{\sm c2})$ of critical values of the applied magnetic field, so that the total average magnetic field $\bar H$ in the interior increases {\it continuously,} rather than abruptly, from zero below the lower bound, i.e.~$\bar H=0$ for $H_0<H_{\sm c1}$, to $\bar H=H_{\sm c2}$ at the upper bound, i.e.~$H_0=H_{\sm c2}$.\footnote{For a calculation of $H_{\sm c2}$ and $H_{\sm c1}$ from the Ginzburg-Landau model, see Saint-James et al.~(1969:~pp.~42-43 and 51, respectively).} As we will see, this mixed state is due to the appearance of vortices with quantized magnetic flux, where the core of each vortex is surrounded by a vortex of electrons, dissociated from each other because of the presence of the magnetic field, which pulls electrons of opposite spin in different directions.

Thus type II superconductors have two {\it superconducting phases}:\footnote{The boundary between types I and II lies at the value $\k=1/\sqrt{2}$, i.e.~when the penetration depth and the coherence length are of the same order. For $\k<1/\sqrt{2}$, the superconductor is of type I. For $\k>1/\sqrt{2}$, the superconductor is of type II. It was already noticed by Ginzburg and Landau (1950:~p.~555) that for $\k>1/\sqrt{2}$ there is a `peculiar instability ... with respect to the formation of thin layers of superconducting phase'. Abrikosov (1957:~pp.~1178-1180) found that type II superconductors have vortex solutions. See also Abrikosov (2004:~p.~977).}

(i)~~A {\bf Meissner phase}, where the external magnetic field is smaller than $H_{\sm c1}$, i.e.~$H_0<H_{\sm c1}$, so that the macroscopic internal magnetic field is zero and the behaviour is like the type I superconductor, i.e.~a perfect diamagnet. 

(ii)~~A {\bf Shubnikov phase}, between the values $H_{\sm{c1}}<H_0<H_{\sm{c2}}$, where the superconductor is {\it in part} diamagnetic, and the average magnetic field in the interior is non-zero, i.e.~$\bar H\not=0$. 

For stronger magnetic fields, $H_{\sm{external}}>H_{\sm c2}$, the superconductor undergoes a phase transition (of the second kind) to its normal, non-superconducting, state.

Alternatively, we can go in the other direction. We can consider reducing the applied field towards $H_{\sm c2}$, regions of superconducting material will begin to nucleate, and the superconductor enters the mixed state (ii) (see Cyrot, 1973:~p.~118).

The Shubnikov phase is characterised by its {\it Abrikosov vortices}. These are long filaments transverse to the surface of the superconductor, arranged in a two-dimensional periodic array on the $x-y$-plane.\footnote{Abrikosov (1957:~p.~1178) stressed the resemblance of his vortices with the distribution of superfluidity in helium II by the creation of vortex filaments, described by Onsager (1949:~p.~280) and Feynman (1955:~p.~31).} 
The core of each vortex carries quantized magnetic flux, and is surrounded by a vortex of electrons, now dissociated from each other because of the presence of the magnetic field, which pulls electrons of opposite spin in different directions. When the external magnetic field is increased to $H_{\sm c1}$, vortices begin to appear, indicating the beginning of the instability of the Cooper pairs as the field increases towards $H_{\sm c2}$.\footnote{The condition for the superconducting state to go, at large magnetic field, into the polarized normal state is that the polarization energy (given by the spin susceptibility) is equal to the condensation energy, i.e.~the formation energy of the Cooper pair, given by the energy gap. For details, see Saint-James et al.~(1969:~p.~158).}

Abrikosov (1957:~pp.~1175-1177) discovered these vortices as solutions of the free energy equations, Eq.~\eq{GLfreeE}, including a non-trivial magnetic contribution from the gauge potential. The quartic term, with a positive coefficient, in effect gives a repulsive force pulling the vortices apart:\footnote{We will see that the free energy is best seen in analogy with the Lagrangian, rather than with the Hamiltonian.} 
hence the tendency of the vortices, which because of the quantization condition all have the same amount of magnetic flux, to form a regular lattice: the minimization problem\footnote{For a discussion of the boundary conditions used in the variational procedure, see e.g.~Tinkham (1996:~pp.~117-118, 121).} 
determines that, for a superconductor infinitely extended in the $x-y$-plane, the most favourable configuration is a honeycomb or hexagonal lattice, i.e.~a configuration of vortices in regular (i.e.~equilateral) triangles, where triangles and inverted triangles are juxtaposed on the plane.\footnote{To write down his vortex solutions, Abrikosov took advantage of the fact that the free energy equation takes the precise form of a harmonic oscillator equation, whose solutions are well-known, and could be suitably superposed to form the desired lattice. Although Abrikosov (1957:~p.~1177) thought that the most favourable lattice was the square one, with minimal value of the free energy, work by Kleiner et al.~(1964):~p.~1227) later found that this solution is unstable, and that the honeycomb lattice has the minimum value of the `Abrikosov parameter' that determines the minimum of the free energy. See also the discussion in Tinkham (1996:~pp.~145-147).}

At the core of a vortex, the order parameter $\psi=\r\,e^{i\chi}$ takes zero value, i.e.~$\rho=0$, which indicates that the superconductivity is lost there and there are no Cooper pairs, whereas the field $H$ is non-zero and has its maximum value there, again indicating the local disappearance of the Meissner effect. 

In Section \ref{analo2}, Abrikosov-type vortices in a relativistic field theory will model strings, as shown by the middle row in Figure \ref{dualSuperC}. To this end, the next Section first introduces the analogy between superconductivity and the Higgs mechanism.

\section{Superconductivity and the Higgs mechanism}\label{analogies}

This Section demonstrates two crucial analogies between superconductivity and particle physics. The first (Section \ref{analo1}) is the analogy between the formation of Cooper pairs in a superconductor and the Higgs mechanism in the electroweak model, i.e.~the model that unifies the electromagnetic and the weak interactions. The second (Section \ref{analo2}) is the analogy between vortices in type II superconductivity and relativistic strings in Higgs models.

These analogies played a pivotal role for the development of the Higgs mechanism, spontaneous symmetry breaking, and models of colour charge confinement that employ the idea of duality (Section \ref{cmds}).

\subsection{Analogy between superconductivity and Higgs mechanism}\label{analo1}

Today, the analogy between superconductivity and the Higgs mechanism in particle physics remains relevant.\footnote{For accounts of this history, see Borrelli (2015) and Karaca (2013).}
For, by the Higgs mechanism, the $W^\pm$ and $Z$ gauge bosons of the electroweak model can have an effective mass.\footnote{The gauge group of the electroweak theory is $\mbox{SU}(2)\times\mbox{U}(1)$, where SU(2) is the gauge group of the weak interactions, and U(1) is the gauge group of the electromagnetic interactions. For simplicity, we will focus on the SU(2) sector, which was historically considered first. Glashow (1959, 1961), Salam (1962; see also Salam and Ward, 1959) and Weinberg (1967) developed the electroweak model with gauge group $\mbox{SU}(2)\times\mbox{U}(1)$.}
And it would be incorrect to take the analogy to be `one-way', so that one learns about particle physics by considering the ``simpler'' setting of condensed matter physics. The analogy also gets used in the other direction, so as to learn about gauge invariance in condensed matter physics (see the double arrow in Figure \ref{dualSuperC}). 

Briefly, the analogy is between the Cooper pairs which condense in the Meissner phase, and the Higgs field taking a non-zero expectation value in the Higgs phase. Both condensates spontaneously break the symmetry, and thereby give the gauge bosons their effective mass. London's exponential decay of the magnetic field near the boundary (see Eq.~\eq{Londoneq}) is analogous to the effective Yukawa potential of the electroweak interactions: both of which give short-ranged forces.

There are important disanalogies in the physical properties, and nature, of the phases. The major one is that, for a superconductor, physicists have good experimental control of external parameters such as temperature: which is much more difficult to attain in particle physics, because increasing the temperature requires collisions at much higher energies, or in cosmology, where observations are indirect (other differences include e.g.~the quantum behaviour of the models). 

But note that the experimental realization of {\it both} phase transitions requires non-zero temperature and real time evolution. In other words, the Higgs mechanism, as presented in standard textbooks, is simplified and idealised. For the coefficients of the effective action in the electroweak theory, including the value of minimum of the Higgs potential, in general depend on the temperature, just as they do in the Ginzburg-Landau model.\footnote{See Ramsey-Musolf, 2020:~p.~4; Huang et al.~2016:~p.~2; Csikor et al.~1999:~p.~22; and Section \ref{GLtheory} in this Chapter.} 

In view of this, one major limitation of the analogy with superconductivity is in the differences between the phase transitions: lattice calculations in the standard model indicate cases where, rather than a phase transition, cross-over behaviour can take place,\footnote{See Csikor et al.~(1999:~p.~21) and Fodor (2000:~p.~1).} 
although including higher-order operators changes this to a first-order or to a continuous phase transition, including in models beyond the standard model.\footnote{See Huang et al.~(2016:~p.~16) and Ramsey-Musolf (2020:~p.~1).} 
Here, `cross-over' means that the transition between phases is smooth and there are no discontinuities or singularities in the free energy or its derivatives. Note that in the BCS model there can also be cross-over behaviour to Bose superconductivity,\footnote{See S\'a de Melo et al.~(1993:~p.~3202).} 
and also that in type II superconductivity, because there is an intermediate phase of mixed normal and superconducting properties, there is not a pure continuous phase transition. Clearly, the differences here concern the details of the different models. Nevertheless, the Ginzburg-Landau model remains a powerful tool in the electroweak theory, and the analogy with superconductivity is a useful one. 

Thus, despite the differences, we disagree with Fraser and Koberinski (2016:~p.~72) that the analogies are {\it purely} formal, since there are substantive similarities in the physical mechanisms involved (the generation of a mass gap, the existence of a Higgs mode, and the possibility of topological defects), and a full account of the Higgs mechanism requires the considerations of the electroweak phase transition of the type just given.\\
\\
{\bf Gauge invariance in symmetry breaking: a two-way analogy.} By using quantum field theory techniques to address the problem of gauge invariance in the BCS model of superconductivity, Nambu introduced the idea of spontaneous symmetry breaking into particle physics.\footnote{See Nambu (1960:~p.~648) and Karaca (2013:~p.~4). Two years earlier, Anderson (1958:~p.~829) had shown that terms that expressed non-gauge-invariant results in the BCS model are negligible. Nambu showed that the BCS model is {\it strictly} gauge invariant.} Namely, he showed that the calculation of the Meissner effect by Bardeen et al.~(1957), which was apparently not gauge-invariant with respect to local U(1) transformations of the fermionic fields inside the Cooper pairs, {\it was} in fact gauge invariant, due to the interactions of the electrons with phonons in the material. The proof was essentially quantum field-theoretic: it used the Ward-Takahashi identity, which expresses the conservation of a current as a consequence of local gauge invariance. Thus in establishing the all-important gauge invariance of the BCS model, he made crucial use of the analogy with quantum field theory, where gauge invariance can be tested through the Ward-Takahashi identity.\footnote{If one can show that the Ward-Takahashi identity holds, as Nambu did, then the current is conserved, and the theory is thereby shown to be gauge invariant.}

In futher work of his and with Jona-Lasinio, he went on to explore the other side of this relation.\footnote{See Nambu (1960a:~p.~382) and Nambu and Jona-Lasinio (1961:~p.~345).} 
Indeed, the analogy with the Meissner effect allowed him to solve an important problem in particle physics: to introduce mass without breaking the {\bf chiral symmetry}, through spontaneous symmetry breaking. Even though the nuclear interactions were still unknown, he derived the nucleon mass (i.e.~the mass of a neutron or a proton) by a method analogous to the derivation of the mass gap for a superconductor.\footnote{He showed this from very general assumptions: (i) the classical concepts of attraction and repulsion between particles, (ii) the chiral symmetry. See Nambu and Jona-Lasinio (1961:~p.~357). In 2008 he was awarded the Nobel Prize in physics for the discovery of the mechanism of spontaneous broken symmetry in subatomic physics, which he shared with Kobayashi and Maskawa.} 

But nucleons are fermions, and the question remained how {\it bosons}, such as appear in the electroweak (rather than strong) interactions, could be massive.\footnote{Glashow (1961) had already tried, and failed, to unify the weak and electromagnetic interactions.}
Because the weak interactions are short range, any vector bosons mediating them must be massive: but this is prima facie incompatible with the {\it gauge invariance} of the Yang-Mills theory, which is required for renormalizability.\footnote{See Komar and Salam (1960).} \\
\\
{\bf Unobserved Goldstone bosons.} Even if Nambu's idea of spontaneous symmetry breaking might be a way to give mass to gauge bosons, it had a problem: Goldstone's theorem predicts that, in the absence of long-range interactions, the spontaneous breaking of a global symmetry entails the existence of massless (scalar) bosons, which are not observed in nature.\footnote{See Goldstone (1961) where he conjectured the existence of such massless bosons. This was proven in Goldstone, Salam and Weinberg (1962).} 
Thus it was widely {\it believed} that, even though the symmetry is local rather than global, spontaneous symmetry breaking is not a promising way of introducing mass.

Without invoking symmetry breaking, Schwinger (1962:~p.~397)\footnote{See also Karaca (2013:~p.~13).} 
had defended the idea that the gauge invariance of a vector field need not entail that it has zero mass. He proposed that vector bosons could acquire a mass by interacting with (i.e.~by their coupling to) a conserved current. 

Anderson (1963) showed that Schwinger's idea is realised in a superconducting plasma: as a result of the coupling of the external magnetic field to the conserved current associated with the Cooper pairs (see footnote \ref{mfq}), the photon becomes effectively massive, because it acquires a longitudinal polarization state, which a massless photon does not have, since a massless photon has only two polarizations, both transverse to its momentum vector.\footnote{For an explanation of the origin of the longitudinal polarization state of the photon in the BCS model, see Karaca (2013:~p.~5).} 
Thus the magnetic interaction is short-ranged in the interior of the superconductor---this is the Meissner effect. The longitudinal polarization state that was acquired by the photon is the result of the spontaneous breaking of the U(1) phase symmetry of the electrons in the Cooper pairs (in our sense of `symmetry breaking' in Section \ref{ssp}), i.e.~it is the massless Goldstone boson that results from the spontaneous symmetry breaking. 

Thus the superconductor analogy resolves two problems at once: 

(i)~~The photon is massive, thus giving the short-ranged magnetic field of the Meissner effect, without breaking the gauge symmetry, i.e.~through spontaneous symmetry breaking.

(ii)~~Goldstone bosons are avoided, because the would-be massless, scalar, Goldstone boson ``recombines'' with the photon, as a longitudinal mode of a {\it massive} photon.

Anderson suggested that this mechanism could also solve the analogous problem of the weak interactions: massless vector fields in Yang-Mills theory can become massive without breaking the gauge invariance, and without creating additional unobserved massless scalar particles. Anderson's analysis of superconductivity is a ``skeleton'' of what would become the Higgs mechanism.

We take the {\bf Higgs mechanism} to be the following phenomenon. Consider a field theory $T_1$ with a global symmetry and degenerate vacua, such that coupling it (minimally) to gauge fields gives another theory $T_2$ which has a local (gauge) symmetry. The Goldstone modes in the vacuum spectrum of $T_1$ correspond to unphysical modes in $T_2$, i.e.~$T_2$ has, in its perturbative spectrum about the vacuum, (physical) massive vector modes in the longitudinal directions.\footnote{We thank Nicholas Teh for a discussion of this point.} 
We illustrate this phenomenon below.\\
\\
{\bf The Higgs mechanism in particle physics.} Higgs (1964:~p.~508) developed the idea of {\it spontaneous symmetry breaking in relativistic field theory}, for a (classical, abelian) vector field $A$ coupled to two real scalar fields, $\f_1$ and $\f_2$, with an arbitrary (rotation-symmetric) potential.\footnote{If we think of the two real scalar fields as a single complex field $\f=\f_1+i\f_2$, then Higgs' theory is the four-dimensional analogue of Eq.~\eq{Lvortex}, for an arbitrary real potential.} 
Thus he used the four-dimensional abelian Higgs model with an arbitrary potential (see Eq.~\eq{Lvortex}).

This model has a local U(1) symmetry that is spontaneously broken (in the sense of Section \ref{ssp}) by the choice of a solution that minimizes the potential: where one of the two scalar fields is zero, $\f_1=0$, and the other one is a constant, $\f_2=\f_0$.\footnote{The U(1) symmetry acts as SO(2) transformations on the scalars, and so this solution is a representative, among an SO(2)-worth of equivalent representatives, of the ground state of the model.}
This choice is analogous to the minimization of the potential, Eq.~\eq{psimin}, in the Ginzburg-Landau model. 

Higgs showed that, in the effective field theory that one obtains by {\it expanding about this solution} (recall, from Section \ref{ssp}, that the broken symmetry model is a linearized model), the equations of motion are equivalent to those of a massive vector model: 
\bea\label{Gmn}
\pa_\n G^{\m\n}+e^2\f_0^2\,B^\m&=&0\,,
\eea
where the new massive vector field $B$ (and thereby the curvature $G$) is defined in terms of the old one and the fluctuation of the scalar field, as follows:\footnote{$B$ also automatically satisfies the condition $\pa_\m B^\m=0$, which follows from Eq.~\eq{Gmn}.}
\bea\label{BG}
B_\m&:=&A_\m-(e\f_0)^{-1}\pa_\m\f'\nn
G_{\m\n}&:=&\pa_\m B_\n-\pa_\n B_\m\,,
\eea
where $\f'$ is the fluctuation around the zero value of first scalar field, i.e.~$\f'=\d\f_1$.\footnote{There is also a massive equation for the small fluctuation of the other scalar field, $\f_2$.} 
For reasons that will become clear in a moment, we will call this approximate set of equations of motion, i.e.~Eqs.~\eq{Gmn} and \eq{BG}, the {\bf effective Higgs model}.

The model defined by Eqs.~\eq{Gmn} and \eq{BG} is the Proca model, i.e.~a model with a massive vector field. Unlike the massless gauge field $A$, which has only two transverse polarizations, the massive photon $B$ has two transverse and one longitudinal polarization. This is because the would-be Goldstone boson $\f'$\footnote{Note that if the U(1) symmetry is local, so that $e=0$ and the scalar fields are decoupled from the vector field, then $\f'$ is indeed a (clasically massless) Goldstone boson. Because of the coupling, it can be reabsorbed into the longitudinal polarization the massive vector field $B$, through Eq.~\eq{BG}.} 
has been reabsorbed as one of the longitudinal polarizations of the vector field.\footnote{Namely, the Fourier transform of the extra term that the photon $B$ acquires compared to $A$ in Eq.~\eq{BG}, i.e.~$\pa_\m\f'$, is proportional to the momentum vector $k_\m$, which is precisely in the direction of the four-momentum, hence longitudinal. This is now a {\it physical} component of the vector $B$, unlike the usual longitudinal components of the {\it massless} vector $A$, which under normal circumstances are unphysical because they can be removed by a gauge transformation.} 
The mass of this photon is $\m=e\f_0$. 

As we mentioned at the beginning of this Chapter, a recent line of philosophical work has worried about the gauge-dependence of the Higgs mechanism (see footnote \ref{SBphil}). There are two key points to emphasise:

(A)~~{\it Gauge invariance is not broken:} it is merely hidden in Eq.~\eq{BG}, which is invariant under the combined gauge transformations of $A$ and $\f'$. (The Proca equation for $B$ is not gauge invariant, but there is no reason to expect that it should be.)

(B)~~{\it The two models are not equivalent:} the phrase `expanding around this solution' indicates this important aspect of the phenomenon of symmetry breaking, which often remains in the background. Namely, the manifestly gauge-invariant abelian Higgs model we started with, and the effective Higgs model in Eq.~\eq{BG}, are not equivalent: the latter is merely an {\it effective model} obtained by a perturbative expansion of the former about a particular state. An important difference regards their quantum properties: the former model is renormalisable, while the latter in general is not.\footnote{The non-abelian case is non-renormalizable: see Shizuya (1977:~pp.~137-139).}

Thus it is best to think of the effective Higgs model as a tool used (a) to exhibit the physical degrees of freedom that are observable at low energies, which are not manifest in the fundamental model, and (b) to calculate their masses. For in the fundamental model it is only the perturbative degrees of freedom around $\f_1=\f_2=0$ that are manifest: but this is an unstable vacuum, and to make the low-energy degrees of freedom manifest one expands around the absolute vacuum, leading to the effective Higgs model. 

In other words, there is a trade-off of virtues between the two models: the fundamental abelian Higgs model is manifestly gauge invariant and is well-defined quantum mechanically, but it does not make explicit the low-energy degrees of freedom at the absolute minimum of the potential explicit. By contrast, the effective Higgs model is not well-defined as a quantum theory, but it does make the low-energy degrees of freedom manifest.\\
\\
{\bf Back to the analogy with superconductivity.} The Higgs mechanism illustrates, for the electroweak model in particle physics, both aspects of Anderson's suggested solution of the mass problem for superconductivity (i.e.~(i) and (ii) above).\footnote{Englert and Brout (1964:~p.~322) offered a {\it quantum} treatment of spontaneous symmetry breaking, including a first treatment of the non-abelian case. They established the Ward-Takahashi identity, and so local gauge invariance, after spontaneous symmetry breaking. As in Higgs's case, the gauge field acquires a mass $\m=e\f_0$ (where $\f_0$ is the {\it quantum} expectation value of one of the scalar fields, i.e.~$\bra\f\ket$), and the longitudinal term of the vector field originates in the would-be Goldstone boson. They also extended the analysis to non-abelian gauge theories, which Higgs (1964) had not done, and calculated the mass in that case. Further pivotal work, among others, is by Guralnik, Hagen, and Kibble (1964), and Higgs (1966), in part itself a response to criticism by Gilbert (1964).}
We already noted the close analogy between the Lagrangian, used for spontaneous symmetry breaking in the electroweak model,\footnote{See Higgs (1964:~p.~508) and Englert and Brout (1964:~p.~321).}
and the free energy of the Ginzburg-Landau model, Eq.~\eq{GLfreeE}.\footnote{As we discussed in Section \ref{GLtheory}, the latter was derived from the BCS model by Gor'kov (1959:~p.~1366).} 
Not only are these minimization principles similar: as we have seen, much of the physics involved in the spontaneous breaking of gauge symmetry, the appearance of a mass gap, and the identification of physical degrees of freedom,\footnote{The formal correlate of this is the absorption of would-be Goldstone bosons into longitudinal components of vector fields} 
are close analogues.\footnote{We discussed some differences earlier. Another important difference is that the Ginzburg-Landau model is non-relativistic. We will return to this in the next Section.}

Thus the Higgs model came to be seen as a relativistic generalisation of the Ginzburg-Landau model.\footnote{See Nielsen and Olesen (1973:~p.~46).} Indeed, for static solutions, the abelian Higgs Lagrangian is {\it identical} to the Ginzburg-Landau free energy for type II superconductors. This prompted a second key analogy (see Figure \ref{dualSuperC}) to which we now turn.

\subsection{Strings from vortices in superconductors}\label{analo2}

As we discussed in the preamble of this Chapter, the analogy with superconductivity is well-suited not just for understanding electroweak phenomena, but also the strong force that keeps quarks together in hadrons. 

A key intermediate step towards contemporary models of colour confinement and string theory is Nielsen and Olesen's (1973:~p.~45)\footnote{The work by Nielsen and Olesen was influenced by Nambu's unpublished notes prepared for the Copenhagen High Energy Symposium (Nambu, 1970:~pp.~289, 295), which, as it turns out, he never delivered! (p.~280). It is also in these lecture notes that Nambu wrote down the fundamental bosonic string Lagrangian (p.~287), i.e.~the Nambu-Goto action (see also Goto, 1971:~p.~1563). For more on Nambu's work in string theory during this early period, see Ramond's (2016:~pp.~3-5) {\it Hommage \`a Nambu}.\label{Nambu}} 
and Nambu's (1974:~p.~4262) relativistic interpretation of vortices in type II superconductors (see Figure \ref{dualSuperC}). Although the idea was phenomenologically and conceptually unsuccessful as a model of confinement, it is an important step towards the more mature models in Section \ref{cmds} and later Chapters.

The analogy is between the Shubnikov phase of type II superconductivity and {\it confinement of colour charge} in mesons, i.e.~hadrons that are pairs of quarks and anti-quarks. Recall that, in the Shubnikov phase, Abrikosov vortices form where the Cooper pairs dissociate (and the order parameter goes to zero), so that the external magnetic field can penetrate into the vortex core, in quantised amounts (see Eq.~\eq{fquant}). The magnetic field outside the vortices gets expelled from the material. Thus, in effect, the magnetic field is {\it confined} to the vortices. 

The analogy with mesons is better if magnetic monopoles of opposite charge are attached to the ends of vortices with finite length: so that pairs of magnetic monopoles are connected by magnetic vortices, and the magnetic field is confined to the interior of the vortices. This is then analogous to a model of a meson as a pair of a quark and and anti-quark, connected by a tube of colour charge, with the colour field confined to the interior of the hadrons. The major difference is of course that {\it quarks are electrically, rather than magnetically, charged}.

Nevertheless, this analogy is a key intermediate step. Making it precise requires three points (see the discussion of Abrikosov vortices in the Ginzburg-Landau theory at the end of Section \ref{GLtheory}):

(i)~~{\it Vortices as solutions of the abelian Higgs model}: Nielsen and Olesen used the fact that solutions of the (three-dimensional, $D=3$) Ginzburg-Landau equations are {\it static} solutions of the four-dimensional (i.e.~$D=3+1$) Higgs model, to show that the Higgs model has vortex-like solutions of the Abrikosov type, as at the end of Section \ref{GLtheory}. Their vortex solution has an elongated core, where the magnetic field has a maximum and the Higgs field is zero: while the magnetic field goes to zero outside the core, and the Higgs field goes to a constant value (namely, the minimum of the potential). 

\begin{figure}
\begin{center}
\includegraphics[height=5cm]{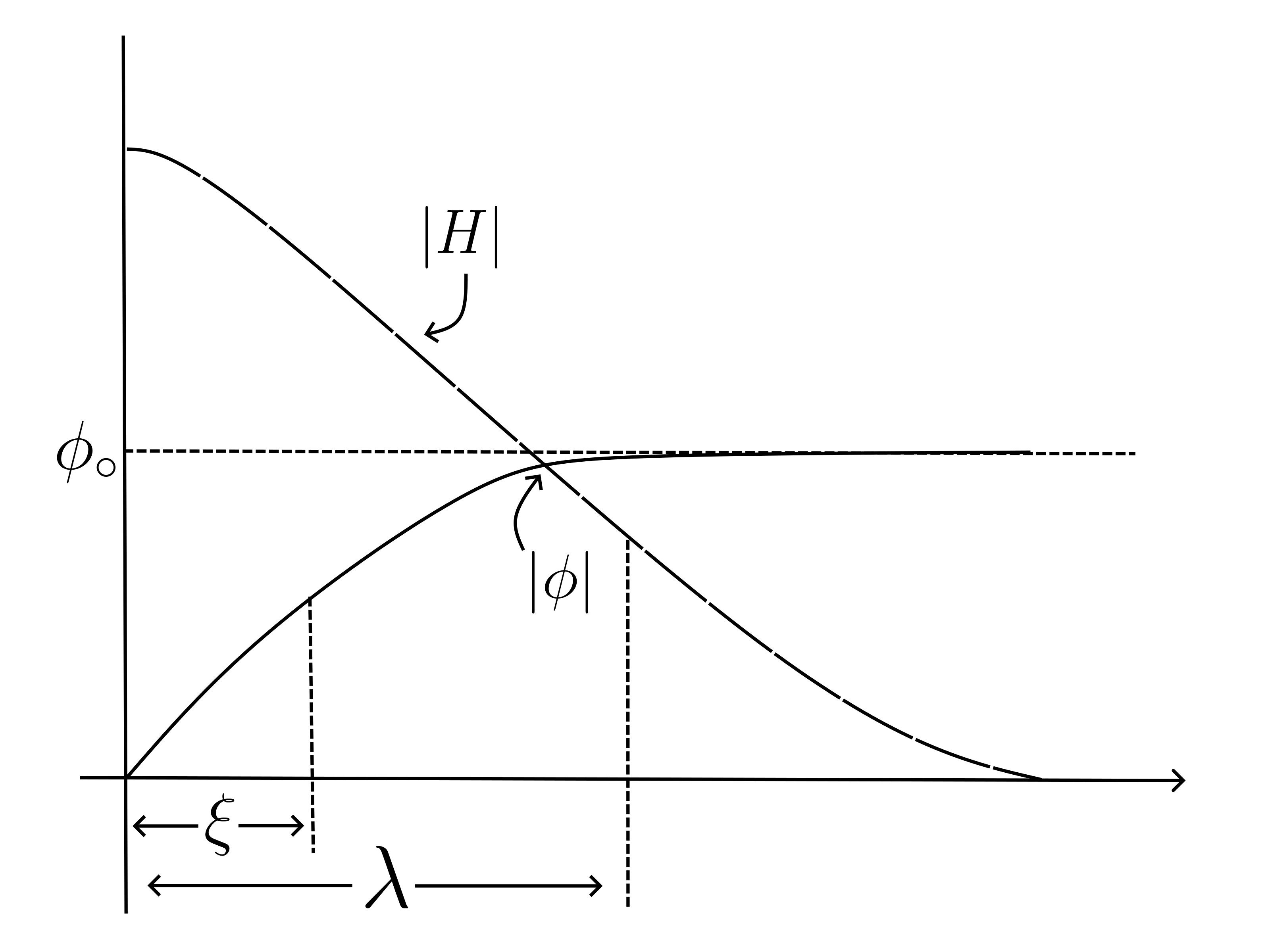}
\caption{\small A cross section of an Abrikosov vortex. The Higgs field goes to a constant expectation value at large distances, and the magnetic field goes exponentially to zero outside the vortex core, so that it is {\it confined} to the core. In the interior of the core, the Higgs field goes to zero, and the magnetic field is non-zero.}
\label{vortex-core}
\end{center}
\end{figure}

This is the behaviour that we discussed for a type II superconductor (see Figure \ref{vortex-core}): the Cooper pairs have dissociated at the vortex core, and so the order parameter takes zero value there, while it goes to a constant at infinity. In other words, the solution interpolates between the two extrema of the Ginzburg-Landau potential: the unstable one at $\psi=0$, and the absolute minimum Eq.~\eq{psimin}. 
The magnetic field is indeed confined to the vortex: it decays exponentially outside of its core.

The effective Lagrangian describing such solutions was the geometric Lagrangian for a relativistic string in spacetime, i.e.~the Nambu-Goto action (see also footnote \ref{Nambu}, and (iii) below). 

(ii)~~{\it Matching parameters}: the two length scales in the Ginzburg-Landau model of superconductivity are the London penetration depth, $\l$, and the correlation length, $\xi$. These set the scales for the masses of the mesons in the model: namely, a vector and a scalar meson, respectively. The vector meson is the meson that mediates the U(1) interaction, and its mass is responsible for the confinement of the U(1) field. The scalar meson is the Higgs, and so its mass is the Higgs mass.
Up to overall constants, and with the usual field theory convention $\hbar=c=1$, the masses for the vector and scalar mesons, respectively, are given by: $m_V=\l^{-1}=e\f_0=e\sqrt{-\a/\b}$, where $\a$ and $\b$ are as in the Ginzburg-Landau model (Eq.~\eq{GLfreeE}), and $m_S=\xi^{-1}=\sqrt{-2\a}$ (cf.~Eq.~\eq{corrL}, and $\a<0$). As in a type II superconductor, we have $\xi\gg\l$.\footnote{A problem of this approach was that, in order for this theory to describe ``thin strings'', $\l$ and $\xi$ have to be much smaller than the typical string length. Since mass and length are inversely related, this implies that the vector and scalar meson masses have to be very large (much larger than the string mass), much larger than typical hadron masses. Thus the model could not be a realistic model of hadrons, and a good string picture only emerges at strong coupling, where the theory is highly quantum (see Nielsen and Olesen, 1973:~p.~55). Nevertheless, the idea is influential in QCD, as we will see.} In the words of Nambu (1974:~p.~4263), `the string is actually two things, a flux of a magnetic field and a hollow vortex line in the Higgs field'. 

Thus the deep {\it disanalogy} with QCD becomes clear here: not just because the analogy is with a {\it magnetic}, rather than electric, version of colour charge, but more important, because the gluon in the standard model (i.e.~so far as we now know) is massless, while here it is massive. In other words, if one wishes to uphold the analogy with QCD, one needs to say that Nielsen and Olesen give a confined version of QCD, but a {\it Higgs phase} of it, i.e.~where the gluon is also ``Higgsed'', and the SU(3) symmetry is spontaneoulsy broken to U(1). This then resembles the electroweak theory, rather than QCD.\footnote{Therefore, 't Hooft and Mandelstam made some crucial changes to this kind of approach to quark confinement (as we shall discuss in Section \ref{cmds}). Nevertheless, it is believed that Bose-Einstein condensation of magnetic monopoles takes place in QCD at the {\it deconfinement temperature}. Above this temperature, the state is a quark-gluon plasma: below, there is confinement of colour charge. See Shuryak (2021:~p.~4).}

(iii)~~{\it Vortices as relativistic open strings}: Nambu (1974) argued that, since the vortices carry magnetic flux, finite (as opposed to infinite, or semi-infinite) vortices can form {\it provided} there are magnetic monopoles at the end-points, where the magnetic flux can terminate. The monopoles are identified with (magnetic analogues of) quarks, which are sources of the magnetic flux of the vortex. Thus there can be {\it open strings} in the Ginzburg-Landau model, with magnetic quarks at the two ends of the string that are bound in pairs, and show a behaviour very similar to that of a superconductor.\footnote{Thus in this picture, the vortex is regarded as a Dirac string connecting two monopoles. Unlike for Dirac's monopole, the vortex line connecting two monopoles is here regarded as physical (cf.~Section \ref{Dqc}). The criterion for the reality of the dual string was that the vector meson mass, which is proportional to the expectation value of the Higgs (cf.~$m_V=e\f_0$) and thus associated with the spontaneous breaking of the symmetry, is non-zero. If the mass is zero, then the Higgs field is everywhere zero, the magnetic field is not screened, and one recovers monopoles interacting via a long-range Maxwell field, i.e.~the interaction is of the Coulomb type, rather than Yukawa-type (the latter is the interaction one obtains when the electromagnetic field is massive, so that there is a non-zero London penetration depth). In that case, the string was unphysical, because---like the Dirac string---it carries no energy or momentum. But when the vector meson mass is non-zero, the string carries energy, and it `acquires physical reality' (p.~4264). The vortex model with magnetic monopoles allowed Nambu to identify the flux quantization condition of the superconductor, which applies to the magnetic flux of the vortex, Eq.~\eq{fquant}, with the Dirac quantization condition of the associated monopoles. } 

Despite the model's not giving a good description of QCD, it did make an important prediction:\footnote{Of course we no longer think that quarks are magnetically charged: rather, their colour charges are electric. Furthermore, since magnetic monopole charge is in general large, assigning quarks a magnetic charge seemed incompatible with the fact that the forces between quarks at small distances are {\it weak} (a phenomenon called `asymptotic freedom'): see 't Hooft (2007:~p.~723). Also, there were problems with the non-Abelian generalization of the model. (Mandelstam (1975a:~p.~476) did attempt to generalise the model to the case of a non-abelian gauge group $\mbox{SU}(N)$.)} 
if the quarks at the end of the string are pulled apart, the force required to pull the quarks apart grows linearly with the distance: an apt qualitative feature for describing the confinement of quarks.\footnote{See Nambu (1974:~p.~4267). Ultimately, Nielsen, Olesen, and Nambu's models, more than models of confinement, may have been more important for {\it string theory}: Nambu's (1974) model was complete with a Lagrangian for both the string and the magnetic monopoles sitting at its two ends. The effective string action that he obtained (and which Nielsen and Olesen had derived using different techniques) included a coupling to the dual of the Faraday tensor, thus expressing the {\it magnetic} nature of the string and its charges. The equations of motion were identical with those of the `dual string', as formulated in Nambu (1970). Thus the model gave new life (although it would be short-lived, since a new transmutation was yet to come!) to an old idea that had been developed to explain the strong interactions, namely: {\it (dual) resonance models}. Veneziano, Nambu, Virasoro, Fubini, Susskind, and others, had developed the `old string theory' around 1968-1970 for different reasons, namely to explain the behaviour of the S-matrix of hadrons, including their approximate duality properties and Regge trajectories (i.e.~the typically string-like, linear dependence between the mass squared and the spin of hadrons). The idea of dualities in the S-matrix of hadrons was introduced by Dolen, Horn, and Schmid (1968) for pion-nucleon scattering. The idea was that there was an equivalence, called `crossing symmetry', between the exchange of resonance particles between $t$- and $s$-channels where $t$ is the square of the momentum transfer, and $s$ is the square of the energy transfer. That is, because of this equivalence, processes that involved intermediate particles emitted through the $t$-channel or through the $s$-channel were equivalent. This means that only summation over a single channel is required. For a discussion of the birth of string theory, see Rickles (2014:~pp.~1-3, 51). We will return to string theory in Chapter \ref{String}, where Nambu's idea of coupling charges to the endpoints of strings plays an important role, especially for D-branes.}

\section{Confinement as dual superconductivity}\label{cmds}

In this Section, we discuss the phenomenon of colour confinement in particle physics as the ``magnetic dual'' of type II superconductivity, i.e.~the third and last row in Figure \ref{dualSuperC}. The scare quotes indicate that this is not a duality, but a {\it quasi-duality} that combines a (conjectured) electric-magnetic {\it effective duality}, and the {\it analogy} between condensed matter physics and quantum field theory. Nevertheless, the idea has been heuristically fruitful, and illustrates the heuristic power of duality. (For simplicity in the rest of this Section, we use the word `duality', even when `quasi-duality' is meant: this will be clear from the context.)

In Section \ref{condm}, we will first give the idea of the 't Hooft-Mandelstam mechanism of confinement, using the analogy with both superconductivity and with the Ising model from Chapter \ref{Simple}. In Section \ref{WilsonL}, we will then discuss some evidence from lattice calculations that confirm that the 't Hooft-Mandelstam mechanism of confinement plays a role. To this end, we will introduce a well-known order parameter for the confining phase, namely the Wilson loop. In Section \ref{phasesGT}, we will discuss more generally the notion of the phase of a quantum field theory, which will be use in the next Chapter when we discuss the Seiberg-Witten theory, and will be an important notion in our geometric view of theories, in Chapter \ref{Heuri}.

Recall how type II superconductivity exhibits {\it confinement of the magnetic field}: the magnetic field is everywhere zero in the interior of a superconductor, except at the core of magnetic vortices. But confinement of quarks in hadrons of course requires {\it confinement of electric colour charge}. This is the idea that we discuss in the next Section.

\subsection{Condensation of monopoles}\label{condm}

't Hooft (1975) and Mandelstam (1976) proposed that quark confinement is the magnetic dual of the disordered phase of type II superconductivity.\footnote{We here follow 't Hooft's (1975:~pp.~3-4) argument, in particular about the expectation value of the monopoles in the magnetic vacuum. Mandelstam (1976:~p.~247) gives a version of the argument that does not introduce monopole fields explicitly, but rather replaces this by an operator that creates 't Hooft-Polyakov monopoles in the vacuum.}
Namely, to get confinement of electric colour charge, we require electric, rather than magnetic, flux tubes. And at the end of the tubes, rather than magnetically charged magnetic monopoles, we have electrically charged quarks. So the question is: how to {\it dualize} a superconductor?

The answer is not known for four-dimensional QCD or other non-abelian models that are used as toy models for QCD (such as the Yang-Mills-Higgs model). After all, confinement is an outstanding open problem in physics!\footnote{However, there is excellent evidence from lattice QCD that 't Hooft-Mandelstam plays a role in the deconfinement phase transition of QCD, i.e.~from a quark-gluon plasma above the critical temperature, to a monopole condensate below the critical temperature (and this might be described either as a superconductor state or as a Bose-Einstein condensate of monopole pairs). See Shuryak (2021:~pp.~68-69), D'Alessandro et al.~(2010:~p.~8), and Bali (2000:~p.~18). In QCD, besides the confinement phase transition, there is also a chiral phase transition: see Bali (2000:~p.~18).}

But the analogy with the superconductor does give us guidance: recall Chapter \ref{Advan}'s discussion of particle-vortex duality, which involves both electric and magnetic versions of the abelian Higgs model. Also the analogy with the duality, in Section \ref{IMD}, between a disordered ferromagnet at high temperature and an ordered phase with dislocations at low temperatures, is useful.\footnote{For the Helmholtz free energy of an {\it ensemble of monopoles}, and entropy considerations analogous to the Berezinskii-Kosterlitz-Thouless phase transition in Section \ref{PVD}, see Shuryak (2021:~pp.~45, 478).} 

't Hooft (1975:~pp.~3-4) noted that the symmetry-breaking phase of the Yang-Mills-Higgs model (i.e.~the right-hand potential in Figure \ref{Phi4potential} above) has, in addition to vortices, magnetic monopole solutions. This is because of the model's non-abelian nature. However, this Higgs vacuum seems the wrong vacuum, because the monopoles appear to have zero expectation value there: they are not visible in terms of the Higgs expectation value, which is an electric order parameter. Thus we need to require that the {\it monopoles} have a non-zero expectation value. 

't Hooft's proposal was to return to the single-well potential (i.e.~the left-hand potential in Figure \ref{Phi4potential}), by giving the scalar field a positive mass squared (i.e.~by, in effect, taking $\a>0$). He conjectured that, in the normal phase of a non-abelian model written in dual variables, magnetic monopoles condense, and {\it colour-electric charge is confined} to the interior of electric flux tubes. This is sometimes called a {\it dual superconductor} mechanism.\footnote{It is important to note that 't Hooft and Mandelstam's proposals do {\it not} entail that confinement by monopole condensation should occur, as in a dual superconductor, through the (dual) Higgs mechanism. Rather, the Higgs mechanism is a {\it toy model} for modelling the phase in which the monopoles condense. Especially Mandelstam (1976) does not use the Higgs mechanism at all. A discussion of the evidence for the 't Hooft-Mandelstam proposal in QCD is in Shuryak (2021).}
(This does not happen in a type II superconductor, because the magnetic monopole solutions require non-abelian fields.) Thus the magnetic monopoles are the magnetic analogues of electrons in a Cooper pair: adding quarks, the electric colour charge is screened at large distances by a dual Meissner effect, and the colour-electric field lines are confined to a tube (a dual Abrikosov vortex) connecting a quark and an anti-quark, which acquire an attractive linear {\it electric} potential. 

The {\it magnetic (dis)order} of this state is not visible in terms of the electric order parameter (and this parameter indicates that there is electric disorder).\footnote{Recall that a type II superconductor is a case of mixed order: due to the Meissner effect, there is order in the superconductor's bulk, and disorder in the interior of magnetic vortices. Thus in the dual case, one expects magnetic order outside electric flux tubes due to the dual Meissner effect, and magnetic disorder in the interior.}
To make the magnetic order visible, the idea is to write an effective action for the {\it dual or magnetic order parameter}, which measures the magnetic order. 

One candidate order parameter for the magnetic phase is (by analogy with pairs of electrons in Cooper pairs) the expectation value of pairs of magnetic monopole and anti-monopole fields. Monopoles are topological configurations (we will discuss their topology in Chapter \ref{EMYM}) and, if they condense, the vacuum has a topological or magnetic order. By analogy with a type II superconductor, the magnetic order parameter has a non-zero value far from an electric flux tube, and is zero in its interior. And the electric field is screened far from the flux tube, and is non-zero in its interior. Thus in effect, the electric field is confined to the interior of the flux tubes.

To better understand this idea, let us rephrase it, using the analogy between superconductivity and the Ising model. In the ordered phase with broken symmetry, i.e.~with the double-well potential, the quadratic term has a negative coefficient $\a<0$, which as we see from the Ginzburg-Landau model corresponds to $T<T_{\sm c}$ (cf.~Eq.~\eq{alphaT}). In the Ising model, this is the ferromagnetic phase, where the magnetisation is non-zero (see Figure \ref{DisorderP}). But, for the reasons discussed above, this is the wrong phase, because if one realized confinement in this phase, it would be electric. Thus one seeks to describe confinement in the disordered, symmetric phase, where $\a>0$ and therefore $T>T_{\sm c}$. By analogy with the Ising model, one looks for a disorder parameter analogous to that in Eq.~\eq{Mtilde}, that describes the consensation of monopoles (as analogues of Cooper pairs in the type II superconductor): namely, as discussed, the expectation value of pairs of magnetic monopole and anti-monopole fields. By analogy with the superconductor, one expects that the electric field in this phase is confined by the dual Meissner effect, except in the interior of electric flux tubes, so that flux tubes of finite length end on quarks and have a linear potential.

Testing this idea analytically turns out to be extremely difficult (we will briefly discuss how it was verified analytically in the Seiberg-Witten theory, in Section \ref{moncond}). Therefore, one first resort is lattice gauge theory, where the idea can be verified. We will discuss this in the next two Sections: in the next Section, we first discuss how to define an order parameter that can be used in practice.

\subsection{Wilson loops as magnetic order parameters}\label{WilsonL}

Although one can find magnetic monopole solutions of gauge theories (see Section \ref{mmYM}), it is difficult to use them to construct order parameters of the type envisaged by 't Hooft and Mandelstam, i.e.~expectation values of pairs of monopole and anti-monopole fields. 

For this reason, it is more practical to first construct another order parameter that also indicates the disordered, symmetric, phase. Then one can try to see whether monopoles do contribute to its being non-zero. Such an alternative order parameter for confinement is the Wilson loop. 

Wilson's (1974:~pp.~2448, 2454) (now classical) criterion is that a system is in a confining phase if the expectation value of a Wilson loop enclosing a surface decays exponentially with the area of the surface. (This contrasts with the normal perimeter-law fall-off of a Wilson loop in a non-confining phase.) We say that the Wilson loop obeys an {\bf area law}:
\bea\label{PolWil}
\bra e^{i\oint{\sm d}x^\m A_\m}\ket=e^{-\g A}\,,
\eea
where $A$ is the minimal area of the surface whose boundary is the loop, and $\g$ is a constant that depends on the model's parameters. The contour integral is over a closed loop enclosing the surface with area $A$, and so using the single-valuedness of gauge transformations, the Wilson loop is gauge-invariant.\footnote{In the non-abelian case, one takes the group trace of the exponential, and also needs to take the path-ordering along the curve.}

Consider a pair of electric charges that travel a distance $R$ before they reunite at time $T$, where $A=RT$, so that their trajectories follow a rectangular Wilson loop. It follows from the area law of the Wilson loop Eq.~\eq{PolWil} that the effective potential between the charges is linear in the distance $R$, i.e.~$V(R)=\g R$, and the force is constant, $F=\g$. This means that the electric charges are confined, because the potential between them increases linearly with the distances.\footnote{This is only a sufficient, and not a necessary, condition for confinement. For a broader definition of colour confinement, see Greensite (2020:~p.~21) and footnote \ref{quarkC}.}
This is Wilson's criterion of confinement.\footnote{In the abelian case, the additivity law of the area implies that the magnetic fluxes through independent but adjacent surfaces is uncorrelated, which indicates the magnetic disorder in the confined phase: see Greensite (2020:~pp.~15-16).}\\

With this in hand, we return to the evidence for the 't Hooft-Mandelstam dual model of superconductivity, which uses the Wilson loop. This model is an idealization and, as an explanation of confinement in four-dimensional QCD, it is speculative.\footnote{In QCD, it seems well-established that monopole condensation {\it contributes} to the deconfinement phase transition, but the details are not settled. According to Shuryak (2021:~p.~68) (see also D'Alessandro et al.~2010), the evidence is consistent with two different analogies, both of which involve the condensation of monopoles: the condensation of pairs of monopoles and anti-monopoles as in a dual superconductor, and Bose-Einstein condensation of monopoles. Also, the flux electric tubes persist above the critical temperature up to at least $T=1.5\,T_{\sm c}$, which one would not expect in a pure dual superconducting model (Shuryak, 2021:~pp.~67-68, 403-404). Finally, the value of the parameter $\k$ which determines the type of superconductor (see Eq.~\eq{kappa} suggests the type we value (Bali, 2000:~p.~34; Shuryak, 2021:~p.~397). However, the profile of the electric field in the condensate is like that of the Abrikosov vortex of a type II superconductor (see Figure \ref{vortex-core}).} 

But there are two main cases where there is good theoretical evidence for it. The first case is a series of {\it lattice gauge models}, sometimes including their continuum limits, in three and four dimensions. Polyakov (1975, 1977) and Banks et al.~(1977) verified the area law, Eq.~\eq{PolWil}, in a number of models, including quantum electrodynamics with gauge group U(1), the SU(2) Georgi-Glashow model (i.e.~SU(2) Yang-Mills-Higgs: see Section \ref{tHPmon}), and other closely related lattice models.\footnote{Polyakov (1977:~pp.~447-449) used the Euclidean version of the model, so that the result for the Wilson loop is positive-definite. In that case, the Wilson loop is often interpreted as an {\it electric} loop, giving the effective potential between electrically charged particles. As the state for the calculation of the expectation value in Eq.~\eq{PolWil}, he chose the one-loop correction to a background of monopole-like solutions, similar to Eq.~\eq{HPmonopole}, which he called `pseudo-particles' (see Polyakov, 1977:~p.~439). Such solutions do exist for the three-dimensional Georgi-Glashow model with gauge group U(1). For the verification of the area law, and the calculation of $\g$ in SU(2) and SU(3) Yang-Mills theory in four dimensions using Monte carlo methods in lattice gauge theory, see Creutz (1980:~p.~314) and Shuryak (2021:~pp.~32-34).} 
Some results are rigorous, especially quantum electrodynamics in $D=3$, while other results have been disputed.\footnote{For a rigorous treatment of $D=3$ lattice QED, see G\"opfert and Mack (1982). For a critique of $D=4$ QED and the Georgi-Glashow model, see Greensite (2020:~pp.~122, 128-129). According to Bali (2000:~p.~21), confinement has been satisfactorily proven for (compact) U(1) quantum electrodynamics (with `Villain' action), the three-dimensional Georgi-Glashow model, and supersymmetric Yang-Mills theory.}

In these models, the area-law fall-off in Eq.~\eq{PolWil} is due to the interaction with a plasma of monopoles. The constant $\g$ is model-dependent and can be calculated in the different models (for example, in the Georgi-Glashow model, it is an analytic function of the parameters $e$, $\l$, and $v$ of the action, Eq.~\eq{LYMH}). 

Polyakov (1977:~pp.~444-445) calculated the {\it contribution of the monopoles} to the expectation value of the Wilson loop:\footnote{More specifically, Polyakov used the Debye, or mean-field, approximation for the monopoles, where the monopoles are well-separated, and he calculated the average contribution of the other monopoles to the magnetic potential between any two of them. This approximation has been criticised in Greensite (2020:~p.~128).} 
it takes the form of the partition function of a magnetic Coulomb gas, i.e.~a gas of monopoles that attract or repel each other by a magnetic Coulomb potential. Thus in this case, the mechanism can be verified.

To consider more realistic cases, especially $D=4$, one has to take into account the various phases.\footnote{In the three-dimensional Georgi-Glashow model, the effective potential is an analytic function of the parameters of the action Eq.~\eq{LYMH} (i.e.~$e$, $\l$ and $v$), and so there is a single phase.} 
Thus in our last Section, we will discuss more generally:

\subsection{Phases of gauge theories}\label{phasesGT}

By analogy with superconductivity, Section \ref{analogies} discussed phase transitions in gauge theories between the normal (i.e.~massless) phase and the Higgs (i.e.~massive) phase. And this Section has so far discussed the confining phase as the {\it magnetic dual} of the Higgs phase. We will now discuss the phases of gauge theories more generally, since this will play an important role in Chapters \ref{EMYM} and \ref{Heuri}.

The {\it phase} of a gauge theory is determined by the values of parameters such as masses, couplings, etc. And, although in a classical field theory, these numbers are fixed experimentally (as in, for example, Millikan's oil-drop experiment to measure the electron charge), in general in ``real life'' effective quantum field theories, these parameters are functions of other variables, such as an external temperature or the renormalization group scale.\footnote{An effective field theory can be obtained, using the renormalization group, by integrating out massive fields that are only visible at high energies. For a philosophical discussion, cf.~Crowther (2016:~pp.~64-71).} 
Thus, while exploring the phase diagram of a gauge theory is in part exploring the `space of possible gauge theories', it is also exploring the `space of possibilities for a single theory' under various external conditions.

One might argue that the renormalization group scale is an aid that we use to describe a system at some level of detail, and that it has no physical meaning. But this is not entirely correct, because while it is true that the scale at which we fix the renormalization scale is arbitrary, changing the renormalization group scale can have a physical correlate, e.g.~when we compare experiments that take place at different energies. The coupling parameters then change with the renormalization group scale (we say that the `couplings run'), and this change expresses the way in which experiments are able to probe different physics at different scales. Thus the running of the effective coupling expresses the way in which different physical processes take place at different scales.\footnote{For an introduction to the `modern approach' to renormalization, see Butterfield and Bouatta (2015:~pp.~456-474).}

These three phases, and the kinds of potentials between matter particles that they in general give rise to, can be summarised as follows:\footnote{See for example Fradkin and Shenker (1979:~p.~3684) and Greensite (2020:~pp.~13-14).}

(i)~~A {\it massive, or Higgs, phase,} where the potential $V(R)$ is of the Yukawa form, i.e.~up to an overall additive constant: $V(R)=-g^2\,e^{-m R}/R$ ($m$ is the gauge boson's effective mass). 

(ii)~~A {\it massless phase,} where $V(R)=-g^2(R)/R$, with $g(R)$ constant or logarithmic. In the rest of this discussion, we will assume that $g(R)$ is constant: the phase is then called the {\it Coulomb phase}.

(iii)~~A {\it confining, or magnetic disorder, phase,} where the potential is linear.

Fradkin and Shenker (1979) did the first systematic search for phases of gauge theories, and for illustration of these ideas we will give one of their examples. They considered abelian and non-abelian lattice gauge theories (including Higgs fields) of various types and in different dimensions.\footnote{For introductions to lattice gauge theory, see Aitchison and Hey (2013:~pp.~158-161), Rothe (2005), and Creutz (1983).} 

\begin{figure}
\begin{center}
\includegraphics[height=5cm]{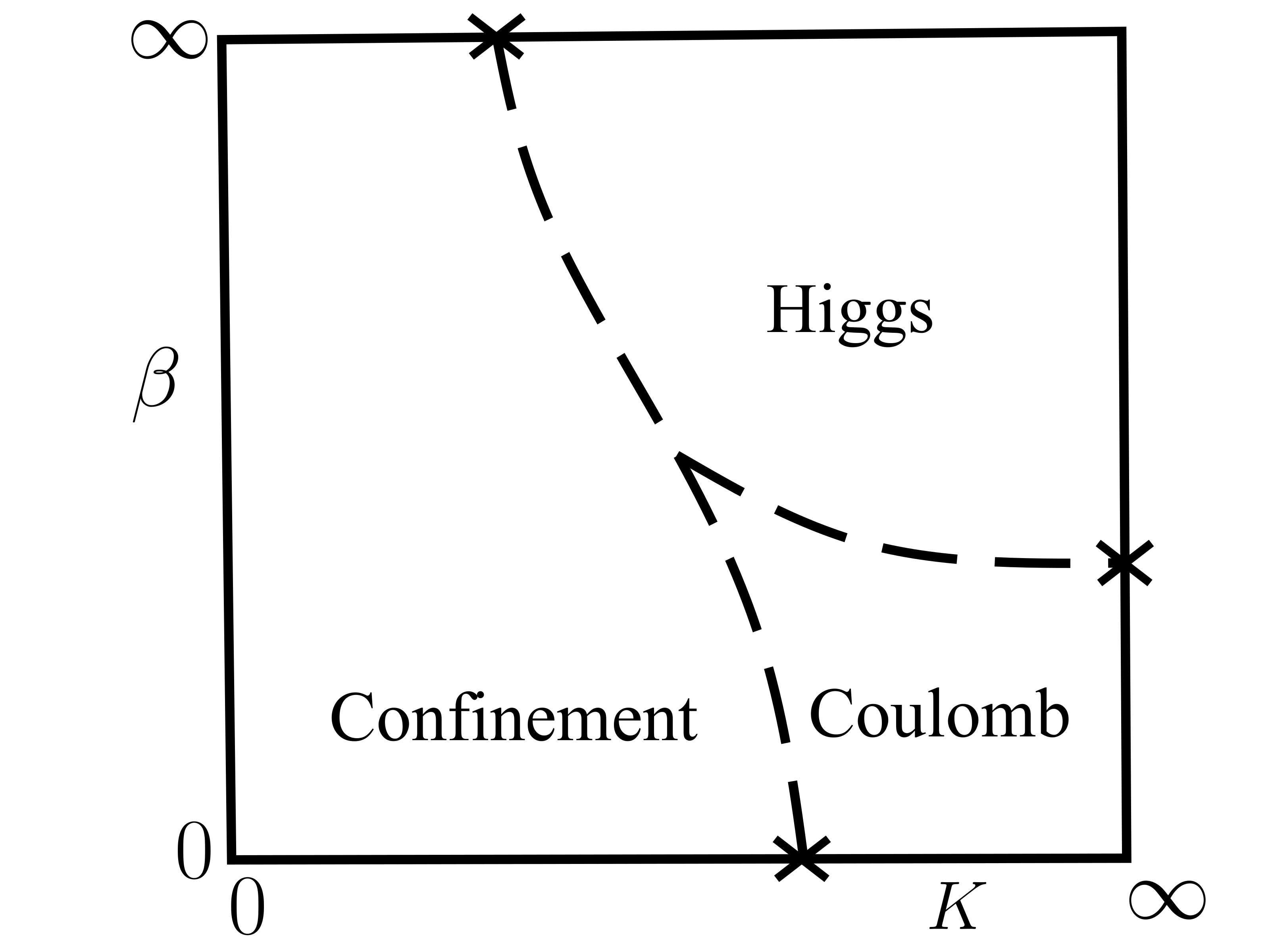}
\caption{\small Phases of the D=4 U(1) Georgi-Glashow model.}
\label{phasesbK}
\end{center}
\end{figure}

We will here focus on the {\it Georgi-Glashow model with U(1) gauge group}. It has two couplings, $\b$ and $K$, where $K$ is the (inverse) electric coupling $1/e^2$, and $\b$ is the inverse coupling of the Higgs part of the Lagrangian, which (as in the analogy with superconductivity) is analogous to the inverse temperature.\footnote{This diagram is for matter (Higgs) fields whose charge is higher than the fundamental electric charge, i.e.~$q>e$. If the Higgs field has the fundamental electric charge, i.e.~$q=e$, then there is no phase separating the Higgs and confining regimes, although there is a distinct Coulomb phase (thus there are a total of two phases). See Fradkin and Shenker (1979:~pp.~3691-3693). Notice that, when the Higgs field has the fundamental charge, the conventional condition for confinement (i.e.~the area-law of the Wilson loop) no longer works, because a particle-anti-particle pair of charged particles can be created in the vacuum, which then shield the gauge of the sources, so that the effective potential no longer rises. In other words, it becomes energetically favourable to break up a pair of electrically charged particles into sub-pairs (where the energy is provided by the Higgs field), without losing the confinement of electric charge.}
The phase diagram is given in Figure \ref{phasesbK}. It shows how, for various values of these parameters, we have one of the three phases, (i) to (iii).

In the Higgs phase (with $K>K_{\tn c}$ and $\b>\b_{\tn c}$) the gauge boson is massive, and electric sources interact through an exponentially damped Yukawa potential: thus the force is short-ranged, and the Wilson loop decays with the perimeter of the surface, rather than with the area.

In the Coulomb phase (with $K>K_{\tn c}$ and $\b<\b_{\tn c}$) the gauge boson is massless, and it gives a Coulomb force between electric charges: there is no confinement either. 

In the confining phase (with $K<K_{\tn c}$, and any $\b$), the gauge boson is massive, and all charges are confined, i.e.~electrically neutral. The Wilson loop decays exponentially with the area. 

In the $D=4$ U(1) Georgi-Glashow model, the explanation of confinement is analogous to the one we discussed in $D=3$, which only had a confining phase. For example, consider the transition between the confining and the Coulomb phase, i.e.~$\b$ low (see Figure \ref{phasesbK}). For $K>K_{\tn c}$, i.e.~small coupling $e<e_{\tn c}$, the monopole loops are small, and there are few of them. Beyond the critical coupling, the density of monopole loops rises with their size, and the vacuum of the theory is a gas of monopoles, analogous to what we discussed in the three-dimensional case. If electric charges are introduced in this vacuum, the electric flux between them has to circumvent the monopole-anti-monopole pairs. The greater the distance between the electric charges, the more difficult it is for the electric flux to do that. This gives a linearly rising potential between the electric charges, and the phenomenon is the dual of the Meissner effect.\footnote{For a discussion, see Banks et al.~(1977:~pp.~500-501).}

These results suggest that, in so far as monopoles play this role in confining electric charge, something like the {\it dual} Meissner effect is indeed the mechanism of confinement in these gauge theories.\footnote{The presence of monopoles was checked early on, using Monte Carlo methods, by DeGrand and Toussaint (1980:~p.~2481).}

Peskin (1978) gave a lattice derivation of the particle-vortex duality from Section \ref{pvdah}, including in four dimensions. He used a discretized version of (a cousin of) the Georgi-Glashow model. The free energy is a function $F(e,g)$ of two couplings, $e$ and $g$, which relate to our earlier couplings (up to numerical factors) as: $e=1/\sqrt{K}$ and $g=1/\b$. The duality relation takes the interesting form:
\bea\label{dualF}
F(0,g)=F({2\pi\over g},0)\,.
\eea
It relates points on the $K=\infty$ line to points on the $\b=\infty$ line of Figure \ref{phasesbK}. On the left-hand side, because the electric charge is zero, $e=0$, the gauge field is decoupled, and we have a pure Higgs model coupled to an external (i.e.~not dynamical) gauge field, analogous to Eq.~\eq{Llin}. On the right-hand side, where the Higgs coupling constant is zero, the Higgs field is decoupled and there is a pure gauge model with a new dynamical gauge field, analogous to Eq.~\eq{jvortex}. In doing this transformation, i.e.~in the process of integrating out the Higgs field to obtain the gauge field, topologically non-trivial configurations appear, with a topological current analogous to the vortex current, Eq.~\eq{vortexj}. In the four-dimensional case, this is a discrete monopole current: namely, the discretization of the conserved magnetic current (see the first equality in Eq.~\eq{magncur}). This current is not zero because of the presence of dislocations along the Wilson loop. Unlike the Dirac theory of monopoles, these monopoles are not put in by hand, but appear, as predicted by 't Hooft, as a result of the non-linearity of the initial Higgs theory. Comparing with Figure \ref{phasesbK}, the duality in effect relates the electric Higgs phase to a magnetic Coulomb-confinement phase.\footnote{Peskin (1978) considered Higgs fields with fundamental charge, and so the Higgs and confinement regimes belong to the same phase. One should keep in mind that, in the present duality, we are only relating to each other the $K=\infty$ and $\b=\infty$ {\it edges} of the diagram in Figure \ref{phasesbK}.} 
Peskin (1978:~p.~150) even concludes that this gives a `complete identification [sic] of a strong-coupling quark-confining gauge theory with a superconductor, along lines suggested earlier by Mandelstam and 't Hooft'.\footnote{Other approaches to confinement use instantons rather than monopoles. Polyakov's monopoles in three dimensions are instanton solutions, i.e.~solutions of the Euclidean field equations with finite action, and they were seen to be responsible for confinement in three dimensions. This naturally leads to the idea of instantons in four dimensions. Although standard instanton solutions do not seem to give confinement, because the field strengh falls off too rapidly at infinity, so that they cannot produce magnetic disorder in the vacuum, recently some researchers have considered {\it calorons}, i.e.~instantons at finite temperatures. See Greensite (2020:~p.~139), Kraan and van Baal (1998), Lee and Lu (1998).} 

Duality proves to be a fruitful principle that gives us insight into the phases of gauge theories, but we also see its limitations: in this case, in relating one regime of parameters to another (see Eq.~\eq{dualF}), it does not by itself give a calculation of this function.

\section{Conclusion}\label{conclusion6}

The {\it analogies} between superconductivity and the electroweak model can be used in both directions: in the forward direction, to suggest a mechanism that gives mass to gauge bosons; and in the opposite direction to prove, using quantum field theory techniques, the gauge invariance of spontaneous symmetry breaking in the BCS model. 

The analogy concerns details of physical behaviour, e.g.~about generating mass and obtaining a short-ranged Yukawa potential for the electroweak force, which is analogous to the London exponential decay of the magnetic field on the surface, characteristic of the Meissner effect. The analogy can be extended to type II Abrikosov vortices embedded as static solutions of a relativistic field theory whose time evolution is a relativistic string. Since the tension of a string grows linearly with the distance, this suggests an approach to confinement.

But this generalization requires a magnetic order parameter, and the new ingredient required is 't Hooft-Polyakov magnetic monopoles: one can in principle take the magnetic order parameter to be the density of monopoles. (We say `in principle' because this is not easy to do in practice, except in lattice simulations: the next Chapter will discuss the Seiberg-Witten theory, where this programme can be realized using analytic methods.) 't Hooft proposed to generate a {\it magnetic Higgs mechanism}, through an effective dual of the normal state of the Higgs model.\footnote{'t Hooft-Polyakov monopoles require a Higgs scalar field that, as in the Georgi-Glashow model, is in the adjoint representation of the gauge group, which QCD does not have. Also, these monopoles require breaking the symmetry of a magnetic gauge group, which in the Georgi-Glashow model is breaking from SO(3) to U(1); but doing this for SU($N$) requires a procedure called the `abelian projection', which is not well-understood (see D'Alessandro, 2010:~p.~1; Bali, 2000:~pp.~18,22). Nevertheless, as we have discussed, the mechanism {\it is} realized in QCD.}

The {\it logical and heuristic structure} of the 't Hooft-Mandelstam confinement argument, schematized in Figure \ref{dualSuperC}, involves all of the concepts previously discussed: solitons, phases of gauge theories, order parameters, analogies, and electric-magnetic quasi-duality. The next Chapter gives instantiations of electric-magnetic duality.

This Chapter also highlights the connection between our themes (2) and (4) in Section \ref{themesd}, {\it elementary-composite} and {\it symmetry-breaking}. Spontaneous symmetry breaking in the transition to a phase with a topological order is best described by an electric-magnetic duality transformation that exchanges particles and solitons. 

A novel theme in the physics of this Chapter is the phases of gauge theories with long-range topological order. This no doubt bears on questions of reduction and emergence that are usually thought to arise only in condensed matter physics. We can now understand why they are also relevant in particle physics:

\begin{quote}\small
unlike in previously existing elementary physical theories, it is not possible to reduce everything down to two-body interactions, but collective excitations of quark and gluon states have to be accounted for. For the first time, excitations of the vacuum that are considered to be fundamental do not occur as initial or final states... Therefore, {\it understanding} confinement, in my opinion, is one of the most exciting challenges of modern physics (Bali, 2000:~p.~17).
\end{quote}
By exploring the role of electric-magnetic duality in the analogies between condensed matter physics and quantum field theory, especially for the mechanism of confinement, this Chapter has aimed to illustrate the heuristic power of duality in theory construction. Although we will only return briefly to confinement in Section \ref{moncond}, the concepts that we have here adapted from condensed matter physics to gauge theories (especially phases of systems and their order parameters) are all-important. We will use them again, in the context of the Seiberg-Witten theory, in the next Chapter. Then Chapter \ref{Heuri} will further discuss the philosophical issues involved: especially the uses of duality in theory construction, reduction, and emergence.

\chapter{Electric-Magnetic Duality in Yang-Mills Theories}
\markboth{\small{\textup{Electric-Magnetic Duality in Yang-Mills Theories}}}{\textup{\small{Electric-Magnetic Duality in Yang-Mills Theories}}}\label{EMYM}

The aim of this Chapter is to formulate Montonen and Olive's (1977) {\it electric-magnetic duality conjecture}, and to test it in two examples of interacting four-dimensional quantum field theories.

The first example is the maximally supersymmetric Yang-Mills theory in four spacetime dimensions (`${\cal N}=4$ supersymmetric Yang-Mills', where ${\cal N}$ is the number of independent supersymmetry generators, i.e.~the number of ways there are to exchange a boson and a fermion and still ``get the same theory''). In this example, although there is no proof of the conjecture, there is good evidence that it is true. The duality maps the electric states to the magnetic states (and vice versa), and reproduces the Dirac quantization condition, Eq.~\eq{Diracq}. 

In supersymmetric Yang-Mills, the electric states are elementary particle states, and the magnetic states are solitonic (i.e.~collective) states, along the lines discussed in Chapter \ref{Advan}; so that Montonen-Olive duality indeed exchanges particles and solitons. In this example, we are able to illustrate, more explicitly than in the previous Chapter, that the duality map is of the type discussed in Chapter \ref{Schema}, i.e.~in terms of an isomorphism between the states and the quantities of the two models that is equivariant for the dynamics (for a number of important states and quantities). 

The second example is ${\cal N}=2$ supersymmetric Yang-Mills theory (which has half the amount of supersymmetry of the ${\cal N}=4$ theory), where the Montonen-Olive duality conjecture {\it fails}. The reasons for this failure are insightful, and lead us to establish an {\it electric-magnetic quasi-duality} between the theory's {\it low-energy models}. This leads in to an explicit realization (for the first time, for an interacting four-dimensional quantum field theory) of the ideas of Chapter \ref{EMDuality}, especially 't Hooft and Mandelstam's explanation of colour confinement through monopole condensation and the dual Meissner effect, which we will briefly touch upon.

The explanation of the mass gap and of colour confinement in theories of the Yang-Mills type, in particular QCD, is of course a major open problem in the standard model of particle physics. The two papers by Seiberg and Witten (1994a, 1994b) were epoch-making, because they derived these and other non-perturbative results in ${\cal N}=2$ supersymmetric Yang-Mills theory from a complete (i.e.~fully quantum) solution of the low-energy effective action. This suggested that an effective electric-magnetic duality plays a role in the explanation of confinement, and opened up new perspectives for understanding non-perturbative aspects of four-dimensional quantum field theories. 

Our discussion of the Seiberg-Witten theory, i.e.~the low-energy limit of ${\cal N}=2$ SYM, will focus on two (out of three) low-energy duals, and on the geometry of the space of states thus obtained. This will be one main example from which Chapter \ref{Heuri} will develop a geometric view of theories for quasi-duals, that goes beyond the Schema.

To lead up to these two examples, this Chapter will first introduce two topics whose physical significance makes them indispensable for philosophers of quantum field theory. First we will describe, in Section \ref{mmYM}, the {\it 't Hooft-Polyakov monopole solutions}, i.e.~solitonic solutions of the Yang-Mills-Higgs model with magnetic charge that is associated with a topological, rather than with a Noether, current. These solutions have many interesting and important properties: such as the emergence of an Abelian electromagnetic field from the monopole's non-Abelian core. In Section \ref{M-O}, monopoles play an important role in the Montonen-Olive duality conjecture.

Second, in Section \ref{basicsusy} we will review, at a conceptual but suitably (and minimally) quantitative level, the {\it basics of supersymmetry}: what it is, what its properties are, and how to construct supersymmetric Lagrangians using the superspace formalism. 

This allows us to discuss, in Section \ref{N=4SYM}, the ${\cal N}=4$ supersymmetric Yang-Mills theory and, in Sections \ref{effD} and \ref{moncond}, the Seiberg-Witten theory, whose features are key to the modern non-perturbative understanding of quantum field theories.\footnote{We use the phrase `non-perturbative' in the sense of `cannot be described (and, usually, not even seen) in perturbation theory'. For other uses of the phrase, see Shuryak (2021:~p.~1).} 

One important feature that we will discuss, and that suggests the conceptualisation of theories and of quasi-dualities that Chapter \ref{Heuri} will dub the `geometric view', is the notion of the {\it moduli space} of a theory. It is the space in which the vacuum states of fields (here, the Higgs field) take values. As we will discuss, this space has singularities at special points where monopoles become massless, and so cannot be left out of the effective low-energy description. Recall, from the previous Chapter, that at such points there is a change of phase:\footnote{There is a change of phase, or physical state, even if there may not be a canonical phase transition. Recall that the previous Chapter also mentioned cross-over of phases.}
the new phase can be described by a duality transformation. Since perturbation theory cannot describe these phenomena, non-perturbative methods are required.\footnote{The reason is that, at such points, physical quantities such as the free energy do not have a perturbative expansion: see Shuryak (2021:~p.~2).} 

\section{Magnetic monopoles in Yang-Mills theories}\label{mmYM}

Since electric-magnetic dualities relate electric and magnetic states, it is crucial to identify the states of non-abelian quantum field theories that are magnetically charged: these are magnetic monopoles. They are described by soliton solutions, i.e.~solutions of the non-linear field equations with finite energy. Their properties are interesting in their own right, and will recur in later Sections and Chapters: they satisfy the Dirac quantisation condition (cf.~Eq.~\eq{Diracq}), their mass is bounded by their magnetic charge, and there is a topological current associated with their magnetic charge (analogous to the currents in two and three dimensions, respectively in Eqs.~\eq{curr1d} and \eq{vortexj}).

Nielsen and Olesen's (1973) analogy between relativistic strings and Abrikosov vortices inspired 't Hooft (1975) and Mandelstam's (1976) idea, that confinement in gauge theories can be understood as dual superconductivity (see Section \ref{cmds}). On this view, magnetic monopoles should play an important role.\footnote{'t Hooft (1974b:~p.~276) motivated his own monopole solution with two arguments to the effect that it is impossible for the Nielsen-Olesen vortex to end on a monopole that is made of the same stuff. See also Polyakov (1974).} 
The 't Hooft-Polyakov monopole is a new kind of magnetic monopole: a non-linear soliton with no analogue in superconductivity.

\subsection{Motivating the 't Hooft-Polyakov monopole}\label{tHPmon}

As we discussed in Section \ref{Dqc}, Dirac's (1931, 1948) attempt to introduce monopoles in quantum electrodynamics entails certain oddities, like for example the famous string singularity that bears his name. Our study of solitons, i.e.~smooth solutions of the non-linear field equations with non-trivial topology, shows that solitons have magnetic fields that are quantized due to the non-trivial topology, and so suggests that solitons are natural candidates for magnetic monopole charge. 

The 't Hooft-Polyakov monopole realizes this expectation: at large distances, it looks like a Dirac monopole, but the Dirac monopole's singular ``core'' is replaced by a {\it smooth} configuration of the non-Abelian fields. The Dirac quantization condition arises from the soliton's non-trivial topology, given by the non-trivial boundary conditions of the fields and the properties of the gauge group.\footnote{Yang (1970:~p.~2360) was the first to discuss the role of the compactness of the gauge group in charge quantization.}

The 't Hooft-Polyakov solution also exhibits {\it spontaneous symmetry breaking}: the local SU(2) gauge symmetry is ``broken''  (in our sense of `symmetry breaking' from Section \ref{ssp}), by the non-zero value of the Higgs field, to a local U(1) that is naturally identified with the gauge symmetry of long-distance electrodynamics.

Before we introduce the 't Hooft-Polyakov monopole, we here list some additional motivations for considering this solution:

(1)~~{\it Electric-magnetic duality for interacting QFTs.} Regardless of confinement, magnetic monopole solutions are natural objects to construct if one wishes to extend {\it electric-magnetic duality} (from Sections \ref{EMduality} and \ref{MEMD}) to interacting four-dimensional quantum field theories, which is the main aim of this Chapter. 

(2)~~{\it Smooth soliton solutions.} Monopoles in non-abelian gauge theories have more realistic properties than the singular Dirac monopole, and so they could potentially give a more satisfactory explanation of Dirac charge quantization. For the 't Hooft-Polyakov monopole does not require us to introduce (somewhat ad-hoc) point-like sources: it is a smooth solution of the field equations. Its magnetic charge is spread over a region of space, and its mass is calculable and finite.\footnote{The mass of the 't Hooft-Polyakov monopole is due to self-energy. This contrasts with the Dirac monopole, where the mass of the monopole cannot be calculated in the theory.} 
Thus while Dirac's monopoles are {\it allowed} by quantum electrodynamics (i.e.~one introduces them by appropriate sources), and their existence implies the quantization of electric charge,\footnote{See Dirac (1931:~p.~68; 1948:~p.~818)} 
the 't Hooft-Polyakov monopole is a natural solution of Yang-Mills-Higgs models.\footnote{See Preskill (1984:~p.~471). By `existence', we here mean `existence as a solution of the model', and not `existence in our world'. Thus there is a possible world, described by the theory, where such solutions exist. This is unlike the Dirac monopole, where one needs to change the theory to accommodate the monopole solution.} 
And since this monopole is smooth, dynamical processes such as scattering between monopoles can also be studied.\footnote{See Manton and Sutcliffe (2004:~pp.~309-314).}

(3)~~{\it Colour confinement.} Magnetic monopoles play a central role in, what is believed to be, one of the most promising candidate mechanisms for the confinement of colour charge, or {\it colour confinement}. The 't Hooft-Polyakov monopole is a natural candidate object for monopole condensation.

\subsection{The Georgi-Glashow model}\label{GGmodel}

We would like to find monopole solutions, and to this end, we use a simple non-Abelian model. This is the Georgi-Glashow (1972, 1974) model with a gauge group of SU(2):\footnote{The name of Georgi-Glashow `model' indicates that it does not give a realistic description of nuclear physics. But the Montonen-Olive conjecture will relate two Georgi-Glashow models, so that these are also `models' in our sense from Chapter \ref{Thies}.}
it is a Yang-Mills-Higgs model where the Higgs field is in the adjoint, i.e.~three-dimensional vector, representation of SU(2).\footnote{This model is different from the Weinberg-Salam model that is the electroweak sector of the standard model (see Glashow (1959, 1961), Salam (1962) and Weinberg (1967)). In the standard model, the (minimal) Higgs is in the fundamental, i.e.~the defining complex two-dimensional, representation of the electroweak SU(2), rather than in a real vector representation, as in the Georgi-Glashow model.} 
This model is the natural non-Abelian analogue of the model considered in Section \ref{simplev}, whose solutions were magnetically charged vortices. 

The Lagrangian has three parts: the Yang-Mills Lagrangian, the Higgs Lagrangian, and the quartic (Mexican hat-type) potential. It looks as follows:\footnote{For an introduction to the Yang-Mills Lagrangian, see Hamilton (2017:~pp.~413-416).}
\bea\label{LYMH}
{\cal L}=-{1\over4}\,F_{\m\n}^aF^{a\m\n}+\half D_\m\F^aD^\m\F^a-V(\F)\,,
\eea
where the index $a=1,2,3$ runs over a basis of the Lie algebra of SU(2), and the Einstein summation convention is used for all the indices. The Yang-Mills field strength and the covariant derivative of the Higgs field are defined in the usual way:
\bea\label{FDPhi}
F^a_{\m\n}&:=&\pa_\m A^a_\n-\pa_\n A^a_\m+e\,\e^{abc}A_\m^b A_\n^c\nn
D_\m\F^a&:=&\pa_\m\F^a+e\,\e^{abc}A^b_\m\,\F^c\,.
\eea
The Levi-Civita $\e$-tensor appears here because both the gauge field $A$ and the Higgs field transform in the adjoint, i.e.~three-dimensional vector, representation of SU(2). In this representation, the infinitesimal generators $T^a$ of the gauge group\footnote{Recall that the infinitesimal generators of a Lie group are Lie algebra elements, i.e.~they parametrize the group elements near the identity element of the group. Exponentiating the infinitesimal generator with suitable coefficients gives the group elements.} are given by the structure constants of the corresponding su(2) Lie algebra, and the structure constants are given (with a suitable choice of basis for the adjoint representation of the Lie algebra) by the Levi-Civita tensor.\footnote{The expression for the generators in the adjoint representation is: $(T^a)_{bc}=-if^a{}_{bc}$, where $f^a{}_{bc}$ are the structure constants of the su(2) Lie algebra, given by the Levi-Civita tensor. In general, the index $a$ and the indices $bc$ are of different kinds: namely, $a$ labels the generator of the representation, and so it runs from 1 to the dimension of the Lie algebra (for $\mbox{SU}(N)$, this is $N^2-1$); while $b,c$ are matrix indices, running from 1 to the dimension of the representation (in the adjoint, this is by definition the same as the dimension of the Lie algebra). For the adjoint representation, these indices are of one kind, and the generators of the gauge group are given by the structure constants.}

Our potential is the Mexican-hat potential, which we parametrize as follows:
\bea\label{MexicanV}
V(\F)={\l\over4}\,\left(\F^a\F^a-v^2\right)^2\,.
\eea
It is of the form we used earlier, in the Abelian case in Eq.~\eq{VMex}, if we set $v^2=m^2/\l$. The Ginzburg-Landau form, Eq.~\eq{GLfreeE}, is obtained by setting $\a=-\l v^2/2$ and $\b=\l/4$. 

Recall that we are interested in {\it soliton} solutions of the equations of motion that follow from Eq.~\eq{LYMH}, and which spontaneously break the symmetry from SU(2) to U(1). As before, we require that the total energy of the soliton be finite. 

Splitting the four-dimensional index $\m$ into its time and space components, i.e.~$\m=(0,i)$ with $i=1,2,3$, the energy density is the 00-component of the canonical stress-energy tensor, given as follows:\footnote{The covariant stress-energy tensor is given in Harvey (1996:~p.~13). It is the non-abelian generalization of Eq.~\eq{Tmunu} that includes the contribution of the Higgs field.}
\bea\label{T00}
T_{00}={1\over2}\left(E^a_iE^a_i+B^a_iB^a_i+D_0\F^aD_0\F^a+D_i\F^aD_i\F^a\right)+V(\F)\,.
\eea
The non-Abelian electric and magnetic fields are given in terms of the curvature as in the abelian case (see Eq.~\eq{EBfields})
\bea\label{EaBa}
E^{ai}&=&F^{ai0}\nn
B^{ai}&=&\half\e^{ijk}F^a_{jk}\,.
\eea
Clearly, the above expression for the energy density is positive semi-definite, i.e.~$T_{00}\geq0$, as it should be. We first discuss, in Section \ref{vacsols}, the {\it vacuum state,} where the energy density is everywhere zero, $T_{00}=0$. Then, in Section \ref{monsol} we discuss the {\it monopole states,} which have non-zero energy.

\subsection{Vacuum solutions}\label{vacsols}

We see from the expression Eq.~\eq{T00} that the energy density is zero everywhere iff the following three conditions hold:
\bea\label{vacsol}
F^a_{\m\n}&=&0\nn
D_\m\F^a=\pa_\m\F^a+e\,\e^{abc}A_\m^b\F^c&=&0\nn
V(\F)&=&0\,.
\eea
Since the field strength is zero, the gauge field $A$ can be zero, but the Higgs field cannot be, since it must minimize the potential. We will call a solution satisfying these conditions a {\it vacuum solution}.\footnote{Solutions that satisfy the second and third equation, but not necessarily the first, are often said to define the `Higgs vacuum'. However, in so far as we are inerested in having zero energy density, we require that the three conditions are simultaneously satisfied. Below we will relax this and require only that Eq.~\eq{vacsol} is satisfied {\it asymptotically}, so that the total energy is finite but not zero: and then we still require that the three conditions are satisfied asymptotically. In other words, we will not use the notion of a `Higgs vacuum'.}

A simple example satisfying the three requirements above is for the Higgs to be {\it constant} and the gauge field to be zero, i.e.~$\F^a=v\,\d^a_3$ and $A^a_\m=0$. 

A solution with a constant Higgs field breaks the gauge symmetry from SU(2) to U(1): namely, those gauge transformations that fix the direction of the Higgs field (here, we think of the Higgs field as pointing in a particular direction, and the remaining U(1) corresponds to those gauge transformations that correspond to rotations around that direction). 

Notice that, since the condition that the energy is everywhere zero is gauge invariant, any solution that is obtained by gauge-transforming this constant solution is also a solution with zero energy, i.e.~it is a {\it vacuum solution}. Any constant solution that minimizes the potential satisfies $\F^a\F^a=v^2$, which (with $a$ taking values from 1 to 3) is the defining equation for a 2-sphere, and so (since $\F^a$ is in the vector representation of SU(2)) SU(2) can be locally identified with the three-dimensional rotations that leave the sphere invariant. Since any such constant solution is not invariant under the whole SU(2), but only under an $\mbox{SO}(2)\simeq\mbox{U}(1)$ sugroup that leaves the particular direction invariant, the symmetry is broken from SU(2) down to U(1).\footnote{As we discussed in Chapter \ref{EMDuality}, the gauge symmetry is not broken by the Higgs mechanism, but is merely {\it hidden}, in the low-energy model that we get by expanding about the vacuum solution. (For a discussion of spontaneous symmetry breaking, see Section \ref{ssp}: and for the Higgs mechanism, see Section \ref{analogies}.) We can {\it quantize} the model around such solutions, by considering small perturbations about a constant solution that minimizes the potential. The steps are similar to those discussed in Section \ref{analogies} for the analogous case of the Ginzburg-Landau model, and we will not go through the detail (see Goddard and Olive (1978:~p.~1383) and Figueroa-O'Farrill (1998:~pp.~19-20)). One finds that the spectrum of linearized fields contains a massless, abelian, vector field, a massive, neutral, scalar (namely, the fluctuation of the Higgs, with mass $m_{\tn{H}}=v\sqrt{2\l}\hbar$), and two massive vector bosons (with masses $m_{\tn W}=ve\hbar$ and charges $\pm e\hbar$). The massless vector field is gauge invariant with respect to the U(1) that is unbroken, and it has all the characteristics of the Maxwell electromagnetic model, so that it is a photon.} 

\subsection{The 't Hooft-Polyakov monopole solution}\label{monsol}

We have so far discussed classical solutions of the field equations that minimize (i.e.~have zero) energy. For solutions with {\it non-zero energy density}, the conditions Eqs.~\eq{vacsol} need not hold everywhere. However, the requirement that the total energy be {\it finite} implies that the local energy density must go to zero sufficiently rapidly at infinity, so that Eqs.~\eq{vacsol} hold {\it asymptotically}, at spatial infinity.\footnote{This is analogous to our analysis of Eq.~\eq{Venergy}, and especially of the connection, discussed for the abelian Higgs model after Eq.~\eq{Dphi}, between the asymptotic behaviour of the gauge field and the scalar field.}

Solutions with these properties can be readily found if we make the following two motivated assumptions: 

(1)~~We expect that the solutions with lowest energy are {\it static}, since, in addition to carrying potential energy, time-dependent solutions will have non-zero kinetic energy.

(2)~~Since large local variations would increase the energy, we expect that solutions with low energy are highly {\it symmetric}. In particular, one expects the lowest-energy solution to have the highest symmetry compatible with having non-zero topological charge.\footnote{For a detailed discussion of the desired symmetries, see Goddard and Olive (1978:~pp.~1384-1386), who parametrize the most general class of solutions, including with angular momentum, that satisfy these conditions. The solution given here is the 't Hooft's (1974b:~p.~280) original, where the angular momentum is zero. See also Manton and Sutcliffe (2004:~p.~253).} 
However, the solution cannot be rotationally invariant, i.e.~invariant under the spatial SO(3), because the vacuum expectation value of the Higgs field is non-zero: but it can be invariant under an SO(3) that is a combination of spatial rotations and gauge transformations.

We here give 't Hooft's original solution satisfying these assumptions:
\bea\label{HPmonopole}
\F^a&=&{r^a\over er^2}\,H(ver)\nn
A^a_i&=&-\e^a{}_{ij}\,{r^j\over er^2}\left(1-K(ver)\right),
\eea
where $r^a$ and $r^j$ are the components of the three-vector ${\bf r}$ in the $a$- or $j$-directions in space (both types of indices run from 1 to 3, and so they can be identified with the spatial part of a spacetime index). $H$ and $K$ are functions of the radial distance $r$ (and of the parameters $v$ and $e$). Their boundary conditions follow from the finiteness of the total energy.\footnote{The boundary conditions are: $K\rightarrow1$, $H\rightarrow0$ as $r\rightarrow0$; and, at infinity, $K\rightarrow0$, while $H$ grows linearly with $r$: $H\rightarrow ver$. Notice that $H$ and $K$ can only be given numerically.} 
The fourth component of the gauge field is zero, $A^a_0=0$, and so it follows (since the solution is static) that the electric field, $E^{ai}$, is zero (see Eq.~\eq{EaBa}). 

Substituting the solution, Eq.~\eq{HPmonopole}, into the magnetic field in Eq.~\eq{EaBa}, we find that the leading behaviour, for large $r$, is as follows: $B^{ai}={r^ar^i\over er^4}$. However, since the SU(2) symmetry is broken by the non-zero Higgs, there is no conserved non-abelian charge associated with this field. There is only an {\it abelian charge}. The result is, to leading order in $1/r$:
\bea\label{Bi}
B^i={r^i\over er^3}\,,
\eea
which is indeed the expected magnetic field for a monopole. Calculating the magnetic charge as the surface integral of this expression at infinity, one finds that the monopole solution has charge $g=4\pi/e$. 

The mass, $M$, of such a solution can in principle be calculated, by integrating the energy density, Eq.~\eq{T00}, over space. One can prove that it satisfies the following inequality, or {\bf Bogomol'nyi bound}: $M\geq |vg|$.\footnote{The case $M=|vg|$ holds iff the potential is zero, $V(\F)=0$ and the Bogomol'nyi equation holds: $B^a_i=D_i\F^a$. See Harvey (1996:~p.~19).} In the limit of small coupling $\l$, the 't Hooft monopole satisfies the minimum of the Bogomol'nyi bound, i.e.~$M=vg$.\footnote{The mass, for other values of the coupling $\l$, are given in Goddard and Olive (1978:~p.~1387).}

In evaluating the magnetic field of the 't Hooft-Polyakov monopole, we used the given boundary conditions for the functions $H$ and $K$ at infinity. These boundary conditions are required for the total energy of the monopole to be finite. By using the field equations that follow from the Lagrangian Eq.~\eq{LYMH}, one can calculate the sub-leading corrections to these functions, which turn out to be exponential in $-r/R_0$, where $R_0=\sqrt{2\l}/\hbar$, i.e.~for large but not infinite $r$, the functions $H$ and $K$ have their boundary values up to exponentially suppressed terms.\footnote{This is analogous to the analysis of the {\it size} of the core of an Abrikosov vortex, in Section \ref{analo2}, although the details are very different.} 

This means that, for $r\gg R_0$, the corrections are exponentially suppressed, and the magnetic field is, to a very good approximation, given by Eq.~\eq{Bi}, so that the electric field is zero and the magnetic field of the 't Hooft-Polyakov monopole in effect looks like a Dirac monopole.\footnote{This is only with respect to the behaviour of the Abelian gauge field corresponding to the unbroken U(1), since there are still non-Abelian fields that are non-zero.} 

However, when we reach $r=R_0$, the corrections are no longer suppressed, and Eq.~\eq{Bi} does {\it not} give a good approximation to the solution. In that case, the Abelian fields are not distinguished from the non-abelian ones, and we need to consider the full, non-linear, set of equations for the non-Abelian fields. This argument indicates that the length $R_0$ can be interpreted as the typical size of the 't Hooft-Polyakov monopole. For $r\gg R_0$, the monopole looks like a Dirac monopole. The region $0\leq r\leq R_0$ is the core of the soliton, where we have a smooth configuration of the interacting non-Abelian fields. Thus the vacuum equations, Eq.~\eq{vacsol}, hold, up to exponentially suppressed corrections, outside a region of size $R_0$.

At large $r\gg R_0$, $H(r)\sim ver$, and the Higgs field has the following asymptotic form (up to exponentially suppressed terms, which we will discuss in the next Subsection):
\bea\label{phia}
\F^a({\bf r})=v\,{r^a\over r}=:v\,\hat r^a\,,
\eea
sometimes called a ``hedgehog'' configuration, for obvious reasons. 

Notice that the Higgs field is taken to minimize the potential at infinity, i.e.~$\F^a\F^a=v^2$. This is the defining equation for a {\it two-sphere} of radius $v$, defined on the Lie algebra of SU(2). Since, for large $r$, $\Phi^a({\bf r})$ is a function of $\hat r:={\bf r}/r$, we can define the asymptotic value of the Higgs field as follows:\footnote{In usual spherical coordinates, $\hat r=(\sin\th\cos\vf,\sin\th\sin\vf,\cos\th)$.}
\bea\label{Finfinity}
\F^a_\infty(\hat r):=\lim_{r\rightarrow\infty}\F({\bf r})\in S_v^2\,.
\eea
Thus $\F^a_\infty$ is the value of the Higgs field at infinity, and as we discussed above it takes values on the two-sphere $S^2_v$ of radius $v$, defined by $\F^a_\infty\F^a_\infty=v^2$. Since $\F_\infty$ is itself only a function of $\hat r\in S^2_\infty$, i.e.~the sphere at infinity, $\F_\infty$ is a map between two spheres: 
\bea
\F_\infty:S^2_\infty\rightarrow S^2_v\,.
\eea
Rescaling the Higgs field by its modulus $v$, $\f^a:={1\over v}\,\F_\infty^a$, we get a map from the two-sphere at infinity to the two-sphere of {\it unit radius}:
\bea\label{spheremap}
\f:S^2_\infty\rightarrow S^2\,.
\eea
In the case at hand, i.e.~for Eq.~\eq{phia}, this ``hedgehog'' configuration is the identity map on the sphere:
\bea\label{idmap}
\f^a(\hat r)=\hat r^a\,.
\eea

Let us quickly anticipate the role of the topology of the Higgs field that will play a central role in determining the Dirac quantization condition later on. The hedgehog vacuum solution Eq.~\eq{idmap}, i.e.~the {\it identity map}, has winding number one (since it is a one-to-one map between the two spheres that preserves their topology), while the {\it constant map} of the previous Section mapped the sphere to a single point, and so it has winding number zero. Since an integer cannot change continuously, the ``winding number'' (more precisely, the degree of the map) is preserved by continuous transformations, and so the hedgehog configuration cannot be continuously deformed to the constant one while keeping the energy finite, i.e.~while preserving the minimum of the potential at infinity.

\subsection{The effective Abelian gauge field and the topological current}\label{eagf}

Let us better understand the nature of the effective U(1) gauge field associated with the monopole: we will argue that, far from the monopole, the magnetic field is in effect given by an {\it abelian gauge field}, and that the {\it Dirac quantization condition} arises from a topological property of the Higgs field. 

To understand the Abelian degrees of freedom of monopole configurations of this type, it is helpful to consider slightly more general solutions, where the vacuum equations Eq.~\eq{vacsol} indeed hold {\it outside some region of size $R_0$}, $r\gg R_0$, but the solutions do not necessarily describe a single monopole, i.e.~they are not of the form Eq.~\eq{HPmonopole}. Thus we drop the two symmetry assumptions, (1)-(2) at the start of this Section, but require finite total energy with its bulk localized in a region of size $R_0$.\footnote{For example, one could consider non-static solutions with several monopoles inside the region. For a brief discussion of the generalization to non-static solutions, see Goddard and Olive (1978:~p.~1388).} The middle equation determines the two ``longitudinal components'' (in the three-dimensional space of the Lie algebra) of the gauge field $A_\m^a$, i.e.~the components that are transverse to $\F^a$.\footnote{This is because the second term of this equation can be read as the vector product between $A$ and $\F$.} 
After some algebra, we can solve for them:\footnote{One multiplies the equation by $\e^{abc}\F^b$ and uses the value of the Higgs at the minimum of the potential.}
\bea\label{Aae}
A^a_\m&=&-{1\over ev^2}\,\e^{abc}\,\F^b\pa_\m\F^c+{1\over v^2}\,A_\m\F^a\,,
\eea
where we have defined $A_\m:=A_\m^a\,\F^a$, i.e.~it is the longitudinal component of the gauge field, along the direction of the Higgs field. Notice that, for the 't Hooft-Polyakov solution, Eq.~\eq{HPmonopole}, this longitudinal component is identically zero, i.e.~$A_\m=0$. 

The non-Abelian field strength, calculated from Eq.~\eq{Aae}, is given in terms of an Abelian antisymmetric two-tensor, as follows:
\bea\label{Fabelian}
F_{\m\n}^a&=&{1\over v}\,\F^a\,F_{\m\n}\nn
F_{\m\n}&:=&-{1\over ev^3}\,\e^{abc}\F^a\,\pa_\m\F^b\pa_\n\F^c+\pa_\m A_\n-\pa_\n A_\m\,.
\eea
The first equation implies that, in our approximation, i.e.~for $r\gg R_0$, the non-Abelian field is proportional to the Higgs field $\F^a$, i.e.~the direction of the unbroken U(1). In the second equation, the thus defined Abelian field strength contains two pieces, one associated with the gauge potential $A_\m$, and a contribution from the Higgs field. Although these two pieces are separately not gauge invariant under the original SU(2) gauge transformations, the field strength $F_{\m\n}$ {\it is} invariant.\footnote{For a more detailed discussion of gauge invariance, see Arafune et al.~(1975:~p.~435).}

The middle equation in Eq.~\eq{vacsol}, and the equation of motion for the Higgs, secure that this field strengh satisfies the equations of motion of a vector field in the absence of sources, i.e.~$\pa_\m F^{\m\n}=0$. However, $F_{\m\n}$ does {\it not} satisfy the Bianchi identity, Eq.~\eq{Max2}, because it is not an exact two-form. Thus, at large distances, an {\it effective} Abelian field arises, whose magnetic monopole charge is carried by a topological current. We can say that the Abelian electromagnetic field {\it emerges}, as a novel and robust long-distance phenomenon, from the non-abelian configuration in the interior of the monopole core (for a philosophical discussion of emergence, see Section \ref{emergence}).

We can define the magnetic current $J^\m_{\tn m}$ (see the analogous case in Eq.~\eq{elmagn}) in terms of the Hodge dual of the Faraday tensor (cf.~Eq.~\eq{Gdual}):
\bea\label{magncur}
J^\m_{\tn m}=\half\e^{\m\n\s\r}\,\pa_\n F_{\s\r}={1\over2ev^3}\,\e^{\m\n\r\s}\,\e^{abc}\pa_\n\F^a\pa_\r\F^b\pa_\s\F^c\,.
\eea
This is a {\it conserved magnetic current}, i.e.~it satisfies $\pa_\m J^\m_{\tn m}=0$, even though this conservation does {\it not} follow from Noether's theorem, nor does it generate a symmetry of the Lagrangian. The conservation of the magnetic current follows from its {\it definition}, regardless of the dynamics of the Higgs field.\footnote{For a discussion, see Arafune et al.~(1975:~p.~433).} Thus there is no Dirac string in the Maxwell equations, but there is nevertheless magnetic monopole charge coming from the contribution of the Higgs to the Abelian field strength Eq.~\eq{Fabelian}. We next compute this charge.

The magnetic flux is obtained by integrating the zeroth (i.e.~time-like) component of the current over space:
\bea
g&=&\int_V\dd^3x~J^0_{\tn m}={1\over2ev^3}\int_V\dd^3x~\e^{ijk}\e^{abc}\pa_i(\F^a\pa_j\F^b\pa_k\F^c)\nn
&=&{1\over2ev^3}\int_{S^2_\infty}\dd S_i~\e^{ijk}\e^{abc}\F^a_\infty\,\pa_j\F^b_\infty\,\pa_k\F^c_\infty\,,
\eea
where we used the identity $\e^{0ijk}=\e^{ijk}$. In the last line, we used Stokes' theorem to rewrite the flux as a boundary integral over the two-sphere $S^2_\infty$, i.e.~a two-sphere (or other closed two-surface) enclosing the region $r\leq R_0$ sufficiently far away out from this region, the volume vector $\dd S_i$ points in the outward-normal direction to the surface of integration, i.e.~it is proportional to the outward-pointing unit vector $n_i$. As in the previous Section, the notation $\F_\infty$ indicates that the integral only depends on the value of the Higgs field at infinity, defined in Eq.~\eq{Finfinity}, and its derivatives in the directions tangential to the surface $S_\infty^2$. 

Using the expression for the magnetic field, Eq.~\eq{Bi}, we reproduce the usual expression for the magnetic flux, Eq.~\eq{gcharge}, in terms of the magnetic field, which is now given by the asymptotic value of the Higgs field:
\bea\label{gmap}
g=\int_{S^2_\infty}\dd{\bf S}\cdot{\bf B}={1\over2e}\int_{S^2_\infty}\dd S_i~\e^{ijk}\e^{abc}\f^a\,\pa_j\f^b\pa_k\f^c={4\pi n\over e}\,,
\eea
where we have used the rescaled map, $\f$ in Eq.~\eq{spheremap}, which maps from the sphere at infinity to the {\it unit} sphere. 

It is not hard to show that $n$ is an integer: namely, the degree of the map, i.e.~the number of times that $\f$ covers the unit sphere, $S^2$, which must be an integer in order for the map to be single-valued.\footnote{In more detail:---the integrand in the last integral of Eq.~\eq{gmap} can be written as $n$ times the volume factor, $\sqrt{g}$, on the sphere. This gives the factor of $n$ times the area of the unit sphere, i.e.~$4\pi$, that is the result of the calculation in Eq.~\eq{gmap}. One can show this by introducing intrinsic, two-dimensional, coordinates $\xi_\a$ tangential to the sphere at infinity, and rewriting the integrand in terms of the unit normal vector to the surface, $n_i\,\e^{ijk}\pa_j\f^b\pa_k\f^c =\e^{\a\b}\pa_\a\f^b\pa_\b\f^c$, which is possible because $n_i$ is perpendicular to the sphere. In these two-dimensional coordinates, the product of the charges is $ge=\int\dd^2\xi~\half\e^{\a\b}\e^{abc}\f^a\pa_\a\f^b\pa_\b\f^c$. It is then readily checked that the square of the integrand in the last expression is precisely the Jacobian of the map, $g=\det(\pa_\a\f^a\pa_\b\f^a)$ (not to be confused with the magnetic charge $g$!), so that $ge=n\int\dd^2\xi\,\sqrt{g}=4\pi n$, where $n$ is the degree of the map, as in the main text (namely, $\xi$ covers the sphere once, while $\f$ covers it $n$ times). For more details, see Arafune et al.~(1975:~p.~434). For example, choosing spherical coordinates $\xi^\a=(\th,\varphi)$ and $\f^a=\hat r^a=(\sin\th\cos\varphi,\sin\th\sin\vf,\cos\th)$, one easily checks that the degree of this map is 1, and the value of the resulting integral is $4\pi$. An example of a map of degree $n$ is given in Goddard and Olive (1978:~p.~1390).} 
The degree of this map is called {\it Kronecker's index}. Higgs fields that are in the same homotopy class, i.e.~that can be continuously and smoothly deformed into each other, have the same Kronecker index, and thus the same magnetic charge.\footnote{The Brouwer degree of the map and the (sum of) Poincar\'e-Hopf indices give alternative characterizations of the Kronecker index. See Arafune et al.~(1975:p.~435) and Milnor (1965:~pp.~27-28, 51).} Since the vacuum manifold (in our case, the two-sphere, $S^2$) is the coset $G/H$, where $G$ is the gauge group and $H$ the unbroken gauge group, the Higgs field is a map to $G/H$, and so the relevant homotopy class is the relative homotopy class of maps $S^2\rightarrow G/H$, i.e.~$\pi_2(G/H)$. A theorem in homotopy tells us that, for a simply-connected gauge group, this is isomorphic to: $\pi_2(G/H)\simeq\pi_1(H)$. In our case, $G=\mbox{SU}(2)$ and $H=\mbox{U}(1)$, and so the homotopy class is $\pi_1(U(1))=\mathbb{Z}$, i.e.~the additive group of integers, as one expects from the Dirac quantisation condition.\footnote{A proof of the theorem is given in Goddard and Olive (1978:~pp.~1432-1433). This paper discusses other relevant aspects of the topology of the Higgs field, on pp.~1403-1420. See also Preskill (1984:~pp.~478-484).}

The magnetic flux $g$ is thus quantized, and it can be shown that it is additive under the union of regions with different charge. Notice that the quantization condition in Eq.~\eq{gmap} is $ge=4\pi n$: and {\it if} one puts $q=e\hbar$, this appears to differ from the Dirac quantization condition, Eq.~\eq{Diracq}, by a factor of 2. (In other words, the quantization condition of the non-Abelian Yang-Mills-Higgs model, with the assumption $q=e\hbar$, reproduces the Dirac quantization condition in Eq.~\eq{Diracq} only for even $n$.) As Goddard and Olive (1978:~p.~1390) advocate, this can be dealt with by putting $q=\half e\hbar$, which reproduces the Dirac quantization condition. For, although half-integer charges do not appear in the vector representation of SU(2) with isospin 1, this value would appear for fields in the fundamental, i.e.~two-dimensional, representation {\bf 2} of SU(2), with isospin $\half$. Fields in the fundamental representation do carry half the charge, and so the Dirac quantization condition for these charges requires the condition Eq.~\eq{gmap}. In other words, the apparent mismatch with the Dirac quantization condition is explained by the fact that the fields considered are in the vector representation.\footnote{For details, see Goddard and Olive (1978:~p.~1416).}

It is important to emphasize the topological nature of the magnetic current, Eq.~\eq{magncur}: the quantization of the charge depends only on the homotopy class of the Higgs field at infinity, i.e.~of the map $\f$.\footnote{For a simple proof, see Goddard and Olive (1978:~p.~p.~1389), who show that the magnetic charge is invariant under continuous deformations of the Higgs field, $\F^a\mapsto\F^a+\d\F^a$, subject to the asymptotic conditions Eq.~\eq{vacsol}.} Thus the conservation of magnetic charge, established by this current for a non-Abelian gauge model, does {\it not} originate from the ``Noether method'' applied to a continuous symmetry of the Lagrangian. Rather, it is a topological invariant of the mapping, $\f$, between the two spheres, i.e.~the sphere at infinity and a sphere in field space (at the minimum of the Higgs potential). Thus this topological invariant can only be defined near the minimum of the potential, i.e.~if the local gauge symmetry is spontaneously broken. 

The formal similarity with the topological current of the one-dimensional sine-Gordon model is not a coincidence: since, as Arafune et al.~(1975:~p.~434) argue, the number of scalars required to build a topological current in $d$ dimensions of space is $d$. Thus for $d=1$ we require one scalar, and for $d=3$ we require three scalars (i.e.~the three components of the Higgs field, $\F^a$).\footnote{We discussed some aspects of the $d=2$ case in Section \ref{pvdah} but the current found in Eq.~\eq{vortexj} was not a current like the one discussed here, but rather originated in a singularity of the scalar field---hence that discussion is slightly different from the present solitons, which are regular. Arafune et al.~(1975:~p.~434) also discuss models with fewer scalars, which can possess conservation laws on submanifolds.} 

\section{The Montonen-Olive duality conjecture}\label{M-O}

By explicitly constructing monopole solutions in a non-Abelian gauge model, and by characterizing monopoles---as against the Dirac monopole---as {\it solitonic} objects, the previous Section has laid the groundwork for this and the next Sections. In this Section, we formulate a general electric-magnetic duality conjecture for non-Abelian gauge theories. 

This conjecture was put forward, for the Georgi-Glashow model, by Montonen and Olive (1997). There are three chief elements of the 't Hooft-Polyakov monopole that their conjecture naturally adopts: 

(i)~~electric-magnetic duality exchanges conserved {\it electric and magnetic charges,} i.e.~it is a map of the type: $q\mapsto g$ and $g\mapsto-q$;\footnote{In Eq.~\eq{gmap}, $n$ is an arbitrary integer. The map $e\mapsto g$ and $g\mapsto-e$ maps the Dirac quantization condition to itself, with an arbitrary integer $n'=-n$.} 

(ii)~~{\it the Dirac quantization condition,} Eq.~\eq{gmap}, is invariant under this duality;

(iii)~~{\it the solitonic nature of the monopoles} is an important aspect of the duality: for the conserved currents, that correspond to the charges exchanged by the duality, are of different kinds. The electric charge corresponds to the {\it Noether current}, while the magnetic charge corresponds to the magnetic current, which is a {\it topological current} (see Eq.~\eq{magncur}).

Montonen and Olive (1977) conjectured that there exists {\bf electric-magnetic duality between two Georgi-Glashow quantum models}, based on Lagrangians of the same type: one with gauge group $G$, and the other with {\it dual gauge group,} $\hat G$ (see below); also the coupling constants are each others' duals.

Although it is known that the Montonen-Olive duality conjecture is incorrect, i.e.~that it is {\it not} true for the Georgi-Glashow model, or for the Yang-Mills-Higgs model, it is nevertheless an important and influential idea, for two reasons. 

First, because the Mandelstam-'t Hooft proposal for confinement, which is believed to play a role in QCD, is an approximate version of the proposed electric-magnetic duality. 

This will lead in to the idea of {\it effective electric-magnetic duality}, which could potentially apply to very many different theories. We will study in some detail the paradigmatic example of the Seiberg-Witten theory (Section \ref{effD}).

Second, there is very good evidence that the Montonen-Olive duality conjecture is {\it true} in ${\cal N}=4$ supersymmetric Yang-Mills theory and in string theory. \\

We will now describe the ingredients of the Montonen-Olive conjecture in more detail. The starting point is again the Yang-Mills-Higgs Lagrangian, Eq.~\eq{LYMH}, for an arbitray compact simple gauge group $G$.\footnote{Yang (1970:~p.~2360) was the first to argue from gauge invariance that, if all electric charges are integral multiples of a universal unit of charge $e$, then the gauge group is {\it compact}, i.e.~the parameter space of the generators of the Lie group is compact. Englert and Windey (1976:~p.~2730), Weinberg (1980:~p.~501), and Olive (1995:~p.~8) further assume that the group is {\it simple}, so that its coroot lattice is isomorphic to a root lattice (of the same, or of a different, Lie group): cf.~Humphreys (1972:~pp.~43, 58, 63). Recall that a simple Lie group is connected and has a simple Lie algebra: a simple Lie algebra is one that contains no proper ideals, i.e.~no proper subset of generators that is closed under commutation with all the generators of the Lie algebra. A semi-simple Lie algebra, i.e.~one that is a direct sum of simple Lie algebras, has a non-degenerate Killing form that establishes an isomorphism between the Cartan subalgebra and its dual. See Di Francesco et al.~(1997:~p.~491).\label{LieA}} 
All fields are in the adjoint representation of the Lie algebra, which is of rank $r$ (i.e.~the number of commuting generators) and dimension $d$ (i.e.~the total number of generators). Thus we replace, in the fields that appear in the Lagrangian (cf.~Eq.~\eq{FDPhi}), the structure constants of the SU(2) algebra by the general structure constants, i.e.~$e\,\e^{abc}\mapsto f^{abc}$, where the Lie algebra indices $a,b,c$ run from $1,\ldots,d$. The (non-Abelian) electric and magnetic fields of this model are, as before, defined through Eq.~\eq{EaBa}. 

As before, we require that the energy is finite. Any configuration with finite energy must have zero energy at infinity, and so the potential is zero there. This means that the (vacuum expectation value of the) Higgs field at infinity is Eq.~\eq{Finfinity}, i.e.~it lies on a $(d-1)$-sphere of radius $v$, where the value $v$ minimizes the potential (see Eq.~\eq{MexicanV}):\footnote{In so far as Montonen-Olive duality is a conjecture about the duality of {\it quantum} models, the {\it expectation value} of the Higgs field at infinity must minimize the quantum potential, i.e.~$\bra\F_\infty^a\F^a_\infty\ket=v^2$. We will discuss this minimization problem in Section \ref{effD}.} 
By choosing a vacuum expectation value, the symmetry $G$ is broken to a subgroup $H$. If $G=\mbox{SU}(2)$, this subgroup is $H=\mbox{U}(1)$ and it fixes the direction of the Higgs at infinity (see the discussion following Eq.~\eq{vacsol}).\footnote{This is also true for the asymptotic hedgehog solutions in Eq.~\eq{phia}, where the $\mbox{SO}(2)\simeq\mbox{U}(1)$ rotations do not lie in the $x^1-x^2$ plane (as in the case $\F^a=\d^a_3\,v$), but in the plane tangent to the two-sphere of radius $v$ at the point $v\hat r^a$).} For general $G$, there can be different ways in which the symmetry is broken, but this will not be important for our discussion,\footnote{For details, see Dorey et al.~(1996:~p.~423).} 
since we will take solutions where the scalar field picks out a specific U(1) subgroup, with Abelian field strength as in Eq.~\eq{Fabelian}, for which the electric and magnetic charges are given as follows:\footnote{Since, at infinity, the fields tend to their classical expectation values, to simplify the notation we remove the expectation values from expressions such as the one below, which involve the values of the fields at infinity.}
\bea\label{NAcharge}
q&=&\int_{S^2_\infty}\dd S_i\,F^{i0}={1\over v}\int_{S^2_\infty}\dd S_i\,\F^a E^a_i\nn
g&=&\int_{S^2_\infty}\dd S_i\,\half\e^{ijk}F_{jk}={1\over v}\int_{S^2_\infty}\dd S_i\,\F^a B^a_i\,.
\eea
Recall that, while $q$ is a Noether charge, $g$ is a topological quantum number (the expression for SU(2) is given in Eq.~\eq{gmap}). 

A first hint of electric-magnetic duality is in the generalization of the 't Hooft-Polyakov solution, Eqs.~\eq{HPmonopole}-\eq{Bi}, from SU(2) to general (simple compact) gauge group $G$.\footnote{Even though it is long, it is worth here sketching the argument, in four steps (we follow Weinberg (1980): see also Goddard and Olive (1978) and Olive (1995); for clarity of notation, we keep using $G$ for the {\it unbroken subgroup} of the gauge group). (i) We generalize the {\it 't Hooft-Polyakov monopole}, through the following asymptotic form of the non-Abelian magnetic field: $B_i^a={r_i\over4\pi r^3}\,G^a(\hat r)$, for some function $G^a$ from $S^2_\infty$ into the $d$-dimensional Lie algebra. Since the Higgs is covariantly constant at infinity (see Eq.~\eq{vacsol}), this function can be written in terms of a basis of generators of the Cartan subalgebra, $H_i$, of $G$, as follows: $\sum_{a=1}^dG^aT^a=\sum_{i=1}^rg_iH_i=:{\bf g}\cdot{\bf H}$, in vector notation where ${\bf g}$ and ${\bf H}$ are $r$-dimensional vectors, and $r$ is the rank of $G$. (For more on Lie algebras, see Di Francesco (1997:~pp.~489-503); Humphreys (1972:~pp.~15-72).) (ii) The {\it Dirac quantization condition} follows from the requirement that the magnetic field does not produce a Dirac string: $\exp(i{\bf g}\cdot\ti{\bf H})=\mathbb{1}_{d\times d}$, where $\ti{\bf H}$ is the vector whose components are the Cartan generators of the universal covering group, $\ti G$, of $G$ (rather than the Cartan generators of $G$, i.e.~the components of ${\bf H}$). This topological argument is given in Englert and Windey (1976:~p.~2729); see also Goddard and Olive (1978:~pp.~1415-1416). The relation between $G$ and its universal covering group, $\ti G$, is: $G=\ti G/k(G)$, where $k(G)$ is the kernel of the homomorphism $\ti G\rightarrow G$, which is a subgroup of $Z(\ti G)$, i.e.~the centre of $\ti G$. $\ti G$ is uniquely determined by the Lie algebra of $G$. (iii) The {\it general solution} of the Dirac quantization condition is obtained from Lie algebra theory: ${\bf g}=2\pi\sum_{i=1}^rm_i\, \hat{\mbox{\boldmath$\a$}}_i$, with $m_i\in\mathbb{Z}$, and $\hat{\mbox{\boldmath$\a$}}_i$ are the {\it duals} (in the sense for vector spaces!) of the simple roots, or {\bf coroots}: $\hat{\mbox{\boldmath$\a$}}_i:=2\mbox{\boldmath$\a$}_i/\mbox{\boldmath$\a$}_i^2$ (with no summation over $i$). This solution was first derived in Englert and Windey (1976:~pp.~2729-2730); cf.~also Bais (1978:~p.~1207). The total magnetic charge, Eq.~\eq{NAcharge}, does come out {\it quantized}. Goddard and Olive (1978:~pp.~8-9) notice that the set of dual roots $\hat{\mbox{\boldmath$\a$}}_1,\ldots,\hat{\mbox{\boldmath$\a$}}_r$ is defined only up to Weyl reflections. This equivalence class of magnetic coroots, defined by the action of the Weyl group, contains all the gauge-invariant information carried asymptotically by the expectation value of the magnetic field. (iv) The significance for {\it electric-magnetic duality} is that the coroots $\hat{\mbox{\boldmath$\a$}}_1,\ldots,\hat{\mbox{\boldmath$\a$}}_r$, whose coefficients give the magnetic charge (as in (iii)), can be seen as {\it roots of a dual group}, $\hat G$ (cf.~footnote \ref{LieA}). $\hat G$ is the group whose roots are the coroots of $G$, and whose global structure is given thus: $Z(\hat G)=Z(\ti G)/Z(G)$ (Olive (1995:~p.~8); Goddard et al.~(1977:~pp.~13-14)). An exceptionally clear description of the relations between a Lie group and its Langlands dual is in Gukov and Witten (2006:~pp.~168-175).}
One finds that the set of magnetic charges spans a lattice of {\it dual roots}, also called the {\it coroots of the Lie algebra} (the coefficients of the coroots are the magnetic charges),\footnote{The weight and coroot lattices are also dual lattices, and so the duality is sometimes stated in terms of the weight, rather than the root, lattice (see Goddard et al.~(1977:~p.~14)). Note that the roots are the weights of the adjoint representation, and that the root lattice is a subset of the weight lattice. For the relation between the various lattices for simple Lie groups, see Gukov and Witten (2006:~pp.~169-170).}
which can be seen as weights of the dual gauge group, $\hat G$, viz.~the {\bf Langlands dual group} or {\it Goddard-Nuyts-Osborn dual group}, such that {\it its} dual is the original group, i.e.~$\widehat{\hat G}=G$.\footnote{This is assuming an unbroken gauge group. If the gauge group is broken, one replaces $G$ by its unbroken subgroup $H$, and it is the Langlands dual of $H$, i.e.~$\hat H$, that matters. For the signficance of the geometric Langlands programme for quantum field theory and string theory, see Kapustin and Witten (2006), Kapustin (2008) and Schlesinger (2010).}
This suggests that the magnetic charges of the original group can be seen as the electric chages of a dual gauge group (and the electric charges as the magnetic charges of the dual): at least for the right kind of theory, where the quantities remain invariant under the duality map. For example, the dual of $\mbox{SU}(N)$ is $\mbox{SU}(N)/\mathbb{Z}_N$ and the dual of $\mbox{SO}(2N+1)$ is $\mbox{Sp}(N)$, while $\mbox{U}(N)$ is self-dual.

Thus the basic statement of the conjecture is that {\it the magnetic charges of the Yang-Mills-Higgs model with gauge group $G$, which are the coefficients of the coroots of $G$ (or its unbroken subgroup), are the electric charges of the dual model with gauge group $\hat G$, which are the roots of $\hat G$} (and likewise for the electric charges).

A first check of this conjecture concerns the mass of the lowest-lying electrically and magnetically charged states. One can show that states with finite energy satisfy (from Eq.~\eq{T00}) the {\bf Bogomol'nyi-Prasad-Sommerfeld (BPS) bound}. Namely, their mass is bounded from below as follows:\footnote{See Bogomol'nyi (1976:~p.~454), Prasad and Sommerfield (1975:~p.~762), Weinberg (1980:~p.~504), and Coleman et al.~(1977:~p.~545).}
\bea\label{bps}
M\geq v\sqrt{q^2+g^2}\,.
\eea
This spectrum is of course invariant under electric-magnetic duality. States that saturate, i.e.~minimize, the bound, are called `BPS-saturated' or, simply, {\bf BPS states}. These include all the single-particle states, including the solitonic ones: the photon, monopoles, dyons, the Higgs particle, and the heavy gauge particles.\footnote{The mass of the heavy gauge particles was calculated by Bais (1978:~p.~1209), who also showed that the masses of the monopoles and massive gauge particles map into each other under duality.}

The hints of duality that we have seen so far---the invariance of the spectrum of BPS-saturated states and of the mass formula under electric-magnetic duality, the fact that magnetic charges are naturally described in terms of the roots of a gauge group $\hat G$ that is the dual of the electric group $G$---are suggestive. They, together with the additional fact that the force between two well-separated monopoles can be calculated to have the expected Coulomb form (decaying as $1/r^2$, where the electric charge is repalced by the magnetic charge) motivated Montonen and Olive to conjecture that duality remains true when quantum corrections (e.g.~to the mass formula) are taken into account, and the bare parameters are replaced by the renormalized ones. They boldly conjectured that the monopole solutions are the elementary solutions of a dual model, with a Lagrangian of the same form as the original model, but with gauge group $\hat G$ and with the dual coupling. 

The duality conjectured by Montonen and Olive involved only what we call $S$-duality, i.e.~the exchange of the electric and magnetic charges and of the corresponding actions, as in Eq.~\eq{dadb}. It did not include a $\theta$-term, as in Eqs.~\eq{taumintau}-\eq{eth}. This term modifies the electric Noether current of the model, so that the admitted values of the electric charge are shifted by it (Witten, 1979:~p.~284). If the 't Hooft-Polyakov magnetic monopole has mass $M=4\pi/e$, the electric charge in this model is not an integer (or even rational) multiple of the elementary electric charge, but can have values: $q=ne-e\th/2\pi$. 

As we will see in the next Section, by electric-magnetic duality in models containing a $\th$-term we mean a duality that involves the $\theta$-term, as in Eq.~\eq{taumintau} (in fact, the electric-magnetic duality of non-Abelian gauge theories naturally generalizes the duality of the Abelian model considered in Section \ref{quantumD}). In this case, the {\it electric-magnetic duality group}, i.e.~the orbit under the duality map, is not $\mathbb{Z}_2$ but $\mbox{SL}(2,\mathbb{R})$ (which is broken by quantum effects to $\mbox{SL}(2,\mathbb{Z})$, which in most cases is the duality group of the quantum model).

\section{A brief overview of supersymmetry}\label{basicsusy}

Before discussing electric-magnetic duality in supersymmetric Yang-Mills theories (Section \ref{N=4SYM}), this Section gives a conceptual and qualitative overview of basic notions of supersymmetry. This level of understanding is enough for our purposes in this Chapter and the rest of the book, but the interested reader can find more details in the many excellent reviews and books on the subject. (As in the rest of the book, we aim to quote the relevant sources where details are found.)\footnote{A helpful guide to supersymmetry is Ferrara (1987:~Chapters I-III); see also Sohnius (1985) and Wess and Bagger (1992). An introduction to supersymmetric Yang-Mills theory is in D'Hoker and Phong (1999).} \\
\\
{\bf Supersymmetry} is a symmetry that relates bosonic and fermionic states (usually, by relating their corresponding fields). It is an extension of ordinary Lie algebras, such as the Poincar\'e algebra (generated by the spacetime displacement operators, $P_\m$, and Lorentz transformations, $M_{\m\n}$), by the addition of Grassmannian, i.e.~anticommuting, generators.\footnote{Supersymmetry in four dimensions was discovered by Wess and Zumino (1974a:~pp.~39-40). For some history of the field, see Ferrara (1987). For a basic introduction to Grassmann algebras, see Berezin (1966:~pp.~49-77).} 
The algebra one obtains is a graded Lie algebra, i.e.~an algebra that formally satisfies the properties of an ordinary Lie algebra (including the Jacobi identities for the structure constants), but where the Lie algebra product between two algebra-valued elements (in the adjoint representation, their commutator) is {\it graded}, i.e.~it is a commutator if it contains either two bosonic quantities or one fermionic and one bosonic quantity, and it is an anti-commutator if it contains two fermionic (i.e.~Grassmann-valued) quantities.\footnote{The graded commutator is often denoted by $[,\}$, and it is defined as follows: $[X_A,X_B\}:=X_AX_B-(-)^{ab}X_BX_A=f^C_{AB}X_C$. Here, the numbers $a$ and $b$ have the value 0 if the corresponding quantity is bosonic, and 1 if it is fermionic. Thus $(-)^{ab}$ equals $-1$ (so that one gets an anti-commutator) only if $a=b=1$, so that both $X_A$ and $X_B$ are fermionic. Otherwise, one gets a commutator. The generalized structure constants, $f^C_{AB}$, satisfy a number of relations that generalize the Jacobi identities. See e.g.~Ferrara (1987:~pp.~3-4).} Thus the supersymmetry algebra is a (usually unique, up to central charges that we will discuss) graded extension of the Poincar\'e algebra, also called the super-Poincar\'e algebra.\footnote{In apparent contradiction of Coleman and Mandula's (1967:~p.~1251) theorem proving `the impossibility to combine spacetime and internal symmetries in any but a trivial way', i.e.~so that the total group is locally isomorphic to the direct product of an internal symmetry group and the Poincar\'e group, supersymmetry effects a fusion of internal and spacetime symmetries (Coleman and Mandula's work generalized O'Raifeartaigh (1965)). The no-go theorem is circumvented because the algebra is not a Lie algebra, but has anti-commuting generators not considered by Coleman and Mandula (see Wess and Zumino (1974b:~p.~54)). Thus the irreducible representations of the super-Poincar\'e algebra combine fermions with bosons (see Salam and Strathdee (1974:~p.~35)).} 
If we denote by $Q$ the conserved charge that generates the supersymmetry algebra (see below), we can write, symbolically:
\bea\label{susyQ}
Q\,|\mbox{boson}\ket=|\mbox{fermion}\ket~,~~Q\,|\mbox{fermion}\ket=|\mbox{boson}\ket\,.
\eea

The model is supersymmetric if the action is invariant (up to a total derivative term) under the transformations of the supersymmetry algebra. In four spacetime dimensions $D=4$, this implies that the dimensions of the bosonic and fermionic state spaces are the same. (In the rest of this Section, unless we mention that $D$ is general, we discuss the $D=4$ case).

In general, fermions are represented by Grassmann-valued variables (usually fields) that transform in half-integral representations of the Poincar\'e group,\footnote{For models without gravity, only spins up to one are allowed, and fermions can have spin $s=\half$ or $s={3\over2}$: see Ferrara (1987:~p.~33).} 
and so the number of components of such fields is given by the corresponding representation. For example, an irreducible Dirac (i.e.~complex) spinor in $D$ dimensions, for even $D$, has $2^{D/2}$ components ($2^{(D-1)/2}$ if $D$ is odd). This number grows exponentially, and to match the number of fermions with the number of bosons, additional conditions are sometimes imposed on the fermions, which reduce the number of independent components. For example, a Majorana spinor is a real spinor, while a Weyl spinor is a chiral spinor (left-handed or right-handed). Also, to match the number of components, bosonic fields (typically scalar fields) often need to be added.

Let us briefly consider our case of interest, the SU(2) Yang-Mills model in four dimensions, in the simplest possible setting. Thus take the pure Yang-Mills action, and add to it a single (Weyl or Majorana) spinor. In four spacetime dimensions, each gauge boson has two helicity states, while a standard Dirac spinor has four complex degrees of freedom (corresponding to a spin-$\half$ particle, together with its anti-particle): imposing the Majorana condition reduces this by half, so that there are only the two independent helicities of a spin-$\half$ particle, i.e.~two states.\footnote{Just as for bosons, the counting is not completely straightforward, in that the four independent components of a Majorana spinor correspond to only {\it two} helicity states, as befits a spin-$\half$ particle. Also, in four dimensions, the Majorana and Weyl conditions cannot be imposed simultaneously, on pain of getting a spinor that is equal to zero.} 
The action is:
\bea
S=\int\dd^4x\left(-{1\over4}F^a_{\m\n}F^{\m\n a}+i\bar\l^a\g^\m D_\m\l^a\right),
\eea
where $\g^\m$ are the four-dimensional gamma matrices.\footnote{For details, see Brink et al.~(1977:~p.~78).} This action is {\it invariant} under the following infinitesimal {\bf supersymmetry transformations}:
\bea\label{susy}
\d A^a_\m&=&i\bar\ve\g_\m\l^a-i\bar\l^a\g_\m\ve\nn
\d \l^a&=&\s_{\m\n} F^{\m\n a}\ve~~\mbox{(similarly for $\bar\l$),} 
\eea
where $\s^{\m\n}:={1\over4}[\g^\m,\g^\n]$ is the antisymmetrized product of two gamma marices. The parameter $\ve$ is a constant anti-commuting spinor, and it is the infinitesimal parameter that generates the supersymmetry transformation (it is infinitesimal, since we are here working with the algebra, rather than the group). The bar denotes the adjoint and, in addition, right-multiplication by the charge conjugation matrix.\footnote{We have: $\bar\ve=\ve^\dagger C$ and $C$ is the charge conjugation matrix, where one can choose $C=\g^0$.}

Under a suitably defined product of two supersymmetry transformations, supersymmetry transformations form a closed algebra. For Lie algebras, the product on the algebra is given by the Lie bracket, which in the adjoint representation is represented by the commutator of Lie algebra elements. The graded Lie algebra consists of the usual generators of the Lie algebra plus, in addition, a number of fermionic {\it supercharges}, $Q$, that act on fields through the (graded) commutator. The relation between the supercharge and the infinitesimal supersymmetry transformations\footnote{For the exponentiation of the Grassmanian generators of the super-Poincar\'e algebra, to form group elements, see Ferrara et al.~(1974:~p.~240).} defined above, in Eq.~\eq{susy}, is as follows:\footnote{See Haag et al.~(1975:~p.~261), who also shows how to derive, using Noether's theorem, the supersymmetry charge $Q$ from a local conserved (Grassmann and vector-valued) supercurrent. See also Wess and Zumino (1974b:~p.~54) and Ferrara (1987:~p.~71).} 
$\d_Q(\psi(x))=i[Q,\psi(x)\}$. For the graded Lie algebra to close, one defines the (anti-)commutator of the supercharge $Q$ not only with itself, but also with all the bosonic generators of the Lie algebra. The algebra of the supercharges takes the following schematic form (suppressing indices):
\bea\label{susyZ}
\{Q,Q\}=Z\,,
\eea
where $Z$ are central elements of the algebra, called the `central charges', i.e.~they commute with all the generators of the graded Lie algebra.\footnote{See Haag et al.~(1975:~p.~259) for the rationale of how to construct a supersymmetry algebra, its various terms, and the conditions they must statisfy.}\\
\\
{\bf ${\cal N}$-Extended supersymmetry}\\
\\
The supersymmetry algebra can be enlarged by the addition of more spinors, with the corresponding supersymmetry generators $\e_i$ ($i=1,2,\ldots,{\cal N}$) that relate each boson state to ${\cal N}$ independent fermion states. This is called ${\cal N}$-{\it extended supersymmetry}, where ${\cal N}=1$ is the case of simple supersymmetry just discussed, i.e.~each boson corresponds to one fermion. For ${\cal N}=2$, each boson $A^a_\m$ is related to two independent fermions, $\l^a_1$ and $\l^a_2$. For field theories without gravity in four dimensions, the maximum is ${\cal N}=4$.\footnote{A useful way to think of these different supersymmetry algebras is in terms of compactification from higher to lower dimensions, where larger spinors split into smaller spinors, thereby raising ${\cal N}$ (recall that, for even $D$, a Dirac spinor has $2^D$ components). Additional scalar fields arise from the components of the gauge field along the compactified dimensions. For example, the ${\cal N}=2$ supersymmetric Yang-Mills (SYM) model in $D=4$ is obtained from the ${\cal N}=1$ model in $D=6$ (where two scalars in $D=4$ arise from the components of the gauge field along the two compactified dimensions), and the ${\cal N}=4$ SYM model in $D=4$ is obtained from the ${\cal N}=1$ model in $D=10$ (with six scalars that correspond to the components of the gauge field along the six compactified dimensions). See Brink et al.~(1977:~pp.~83-90).} 

A {\bf multiplet} of the supersymmetry algebra, or {\it supermultiplet}, is the set of fields that are related by the supersymmetry transformations, as in Eq.~\eq{susy}.\footnote{See Wess and Zumino (1974c:~pp.~3-5). A useful reference for the properties of the various multiplets is D'Hoker and Phong (1999:~pp.~11-14).} 
In the ${\cal N}=1$ case reviewed above, this is called the {\bf ${\cal N}=1$ vector multiplet}, and it contains just two members: namely, the gauge field $A$ and the Majorana spinor $\l$, with a total of four states (i.e.~two helicities states of the gauge field, $\pm1$, and two helicity states of the Majorana fermion, $\pm\half$). 

If we wish to construct an ${\cal N}=1$ supermultiplet whose bosonic content is a scalar field, then this scalar must be complex (so that it has two real components), since the smallest spinor that one can form in four dimensions has two independent components, i.e.~two states, rather than one (assuming Lorentz invariance and PCT-symmetry). In this way, one obtains the {\bf ${\cal N}=1$ chiral multiplet}, with a complex scalar and a chiral spinor.\footnote{Alternatively, one can consider a supermultiplet consisting of a real scalar, a pseudo-scalar, and a spinor field. Wess and Zumino (1974b:~pp.~52-53) also considered the renormalizability of this model, and found that some of the divergences cancel. By adding auxiliary fields, the model is, to one-loop order, fully renormalized by a single, logarithmically divergent, wave-function renormalization.}

We can form an {\bf ${\cal N}=2$ vector multiplet} by coupling an ${\cal N}=1$ vector multiplet and an ${\cal N}=1$ chiral multiplet. We have a total of eight states: two helicity states of the gauge field, two real states of the complex scalar field, and four states of the Dirac fermion.

The {\bf ${\cal N}=4$ vector multiplet} has a gauge field, six real scalars, and four Majorana spinors.\footnote{See Osborn (1979:~p.~322). One easily checks that the number of helicity states matches. For the bosons, it is a total of eight: two helicity states for the gauge field and six for the scalar fields. For the fermions, it is also eight: four spinors with two helicity states each. It can also be seen as an ${\cal N}=2$ vector multipled coupled to an ${\cal N}=2$ hypermultiplet.} 
The bosonic part of the action is again of the Yang-Mills-Higgs form, Eq.~\eq{LYMH}, for a set of six real scalars $\F$, where the potential is quartic, but not of the Mexican hat form Eq.~\eq{MexicanV} (i.e.~$v=0$), and it involves not just a single scalar, but quartic interactions among the six scalars. \\
\\
{\bf Superspace and superfields}\\
\\
A very efficient way to write supermultiplets that automatically satisfy the supersymmetry algebra, and can be used to construct supersymmetric Lagrangians, is using superspace, with superfields living on it.\footnote{Superspace was first constructed by Salam and Strathdee (1974), who showed how to systematically construct fields that correspond to various representations of the supersymmetry algebra on superspace (scalar, vector representations, etc.). For a philosophical study of superspace, see Menon (2021).} 
{\bf Superspace} is an extension of ordinary four-dimensional Minkowski space that adds four coordinates that are Grassmann variables (i.e.~four c-number Majorana fermionic variables), $\th$. A supersymmetry transformation is then an element of a group of motions in superspace, $\th'=\th+\ve$.\footnote{In addition, the ordinary Minkowski coordinates also change, $x'_\m=x_\m+i\bar\ve\g_\m\th$, but we will not need these details.} The supersymmetry generators (the charges) $Q$ can thus be represented as translations in superspace, $-{\pa\over\pa\th}+i\g^\m\th{\pa\over\pa x^\m}$, in much the same way as the generators of translations and Lorentz transformations, $P_\m$ and $M_{\m\n}$, are represented by differential operators acting on fields. Superfields thus give representations of the super-Poincar\'e group on superspace.\footnote{For details about the particle content of the super-Poincar\'e algebra, for massless and massive representations, see Ferrara (1987:~pp.~32-34).}

Superfields have monomial expansions, i.e.~in terms of the superspace coordinate $\th$ and powers of it, that, due to the Grassmannian nature of $\th$, terminate at the fourth order.\footnote{Since $\th$ are four Grassmann variables, any monomial $\th_1\th_2\cdots\th_n$ (i.e.~products of different $\th$'s) with $n>4$ is zero.} Thus the Taylor series in powers of $\th$ of e.g.~a scalar superfield $\F(x,\th)$ stops at the fourth order:
\bea
\F(x,\th)=\f_0(x)+\th\,\f_1(x)+\ldots+\th^4\f_4(x)\,,
\eea
where the five coefficients of this sequence, $\f_0(x),\ldots,\f_4(x)$, are (built out of) ordinary spacetime fields, both bosonic and fermionic, and so contain the physical content of the model. These terms must have the correct statistics, so that $\f_0$ is a bosonic quantity (a scalar field), $\f_1$ is Grassmann-valued, etc. Also, one can see from the supersymmetry algebra that the dimension of $\th$ is $\half$.\footnote{By `dimension', we here mean the engineering dimension i.e.~the inverse length, $1/L$, also called the `mass dimension', in conventions where $\hbar=c=1$. Thus $\pa_\m$ has dimension one.} Thus the various components have different dimensions. For example, the dimension of $\f_4$ is the dimension of $\f_0$ plus two.

Superspace is often conveniently written in terms of chiral spinor coordinates, $\th$ (where $\th$ is chiral, and $\bar\th$ is anti-chiral). This allows the construction of {\bf chiral superfields} with complex scalar fields. Thus take $\th$ to be a set of four c-number spinors, and impose a chirality condition, so that two of the $\th$'s have positive chirality, and the other two have negative chirality (and we denote the latter by $\bar\th$).\footnote{A chirality condition on a spinor $\psi$ is $\g_5\psi=\psi$, whereas an anti-chirality condition is $\g_5\psi=-\psi$, where $\g_5=\g_1\g_2\g_3\g_4$.} 
Then we can define $\F(x,\th)$ to be a chiral superfield, which depends on only two $\th$'s, by requiring that it is independent of the negative chirality variables, $\bar\th$'s, i.e.~we require that: $\bar D_\th\F=0$. The Taylor series of this chiral field then only has three terms:\footnote{We are slightly simplifying here, in that the fields below are functions of $y^\m=x^\m+i\th\s^\m\bar\th$ rather than $x^\m$. However, it is clear how to go back and forth between the two types of expressions.}
\bea\label{chiralF}
\F(x,\th)=\f(x)+\th\sqrt{2}\psi(x)+\th^2F(x)\,,
\eea
where the $\sqrt{2}$ is conventionally chosen. Here, $\f$ is a complex scalar field, $\psi$ is a fermionic field, and $F$ is an auxiliary scalar that can be set equal to zero by its equation of motion (see below; this must be so, because the minimal supermultiplet has only one dynamical scalar). Under supersymmetry, an arbitrary superfield $\F$ transforms as $\d\F=(\ve Q+\bar\ve\bar Q)\F$, where the supercharges $Q$ and $\bar Q$ are represented by translations in the Grassmannian variables (as above).\footnote{See Wess and Bagger (1992:~p.~29).} 
From here, one readily reproduces the ordinary (i.e.~non-superspace) form of the supersymmetry transformations, see e.g.~Eq.~\eq{susy}. Thus, in effect, the physical content of a chiral superfield is that of an ${\cal N}=1$ {\it chiral multiplet}.

The supersymmetric action for the scalar field $\F$ of Eq.~\eq{chiralF} is constructed by integrating over the whole superspace:
\bea\label{N=1scalar}
S_{\tn{scalar}}^{{\cal N}=1}={1\over4}\int\dd^4x\,\dd^2\th\,\dd^2\bar\th~\F^\dagger\F\,.
\eea
This is the simplest real action that one can construct with the chiral superfield $\F$. The integral over the Grassmann variables is readily carried out after using the expansion of the field, Eq.~\eq{chiralF}. The result contains the kinetic term of the scalar field as its first term:
\bea\label{1scalar}
S_{\tn{scalar}}^{{\cal N}=1}=\int\dd^4x\left(\pa_\m\f\pa^\m\f^\dagger-i\bar\psi\bar\s^\m\pa_\m\psi+F^\dagger F\right).
\eea
$F$ has no kinetic term,\footnote{This is determined, by its engineering dimension, from the place where it appears in the expansion of the chiral superfield, Eq.~\eq{chiralF}: namely, as the $\th^2$-term. The difference between the engineering dimensions of $\f$ and $F$ is $1$, which means that the term quadratic in $F$ in the action cannot have any derivatives.} 
and its equation of motion is $F=0$, so that it is indeed an auxiliary. Thus the resulting action is simply the supersymmetric extension of the action of a complex scalar field, i.e.~a scalar supermultiplet.

Likewise one defines a real, {\bf ${\cal N}=1$ vector superfield} $V$ which contains the content of an ${\cal N}=1$ {\it vector multiplet}. One then defines from it a chiral, gauge invariant superfield that contains the field strength as one of its components. This superfield is conventionally denoted $W_\a$, where $\a=1,2$ labels the two components of a chiral spinor (and it is also Lie-algebra valued).\footnote{That $W_\a$ is Lie-algebra valued means that it can be written as: $W_\a=W^a_\a T^a$, where $T^a$ are the generators of the Lie algebra in the adjoint representation. Recall that a four-dimensional spinor has four components, and can be decomposed into two chiral spinors, with two components each. The reason that this superfield acquires this spinor index is that it is obtained by taking derivatives of the vector superfield $V$ with respect to the $\th$'s.} Since $W_\a$ is a chiral superfield, it contains only $\th$'s and no $\bar\th$'s, so that one only needs to integate over the two chiral $\th$'s in the action. The simplest invariant action is again quadratic in $W$:
\bea\label{Waction}
S_{\tn{gauge}}^{{\cal N}=1}&=&{1\over4\pi}\,\mbox{Im}\int\dd^4x\,\dd^2\th~\t\,\Tr\,(W^\a W_\a)\nn
&=&\int\dd^4x\left(-{1\over4e^2}\,F^a_{\m\n}F^{a\m\n}-{\th\over32\pi^2}\,F^a_{\m\n}*F^{a\m\n}\right)+\ldots
\eea
The trace here is taken in the Lie algebra. (The factor of $\th$ in the second line is the usual $\th$-term, see Eq.~\eq{Swtheta}, and should not be confused with the superspace Grassmann variables). The complex coupling constant $\t$ is included so as to get the correct coefficients for the kinetic and $\th$-term, and the imaginary part is taken so as to get a real action. 

Note the similarity with the action Eq.~\eq{SAeth}: this is because the $\th^2$ term of $\Tr(W^\a W_\a)$ in fact contains the non-Abelian analogue of $\mathscr{F}$ (defined in Eq.~\eq{calF}). The omitted terms involve the fermion and an auxiliary scalar field.

Since supersymmetric actions are simple in the superfield formalism, this formalism is useful to construct actions with ${\cal N}$-extended supersymmetry. For example, ${\cal N}=1$ chiral and vector multiplets can be combined into an ${\cal N}=2$ vector multiplet. Since the number of supersymmetry generators doubles, one enlarges the superspace by adding a new set of coordinates, $(\ti\th,\ti{\bar{\th}})$, to the already existing $(\th,\bar\th)$ of ${\cal N}=1$ superspace. The {\bf ${\cal N}=2$ chiral superfield} is then (somewhat schematically) defined as follows:
\bea\label{chirals2}
\Psi(x,\th,\ti\th)=\F(x,\th)+\sqrt{2}\,\ti\th \,W_\a(x,\th)+\ti\th^2G(x,\th)\,.
\eea
The first term in the expansion is the ${\cal N}=1$ chiral superfield, $\F$; the second term is the vector superfield, $W$; and the third term, $G(x,\th)$, is defined as a combination of the chiral and vector superfields. 

Then the simplest superspace action is analogous to the ${\cal N}=1$ gauge action, Eq.~\eq{Waction}:
\bea\label{N=2action}
S_{{\cal N}=2}=\mbox{Im}\int\dd^4x\,\dd^2\th\,\dd^2\ti\th~{\t\over2}\,\Tr\,\Psi^2\,.
\eea
Doing the integration over the Grassmann variables, one sees that this action contains both the ${\cal N}=1$ scalar action, Eq.~\eq{1scalar}, and the ${\cal N}=1$ gauge action, Eq.~\eq{Waction}, added with relative coefficients so that the kinetic terms of the fermions of the two supermultiplets have the same coefficient, and the supersymmetry is extended to ${\cal N}=2$.

Charged matter fields with ${\cal N}=2$ supersymmetry can be constructed by combining an ${\cal N}=1$ chiral multiplet with its complex conjugate, into an {\bf ${\cal N}=2$ hypermultiplet}. And by combining an ${\cal N}=2$ vector multiplet with an ${\cal N}=2$ hypermultiplet, one obtains an ${\cal N}=4$ vector multiplet, which can be used to build the ${\cal N}=4$ super Yang-Mills action. 

\section{Duality of
{\it N=4}
supersymmetric Yang-Mills theory}\label{N=4SYM}

${\cal N}=4$ supersymmetric Yang-Mills (SYM) is one of the theories for which the Montonen-Olive conjecture, from Section \ref{M-O}, is expected to be true. It is called S-duality, and it is a strong-weak duality. That is, we conjecture under S-duality, that {\it ${\cal N}=4$ SYM with gauge group SU(2) and complex coupling $\t$ is dual to ${\cal N}=4$ SYM with gauge group SO(3) at the dual value of the coupling, $\t'=-1/\t$} (cf.~Eq.~\eq{taumintau}; and recall, from Section \ref{M-O}, that the Langlands or Goddard-Nuyts-Osborn dual of SU(2) is isomorphic to SO(3)).\footnote{More precisely, the Langlands dual of $\mbox{SU}(N)$ is $\mbox{SU}(N)/\mathbb{Z}_N$, and so the dual of $\mbox{SU}(2)$ is $\mbox{SU}(2)/\mathbb{Z}_2$. The homomorphism from SU(2) to SO(3) has kernel $\mathbb{Z}_2$, and so $\mbox{SU}(2)/\mathbb{Z}_2\simeq \mbox{SO}(3)$. See e.g.~Jeevanjee (2011:~pp.~106-108).} 
The duality is such that that the magnetic states of one model, as specified by their charges and spins, are the electric states of the dual, and vice versa.

The Montonen-Olive duality is a formidable conjecture to, in full generality, verify: and even to formulate. Thus we will make two simplifying conceptual assumptions that are standard in the physics literature. 

First, we will not explicitly distinguish between a model with gauge group SU(2) and a model with gauge group SO(3). For, although SU(2) is the double cover of SO(3), which is non-simply-connected, while SU(2) is simply connected, their algebras are isomorphic: and so, so long as we do not explicitly require the groups' global properties, the models based on them are in effect the same. 
Second, we will continue to take the duality group to be $\mbox{SL}(2,\mathbb{Z})$, which we discuss further below, even though in some cases only a subgroup of it is the duality group. Both assumptions are indeed justified for the simple set of quantities that we will consider.\footnote{Both points are made by Gukov and Witten (2006:~p.~59), who point out that, on $\mathbb{R}^4$ and without considering Wilson or 't Hooft loop operators, it is correct to, in effect, regard the ${\cal N}=4$ SU(2) and SO(3) SYM models as isomorphic, and that the duality group can be taken to be $\mbox{SL}(2,\mathbb{Z})$. See also Gaiotto and Witten (2022:~Section 2.7).} With these simplifications, the conjecture is that ${\cal N}=4$ SYM with gauge group SU(2) is, in effect, dual to itself at the dual value of the coupling: we consider a single model, and there is no significant difference between the theory and its model. 

Thus, with these simplifying assumptions, we can think of ${\cal N}=4$ SU(2) SYM as our theory. As we know from the previous Section, it contains a spin-1 gauge field, six scalars, and four Majorana spinors, all of them massless: hence eight bosonic and eight fermionic states, i.e.~a total of sixteen states.

To verify the Montonen-Olive duality conjecture, we should first check that the spectrum of lowest-energy states is mapped to itself. The theory contains elementary particle states (i.e.~the sixteen states reported above), but it also contains {\it solitonic} states that are not seen by looking at the linearized spectrum (i.e.~the theory's spectrum in perturbation theory, where interactions are assumed to be small). These soliton states have topological quantum numbers that are the central charges appearing on the right-hand side of the supersymmetry algebra, Eq.~\eq{susyZ}. These central charges come from certain surface terms in the calculation of the right-hand side of the supersymmetry algebra after spontaneous symmetry breaking (i.e.~after the Higgs mechanism), which give topological charges that are non-zero for soliton states.\footnote{This analysis was first done by Witten and Olive (1978).}

Thus, by including central charges in the supersymmetry algebra, the spectrum after spontaneous symmetry breaking can be determined exactly: which is remarkable, since the spectrum of an interacting quantum field theory can usually only be determined non-perturbatively. The supersymmetry algebra allows us to determine the solitonic states after symmetry breaking.\footnote{${\cal N}=4$ supersymmetric Yang-Mills theory was the first non-trivial four-dimensional example of a theory for which this spectrum could be derived.} 

The number of states indeed works out, i.e.~the spins of the elementary particles are identical to those of the monopoles: there is (in the broken symmetry vacuum, where one of the six scalars has acquired a vacuum expectation value) one particle with spin $s=1$ (the massive gauge boson), five with $s=0$, four with $s=\half$, and exactly the same number of monopoles.\footnote{See the discussion in Osborn (1979:~p.~326). For the states in the case of an unbroken symmetry group, see Figure \ref{N=4N=2spectrum} below.} 
This means that, if in addition we invert the coupling, the set of quantum numbers of the states (i.e.~their charges and spins) is invariant under S-duality!

Of course, more is needed for duality: all other properties characterizing the states should also leave the set of quantities invariant. There is at present no full proof, but a number of important quantities have been calculated and found to be suitably invariant.

The first quantity on the states to consider is their energy (their mass). This can also be analyzed from the supersymmetry algebra, from which one can derive the BPS bound for the mass, Eq.~\eq{bps}. This bound is saturated for both elementary particle and soliton states, all of which satisfy: $M=v\sqrt{e^2+g^2}$. Thus {\it any} particle with electric charge $e$ and magnetic charge $g$ (i.e.~photons, Higgs particles, fermions, W-bosons, magnetic monopoles, and dyons) satisfies this bound. Note that this formula is exact, i.e.~it receives no quantum corrections. We already know that this formula is invariant under duality, and so the set of masses is also invariant. We will discuss other quantities below. 

Although the original Montonen-Olive conjecture only considered S-duality, i.e.~which inverts the coupling $e\mapsto4\pi/e$ (cf.~Eq.~\eq{dadb}), this can be extended to $\mbox{SL}(2,\mathbb{Z})$ if the $\th$-term is included in the action, as we did in Eq.~\eq{Waction}.\footnote{The group $\mbox{SL}(2,\mathbb{Z})$ is generated by two transformations: $S$ maps $\t\mapsto-1/\t$, i.e.~it inverts the coupling and so it is the usual S-duality (see Eq.~\eq{taumintau}; it is the same $\t$ as in Eq.~\eq{Waction}) and $T$ maps $\t\mapsto\t+1$, which is a symmetry of the action because the partition function is periodic in $\th$ with period $2\pi$ (this is because the $\th$-term in the action, Eq.~\eq{Waction}, is a topological term, see Witten (1979:~p.~283)). $S$ and $T$ are represented by $2\times2$ matrices as: $S=\left(\begin{array}{cc}0&1\\-1&0\end{array}\right)$, $T=\left(\begin{array}{cc}1&1\\0&1\end{array}\right)$. For the correct action of S-duality on the coupling for an arbitrary simple Lie group, see Gukov and Witten (2006:~p.~174).} 
The $\mbox{SL}(2,\mathbb{Z})$ group then acts on $\t$ as follows:
\bea\label{modular}
\t\mapsto{a\t+b\over c\t+d}~,~~a,b,c,d\in\mathbb{Z},~~ad-bc=1\,.
\eea
These can be combined into an $\mbox{SL}(2,\mathbb{Z})$ matrix with unit determinant:
\bea\label{SL2Z}
\left(\begin{array}{cc}a&b\\c&d\end{array}\right).
\eea

Another important quantity to consider is the free energy, {\it for any excited state} (i.e.~the value of the Wilsonian Euclidean action, at a given scale).\footnote{Girardello et al.~(1994:~p.~335) argued that, because the beta function of ${\cal N}=4$ supersymmetric Yang-Mills theory is zero, the Wilsonian effective action for the monopoles describes the monopole dynamics {\it at all scales}, i.e.~that the physics at any energy is captured by effective action at {\it low energies}. That the beta function is zero follows from the ${\cal N}=4$ supersymmetry algebra.} 
Girardello et al.~(1994:~p.~336) showed that the free energy is $\mbox{SL}(2,\mathbb{Z})$-invariant, which again is a strong argument for the electric-magnetic duality of the theory with gauge group SU(2).

Girardello et al.~(1995:~p.~127) generalized this to an arbitrary gauge group, $G$. They showed that the partition function of the ${\cal N}=4$ supersymmetric Yang-Mills model with gauge group $G$ transforms as required by S-duality, i.e.~it maps to the partition function of a model with dual gauge group, $\hat G$. 

Vafa and Witten (1994:~p.~6) did another non-trivial check, in a vacuum where the non-abelian gauge symmetry is unbroken. They calculated the partition function of the ${\cal N}=4$ supersymmetric Yang-Mills model on a curved manifold.\footnote{It is not straightforward to put a non-abelian supersymmetric quantum field theory on an arbitrary orientable pseudo-Riemannian manifold, unless it has a distinguished time direction. The basic reason is that in verifying supersymmetry, one finds commutators of covariant derivatives, so that the Riemann tensor appears, and the theory is not supersymmetric. Witten (1988a:~p.~359) devised a method of `topological twisting', whereby part of the local rotation group of $\mathbb{R}^4$ is replaced by a diagonal sum of that part of the rotation group, and part of the symmetries of the extended supersymmetry charges. The result of twisting is a ``topological'' quantum field theory is a theory that depends less on geometry (e.g.~depending on the type of twisting, the model is independent of the volume of the space, or it is independent of the complex structure). Vafa and Witten (1994:~p.~6) considered a twisted version of ${\cal N}=4$ supersymmetric Yang-Mills model on a curved manifold that agrees with the usual (non-twisted) ${\cal N}=4$ supersymmetric Yang-Mills model on $\mathbb{R}^4$.} 
This partition function, for the gauge connection in a given topological class, was given by the Euler characteristic of the instanton moduli space.\footnote{The Euler characteristic on a compact oriented four-manifold $M$ is ${1\over96\pi^2}\int_M\mbox{Pf}\left(F\wedge F\right)$, where Pf is the Pfaffian on the tangent space indices, and $F$ is the curvature of the gauge connection $A$.} For a specific class of manifolds, they showed that the partition function is a modular object, i.e.~it has the expected transformation properties under a modular transformation Eq.~\eq{modular}, i.e.~it is not simply modular invariant, but it transforms with a certain weight. Note that the electric-magnetic duality is not a {\it symmetry} of the partition function, in the sense of the partition function's being invariant under it, but rather relates the partition functions of models with dual gauge groups.\footnote{The extension of Montonen-Olive duality to the duality group $\mbox{SL}(2,\mathbb{Z})$ is conjectured to be an exact {\it symmetry} of a certain version of string theory: namely, the `heterotic' string theory compactified to four dimensions; (for more on string theory, see Chapter \ref{String}: we postpone details about string theory until then). The duality seems to be ``universal'', in the sense that it is a symmetry regardless of which internal manifold is chosen for the compactification, so long as some supersymmetry is preserved in four dimensions. For example, if the internal compactification manifold is a six-torus, one gets a four-dimensional theory with ${\cal N}=4$ supersymmetry. The low-energy parameters that appear in the Wilsonian effective action description of string theory include the vacuum expectation value of the `dilaton' supermultiplet, $\l$, i.e.~the supermultiplet associated with the dilaton field: namely, the scalar field that determines the string coupling (which is not a constant, but a scalar field that can vary over the spacetime) and appears in the low-energy four-dimensional model as a scale factor in the action. The dilaton supermultiplet $\l$ transforms, under $\mbox{SL}(2,\mathbb{Z})$ duality, in the same way as the coupling constant $\t$ in a gauge theory, i.e.~as in Eq.~\eq{modular}. }

\section{Effective duality of {\it N=2} 
supersymmetric Yang-Mills theory}\label{effD}

While ${\cal N}=2$ supersymmetric Yang-Mills theory is not known to have electric-magnetic duality-related models, it has approximate (low-energy) models that are quasi-duals: more specifically, they are {\it effective duals} (see theme (5) in Section \ref{themesd}). 

So Section \ref{Wea} will contrast ${\cal N}=2$ with its more symmetric cousin, the ${\cal N}=4$ theory, so as to explain their differences and the sense in which ${\cal N}=2$ SYM instantiates a {\it quasi-duality}. Sections \ref{swt} to \ref{dualm} expound the details of Seiberg and Witten (1994a)'s remarkable development of this quasi-duality. Section \ref{moncond} briefly discusses how further breaking of the symmetry leads to physical phenomena whose existence one wishes one could predict in QCD: a mass gap, monopole condensation, and charge confinement.

\subsection{Effective duality and the Wilsonian effective action}\label{Wea}

It is instructive to understand why, unlike the ${\cal N}=4$ supersymmetric Yang-Mills theory, the ${\cal N}=2$ theory does {\it not} illustrate the Montonen-Olive duality conjecture: since this will lead in to an instantiation of {\it effective duality} (in the sense of Section \ref{featurerole}: Section \ref{dualm} will then explain the sense in which this is a quasi-duality.). \\
\\
{\bf Two conditions for Montonen-Olive duality.} One possible reason why a given pair of models of a non-Abelian gauge theory $T$ does not instantiate the Montonen-Olive duality conjecture is that it does not satisfy two of the required conditions:

(i)~~The particles and the solitons (in particular, the monopoles) must have the {\it same (spin) quantum numbers}, so that their state spaces  can be isomorphic. 

Montonen-Olive duality exchanges the electric and magnetic charges (i.e.~it exchanges the roots of the gauge group with the coroots of the Langlands dual group), thereby exchanging particle and soliton states. Thus if the models have different spin quantum numbers, the Montonen-Olive duality is not instantiated.

\begin{figure}
\begin{center}
\includegraphics[height=1.5cm]{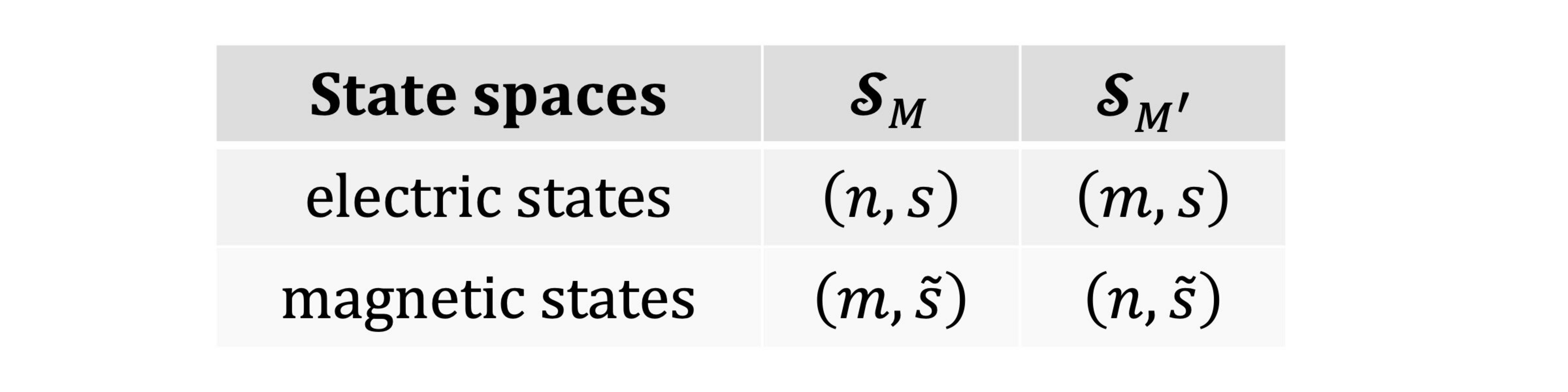}
\caption{\small State spaces of models related by Montonen-Olive duality. The two models satisfy {\it different supersymmetry algebras}, except if $s=\ti s$.}
\label{MON=2}
\end{center}
\end{figure}

Let us schematise this argument for two models, $M$ and $M'$, which have the {\it same supersymmetry algebra}, but different gauge groups (respectively, $G$ and its dual, $\hat G$), couplings (respectively, $\tau$ and its dual), and also different sets of spins. (This is the situation in ${\cal N}=2$ SYM, where $M$ is the usual ${\cal N}=2$ SYM, and $M'$ is the model obtained from $M$ by Montonen-Olive duality, i.e.~they differ in the charges that they call `electric' and `magnetic': since they differ by the exchange of their roots and coroots.) Label $M$'s set of {\it electric} charges (schematically, up to a multiplicative constant) by $n$ (these are elements of the {\it root lattice} of $G$), and their corresponding spins by $s$. Label $M$'s {\it magnetic} charges, up to a multiplicative constant, by $m$ (these are elements of the {\it coroot lattice}), and their corresponding spins by $\ti s$. Thus the state space ${\cal S}_M$ has electric states with quantum numbers $(n,s)$, and magnetic states with quantum numbers $(m,\ti s)$: see Figure \ref{MON=2}. We will now show that the state spaces of the two models, ${\cal S}_M$ and ${\cal S}_{M'}$, are mapped onto each other by the putative duality only if $s=\ti s$.

Montonen-Olive duality maps the roots of $M$ to the coroots of $M'$ and vice versa, so that 
$m$ is the set of {\it electric} charges in ${\cal S}_{M'}$, i.e.~roots of $\hat G$, and $n$ is the set of {\it magnetic} charges in ${\cal S}_{M'}$, i.e.~coroots of $\hat G$ (and the spins do not change). Thus under the duality map there are, in the state space ${\cal S}_{M'}$, {\it electric} states $(m,s)$, and {\it magnetic} states $(n,\ti s)$ (see Figure \ref{MON=2}). But if $s\not=\ti s$, these states have the ``wrong'' spin, and they are incompatible with the supersymmetry algebra which the two models have in common. (Namely, a putative algebra with states $(m,s)$ and $(n,\ti s)$ is different from the algebra of ${\cal N}=2$ SYM, which has states $(n,s)$ and $(m,\ti s)$.) Thus realizing Montonen-Olive duality requires that the spins are the same, i.e.~$s=\ti s$.\footnote{$n=m$ is in general not possible, because the electric and the magnetic charges are related by the inversion of the coupling.}

(ii)~~ {\it Quantum corrections} must respect Montonen-Olive duality, which relates bare couplings in the model $M$ to bare couplings in its dual, $\hat M$, either by inversion of the coupling $e$ (see Eq.~\eq{dadb}), or generalized as an $\mbox{SL}(2,\mathbb{Z})$ transformation (i.e.~Eq.~\eq{modular}). 

But in general, the coupling constant $e$ {\it runs}, i.e.~receives quantum corrections, and this running breaks the duality. So, even if $M$ and $\hat M$ are duals as classical models, quantum corrections will in general spoil the duality. This is of course further complicated by the fact that the inversion of the coupling implied by electric-magnetic duality relates a classical model to a highly quantum one.\\

\begin{figure}
\begin{center}
\includegraphics[height=2.5cm]{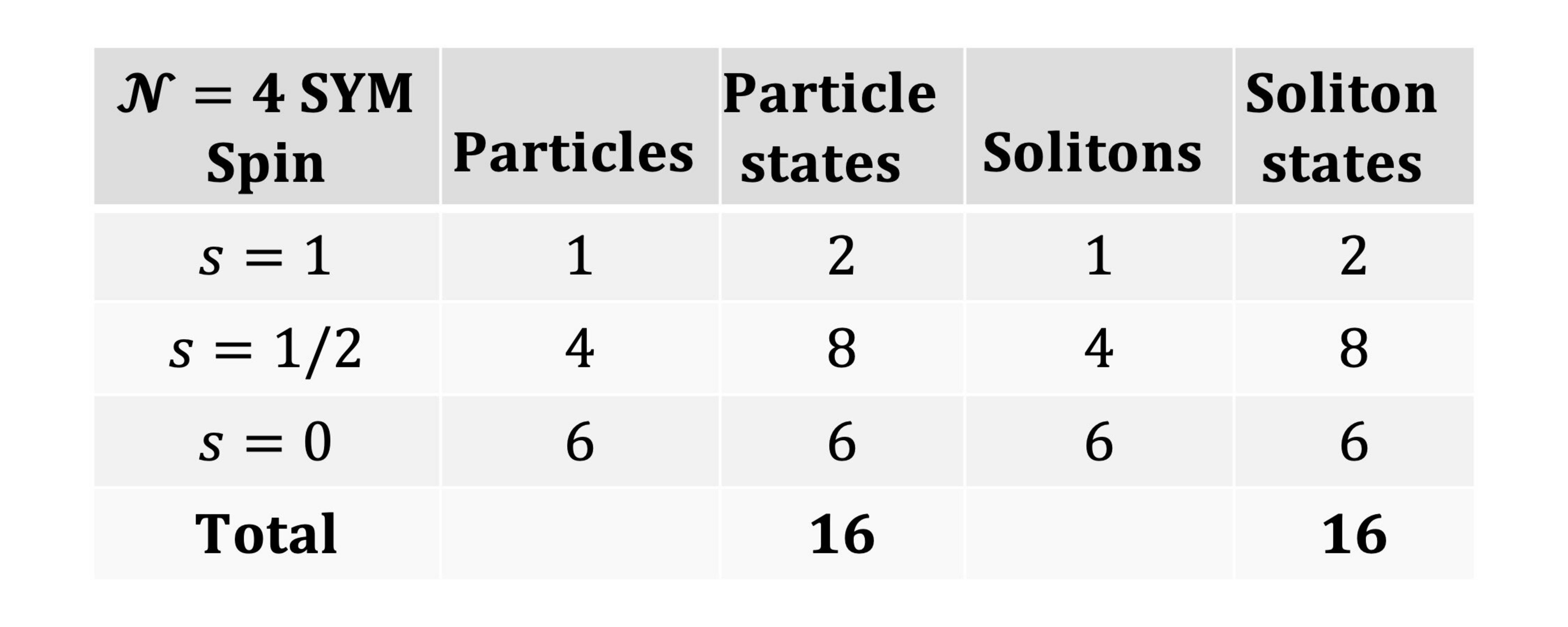}~
\includegraphics[height=2.5cm]{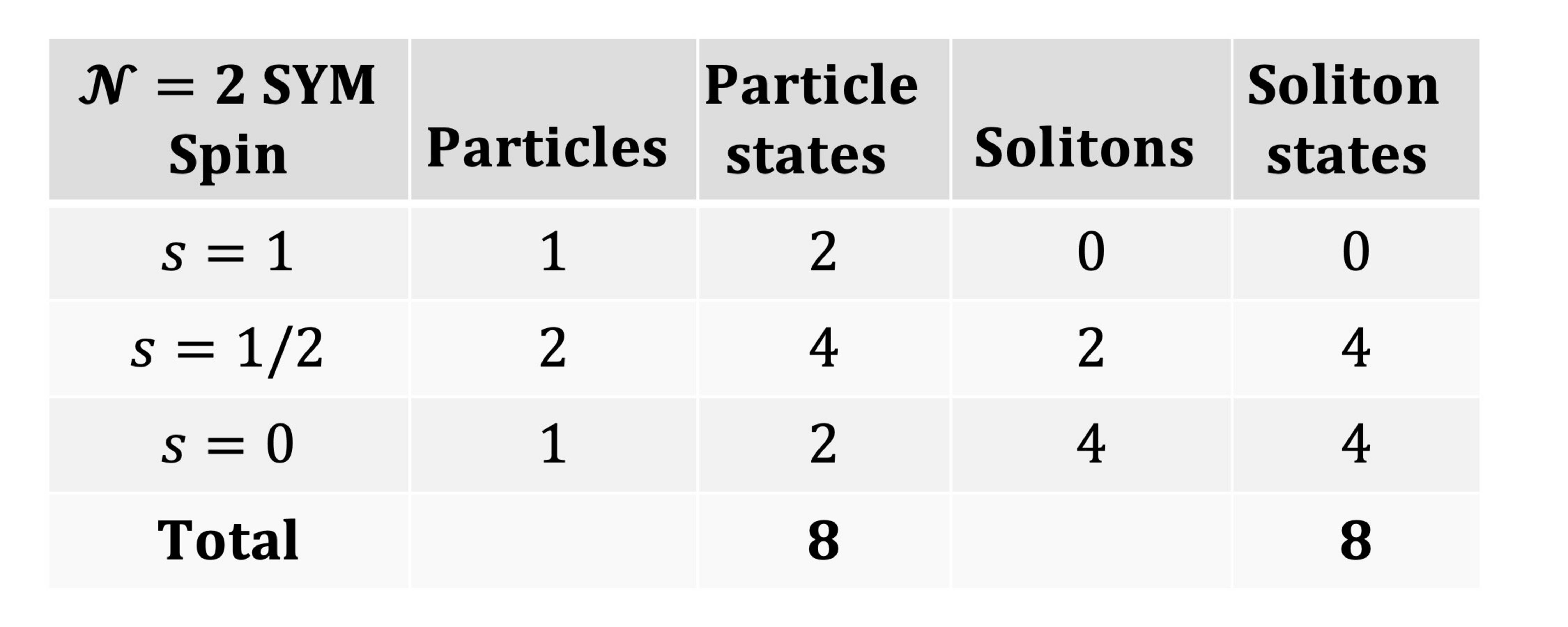}
\caption{\small Particle and soliton spectrum of: {\it left}: ${\cal N}=4$ SYM; {\it right}: ${\cal N}=2$ SYM. The sets of particle and soliton states of the ${\cal N}=4$ theory match, while those of the ${\cal N}=2$ theory do {\it not} match, since there is no soliton with $s=1$. In each theory, the numbers of bosonic and fermionic states match, because they belong to appropriate ${\cal N}=4$ and ${\cal N}=2$ supermultiplets. Note that the most significant vertical columns are the {\it particle states} and {\it soliton states}, because the numbers of {\it particles} and {\it solitons} depend on conventions, such as whether scalars are taken real or complex. (In the broken case, the states are in effect reshuffled vertically within columns.)}
\label{N=4N=2spectrum}
\end{center}
\end{figure}

One easily sees that (i) and (ii) are satisfied in the ${\cal N}=4$ SYM theory: (i) as already discussed in Section \ref{N=4SYM}, the spin quantum numbers of the elementary particles in this model are the same as those of the solitons (see Figure \ref{N=4N=2spectrum}). (ii) The model is conformally invariant: the beta-functions of the couplings $e$ and $\th$ are zero, and the couplings do not run. Since the bare couplings are also the quantum couplings, the modular invariant, low-energy, effective action is {\it valid at all} scales, including in the UV (see Girardello et al., 1994). 

On the other hand, the ${\cal N}=2$ SYM theory does not satisfy (i):\footnote{As we will see, it does satisfy condition (ii): even though the coupling of the ${\cal N}=2$ theory runs, there is nevertheless a suitable function of the expectation value of the Higgs field that plays the role of a coupling, and transforms correctly under $\mbox{SL}(2,\mathbb{Z})$. (Strictly speaking, there is only a section on a fibre bundle, as will become clear in Section \ref{physsig}: although for the purposes of this exposition we will not need to spell out the mathematical details. See Lerche (1998:~p.~176).)} 
for there is a mismatch between the spins of the states in the spectrum of elementary particles and the spectrum of solitons: electrons are in a supersymmetric multiplet with spins $\leq1$, while monopoles are in a multiplet with spins $\leq\half$: see Figure \ref{N=4N=2spectrum} (Osborn, 1979:~p.~326). Thus ${\cal N}=2$ SYM does not realize Montonen-Olive duality.\\
\\ 
{\bf Effective duals.} Even if a theory does not realize Montonen-Olive duality, there can be a {\it quasi-duality} of effective models, where each model is defined by a Wilsonian effective action, valid up to some energy. The models are quasi-duals in the sense that, although their action principles are not isomorphic and there is no isomorphism of state spaces, there is nevertheless an appropriate (bijective) map between the state spaces that preserves a relevant subset of the quantities, up to the scale at which the Wilsonian effective actions are well-defined. 

Wilson (1974) introduced a method of renormalization whereby the momentum cutoff, $\L$, of a theory is not taken to infinity, but is kept finite. The {\bf Wilsonian effective action} is the action that describes the dynamics of the field modes whose momentum is smaller than the cutoff, and the effects of the field modes with momentum higher than the cutoff are included (a) in the renormalization of the coupling constants of the theory, and (b) in terms in the effective action that may have been absent from the theory's original action. Thus the modes with high momentum are ``integrated out'' in deriving the effective action from the path integral.\footnote{For a more detailed description of Wilsonian renormalization, see Fisher (1998) and Binney et al.~(1992). Hartmann (2001), Batterman (2011), Butterfield and Bouatta (2015), and Dieks et al.~(2015) give philosophical discussions.} 
Also, the fields in the effective action are usually rather different from the original fields, for example because fields have been ``lumped together'' when integrating out. Recall, for example, that the order parameter $\psi$ of the Ginzburg-Landau model, Eq.~\eq{GLfreeE}, is an effective bosonic field, quite different from the pair of fermionic fields whose low-energy behaviour it describes.

In practice, to write the Wilsonian effective action one uses symmetry arguments. For example, Ginzburg and Landau formulated their low-energy model of superconductivity based on symmetry and other physical considerations (see Section \ref{GLtheory}). Later Gor'kov (1958:~p.~505), by focussing on the formation of bound states of electron pairs with zero momentum, and neglecting other effects such as scattering, showed that the same model follows from the BCS model.\\
\\
{\bf The Seiberg-Witten duality introduced.} The ${\cal N}=2$ theory has models that are quasi-duals, in the following sense: the electric states of $M$ are mapped to the magnetic states of $M'$, and vice versa,\footnote{In particular, this duality acts in the usual way on the gauge field (cf.~Eq.~\eq{EMd2} for S-duality), and it also relates the quantities and the dynamics of the two effective actions.} 
and their partition functions can be shown to be dual to each other. 

More precisely: the Wilsonian effective model, $M$, is the ${\cal N}=2$ supersymmetric Yang-Mills model (SYM), with gauge group U(1). It is obtained, at low energies, by the Higgs mechanism from the full ${\cal N}=2$ SU(2) SYM theory.

\begin{figure}
\begin{center}
\includegraphics[height=3cm]{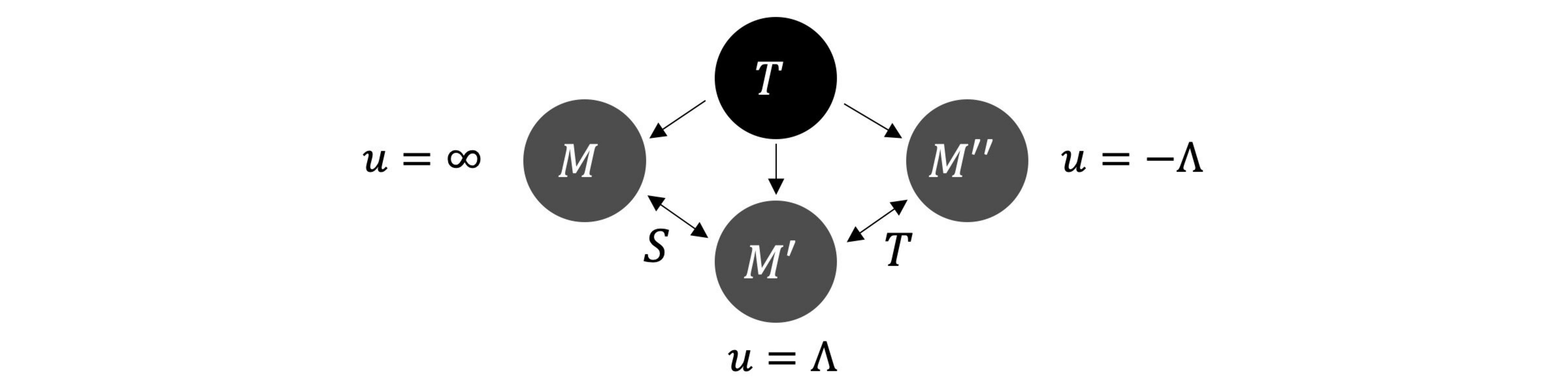}
\caption{\small The Seiberg-Witten theory, i.e.~the quasi-duality for $T=[{\cal N}=2~\mbox{SYM}]$. $M$, $M'$, and $M''$ are low-energy limits of $T$, each defined near one of three singularities in the $u$-plane (at $u=\infty,\pm\L$). An effective $S$-duality maps $M$ to $M'$, and the action with the $T$-generator of $\mbox{SL}(2,\mathbb{Z})$ maps $M'$ to $M''$.}
\label{SWduality}
\end{center}
\end{figure}

The dual effective model, $M'$, is ${\cal N}=2$ quantum electrodynamics (SQED), i.e.~a massive ${\cal N}=2$ BPS vector multiplet coupled to a light ${\cal N}=2$ hypermultiplet (see Section \ref{basicsusy}).\footnote{As the next Section will argue, the mass of this vector multiplet, in the model's region of weak coupling, is close to zero.}
Thus the duality is as follows:\\
\\
{\bf Seiberg-Witten quasi-duality:} {\it ${\cal N}=2$ SYM theory, with gauge group SU(2) broken by the Higgs mechanism to U(1), has two low-energy models, $M$ and $M'$, which are quasi-duals under the exchange of electric and magnetic charges and under inversion of the running coupling} (see figure \ref{SWduality}):\footnote{The low-energy models are dual under any element of $\mbox{SL}(2,\mathbb{Z})$, where the fields are appropriate linear combinations of the supermultiplets. This duality group is generated by the generators $S$ and $T$ in Figure \ref{SWduality}.}
\bea
M&=&\left[{\cal N}=2~\mbox{SYM with gauge group U(1)}\right],\nn
M'&=&\left[{\cal N}=2~\mbox{SQED with light ${\cal N}=2$ hypermultiplet}\right].\nonumber
\eea
The next Section will discuss the matching of states, quantities, and symmetries of these two models (for brevity, we will not give detail about the third dual model in Figure \ref{SWduality}).\footnote{The types and numbers of states of $M$ are, in effect, as on the right Figure \ref{N=4N=2spectrum}, but they are in an ${\cal N}=2$ BPS massive vector multiplet, rather than a massless ${\cal N}=2$ vector multiplet (and they are states of U(1), rather than SU(2)). For the states of $M'$, see Section \ref{dualm}.}

\subsection{Details of the Seiberg-Witten theory}\label{swt}

Seiberg and Witten (1994a, 1994b) wrote two influential papers that contained the {\it exact} quantum effective action of ${\cal N}=2$ supersymmetric Yang-Mills theory with gauge group SU(2), for vacua where the SU(2) gauge symmetry is broken by the Higgs mechanism to U(1).\footnote{Useful reviews of this work, in increasing order of sophistication, include: Bilal (1997), Alvarez-Gaume and Hassan (1997), and Lerche (1998).} 
`Exact' here means that their quantum effective action includes {\it all} the quantum corrections, both perturbative (roughly, expressible in ordinary Feynman diagrams) and non-perturbative (not so expressible).

As usual in Yang-Mills-Higgs models, the Higgs mechanism gives mass to two of the gauge modes, while a third mode remains massless, and corresponds to the low-energy photon whose gauge group is the unbroken U(1). Seiberg and Witten assumed that ${\cal N}=2$ {\it supersymmetry is preserved} in the effective theory, so that also the fermionic superpartners of the photon remain massless, as do the fluctuations of the Higgs field itself, since all of these fields belong to the same ${\cal N}=2$ vector multiplet (see Section \ref{basicsusy}). The low-energy effective action describes the physics of these massless modes. \\
\\
{\bf Wilsonian effective action.} The effective action is constrained by ${\cal N}=2$ supersymmetry to be of the following form:
\bea\label{Seff}
S_{\sm{eff}}={1\over16\pi}\,\mbox{Im}\int\dd^4x\,\dd^2\th\,\dd^2\ti\th~{\cal F}(\Psi)\,,
\eea
where ${\cal F}(\Psi)$ is called the {\bf prepotential}, and it is locally a function of the ${\cal N}=2$ chiral superfield $\Psi$ but not its complex conjugate, $\Psi^\dagger$ (the notion of a chiral superfield was introduced in Section \ref{basicsusy}). $\Psi$ is the Abelian chiral superfield that contains the photon, Higgs, and fermion fields that are unbroken at low energies. Due to the gauge symmetry breaking, this Wilsonian effective action is Abelian. 

The prepotential ${\cal F}$ has an infinite series expansion whose first term is the quadratic term of the classical action, Eq.~\eq{N=2action}, that appears in the path integral. Thus the leading term of the effective action, Eq.~\eq{Seff}, has the same form as the original (microscopic) action, but with a {\it single} unbroken Abelian ${\cal N}=2$ supermultiplet $\Psi$, rather than the full non-Abelian supermultiplets. 

The first term in the expansion of the superfield $\Psi$ is the expectation value, $a$, of the Higgs field at weak coupling, $\Psi=a+\ldots$ (see Eqs.~\eq{chirals2} and \eq{chiralF}), because the Higgs potential $V(\f)=\Tr[\phi,\phi^\dagger]^2$ is minimized, classically, by $\f=a\s_3$ (i.e.~the third Pauli matrix). This is analogous to the discussion, in Section \ref{vacsols}, of the minimization of the Mexican hat potential, Eq.~\eq{MexicanV}, in the non-supersymmetric case. However, a major difference is that, while the parameter $v$ of Section \ref{vacsols} was fixed by the Mexican hat potential, the expectation value of the Higgs is here {\it not} determined by the action: thus there is a space of possible vacua, parametrized (at weak coupling) by $a$. \\
\\
{\bf Moduli space of vacua.} To parametrize the vacua in a gauge-invariant way, we define our {\it order parameter} $u$ as follows:
\bea\label{Higgsvev}
u:=\bra\Tr\,\f^2\ket=2a^2\,.
\eea
$u$ is a good order parameter not only because it is gauge invariant, but also because it can in principle be calculated in the quantum theory, where the classical relation $u(a)=2a^2$ will be modified by quantum effects. By analogy with the {\it order parameters} discussed in the previous Chapter, we will use $u$ to characterize the validity of the Wilsonian effective action in various regimes. 

The space of vacua parametrized by the complex variable $u$ is a two-dimensional complex manifold called the {\bf moduli space}, ${\cal M}$, i.e.~the space of possible vacuum values of the (Higgs) field: it is, under the Higgs mechanism, the space of the theory's vacua at low energies.\footnote{The {\bf moduli} are a distinguished set of quantities or variables (here, the Higgs field) whose values characterise the properties of states, e.g.~by distinguishing different macroscopic phases. Note that the use of the word `moduli' is somewhat muddled in the literature, since it is sometimes used for the {\it values} of the moduli, rather than for the moduli themselves. Thus a moduli space is not a space of {\it values} of moduli, but rather an abstract space whose {\it coordinates} are moduli. The values of the moduli are typically interpreted as order parameters. For more details, see Section \ref{mvd}.} 
And, as in Figure \ref{SWduality}, different regions of the moduli space will be described by different {\it models}.

One main ingredient of Seiberg and Witten's analysis is the following two statements about the moduli space, which we will discuss below:\footnote{Other properties of the moduli space are as follows: it is a K\"ahler manifold, where the K\"ahler metric is given by the prepotential, which is a meromorphic function (see Eq.~\eq{SWmetric}). Also, the moduli space has a (metric-compatible) symplectic structure that is induced by a non-degenerate two-form called the `Seiberg-Witten differential'.\label{Kmetric}}

(i) the moduli space, ${\cal M}$, is endowed with a {\it metric};

(ii) the metric has at least three {\it singularities} (and a low-energy model will be associated with the region near each of the singularities), which are not part of the moduli space. 

As we will see, the metric is flat except for an overall scale factor, which is given in terms of the prepotential. And since the metric has singularities (cf.~(ii)), it is not everywhere well-defined, and we will look for coordinates that render it well-defined in various regions. As it turns out, a choice of {\it dual} variables (defined by the Legendre transform) renders the prepotential and the metric non-singular in ``dual regions''. 

Thus the (perhaps surprising) picture that we will argue for is as follows: an electric-magnetic (quasi-)duality map is a transition function between two open sets of the Riemann surface (i.e.~between two regions, each with a locally defined metric). This transition function induces a change of coordinates of the moduli space (the space of vacua) of the theory. As we will see, the duality group $\mbox{SL}(2,\mathbb{Z})$ arises as the group of transformations that preserves the form of the metric on the moduli space.

Seiberg and Witten (1994a) make (and partly motivate) the minimal assumption for (ii), i.e.~that there are only {\it three} singularities. Since the moduli space is the space spanned by expectation value of the Higgs, Eq.~\eq{Higgsvev}, the classical value $a$ turns out to give a good description of the semi-classical region, where $a$ is large, while the dual Higgs variable, $a_D$, describes its S-dual region (and there is yet another variable for the third region, which is a linear combination of the other two). \\
\\
{\bf Metric on moduli space.} Let us first see how the metric on moduli space appears. To see this, we do the $\ti\th$-integrals in Eq.~\eq{Seff}, which gives the following effective action, written in terms of the ${\cal N}=1$ superfields:
\bea\label{SeffF}
S_{\sm{eff}}={1\over16\pi}\,\mbox{Im}\int\dd^4x\left(\int\dd^2\th\,\dd^2\bar\th~\F^\dagger{\cal F}'(\F)+\int\dd^2\th~{\cal F}''(\F)\,W^\a W_\a\right),
\eea
where $\F$ is the ${\cal N}=1$ chiral superfield respecting the unbroken U(1) symmetry, and whose expansion begins as: $\F=a+\ldots$ (see Eq.~\eq{chiralF}). 

The first term is analogous to the ${\cal N}=1$ scalar action Eq.~\eq{N=1scalar}, and the second term is analogous to the ${\cal N}=1$ gauge action, Eq.~\eq{Waction}. Thus the effective action written as in Eq.~\eq{SeffF} is analogous to the classical ${\cal N}=2$ action Eq.~\eq{N=2action}, which gave the sum of the two ${\cal N}=1$ scalar and gauge actions. Indeed, one easily sees that if the prepotential is a quadratic function, then one simply gets the same quadratic terms as before.\footnote{There are of course crucial differences with respect to the two ${\cal N}=1$ multiplets involved: first, the effective action now depends on the chiral superfield in the broken Higgs vacuum, $\F=a+\ldots$, rather than on the fundamental Higgs field. Second, the gauge superfield $W_\a$ is abelian and transforms under the unbroken U(1).} 

If one does the remaining $\th$-integrals in the above action, one gets the usual scalar and kinetic terms for the fields $a$, $F_{\m\n}$, etc.~(as in Eqs.~\eq{1scalar} and Eq.~\eq{Waction}), but each of them comes with an overall multiplying coefficient, namely: $\mbox{Im}\,\t(a):=\mbox{Im}\,{\cal F}''(a)$. This additional multiplying factor is the scale factor of the {\bf metric in the space of field values}, written in complex coordinates $(a,\bar a)$:\footnote{It is sometimes called a {\it sigma-model metric}, written (classically) as $\dd s^2=\mbox{Im}\,{\cal F}''(\f)\,\dd \f\,\dd\bar \f$.}
\bea\label{SWmetric}
\dd s^2=\mbox{Im}\,\t(a)\,\dd a\,\dd\bar a\,.
\eea
Seiberg and Witten proposed that this is the local metric on an appropriate coordinate patch on (more precisely, an open set of) the moduli space: namely, around $a=\infty$ i.e.~$u=\infty$. Comparing the above action with the classical expression Eq.~\eq{deftau}, we see that it contains the effects of renormalization on the couplings:
\bea\label{ta}
\t(a)={\th(a)\over2\pi}+{4\pi i\over e^2(a)}\,.
\eea
This is the effective coupling that appears in the Wilsonian effective action, and it is valid up to scales set by the dimensionful parameter $a$, i.e.~the expectation value of the Higgs, compared to the Wilsonian cut-off $\L$. Classically, $\t=\t_0$ is the bare coupling constant, and the prepotential is: ${\cal F}(a)=\half\t_0\, a^2$. The first main point of the Seiberg-Witten calculation is to find the full expression for the prepotential, with all the quantum corrections, in this region:
\bea\label{Fsum}
{\cal F}(a)=\half\t_0\,a^2+{i\over\pi}\,a^2\ln{a^2\over\L^2}+{a^2\over2\pi i}\sum_{k=1}^\infty c_k\left({\L\over a}\right)^{4k}\,,
\eea
where $\L$ is the cutoff. The first term, quadratic in the field, is the classical term (which reproduces the classical action, Eq.~\eq{N=2action}). The logarithmic term is a one-loop correction to the action, and the infinite sum contains the non-perturbative contributions.\footnote{See Lerche (1998:~p.~172). For a general discussion of non-perturbative contributions to amplitudes, see Shuryak (2021:~p.~1).}
The latter are sub-leading in the perturbative regime, where the expectation value of the Higgs is large, $a\gg\L$, and the theory is asymptotically free. In other words, Eq.~\eq{Fsum} gives a good description of the moduli space near $a=\infty$, i.e.~$u=\infty$.

The non-perturbative terms in Eq.~\eq{ta} are important at strong coupling, where $a$ is not large.\footnote{Note that $a$ has dimensions of momentum, and so the perturbative approximation is good for large momenta, where the coupling is small, due to the asymptotic freedom of the theory, as seen in the negative sign of the beta function.\label{largeHiggs}} The determination of the exact form of the prepotential (i.e.~of the coefficients $c_k$), and of how it transforms to other regions of the moduli space (i.e.~to regions of strong coupling, where the series in Eq.~\eq{Fsum} does not converge), is Seiberg and Witten's main technical achievement.\footnote{For a more detailed description of the {\it mathematical} problem in the Seiberg-Witten theory, see Lerche (1998:~pp.~178-184; 186-189). $\t$ and the prepotential are not functions on the moduli space, but rather multi-valued sections. The monodromy problem then allows one to solve for the prepotential in each of the regions of the moduli space.}\\
\\
{\bf The metric is not globally defined.} One can argue that the metric on the moduli space, in a region with scale factor $\t(a)$, is {\bf not globally defined}. For, requiring that the low-energy theory is unitary and the fields have low-energy interpretations as normal particles, the scale factor $\t(a)$, which multiplies the kinetic terms of all the fields of the Wilsonian effective action, must be everywhere positive. However, the positivity of $\t(a)={\cal F}''(a)$ is incompatible with its being a harmonic function of the fields. Thus the metric cannot be positive definite everywhere, and (since it is not a constant: cf.~Eq.~\eq{ta}) it is zero somewhere on the moduli space.\footnote{$\tau(a)$ is harmonic because it depends on $a$ but not $\bar a$, so that ${\pa\t\over\pa\bar a}=0$. A non-constant harmonic holomorphic function cannot have a maximum nor a minimum in its region of definition: see Ahlfors (1979:~pp.~134, 166). In the case at hand, zero cannot be a global minimum of the function, and so we cannot have $\t(a)>0$ everywhere.} 

But near these singularities a dual description exists, that is good in the region of small $a/\L$, where the theory is highly quantum and the expansion Eq.~\eq{Fsum} does not converge.

The above statement, i.e.~that the prepotential ${\cal F}$ depends on $a$ but not $\bar a$, is implied by ${\cal N}=2$ supersymmetry, and is one of the premisses of the argument that the prepotential must have singularities, and that in the region near these singularities a new prepotential, ${\cal F}_D$, must be found that is related to the normal prepotential by an appropriate change of coordinates.\footnote{This argument is reminiscent of the Ising model, where the assumption that the free energy has a single singularity allowed one to determine the critical temperature, thereby allowing one to ``patch together'' the free energy and its dual (see the end of Section \ref{dualpf}).} As we will see, the singularities are interpreted in terms of monopoles becoming massless.

What is the physical interpretation of a {\bf singularity} in moduli space? According to 't Hooft (1981:~pp.~463-465, 456), singularities in the space of fields indicate that states that were neglected in the effective description become relevant at the value of the modulus that gives the singular behaviour. This means that one begins with a description in which all particles whose mass is above the Wilsonian cut-off are integrated out, and that, as one moves in the moduli space, one reaches a point where a particle becomes massless (e.g.~where the expectation value of the Higgs field is zero). At this point, the Wilsonian effective action is not valid, and this could again be indicated by the moduli being outside the radius of convergence of the approximation used to derive the Wilsonian effective action---this situation would then result in a singularity of the Wilsonian effective action. As we will see, the Seiberg-Witten theory realizes this idea.\footnote{For a discussion of the expansions of the prepotential near each of the singularities, see Lerche (1998:~p.~179).}

\subsection{Dual effective action}\label{dualpro}

The main innovation by Seiberg and Witten (1994a) was to realize that the {\it Legendre transform} of the effective action Eq.~\eq{SeffF} could give a formulation that is both free of singularities in the appropriate (small $a$) region and that is the electric-magnetic (quasi-)dual of this action. Although the argument that follows does not amount to a {\it proof} of the electric-magnetic (quasi-)duality of the effective action because of certain details that we will mention, it is good evidence for it.\footnote{The current presentation, in terms of the Legendre transform, follows Bilal (1997:~pp.~100-101), who also mentions that the present duality transformation is a canonical transformation with a Jacobian that is equal to one, i.e.~it is a symplectic transformation that preserves the area form of the moduli space of vacua.} 
In any case, the point of view of Seiberg and Witten is that, if one assumes that the effective action transforms in the way that the manipulations below show (these are similar to the steps in Section \ref{quantumD}), important consequences of this (quasi-)duality follow.

The Legendre transformation is done in the usual way. Introduce the {\it dual} variable, $\F_D$ of $\F$, and a {\it dual prepotential}, ${\cal F}_D$ of ${\cal F}$, as follows:
\bea\label{LegendreF}
\F_D&:=&{\cal F}'(\F)\nn
{\cal F}'_D(\F_D)&:=&-\F\,.
\eea
Using this definition, one checks that the first term in the effective action involving the chiral superfield, Eq.~\eq{SeffF}, takes the same form in both sets of variables:
\bea
\mbox{Im}\int\dd^4x\,\dd^2\th\,\dd^2\bar\th~~\F^\dagger{\cal F}'(\F)&=&-\mbox{Im}\int\dd^4x\,\dd^2\th\,\dd^2\bar\th~({\cal F}'_D(\F_D))^\dagger\F_D\nn
&=&\mbox{Im}\int\dd^4x\,\dd^2\th\,\dd^2\bar\th~~\F_D^\dagger\,{\cal F}'_D(\F_D)\,,\nonumber
\eea
and in the last line we used that, for any complex $A$ and $B$, $\mbox{Im}(A^\dagger B)=-\mbox{Im}(B^\dagger A)$. 

One can do similar manipulations for the second term of the effective action Eq.~\eq{SeffF}, i.e.~the one including the gauge fields. The manipulations required here are the superspace analogues of the ones we did in Section \ref{quantumD}, and so we will not repeat them in detail.\footnote{One adds to the path integral, over the low-energy effective action for the gauge field, a Lagrange multiplier superfield that implements the Bianchi identity for the gauge field (because the path integral over this Lagrange multiplier superfield is a delta function whose argument is the Bianchi identity, see Eq.~\eq{intB}). Changing the order of the integrations, i.e.~leaving the Lagrange multiplier in the path integral and carrying out the integral over the gauge field (which can now be carried out, since it is a quadratic field that no longer satisfies the Bianchi identity, and the resulting path integral is a Gaussian functional integral: see Eq.~\eq{DFFAF}: such integrals can be performed by lattice regularization). One gets an effective action for the Lagrange multiplier superfield that is not linear, but quadratic in the superfield, which can therefore be seen as a dynamic gauge superfield, since it includes derivatives. In fact, the effective action depends on the field strength of the dual gauge superfield, but with a {\it dual coupling}, $-1/{\cal F}''(\F)$ rather than ${\cal F}''(\F)$. Expressing the dual coupling in terms of $\F_D$, i.e.~using the Legendre transform, Eq.~\eq{LegendreF}, one finds the S-dual of the coupling, $\t_D(a_D)=-{1\over\t(a)}$. This is analogous to the Abelian case in Section \ref{quantumD}, where the model is the Maxwell model, Eq.~\eq{Maxwth}, with dual model, Eq.~\eq{Maxwth2}, obtained by the addition of such a Lagrange multiplier, and dual couplings Eq.~\eq{taumintau}. Thus also this second term of the Wilsonian effective action, Eq.~\eq{SeffF}, containing the gauge fields, can be rewritten, in terms of the thus defined dual fields, in the form ${\cal F}'_D(\F_D)\,W^\a_DW_{D \a}$, where $W_D$ is the dual gauge superfield that contains the dual field strength (obtained from the Lagrange multiplier superfield). } 

If the above manipulations of the path integral are correct (as one expects that they are, since this model is a superspace analogue of the model in Section \ref{quantumD}, which can be put on a lattice),\footnote{See for example Aitchison and Hey (2013:~pp.~158-161).} 
then the form of the effective action indeed remains the same under electric-magnetic duality, where the {\it dual coupling} is related to the original ones in the already familiar form: $\t_D(a_D)=-{1\over\t(a)}$.

It is useful to rewrite the moduli space metric, Eq.~\eq{SWmetric}, using a mix of the original and the dual variables, Eq.~\eq{LegendreF}:
\bea
\dd s^2=\mbox{Im}\left(\dd a_D\,\dd\bar a\right)={i\over 2}\left(\dd a\,\dd\bar a_D-\dd a_D\,\dd\bar a\right).
\eea
Here, $a_D$ is the expectation value of the dual Higgs (defined from Eq.~\eq{LegendreF}), $\bra\f_D\ket=a_D\,\s_3$. This metric is invariant under $\mbox{SL}(2,\mathbb{Z})$ duality, which acts as follows:
\bea\label{dualHiggs}
\left(\begin{array}{cc}a_D\\a\end{array}\right)\mapsto\left(\begin{array}{cc}\a&\b\\\g&\d\end{array}\right)\left(\begin{array}{cc}a_D\\a\end{array}\right),~\a\d-\b\g=1\,.
\eea
By rewriting the effective action using a mix of the Higgs field and its dual, i.e.~$\F$ and $\F_D$, one can show that this is the duality group of the effective action.

\subsection{Physical signficance of the singularities}\label{physsig}

The duality map acts on the Higgs field and its dual according to Eq.~\eq{dualHiggs}. This allows us to determine how the {\it set of (lowest-energy) states} of the model is mapped by the quasi-duality, along the lines of the Schema's discussion in Section \ref{isomdef}. The particle spectrum is best seen from the supersymmetry algebra (see also the discussion of the ${\cal N}=4$ case in Section \ref{N=4SYM}). We are interested in the states with minimum energy, i.e.~the BPS states. For such states, the BPS condition, which minimizes the mass in Eq.~\eq{bps}, is exact, i.e.~it receives no quantum corrections. Also, the BPS relation between the mass and the charges is invariant under electric-magnetic duality. This can be checked from the supersymmetry algebra (indicated schematically in Eq.~\eq{susyZ}) for the duality map, Eq.~\eq{dualHiggs}, that we have just discovered: since, for states with SU(2) symmetry broken to U(1), the expectation value of the Higgs field appears on the right-hand side of the supersymmetry algebra. 

An {\it electric state} has central charge $Z=an_e$, where $n_e$ is the integral electric charge. Such a state is dual, under an S-duality transformation $\left(\begin{array}{cc}0&1\\-1&0\end{array}\right)$, to a {\it magnetic state}, with central charge $Z=a_Dn_m$. A general duality transformation, Eq.~\eq{dualHiggs}, gives a {\bf dyon}, i.e.~a state with both electric and magnetic charge, with central charge $Z=an_e+a_Dn_m$. These states are all BPS states, and one can show that their mass is invariant under $\mbox{SL}(2,\mathbb{Z})$ transformations.

Since the Higgs field and its dual are related by the Legendre transformation Eq.~\eq{LegendreF}, the physical consequences of the duality relation are encoded in the form of the prepotential ${\cal F}$, i.e.~in Eq.~\eq{Fsum}. To determine this prepotential is to determine, through Eq.~\eq{LegendreF}, the expectation value of the dual Higgs (at strong coupling), as a function of the original Higgs at weak coupling, i.e.~$a_D(a)$. 

We already know that the moduli space metric (and thus the prepotential) is not globally well-defined because of singularities, which we interpret as massless particles that had been integrated out in regions of moduli space where they are heavy, but need to be taken into account in regions of the moduli space where they are light. We can see this from the general form of the prepotential, Eq.~\eq{Fsum}, which we expect to have singularities at $u=\infty$, $u=\pm\L^2$, and $u=0$ (but whether these are true singularities of course depends on the yet undetermined coefficients $c_k$). The point $u=0$ turns out to not be a singularity of the quantum model, and so we will discard it.\footnote{See e.g.~Bilal (1997:~p.~107) and Seiberg and Witten (1994a:~p.~37).} 
This leaves us with three singularities, at $u=\infty$ and $u=\pm\L^2$ (recall that $\L$ is the Wilsonian cutoff). In principle there could be more, but Seiberg and Witten proceeded on the assumption that there are just three. The point will be that, {\it although the effective action is not valid at the singular points, duality allows us to recoordinatize the moduli space so that a new effective action is valid near (though not at) the singular point, and the whole moduli space, which excludes the singularities, is thus covered by three dual models.}

The region near $u=\infty$ is where the expectation value of the Higgs field is large, and so it corresponds to weak coupling, where the semi-classical approximation is good (see footnote \ref{largeHiggs}). In that region, the leading result is given by the classical and one-loop terms of the prepotential, Eq.~\eq{Fsum}, and the instanton terms are sub-leading. Thus our effective model $M$, i.e.~${\cal N}=2$ U(1) SYM, is valid in this region.

The region of moduli space away from $u=\infty$ lies outside the domain of convergence of the instanton sum in Eq.~\eq{Fsum}, where the prepotential is not well-defined. To find coordinates in terms of which the prepotential is well-defined, we are physically motivated to use the dual variables $a_D$ and ${\cal F}_D$, from Eq.~\eq{LegendreF}, because the model is weakly coupled in these variables.\footnote{We have stated the definition of the dual variables as a {\it physically motivated assumption}, rather than as an established mathematical fact, because Seiberg and Witten (1994a) juxtapose heuristic physical assumptions, like this one, with detailed mathematical methods.}

These assumptions---that there are three singularities, and that the description in terms of the dual Higgs variable, $a_D\simeq0$, is good near the strong-coupling singularity $u=\L^2$---allow us to find the prepotential ${\cal F}_D(a_D)$, which amounts to resumming the instanton series in Eq.~\eq{LegendreF} by using magnetic variables. 

The {\bf monodromy} of $(a_D,a)$ around a singularity characterizes the singularity precisely. Briefly, if we go around a singular point, the logarithmic term of the prepotential (in its $a$-dependent form Eq.~\eq{Fsum}, or in its dual reformulation ${\cal F}_D$) shifts the pair $(a_D,a)$ by an element of $\mbox{SL}(2,\mathbb{Z})$, Eq.~\eq{SL2Z}.\footnote{This logarithmic term is the one-loop contribution to the prepotential. See Lerche (1998:~p.~172).} 

Assuming that the best interpretation of the strong-coupling singularities at $u=\pm\L^2$ is that massive particles become massless near the singularity, there are two ways to argue that these particles are {\it magnetic monopoles}: in the original model $M$, and from its dual, $M$'.

In $M$, there are two candidates for massive particles becoming massless: (i) the particles in the vector multiplet of $T$, i.e.~${\cal N}=2$ SU(2) SYM, which are massive by the Higgs mechanism and are integrated out, and (ii) the soliton states whose charges appear in the Bogomol'nyi bound derived from the supersymmetry algebra, Eq.~\eq{susyZ}. Since the particle states cannot be responsible for the singularity, only the soliton states remain.\footnote{The argument that the massless particles cannot be gauge bosons, and thus also not Higgs particles, which are in the same ${\cal N}=2$ supermultiplet, is spelled out in Bilal (1997:~p.~107). Since there are no other elementary ${\cal N}=2$ multiplets in the theory, only soliton states remain.} 
The Bogomol'nyi bound, $M=|an_e+a_Dn_m|$, implies that, at the point $a_D=0$, the state that becomes massless must be a {\it magnetic monopole}, i.e.~a soliton with charge $n_e=0$. These are supersymmetric 't Hooft-Polyakov monopoles of $T$, which at long distances are charged under the unbroken U(1).

\subsection{The dual model}\label{dualm}

We can use the above argument to find the dual model, $M'$ (i.e.~Eq.~\eq{LegendreF}), and show that the assumption that the extra states are monopoles is a coherent assumption that gives well-defined physics. First, from the Legendre transform of the effective action in Section \ref{dualpro}, it follows that $M'$ has a dual photon, $W_{D\a}$, and a dual Higgs, $\F_D$.\footnote{It follows from ${\cal N}=2$ supersymmetry that the dual photon is in an ${\cal N}=2$ BPS vector multiplet. The dual Higgs is in the same hypermultiplet as the monopoles: see below.}

Second, magnetic monopoles are charged particles with spin 1/2:\footnote{This follows from the supersymmetry algebra Eq.~\eq{susyZ}, see also the discussion in Section \ref{Wea}.} 
a monopole belongs to an ${\cal N}=2$ hypermultiplet that couples electrically to the dual gauge field.\footnote{There are additional states, but at low energies these are the main ones. The helicity states of $M'$ are as on the right Figure \ref{N=4N=2spectrum}, with `particle' and `soliton' states exchanged (and the vector and hypermultiplets are massive BPS, rather than massless, multiplets). (The hypermultiplet contains a dyon state, which plays a role in $M''$, which we do not consider in detail here.)}
Thus $M'$ has a gauge vector boson $W_{D\a}$, a Higgs field $\F_D$, and a charged spin-1/2 particle. This model is massive {\it ${\cal N}=2$ quantum electrodynamics} (SQED), where the mass of the electrically charged particles is small (note that the spin-1/2 particles are monopoles in $M$, but are {\it electrically charged} in $M'$, because they couple electrically to the gauge field!). 

Unlike $T$, this model is known to {\it not} be asymptotically free; its beta-function, to one loop, is known and positive.\footnote{The RG equation is given by ${\m\,{\dd\over\dd\m}}\,e_D={e^3_D\over8\pi}$, where $e_D$ is the dual electric coupling, and the scale $\m$ is proportional to $a_D$, just as $a$ played the role of the renormalization group scale in the original model.} 
This beta-function equation allows one to determine the complex coupling $\t_D$ as a function of the expectation value of the dual Higgs, $a_D$, and to thus find the relation between the expectation value of the Higgs of $M$ and its dual in $M'$, i.e.~the duality relation, $a(a_D)$, near $a_D\simeq0$. From there, one can calculate the monodromy matrix of the pair of variables $(a_D,a)$ in the complex $u$-plane, around $u=\L^2$. One finds that the magnetic monopole state $(n_m,n_e)=(1,0)$ is an eigenstate of this monodromy matrix, as it should if the singularity corresponds to a magnetic monopole becoming massless---namely, it is the state left invariant by the monodromy.

The three monodromies thus calculated allow one to solve for the geometry of the moduli space, which can be rewritten in terms of a two-sheeted Riemann surface with two branch cuts.\footnote{The appearance of a Riemann surface is closely related to other ideas string theory. See for example Dijkgraaf and Vafa (2002:~pp.~24-26) and Chapters \ref{String} to \ref{HABHM}.} 
If one adds the point at infinity, this Riemann surface becomes a torus. The expectation values of the Higgs and its dual, $a$ and $a_D$, can then be reinterpreted as the periods around the two cycles of the torus of a certain meromorphic one-form on the Riemann surface. These integrals then allow one (through Eq.~\eq{LegendreF}) to find the prepotential ${\cal F}$ and its dual, ${\cal F}_D$, and to thus find the effective action in any region of the moduli space. Thus, by finding the three models, which are valid in three different regions, the ${\cal N}=2$ low-energy effective theory is solved.\\
\\
{\bf Duality or quasi-duality?} Recall, from Section \ref{Wea}, our two main arguments explaining why electric-magnetic duality is not a duality of ${\cal N}=2$ SYM: (i) the models are only valid at low energies; and (ii) they have different quantum numbers. While (i) does not in principle prevent there being a duality {\it of low-energy models}, i.e.~of the Seiberg-Witten theory: as we also discussed, (ii) does prevent it, because it implies that there is no isomorphism of state spaces. This is indeed confirmed by the analysis of the dual model that we just reported. For 
the appearance of a hypermultiplet in $M'$, there being none in $M$ (or in $T$), highlights that there is {\it no isomorphism} of models, and so that $M$ and $M'$ are not duals: in $M'$, the monopoles are in supermultiplets with spins up to 1/2, while in $M$ they are solitons in a supermultiplet with spins up to 1. Although the state-spaces are equinumerous, the quantum numbers of the states are different: `the fact this spectrum is not duality invariant is precisely the reason that it was concluded many years ago that Olive-Montonen duality did not hold for ${\cal N}=2$ super Yang-Mills theory'.\footnote{Seiberg and Witten (1994a:~p.~50). Although the values of some of the quantities (for example, the masses of BPS states) are the same in the three models, not all quantities have the same values. For if one evaluated the free energies of $M$ and $M'$, these would not be the same, not even up to overall factors as in e.g.~Eq.~\eq{Zmod} in the Ising model (they might agree at special points, but not on large regions of, the moduli space). Rather, what is required is a model-independent calculation of the partition function of $T$ (and here, the model-independent information given by the models, such as the moduli space, is surely useful). Such partition functions have been evaluated in Pestun (2012:~p.~112) and Hollowood and Kumar (2015:~p.~30) on $S^4$, and their transformation properties under duality are anomalous. This can be cured by adding gravitational curvature terms along the lines of Vafa and Witten (1994), and ultimately there may be a cousin of $T$ that implements duality as some kind of symmetry: but this requires considering the whole theory, rather than its low-energy models.}

 Even so, one might have thought that the prepotential, Eq.~\eq{Fsum}, could be used to define a common core: and that the Legendre transformation discussed in Section \ref{dualpro} is a duality between two representations of this common core. 

We argue that this conclusion does not follow: first of all, the prepotential does not by itself define a theory; but second, and more important, the prepotential has a {\it limited radius of convergence}, so that there is a structure-preserving map between $M$ and $M'$ {\it only in the region where the two models overlap}. And that region is not the whole moduli space. (In particular, the Legendre transformation Eq.~\eq{LegendreF} only makes sense where the models overlap.) This again means that the state spaces, as thus defined by Seiberg and Witten, are not isomorphic, and we have a quasi-duality.

As we will discuss in the next Section, the correct way to think of the quasi-dual models here defined is not as representations of a common core, but as overlapping models that together define a single theory. This will lead in to our {\it geometric view of theories}.
\\
\\
{\bf Stipulated and proper symmetrties.} The theory $T$ has a stipulated ${\cal N}=2$ supersymmetry, which includes the Poincar\'e group, and is represented in both $M$ and $M'$ (as we have discussed, these representations are {\it not} isomorphic). And each of $M$ and $M'$ has its own proper symmetries, in the sense of Section \ref{dualsym}. For example, although $M$ and $M'$ both have local U(1) symmetries, these U(1)s are not the same. The U(1) in $M$ is what remains of SU(2) after symmetry breaking, while the U(1) of $M'$ is not related to SU(2), but emerges in the dual phase, where the monopoles are light and couple electrically to the dual gauge field (this is similar to Section \ref{MEMD}, where the two U(1)s are also not related). 

\subsection{Monopole condensation, confinement and supersymmetry breaking}\label{moncond}

So far, the states in the low-energy models were very light, i.e.~the massive modes are integrated out, and singularities appear at points where one of these modes is massless. Thus the models were in Abelian Coulomb phases (see Section \ref{cmds}).\footnote{For more detailed discussions of the phases of supersymmetric gauge theories and how they are related by dualities, see Intrilligator and Seiberg (1996:~pp.~1-2), Alvarez-Gaum\'e and Zamora (1998:~pp.~54-56, 65-68).}

One can go into a {\it confining phase} by adding a mass term for the ${\cal N}=1$ chiral superfield in the bare Lagrangian, i.e.~$m\,\Tr\,\F^2$. Adding this term breaks the ${\cal N}=2$ supersymmetry to ${\cal N}=1$. The minimization of the low energy Wilsonian effective action removes the vacuum degeneracy (since it fixes $a_D=0$, and so the perturbation is at the point where the monopoles would have been massless),\footnote{There are two vacuum states, with a $\mathbb{Z}_2$ symmetry, rather than a whole moduli space of vacua. See Seiberg and Witten (1994a:~p.~41).} and gives a vacuum expectation values to the hypermultiplets, i.e.~the monopoles. Expanding around this vacuum, the results can be summarized as follows (see Alvarez-Gaume and Zamora (1998:~p.~75)):

(i)~~There is a mass gap of order $\sqrt{m\L}$, which is the mass of the lightest particle.

(ii)~~The objects that condense are magnetic monopoles, and this is how the gauge field, under (i), gets a mass: namely, through a magnetic Higgs mechanism, whereby pairs of monopoles, rather than the Higgs field, acquire a vacuum expectation value. This is seen in the non-zero expectation value of the (dual) hypermultiplet containing the monopoles. There are electric flux tubes with a non-zero string tension of the order of the mass gap, that confines electric charges of the unbroken U(1) gauge group. 

Thus this is a precise realization of the Mandelstam-'t Hooft mechanism for confinement, from Section \ref{cmds}, through a dual (i.e.~magnetic) Meissner effect. It seems to be the first four-dimensional relativistic quantum field theory `in which the confinement of charge is explained in this long-suspected fashion' (Seiberg and Witten, 1994a:~p.~21).\footnote{Supersymmetry can be completely broken in such a way that the condensation of magnetic monopoles remains a valid mechanism of confinement. For a discussion, see Alvarez-Gaume and Zamora (1998:~p.~76).}

\section{Illustrating the Schema, and the geometric view}\label{SWmanifold}

We have discussed how Seiberg-Witten quasi-duality maps states, quantities, and dynamics, and it is worth now adding a few words about how it illustrates the Schema, and about what we are going to call the `geometric view' of theories: which goes beyond the Schema, and will be taken up in Section \ref{mvd}.

Although ${\cal N}=4$ SYM theory illustrates the Schema's conception of duality directly, we could not present it in full detail here. This is both due to the limitations of our discussion, and to the lack of a full proof of the duality. 

We have focussed on the subset of supersymmetric BPS states, whose mass equals their charge. They are tabulated in Figure \ref{N=4N=2spectrum}, and for each value of the total spin, the number of electric (particle) states and magnetic (soliton) states match. The duality map exchanges the gauge group with its Langlands dual, and thus exchanges the electric and magnetic charges, so that it exchanges these states one by one (see also the discussion in Section \ref{Wea}).

Up to an overall normalization, the values of the physical quantities such as the partition function remain the same under this exchange.\footnote{As we already saw in Section \ref{MEMD}, quantities like the partition function transform like modular forms under duality.}
This is similar to other dualities that we have been able to discuss in more detail, such as Kramers-Wannier duality (see Eq.~\eq{Zmod}), and electric-magnetic duality for the Maxwell theory (see Section \ref{quantumD}). (We do not need to discuss the dynamics separately here, because the match of the partition function and the other correlation functions in a quantum field theory implies the matching of the dynamics.)

As Section \ref{Wea} discussed in detail, ${\cal N}=2$ SYM does not instantiate Montonen-Olive duality, because the quantum numbers of the electric and the magnetic states do not match (see the right table in Figure \ref{N=4N=2spectrum}). Rather, the low-energy models are quasi-duals, and the quasi-duality is a bijection that preserves some of the structure. It also maps the Wilsonian effective actions through a Legendre transformation.

This quasi-duality can be reformulated more generally as a {\bf geometric view} of theories. At low energies, the moduli space of vacua of the ${\cal N}=2$ theory is a two-dimensional manifold: the space of vacuum expectation values of the Higgs field (see Section \ref{swt}). Two comments are in order, both of which aim to clarify the matching of states and quantities between the models: about the interpretation of the vacuum expectation value of the Higgs field, and about the properties of the moduli space.\\
\\
(i)~~{\it Expectation value of the Higgs field and classical states}: in the low-energy approximation, the Higgs field is treated as a background field, whose classical expectation value minimizes the potential (see the discussion in Section \ref{vacsols}). This enables an effective description, by considering the fluctuations around this value (but, as we have discussed, this effective description has a limited region of validity). The Higgs field is a complex coordinate on the moduli space, with each point in the moduli space being a semi-classical state of the theory, i.e.~a solution of the vacuum equations. The vacuum expectation value of the Higgs depends on the theory's parameters, such as the coupling and mass (see the paragraph following Eq.~\eq{MexicanV}). (In a laboratory experiment, these values can be varied by changing the energies and momenta of the particles, and other external parameters, such as the temperature or the value of the externally applied magnetic field.) The three regions of the moduli space correspond physically to three phases of the theory (Coulomb, monopole, and confining phases), each of which has a different qualitative behaviour, that is indicated by an appropriate order parameter (see the discussion in Section \ref{cmds}). However, these classical states of the system are not the system's {\it quantum} states. The information about the quantum states is encoded in the:\\
\\
(ii)~~{\it Properties of the moduli space and quantum states}: the moduli space is a two-dimensional complex plane with three punctures, equipped with a metric and a complex structure that is metric-compatible.\footnote{For more details, see footnote \ref{Kmetric}.}
These geometric quantities on the moduli space contain all the physical information about the states, quantities, and dynamics of the theory in the low energy limit. The expectation value of the Higgs field is a coordinate on the moduli space (i.e.~a map from the three-punctured plane into $\mathbb{C}$; and in the other regions, a dual field plays the role of a coordinate). The metric on the moduli space (which is derived from the prepotential) encodes the information about the BPS states: namely, the monodromies of the coordinates around the singularities are elements of $\mbox{SL}(2,\mathbb{Z})$, and the corresponding integers give the values of the electric and magnetic charges in each state (see Eq.~\eq{dualHiggs}). Of the 16 BPS states shown in Figure \ref{N=4N=2spectrum} (right), eight are electric and eight are magnetic.\footnote{By a change of variables, one can associate with the moduli space a torus with a one-form defined on it; the two periods of this one-form are integers that give the charges of the various states.}

Section \ref{mvd} will further discuss how the idea of a moduli space of a quantum theory, as the space of a theory's classical vacua, whose geometry encodes the relevant information about the quantum states and quantities at low energies, is relevant more generally for a view of dualities and quasi-dualities.

\section{Conclusion}\label{disc7}

Montonen-Olive duality is realized for ${\cal N}=4$ SYM, but not for ${\cal N}=2$ SYM: and this failure is a source of interesting physics, including a confining phase that, as predicted by 't Hooft and Mandelstam, is triggered by the condensation of monopoles. 

The ${\cal N}=4$ theory with gauge group SU(2) is invariant under the duality group $\mbox{SL}(2,\mathbb{Z})$ that mixes the electric and magnetic states. The masses of the lowest-energy states (the BPS states) can be calculated and are invariant under the duality, as are the values of other quantities.

In ${\cal N}=2$ SYM with gauge group SU(2), there is an {\it effective duality} of low-energy models. The Higgs field is a complex coordinate on a two-dimensional moduli space of vacua. This moduli space comes equipped with a metric that has three singularities, i.e.~points where the conformal factor blows up. At these points, dual variables defined by Legendre transformation (Eq.~\eq{LegendreF}) render the metric regular. This Legendre transformation exchanges electric and magnetic states. (As we will discuss in Chapters \ref{STII} and \ref{HABHM}, the appearance of moduli spaces with non-trivial metrics is closely related to black hole microstates and holographic dualities.)

The above discussion is in the Coulomb phase, where the low-energy particles are massless: there is also a confining phase (obtained, as anticipated in Chapter \ref{EMDuality}, by giving a mass to a scalar field), in which monopoles condense. 

't Hooft-Polyakov {\it monopoles} illustrate, in interacting four-dimensional quantum field theories, aspects of solitons from previous Chapters. The monopole has a smooth core of SU(2) fields that, from the outside, looks like a Dirac monopole. The long-range abelian gauge field emerges from this non-abelian core. Its magnetic charge is the integral of a topological current (not a Noether current), and the degree of a map implies the Dirac quantization condition.

The {\it quasi-duality} of the Seiberg-Witten theory shares features with dualities: it is a bijective map, between state spaces and quantities of effective models, that preserves some, {\it but not all}, of the values of the quantities. The models describe regions of the moduli space of $T$, including the metric and the prepotential. Thus a characterization of $T$ emerges as a {\it successor theory}. This will be one of the guiding ideas in Chapter \ref{Heuri}.

The quasi-duality also illustrates our theme from Section \ref{featurerole} of {\it hard-easy}: namely, the singularities indicate the lack of convergence of the effective action (i.e.~the infinite series in Eq.~\eq{Fsum}). But near a singularity, the series can be defined outside the radius of convergence of the original variable, by using a dual variable. Thus quasi-duality gives us access to the theory's domain in a range of parameters that is intractable in one model, but tractable in a dual model. 

\chapter{Elements of String Theory Dualities}\label{String}
\markboth{\small{\textup{Elements of String Theory Dualities}}}{\textup{\small{Elements of String Theory Dualities}}}

This Chapter and the next illustrate the profound implications of dualities in both physics and philosophy in the context of string theory. This is not only because some of these dualities are very surprising, and unexpected; but also because duals purport to describe the whole universe, including gravity. Thus we will pick up these examples in Part III, when we discuss what it is for two models to make `the very same claims about the world' (what we call `physical equivalence'), and how dualities bear on our view of scientific theories.

The first two Sections are introductory: Section \ref{DSTov} gives an overview of dualities in string theory (summarized in Figure \ref{Mthfig}), and Section \ref{stringth} briefly introduces relativistic strings. Section \ref{T-d} then discusses T-duality (i.e.~`target space' duality). Sections \ref{Dbopen}, \ref{Dcst}, and \ref{dynD} discuss the interaction between strings and D-branes, which are non-perturbative objects that are required for string models to be dual, as well as the dynamics of D-branes. (Appendix 8.A discusses the generalization of classical electric-magnetic duality to D-branes.)

\section{Duality in string theory: an overview}\label{DSTov}

\subsection{Five different superstring theories}\label{5diff}

The foundational idea of string theory is that point particles, whose world-lines are (0+1)-dimensional curves in spacetime, are replaced by strings, i.e.~objects that are spatially extended along one dimension: they sweep out (1+1)-dimensional world-sheets as they move in spacetime. The main motivation for this is that such theories contain gravitational excitations in their spectrum {\it and} have optimal ultraviolet behaviour, i.e.~ultraviolet divergences can be renormalized. (This is not the case if one attempts to quantize objects that are extended along more than one dimension: thus the string is an ``optimal'' compromise between point particles and higher-dimensional objects).

The parameters and field content of string theories are constrained by the rules of quantum mechanics: avoiding unphysical results (such as the presence of `tachyons', or excitations of the string spectrum with negative mass-squared, and preserving unitarity so that the theory has the usual Born-rule interpretation) imposes a series of constraints. We will say that these constraints arise from the {\bf quantum consistency} of the theory: quantum consistency determines that (so far as we know) only five kinds of conventional string theories can be defined, and also the number of spacetime dimensions in which they move is not arbitrary: this number can only be ten. With the aim of understanding the diagram in Figure \ref{Mthfig}, we briefly discuss these two issues.

The first ingredient that we require to secure quantum consistency is supersymmetry, i.e.~in addition to the bosonic fields that describe the movement of the string in spacetime, we add matching fermionic fields (see Section \ref{basicsusy}): thus is born a {\it superstring}.

There are two kinds of detail that needs to be settled for a superstring. First, a string is either open or closed, depending on whether it has endpoints or forms a closed loop, so that its spatial topology is a line segment (i.e.~an interval) or a circle. The theory of {\it open} superstrings is called {\bf Type I}. (It has both open an closed strings because, as we will see, {\it all} five string theories have closed strings, which describe gravitational excitations.) Theories with only {\it closed} strings were (initially) called {\bf Type II} theories.

While it is important to note that there are two Type II theories, depending on the chirality of the fermions: Type IIA and Type IIB,\footnote{In the Type IIA theory, left- and right-chiralities are equally distributed among the fermions, while in the Type IIB theories all fermions have the same chirality.} 
we will not need to use details about the fermions here. For we will alternatively distinguish Type IIA and Type IIB theories through the different kinds of {\it bosonic} spacetime fields and D-branes that they contain.

We said that closed string theories were {\it initially} called Type II, because two more closed string theories were subsequently discovered.\footnote{For a brief history of superstring theory, see Rickles (2014).} 
These were dubbed `heterotic' theories, because both of them are hybrids of a superstring and a bosonic superstring. Although they are important theories, we will not need their details in this Chapter, and it will be enough for us to know that they can be defined, because they satisfy the {\it well-definedness} condition mentioned above in a different way than the usual superstring, indeed by admixing properties of the purely bosonic string and the superstring. These two theories are called Heterotic $E_8\times E_8$ and Heterotic SO(32), depending on their gauge group.

Summarizing, we have {\it five} superstring theories: the Type I theory (with both open and closed strings), the Types IIA and IIB theories (with closed strings only), and the two Heterotic string theories, which also have closed strings only, and are constructed as hybrid theories. The mathematical well-definedness of superstring theories also requires the number of dimensions of spacetime to be ten, i.e.~nine space and one time dimension. See Figure \ref{Mthfig}.

There is a cautionary note about the above, which will later reconnect to our discussion of non-perturbative physics in the previous Chapter: string theories, as thus defined through their elementary excitations, are only {\it perturbatively} well-defined, i.e.~for interactions between the strings that are weak, so that a stringy analogue of the perturbative expansion of quantum field theory applies. The string diagrams are actually ultraviolet finite, i.e.~there are no short-distance divergences, up to two loops, basically because there is no contribution from the high-energy region.\footnote{One-loop finiteness is discussed in e.g.~Green et al.~(1987:~p.~17) and Polchinski (1998a:~p.~219). Two-loop finiteness is discussed in D'Hoker and Phong (2002:~p.~252). It has also been argued, e.g.~by Sen (2015:~p.~5), that ultraviolet divergences are absent at higher loops.} 
However, the perturbative expansion is a divergent asymptotic power series in the {\bf string coupling} constant, $g_{\sm s}$.\footnote{Thus, even though the individual terms in the perturbative expansion are finite at each order, the whole series diverges. Asymptotic expansions nevertheless can give a reliable approximation for physical quantities. For a discussion of asymptotic series, see Paulsen (2014:~p.~14) and Dingle (1973:~pp.~2, 19). We will see later that the string coupling constant is the asymptotic value of a field, the dilaton field.}
Thus, although the theory gives good, finite, results for small values of $g_{\sm s}$ (where `good' here means `to the desired accuracy in the perturbative expansion'), the series does not converge for large values of the coupling. In fact, the perturbative definition of the superstring does not give us a way to treat non-perturbative effects. As we will see, dualities give crucial clues about non-perturbative effects in string theory. 

\subsection{Dualities between string theories}\label{dbst}

In the early 1990s, there were strong hints that these five superstring theories were related by dualities. And it was known that, at least classically, ten-dimensional superstrings can be obtained from membranes in eleven dimensions, by dimensional reduction of the eleventh dimension, i.e.~this eleventh dimension is a circle so small compared to the rest of the spacetime, that the spacetime effectively looks ten-dimensional. Also, the membrane is wrapped on this circle, so that it looks like a {\it string} that is moving in a ten-dimensional spacetime. This is called `double-dimensional reduction'. The theory that describes such a configuration in eleven dimensions is {\it eleven-dimensional supergravity}.\footnote{This is a classical theory, because membranes are not straightforwardly quantized. See De Wit et al.~(1989:~p.~141) and Dasgupta et al.~(2002).}

\begin{figure}
\begin{center}
\includegraphics[height=5cm]{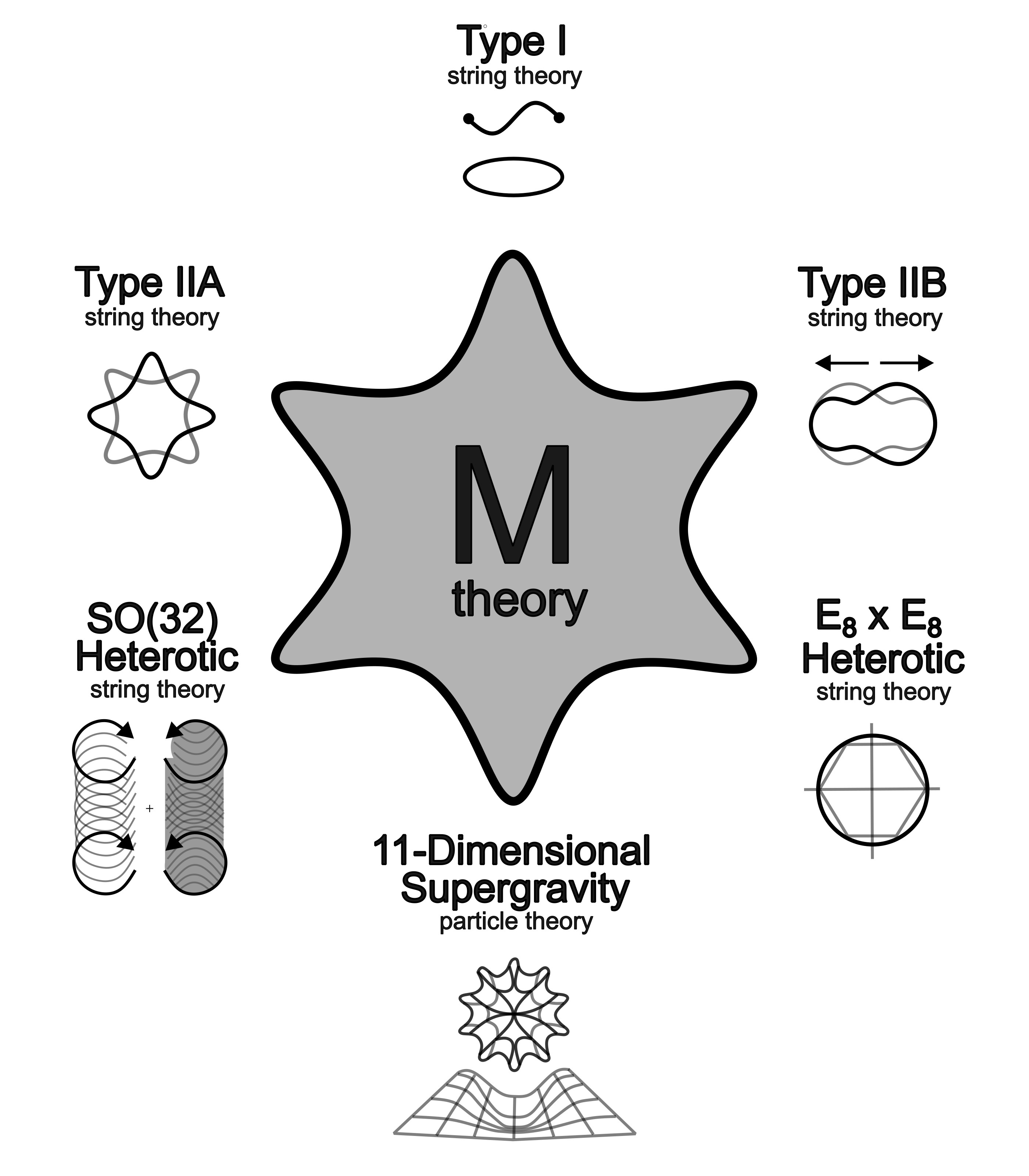}~~~
\caption{\small The space of M-theory, with its five corners of Type I, Type IIA, Type IIB, Heterotic SO(32) and Heterotic $E_8\times E_8$ string theories, and eleven-dimensional supergravity.}
\label{Mthfig}
\end{center}
\end{figure}

In 1995, papers by Hull and Townsend (1995) and Witten (1995a) gave evidence for the existence of a series of dualities that {\it relate the five string theories} to each other and to eleven-dimensional supergravity. Combined with Polchinski's (1995) work on D-branes, this gave a new understanding of the physics underlying these dualities: Witten conjectured that a yet unknown eleven-dimensional `M-theory' exists that explains the web of dualities (see Figure \ref{Mthfig}). These developments are called the {\it second superstring revolution}.\footnote{For a discussion of the first and second superstring revolutions, in 1984 and 1995 respectively, see Rickles (2014:~Chapters 8 and 10).} 

In a nutshell, Witten realized that the various hints of duality previously discovered, as well as the new dualities that he conjectured, required the existence of a new class of lowest-energy states in string theory that are noticeable when the string coupling is large. These are BPS-saturated states, i.e.~they are massive (supersymmetric) states, and their mass saturates a BPS-bound of the type Eq.~\eq{bps}, namely:\footnote{This is analogous to our discussion, in Chapter \ref{EMYM}, of electric-magnetic duality for supersymmetric Yang-Mills, which requires the existence of magnetic monopoles, i.e.~solitons that satisfy the BPS condition. However, the detail of the way the BPS formula in string theory depends on coupling constant is different.}
\bea\label{Mcharge}
M=c\,{|n|\over g_{\sm s}}\,,
\eea
where $n$ is a unit of elementary electric charge of these new objects, and $c$ is a numerical constant (we will fix it later, in Section \ref{dynD}). The charge $n$ of these novel BPS states is {\it not} the electric charge of the superstrings.\footnote{Superstrings are themselves also sources of electric charge because, as we will see later, one can couple a suitable gauge field to their world-sheet coordinates, analogously to how a point particle is coupled to a gauge field.}
It is a new type of charge that is {\it invisible} to the string world-sheet, because, when the string coupling is zero, the BPS states are infinitely massive (see Eq.~\eq{Mcharge}). Thus they are non-perturbative states: they are only visible, because light, when the coupling is strong, i.e.~when $g_{\sm s}\gg1$. 

This prediction of dualities, i.e.~the requirement that a set of new BPS objects exists, is possible because some of these duality transformations involve the inversion of the coupling, i.e.~$g_{\sm s}\mapsto1/g_{\sm s}$ (and-or other parameters). In a way analogous to previous Chapters, the {\it hard-easy} feature of dualities enables their {\it unifying} seemingly different string models (see Section \ref{featurerole}). If the duality conjectures are true, what appeared to be different theories are found to be models of a single theory: namely, string theory, or M-theory.

\section{A brief introduction to string theory}\label{stringth}

Before we go on to study the first of these dualities (viz.~T-duality, in Section \ref{T-d}), we here give a lightning review of the aspects of string theory that we will use in this Chapter and the next.

\subsection{Rudiments of the world-sheet theory of strings}\label{rudiments}

Strings are best understood as analogues of massless point particles moving in spacetime. The geodesic equation for a massless point particle in Minkowski space is the free equation ${\sm d^2x^\m\over\sm d\s^2}=0$, where $x^\m(\s)$ is the spacetime coordinate of the particle as a function of the affine parameter, $\s$, along the null geodesic. That the geodesic is null is expressed by the condition that the four-velocity lies on a light-cone: ${\sm d x_\m\over\sm d\s}{\sm d x^\m\over\sm d\s}=0$ (contracting the indices using the Minkowski metric). 

The particle's geodesic equation and the null or light-cone condition are both readily derived by extremizing the path length along the particle's trajectory. This latter quantity can be conveniently rewritten as the following action:
\bea\label{mpa}
S_{\tn{mp}}=\int{\dd\s\over e(\s)}~{\dd x_\m\over\dd\s}{\dd x^\m\over\dd\s}\,,
\eea
where $e$ is a Lagrange multiplier that can be set to one by a convenient reparametrization of the integral. Extremizing this action with respect to both $x$ and $e$, and then setting $e=1$, gives the geodesic equation for $x$ and the light-cone condition for the four-velocity. \\
\\
{\bf Classical theory.} The {\it classical action for a string} moving in Minkowski space is a straightforward analogue of the above expression for the point-particle. As in the point-particle case, it is the action whose variation extremizes the area in spacetime swept out by the string, i.e.~it is a {\bf world-sheet action}, often called the {\bf Polyakov action}:\footnote{Also Brink-Deser-Di Vecchia-Howe-Polyakov-Zumino action. The {\bf Nambu-Goto action} that we discussed in Section \ref{analo2} is the integral of the area of the world-sheet. Since the area contains a square root, the Nambu-Goto action is difficult to quantize. The Polyakov action is quadratic in the fields, which it achieves by introducing the auxiliary metric $h^{\a\b}$, and one obtains a conformal field theory. The two actions are related by varying the Polyakov action with respect to $h^{\a\b}$ and substituting the equation of motion back into the action.\label{NambuGoto}}
\bea\label{WSa}
S_{\tn{WS}}=-{T\over2}\int\dd^2\s\,\sqrt{h}\,h^{\a\b}\,\pa_\a X_\m\,\pa_\b X^\m\,.
\eea
The fields $X^\m(\t,\s)$ are the {\bf target space coordinates} of the string in the target Minkowski spacetime, and they are the analogues of the coordinates $x^\m(\s)$ of a point particle. The variables $\t,\s$ are, respectively, the time and space coordinates on the {\bf world-sheet} of the string. Just like the affine parameter for the massless particle, the world-sheet coordinates are arbitrary, and the action is invariant under reparametrizations.\footnote{One can identify these two coordinates with two of the target space coordinates (e.g.~so that the Minkowski time is the time coordinate on the world-sheet, $X^0=\t$), but we will not need this since we prefer to work with the general expressions.} 
Latin indices $\a,\b$ take values $1,2$ and label the world-sheet coordinates $\s^\a$, i.e.~$(\s^1,\s^2)=(\t,\s)$. The metric $h^{\a\b}$ on the world-sheet secures the reparametrization invariance of the action. Like for a point particle, the reparametrization invariance of the world-sheet can be used to simplify the world-sheet action so that this metric is flat:\footnote{Comparing the world-sheet action Eq.~\eq{WSa} to the point-particle action Eq.~\eq{mpa}, we see that the Lagrange multiplier $e$ can be interpreted as an {\it einbein}, i.e.~a $1\times1$ metric on the trajectory of the particle. For the string this is more complicated, because $h^{\a\b}$ is a $2\times2$ matrix. It has three independent components; using reparametrization invariance to bring it to a diagonal form leaves us with a scale factor that is non-zero, but which nevertheless drops out of the action because of conformal invariance. This conformal symmetry of the string action is an important property of the string action.}
\bea\label{WS2}
S_{\tn{WS}}=-{T\over2}\int\dd^2\s\,\pa_\a X_\m\pa^\a X^\m\,.
\eea
By dimensional analysis, we see that the overall constant $T$ has dimensions of mass squared (or one over length squared): it is the string tension, and for historical reasons, it is often written as $T=1/2\pi\a'$, where $\a'$ is the `Regge slope'.\footnote{In the 1960s, the Regge slope gave the relation betwen the mass and the spin of hadron resonances, $m^2=J/\a'$, and it was one of the reasons for the introduction of string theory, then called `dual models'. See Green et al.~(1987:~p.~1) and Rickles (2014:~p.~32-33).} 
$\a'=\ell_{\sm s}^2$ is a fundamental constant in string theory, with dimensions of length squared: it sets the fundamental {\bf string length}, $\ell_{\sm s}$, i.e.~its typical length scale. Since it multiplies the whole action, it can be taken as the coupling constant of the world-sheet theory. However, the string length does {\it not} play the role that a coupling constant plays in the Feynman diagram expansion of a quantum field theory, where each loop comes with a power of the coupling constant (in string theory, that role is played by the string coupling constant $g_{\sm s}$): rather, the string length governs the ``stringy effects'' at short distances, which deviate from the point-particle behaviour.\footnote{Thus in string theory there are {\it two different coupling expansions}: the first is related to the fact that at short distances the string is an extended object. This is the expansion in $\a'$, where at lowest order in $\a'$ only massless the states contribute (thus corrections in $\a'$ are short-length or high-energy corrections, where the massive string modes also contribute). The second is the expansion in the string coupling $g_{\sm s}$, which is the loop expansion of the diagrams, with higher loops contributing when the coupling between the strings is large. However, this is not an ordinary loop expansion, because $g_{\sm s}$ is the expectation value of the dilaton field, which varies with the location in spacetime.}
Thus the corrections in $\a'$ indicate that the string is an extended object (see Section \ref{cea}).

The equation of motion derived from the above world-sheet action is, as one would expect, a two-dimensional wave equation, i.e.~a two-dimensional generalization of the geodesic equation:\footnote{This is the equation of motion that arises from the variation of $X$. As in the point-particle case, one in addition needs to vary the metric $h^{\a\b}$, which then imposes the vanishing of the stress-energy tensor on the world-sheet.} 
\bea\label{boxX}
\left(\pa_\t^2-\pa_\s^2\right)X^\m=0\,.
\eea
This is a wave equation, whose general solution is a sum of arbitrary functions of $\t\pm\s$. Such functions can be written as superpositions of sines and cosines or, alternatively, exponentials of these variables.\footnote{One can see this by rewriting the equation thus: $(\pa_\t+\pa_\s)(\pa_\t-\pa_\s)X^\m=0$. The general solution is: $X^\m=X^\m_+(\t+\s)+X^\m_-(\t-\s)$, where $X^\m_{\pm}$ are {\it arbitrary functions}, which we then expand in an appropriate basis of sines and cosines, or of exponentials. The periodicity or boundary conditions then limit the coefficients, as in Eq.~\eq{Xclosed}.} 
The exponential form is useful, because (by analogy with ordinary quantum field theory) we can identify positive energy, and make a choice of constants that, upon quantization, will give us {\it creation and annihilation operators} below.\\ 
\\
{\bf Closed and open strings.} In deriving the above wave equation, one sets a {\it boundary term} to zero.\footnote{This boundary term appears upon doing a partial integration of the action that gives the equation of motion.} 
That this term is zero requires appropriate boundary conditions.\footnote{The coordinate $\s$ along the string's spatial direction is often conventionally taken to run from $0$ to $\pi$ for open strings, and from $0$ to $2\pi$ for closed strings. On this convention, the expression Eq.~\eq{osbc} is for open strings: the expression for closed strings is obtained by replacing $\pi\mapsto2\pi$. For general $\ell$, see Polchinski (1998:~pp.~18-19).} 
The boundary term is:
\bea\label{osbc}
\int_{-\infty}^\infty\dd\t\left(\pa_\s X_\m\d X^\m|_{\s=0}-\pa_\s X_\m\d X^\m|_{\s=\pi}\right)=0\,.
\eea
We have followed the convention of parametrizing the two {\it endpoints} of the (open) string by $\s=0$ and $\s=\pi$. 

Three distinct ways to satisfy this boundary condition will give us different types of strings: 

(i)~~We require that the two boundary terms cancel each other out, through $X$'s (and its variation, $\d X$) being {\it periodic} in the world-sheet coordinate $\s$: this gives {\it closed strings}. 

(ii)~~We keep the variation $\d X$ arbitrary: then the boundary term is zero iff we impose a {\bf Neumann boundary condition}:\footnote{It is possible to choose special variations that are zero at the endpoints, i.e.~$\d X^\m|_{\s=0,\pi}=0$. This requires special care, since it means that the endpoints of the string are {\it fixed in space(time)}. This is how D-branes appear on the world-sheet.} 
\bea\label{Neumann}
\pa_\s X^\m|_{\s=0,\pi}=0\,.
\eea
This gives an {\it open string}, whose endpoints move freely, but {\it no momentum flows} through them. (For {\it arbitrary} world-sheet coodinates, we can write this condition as $n^a\pa_a X^\m|_{\pa M}=0$, where $n^a$ is normal to the boundary of the world-sheet, $\pa M$.) 

(iii)~~We require that the variations $\d X$ are zero at the boundary. This is a Dirichlet boundary condition that also gives open strings, and leads in to the idea of {\it D-branes}, which we will discuss in Section \ref{Dbopen}.

We here discuss in more detail the closed and open string solutions, (i) and (ii) above.

(i)~~The most general {\it periodic} solution of Eq.~\eq{boxX} is a {\bf closed string}:\footnote{As we announced earlier, and as is conventional for closed strings, we now take the periodicity of $\s$ to be $2\pi$.}
\bea\label{Xclosed}
X^\m=x^\m+2\pi\a'\,p^\m\,\t+i\sqrt{\a'\over2}\sum_{n\not=0}{1\over n}\left(\a^\m_n\,e^{-2in(\t+\s)}+\ti\a^\m_n\,e^{-2in(\t-\s)}\right).
\eea
Here, $x^\m$ and $p^\m$ are arbitrary constants that have the obvious interpretation as the initial centre-of-mass position of the string and four-momentum (analogous quantities are present for a point particle). In fact, $p_\m={\d S\over\d(\pa_\t X^\m)}={1\over2\pi\a'}\,\pa_\t X^\m$ is the conserved momentum obtained from the action Eq.~\eq{WS2} through Noether's theorem. There is no linear term in $\s$ in Eq.~\eq{Xclosed} because, for a closed string, the string coordinate $\s$ is periodic (i.e.~$\s\in[0,2\pi)$) while the target space coordinate $X$ has infinite range, and a linear term in the periodic coordinate $\s$ would make the target space coordinate $X$ not singly-defined. The remaining terms represent the {\bf string oscillations}: the set of constants $\a^\m_n$ and $\ti\a^\m_n$ are the higher Fourier modes. Since $X$ is real, the positive and negative Fourier modes are related by: $\a^\m_{-n}=(\a^\m_n)^\dagger$, $\ti\a^\m_{-n}=(\ti\a^\m_n)^\dagger$. For a classical string, the dagger is simply complex conjugation, but for a quantum string this gives us the natural separation between {\bf creation and annihilation operators}.\footnote{As usual in quantum field theory, the annihilation operators are those that multiply the mode $e^{-i\om t}$, for $\om>0$, where in these units the frequency is quantized: $\om=n$.} The two types of modes, the $\a^\m_n$ and $\ti\a^\m_n$, are independent of each other. The $\a^\m_n$ are left-moving modes,\footnote{These are modes that propagate at the speed of light towards positive $\s$ on the world-sheet, i.e.~$\s=\t$.} and the $\ti\a^\m_n$ modes are right-moving.\footnote{These modes propagate at the speed of light towards negative $\s$, i.e.~$\s=-\t$.}

(ii)~~For the Neumann boundary conditions (i.e.~Eq.~\eq{Neumann}), the solutions are {\it open strings}. The solutions are of the same type as under (i), but the boundary condition selects out half of the modes: we keep all the cosines in $\s$, and set the coefficients of all the sines to zero.\footnote{This is because $\sin n\s|_{\s=0,\pi}=0$, and we require that the derivative of $X$ is zero at the boundary. Thus $X$ itself can only contain cosines $\cos n\s$. In terms of the exponentials in Eq.~\eq{Xclosed}, this in effect means that the left- and right-moving oscillators are tied together (i.e.~$\a^\m_n$ and $\ti\a^\m_n$ are not independent, but are related to each other), and the left- and right-moving oscillators combine to standing waves.} 
The general open string solution is: 
\bea\label{opens}
X^\m =x^\m+2\pi\a'p^\m\t+i\,\sqrt{2\a'}\sum_{n\not=0}{1\over n}\,\a^\m_n\,\cos n\s\,e^{-in\t}\,.
\eea
Again, the condition that this is a self-adjoint quantum field gives: $\a^\m_{-n}=(\a^\m_n)^\dagger$.\\ 
\\
{\bf Quantization: states and operators.} There are several ways to quantize the world-sheet theory.\footnote{Another method is path integral quantization: see Green et al.~(1987:~Chapter 3) and Polchinski (1999:~Chs.~2 and 3), where Polchinski uses field theory techniques.} 
In covariant quantization, one replaces Poisson backets by commutators, and takes the fields $X$ and their canonically conjugate momenta to satisfy standard equal-time commutation relations, from which one derives the algebra of the Fourier modes $x,p,\a,\ti\a$.\footnote{There are two main ways to do this: in covariant quantization, {\it all} of the $X^\m$'s are taken to satisfy the commutation relations. The Lorentz invariance of the algebra is then manifest, but the unitarity of the S-matrix is not: there are unphyiscal ghost states (i.e.~states of negative norm squared) that only disappear if the spacetime dimension is $D=26$ ($D=10$ in the case of a superstring). In light-cone gauge quantization, one gauge fixes the diffeomorphism invariance of the world-sheet so that, in effect, one only quantizes the physical states (i.e.~$D-2$ $X$'s, transverse to the light-cone, among all $D$ of them). In this non-covariant treatment there are no ghosts and unitarity is manifest because one only quantizes physical states, but the Lorentz symmetry of the algebra requires (again) $D=26$ for the bosonic string, and $D=10$ for the superstring.\label{ghostsuni}} The physical conditions on the algebra (see footnote \ref{ghostsuni}) require that the spacetime dimension be $D=26$ for bosonic strings, and $D=10$ for superstrings (i.e.~fermionic fields are added to the $X$'s to make the action Eq.~\eq{WSa} supersymmetric). 

A distinguished operator in the algebra of operators is the mass-squared operator, $M^2$. The mass levels of the states are quantized (this operator is quadratic in the $\alpha$'s: roughly speaking, it is proportional to the sum of number operators $\alpha^\dagger_n\alpha_n$ of each Fourier mode). The eigenvalues of this operator give the mass levels of the string (the following discussion focusses on the closed string spectrum).\footnote{Considering open strings gives additional states that need not concern us here.} The ground state of the bosonic closed string is a {\bf tachyon}, i.e.~it has negative mass-squared, $\a'\,M^2=-4$. Let us call this state $|0,0;k\ket$, where the zeroes indicate that this state is annihilated by both $\a_n$ and $\ti\a_n$, and $k$ is the centre-of-mass momentum of the string (i.e.~the eigenvalue of $p$). The fact that the ground state has negative mass-squared is a problematic property of the bosonic string: this state disappers in the superstring (and this is strong motivation to introduce {\it superstrings}, which are free of tachyons). 

The {\bf first excited state} is more interesting, and it is the state obtained by acting on $|0,0;k\ket$ with the creation opertors of the $n=1$ left- and right-moving modes, i.e.~$(\a^i_1)^\dagger(\ti\a^j_1)^\dagger|0,0;k\ket$ ($i$ and $j$ are transverse spatial indices, and range from $2$ to $D-1$).\footnote{The argument is clearest in light-cone gauge quantization, where only the transverse modes are quantized; one uses the diffeomorphism invariance of the world-sheet theory to set $X^0=\t$ and $X^1=\s$, so that there are no $\m=0$ and $\m=1$ physical oscillations.} This state is {\it massless}, i.e.~$M^2=0$. Since it has two indices $i$ and $j$ ranging from $1$ to $D-1$, it is in a reducible representation of the relevant part of the Lorentz group, namely $\mbox{SO}(D-2)$: one can think of this state as a $(D-2)\times(D-2)$ matrix. The irreducible representations consist of a massless symmetric state of spin two (symmetric in $i$ and $j$), a massless scalar (i.e.~the trace of the matrix), and an antisymmetric rank-two tensor (antisymmetric in $i$ and $j$). 

The importance of these states is that they are the {\bf massless string states} (also present in the superstring). They represent a {\bf graviton} (i.e.~a symmetric, massless spin-two excitation: more on this below), a {\bf dilaton field} (a massless scalar excitation with a specific coupling to the string, usually denoted by $\Phi$) and an {\bf antisymmetric rank-two tensor field}, $B_{\m\n}$. These are called the {\bf NS-NS states} (for Neveu and Schwarz).\footnote{For the significance of this nomenclature one needs to go to the superstring, where these are bosonic closed string states whose left- and right-moving parts both have anti-periodic boundary conditions on the fermion partners; this then results in bosons.} 
In general, the NS-NS states in Type II superstring theories are the bosonic closed string states whose left- and right-moving parts are bosonic. 

While a systematic construction of the physical states along the above lines in principle allows one to construct the whole spectrum of the perturbative string theory, in practice we are more interested in scattering between strings, for which we require analogues of plane wave states in quantum field theory, which allow the construction of classical fields localized in spacetime. 

The corresponding states in string theory are obtained as {\bf coherent states}, which are superpositions of states with {\it arbitrary} mode numbers, i.e.~not only the lowest ones. These coherent states generalize plane waves and are obtained, in the path integral formalism, from {\bf vertex operators}, i.e.~local operators that can be inserted into a state to represent the absorption or emission of string states or, equivalently, local vertex functions added to the path integral to represent such absorption or emission.\footnote{In conformal field theories, the connection between the canonical formalism and the vertex operator formalisms is underpinned by the correspondence that exists between on-shell physical states and vertex operators, or {\bf state-vertex correspondence}.}

Specifically, coherent states can be used to show that the state of the massless closed string spectrum that we found above deserves the name `graviton', i.e.~that a {\it macroscopic superposition of graviton states}, obtained as a coherent state of them, gives a curved metric that satisfies the Einstein field equations. Using the vertex operator formalism, this is readily shown when the resulting target space metric differs from the Minkowski metric by a small amount that is given by the coherent state. One finds the vertex function that (through the vertex-state correspondence) describes a coherent state of such massless strings (i.e.~not just the graviton state, but also the dilaton and the antisymmetric tensor). The vertex function modifies the path integral in such a way that the original world-sheet action, Eq.~\eq{WSa}, is in effect replaced by the following:
\bea\label{fullWS}
S_{\tn{WS}}=-{T\over2}\int\dd^2\s\,\sqrt{h}\left(h^{\a\b}\pa_\a X^\m\pa_\b X^\n g_{\m\n}+\e^{\a\b}\pa_\a X^\m\pa_\b X^\n B_{\m\n}+\a'R(h)\,\F\right).
\eea
The first term is the generalization of the world-sheet action Eq.~\eq{WSa} to a curved target space metric, where in this case the metric is a small perturbation of the Minkowski metric, i.e.~$g_{\m\n}(X)=\eta_{\m\n}+f_{\m\n}(X)$, and $f_{\m\n}$ is the perturbation coming from the vertex function that describes the interaction of the string with a coherent state of gravitons with wave-function $f_{\m\n}$ (see Green et al.~(1999:~p.~165)).

Likewise, the second and third terms are the contributions of the vertex operators for an interaction of the string with a coherent state of antisymmetric two-tensor states with wave-function $B_{\m\n}(X)$, and a coherent state of scalar states with wave-function $\F(X)$. $R(h)$ is the curvature in the world-sheet metric $h$. The full bosonic string action, Eq.~\eq{fullWS}, is the most general renormalizable, two-derivative, action in $X$. 

The second term is called the {\bf Wess-Zumino term}, and it is the unique way to couple an antisymmetric two-tensor $B_{\m\n}$ to the world-sheet of a string in a gauge-invariant way (see below), while preserving the other symmetries of the string. This term generalizes the gauge-invariant coupling, $e\int\dd\t\,\dot x^\m A_\m(x(\t))$, of a charged point-particle to the electromagnetic four-vector potential.\footnote{The coupling of the antisymmetric two-tensor to the world-sheet scalar fields $X$ is a coupling of the antisymmetric two-tensor to the volume form on the world-sheet of the string, $\dd\s^i\wedge\dd\s^j$, {\it pulled back} to the spacetime through the scalar fields $X^\m(\s)$ (which give an embedding map of the string onto the spacetime). The coupling on the world-line of a point particle can be understood as coming from the field-theory coupling $\int\dd^4x\,J^\m A_\m(x)$, where $J^\m$ is a conserved four-current for the point-particle's trajectory, $y^\m(\t)$: namely, $J^\m=e\int\dd\t\,{\sm dy^\m\over\dd\t}\,\d^{(4)}(x-y(\t))$. Substituting this into the field-theory coupling, and performing the integral over the Dirac delta function, we get the world-line coupling: $e\int\dd\t\,\dot y^\m A_\m(y(\t))$. The gauge-invariance of this term can be derived in two ways: either from the conservation of the four-current $J^\m$ in the field theory coupling, or directly in the world-line coupling: $\int\dd\t\,\dot y^\m\,\pa_\m\l=\int\dd\t\,\dot\l$, which is merely a boundary term at the end-points of the particle's world-line.\label{wlp}} 

Thus the antisymmetric two-tensor $B_{\m\n}$ is a two-tensor generalization of a vector potential. There is an antisymmetric three-tensor $F_{(3)}$ corresponding to it:
\bea\label{H3}
F_{\m\n\l}=\pa_\l B_{\m\n}+\pa_\m B_{\n\l}+\pa_\n B_{\l\m}\,,
\eea
which is invariant under generalized gauge transformations, whose gauge parameter is a vector $\L_\m$, i.e.~$\d B_{\m\n}=\pa_\m\L_\n-\pa_\n\L_\m$ (and the above action is also invariant under this gauge transformation).

The above shows how the string couples to the {\bf NS-NS fields} $(g_{\m\n},B_{\m\n},\F)$, which are obtained from coherent states of massless NS-NS states (and therefore they are called `NS-NS fields'). Since these fields are functions of $X$, the theory Eq.~\eq{fullWS} is no longer quadratic in $X$, but is non-linear. From the perspective of the theory on the world-sheet, the fields $(g_{\m\n},B_{\m\n},\F)$ play the role of {\it coupling functions} for the world-sheet fields $X$. 

\subsection{Couplings and effective actions}\label{cea}

The classical world-sheet action is invariant under {\it local Weyl rescalings}, i.e.~arbitrary rescalings of the world-sheet metric $h$. Since $h$ is a field that is integrated over in the path integral, the consistency of the formalism is maintained by requiring that there is no anomaly, i.e.~the path integral measure must also be invariant under local Weyl transformations.\footnote{This secures that the path integral is independent of the metric $h$ on the world-sheet. Dependence on $h$ would break the diffeomorphism invariance of the world-sheet theory: only those diffeomorphisms that leave this (arbitrarily chosen) metric invariant, i.e.~its isometries, would be symmetries. (For the conception of `diffeomorphism invariance', see Pooley (2017:~p.~117).) For a philosophical discussion of conformal invariance in string theory, see Huggett and Vistarini (2015) and Dougherty (2021).}

However, this requirement is in general not satisfied, unless the beta-functions of the coupling functions $(g_{\m\n},B_{\m\n},\F)$ are zero. The beta functions describe the changes of the couplings under changes of scale, in particular under Weyl rescalings. And so, having a string theory that is free of a Weyl anomaly requires that {\it the beta functions are zero}.

The beta functions for the couplings $(g_{\m\n},B_{\m\n},\F)$ can be shown to be (to lowest order in $\a'$) the full set of Einstein field equations for these fields, where $B_{\m\n}$ and $\F$ are tensor and scalar contributions to the stress-energy tensor. (The vanishing of the beta function of $\F$ requires that $D=26$ for the bosonic string and $D=10$ for the superstring: this is the quantum consistency condition announced in Section \ref{5diff}). To see the dependence of Newton's constant on the string coupling below, it is useful to shift the dilaton by its constant vacuum expectation value $\f_0$, i.e.~$\F=\f_0+\f$, where $\bra\F\ket=\f_0$ and $\bra\f\ket=0$. The Einstein field equations are then derived from the following action:
\bea\label{betaf}
S=-{1\over16\pi G_{\tn N}}\int\dd^Dx\,\sqrt{-g}\,e^{-2\f}\left(R(g)-4(\na\f)^2+{1\over12}\,H_{\m\n\l}H^{\m\n\l}\right),
\eea
where Newton's constant is given in terms of the parameters of the string: $G_{\tn N}:={g_{\tn s}^2\over2\pi}\,(2\pi\ell_{\sm s})^{D-2}$, and $g_{\sm s}=e^{\f_0}$ is the {\bf string coupling constant} (or, simply, the `string coupling').\footnote{If the redefined dilaton field $\f$ goes to a constant value at spatial infinity, then this value must be zero, and the string coupling constant is given by the value of the dilaton field (in string frame) at infinity.}

After including the fermionic fields, this action is equivalent to the supergravity action.\footnote{Notice that, because of the multiplicative factor of the dilaton, the action is not in the standard Einstein-Hilbert form: hence the phrase `is equivalent to'. We obtain the supergravity action by a Weyl rescaling of the metric (see Eq.~\eq{S10} below).} 
It is the Wilsonian effective action for the NS-NS massless modes of string theory, described at low energies by coherent states of the graviton, the dilaton and the antisymmetric tensor, where the massive states that are relevant at high energies have been integrated out. This underpins the claim that {\it supergravity theory is the low-energy limit of string theory}. Including the massive modes, i.e.~going beyond the low-energy approximation, adds terms to the integral of order $\a'$ and higher.

The above action also gets corrections proportional to powers of the string coupling constant, i.e.~$(g_{\sm s}^2)^{g-1}$, where $g$ is the genus of the closed string wold-sheet, i.e.~the number of holes (the above action is proportional to $1/g_{\sm s}^2$, and so has $g=0$, which gives a spherical world-sheet with no holes; a torus has $g=1$, etc.). Thus the expansion in the string coupling constant is the analogue of the Feynman diagram expansion for strings, but here it is simpler, because loops are counted by the genus $g$ of the world-sheet, i.e.~it is a topological expansion.

Thus our effective field theory has a {\it double expansion}: corrections in the dimensionful constant $\a'$ are high-energy corrections, and corrections in the dimensionless string coupling constant $g_{\sm s}$ are corrections by string world-sheets of higher genus. Such a double expansion is not new: recall that, in the Seiberg-Witten theory, we had a low-energy Wilsonian effective action, and in addition an expansion in the expectation value of the Higgs field, which played the role of a coupling constant. There, the expansion in the expectation value of the Higgs could be determined using quasi-duality. (We will see in the next Chapter that S-duality also sometimes fixes the dependence on the dilaton field.)

\section{When small is large: T-duality}\label{T-d}

Our first example of a duality in string theory is T-duality, which relates Type IIA strings on a circle of radius $R$ to Type IIB strings on a circle of radius $1/R$, i.e.~it relates a model on a large circle to a model on a small circle, and vice versa. T-duality is believed to be a {\it perturbative} duality of string theory, in the sense that it can be demonstrated order by order in perturbative expansion. It involves the inversion of the {\it radius}, but it does not require the inversion of $\a'$ or the string coupling constant. Nevertheless, it is a very non-trivial duality, because it does involve relating terms with different powers of $\a'$, as well as changing the string coupling constant. 

Section \ref{cswc} first discusses the simplest example of T-duality: a classical closed string winding a circle of radius $R$. Section \ref{SchemaT} then discusses the interpretation of T-duality according to the Schema.

\subsection{A closed string winding a circle}\label{cswc}

Since the bosonic parts of the world-sheet Type IIA and Type IIB string models are the same, i.e.~they are both given by the action Eq.~\eq{WS2}, we may use the closed string solutions Eq.~\eq{Xclosed}, regardless whether we are in Type IIA or Type IIB. Thus our task is to show that a solution of the world-sheet theory in which one of the target space dimensions is a circle of radius $R$, is isomorphic to the solution in which the radius is $1/R$.

To this end, we take the target spacetime, in which the string is embedded, to be a nine-dimensional flat space times a circle of radius $R$, i.e.~$\mathbb{R}^9\times S_R$. Notice that the target space coordinates $X^\m$ ($\m=0,\cdots,9$) of the closed string take values in this space: therefore, we choose one of the nine spatial coordinates to take values on the circle $S_R$. Conventionally, we take this special coordinate to be $X^9$. $X^9$ is then a map from the closed string world-sheet, which contains a spatial circle, into the target space circle, i.e.~it maps one spatial circle onto another, so that it is a multi-valued function of $\s$, where the multi-valuedness corresponds to a possible {\it winding} of the closed string around the {\it compact} ninth coordinate. Thus under a translation of the world-sheet coordinate $\s\mapsto\s+2\pi$, which maps the world-sheet circle onto itself, the target space coordinate $X^9$ can jump by $X^9\mapsto X^9+2\pi mR$, where $m\in\mathbb{Z}$ is the {\bf winding number} of the string, i.e.~the number of times the string winds around the circle.

Back in our expansion of the closed string coordinates with infinite range, i.e.~Eq.~\eq{Xclosed}, we ruled out a linear term in $\s$ in order that the spacetime coordinates be single-valued. However, when one of the target space dimensions is compact, we require a multi-valued scalar, and we {\it do} get a term linear in $\s$:
\bea\label{x9}
X^9=x_0+2\pi \a' p\,\t+mR\,\s+\mbox{oscillators}\,,
\eea
where $x_0$ is the centre-of-mass coordinate in the 9th direction, and $p$ is shorthand for $p^9$. The above indeed satisfies: $X^9(\t,\s+2\pi)=X^9(\t,\s)+2\pi mR$. Thus the range of $X^9$ is still infinite, but if the string wraps $m$ times around the circle, then $X^9$ jumps by an integer multiple of the diameter of the circle. 

As we noticed earlier, the constant $p$ is the canonical (Noether) momentum associated with $X^9$, i.e.~$p=\pa_\t X^9/2\pi\a'$. Since $X^9$ is a multi-valued function on a circle of radius $R$, the corresponding conjugate momentum is quantized in units of $1/R$ (for any value of $m$). This is a standard result of quantum mechanics on a circle: the single-valuedness of the wave-function around this circle implies that the momentum is quantized as follows (e.g.~consider a plane wave $e^{ipX^9}$):
\bea
p={n\over R}\,,
\eea
where $n$ is any integer, independent of $m$. Thus we can write the above expansion as:
\bea\label{X9}
X^9=x_0+2\pi\a' {n\over R}\,\t+mR\,\s+\mbox{oscillators.}
\eea
There are two independent kinds of modes here: {\it momentum modes}, quantized in units of $1/R$ and characterized by the quantum number $n$, and {\it winding modes}, which are classical and topological in origin, and are characterized by the integer $m$.

As we will now argue, T-duality exchanges these two kinds of modes, if at the same time the radius goes to its inverse, while all the physical quantities remain invariant. 

In the ground state, the oscillator modes are all unoccupied: thus these states are characterized by the zero modes, i.e.~the momentum $p^\m$ and the winding, and they can be written in the form $|\{p\};n,m\ket$, where $\{p\}$ indicates the momentum in the other spatial directions, perpendicular to $X^9$. Further (excited) states are obtained by acting on these states with creation operators, $(\a_n)^\dagger$ and $(\ti\a_n)^\dagger$. 

The mass formula is as follows:
\bea\label{M2}
M^2={n^2\over R^2}+{m^2R^2\over\a'{}^2}+\mbox{oscillations.}
\eea
Clearly, this formula is invariant under the simultaneous interchange:
\bea\label{Tduality}
n\leftrightarrow m~~\mbox{and}~~R\leftrightarrow \a'/R\,.
\eea
In other words, the {\it closed string spectrum} of a bosonic string on a circle of radius $R$, i.e.~the set of states, is identical to the closed string spectrum on a circle of dual radius, $\ti R:=\a'/R$, provided we also interchange the quantum numbers of momentum and winding. Notice that $\a'=\ell_{\sm s}^2$, and so $\ti R$ is indeed a length.\footnote{As we mentioned in the preamble of this Section, the string coupling also changes under T-duality, as follows: $\ti g_{\tn s}=g_{\tn s} \ell_{\tn s}/R$. See e.g.~Alvarez et al.~(1995:~p.~2). This paper (p.~11) also gives the {\it Buscher transformation rules} of the low-energy supergravity action under T-duality.}

Up to terms that are independent of $n$ and $m$, the mass-squared Eq.~\eq{M2} is the (value of the) Hamiltonian of the closed string theory. Thus the duality is equivariant for the closed string dynamics, which is our condition (ii) in Section \ref{isomdef} for the structure that the isomorphism must preserve.

But one should also show that T-duality preserves the values of all the quantities, i.e.~our condition (i). To find how the duality map in Eq.~\eq{Tduality} maps quantities, it is best to see how it acts on the field $X^9$ in Eq.~\eq{X9}, which, along with the remaining $X^\m$'s ($\m=0,\ldots,8$, and these fields are invariant under T-duality since they do not wind around the circle), is the fundamental field in the world-sheet action Eq.~\eq{WS2}, from which the operators in this spacetime are constructed. 

Acting with the duality map gives the following {\it dual target space coordinate}:
\bea\label{tiX9}
\ti X^9=\ti x_0+2\pi\a'{n\over R}\,\s+mR\,\t+\mbox{oscillators.}
\eea
This dual target space coordinate is a multi-valued map from the world-sheet closed string into the target space circle of {\it dual radius}, Eq.~\eq{Tduality}. While, from the comparison with Eq.~\eq{Tduality}, it might seem like T-duality merely exchanges the world-sheet coordinates $\s\leftrightarrow\t$, this is not the case, because, in order for T-duality to preserve the canonical commutation relations, it must also map to a new centre-of-mass coordinate, $x_0\mapsto\ti x_0$. Indeed, $x_0$ is canonically conjugate to the momentum operator $\hat p$ (i.e.~they satisfy $[x_0,\hat p]=i$), and since duality maps momentum to winding, i.e.~$\hat p\mapsto \tilde{\hat p}=\hat w$, the dual centre-of-mass coordinate $\ti x_0$ must be conjugate to the winding operator rather than the momentum operator, i.e.~$[\ti x_0,\hat w]=i$.

The above can be implemented systematically by a change of sign in the right-movers of the dual target space coordinate, $\ti X^9$. Namely, recall from the discussion after Eq.~\eq{Xclosed} that $X^9=X_L(\t+\s)+X_R(\t-\s)$, where $X_L$ are the left-moving modes (with the normal oscillators $\a_n$) and $X_R$ are the right-moving modes (with tilded oscillators, $\ti\a_n$). Then the dual field Eq.~\eq{tiX9} can be defined as:
\bea\label{dualX9}
\ti X^9=X_L-X_R\,.
\eea
This is a convenient reformulation of the previous discussion (including not only the zero modes but also the oscillators), where T-duality is realized as a {\it parity operation on the right-moving string modes}. 

The dual string theory has a world-sheet formulation like the original one, with world-sheet action Eq.~\eq{WS2} written in terms of the dual field, and its commutation relations have the same canonical form. Because the actions of the two theories have the same form, as do the commutation relations, these theories are isomorphic. 

\subsection{Illustrating the Schema and interpreting T-duality}\label{SchemaT}

The previous Section already argued that a string on a circle of radius $R$ and its dual are isomorphic models, in the Schema's sense. The main point is that the quantized scalar $X^9$ (and its dual, $\ti X^9$) give the states, and then the operators are constructed from them (as well as from the other fields in the model). In particular, $|\{p\};n,m\ket$ is the ground state. The duality map is then as in Eq.~\eq{Tduality}. Also the Hamiltonians of the two models match: the values are given by Eq.~\eq{M2}, and they are the same for the two models. In this Section, we discuss the Schema's interpretation of T-duality in more detail. To have a more precise discussion of the interpretation, we first introduce a simpler notation:\\
\\
{\bf Notation for models:} we will denote our T-duals of the closed string on a circle by $M$ and $\ti M$. In $M$, the states are labelled by pairs of integers $(n,m)$, so that the momentum and winding operators, $\hat p$ and $\hat w$, have eigenvalues $p=n/R$ and $w=mR/\alpha'$, with canonically conjugate variables $\hat x$ and $\hat y$ that are periodic, i.e.~have values in the ranges $x\in[0,2\pi R)$ and $y\in[0,2\pi\a'/R)$.\footnote{Give the wave functions for $x$ and $y$.} 
(Here, $x$ and $y$ are the variables that we called $x_0$ and $\ti x_0$ earlier: $x_0$ is a centre-of-mass variable in target space, and $\ti x_0$ is a centre-of-mass variable in an auxiliary space called `winding space'.)\footnote{Thus for simplicity we will, from now on, not consider the oscillator modes, or the momentum modes in the other spatial directions, denoted by $\{p\}$.}

In $\ti M$, we have tilded states and quantities: the states are labelled by pairs of integers $(\ti n,\ti m)$, so that the operators $\hat{\ti p}$ and $\hat{\ti w}$ have eigenvalues $\ti p=\ti n/\ti R$ and $\ti w=\ti m\ti R/\a'$, with canonically conjugate quantities with values $\ti x\in[0,2\pi\ti R)$ and $\ti y\in[0,2\pi\a'/\ti R)$, and the dual radius is defined by: $\ti R:=R/\a'$. 

The duality map thus exchanges the tilded and untilded states:\footnote{Comparing to Eq.~\eq{Tduality}, the stipulation that the radius is exchanged with its dual is now being implemented through the exchanges $x\leftrightarrow\ti x$ and $y\leftrightarrow\ti y$, and so is not an additional requirement.}
\bea\label{mapTd}
(n,m)\,\mapsto\,(\ti n,\ti m)&=&(m,n)~~~\Leftrightarrow~~~
(p,w)\,\mapsto\,(\ti p,\ti w)=(w,p)\nn
(x,y)\,\mapsto\,(\ti x,\ti y)&=&(y,x)\,.
\eea
Here, the states $(n,m)$ are in $M$'s state space ${\cal S}$ in the momentum-winding representation, while the states $(x,y)$ are in the position representation in target-winding space (and likewise for the dual state space, $\ti{\cal S}$, of $\ti M$). Below, we will usually write the states in the momentum-winding representation, i.e.~as $(n,m)\in{\cal S}$ and $(\ti n,\ti m)\in\ti{\cal S}$. With this notation, we can now discuss the Schema's two kinds of interpretation maps:\\
\\
{\bf Internal and external interpretations.} Recall, from Section \ref{intext}, that internal interpretations only interpret the common core of duals, while external interpretations also interpret the specific structure. 

The specific structure is here an assignment, to the pairs of integers $(n,m)$, of the {\it classical} structure typically associated with either momentum or winding (and to the pairs of real numbers $(x,y)$ an assignment of the classical structure associated with either target space or winding space). By an `assignment of the classical structure', we mean the classical structure that we use for quantization: namely, the classical string fields and their action. This classical structure gives $p$ its {\bf standard interpretation} as momentum and $w$ its standard interpretation as winding (and consequently, $n$ is a momentum quantum number, and $m$ is a winding number). This standard interpretation is based on the understanding that our classical string moves on, and wraps, a circle of radius $R$. Thus the standard {\it external interpretation} of $M$ says that there are $n$ units of momentum and $m$ units of winding, while the standard external interpretation of $\ti M$ says that there are $\ti n=m$ units of momentum and $\ti m=n$ units of winding. Since $\ti n\not=n$ and $\ti m\not=m$, the duals disagree: their domains of application are different.

By contrast, an {\it internal interpretation} maps what is common to the two models, and does not map the specific structure. The duals agree that the string moves on a circle and that it has both momentum and winding properties, but they are silent about the radius of the circle and about the values of the momentum and the winding, since it is only their combination that is common. The reason is that, on an internal interpretation, some aspects of the geometry cannot be probed by the objects of the theory, i.e.~by the closed strings. There is no independent definition of `large' and `small' radius (or of `momentum' and `winding': there is only a definition of states with both properties). 

This need not be as strange as it appears: it is quite reasonable for a theory to postulate the shape of a space (i.e.~a circle), and not to specify all of its metric properties (e.g.~the radius of the circle).\footnote{For example, conformal field theories are independent of local scales. Also, by a procedure called `twisting', one can obtain a string model that is independent of part of the geometric structure of the target space (i.e.~it is independent of the K\"ahler structure, or independent of the complex structure), depending on the type of twisting. These two models are related by mirror symmetry: see Witten (1991). In other words, one can have well-defined quantum field theories or string theories that depend on less geometric structure than other relativistic theories do. For the relation between mirror symmetry and T-duality, see Strominger, Yau and Zaslow (1996).} 

Section \ref{EmdS}'s analogy with electromagnetism is helpful here: while, for the most part and prior to the discovery of the Maxwell theory, electricity and magnetism were seen as independent phenomena, they are related by the relativistic Maxwell theory, where the electromagnetic field is the Faraday tensor field, which gets an internal interpretation that is independent of particular choices of coordinates. What an observer in a particular frame of reference takes to be a purely `magnetic' force or a purely `electric' force is part of an external interpretation: the rods and clocks in that frame of reference are specific structure that is added to the common core, i.e.~to the relativistically invariant theory, which is itself formulated without reference to particular choices of coordinates, observer's trajectories, or sets of rods and clocks. On the theory's internal interpretation, there are no independent electric and magnetic forces defined, but an {\it electromagnetic field} that transforms covariantly under changes of coordinates.

By analogy, our states have both momentum and winding, and under a duality map they transform as indicated in Eq.~\eq{mapTd}, but on an internal interpretation there are no states with `pure momentum' or `pure winding'.\footnote{This has given rise to the research programme of {\it double field theory}, which is based on a space that has twice as many coordinates as the target space, because it includes both momentum and winding modes, and incorporates duals into a single theory. Despite the doubling, there are no new degrees of freedom introduced, because of a constraint on the fields and gauge parameters that arises from a level-matching condition in closed string theory (usually called a `physical section condition'). This theory has a generalized diffeomorphism invariance, and a generalized metric, which encodes the usual metric, the two-form field, and the dilaton field. Berman and Thompson (2015:~p.~3) argue that double field theory is not strictly speaking a low-energy limit of string theory, but rather a truncation of it, in which both the momentum and the winding modes are kept, but not the oscillations. There is also an M-theory version of double field theory that incorporates dualities of M-theory.\label{dft}} 
Such interpretations are only obtained by adjoining specific structure, which as we explained above is defined by relating the states with a specific limit of e.g.~large $R$, in which the semi-classical world-sheet description is valid.

\section{D-branes from open strings}\label{Dbopen}

In this Section, we discuss T-duality for open strings. Unlike T-duality for closed strings, which relates strings on a large circle to strings on a small circle (cf.~Section \ref{T-d}), T-duality relates open strings to a novel class of objects: {\bf D-branes} are introduced into the wordsheet perspective as higher-dimensional objects on which open strings can end. Thus this Section will argue that {\it the T-dual of an open string is another open string ending on a D-brane}.

In Section \ref{rudiments}, open strings were introduced by using Neumann boundary conditions (see Eq.~\eq{Neumann}). For an open string on a circle of radius $R$, this implies that the open string has no winding. If $X^9$ is the target space coordinate of the string along this circle, then the Neumann boundary condition, Eq.~\eq{Neumann} applied to the scalar $X^9$ in Eq.~\eq{x9}, implies that the winding number is zero: $m=0$. 

This can be understood as follows: while a closed string with non-zero winding around the circle cannot shrink to a point because of its winding, if an open string whose endpoints can move freely on the $S^1$ shinks to a point, it does so at no extra cost in energy. By contrast to the winding of the closed string, there is no topological obstruction against shrinking an open string to a point.

But, as we will now show, an open string {\it can} wind around the circle if its endpoints are fixed on a D-brane transverse to the circle. T-duality will relate the two kinds of situations: the Neumann and the Dirichlet boundary conditions for the open string are maped onto each other under T-duality.

It is straightforward to see how T-duality along the circle direction acts on the Neumann boundary condition:\footnote{This was discovered by Dai et al.~(1989:~p.~2076).} 
it is mapped into a
\bea\label{Dirichlet}
\mbox{{\bf Dirichlet boundary condition} on $X^9$:}~~~~\pa_\t X^9|_{\s=0,\pi}=0\,.
\eea
This can be verified by constructing the dual variable, $\ti X^9$, as in Eq.~\eq{dualX9}: the momentum term in the expansion Eq.~\eq{opens} turns into a winding term, and in the oscillator term the cosine transforms into a sine, which is zero at $\s=0$ and $\s=\pi$. An open string satisfying a Dirichlet boundary condition is an open string whose endpoints do not move: they are fixed to lie somewhere on the circle.

Thus if we begin with an open string that has momentum $p=n/R$ and zero winding, we get, after a T-duality transformation along the circle direction, an open string with zero momentum and non-zero winding, i.e.~$\ti p=0$ and $\ti w=n\a'/R=n\ti R$. But this is a (dual) open string that stretches between two fixed points, viz.~$\ti X^9|_{\s=0}=0$ and $\ti X^9|_{\s=\pi}=2\pi n\ti R$. We interpret this as there being a {\it defect} in the eight spatial directions transverse to the circle, such that the endpoints can move freely along those directions, but not along the circle: the open string is stuck to move on an 8-dimensional hyperplane. This is what we call a D8-brane, where the `D' stands for the `Dirichlet' boundary condition for the open string.\footnote{Thus we have a Dirichlet boundary condition on the circle, and Neumann boundary conditions on the eight other directions. This is because we have T-dualized the circle coordinate, $X^9$. As we will discuss below, if we have Dirichlet boundary conditions along other directions, then we have a D-brane of lower dimensionality, because the endpoints are fixed along more dimensions.}

Dirichlet boundary conditions are allowed in the variational problem of the open string, in Eq.~\eq{osbc}. Making the boundary terms $\pa_\s X_\m\d X^\m$ to be zero can be achieved either by the Neumann boundary condition Eq.~\eq{Neumann}, or by the alternative, Dirichlet, boundary condition that we have just discovered (Eq.~\eq{Dirichlet}), because the latter amounts to fixing the value $X^\m$ at the endpoints of the string, so that $X^\m$ cannot change in time and the endpoint is fixed: this is precisely the condition $\d X^\m|_{\s=0,\pi}=0$. And the interpretation of such boundary conditions is in terms of the directions normal to the motion of the D-brane. If $p$ out of ten coordinates $X^\m$ satisfy the Dirichlet boundary condition Eq.~\eq{Dirichlet}, the D-brane has $p$ normal directions in spacetime. Thus a {\bf D$p$-brane} is a {\it spatially} extended $p$-dimensional object on which open strings can end (its world-volume is $p+1$-spacetime-dimensional).

\begin{figure}
\begin{center}
\includegraphics[height=5cm]{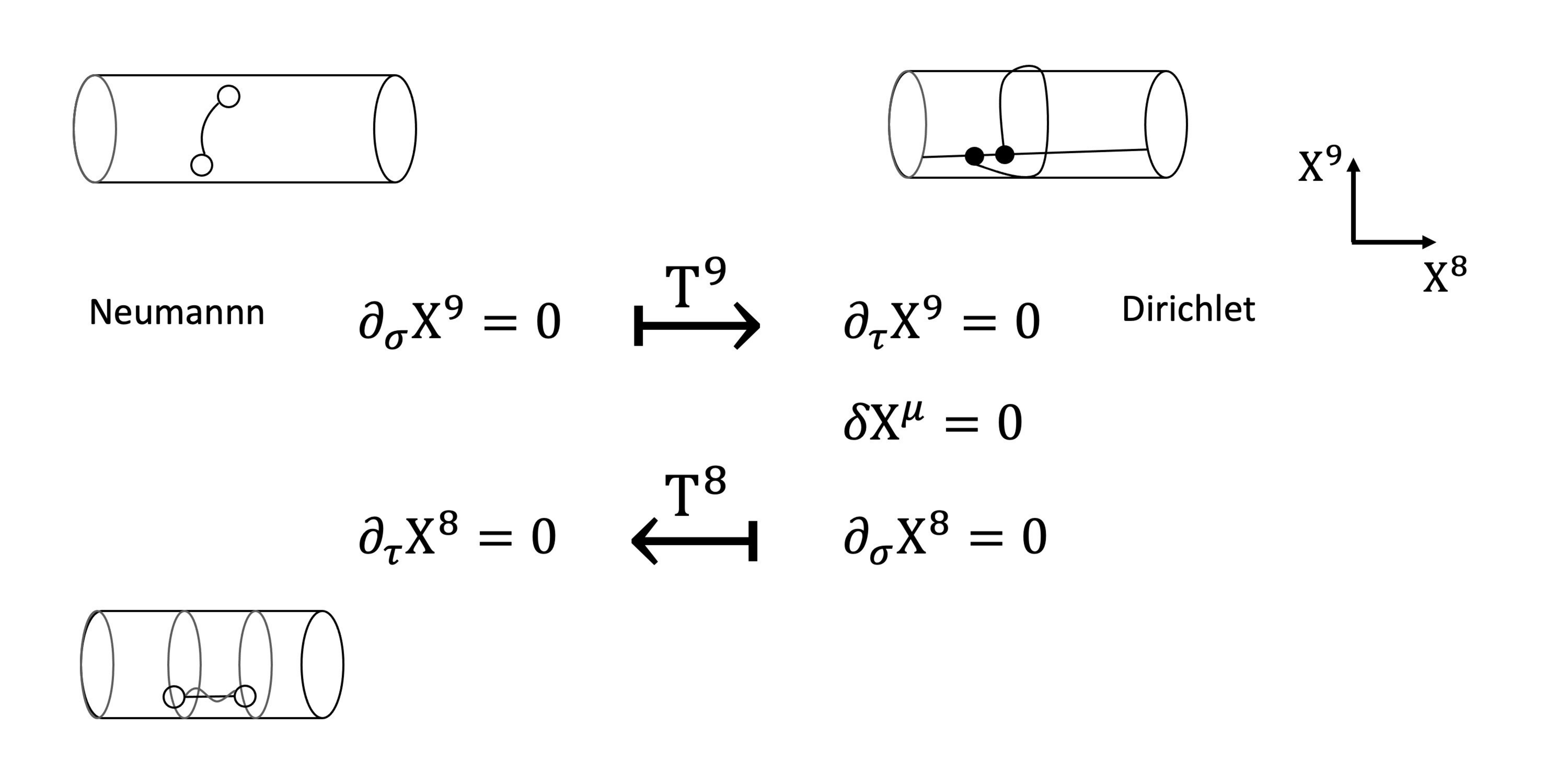}~~~
\caption{\small T-duality for open strings maps an open string with Neumann boundary conditions (left) to an open string with non-zero winding number and Dirichlet boundary conditions (right). The endpoints of the string on the right are fixed on the world-volume of a D-brane.}
\label{Tdualcircle}
\end{center}
\end{figure}

In the example above, we assumed that the original open string had Neumann boundary conditions in all ten spacetime dimensions. We can reformulate this in terms of D-branes by saying that in the original model there is a D10-brane, i.e.~a D-brane filling the whole space: thus there are no directions transverse to the D-brane, and the string endpoints can move freely in all directions. After T-dualizing along the circle, the open string is attached to a D9-brane (i.e.~one Dirichlet boundary condition on $X^9$, so that the open strings are free to move in any direction except along the circle direction). Thus T-duality maps a state with an open string on $S_R$ and a D10-brane to a state with an open string on $S_{\ti R}$, and with the endpoints attached to a D9-brane that is transverse to the circle. See Figure \ref{Tdualcircle}.

This pattern generalizes. If the spacetime contains several circles, we can T-dualize along the other circles (i.e.~we dualize the corresponding target space coordinates, e.g.~$X^8$, etc.). Each T-duality transformation reduces the dimensionality of the D-brane by one, so that we first get a D8-brane, then a D7-brane, etc. In other words, if we dualize a D-brane along a direction that is parallel to the D-brane (so that a Neumann boundary condition is mapped to a Dirichlet boundary condition on that direction), we map a D$p$-brane to a D$(p-1)$-brane. The inverse operation is T-dualizing along a circle {\it transverse} to the D-brane (so as to map a Dirichlet boundary condition to a Neumann boundary condition along that direction), thus mapping a D$p$-brane to a D$(p+1)$-brane.\footnote{To see the boundary conditions for the coordinates of a D$p$-brane in $D$ dimensions, we can split the indices as follows: $\m=0,\ldots,p\,||\,p+1,\ldots,D-1$, where the first block are the directions {\it parallel} to the world-volume of the D-brane, and the second block are the remaining directions (i.e.~the directions perpendicular to the world-volume, also called `transverse directions'). Then the D$p$-brane has Dirichlet boundary conditions $\d X^\m|_{\s=0,\pi}=0$ for $\m=p+1,\ldots,D-1$ i.e.~it on the $D-p-1$ transverse directions, and Neumann on the $p+1$ directions (including time) parallel to the world-volume, i.e.~$\m=0,\ldots,p$.}

The existence of D-branes on which open strings end breaks (spontaneously) translation invariance in the directions transverse to the D-branes. This reflects the fact that, in the original string model with Neumann boundary conditions, the winding number is not conserved: a closed string can break up by attaching its endpoints to a D-brane. Since, under T-duality, winding number is dual to momentum, the momentum of the open string in the direction transverse to the D-brane is not conserved either, and so translation invariance is broken.

However, string theory is a theory of gravity, and so the present picture is an approximation: D-branes are dynamic objects that e.g.~can recoil when they emit a string, so that the total momentum {\it is} conserved.\footnote{Near the D-branes, the excitations of the open strings ending on D-branes are described by non-abelian gauge theories defined on the world-volumes of the D-branes. For example, for $n$ parallel D3-branes that are coincident in space, one gets ${\cal N}=4$ $\mbox{U}(n)$ SYM.}

We will discuss D-brane dynamics in Section \ref{dynD}, after we discuss D-branes in closed string theory, and the more general significance of D-branes, in the next Section.

\section{D-branes in closed string theory}\label{Dcst}

The crucial significance of D-branes for string theory remained obscure until the series of developments in 1995 known as the {\it second superstring revolution}. 

D-branes, which the previous Section discussed from the point of view of open strings, had been discovered in 1989 by Dai et al.~and were rediscovered in 1995 in the closed string spectrum. Although we discuss this more fully below, this can here be understood from the fact that, since D-branes have mass, they curve the spacetime around them---and this curvature is visible to the closed strings, which carry the graviton excitations. Thus the presence of D-branes affects not only the open, but also the closed, strings.

In extreme brevity, Witten (1995a) announced a threefold discovery:\\
\\
(1)~~There are {\it states of closed string theory} that are only seen at strong coupling, and whose properties can be readily described: they are BPS states, i.e.~their mass saturates the BPS bound Eq.~\eq{Mcharge} (for a discussion of BPS states, see Section \ref{M-O}).\footnote{Each D-brane comes with one unit of elementary charge, but if there are many D-branes on top of each other, $n$ can be large.}\\
\\
(2)~~These states are important when the string interactions are strong, i.e.~at {\it strong string coupling}. D-brane states are required in order to get a full picture of the dualities relating various string theories, i.e.~the duality web in Figure \ref{Mthfig}.\footnote{This Chapter focusses on the {\it ten-dimensional} theories and their D-branes. Chapter \ref{STII} will discuss the {\it eleven-dimensional} origin of the dualities conjectured by Witten and others.}\\
\\
(3)~~Being states of {\it closed string theory}, D-branes have a natural spacetime interpretation as {\it extremal black holes} (i.e.~black holes whose mass satisfies the BPS bound).\footnote{This will be discussed further in Chapter \ref{HABHM}.} 
They correspond to the {\it black $p$-brane solutions} of supergravity theories that had been known since the 1980s.\footnote{Black $p$-branes are solutions of the supergravity field equations that describe black holes with higher-dimensional event horizons. See e.g.~Horowitz and Strominger (1991:~p.~198). For more on this, see Sections \ref{dynD} and Appendix 8.A below.}\\
\\
Furthermore, Polchinski (1995) understood a fourth point, which connects the D-branes in the closed string spectrum to D-branes in the open string spectrum:\\
(4)~~An alternative (i.e.~dual) description of Witten's closed string D-brane states is as the {\it open string} D-branes of Section \ref{Dbopen}, introduced through the Dirichlet boundary conditions at the endpoints of strings.

\begin{figure}
\begin{center}
\includegraphics[height=5cm]{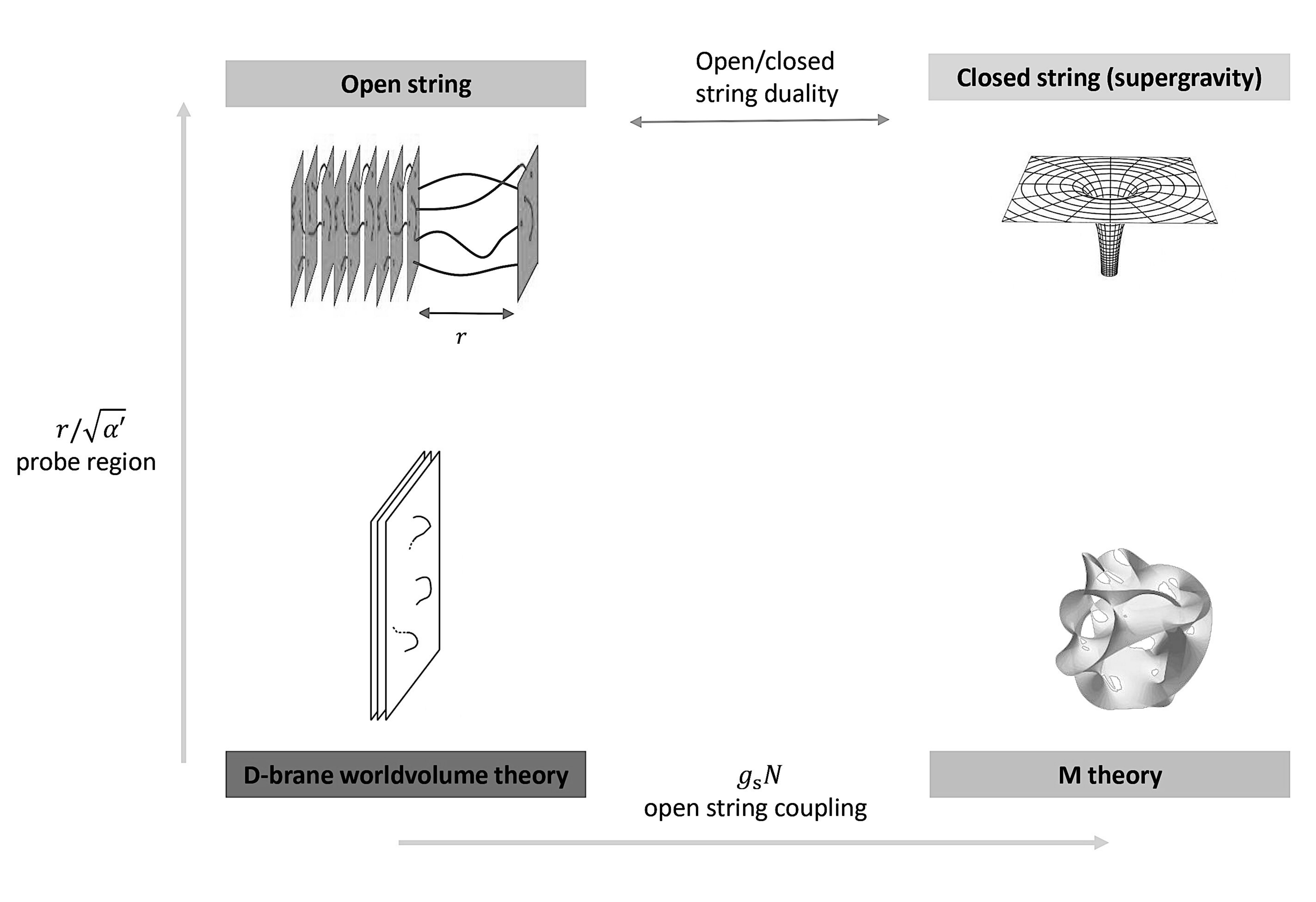}~~~
\caption{\small Summary of the relations between the states discovered during the second superstring revolution, in terms of two parameters: the probing distance and the string coupling. {\it Top left}: D-branes in open string theory at weak coupling, probed at large distances, $r$. {\it Bottom left}: world-volume description of D-branes in open string theory at weak coupling, probed at short distances. {\it Top right}: closed string description of D-branes at long distances and strong string coupling, as black holes or other curved spacetime solutions of supergravity. {\it Bottom right}: (largely unknown) M-theory description, at short distances and strong coupling.}
\label{SV}
\end{center}
\end{figure}

The interrelations (1) to (4) are summarized in Figure \ref{SV}. Points (1) and (2) have the following consequence: according to the mass formula for D-branes, Eq.~\eq{Mcharge}, when the string coupling constant is {\it weak}, i.e.~$g_{\sm s}\ll1$, D-branes are heavy, and so they in effect decouple from the perturbative string theory on the world-sheet: they are too massive to be ``seen''. However, at {\it strong coupling} (or if there are a large number of them), the D-branes back-react on the spacetime, which gets highly curved: we get a black hole (i.e.~point (3) is the top-right corner of Figure \ref{SV}). 

The correspondence, in point (4), between Witten's closed-string picture and the open-string description of D-branes from Section \ref{Dbopen}, is a {\it quasi-duality} between the open string model with D-branes and the black $p$-brane supergravity model. (We will return to this when we discuss black hole microstates, in Section \ref{ocs}.)\footnote{For a detailed discussion of this point and its significance for the microscopic calculation of black hole entropy in string theory, see De Haro et al.~(2020:~pp.~85-93), Van Dongen et al.~(2020:~pp.~117-118), and Strominger and Vafa (1996).}

The $1/g_{\sm s}$-dependence of the mass of the D-brane contrasts with the masses of perturbative string states that we discussed in Sections \ref{stringth} and \ref{T-d}. Those were {\it independent} of the string coupling (and usually proportional to $1/\a'$). It also differs from the masses of solitons in field theory, such as monopoles, which are usually inversely proportional to the {\it square} of the coupling constant (see Eq.~\eq{Eresult}, where $\b$ is the coupling constant of the sine-Gordon model).\footnote{For the one over coupling-squared dependence of the mass for magnetic monopole solutions, see 't Hooft (1974b:~p.~283) and Prasad and Sommerfield (1975:~p.~762).} 

This unusual dependence on the string coupling is explained by the nature of the D-brane charge, and how it couples to the dilaton field: it is Ramond-Ramond (RR) charge, and D-brane states do not belong to the fields in the usual (perturbative) NS-NS sector on the string world-sheet. Rather, the fields in the RR sector are non-perturbative and, accordingly, their coupling to the dilaton field differs from the usual coupling in the NS-NS sector.\footnote{This can be seen from the effective supergravity actions for D-branes, under point (3).} (In Section \ref{S-d}, this specific coupling of RR fields will be crucial to get correct transformation properties under S-duality.)

\section{The dynamics of D-branes}\label{dynD}

Section \ref{Dbsugra} first introduces the low-energy (i.e.~supergravity) description of D-branes in closed string theory. Section \ref{elemsolit} then discusses two types of D-branes: elementary, and solitonic. These two types of D-branes will be used when we discuss various string dualities in later Sections.

\subsection{D-branes in supergravity}\label{Dbsugra}

The previous Sections discussed two different descriptions of D-branes, in terms of open and closed strings, that are believed to be quasi-duals (see Figure \ref{SV}): the contrast is between $p$-dimensional hyperplanes on which open strings can end, and {\it black $p$-brane} solutions of supergravity theories that are low-energy limits of closed string theories. In both cases, we will continue to speak about `D-branes' (or, to emphasize their dimensionality, D$p$-branes), since there will be no confusion and it will be clear from the context when we are discussing the hyperplanes on which open strings can end, and when we are discussing solutions of supergravity. In this Section, we will give the details about the supergravity (i.e.~the low-energy limit of the closed string) description that, in the rest of this Chapter, will allow us to understand string dualities.

We first need to understand how D$p$-branes modify the geometry. Since $p$-branes are charged, the energy-momentum tensor in the Einstein field equations will include the gauge fields that carry the D-brane's charge. Thus we ask: what kind of gauge field carries the charge for which a $p$-dimensional object is a source? This question is best answered by analogy with the point particle ($p=0$) and the string ($p=1$) that we already encountered. A charged point particle is a source for an electric field described by the gauge field $A_\m$ (if there is a current, there is also a magnetic field). We say that the point particle `couples' to the gauge field, through a coupling $A_\m\,\dot x^\m$ in the world-line action of the particle. (This term contributes a delta function, along the particle's world-line, to the four-current $J^\m$: see footnote \ref{wlp}.) The electric field contributes to Einstein's equations through the energy momentum tensor (Eq.~\eq{Tmunu}). 

We have also seen that strings, i.e.~one-dimensional objects, naturally couple to antisymmetric two-index potentials (see Eq.~\eq{fullWS}), with corresponding antisymmetric three-index field strength (Eq.~\eq{H3}), which have a generalized gauge symmetry. In general, the number of indices that we need to couple the gauge field to the world volume of the D-brane, so that all of the dimensions of the D-brane couple to the charge, equals the number of spacetime dimensions of the $p$-brane.\footnote{From the middle term of Eq.~\eq{fullWS}, one sees that the antisymmetric tensor $B_{\m\n}$ couples to the world-sheet of the string by the natural pull-back of the string's world-volume element to the spacetime, i.e.~$\e^{\a\b}\pa_\a X^\m\pa_\b X^\m$. This generalizes to $p$-branes of higher dimensions.} 
Thus for a 2-brane, i.e.~a membrane (with two space and one time dimensions), we need a three-index gauge potential $A_{(3)}$, from which we construct a four-form field strength $F_{(4)}$, etc. In general, a $p$-brane couples to (or is a source for) a $(p+1)$-form $A_{(p+1)}$, with field strength $F_{(p+2)}$. These are conveniently written in form notation:
\bea
A_{(n)}=A_{\m_1\cdots\m_n}\,\dd x^{\m_1}\wedge\cdots\wedge\dd x^{\m_n}\,,
\eea
where the field strength is $F_{(n+1)}=\dd A_{(n)}$, with gauge transformations $\d A_{(n)}=\dd\L_{(n-1)}$, and $n=p+1$. In $D$ spacetime dimensions (the relevant case will be $D=10$), the charge of the D-brane can be calculated by a generalization of Gauss' law:\footnote{The details are as follows. The gauge field $A_{(p+1)}$ couples to the conserved current, $J_{(p+1)}$ on the world volume of the D-brane, through the term $\int\dd V_D\,A_{\m_1\cdots\m_{p+1}}\,J^{\m_1\cdots\m_{p+1}}$ (where $\dd V_D$ is the $D$-dimensional infinitesimal volume of integration), and the integrand is more conveniently written in form notation as $A\wedge*J$. The Maxwell action then contains two terms: $\half\dd A\wedge*\dd A$ and $A\wedge*J$ (where, analogously to the point particle case, the current $J$ contains appropriate delta functions that effectively restrict the integral to the world volume of the D-brane). By varying the gauge field $A$, one obtains the Maxwell equations: $\dd*F=*J$, where $*J$ is a $D-p-1$-form, since $J$ is a $(p+1)$-form. Integrating $*J$ over the a $(D-p-1)$-dimensional volume containing the charge, gives: $Q=\int_{V_{D-p-1}}*J=\int_{V_{D-p-1}}\dd*F=\int_{\Si_{D-p-2}}*F$, where $\Si:=\pa V$ is the $(D-p-2)$-dimensional boundary of the volume enclosing the charge (and in the last step we used Gauss' law). Notice that, if the Maxwell term contains a dilaton coupling, as in Eq.~\eq{S10} below, then this will introduce a dilaton dependence in the calculation of the charge.}
\bea\label{Qcharge}
Q=\int_{S^{D-p-1}}*F\,,
\eea
and $S^{D-p-1}$ is a $(D-p-1)$-dimensional sphere encircling the $p$-brane, and $*F$ is the Hodge dual of $F_{(p+2)}$, i.e.~a $(D-p-2)$-form obtained by contracting the totally antisymmetric epsilon $D$-tensor with the antisymmetric $(p+2)$-tensor $F_{(p+2)}$ (i.e.~it generalizes Eq.~\eq{Gdual}). Such a tensor can indeed be integrated over a $(D-p-2)$-dimensional spatial surface. 
The Hodge dual of a tensor of rank $n$ is a tensor of rank $m$ with components:
\bea
(*F)^{\m_1\cdots\m_m}:={1\over n!}\,\e^{\m_1\cdots\m_m\n_1\cdots\n_n}F_{\n_1\cdots\n_n}\,,
\eea
where $m=D-n$. 

By analogy with the $p=0$ and $p=1$ cases, the bosonic part of the supergravity action is:\footnote{For D-branes that are non-solitonic we would also add the D-brane's world volume action: see the discussion below.}
\bea\label{S10}
S_{\tn{spacetime}}^{(D,p)}={1\over2\k_D^2}\int\dd^Dx\,\sqrt{-g}\left(R-\half(\pa\f)^2-{1\over2(p+2)!}\,e^{-a\f}\,F_{(p+2)}^2\right),
\eea
where $\k_D^2=8\pi G_{\tn N}$, and the value of $a$ depends on the values of $D$ and $p$, and on the type of $p$-brane one considers.\footnote{For elementary $p$-brane solutions, the value of $a$ is given by: $a^2=4-2(p+1)(D-p-3)/(D-2)$. See Duff et al.~(1995:~p.~254).} This coupling of the D-brane fields to the dilaton field differs from that of the Ricci scalar (recall that the metric is a field in the NS-NS sector), and indicates that the gauge field $A_{(p+1)}$ is a RR field. 

In supergravity, the above action can be made supersymmetric by adding fermionic terms. Note that, just like the string theories, not all supergravity theories contain D-branes for all values of $p$. Here we will focus on the Type IIA and Type IIB models. While all supergravity theories that are low-energy limits of string theories allow the value $p=1$ (as they should, since there should be a string in the low-energy spectrum), the other values are divided as follows: the Type IIA model only allows the even values $p=0,2,4,6,8$, while the Type IIB model only allows the odd values $p=1, 3, 5,7,9$.\footnote{As we will discuss below, D-branes come in electric-magnetic pairs, so that D-branes with $p<3$ are electric, the $p=3$ D-brane is self-dual, and D-branes with $p>3$ are magnetic. Polchinski (1995:~p.~4726) mentions that, even though the $p=8$ and $p=9$ cases do not correspond to propagating string states, they are still present in the Type IIA and Type IIB models. A 9-brane is a D-brane that fills the whole space, i.e.~it is a Neumann boundary condition. The 8-brane is the one that we began from above when we discussed the Dirichlet boundary condition. There is also a $p=-1$ D-brane, which is interpreted as a D-instanton, but we do not discuss here. Zwiebach (2009:~p.~324, 371) explicitly constructs the massless states in the RR sector of the Type IIA and Type IIB models.}

For various values of $p$, the above action has solutions describing spacetimes that are curved by the presence of charged D-branes. The core of such a solution extends along $p$ spatial and one time dimensions. Because D-branes are BPS solutions, they preserve half of the spacetime supersymmetry (i.e.~the condition that the solutions are supersymmetric in effect gives the BPS condition Eq.~\eq{Mcharge}), and hence those properties that are defined by the supersymmetry algebra cannot change between one regime of couplings and another. This is the case with, for example, the Hawking entropy of these black holes, and lies at the basis of Strominger and Vafa's successful counting of black hole microstates (see Section \ref{ocs}).

The BPS mass of a D$p$-brane with RR charge is (i.e.~including the value of the overall constant in Eq.~\eq{Mcharge}):\footnote{The mass follows from the tension of the D-brane, which is the mass per unit volume: $T_p=1/((2\pi\ell_{\tn s})^pg_{\tn s}\ell_{\tn s})=M_p/V_p$ (the tension of the D-brane is calculated in Becker et al.~(2007:p.~234)). The unitless charge is: $n_p=V_p/\ell_{\tn s}^p$, which gives the value of the mass reported in Eq.~\eq{Dbmass}. Since the tension is independent of the volume of the D-brane, from now on it will be clearest to work with the tension, i.e.~the mass density, rather than the mass.\label{DbT}}
\bea\label{Dbmass}
M_p={1\over(2\pi)^p\,\ell_{\sm s}}\,{|n_p|\over g_{\sm s}}\,,
\eea
where $n_p$ is the RR charge of the D$p$-brane. The next Section will contrast this expression with the mass of a macroscopic fundamental string.

An argument by Nepomechie (1985:~p.~1923)\footnote{See also Teitelboim (1986:~p.~71).} 
involving a gauge potential $A_{(p+1)}$ (with field strength $F_{(p+2)}$) on the northern and southern hemispheres of a $S^{p+2}$-sphere implies that the electric and magnetic charges of the gauge field, independent of the dimension of the spacetime, satisfy the Dirac quantization condition. Applied to D-branes that are the sources for the electric and magnetic charges, note that the magnetic dual of a D$p$-brane is a D$\ti p$-brane, with $\ti p=D-p-4$ (see Eq.~\eq{dualitybranes} in Appendix 8.A). Thus in $D=10$, $\ti p=6-p$. This means that D1-branes and D5-branes are dual, and D2-branes and D4-branes are dual, and the D3-brane is self dual. Their corresponding charges satisfy: $e_pg_{6-p}=2\pi n$ (for integer $n$). 

\subsection{Elementary and solitonic D-branes: two examples}\label{elemsolit}

We have characterised solitons in field theory as non-linear, finite-energy, solutions that are stable due to a conservation law that is unrelated to Noether's theorem, and has a topological charge (i.e.~one that is quantized by a topological condition). Physically, solitons are collective oscillations of the field that look like localized lumps of energy. On this notion, many of the above D-branes {\it are} solitons of general relativity, and for this reason we did not include the soliton's world volume action into Eq.~\eq{S10}: for solitons we are not supposed to, since they ``emerge as non-trivial solutions'', rather than being inserted as fundamental degrees of freedom to be added to the action.

However, while the solitonic nature of D-branes is indeed important to string theory dualities, not all D-branes are solitonic: many are singular, in the sense that they are solutions of the equations only if an appropriate Dirac delta-function for their world volumes is added. (For this reason, some authors call these singular D-branes `elementary', rather than `solitonic').

Yet other authors argue\footnote{See Schwarz (1997:~p.~8).} 
that the distinction between `elementary' and `solitonic' D-branes should not be taken too literally, and their argument proceeds in two steps, of which the first is general, and the second is a further specification: (1) The classical supergravity description, including the singularities, being an effective (low-energy) description, is not guaranteed to be significant in string theory when quantum effects are included. (2) Various electric-magnetic dualities (see below) relate the singular and the non-singular solutions, including in cases where these dualities are taken to be fundamental for the underlying string theory (i.e.~more fundamental than the specific forms of the low-energy solutions). If such dualities relate singular and non-singular solutions, while the physics is believed to in effect be the same, then there is no reason why the classical solutions' being singular or non-singular should be significant.\footnote{Note that this approach towards singularities is what Crowther and De Haro (2022) call `attitude 4', i.e.~`indifference to singularities---they do not point to quantum gravity'.} 

Even assuming that these two points (1) and (2) are correct, the argument seems inconclusive. For some dualities {\it do} relate elementary and solitonic objects (see the examples in Chapter \ref{Advan}). Thus it is indeed possible that the `elementary vs.~solitonic' distinction has a correlate in the quantum models, and that these roles are exchanged by dualities. For example, bosonization exchanges an elementary fermionic operator and a composite bosonic one (see Eq.~\eq{expf}: we will return to this in Part III.) This suggests that, even if the common core theory may not make this distinction, it is an important ingredient in formulating dual models, even after quantization. For this reason, and because this distinction {\it is} significant to the supergravity models, we will uphold it below.

\subsubsection{The elementary superstring}

Since the form of the solutions for general D$p$-branes are complicated, we here focus on the string solution in ten dimensions (i.e.~$p=1$ and $D=10$) and its dual, namely the fivebrane (for the duality between the two, see Appendix 8.A). 

Such a string is a solution of supergravity, and so it is a {\it macroscopic}, rather than a microscopic, string (Dabholkar et al.~1990:~p.~34). A macroscopic string can be seen as a coherent state of microscopic strings at low energies. 

The action of such an elementary string is the spacetime action Eq.~\eq{S10}, with the value $p=1$, coupled to the string's world-sheet action, Eq.~\eq{fullWS}, i.e.~$S_{\tn{spacetime}}^{(D=10,\,p=1)}+S_{\tn{WS}}$. (This is analogous to coupling the Maxwell equations to a point-particle source for an electron.) Thus there are four kinds of fields, $(g,B_{(2)},\f,X)$: the spacetime metric, the two-form gauge field (with associated field strength $F_{(3)}$), the dilaton, and the scalar fields that embed the string into the target spacetime. 

We look for a judicious Ansatz for these fields that solves the set of four coupled equations and describes a static, flat, infinitely extended, supersymmetric string. Thus we have three requirements: (i) the solution should preserve the symmetries of a flat, static, infinite extended string, (ii) it should preserve supersymmetry, and (iii) it should satisfy the four coupled field equations.\footnote{Taking the string, under (i), to be {\it static} amounts to solving for the equation of motion of the target space coordinates, $X$, on the world-sheet. Thus, under (i) we have already fully specified the kind of solution of the world-sheet equations of motion that we are looking for (and we will use this choice in (ii)). Since in (ii) we use this choice of string solution, it is clearest to specify the string solution in step (i). In step (iii), we of course check that we get a solution of all the equations of motion.}

As we will discuss below, condition (i) determines the form of the metric and the gauge two-form in terms of three unknown functions $A,B,$ and $C$. Condition (ii) then gives these three functions in terms of the dilaton field. (iii) For a static solution, i.e.~a string at rest, the field equations fix the value of the dilaton field. We discuss these three steps in turn:

{\bf (i) Spacetime symmetries:} the world-sheet of a flat, infinitely extended, string in spacetime is Poincar\'e invariant. Furthermore, since there is no other matter anywhere in the embedding (i.e.~target) spacetime, the latter is rotationally symmetric in the eight other directions, transverse to the string. 

Thus the symmetry group of the target space is $\mbox{Poincar\'e}_2\times\mbox{SO}(8)$, i.e.~there is two-dimensional Poincar\'e symmetry for the longitudinal directions (with $\m=0,1$) and SO(8) rotation symmetry for the remaining eight transverse directions (i.e.~$\m=2,\ldots,9$). Thus we naturally split the coordinates as follows: $x^\m=(x^\a,y^m)$, where $\a=0,1$, $m=2,\ldots,9$. The most general line element with this symmetry is:
\bea\label{stringmetric}
\dd s^2=e^{2A}\eta_{\a\b}\dd x^\a\dd x^\b+e^{2B}\d_{mn}\dd y^m\dd y^n\,,
\eea
where $A$ and $B$ are yet to be determined functions, with the same symmetry. 

The Ansatz for the two-form gauge field $B_{(2)}$ (whose three-form field strength is $F_{(3)}$) follows by considering that the source for $B_{(2)}$ is the string's world-sheet (through the coupling in Eq.~\eq{fullWS}), and so we may take it to be longitudinal to the world-sheet, and we are left with a single non-zero component, viz.~$B_{01}$, which we write in terms of an arbitrary function $C$: $B_{01}=-e^C$.\footnote{At this point, writing $C$ in the exponential is a mere convenience, and no assumption about the spatial orientation of the gauge field is required. However, under (iii), $e^C$ will be seen to satisfy the Poisson equation with a delta function source on the right-hand side. $C$ is then a real function, and the orientation is indeed $B_{01}<0$.}

{\bf (ii) Supersymmetry:} a putative solution of the field equations is invariant under supersymmetry iff a supersymmetry transformation does not modify the putative solution, i.e.~the supersymmetry variation of each of the fields for that putative solution is zero. For example, the supersymmetry variation of a (non-abelian) gauge field and its supersymmetric partner is given in Eq.~\eq{susy}. 

The superpartners of the three bosonic fields of interest in our example, i.e.~$(g, B_{(2)},\f)$, are a spin-${3\over2}$ field, $\psi_\m$, and a spin-${1\over2}$ field $\l$.\footnote{Two comments regarding supersymmetry: (1) Recall that, in (i), we required that the string be static. Such a solution does preserve supersymmetry: see Duff (1996:~p.~39). (2) We assume minimal, unextended supersymmetry (i.e.~${\cal N}=1$, see Section \ref{basicsusy}), which corresponds to Type I supergravity. In Type II supergravities, the current solutions are still correct, but the supersymmetry analysis includes a second set of fermionic fields. The fermionic part of the supergravity action is given in Dabholkar et al.~(1990:~p.~45).} Recall that, from Pauli spin statistics (see Eqs.~\eq{susyQ}-Eq.~\eq{susy}), the supersymmetry variation of a bosonic field contains {\it at least one fermionic} field (namely, either $\psi_\m$ or $\l$), and the supersymmetry variation of a fermionic field (i.e.~$\psi_\m$ or $\l$) contains fields in {\it bosonic combinations}, as follows:
\bea\label{varf}
\d(\mbox{bosonic field})&=&\mbox{combination of fermionic fields}~(\psi_\m,\,\l)\overset{\psi_\m,\l=0}{=}0\,,\nn
\d(\psi_\m,\l)&=&\mbox{bosonic combination of fields.}
\eea
The first line is zero because, for the (putative) bosonic solution that we are considering, all the fermions are zero, i.e.~$\psi_\m=\l=0$, which automatically sets the variations of all the bosonic fields to zero. The same argument does {\it not} apply to the variations of the {\it fermionic} fields in the second line of Eq.~\eq{varf}, since the bosonic combinations of fields that appear on the right-hand side are in general not zero. Thus the supersymmetry of the solutions of the type we are considering (i.e.~with bosonic fields only, and all fermionic fields set to zero) requires setting the right-hand side of the second equation to zero. This is what our condition (ii) comes down to, and it is a non-trivial condition. In effect, this condition relates the three unknown functions $A,B$, and $C$ to the dilaton field, with values: $A={3\over4}\f$, $B=-{1\over4}\f$, and $C=2\f$. The boundary conditions are chosen such that the metric is asymptotically Minkowskian, i.e.~$\f$ goes to zero at infinity (recall that the constant expectation value of the dilaton at infinity was subtracted from $\f$ in Eq.~\eq{betaf}), and the gauge field strength, $F_{(3)}$, is asymptotically zero. Thus, except for the arbitrary function $\f$, the solution is thus fully determined, as we claimed above.\footnote{There is also a chirality condition on the spinors $\e$ that can be constructed for such a solution. Because chiral spinors have half the number of components, the jargon says that the solution preserves `half the supersymmetry'.}

{\bf (iii) Field equations:} there are four types of field equations that follow from the action, $S_{\tn{spacetime}}^{(D=10,p=1)}+S_{\tn{W}}$, one for each of our fields, $(g,B_{(2)},\f,X)$: (1) Einstein's gravitational field equations, where all the matter fields contribute to the energy-momentum tensor; (2) the Maxwell-type equation for the field strength, $F_{(3)}$; (3) the scalar equation for the dilaton; (4) the world-sheet equations of motion, derived from the string world-sheet action. 

Since the world-sheet equations of motion, (4), were previously solved by the requirement that the string is static, we just need to check (1) to (3). And, as it turns out, with our Ansatz (where the dilaton is the only unknown function), these three equations boil down to a {\it single} equation: namely, a Poisson-type of equation for the dilaton (with a source at the locus, $\{y^m=0\}_{m=2,\ldots,9}$, of the string), with solution (unique up to the asymptotic boundary value of the dilaton, $\f_0$):
\bea\label{ec2}
e^{-2\f}=1+{c_2\over r^6}\,,
\eea
where $r=\sqrt{\d_{mn}y^my^n}$ is the transverse distance to the string, and the constant $c_2$ is given in terms of Newton's constant, the string tension, and other numbers (see Duff et al.~(1999:~p.~229)).

The elementary superstring satisfies the BPS bound, Eq.~\eq{bps}, for the mass per unit length, where on the right-hand side we have the electric, Noether charge of the string, i.e.~essentially the integral over the flux of the NS-NS three-form, $F_{(3)}$ (see Eq.~\eq{Qcharge}).\footnote{The Noether electric charge is: $e_2=\sqrt{2\k^2}\int_{S^3}e^{-\f}*H=2\sqrt{2}\k\,T$, where $T=1/2\pi\a'$ is the string tension. The charge is conserved in virtue of the equation of motion of the two-form.}
The mass density of the macroscopic elementary supestring is equal to the string tension, $T=1/2\pi\a'$.\footnote{See Dabholkar et al.~(1990:~p.~4). Duff et al.~(1995:p.~231) give the mass in the Einstein frame.}

The mass and the tension of an elementary macroscopic string differ in an important way from the mass and the tension of a D1-brane with RR charge, because for the elementary string the mass and the tension are independent of the string coupling, while for the D1-brane they are multiplied by the inverse of the string coupling: $T_1=1/(2\pi\a'g_{\sm s})$.\footnote{This follows by setting $p=1$ in the tension given in footnote \ref{DbT}. Alternatively, Eq.~\eq{Dbmass} gives: $M_1=|n_1|/(2\pi\a'g_{\sm s})$, where the charge is the length of the macroscopic string divided by the fundamental string length $\ell_{\tn s}$. Therefore, the tension, i.e.~the mass per unit length, has the value reported.}
Even though both objects are stringlike, they have different types of charges, and so their dependence on the coupling is different. (The next Chapter will discuss that they are related by S-duality.) This difference reflects the fact, already discussed in Section \ref{dbst}, that the Ramond-Ramond charge of D-branes is non-perturbative in the string coupling, and so their mass depends on the {\it inverse} of the string coupling. By contrast, the mass of an elementary macroscopic string is independent of the coupling.\footnote{The BPS equation for the mass of the macroscopic string follows from its supersymmetry condition, i.e.~from step (ii) above, and it does not require all the details from step (iii). It also expresses a condition that the force between parallel strings with the same orientation is zero. For a stationary test string at some distance from the elementary string solution, with the same orientation, experiences a zero net force: the gravitational and electric force cancel each other out. One can see this from the geodesic equation in the above string geometry, where the contribution of the Christoffel symbols is cancelled by the three-form field strength, $F_{(3)}$. If the orientation is opposite, there is a net attractive force.} 

\subsubsection{The solitonic fivebrane}

We have seen that we can obtain an elementary string solution of supergravity from the equations of motion of the action with a source at the location of the string, i.e.~$S_{\tn{spacetime}}^{(D=10,p=1)}+S_{\tn{W}}$. This is analogous to getting the electric Coulomb field by coupling the Maxwell equations to a point-particle source for an electron.

Remarkably, we can get the {\it magnetic dual} of an elementary string by considering the {\it source-free} supergravity equations, following from the action $S_{\tn{spacetime}}^{(D=10,p=1)}$ with {\it no source}, i.e.~as a soliton. This is somewhat analogous to the 't Hooft-Polyakov magnetic monopole, which is not obtained by adding a magnetic source, but as a finite-energy solution of the non-linear field equations, Eq.~\eq{HPmonopole}, with a charge that is topological, rather than conserved by Noether's theorem (see Eq.~\eq{gmap}). In the following Section, we will explain why the electric-magnetic dual of a {\it string}, in ten dimensions, is a {\it fivebrane}, instead of just another string.

The steps are the same as in the previous Section, with two differences: there are three field equations to be satisfied rather than four (since for a soliton there are no target space fields $X$), and the symmetries are different. A fivebrane has a six-dimensional world-volume and four transverse directions. Hence we split the spacetime coordinates as: $x^\m=(x^\a,y^m)$, where $\a=0,\ldots,5$ and $m=6,\ldots,9$, and the symmetry group is $\mbox{Poincar\'e}_6\times\mbox{SO}(4)$. The Ansatz for the three-form field strength with these symmetries is:\footnote{This Ansatz can be obtained as the Hodge dual of the electric field of the string, using the duality worked out in the next Section.}
\bea
F_{(3)}=C\,\e_3\,,
\eea
where $C$ is a constant containing the magnetic charge (see below), and $\e_3$ is the volume form (i.e.~the appropriately normalized three-dimensional Levi-Civita tensor) on the $S^3$ surrounding the fivebrane, in the transverse Euclidean space characterized by the coordinates $y^m$ ($m=6,\ldots,9$). The rest of the solution is then given by:
\bea\label{ec6}
\dd s^2&=&e^{-\f/2}\,\eta_{\a\b}\,\dd x^\a\dd x^\b+e^{3\f/2}\d_{mn}\dd y^m\dd y^n\,,\nn
e^{2\f}&=&1+{c_6\over r^2}\,,
\eea
where $r=\sqrt{\d_{mn}y^\m y^\n}$ ($m=6,\ldots,9$), and $c_6$ is a constant of integration that, up to constants, is equal to the magnetic charge, $g_6$.\footnote{The constants, in the Einstein frame, are: $c_6=g_6/\sqrt{2\k^2}\,\Om_3\,e^{-\f_0/2}$ and $C=2c_6e^{\f_0/2}$, where $\Om_3$ is the volume of the unit three-sphere.}

This solution is again supersymmetric, and its tension saturates the BPS bound, i.e.~it is given by Eq.~\eq{Dbmass} with $p=6$. At weak string coupling ($g_{\sm s}\rightarrow0$), fivebranes are very massive, as we expect from solitons.

\section{Conclusion}

This Chapter has done two things: it has introduced the basics of string theory, and it has discussed conceptual aspects of dualities that Chapters \ref{physeq} and \ref{Heuri} will develop further. Among these conceptual aspects, we here highlight three ideas: namely, {\it augmentation}, the contrast between {\it internal and external interpretations}, and {\it abstraction}. That these three aspects are not specific to dualities in string theory can be seen from the analogies with other dualities, where they also play an important role.

A key aspect of dualities in string theory, and in particular T-duality, is the {\it augmentation} of the sets of states and quantities that is required for a full formulation of duals. Dirichlet boundary conditions for open strings require us to introduce D-branes: and D-branes are found to be dynamical objects themselves, with dynamics both on their world-volumes, where they interact with open strings, and on the surrounding spacetime, which they curve and can become a black hole, as the next two Chapters will discuss. 

The augmentation of the sets of states and quantities reminds us of bosonization duality, which requires having fermionic states and operators that are coherent states of bosons (see Eq.~\eq{expf}). It is also analogous to the discussion of soliton states in Chapter \ref{EMYM}, and especially in the Seiberg-Witten theory, where the electric states of the dual model, $M'$, come from magnetic solitons that are only seen at strong values of the Higgs field. While the low-energy models $M$ and $M'$ only contain one type of states, the spectrum of the full ${\cal N}=2$ SYM theory contains both types. By analogy, perturbative string states and non-perturbative D-brane states are both in the spectrum of string theory. 

A counterpart of this idea, purely within the closed string sector, is T-duality's relating momentum states to winding states. On a common core interpretation, these states are all of one kind, and spacetime has a definite shape, but not a definite radius. (This has also prompted the research programme in physics called `double field theory': a kind of truncation of string theory that is explicitly invariant under T-duality, and can be seen as a stringy version of effective field theory: cf.~footnote \ref{dft}.) This can be understood by analogy with the Maxwell theory, where an {\it internal interpretation} mentions an electromagnetic field but no purely electric or magnetic fields, while {\it external interpretations}, given by formulations of the theory in various systems of coordinates, do mention `electric' and `magnetic'. By analogy, `momentum', `winding', and the radius of space, are parts of external interpretations that are constructed in a semi-classical limit of the theory. 

In these cases, the external interpretations of duals are not incompatible with an internal one, but rather mention aspects about which the common core is silent. Chapter \ref{physeq} will develop this idea further, in terms of {\it abstraction}, and of the logico-semantic relations between a theory and its models.

Among the surprising aspects of T-duality is the variety of situations that it relates: it maps a D-brane of some dimension to a D-brane of a different (higher or lower) dimension. Also, we have seen that D-branes come in electric and magnetic pairs, which are mapped into each other by Hodge duality (see Appendix 8.A).

Regardless of dualities, we have seen that string theory satisfies a basic hallmark of a more fundamental theory: namely, Newton's constant is given in terms of the string length and the string coupling. This is analogous to other examples in physics, where experimental constants such as the Bohr radius, the Rydberg constant, or the Stefan-Boltzmann constant, could be predicted in terms of the more fundamental constants of quantum mechanics. But there is of course more than this, since as we have seen, the low-energy limit of string theory is supergravity theory, and the world-volume theories of D-branes are supersymmetric versions of quantum field theories. Thus, for appropriate regimes of parameters, string theory gives supersymmetric versions of general relativity and quantum field theory: which is of course a requirement of any theory of quantum gravity. The next Chapter will further elaborate on these topics.

By contrast with quantum theory, where the values of constants can be determined with increased accuracy, a key limitation of the current understanding of string theory is that the numerical value of e.g.~Newton's constant cannot be predicted, because the values of the string length and the string coupling are not known. Also, there is believed to be a vast landscape of vacua where string theory reproduces a dazzling variety of low-energy models, without yet a principled explanation for why, out of this vast landscape, our universe occupies one particular vacuum.

\section*{Appendix 8.A. Electric-magnetic duality for D-branes}\label{EMDb}
\addcontentsline{toc}{section}{Appendix 8.A. Electric-magnetic duality for D-branes}

Our aim in this Appendix is to argue that the classical electric-magnetic dual of a D$p$-brane is a D$\ti p$-brane, where $\ti p=D-p-4$ (where $D$ is, as usual, the spacetime dimension), and the electric charge of the D$\ti p$-brane (i.e.~defined, through Eq.~\eq{Qcharge}, for the {\it dual} Faraday tensor) is the magnetic charge of the original D$p$-brane (and the magnetic charge of the dual D$\ti p$-brane is the electric charge of the D$p$-brane).

This version of electric-magnetic duality, that holds for D-branes, is suggested by the Hodge dual that we used earlier, and generalizes the point-particle case. The structure is analogous to the discussion of the Maxwell theory in Section \ref{EmdS}, for $D=4$ and $p=0$. Even though not all of these dualities generalize to quantum dualities, several important examples do, in string theory (and the next Chapter will illustrate their significance for M-theory). Thus the current Section illustrates how electric-magnetic duality is realized classically though Hodge duality.

\subsection*{8.A.1. Electric-magnetic duality for point particles, revisited}
\addcontentsline{toc}{subsection}{8.A.1. Electric-magnetic duality for point particles, revisited}

In four dimensions, Hodge duality maps a two-form to another two-form, and it maps one-forms and three-forms onto each other. Thus for a point-particle, the dual (magnetic) theory is based on the Hodge dual of the Faraday tensor (see Eq.~\eq{Gdual}), itself again a two-index antisymmetric tensor. Electric-magnetic duality then interchanges the equation of motion and the Bianchi identity (see Eq.~\eq{Max2}): the equation of motion of the original (electric) theory becomes the Bianchi identity of its dual, and vice versa. However, since we wish to consider sources rather than vacuum solutions, we require a slight generalization of the electric-magnetic duality from Section \ref{EmdS}. 

Recall, from Eq.~\eq{Max2}, the form of the field equations of the Maxwell theory using differential forms. We now add a conserved electric current $J$, that is a one-form, on the right-hand side of the Maxwell equations:\footnote{We follow the usual conventions in the high-energy physics literature of, in addition to setting $c=1$, also rescaling the Faraday tensor so that $\m_0$ does not appear in the Maxwell equations (cf.~Eq.~\eq{elmagn}). Note that, in order for the second equation in form notation to give back the equation of motion, the dual of the source must appear on the right-hand side (compare this with footnote \ref{Jdual}).}
\bea
\mbox{Bianchi identity:}~~~~~~~\dd F&=&0\nn
\mbox{Equation of motion:}~~~~~\dd*F&=&*J\,.
\eea
The Hodge dual of the current, $*J$, is a three-form, which for a point-particle sitting at the origin is a Dirac delta function for the particle's world-line, $x=0$. In components: $(*J)_{123}=e\,\d^{(3)}(x)$.\footnote{For a point-particle at the origin at rest, the conserved current is (see footnote \ref{wlp}): $J^\m=e\int\dd\t\,\dot y^\m\d^{(4)}(x-y(\t))$, where $y^\m(\t)=(\t,{\bf 0})$, so that $J^\m=e\int\dd\t\,\d^\m_0\d^{(4)}(x-y(\t))=e\,\d^\m_0\,\d^{(3)}(x)$. The dual is: $(*J)^{123}=\e^{123\m}\,J_\m=\e^{1230}J_0=e\,\d^{(3)}(x)$, with all components along the $0$-direction being zero.} 
Filling this into Eq.~\eq{Qcharge}, we indeed get the electric charge: $e=\int_{V_3}*J=\int_{S^2}*F$, where $V_3$ is a three-volume containing the particle. 

The conservation of the current $J$ follows as a consequence of the equations of motion: namely, $\dd* J=\dd^2*F=0$. 

We know from Section \ref{Dqc} that the presence of an electric current but no magnetic current breaks the electric-magnetic duality. Thus to get a set of equations that illustrate electric-magnetic duality, we include an electric {\it and} a magnetic current in the original equations, so that the dual equations have the same form as the original ones, under Hodge duality, $F\mapsto\ti F:=*F$, while also exchanging electric and magnetic currents. In the presence of {\it both} electric and magnetic currents, the Maxwell equations take the following form:\footnote{The first Maxwell's equation implies that the Faraday tensor is not exact, as it is in the Maxwell theory without magnetic currents. Thus to get a non-zero magnetic current on the right-hand side of the first equation, we allow the Faraday tensor to be of the following form: $F=\dd A+\om$, where $\om$ is a {\it non-exact} two-form (recall, from Section \ref{EMduality}, that an exact Faraday tensor cannot give magnetic charge). The magnetic current then is the exterior derivative of the non-exact piece, i.e.~$J^{\tn m}:=\dd\om$. Thus, by construction, and unlike the electric current, the magnetic current is exact and thus {\it identically conserved}: $\dd J^{\tn m}=0$. For this reason, it is often called a `topological current'.\label{defom}}
\bea\label{Mx12}
\mbox{Maxwell}_1:~~~~~\dd F&=&J^{\tn m}\nn
\mbox{Maxwell}_2:~~\dd*F&=&*J^{\tn e}\,,
\eea
where the magnetic current, $J^{\tn m}$, is a three-form, and the electric current, $J^{\tn e}$, is a one-form (as one can read off from the left-hand side of the equations, which are both three-forms).\footnote{There are several ways to see that we must have $J^{\tn m}$ in the first equation, but the {\it Hodge dual} of $J^{\tn e}$ in the second. This follows from writing out the components of these equations and comparing with the sources in the usual form of the Maxwell equations with the electric and magnetic fields, Eq.~\eq{elmagn}. We will also see, in Eq.~\eq{dualM} below, that this form gives a natural duality map between the electric and magnetic currents.} 

A magnetic monopole sitting at the origin gives the following non-zero components for the magnetic current: $J^{\tn m}_{123}=g\,\d^{(3)}(x)$. The magnetic charge is recovered by integration over space: $g=\int_{V_3}J^{\tn m}=\int_{S^2}F$. 

While the electric current is conserved by {\it Noether's theorem}, and requires the use of the (second) Maxwell equation, the conservation of the magnetic current is an {\it identity}, and does not require the use of the Maxwell equations (this is because the magnetic current is an exact three-form: see footnote \ref{defom}).\\
\\
{\bf Proposition: Duality of the Maxwell theory.} The Maxwell equations with both electric and magnetic sources, Eq.~\eq{Mx12}, are invariant under the following duality map:
\bea\label{dualM}
F&\mapsto&\ti F:=*F\,,\nn
J^{\tn m}&\mapsto&\ti J^{\tn m}:=*J^{\tn e}\,,\nn
J^{\tn e}&\mapsto&\ti J^{\tn e}:=*J^{\tn m}\,,
\eea
where by `invariant' we mean that the equations are mapped one into the other, so that the set of equations is invariant:
\bea\label{MX}
\mbox{Maxwell}_1&\mapsto&\mbox{Maxwell}_{\ti 2}\,,\nn
\mbox{Maxwell}_2&\mapsto&\mbox{Maxwell}_{\ti 1}\,.
\eea
{\it Proof.} Although this proposition is trivial to prove, it is insightful to give a proof that assumes only the first item in Eq.~\eq{dualM}, namely the Hodge duality of the field strength, and derives the maps on the electric and magnetic currents from the form preservation of the equations. In this way, we use what we ``already know'' (i.e.~that duality involves the Hodge duality of the field strength) and derive, from the form preservation, what we ``do not yet know'' (i.e.~how the currents in the dual model are related to those in the original model).

Thus we first rewrite the two Maxwell equations, Eq.~\eq{Mx12}, in terms of the {\it Hodge dual} of the field strength, i.e.~$\ti F:=*F$.\footnote{The converse of this equation, $F=-*\ti F$, gives the minus sign in the first equation in Eq.~\eq{Max3}, and it follows from the property of the Hodge dual in four-dimensional Minkowski space, $* *F=-F$.} 
Substituting, we get:
\bea\label{Max3}
\mbox{Maxwell}_1:~~-\dd*\ti F&=&J^{\tn m}\,,\nn
\mbox{Maxwell}_2:~\,~~~~~~~\dd\ti F&=&*J^{\tn e}\,.
\eea

The idea of the Hodge {\it duality} is that the original Maxwell equations, Eq.~\eq{Mx12}, and the equations written in the dual variables, should have {\it the same form}. Since the duality map maps $F\mapsto\ti F$, we should compare the left-hand side of the first Maxwell equation in Eq.~\eq{Mx12} with the left-hand side of the {\it second} equation in Eq.~\eq{Max3}, and vice versa. We see that the two sets of equations have {\it the same form} under the replacement of tilded and untilded variables, iff we define:
\bea
\ti J^{\tn m}&:=&*J^{\tn e}\nn
{}*\ti J^{\tn e}&:=&-J^{\tn m}~\Rightarrow~J^{\tn e}=*J^{\tn m}\,.
\eea
Given these unique definitions, the dual Maxwell's equations have the same form as the original equations, under the duality map, Eqs.~\eq{dualM}-\eq{MX}:
\bea\label{Maxt21}
\mbox{Maxwell}_{\ti2}:~~\dd*\ti F&=&*\ti J^{\tn e}\,,\nn
\mbox{Maxwell}_{\ti1}:~~~~~\dd\ti F&=&\ti J^{\tn m}\,.~\Box
\eea

Under the duality, the magnetic monopole charge gets mapped to (minus) the electric charge of the dual. By definition, the electric charge of the dual theory is defined by Eq.~\eq{Qcharge}, with $F$ replaced by $\ti F$:
\bea
\ti e:=\int_{S^2}*\ti F=-\int_{S^2}F=-\int_{V_3}\dd F=-\int_{V_3}J^{\tn m}=-\int_{V_3}g\,\d^{(3)}(y)=-g\,.
\eea

\subsection*{8.A.2. Hodge duality for D-branes}
\addcontentsline{toc}{subsection}{8.A.2. Hodge duality for D-brane forms}

The previous generalizes to the higher forms of D-branes. In $D$ dimensions, the Hodge dual of an $n$-form is a $(D-n)$-form. As we saw in Section \ref{dynD}, the Faraday tensor that is associated with the electric field of a D$p$-brane, is a $(p+2)$-form, i.e.~$F_{(p+2)}$, where the subindex indicates the kind of form. Thus its Hodge dual is a $(D-p-2)$-form:\footnote{Because of the dilaton coupling in the Einstein action Eq.~\eq{S10}, mapping one unit of electric charge to one unit of magnetic charge requires including a coupling to the dilaton, in addition to taking the Hodge dual. We will gloss over this point. For details, see Duff (1996:~pp.~250-252).}
\bea\label{Hdual}
\ti F_{(D-p-2)}:=*F_{(p+2)}=(*F)_{(D-p-2)}\,
\eea
The subindex always refers to the last form in a product or, if there is a bracket, to the whole expression in the bracket, so that $*F_{(p+2)}$ has subindex $p+2$ because $F$ is a $(p+2)$-form, and $(*F)_{(D-p-2)}$ has subindex $D-p-2$ because the Hodge dual maps $(p+2)$-forms to $(D-p-2)$-forms.

With this notation, the two Maxwell equations with sources take the following form:
\bea\label{Mx12D}
\mbox{Maxwell}_1:~~~~~~~~~~\dd F_{(p+2)}&=&J^{\tn m}_{(p+3)}\nn
\mbox{Maxwell}_2:~~\dd(*F)_{(D-p-2)}&=&(*J^{\tn e})_{(D-p-1)}\,,
\eea

Recall that D-branes are here the {\it sources} of electric and magnetic charge. Thus we ask not just what is the electric-magnetic duality that preserves these equations, but also what is the dimension of the {\it dual D-brane} that is the source of the dual charge. We will prove the following proposition:\\
\\
{\bf Proposition: Electric-magnetic duality for D$p$-branes.} The electric-magnetic dual of a D$p$-brane that is a source of electric and-or magnetic currents in the Maxwell equations, Eq.~\eq{Mx12D}, is a D$\ti p$-brane that is a source of electric and-or magnetic currents in the Maxwell equations of the same form, with the following duality map:
\bea\label{dualitybranes}
F_{(p+2)}&\mapsto&\ti F_{(\ti p+2)}:=*F_{(p+2)}\nn
J^{\tn e}_{(p+1)}&\mapsto&\ti J^{\tn e}_{(\ti p+1)}:=*J^{\tn m}_{(p+3)}\,,\nn
J^{\tn m}_{(p+3)}&\mapsto&\ti J^{\tn m}_{(\ti p+3)}:=*J^{\tn e}_{(p+1)}\,,\nn
p&\mapsto&\ti p:=D-p-4\nn
\mbox{Maxwell}_1&\mapsto&\mbox{Maxwell}_{\ti 2}\,,\nn
\mbox{Maxwell}_2&\mapsto&\mbox{Maxwell}_{\ti 1}\,.
\eea
{\it Proof.} As before, the proof of this proposition is straightforward, but we will here take a ``constructive'' approach that only assumes Hodge duality and the relevant facts about how sources in the Maxwell equations represent D$p$-branes, and derives the remaining items on the list of duality transformation rules, Eq.~\eq{dualitybranes}. In this way, we will be able to understand why the dual $D$-brane has the dimension $\ti p=D-p-4$, and why the sources must transform in the way indicated in Eq.~\eq{dualitybranes}.

First, the physical idea of the proposition is that, if the original D$p$-brane is a source for the field strength $F_{(p+2)}$ (in the sense that its electric flux integrates to an electric charge as in Eq.~\eq{Qcharge}), then the dual D$\ti p$-brane must be a source for a dual field strength, $\ti F_{(\ti p+2)}$. 

Second, we assume the first item in the duality transformation rules, i.e.~we assume that the dual field strength is the Hodge dual of the original field strength, i.e.~it is a $(D-p-2)$-form: $\ti F_{(D-p-2)}=*F_{(p+2)}$. Combining this with the first step, we get the value: $\ti p+2=D-p-2\Rightarrow \ti p=D-p-4$ (listed in the fourth duality rule, Eq.~\eq{dualitybranes}).\footnote{Another way to see this is by considering the Wess-Zumino coupling (cf.~the discussion after Eq.~\eq{fullWS}) of the gauge field to the D$\ti p$-brane world-volume. The gauge field corresponding to the dual form $\ti F_{(D-p-2)}$ is a $(D-p-3)$-form. This gauge field couples, through the Wess-Zumino term, to the pull-back of the world-volume form to the spacetime, which is a $(\ti p+1)$-form (namely, $\ti p$ spatial dimensions plus time). Thse two forms are contracted with each other, and so they must be of the same rank, which gives: $D-p-3=\ti p+1\Rightarrow \,\ti p=D-p-4$.}

Third, like in the point-particle case, we rewrite the left-hand side of the two Maxwell equations in terms of the dual field strength, i.e.~$F_{(p+2)}=-*\ti F_{(\ti p+2)}$ in the first equation, and $(*F)_{(D-p-2)}=*F_{(p+2)}=\ti F_{(\ti p+2)}$ in the second (and we also multiply the first equation with an overall minus sign):
\bea\label{tiMx2}
\mbox{Maxwell}_1:~~\dd*\ti F_{(\ti p+2)}&=&-J^{\tn m}_{(p+3)}\,\nn
\mbox{Maxwell}_2:~~~~\dd F_{(\ti p+2)}&=&(*J^{\tn e})_{(D-p-1)}\,.
\eea
Requiring that these two equations have the same form as the original Maxwell equations, Eq.~\eq{Mx12D}, where the two equations are interchanged, implies that the dual currents on the right-hand side must be defined, in terms of the original ones, as follows:
\bea
(*\ti J^{\tn e})_{(D-\ti p-1)}:=-J^{\tn m}_{(p+3)}~~\,\,~~~~~\Rightarrow~~\ti J^{\tn e}_{(\ti p+1)}&:=&*J^{\tn m}_{(p+3)}\nn
*\ti J^{\tn m}_{(\ti p+3)}:=(*J^{\tn e})_{(D-p-1)}~\Rightarrow~~\ti J^{\tn m}_{(\ti p+3)}&:=&*J^{\tn e}_{(p+1)}\,.
\eea
This guarantees that, under the duality map Eq.~\eq{dualitybranes}, the Maxwell equations have the {\it same form} as the original Maxwell equations, Eq.~\eq{Mx12D}, where the two equations are interchanged:
\bea\label{tiMx21}
\mbox{Maxwell}_{\ti2}:~~\dd(*\ti F)_{(D-\ti p-2)}&=&(*\ti J^{\tn e})_{(D-\ti p-1)}\nn
\mbox{Maxwell}_{\ti1}:~~~~~~~~~~\dd F_{(\ti p+2)}&=&\ti J^{\tn m}_{(\ti p+3)}\,,
\eea
as is obvious by comparing with the Maxwell equations, Eq.~\eq{Mx12D}. $\Box$

As in the previous Section, we can calculate the electric and magnetic charge of a static, electrically or, respectively, magnetically charged D$p$-brane at the origin:
\bea\label{sourcesD}
(*J^{\tn e})_{12\cdots D-p-1}&=&e_{p+1}\,\d^{(D-p-1)}(x)\,,\nn
J^{\tn m}_{12\cdots p+3}&=&g_{D-p-1}\,\d^{(p+3)}(x)\,.
\eea
A D$p$-brane with the first kind of (electric) charge is extended along $p+1$ dimensions, and has $D-p-1$ transverse directions that define its location, by the set of $D-p-1$ equations $y=0$. The second kind of D-brane is extended along $D-p-1$ longitudinal dimensions, and has $p+3$ transverse directions that define its location, by the set of $p+3$ equations $z=0$. Their charges are indeed given by:
\bea
Q_{p+1}&\overset{\sm{Eq.~}\eq{Qcharge}}{=}&\int_{S^{D-p-2}}(*F)_{(D-p-2)}\overset{\sm{Gauss' th.}}{=}\int_{V_{D-p-1}}\dd(*F)_{(D-p-2)}\\
&\overset{\sm{Eq.~}\eq{Mx12D}}{=}&\int_{V_{D-p-1}}(*J^{\tn e})_{D-p-1}\overset{\sm{Eq.~}\eq{sourcesD}}{=}\int_{V_{D-p-1}}e_{p+1}\,\d^{(D-p-1)}(x)=e_{p+1}\,.\nonumber
\eea
In the dual theory, the electric charge (which is magnetic in the original theory) is defined in the same way, but in terms of the dual variables:
\bea
\ti Q_{\ti p+1}&\overset{\sm{Eq.~}\eq{Qcharge}}{=}&\int_{S^{D-\ti p-2}}(*\ti F)_{(D-\ti p-2)}\overset{\ti p=D-p-4\,\&\,*\ti F=-F}{=}-\int_{S^{p+2}}F_{(p+2)}\nn
&\overset{\sm{Gauss' th.}}{=}&-\int_{V_{p+3}}\dd F_{(p+2)}\overset{\sm{Eq.~}\eq{Mx12D}}{=}-\int_{V_{p+3}}J^{\tn m}_{(p+3)}\nn
&\overset{\sm{Eq.~}\eq{sourcesD}}{=}&-\int_{V_{p+3}}g_{D-p-1}\,\d^{(p+3)}(x)=-g_{D-p-1}\,,
\eea
thus generalizing the point-particle case. The two quantities are related by the Dirac quantization condition (see Duff et al.~(1995:~p.~220)):
\bea
e_{p+1}\,g_{D-p-1}=2\pi n\,.
\eea
{\it String-fivebrane duality}. Section \ref{elemsolit} showed (and this holds for general $p$-branes) that elementary strings, i.e.~$p=1$ branes, are singular solutions of the equations of motion of the supergravity action coupled to the action of a $p=1$ string {\it source}, while magnetically charged solitons are non-singular solutions of the supergravity equations that arise as $\ti p=5$ fivebrane {\it solitons}. 

As we have discussed in this Section, this classical duality is Hodge duality between the $(p+2)$-form field strength of an electric $p$-brane and the $(\ti p+2)$-form field strength of a magnetic $\ti p$-brane, where $\ti p=D-p-4$.

The exchange of elementary and solitonic solutions combines with Hodge duality in a more complete picture of string/fivebrane duality (and again, the picture generalizes to other $p$-branes).\footnote{See Duff and Lu (1993:~pp.~313-316).}
We reinterpret a solitonic fivebrane as an elementary, electric, solution of the {\it Hodge dual} theory (based on the dual seven-form $\ti F_{(7)}$). This is done by replacing the action, Eq.~\eq{S10}, with its dual, and by adding to it a fivebrane source---thus getting an elementary solution. Likewise, the elementary string can be reinterpreted as a magnetic (solitonic) solution of the dual theory, with no source. This works out as expected, i.e.~the exchange of actions, and of their corresponding equations of motion, leads to a reinterpretation of the solutions: fivebranes are elementary electric solutions of the dual supergravity-fivebrane action, and strings are solitonic, magnetic, solutions of the dual supergravity action, with no source, and so electric-magnetic duality indeed {\it exchanges the roles of elementary vs.~solitonic solutions}. This is of course a theme that we have encountered in several Chapters in this book.\footnote{The details of these new solutions are in Duff et al.~(1995:~pp.~235-240). See also Duff and Lu (1993:~pp.~313-316).}\\
\\
{\it Some more examples.} For illustration, we discuss a few more examples in type IIA and type IIB supergravity in ten dimensions $D=10$ (in Chapter \ref{STII} we discuss $D=11$).\footnote{Many of the early investigations of duality in string theory focussed on lower dimensional cases, of string theories compactified to e.g.~$D=6$. This is not the approach we will take here: he look at the uncompactified theories.}
Recall from Section \ref{dynD} that, in type IIA supergravity, D$p$-branes in the RR sector have values $p=0,2,4,6,8$. 

$p=0$ is the point particle, and in this case the field strength $F_{(p+2)}$ is a two-form: namely, the Faraday tensor. Under the duality map, Eq.~\eq{dualitybranes}, for $D=10$, this maps to $\ti p=10-4=6$ i.e.~a D$6$-brane, with eight-form field strength, $\ti F_{(8)}$. $p=2$ is the membrane, with four-form field strength, $F_{(4)}$. This gets mapped to a D4-brane, with six-form field strength, $\ti F_{(6)}$.

In type IIB, $p$ can take the values $p=1,3,5,7,9$. $p=3$ is a D$3$-brane with field strength $F_{(5)}$, which is self-dual in ten dimensions, i.e.~it maps to another D$3$-brane (just like a point particle in $D=4$ maps under duality to another point particle). In fact, the five-form is self-dual in this case: $F_{(5)}=\ti F_{(5)}$. This will be important for the discussion of S-duality in the next Chapter.

\chapter{M-theory and Gauge-Gravity Duality}\label{STII}
\markboth{\small{\textup{M-theory and Gauge-Gravity Duality}}}{\textup{\small{M-theory and Gauge-Gravity Duality}}}

This Chapter discusses three central examples of non-perturbative dualities in string theory that relate models in different numbers of spacetime dimensions. The dualities discussed here have been part of the main reason for the interest in dualities in string theory over the past few decades. These dualities also feed into the philosophical views that we will develop in Part III, especially the idea of theoretical equivalence (in Chapters \ref{Theor} and \ref{physeq}) and the ideas of successor theories and of emergence and fundamentality (Chapter \ref{Heuri}). 

Section \ref{11DM} discusses the effective duality between Type IIA string theory in 10D and M-theory in 11D. Section \ref{S-d} discusses the S-duality of Type IIB string theory in 9D, which can be understood as a symmetry of the torus in 11D. Section \ref{ggd} discusses gauge-gravity duality, which is a duality between a theory of gravity in $D$ dimensions and a quantum field theory without gravity in $D-1$ dimensions. 

\section{Eleven dimensions and M-theory}\label{11DM}

This Section expounds the eleven-dimensional origin of ten-dimensional Type IIA string theory, which was conjectured by Witten (1995a:~pp.~92-93; 1996:~p.~383) to explain the existence of various other string dualities (see Section \ref{Dcst}). Witten's key realization was that, at strong coupling, the coupling constant of Type IIA string theory can be reinterpreted as the {\it radius of a hidden dimension}: an eleventh (or tenth spatial) dimension that opens up at strong coupling. The radius is (in dimensionless units: we will introduce length scales below):
\bea\label{R10}
R_{11}=e^{2\f_0/3}=g_{\tn{s}}^{2/3}\,.
\eea
From the dimensional reduction, one can calculate the masses of particles that arise upon compactification on this circle (these particles are called `Kaluza-Klein': see Section \ref{IIAsugra}). Their mass saturates the BPS bound, Eq.~\eq{Dbmass}, and the $1/g_{\sm s}$-dependence, which indicates their non-perturbative nature, follows from the eleven-dimensional derivation, as we will show below. Thus eleven-dimensional supergravity gives a derivation and a natural reinterpretation of the existence of the RR states appearing in the spectrum at strong coupling.

Thus Type IIA string theory, including the non-perturbative information about the masses of RR states, is reinterpreted as the low-energy limit of eleven-dimensional supergravity, compactified on $\mathbb{R}^{10}\times S^1$. This is an effective duality between Type IIA string theory and eleven-dimensional supergravity.

In the next Section, we will give two pieces of evidence for this effective duality:\footnote{Witten's (1995a:~Sections 3-4) original paper contains much more detailed evidence, which involves compactifying string theory to dimensions $D<10$, thus making contact with (and explaining) a series of previously known dualities, known as U-duality.} 
the first is the matching of the low-energy effective actions of the two models, i.e.~eleven-dimensional supergravity with ten-dimensional type IIA supergravity, which is the low-energy limit of Type IIA string theory. The second is the derivation of the world-sheet action of the superstring from the world-volume action of a supermembrane in eleven dimensions. 

\subsection{10D supergravity from 11D supergravity}\label{IIAsugra}

We here discuss three aspects of the matching between 11D and 10D type IIA supergravity: the dimensional (i.e.~Kaluza-Klein) reduction of the action for the massless modes, the relations between the fundamental constants, and the correspondence between the massive modes.

\subsubsection{Kaluza-Klein reduction of massless modes}

Our basic task is to derive the ten-dimensional supergravity action, Eq.~\eq{S10} (for $D=10$), from eleven dimensions, including not just the massless fields from the NS-NS sector, but also the RR-sector. To this end, we use the string-frame form of the action, Eq.~\eq{betaf}. The full action is then the {\it sum} of the following Neveau-Schwarz and Ramond actions:
\bea\label{10Daction}
S_{\tn{NS}}&=&{1\over2\k^2}\int\dd^{10}x\,\sqrt{-g}\,e^{-2\F}\left(R+4(\na\F)^2-\half H_{(3)}^2\right),\\
S_{\tn{R}}&=&-{1\over4\k^2}\int\dd^{10}x\,\sqrt{-g}\left(F_{(2)}^2+F'{}^2_{\!\!\!(4)}\right)\nonumber
-{1\over4\k^2}\int B_{(2)}\wedge F_{(4)}\wedge F_{(4)}\,,
\eea
where the massless bosonic fields of the NR-NS sector are the metric tensor, the antisymmetric tensor $B_{(2)}$, and the dilaton. Here, $2\k^2=(2\pi)^7\ell_{\sm s}^8$ is the ten-dimensional Newton's constant. In the RR sector of type IIA, there are a one-form $A_{(1)}$ and a three-form $A_{(3)}$, with corresponding field strengths $H_{(3)}=\dd B_{(2)}$, $F_{(2)}=\dd A_{(1)}$, $F_{(4)}=\dd A_{(3)}$ (see the discussion in Section \ref{Dbsugra}). In addition, we have defined a field strength that couples the two fields: $F'_{(4)}=\dd A_{(3)}+A_{(1)}\wedge H_{(3)}$.\footnote{Here, and below, the Chern-Simons term contains no explicit volume element, because this is included in the wedge products of the forms: the integrand is an 10-form.}

The NS-NS and RR sectors have a different dependence on the dilaton field: the terms in the NS-NS sector couple to the dilaton by an overall power of $e^{-2\F}$, while the terms in the RR sector, in the string frame here used, do not couple to the dilaton.\footnote{One can redefine the RR fields to get the same dilaton coupling as for the NS-NS fields, but the action then looks more complicated; the present form is easier.}

Unlike the Type II string theories, which contain several higher-rank gauge fields, eleven-dimensional supergravity has a single gauge field: namely, a three-form potential $A^{11}_{(3)}$. This is required by the matching of bosonic and fermionic degrees of freedom in eleven dimensions.\footnote{In $D$ dimensions, the gravitational field has $D(D-3)/2$ on-shell degrees of freedom: namely, the components of a transverse, symmetric, traceless tensor. A three-form gauge field has $(D-2)(D-3)(D-4)/3!$ transverse degrees of freedom. For $D=11$, this gives $44$ degrees of freedom for the graviton, and 84 for the three-tensor, i.e.~a total of 128 propagating (on-shell) degrees of freedom. As for the fermionic degrees of freedom: these are the polarization states of the gravitino in eleven dimensions, which is a 32-component Majorana spinor. Such a spinor has 128 components, and so the number of bosonic and fermionic degrees of freedom match. For a discussion, see Becker et al.~(2007:~pp.~303-304).} Because the requirement of supersymmetry fixes the maximum number of degrees of freedom in eleven dimensions, there can be no dilaton field in a supersymmetric theory in eleven dimensions.\footnote{This also implies that, in eleven dimensions, there is no obvious coupling constant that can be used for perturbation theory: see Townsend (1995:~p.~185) and Dasgupta et al.~(2002:~Section 2).} The bosonic part of this unique action is:\footnote{First written down by Cremmer et al.~(1978:~p.~410).}
\bea\label{11sugra}
S_{\tn{11D}}={1\over2\k_{11}^2}\int\dd^{11}x\,\sqrt{-G_{11}}\left(R-\half|F^{11}_{(4)}|^2\right)-{1\over6}\int A^{11}_{(3)}\wedge F^{11}_{(4)}\wedge F_{(4)}^{11}\,,
\eea
where, as usual, $F^{11}_{(4)}=\dd A_{(3)}^{11}$. 

We will now dimensionally reduce this action to ten dimensions. Without loss of generality, we write the eleven-dimensional metric in the following {\bf Kaluza-Klein form}:
\bea\label{KKmetric}
\dd s^2_{11}=e^{-2\F/3}g_{\m\n}\,\dd x^\m\dd x^\n+e^{4\F/3}(\dd x^{11}+A_\m\dd x^\m)^2\,,
\eea
in terms of the effective ten-dimensional metric $g_{\m\n}$, the ten-dimensional vector $A_{(1)}$, and the scalar field $\F$, which will be identified with the dilaton field. Note that the numerical factors in the exponentials are chosen such that the fields can be compared with the ten-dimensional action, Eq.~\eq{10Daction}.\footnote{To see that this is possible, first write the eleven-dimensional metric in the general form $\dd s^2=G_{\m\n}^{10}\,\dd x^\m\dd x^\n+e^{2\g}(\dd x^{11}-A_{(1)\m}\dd x^\m)^2$, where $\g$ is any scalar field. Then do a Weyl rescaling of the ten-dimensional metric $G_{10\m\n}:=e^{-\g}g_{\m\n}$, so that the kinetic terms of the Ramond-Ramond fields do not depend on $\g$ (i.e.~as in Eq.~\eq{10Daction}, where the Ramond-Ramond fields are independent of the dilaton). The other terms in the action then still have the following dependence on $\g$: $e^{-3\g}(R+|\na\g|^2+|\dd B_{(2)}|^2)$. To match this with the NS-NS part of the ten-dimensional action, Eq.~\eq{10Daction}, we redefine $\g:=2/3\F$. Filling this value into the general form of the metric above, we recover Eq.~\eq{KKmetric}.\label{KKred}}

Likewise, the eleven-dimensional three-form gauge field, $A^{11}_{(3)}$, decomposes into two kinds of components: (1) Components whose indices are all along the ten dimensions $\m=0,\ldots,9$, which gives, from the ten-dimensional perspective, a three-form with indices $A_{\mu\nu\l}:=A^{11}_{\m\n\l}$, with field-strength $F_{(4)}$. (2) Components with one index along the eleventh direction, i.e.~$A^{11}_{\m\n10}$. From the ten-dimensional perspective, these are the components of a ten-dimensional two-form, which we will identify with the ten-dimensional $B_{(2)}$, whose field-strength is $H_{(3)}$. Because $A^{11}_{(3)}$ is antisymmetric, there are no non-zero components with more than one index along the 11th direction. In other words, the decomposition can be written as: $A^{11}_{(3)}=(A_{(3)},B_{(2)})$.

Under the above two decompositions, Kaluza-Klein reduction takes (in the leading approximation) all the fields to be independent of the 11th coordinate. This corresponds to neglecting the massive modes, and this is a good approximation when the radius of the eleventh dimension, $R_{11}$, is small (these modes will be considered below: see Eq.~\eq{D0mass}). The eleven-dimensional supergravity action is then readily seen to dimensionally reduce to the ten-dimensional type IIA action,\footnote{For some of the details, see footnote \ref{KKred}, and Becker et al.~(2007:~pp.~309-310).} Eq.~\eq{10Daction}. 

\subsubsection{Relation between fundamental constants}

To complete the dimensional reduction, we need to match the fundamental constants. Thus we first introduce the following widely-used convention for {\bf Newton's constant}, $G_D^{\tn N}$, in terms of the {\it Planck length}, $\ell_D$, in $D$ dimensions:
\bea\label{GNewton}
16\pi\,G_D^{\tn N}={1\over2\pi}\,(2\pi\ell_D)^{D-2}=:2\k_D^2\,.
\eea

Second, we relate the ten-dimensional Newton constant to the {\it gravitational constant}, $\k$, in the type IIA supergravity action, Eq.~\eq{10Daction}. This constant is Newton's gravitational constant written {\it in the string frame}, i.e.~$\k$ is not defined with respect to the canonical Einstein metric (the one in the Einstein-Hilbert action), but with respect to the string frame (which contains the dilaton field): it is the effective Newton's constant as seen on the worldsheet of the string. 

But just like for the Einstein-Hilbert action, the Newton gravitational constant $\k$ is given in terms of a length (see Eq.~\eq{GNewton}), namely the string length:
\bea\label{GNstring}
2\k^2={1\over2\pi}(2\pi\ell_{\sm s})^8\,.
\eea
The two are related through the string coupling:\footnote{This is because the string-frame Newton gravitatial constant comes with a factor of $e^{-2\F}$ in the integrand of the action. The asymptotic value of the dilaton, i.e.~the value of the dilaton in an otherwise empty region of spacetime, is the expectation value of the microscopic dilaton field, which is the string coupling constant.}
\bea
\k_{10}=g_{\sm s}\k\,.
\eea
This gives us the relation between the ten-dimensional Planck length and the string length: $\ell_{10}=g_{\tn s}^{1/4}\,\ell_{\tn s}$. 

Third, we relate the eleven- and ten-dimensional {\it fundamental lengths}. First, from the Kaluza-Klein metric we find that the eleven-dimensional Planck length and the string length are related as follows:
\bea\label{ell11}
\ell_{11}=g_{\sm s}^{1/3}\ell_{\sm s}\,.
\eea
This follows from the requirement that the lengths, regardless of whether they are measured in eleven-dimensional and in ten-dimensional units, have the {\it same invariant length}.\footnote{Considering only the first term in Eq.~\eq{KKmetric} (i.e.~setting the variations in the 11th dimension to zero, and setting the gauge field to zero), the invariant length along some path is: $\ell=\int\dd\s\sqrt{G_{\m\n}\dot X^\m\dot X^\n}=\int\dd\s\sqrt{g_{\m\n}\dot x^\m\dot x^\n}$, where $X$ are eleven-dimensional distances and $x$ are ten-dimensional distances measured in the string-fame metric $g_{\m\n}$. The two are related by: $\dd X^\m=\ell_{11}/\ell_{\tn s}\,\dd x^\m$. Substituting this into the length, and using $G_{\m\n}=g_{\tn s}^{-2/3}g_{\m\n}$ (in an empty region of spacetime where the dilaton is constant) gives Eq.~\eq{ell11}. Alternatively, we can consider the second term in the Kaluza-Klein metric, Eq.~\eq{KKmetric}, i.e.~$e^{4\F/3}\dd X_{11}^2$. In eleven-dimensional units, this expression gives the square of the perimeter of the circle, i.e.~$(2\pi R_{11})^2$, which measured in terms of the eleven-dimensional Planck length is $g_{\tn s}^{4/3}\ell_{11}^2$. This gives $2\pi R_{11}=g_{\tn s}\ell_{\tn s}$, which is equivalent to Eq.~\eq{ell11} combined with Eq.~\eq{k10}.\label{11210}}

Fourth, the {\it eleven-dimensional and ten-dimensional Newton gravitational constants} are related as follows:
\bea\label{k10}
{1\over\k_{10}^2}={2\pi R_{11}\over\k_{11}^2}~\Rightarrow~G_{11}=2\pi R_{11}\,G_{10}\,.
\eea
This follows by integration of the Einstein-Hilbert action over the 11th dimension, in the Kaluza-Klein approximation that all the fields are constant in the 11th dimension.

With these identifications, the eleven-dimensional supergravity action, Eq.~\eq{11sugra}, {\it dimensionally reduces} (upon doing the integral over the circle and using Eq.~\eq{k10}) to the Type IIA superstring action, Eq.~\eq{10Daction}. While we have here matched the bosonic actions, the fermions also match (see Becker et al.~(2007:~p.~312)).

\subsubsection{Including the massive modes}

As a first check of our claim that the eleven-dimensional perspective allows us to derive non-perturbative results about D-brane states in Type IIA string theory, let us check the formula for the D-brane mass, Eq.~\eq{Mcharge}, for BPS states of D0-branes (i.e.~point particles) in Type IIA string theory.

The RR charge in Type IIA that couples to D0-branes is the charge of the U(1) field $A_{(1)}$ (see Eq.~\eq{KKmetric}). This one-form couples to the world-line of a point particle, i.e.~a D0-brane with momentum $p_\m$. A supersymmetric massive particle in ten dimensions can be obtained from a supersymmetric massless particle in eleven dimensions, i.e.~one that satisfies: $P_MP^M=0$, where $M=0,\ldots,10$ is an the eleven-dimensional index. By expanding $P_M=(p_\m,p_{11})$, where the Greek index is ten-dimensional, $\m=0,\ldots,9$, and the eleventh coordinate is the momentum along the circle, i.e.~$P_{10}=:p_{11}$, this eleven-dimensional massless equation can be written as: $M^2=-p_\m p^\m$, where $M:=|p_{11}|$. From the ten-dimensional perspective, this is the equation for the momentum of a massive particle, and so $M$ is identified as the ten-dimensional mass. 

Since the eleventh dimension is a circle of radius $R_{11}$, the momentum $p_{11}$ is quantized in units of the radius: $p_{11}=n/R_{11}$, where $n$ in an integer (namely, the charge that couples to the RR U(1) field $A_{(1)}$). Thus we get the following mass:
\bea\label{D0mass}
M={|n|\over R_{11}}={|n|\over g_{\sm s}\ell_{\sm s}}\,,
\eea
where in the last equality we used $R_{11}=g_{\tn s}\ell_{\tn s}$.\footnote{This follows by by inserting the Planck length into the expression for the radius in terms of the string coupling, Eq.~\eq{R10}, and using the relation between the Planck length and the string length that we have just derived, i.e.~Eq.~\eq{ell11}.}
This agrees with the independent calculation for D0-branes (i.e.~Eq.~\eq{Dbmass} for $p=0$). The properties of these particles are as we expect: they are are very massive at weak coupling (i.e.~when the eleventh dimension is small), and are light at strong coupling, i.e.~when the eleventh dimension is large. This is a non-trivial test of the effective duality between eleven-dimensional supergravity and Type IIA string theory.\footnote{The ten-dimensional type IIA supergravity solutions with this mass and U(1) charge are extremal black hole solutions. See Horowitz and Strominger (1991:~pp.~199-203) (setting $D=4$ and choosing the extremal case in their formulas).} 

To summarize: we have shown that a massless particle in eleven dimensions, whose momentum along the eleventh dimension is quantized, gives a configuration that from the ten-dimensional perspective looks like a D0-brane, with the correct dependence of the D0-brane tension on the string coupling. 

An elementary D0-brane has one unit of RR charge under the field $A_{(1)}$, just like an electron has unit elementary charge. This means that, for $n>1$, the mass calculated in Eq.~\eq{D0mass} is the mass of a {\it bound state} of $n$ D0-branes. Note that, even though these D0-branes all have unit RR charge, they do not repel each other, and their binding energy is zero, since all their mass is given by the sum of their individual masses. This is yet another manifestation of the fact that this is a BPS-state, where the attractive and repulsive forces cancel each other out. And since this formula follows from the supersymmetry algebra, it is exact for all values of the string coupling, i.e.~it does not get any corrections.

\subsection{What is fundamental: superstrings or supermembranes?}\label{sssm}

Given the evidence for M-theory just discussed, a pressing question is whether supermembranes, or perhaps some other objects, replace strings as fundamental degrees of freedom.\\
\\
{\bf Hopes for a fundamental membrane theory in eleven dimensions.} As we discussed in the previous Section, eleven-dimensional supergravity has a single gauge field: namely, the 
three-form $A^{11}_{(3)}$. The charged object to which this field couples, through a Wess-Zumino term (see the discussion after Eq.~\eq{fullWS}), is {\it three-dimensional}: it is a 2-brane, or {\it supermembrane}. 

When supermembrane theory was being developed in the 1980s, it was hoped that, by analogy with string theory, which reproduces general relativity in the low-energy limit (as ten-dimensional supergravity), the supermembrane could give a fundamental starting point of a theory of quantum gravity in eleven dimensions: perhaps even a more fundamental theory than string theory, because eleven is the highest dimension in which conventional supersymmetric theories can be formulated (and, as we will see below, strings can be obtained as rolled-up membranes).\footnote{For expressions of this hope, see, for example, Duff et al.~(1987:~p.~73) and de Wit et al~(1988:~p.~545). For why eleven is the highest dimension that admits conventional supersymmetric theories, see Nahm (1978:p.~162) and Cremmer et al.~(1978:~p.~409).} 
This conjecture seemed to enjoy support from both the uniqueness of supermembranes and supergravity in eleven dimensions, and from the discovery that the equations of motion of eleven-dimensional supergravity follow as a requirement of the local fermionic symmetry (called $\k$-symmetry) of the world-volume action of the membrane.\footnote{See Bergshoeff et al.~(1987:~pp.~76-77; 1988:~pp.~335-338). In Hughes et al.~(1986:~p.~373), $\k$-symmetry is introduced as a consequence of the requirement of local supersymmetry of the world-volume action of the supermembrane. Supermembranes can only be formulated in $D=4, 5, 7$ and $11$ (see de Wit et al.~(1988:~p.~548).}

However, for many physicists these hopes were dashed when, in the late 1980s, de Wit et al.~(1989:~p.~141) showed that supermembranes have a continuous spectrum and no mass gap between the zero-mass states and the rest of the spectrum. This is seen as a manifestation of an {\it instability} of the supermembrane, which is prone to form ``spikes'' of arbitrary length but zero area, without increasing its mass.\footnote{The physical reason for this instability is that the potential has valleys, through which some of the membrane coordinates can escape to infinity (without increasing the mass), thus acquiring very large values, and giving string-like configurations of arbitrarily long length.} To put it differently, an obvious {\it sufficient} condition (although not a necessary condition) for the eleven-dimensional supergravity to be the low-energy limit of the supermembrane does {\it not} appear to be satisfied: namely, there is no supersymmetric ground state with a mass (or energy) gap, relative to the first excited state. Because there is a {\it continuum} of states at the bottom, rather than a mass gap, it takes an arbitrarily small amount of energy to excite extra states, and so there does not appear to be a low-energy theory with a fixed set of fields for which one could write the equations of motion. Thus membranes are not good as fundamental objects, basically because they decay into strings of arbitrarily high length, and so it is not clear that their massless spectrum is anything useful, like a supergravity theory.

Yet not all was lost: for, in a highly-cited paper, Banks et al.~(1997) proposed a rebranding of membrane theory that could serve as a fundamental theory. The telling title was `M-theory as a Matrix Model: A Conjecture'. The idea was that membranes are made of a web of D0-branes whose positions in spacetime are not c-numbers, but non-commuting matrices. These matrices can be obtained by discretizing the world-volume of the membranes (which cures the problem of the continuous spectrum), and they are candidates for the fundamental M-theory variables to be quantized. Thus M-theory was conjectured to be a theory of matrices that are the quantum analogues of a set of D0-branes that make up a membrane. While we will not go further into these developments here, the matrix proposal is still a live proposal.\\
\\
{\bf Elementary membranes, dimensionally reduced to strings.} Back to the relation between eleven-dimensional supermembranes and strings in 10D: our job is to show that a supermembrane, compactified on a circle, is a fundamental superstring---so that the supermembrane could replace the string as a fundamental object.\footnote{If eleven dimensions are taken to be fundamental, then one should be able to reproduce not just the superstring, but also the other D-branes in lower dimensions. This can indeed be done: the D-branes of Type IIA string theory are obtained by dimensional reduction on a circle, and the D-branes of Type IIB string theory are obtained by T-duality (and by further compactifications one obtains D-branes in dimensions lower than ten). Very briefly: in eleven dimensions there is, in addition to the membrane, also its magnetic dual, namely a fivebrane. By double dimensional reduction of these two solutions, one obtains the fundamental Type IIA string (as discussed in the rest of this Section) and a D4-brane, respectively. On the other hand, by a single dimensional reduction, i.e.~by a dimensional reduction along a direction {\it transverse} to the membrane or the fivebrane, we get a D2-brane or a D5-brane, respectively. Finally, D0-branes are obtained as Kaluza-Klein momentum modes, as in Eq.~\eq{D0mass}. For details, see Polchinski (1998b:~pp.~201-203), Townsend (1995), and Becker et al.~(2007:~pp.~325-326). The D$p$-brane solutions of type IIA and type IIB supergravity with RR charge for various values of $p$ are in Townsend (1995:~pp.~186-187) and Horowitz and Strominger (1991:~pp.~203-206). For the eleven-dimensional origin of various solutions in Type IIB string theory, see Russo and Tseytlin (1997:~pp.~135-143).} 
It is a plausible conjecture that, at least classically, this must be the case, if supermembranes are to have a fighting chance as fundamental objects in eleven dimensions (whether by themselves, as the proper objects to be quantized, or as the classical starting point of a more fundamental theory of D0-branes or of something else). 

There are two main ways to dimensionally reduce a supermembrane to a string: one is to reduce to the Type IIA string in 10D, which is the route we will take here. (We will follow the alternative route, of reducing to nine dimensions and, by T-duality, getting the Type IIB string, in Section \ref{S-d}.) Thus we require {\it double-dimensional reduction}, i.e.~we dimensionally reduce both the target space and the world-volume of the membrane, both of which have a circle. Thus the supermembrane is rolled around the 11th dimension to form a spatial tube, and one of the spatial dimensions of the membrane is a circle of the same radius as the 11th dimension. At distances large compared to the eleventh dimension, the supermembrane in effect looks like a string.\footnote{A detailed treatment is in Duff et al.~(1987:~pp.~205-208).} 
In doing this type of dimensional reduction, the main caveat is the instabilities that we mentioned above. 

But, despite the instability, there are two reassuring facts and a surprise. First, it is reassuring that the {\it same} Kaluza-Klein Ansatz for the decomposition of the metric, Eq.~\eq{KKmetric}, that we used to dimensionally reduce the supergravity action to the type IIA supergravity action, also dimensionally reduces the supermembrane action to the superstring action. Second, the equations of motion of the supermembrane indeed dimensionally reduce to those of the superstring, complete with all the fermionic terms that are required for type IIA supersymmetry.\footnote{The dimensional reduction works clearest at the level of the equations of motion. If one dimensionally reduces the world-volume action of the supermembrane directly, one finds a ten-dimensional action that is not quite the Polyakov form of the string action, Eq.~\eq{WSa}, but the Nambu-Goto form, which is equivalent to it. See Duff et al.~(1987:~p.~72) and footnote \ref{NambuGoto}.} 
This is satisfying, and indeed not a priori guaranteed. The dimensional reduction gives the following relation between the string tension, $T_2=1/2\pi\a'$, and the membrane tension, $T_3$:
\bea
T_2=2\pi R_{11}\,T_3\,,~\mbox{i.e.}~~~~~2\pi\a'={1\over2\pi R_{11}\,T_3}\,.
\eea
In the approximation in which we only keep the massless modes and throw away the massive modes, Eq.~\eq{D0mass}, $R_{11}$ is small while the string tension, $T_2$, is finite. This requires that we take the tension of the supermembrane, $T_3$, to be large. 

The surprise is that the three-form itself {\it decouples} from the bosonic part of the action. This is the unique three-form, $A_{(3)}^{11}$ (cf.~Eq.~\eq{11sugra}), of eleven-dimensional supergravity, with a natural Wess-Zumino coupling to the membrane (that is how we motivated the existence of the membrane to begin with: namely, as the $p$-brane that is charged with respect to this unique $(p+1)$-form). And recall, from Section \ref{IIAsugra}, that the three-form decomposes, under dimensional reduction, into a three-form and a two-form, i.e.~$A_{(3)}^{11}=(A_{(3)},B_{(2)})$. The fact we called `surprising' is that, upon Kaluza-Klein reduction (i.e.~keeping only the massless modes), this ten-dimensional three-form does not appear in the bosonic part of the action. 

But, from the superstring perspective, if the dimensional reduction is to work, this decoupling is {\it required}: since the superstring does not have a three-form in its NS-NS sector Eq.~\eq{fullWS}, no three-form should appear. (It does have it in the RR sector, where it couples to the fermions.)

Likewise for the ten-dimensional one-form and the dilaton fields, $A_{(1)}$ and $\F$, from the Kaluza-Klein Ansatz Eq.~\eq{KKmetric}: these also disappear from the bosonic sector of the string, although they survive in the fermionic sector. This is also to be expected, but for different reasons: for, while $A_{(1)}$ is a RR field that cannot appear in the NS-NS sector of the superstring, the dilaton field is one of the three fields in the NS-NS sector. That it disappears from the action appears to be a consequence of the quantum nature of this term vs.~the low-energy required for Kaluza-Klein reduction: notice that this term appears with the coefficient $\a'$ in the world-sheet action, Eq.~\eq{fullWS}, i.e.~the square of the string length. And, from Eq.~\eq{GNstring}, the string length sets the length of the gravitational interactions. This seems to imply that the present approximation is only valid at energies much smaller than the string mass, i.e.~$E\ll1/\ell_{\sm s}$. This is {\it in addition} to the condition for the validity of the Kaluza-Klein approximation, which requires that massive modes can be discarded, and so $E\ll1/R_{11}$. All in all, the low-energy condition required to get the superstring from the supermembrane is $E\ll1/\ell_{\tn s}\ll1/R_{11}$. This further restriction of the energy spectrum appears related to the instability, and the absence of a mass gap, that we mentioned above. Indeed, this is precisely the same restriction, required by Russo (1997:~p.~206), to decouple extra supermembrane states that spoil the low-energy limit (these extra states are ``frozen'' in this limit).\\
\\
{\bf Solitonic membranes as fundamental?} To resolve the problem of the continuity of the membrane spectrum, Townsend (1995:~p.~184) proposed that the elementary supermembrane is {\it not} the fundamental object of M-theory. Instead, the fundamental degrees of freedom of M-theory are the degrees of freedom of the {\it solitonic supermembrane} of eleven-dimensional supergravity. The difference can be understood as follows: the action of the elementary supermembrane is its world-volume action (analogous to the string theory action, Eq.~\eq{fullWS}). The supergravity solution for a macroscopic supermembrane consequently has a delta function source. After quantization, the spectrum is continuous and does not have an energy gap.

By contrast, the {\it solitonic} supermembrane is a solution of the supergravity (Einstein) equations with no supermembrane source. Locally, this solution takes the same form as the elementary membrane solution, but it differs from it in that the solution is extended beyond the purported singularity, which is a {\it coordinate singularity} of the extended solution at an event horizon, with a curvature singularity in the extended solution hiding behind the horizon.\footnote{Duff, Gibbons, and Townsend (1994) showed that the singularity of the (elementary) eleven-dimensional supermembrane solution of supergravity of Duff and Stelle (1991:~p.~115) can be analytically extended to give a {\it non-singular} solution. This is consistent with the fact that $k$-symmetry requires that the source-free equations of eleven-dimensional supergravity, rather than the solutions with a source, are satisfied.}

In its extended form, the core of the supermembrane has, due to its gravitational field, {\it finite size}, and it is hidden behind a horizon.\footnote{As with all supersymmetric solutions, this black hole is extremal. For a philosophical discussion of extremal black hole solutions, see De Haro et al.~(2020).} 
Townsend argued that one does not expect the spectrum of such a solution to be continuous: he ascribed the spikes of the elementary supermembrane to its core of zero size, which is an idealization. Instead, the solitonic membrane is extended, and so no spikes should be present, and the supermembrane should be stable. 

Although Townsend quoted additional corroborating evidence for his conjecture, to our knowledge this problem has been open for almost thirty years (perhaps because research moved on to the related matrix theory proposal,\footnote{I.e.~the matrix theory proposal, discussed at the beginning of this Subsection, to the effect that the fundamental objects of M-theory are D0-branes, whose degrees of freedom are described by non-abelian matrices. Matrix theory reproduces some aspects of supergravity at low energies. See Banks et al.~(1997) and Nicolai and Helling (1999).} 
and the proposals discussed in the next Sections). Indeed, Townsend's conjectures is one of several extant conjectures for the nature of the fundamental degrees of freedom of M-theory.\footnote{In M-theory, two fundamental objects are supposed to be M2-branes and M5-branes: these are extended objects that arise as solitonic solutions of 11D supergravity and carry electric and magnetic charges under the 3-form potential of M-theory2. M2-branes and M5-branes are expected to play a fundamental role in M-theory, but their quantum dynamics are not fully understood. Some attempts to describe them include the ABJM theory and the BLG model for M2-branes, and the (2,0) superconformal field theory for M5-branes. In eleven dimensions, the only D-branes allowed by supersymmetry have $p=2$ and $\ti p=11-2-4=5$. Thus these are the supermembrane and its dual, the fivebrane.}

\section{When weak is strong: S-duality}\label{S-d}

The previous Section reviewed some of the proposals and evidence for M-theory. Another major conjecture by Witten (1995a:~p.~87) is that Type IIB string theory is {\it self-dual} under a non-abelian $\mbox{SL}(2,\mathbb{Z})$ group of dualities.\footnote{This conjecture built on ealier work on strong-weak couping duality in string theory, especially by Sen (1994).} 
This is analogous to Montonen-Olive duality (see Eq.~\eq{modular}), which also exchanges weak and strong coupling regimes. 

\subsection{S-duality as a symmetry of the torus}

While the S-duality of Type IIB can be studied directly in the supergravity limit in ten dimensions, where it extends to $\mbox{SL}(2,\mathbb{R})$, it is illuminating to briefly discuss its eleven-dimensional origin. The basic idea is that, when eleven-dimensional supergravity is compactified on a two-torus $T^2$ down to {\it nine} spacetime dimensions, the result is Type IIB string theory in (in effect) ten dimensions, where the diffeomorphisms of the two-torus (namely, the modular group $\mbox{SL}(2,\mathbb{Z})$) is identified with the $\mbox{SL}(2,\mathbb{Z})$ duality group of Type IIB string theory.

This can be understood in two steps: from eleven to ten, and from ten to nine dimensions. The torus $T^2$ is the product of two circles, $S_{11}$ and $S_{10}$. The first circle has radius $R_{11}$, and brings us from eleven to ten dimensions; the second circle has radius $R_{10}$, and brings us from ten to nine dimensions. From the previous Section, we know that the first step gives an effective Type IIA string theory in ten dimensions, which is weakly coupled when the radius of the eleventh dimension is small (recall, from footnote \ref{11210}, that the radius of the eleventh dimension is $R_{11}=g_{\tn s}\ell_{\tn s}/2\pi$). 

In the second, we compactify a second coordinate: this gives Type IIA string theory on $\mathbb{R}^9\times S_{10}$, which is T-dual to Type IIB string theory on $\mathbb{R}^9\times S'_{10}$, with a circle of inverse radius, $R_{10}':=\a'/R_{10}$. If $R_{10}$ is small, so that we only keep the massless modes, then the dual radius, $R_{10}'$, is large, and in effect we have Type IIB string theory in ten (large and flat) dimensions.

Although the duality of Type IIB string theory is very non-trivial, because it involves exchanging fundamental strings and D1-branes (also called D-strings: see the next Section), it is straightforward to check that S-duality corresponds to a symmetry of the two-torus: namely, exchanging the two torus circles, $T^2=S_{11}\times S_{10}=S_{10}\times S_{11}$, i.e.~$R_{11}\leftrightarrow R_{10}$, permutes two of the spatial coordinates.\footnote{The self-duality group $\mbox{SL}(2,\mathbb{Z})$ of Type IIB string theory corresponds, through an effective duality, to the diffeomorphism group of the torus. For the details about how compactification from eleven to nine dimensions gives the duality group of Type IIB string theory, see Bergshoeff et al.~(1995:~pp.~563-564). See also Witten (1995c:~pp.~502-203).} 
To see this, first note that T-duality relates the Type IIA and Type IIB string couplings as follows:\footnote{See Witten (1995c:~p.~503).}
\bea
g_{\tn{IIB}}={g_{\tn{IIA}}\over R_{\tn{IIA}}}\,,
\eea
where $R_{\tn{IIA}}$ is the radius of the eleventh dimension, $R_{11}$, measured in ten-dimensional Type IIA string theory units. The two are related by $R_{\tn{IIA}}=\sqrt{R_{11}/\ell_{11}}\,R_{10}$.\footnote{Recall, from the Kaluza-Klein form of the eleven-dimensional metric, Eq.~\eq{KKmetric}, that the eleven-dimensional and the ten-dimensional (Type IIA string theory) metrics are related by the following power of the dilaton: $G_{\m\n}^{11}=g_{\tn s}^{-2/3}g_{\m\n}^{10}$. Since, by the duality between eleven-dimensional supergravity and ten-dimensional string theory, the eleven-dimensional radius $R_{11}$ grows with the same power of the dilaton, i.e.~$g_{\tn s}^{2/3}$, we find that the eleven-dimensional metric and the ten-dimensional one are related by the radius: $G_{\m\n}^{11}={\ell_{11}\over R_{10}}\,g_{\m\n}^{10}$, and thus the lengths are related by the square root of this factor, as claimed in the main text.}
Using $g_{\tn{IIA}}=(R_{11}/\ell_{11})^{3/2}$, from Section \ref{11DM},\footnote{For clarity, we reinsert the 11th-dimensional Planck length, which had been left out from Eq.~\eq{R10}, and nevertheless drops out.}
we get:
\bea
g_{\tn{IIB}}={R_{11}\over R_{10}}\,,
\eea
i.e.~the string coupling constant of the Type IIB string theory is given by the ratio of the two radii. From here, we see that exchanging the two circles, $R_{11}\leftrightarrow R_{10}$, gives the S-duality transformation of the Type IIB string theory:
\bea\label{Sdual}
g_{\tn{IIB}}\mapsto 1/g_{\tn{IIB}}\,,
\eea
which is the pure S-duality component of the modular transformation Eq.~\eq{modular}, i.e.~the one that inverts the coupling (i.e.~it is the analogue of $e\mapsto4\pi/e$, where the value of the $\theta$-angle does not change). This relates weakly coupled Type IIB string theory to its strongly coupled version, and vice versa. More generally, the full S-duality group $\mbox{SL}(2,\mathbb{Z})$ can be identified with the modular group of the two-torus on which eleven-dimensional supergravity is compactified.

This effective duality between eleven-dimensional supergravity and Type IIA string theory, combined with T-duality between Type IIA and Type IIB string theories, amounts to a {\it geometrization} of the S-duality of Type IIB string theory. While, from the perspective of ten-dimensional Type IIB string theory, S-duality looks like a duality involving the coupling constants of the theory, from the eleven-dimensional perspective it is the diffeomorphism symmetry of the compactification torus.

\subsection{S-duality of Type IIB string theory}\label{S-dIIB}

To further check the S-duality conjecture for Type IIB string theory, we verify the action of S-duality on various states. (We here give several results without proof, and refer to the literature for details.)\footnote{For a brief account of the history of the discovery of (various forms of) S-duality in string theory, see Schwarz (1996:~pp.~2-3).}
We first discuss the massless states and then the massive states.

The massless bosonic fields of Type IIB string theory, in the NS-NS sector, are the same as those of the Type IIA theory, but they differ in the RR sector. As we discussed in Section \ref{Dbsugra}, the RR sector of the Type IIA theory has even-form gauge field strengths, and Type IIB has odd field strengths, i.e.~$F_{(n)}$ with $n=1,3,5$ (and their electric-magnetic duals with $n>5$). Therefore, the corresponding gauge field, $A_{(n-1)}$, has an even number of indices. 

We will treat these cases one by one: the aim is to make plausible that $\mbox{SL}(2,\mathbb{Z})$ duality in Type IIB can be seen as the ``mirroring'' of the fields of the NS-NS sector by fields from the RR sector, i.e.~they come in pairs that get exchanged by the duality (i.e.~as `doublets'): this does not happen in Type IIA, where the fields in the RR sector do not match those of the NS-NS sector.\\
\\
{\bf Case $n=1$.} This is a scalar field $A_{(0)}$, called the `axion', usually indicated by $\chi$. It is ``mirrored'' by the dilaton from the NS-NS sector, in that they transform into each other under duality. Together, they form a complex scalar field:
\bea\label{scalarF}
\t=\chi+ie^{-\f}\,.
\eea
This complex scalar field transforms under $\mbox{SL}(2,\mathbb{Z})$ as a modular parameter, i.e.~as in Eq.~\eq{modular}.\\ 
\\
{\bf Case $n=3$.} This is the three-form field strength, $F_{(3)}$, of a gauge two-form $A_{(2)}$. This RR three-form ``mirrors'' the NS-NS three-form $H$ (see Eq.~\eq{betaf}), i.e.~they transform into each other under S-duality, and together they form an $\mbox{SL}(2,\mathbb{Z})$ doublet. Like for the dilaton and the axion, the NS-NS and RR gauge fields can be conveniently written in terms of a single complex field, whose field strength is analogous to Eq.~\eq{calF}.

As we discussed in Section \ref{Dbsugra}, the RR three-form $F_{(3)}$ couples to a D1-brane or Dirichlet string, i.e.~the solitonic solution, and the NS-NS three-form $H$ couples to a fundamental string (i.e.~the D-string and the fundamental string are charged under the three-forms to which they couple). Since $\mbox{SL}(2,\mathbb{Z})$ exchanges the two three-forms, the duality exchanges the fundamental string and the D-string, including not just the pure S-duality case, but all also ``in-between'' states, i.e.~mixtures of D-strings and fundamental strings (see below).

As we discussed in Section \ref{elemsolit}, the tensions of the fundamental and Dirichlet strings are related by a factor of the string coupling: $T_{\sm{D1}}=T/g_{\sm s}$, where $T=1/2\pi\a'$ is the fundamental string tension. This follows from supersymmetry, and so it is an exact relation. 

We have discussed that $\mbox{SL}(2,\mathbb{Z})$ duality exchanges the roles of the fundamental string and the solitonic string, including linear combinations of them.\footnote{For a discussion, see Polchinski (1998a:~p.~181).}
The states with three-form charge can indeed be found, from the supersymmetry algebra: they are bound states of fundamental strings and Dirichlet strings, with charges $(q_1,q_2)$, where $q_1$ is the fundamental string charge, obtained from $H_{(3)}$ ($q_1=1$ for a single fundamental string) and $q_2$ is the Dirichlet string charge, obtained from $F_{(3)}$ ($q_2=1$ for a single Dirichlet string). Under $\mbox{SL}(2,\mathbb{Z})$, a state with charges $(q_1,q_2)$ transforms into a state with charges $(aq_1+bq_2,cq_1+dq_2)$, where $ad-bc=1$.\footnote{The constraint $ad-bc=1$ implies that $a$ and $c$ are relatively prime.} 
In particular, an electric state $(1,0)$ transforms into a dyon $(a,c)$, i.e.~a state that is both electrically and magnetically charged. 

The $\mbox{SL}(2,\mathbb{Z})$ symmetry of Type IIB string theory can be made explicit for the low-energy supergravity action, Eq.~\eq{betaf}, with the massless fields of the RR sector added. This action is written in terms of appropriate doublets $(\f,\chi)$ and $(H_{(3)},F_{(3)})$, on which $\mbox{SL}(2,\mathbb{Z})$ act as above, i.e.~as in Eq.~\eq{SL2Z}. 
This supergravity action has a family of macroscopic string solutions, with charges $(q_1,q_2)$, that are mapped onto one another under $\mbox{SL}(2,\mathbb{Z})$.\footnote{Russo and Tseytlin (1997:~p.~121) identify eleven-dimensional counterparts of such solutions, as pure gravitational waves propagating along a cycle of a two-torus.} 

Even though the three-form fields are massless, the bound states of D1 strings and fundamental strings on a circle, with charges $(q_1,q_2)$, are massive, i.e.~they have a Kaluza-Klein mass. Schwarz (1996:~pp.~187-188) calculated the mass spectrum of such $(q_1,q_2)$ Type IIB strings, compactified to nine dimensions, on a circle of radius $R_{10}$, and compared it with the Kaluza-Klein modes of supergravity on a torus with modular parameter $\t=(x+iy)/2\pi R_{11}$. These modes have the following wave-function:
\bea
\psi_{\ell_1,\ell_2}(x,y)=\exp\left({i\over R_{11}}\left(x\ell_2+{1\over\t_2}\,y(\ell_1-\ell_2\t_1)\right)\right),~~\ell_1,\ell_2\in\mathbb{Z}\,,
\eea
where $x$ and $y$ are the spatial coordinates of the torus, and $\ell_1$ and $\ell_2$ are integers that characterize the (Kaluza-Klein) momentum. A convenient complex coordinate on the torus is $z=(x+iy)/2\pi R_{11}$.\footnote{In terms of the radius of the eleventh dimension and the complex structure, the two periods of the torus are $2\pi R_{11}$ and $2\pi R_{11}\t_2$. The area of the torus is $(2\pi R_{11})^2\t_2$, i.e.~$\t_2=R_{10}/R_{11}$.} In terms of this coordinate, the wave-function is invariant under the two transformations that generate $\mbox{SL}(2,\mathbb{Z})$: namely, $z\mapsto z+1$ and $z\mapsto z+\t$. Calculating the mass of a supermembrane that wraps an integer number of times around the torus, he found precise agreement with the formula from Type IIB string theory.\footnote{The matching of the two formulas indeed requires that the modular parameter $\t$ of the torus in eleven dimensions is identified with the (constant) value of the scalar field, Eq.~\eq{scalarF}. The matching also gives the value of the tension of the membrane in terms of the ten-dimensional parameters.} Schwarz's results are valid for the zero-mode part of the mass of the BPS states of Type IIB $(q_1,q_2)$ strings on a circle, as compared to the zero-mode states of a fundamental supermembrane wrapped around a two-torus. This result was generalized, by Russo and Tseytlin (1997:~pp.~135-143), to the oscillator part of the mass (see Eq.~\eq{M2}).

This is further evidence for the effective duality between M-theory on a torus and Type IIB on a circle: `the duality is really between a torus and a circle rather than between two circles' (Schwarz (1996:~p.~188)).\footnote{This duality can be generalized to other two-dimensional manifolds, such as the cylinder, $I\times S^1$, where $I$ is an interval.} \\
\\
{\bf Case $n=5$.} This is the RR five-form $F_{(5)}$, and it does not have a ``mirror'' in the NS-NS sector. For it is self-dual, i.e.~it transforms into itself under S-duality.\footnote{Since a five-form field strength with four-form gauge field couples to the world-volume of a three-brane, S-duality takes the D3-brane into itself. We will not discuss the self-dual case here; see Polchinski (1998a:~pp.~182-183), who also discusses D5-branes and NS5-branes (pp.~182-185). In general, $(p,q)$ 5-branes are the magnetic duals of $(p,q)$ strings (Becker et al.~2007:~p.~329).} In fact, this conjectured duality, when appropriately generalized, is closely related to the Montonen-Olive duality of ${\cal N}=4$ supersymmetric Yang-Mills theory, from Section \ref{N=4SYM}, as we will see in the next Section.

\section{Gauge-gravity duality}\label{ggd}

Gauge-gravity duality is, even more than other dualities, a huge topic, with ramifications in several other fields, of which we can here only give a brief indication.\footnote{An accessible introduction that discusses a number of philosophical issues is in De Haro, Mayerson and Butterfield (2016). Book-length treatments and reviews are in Ammon and Erdmenger (2015) and Aharony et al.~(2000).} 
Our exposition will limit itself to three aspects: motivating the duality so as to make it plausible (Section \ref{motgg}), matching the partition functions so as to illustrate the Schema (Section \ref{matchpf}), and discussing the themes that are illustrated by gauge-gravity duality (Section \ref{themesgg}).

\subsection{Motivating the duality}\label{motgg}

For more than twenty-five years, gauge-gravity dualities have been a central topic of research in string theory. Although several precursors existed,\footnote{See Brown and Henneaux (1986), which is widely regarded to be a ``precursor'' of AdS$_3$/CFT$_2$, and also Ba\~nados et al.~(1992), who discovered the three-dimensional `BTZ' black hole. Witten (1991) is an early description of a black hole using conformal field theory. For a history of string theory, see Rickles (2014).} 
it was Maldacena's (1997) paper on AdS-CFT duality that transformed the field, quickly followed by Witten (1998), who, by putting it in the elementary language of quantum field theory, clarified basic aspects of the duality.

Maldacena gave evidence for a series of conjectured dualities between: (i) string theories in an anti-de Sitter space\footnote{Anti-de Sitter space is the maximally symmetric spacetime with a negative cosmological constant. For introductions, see Aharony et al.~(2000:~pp.~214-216) and De Haro, Mayerson and Butterfield (2016:~pp.~1383-1385).} 
of dimension $d+1$ (with an internal manifold of positive curvature, to a total or 10 or 11 dimensions), and (ii) conformal field theories\footnote{A conformal field theory is a quantum field theory that is invariant under local scale transformations, i.e.~transformations that change the volume by a spacetime-dependent factor, such that the angles are preserved. Quantum mechanically, this implies that the beta-functions of such a quantum field theory are zero. See Section \ref{bosoniz}.} 
in various dimensions (denoted by $d$).

Maldacena's argument for these dualities involved considering a large number $N$ of parallel D-branes (see the left side of Figure \ref{SV}), in the limit in which the D-branes are on top of each other. He argued that, in a certain low-energy limit that he dubbed the {\bf decoupling limit}, the D-branes effectively decouple from the rest of the geometry, i.e.~they do not interact with it. 

In the string theory describing this system, he distinguished two alternative points of view, or descriptions: (i) that of the ambient geometry surrounding the D-branes, which is an anti-de Sitter (AdS) geometry; (ii) the interactions between the D-branes, described by a non-abelian conformal field theory on the world-volume of the D-branes. Maldacena argued that, in the decoupling limit, the systems (i) and (ii) are dual to each other.\footnote{It would take long to reproduce the whole argument for this conjecture. For an excellent explanation, see Zwiebach (2009:~pp.~537-541), and also Aharony et al.~(2000:~pp.~225-227).}

There are two related questions that require clarification, and that are crucial for Maldacena's argument: (1) What are the `two alternative points of view'? (Maldacena et al., 2000:~p.~228). From Chapter \ref{String}, we already know that (i) is the description of the D-branes given by closed strings, and (ii) is the description given by open strings, with the difference that we now take a decoupling limit. (2) What is the argument that these are descriptions of the {\it same} system? We will postpone the answers to these two questions, since to appreciate them it is best to first give the conjecture, as follows:\\
\\
{\bf Duality: the standard case.} Let us consider the case of five-dimensional anti-de Sitter space (AdS$_5$) and a four-dimensional conformal field theory (CFT$_4$), i.e.~$d=4$. The basic conjecture is that we have a duality between:

(i)~~Type IIB string theory on $\mbox{AdS}_5\times S_5$, i.e.~five-dimensional anti-de Sitter space times a five sphere (see Section \ref{S-d} on Type IIB string theory);

(ii)~~Four-dimensional ${\cal N}=4$ supersymmetric Yang Mills (SYM) field theory (see Section \ref{N=4SYM}).\\
\\
{\bf Relations between the parameters.} The two models have two {\it dimensionless parameters} that are related by the duality: (i) the Yang-Mills model has a coupling constant $g_{\tn{YM}}$\footnote{In the action of the ${\cal N}=1$ SYM model, i.e.~Eq.~\eq{Waction}, the coupling constant $g_{\tn{YM}}$ was called $e$. Thus we do not here consider the $\theta$-angle defined in Eqs.~\eq{deftau} and \eq{modular}.} 
and an integer $N$ which is the rank of its non-abelian gauge group, namely $\mbox{U}(N)$. (ii) The parameters of the Type IIB string theory are the string coupling, $g^{\tn{IIB}}_{\tn s}$ (see Section \ref{S-d}), and the cosmological constant scale, i.e.~the radius of curvature of the $\mbox{AdS}_5$, measured in units of the string length, i.e.~$R/\ell_{\sm s}$.\footnote{The radius of curvature of $\mbox{AdS}_5$ is required to be the same as that of the $S^5$, so that the total curvature is zero. Also, one can view the ratio $R/\ell_{\sm s}$ as the product of the cosmological constant and Newton's constant (see Eq.~\eq{GNstring}).}
The parameters are related as follows: the Yang-Mills coupling is directly related to the closed string coupling:\footnote{The Yang-Mills coupling is the coupling strength between the D-branes, which interact by the exchange of open strings. Therefore, the Yang-Mills coupling is the coupling strength of the open strings. In turn, the square of the open string coupling is the coupling between the closed strings, i.e.~$g_{\tn{IIB}}$ (see Section \ref{ocs}). A Feynman-diagram heuristic to understand this relation is that the union of two open strings forms a closed string, and so a closed string amplitude with coupling strength $g_{\tn{IIB}}$ can be ``broken up'' into a product of two open string amplitudes, each with coupling strength $g_{\tn{YM}}$.}
\bea\label{gYMIIB}
g_{\tn{YM}}^2=4\pi g_{\tn{IIB}}\,,
\eea
and the radius of curvature is related to the following combination of the Yang-Mills coupling and the number of D-branes:\footnote{This relation follows by considering the gravitational radius of a system of $N$ coincident D-branes, and the radius of curvature that they will produce. This can be seen from the analogues of the supergravity solutions for the one-brane (superstring) and five-brane, Eqs.~\eq{ec2} and \eq{ec6}, for the 3-brane. In place of the constants $c_2$ and $c_6$, there will be a constant $c_4$ with length dimensions $R^4$ that sets the radius of curvature $R$ of the $N$ D3-branes. It is given by $R^4=4\pi g_{\tn s}^{\tn{IIB}}N\ell_{\tn s}$, which combined with Eq.~\eq{gYMIIB} reproduces Eq.~\eq{RAds}. See also Aharony et al.~(2000:~p.~226) and Zwiebach (2009:~p.~537).}
\bea\label{RAdS}
{R\over\ell_{\tn s}}=\left(g_{\tn{YM}}^2N\right)^{1/4}=:\lambda^{1/4}\,.
\eea
$\l:=g_{\tn{YM}}^2N$ is called the {\bf 't Hooft coupling}, and it can be seen as the effective coupling of a system of $N$ D-branes, each of which contributes a factor of $g_{\tn{YM}}^2$ to the total coupling.\footnote{The 't Hooft coupling is very useful to understand the perturbative expansion of a non-abelian quantum field theory (this will be one of the examples of scientific understanding, in Section \ref{UEwST}).} \\
\\
{\bf Two regimes.} Both in the string theory and in the CFT, the 't Hooft coupling is the effective coupling that determines the validity of the perturbative approximation. We have two cases:

(I)~~At {\it strong} 't Hooft coupling, i.e.~$\l\gg1$, the radius of curvature $R$ is large: thus the spacetime curvature is small, and the supergravity model is a good approximation to the string model (see Section \ref{cea}). From Eq.~\eq{RAdS}, we see that the perturbative approximation in this limit is geometric, i.e.~higher-order terms are corrections in powers of the small number $\ell_{\tn s}/R$, which appear in the Einstein-Hilbert action as higher-order curvature corrections coming from the string theory.

(II)~~At {\it weak} 't Hooft coupling, i.e.~$\l\ll1$, the diagrammatic expansion of the SYM model is a good approximation (this is a double-line diagrammatic, also called `planar', expansion: see Section \ref{UEwST}).\footnote{For an overview of results available in ${\cal N}=4$ SYM at all values of the 't Hooft coupling, see Beisert et al.~(2012).} \\
\\
{\bf The argument, in more detail.} We are now in a position to unpack Maldacena's argument leading to the duality, and to answer questions (1) and (2) above. One begins with Type IIB string theory, with a set of $N$ D-branes and two kinds of excitations: open strings attached to D-branes, and closed strings. These excitations are relevant in two distinct regimes: (I) At strong 't Hooft coupling, the gravitational effects of the $N$ D-branes are important, while the short-distance interactions between the D-branes due to the open strings are relatively unimportant. In this regime, in effect we have a model of closed strings, which can be approximated by Type IIB supergravity. (II) At weak 't Hooft coupling, the gravitational effects are negligible so that the space is effectively flat, and the physics of the D-branes is effectively described by the SYM model. Thus Maldacena's `two perspectives' means focussing on the relevant physical degrees of freedom, i.e.~closed vs.~open strings, in two different regimes of parameters.\footnote{In both regimes of parameters, there is another system, common to both systems, that consists of the closed string excitations of flat space, i.e.~far away from the D-branes. However, this is the same from the D-brane system, which effectively decouples.}

So far this argument only establishes that the same system of $N$ D3-branes can be described by two different models in two different regimes, i.e.~theories (i) and (ii) give a good approximation to the physics of the D3-branes in Type IIB string theory, in two different regimes (I) and (II). It does not establish {\it duality}, and also it does not establish an effective duality, which requires arguing that the two are models of a single theory {\it for the same regime of parameters}.

This last bit is the {\it conjecture} in AdS-CFT duality. The conjecture is motivated by the fact that both models, i.e.~Type IIB closed strings in $\mbox{AdS}_5\times S^5$ and SYM, are believed to be well-defined {\it away from the regimes} (I) and (II) of couplings. Thus {\it if} there are values of the 't Hooft coupling where the models overlap, in the sense of both giving good descriptions of the D3-branes: since the D3-branes are in effect isolated from the rest of the spacetime (i.e.~in the jargon mentioned earlier, at low energies they are {\it decoupled}), then these are descriptions of the same system, and are physically equivalent. And {\it if} the two descriptions remain valid for {\it all} values of the 't Hooft coupling, then they are physically equivalent for all the values.

The crucial `ifs' prompt two possible forms of the correspondence: as a quasi-duality (if the two models are not equally good descriptions of the physics, but e.g.~only by approximation) or as a duality (if they give equally accurate descriptions of the physics).\footnote{Also, the above argument only says that the descriptions are physically equivalent, but does not specify that they should be duals, i.e.~isomorphic models. However, since they are both quantum theories, it seems very unlikely that two physically equivalent descriptions would be anything but duals. Nevertheless, this is of course also part of Maldacena's conjecture: that there is a duality that in particular involves an isomorphism of the Hilbert spaces.}
Maldacena conjectured the strong form: namely, the two models are duals, and give different descriptions of the same physical system.\footnote{See Aharony et al.~(2000:~p.~229) and Maldacena (1997:~p.~1117).}

Although Maldacena's conjecture appears to be formulated on a fixed $\mbox{AdS}_5\times S^5$ spacetime, it in fact allows for generalizations where the spacetime has AdS-like boundary conditions at infinity, but is allowed to fluctuate in the bulk.\footnote{The significance of boundary conditions was explained by Witten (1998) and Gubser et al.~(1998), and the full form of the conjecture, where the interior of the space need not be pure AdS, was given in Aharony et al.~(2000:~p.~230). However, even this form of the conjecture is not general enough, since the boundary conditions need not be AdS: it is sufficient that the space is asymptotically a solution of the Einstein field equations with a negative cosmological constant, including with non-zero matter fields. Thus no anti-de Sitter space needs to be involved, but only an asymptotically locally AdS. This form of the conjecture was developed in De Haro et al.~(2001).}

Additional evidence for gauge-gravity duality exists. One basic point is that the symmetries of the two models agree: in the basic case of a fixed AdS spacetime, the spacetime symmetry is $\mbox{SO}(d,2)$, which is the conformal group of a $d$-dimensional conformal field theory. Furthermore, the internal five-sphere, $S^5$, has a symmetry group $\mbox{SO}(6)$, which is the internal symmetry group of the six scalars of the ${\cal N}=4$ SYM theory (see Section \ref{N=4SYM}).\footnote{The six scalars are listed in Table \ref{N=4N=2spectrum}. For more details on the symmetries, including in other dimensions, see Aharony et al.~(2000:~pp.~221-222).}
In the nomenclature from Section \ref{dualsym}, these are the stipulated symmetries of the theory behind the duals.\footnote{For a review on the progress on relating SYM to string theory at all values of the 't Hooft coupling, see Beisert et al.~(2012).}\\
\\
{\bf The holographic principle.} Witten established a quasi-local form of gauge-gravity duality, i.e.~he explained that, although there cannot be a local correspondence between the gravity and the gauge theory quantities, there is a correspondence of quantities defined on spacetime {\it surfaces}, including the asymptotic boundary. Thus he argued that Maldacena's gauge-gravity duality is {\it holographic}. 

The {\it holographic principle} had been formulated by 't Hooft (1993), and in the context of string theory by Susskind (1995), as a general principle that models of quantum gravity ought to satisfy: it is a principle about the nature of the degrees of freedom of any quantum theory of gravity.\footnote{Hologrpahic relations of duality and quasi-duality between $d$-dimensional quantum systems and $(d+1)$-dimensional quantum gravity theories appear much more generally, beyond string theory (see e.g.~Qi (2013)); and have also been conjectured in loop quantum gravity, see Han and Hung (2017).}

By reinterpreting the Bekenstein-Hawking black hole entropy formula as a statement about the dimension of the Hilbert space of a black hole, and arguing that a black hole with horizon area $A$ contains the maximum amount possible in the interior of the volume bounded by the area $A$, 't Hooft argued that any theory of quantum gravity in 3+1 dimensions can be formulated as a quantum theory (with no gravity) in 2+1 dimensions. Thus quantum gravity is like a hologram, where the two-dimensional surface contains all of the information about a three-dimensional volume. More specifically, 't Hooft (1993:~p.~6) argued that quantum gravity can be formulated as {\it topological} quantum field theory, with all of its degrees of freedom projected onto the boundary. 

Thus Witten proposed that AdS-CFT duality realizes 't Hooft's holographic principle, with the asymptotic boundary of the AdS-type spacetime playing the role of boundary onto which the degrees of freedom are projected. In other words, 't Hooft's idea of `quantum gravity as a topological quantum field theory' is realized through a quasi-locally defined Hilbert space and set of quantities.

\subsection{Illustrating the Schema: matching of partition functions}\label{matchpf}

The Hilbert space of gauge-gravity duality is not known beyond various limits and special cases. But if one is willing to enter a non-rigorous discussion, then the conjecture can be reformulated as follows:\footnote{We here follow closely the discussion of De Haro, Teh and Butterfield (2016:~pp.~1395-1396).} 

(A) first, cast the two models as triples of states, quantities, and dynamics; 

(B) when this is done, they are duals in the Schema's sense.\footnote{Discussions in the physics literature of the duality as a unitary map between the Hilbert spaces include Giddings (2015) and Aharony et al.~(2000).}

Here we just emphasise that the main conceptual point, as regards (A), is that the partition function gives correlation functions of boundary operators (i.e.~of canonical momenta) that agree with those of the dual CFT. And so, the duality is best formulated as a map between two partition functions, as follows:

(i)~~{\it Type IIB string partition function.} The partition function of the string model on the AdS$_{d+1}$ spacetime is a function of the $d$-dimensional metric, $g$, induced on the conformal boundary of AdS$_{d+1}$.\footnote{For simplicity, we do not consider matter fields here.}
In the leading semi-classical approximation, i.e.~case (I) above, we can approximate this partition function, to leading order, by the supergravity action, evaluated on the relevant solution of the Einstein field equations with the boundary condition $g$:
\bea
Z_{\tn{string}}[g]\simeq e^{-S_{\tn{grav}}[g]}\,.
\eea
By taking functional derivatives with respect to the boundary metric, one obtains the correlation functions of the canonical momentum. For the one-point function, the result is the Brown-York quasi-local stress-energy tensor.\footnote{For a conceptual discussion, see De Haro, Teh and Butterfield (2016:~p.~1379).}

(ii)~~{\it The CFT partition function:} this is the generating functional for (dis-) connected correlation functions, whose logarithm is (minus) the generating functional for connected correlation functions in the CFT, on a curved background $g$:
\bea
Z_{\tn{CFT}}[g]=e^{-W_{\tn{CFT}}[g]}\,.
\eea
Taking functional derivatives, one obtains expectation values of the stress-energy tensor of the CFT, i.e.~$\bra T\ket_{\tn{CFT}}$.\footnote{For a discussion of the definition of energy and Noether's theorems in general relativity, and its relation to the quantities of the CFT, see De Haro (2022:~pp.~219-227). The duality map between the quasi-local stress-energy tensor in the bulk and the stress-energy tensor of the CFT vindicates the relevance of Penrose (1982, 1988)-type quasi-local quantities in general relativity. For a discussion, see De Haro (2022:~pp.~239-246).}

The duality conjecture, i.e.~valid in the regimes (I) and (II) and in between, is as follows:
\bea
Z_{\tn{string}}[g]\equiv Z_{\tn{CFT}}[g]\,.
\eea
In the regime (I) where supergravity is valid, this simplifies to:
\bea\label{AdSCFTdict}
S_{\tn{grav}}[g]\simeq W_{\tn{CFT}}[g]\,,
\eea
and this is the {\it effective duality} (with its generalizations including matter fields in supergravity) that the literature has checked in great detail.

Gauge-gravity has thus been conjectured to be a {\it quantum} equivalence between (i) Type IIB strings in $d+1$ dimensions with a negative cosmological constant, and (ii) a $d$-dimensional conformal field theory model. For example, Maldacena (1997:~p.~1117) and Aharony et al.~(2000:~pp.~231, 237-247, 254-265) explicitly speak about the correspondence between gravity operators (i.e.~asymptotic fields) and boundary operators and matching of algebras and correlation functions, and about the isomorphism of the Hilbert spaces. In other words, if gauge-gravity duality is true, then it illustrates the Schema. 

\subsection{Themes illustrated by gauge-gravity duality}\label{themesgg}

{\bf Quantum duality.} Since gauge-gravity duality is an isomorphism between quantum models that are classically very different, it is an example of a quantum duality (see Section \ref{std}): which also explains why this duality is so `surprising'. The explanation is in the two regimes (I) and (II) of the 't Hooft parameter $\l$ considered above: namely, there are {\it two limits where the common core theory shows very different semi-classical behaviour}:

(I)~~At strong 't Hooft coupling, the quantum theory is well-approximated by supergravity in an AdS-type space of dimension $d+1$. Thus the spacetime develops an {\it effective} extra dimension at strong coupling---a phenomenon that is itself reminiscent of the appearance of the eleventh dimension in string theory, in Section \ref{11DM}. 

(II)~~At weak 't Hooft coupling, there is a second semi-classical limit, where the theory is effectively a free CFT in $d$ dimensions.

Since the classical descriptions are not valid for other values of the coupling, there is no sense in which the two classical limits are dual to each other. Thus we have a quantum duality.

Witten (1980:~\S III) has explained the stringlike behaviour of gauge theories in the 't Hooft limit in terms of the appearance of a `master field'. The idea is that in the 't Hooft limit, the gauge theory in effect becomes classical, but still describes quantum effects relevant to the chosen regime of the parameters. In this approximation, the theory is governed by a master field which represents a closed string worldsheet (i.e.~in AdS-CFT, the master field configuration of the fields in the CFT represents a gravitational field in AdS). 

A {\bf master field}, in a quantum field theory, is a configuration of the field(s) that, in an appropriate limit, dominates the path integral (see Coleman (1985:~p.~392) and Witten (1980:~\S IV)). In this limit, the path integral (and the quantities evaluated from it) gets a {\it single} contribution, namely, from the master field (if there are several fields in the theory, then the master field is the single contribution from this set of fields). The theory in effect becomes classical, and the master field satisfies a set of classical field equations. In the state-operator formalism, this implies that operators have zero standard deviation from their classical values, i.e.~their expectation values satisfy, $\bra(Q-\bra Q\ket)^2\ket\rightarrow0$ as $N\rightarrow\infty$, as well as factorization properties characteristic of c-numbers, such as $\bra Q_1\,Q_2\,Q_3\ket=\bra Q_1\ket\,\bra Q_2\ket\,\bra Q_3\ket$.\footnote{For an example of the factorization of a correlation function, see the two-point function of the disorder parameter of the Ising model above the critical temperature, in Eq.~\eq{factorize2}. There are many similarities between the master field in a quantum field theory and coherent states in ordinary quantum mechanics, say for the harmonic oscillator: the expectation values of operators satisfy `classical relations' (the harmonic oscillator's equation of motion), uncertainty is minimised, etc. However, the master field as here defined seems to explain more generally the appearance of semi-classical behaviour.}\\
\\
{\bf Hard-easy:} the semi-classical regimes (I) and (II) are {\it incompatible} (namely, the ranges of values of the 't Hooft coupling in the two regimes do not overlap), so that the regime that allows us to do perturbation theory in one model, maps to a non-perturbative regime (highly quantum) over in the other model, and vice versa.\\
\\
{\bf Unification:} the duals differ in the dimension that they ascribe to space, in the curvature of spacetime, and much more. Thus gauge-gravity duality is perhaps the most non-trivial example of a duality we have so far encountered.

In the context of particle physics, `unification' is often understood as the integration of different forces into a single unified force: e.g.~electricity and magnetism becoming part of a single field, namely the electromagnetic field. 

But the unification brought about by gauge-gravity duality does not seem to be of this type: so far as we now know, the common core theory is not a `bigger framework' incorporating the gravitational force in the interior of AdS and the non-gravitational forces on the CFT of the boundary into a single framework (as e.g.~some of the approaches to T-duality do advocate: for example, double field theory). Rather, we find that these two different types of forces are {\it different}, viz.~mutually exclusive, manifestations of a single common core. Namely, the regimes (I) and (II) exclude one another: and, where they might coincide, one is a {\it reformulation} of the other, rather than meshing with it as part of a single description. Thus there is here an echo of the `incompatibility' condition of complementarity, i.e.~(c) in Section \ref{complement} (as is often the case with quantum dualities: for a discussion, see Section \ref{iob}).\\

Part III will discuss the practical functions of duality. Thus Chapter \ref{Heuri} will discuss the interplay between duality and {\it emergence}, including for gauge-gravity duality, and Chapter \ref{Understand} will discuss the fruitfulness of 't Hooft's large $N$ expansion, and the kind of geometric reformulation that gauge-gravity dualities invite, for the scientific aims of {\it explanation} and {\it understanding}. Here we will discuss one remaining practical function where gauge-gravity duality stands out among other dualities in string theory:\\
\\
{\bf Analogies and experimental approaches.} Compared to other string dualities, gauge-gravity dualities and quasi-dualities are extremely general: on the boundary, one can have any one of a large class of quantum field theories, of various dimensions and with many types of fields and other background conditions such as finite temperature. This means that one can, in principle, find duals of systems of experimental interest, as has indeed already been achieved. This includes hydrodynamic systems (including analogues of the quark-gluon plasma observed in collisions between heavy ions), quantum chromodynamics (including aspects of quark confinement, Wilson loops, and chiral symmetry breaking already discussed in Chapters \ref{EMDuality} and \ref{EMYM}) and strongly coupled condensed matter systems (superfluids and superconductors, Fermi liquids, etc.).\footnote{For introductions, see Ammon and Erdmenger (2015:~Part III) and Zaanen et al.~(2015).}
A more recent development is the analogue experimental simulations on a quantum computer of communication through a wormhole, by applying unitary operations to the qubits.\footnote{See Jafferis et al.~(2021).}

\section{Conclusion}

This Chapter has discussed three main examples, which illustrate various aspects of dualities and quasi-dualities: (i) the effective duality between Type IIA string theory and 11D supergravity; (ii) S-duality of Type IIB string theory (and its eleven-dimensional origin); (iii) gauge-gravity duality.

The effective duality between type IIA string theory and 11D supergravity was instrumental in suggesting the existence of a {\it successor theory} dubbed `M-theory': and in various proposals for identifying its fundamental degrees of freedom. (We will discuss successor theories, in Chapter \ref{Heuri}, when we discuss the heuristic function of dualities.) M-theory, which is believed to be more fundamental, purports to describe the strong-coupling, high-energy, regime of string theory. 

The extant proposals for M-theory (elementary and solitonic supermembranes, Matrix theory, and also special cases of AdS-CFT where the spacetime dimension of the gravity model is 11)\footnote{Some other programmes that also purport to describe various parts or limits of M-theory include M5-branes and the M5-brane/M2-brane correspondence (i.e.~D-branes in eleven dimensions). For an overview, see Berman (2007); also Maldacena et al.~(2003).}
reproduce many of the features of Type IIA and Type IIB string theories at low energies, but a major obstacle is quantising the degrees of freedom in eleven dimensions. 

S-duality in Type IIB string theory is analogous to the electric-magnetic dualities discussed in previous Chapters, in that the Type IIB theory includes both elementary and solitonic degrees of freedom in its spectrum: notably, fundamental and D-strings, which transform into each other under S-duality. In the ${\cal N}=4$ and ${\cal N}=2$ SYM theories, electric-magnetic (quasi-)dualities relate elementary (electric) and solitonic (magnetic) states. The S-duality of ${\cal N}=4$ SYM is conjectured to be mapped by gauge-gravity duality to the S-duality of Type IIB strings on $\mbox{AdS}_5\times S^5$. 

S-duality is a {\it self-duality} of the Type IIB string theory, i.e.~there is a single model whose states are mapped onto one another by the duality. This symmetry-aspect of the self-duality is particularly clear in the geometrization of S-duality in eleven dimensions, where the $\mbox{SL}(2,\mathbb{Z})$ duality group of the Type IIB on $\mathbb{R}^9\times S$ is identified with the diffeomorphism group of the torus of type IIB supergravity on $\mathbb{R}^9\times T^2$.

As we have discussed, gauge-gravity dualities are very rich dualities, in that they map gravitational models into non-gravitational models in various dimensions. It is worth emphasising how AdS-CFT dualities illustrate {\it quantum dualities}, in which a quantum common core theory can have two (or more) limits in which the theory in effect becomes classical. The semi-classical behaviour in these two limits can be very different. One limit is the CFT's semi-classical limit in $d$ dimensions that is the starting point of the quantization of the CFT, with usual perturbation theory for that CFT being applicable around that point. The other limit is the 't Hooft limit, where the common core theory has a semi-classical limit as a $(d+1)$-dimensional supergravity model. 

All four string theory dualities that we have discussed in extenso (namely, T-duality, M-theory/Type IIA quasi-duality, S-duality, and gauge-gravity dualities) relate models in different dimensions, and-or involve the geometrization of field-theoretic properties. This is again illustrated particularly vividly by gauge-gravity duality, which relates a gravitational and a non-gravitational model. This may suggest that {\it spacetime and gravity are emergent} in string theory, and this has naturally been a main topic of research in high-energy physics over the past twenty years. We will return to this topic in Chapter \ref{Heuri}.

\chapter{A Tale of Two Holes in Spacetime}\label{HABHM}
\markboth{\small{\textup{A Tale of Two Holes}}}{\textup{\small{A Tale of Two Holes}}}

This Chapter addresses two spacetime topics in philosophy of physics on which dualities bear: the first (the hole argument) is well-trodden ground for philosophers of physics: but we will here argue that there is still a vast and unexplored territory; the second (black hole microstates) is nearly virgin territory for philosophers. While both feature holes in spacetime, they are in fact very different topics. But they both involve foundational discussions about the nature of spacetime: the hole argument for classical spacetime theories, and black hole microstates for theories of quantum gravity.

We address these two `hole arguments' here, in the final Chapter of Part II, because the philosophical issues that they raise build on the preceding Chapters and complete our discussion of Part II. Part III will then take up the implications of dualities for philosophy of science more generally, especially for inter-theoretic relations such as equivalence, scientific theories, emergence, and fundamentality.

Thus Section \ref{holeA} discusses {\it the} hole argument, and three kinds of responses that have been given to it: relationism, sophisticated substantivalism, and mathematical responses. The reason why the hole argument has become such a central issue in philosophy of physics surely has to do with its addressing one of the central---perennial---questions confronting space-time theories since the days of Newton and Leibniz (going back to Aristotle): namely, whether the spacetime described by our theories (especially general relativity) is a substance, or whether we should construe it some other way---for example, relationally. While the hole argument reflects Einstein's own struggles with this question as he made his way towards the theory of general relativity, Earman and Norton (1987) showed that the argument has a significance for contemporary spacetime theories well beyond Einstein's original discussion. And so, the hole argument remains, even up to this day, an important and lively topic of debate.

We also discuss a version of the hole argument where the cosmological constant is non-zero (positive or negative), on which gauge-gravity duality bears: for many philosophers, this connection will be novel. The overall aim is to argue that some diffeomorphisms change the physical possibility being represented, and thus that Leibniz equivalence is not always satisfied. Using gauge-gravity duality, these diffeomorphisms that change the physical possibility can be argued to be {\it visible} in the dual boundary theory. 

By showing how different types of diffeomorphisms are in different ``parts'' of our Schema, our results will illustrate one main theme of this book: namely, the fruitfulness of the distinction between a common core theory and its dual models. It will also make concrete, in the context of our Schema, Belot's (2018:~p.~968) remark that one should think of the solutions of general relativity as forming a structured space. (Chapter \ref{Heuri} will further develop the idea of a theory as being {\it geometrically} structured.) In addition, these results clarify how Leibniz equivalence depends on the interpretation that is given to the bare theory and its models.

Section \ref{ocs} discusses the counting of black hole microstates in string theory, following the ground-breaking work by Strominger and Vafa (1996), who gave the first microscopic account of the Bekenstein-Hawking entropy in string theory. By drawing from several previous Chapters in Part II, this Section brings together several major developments in quantum field theory and string theory around the single question of the microscopic origin of black hole entropy: in particular, dualities are a major ingredient of the putative explanation of black hole entropy in string theory. As such, the derivation illustrates several previous physical and philosophical themes and roles of dualities: especially the heuristic function of dualities within the M-theory programme, and the emergence of spacetime. These philosophical themes will be taken up in Chapters \ref{Heuri} and \ref{Understand}, when we discuss the practical functions of duality.\footnote{For philosophical discussions of further aspects of entanglement in gauge-gravity dualities, see Bain (2020, 2021), Jaksland (2021), and Cinti et al.~(2022).}

\section{The hole argument}\label{holeA}

This Section discusses Einstein's 1913 hole argument and how it interacts with gauge-gravity duality. After reviewing, in Section \ref{historyH}, the hole argument itself and its role in contemporary philosophy of spacetime, Section \ref{responsesH} discusses some salient recent responses to it. Section \ref{holecosm} then goes on to give a version of the argument in a space-time with a cosmological constant, and Section \ref{holegg} discusses how the hole argument is mapped by gauge-gravity duality: in particular, hole diffeomorphisms are argued to act on the specific structure, while diffeomorphisms that are boundary symmetries are symmetries of the common core.\footnote{Some of the issues that we discuss in this Section, especially in the context of gauge-gravity dualities, also bear on the question of background-independence. For discussions, see Read (forthcoming) and De Haro (2017a:~pp.~113-116).} 

\subsection{The hole argument and its legacy}\label{historyH}

In 1913, as he struggled to formulate a general theory of relativity, Einstein developed the {\it hole argument}, which was aimed at defending his draft version, or {\it Entwurf} theory, with Marcel Grossman.\footnote{For the early history of the hole argument, see Janssen and Renn (2022:~pp.~30-33). A brief account is in Kosmann-Schwarzbach (2011:~p.~38).}
Although Einstein would later resolve the hole argument and reject its main conclusion, Earman and Norton (1987) reformulated the argument and redirected it towards a different target: not general covariance, but spacetime substantivalism. Their version of the argument has become a central focus of debates in the philosophy of spacetime about, in short, the question {\it whether spacetime is a substance}: with the associated topics of substantivalism, relationism, and determinism.\footnote{For the recent history of the hole argument in the hands of philosophers of physics, see Roberts and Weatherall (2020:~pp.~218-221). This special issue contains a collection of papers that develop new and old perspectives on the argument. See also Gomes and Butterfield (2023:~pp.~1-15).}
At stake is not only the nature of spacetime, but also how we ought to interpret well-established scientific theories. 

While the field equations of Einstein's {\it Entwurf} theory with Marcel Grossman were second order differential equations that generalized the Poisson equation of Newtonian gravitation, they would turn out to disagree with the field equations that Einstein published on 25 November 1915 for his general theory of relativity.\footnote{See Einstein and Grossman (1913:~p.~165).}
And while Einstein found the {\it Entwurf} theory's lack of general covariance unsettling, he became resigned to it. For, although in general Einstein (1914a:~p.~7) thought it desirable for the field equations to be generally covariant, in 1913 he concocted a clever argument to the effect that generally covariant field equations are physically undesirable, so that the class of admissible diffeomorphisms must be restricted.\footnote{See Einstein (1914a:~p.~8; 1914b:~p.~66). Einstein uses the phrase `coordinate transformations', here replaced by diffeomorphisms. But we will still use the common phrase `general covariance' rather than `covariance under diffeomorphism', which should not lead to confusion. For a discussion of this distinction, see e.g.~Pooley (2017:~pp.~115-116).} 
This was his hole argument. Namely, in a generally covariant theory, one can consider a region of spacetime with no matter in the interior (`the hole'), and conclude that the fields outside do not determine the fields in the interior, because a diffeomorphism that is the identity at the boundary of the hole and outside of it, but not in the interior, produces {\it two different gravitational fields on the same manifold}: namely, any metric tensor that solves the field equations in the interior of the hole, and one that is diffeomorphic to it by a hole diffeomorphism. This means that the gravitational field in the interior cannot be uniquely determined from the given initial or boundary data outside the hole. Thus a generally covariant theory violates determinism (or, as Einstein (1914b:~p.~65) called it, `the law of cause and effect'). Thus one should search for theories with the right kind of non-covariance---which his {\it Entwurf} theory did possess.

Earman and Norton (1987:~p.~521) reformulated the hole argument, not as an argument against generally covariant theories, but as an argument against spacetime substantivalism, which they defined as the doctrine that the spacetime manifold `can exist independently of any of the things in it' (and so, the view is better dubbed `manifold substantivalism': see below). In particular, the manifold's {\it points} exist, and are not reducible to other structures, geometric or otherwise (e.g.~matter). They argue that such substantivalism faces a radical form of indeterminism, of the kind that Einstein had formulated.\footnote{They dub the indeterminism `radical', because even for a very small neighbourhood (the hole) the fields within the neighbourhood cannot be determined from the fields just outside (Earman and Norton, 1987:~p.~516). Thus it is not only indeterminism about `large regions' of spacetime. In the next Section, we will argue that it is worth exploring other versions of the hole argument that involve holes of different size and global topology, whose boundary is timelike, or is located at future timelike infinity.}
Thus one must either reject substantivalism, or accept indeterminism.\footnote{Earman and Norton (1987:~p.~524) choose the former, not because they are in principle opposed to indeterminism, but rather because if determinism fails, it should fail for a reason of physics, and not because of a commitment to a substantivalism without additional empirical advantages.}
They conclude that substantivalism is {\it untenable}.

Part of the reason why Earman and Norton (1987:~p.~524) think that substantivalism must be rejected is because there is a readily available principle that the substantivalist must deny, and which resolves the indeterminism problem: `indeterminism can be escaped easily by just accepting Leibniz equivalence'. They define the latter as follows:\footnote{Earman and Norton (1987:~p.~520) envisage that other structure (such as e.g.~metric structure), that may be defined on the manifold, is also ``carried along'', using the pull-back of the diffeomorphism.}\\
\\
{\bf Leibniz equivalence:} {\it diffeomorphic models represent the same physical situation.}\footnote{Roberts (2020:~p.~250) also concludes, for different reasons, that Leibniz equivalence is `strictly false'. He distinguishes between `weak' and `strong' Leibniz equivalence, where the difference is whether the manifold is used to represent the same physical situation `not necessarily at once' vs.~representing it `at once'. At p.~255, he argues that only the strong version is relevant to the hole argument: and that this version is false. For a discussion of Roberts' paper, see Pooley and Read (2021:~Section 5), who also discuss other versions of Leibniz equivalence.} 

The substantivalist must deny this, because the substantivalist's assumption is precisely that models of spacetime that differ by a diffeomorphism represent different spacetimes.

There can be a temptation to dismiss the hole argument as a philosophical confusion, especially: (a) on the grounds that Leibniz equivalence is well-entrenched in mathematical and physical practice; and (b) on the grounds that the focus on the nature and-or existence of the manifold (with its set of points) is a metaphysical question that one ought to avoid, since it is not just the manifold that matters, but rather the manifold together with the other structures (the metric, and other, fields) defined on it.

There are three reasons to reject this quick dismissal, of which the first is specific (and responds to (a) above), and the second and third are general (together they respond to (b) above):

(A)~~The specific reason is that Leibniz equivalence, as formulated above, is demonstrably {\it false}: not on philosophical, but on physical, grounds. This will be the task of Section \ref{holecosm}, which will show that some diffeomorphisms do not change the physical situation, while others {\it do}. \footnote{This point was made in De Haro, Teh, and Butterfield~(2016:~pp.~1067-1069; 2017:~pp.~78-79), De Haro (2017; 2022:~pp.~227-231), and Belot (2018:~p.~965).} 
And Section \ref{holegg} will argue that gauge-gravity duality gives additional support to this distinction.

To justify something like Leibniz equivalence, one must do a great deal more: one must investigate which diffeomorphisms are true `gauge' symmetries of general relativity, and which are not (here, we take `gauge' symmetries in the philosophical sense of `redundancies', not the physical sense of `local or global reparametrizations of the dependent variables'). And to that end, one must define the states and quantities of general relativity that it is appropriate to preserve. Section \ref{holecosm} will argue that verdicts of Leibniz equivalence depend on both the structure of a theory and its models (in a way that is not normally considered in discusions of the hole argument), and on the interpretation that is given to them.

(B)~~Although it is of course true that, in general relativity, one ought to consider not just the {\it manifold}, but especially also the {\it metric tensor} defined on it (and this leads to one of the possible replies to the hole argument, as in sophisticated substantivalism: see Section \ref{responsesH}), the question about the semantics of the models of general relativity, and in particular about the existence of the manifold, cannot be so easily dismissed on either a scientific realist position, or an anti-realist position: (see Chapter \ref{Realism}, especially Section \ref{scirer}, for details of these positions, especially van Fraassen's anti-realism, dubbed `constructive empiricism'). In other words, the question of whether, by formulating general relativity using a manifold, we are asserting that the manifold exists (including the set of points and structures on it, such as a family of open sets that cover the whole manifold and satisfy usual differentiability conditions etc.) is, for both realists and anti-realists such as constructive empiricists, a pertinent question about the interpretation of general relativity.

(C)~~Focussing on the semantics of the {\it manifold}, rather than of the manifold equipped with further structures, is entirely appropriate if one is interested in the question whether there is a common (core) ontology underlying what Earman and Norton (1987:~p.~515) call a `very broad class of spacetime theories'.\footnote{At p.~517, they also dub them `local spacetime theories'.} 
Apart from general relativity, this class of theories includes theories that postulate a manifold but less geometric structure: topological theories (of gravity and-or gauge fields),\footnote{For example, gravity in three dimensions is a topological theory with no local degrees of freedom (in particular, no gravitational waves), but only topological degrees of freedom. Also Chern-Simons theory is a topological theory of a gauge connection on a three-manifold. See Witten (1988c, 1989).}
theories based on geometric structures that do not require a metric tensor (such as a complex structure or a K\"ahler form),\footnote{In the topological string theories called the A- and B-models, the A-model is independent of the K\"ahler structure, while the B-model is independent of the complex structure. See Witten (1988b), Vonk (2005) and Alim (2012).} 
conformal theories which possess only a conformal class of metrics, etc.\footnote{See e.g.~Di Francesco et al.~(1997) and Ginsparg (1988).} 
All of these theories study phenomena whose description requires the notion of a manifold, but they do not postulate the metric tensor of general relativity. And so, it is a sensible question to ask whether these theories have a common spacetime ontology that they share with each other and with general relativity: and whether the {\it manifold}, regardless of these other structures, can be understood as a fundamental entity that underlies the various phenomena described by these theories, i.e.~as a substance. (If this were the case, it would be interesting: for it would suggest that there is a common ontological basis that all of these theories share, and so that they can be seen as different plays, with different plots and dramatis personae, all written to be performed on the same stage.)

The hole argument prompts one to say that, unless one is prepared to embrace radical indeterminism, this hope is dashed: the manifold cannot by itself be part of the ontology of such a common core,\footnote{We here use the term `common core' informally, i.e.~not strictly the Schema's sense, because there is presumably no common dynamics to these theories: so it does not deserve the name `theory'.}
and so spacetime theories are less unified than one might have expected or hoped, given their common use of manifolds and topology. (And, given that this hope is dashed, the question remains how one {\it does} make sense of the spacetime ontologies of each of these theories.)

But the question could not be easily dismissed before it had been asked, since it is the analysis of the hole argument that allowed us to conclude, in (C), that `this hope is dashed'.\footnote{Agreed, there is also a mathematical way to start addressing this question, by asking which of these theories are `mutually compatible', e.g.~in the sense that their Lagrangians can be coupled in a non-trivial way on the same manifold. Solutions to such generalized theories would then describe possible worlds in which the various structures coexist and interact with each other. But the hole argument has the virtue that it asks a precise interpretative question in general, regardless of the technical details of the theories in question.} 
In other words, general relativity has in its ontology spacetimes which are manifolds equipped with a metric tensor: but it cannot answer the question whether, for a variety of theories that may or may not use a metric, it makes sense, independently of the metric, to talk about the manifold.

We take the main moral of Earman and Norton's formulation of the hole argument to be like Lewis's (1969:~pp.~1-2) moral about conventions: 
\begin{quote}\small when a good philosopher challenges a platitude, it usually turns out that the platitude was essentially right; but the philosopher has noticed trouble that one who did not think twice could not have met. In the end the challenge is answered and the platitude survives, more often than not. But the philosopher has done the adherents of the platitude a service: he has made them think twice. ... The platitude ... is no dogma of any school of philosophy, but commands the immediate assent of any thoughtful person---unless he is a philosopher.
\end{quote}

\subsection{Recent responses to the hole argument}\label{responsesH}

This Section will discuss three of the most relevant recent responses to the hole argument: relationism, sophisticated substantivalism, and mathematical responses. But before we do so, we need slightly more precision on the formulation of the hole argument.\footnote{Below, we mostly follow Pooley's (2022) analysis of the hole argument, appropriate to general relativity. In view of our point (C) of the previous Section, the notion of a spacetime could in general be weakened to a pair $(M,X)$, where $X$ is an appropriate topological or geometric structure on $M$ (an orientation, a conformal class of metrics, a complex structure, etc.). However, since in the rest of this Section we will be interested in general relativity, it will be appropriate to take $X$ to be the metric tensor.}\\
\\
{\bf Spacetime substantivalism} is the view that spacetimes, which are described by solutions $(M,g)$ of a spacetime theory (e.g.~general relativity), i.e.~a manifold $M$ with a pseudo-Riemannian metric field $g$, are in the theory's domain of application, and are not reducible to other elements of the domain, such as the matter fields. 

A diffeomorphism of the manifold to itself, $\f:M\rightarrow M$, acts on the manifold's points, i.e.~$\f:p\mapsto \f(p)=q$, where $p$ and $q$ are two (in general distinct) points on the manifold, i.e.~$p,q\in M$. Since the manifold is equipped with a metric tensor field, the diffeomorphism thereby induces an action on the metric tensor, through the pull-back map, written as: $\f^*g(p)=g(\f(p))$; and likewise for other tensor fields, denoted by $\F$. 

The hole argument can then get off the ground when the substantivalist regards two spacetimes related by a diffeomorphism, i.e.~${\cal M}=(M,g,\F)$ and $\f^*{\cal M}=(M,\f^*g,\f^*\F)$, as representing {\it distinct} possibilities. 

What could motivate such a position? Pooley (2022:~p.~147) identifies two assumptions that are close to the core of many substantivalist positions, and that amount, in effect, to regarding diffeomorphism-related spacetimes as distinct spacetimes: 

(i)~~The points of the manifold $M$ represent genuine entities. This is motivated by the substantivalist's commitment to the existence of the manifold, which is defined in terms of a set of spacetime points. And so, it seems reasonable that the points are the fundamental ingredients of the manifold (if the spacetime points do not exist, then the set of them does not exist either).

(ii)~~Two spacetimes that differ by a diffeomorphism, i.e.~${\cal M}$ and $\f^*{\cal M}$, assign different {\it properties} to the same points. This is because, while the diffeomorphism $\f$ maps points to points, the pullback $\f^*$ maps properties to properties (namely, the metric relations and distribution of matter fields over points). Thus, since the spacetimes ${\cal M}=(M,g,\F)$ and $\f^*{\cal M}=(M,\f^*g,\f^*\F)$ share the same manifold $M$ (i.e.~the same set of points), but define different fields on them, they assign {\it different properties} to the same points.

If we then take $\f$ to be a hole diffeomorphism, i.e.~the identity outside the hole but different from the identity in the interior, we have spacetimes that agree outside the hole, but are distinct in the interior. Thus the spacetime in the hole's interior cannot be determined, or reconstructed, from the data outside.\footnote{Pooley (2022:~p.~147) spells out the assumptions (i) and (ii) that go into the hole argument in terms of two premisses and a definition, namely: (1) The spacetimes ${\cal M}$ and $\f^*{\cal M}$ describe possible physical situations that are {\it distinct spacetimes}, i.e.~that: (A) have the same pattern of spatiotemporal properties; but (B) differ solely over which properties they assign to which spacetime points. (2) Furthermore, the two spacetimes are equally physically possible, because they are both solutions of general relativity. (3) Therefore, the spacetime in the interior of the hole is undetermined. Notice that assumption (1) is substantivalism, assumption (2) is general covariance, and (3) involves a (standard) definition of determinism for spacetime theories.}

To respond to the hole argument, one must either reject the substantivalist views (i) or (ii), or embrace them and give up spacetime determinism. We will here give three distinct recent responses that do the former.\footnote{Given the body of philosophical literature built around the hole argument, these responses are of course not exhaustive. Another possible response is a structural realist one: see Dorato (2000) and Esfeld and Lam (2008).}\\
\\
{\bf Relationism} rejects the substantivalist assumption (i) that the manifold represents a genuine entity. It postulates that facts about spacetime can be reduced to facts about spatiotemporal properties and relations of {\it matter}: for example, that all motion is motion of bodies relative to one another, and not relative to some background space(time).\footnote{Barbour (1982, 1999) and Barbour and Bertotti (1982) are ardent advocates of relationism. They propose a Machian treatment of motion and of time that involves reformulating relativistic dynamics.} 

Since, for a relationist, spacetime points have no independent existence, two spacetimes ${\cal M}$ and $\f^*{\cal M}$ related by a hole diffeomorphism cannot differ over the properties that the fields assign to spacetime points: since there exist no independent spacetime points, two spacetimes that differ solely over the properties of individual spacetime points do not differ at all.\footnote{Even more than in the case of substantivalism, relationism is not a single position, but an umbrella term for various positions, with at their core the rejection of spacetime as a substance. In Earman's (1989:~p.~12) words, `there are almost as many versions of relationism as there are relationists'. Earman goes on to give a helpful characterization of relationism in terms of three themes, which boil down to Leibniz's insistence, in his debate with Clarke, on all motion as being relative motion of bodies, and his opposition to space being a substance. In the context of general relativity, relationism is often associated with Mach's views: see Pooley and Brown (2002:~p.~188) and Earman (1989:~pp.~43, 67, 82).}

Note that, unlike the various substantivalisms, which aim to give literal\footnote{They aim to, at least, give as literal as possible readings of the formalism: see the discussion in Section \ref{ncsr}.} readings of the formalism of general relativity, relationism, in order to be precise and allow a straightforward referential semantics, seems to require a {\it revision} of this formalism, that implements the idea of reducing facts about spacetime to facts about matter.\footnote{For example, Earman (1989:~p.~135) admits that `the relationist ... must meet the challenge of formulating a relationally pure physics. The history of relationism is notable for its lack of success in meeting this challenge'. For a defence of relationism, including a discussion of historical approaches like the Hofmann-Reissner-Schr\"odinger theory and the Baierlein-Sharp-Wheeler reformulation of general relativity, see Barbour (1999). Vassallo and Esfeld (2016:~p.~102) seem to reject the project of construing the claims of physical theories literally, which we endorsed at the start of Chapter \ref{Thies}, where we also endorsed referential semantics (we will return to this in Chapter \ref{Realism}, and especially Section \ref{ncsr}, where we will add some qualifications to the phrase `literal reading'): `The formalisms are not the guide to ontology, but parsimony and coherence together with empirical adequacy are'. Thus, since for them, relationism is a matter of ontology not formulation of physical theory, they reject the requirement that the relationist should formulate their physical theory `relationally'.} 
Also, relationism needs to explain the possibility of the existence of non-trivial vacuum solutions of general relativity, with no matter fields whatever.\\
\\
{\bf Sophisticated substantivalism} (i) accepts that spacetime points are genuine entities, but (ii) denies that diffeomorphism-related spacetimes can differ solely by the properties that are assigned (by the metric) to the same spacetime points.\footnote{This position is called {\it anti-haecceitistic}, because it denies that spacetimes can differ merely by haecceitistic differences, i.e.~by which properties are assigned to which points. For a discussion of haecceitism and the arguments in favour and against it, see Cowling (2022). Early statements of sophistication are in e.g.~Mundy (1992), Brighouse (1994), Rynasiewicz (1994), Bartels (1996), Hoefer (1996), and Pooley (2001, 2006, 2013). For a generalization of the idea of sophistication, and a treatment of it that modifies the model-theoretic semantics, see Dewar (2019:~pp.~498-503). Martens and Read (2021:~p.~323) question it, on the grounds that it does not come with an adequate metaphysical analysis. For a discussion of external vs.~internal sophistication, see (ibid, p.~324).} 

We can gloss the above by saying that, for the sophisticated substantivalist, spacetime exists, but not as an independently existing manifold: spacetime points are individuated by the metric field, i.e.~by the metric distances between them. In the words of Pooley (2022:~p.~153), `spacetime is exhaustively specified by a complete catalog of the qualitative facts concerning the full pattern of spatiotemporal relations that are instantiated by its points ... there simply are no further ``individualistic'' facts concerning which objects [i.e.~points] possess which properties. At a fundamental level, reality, at least concerning spacetime points, is purely qualitative'.

Read (2016:~p.~224) points to an analogy between sophisticated substantivalism and the internal interpretation of dualities, which we will discuss in Section \ref{cco}.\\
\\
{\bf Mathematical responses:} Weatherall (2018:~p.~330) has argued that `the mathematical argument that allegedly generates the interpretational problem is misleading'. In short, his point is that isomorphism is the mathematical standard of sameness for spacetimes of general relativity: and so, isomorphic models can be taken to represent the same possible worlds (they `should be taken to have the same representational capacities', ibid, p.~332).\footnote{Earlier formal responses to the hole argument include Mundy (1992) and Leeds (1995) (for a reply, see Rynasiewicz (1996)). These are syntactic approaches, that aim to state precisely, in a logical language, the physical content of the theory of general relativity: and to thus avoid indeterminism. Weatherall (2018) distances himself from this kind of response, since he takes himself to be making a conceptual point about how mathematics is used in physics.} 
For example, manifolds are defined only up to isomorphism. Thus good mathematical practice teaches us to take isometric spacetimes as being the same:\footnote{Weatherall's point is that, having defined general relativity in terms of pseudo-Riemannian spacetimes, which are {\it pairs} $(M,g)$, it goes against good mathematical practice to then compare spacetimes using an isomorphism of only the {\it first member} of the pair that does not fix the second member: namely, a diffeomorphism $\f:M\rightarrow M$ is an automorphism of the manifold, but it is not an automorphism of the metric tensor (unless it is in fact an isometry).} 
if one follows correct mathematical practice, then, regardless of metaphysics, the hole argument is blocked.\footnote{A detailed reply to Weatherall has been given by Pooley and Read (2021). One main point of debate is whether it is appropriate to use two different standards to compare spacetimes: one to compare manifolds regardless of the metric (thus concluding that the manifolds are the same), and one to compare spacetimes including the metric (thus concluding that the spacetimes are different, hence the indeterminism). Landsman (2022:~Section 1) has argued that it is standard mathematical practice to compare manifolds in the way that Weatherall criticizes, and that the notion of an isometry rests on being able to make this comparison. For more details, see Pooley and Read (2021), who argue that it is legitimate to use different maps for different types of comparisons. For a summary of responses to Weatherall, and to Halvorson and Manchak, see Gomes and Butterfield (2023:~pp.~15-19).} 

A similar point by Halvorson and Manchak (2022) is that manifold substantivalism adds a new axiom to general relativity that does not appear in the standard formulation of the theory used in physics.\footnote{The axiom adds Zermelo-Fraenkel set theory read in `material mode'. The axiom says that, roughly speaking, by naming the points of the manifold across different possible worlds, the physical possibilities represented by different points, that are assigned isometric metric properties, are {\it distinct}.} 
This introduces into general relativity a metaphysics that is incompatible with physical practice.\footnote{The main technical claim of Halvorson and Manchak's (2022) paper is a different one, and revolves around the non-existence of hole isomorphisms, but we will not discuss it here. See the sustained critique by Menon and Read (2023).}

Similarly, Bradley and Weatherall (2022) point out that the facts about models that the manifold substantivalist wishes to establish (e.g.~distinguishing the individual points by giving them names) cannot be expressed in the language of general relativity, but only in the metalanguage, i.e.~using the semantics that we use to talk about the models. And assertions in the metalanguage, Bradley and Weatherall (2022:~p.~1231) claim, do not have a physical significance. In other words, there is no mathematically rigorous way to formulate the hole argument. Thus they demand that `anyone who wishes to claim that GR is incomplete in this regard needs to provide some justification for extending the theory'. 

We submit that our first reason for engaging with the hole argument, i.e.~point (C) in Section \ref{historyH}, justifies---not the claim that general relativity is formally incomplete, but rather---our taking the substantivalist arguments seriously. Second, as we will discuss in the next Section, there is an interesting interaction between gauge-gravity dualities, the hole argument, and different kinds of diffeomorphisms, that casts light on the conditions under which Leibniz equivalence obtains. Furthermore, there is an interesting contrast between the analysis of equivalence given in Part III and Weatherall's two main points: (i) isomorphism is the preferred mathematical standard of equivalence for spacetime theories; (ii) assertions in the metalanguage do not have a physical significance. 

About (i): Section \ref{defencei} will defend isomorphism as a preferred formal standard of equivalence, thus agreeing with Weatherall's first point. 

But Chapter \ref{physeq} will also say that isomorphism is a necessary, but not sufficient, condition for theoretical or physical equivalence, because an interpretative criterion of equivalence is also required (and this ties into our response to (ii)).

About (ii): there is further motivation, additional to (C) above, to allow the substantivalist to make the hole argument. For, as we discussed in Chapter \ref{itm}, a bare theory requires an interpretation to describe the physical world. The mathematics of general relativity only describes something physical if it is supplied with an interpretation. Thus one cannot avoid engaging in an interpretative project.\footnote{Bradley and Weatherall do not deny that {\it some} interpretation may be required: `We agree that purely formal considerations do not settle the issue' (p.~1229). They seem only to deny that the `metalanguage' used to formulate general relativity should be taken seriously. However, they do go on to affirm approvingly that `the advocate for the mathematical response would presumably deny that there is any problem of reference in the first place ... to generate the problem for any given theory, one must move from the formal theory under consideration to the metatheory. And the metatheory is representationally irrelevant' (p.~1231). Rather than metatheory, what we think is important for how general relativity describes a domain of application, using a conception of reference like the one we defended in Section \ref{itm}, is semantic interpretation of the type that we will discuss in Part III.}

Furthermore, as Section \ref{theoreq} will argue, the semantics of scientific theories is not limited to a structuralist semantics, with models understood as structures. A structuralist semantics does not capture those aspects of the ontology and epistemology of physical theories that do not reduce to mathematical structure. For verdicts of theoretical equivalence also require an account of how models describe a domain of application. (Part III will give some principles for this interpretative project.)

It is worth pointing out that, setting aside the technical details, the differences between opponents and proponents of the hole argument are smaller than they appear. For Weatherall (2018), and Halvorson and Manchak (2022), all, in effect, defend something like sophisticated substantivalism, as indeed do Pooley and Read (2021), and many others.\footnote{Weatherall (2018:~346-347) explicitly says that, in spite of his not being committed to a Lewisian metaphysics, his views are close to those of Butterfield (1989) and Brighouse (1994). And Halvorson and Manchak (2022:~Section 7) endorse `metric realism, that is, the claim that spacetime has metric structure'. And so, they seem to also commit to the existence of the {\it spacetime} that is individuated by its metric structure.} 
The difference is that Weatherall, Halvorson and Manchak do not view sophisticated substantivalism as a response, or as a resolution of, the hole argument through {\it metaphysical} reasoning. Rather, they take themselves to be providing a correct interpretation of the {\it mathematics} of general relativity. By contrast, authors such as Pooley and Read {\it do} defend sophisticated substantivalism as a metaphysical response to the hole argument. Thus their responses are different ways to reach the same, or at least a very similar, conclusion about substantivalism. (Here we again recall the quote from Lewis, at the end of the previous Subsection.)

\subsection{The hole argument with a cosmological constant}\label{holecosm}

This Section discusses a version of the hole argument for spacetimes with a non-zero (positive or negative) cosmological constant, $\L$. The overall aim is to argue that some diffeomorphisms do change the physical possibility being represented, so that Leibniz equivalence is not always true (see Section \ref{historyH} (A)).\footnote{This Section is based on De Haro (2017; 2022:~pp.~227-231) and De Haro, Teh, and Butterfield~(2016:~pp.~1067-1069; 2017:~pp.~78-79). The analysis has been explicitly done for a negative cosmological constant, but the results can be analytically continued to the case of positive $\L$. See also Balasubramanian et al.~(2001:~p.~3) and Anninos et al.~(2011:~pp.~3-6).} 

Considering non-zero $\L$ allows for the hole argument to interact with gauge-gravity dualities. We will see that different kinds of diffeomorphisms correspond to different symmetries of the common core theory and of its models: 

(1) `Hole diffeomorphisms' will be argued to be part of the specific structure of the gravity model, and so these are {\it proper symmetries of the gravity model} (see our classification of symmetries in Section \ref{dualsym}). These diffeomorphisms are `invisible' to the other dual, and to the common core. 

(2) Diffeomorphisms that generate conformal symmetries on the conformal boundary are {\it stipulated symmetries} of the common core. Thus these diffeomorphisms are `visible' to the other dual, and to the common core. If these symmetries are defined and interpreted in the common core theory as `gauge' redundancies (as against their being physical symmetries), then these diffeomorphisms are also candidates for satisfying Leibniz equivalence.\footnote{Diffeomorphisms of type (2) are the analogues of the boosts, rotations, and translations at infinity, that Belot (2018:~p.~966) discusses in the asymptotically flat case. For non-zero $\L$, one gets the whole conformal group, i.e.~there are additional generators, of special conformal transformations and scalings.}

But perhaps surprisingly, these results highlight that Leibniz equivalence with non-zero $\L$ depends on both the formal and the interpretative aspects of the Schema, as follows:

(A) {\it Different types of diffeomorphisms are in different ``parts'' of the Schema}: diffeomorphisms of type (1) are symmetries of one of the duals, while diffeomorphisms of type (2) are symmetries of the bare theory. This shows how different verdicts of Leibniz equivalence have to be considered in different ``parts'' of the Schema, i.e.~in the bare theory vs.~in one of its representations.

(B) {\it Leibniz equivalence depends on the interpretation}: whether Leibniz equivalence obtains depends on how the diffeomorphisms are interpreted within the common core, and within each of the duals. For example, whether stipulated symmetries satisfy Leibniz equivalence depends on how the common core is interpreted.

Aspect (A) will deepen, because it will make concrete in the context of our Schema, Belot's (2018:~p.~968) remark that one should think of the solutions of general relativity as forming a structured space. Namely, the Schema gives us a natural classification of the different types of diffeomorphisms, in terms of (1) and (2) above (and one should remember that the common core here has a state-space of one dimension lower!). (Aspect (A) of course ties into our own geometric view of theories, which we shall develop in Chapter \ref{Heuri}.)\footnote{Aspect (B) seems to differ from Belot (2018), since it is not clear whether he recognizes that, even within a given type of symmetry like (2), Leibniz equivalence depends on the interpretation that one gives to the common core and to the models.}

Regardless of the hole argument, the argument below has its own significance for: (a) the empirical significance of diffeomorphisms; (b) the status of diffeomorphisms vis-\`a-vis gauge-gravity dualities, i.e.~as being part of the common core or part of the specific structure.

There are two important additional reasons for considering the hole argument with a cosmological constant. First, the cosmological constant should be brought into the hole argument. For, while Landsman (2022:~Section 1) correctly argues that including {\it matter fields} into the hole argument is unnecessary, his argument does not go through for the cosmological constant---which, unlike the matter fields, {\it cannot} be set to zero by focussing on a region of ``empty space'': and it does affect the global structure of the spacetime (see Penrose and Rindler (1986:~p.~353)). After all, evidence points to the cosmological constant in our universe being positive. And, although infinitesimal regions can always be approximated by a Lorentzian spacetime, this is of course not true for large regions. Also, it is interesting in its own right to consider what kind of trouble the hole argument spells when we let it interact with the global structure of the spacetime, in the way that we will discuss below. 

Second, standard discussions of the hole argument, including our exposition in Section \ref{holeA}, require that the {\it hole diffeomorphism}, $\phi$, is the identity mapping outside the hole. As this Section will explain, in the context of Einstein's field equations with a non-zero cosmological constant, this is not the correct condition: in one sense it requires a strengthening, and in another sense it allows a weakening.\footnote{For a full discussion, see De Haro (2017; 2022:~pp.~227-231). For a comparison with gauge symmetries, see De Haro, Teh, and Butterfield~(2016:~pp.~1067-1069; 2017:~pp.~78-79).}
Although our argument will be set in a specific cosmological setting, its consequence---that more discrimination is required about the relevant class of hole diffeomorphisms---applies more generally, even though the details may differ.\footnote{In the case of asymptotically flat spacetimes, our point below, that one must distinguish diffeomorphisms that are isometries outside the hole, and those that only fix part of the boundary structure (i.e.~our cases (1), respectively (2) and (3), below), has also been made, in general terms, by Belot (2018:~p.~965).}\\

We will adopt the following cosmological setting: the region outside the hole is the asymptotic boundary of the spacetime, and the hole is the interior region, of which we will only need to consider a neighbourhood. The `asymptotic boundary' is, in the case of a positive cosmological constant, future timelike infinity; in the case of a negative cosmological constant, it is the asymptotic boundary at spatial infinity. (Although an asymptotically locally anti-de Sitter space does not in general have a Cauchy surface, the asymptotic boundary in effect plays the role of the Cauchy surface, because it defines a unique boundary problem for the fields, including the metric tensor.)\footnote{See Avis et al.~(1978:~p.3567), Witten (1998:~pp.~4-7), and Skenderis (2002:~pp.~5853-5855). For a careful discussion of the initial and boundary conditions in the case of Lorentzian signature, see Skenderis and van Rees (2008, 2009).}

This means that, when the cosmological constant is non-zero, the hole argument implies the incompatibility of substantivalism and indeterminism, with two different examples of indeterminism, with the nature of the example tied to the structure of the region that is the candidate to do the determining: namely, the possibility to postdict the past (if the cosmological constant is positive) or to predict a unique spatial development of the geometry given the boundary data (if the cosmological constant is negative). Although the former seems to give a metaphysically more severe problems than the latter, both cases give one reasons to worry. For in both cases, substantivalism implies the {\it lack of uniqueness} of the geometry described by the solutions of the Einstein field equations, given data that are normally sufficient to solve a well-defined boundary problem. And, following Earman and Norton (1987:~p.~524), lack of uniqueness is a high price to pay for a substantivalist metaphysics. Thus, also if the cosmological constant is non-zero, substantivalism is prima facie faced with a problem that it needs to address.\\

As we said in our second `additional reason' at the beginning of this Section, hole diffeomorphisms must be defined using a stronger condition, since the requirement that the diffeomorphism goes to the identity is too weak. The reason is that a diffeomorphism that `is the identity at the boundary' does not in general fix the {\it quantities} (in the Schema's sense) nor the {\it asymptotic states} of general relativity. Thus a diffeomorphism may be the identity outside the hole, and still relate distinct physical situations.

Thus we will consider both strenghtenings and weakenings of this requirement, by distinguishing three properties below that a diffeomorphism can have. Property (a) is the one just discussed, i.e.~that the diffeomorphism is the identity at the hole's boundary, while properties (b) and (c) fix other (boundary) structure:

(a)~~{\bf Identity:} the diffeomorphism $\f$ is the identity at the boundary of the hole, i.e.~$\f|_{\pa M}=\mbox{id}_{\pa M}$, where $\pa M$ is the asymptotic boundary of $M$. This condition can be understood as the diffeomorphism's fixing the points of the boundary manifold, i.e.~it maps each boundary point to itself.

(b)~~{\bf Boundary isometry:} the diffeomorphism $\f$ is an isometry of the boundary metric\footnote{This is in fact a {\it conformal} metric, i.e.~it is only defined up to a conformal factor. The general notion is in Penrose and Rindler (1986:~Chapter 9). For the conformal properties of asymptotically locally anti-de Sitter spaces, as well as more general geometric aspects, see Fefferman and Graham (1985, 2012) and Biquard (2005).\label{confP}} 
$g$ defined on the boundary manifold, i.e.~for all boundary points $p$, $(\f^*g)(p)=g(p)$.\footnote{This condition is distinct from (a), because it is defined with respect to the conformal boundary metric, and not with respect to the full metric in the interior of the spacetime. Thus the condition under consideration is $(\f^*|_{\pa M}g)(p)=g(p)$, which is {\it not} given by $(\f^*|_{\pa M}g)(p)=g(\f|_{\pa M}(p))$, since this last condition only holds for the metric in the interior and {\it not} for the metric on the conformal boundary.}

(c)~~{\bf Normal form:} the diffeomorphism $\f$ fixes the normal form of the spacetime metric, i.e.~the metric written in (Poincar\'e normal form) coordinates $(r,x)$ that are adapted to the geometry of the boundary. In terms of such coordinates, the boundary surface is parametrized by $r=0$, and it is spanned by the remaining coordinates $x$.\footnote{For details, and proofs of the various results below, see De Haro (2017); and (2022).}

Note that, in the boundary problem where the spacetime metric is fully determined by the boundary metric $g$, conditions (b) and (c) together imply that the diffeomorphism $\f$ is an isometry of the full spacetime metric (and not only an isometry of the boundary metric).\footnote{There are different uses of the word `isometry' in the literature. We use isometry in the sense of Hawking and Ellis (1973:~p.~43), i.e.~as preservation of scalar products of tangent space vectors under a diffeomorphism. A vector field that generates a one-parameter group of isometries is always a Killing vector field: see also Malament (2012:~pp.~74-75) and Nakahara (2003:~pp.~273, 279).} 
But we do not need to require such a strong condition.

Given the above general properties of a diffeomorphism $\f$ with respect to its behaviour outside the hole (i.e.~on the boundary of the spacetime), we have the following four jointly exhaustive, and mutually exclusive, classes of diffeomorphisms:

(1)~~{\bf Hole diffeomorphisms:} these diffeomorphism go to the identity, and they are also isometries, outside the hole. Thus they satisfy conditions (a) and (b), i.e.~they fix the boundary manifold and the boundary metric.

(2)~~{\bf Boundary symmetries:} these diffeomorphisms fix the normal form of the metric plus one other boundary structure (i.e.~they either go to the identity at the boundary, or they fix the boundary metric), hence they have properties (b) and (c) or (a) and (c). The reason for the name `boundary symmetries' is that these diffeomorphisms generate the {\it conformal group} of transformations on the boundary manifold.\footnote{This is the analogue, for the case of non-zero $\L$, of the BMS group that one finds when $\L$ is zero. The group of asymptotic conformal symmetries for three-dimensional spacetimes with negative $\L$ was worked out in Brown and Henneaux (1986:~pp.~213-218), and in any dimension by Imbimbo et al.~(2000:~p.~1130-1132). De Haro (2017a:~pp.~1477) pointed out that the class of boundary diffeomorphisms is larger than the ones derived by Imbimbo et al., and derived the general conformal Killing equation satisfied by the boundary diffeomorphisms. The analysis was extended to positive $\L$ in Strominger (2001:~pp.~4-5), Balasubramanian et al.~(2001:~pp.~7-8) and Anninos et al.~(2011:~pp.~3-4). See also Penrose and Rindler (1986:~Section 9.8), and the references in footnote \ref{confP}. For the notion of asymptotic symmetries in general relativity, see also Arnowitt et al.~(1959, 2008), Sachs (1961, 1962), Bondi et al.~(1962), Penrose (1963, 1964), Newman et al.~(1966), Geroch (1972), and Ashtekar et al.~(1978).}

As we will argue below, the above two types of diffeomorphisms are the relevant ones for discussions of the hole argument. Each of them has {\it two} of the properties defined above, i.e.~they each preserve two boundary structures, and so they are `intermediate' cases. In addition, there are two `extreme' cases (3) and (4): we will argue that (3) is irrelevant to the hole argument, and (4) is an empty class:

(3)~~{\bf Large diffeomorphisms:} these diffeomorphisms either have just {\it one} of the three properties above, or {\it none}. 

(4)~~{\bf Hole isometries:} these diffeomorphisms have all {\it three} properties above, and so (assuming that the spacetime metric is fully given in terms of the boundary metric) they are spacetime isomorphisms that are the identity at the boundary.

We now comment on the physical interpretation of these four types of diffeomorphisms, which exhaust all the possibilities with respect to the given boundary structures.

\begin{figure}
\begin{center}
\includegraphics[height=4cm]{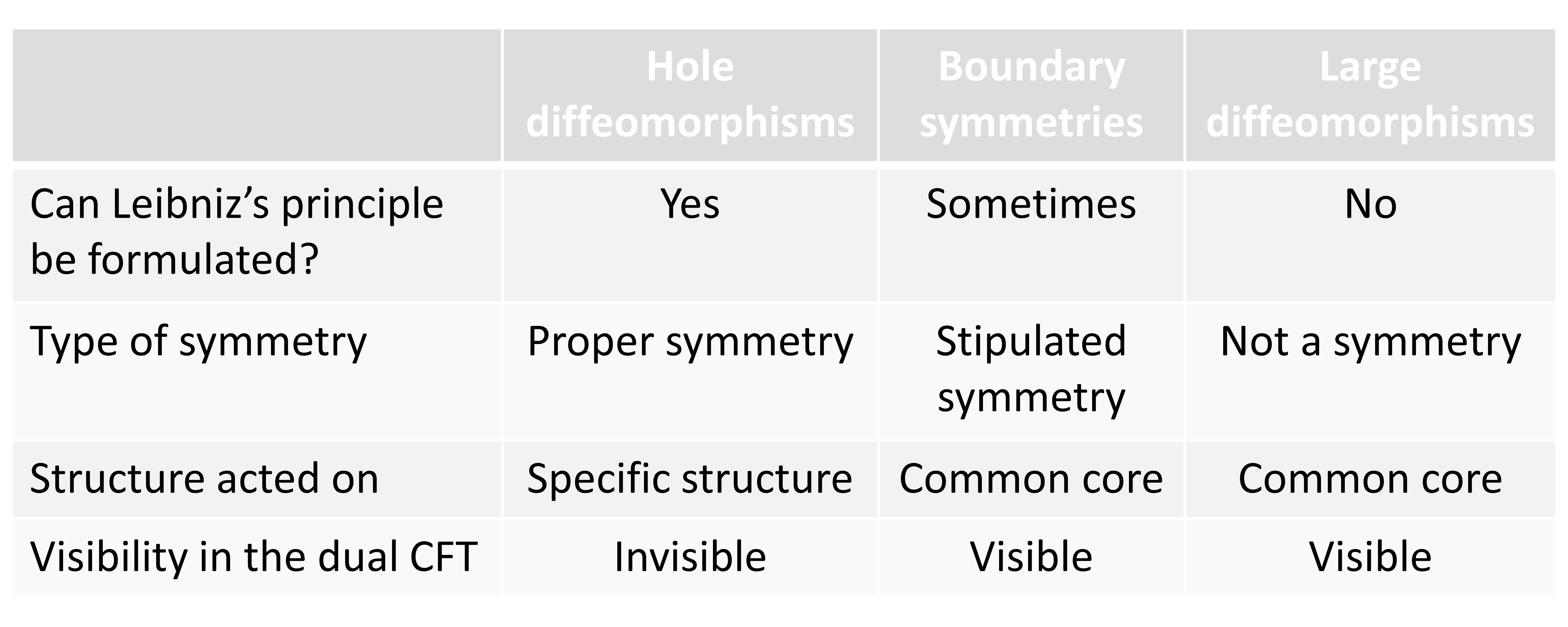}~~~
\caption{\small Summary, for each non-empty class of diffeomorphism (1) to (3), of the analysis with respect to Leibniz equivalence, type of symmetry, and their relation to the common core and specific structure.}
\label{Hole_diffeos}
\end{center}
\end{figure}

(1)~~{\it Hole diffeomorphisms} are the relevant ones in discussions of the hole argument, as we will also discuss below. These diffeomorphisms do not change any of the asymptotic physical properties that characterize a spacetime, such as its ADM or Bondi mass and angular momentum, its quasi-local energy and momentum (as defined e.g.~by Brown and York (1993)), etc. When such a diffeomorphism acts on the spacetime, it generates a `distinction without a difference' (assuming one of the responses to the hole argument discussed in Section \ref{responsesH}): they express specific structure that is used in the description of the spacetime, and they relate {\it physically equivalent} situations. Hole diffeomorphisms can be taken to be `gauge', in the philosophical sense of the word, i.e.~redundancies of the description.\footnote{For a discussion of this point, see De Haro et al.~(2017).}

(2)~~{\it Boundary symmetries} are a very important class of diffeomorphisms, since they characterize the asymptotic symmetries of the spacetime: namely, conformal symmetry.\footnote{For AdS$_{d+1}$, the asymptotic symmetry group is $\mbox{SO}(2,d)$. For de Sitter$_{d+1}$, it is $\mbox{SO}(1,d+1)$.}
In the case of a negative cosmological constant, this conformal symmetry is mapped, by gauge-gravity duality, to the conformal group of the dual CFT. In general, these diffeomorphisms {\it change the physical state}, since they change the asymptotic metric by a conformal factor, and thus also change the values of other physical quantities that characterize the spacetime, such as the quasi-local energy-momentum tensor defined on the boundary.\footnote{See Skenderis (2001:~Eq.~(20)). For a philosophical discussion, see De Haro (2022:~p.~229).}

However, as we suggested in De Haro et al.~(2017:~p.~79), in some cases there can be an internal perspective, in which only conformal classes of observables are considered physical. (This then agrees with the situation in the CFT, where conformal symmetry is part of the definition of the theory in some, but not all, cases: for example, because of the conformal anomaly, conformal symmetry is not a symmetry of every CFT on a curved spacetime.) Then, on an internal interpretation of the conformal symmetry under which conformal symmetry-related states are physically equivalent, these diffeomorphisms are also candidates for hole diffeomorphisms, i.e.~the notion of hole diffeomorphisms in (1) can be weakened to admit boundary symmetries: the diffeomorphisms (2) are then also `gauge'.\footnote{This is only possible if the conformal anomaly is zero, which is the case if $d$ is odd.}

(3)~~{\it Large diffeomorphisms} are physically relevant, but they are irrelevant to the hole argument. For in general they change the physical properties of the spacetime, such as its ADM mass, and so they are not the kinds of diffeomorphisms considered in the hole argument: they change the properties of the spacetime not only in the interior of the hole, but also outside. Thus they relate physically inequivalent situations.

(4)~~{\it Hole isomorphisms} form an {\it empty} class, i.e.~there exist no hole diffeomorphisms that have all three properties (a) to (c).\footnote{This is a consequence of Theorem 3 in De Haro (2017b:~p.~1478). See also the discussion in De Haro, Teh and Butterfield (2016:~p.~1067; 2017:~p.~78).}
This means that the class of non-trivial hole diffeomorphisms is smaller than is sometimes expected.

Most analyses of the hole argument proceed by general arguments regarding spacetimes and diffeomorphisms. In particular, they take models of general relativity to be pairs of a manifold and a metric: in the familiar notation, ${\cal M}=(M,g)$. But on the Schema's analysis, our discussion here takes into account the set of {\it quantities} and their values. For example, ADM and Bondi mass and momentum, and the quasi-local energy and momentum defined at the boundary of the hole, need to be taken into account.\footnote{One might think that it suffices to have a manifold $M$ and a smooth metric $g$ that solves the Einstein field equations. But this does not suffice for a well-defined solution of general relativity as a {\it theory}, because there are smooth metrics that do not yield finite quasi-local energy. For a discussion, also in comparison with other notions of energy and momentum in general relativity, see De Haro (2022).}

This analysis shows that there is a rich landscape, with three non-trivial classes of diffeomorphisms, one of which is the set of genuine hole diffeomorphisms, another the set of large diffeomorphisms (which are irrelevant to the hole argument), and another the intermediate class of boundary symmetries.

This proves that Leibniz equivalence, as normally stated to obtain between `diffeomorphic models' (see Section \ref{historyH}), is false, because it does not obtain for the class (3) of large diffeomorphisms (even if such a diffeomorphism might go to the identity outside the hole). Rather, Leibniz equivalence can be upheld for the class (1) of hole diffeomorphisms---e.g.~on a sophisticated substantivalist view. And, in some cases, it can also be upheld on an internal interpretation of class (2) of boundary symmetries.\footnote{For a detailed discussion of this point, see De Haro et al.~(2017:~p.~79).}

\subsection{The hole argument, duality and Leibniz equivalence}\label{holegg}

If the gravitational theory has a dual, and in particular in the case of negative cosmological constant, the analysis in the previous Section can be used to study how the diffeomorphisms are mapped across gauge-gravity duality. Thus we can verify whether this mapping agrees with our expectations about the common core theory of gauge-gravity duals (in Section \ref{matchpf}), also based on the known spacetime properties of the dual CFT. We will here focus on the two classes of diffeomorphisms that are relevant for the hole argument: namely, (1) and (2) in the previous Section.

Since the duality maps only the triples of states, quantities and dynamics but not the specific structure, we expect that there are no diffeomorphisms in the CFT that correspond to the class (1) of hole diffeomorphisms, since these are part of the specific structure. This is indeed the case: using the duality map in Eq.~\eq{AdSCFTdict}, it follows that the CFT model is inert under hole diffeomorphisms carried out in the gravity model, i.e.~there are no transformations that correspond to these diffeomorphisms in the CFT. For this reason, these diffeomorphisms are called {\it invisible}, since they are not seen by the dual CFT (De Haro (2022:~p.~229; 2017:~p.~1468)). In the terminology of Section \ref{dualsym}, hole diffeomorphisms i.e.~class (1), are {\it proper symmetries of the gravity model}, i.e.~they are symmetries of the specific structure of the gravity model.\footnote{There is therefore a sense in which these diffeomorphisms are emergent. We will discuss this in Section \ref{emergence}.}

The boundary symmetries, (2), are {\it visible} in the CFT, since they are mapped to conformal rescalings of the metric of the CFT, so that the quantities are also modified. The conformal group that was found in the previous Section for this class of diffeomorphisms is then mapped, by the duality map, onto the conformal group of the CFT. In fact, this correspondence between symmetries was one the insights leading up to AdS-CFT.\footnote{See Maldacena (1997:~pp.~1114-1115) and Brown and Henneaux (1986:~pp.~213-218).} 
In the terminology of Section \ref{dualsym}, boundary symmetries i.e.~diffeomorphisms of class (2), are {\it stipulated symmetries} of the common core, because they are represented by both duals (even though in different ways: as the asymptotic symmetries of a spacetime, or as the conformal symmetry of a CFT).

That these two classes of diffeomorphisms (i.e.~the hole diffeomorphisms and the boundary symmetries) are mapped in different ways by the duality map (namely, one class is not mapped at all, while the other is mapped onto the conformal symmetries of the CFT) is consistent with, and thereby supports, our statement that the usual formulation of Leibniz equivalence, for diffeomorphisms {\it tout court}, is false, because one must distinguish different types of diffeomorphisms. As follows:

{\it Hole diffeomorphisms} are not mapped by the duality, since they are part of the specific structure. Thus, on an internal interpretation of the duality, these diffeomorphisms are not in the ontology of the common core theory. This is consistent with Read's (2016:~p.~224) analogy, mentioned in Section \ref{uds}, between anti-haecceitism and internal interpretations: on an anti-haecceitistic position, Leibniz equivalence obtains for hole diffeomorphisms. And on an internal interpretation of the duality, hole diffeomorphisms are not in the ontology of the common core theory.\footnote{We will discuss the comparison between an internal interpretation of dualities and sophisticated substantivalism further in Section \ref{cco}: see also De Haro, Teh, and Butterfield (2016, 2017).} 

For {\it boundary symmetries}, the question whether Leibniz equivalence obtains depends on the interpretation adopted: and this corresponds to the question whether the common core theory's symmetries are redundancies (i.e.~`gauge' symmetries, in the philosophical sense) or physical symmetries.\footnote{For a more detailed discussion of this, see De Haro, Teh, and Butterfield (2016, 2017) and De Haro (2017a).}

\section{Black hole microstates}\label{ocs}

Ever since Hawking's 1974 discovery that black holes radiate, black holes have become standard laboratories for studying the quantum nature of spacetime. Thus the black hole's event horizon, more distinctly than the singularity, became the semi-permeable membrane where to peer into yet unknown quantum gravity effects.\footnote{This Section has benefitted from collaborations in various projects with Jeroen van Dongen and with Manus Visser. It is largely based on De Haro et al.~(2020) and van Dongen et al.~(2020).}
There were two main reasons for this.

First, Hawking radiation realizes a physical process that is classically apparently impossible, analogous to the quntum tunneling of particles through a potential barrier, as a purely gravitational effect. Since this process can decrease the mass of the black hole, turning the mass into kinetic energy for the outgoing matter particles, this suggests that quantum gravity effects can convert geometry (back) into matter. To a number of researchers, this indicated that black holes are like other particles, elementary or otherwise: they can form and decay.

Second, less than two years after his discovery of black hole radiation, Hawking speculated that black holes {\it destroy} information (rather than just storing it, or redirecting it elsewhere in spacetime), and so that, in the presence of black holes, the future cannot be predicted.\footnote{Note that, for Hawking (1976:~p.~2460), the presence of the singularity was essential in producing the unpredictability: `Because all known laws of physics are formulated on a classical space-time background, they will all break down at a singularity. This is a great crisis for physics because it means that one cannot predict the future: One does not know what will come out of a singularity'.}
(Thus there {\it is}, even more than we discussed in the preamble of this Chapter, an analogy betwen Hawking's quantum unpredictability argument and Einstein's hole argument.) He then went on to speculate that quantum gravity brings in an uncertainty that is complementary to Heisenberg's uncertainty principle. In the presence of black holes, the Schr\"odinger equation needs to be replaced by a more general equation that describes, through what he dubbed the `superscattering operator', the time evolution of pure density matrices into mixed density matrices.\footnote{In the philosophical literature, this has motivated the `open systems view': see Cuffaro and Hartmann (2023:~Section 1).}

Thus Hawking radiation promises to hold the key to the mysteries of spacetime in quantum gravity. Accordingly, a great deal of work in physics has been done on understanding the quantum properties of black hole horizons. Prominent among these is the formulation of the holographic principle, which in turn galvanised---through the studies of black hole entropy that we will review in this Section---gauge-gravity duality (see Section \ref{ggd}).

The philosophical discussion of these matters has just hardly begun. For the most part, the philosophical discussion has so far focussed on the derivation of the Unruh effect,\footnote{For a philosophical discussion, see Earman (2011).}
the relation between black hole thermodynamics and ordinary thermodynamics,\footnote{See Dougherty and Callender (2016) and Wallace (2018, 2019).}
and the significance of analogue experimetns for the confirmation of Hawking radiation of black holes.\footnote{See Dardashti et al.'s~(2019) Bayesian analysis vindicating that analogue experiments are confirmatory. Crowther et al.~(2021) counter that analogue experiments cannot confirm the existence of Hawking radiation, because they presuppose the validity of quantum field theory in curved spacetime and the universality of Hawking radiation (i.e.~its independence of high-energy effects), which according to them are the real issues. For a discussion of the universality of Hawking radiation, see Gryb et al.~(2021). For the epistemic function of Wheeler's and Geroch's thought experiments on black hole thermodynamics, see El Skaf and Palacios (2022).}
The attention that the information paradox and some of its proposed resolutions have received has so far been scattered.\footnote{See e.g.~Belot et al.~(1999) and van Dongen and De Haro (2004). A historical and philosophical discussion of the black hole information paradox is in van Dongen, De Haro and Landsman, {\it History and Philosophy of the Black Hole Information Paradox} (forthcoming). For a recent discussion of black hole complementarity, see Muthukrishnan (2022).}

\subsection{Statistical mechanics of Bekenstein-Hawking entropy}

A central question regarding Hawking radiation is its statistical mechanical underpinning. Bardeen, Carter and Hawking (1973) had discovered a suggestive analogy between the laws of black hole mechanics and thermodynamics, with the horizon area $A$ in the role of entropy, the surface gravity of the event horizon playing the role of temperature, and the stationarity of the event horizon playing the role of thermal equilibrium. Hawking's (1974, 1975) discovery, in the following year, of black hole radiation, was the ``missing link'' that would assign the black hole a {\it non-zero} `Hawking' temperature, thereby completing the analogy between black hole mechanics and thermodynamics:
\bea
T_{\tn H}={\hbar c^3\over8\pi G_{\tn N}k_{\tn B}M}\,.
\eea
This also enabled the precise calculation of the coefficient linking entropy to area in the Bekenstein-Hawking entropy formula:
\bea
S_{\tn{BH}}={c^3A\over 4G_{\tn N}\hbar}\,.
\eea
Thus Hawking's discovery confirmed Bekenstein's (1972, 1973, 1974) earlier work on black hole entropy and the generalized second law, which Hawking had previously opposed, that the entropy in the black hole exterior plus the black hole entropy never decreases. Thus the analogy between black holes and thermodynamics was complete: black hole thermodynamics could now be seen as a special case of thermodynamics.

But to understand black holes as thermodynamical systems was not yet to understand the microphysical undepinning of Hawking radiation and of the states that are counted by the Bekenstein-Hawking entropy formula. For most felt that identifying the microscopic states is bound to cast light on the nature of the quantum degrees of freedom underlying a classical black hole, and the quantum nature of spacetime. 

Following the analogy between black holes and the statistical underpinning of thermodynamics, it seems natural to suggest that, just as thermodynamic systems are often considered to emerge from the underlying microscopic statistical mechanical systems, general relativity should be regarded as an effective field theory that emerges from the microscopic degrees of freedom. And thus, spacetime itself is emergent from the underlying quantum structure. 

At this juncture, there are two basic options: the underlying fundamental degrees of freedom could themselves be gravitational degrees of freedom (in some appropriate sense: they could be geometric degrees of freedom, of quantum versions of the gravitational degrees of freedom) or they could be degrees of freedom of matter fields. 

While these two options are often associated with competing quantum gravity programmes (loop quantum gravity and spin foams favouring gravitational degrees of freedom, and string theory favouring matter), one might take gauge-gravity duality to suggest that this distinction may not be so fundamental. 

The following two Sections focus on the microscopic origin of black hole entropy in string theory, and in particular on the class of black hole entropy countings advanced by Strominger and Vafa (1996). The title of their paper lays a claim to the `Microscopic Origin of the Bekenstein-Hawking Entropy'. Accordingly, their aim is not just to {\it count} the microscopic degrees of freedom underpinning black hole entropy, but also to {\it explain} its origin. They do this by using {\it matter}: namely, sets of D-branes. However, their work crucially also depends on the idea of open-closed string duality, an effective duality in string theory at the origin of gauge-gravity duality.

\subsection{Microscopic origin of black hole entropy in string theory}

Our topic in this Section is the microscopic origin of black hole entropy for an important class of black holes in string theory: their mass, angular momentum, and electric charges satisfy extremality or near-extremality conditions.\footnote{The black hole considered by Strominger and Vafa (1996) has two electric charges, corresponding to two independent gauge fields (see Section \ref{dynD}). Subsequent versions by Breckenridge, Myers, Perry and Vafa (1997:~p.~93) have three charges. See also Maldacena (1996), Peet (2001) and Johnson (2003).}
`Extremal' means having maximum possible electric charge, i.e.~the black hole saturates the BPS bound for the mass, Eq.~\eq{bps} (See Section \ref{Dcst} (3)).\footnote{For a discussion of the physical significance of these solutions, see De Haro et al.~(2020:~pp.~97-99).}

The calculation reported in Strominger and Vafa's (1996) brief (6-page!) paper is easy to summarize but it is not straightforwardly unpacked. (Fortunately for us, the required footwork was done in Section \ref{Dcst}.) They calculated the degeneracy of the ground state of a set of D-branes (which is a bound state with the D-branes on top of each other) with given mass and electric charges. The result agreed, in leading order, with the Bekenstein-Hawking entropy of a black hole with the same mass and electric charges (in units where $c=\hbar=1$):
\bea\label{SVentropy}
S_{\tn{BH}}={A\over4G_{\tn N}}=2\pi\sqrt{Q_1Q_2^2\over2}\,,
\eea
where $Q_1$ and $Q_2$ are the two independent electric charges. (Subsequent work generalized this to black holes with non-zero angular momentum and a with third electric charge, and found again agreement, also for the sub-leading corrections to the Bekenstein-Hawking formula.)\footnote{For an overview, see De Haro et al.~(2020:~pp.~99-101).}

In short, this was the first successful microscopic calculation of black hole entropy in string theory. But how to understand the calculation and its significance?\\

\begin{figure}
\begin{center}
\includegraphics[height=5cm]{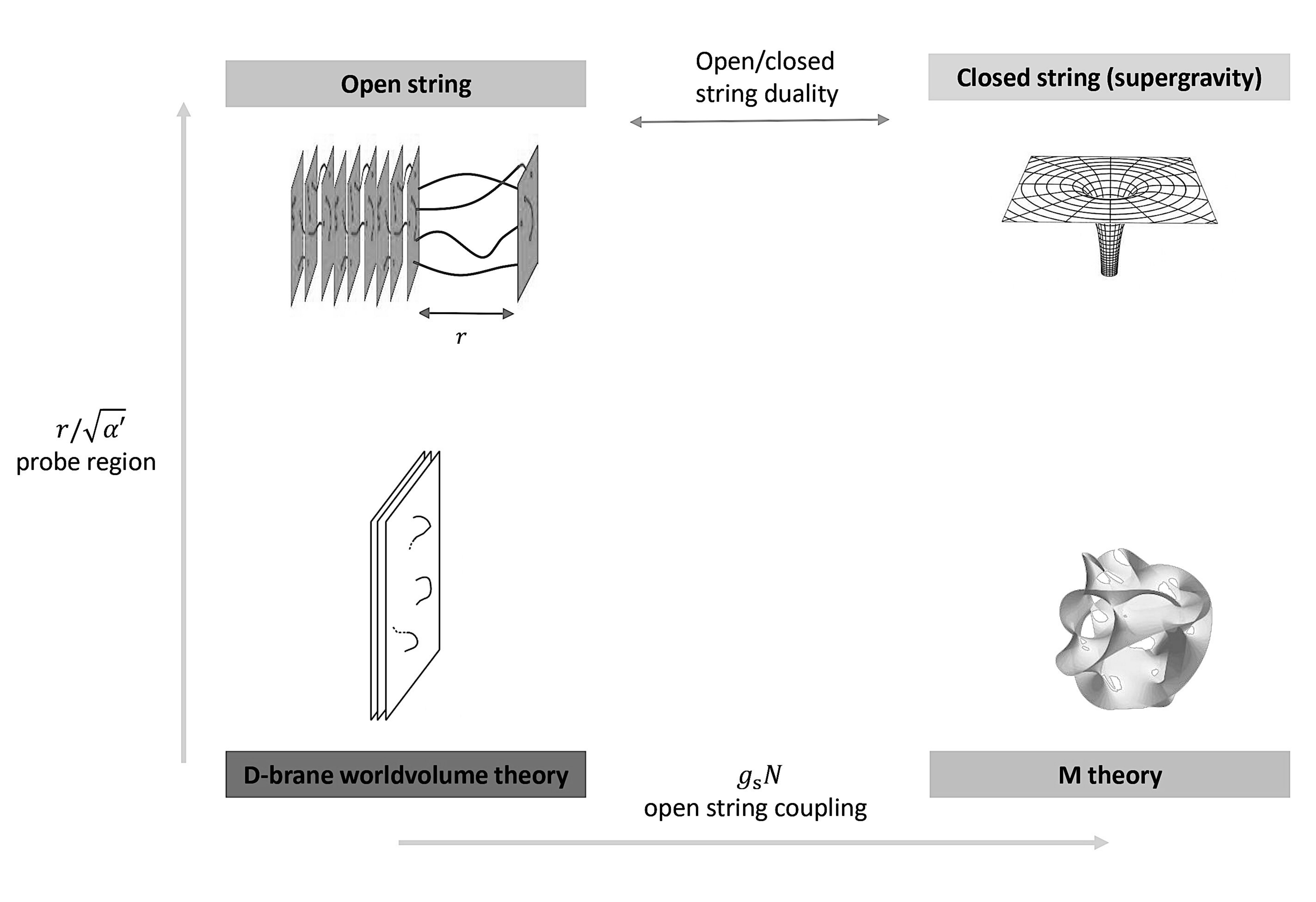}~~~
\caption{\small Anatomy of the Strominger-Vafa entropy counting argument. {\it Top right}: five-dimensional black hole solution, whose entropy is given by the Bekenstein-Hawking formula. {\it Bottom left}: microscopic system i.e.~quantum field theory on the world-volume of the D-branes, where the microscopic counting is done. {\it Top left}: open string theory description, used by the argument to relate the top-right and the bottom-left corners. {\it Bottom right}: the argument does not use this corner, but the microscopic counting is thought to be relevant for the formulation of a full quantum theory of gravity.}
\label{SV2}
\end{center}
\end{figure}

The basic structure of the Strominger-Vafa argument was laid out in Figure \ref{SV}. Here we fill in the details, using the new Figure \ref{SV2} (adapted from De Haro et al.~(2020:~p.~95)). The black hole is in the top-right corner: like any black hole, it is uniquely defined by its mass, angular momentum, electric charges, and the dimension of spacetime. The black hole is a macroscopic physical object, described by type IIA supergravity (see Section \ref{dynD}).\footnote{Recall, from Section \ref{Dbsugra}, that type IIA supergravity has $p$-brane solutions for $p=0,2,4,6$. The Strominger-Vafa black hole is a solution for a system of D0-D2-D4 branes intersecting on the world-volume of a fundamental string on $S^1$.}
It is in effect five-dimensional: it is obtained by compactifying type IIA supergravity from ten dimensions to five.\footnote{For a discussion of compactification, see Section \ref{11DM}. The five-dimensional manifold on which the solution is compactified is $K3\times S^1$, where $K3$ is a compact, Ricci-flat, hyperk\"ahler manifold, and the string of the previous footnote is wound around this $S^1$.\label{K3S1}} 
(Subsequent work has performed similar calculations for four-dimensional black holes.)

The microscopic system is in the bottom-left corner of Figure \ref{SV2}. It is described by a quantum field theory: namely, the world-volume theory of the set of D-branes with the same mass, angular momentum, and charges as the macroscopic black hole. This world-volume theory is a (1+1)-dimensional CFT on $S^1\times\mathbb{R}$, where $S^1$ is a spatial circle and $\mathbb{R}$ is time.\footnote{More precisely, the theory is a supersymmetric sigma model, i.e.~a quantum field theory, with as fields maps from $S^1\times\mathbb{R}$ into the moduli space of the D-branes, i.e.~the space of possible field values that describe the degrees of freedom of the bound states of the D-branes (we already encountered the notion of moduli space in Sections \ref{EMYM} and especially \ref{swt}). This moduli space describes the possible motions of the D0-branes in the K3 in the T-dual picture in Type IIA, up to permutations of the D-branes. See De Haro et al.~(2020:~p.~92).}
This (1+1)-dimensional world-volume is the place where a set of D1-D3-D5 branes in Type IIB string theory intersect: namely, the world-volume of the D1 (and D1-D3-D5 are the types of D-branes required to reproduce the black hole charges).\footnote{There is also a contribution from the momentum $P$ along the $S^1$, which gives a charge equal to the fundamental string charge in the Type IIA system.}
The entropy is the logarithm of the degeneracy of states in the ground state, which is calculated using a standard result about the asymptotic growth of the number of states for fixed energy in a CFT: namely, the Cardy formula.\footnote{For a discussion, see De Haro et al.~(2020:~pp.~92-93).}

\subsection{Structure of the argument and inference relations}\label{strar}

The relation between the macroscopic black hole and the microscopic D-branes, that justifies the microscopic origin of the entropy, Eq.~\eq{SVentropy}, is best characterized as one of `correspondence'.\footnote{For a discussion of, and a comparison with, Bohr's correspondence principle in this context, see van Dongen et al.~(2020:~pp.~122-124). For a more general discussion, see Radder (1991) and Post (1971).} 
In this Section, we unpack this correspondence by identifying three chief assumptions that the Stominger-Vafa argument depends on. These three assumptions are, in effect, about the models, the states, and the quantities involved, as follows.

{\bf (i) (Effective) duality:} the two physical systems, namely the macroscopic and the microscopic one, are described by two different models (viz.~types IIA and IIB string theories). And these models are related by {\it T-duality} (see Section \ref{T-d}).\footnote{The T-duality is along the $S^1$ on which the fundamental string in Type IIA is compactified (thus a small circle). Under T-duality, this corresponds to the large circle that (with the time direction) is the world-volume of the D1-branes of Type IIB. T-duality is a technical assumption that later work dispensed with, by doing all the calculations with the D1-D5 system.} 

But this does not suffice to match the states and quantities of the two systems. And so, one assumes what, in Section \ref{Dcst} (points (3) and (4)), we called the `alternative i.e.~dual description' of closed string states as open string states: namely, the right-hand side of Figure \ref{SV2} is the {\it closed string sector} of string theory, while the left-hand side is the {\it open string sector}. These two subsectors (i.e.~subsets of states and quantities) are part of a {\it single string theory}. Thus, building on Polchinski (1995), Strominger and Vafa assumed that these two sectors give, for a regime of parameters, different descriptions of the same type of system (more on this below). Here one does not expect a full duality between the two sectors, but only an {\it effective duality} for selected sets of states and quantities.

{\bf (ii) Regimes:} matching of states and quantities (and the systems they describe, namely the black hole and the set of D-branes) can only be expected in appropriate regimes. These regimes are marked in the horizontal and vertical axes of Figure \ref{SV2}: the supergravity description of the black hole is valid when the black hole horizon is much larger than the string scale, so that the black hole is indeed macroscopic (and we probe it at large distances: hence the large distance on the vertical axis). Furthermore, the string coupling should be small. Both conditions jointly imply that the 't Hooft coupling is large, as shown by the horizontal axis (see also the discussion of the 't Hooft coupling for gauge-gravity duality in Section \ref{ggd}, after Eq.~\eq{RAdS}). 

By contrast, the CFT calculation is valid for small 't Hooft coupling, which is the effective expansion parameter used in perturbation theory in the CFT. And the D-branes are probed at short distances: they are on top of each other, and the fact that they are probed at short transverse distances means that one can approximate the system using the world-volume of the D-branes.\footnote{One then in effect has a theory with only massless fields, where the massive modes have been integrated out, and their effects have been incorporated in the couplings.}

Thus the two regimes of validity are perfectly incompatible: this goes under the name of the `UV/IR connection', because the effective duality relates the long-distance behaviour of the black hole to the short-distance behaviour of the D-branes (we already encountered it, in the context of gauge-gravity duality, in Section \ref{ggd}). It is indeed one of the surprising facts about D-branes that they describe the short-distance behaviour of string theory very well.

{\bf (iii) Supersymmetry protection:} since the match of quantities in Eq.~\eq{SVentropy} (namely, the Bekenstein-Hawking entropy and the logarithm of the density of D-brane ground states) assumes calculations in different regimes of parameters (as per point (ii)), the match relies on these quantities' being (in physics jargon) `protected by supersymmetry'. In short, the supersymmetry of the states, reflected by the BPS condition, secures that the quantities are {\it independent of the 't Hooft coupling}. For the extremal black hole, the area of the horizon is indeed independent of the 't Hooft coupling. And for the system of D-branes, the degeneracy of the ground state is also independent of the 't Hooft coupling.\footnote{In the world-volume theory of the D-branes, supersymmetry protection manifests itself as the invariance of a certain topological quantity, independent of the string coupling: namely, the elliptic genus (also called the `Witten index'), which is a topological invariant of non-linear sigma models. It is an index associated with the supercharge of the sigma model, anlogous to the Dirac index associated with the Dirac operator. See Witten (1987:~p.~535) Elliptic Genera.}
(Note that, even though supersymmetry is broken by the near-extremal solutions, there is nevertheless a perfect match.)

This calculation was then generalized to other cases, such as near-extremal black holes, spinning black holes, and black holes in M-theory. Indeed, this type of match between the properties of black hole horizons and the properties of quantum field theories led up to Maldacena's AdS-CFT.\footnote{For a discussion of the generalisations of the Strominger-Vafa result, see De Haro et al.~(2020~:~pp.~99-101). For its historical role in leading up to AdS-CFT, see (idem, pp.~101-105).}

\subsection{On the explanatory and ontological status of the entropy counting}

The Strominger-Vafa calculation does not require that the black hole and the set of D-branes are the same system---not even the same system at different values of the parameters. It only requires that the relevant subsectors are effective duals (i.e.~that a subset of the states and quantities match: cf.~points (ii) and (iii) above).

However, the paper's more ambitious claim (also in other work like it), that it provides the {\it microscopic origin} of the entropy of a particular system, involves an ontological claim: namely, about the {\it ontological relation} between the two systems. This ontological relation is then put forward as an {\it explanation} for the system's entropy. 

Thus the question is: what is the correct ontological relation between the two systems?

The consensus in string theory is that the systems described by the two models (the supergravity model and the D-brane model) are one and the same. This condition, that the two models must describe the same system, is both assumed by proponents of the argument, and strongly doubted by its critics, who require that an explanation of black hole entropy should involve {\it only the black hole}, and not some `other' system that is not a black hole.

We shall argue against both views, which agree that `sameness' is a requirement for explanation: thus we shall argue that it is not. Our argument goes in two steps. First, we will argue that a good explanation does not require that the systems are the {\it same}: and thus that the arguments brought in by the critics (namely, that black hole entropy has not been explained, because the set of D-branes {\it is not} a black hole) are irrelevant for the questions of explanation and understanding that are addressed in the philosophy of science. Then we will argue that, to the best of our knowledge, the systems described by the two models are not the same, but are in an appropriate ontological relation of {\it correspondence}: more precisely, one is emergent with respect to the other. The two points are as follows:

(1)~~{\it Regardless of whether the supergravity and D-brane models describe the same system or not, the Strominger-Vafa microphysical account gives, for extremal and near-extremal black holes, a good explanation, and provides understanding, of Bekenstein-Hawking entropy}.

What is a good explanation, and what is scientific understanding of a phenomenon? We will discuss this question in detail in Chapter \ref{Understand}, where our answer will be mainstream: namely, we will endorse a version of Hempel's deductive-nomological model of explanation, which defines explanation as an appropriate relation (namely, a {\it deductive argument}) between a theory or model that does the explaining, and (the occurrence of) the phenomenon to be explained.\footnote{This notion admits the usual weakenings of the more general covering-law model; also, the argument need not be deductive in the sense of formal logic, but in the sense of the relevant science---in this case, string theory, and the broader field of high-energy physics.}
Such deduction allows the use of {\it bridge laws} that link the phenomenon to the underlying theory and enable the logical deduction. (Furthermore, there are the usual requirements of adequacy: see Section \ref{dRU}).\footnote{The deductive-nomological model assumes a syntactic formulation of a theory. But this is not a problem, because the important point is that we can point to the deductive arguments that do the explaining: and this we will do below.}

Although the deductive-nomological model is not (or no longer) universally accepted across all philosophy of science, this is for reasons mostly unrelated to high-energy physics.\footnote{One main motivation for the critiques of the deductive-nomological model was that it does not say what understanding consists in. But, as we will lay out in Chapter \ref{Understand}, there is a wide consensus that the deductive-nomological model can be augmented by an additional condition (condition (3) in Section \ref{dRU}) that spells out how explanations provide understanding. In short, an explanation is a form of knowledge, while understanding has a pragmatic dimension: it requires the ability to construct and use explanations.}
And it remains an excellent---indeed mainstream---choice, especially in physics, and in particular in the more formal branches of physics. Furthermore, even if one is a pluralist about explanation (as we are), in that one allows the use of different models of explanation in various contexts, we will argue below that this is irrelevant for our question. For, by any of the main models of explanation that are relevant in high-energy physics, the statement (1) is true.

The Strominger-Vafa account is a good explanation, in the context of string theory, {\it because it is a deductive argument} predicting the properties and the value of the Bekenstein-Hawking entropy, as well as the occurrence of other phenomena associated with it (such Hawking radiation, the black hole's greybody factor, the Hawking temperature, the specific heat, the second law, etc.). Namely, the deduction proceeds from the D-brane model, using the inference relations (i) to (iii) from Section \ref{strar}.

The deduction is enabled by the use of the effective duality (see (i) in Section \ref{strar}): which, in the syntactic formulation that we are adopting here,\footnote{For a linguistic formulation of dualities, see Section \ref{mtce}.}
provides the bridge principles that connect the languages of D-branes and black holes (for example, the bridge principles relate D-brane charges to black hole charges). These bridge principles have physical content, because the derivation is not merely formal: the relation between the two models is also physical (see point (2) below). Thus it predicts the occurrence of phenomena like Hawking radiation. (By the relation being `physical', we mean that, under changes of the parameters, the systems that the two models describe, even if they differ, do mutually correspond.) Thus, on the deductive-nomological model, and in the context of string theory, the Strominger-Vafa derivation is a good explanation.

Hempel's model of explanation is of course theory-relative, which implies that the Strominger-Vafa account is a good explanation of Bekenstein-Hawking entropy for extremal and near-extremal black holes, {\it only if one takes string theory to be the kind of theory that can give such explanations}. But, while this is in the neighbourhood of a substantive philosophical point, we argue that it is not a problem. For the explanation is of the type: {\it if the world were as string theory says it is, then there would be black hole radiation with the properties that Bekenstein and Hawking ascribe to it}. And, besides the fact that explanations are always modal (i.e.~to explain a phenomenon, one must consider both conditions under which the phenomenon will occur, and conditions under which it will not occur), this kind of explanation is of course standard practice in studies of ontologies of scientific theories that are idealisations, rather than full descriptions of the actual world (see our discussion of the `ideal of pristine interpretation', and related views, in Section \ref{ncsr}).

We will be brief about alternative models of explanation. One of the major accounts, namely the unificationist accounts of Friedman (1974:~pp.~15, 18) and Kitcher (1981:~pp.~512, 519), says that an explanation is a reduction of some phenomena to other phenomena, so that the number of unexplained phenomena that we need to accept is smaller; alternatively, an explanation is the set of acceptable arguments that best unifies a theory, i.e.~systematizes and simplifies it, by deriving some sentences from others. On either account, the Strominger-Vafa argument is surely an explanation.\footnote{Other accounts of explanation, such as the causal mechanical, manipulationist, and mechanistic accounts, seem less relevant in the context of high-energy physics.} 
Thus the Strominger-Vafa derivation satisfies usual standards for good explanations, on two broad classes of mainstream accounts. 

After endorsing one or several models of explanation, and establishing that the Strominger-Vafa argument is indeed an explanation by those standards, there is no room for ad-hoc claims that purport to deprive the argument from its explanatory power, i.e.~claims to the effect that it is not a {\it genuine}, and thus not a good, explanation.\footnote{See e.g.~Smolin (2006:~p.~140), who asks whether string models `provide a genuine explanation of the entropy and temperature of black holes', and distinguishes between an `optimistic' and a `pessimistic viewpoint' on the physical equivalence of the two sides of the duality.} 
For it is the job of the models to give criteria of explanation. Thus one would be liable to commit a `No true Scotsman' fallacy. Instead, the correct procedure is to give a model of explanation and to verify whether the argument satisfies the model or not. In other words, one should not conflate `explanation' with `genuine explanation', just as in van Fraassen's (1980:~pp.~205-206) delightful joke attempting to prove scientific realism---a clever joke that Musgrave (1985:~p.~215) self-reportedly spoils, by pointing out the tacit conflation of `explanation' with `ultimate explanation'.

Does the Strominger-Vafa argument give understanding of black hole entropy? We will not discuss this point in detail here, but only note that on the standard account of scientific understanding from Chapter \ref{Understand}, it surely qualifies as giving understanding of black hole entropy and the related associated phenomena of Hawking radiation etc. For not only is the explanation surely extremely useable within string theory, but as subsequent developments have demonstrated, this argument has had enormous heuristic value in spurring further advances in the field: notably, gauge-gravity duality.\footnote{See De Haro et al.~(2020:~pp.~101-105).} 

(2)~~{\it The systems described by the supergravity and D-brane models, for different values of the parameters, are in a relation of correspondence.} There are two main arguments for this. The first is from the derivation, starting with the full (perturbative) string theory, of how the supergravity and D-brane equations come to model the black hole in the first place: namely, the two `points of view' that we discussed in Section \ref{ggd} for gauge-gravity duality and which appear at low energies, at different values of the coupling (see Aharony et al.~2000:~p.~226-227; Zwiebach, 2009:~pp.~538). This argument has admittedly a limited validity, since there is only a duality between the two points of view for the excitations near the horizon: in general, the two models give different results. Nevertheless, they give approximate descriptions of, if not numerically the same system, systems at different values of the parameters (especially the 't Hooft coupling) that go into each other when we change the coupling. Here, by `going into each other', we mean that one system transforms into the other under the change of coupling: for there is just one underlying string theory modelling a single class of systems, of which two different limits are being considered.

We avoid using the phrase `the same system, at different values of the coupling' since that would, by analogy, suggest that ice and water are `the same system, at different values of the temperature': while many would agree that, while the underlying molecules are in themselves the same, ice and water are in {\it different states}. Ice is not simply `water at low temperature', but rather a state that emerges from water, or that water {\it transforms into}, when it freezes.\footnote{A more precisely worked out example is the ferromagnetic phase transition, from Section \ref{dualpf}. This example has been worked out as an example of emergence in De Haro (2019) theory of emergence. In the language of De Haro (2020:~pp.~37-39) and Radder (1984:~pp.~131-134), there is material correspondence between ice and water.For a detailed discussion of the role of the correspondence principle in the Strominger-Vafa case, see van Dongen et al.~(2020:p.~122-125).} 

The second argument follows our treatment, in Chapter \ref{String}, of the non-perturbative BPS states of string theory with Ramond-Ramond charge. Namely, the main question that the work by Strominger and Vafa addresses is how to identify the microscopic states underlying a black hole in string theory of given mass, angular momentum, and charge (where the electric charges are Ramond-Ramond and fundamental string charges: see Section \ref{Dcst}). Given what is known about BPS states with Ramond-Ramond charges satisfying Eq.\eq{Dbmass}, the answer is that no other such states are available in string theory. 

Could there be other states that have not yet been discovered and that do the job? While this logical possibility cannot be fully excluded until a non-perturbative formulation of string theory or M-theory is known, it is very implausible. For if such states exist, they should be visible in the low-energy limit of the theory, which is now well understood. The BPS states with Ramond-Ramond charges carried by D-branes are just what is predicted by the supersymmetry algebra, Eq.~\eq{susyZ}, and there is no indication whatever of the existence of putative additional states with the same properties (nor are other BPS states allowed by the supersymmetry algebra).

A different question in the neighbourhood of this one is as follows. Since the characterization of a black hole by the no-hair theorem as a macroscopic object with mass, angular momentum and charges, is very coarse, could such a black hole not be made up of very different objects---basically, any old lump of matter with the right properties might come to collapse to a Strominger-Vafa black hole, without using D-branes? In other words, the objection is that the microstates are extremely under-determined by the macrostate of the black hole.

The answer to this objection, given the discussion of the low-energy spectrum of string theory above, is that: No, this is not possible---and this makes BPS states special. For building an extremal black hole out of smaller building blocks requires us to either: (a) use building blocks that satisfy the BPS condition, or (b) use some building blocks of ``ordinary'' matter exceeding the BPS bound (i.e.~their mass exceeds their charge, as in Eq.~\eq{bps}), but respect the extremality condition by adding in some building blocks that are {\it below} the BPS threshold and supply the remaining charge (i.e.~their mass is lower than their charge). Option (a) brings us back to D-brane states. And option (b) is not possible: not just because such building blocks violate supersymmetry and do not exist in the low-energy spectrum of string theory, but also crucially because such matter would have pathological properties: for example, in general relativity, it would give naked singularities. In other words, there seems to be no way in string theory to build extremal black holes with the required RR charge other than using D-branes: as ground states of the theory, they are indeed unique.

The picture of low-energy physics that string theory offers is wonderfully coherent and closely knit. While many things are not well understood (e.g.~the main string dualities remain unproven), accounting for black hole entropy is a problem of computational tractability, rather than a problem of putative states that are unknown. For the reasons explained above, adding new states to the low-energy spectrum would seem to compromise the whole theory.

\section{Conclusion}

This Chapter has addressed two hole arguments that are widely thought to bear on the nature of spacetime, in classical spacetime theories and in theories of quantum gravity.

Section \ref{holeA} argued that, despite some claims in the literature, to the effect that the hole argument dissolves when properly formulated mathematically, the hole argument remains an important question to which both realists and anti-realists \`a la van Fraassen must give an appropriate answer. 

Where usual discussions of the hole argument and of the Leibniz equivalence principle use the generic word `diffeomorphism' (requiring, in the case of the hole argument, that it goes to the identity outside the hole), we have argued that, in general, more discrimination about the relevant class of diffeomorphisms is required.

To illustrate this, we gave a version of the hole argument for a spacetime with a non-zero cosmological constant. The conclusions regarding Leibniz equivalence, symmetries, and dualities are summarized in Figure \ref{Hole_diffeos}. 

We distinguished three types of diffeomorphisms: hole diffeomorphisms, boundary symmetries, and large diffeomorphisms. Leibniz equivalence and the hole argument can only be formulated for the first two classes, and not for the third. So where usual formulations of the Leibniz equivalence principle (see Section \ref{historyH}) do not distinguish between different types of diffomorphisms, they are incorrect.

The hole argument is consistent with gauge-gravity duality and the Schema's treatment of it. Namely, hole diffeomorphisms are proper symmetries of the gravity model and are not `seen' by the dual CFT, i.e.~the CFT carries no representation of them. By contrast, boundary symmetries are stipulated symmetries: they are visible both in the gravity model and in the CFT on the boundary of the hole. They represent the symmetries of the common core. 

Section \ref{ocs} presented the anatomy of Strominger and Vafa's black hole microstate counting. In this anatomy, three arguments play a principal role: effective duality, the properties of the regimes which allow the matching of quantities, and supersymmetry. We argued that, by two mainstream accounts of explanation in philosophy of science, the Strominger-Vafa argument is explanatory and provides understanding of black hole entropy, for a class of extremal and near-extremal black holes in string theory. Strominger and Vafa count the degeneracy of the ground state of a set of D-brane BPS states, described in open string theory at weak 't Hooft coupling, which turn into the $p$-branes that make up the black hole at strong coupling. The relation between them is one of correspondence, where the black hole geometry emerges as one goes from weak to strong 't Hooft coupling.

\newpage
\thispagestyle{empty}
$ $
\begin{center}
\vskip7cm
{\bf\Huge Part III. Theoretical and Practical Functions of Duality}
\end{center}
\addcontentsline{toc}{chapter}{Part III. Theoretical and Practical Functions of Duality}
\newpage
\thispagestyle{plain}

\chapter{Equivalence in Logic and Philosophy of Science}\label{Theor}
\markboth{\small{\textup{Equivalence in Logic and Philosophy of Science}}}{\textup{\small{Equivalence in Logic and Philosophy of Science}}}

Part III of this book explores the theoretical functions, and practical functions (or `roles'), of dualities. As we explained in Section \ref{featurerole}: the theoretical functions of a duality are closely tied to satisfying the Schema: i.e.~a common core theory being represented by two isomorphic models. 

The practical functions ``look beyond'' the satisfaction of our Schema; cf.~Chapter \ref{Heuri}. Of course, these two types of function overlap a good deal. But just as philosophers of science broadly distinguish issues of ontology, semantics and logic from issues of epistemology or methodology: so we broadly distinguish these two types of function for dualities.

In this first out of three Chapters on the theoretical function of duality, we set the stage by discussing two issues in logic and philosophy of science that will mould our discussion of the theoretical functions of duality in the next two Chapters. (Since this Chapter is preparatory, it is largely independent of our Schema for dualities: the next Chapter will return to the detail of the Schema.)

The first issue in this Chapter is equivalence (Sections \ref{eil} and \ref{theoreq}): both in logic and in philosophy of science, where equivalence involves whole scientific theories, and so is called `theoretical equivalence'. These Sections also explain in more detail the syntactic and semantic conceptions of theories, introduced in Section \ref{thsmodels}.

In this Chapter, dualities enter in Section \ref{mtce}, which develops a syntactic conception of the idea of duality, thus clarifying the relationship between dualities and the criterion of isomorphism of models, which is the natural criterion of equivalence in the semantic conception of theories. The isomorphism criterion has been criticized as being too strict, i.e.~too logically strong. 

Thus the second issue, in Section \ref{defencei}, is to discuss the desired logical strengths of a formalism and its interpretation, and defend the isomorphism criterion against the accusation of strictness. This Section clears the ground for the discussion of theoretical equivalence in the next Chapter by first giving such a defence: and then turns the tables and argues against (what are often called) more `liberal', i.e.~logically weaker, criteria of equivalence, through a natural requirement of {\it univocality} of the interpretations of physical theories. This is cashed out as what we dub a `No {\it deus ex machina}' argument about the desired relative logical strength of a mathematical formalism and its interpretation: in short, interpretations of physical theories ought not to make distinctions that are not made by the formalism. 

The first Section sets up the discussion by introducing different conceptions of equivalence and definability in logic. Given that both physicists and philosophers often use the phrases `different languages', `translations', and `equivalence' (especially in the context of dualities), it is worth recalling how logicians think about equivalence.

\section{On equivalence in logic}\label{eil}

In logic, the standard criterion of equivalence is of course {\it logical equivalence}: two sentences, $\phi_1$ and $\phi_2$, are logically equivalent if they have the same set of models, i.e.~if $M\models\phi_1$ iff $M\models\phi_2$, where $M$ is a model. Here, the notation $M\models\phi$ means that the sentence $\f$ is {\it satisfied} in $M$, or, in other words, that $\f$ is {\it true} in $M$.\footnote{For an introduction to model theory and the use of this notation, see Hodges (1997:~p.~12). We here gloss over several details, such as the distinction between atomic and non-atomic sentences, that are comparatively unimportant for our aims in this Chapter. For a full treatment of model theory, see Hodges (1993:~p.~12).} (More details on models below, and in Section \ref{theoreq}).

In Section \ref{theoreq}, we will argue that, although logical equivalence can be applied to whole theories, this criterion is in general too strict, i.e.~logically too strong, for scientific theories. And it is also too strict in logic and in computer science, where one is often interested in translating (fragments of) one logic into another, or of one database into another. 

This Section aims to give a breezy review, for philosophers and physicists, of some of these important developments that will inform our discussion of theoretical equivalence in the philosophy of science.\footnote{This Section has benefitted from discussions with Johan van Benthem and Sonja Smets, to both of whom we are grateful to for comments.}

In this Chapter, the word `model' will be used in the usual model-theoretic sense, which we will distinguish from the Schema's sense of `model' (see Section \ref{ThisB}; the relation between the two senses will be explained in Section \ref{mtce}).

\subsection{Correspondence between modal and classical logic}

Dutch logician Johan van Benthem (1976, 1985, 2001) is one of the founders of (what he dubbed) the `correspondence theory' that relates modal logic and classical logic.\footnote{There are of course precursors: see for example Lewis' (1968) counterpart theory. For an overview of recent developments, see Baltag and Smets (2014) and van Benthem (2010a). A recent introduction is van Benthem (2010b).} 
This word, `correspondence', appears to nod at the `correspondence rules' of twentieth-century philosophy of science, and-or at the `correspondence principle' that relates quantum mechanics to classical mechanics.

In logic, correspondence theory aims to make precise the analogies that exist between apparently different logical languages and systems, and enable the transfer of results across both sides of the correspondence relation. Given the proliferation of languages and logical systems considered by logicians, correspondence theory promises to bring their underlying unity to light, especially when the correspondence relation is surprising and difficult to find. (The notions introduced below, rather than being part of our core proposals in this Chapter, provide relevant background about the notions of equivalence and translation.) 

Van Benthem proved that relevant fragments of modal logic (including modal propositional logic, modal predicate logic, and temporal logic) can be translated into first-order or into second-order languages. The notion of `translation' is informal here: it means a replacement of standard modal operators, such as $\Box$ (written $\Box p$, i.e.~`it is necessary that $p$') and $\Diamond$ (written $\Diamond p$, i.e.~`it is possible that $p$'), by sequences of first-order symbols.\footnote{For example, if $\f$ is a formula in modal propositional logic, $L_{\tn m}$, then $\Box\f$ has a `standard translation', denoted by a map, dubbed ST, from $L_{\tn m}$ into $L_1$ (where $L_1$ is a first-order language with a binary predicate constant $R$, called the `accessibility relation', and unary predicate constants). The map is: $\mbox{ST}(\Box\f)=\forall y\,(Rxy\rightarrow[y/x]\,\mbox{ST}(\f))$, where $y$ does not occur in $\mbox{ST}(\f)$, and $[y/x]$ means substitution of $x$ by $y$ in the expression $\mbox{ST}(\f)$, which depends on the fixed variable $x$. The idea behind this translation is of course that: (i) whatever the translation of $\f$ says about $x$, it will say about any variable $y$ bearing the relation $R$ to $x$, (ii) the variables $x$ and $y$ will be assigned possible worlds. See also Lewis (1968:~p.~118).} 

For example, the axiom $\Box p\rightarrow p$ of the modal logic system T is true in a Kripke frame, $\bra W,R\ket$, iff the accessibility relation $R$ is {\it reflexive}. Here, $W$ is the {\it domain} of the model, i.e.~a set (whose elements are thought of as possible worlds), and $R$ is a binary relation on this domain. The idea behind the introduction of an accessibility relation $R$ in the interpretation of the modal operator is that, if $\Box p$ is true at some possible world $w\in W$, then $p$ does not need to be true at {\it all} possible worlds, but rather at those possible worlds $w'$ that are related to $w$ by an appropriate relation $R$, i.e.~$Rww'$ (namely, those worlds that are possible given the facts of $w$). $\Box p\rightarrow p$ then translates into the {\it reflexivity} condition of $R$, because the implication that $p$ is true at $w$ follows from the assumption that $\Box p$ is true at $w$, iff (by the interpretation of $\Box$) we require that $w$ is related to itself by $R$, i.e.~$Rww$.\footnote{For a proof, see van Benthem (2001:~p.~326).} 

By a similar argument, the axiom $\Box p\rightarrow\Box\Box p$ of system S4 is true iff $R$ is {\it transitive.} All in all, a modal formula is a theorem of S4 iff it is true in all reflexive and transitive Kripke frames.\footnote{For philosopher-friendly overviews of the basic modal systems, including T and S4, see Hughes and Cresswell (1968) and Lowe (2002:~pp.~116-117). An explanation of the modal counterparts of the {\it symmetry} and {\it transitivity} conditions on the accessibility relation are on pp.~119-120. For an explanation of the transitivity and seriality conditions based on the interpretation of the modal operator $\Box$ in temporal logic, see Garson (2018:~Section 7).} 

In this way, modal systems can be seen as axiomatizations of Kripke frames satisfying ``classical'' constraints of varying strength on the accessibility relation $R$. Using this translation from modal into classical logic, established results of classical logic can be reformulated as results about modal logic.\footnote{The L\"owenheim-Skolem theorem is one example of this: see van Benthem (2010b:~p.~77). More generally, translations of one logic into another can be used to establish results about decidability and complexity: for example, about the decidability of the satisfiability problem for fragments of a guarded logic: see Gr\"adel and Otto (2014, pp.~16-18). See also van Benthem (2010b:~pp.~82-83), who furthermore discusses increasing the richness of the logic, by increasing the expressive power of the language (Chapter 7), its deductive power (Chapter 8), and its descriptive power (Chapter 10).}

Correspondence theory can be generalized to many other logics, for example intuitionistic logic and tense logic (van Benthem, 2010b:~pp.~234-243; 206-217). There is also a correspondence theorem between modal predicate logic and modal propositional logic.\footnote{Van Benthem (2010b:~pp.~78-81) discusses extended modal languages. For modal predicate logic, see (pp.~117-123).}

But it is important to note that the above {\it correspondences} are not cases of {\it equivalence}. For we have simply mapped various modal logics to {\it fragments of} classical logic. Thus one way to put the question about equivalence is to ask: {\it what are the first-order accessibility conditions that are modally definable?} This brings us from correspondence theory to {\it definability theory}. 

\subsection{Definability theory: bisimulation}\label{synsemeq}

Note that, although the correspondence theory just discussed could, in principle, be seen as purely syntactic, i.e.~as a correspondence between different linguistic forms and rules, regardless of meaning, it of course makes sense only in the light of the semantic considerations used: like, for example, when we argued that the system T's axiom $\Box p\rightarrow p$ should correspond to the {\it reflexivity} requirement on the accessibility relation. For that argument involved considering the truth of this axiom at various possible worlds.\footnote{In fact, the {\it proofs} of the correspondence of three of the systems T and S4's axioms with the requirements of reflexivity, transitivity, and directedness of the accessibility relation require considering the {\it truth} of the axioms at various worlds (see e.g.~van Benthem, 2001:~p.~326). Thus they involve {\it semantic entailments} rather than just formal derivations, in the sense of Beth. Beth (1969:~pp.~14-15) is a lucid discussion of the contrast between formal derivability and semantic entailment.}

As we mentioned in Section \ref{thmscph}, a semantics is an assignment, to linguistic items such as $p$, of referents such as a set of worlds (namely, the set of possible worlds where $p$ is true). Specifically, the model-theoretic notion of a model, $M$, is a frame together with a {\it valuation map}, $V$, that assigns a set of worlds, $V(p)\subset W$, to a proposition letter $p$: $M=\bra W,R,V\ket$. Thus the sentence `$\phi$ is true in $M$ at $w$' is written as $w\models_M\phi$.

Thus imagine two languages (e.g.~a modal logic and a first-order language), each of them equipped with a semantics so defined. Our considerations above suggest that the correspondence between the two languages is established by looking at fragments of the semantics of these two languages that are indistinguishable. Van Benthem (2001:~p.~340; the first emphasis is ours) defines the {\it first question of model theory} thus:

\begin{quote}\small
[T]he `first question' of any model theory is that concerning the relation between linguistic indistinguishability (equality of modal theories) and structural indistinguishability (isomorphism) of semantic structures. {\it How far do the webs of language and ontology diverge?} In classical logic, we know that (first-order) elementary equivalence coincides with isomorphism on the {\it finite} structures, but no higher up: isomorphism then becomes by far the finer sieve.
\end{quote}

\noindent Van Benthem's quote is important because it shows that one of the main questions about inter-theoretic relations in philosophy of science (namely, that formal inter-theoretic relations such as equivalence and reduction need not amount to equivalences or reductions between ontologies) is already relevant in logic, where syntax and semantics also ``come apart''. Thus a major question for inter-theoretic relations is how far one should go, and what criteria one can use, to `hold together the webs of language and ontology'. (We will return to these issues in Sections \ref{lsps} and \ref{lessons}, and in the next two Chapters.)

A first important notion that will allow us to state a theorem that makes precise how to define one logic in terms of another (while ``holding the webs of language and ontology close together'') is the notion of a bisimulation. The heuristic idea behind bisimulation is that ``systems behave in the same way'', so that each system can be used to simulate the other. As follows:

A {\bf bisimulation} (also called a `p-relation' or a `zigzag connection') is going to be a relation between models, $M_1=\bra W_1,R_1,V_1\ket$ and $M_2=\bra W_2,R_2,V_2\ket$, that is required to preserve the accessibility relations. Namely, it is defined as a relation $C$ that has $W_1$ in its domain and $W_2$ in its range, and satisfies the following two natural conditions (these two conditions jointly work towards ``holding the webs of language and ontology close together''):\footnote{There are of course several variations on this definition in the literature. Here we follow van Benthem (2001:~p.~341). See also van Benthem (1976:~p.~10) and Andr\'eka et al.~(1998:~p.~220).}

(i)~~~$C$ forms a {\it commuting diagram} with the accessibility relations, with specific conditions on the vertices: see Figure \ref{Ccommdia};

(ii)~~$C$ {\it preserves the semantics}: namely, if $C$ relates $w_1\in W_1$ and $w_2\in W_2$ (i.e.~$Cw_1w_2$), then $w_1$ and $w_2$ satisfy the same proposition (letter) $p$, i.e.~$\forall p$, the following holds: $w_1\in V_1(p)$ iff $w_2\in V_2(p)$. 
\begin{figure}
\begin{center}
\begin{tikzcd}
w_1\in W_1~~ \arrow[mapsto,color=black]{r}{C} 
\arrow[mapsto,color=black]{d}{R_1}
& ~~w_2\in W_2\arrow[mapsto,color=black]{d}{R_2} \\
w_1'\in W_1 ~~\arrow[mapsto,color=black]{r}[color=black]{C}
& ~~w_2' \in W_2
\end{tikzcd}
\caption{\small Commuting diagram for $C$ and the accessibility relations. Condition (i) for a bisimulation requires that, given $w_1,w_2,w_1'$ satisfying the relations shown in the diagram, some $w_2'\in W_2$ exists, satisfying the relations in the diagram; and, conversely, that given appropriate $w_1,w_2,w_2'$, some such $w_1'\in W_1$ exists.}
\label{Ccommdia}
\end{center}
\end{figure}

Second, we say that a modal formula $\f$ is {\bf invariant under a bisimulation} $C$ if, for all models $M_1$ and $M_2$ related by $C$, and for all $w_1\in W_1$ and $w_2\in W_2$ pairwise related by $C$ (i.e.~$Cw_1w_2$), the following holds (van Benthem, 1976:~p.~10): 
\begin{center}\label{modelpres}
$w_1\models_{M_1}\f~~~$ iff $~~~w_2\models_{M_2}\f$ .
\end{center}

With these definitions, we have the following important theorem that expresses ``how the webs of language and ontology can be held close together'':\footnote{See van Benthem (1976:~p.~10; 2001:~p.~341).} 

{\bf van Benthem's zigzag theorem:} if two models, $M_1$ and $M_2$, are related by a bisimulation $C$, then all the modal formulas $\f$ are invariant under this bisimulation. 

Furthermore, there is a condition that states when a modal formula can be translated to a first-order formula: namely, a first-order formula $\f(x)$ is logically equivalent to a translation of a modal formula iff that translation is invariant under all bisimulations.\footnote{See van Benthem (2001:~p.~342) and Gr\"adel and Otto (2014:~p.~10). For an account of the significance of this result and its generalizations, see Venema (2014:~pp.~33-34).}
This gives a nice correspondence between first-order and modal logic: namely, it characterizes the fragment of first-order logic that has a translation from modal formulas, as that fragment that is invariant under all bisimulations.

Bisimulations can be generalized to obtain between {\it frames}, rather than models. The resulting notion is that of a {\bf p-morphism}, $f$, from a frame $F_1=\bra W_1,R_1\ket$ to a frame $F_2=\bra W_2,R_2\ket$, satisfying the analogue of the condition (i) above, that the diagram commutes: but without the condition (ii) that the semantics is preserved. The generalization of van Benthem's zigzag theorem to this case then says that, if $f$ is a p-morphism from $F_1$ to $F_2$, then, for all modal formulas $\f$, $F_1\models\f$ implies $F_2\models\f$. In this case, a modal formula is in general translated to a second-order formula, although for many modal formulas translations also exist to first-order formulas.\footnote{See van Benthem (2010b:~pp.~99-106).}

A bisimulation or p-relation (more generally, a p-morphism between frames) thus gives an analogue of the notion of partial isomorphism for modal logic.\footnote{Partial isomorphisms are thought to play a (somewhat different) role in the relation between a theory and the world: see Sections \ref{lessons} and \ref{intext}.} 
Furthermore, bisimulations give appropriate conditions under which formulas in a modal language {\it say the same thing}: namely, if the models are bisimilar (i.e.~if a bisimulation exists between them), then the modal formulas are invariant under this bisimulation, in the sense that their meaning is preserved under the translation. 

The study of bisimulations is one of the leading themes in modal logic, and also a thriving field in database science.\footnote{For an elementary introduction to bisimulation in the (different) context of set theory for non-wellfounded phenomena, see Barwise and Moss (1996:~pp.~78-81). A further generalization of bisimulation to decidable logics is the notion of guarded bisimulation, and further for undecidable logics in simulations that are not guarded. See van Benthem (2010a:~pp.~122-123), Gr\"adel and Otto (2014:~pp.~12) and Baltag and Smets (2014). In database science, one is interested in `schema mappings', i.e.~high-level mappings that specify the relationships between two database schemas (not our sense of `Schema'!). Schema mappings are syntactic objects that help solve inter-operability problems. They can be assigned a semantics to solve problems such as: given a source of data that is an instance of a schema, how to transform it to a solution with respect to another schema. See e.g.~ten Cate and Kolaitis (2014). } 
According to Gr\"adel and Otto (2014:~p.~3; our emphasis):

\begin{quote}An appropriate notion of bisimulation for a logic allows us to study the {\it expressive power} of that logic in terms of semantic invariance and logical indistinguishability. As bisimilar nodes or tuples in two structures cannot be distinguished by formulae of the logic, bisimulations may be used to control the complexity of the models.\end{quote}

To summarize, we have argued the following three points regarding translations between logics: 

(i)~~~There is a natural question of how to {\it translate} a given (modal) logic into another (classical) logic. 

(ii)~~\,The construction of such translations is not regardless of meaning, but uses the {\it interpretation} of the modal operators and the {\it truth conditions} for the sentences in the modal language. More generally, `semantic invariance' is a requirement on such translations. 

(iii)~~Bisimulations give the appropriate notion of {\it partial isomorphism} in modal logic: if two models are bisimilar (i.e.~if a bisimulation exists between them), then the meaning of their formulas is preserved under the translation.

The next two Sections discuss these three issues, i.e.~translations, semantics, and isomorphism, in philosophy of science.

\section{Theoretical equivalence in philosophy of science}\label{theoreq}

Each of the logical criteria of equivalence that we discussed in the previous Section has a linguistic and a semantic component: as we saw, according to van Benthem, the `first question of model theory' is `how far the webs of language and ontology diverge'---to which his zigzag theorem provides an answer in modal logic. 

In philosophy of science, the two basic conceptions of a scientific theory, i.e.~the syntactic and the semantic conceptions (see Section \ref{thmscph}) naturally give different criteria of theoretical equivalence. (As we noted in Section \ref{refsr}, we endorse the recent literature in thinking that the contrast between the syntactic and the semantic conceptions of theories has often been over-emphasised, and in the following pages we are going to see that each conception relies on the other.)\footnote{Thus we agree with Halvorson's (2012, 2013) critique of a literal interpretation of the semantic conception of theories, and with the replies to it by Lutz (2013), Glymour (2013), and van Fraassen (2014). For a summary of the discussion, see Frigg (2023:~p.~167).} 
And, since scientific theories are always formulated using both a language or calculus and an interpretation that assigns referents, i.e.~items in the world, to the items of that language or calculus, all the criteria of theoretical equivalence that we will here discuss have both linguistic and semantic components. 

This Section discusses various proposals, both traditional and recent, for understanding theoretical equivalence, and relates them to our account of duality. 

\subsection{Language and semantics in the philosophy of science}\label{lsps}

Although, for the question of theoretical equivalence in the philosophy of science, we are interested in translating not logics, but scientific theories, the larger aim is closely related: namely, to find both formal and interpretative criteria under which two theories say the same thing.

For this reason, it is important to here point out some salient differences between the use of language and semantics, in logic and model theory and in the philosophy of science, that will guide the rest of our discussion:

(i)~~~As we noted in Section \ref{thmscph}, the language of scientific theories is not always a formal language, but is built on the language of mathematical physics, which e.g.~includes pieces of natural language.\footnote{It may even include non-linguistic forms of communication: see e.g.~M\"ossner (2018:~Chapter 4), although this particular point will not play any role in the next three Chapters. See also Frigg (2023:~pp.~172), and the discussion in Section \ref{refsr}.} 

(ii)~~The semantics of scientific theories is not limited to the {\it structuralist}, model-theoretic kind of semantics that we have so far discussed, with interpretations given by models understood as {\it mathematical structures}, e.g.~in modal logic $M=\bra W,R,V\ket$. 

About (ii): although we do use model-theoretic semantics in physics (and this will be the main point of our comparison of the Schema with the traditional views in philosophy of science in Section \ref{mtce} (A)): in physics, {\it model-theoretic semantics, with models understood as structures, is not enough}. There are two reasons for this:\footnote{There is of course a long tradition of distinguishing formal semantics and physical semantics (distinguished below, and in Section \ref{defencei}): see Suppe (1974:~p.~102). For example, Nagel (1961:~p.~349) writes: `We shall ... be primarily concerned with the so-called `{}`descriptive expressions', signifying what are generally regarded as ``empirical'' objects, traits, relations, or processes, rather than purely formal or logical identities.' And he notes that formalization comes in various degrees that are appropriate to each specific field of empirical science, adding---in his discussion of reduction---appropriate non-formal conditions (pp.~349, 359, 366). Frigg (2023:~pp.~198-200) motivates the development of a linguistic formulation, {\it in addition} to the structural semantics, through Newman's problem. For target systems do not have unique structures, and so `the attribution of a structure to a target system is always relative to a substantive---non-structural---description', in terms of a `fully interpreted physical language'.} 

(1)~~By definition, a semantics in which models are mathematical structures (of a certain type) cannot capture those parts of the ontology that involve certain aspects that, despite being realized, or instantiated, in the mathematical structure, are not reducible to structural properties (e.g.~about causation, or about the natures of the entities that appear in physical theories). In other words: {\it unless one adopts a structural realist view, a structural analysis of models does not exhaust a (realist) semantics}. And structural realism is of course a disputed position in the philosophy of science. Note that this point is not only relevant for the scientific realist, but also for the anti-realist who, like van Fraassen, endorses a referential semantics (see the discussion in Section \ref{scirer}). 

(2)~~Since physics is an empirical science, materiality and even the experimenter's actions may enter into the epistemology and ontology of a scientific theory in ways that are not accounted for by mathematical structuralism. For example: `by applying a solvent to an amount of soot, the organic chemist extracted a sample of fullerene that he then brought over for analysis to the spectroscopy lab'. The important fact that both labs worked with the {\it same} sample of fullerene is not determined by the chemical or physical structure of fullerene, but by the fact that the organic chemist brought the sample (a token) over to the spectroscopy lab.

While these two points are substantive, they are also well-known: and, luckily for us, we can deal with them in {\it two steps}. In the first step, we discuss the model-theoretic semantics, i.e.~we focus on formal and structural matters (in this Section). In the second step, we also discuss other aspects of the ontology and epistemology of scientific theories that may not be so modelled (Chapter \ref{physeq}). Thus, at least conceptually and methodologically, we distinguish two aspects of the semantics of scientific theories: the {\it model-theoretic semantics} and the {\it physical semantics}, where the latter of course uses the former. In the model-theoretic semantics, a model is a structure like the models $M=\bra W,R,V\ket$ of modal logic, and a formula $\f$ acquires mathematical meaning by its being mapped into an appropriate item in the model, e.g.~in modal logic into a subset of the set of possible worlds, $V(\f)\subset W$, understood as structures. The {\it physical semantics} then uses these structures to map the theory into the physical world: as already structured by the model-theoretic semantics, but furthermore taking into account the all-important aspects (1) and (2). (Recall our discussion in Section \ref{srpost}.)

We admit that the distinction between structure and physical interpretation that we are going to advocate is shiftable. For we expect that, as our theories mature, more of the physical content will be described by formal structures (in our example above, we may choose coordinates for the organic chemistry and the spectroscopy labs, and we may write down the worldline of the sample of fullerene etc.). And we also welcome the formalization, where possible, of the epistemological and metaphysical work required to explicate how a physical theory describes its domain of application. Indeed, the more explicit we are about what our theories mean, the more we use some type of formalization (as our principle in Section \ref{lessons}, of the `univocality of the interpretation of physical theories', attests). But this does not mean that this distinction is irrelevant: only that it is not absolute.\footnote{We thank Hans Halvorson for bringing this point to our attention.} 
Nor does it imply that the physical content of a theory shrinks: quite at the contrary, formalization helps articulate and develop both form and physical interpretation.\footnote{Dualities are a case in point, where some early philosophical commentators claimed that dualities pose a problem for standard scientific realism, and that dualities favour an ontologically ``lighter'' structural realist view. With the development of the field, it is clear that dualities are compatible with scientific realism, and many philosophers of dualities now reject the former view (for a discussion, see Section \ref{srr}). The hole argument (see Section \ref{holeA}) is another example: although, after the rediscovery of the argument, one might have been tempted to say that a metaphysically ``lighter'' relationism is the preferred response to it, the development of sophisticated substantivalism shows that there can be a realist position that is in full agreement with physical practice. Thus the distinction between relationism and substantivalism is not primarily a difference of form, but a difference of interpretation.}\\ 
\\
There is a second distinction that is in the neighbourhood of the first one, but is also more general, because it includes both language and semantics. The distinction is between two different aspects of the notion of theoretical equivalence, which is usually seen to be a conjunction of two criteria:

(A)~~A {\it formal, i.e.~non-interpretative, criterion} of equivalence (e.g.~logical or---yet to be introduced---definitional equivalence, or the isomorphism criterion of models).

(B)~~An {\it interpretative criterion} of equivalence (almost always: sameness of the domains of application, i.e.~that the interpretation maps have the same range, i.e.~the same image; see Section \ref{itm}).

We believe there is a consensus that this distinction is mandatory (see for example the distinction between (i) and (ii) at the end of Section \ref{contcons}). And we endorse this consensus,\footnote{For more on this, including a discussion of some authors who do not (fully) endorse this consensus, see De Haro (2021:~pp.~5153-5155).} since these are indeed two main aspects of scientific theories, and correlate with our distinction, in Section \ref{ThisB}, between a {\it bare theory} and an {\it interpretation}.\footnote{This distinction follows the `two-step procedure' that is often used to conceptualise the formulation of a scientific theory: see Section \ref{ThisB}, especially footnote \ref{Ruetsche2002}.}

Note that the formal, i.e.~non-interpretative, criterion of equivalence (A) can be either linguistic or semantic, as we have already discussed. And note that the interpretative criterion of equivalence can be model-theoretic, although for physical theories it will always be the equivalence of the physical semantics.

Sections \ref{sse} and \ref{mtce} discuss linguistic and model-theoretic criteria of equivalence that emphasize the formal i.e.~non-interpretative aspect, (A). We will return to a discussion of the physical semantics in Section \ref{defencei}.

\subsection{Linguistic criteria of equivalence}\label{sse}

In this Section, we first discuss logical equivalence of scientific theories, and then two weaker inter-translatability criteria, due to Glymour and Quine.

\subsubsection{(A)~~Logical equivalence}

As we discussed in the preamble of Section \ref{eil}, although logical equivalence is the natural criterion of equivalence of sentences in logic, this criterion is too strict, i.e.~too logically strong, when applied to whole theories. 

To see that logical equivalence is too strong, consider a theory, $T_1$, that is presented syntactically, as a set of sentences and a set of rules of deduction, or syntax, relating the sentences. The sentences of $T_1$ have two types of symbols: {\it non-logical} and {\it logical}. {\it Non-logical} symbols are of four types: function symbols ($+,\times$, etc.), relation symbols (e.g.~$>,<$), constant symbols (e.g.~$e,\mathbb{1}$,0) with fixed values in the domain, each with a given arity, and variables (e.g.~$x,y,F,m,a$) that can take on any value in the domain. Thus, roughly speaking, the non-logical symbols (in particular, variables) ``stand ready'' for interpretation: in a mathematical structure, and also as things or physical systems (e.g.~$S$ stands ready for interpretation as a system comprising a single point particle) or as properties of things. For example, $F$, $m$, and $a$, are non-logical terms that stand ready for interpretation as the force on, mass of, and acceleration of a given body. The set of non-logical terms of a theory is also called the theory's {\it signature}; $\Si$ (we only consider single-sorted signatures here). 

The {\it logical vocabulary} (`not', `and', `or', `if', `any', `some' etc.) does not itself ``stand ready'' for interpretation in terms of things and properties of things. Logical equivalence, between $T_1$ and another theory $T_2$ with the same signature, is then defined as follows.

Two theories, $T_1$ and $T_2$, are {\bf logically equivalent} iff they have the same class of models, i.e.~iff $\{M_1|\,\forall \f\in T_1\,(M_1\models\f)\}=\{M_2|\,\forall\varphi\in T_2\,(M_2\models\varphi)\}$. 

Note that a necessary condition for logical equivalence is that the two theories use the {\it same signatures}, or non-logical vocabularies.\footnote{For an example, see Barrett and Halvorson (2016a:~p.~2).} This is because models are mathematical structures that satisfy the corresponding sentences, and each structure has a signature (an interpretation of the signature of the corresponding theory), in terms of which we can know whether the model makes the sentence true or false. And so, one sees immediately that the above condition is too strong. For if the vocabularies of the two theories are different, so that one theory e.g.~uses the vocabulary $\{F,m,a,\times\}$ and contains the formula $F=m\times a$, while the other one uses the vocabulary $\{G,n,b,\cdot\}$ and contains the formula $G=n\cdot b$, then the two theories cannot have the same models. But formulating Newton's second law as $F=m\times a$ or as $G=n\cdot b$ should not count as a difference. 

While this might look like a minor point, easily fixed by an appropriate translation between the two vocabularies: as we saw in the Section \ref{eil}, translations are far from trivial. For once one weakens one's notion of equivalence, the following question is unavoidable: how {\it weak} should a good translation be? Should we just allow relabellings of symbols (which suffices to handle the previous example of Newtonian mechanics) or should we also allow more general mappings (e.g.~mappings not only between individual symbols, but also mappings from individual symbols to expressions with multiple symbols, whilst ``preserving the arity'' of the expressions)? Our discussion of bisimulation in the previous Section illustrates how translations even between fragments of elementary logics involve mappings that are not surjective, i.e.~do not have a unique inverse. And for translations between {\it scientific theories}, Glymour (1970:~p.~280) and Quine (1975:~p.~320) argued, along the same lines, that a linguistic criterion of equivalence should be rather more liberal, i.e.~logically weak. Thus we will summarize their criteria of theoretical equivalence. 

\subsubsection{(B)~~Glymour and Quine's inter-translatability criteria}

Glymour and Quine's linguistic accounts of equivalence have been very influential in discussions of theoretical equivalence and under-determination. Glymour's (1970:~p.~279) criterion of theoretical equivalence uses the notion of a (common) definitional extension:\footnote{This formulation of the criterion draws on Barrett and Halvorson (2016a:~pp.~469-470), to which we refer for more details. In the context of dualities, this criterion seems to have been first discussed by Rickles (2017).
}\\
\\
{\bf Definitional extensions}.\footnote{The problem of the definability of new concepts was discussed and formalized in Tarski (1983:~p.~301) [1934]. For recent accounts of definability in many-sorted logic, see Andr\'eka et al.~(2008:~pp.~20-21) and Barrett and Halvorson (2016b:~pp.~563-564), who dub their generalization `Morita equivalence', and show that their criterion is weaker than definitional equivalence, but stronger than categorical equivalence (p.~574); see also Halvorson and Tsementzis (2017:~pp.~406-407).} 
Take two theories, $T_1$ and $T_2$, with disjoint signatures (i.e.~non-logical vocabularies) $\Si_1$ and $\Si_2$, respectively. (Recall that a signature, or vocabulary, is the set of non-logical symbols in which the theory's sentences are written.) Then extend the respective signatures of each of these theories to include the symbols of the other theory: and then add  to each of these theories appropriate {\it explicit definitions} of each of the non-logical symbols of the other, so as to obtain new theories that we denote by dashes, $T_1'$ and $T_2'$. So these dashed theories have a common signature, $\Si_1\cup\Si_2$; and we say that  $T_1'$ is a {\it definitional extension} of $T_1$, and similarly  $T_2'$ is a definitional extension of $T_2$. 

But suppose also that in each such augmented theory, you can deduce all the theorems of the original version of the other theory. That is: suppose that in $T_1'$ you can deduce all the theorems of $T_2$ (as well, of course, as all the theorems of $T_1$ itself, which is by definition contained in $T_1'$); and similarly, in $T_2'$ you can deduce all the theorems of $T_1$ (as well as all of $T_2$). So in effect, the supposition is that the explicit definitions that we added so as to define the new augmented theories are appropriate, indeed `powerful', in that each such theory `recovers' or `encompasses' the original version of the other.  (See an example below.) Then if these new theories, $T_1'$ and $T_2'$, are {\it logically equivalent}, we say that the original theories have a {\it common definitional extension}. We speak of a {\it common} extension just because the two extensions are, up to logical equivalence, the same theory. Then Glymour's criterion is as follows.\\
\\
{\bf Glymour inter-translatability}: two theories, $T_1$ and $T_2$, are {\it definitionally equivalent} iff they have a {\it common definitional extension}\footnote{De Bouvere (1965:~pp.~404-405) defines the semantic notion of `coalescence' of theories, regarded as sets of models, as a counterpart of linguistic synonymy, and proves a theorem according to which two theories are synonymous iff their sets of models are coalescent.}

We will first make two comments about this criterion: one about its relation to logical equivalence and one about its being only a necessary condition for theoretical equivalence; and then we will give an example. First of all, this notion of inter-translatability seems promising, because it is a welcome weakening of logical equivalence, which in Section \ref{sse} we saw to be too strict. The idea of definitional equivalence is indeed that, when the two theories do not have all terms in common, one may enlarge their vocabularies by adding definitions of new terms {\it in terms of old ones},\footnote{This is an important restriction, because in the context of equivalence we should not be allowed to simply augment a theory with arbitrary new terms. Thus Glymour (1977:~p.~229) rightly says that one may add `a predicate, but no new axioms' to the theory's language. In the context of the reduction of one theory to another, the new definitions are usually called `bridge laws'. A standard example is: to reduce the thermodynamics of an ideal gas to the statistical mechanics of massive particles, one defines the temperature in terms of the average kinetic energy of the particles. (See the discussion of the Nagelian idea of reduction of one theory to another, in Section \ref{eandr} and Chapter \ref{Understand}.)} 
and one does this for the two theories in such a way that the two theories that result {\it are} logically equivalent.

Thus the definability criterion {\it is} the logical equivalence criterion, applied to a new pair of theories that are obtained from the original ones by adding appropriate definitions. (The idea of a `standard translation', in Section \ref{eil}, can be understood as being of this type: for example, a modal symbol such as $\Box$ is defined in terms of an open formula in the language of an extensional predicate logic.)

Second, it is important to note that Glymour (in agreement with our discussion in Section \ref{lsps}, and as against Quine whom we will discuss shortly) sees definitional equivalence as only a {\it necessary condition} for theoretical equivalence, and not as a sufficient condition.\footnote{Thus Glymour and Quine agree that empirical equivalence is a necessary condition for theoretical equivalence, but construe the formal requirement differently (as inter-translatability vs.~a reconstrual of predicates). And both do seem to have in mind a notion of theoretical equivalence in which both theories in some sense ``say the same thing about the world'', i.e.~a notion that is also interpretative. However, Quine's and Glymour's interpretative attitudes are quite different, because Quine was sceptical about meaning and ontology (see Quine's (1960:\,\S2) discussion of referential indeterminacy) while Glymour (1977:~p.~228) is not so sceptical.} 
The reason is that purely linguistic criteria {\it never} are, by themselves, sufficient conditions for the theoretical equivalence of scientific theories, since there are further epistemic and ontological requirements for two theories to count as the same. (In other words, Glymour implicitly appeals to the distinction between the model-theoretic semantics and the physical semantics, that we discussed in \ref{lsps}.)\footnote{Glymour (1977:~pp.~227-228) emphasizes that the required additional criteria are not {\it a priori} criteria, but rather have to do with the body of evidence with which a theory is tested. Thus he mentions differences in confirmation by the evidence, testability by the body of evidence in question, and support for the theories. Among these additional criteria are also ontological criteria, e.g.~about the different number of entities that one needs to assume for the two theories (p.~237). See also Glymour's (1980:~pp.~12, 52, 131) `bootstrap theory of confirmation', which proposes that a hypothesis is confirmed by the evidence, relative to some theory, if a positive instance of the hypothesis can be deduced from the observations and the theory. Christensen (1983:~pp.~478-481) crititicized it for allowing cases of evidential irrelevance. For Glymour's reply and revised bootstrap theory, see Glymour (1983:~p.~627).}\\
\\
{\bf Example: definitional equivalence of SO(2) and U(1).} Maybe the simplest non-trivial example that is relevant for physical theories and their symmetries is the isomorphism between the groups U(1) and SO(2) (see e.g.~Section \ref{EM-duality}). We will argue that, while these groups are of course isomorphic, once written as a theory in a formal logical language, they would turn out {\it not} to be logically equivalent (this is because their signatures are different).\footnote{We are not alone in thinking that groups and their representations, axiomatized within set theory by defining an appropriate set-theoretic predicate, rather than directly within e.g.~first-order logic, are good illustrations of the relation between theories and models: see Suppes (1962:~p.~252).} 
Nevertheless, they are {\it definitionally} equivalent, i.e.~one can construct appropriate definitional extensions that are logically equivalent. We will first remind the reader how these groups are isomorphic, and then give the explicit definitions that show the theories of the two groups to be definitionally equivalent.\footnote{Thus, for the sake of illustration, we take these groups as ``bare theories''. Of course, there is only a space of states here, and no dynamics. Barrett and Halvorson (2016) consider similar examples.}

A natural way to find the isomorphism between U(1) and SO(2) is by considering their actions on a vector space: namely, on the complex numbers, $\mathbb{C}$, and on the plane, $\mathbb{R}^2$ respectively. The natural isomorphism $\mathbb{C}\simeq\mathbb{R}^2$ is of course given by $z:=x+iy\in\mathbb{C}\mapsto {\bf v}:=\left(\begin{array}{c}x\\y\end{array}\right)\in\mathbb{R}^2$. Then the action of counter-clockwise rotations of a vector ${\bf v}$ on the plane, i.e.~by an element of SO(2), induces a counter-clockwise rotation of the complex vector $z\in\mathbb{C}$ on the plane (and vice versa), and establishes the obvious isomorphism between U(1) and SO(2).

U(1) is of course defined as the set of complex numbers $u\in\mathbb{C}$ that preserve the norm of vectors on the complex plane, as follows:
\bea\label{U1}
\mbox{U}(1):=\{u=u_1+iu_2\in\mathbb{C}\,|\,uu^*=1\}\,,
\eea
where we have simplified the notation by not using an explicit dot for complex number multiplication and addition, since this follows from the corresponding operations of the real numbers. In effect, the axiom $uu^*=1$ requires that $u_1$ and $u_2$ are both real numbers with range $[-1,1]$, and that they satisfy the condition $u_1 u_1+u_2 u_2=1$, as they of course should, since they span the unit circle in the complex plane.

Likewise, SO(2) is defined as the set of $2\times2$ matrices that preserve the norm and orientation of vectors on the real plane:
\bea\label{SO2}
\mbox{SO}(2):=\{A:=\left(\begin{array}{cc}a&b\\c&d\end{array}\right)\in\mbox{SL}(2,\mathbb{R})\,|\,A\cdot A^{\tn T}=A^{\tn T}\cdot A=\mathbb{1}_{2\times2}\}\,.
\eea
Given the action of SO(2) on the plane, we can induce the action of U(1) on the complex plane and obtain the corresponding element of U(1) that performs that action, and vice versa, thus establishing an isomorphism between the two groups.

We can also see that these two isomorphic groups, if we view them as theories $T_1=\mbox{U}(1)$ and $T_2=\mbox{SO}(2)$ in a formal language, are {\it not} logically equivalent. For the models of U(1) are given by a specification of a subset of the complex numbers (with the operations of addition and multiplication and constants 0 and 1 naming the corresponding numbers), of the form $u=e^{i\th}$ (where $\th\in[0,2\pi)$), while the models of SO(2) are a specification of a subset of the $2\times2$ special linear matrices, of the form $R(\th)=\left(\begin{array}{cc}\cos\th&-\sin\th\\\sin\th&\cos\th\end{array}\right)$. And although these models are isomorphic, they are not the same, because a complex number is not a $2\times2$ matrix. (Namely, they are written in different signatures: $2\times2$ complex matrices with real entries vs.~complex numbers.)

To show that these two theories are definitionally equivalent, we need to construct a common definitional extension, and so we first need to write out their signatures. Note that the signatures of mathematical and physical theories normally share standard mathematical symbols (e.g.~for the set of reals, $\mathbb{R}$, and the operations defined on it, etc.). Although, strictly speaking, one should rewrite these shared symbols in disjoint signatures, this can always be done by a simple relabelling. And so, for the simplicity of the notation, we will not include this common part of the signature explicitly in the signatures $\Si_{\sm{U}(1)}$ and $\Si_{\sm{SO}(2)}$ below, which in this way {\it are} disjoint. 

A more substantial comment regards formalization: we will here formalize only the {\it new} definitions that are required for the definitional extensions. Thus our discussion of Eqs.~\eq{U1}-\eq{SO2} assumes that the field of complex numbers $\mathbb{C}$, and the matrix group $\mbox{SL}(2,\mathbb{R})$, have been previously defined in a formal language. They {\it can} be so expressed using predicate logic, but continuity is an irreducibly second-order notion. But we will take this in our stride, and also assume we have secured some axiomatic distinction between $\mathbb{R}$ and $\mathbb{C}$. (Thus, in our discussion above, one envisages variables ranging over the real numbers, and an individual constant $i$, with the operations of addition and multiplication, both for the real and the complex numbers, and e.g.~a specification of a subset of the complex numbers, $e^{i\th}\in\mathbb{C}$, by the axiom in Eq.~\eq{U1}, under complex multiplication.)

With these remarks in hand, the signatures can be written as follows: 
\bea
\Si_{\sm{U}(1)}&=&\{i,*\}\nn
\Si_{\sm{SO}(2)}&=&\Big\{\left(\begin{array}{cc}-&-\\-&-\end{array}\right),
|\cdots|,\,\cdot~\Big\}\,,
\eea
where $i^2:=-1$, and $*$ is the complex conjugation symbol, $z^*$. The four horizontal lines are places for $2\times2$ matrix components, the triple dots are a place for an arbitrary vector whose norm is being calculated, i.e.~$|{\bf v}|$, and the single dot $\cdot$ indicates matrix multiplication (including multiplication between a matrix and a vector, i.e.~$A\cdot{\bf v}$).

Explicit definitions of the elements $i$, $*$, of the signature $\Si_{\sm{U}(1)}$, in the signature $\Si_{\sm{SO}(2)}$, can be given along the following lines. Namely, we have the following sentences, where $\mbox{Def}_s$ is an explicit definition of the symbol $s\in \Si_{\sm{U}(1)}$ (as it occurs in the context of any complex complex number $u_1 + iu_2$) in terms of the signature $\Si_{\sm{SO}(2)}$:\footnote{For simplicity in the notation, we here only give the translation for the symbols themselves, rather than giving explicit definitions containing formulas for the relations, constants, and function symbols. For example, the constant symbol $i$ of $\mbox{U}(1)$ can be explicitly defined as: $\forall u_1\forall u_2\left(u_1+i\,u_2=i~~\leftrightarrow~\left(\begin{array}{cc}u_1&-u_2\\u_2&u_1\end{array}\right)=\left(\begin{array}{cc}0&-1\\1&0\end{array}\right)\right),$ and so on. For details, see Hodges (1997:~pp.~52-53).} 
\bea
\mbox{Def}_i:&&\forall u_1\forall u_2\in[-1,1]\left(u_1+iu_2~:=\,\left(\begin{array}{cc}u_1&-u_2\\u_2&u_1\end{array}\right)\right)\nn
\mbox{Def}_*:&&\forall u_1\forall u_2\in[-1,1]\left((u_1+iu_2)^*:=\, \left(\begin{array}{cc}u_1&u_2\\-u_2&u_1\end{array}\right)\right).
\eea
We here take $u_1$ and $u_2$ to range over the interval $[-1,1]$, so as to match the defining axiom of $\mbox{U}(1)$ in Eq.~\eq{U1}, which requires that any $u_1$ and $u_2$, that enter into sentences in $T_1$ have that range (and likewise for the definitions of the symbols of $\Si_{\sm{SO}(2)}$ below). Note that, from the defining axiom of $T_1$ in Eq.~\eq{U1}, $u_1$ and $u_2$ span the unit circle, and so there is the additional condition that $u_1$ and $u_2$ satisfy the equation $u_1u_1+u_2u_2=1$. However, in order to simplify the notation, we suppress this condition, which is common to the two sets of explicit definitions. (Another simplification: we write sentences containing $u_1$ and $u_2$, rather than writing open sentences with arbitrary variables $x,y,\ldots$, and $u_1$ and $u_2$; and likewise for $\mbox{SO}(2)$, below.)

Adding these two definitions to the signature $\Si_{\sm{SO}(2)}$ of $T_2$ gives the signature $\{i,*\}\,\cup\,\Si_{\sm{SO}(2)}=\Si_{\sm{U}(1)}\,\cup\,\Si_{\sm{SO}(2)}$.\footnote{To translate {\it all} of the statements of $T_1$ into $T_2$, and not just the group elements as in the main text, we should strictly speaking also add definitions that translate complex numbers into numbers on the real plane and vice versa. As we mentioned before, we omit this for simplicity.} Likewise, the explicit definitions of the symbols of $\Si_{\sm{SO}(2)}$ in terms of $\Si_{\sm{U}(1)}$ are as follows:
\bea\label{defs2}
\mbox{Def}_{\left(\begin{array}{cc}-&-\\-&-\end{array}\right)}
:&&\forall u_1\forall u_2\in[-1,1]\left(\left(\begin{array}{cc}u_1&-u_2\\u_2&u_1\end{array}\right):=\,u_1+iu_2\right)\nn
\mbox{Def}_{\tn T}:&&\forall u_1\forall u_2\in[-1,1]\left(\left(\begin{array}{cc}u_1&-u_2\\u_2&u_1\end{array}\right)^{\!\sm T}:=\,u_1-iu_2\right)\nn
\mbox{Def}_.:&&\forall\th_1\forall\th_2\in[0,2\pi)\left(R(\th_1)\cdot R(\th_2)\,:=\, e^{i(\th_1+\th_2)}\right),
\eea
where $R(\th)$ is a counter-clockwise rotation matrix by an angle $\th$. Adding these definitions to the signature $\Si_{\sm U(1)}$ of $T_1$ gives again the signature $\Si_{\sm{U}(1)}\,\cup\,\Si_{\sm{SO}(2)}$.

The dashed theories thus defined, i.e.~$T_1'=T_1\cup \Si_{\sm{SO}(2)}$ and $T_2'=T_2\cup\Si_{\sm{U}(1)}$, are logically equivalent. This is because their models are {\it the same}: namely, any model, i.e.~the set of numbers $u=e^{i\th}=\cos\th+i\sin\th$, of $T_1$ with the operations of multiplication and complex conjugation on it, provides, through the added explicit definitions, a 
model whose domain {\it is} a set of matrices (as defined by Eq.~\eq{defs2}) $R(\th)$ of $T_1'$, which is also a model of $T_2$ and hence also of $T_2'$. And the procedure is symmetric: any model of matrices $R(\th)$ of $T_2$ provides, through the added explicit definitions, a model whose elements are complex numbers $\cos\th+i\sin\th$ of $T_2'$, which is a model of $T_1$ and hence also of $T_1'$. In other words, SO(2) and U(1) are definitionally equivalent.\\
\\
{\bf Quine's criterion of theoretical equivalence:} Rather than adding definitions to reach a common definitional extension of the two theories, Quine (1975:~p.~320) in effect leaves the two theories intact, but maps one into the other by what he calls a {\bf reconstrual} of the predicates. A reconstrual is any mapping of $n$-place predicates of one theory into $n$-variable open sentences of the other. In other words, every $n$-ary symbol of $\Si_1$ is mapped to an open formula of $\Si_2$ with the corresponding $n$-arity.\footnote{Quine's criterion has been formalized by Barrett and Halvorson (2016a:~pp.~471-473).}

The criterion is then that {\it two theory formulations express the same theory if they are empirically equivalent and if there is a reconstrual of predicates that transforms one theory into a logical equivalent of the other}. 

Quine's criterion of reconstrual of predicates is too weak, since it counts as theoretically equivalent theories that one does not expect to count as equivalent. For example, it can map an incomplete theory (i.e.~one in which it is {\it not} true that every well-formed sentence is either a proven theorem, or its negation is a theorem) to a complete theory (in which all such sentences are either proven or refuted), rendering them theoretically equivalent:\footnote{This defect is exhibited in an example given by Barrett and Halvorson (2016a:~p.~475).} which is surely not a correct verdict, according to the requirement that theoretical equivalence should be truth-preserving (see (B) in Section \ref{lsps})).

More broadly, the problem is that, despite Quine's criterion mapping $T_1$ to a theory that is logically equivalent to $T_2$, the converse is not necessarily true, and so the mapping is not truth-preserving in both directions. Quine's criterion can be appropriately strengthened by adding a map in the other direction, i.e.~having mappings both ways (although these mappings need not be each others' inverses: the combination of the two maps might map a sentence of $T_1$ to another sentence that is equivalent to it according to $T_1$). We will call this strengthened criterion {\bf Quine inter-translatability}.

Barrett and Halvorson (2016a:~p.~478) then prove a theorem to the effect that, if two first-order theories are Glymour inter-translatable (i.e.~they have a common definitional extension), then they are also Quine inter-translatable. And if the signatures of the two theories are disjoint (which can always be achieved by a mere relabelling of the symbols), then the converse is also true. Thus, at least in first-order logic, Quine inter-translatability and a natural strengthening of Glymour's account of theoretical equivalence amount to the same inter-translatability condition. 

Both kinds of inter-translatability hinge on rendering two theories logically equivalent, in the sense of their having the same models. Thus it is clear that Glymour and Quine's notions of inter-translatability are not purely syntactic in the sense of `formal i.e.~non-interpretative', but are also semantic: in spite of their formalistic appearance, making the same sentences true is in the end what matters. And thus they agree with the three points, (i)-(iii), that we mentioned at the end of Section \ref{eil} as general characteristics of definability theories (note that those conditions, especially (iii), are much weaker, i.e.~inter-translatability is a strong requirement).

\subsection{Model-theoretic criteria: a syntactic conception of duality}\label{mtce}

After having reviewed two main linguistic criteria of equivalence in the philosophy of science, we now turn to the semantic conception of theories, where a theory is seen as a collection of models, i.e.~of mathematical structures (see Section \ref{eil}).

When we compare one mathematical structure to another, isomorphism is the natural criterion of equivalence. For an isomorphism is, by definition, a bijective map that preserves the relevant structure (i.e.~a bijective homomorphism or surjective embedding: see Hodges, 1993:~p.~5). By `structure', we usually mean a set with relations between the elements of the set, e.g.~a Kripke frame. The structure is preserved just in case the relations between elements of the one set obtain iff the corresponding relations also obtain between elements of the other set.\footnote{The relations are in general not the same, because, unlike in the case of logical equivalence, where the two theories were required to use the {\it same} signature (see Section \ref{sse} (A)), there is no requirement in model theory that isomorphic models should be written using the same signature. For a brief philosophical discussion, see Burgess (2016:~p.~109). See also van Benthem (2010b:~p.~26).}

Thus also in model theory, isomorphism of models is a very natural criterion of equivalence, here understood as the `structural indistinguishability' of semantic structures (van Benthem, 2001:~p.~340: see Section \ref{synsemeq}).\footnote{Isomorphism is of course not the only structural criterion of equivalence. In a recent series of papers, Weatherall (e.g.~2016a, 2016b) has advocated {\it categorical equivalence} as a standard of equivalence, and has applied it to examples in physics. As Barrett and Halvorson (2016b:~p.~566) and Hudetz (2019:~p.~51) have argued, this criterion is too weak (and our own reasons for this judgment follow from our analysis of the Maxwell theory: De Haro, 2020a:~p.~5166)). So far as we are aware, the only example from Part II of this book where the philosophical literature has used category theory is the elementary duality of the classical Maxwell theory (Section \ref{MEMD}). And so, it is an interesting open question whether the category-theoretic criterion of equivalence can be fruitfully strengthened and applied to examples, such as sine-Gordon-Thirring duality, where duality is already known to be a good formal criterion of equivalence. In the physics literature, monoidal categories have been used extensively in statistical mechanics, conformal field theory, and topological quantum field theory: see e.g.~the excellent Turaev (1992, 2010); also Dijkgraaf (1997:~pp.~20-27), Atiyah (1988:~pp.~177-182) and Kock (2003). In particular, they have been used to axiomatize two and three-dimensional quantum field theories. So far as we know, these categories have not yet been used by philosophers of physics. Another discussion of category-theoretic criteria of equivalence is in Halvorson and Tsementzis (2017:~pp.~406-413).\label{categoryT}} The isomorphism drags along the extension in the domain of one model to the codomain, where it coincides with the relation over there.

But also in natural science, especially physics, isomorphism is a privileged criterion of equivalence.\footnote{When the criterion is weakened to {\it partial} isomorphism, one is simplifying one's problem by considering only parts of the structure, or by abstracting from the details of one of the structures, or by considering only the aspects that are relevant to the problem at hand. In other words, one is happy to consider structural {\it similarities} only; for structural {\it equivalence} requires isomorphism.} And we already saw that {\it duality} is an isomorphism between theories, as physicists usually define their theories. Thus in this Section, we will discuss whether the model-theoretic notion of isomorphism meshes with the conception of duality in the practice of physics.

In order to see the close connection between the syntactic and semantic conceptions of theories (and to be able to contrast them later),\footnote{Although it is true that the semantic conception of theories was in part developed as a rival, or as a replacement, of the syntactic conception, there are relations between them that are worth emphasizing. See point (i) in Section \ref{refsr}.} it will be convenient to define the model-theoretic criterion of isomorphism in terms of a weakening of logical equivalence, which---recall from Section \ref{sse} (A)---requires the {\it identity} of models, and can be conveniently rewritten as follows:
\bea\label{LE}
T_1\overset{\sm{LE}}{\sim}T_2~~\Leftrightarrow~~\textfrak{M}_1=\textfrak{M}_2\,.
\eea
Here, $\overset{\sm{LE}}{\sim}$ means `logical equivalence', and we have introduced the notation $\textfrak{M}_1$ and $\textfrak{M}_2$ for the sets of models of the {\it syntactically} defined theories $T_1$ and $T_2$, respectively:
\bea\label{Mmodels}
\textfrak{M}_1&:=&\{M|\,\forall \f\in T_1~M\models\f\}\nn
\textfrak{M}_2&:=&\{N|\,\forall\varphi\in T_2~N\models\varphi\}\,.
\eea
Then the {\bf isomorphism criterion} is a {\it weakening} of the above condition, as the {\it pairwise isomorphism} between models, $M$ and $N$, that are elements of $\textfrak{M}_1$ and $\textfrak{M}_2$, as follows:
\bea\label{isoT}
T_1\overset{\sm{isom}}{\sim}T_2~~~\Leftrightarrow~~\left\{\!\!\begin{array}{c}
\mbox{(i)~~$\forall M\in\textfrak{M}_1~\,\exists N\in\textfrak{M}_2$ such that $M\simeq N$}~~\\
\mbox{(ii)~~$\forall N\in\textfrak{M}_2~\,\exists M\in\textfrak{M}_1$ such that $M\simeq N$}\,,
\end{array}\right.
\eea
where $\simeq$ is the isomorphism between two models. This isomorphism criterion is weaker than logical equivalence, because $M_1 \simeq M_2$ is weaker than identity, i.e.~$M_1 = M_2$, in two different ways. 

These two ways reflect that models are finely individuated, both by (i) which set of objects is their domain of quantification; and (ii) which elements, i.e.~non-logical vocabulary items, appear in their signatures. Namely, two models can differ by: 

(i)~~having different domains of quantification (even though they are equinumerous sets, and so bijection is possible);

(ii)~~having different signatures (even if the number of items of a given arity, in each of the two signatures, is the same).

Thus the objects $\textfrak{M}_1$ and $\textfrak{M}_2$, perhaps endowed with additional structure, correspond to what one calls `theories' in the {\it semantic conception} of scientific theories: namely, according to this view, a scientific theory is defined directly as a {\it collection of models} (usually endowed with additional structure, depending on the authors).\footnote{For recent discussions, see Frigg (2023:~Chapter 5) and Halvorson (2012:~pp.~201-202). See also van Fraassen (1970:~p.~337) and Suppes (1962:~p.~252).} 
And it is important to note that, although the above criterion is obtained within a `syntactic' formulation of a theory, it is really the official model-theoretic criterion of isomorphism of models, including its appeal, shown in Eq.~\eq{Mmodels}, to a particular language.

In the philosophical literature, the view that a scientific theory is not flat, i.e.~a collection of models, but is structured has been advocated by Halvorson (2012:~pp.~204-205), who asks when a set of models contains all the information about a syntactically formulated theory. He suggests that introducing a topology on the set of models gives the required information, and Halvorson and Tsementzis (2017:~pp.~413-414) discuss various candidate categories to play the role of a `category of theories'. In the context of classical mechanics, Curiel (2014:~p.~275; cf.~p.~318) remarks that `quantities determine the topological and differential structure of the space of states'.\footnote{An excellent brief exposition of this point is in Wald (1994:~pp.~11-14). Arnold (1989:~Chapter 8) is a detailed discussion. In Section \ref{mvd}, we will endorse this view for other theories than classical mechanics, since what Chapter \ref{EMYM} called `moduli' are physical quantities and variables that can be used to coordinatize a moduli space equipped with symplectic and metric structure.}
Also Fletcher (2016:~p.~366) has adverted to the use of topology for models of general relativity, especially in connection with the question of the stability of properties.\footnote{See also Hawking and Ellis (1973:~Section~7.6), Belot (2018:~p.~968), and Geroch (1969:~p.~182).}\\
\\
Regarding dualities: we see that, if $\textfrak{M}_1$ and $\textfrak{M}_2$ correlate with what, in Section \ref{ThisB}, we called `models' (and which we denoted by $M_1$ and $M_2$), we recover our notion of {\it duality}. More precisely, it seems that we have arrived at a version of the idea of duality {\it within a syntactic formulation} of theories. 

It is worth spelling out the relation between these two formulations of the idea of duality, since there are also some differences---as we expect there to be, since a syntactic and a semantic formulation of a theory are in general different. 

First, about the terminology: we should stress the different {\it objects} that the syntactic and semantic conceptions call `theories'. The syntactic conception calls the sets of {\it sentences}, $T_1$ and $T_2$, its `theories': while the semantic conception calls the sets of {\it models}, $\textfrak{M}_1$ and $\textfrak{M}_2$, its `theories'. On its side, the Schema uses a duality-motivated terminology that is intermediate between the two: it agrees with the syntactic conception that $\textfrak{M}_1$ and $\textfrak{M}_2$ are not themselves theories, but rather representations or instantiations, of theories: but it also agrees with the semantic conception that, in scientific practice, a theory is most conveniently defined---not as a set of sentences, but semantically---as a structure: in our notation, a triple, $T=\bra{\cal S},{\cal Q},{\cal D}\ket$.\footnote{For a brief discussion of the semantic conception of theories, see the discussion at the end of Section \ref{thmscph}.} 
(Having said that, in the next Chapter we will find that the syntactic reformulation of a duality discussed here gives a very useful way to understand serveral aspects of dualities.)

Note that, just like the two sets of models, $\textfrak{M}_1$ and $\textfrak{M}_2$ in Eq.~\eq{Mmodels}, the Schema's `models' (in the Schema's sense!), $M_1$ and $M_2$, are themselves {\it sets of models}. That is, they are {\it sets of structures}: in particular, a structured set of {\it states} (typically, either states in a Hilbert space, or points of a manifold) and a set of {\it quantities}, dual (in the mathematical sense) to these states and equivariant with the dynamics. 

A second similarity between the sets $\textfrak{M}_1$ and $\textfrak{M}_2$ and the Schema's models is that they both contain {\it specific structure}: $\textfrak{M}_1$ and $\textfrak{M}_2$ are each a set of models that instantiate a syntactically defined theory (i.e.~respectively, $T_1$ or $T_2$), but $\textfrak{M}_1$ and $\textfrak{M}_2$ are in general very different (in both our ways (i) and (ii) above), and so there can be structure that is not mapped across in the isomorphisms between members of $\textfrak{M}_1$ and $\textfrak{M}_2$. Likewise, the Schema's models contain specific structure that we denoted by an overbar (i.e.~$\bar M$). This specific structure is used to construct the models, but it is not mapped by the duality.

For example, in the sine-Gordon-Thirring duality from Section \ref{SGT}, there are two ways of representing a conserved current $J^\m$: (1) as a topological current for the soliton number of a bosonic field (Eq.~\eq{curr1d}); or (2) as the Noether current for a global U(1) symmetry, which is a fermion bilinear (Eq.~\eq{ThirrJ}). In fact, {\it if} we had independent syntactic formulations of the quantum theories of the boson and the fermion in two dimensions (for example, as sets of axioms in an axiomatic quantum field theory of bosons and of fermions), then those formulations would be candidates for the syntactically formulated theories $T_1$ and $T_2$. We will give a toy version of such an axiomatic formulation, using conformal field theory, in Section \ref{lsrr}, for which the example below will give some preparatory comments.

Likewise for the other dualities that we have encountered: if we had pairs of axiomatically formulated theories whose respective sets of models $\textfrak{M}_1$ and $\textfrak{M}_2$ are the Schema's dual models, then these axiomatic theories would be candidates for the syntactically formulated theories $T_1$ and $T_2$, and this would explain the differences in the specific structure of duals.

Thus we conclude that, according to a syntactic conception of theories, there is, at least in principle, a natural way to interpret the Schema's conception of duality as an isomorphism of models of the type Eq.~\eq{isoT}, where $T_1$ and $T_2$ are axiomatic theories whose models carry specific structure.\footnote{An important difference between the Schema's models and $\textfrak{M}_1$ and $\textfrak{M}_2$ is that the former sets are themselves almost always {\it structured} (e.g.~they are endowed with symmetries, and in some cases with a measure or other geometric quantities defined on the set of states), while the latter have been defined as sets with no further structure. This then appears to be the {\it main} difference between the Schema's syntactic reinterpretation here discussed, and the official treatment of the set of models in the syntactic conception of theories. We shall return to this point in Section \ref{mvd}. This distinction is also the springboard of Halvorson's critique of the semantic conception of theories and his advocacy of category-theoretic notions, e.g.~in Halvorson and Tsementzis (2017:~pp.~406-413).\label{Mstructured}} 
Needless to say, finding axiomatic $T_1$ and $T_2$ in the examples from Part II is a task that is well beyond the scope of this book.

There is an important new element that the Schema adds to the syntactic conception of theories and that will recur in the practical uses of dualities in the following Chapters: namely, the notion that the models (in the Schema's sense) are mathematical representations or instantiations, not of syntactically defined theories, but of a semantically defined theory $T=\bra{\cal S},{\cal Q},{\cal D}\ket$, i.e.~a theory that is formulated in the same way as the models themselves: as a structured set of states, quantities, and a dynamics.\\
\\
{\bf A syntactic formulation of a toy quantum duality.} We illustrate the previous ideas in a schematic or toy example that is relevant to quantum dualities. (Section \ref{lsrr} will fill in details for the example of bosonization.)

We consider two theories, $T_1$ and $T_2$, each defined by a set of $n+1$ axioms, of which $n$ are common to the two theories.\footnote{As in the rest of this Chapter, we adopt the normal usage, calling the (here, syntactically formulated) duals, $T_1$ and $T_2$, `theories'. We will return to our jargon of `theory above' and `dual models below' in the next Chapter.}
We dub these common axioms $A_1,\ldots,A_n$, and they are written using a signature $\Sigma$ that is also common to the two theories. In addition, each theory has a specific $(n+1)$th axiom  that uses symbols that are added to the common signature: these sets of extra symbols are $\Sigma_1$ for $T_1$ and $\Sigma_2$ for $T_2$ (see Section \ref{sse}'s definitional extensions). Thus the two theories are written as:
\bea\label{T1T2}
T_1&:=&\bra A_1,\ldots,A_n,B_1;\Sigma,\Sigma_1\ket\nn
T_2&:=&\bra A_1,\ldots,A_n,B_2;\Sigma,\Sigma_2\ket\,,
\eea
where $B_1$ and $B_2$ are the specific axioms. Since the two theories share a common core $T=\bra A_1,\ldots,A_n;\Sigma\ket$, we can rewrite them as:
\bea\label{T1T2b}
T_1&=&\bra T;B_1;\Sigma_1\ket\nn
T_2&=&\bra T;B_2;\Sigma_2\ket\,.
\eea
To secure that there is an isomorphism, according to the condition in Eq.~\eq{isoT}, we further stipulate that the specific axioms $B_1$ and $B_2$ only change the models of the common core $T$ in the senses of (i) and (ii) above. 

With an eye on examples of quantum dualities, we take each additional axiom, $B_1$ and $B_2$, jointly with its additional pieces of signature, $\Sigma_1$ and $\Sigma_2$, to be a {\bf bridge law} between the common core theory $T$ and some other (familiar) theory, that we will think of as a semi-classical theory. The idea (which we will discuss a bit more for bosonization in Section \ref{lsrr}) is that $T$ is a quantum field theory, and that the additional axioms $B_1$ and $B_2$ specify two limits, in each of which the common core theory reproduces a different semi-classical theory. Thus in particular, each of the additional axioms specifies a notion of convergence to a classical limit. The additional pieces of signature, respectively $\Sigma_1$ and $\Sigma_2$, enable the definitions required to rewrite the quantum theory, in the limit, as a semi-classical theory. This in effect secures a Nagelian reduction (cf.~Section \ref{eandr} and Chapter \ref{Understand}) of each of the semi-classical theories to the common core theory $T$.

To illustrate, consider a simpler example, where the common core theory $T$ specifies the set of axioms and the signature for a vector space of dimension $m$ (see the `example of a representation map' in Section \ref{lsr}), and $B_1$ and $B_2$ each specify a surjective homomorphism onto two other vector spaces of dimension $m-1$ (and these lower-dimensional vector spaces are also different from each other). Thus in effect, the additional axioms project the vector space onto a lower-dimensional vector space along two different directions, and the additional pieces of signature secure an embedding of the lower-dimensional vector spaces into the one $m$-dimensional vector space (in two different ways). 

The sets, $\textfrak{M}_1$ and $\textfrak{M}_2$, of models of the theories $T_1$ and $T_2$, are the sets whose elements are vector spaces of dimension $m$ (e.g.~$\mathbb{R}^m$), and they are pairwise isomorphic. Each model comes equipped with a projection onto a specified subspace, and each model is written using its own signature. The $m$-dimensional vector spaces are representations of the common core theory $T$, while the projections (and the changes of signature required to link the higher-dimensional to the lower-dimensional vector spaces) are specific structure.

Returning to the quantum case, the above toy duality illustrates that we have two isomorphic theories, in the sense of Eq.~\eq{isoT}, with pairwise isomorphic models, which are identical as quantum theories (i.e.~they have the same common core quantum field theory $T$). However, each theory has a different semi-classical limit, with its own signature, and so the models of each of the theories can look very different.

\section{In defence of isomorphism; against weaker criteria of equivalence}\label{defencei}

In the previous Section, we presented the model isomorphism criterion as a {\it weakening} of logical equivalence (i.e.~Eq.~\eq{isoT} as against Eq.~\eq{LE}). And recall, from Section \ref{synsemeq}, van Benthem's `first question' of model theory about how far the `webs of language and ontology diverge': and his judgment that, compared to the linguistic criteria, `isomorphism [is] by far the finer sieve'. Thus the isomorphism criterion strikes a balance between logical equivalence and the weaker linguistic criteria discussed in Section \ref{sse} (B).

However, a number of philosophers have criticized the isomorphism criterion of equivalence for being---just like the criterion of logical equivalence---{\it too strict}. 

In this Section, we argue that such criticisms---often given in just a few lines---usually rest on misconceptions that are best exposed by distinguishing between the bare theory and the interpretation maps from Section \ref{itm}, i.e.~by distinguishing aspects (A) and (B) (in Section \ref{lsps}) in our notions of equivalence. (We do not deny that the isomorphism criterion is too strict in e.g.~modal logic: but we will argue that it is not too strict for physical theories, e.g.~in the examples of dualities discussed in Part II.) 

Sections \ref{bsa} and \ref{incorr} illustrate how one main example has been incorrectly taken to show that the isomorphism criterion is too strict.\footnote{The example is in Barrett (2015:~p.~826; 2019:~p.~1171; 2020:~p.~1187). In De Haro (2021:~pp.~5150-5152), we criticized counter-examples that fail to adequately make the case against the isomorphism criterion, because they make appeals to vague and-or unqualified notions of isomorphism (e.g.~a criticism of the equivalence between Heisenberg's matrix and Schr\"odinger's wave mechanics that compares representations of matrix algebras and of wave-functions directly, rather than comparing the functions on these two spaces, i.e.~sequences which are functions on the discrete configuration space, $l^2$, and wave-functions on the continuous configuration space: and these {\it are} indeed isomorphic---see also Section \ref{wpd}). It is unfortunate that these examples are sometimes given without a detailed critical analysis, and then repeated as reasons to dismiss the isomorphism criterion. \label{isocritique}} 
Section \ref{counterA} then turns the tables, and argues that the mistakes thus revealed undermine how the logically weaker criteria of equivalence themselves (i.e.~logically weaker than the isomorphism criterion) apply to natural science. Then, in Section \ref{lessons}, we further substantiate our argument by giving two general interpretative principles that, in effect, mean that the stricter formal criteria of equivalence (i.e.~isomorphism, Eq.~\eq{isoT}) are favoured over the logically weaker ones.\footnote{Section \ref{defencei} has benefitted from discussions with Thomas Barrett,  Jill North and James Weatherall, all of whom we thank. Especially we thank Barrett for a discussion of some of the content of Section \ref{counterA}. However, disagreements remain, as we are about to explain.}
These two interpretative principles constrain the desirable relative logical strength between a formalism and its intepretation: we will use this in Chapter \ref{physeq}, where we will more fully develop the topic of the `desirable logical strength'. 

\subsection{Barrett's signature argument}\label{bsa}

In this literature, formal standards of equivalence are taken as necessary conditions for theoretical equivalence (we will endorse this in Chapter \eq{physeq}). Furthermore, the rhetoric of this literature appeals to theoretical equivalence as motivation to construct logically weak criteria of equivalence (i.e.~weaker than the isomorphism criterion).\footnote{Elsewhere, we have described the project, in this recent literature on theoretical equivalence by Barrett, Weatherall, and others, as one of `explicating equivalence in formal terms': and we have qualified it as a {\it quietist} position about interpretative equivalence: it is `a minimal requirement that their project needs, but on which they do not wish to focus' (see De Haro (2021:~p.~5154)). While we endorse their project of formal explication of equivalence, which we understand in terms of finding a {\it formal correlate of theoretical equivalence}: we reject quietism, because, as we will argue, not focussing on interpretation in the natural sciences is prone to lead to error. For another critique, see North (2021:~pp.~212-214), who argues that this literature attempts to capture `informational equivalence between theories, and they take this to be wholesale theoretical equivalence'. This focus on informational equivalence comes at the expense of what she calls `metaphysical equivalence'.}
In particular, the rhetoric of Barrett's critique of the model isomorphism criterion is that this criterion does not count as theoretically equivalent, situations that are supposed to be the same: i.e.~the difference is said to be a ``mere convention'', that one does not wish to count as a real difference---as we will now discuss.

Barrett's (2015:~p.~826) argument against the model isomorphism criterion of equivalence uses as an example the choice of a signature convention in general relativity: 
\begin{quote}\small
one can define a relativistic space-time as a pair $({\cal M},g)$ where $g$ has signature $(1,-1,-1,-1)$, or as a pair $({\cal M},g')$, where $g'$ has signature $(-1,1,1,1)$. The only difference between these two formulations of general relativity is a sign convention.
\end{quote} 
Barrett argues that, since these two formulations of general relativity differ only by a sign convention for the signature of their metric tensors, we should consider them to be {\it equivalent}, in the sense that they ascribe the {\it same structure to the world}. In other words, this is supposed to be a case of theoretical equivalence.

Yet, according to the isomorphism criterion, two spacetimes with different signature are not isomorphic, and so these two formulations of general relativity are not equivalent. He goes on to take this example (and another one similar to it) to
\begin{quote}\small
show that the isomorphism criterion of equivalence is too strict a standard of equivalence (Barrett, 2020:~p.~1187).
\end{quote} 

Barrett's argument is an {\it enthymeme}, i.e.~it has unstated premisses: and to be able to analyse it, we need to unpack it and explicitly formulate its assumptions. Thus we reconstruct Barrett's argument, as given in (2015:~p.~826; 2019:~p.~1171; 2020:~p.~1187), as follows:

(1)~~Two formulations of general relativity, one whose models are manifolds with $(p,q)$ metric signature,\footnote{Here, and in what follows, $p$ is the number of positive eigenvalues of the metric tensor, and $q$ its number of negative eigenvalues. $p$ and $q$ can be any positive numbers that add up to the dimension of the spacetime. The most familiar cases are of course $(1,3)$ and $(3,1)$, but the argument benefits from this more general formulation. For more on the interpretation of this notation, see Eq.~\eq{pqcon} below.} and another with $(q,p)$ metric signature, differ by a mere {\it sign convention} for the metric tensor (i.e.~a distinction that is not a real difference).

(2)~~If two formulations of general relativity differ only by a sign convention for the signature of the metric tensor, a good formal criterion of equivalence should judge them to be {\it equivalent}.

(3)~~(From (1) and (2)): A good formal criterion of equivalence should judge two formulations of general relativity, whose models are manifolds with metric signatures $(p,q)$, respectively $(q,p)$, to be {\it equivalent}.

(4)~~According to the {\it model isomorphism criterion} of equivalence, manifolds with different metric signature are not isomorphic.

(5)~~(From (4)): According to the {\it model isomorphism criterion} of equivalence, two formulations of general relativity whose models are manifolds with metric signatures $(p,q)$, respectively $(q,p)$, are {\it inequivalent}.

(6)~~(From (5) and (3)): {\it The model isomorphism criterion of equivalence is not a good formal criterion of equivalence. }

Barrett's argument should be understood\footnote{The argument does not appear to depend on the theory being pure general relativity (i.e.~the Einstein field equations in vacuum), since it only requires that one can adopt a different convention for the spacetime signature. Thus the argument could, and even {\it should}, be formulated generally, for theories based on a relativistic spacetime (including e.g.~special relativity and quantum field theories). In other words, besides a metric tensor field (dynamical or fixed), the relativistic manifold might also have other fields on it. However, in order to not depart too much from Barrett's literal words, we do not do this in the list (1)-(6): but we will consider appropriate generalizations below.} as in Figure \ref{barrett}, where the two formulations of general relativity (both of them understood as a set of models of spacetimes) are related by a map ${\cal C}$ (a mnemonic for `change of signs' or `convention', or `criterion of equivalence') that changes the metric signs of the models. The map ${\cal C}$ maps one theory into the other, and thereby changes the sign convention for the metric tensor, i.e.~it changes a model with  the set of signs $(p,q)$ into a model with the set of signs $(q,p)$. 

\begin{figure}
\begin{center}
\bea
\begin{array}{ccccc}
~~~~~~~~~~~T_1\!\!\!\!\!\!\!\!\!\!\!\!&\xrightarrow{\makebox[.6cm]{$\sm{${\cal C}$}$}}&\!\!\!\!\!\!\!\!\!\!\!T_2~~~~~~~~~~\\
~~~~~~~~~~~~~~~~~{\sm{$i$}}\!\!\searrow\!\!\!\!&&\!\!\!\!\swarrow\!{\sm{$i'$}}~~~~~~~~~~~~~~~~~\\
&W&\end{array}\nonumber
\eea
\caption{\small Two formulations of general relativity, $T_1$ and $T_2$, that differ only in their convention for the  signs of the metric. ${\cal C}$ maps one formulation to an equivalent one, i.e.~$T_2\sim T_1$, under the equivalence relation derived from ${\cal C}$. Both theories describe the same (class of) spacetimes, i.e.~the set of possible worlds $W$.}
\label{barrett}
\end{center}
\end{figure}

The signature-changing map ${\cal C}$ can be used to give an {\it equivalence relation} between the two formulations of general relativity with the two sets of signs $(p,q)$ and $(q,p)$, i.e.~$T_1\overset{{\cal C}}{\sim}T_2$: and so, it satisfies the above requirement in (3), in our reconstruction of Barrett's argument, for a `good formal criterion' of equivalence. 

But what motivation or justification could one have for the requirement in statements (2) and (3), i.e.~that a map like ${\cal C}$ makes the two formulations {\it formally} equivalent? Why should solutions of general relativity with  the two sets of signs $(p,q)$ and $(q,p)$ be {\it formally equivalent}, in any substantive sense of the word ``formal''? 

Barrett's answer is in (1): namely, a sign convention for the signature of the metric is not a `real difference', and is expressed in Figure \ref{barrett}: the argument seems to be that, since we are conjoining the ``same reference'' requirement (i.e.~same set of worlds) with a formal requirement of equivalence (see Section \ref{lsps}), we are, in effect, requiring that the two formulations of general relativity are ``theoretically equivalent''. 

Another reason for the formal equivalence could of course be that, although ${\cal C}$ is not an isometry of the given metric manifold (i.e.~it does not preserve the metric), {\it ${\cal C}$ is a diffeomorphism}, i.e.~a bijective, differentiable map between manifolds (and its inverse is also differentiable). 

Reasonable though this might at first sight seem, we will now argue that Figure \ref{barrett} is {\it not} a case of theoretical equivalence: and that, in fact, each of the statements (1) to (3), and (5), are problematic.

\subsection{An incorrect critique of the isomorphism criterion}\label{incorr}

Barrett errs in claiming to have shown that the model isomorphism criterion is too strict. For, as we will now argue, this is not what his examples show. And, although to this end we will use our reconstruction of his (very brief) argument, i.e.~statements (1) to (6), we claim that the problems that we will expose are quite independent of the precise form of this reconstruction. Indeed, there are three main errors, where (A) is the main one that leads in to (B), and (B) and (C) are errors in the application of the isomorphism criterion:

(A)~~The first error is in statement (1), which says that two models (i.e.~two solutions) of general relativity that differ only in their metric signature (i.e.~$(p,q)$ as against $(q,p)$) differ by a `mere sign convention' for the signature of the metric tensor.\footnote{It is sometimes said that a spacetime ``has signature $(-1,1,1,1)$''. But strictly speaking, the signature $s$ is defined as $s=p-q$, and it is this quantity that, if the metric is non-degenerate and continuous, is {\it constant} throughout the manifold (Hawking and Ellis, 1973:~p.~38). Furthermore, since the sum $p+q$ is the dimension of spacetime, $p$ and $q$ are separately also constant throughout the spacetime. On the other hand, in the usual coordinates, a change from e.g.~$(-1,1,1,1)$ to $(1,-1,1,1)$ takes place at the horizon of a black hole: and so, $(-1,1,1,1)$ is not constant throughout the spacetime, and it is {\it not} a signature of the spacetime. In other words, at the horizon of a black hole there is no change of signature, while the signs of the eigenvalues of the metric {\it do} change. See also the example of the two-dimensional black hole, below.} 
This confounds the formal and interpretative aspects of the words used. 

The point is that a {\it convention} for the  set of signs of the metric tensor is an (arbitrary) assignment of a specific number of space and a specific number of time dimensions to a {\it formal} signature $(p,q)$ (i.e.~the number of positive and negative eigenvalues of the metric tensor field), and accordingly also an assignment of which vectors we define to be timelike and spacelike, depending on the sign of their norm. In other words, a sign convention for the signature of the metric tensor is a choice between two possible interpretation maps, one where the set of signs $(p,q)$ is interpreted as the manifold's having $p$ dimensions of space and $q$ of time, and the other where $p$ and $q$ are exchanged (see Eq.~\eq{pqcon} below). Thus a convention for the signature of the metric tensor is only fixed when an {\it interpretation map} is fixed, because the convention is not about the formal formulation of the theory, but about how the formulation of the theory relates to the world.\footnote{For this reason, conventions in physical theories are sometimes called `meaning stipulations' (Putnam, 1996:~p.~36), since they are theory-world relations, rather than formal i.e.~non-interpretative stipulations.} 

Agreed, one sometimes hears this simple view stated about conventions: that one may choose one convention or one may choose another, and that the choice is ``irrelevant for the physics''. And this is true, so long one also agrees that thinking about conventions in this way is an {\it enthymeme} with unstated assumptions that do not cover all cases (we will make these unstated assumptions explicit below). We will argue that {\it when we change a convention, we also change our interpretation: so that we should not require our formal criteria of equivalence to judge as equivalent situations that come out to be equivalent only if we change our interpretation.} To avoid making errors, one should be explicit about one's interpretative choices, i.e.~one must distinguish between bare theory and interpretation. Namely, what we argue is that it is best to think of theories in terms of bare theories and interpretations, and to be explicit about which interpretation one is using.\footnote{We think that there are few philosophers who disagree that the distinction between a bare theory and its interpretation is mandatory. We discuss here potential candiates, and argue that they do not militate against this distinction: (1) Section \ref{refsr} already argued for a reconciliation between our advocacy of referential semantics, and various lines of work that downplay reference: under `Theory' in Section \ref{Ourthm} we mentioned de Regt's observation that some of the claims of theory-independence are not as radical as they appear. (2) We endorsed Halvorson's response to French's eliminativism about scientific theories. (3) One might think that Williams' (2019:~p.~209) realism about effective field theories, especially his rejection of what he calls the ``standard account'' of interpretation (which is close to the `ideal of pristine interpretation'), is a reason to reject our interpretative analysis, and might block our reply to Barrett. There are two comments to make: the first is general, and the second is our reply to the worry. (i) We {\it in part} endorse Williams' (2019:~p.~215) critique of the ideal of pristine interpretation, especially what he calls its two main `vices': and we are sympathetic to his attitude towards effective field theories. However, we argue that this is not a reason to reject referential semantics: we will discuss this in Section \ref{ncsr}, which will contrast other scientific realist views to our own. (ii) Although a pragmatic view along the lines of Williams' might say that it is fruitless to try to get the kind of precision into the interpretative discussion that we aim for here: it seems that such a view is also committed to rejecting the relevance of formal criteria of equivalence of the type that Barrett advocates. Namely, on Dougherty's (forthcoming:~p.~2) analysis, this pragmatic view `drops the ... logicist policy on mathematics, replacing it with case-by-case analysis of the semantic content of mathematical representations'. Thus such a view must be {\it deflationary} about formal criteria of equivalence for scientific theories in general: there is no physically significant formal question of equivalence, and so no issue about the isomorphism criterion either. Here, we agree with the literature on theoretical equivalence that mathematical questions about the equivalence of theories in general {\it are} physically significant.}

Thus it is incorrect to say that a difference in the sign of the metric signature is a {\it mere difference of convention}, i.e.~a distinction without a difference, since this claim just depends on how the sign difference is {\it interpreted}, i.e.~according to which interpretation map. Namely, solutions of the Einstein field equations that differ in the sign of the metric signature can be interpreted as {\it either} using different conventions (and thereby describing the {\it same} spacetime) {\it or} as describing {\it different spacetimes}---and Barrett's enthymeme misses this crucial disjunction. 

This leads in to our next point, (B).\footnote{There is a Reichenbachian echo in the statement that decisions about conventions are choices between equivalent conceptions: see Section \ref{eronos}.}

(B)~~In {\it defence} of the isomorphism criterion, we will explain why statement (3), and therefore (6), are untrue. Namely, there is a very good reason why a spacetime, whose metric has eigenvalues with signs $(-1,1,1,1)$, is not isomorphic, and should not be equivalent, to a spacetime with signs $(1,-1,-1,-1)$: namely, these signs represent {\it very different spacetimes.} For, given an interpretation map for the metric tensor (along the lines of Section \ref{intext}, i.e.~the usual referential semantics of mathematical objects by a mapping into a suitable domain of application), the metric tensor field whose eigenvalues have signs $(1,-1,-1,-1)$ is interpreted as describing the metric of a spacetime with one timelike and three spacelike dimensions, while the metric tensor field with signs $(-1,1,1,1)$ is interpreted as a spacetime with one spacelike and three timelike dimensions; (or, if one adopts an interpretation map that swaps the coordinates it takes to represent space and those it takes to represent time, it is the other way around). These are indeed very different spacetimes, and a sensible criterion of equivalence should {\it not} say that they are the same.\footnote{An elementary argument shows that a change of the spacetime signature, $g_{\m\n}\mapsto-g_{\m\n}$, is {\it not} a symmetry of the Einstein field equations with a non-zero cosmological constant, or with a stress-energy tensor for a scalar field with a non-zero potential, or other types of matter fields (Gibbons, 2012:~pp.~120-121).\label{Mspin}} 
Thus (3), and therefore (6), is not true. (If you thought that spacetimes with one time and three space dimensions are the only relevant ones, see the next Section.)\footnote{See also Appendix 11.A.} 
The point is not about three time dimensions being physically sensible or not, but rather that, if at the outset the $(1,3)$ and $(3,1)$ metrics are said to differ by a `mere sign convention', i.e.~they describe one time dimension and three space dimensions, then we do not know how to even {\it begin} to discuss whether a different case, e.g.~three timelike and one spacelike dimensions, is physically meaningful.

(C)~~The inference from (4) to (5) relies on a (too) quick jump from the isomorphism criterion, applied to pairs of models of general relativity (which judges manifolds to be equivalent iff they are isomorphic, i.e.~isometric), to the {\it model isomorphism criterion} for two versions of general relativity, i.e.~for a whole set of models.

Barrett's idea is surely that the first theory, i.e.~$T_1$, is the set of solutions of general relativity with {\it fixed}  set of signs, i.e.~$(p,q)$ for fixed $p$ and $q$ (signs $(1,3)$, say). Then the other theory, i.e.~$T_2={\cal C}(T_1)$, is the set of solutions with the opposite  set of signs, i.e.~$(q,p)$. And thus the conclusion (5) relies on having fixed the relevant sign convention for the signature of each set of models, and having {\it already discarded} the solutions with a different set of signs, for being unphysical: and thus the argument begs the question. For if we fix the signature {\it convention} at the outset, by dropping the subset of models that we deem unphysical, we will {\it of course} get that the isomorphism criterion judges the complementary set of solutions, with the opposite signature, to be inequivalent to them: since their being `unphysical' was the motivation to drop them in the first place!

Note that (B) already taught us that fixing a convention is an interpretative move, and that the motivation to keep the class of solutions with the set of signs $(p,q)$, and discard all others, is a substantive interpretative move. 

If we, correctly, do {\it not} fix the signature before we go on to interpret the models, then we get the set of {\it all} models, i.e.~all smooth solutions of the Einstein field equations, with all possible signatures for a given dimension. And this set is invariant under the signature-changing map ${\cal C}$, which is an automorphism of this set! In particular, this set of models contains all {\it pairs} of solutions with  the sets of signs $(p,q)$ and $(q,p)$, and so ${\cal C}$ maps these solutions pairwise into each other.\footnote{For example, Blencowe and Duff (1988:~p.~396; our emphasis) consider supermembrane solutions in supergravity, and find that they come in {\it pairs of in principle distinct} solutions: `For every supermembrane with $(S,T)$ signature, there is {\it another} with $(T,S)$'. They also correctly note that physical boundary conditions that treat time differently from space might break this symmetry. Solution spaces that contain both cases will be excluded if one fixes the signature beforehand.} 
(As we mentioned before, we are are here considering general relativity in vacuum, as in Barrett's argument: see footnote \ref{Mspin}.) Since (4) and (5) (just like (6)) are about the {\it formal criterion} of equivalence rather than the interpretative criterion, the point here is, as in the contrast between (A) and (B) in Section \ref{lsps}, to distinguish clearly between the bare theory and the interpreted theory. Only in a next step (i.e.~in the interpretative step (B) from Section \ref{lsps}) do we drop the spacetimes with signatures that our interpretation deems unphysical---and then the model isomorphism criterion judges the two resulting versions of general relativity to be inequivalent, since the signature-changing map ${\cal C}$ is not an automorphism of the remaining subsets of models.\footnote{In practice, we of course do not always require to first write solutions with every signature, to then only keep the ones that our interpretation map deems physical (but see the previous footnote!). But the point is that, if we wish to avoid the muddle of mixing the formal and interpretative parts of our actual methods (as we have argued that we must, see e.g.~points (A) and (B) in Section \ref{lsps}), especially when it is the interpretation of the {\it signature} that is at stake, then we must reconstruct our methods thus. This is the `two-step procedure' that Section \ref{ThisB} argued is a widespread conceptualisation of the formulation of scientific theories.}

To sum up this rebuttal of Barrett's critique of the isomorphism criterion: Barrett's (2015:~p.~826) assumption leading into his incorrect verdict about the isomorphism criterion is that a formal criterion of equivalence should not count two formulations of spacetime theories as inequivalent if they only differ by what he calls a ``sign convention''. He appears to overlook the fact that fixing a ``sign convention'' for the signature of the metric tensor requires fixing an {\it interpretation}, and that once we have fixed such an interpretation map, our interpretation map will judge other signatures to be {\it inequivalent}---as the isomorphism criterion also correctly judges, and ${\cal C}$ incorrectly. Thus isomorphism, and not ${\cal C}$, meshes with the interpretation map and is the formal correlate of theoretical equivalence. 

\subsection{Consequences for proposed logically weaker criteria}\label{counterA}

The three problems expounded above are not just weaknesses of Barrett's specific argument against the isomorphism criterion. For as we will now argue, the exposed errors backfire, and erode some of the main {\it motivations} given in the literature, to find logically weaker formal criteria equivalence (weaker than the isomorphism criterion, that is), such as ${\cal C}$, for theories in the natural sciences. 

To spell this out, we will use the following notation as a shorthand for the two relevant interpretation maps:
\bea\label{pqcon}
i(p,q)&=&\mbox{`$p$ space and $q$ time dimensions'}\nn
i'(p,q)&=&\mbox{`$p$ time and $q$ space dimensions'}\,.
\eea
The isomorphism criterion meshes with these interpretations, as any good criterion must: namely, it distinguishes a spacetime with the set of signs $(p,q)$ from one with the set of signs $(q,p)$---as each interpretation of course also does. Requiring that a formal criterion of equivalence be weaker like ${\cal C}$, so that it maps $(p,q)$ to $(q,p)$, spells trouble, because this requirement is incoherent with {\it either} interpretation map. In short, the trouble is that $(p,q)\overset{{\cal C}}{\sim}(q,p)$, while $i(p,q)\not=i(q,p)$ and $i'(p,q)\not=i'(q,p)$.

Adopting the above notation, it should be clear how we can change our sign convention for the signature: namely, by changing our model from $(p,q)$ to $(q,p)$ {\it and}, crucially, also changing our interpretation map from $i$ to $i'$, or vice versa (see Figure \ref{barrett}).\footnote{The interpretation map $i'$ is what, in Section \ref{pud}, we will call a `peculiar', `Pickwickian' or `perverse' interpretation map, since it exchanges what we mean by the meanings of the words `space' and `time'. For a closely related discussion, see Butterfield (2021:~pp.~55-56).} In other words, the theory $T_1$ with interpretation map $i$, and the theory $T_2$ with interpretation map $i'$, have the same set of spacetimes, $W_1$, in their range. Thus to {\it change our convention}, we must {\it change both the signature and the interpretation map}. This is the correct way to change the signature convention.

We will now describe three steps in which the above argument erodes the motivation for the logically weaker criteria of equivalence.\\ 
\\
{\bf Step 1. The general moral from Section \ref{incorr}.} On further scrutiny, the (incorrect) critique of the isomorphism criterion of equivalence backfires on the logically weaker (more ``liberal'') criterion of equivalence, ${\cal C}$. For this criterion was constructed to make the spacetimes with  the two sets of metric signs $(p,q)$ and $(q,p)$ equivalent---which we just concluded are {\it not} equivalent unless we change our interpretation! Indeed, the criterion ${\cal C}$ is perverse because, on any given interpretation $i$ or $i'$, it mixes space and time: it {\it does not commute with either of the two interpretation maps (taken individually---as they should be), and so it is not a good criterion of theoretical equivalence.}

Note that the problem is not that the formal criterion ${\cal C}$ does not commute with some {\it specific} interpretation: none of the reasonable interpretations that we can construct mesh with the criterion of equivalence ${\cal C}$, because ${\cal C}$ does not commute with any of the interpretation maps: $T_1\overset{{\cal C}}{\sim}T_2$, while $i_k(T_1)\not=i_k({\cal C}(T_1))$ (where $k=1,2$ for the two interpretation maps), because $i(T_1)=W_1$ while $i({\cal C}(T_1))=i(T_2)=W_2$. This ``not meshing'' is illustrated in Figure \ref{noncomm}. Namely, {\it the formal criterion equivocates between solutions that all the relevant interpretation maps say correspond to inequivalent spacetimes}: and so, it is too weak, and not sufficiently fine-grained.

\begin{figure}
\begin{center}
\bea
\begin{array}{ccccc}
~~~~~~~~~~~T_1\!\!\!\!\!\!\!\!&\xrightarrow{\makebox[.6cm]{$\sm{${\cal C}$}$}}&\!\!\!\!\!\!\!T_2~~~~~~~~~~\\
~~~~~~~~~~~~~{\sm{$i$}}\big\downarrow\!\!&&\!\!\big\downarrow {\sm{$i$}}~~~~~~~~~~~~~\\
~~~~~~~~~~~~~~~~~W_1&\not=&W_2~~~~~~~~~~~~~~~\end{array}\nonumber
\eea
\caption{\small The equivalence map, ${\cal C}$, does not commute with the interpretation map, $i$.}
\label{noncomm}
\end{center}
\end{figure}

In other words, the isomorphism criterion of equivalence gets things {\it exactly} right both formally and physically (i.e.~it commutes with the interpretation maps, and a change of sign convention for the signature of the metric tensor is a change of interpretation map), while the criterion ${\cal C}$ is too weak, and incompatible with the physical interpretation (i.e.~it does not commute with the interpretation maps). As a result of its logical weakness, the criterion ${\cal C}$ does not make distinctions that it {\it should} be making. It is not a criterion of {\it equivalence}.\\
\\
{\bf Step 2. Considering a possible reply by Barrett.} What could Barrett reply to this? It might seem that to the above line of thought, he could reply that we should distinguish between a spacetime with `$p$ space and $q$ time dimensions' and a spacetime with `$p$ time and $q$ space dimensions' by {\it using the interpretation maps}, and this {\it in spite of the fact that the equivalence map ${\cal C}$ does not make this distinction}, since it tells us that the $(p,q)$ and $(q,p)$  metric signs are {\it formally} equivalent. (Indeed, Barrett has thus replied (personal communication, 2018).) 

But we will argue that this reply stumbles: in short, because the motivation for developing more liberal criteria was the wish to repair what is seen as a defect of the isomorphism criterion: but, as the reply admits, the criterion ${\cal C}$ does not give a criterion of theoretical equivalence either, and so it is {\it no better than the isomorphism criterion.} 
(Step 3 will also argue that ${\cal C}$ is not a desirable mathematical criterion of equivalence of spacetimes.) 

Thus Barrett explicitly says that ${\cal C}$ is a formal criterion, and then goes on to admit that it does not commute with the interpretation. In other words, Barrett admits that---unlike all the other criteria we have seen in previous Sections---the criterion ${\cal C}$ is not a criterion of {\it theoretical equivalence}, but only a criterion of {\it formal or syntactic equivalence} (and perhaps this is the way the weaker criteria are meant more generally).\footnote{In anticipation of this reply, we already made this distinction in statements (1) to (6), i.e.~so that ${\cal C}$ is a formal equivalence map, which is the correct reading of Barrett.}

This admission is puzzling, because it contradicts the {\it rhetoric} of this literature, mentioned at the start of Section \ref{bsa}, with its appeal to ``theoretical equivalence''. For despite Figure \ref{barrett}'s appearance, the situation is really a {\it non-commuting diagram}, as is clearly illustrated by Figure \ref{noncomm}!

Specifically, this admission runs against the critique of the isomorphism criterion, which was based on the alleged failure of the isomorphism criterion to judge theoretically equivalent situations to be theoretically equivalent (a critique which Section \ref{incorr} already established is incorrect). This alleged failure can hardly be taken to favour a logically weaker criterion, since the weaker criteria are neither criteria of theoretical equivalence {\it nor} their formal correlates, as Barrett's own reply and the above arguments establish: for ${\cal C}$ judges theoretically {\it in}equivalent spacetimes to be formally {\it equivalent}.

In fact, the isomorphism criterion follows closely (i.e.~is the formal correlate of) theoretical equivalence, while a formal criterion with the properties advocated by Barrett is not a formal correlate of theoretical equivalence, because it does not mesh with the interpretation maps (see step 1 above). In other words, weaker criteria of equivalence like ${\cal C}$ do not do what the rhetoric advertises them to do: while the isomorphism criterion does the correct jobs, both of straightforwardly securing theoretical equivalence, {\it and} of allowing a change of sign convention for the signature of the metric tensor in the way that one expects: namely, by a change of interpretation map.\footnote{Other oft-repeated examples against the isomorphism criterion can be dealt with similarly. See footnote \ref{isocritique}, and also De Haro (2020a:~pp.~5150-5152).}

This admission is puzzling also because the reasons why we wish to distinguish between the signatures $s=2$ and $s=-2$ are not only interpretative or empirical, but also {\it formal}: namely, the signature-changing transformation $g_{\m\n}\mapsto-g_{\m\n}$ is really {\it not} a symmetry of the general Einstein field equations (see footnote \ref{Mspin}). Furthermore,  on a spacetime that is space, time or spacetime non-orientable, the former signature allows the definition of Majorana spinors, while the latter does not.\footnote{On a spacetime that is space, time, or spacetime non-orientable, the signature-changing transformation leads to physical differences between fermions defined on two such spacetimes. For the Clifford algebra on a manifold of signature $s=2$ is isomorphic to the algebra of $4\times4$ real matrices, while the Clifford algebra on a manifold of signature $s=-2$ is isomorphic to the algebra of $2\times2$ quaternionic matrices. The former admits Majorana spinors, while the latter does not (Gibbons, 1994:~p.~65; 2012:~pp.~118-119).\label{Mspin2}} 
This kind of critique leads in to our next step:\\
\\
{\bf Step 3. Extending the discussion to substantive questions in physics.} An important problem in physics, both classical and quantum, is that of signature change. So, forget theoretical equivalence, and ask: how does the formal criterion ${\cal C}$ fare with physical situations of {\it signature change} in general relativity and related relativistic spacetime theories? We will argue that it does not fare well and that this reinforces the critique of it, established in steps 1 and 2.

For example, how does the equivalence criterion ${\cal C}$ fare with changes of signature between $(1,3)$ and $(3,1)$  metric signs in cosmological situations?\footnote{Gibbons (1994:~p.~61) considers topology change and the closely related notion of signature change, for various  sets of signs, including $(2,2)$. See also Alty (1994:~p.~2523). Topology change in the context of brane collisions is discussed in Gibbons (2012:~p.~122).} 
Can it distinguish between spacelike and timelike dimensions on manifolds with Kleinian $(2,2)$ signature, which, for a variety of reasons, are routinely considered in several areas of relativistic physics?\footnote{This signature is routinely considered, by an analytic continuation of one of the spatial coordinates, in twistor theory (Penrose, 1967:~p.~354; 1968:~p.~64). For gauge theories with this signature, see Witten (2004:~p.~192). For string theories with this signature, see Ooguri and Vafa (1991:~p.~469). While in these papers the $s=0$ signature is considered for reasons of mathematical convenience, Heckman et al.~(2022:~p.~2), while of course avoiding getting closed timelike curves, also emphasize other physical motivations for considering this signature.} 
How does it fare with physical situations where the lapse function has a zero?\footnote{See Hayward (1992:~p.~1853). For signature change on a cosmological scale, see Gibbons and Hartle (1990:~p.~2459). Sakharov (1991) [1984] is an early version of the idea of signature change in cosmological transitions.} 
Or with black hole solutions of $(1+1)$-dimensional stringy-inspired versions of general relativity coupled to matter, where the space and time coordinates are basically swapped at the horizon?\footnote{There is a large literature on such two-dimensional black holes. The first solution discovered is in Witten (1991:~p.~317). The point about these examples is, of course, that they do not require exotic signatures, and are straightforward relativistic solutions.}

The criterion ${\cal C}$ does not appear to deal with these situations well. For, if ${\cal C}$ were a good formal or mathematical criterion of equivalence of solutions, i.e.~$T_1\overset{{\cal C}}{\sim}T_2$, in an appropriately substantive sense, its adoption would allow us to remove the redundancy and, in effect, reduce the relevant space of solutions, by dropping half of the solutions out of pairs of ${\cal C}$-related solutions. But in the presence of signature changes in the metric, like the examples just mentioned, we must {\it keep} different ${\cal C}$-related solutions (as already advocated, for formal reasons, in (C) of Section \ref{incorr}).\footnote{We get no guidance from the formal criterion ${\cal C}$ as to whether we should keep the ${\cal C}$-related solutions, or throw them away: taking the criterion seriously says that we {\it may} thrown them away to reduce the redundancy. Other relevant examples involve spacetimes of the form $(p,q)\otimes(m,n)$, i.e.~tensor products of spacetimes with sets of signs $(p,q)$ and $(m,n)$. It is clear that one cannot apply the formal equivalence to only one of the factors (even though the corresponding map, ${\cal C}\,\otimes\,\mathbb{1}$, {\it is} a diffeomorphism), because the resulting spacetime would be physically distinct. But the formal criterion does not seem to be able to answer the question (other than by appealing to physical equivalence) as to why this is not an equivalence map.} For reducing the space of solutions through the formal equivalence criterion ${\cal C}$, as we would naturally be allowed to do if the solutions we drop were mathematically equivalent to their ${\cal C}$-related counterparts, means throwing away solutions that are in principle physically relevant, and distinct from the ones we keep (note that the issue at stake here is not moving to a reduced formalism, but rather dropping the solutions that are equivalent to other solutions). Thus taking a weaker criterion of equivalence like ${\cal C}$ {\it mathematically seriously} as a criterion of equivalence, in a more general setting for general relativity of this type, seems to lead to disaster.\footnote{The reason to not drop these solutions is of course that they have different interpretations. But the point is that, if an equivalence map like ${\cal C}$ really gave us a criterion for when two solutions have the {\it same structure}, as Barrett (2020:~p.~1188) claims, then we could use it to reduce the excess structure by dropping half of the solutions, in pairs of ${\cal C}$-related solutions. But we see that this leads to incorrect results.} In the words of Gibbons (1994:~pp.~61-62):
\begin{quote}\small
there is now an important distinction between signature $(+++-)$ and $(---+)$. In some cases one signature may be excluded and the other allowed ... As our understanding of physics progresses one may expect to find increasingly that ... even in some cases what seem to be merely arbitrary conventions, have arisen, perhaps even accidentally, as a consequence of present location in spacetime.
\end{quote}

\subsection{A `No deus ex machina' argument}\label{lessons}

We will now use the above discussion to formulate two general principles, one about good physical theories, and the other about good criteria of equivalence of such theories, where the latter follows from the former. The first criterion is as follows:\\
\\
{\bf Univocality of the interpretation of physical theories:} {\it the semantics of physical theories should not cut any finer than their bare theories (or models)}, i.e.~we should {\it aim} to formulate theories (or models) where any distinction made by a physical interpretation has a counterpart in a distinction in the bare theory (or model).\footnote{Like in the rest of this Chapter, our use of `theory' and `model' here is the standard one, i.e.~not the Schema's `theory' and `model', to which we will return in the next Chapter.} In other words, an interpretation of a bare theory should be {\it univocal}.

This is of course a very general principle about how to construct physical theories: we wish the formalism to be as detailed as possible,  so that we can adapt it to the features of our physical system, and do not need to make any interpretative distinctions that do not have a counterpart in the formalism. In other words, in the interpretation of physical theories we do not make distinctions in the world where there is none in the bare theory or model. (This is not to deny that nature is of course richer than our theories. Further interpretative distinctions could be made, perhaps in another theory.)

Notice that this principle is about the {\it semantics} of a scientific theory, and not about its metaphysics: namely, it is concerned with what Section \ref{srpost} called a `literal construal' of a scientific theory, in the sense of giving a scientific theory a referential semantics (more specifically, in Section \ref{ints}: giving it an intensional semantics). For, while there might be metaphysical reasons to make a distinction that does not have a counterpart in a distinction in the bare theory (and we agree that, in a process of theory construction, metaphysical analysis may suggest modifications of the bare theory), this requires separate metaphysical argumentation, beyond the engagement with the logic and semantics of a theory that we have discussed in this Chapter. Thus a realist and an anti-realist like van Fraassen, while disagreeing about metaphysics, can agree about the empirical evidence and about a theory's literal construal. (We will discuss scientific realism, literal construals, and questions about metaphysics, in Chapter \ref{Realism}.)

This principle is of course not just interpretative: it is as much a {\it formal requirement} to allow an interpretation map that is {\it single-valued}, and does not assign more than one referent to the same term, or to terms that are in the same equivalence class (where the equivalence class is here defined by the formalism). If the interpretation map is not single-valued, then the theory is {\it ambiguous}. And, in general, this is an undesirable property for a theory in the physical sciences. 

But this is precisely what the logically weaker equivalence criteria, such as ${\cal C}$, do. For two metric tensors, related by a ``formal equivalence'' map ${\cal C}$, get assigned {\it different} referents by the {\it same} interpretation map (see Figure \ref{noncomm}, where $T_1$ and $T_2\overset{{\cal C}}{\sim}T_1$ have {\it different} domains of application). Furthermore, none of the relevant interpretation maps (in Eq.~\eq{pqcon}) map ${\cal C}$-related solutions to the {\it same} domain of application. (This is why, in Barrett's `possible reply' in step 2, {\it two} interpretation maps, rather than one, were required to distinguish theories with  metric signs $(p,q)$ from theories with metric signs $(q,p)$.)

Clearly, the problem here is not with the interpretations or with the isomorphism criterion, which does this correctly, but with the logically weaker criteria of equivalence: {\it logically weaker criteria of equivalence such as ${\cal C}$, if implemented as equivalence relations within a single theory, do not allow for single-valued intepretation maps}, and so they are not good formal criteria of equivalence in the natural sciences.

The above principle has the following important corollary for formal criteria of equivalence, which holds {\it all other things being equal}:\\
\\
{\bf A stronger criterion of formal equivalence is to be preferred to a weaker one:} this is because an interpretation map need of course not be injective (see the discussion in Section \ref{srpost}), and so different objects of the bare theory can be mapped to the same object in the world. In other words, distinctions that are made by a bare theory can be ``erased'', or ``forgotten'', by the interpretation (recall, from Section \ref{intext}, that in referential semantics an interpretation is standardly defined as a {\it partial} function, i.e.~a function that, for some element in its domain, can fail to provide a value). In particular, terms that are inequivalent (e.g.~not isomorphic) might get the same interpretation, because the interpretation map makes fewer distinctions than the formal criterion of equivalence. 

What the principle says is that, other things being equal, it is better for an interpretation to {\it forget} a formal distinction, than for it to {\it introduce} a distinction where the bare theory (with its equivalence relations) says there is none. We should introduce new elements, or new distinctions, into our theory by enriching the bare theory, rather than by enriching the interpretation. For a theory thus constructed gives a {\it more detailed description of its domain of application}.

Thus once one gives up the strong requirement that ``interpretation is an isomorphism between a model and the world'', and accepts that interpretations are more general (partial) structure-preserving maps,\footnote{Although the notion that models are {\it isomorphic} to the world is sometimes cited in the context of the semantic conception of theories (see e.g.~Halvorson, 2012:~p.~185; van Fraassen, 1989:~pp.~219, 226), most recent advocates favour weaker theory-world relations: for example, {\it partial} isomorphism, or isomorphism of substructures: see e.g.~van Fraassen (1980:~pp.~43, 64-68; 1989:~p.~227), Da Costa and French (1990:~pp.~254-256) and Ladyman (1998:~p.~416). Indeed, the idea that a theory is given a {\it partial} interpretation was common in the syntactic conception of theories: see Suppe (1974:~pp.~3, 86-94). Other authors favour accounting for the theory-world relation as a homomorphism: Frigg (2023:~p.~200) dubs this the `morphism account', and discusses five challenges that any account of scientific representation must be able to answer, and five conditions of adequacy that the answers must satisfy (pp.~185-186). As we discussed in Section \ref{srpost}, we do not need to provide an answer to these challenges, or even take a stand about the question of representation, since our account of duality does not depend on it. That is, we can give an account of dualities and of how they contribute to questions of theoretical equivalence, without having to answer (perennial!) questions about the objectivity of representation.} 
in general not injective but (where possible!) {\it well-defined}, i.e.~single-valued, a formalism that makes more distinctions is better, because scientifically more detailed or accurate, than one that does not make them, if those distinctions are physically significant (as the ones we have discussed indeed are). Such a formalism is better suited to avoid ambiguity, while the partiality of the interpretation makes up for differences in the formalism that do not have a correlate in the world.

The principle of univocality of physical interpretations and the single-valuedness of interpretation maps thus warn us against making distinctions that are not there in the bare theory, but that only appear in the interpretation like a {\it deus ex machina}. 

\section{Conclusion}

Important developments in logic since the 1960s and 70s have focussed on finding {\it definability conditions}, which identify fragments of one logic that can be defined in terms of another. Van Benthem's zigzag theorem, from Section \ref{eil}, identifies a fragment of first-order logic that is logically equivalent to a translation of modal logic: namely, the fragment that is invariant under bisimulations (and this result generalises in multiple ways, with numerous applications in e.g.~computer science). The fact that the definability conditions involve logical equivalence means that besides having an appropriate formal translation, they also must have the same models.

The discussion in philosophy of science builds on this understanding: namely, the inter-trans\-la\-tability criteria of Glymour (i.e.~definitional equivalence) and Quine (what he dubs a `reconstrual of the predicates') require logical equivalence, and so sameness of models. Thus, also in philosophy of science, there are both (A) formal i.e.~non-interpretative, and (B) interpretative criteria of theoretical equivalence, which must be of the ``right'' logical strength, and must mesh with one another. By `meshing' we here mean, along the lines of van Benthem's theorem, that the strength of the formal criterion (definitional equivalence, bisimilarity, etc.) is compatible with the sameness of the models. 

In philosophy of science, this `meshing' is complicated by the semantics' not just being a formal semantics, but what in Section \ref{lsps} we dubbed a `physical semantics'. In the two Chapters that follow, this physical semantics will take centre stage.

In the semantic conception of scientific theories, the natural criterion of equivalence is the isomorphism criterion. As such, this criterion does not suffice, because---like any other formal criterion---it needs to be complemented with a criterion of equivalence for the physical semantics. And, while the model isomorphism criterion has been criticized for being too strict, we have shown that some of these criticisms are incorrect. (In fact, we never saw a convincing case against the isomorphism criterion in an example of a scientific theory formulated in set theory.) Indeed, as we argued in Section \ref{defencei}, the logical strength of the isomorphism criterion, as a formal criterion, is just right. And there are good physical reasons why spacetimes with $(1,3)$ and $(3,1)$ metric signatures are judged to be neither isomorphic nor equivalent.

Thus we proposed a minimal condition for the desired logical strength, and meshing, of the criteria (A) and (B). We dubbed it the `principle of univocality': namely, a formal criterion should not be so weak as to require the interpretation map to make distinctions that are not there in the formalism. In other words, one's formalism must be adequately strong: a different way of proceeding, by involving an unwarranted {\it deus ex machina}, makes a scientific theory less descriptive. Namely, if a criterion is too weak, the relevant interpretation map cannot be single-valued, and so not well-defined. Hence the requirement that formal criteria, if they are to be useful in natural science, ought to mesh with the relevant interpretation. This will be the starting point of our discussion, in the next Chapter, of the duality-based criterion of theoretical equivalence.

\section*{Appendix 11.A. On the physical relevance of non-standard signatures}
\addcontentsline{toc}{section}{Appendix 11.A. On the physical relevance of non-standard signatures}

In Section \ref{defencei}, we discussed spacetimes with an arbitrary set of signs $(p,q)$, other than the familiar $(1,3)$ and $(3,1)$ ones. From now on, we fix the interpretation map $i$ from Eq.~\eq{pqcon}, so that the first factor of the pair $(p,q)$ is space and the second factor is time. Thus to remind ourselves of this, we use the suggestive notation $(S,T)$, where $S$ is the number of space dimensions and $T$ the number of time dimensions.

Although the point of the argument was not about the spacetimes with $T>1$ having {\it actual} physical significance, but rather about {\it the possibility of discussing} their physical significance, such an argument is of course much more vivid if we do find physical systems where these other signatures are of interest.

The answer is straightforward: yes, these signatures are of physical interest. We will now give some examples from the physics literature, in addition to the examples already given in Section \ref{defencei}. 

First of all, let us note that there is {\it no} formal result in general relativity that requires that $T\leq1$. While the existence of only one time dimension is a given of experience, and there are physical arguments that limit $T\leq1$ (like the avoidance of closed timelike curves),\footnote{For a related argument, about the loss of the classical stability of particles if time were multi-dimensional, see Dorling (1970:~p.~540). But note that this argument does not exclude the possibility of a spacetime with $T>1$ in the models discussed below, either because the additional time dimensions are `gauge', or because these models do not consider pointlike particle processes of the type that Dorling discusses.} 
the theory can be consistently formulated in {\it any} signature, and the problems of closed timelike curves are bypassed in the relevant examples.

For example, one of the standard ways to define a $D$-dimensional anti-de Sitter spacetime is by treating it as a hyperboloid inside a $(D+1)$-dimensional flat spacetime with metric signs $(D-1,2)$.\footnote{See Hawking and Ellis (1973:~p.~131) for a treatment of the geometry of the spacetime, and Avis et al.~(1978:~p.~3566) for a treatment of the quantum field theory on this spacetime, both of them using this representation.}

Also, there is a substantive line of research in theoretical physics into two-time physics (i.e.~$T=2$; cf.~Bars, 2001). This kind of two-time theories are reformulations of one-time theories (i.e.~$T=1$) that make hidden symmetries explicit. The simplest idea is that a particle's position and momentum variables, $(x,p)$, can be treated as a doublet in a phase space with two times and a $\mbox{Sp}(2,\mathbb{R})$ symmetry (Bars et al., 1998:~p.~3114). The spacetime symmetry of such a particle is $\mbox{SO}(S,2)$. This work also appears to give insight into gauge-gravity duality: because $\mbox{SO}(d,2)$ is the conformal group in $d$ spacetime dimensions, the conformal group of a theory in $d$ dimensions (in particular, of a conformal field theory) can be realized as the Lorentz group of a gravity theory with metric signs $(d,2)$. (And, as already discussed in Section \ref{defencei}, there is a vast literature, with various motivations, on relativistic theories with metric signs $(2,2)$.)

Note that, in these theories, it is still true that only one of the time coordinates is a time coordinate in the ordinary way---the other one can be gauged away. But which coordinate is to be interpreted as ordinary time depends on a choice of gauge. 

Also, Blencowe and Duff (1988) and Duff (1996) give systematic treatments of the set of signs $(S,T)$ that are allowed by the consistency of supergravity theories, including the case $T>1$. In the context of string theory, Blencowe and Duff (1988:~p.~388) write: 
\begin{quote}\small
one might hope that a Theory of Everything should predict not only the {\it dimensionality} of spacetime, but also its {\it signature}. For example, quantum consistency of the superstring requires 10 spacetime dimensions, but not necessarily the usual $(9,1)$ signature. The signature is not completely arbitrary, however, since spacetime supersymmetry allows only $(9,1)$, $(5,5)$ or $(1,9)$.
\end{quote}
In other words, string theories are rather liberal in the kinds of signatures that they allow as being in principle physically significant.

\chapter{Duality and Theoretical Equivalence}\label{physeq}
\markboth{\small{\textup{Duality and Theoretical Equivalence}}}{\textup{\small{Duality and Theoretical Equivalence}}}

This Chapter discusses the conditions for two dual theory formulations, which we call models, to make the very same claims about the world. (One might instead say, with equal naturalness: `interpretative equivalence'.)

The discussion builds on our account of duality and interpretation in Section \ref{dualint} in the light of the examples in Chapters \ref{Simple} to \ref{HABHM}. In particular, the notion of internal and external interpretations of a theory, and of the common core of two duals, are further developed: which sets the stage for the discussion of scientific realism in the next Chapter.\\
\\
In the previous Chapter, we discussed logico-semantic criteria of theoretical equivalence. Our approach was model-theoretic, i.e.~the semantics was given by a set of models, construed as mathematical structures. This model-theoretic semantics was our `first step' in constructing the semantics of scientific theories (Section \ref{lsps}). The `second step' we dubbed the `physical semantics': which incorporates, and builds on, the model-theoretic semantics, but is not reduced to it, because it refers to worldly items, which are not always best understood solely structurally. Thus we need to discuss this realist physical semantics that both realists and anti-realists use. 

This Chapter takes on this second step. Thus specifically, our main question, in Section \ref{semanticeq}, is: given duality as a semantic, model-theoretic criterion of equivalence between theories, and an internal interpretation of duals, {\it what are necessary and sufficient conditions for theoretical equivalence?} We will dub the duality-based criterion of theoretical equivalence `physical equivalence', since it is motivated by physical theories, and their semantics and epistemology. 

Although, in answering this question, our focus will be on constructing a duality-based criterion of theoretical equivalence, we will discover {\it general interpretative constraints} on theoretical equivalence that are sensible regardless of dualities, i.e.~which other formal accounts of equivalence also ought to satisfy.\footnote{Our `No deus ex machina' argument in Section \ref{lessons} already constrained the desirable relative logical strength between a formalism and its interpretation: namely, through the criterion of univocality of the interpretation of physical theories, which in particular requires that interpretation maps are single-valued i.e.~well-defined.}
(These constraints are not only about the semantics of theoretical equivalence, but also about its epistemology. This Chapter sets aside the epistemological aspects: the next Chapter will discuss them.)

Section \ref{abstraction} goes on to develop a complementary aspect of this discussion of dualities: namely, the common core theory behind two duals, and how it can be constructed and interpreted, by abstraction, such that it is compatible with the two duals. 

Section \ref{lsr} then points out that, in a syntactic formulation of theories, the relation between a theory and its models is best interpreted as an entailment relation that goes from the (logically stronger) set of models to the (logically weaker) common core theory. The next Chapter will then argue that these logico-semantic relations bear on the issue of scientific realism.

Section \ref{lsrr} illustrates the logico-semantic relations between a theory and its models, and the ideas of internal and external interpretations, in an example familiar from Chapter \ref{Advan}: namely, bosonization.

\section{Physical equivalence}\label{semanticeq}

`Equivalence' is of course a term of art: that is, a vague term that is made precise by different authors in different ways, according to their various different purposes. So also, therefore, is `theoretical equivalence'; and we shall need to regiment our usage, along the lines of our conditions (A) and (B) from Section \ref{lsps}. One may indeed use `theoretical equivalence' in such a way that dual models (in the sense of our Schema) are theoretically equivalent. Section \ref{nscte} gives our regimentation of theoretical equivalence, regardless of the detail of our Schema for dualities. Section \ref{meshdi} then fills in the conditions for theoretical equivalence for theories and models that are formulated according to our Schema: we call our dualities-based proposal for theoretical equivalence `physical equivalence', since it is based on the examples of theories in physics discussed in Part II.

\subsection{Theoretical equivalence: necessary and sufficient conditions}\label{nscte}

For two scientific theories to be theoretically equivalent is for them to ``say the same thing''. And, taking up again our two criteria (A) and (B) for theoretical equivalence from Section \ref{lsps} (see also the `consensus' at the end of Section \ref{contcons}), we will in this Chapter argue that they are both necessary and jointly sufficient for the theoretical equivalence of physical theories. Recall that these two criteria are:\\
\\
{\bf (A)~~A formal (syntactic and semantic) criterion of equivalence:} we already discussed various formal criteria in the previous Chapter. And, in Section \ref{defencei}, we argued that duality is the correct {\it model-theoretic criterion of equivalence} in physics, with the right logical strength. (A syntactic version of it was discussed in Section \ref{mtce}.)\\
\\
{\bf (B)~~An interpretative criterion of equivalence:} this will be the focus of the present Chapter. The leading idea will be that---extending the notion that equivalent theory formulations must have the same model-theoretic semantics---they have the same physical semantics (cf.~Section \ref{lsps}): namely, they have {\it the same domain of application}, as defined in Sections \ref{thsq} (where we also called it `regime') and \ref{intext}. We will also say that they have the `same interpretation', i.e.~that their interpretation maps have the same range, or image.\\
\\
Criterion (A) formalizes the intuition that, if two theories are theoretically equivalent, they should not just have the same physical semantics, i.e.~say the same things about the physical world, as in (B), but that they should do so using mathematical structures that {\it match}, i.e.~that are appropriately isomorphic. This intution was e.g.~expressed in the quote by Gr\"adel and Otto, in Section \ref{synsemeq}, who notice that an appropriate notion of bisimulation allows us to study the expressive power of a theory.\footnote{For more on the varying degrees of complexity of a logic, see van Benthem (2010a:~pp.~122-123).}
This was also the gist of our critique, in Section \ref{defencei}, of more liberal criteria of theoretical equivalence.

But one can still ask: why is formal equivalence a requirement for theoretical equivalence? That is, if theoretical equivalence means `saying the same things {\it about the world}', then why is it not sufficient to require that two models have the {\it same domain of application}? (This question is pertinent even recalling, from Section \ref{refsr}, that the interpretation map usually {\it requires a language} to describe, not just isolated properties or objects, but also e.g.~physical events. Thus an interpretation is not always a map from numbers to properties or objects, but admits descriptions like `an electron and a positron collide and produce a photon' or `the probability for phenomenon $P$ to occur is $1/2$'. In other words, mapping two models into the same codomain of events is not enough for theoretical equivalence.)

The reason why we require formal equivalence (and we will also require `meshing of equivalence and interpretation': see below) is that the condition of `same domain of application' is merely a condition for `saying things about the {\it same world}': it is not a condition for `saying the {\it same things} about the world'. As an example, imagine a model with three linear algebraic equations determining three real numbers, and a model with only two. And imagine mapping both models into a world with three properties, of which two are directly measurable and the third is not (this third quantity might be an unobservable quantity, or a theoretical quantity, in the sense of `physically significant but not directly measurable'---e.g.~entropy). In such a simple world, both models are empirically adequate in the official sense. They also both map to the same world, and yet they are not equivalent. For example, the model that determines three numbers, if true at this world, has a richer semantics and may give deeper explanations: e.g.~it may give an entropic explanation that the model with two numbers lacks. Clearly, mapping to the same domain, even taking into account the uncontroversial requirements of empirical adequacy and predictive power, is too weak a requirement. For the two models do not describe the given domain of application at the same level of detail.

What is required is a formal relation between the models that forms a commuting diagram with the interpretation maps, as in Figure \ref{interp1}: the models must be suitably compatible, in the sense that their structures match. Only then can we say, by requiring commutation, that the interpretations are the same. For the representing side of the interpretation relation (i.e.~of the interpretation map), namely the model (and in particular the model root), determines the level of detail at which a model can describe the domain of application. 

The above example, of a world with two measurable quantities and one additional quantity that is not directly measurable but provides an explanation for the other two, motivates the formal requirement of duality from broader requirements in philosophy of science: namely, for theories formulated as triples of states, quantities, and dynamics, this mathematical structure is used in e.g.~giving explanations: and for the two models to give the same explanations, this mathematical (one could say: logico-semantic) match is required.\footnote{Thus Coffey (2014:~p.~823) is incorrect to say that `claims of theoretical equivalence are normative claims about how theoretical formalisms ought to be interpreted', because the question about theoretical equivalence cannot be reduced or limited to just the problem of interpretation. Although we agree that there are cases where formally inequivalent theories have the same domain of application, these are not cases of {\it theoretical equivalence}, because the {\it theories} are not appropriately related or systematically inter-translatable: there is no proper inter-theoretic relation. They are simply cases of `having the same domain of application', as in our example of a world with three properties, of which one model only describes two.}

As Sklar (1982:~p.~90) aptly puts it, an account of theoretical equivalence should `bring ... into a harmonious whole a theory of meaning, of evidence, of ontology, of truth, of explanation and of equivalence itself'.\footnote{For example, Putnam (1996:~pp.~39-40) requires equivalent descriptions, under translation of one theory into the other, to preserve the relation of explanation, so that the same phenomena are explained by both theories. Furthermore, he argues that `to say that they are equivalent {\it because} they are notational variants is to put the cart before the horse'.} 
We agree, while we of course admit that this is a tall order (and we do not claim to have finished all these tasks in this book, although we do hope to give what may be useful sketches of each of them).\footnote{For more discussion of why criterion (B) is required, see the end of Section \ref{lsps}. This criterion has often been neglected, or discussed superficially, in some of the recent literature on theoretical equivalence. But, as we already saw in Section \ref{lessons}, and will discuss further in this Chapter, there are substantive general things to say about it.} 

Criteria (A) and (B) are jointly sufficient for theoretical equivalence, but as we just argued they are also necessary.\footnote{That criterion (A) gets filled in as `duality', i.e.~as the model isomorphism criterion of equivalence, anyway depends on having a semantic formulation of a theory (however, Section \ref{mtce} of course proposed a syntactic version of duality), or in any case on having a specific presentation of it that is apt for applying the isomorphism criterion. Thus what we argue is that, {\it given} a theory thus presented and interpreted, duality is {\it the} correct criterion, i.e.~has the right logical strength (see the discussion in Section \ref{defencei}).} (A) and (B) are what our Schema means by the phrase `theoretical equivalence': namely, that the two theories have the same logico-semantic structure, and the same interpretation.

(Note that, assuming that two models give sufficiently good descriptions of a given domain of application, \`a la Sklar, i.e.~satisfying whatever might turn out to be the agreed requirements in philosophy of science, of prediction, empirical verification and explanation, the question of theoretical equivalence only regards whether the models do this in {\it equivalent} ways. That is, the present question is {\it not} whether there is or can be a pair of interpretations of dual models that are also equivalent and give {\it better} (more detailed or more accurate by some additional standard) explanations about a given domain than some other pair. The question about theoretical equivalence is concerned solely with the equivalence of models whose interpretations satisfy whatever turn out to be the agreed requirements in philosophy of science. That additional question, of `level of detail or depth of the description', will be addressed in Section \ref{abstraction}, when we discuss abstraction.)\\

Having argued that criteria (A) and (B) are both necessary, and jointly sufficient, conditions for theoretical equivalence, an important epistemic question is whether one is {\it warranted in making a verdict} of theoretical equivalence.\footnote{Recall, from Sections \ref{srpost} and \ref{substd}, that we had postponed this epistemic question until Part III. Notice that there is an analogy here with our discussion of scientific realism in the preamble of Chapter \ref{Realism}, and with the analysis of knowledge as true, justified belief. A correct verdict of equivalence is analogous to a true belief, but this belief may require justification, as we discuss below. However, there do not seem to be any Gettier cases, because of the different kind of justification here required, compared to the Gettier cases.} 
This requires what we shall call {\it specifying the ontology} and making the ontological commitments of theories explicit.\footnote{Read and M\o ller-Nielsen (2020:~p.~282) talk about `explicating the ontology'. Their preferred view, which we endorse, is called `motivationalism'. For a discussion, see Section \ref{comparison}, and De Haro (2021:~pp.~5163-5164).} 
Furthermore, besides specifying, it requires {\it justifying} the ontology, i.e.~giving appropriate epistemic and metaphysical reasons for the choice of a common ontology between the two models.\footnote{Since this last requirement is not specific to cases of theoretical equivalence, but to any scientific theory, we will not discuss it separately, although our comments on scientific realism will bear on it. For example, the hole argument is a well-known debate between substantivalists and anti-substantivalists over the appropriate ontology of general relativity (see Chapter \ref{HABHM}).}
These questions will be discussed in Chapter \ref{Realism}.

In the next Section, we begin by specifying our two criteria of equivalence, (A) and (B), for theories that are formulated according to our Schema from Section \ref{isomdef}.

\subsection{Physical equivalence as meshing of duality and interpretation}\label{meshdi}

The question in this Section builds our Schema from Section \ref{wde}, and is analogous to the question in Section \ref{mtce}, i.e.~{\it what is an appropriate semantic criterion of equivalence for dual theories?} But now we ask this question in the context, not of a strictly model-theoretic semantics (since duality is itself the preferred model-theoretic standard of equivalence!), but in the context of our {\it physical semantics}, or of the physical interpretation of the theory. Thus we have two dual models, $M_1$ and $M_2$, of a bare theory, $T$, and their physical semantics is provided by an appropriate mapping of the models to a domain of the world, which is not a piece of language, but a hunk of reality---what we call the {\it domain of application} (see Section \ref{itm}).

The required criterion of theoretical equivalence, i.e.~our regimentation of this phrase, is what we shall call `physical equivalence': and it is a kind of ``meshing condition'' between duality and interpretation, as follows:\\
\\
{\bf Physical equivalence:} two models, $M_1$ and $M_2$, of a bare theory $T$, with interpretations $i_1$ and $i_2$, are physically equivalent iff: \\
\\
(A)~~they are duals; and \\
(B)~~their domains of application are the same: and, more strongly, the interpretation maps form a commuting diagram with the duality map, i.e.~$i_1=i_2\,\circ\,d$ (see Figure \ref{interp1}).\\
We shall comment on these two conditions in order:\\
\\
{\it About (A)}: we have already extensively commented, in Chapter \ref{Theor} and earlier Chapters, on this condition, so that a brief discussion suffices. First, our condition (A) of having duals assumes bare theories and models that are as discussed in Section \ref{ThisB}: i.e.~the models are representations of a bare theory, and within each model we distinguish the model root from the specific structure (see Section \ref{modelrootss}), where the model root is the structure that realizes the common core. (Model roots and bare theories are usually formulated as triples, as we discussed in item (3) of `Theory' in Section \ref{Ourthm}: at the end of Section \ref{isomdef}, we discussed alternative formulations.) 

Condition (A) says that, {\it for theories and models thus formulated}, duality is a privileged formal criterion of equivalence: namely, it is an isomorphism criterion, which we argue is a privileged criterion for {\it all} physical theories thus formulated. Our evidence for this comes, first, from the examples: especially the ones that we discussed in Chapters \ref{Simple} to \ref{HABHM} (and there are many others), which indeed cover various areas of physics. It is also confirmed by the fact that no better alternative is known that applies fruitfully to physical theories.\footnote{There are two reasons why we do not currently count category theory as an alternative to duality. First, since physical theories are not usually formulated in category theory, applying category theory to physical theories may require a certain amount of reformulation of physical theories (what we call `revising' physical theories below). Second, we agree with other authors that categorical equivalence is too weak, and thus not (yet) a good criterion of equivalence in physics: see Barrett and Halvorson (2016b:~p.~566), Hudetz (2019:~p.~51), and De Haro (2020a:~p.~5166).  Of course, we welcome further research in this direction.}
And, as we discussed in the previous Chapter, it is the preferred formal criterion of equivalence for theories formulated in set theory: and two of the major critiques against the isomorphism criterion in natural science stumble. 

This agrees with what what we said at the end of Section \ref{Intro}, that we propose our Schema for duality in an undogmatic spirit: we accept that our Schema maybe could be improved, and that there may be more mathematically precise formal criteria of equivalence that could replace (A). But such improvements will also involve a certain amount of {\it revision} in the way we formulate our theories, and in our usual assumptions about what bare theories and models are (i.e.~structured triples, of which the models are representations, etc.): thus such revisions do not satisfy our clause `thus formulated', and they do not contradict our conception of physical equivalence. In other words, we argue that, if we take scientific theories as they are {\it given} to us by mathematical physicists within set theory, and as we can ourselves also formulate them using our Schema, then the isomorphism criterion is {\it the} correct formal condition (A).\footnote{Recall that Section \ref{mtce} discussed a syntactic version of a duality: so that there is no need for us to work ``free of language''.}\\
\\
{\it About (B)}: to further clarify this interpretative condition, and its meshing with (A) through an appropriate commuting diagram, we will now make the following four comments: (1) about `sameness', (2) about the undesirability of weakening the interpretative criterion (B), (3) about internal interpretations and explanation, and (4) about the domain of application and possible worlds.\\
\\
{\it (1)~~Comparison with the logico-semantic criteria, and `sameness'.} It is worth comparing the interpretative criterion of physical equivalence (i.e.~part (B) of the criterion) with the model-theoretic criteria of equivalence from Section \ref{mtce}: logical equivalence (which requires that the models of the two theories are the same) and isomorphism (which requires only isomorphism of the models). Clearly, physical equivalence is analogous to the stronger criterion of logical equivalence, since it requires that dual models have {\it the same domain of application} (as against the domains being isomorphic). This is a natural criterion: it is simply what we mean by `the same': namely, {\it numerical identity}, or, in a weaker sense, the {\it lack of a difference} between the domains of application (and from now on, we will use `the same domain of application' without distinguishing between these two senses).

But physical equivalence is free of the problems of logical equivalence, which requires that the sentences of two theory formulations that share their models are formulated in the {\it same language}, and therefore is too strong a criterion of equivalence. For the notion of interpretation from Section \ref{intext} as a partial, structure-preserving, map to a domain of application is largely independent of language (even if language is almost always used), and so there is no question about the satisfaction of the theories' formal sentences in the world, and no requirement that they should be formulated using the same signature. Language plays a quite different role here: rather than defining the theory, it aids interpretation in the world.\footnote{See the discussion in Section \ref{refsr}, especially points (i) and (ii).}\\
\\
{\it (2)~~Weakening the equivalence criterion?} It is of course tempting, by analogy with the model isomorphism criterion of equivalence mentioned in (1), to weaken the criterion (B) of physical equivalence, and require only that the domains of application are {\it isomorphic}. This would amount to, in the example from Section \ref{defencei}, saying that the theories that we get by interpreting general relativity according to the interpretation map $i_1=i$, or according to the interpretation map $i_2=i'$, where the two maps differ in their swapping the number of space and time dimensions that they assign to the manifold, are the same theory (alternatively, this would be like saying that the non-commuting diagram in Figure \ref{noncomm} is a case of physical equivalence). Or to saying, in Kramers-Wannier duality (see Section \ref{dualpf0}), that a lattice at high temperature is physically equivalent to a lattice at low temperature. 

But such weakenings, even though useful for purposes other than discussing physical equivalence, should be resisted. For dual models are {\it not} physically equivalent if they describe isomorphic but distinct domains of application,\footnote{This would also require a standard of `isomorphism of domains of application', while physical equivalence simply requires that the domains are the same: in any case, that they lack any difference.} but only if they describe the same domain of application. In other words, we do not want to say that a lattice at high temperature is physically equivalent to a lattice at low temperature, since they are not---as one quickly finds out when burning one's fingers. (We will return to this in our discussion, in Section \ref{srr}, of under-determination and scientific realism.)\\
\\
{\it (3)~~Internal interpretations of the models and explanation.} The condition that the two interpretations and the duality map commute, i.e.~$i_1=i_2\,\circ\,d$, implies that the two interpretations are {\it internal}, because they do not map the specific structure of the models into the common domain $D$, i.e.~they only map the common structure.\footnote{To show this, note that (using the notation from Section \ref{modelrootss}) if $x\in\bar M_1$, then by definition the duality map $d$ does not map $x$ to anything (i.e.~it is a partial map that does not map $\bar M$). Thus by the commutativity condition, $i_1$ also does not map $x$ to anything (and likewise if we start from the second model).} Or, in other words, the domain of application $D$ is ``independent'' of which interpretation we use: it can be reached from either model, by using the duality and the commutativity of the diagram. 

This is also the reason that, relative to internal interpretations, we can argue that both duals are {\it equally explanatory}.\footnote{We here anticipate Hempel's deductive-nomological model of explanation, to be discussed in more detail in Chapter \ref{Understand}. Hempel defines an explanation as an objective relation between a theory or model and a phenomenon: namely, as a deductive argument in which the occurrence of the phenomenon to be explained (the `explanandum') is deduced from the theory or model, for particular circumstances (i.e.~for an appropriate state and-or value of a quantity).}
For explanations that use the model's internal interpretation are preserved by the duality map, since they commute with it. In the Schema's formulation of theories and models, an explanation (in Hempel's sense, which will be explained in Chapter \ref{Understand}, of `deducing a phenomenon from a theory or model') is a sentence in natural language including one or more values of the interpretation map: such a sentence typically relates the internal interpretation's output (i.e.~the event in the world that is the explanans, e.g.~the occurrence of phenomenon $P$) to either the input (i.e.~the conditions under which the occurrence of $P$ can be predicted) or to other outputs of the interpretation map (e.g.~an earlier state of affairs in the world prior to $P$'s occurrence), which is the explanandum. But these values are independent of the specific structure, and the same sentence in natural language is valid for either model, by relating the inputs that are involved using the duality map: namely, the commutation condition secures that the corresponding explanans is mapped to the same explanandum.\footnote{This will not be the case for understanding, which, as we will discuss in Chapter \ref{Understand}, is relative to an epistemic community, which may use {\it tools} that are not part of the model's ``official interpretation'', i.e.~not part of the internal interpretation picked: thus they may of course use the model's specific structure, and a given epistemic community may have better epistemic access to one specific structure than to the other. But this is irrelevant for theoretical equivalence. Explanations, in so far as they claim to be objective, may not make essential use of tools that are not in the internal interpretation. This is because the interpretation chosen sets the standard of objectivity to be used. Thus once an internal interpretation is chosen, the specific structure should not do any essential explanatory work---on pain of not being objective by the interpretation's own standards.}\\
\\
{\it (4)~~The domain of application and possible worlds.} One important difference, between the model-theoretic semantics that we discussed in the previous Chapter and the physical semantics that we are now discussing, is that the model-theoretic semantics in Eq.~\eq{Mmodels} considers {\it all structures} where all of the theory's sentences are satisfied, i.e.~all possible sets with relations on them satisfying all of the theory's sentences. By contrast, the domain of application $D$ is an appropriate subset of the set of physically possible worlds, and this requires a brief discussion.

In philosophy of physics, a contrast is often made between the set of {\it kinematically possible worlds} and the set of {\it dynamically possible worlds}, where the latter is a subset of the former.\footnote{For an elementary discussion, see Section \ref{thsq}.} 
Since the state spaces of the Schema's model triples (see Section \ref{ThisB}) are ``off-shell'', i.e.~regardless of dynamics, the set of `physically possible worlds', $W$, is here to be understood as the set of kinematically possible worlds (the dynamics in the triple then further specifying the subset of dynamically possible worlds, which are usually distinguished by their initial and-or boundary conditions). And this played a role in Section \ref{MEMD}, where the dynamically possible states of the common core were isormorphic to those of the duals, but the kinematically possible states were not isomorphic.

But this identification requires the following remark: namely, the kinematics of $M_1$ and $M_2$ (where `kinematics' here means the variables of the models that describe states in the state space, and the quantities on it), even if isomorphic, may be very different, because it can advert to the specific structure. Thus $W$ is the set of kinematically possible worlds ``up to duality'', i.e.~it is the set of possible worlds that is compatible with the kinematics as formulated by both $M_1$ and $M_2$. In fact, (3) already showed that the interpretation maps are internal, and so that the domain of application $D$, and a fortiori the set of kinematically possible worlds $W$, is independent of the models. 

This leads in to an analysis, in the next Section, about how to construct and interpret the bare theory, $T$.

\section{Getting the common core by abstraction}\label{abstraction}

In this Section, we consider the endeavour of defining a theory by abstraction from its models. For a duality, this means defining the bare theory, $T$, from its set of models: some of which are isomorphic with respect to the bare-theoretic structure we intend to define---and which are therefore duals, according to our Schema. In Section \ref{dpp}, we called this the {\it common core} theory. But of course, the general idea of defining a theory by abstraction from its models does not depend on there being a duality. 

\subsection{Two ways to get a common core theory}\label{aa}

Suppose that we are given two dual models, $M_1$ and $M_2$. Then one way to obtain a (bare) common core theory $T$, that the two models represent or instantiate, is to ``cross out'' the structure of the models that is not common under the isomorphism. As a result, $T$ is a theory with less structure: it is a structured set of states (including stipulated symmetries, cf.~Section \ref{dualsym}), a structured set of quantities, and a dynamics. In other words, the structure of the common core theory is precisely the structure that we discussed in Section \ref{ThisB}, and that we required was preserved by the duality.

\begin{figure}
\begin{center}
\includegraphics[height=4cm]{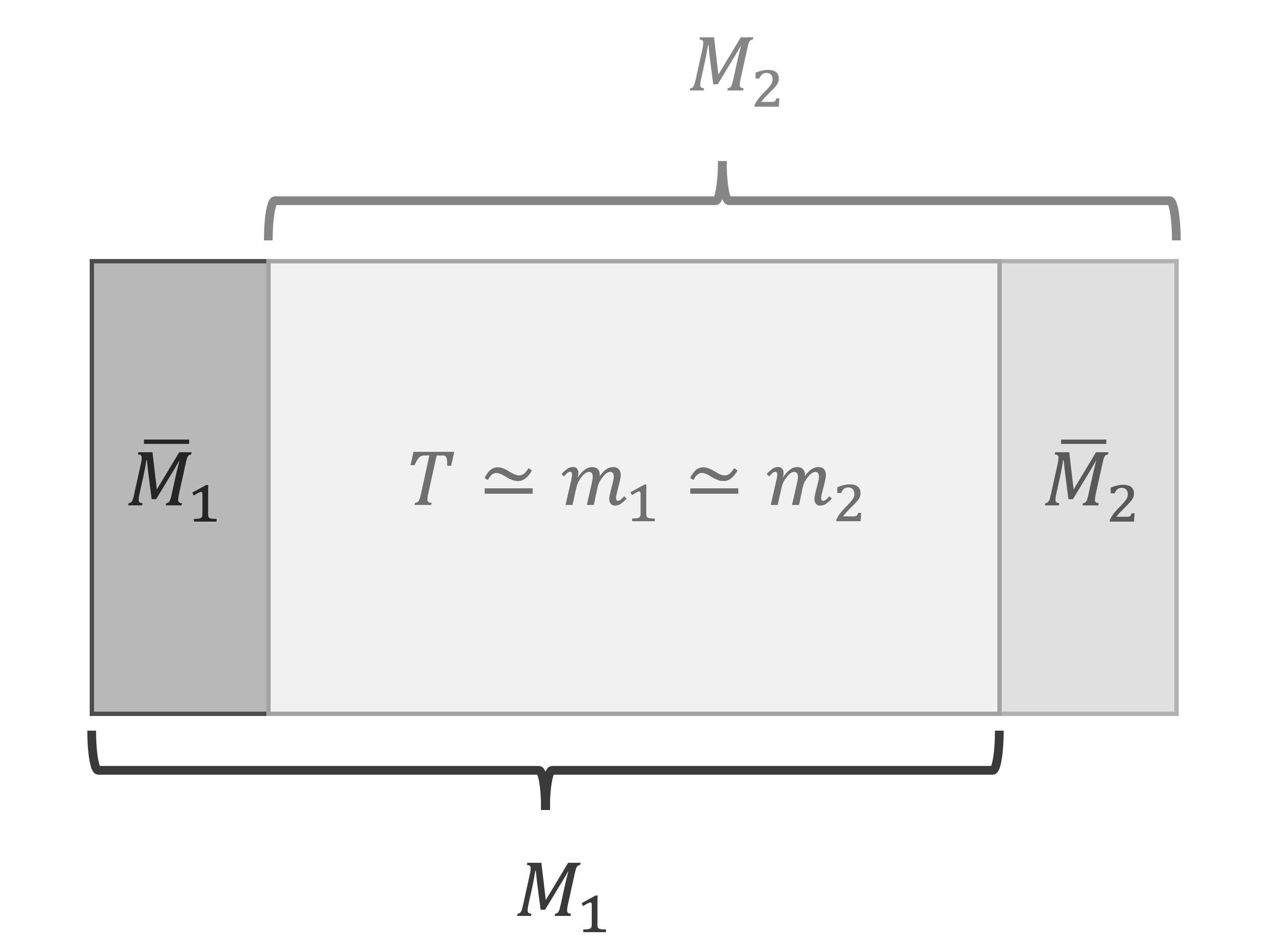}~~~
\caption{\small A common core theory, $T$, of ``desirable strength'', i.e.~isomorphic to its dual models, $M_1$ and $M_2$. The overlapping regions are isomorphic to each other. $\bar M_1$ and $\bar M_2$ are specific structure, and they lie outside the overlap.}
\label{commonc}
\end{center}
\end{figure}

Because the common core theory has less structure, it ``says fewer things'', and so it is {\it compatible} with either of the two duals, since it only contains structure that is common to $M_1$ and $M_2$---as per the isomorphism of these models. This is a well-known way to realize {\bf abstraction}: by being less detailed and more general.\footnote{Lewis (1986:~p.~84) sums up the idea of abstraction as `subtracting specificity, so that an incomplete description of the original entity would be a complete description of the abstraction'. Frege's (1956:~pp.~74-75) classic example is the concept of `direction', which is abstracted from the relation between parallel lines. Given two lines related by the (reflexive, symmetric, and transitive) relation of `being parallel', this relation is replaced by an {\it identity}, thereby formulating the new concept of `direction'. See also Falguera et al.~(2021:~Section 3.4).\label{Fregeex}} 
See Figure \ref{commonc} (for the case to be discussed in Section \ref{ica}, where the common core theory is itself {\it isomorphic} to its dual models).

Before proceeding, let us announce two topics that will crop up in the sequel, and then give an example of abstraction:

(1): It is often very non-trivial how to ``cross-out'' the non-common structure, i.e.~structure that is not common to the models: this usually requires a certain amount of theoretical reformulation. The obvious illustration is gauge variables: it is hard to do physics in a wholly gauge-invariant formalism! This is clear from the simple examples in Chapter \ref{Simple}, especially duality as a Fourier transformation, and electric-magnetic duality.
We also saw, in the examples of Part II, that in practice it can be difficult to extract the common core theory behind duals in quantum field theory and string theory: however, this is also often due to our lack of exact formulations of such theories. In fact, precision about the states, quantities, and dynamics helps theory construction: we will discuss examples of this in Chapter \ref{Heuri}. (Section \ref{lsrb} will illustrate abstraction in a non-trivial example: we will see that an axiomatic presentation of a theory can be useful for this aim.)

(2): Even apart from (1), there is a question of `how far to go' in the process of abstraction: this will be discussed in Section \ref{ica}.\footnote{As always, interpretation constrains mathematical formulation: thus, here, one wishes the common core theory to be referentially successful and capture what the models have in common. We will discuss these interpretative issues in the next Chapter, where we will endorse motivationalism (see especially Section \ref{cco}). As we already noted under (1), and as the examples in Chapter \ref{Simple} illustrated, abstraction is usually not as simple as literally ``crossing out'' structure from the given models, but often requires a more unified and sophisticated reformulation of the models that allows us to excise adequate structure.}

To illustrate the notion of abstraction, we will use the obvious example, of the representations of a group, $G$, that is defined by a set of (abstract) axioms, compatible with the definition of a group. Since the group representations in a structure $M$ are homomorphisms from the group to the structure, i.e.~$h:G\rightarrow M$, the range of the representation map $h$ satisfies the axioms of the group, as represented in the structure. Thus the structure defines a model where the group axioms are satisfied, i.e.~made true (see Section \ref{eil}). 

An elementary example is the cyclic group, $C_n=\{c^r\,|\,c^n=e\,\wedge\,r=1,\ldots,n\}$ (where the powers of $c$ are defined by the group binary or multiplication operation in the obvious way), whose only real one-dimensional representation is the trivial one, with a two-dimensional representation given by the complex numbers $e^{2\pi r/n}$ i.e.~$2\times2$ real matrices, and with general $d$-dimensional representations given by the $d\times d$ matrices $(h(c))^r$, where $(h(c))^n=\mathbb{1}_{d\times d}$, and the relevant multiplication is of course matrix multiplication. The group $C_n$ has the same (i.e.~homomorphic) basic structure as the representations on $M$, but it is more abstract, because it lacks structure that the representations have. For example, $C_n$ has both representations where $h(c)$ is a $d\times d$ real matrix, and representations where it is a complex matrix: but this additional specificity (being defined over a real or over a complex field) is lacking in $C_n$.\footnote{For a more detailed illustration of such a homomorphism, see the example of a representation map in Section \ref{lsr}.} 

The general idea of abstraction is, of course, that we go from a logically stronger system (the set of dual models) to a weaker one (the common core theory). In general, if presenting the theory syntactically, the logically weaker system contains fewer (atomic) sentences, and therefore applies to more cases, while the logically stronger system contains more (atomic) sentences, and thereby allows us to prove more.\\

A second way to construct the bare theory is by {\bf augmentation}. The assumption here is that we begin with two models, $M_1$ and $M_2$, that are {\it not duals} (or not evidently so), but that we can construct from them augmented models, $M_1'$ and $M_2'$, that are duals in the usual sense. 

\begin{figure}
\begin{center}
\includegraphics[height=3.5cm]{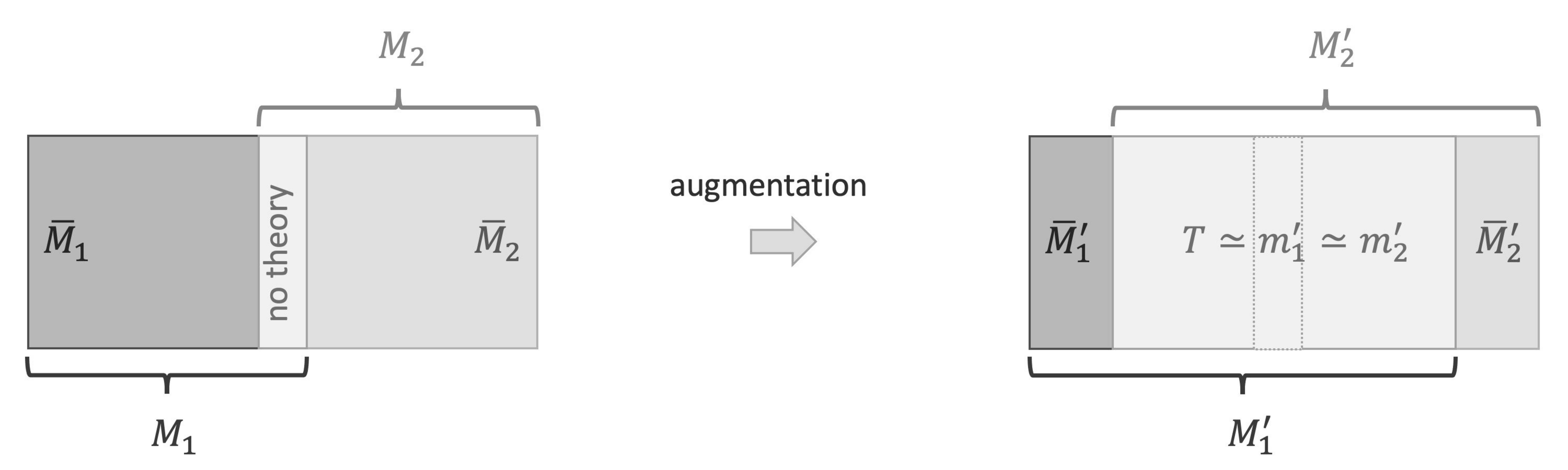}~~~
\caption{\small Left: the region where the two initial models overlap, i.e.~where $M_1$ and $M_2$ are isomorphic, is not a theory (i.e.~not an appropriately structured set of states, quantities, and dynamics). Right: the initial models, $M_1$ and $M_2$, are augmented to models $M_1'$ and $M_2'$ that correctly overlap on a theory, $T$.}
\label{augmentation}
\end{center}
\end{figure}

We have discussed two types of examples of this: (i) we {\it changed the common core} by adding quantities to one of the models that match quantities over in the other model, and (ii) we merely added {\it specific structure}, through Lagrange multipliers.\footnote{There is here an echo of Lewis' (1986:~p.~85) second way of making `good sense' of abstractions, i.e.~his reinterpretation of Frege's example of the concept of direction in footnote \ref{Fregeex}, `by the stratagem of taking equivalence classes ... There is no genuine subtraction of specific detail, rather there is multiplication of it; but by swamping if not by removal, the specifics of the original line get lost.'} 

(i)~~In Section \ref{bosoniz}, we saw that the fermionic and bosonic models, as sometimes defined, are not dual, because, although there is a map between the single fermion $\psi$ and the bosonic operator $:e^{i\f}:\,$, this bosonic operator is usually not in the set of quantities, since it is not self-adjoint, and it has fermionic rather than bosonic statistics. The two models can of course be made dual, by {\it adding quantities} (i.e.~new operators, defined in terms of the other operators), thus obtaining models with more quantities and states. In other words, although the initial (historically given) models, $M_1$ and $M_2$, are not duals, there is a systematic inter-translation that allows us to augment each of the models with definitions from the other model (such as $\psi= :e^{i\f}:$), to get new models, $M_1'$ and $M_2'$, that {\it are} dual to each other. 

(ii)~~In the example of electric-magnetic duality in Section \ref{MEMD}, to get the common core theory we introduced Lagrange multipliers (in the model with gauge field $A$, we added $B$ as a Lagrange multiplier; in the model with gauge field $B$, we introduced ${\cal F}$ as a Lagrange multiplier).\footnote{We here describe how to go from the models to the theory, which is ``inverse'' to how the calculation was presented in Section \ref{MEMD}. This is because this calculation can be run in either direction.}
This amounts to augmenting the models with {\it specific structure}, such that the augmented models, $M_1'$ and $M_2'$, are obviously dual, i.e.~they are the same common core theory $T$.

Both cases are of course analogous to the definitional equivalence discussed in Section \ref{sse} (i.e.~roughly: the logical equivalence of two syntactically formulated theories, obtained by adding appropriate definitions).

An augmented pair of models thus obtained, $M_1'$ and $M_2'$, are duals, and the bare theory is obtained in the above way by abstraction, and so deserves the name {\it common core theory}. This case differs from other cases only in that the bare theory is obtained from the {\it augmented models}, and not from the original ones. The idea of augmentation is illustrated in Figure \ref{augmentation} (for the case discussed in Section \ref{ica}, i.e.~of a common core theory itself isomorphic to its dual models).

\subsection{Internal interpretation of the common core}\label{iicc}

So far, the interpretation of the bare theory, $T$, has not played any role: in particular, its interpretation does not enter into the condition (B) of physical equivalence. So far, $T$'s role has been formal: namely, $M_1$ and $M_2$ are representations of $T$, i.e.~$T$ expresses the common structure of the dual models. (In Section \ref{abstraction}, we will discuss how to get such a bare theory from two given models.)

\begin{figure}
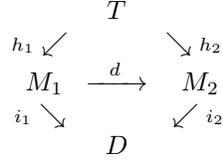

\begin{center}
\bea
\begin{array}{ccccc}&T&\\
~~~~~~~~~~~~~{\sm{$h_1$}}\swarrow\!\! &&\!\!\searrow{\sm{$h_2$}}~~~~~~~~~~~~~~
\\
~~~~~~~~~~~M_1\!\!\!\!\!\!\!\!&\xrightarrow{\makebox[.6cm]{$\sm{$d$}$}}&\!\!\!\!\!\!\!M_2~~~~~~~~~~\\
~~~~~~~~~~~~~{\sm{$i_1$}}\searrow\!\!&&\!\!\swarrow {\sm{$i_2$}}~~~~~~~~~~~~~\\
&D&\end{array}\nonumber
\eea
\caption{\small Interpretative commuting diagram. The bare theory, $T$, has dual models, $M_1$ and $M_2$, that are representations of it, given by maps $h_1$ and $h_2$. $d$ is the duality map between the models. The models are mapped by their respective interpretation maps, $i_1$ and $i_2$, to the same domain of application. The commutation conditions are on the (sub) diagrams that contain interpretation maps, i.e.~the diamond ($i_1\,\circ\,h_1=i_2\,\circ\,h_2$) and the bottom triangle ($i_1=i_2\,\circ\,d$).}
\label{Physeq}
\end{center}
\end{figure}

In this Section, we will require a natural condition for the bare theory $T$ to have an internal interpretation:\\
\\
{\bf Internal interpretation of the bare theory:} if the representation and interpretation maps commute (i.e.~the maps $h_1, h_2$ and $i_1, i_2$ form the commuting {\it diamond diagram} in Figure \ref{Physeq}), then the bare theory, $T$, has a {\it unique internal interpretation map}, $i:T\rightarrow D$, defined in terms of the interpretation maps of the duals. This internal interpretation map is given as follows:\footnote{In Section \ref{giantS}, we said that dual models are `formulations, or versions, or realizations, of the common core theory ... they are almost always representations of it'. Thus to define our notion of physical equivalence, which requires a full match between two models, both structurally and in terms of the interpretation, we now require that the duals $M_1$ and $M_2$ are {\it representations} of the bare theory, $T$, in the sense of representation theory, i.e.~that they are homomorphic copies of it (and in Figure \ref{Physeq} we denote the corresponding homomorphisms by $h_1$ and $h_2$).}
\bea\label{ii1i2}
i=i_1\circ h_1=i_2\circ h_2\,.
\eea
This map is appropriately structure-preserving (as any interpretation should be), because both the partial interpretation maps $i_1$ and $i_2$, and the two representation maps, $h_1$ and $h_2$ (which are homomorphisms between the bare theory and the models) are structure-preserving. In other words, $i$ is indeed an interpretation map, i.e.~an appropriately {\it structure-preserving partial map}.\footnote{In general, despite the commutativity of the diagram, the range of the representation map is a (proper) subset of the domain of the interpretation map of the corresponding model, i.e.~$\mbox{ran}(h_1)\subset\mbox{dom}(i_1)$ (and likewise for the second model). This is because, while the representation map $h_1$ maps the elements of $T$ into the model $M_1$, not every element of $M_1$ is mapped to by a corresponding element of $T$: in other words, the representation maps are not required to be surjective. This means that, in general, the domain of application of the bare theory, $T$, is a {\it subset} of the domain of application, $D$, that is common to the two models. In Section \ref{ica} we will discuss a natural duality-based criterion to define a (common core) bare theory $T$, which will be precisely that the representation maps $h_1$ and $h_2$ are surjective. In that case, the domains of application of the bare theory and of the dual models are the same.} 

All in all, the three internal interpretation maps, i.e.~$i$, $i_1$, and $i_2$, for the theory and its dual models, form the interpretative commuting diagram in Figure \ref{Physeq}.\footnote{By an `interpretative commuting diagram', we mean that the two (sub)diagrams for the interpretation maps commute. Thus we require that the {\it diamond} and the {\it bottom triangle} are commuting (sub)diagrams, i.e.~$i_1\,\circ\,h_1=i_2\,\circ\,h_2$ (diamond) and $i_1=i_2\,\circ\,d$ (bottom triangle). And we do not need to require the stronger condition that the {\it top triangle} commutes, i.e.~$h_2=d\,\circ\,h_1$, because it does not contain any interpretation maps---duality is the only formal requirement, and this additional requirement on the bare theory and the models is not needed. Note that the commutativity of the top and bottom triangles implies the commutativity of the diamond (and the bottom triangle), while the converse is not true. To see that the converse is not true, note that we can derive the commutativity of the diamond diagram from the commutativity of the bottom triangle together with the following weakening of the commutativity of the top triangle diagram: take $h_2$ and $d\,\circ\,h_1$ to differ by a non-zero map, $\epsilon:T\rightarrow M_2$, that maps the elements of $T$ into the kernel of $i_2$ (and we can take it that this `taking the difference' here is allowed, because we can consider quantum theories where the sets of states and quantities are both linear spaces subject to superselection rules). Then, using the commutativity of the bottom triangle diagram, we still derive the commutativity of the diamond (and the argument is symmetric with respect to $M_1$, i.e.~$h_1$ and $d^{-1}\,\circ\,h_2$ differ by a map $d^{-1}\,\circ\,\epsilon$ that maps $T$ into the kernel of $i_1$).}

The commutativity of the diamond diagram in Figure \ref{Physeq} secures that the above expression for the interpretation map, $i$, is independent of the model that we use---hence the second equality sign---i.e.~that the interpretation of the bare theory, that is induced by the interpretation of the dual models, is unique, or ``duality-invariant'', i.e.~independent of the model that we use to define it.\footnote{While $h_1$ and $h_2$ of course map to the specific structure of the models, the interpretation maps $i_1$ and $i_2$ ``wash out'' this model-dependence, because it is in their kernel.} In other words, the bare theory's interpretation map, $i$, is independent of the specfic structure of the models: it is what we dubbed an {\it internal interpretation}. Thus for example, the state space, ${\cal S}$, as formulated in the bare theory $T=\bra{\cal S},{\cal Q},{\cal D}\ket$, is independent of the models' specific structure, and gives a correct (i.e.~suitably duality-invariant) characterization of the states, and thus of the relevant set of worlds, $W$ (see also our comment (4) at the end of Section \ref{meshdi}). This underpins the conceptual and physical significance of the bare theory, $T$. 

Thus such an interpretation map, $i$, gives us an internally {\it interpreted} common core theory, $T$. The next two Sections first discuss the desired logical strength of this interpreted theory, and then its logico-semantic relations with its dual models: which the next Chapter shows have a bearing on scientific realism.

\subsection{Desired logical strength of the common core}\label{ica}

We have seen that the common core theory, $T$, is always related to its set of models by abstraction (which, as we will discuss in the next Section, is the logico-semantic ``opposite'' of representation). In a process of abstraction, the structure defined at each next step is more general, whilst less detailed, than the previously defined structure. But Section \ref{aa} did not address the question: Is there a stopping-point for abstraction? In other words, what is the ``right'' balance between generality and detail?

Although {\it in general}, there is of course no right answer to this question, we will argue that there is a natural standard for the logical strength of the common core theory $T$ between duals---what we shall dub the `desirable logical strength'.

As we will now argue, the standard, stated in general form, is that {\it the common core theory, $T$, should ``say no more, but also no less'', than its dual models}. This standard is natural because it has a balance between strength and generality that is desirable for scientific theories: it is as logically strong as the duals (and so, intuitively, it is as detailed as it should be), but not stronger than them, so that ``it does not miss any important facts''. If the dual models are empirically well-tested and semantically suited, then it is natural for the bare theory to be isomorphic to them, so that the structure of the duality carries over to the common core at the same level of detail.\footnote{Another reason why this standard is natural is because, when we focus on dualities and on finding the theory behind two duals, our main interest is in reproducing the duals, and not in finding a theory that represents other (non-dual) models. In other words, abstraction from duals aims to get a common core theory, and not a theory that duals and non-duals are both representations of. (Even if our interest were finding a theory that has as its representations both dual and non-dual models, finding a logically strong common core theory between the dual models can be a preliminary step for finding a theory whose representations include both dual and non-dual models. Thus, even in that case, the natural standard of duality is useful.)} 
Thus we require that $T$ is logically as strong as the models themselves, ``up to duality'': see Figure \ref{commonc}.

As we already signalled in Section \ref{meshdi}, it can be non-obvious how to cross out all the specific structure, and so it can be non-obvious how to formulate $T$. Also, in the sequel the discussion specializes to a pair of dual models that are isomorphic to $T$.

This natural standard immediately gives us two desirability criteria: a formal (i.e.~non-interpretative) criterion and an interpretative criterion. Both desirability criteria turn out to agree: they say that abstraction should stop at {\it isomorphism}.

(1)~~The formal i.e.~non-interpretative criterion is as follows: given two dual models, $M_1$ and $M_2$, a preferred way to obtain a (bare) common core theory $T$ is by constructing a theory that is isomorphic to its dual models (model roots), i.e.~$T\cong m_1\cong m_2$. For this theory has the same amount of structure as the model roots do, and it does not depend on the models' specific structure. In other words, $T$ is the structure that is by definition preserved by the duality, and which one gets by keeping the common structure of the model roots, $m_1$ and $m_2$, while dropping the specific structure, $\bar M_1$ and $\bar M_2$ (however difficult that may be to do in practice: see Section \ref{lsrr}). This is illustrated in Figure \ref{augmentation}. Other things being equal, this is the natural {\it structural criterion}, because it recommends that we drop the specific structure, but not the structure that is left invariant by the duality map. In this case, the common core theory then simply {\it is} the common structure that the models share.

(2)~~A natural {\it interpretative} criterion also arises for internal interpretations (which do not map the specific structure, and give cases of physical equivalence: see Figure \ref{Physeq}). As we saw in Eq.~\eq{ii1i2}, such an internal interpretation of the common core theory, $T$, is induced by the internal interpretations of its dual models, and it is unique (i.e.~independent of the models), so that we get the internally interpreted common core theory, $T$ (see the end of Section \ref{iicc}). The desirable criterion is that, if the common core theory is to ``say the same thing'' as the dual models, then it should be possible to {\it reconstruct} the interpretations of the dual models, $i_1$ and $i_2$, from the internally interpreted common core theory, i.e.~from $i$, together with the representation maps. And this can be done just in case the common core theory is isomorphic to its dual models, i.e.~$T\cong M_1\cong M_2$ (thus also $T\cong m_1\cong m_2$). Namely, it follows from the diamond commutativity condition, Eq.~\eq{ii1i2}, that the interpretation maps of the models, $i_1$ and $i_2$, can be recovered iff the representation maps $h_1$ and $h_2$ are invertible, i.e.~if $i_1=i\,\circ\, h_1^{-1}$ and $i_2=i\,\circ\, h_2^{-1}$. Since the representation maps are homomorphisms, this is an {\it isomorphism} condition: and so, we find that the common core theory is isomorphic to its dual models.\footnote{This argument is not restricted to {\it two} dual models, but holds for any number of dual models.} 

Thus, although the common core theory, $T$, can in principle have any logical strength, natural structural and interpretative conditions agree on the criterion that they give for the ``desirable strength'' of the common core theory, $T$: namely, that it should be isomorphic to its dual models.

\section{Logico-semantic relations between a theory and its models}\label{lsr}

The aim of this Section is to argue for an important Remark that will help clarify the under-determination argument in the next Chapter. The Remark concerns the logico-semantic relations between the common core bare theory, $T$, and its models, $M_1$ and $M_2$, {\it for the general case of both internal and external interpretations}: and thus throughout this Section, the bottom triangle diagram in Figure \ref{Physeq} need not commute. 

The Remark uses the notion of an {\it induced interpretation} $i$, namely a unique internal interpretation that is obtained from the external interpretations of dual models (we will define the induced interpretation below).\footnote{The Remark is a mandatory consequence of the supposition, stated in {\it About (i)} below, that there are proper subsets of the dual models, $M_1$ and $M_2$, on which the restrictions of the interpretation maps, $i_1$ and $i_2$, agree.} The Remark comes in two different versions: one semantic and one syntactic.\\
\\
{\bf Remark:}\\
\\
(i)~~{\bf Semantic version:} Under the induced interpretation $i$, the common core theory, $T$, is {\it compatible} with its dual models, $M_1$ and $M_2$. The common core theory {\it agrees} with the duals, and says less than they do. (This is regardless of whether $M_1$ and $M_2$ are mutually compatible or incompatible.)\\
\\
(ii)~~{\bf Syntactic version:} Under the induced interpretation $i$, the common core theory $T$ is {\it entailed} by its models.\footnote{The question of entailment of a theory by its models, in the sense of the homomorphism $h$ respecting the truth values (see the discussion below), so that the theory is logically compatible with its models, should not be confused with the question of the ability to mathematically {\it reconstruct} a theory from a given model, which in general is impossible unless one is given the homomorphism.}
The syntactic entailment relations are as follows:\footnote{We use the double turnstile $\vDash$ to indicate semantic entailment between sets of sentences, not syntactic derivability. We of course construe such a set of sentences  conjunctively: the conjunction of elements of $M_2$ semantically entails the conjunction of elements of $T$, and so on.}
\bea\label{matim}
M_1\vDash\,T~&\mbox{and}&\!M_2\,\vDash\,T\,,
\eea
and the models can be either theoretically equivalent or {\it inequivalent} (the next Chapter will apply this Remark to theoretically inequivalent models).\footnote{Agreed, the two dual models could be about wholly different subject-matters or topics, and in that sense not contradict each other, and still be theoretically inequivalent (Butterfield 2021:~p.~42). But this is not the relevant case for our discussion of under-determination in the next Chapter, because we will discuss empirically equivalent models (which agree about all the observables---are {\it a fortiori} about the same topic---and disagree about the unobservables).} 
In the latter case, where the models say different things about the world, we have: $M_1\vDash\neg\,M_2$, where $\neg\,M_2$ denotes the disjunction of all the negations of the elements (construed conjunctively) of $M_2$.

We now explain and prove these two versions of the Remark in order.\\
\\
{\it About (i)}: while the external interpretation maps of two duals (in general) do not commute with the duality map, i.e.~$i_1\not=i_2\,\circ\,d$, it may happen that the two interpretations still substantially agree, i.e.~there are proper subsets of the dual models, $M_1$ and $M_2$, on which the restrictions of the interpretation maps, $i_1$ and $i_2$, agree. 

To implement this `agreement', we will consider the case that the common core theory, and the representation maps, $h_1$ and $h_2$, are such that the ranges of the representation maps are in these proper subsets, i.e.~we take $T$ such that $h_1$ and $h_2$ map elements of $T$ into subsets of the models on which the interpretation maps agree (i.e.~they have the same range). Namely, we require that the restrictions satisfy: $i_1|_{\sm{Im}(h_1)}=i_2|_{\sm{Im}(h_2)}\,\circ\,d|_{\sm{Im}(h_1)}$. (These representation maps are of course not guaranteed to exist; but if they do, then this procedure delivers a unique interpreted theory $T$.) 

This condition secures the diamond diagram commutativity condition, i.e.~$i_1\,\circ\,h_1=i_2\,\circ\,h_2$ (even though the bottom triangle in Figure \ref{Physeq} does not commute, i.e.~on their entire domains, we have $i_1\not=i_2\,\circ\,d$). And since the interpretation maps agree, this gives a unique interpreted theory. For this gives an internal interpretation of the common core theory, as in Eq.~\eq{ii1i2}, but here constructed out of two {\it external} interpretations. We will dub the internal interpretation thus constructed $i$ (in Eq.~\eq{ii1i2}) of the common core theory an {\bf induced interpretation}, i.e.~induced by the requirement $i_1|_{\sm{Im}(h_1)}=i_2|_{\sm{Im}(h_2)}\,\circ\,d|_{\sm{Im}(h_1)}$. Since the induced interpretation has as its image a subdomain of application of the two external interpretations, the common core theory agrees with the two duals, and says less than they do. This establishes (i). $\Box$


What is substantive about the above construction of the induced interpretation is the assumption that the common core theory $T$ thus obtained, is indeed a theory, i.e.~a structured set of states, quantities and dynamics {\it for some physical system}. And, while this is not true in general, we argue that it {\it is} true in the examples of dualities from Parts I and II for which the common core theory, $T$, is more than a formal structure, i.e.~for which there {\it is} an interpreted common core theory. (These examples include the dualities for which a quantum version is available, and in so far as the common core has been worked out: see the discussion in Section \ref{cco}; Section \ref{lsrr} gives a detailed example, of bosonization duality.)\footnote{For position-momentum duality in elementary quantum mechanics, the common core theory is an algebra of operators on Hilbert space, and abstracts the choice of a position or a momentum basis: see Section \ref{pmd}. For electric-magnetic duality, see Sections \ref{EmdS} and \ref{MEMD}. For T-duality, see the discussion in Section \ref{SchemaT}. For gauge-gravity duality, see De Haro (2023:~Section 4.4).} 
In other words, although one cannot always construct, from the common structure of the duals, an interpreted common core theory: in the examples in which such a theory is {\it given}, its interpretation can be constructed from the external interpretations of the dual models in the way just outlined. \\
\\
{\it About (ii)}: we consider a syntactic reformulation of the bare theory and its models (see Section \ref{mtce}). The language will be the uninterpreted language of mathematical physics. We will first define and give examples of representation and interpretation maps; then we will clarify the entailment relations used in the Remark, and then we will prove the Remark.\\
\\
{\it Defining representation maps}: the representation map between a theory and one of its models, $h:T\rightarrow M$, is a homomorphism. Thus we require that it satisfies three properties:

(1) $h$ is a well-defined map;

(2) $h$, defined on a set of sentences, preserves the logical connectives, e.g.~it distributes over $\vee$, so that with the codomain of $h$ with its own $\vee$, we require: if $p$ and $q$ are sentences in the domain of $h$, then $h(p\vee q) = h(p)\vee h(q)$ in the codomain, and likewise for other connectives;

(3) truth-preservation by $h$ is defined as follows: if $p$, in the domain of $h$, is true, then $h(p)$ is true.\\
\\
{\it Example of a representation map}: let our theory be a state space ${\cal S}$ defined as an $n$-dimensional vector space in a two-sorted language, with one sort for vectors and another for scalars. We have $n$ linearly independent basis vectors with constant symbols $s_1,\ldots,s_n$. We use binary functions, $o_1,\ldots,o_4$, for the operations of vector addition, scalar addition, scalar multiplication, and multiplication of vectors and scalars, and a binary relation $=$ for arguments of the same sort. We use constant symbols for the zero vector and scalar, and for the unit scalar. We have axioms for vector addition and vector and scalar addition and multiplication, and for every vector in the vector space being a linear combination of the basis vectors with scalar coefficients. (We can have more relations, such as orthogonality or parallelism of vectors. And we can add other functions, e.g.~assigning a norm to vectors.) A {\it representation} is a map $h:{\cal S}\rightarrow {\cal S}_M$ to a vector space ${\cal S}_M$ of any dimension, i.e.~$v\in{\cal S}\mapsto h(v)\in{\cal S}_M$. ${\cal S}_M$ has $l\geq4$ functions $o_1^M,\ldots,o_l^M$, and $m\geq1$ relations $r_1^M,\ldots,r_m^M$ ($o_1^M,\ldots,o_4^M$ and $r_1^M,\ldots,r_m^M$ are binary relations and functions that respect the homomorphism). For example, mapping ${\cal S}$ into a vector space ${\cal S}_M$ of dimension higher than $n$ gives a representation that is not surjective, because not every vector of ${\cal S}_M$ has a pre-image in ${\cal S}$ (and it is in general not injective, e.g.~if some coordinates are projected). Thus for every sentence in ${\cal S}$ there is a sentence in ${\cal S}_M$ with the same truth values, but not the other way around. \\
\\
{\it Defining interpretation maps}: while on a semantic formulation, an interpretation map maps theories or models into domains of application, on a syntactic formulation, an interpretation map $i$ outputs a 1 if a sentence is true on the corresponding domain of application, and 0 if it is false (these are often called the `semantic values'). Thus interpretations are maps from sets of sentences into Boolean algebras.\footnote{See Button and Walsh (2018:~p.~299). For uses of Boolean algebras in quantum mechanics, see e.g.~van Fraassen (1974:~pp.~203-204) and Bub (1982), which draw on the classic Birkhoff and von Neumann (1936).}
For example, with the domain of $i$ being the theory $T$ and the codomain being the 2-element Boolean algebra of classical truth values, if $p$ is a sentence in $T$, then $i(p)=x$, and $x=1$ if $p$ is true in the theory's domain of application, and $0$ if $p$ is false.

Since an interpretation map is well-defined (i.e.~by the discussion in Section \ref{lessons}, it is single-valued), it satisfies (1). And since it is a partial homomorphism, it satisfies (2) only partially, i.e.~it may not be defined on some sentences.\footnote{For quantum logics, the interpretation map does not preserve the logical connectives for propositions that are not simultaneously determinate. For example, for an electron in a state with spin up along the $z$-direction, the probability of measuring spin up along the $z$-direction is 1, and the probability of measuring spin up along the $x$-direction is 1/2 (on the same state), but the probability of measuring spin up along the $z$- and along the $x$-direction is undefined, since the two measurements involve incompatible operators. Also, the case of quantum logic is more complicated and requires separate consideration, because the relevant set of propositions is the orthocomplemented lattice of projection operators, which is not a Boolean lattice. We will here restrict ourselves to classical logical interpretations.}\\
\\
{\it Entailment.} Here, we clarify what we mean by the entailment relations $M_1\vDash T$ and $M_2\vDash T$ in Eq.~\eq{matim}. These mean that the common core theory has been defined such that all its sentences are true whenever the corresponding sentences of the models are true, but, in addition, the models in general have sentences that do not have a counterpart in $T$, and to which no truth value is assigned (thus if a model is false, the common core can be either true or false).\footnote{A familiar example where such entailment relations are considered for scientific theories is Ramseyfication: for some theory $T$, the Ramsey sentence $R$ obtained from $T$'s postulate sentence $T'$ (i.e.~the conjunction of all of $T$'s sentences) by Ramseyfication, i.e.~by replacing terms in the theoretical or unobservable vocabulary by existentially quantified variables, is strictly weaker that the postulate sentence $T'$ itself. The Ramsey sentence is true if the postulate sentence is true, but not vice versa. Thus we say: $T'\Rightarrow R$. However, we will not work with postulate sentences here, but rather with sets of sentences.}

More precisely, there are two ways in which the models can ``say more'' than the common core theory:\footnote{We thank Hans Halvorson for a discussion of this point.} 
(i) by their being extensions of $T$ as sets of sentences, i.e.~by their having sentences written in $T$'s vocabulary that either do not have a truth value in $T$, or are false in $T$; (ii) by their having new vocabulary or concepts that do not occur in $T$, and which therefore have no truth-value in $T$. As we discussed in our `Syntactic formulation of a toy quantum duality' in Section \ref{mtce}, we take (ii) to be the relevant case for dualities: namely, the models augment the theory $T$ by adding new sentences that use new bits of signature, {\it without} implying more sentences in the original signature of $T$ (or, at least, they do not imply any sentences that are false in $T$).\footnote{Note that sentences in the signature of $T$ that do not have a truth value in $T$ would be admissible. This requirement on augmentations of type (i) follows from our truth-preservation condition (3) on the representation map $h$ above. We require the relation between the bare theory and its models to be such that the models do not add any sentences that are false in $T$ (else they are not representations of $T$). Although a model may ``make a choice from among a number of options'', e.g.~by fixing a boundary or an initial condition, such choices do not {\it contradict} the theory, because the theory leaves the boundary or the initial conditions {\it open}. In other words, adjoining to the theory a sentence that specifies a boundary condition gives a consistent set of sentences. Otherwise the choice of boundary condition would be incompatible with the theory, which is clearly an undersirable way to formulate the relation between a theory and its models.}\\
\\
{\it Unpacking the syntactic version, (ii).} In the syntactic formulation of the common core theory and its models, the representation map $h_1:T\rightarrow M_1$ maps sentences to sentences, i.e.~$\forall p\in T~\exists q\in M_1\,(h_1(p)=q)$, as it should.\footnote{A representation map of course maps different sentences of the common core theory into the model, so that e.g.~two sentences, $p,p'\in T$, have the same truth value as a sentence $q\in M_1$, i.e.~$i(p)=i(p')=i_1(q)$. Thus for $M_1$ to be a {\it representation} of $T$, the sentences of the common core theory that are mapped to the same sentence of the model must be {\it compatible} with each other and with the corresponding sentence of the model. In other words, in a {\it syntactic} formulation of a theory, a representation map is appropriately truth-preserving for individual sentences. And, as we argued at the beginning of Section \ref{aa}, this is precisely the idea of a representation: it makes the axioms of a theory (e.g.~a group) true in a structure (viz.~the representation of the group).\label{hrep}}

The syntactic formulation of the induced interpretation, $i$, is again given by the diamond diagram commutativity condition, Eq.~\eq{ii1i2}. Namely, suppose that $i(p)=x$, where $x=1$ or $0$ is the truth value of the sentence $p\in T$ in the theory's domain of application. The condition that the interpretation is induced from the model's interpretation, i.e.~$i=i_1\,\circ\,h_1$ (see Eq.~\eq{ii1i2}), gives: $x=i(p)=(i_1\circ h_1)(p)=i_1(h_1(p))=i_1(q)$, where in the last equality we used the representation map, $h_1:p\mapsto q$. Thus we have derived the truth value $i_1(q)=x$, which is what we expect: namely, the truth value of the sentence $q\in M_1$ is the truth value of the sentence $p\in T$ that it corresponds to under the representation map. (Note that not all of the model's sentences are thus obtained from sentences of the common core theory under the representation map.) This establishes that the common core theory $T$, thus interpreted is compatible with its model, $M_1$, i.e.~we have established the compatibility relation $M_1\vDash\,T$ in Eq.~\eq{matim}. The same argument of course applies to the other model under its representation map, so that $M_2\vDash\,T$, and a fortiori $M_1\vee M_2\vDash\,T$. Thus we have established, under an induced interpretation, $i$, the entailments in Eq.~\eq{matim}, i.e.~(ii). $\Box$\\
\\
{\it Generality of the proof and the direction of the entailment.} First, note that the above proof is {\it regardless of theoretical equivalence and duality}, so that it holds good for any set of models that are representations of a theory, so long as they have an induced interpretation as here discussed (needless to say, an induced interpretation is easier to come by if the models are duals, so that their external interpretations are in sufficient agreement).

It is worth adding a brief comment on the direction of the entailment relation: this depends on $h$'s being a {\it representation}, i.e.~on $h$'s properties (1) to (3). That the map $h$ is well-defined gives the correct direction of the arrow, i.e.~from a model to the theory, $M\vDash\,T$. Namely, because the map $h:T\rightarrow M$ is single-valued, every sentence of the bare theory gets mapped to {\it some} (exactly one!) sentence of the model, where the truth values are preserved by the homomorphism, and so all the truth values of the bare theory are determined by the truth values of the model, and the arrow {\it can} go from $M$ to $T$.\footnote{$h$ is not one-to-one, i.e.~it may map multiple elements of $T$ into $M$, which is compatible with the entailment arrow going from $M$ to $T$: see footnote \ref{hrep}.}
But it {\it cannot} go from $T$ to $M$, because $h$ need of course not be surjective, and so the truth values of $T$ do not give the truth values of $M$, since the model has sentences that lack a corresponding sentence in the bare theory. Thus the direction of the arrow is from $M$ to $T$. (And this is what we expect: the logically stronger set of sentences implies the weaker one, and not the other way around.)

\section{Theoretical equivalence illustrated by bosonization}\label{lsrr}

This Section illustrates the development of the Schema in the previous Sections, in an example of a duality that is both relatively simple and of scientific importance: namely, bosonization (see Section \ref{bosoniz}). The Section aims to illustrate the entailment relations of the syntactic formulation of duality (i.e.~Eq.~\eq{matim}). To do so, we will build on the syntactic formulation of a quantum duality, at the end of Section \ref{mtce}.

First, Section \ref{ccb} gives a (simplified!) syntactic formulation of the {\it common core theory} behind bosonization, in five axioms. Section \ref{mmcs} then shows that the bosonic and fermionic {\it models} can be constructed by adding, to this common core, a sixth axiom. Section \ref{lsrb} illustrates the logico-semantic relations between the theory and its models in terms of the relation between the axioms. Finally, Section \ref{iob} discusses the relation between the internal and external interpretations of the models.

\subsection{The common core theory of bosonization}\label{ccb}

In order to avoid unnecessary clutter and detail about conformal field theory, we will take as our common core theory just the set of holomorphic quantities that can be constructed from the two currents in bosonization, i.e.~the local algebra of the holomorphic currents $T(z)$ and $J(z)$ that we discussed in Section \ref{bosoniz} (i.e.~we temporarily disregard the state-space and the dynamics, and also the anti-holomorphic part of the theory; and `local' here means `in the neighbourhood of a given point').\footnote{Note that this is correct, because the holomorphic and the anti-holomorphic quantities `decouple': the representations of the enveloping Virasoro algebra from Section \ref{bosoniz} can be reconstructed from the tensor products of holomorphic and anti-holomorphic representations: see Ginsparg (1988:~pp.~7, 17, 44, 144).} 
$T(z)$ is the stress-energy tensor i.e.~the conserved current for the conformal transformations, and $J(z)$ is the conserved current for holomorphic transformations on the fields. Recall that these two currents gave us grounds, in Eq.~\eq{currco}, to view bosonization as a unitary equivalence of representations of algebras.

These currents satisfy what we shall call our five axioms of the common core theory,\footnote{These five axioms are special cases of more general expressions that, in conformal field theory, can be derived for primary fields (see footnote \ref{primaries} in Chapter \ref{Advan}). For a proof of the short-distance behaviour of products of the stress-energy tensor with itself and with primary fields, see Ginsparg (1988:~pp.~18-20, 27) and L\"ust and Theisen (1989:~pp.~60-64). For the remaining four axioms of any conformal field theory, see Ginsparg (1988:~p.~10).} 
i.e.~as envisaged at the end of Section \ref{mtce}. (Thus in particular, our aim is {\it not} to describe the full conformal field theory, but rather the subset of quantities constructed from the currents; the operator product expansions given below guarantee that this subalgebra is closed, and thus is a good illustration of the Schema.)\\
\\
{\bf Common core theory:} ${\cal Q}={\cal A}\{T(z),J(z)\}$, where ${\cal A}$ is the local algebra spanned by the currents $T(z)$ and $J(z)$ (defined by their normally-ordered product operator expansions),\footnote{Normal ordering in real coordinates becomes, in the present coordinates on the complex plane, `radial' ordering: see the operator ordering conventions in Ginsparg (1988:~p.~19). The operator product expansion is the short-distance expansion of a product of operators (i.e.~$z$ and $w$ are very close to each other), where in a conformal field theory the operators that can appear on the right-hand side are determined by the dimensions (i.e.~their `conformal weight', i.e.~the eigenvalue of an operator under dilatations: see footnote \ref{primaries}) of the operators of the left-hand side. In a conformal field theory, the operator product expansion is fixed by a set of operator coefficients which then fix the three-point Green's function of the model (in other words, the two- and three-point Green's functions in a two-dimensional conformal field theory are fixed by conformal invariance). The limit needs to be taken with care, and is usually defined by radial ordering: see Ginsparg (1988:~p.~19). Also, $c$ is the anomalous charge, with value $c=1$.	Operator product expansions not only provide an axiomatic basis for conformal field theory, but more generally also for quantum field theory in a curved spacetime: recently, Hollands and Wald (2010:~pp.~85, 117) have proposed an axiom-by-axiom alternative to the Wightman axioms for quantum field theory in curved spacetime, in which the operator product expansion plays a fundamental role. For the relation between the axioms of conformal field theories and the Wightman axioms, see Kac (1998: Sections 1.1, 1.2), who derives the former from the latter. In particular, the operator product expansion follows from Wightman's locality axiom (ibid, pp.~11, 25-26).} 
subject to the following four axioms:

{\it A1.}~The Sugawara construction obtains, i.e.~$T(z)=-{1\over2}:J(z)J(z):$.\footnote{For the Sugawara construction, see Section \ref{bosoniz}.}

{\it A2.}~The operator product expansion of the stress-energy tensor with itself has $c=1$,
i.e.~$T(z)\,T(w)={c/2\over(z-w)^4}+{2\over(z-w)^2}\,T(w)+{1\over z-w}\,\pa T(w)$.

{\it A3.}~The operator product expansion of the stress-energy tensor with the affine current has weight $h=1$, i.e.~$T(z)\,J(w)={h\over(z-w)^2}\,J(w)+{1\over z-w}\,\pa J(w)$.\footnote{The weight that appears in this operator product expansion is the conformal weight of the current $J$. See footnote \ref{primaries}.}

{\it A4.}~The operator product expansion of the affine current with itself has level $k=1$ and is a $c$-number, i.e.~$J(z)\,J(w)={1\over(z-w)^2}+\mbox{finite terms},~~(z\rightarrow w)$.

{\it A5.}~The stress-energy tensor and the affine current are currents for a chiral primary field of conformal weight $h=\half$, i.e.~a chiral fermion $\hat\psi$: $J=:\hat\psi^\dagger\hat\psi$.

The first axiom specifies the relation between the two currents in the algebra (i.e.~Eq.\eq{TTbar}). The next three axioms have been added for completeness: they specify the three possible products between elements of the algebra (namely, $TT$, $TJ$, and $JJ$). But the details of these axioms are not important for our argument in this Section. The fifth axiom stipulates that the currents are given by a specific chiral primary field in the conformal field theory: namely, a fermionic field.\footnote{This then specifies the irreducible representation, because the highest-weight vector of the representation is the one that is obtained by applying a chiral primary of weight $h$ to the vacuum, in this case applying the fermionic field $\F$ at $z=0$ to the vacuum.}

Agreed, these axioms assume background knowledge about conformal field theory: for example, they assume knowing that the `operator product expansion' is the expansion of a product of operators, evaluated at short distances, i.e.~in the limit $z\rightarrow w$, and the kinds of terms that, depending on the dimension of the fields, can appear in such an expansion. But we will take this in our stride as being {\it background information}, that does not require formalization by additional axioms, analogous to how Section \ref{theoreq} took the axiomatizations of the reals and complex numbers as background information.

Although the fifth axiom might give the impression that a fermionic model is more fundamental than a bosonic model, neither model is in fact more fundamental than the other. This is because in a two-dimensional conformal field theory (restricting, as we have been doing, to the holomorphic sector) the operators are uniquely determined by their conformal weight.\footnote{For more details, see footnote \ref{primaries}.} Thus a formulation in terms of a chiral primary operator $\hat\psi$ of conformal weight $h=\half$ (as in the fermionic model), and a formulation in terms of a bosonic operator\footnote{This operator is defined such that its derivative, $\pa\hat\f$, is a chiral primary operator of conformal weight $h=1$.}
$\hat\f$ (as in the bosonic model) are, in virtue of Eq.~\eq{expf}, one and the same. In other words, writing a conformal field theory using one variable or the other is a trivial reformulation: the content of the two theories---the states and the quantities---is the same. This is because the content of a conformal field theory, for a given conformal family, is exhausted by the set of operators of given conformal weights. Thus as in the case of quantum mechanics, where the reformulation in terms of Hilbert space shows that writing the wave-function in terms of $x$ or $p$ is merely a different expansion of the state (see Section \ref{pmd}), the reformulation in terms of conformal field theory, which does not know about bosonic or fermionic fields, but only about operators of a certain conformal weight, shows that the two (bare) models are the same. 

Thus this argument says that, {\it augmenting} the models by adding fermionic operators into the bosonic model, as in Eq.~\eq{expf} (see Figure \ref{augmentation}), the two {\it quantum mechanical} models are trivially the same when formulated in the framework of a conformal field theory with axioms {\it A1} to {\it A5}.

But we need to say more about the {\it classical} models that we started from in Section \ref{SGT}: namely, the free massless boson and free massless Dirac fermions, with actions given by the kinetic terms of Eqs.~\eq{sineGact} and \eq{Thirrac}. This will lead us back to our theme of {\it quantum duality}, from Section \ref{std}, in the syntactic formulation from Section \ref{mtce}.

\subsection{Two models with additional classical structure}\label{mmcs}

We will now add two axioms, {\it B6} and {\it F6} below, that relate the theory to two of its classical limits, and that, in the next Section, will give an illustration of the logico-semantic relations from Section \ref{lsr}. Thus we fill in details in our `toy quantum duality' in Section \ref{mtce} (especially Eq.~\eq{T1T2}), for the case of bosonization. We do this using the Schema's jargon, where there is a common core theory $T$ with two models, $M_1$ and $M_2$. (Thus, compared to Section \ref{mtce}, we write $M_1$ and $M_2$ rather than $T_1$ and $T_2$.)

The first axiom, {\it B6}, that we will introduce, defines the type of model.\\
\\
{\bf (1)~~Bosonic model:} The bosonic model is obtained when we stipulate that the theory's currents, $T$ and $J$, proceed from quantizing the corresponding classical currents of a bosonic scalar field (i.e.~as in Section \ref{bosoniz} (A), especially Eq.~\eq{affJ}). Thus we add to {\it A1} to {\it A5} the following axiom:

{\it B6.} The affine current $J$ is obtained by quantizing the classical current $J_{\tn B}$ given by the derivative of a bosonic scalar field, i.e.~$J_{\tn B}=\pa\f$.\\
\\
{\bf (2)~~Fermionic model:} The fermionic model is obtained when we stipulate that the theory's currents, $T$ and $J$, proceed from quantizing the corresponding classical currents of a chiral fermionic field (i.e.~as in Section \ref{bosoniz} (B), especially point (c)). Thus we add to {\it A1} to {\it A5} the following axiom:

{\it F6.} The affine current $J$ is obtained by quantizing the classical current $J_{\tn F}$ given by a chiral symmetry-invariant fermion bilinear, i.e.~$J_{\tn F}=\,\psi^\dagger\psi$.\\
\\
We will make six comments about these two specific axioms, of which (a) to (c) are about the axiomatization itself, and lead in to (d) and (e) about bridge laws and about the isomorphism, and the last comment is about the difference between the two models.\\

(a)~~As we stipulated in Section \ref{mtce}, each specific axiom {\it B6} and {\it F6} specifies a different {\it semi-classical limit of (a subsector of) the theory}. The bosonic model gives an accurate description of the semi-classical limit of the subset of coherent states where the bosonic operator $\hat\f$ satisfies the equation of motion of the classical bosonic model (these are states where fermions are condensed into pairs, i.e.~coherent states of the fermion bilinear operator $\hat\psi^\dagger\hat\psi$). The fermionic model gives an accurate description of the semi-classical limit of the subset of coherent states where the expectation value of the fermionic operator $\hat\psi$ satisfies the equation of motion of the classical fermionic model.

(b)~~To formulate its semi-classical limit, each axiom uses additional symbols: namely, elements $\Sigma_1=\{J_{\tn B},\phi\}$ and $\Sigma_2=\{J_{\tn F},\psi\}$ of the two signatures. These elements should be included in the signatures, as in Eq.~\eq{T1T2}. (However, in this Chapter, to avoid clutter, we do not separately keep track of signatures.)

(c)~~As before, each of the specific axioms comes with additional background knowledge about classical bosonic and fermionic fields and about how to quantize them, for which we will not require additional axioms. 

(d)~~Points (a) and (b) imply that, as we discussed in Section \ref{mtce}, these axioms are {\bf bridge laws} that reduce the semi-classical bosonic and fermionic models to the quantum theory.\footnote{Their role is like Schaffner's (1969:~p.~283) `type III' correspondence rules (really more like bridge laws), which link the meanings of new terms to `antecedent theoretical meanings'.}

(e)~~Points (a) and (b) allow the models to be isomorphic in the sense of Section \ref{mtce} (i.e.~Eq.~\eq{isoT}), because the structures that satisfy these axioms only differ through their having (i) different domains of quantification, and (ii) different signatures (see the `two ways that models are finely individuated, after Eq.~\eq{isoT} in Section \ref{mtce}). 

(f)~~The two additional axioms express a {\it real difference} between the two semi-classical limits of the theory, i.e.~the two semi-classical (bosonic and fermionic) models that reduce to the theory, but not necessarily a real difference about the two models of the theory; (whether this is a real difference depends on having an external or an internal interpretation). It is a real difference of the semi-classical bosonic and fermionic models because there is no simple relation between the actions or the equations of motion of the classical boson and the classical fermion, since duality entails the {\it quantum} relation Eq.~\eq{expf}. This is our theme of {\it quantum duality}, from Section \ref{std} (which we also encountered in the string theory examples in Part II): the duality is realized only at the quantum level, i.e.~the classical models are not duals. The two models are isomorphic to each other, in the sense that the bare theory is indeed their common core, but in addition each of the models has its own specific structure: namely, the classical structure (including the symbols in the signature) that gives the operators $\f$ and $\psi$ their specificity as, respectively, massless free bosons or as free chiral fermions. 

\subsection{Illustrating the Schema by bosonization: logico-semantic relations}\label{lsrb}

We are ready to illustrate the logico-semantic relations between a theory and its models, from Section \ref{lsr}. In the example of bosonization, the formulation of the theory and its two models can be summarized as follows. The common core theory is the set of quantities ${\cal Q}$, subject to the axioms {\it A1} to {\it A5} (again for simplicity of the notation, we avoid a separate mention of the signature): 
\bea
T:=\bra{\cal Q},A1,\ldots,A5\ket\,.
\eea
Each of the two models is obtained by {\it adding a sixth axiom as specific structure} (and each axiom comes with additional symbols in its signature):\footnote{Although this notation is reminiscent of our earlier notation for models (see Eq.~\eq{eqmodel}), the specific structure here are additional axioms, and so we have refrained from indicating representations by subscripts, i.e.~we have included the theory $T$ as the model root. This is because the models must be written in the same language as the theory. In this syntactic formulation, the additional axioms play the role of the specific structure (see Section \ref{mtce}).}
\bea\label{MbMf}
M_{\tn B}&:=&\bra{\cal Q},A1,\ldots,A5;B6\ket=\bra T;B6\ket\nn
M_{\tn F}&:=&\bra{\cal Q},A1,\ldots,A5;F6\ket=\bra T;F6\ket\,.
\eea
Apart from the fact that we lifted our talk of theories ``one level up'' (so that Eq.~\eq{MbMf} are now models, rather than theories), and that our notation does not keep track of the signatures, this is the form that duals took in our discussion of the syntactic version of duality, in Section \ref{mtce}. 

There are two points to be made about this presentation of the models. The first is about how it illustrates the Schema, and the second is our main point about the logico-semantic relations implied by the Schema.

First, the two models have the bare theory $T$ in their model root. Thus in this syntactic formulation, the bare theory is the `common core' of the two models in the literal sense of being a common conjunct in the definitions of the models. And in both models, the sixth axiom, which relates the model to its classical limit, is `specific structure'. Since bosonization is a quantum duality, the additional axioms are only required as bridge laws that relate our quantum theory to a specific classical limit (in the next Section, they will also enable a classical interpretation for each of the models): these classical limits are not part of the structure mapped by the duality.

Second, this syntactic formulation of the theory and its two models, in terms of structures satisfying a list of axioms, illustrates the logico-semantic relations in Eq.~\eq{matim}. For the common core theory $T$ is obtained from either model by dropping the model's sixth axiom and its signature, i.e.~by dropping the specific structure. And so, whenever either model is true, i.e.~whenever ${\cal Q}$ satisfies the five axioms of either model, the common core theory is also true, since in either model ${\cal Q}$ evidently satisfies the subset of five axioms, {\it A1} to {\it A5}. This establishes that $M_{\tn B}\vDash T$ and $M_{\tn F}\vDash T$, and {\it a fortiori} $M_{\tn B}\vee M_{\tn F}\vDash T$. Also, in so far as there is no duality relation between the models' specific axioms {\it B6} and {\it F6}, because there is no change of variables that renders their classical actions or their equations of motion equivalent, the two models are {\it incompatible}, i.e.~$M_{\tn B}\vDash \neg M_{\tn F}$.

Notice that the common core theory is {\it silent} about the additional axioms {\it B6} and {\it F6}, since these axioms cannot be formulated in its signature, which does not contain classical fields or currents (these axioms correspond to `sentences [of the dual models] that do not have a counterpart in $T$, and about which the common core is silent': see Section \ref{lsr}). Thus these axioms do not have a truth value in $T$, which is consistent with the common theory being implied by its dual models, but not the other way around.

\subsection{Interpretation of bosonization}\label{iob}

The previous Sections illustrated how bosonization's common core bare theory follows from the dual models by abstraction, i.e.~by dropping their specific structure (see Figure \ref{commonc}). In the syntactic reformulation of bosonization, the common core follows by dropping the model-specific axioms, {\it B6} and {\it F6}. 

We now discuss the interpretation of the theory and of its models---and this will cast further light on the role of models in the Schema. In particular, we distinguish the internal and external interpretations (see Section \ref{intext}):

An {\it internal interpretation}, by definition, maps only the model roots, i.e.~in this case, the axioms {\it A1} to {\it A5} of the bare theory. This means that this internal interpretation is a map from the common core theory to the domain of application. By construction, it satisfies the condition Eq.~\eq{ii1i2}, i.e.~the commuting diagram in Figure \ref{Physeq}, because the range of the representation maps $h_{\tn B}$ and $h_{\tn F}$ for the bosonic and the fermionic models, respectively, is the theory itself. In other words, for any interpretation of the common core theory $T$, there are internal interpretations of the two models with the same domain of application. On such an interpretation, there are bosonic and fermionic states in the theory, but they are not associated with the quantization of the classical actions of the massless free boson and massless chiral fermion.

Also, since the common core theory is a quantum theory that does not mention classical fields, there is no sense in which the classical bosonic or the fermionic field that appears in the classical actions is fundamental or privileged. Since the classical fields are part of the specific structure, they are not mapped by the internal interpretation: thus on an internal interpretation, there are no `elementary boson' or `elementary fermion' {\it fields}, and we speak instead about a `bosonic state' or a `fermionic state' that is characterized by the quantum numbers for the conformal weights (i.e.~its spin and dilatation eigenvalues), and the expectation values of the operators in that state, such as energy and momentum. (This is again the point about additional signatures being required by the bridge laws, in Eq.~\eq{T1T2b}.)

An {\it external interpretation} also assigns referents to the specific structure, i.e.~to the model-specfic axioms {\it B6} and {\it F6}: and so it {\it does} allow us to interpret the theory's states as having specific semi-classical regimes where the bosonic and the fermionic classical descriptions are valid. The bosonic and fermionic interpretations are, in a weak sense, mutually incompatible: namely, as descriptions of the actual world, i.e.~they cannot be simultaneously realized for a single set of initial boundary conditions and range of parameters. For when the bosonic model gives an accurate semi-classical description (i.e.~for a subset of states and operators), the fermionic model in general does not: and this is of course reminiscent of complementarity (see Section \ref{adcom}). This distinction will in general be made by the initial or boundary conditions: the initial asymptotic state (at $z=0$) may allow an approximation of the theory's behaviour by a subset of states and operators that (at least for a certain amount of time) behave semi-classically, as described well by one model but not by the other.\footnote{This will for example be the case with coherent states of one of the two models, where either the bosonic or the fermionic operator acquires an expectation value that satisfies suitable factorization properties and satisfies the classical equation of motion.}
But such semi-classical approximations are sensitive to the initial conditions, and may not hold everywhere in the two-dimensional spacetime.

Note that the internal and external interpretations that we have just discussed, which were literal readings of the axioms, agree with the relations in Eq.~\eq{matim}. This casts light on the relations between internal and external interpretations. For the external interpretations just discussed, i.e.~the literal readings of the two sets of axioms in Eq.~\eq{MbMf}, are {\it compatible} with the internal interpretation. (And the two external interpretations are mutually incompatible.) 

Thus if our target system is quantum mechanical, it is best described by the theory's internal interpretation, and the external interpretation of the sixth axiom gives an {\it approximate} or {\it idealized} description of the target system as a classical system. In this sense, the internal interpretation has priority over the external ones. For the latter add statements that are only approximately true: namely, in the semi-classical approximation.

In our example, the classical structure added by one of the models is a case of {\it idealisation}, because it describes a possible target system, i.e.~there is a possible world associated with the semi-classical regime (namely, the possible world described by the solutions of the classical equations of motion of the model).\footnote{We use Norton's (2012:~p.~219) notion of `idealisation', which we further discuss in Section \ref{Sem}.}
This means that, regardless of the question which target system we use the theory and the models to describe (a quantum system or a classical system), on an external interpretation this can be viewed as a case of emergence of the classical regime from the quantum regime.

While, as we stressed in Section \ref{std}, not all dualities are quantum dualities, where the models add classical structure to a quantum theory, many interesting dualities {\it are} of this type, and so the previous discussion generalizes. 

Namely, dual models describe different classical limits of a given quantum theory, and we can view the richer ontological picture given by an external interpretation of a dual model as emerging from the sparser ontological picture given by the (internal) interpretation of the common core theory. (We will return to the question of emergence in Chapter \ref{Heuri}.)

\section{Conclusion}

This Chapter has developed a view of how dualities bear on two general issues about inter-theory relations: namely, theoretical equivalence, and the relation between a theory and its models. 

First, we discussed physical equivalence as the duality-based criterion of theoretical equivalence: and thus as the answer to the questions, in Chapters \ref{Theor} and \ref{physeq}, about the physical interpretation and how it meshes with duality.

The second aspect we developed is the common core theory, both in general, and more specifically for dualities. Besides its highlighting complementary aspects of dualities (in particular, their external interpretations, which make duals theoretically inequivalent), this aspect bears on the issues discussed in the next Chapters: in particular, on scientific realism and on the practical functions of dualities.

Both aspects are summarized in the commuting diagram in Figure \ref{Physeq}. Physical i.e.~theoretical equivalence is the condition that the bottom triangle diagram commutes (regardless of whether the diamond diagram commutes). In particular, we have argued that the standard of {\it sameness} of, or at least the lack of a difference in, the domains of application should not be weakened, and so that the criterion of theoretical equivalence that results has the right logical strength. 

The second aspect, i.e.~the common core theory, is summarized by the commutativity of the diamond diagram (regardless of whether the bottom triangle diagram commutes, i.e.~regardless of whether the interpretation is internal or external!). The common core theory, including its internal interpretation map, is obtained from the models by abstraction, i.e.~through the commutativity of the diamond diagram (and this is again regardless of dualities, and applies if there are more than two models).

We argued that, from both formal and interpretative points of view on dualities, when the common core theory is isomorphic to its models, and the diamond diagram commutes, the common core theory is of the desired logical strength. This criterion is regardless of non-dual models, which can be added in a next step, and gives the most perspicacious formulation of the theory behind dual models, so that it can be a starting point for the discussion of the practical functions of dualities.

In cases of theoretical equivalence, with a common core of the ``right strength'', i.e.~isomorphic to its dual models, the models are reformulations of one and the same theory.

We also found important logico-semantic relations between a theory and its models, summarized in Eq.~\eq{matim}. As it turns out, the representation map from a theory to its models can be interpreted, in a syntactic formulation, as the entailment of the theory by its models. In other words, the process of abstraction from logically stronger models to the logically weaker common core theory is correctly interpreted as a relation of entailment. This then bears on our discussion of scientific realism, in the next Chapter.

\chapter{Realism and Equivalence}\label{Realism}
\markboth{\small{\textup{Realism and Equivalence}}}{\textup{\small{Realism and Equivalence}}}

In the previous two Chapters on the theoretical function of duality, we explored how duality bears on {\it inter-theoretic relations}, and in particular on theoretical equivalence. (Emergence, which is another important inter-theoretic relation, will be explored in the next Chapter.)

As we announced at the start of Chapter \ref{Thies}, we have been able, so far in this book, to set aside the epistemology of physical theories. The presentation of the Schema, and of its illustrations, in the preceding Chapters depended on construing the claims of theories literally, and more specifically, on giving them a referential semantics. But issues about the scientific community's warrant for believing these claims could be set aside; mainly because the two main rival views about our warrant, viz.~realism and anti-realism, agree in giving the various claims a referential semantics. But from this Chapter on, these epistemological issues will come to the fore. 

This Chapter discusses the question of {\it scientific realism}, and how theoretical equivalence and dualities bear on it. Thus the first issue is realism, both in general philosophy and in philosophy of science (Section \ref{realism}), which is a view about what theories, especially scientific theories, are taken to say about the world. 

Section \ref{srr} then discusses scientific realism and anti-realism. In the light of dualities and our earlier logico-semantic analysis, we propose a {\it cautious scientific realism}, both as a reading of scientific theories and of their inter-theoretic relations, and as a reply to a possible threat of under-determination. We argue that a structured, as against a flat, view of theories is required. Thus we conclude that dualities do not give new ammunition to the anti-realist who wishes to defend a Quinean under-determination thesis. 

Section \ref{ejpe} discusses the {\it epistemic justification} of verdicts of physical equivalence. Here, the notion of an unextendable theory is developed in light of Part II's quantum field theory and string theory examples. 

So we set the scene for these issues by starting this Chapter with a discussion of realism, both in general philosophy and in the philosophy of science. Sections \ref{rip} and \ref{scirer} first introduce our topic for non-experts. Section \ref{eronos} then discusses empirical equivalence, and Section \ref{pud} introduces the anti-realist challange of under-determination.

\section{Realism in philosophy; scientific realism}\label{realism}

Broadly speaking, we are realists, both in general philosophy and in the philosophy of science. This viewpoint sets the tone of what follows: more specifically, it influences our views about theoretical equivalence (in Chapter \ref{physeq}) and emergence (in Chapter \ref{Heuri}). Nevertheless, many of our views on these topics are developed independently of a scientific realist position: thus we will mention realism explicitly when it is required. We begin by introducing it. 

\subsection{Realism in philosophy}\label{rip}

{\bf Realism} is traditionally defined as a conjunction of two claims. The first is metaphysical, saying that the world---or some more specific topic being enquired into---is independent of us and of how our minds work: `the world is mind-independent'. The second is epistemological, saying that we can have knowledge, or at least warranted belief, about the world (the topic). 

Here, we say `or at least warranted belief' because in traditional philosophy (ever since Plato), knowledge has been `put on a pedestal', as a kind of true belief for which one not only has warrant (justification) but which also is indefeasible. That is: philosophy has traditionally maintained that once knowledge is attained, one need not, and should not, ever give it up; nor revise it, even mildly. Also, in traditional philosophy, especially since Descartes, `knowledge' connotes certainty (i.e.~complete 100 percent conviction) and exact, not approximate, truth. But we, and most other realists, in particular scientific realists, do not need to claim knowledge in this strong sense of indefeasible, certain and exactly true belief. (Recall the discussion in Section \ref{itm} about the notion of approximate truth.) 
This independence of the world from our minds does not conflict with the fact that we are ourselves part of the world, and so our knowledge always begins from a specific epistemic situation. Hence our saying `or at least warranted belief'. The main point is that the second thesis is one of cognitive optimism: despite the independence of the world or topic, we can know about it. More precisely, and more modestly: we can have warranted belief about it, where the belief may be defeasible, not certain and only approximately true.\footnote{Our realism about the world (or topic, or objects) contrasts with another kind of realism that we do not endorse, held by Plato, about {\it ideas}, i.e.~the view that ideas, general concepts and-or universals are independent of the mind, and exist independently of our knowledge of them. The traditional rival of realism about ideas is {\it nominalism}, i.e.~the claim that such ideas, general concepts or universals are mere names or concepts generated by the mind. Although nominalism can be a form of scepticism, it need not be, and in any case nominalism is a more specific doctrine. While this realism/nominalism debate is important in e.g.~the philosophy of mathematics, it will not play an important role in this book. See also the discussion of empiricism, below.\label{nomin}}

In the traditional broad debates of philosophy, realism is opposed to two rivals. First: {\bf scepticism}, which agrees that the world or topic is mind-independent; but then denies realism's cognitive optimism, saying that therefore, we cannot know, or have warranted belief, about the world or topic. So in effect, scepticism is cognitive pessimism.\footnote{As we will discuss below, especially when we talk about the philosophy of science, `scepticism' is a vague term that includes both clearly articulated doctrines, such as instrumentalism, and other ways of thinking.} 
Second: {\bf idealism}, which denies that the world or topic is mind-independent. This denial need not be as crazy as it at first appears. For it need not mean that the world or topic is literally ideas of ours, and would not exist without our mental activity. It can mean instead that the way we think of and describe the world or topic is so greatly moulded by us, i.e.~by the biological, or cultural, or individual psychological factors in cognition, that it is pointless, and perhaps impossible, to try and winnow out some notion of the world (or topic), in itself, rather than as moulded by us (see footnote \ref{nomin}, and Section \ref{srpost}). But whether one takes this sort of view, or the first, more simple-minded and apparently crazy, view that the world or topic is literally ideas, the overall intent of idealism is to try and answer scepticism. The intention is, by `bringing the world closer to us' (if not literally, in to our minds), to make it knowable.\footnote{Nowadays, the second, more nuanced, view is often called `anti-realism', to distinguish it from the `literally, ideas' of traditional idealism. Of course, all these `isms' are vague. But it will be useful to us in the sequel to have adumbrated the three core doctrines of realism, scepticism and idealism.} 

As we have said, we will not need to defend realism at this very general level. But we will now in Section \ref{scirer}, say a bit more about the more modest, because specific, doctrine, {\it scientific realism}. It is in effect cognitive optimism of the above sort, about the claims of well-confirmed scientific theories. Although most of our views do not depend on scientific realism, a short discussion is called for, for two reasons. First, {\it some} of our views in the next Chapters do so depend (see Section \ref{srr}). So we need to clear the path. Second, we need to expound scientific realism and its rivals in sufficient detail to justify our claim (also at the start of Chapter \ref{Thies}) that nowadays, scientific realists and their opponents agree on the semantic aspects of interpreting scientific theories.

\subsection{Scientific realism and its rivals}\label{scirer}

We said that scientific realism is cognitive optimism about well-confirmed scientific theories. More precisely, and confining ourselves from now on to physics, i.e.~physical theories: we will take scientific realism as the conjunction of (i) a claim about the semantics or interpretation of physical theories and (ii) a claim about their epistemology.\footnote{For a clear exposition of the distinction between semantic and epistemic versions of realism, see Earman (1993:~p.~19)}

About (i), the semantics or interpretation of physical theories: the claim is that all the statements of a physical theory should be given a referential semantics in the tradition launched by Frege and consolidated in modern logical theory. But for the moment, the main point lies in the phrase `all the statements'. 

For there is a traditional contrast between two types of statement, within a physical theory: {\it observational statements}, that are about observable objects and states of affairs, and {\it statements about unobservables} (often also called `theoretical statements') that are about unobservable objects and states of affairs. This contrast is usually made a bit more precise along the following lines. The theory's vocabulary, i.e.~the predicates it ascribes to the objects it deals with, divides in to two sets: the observational or observable vocabulary, i.e.~predicates that can be ascribed to objects on the basis of looks (roughly: untrained observation, or observation without instruments); and theoretical vocabulary, i.e.~predicates that cannot be so ascribed (one can also say: `non-observational vocabulary'). The traditional idea is then that for observational statements, we have both a better understanding of what they claim about the world, and a better (at least: less corrigible) warrant for believing them true, than we have for theoretical statements.\footnote{In the usual conception, observational sentences can be verified using the {\it unaided} human senses. While there are deep epistemological reasons for this, having to do with the theory-ladenness of observations that uses scientific instruments, as well as the question of the different nature and degree of certainty of the evidence provided by observation mediated by instruments, it is less natural in the face of contemporary scientific practice, where observation is almost never done with the unaided human senses, but almost always uses measuring or observation apparatusses. The notion of `observation' has accordingly been broadened by Lenzen (1955:~p.~307), who allows observation mediated by scientific instruments, his reason being that such observation `has achieved the practical certainty of direct observation'.\label{observe}}

The rationale for this view is, for almost all authors, empiricism as understood within general philosophy. {\bf Empiricism} is the doctrine that experience is the source of both: (i) our concepts and categories (the meanings of our predicates by which we describe and classify objects), and so the content of our statements about the world; and (ii) our evidence for those statements. So the `closer' a concept or category---the meaning of a predicate---is to experience, the surer our understanding of claims that involve it, and the better (at least: less corrigible) the warrant that experience provides for believing such claims. 

And thus we come to the traditional `decision-point', or branching of the road, that distinguishes scientific realism from its rivals. There will be a broad threefold division, as we saw in general philosophy: which we can again label as `realism', `scepticism' and `idealism'. And in this more specific context of philosophy of science, each will have more specific versions: which, we shall report, have been advocated in twentieth-century philosophy of science. 

As regards the broad threefold division: the first option is realism (now called `scientific realism'). Again, it is a conjunction. As we said above, its claim about semantics and interpretation ((i) above) is that all the statements of a scientific theory, whether observational or non-observational, are to be given a referential semantics of the sort given in modern logical theory. As it is often put: the theoretical statements are to be understood literally, with terms for unobservable objects or states of affairs being assigned a reference in exactly the same way as terms for observable objects. And its claim about epistemology ((ii) above) is that we can have warranted belief in non-observational statements, distant though their claims are from our experience. So here again the view is one of cognitive optimism. We can formulate statements, understood literally, about what is beyond our experience; and we can have warranted belief in them. 

On the other hand, there is scepticism. It accepts (i) but denies (ii). It accepts referential semantics for all statements of a scientific theory: it agrees that we have, somehow or other, invested our theoretical terms with meanings that sustain a literal understanding of them. But it denies that we can have warranted belief in such statements.\footnote{Agreed: in traditional philosophy, scepticism is the denial of the validity of knowledge claims, where `knowledge' connotes various features additional to true belief, especially having an indefeasible justification. But recall that above, we said that indefeasibility was a tall order; and so we settled on warranted (but defeasible) belief.} 

Finally, the third option is to deny (i): theoretical statements are not to be understood literally. But then one advocates some other understanding of them that gives them some useful role in empirical enquiry. This is done in one of two broad ways. Either one reconstrues the theoretical statements as another type of statement, and then allows that we can have warranted belief in these reconstruals. So according to this way, the reconstruals are statements, and have a straightforward referential semantics. Example: `Such-and-such statement about electrons should be reconstrued as a long and complex conjunction of conditional statements about point-readings on ammeters, volt-meters etc.' Or one denies that theoretical statements can be reconstrued as any kind of {\it statement}, i.e.~claim about the world. So according to this way, their role is not representational. (Some authors say that the role is to organize observational statements, and-or as `inference-tickets' that license certain inferences between them.) Here, we do not need to go in to details about these two ways. We just remark that, of course, either way corresponds to the theme in idealism, within general philosophy, of `bringing the world closer to us' (if not literally, in to our minds as a set of ideas), so as to make it knowable.

We should also remark on the jargon. Though `scientific realism' is standard jargon for the first option, `scepticism' and `idealism' are not standard jargon for the second and third. Another often used jargon is:

(a) `anti-realism': which is used as an umbrella-term for both options (scepticism or idealism), i.e.~for any view opposed to realism;

(b) `instrumentalism': is used to denote scepticism and (equally often) as a term for the second way of the third option, i.e.~the view that theoretical statements are not really statements at all, but have some other role---they are `instruments' for organizing observations, or for warranting inferences between observational statements.

In recent decades, debate has focussed mostly on realism and a version of anti-realism advocated by van Fraassen (1980). He calls his view `constructive empiricism'\footnote{Note that the name of the view, `constructive empiricism', has misleading connotations, because empiricism is compatible with realism, i.e.~{\it empiricism is not anti-realism}. This misleading connotation already appears in the opening sentence of {\it The Scientific Image}: `The opposition between empiricism and realism is old', and then van Fraassen goes on to explain the opposition between realists and {\it nominalists}! Indeed, as Salmon (1989:~p.~150) stresses, this `striking' sentence rests on `an [anti-realist] assumption that goes unquestioned throughout the rest of the book---namely, that it is impossible to have empirical evidence that supports or undermines statements about objects, events, and properties that are not directly observable by the unaided normal human senses'.} 
and formulates the debate over scientific realism in much the same terms as we have adopted above. He takes scientific realism and his own view, constructive empiricism, to both endorse a uniform semantic treatment of observational and theoretical statements. Thus he holds that anti-realists,\footnote{Van Fraassen's endorsement of classical referential semantics makes him an epistemic anti-realist but a semantic realist (see the discussion at the beginning of Section \ref{realism}). Salmon (1989:~p.~135) labels him a `sophisticated instrumentalist', because `van Fraassen recognizes that theories, which at least appear to make reference to unobservable entities, have played an indispensable role in the development of modern science ... The most fundamental difference between the instrumentalist and the constructive empiricist is that the former {\it denies} the existence of unobservables while the latter remains {\it agnostic} with respect to their existence.'} 
including he himself, should, like the realist, interpret the theoretical terms---and so the theoretical statements---of scientific theories literally, i.e.~with a classical referential semantics, just as they do the observational statements: rather than reconstruing the theoretical statements, or as `inference-tickets' or something similar (as instrumentalists traditionally did). 

The literature on scientific realism has very largely followed van Fraassen's lead. Of course, agreement on the points above is only to be expected from scientific realists like us. But other critics of scientific realism, such as Fine with his advocacy of what he calls `the natural ontological attitude', also endorse a referential semantics for both the theoretical and the observational terms.\footnote{See Fine (1984, 1986).}

Thus we submit that various forms of anti-realism can endorse our preferred framework for semantics: and we can now see {\it why} our main claims about dualities are independent of scientific realism: because they involve the logic and semantics of dualities, rather than the claim that our use of such semantics is epistemically warranted (which {\it is} the main topic of this Chapter). 

Accordingly, the recent debate over scientific realism has focussed almost entirely, not on semantics, but on the epistemology of scientific theories: to which we now turn.

According to van Fraassen's influential formulation, the scientific realist says that advocacy of a scientific theory requires belief in its theoretical claims, as much as in its observational claims. Besides, the scientific realist says that we (not just the scientist, but also the philosopher or other user of science) can be cognitively optimistic about mature science. More precisely: we should believe that the well-confirmed theoretical claims of mature scientific theories are, in some sense, true, or approximately true.

Famously (notoriously!), van Fraassen's own position, constructive empiricism, is very different. Namely: advocacy of a scientific theory requires, by way of belief, only belief in the theory being empirically adequate---not in its theoretical claims. (Advocacy also involves exhibiting the theory to have other theoretical virtues.) And we (not just the scientist, but also the philosopher or other user of science) are entirely rational in {\it not} believing that the well-confirmed theoretical claims of mature scientific theories are true, or even approximately true.

The contrast, at the beginning of this Section, between observational and non-observational sentences, can be used to give a conception of empirical equivalence that will enable, in Section \ref{pud}, a discussion of {\it under-determination}.\footnote{Empirical equivalence is sometimes used as the {\it interpretative criterion of equivalence}, i.e.~the second aspect in the conception of theoretical equivalence, (B) in Section \eq{lsps}. Although this interpretative criterion is too weak for a conception of theoretical equivalence, it is an important minimal conception of equivalence.}
Thus we first discuss empirical equivalence:

\subsection{Empirical equivalence}\label{eronos}

Two theories or models are {\bf empirically equivalent} if they imply {\it the same observational sentences}, for all possible observations---present, past, and future.\footnote{This formulation follows Quine's (1975:~p.~323) syntactic construal of theory. Glymour (1970:~p.~277) holds a similar view. On van Fraassen's (1980:~p.~4) semantic construal, two theories are empirically equivalent if the empirical substructures of their models are isomorphic to each other. For a discussion, see De Haro (2020b:~pp.~94-95).}
Some logical positivists, in particular Reichenbach and Schlick, argued that empirical equivalence counts as a criterion of theoretical equivalence,\footnote{Thus they took empirical equivalence for the interpretative criterion, (B), in Section \ref{lsps}.} 
because what we really care about is the {\it empirical content} of theories. Thus so long as theories agree on the observational sentences, any differences in their non-observational sentences are no significant differences at all: we would not count them as `different' (theories are e.g.~`mere instruments to derive observable predictions').\footnote{For a critical discussion, see Sklar (1982) and Putnam (1996).}

Reichenbach went so far as to regard `logical equivalence' as interchangeable with `having the same observable facts'.\footnote{`[T]he theories compared are logically equivalent, i.e., correspond in all observable facts' (Reichenbach (1938:~p.~374)).} 
For him, only observational sentences are capable of the usual semantic analysis: only observational sentences are capable of having truth-values, while non-observational sentences are not. For the latter involve {\it decisions} or `volitional resolutions' (p.~9), of which conventions are a special class: `they represent a choice between {\it equivalent} conceptions'. And the difference between different conventions is a matter of convenience: `[t]he greater simplicity ... is really a matter of taste and economy' (p.~374). Thus there is no difference that bears a truth character between a system of coordinates at rest or in motion, or between different notions of simultaneity, or between the choice between Euclidean and non-Euclidean geometries. For different interpretations of non-Euclidean geometry (in terms of a non-Euclidean space, or in terms of an Euclidean space with additional forces or a temperature on it) amount to the ``same thing'', since observers cannot distinguish between the two interpretations. The two viewpoints are {\it equivalent descriptions}.

This conflation of `observability' and `existence' by some of the early logical positivists is now widely rejected. For empirical equivalence is {\it not a sufficient criterion} of theoretical equivalence.\footnote{The word `empirical' often has a broader connotation than the word `observable'. We here follow the philosophical jargon, where `empirical equivalence' is the two theories or models having the same set of observable statements.} 
Namely, two theories that agree on their observational sentences cannot count as the same theory if they {\it contradict one another} about unobservables.\footnote{While this argument is most persuasive for the realist, who thinks that statements about unobservables are true or false, it also matters for the anti-realist who, like van Fraassen, thinks that statements about unobservables {\it can} be true or false, and therefore requires a semantics. Thus, also for the anti-realist, duals can contradict each other: and when they do, they are not theoretically equivalent. See also the discussion in Section \ref{scirer}.} 
Therefore, our account of theoretical equivalence in the previous Chapter required full interpretative equivalence of models.

In the next Section, we discuss one of the arguments against scientific realism, from the existence of empirically equivalent theories that are theoretically inequivalent:

\subsection{The problem of under-determination}\label{pud}

The under-determination argument says that physical theories are under-determined by the empirical data, or that {\it mutually incompatible theories can be empirically equivalent} and thus equally agree with the data.\footnote{The empirical under-determination argument was put forward by Quine (1970:~p.~179; 1975:~p.~322). It should not be confused with other forms of under-determination, such as transient under-determination (see Sklar 1975, p.~380) or Quine-Duhem holism. Laudan (1990:~pp.~269-281) and Lyre (2011:~p.~236) carefully distinguish and compare empirical under-determination, Duhem-Quine holism, and Humean under-determination. {\it Transient under-determination} is under-determination of theory by evidence {\it so far}, while empirical under-determination is under-determination by {\it all possible} evidence (and so, we will call it: {\it in-principle under-determination}). In transient under-determination, the empirical basis is limited, and so it is likely that future evidence will break the under-determination. Transient under-determination, in the context of string theory (including dualities) is discussed by Dawid (2006:~p.~303), who puts forward a `structural uniqueness' thesis to undercut such under-determination, for which he gives several arguments (pp.~310-318). We will discuss the idea of `finding a successor theory behind the duals' in Section \ref{bcc}. Dawid (2007:~p.~8) cogently argues that, in the age of particle accelerators and space telescopes, we cannot limit what we take to be `empirical evidence' to just the observable realm as traditionally construed, but must accept evidence and justification from within the non-observable structure. In other words, the boundary between what is `observable' and `non-observable' is blurred (see also the discussion in Section \ref{scirer}, especially footnote \ref{observe}).\label{DawidUD}}

We can put the anti-realist challenge as follows. It appears that, in the presence of several alternatives that account for the same data, the realist doctrine, that we are justified to accept the claims of our well-confirmed scientific theories, will make us contradict ourselves, i.e.~accept the claims of incompatible theories. Or, at least, the realist doctrine prompts us to accept any out of a set of mutually incompatible claims, without a principled justification of which to accept. And, by putting mutually incompatible options in front of us, scientific theories not only fail to tell us what the world is like, but they also give us conflicting accounts.

According to Quine, two theory formulations (here, models) are under-determined if they are empirically equivalent (in the sense of Section \ref{eronos}) but theoretically inequivalent.\footnote{For Quine's own criterion of theoretical equivalence, see Section \ref{sse}.}
We will call this {\bf in-principle under-determination}.\footnote{One often-used jargon is `empirical under-determination'. However, `empirical' is a very broad term: since we are interested in semantic responses to under-determination (see footnote \ref{semUD}), and we wish to distinguish in-principle underdetermination from transient under-determination, we will use the label `in-principle'.}

There are two broad types of responses to the in-principle under-determination argument: the first is an epistemic response, that uses alternative assessment criteria to argue that, despite the empirical equivalence, one model is better supported by the evidence than the other.\footnote{Thus the epistemic grounds to distinguish two theory formulations are not limited to their empirical equivalence, understood as equivalence of their observation sentences. Laudan and Leplin (1991:~p.~465) and Laudan (1990:~276, 278) cogently argue that purported examples of under-determination are epistemically unsuccessful, because theories that have the same empirical consequences in general have different degrees of evidential support, or indeed they may differ with respect to other virtues. The same argument was, in effect, made much earlier, in the restricted context of spacetime theories, e.g.~by Glymour in (1977:~pp.~228, 235-238). Earman (1993:~pp.~19-20) distinguishes four different responses to the under-determination argument, each of which take on a different attitude towards the semantic and epistemic components of scientific realism. Both he and Norton (2003:~p.~28) defend theses to the effect that the nature of inductive inference is a complex matter that does not support a simple in-principle under-determination thesis. For a discussion of the epistemic and semantic responses to in-principle under-determination, see De Haro (2023:~pp.~119-123).\label{semUD}} 
While the epistemic response is cogent and is almost always available, so that it can solve the under-determination problem of the practising scientist, we submit that it does not do full justice to Quine's under-determination problem, which is primarily a question about the semantics of theories, i.e.~the question of the `literal reading of an ambiguous text', which we will discuss in Section \ref{ncsr}.

The semantic response to in-principle under-determination is to further analyse, or where required modify, the notions of empirical and theoretical equivalence, and thus argue that there is no problem of under-determination after all. Since empirical equivalence and theoretical equivalence ``pull in different directions'', a reduction of the number of cases of under-determination would apparently require either a strengthening of the criterion of empirical equivalence, so as to reduce the number of pairs of empirically equivalent theories (but this is implausible, since traditional empirical equivalence is often already regarded as too strong), and-or a weakening of the criterion of theoretical equivalence, so as to increase the number of pairs of theoretically equivalent theories.

Thus the question is: {\it do dualities give cases of in-principle under-determination? And, if so, is this under-determination a threat to scientific realism?} 

To answer this question, we will set aside the alternative assessment criteria and focus on the semantic response to under-determination. But we will not strengthen our conception of empirical equivalence or modify our conception of theoretical equivalence. (In fact, we will argue that dualities are not examples of under-determination: and, to allow for an as strong as possible case of under-determination, we will allow a weakening of empirical equivalence that is sometimes considered in the literature.) Rather, we will point to the existence of relevant logico-semantic relations between purported under-determined models that break the under-determination. In effect, the argument will be that theoretical inequivalence (combined with empirical equivalence) is not sufficient for vicious under-determination.

In asking this question, we are interested in {\it actual scientific cases} of under-determination, and not in imagined sceptical scenarios. Thus we agree with Stanford's (2006:~p.~17, our emphasis) thesis: 
\begin{quote}\small
[T]he critics of underdetermination have been well within their rights to demand that serious, nonskeptical, and genuinely distinct empirical equivalents to a theory {\it actually be produced} before they withhold belief in it and refusing to {\it presume that such equivalents exist when none can be identified}.\end{quote}

To complete the defence of our views on theoretical equivalence, we discuss, in the next Sections, two epistemic aspects of theoretical equivalence. 

\section{Scientific realism revisited}\label{srr}

This Section addresses how dualities, and our Schema for them, bear on scientific realism, and the possible challenge of under-determination of theory by empirical data, which is thought to be one of the main arguments against scientific realism (as already mentioned in Section \ref{contcons}). 

\subsection{Na\"ive vs.~cautious scientific realism}\label{ncsr}

Section \ref{scirer} discussed the realist's claim that theoretical statements are to be understood {\it literally}, and thus to be treated uniformly with, observational statements. 

To clarify the `cautious realism' that we---and, we submit, dualities in physics---favour, we first comment on this `literal understanding', or what van Fraassen (1980:~pp.~8-11) calls the `literally true story' (also a `literal construal of the language') that scientific theories aim to give us: since these are of course vague expressions.\\
\\
{\bf Literal vs.~na\"ive readings of theories.} The words `story' and `reading' of course remind us that scientific theories have a semantics, which we have here construed as a referential semantics. For us, the main point to emphasise is that one can give a {\it literal reading} of an {\it ambiguous text}. For an ambiguous text admits several legitimate readings, all of them compatible with the words of the text and with their legitimate meanings.

The point is of course that bare theories, even if formulated in mathematically precise language, are almost always ambiguous when applied to the empirical world, and so their interpretations are not always unique. Examples abound: see the different views on the hole argument about the correct physical interpretation of the mathematical objects that we use to formulate general relativity, especially manifolds and their points (Section \ref{holeA}). Another common example is gauge symmetries: the four-vector potential of the Maxwell theory of electromagnetism (Section \ref{EMduality}) is ambiguous because local gauge transformations can change it without changing the state of the system, and so the four-vector potential in general cannot be directly interpreted as a physical field. (And yet in other cases, a gauge transformation does change the state: e.g.~if it changes the boundary conditions at infinity.)\footnote{There is in the philosophical literature an ongoing discussion about the empirical significance of gauge symmetries. While a consensus has emerged that global gauge symmetries applied to subsystems can have direct empirical significance, there is debate whether local gauge symmetries can have direct empirical significance. While Kosso (2000:~pp.~95-95), Brading and Brown (2004:~pp.~656-567), and Healey (2009) argued that local gauge symmetries have no direct empirical significance (i.e.~only indirectly, by their connection with conservation laws through Noether's theorem), Greaves and Wallace (2014) and Teh (2016) argue that, applied to subsystems, local gauge symmetries {\it can} have a direct empirical significance (especially through Galileo's ship scenarios). For the recent discussion whether local gauge symmetries express mere redundancies of the description, or whether they can be empirically significant, see, for example, Friederich (2015:~p.~537), Murgueitio (2021:~p.~1021), Murgueitio and Teh (2020:~p.~3), and Gomes (2021:~p.~11). As we discussed in Section \ref{holeA}, there is in general relativity a class of `visible' diffeomorphisms that are analogues of the class of gauge symmetries that can have a direct empirical equivalence in gauge theories. The latter are called `non-interior, boundary-preserving symmetries' by Greaves and Wallace (2014:~p.~71), and their definition is further refined by Teh (2016:~p.~116).\label{empsym}}

In both cases, giving a `literal reading' does not amount to regimenting the theory in a preferred language (e.g.~first-order logic) and, as it were, reading this language out loud and committing oneself to the existence of the objects that the regimented terms refer to. For, even if we have decided which sentences we take to be true, i.e.~which scientific theory we take to be our best, this does not {\it automatically} lead to unique existential commitments. Regimenting the theory in such a way that we can extract our ontological commitments from it requires physical and philosophical analysis: about the nature and scope of local gauge symmetries and diffeomorphism symmetries, about the correct way to formalize quantities, etc. Constructing the kind of formalism (i.e.~the kind of regimentation) that will give us the right ontological commitments, itself involves interpretation. And even after we have found our preferred formalism, constructing a suitable semantics requires work, and ambiguities may remain. However, we maintain that our goal is to give as literal as possible readings of formalisms. For theories that are formulated as triples (see item (3) in `Theory' in Section \ref{Ourthm}), the literal reading is assigning a referential semantics to the states and the quantities, in the way that we have advocated earlier in this book (see Section \ref{ints}). Thus an interpretation map assigns to them referents in a domain of application (physical states, and physical quantities) such that the structure is preserved, and in particular such that the dynamics is satisfied. 

Our rejection of `na\"ive readings' concurs with North's (2021:~pp.~4, 218) recent emphasis that we should not `naively read off everything about the physical world directly from the mathematics' and that `there is no algorithm that takes us from a theory's formalism to its metaphysics'. Even though we are largely in agreement with North, it will help to clarify our realism, which is cautious, with North's `direct' realism. \\
\\
{\bf Cautious reading of a theory.} What we mean by the word `caution' is as follows. First, while a na\"ive reading takes the theory as given and assumes that there is a unique and, perhaps, fairly obvious semantics in terms of existential quantification in our preferred formalism (viz.~quantification over our fundamental fields, basic quantities, or what have you), a {\it cautious} reading denies that interpreting a scientific theory---even though it requires literal reading---is a matter of {\it semi-automatic} reading. 

The first point is `caution'{}'s negative meaning, viz.~telling us what not to do; and cautious scientific realism is indeed not a specific {\it kind} of realism. But `caution' has two implications for dualities and, more generally for theoretical equivalence, that, we hope, result in a more positive view, in the next three Subsections: the first implication is general, and the second more specific. Then we will contrast our view with North's (2021) `directness criterion'.

The general implication is that, when faced with dualities or similar inter-theoretic relations (including criteria for theoretical equivalence), the cautious scientific realist will naturally take these inter-theoretic relations into account when she determines her realist commitments. In other words, when faced with an inter-theoretic relation she will not jump to ontological conclusions, but first do the required logico-semantic analysis---and this is a substantive business, about which we will say more, in Sections \ref{srr} and \ref{ejpe}.

This bolsters a general view about scientific theories and models, advocated by e.g.~Belot (1998:~pp.~551-554), who holds that theories are not self-contained units, but are situated within {\it networks}: thus the interpretation (what he calls the `content') of a theory depends on its place within that network.\footnote{See also our discussion in Section \ref{mtce}. Furthermore, Hesse (2000:~p.~306-307) has advocated an `analogical conception of theories', according to which theories are historically changing entities, and consist of models that are analogues of reality, and change as new data are obtained and new models are developed. She contrasts her view with the semantic conception of theories, where models and theories are taken as static or `given' (and which we are advocating here). Hesse's conception differs from ours, since our discussion of realism here is concerned with the theoretical, rather than with the practical, functions of dualities. The themes of theory and model development, and of theory succession, will be discussed in Chapters \ref{Heuri} and \ref{Understand}. Also Ruetsche (2002:~p.~351) has advocated that the interpretation of a scientific theory should take into account how well `the theory as interpreted fit[s] with environing theories'.\label{hesse}} 
We will say that a {\it structured view} of theories, as against a flat view, is the correct view. (We will discuss a more specific `geometric view' of theories, in Chapter \ref{Heuri}.) 

Belot's discussion regards {\it transient} under-determination: he argues that, although the Maxwell theory of electromagnetism can have several possible interpretations, the Aharonov-Bohm effect suggests that the vector potential is physically real. Thus the connection of the Maxwell theory with quantum mechanics guides our interpretative stipulations away from the familiar ontology of electric and magnetic fields, towards an ontology in which the four-vector potential is physical.\footnote{At Belot (1998:~pp.~541-543, 550), he discusses two possible ontologies whose starting point is the vector potential, rather than the electric and magnetic fields, as well as their relative merits.}
As in other responses to transient under-determination, this change of ontology is elicited by the discovery of a new phenomenon (the Aharonov-Bohm effect), and is reflected in the kind of constraint on interpretation that Belot uses to choose between ontologies: namely, their practical fruitfulness: `one interpretation provides a more fruitful way of thinking about a certain range of phenomena than the other' (p.~554). 

While we endorse Belot's picture, the aspect of scientific realism that, we argue, dualities cast light on, is not (only) {\it practical fruitfulness}, but logico-semantic relations, independent of the question of new empirical evidence. We will argue that the under-determination is broken by the constitutive relations between the theory and its models, and by the deductive dependence of the induced interpretation of the common core theory on the models: dualities bear on {\it in-principle} under-determination.

The more specific implication is the flipside of this general point, and echoes the analysis in the previous Chapter. Namely: our caution, about drawing quick physical or ontological conclusions from formal inter-theoretic relations, makes us wary of any claims that some special formalisms can provide us with easy methods to decide how to individuate theories. Formalisms are needed and useful, but semantics is equally substantive. This agrees with the `directness criterion':\\
\\
{\bf A comparison with North's `directness criterion'.} North (2021) 
advocates a direct approach to realism: namely, other things being equal, we should {\it prefer the formulation that represents our represented item, or target, most directly}. 

While our cautious realism broadly agrees with this approach, it disagrees about specifics: viz.~about the use of the clause `all other things being equal'. For North's discussion of examples of equivalence emphasises the metaphysical {\it inequivalence} of models, while it seems to envisage almost no role for various forms of equivalence---at least, no role that is either noticeable or substantive. For example, she takes the Maxwell theory, formulated in terms of the electric and magnetic fields, to be formally equivalent, but metaphysically inequivalent, to the formulation in terms of the gauge field: she judges that the electric and magnetic fields give a more direct or more perspicuous representation of the nature of the target (namely, the electromagnetic field). By contrast, we have argued elsewhere that a model based on the Faraday tensor, under a suitable interpretation, is theoretically equivalent to a gauge-invariant model, where such a model is based on a {\it gauge orbit}, i.e.~an equivalence class of gauge-related gauge fields.\footnote{See De Haro (2021:~pp.~5169-5174), where we contrasted our verdicts using the isomorphism criterion with Weatherall's (2016a, 2016b) category-theoretic verdicts. (See our discussion of category theory in Section \ref{mtce}.) For the sake of comparing with the category-theoretic verdicts, the discussion of the Maxwell theory in De Haro (2021) takes the Maxwell theory ``as given'', i.e.~it is regardless of the aspects of extendability that we are here discussing, and will discuss further in Section \ref{ejpe}. A further point seems to be that the direct realist would have to decide whether the electric and magnetic fields count as metaphysically equivalent to the Faraday tensor or not, since (given the direct view's insistence on the {\it inequivalence} of representations that are commonly held to be equivalent) this is not prima facie clear. For the electric and magnetic fields do not themselves transform as tensors under changes of coordinates (they are the components of a tensor in a particular system of coordinates), while the Faraday tensor is an antisymmetric two-tensor.}

While North recognises that there can be genuine cases of theoretical equivalence, these are usually restricted to changes of coordinates (e.g.~Lagrangian mechanics stated in different sets of coordinates, or Newtonian mechanics in different sets of inertial coordinates: see North (2021:~p.~229)). Her general approach, in effect, both reduces the number of cases of theoretical equivalence and {\it understates} the logico-semantic significance of formal, and other forms of, equivalence. We are going to argue that, even in cases of theoretical inequivalence, the presence of a duality, and its correlate in the interpretation, are important for a cautious scientific realism. 

North does of course accept that, short of full equivalence, theories can be equivalent in various respects. However, judgments of equivalence aside, there is a further concern: although the direct approach applies `other things being equal', so that it officially recognises that contextual variations may affect the way in which a scientific theory is applied, these sorts of considerations do not seem to play any role in determining one's realist commitments. `Directness', as a method to determine one's realist commitments, seems to focus on the {\it possible world}  (which could of course be the actual world) described by a scientific theory, regardless of other considerations. 

By contrast, the cautious realist recognises that, among the various literal readings allowed by a model, one of them may be favoured by objective facts that are external to the model and the system that it describes: such as the {\it location of the model within a network or manifold of models}, and how a given physical system of interest {\it interacts with other physical systems}. 

This lack of concern for variations in ontology prompted by external factors is a worrying feature of the `directness criterion', because it does not make relevant distinctions such as e.g.~between internal and external interpretations, and between extendable and unextendable models (which we will do in this Chapter). By focussing on `the possible world described by the theory', in isolation from everything else, it sometimes privileges a reading of the ontology of a theory in some possible world, {\it at the expense} of an equally literal reading of the theory that is more apt as a description of the {\it actual world}. In the actual world, our system of interest interacts with other systems: and these interactions show important features of the nature of our system. 

This is further illustrated by North's (2021:~pp.~215-216) preference for the formulation of the Maxwell theory in terms of the electric and magnetic fields, over the formulation in terms of a gauge field: but the former is only usefully applicable to the Maxwell equations {\it in vacuum}. For, in general, we cannot write a Lagrangian or a Hamiltonian for the Maxwell theory coupled to classical sources (e.g.~a point particle, a spinor, etc.) other than by an appropriate coupling to the gauge field. More generally, the whole development of both classical and quantum field theory (including the examples in Chapters \ref{Advan} to \ref{EMYM}) relies on the notion of a gauge field being available (see also our approval of Belot's view on the Aharonov-Bohm effect above). 

Not taking these kinds of considerations, and their ontological implications, into account, makes North's directness criterion border on the na\"ive realism that she criticizes. The crucial point that we will argue in this Chapter and the next is that a cautious realism should take into account inter-theoretic relations of various sorts (both formal and interpretative) when we determine the theory's ontology: inter-theoretic relations, and the interactions of systems with other systems in the actual world, are necessary to correctly individuate our theories.

If the mathematics of scientific theories was a gift from Athena's hand, their interpretation was not. We need to develop it not only by imagining possible worlds, but also by considering mundane theoretical facts about the actual world.\footnote{There is here an echo of Ruetsche's (2011:~p.~3) critique of the `ideal of pristine interpretation'. However, our reasons for rejecting it are not `pragmatic', in the sense of `subjective', but principled and objective: the ideal of pristine interpretation may give us a set of possible worlds, but more work (including, in some cases, enlarging the set of possible worlds) is required to get at the actual world.}

\subsection{Under-determination for duals?}

Since dualities are a rich and scientifically active field, they could potentially give interesting examples of under-determination. As we discussed in Section \ref{comparison}, whether dualities give genuine cases of under-determination is a controverted point among philosophers of dualities. (As we are about to discuss, this is in part influenced by what we mean by `duality'.)

When duals are theoretically equivalent, as in Figure \ref{Physeq}, there is no question of under-determination, because the interpretations of the models are the same, and also the bare theory's interpretation is the same, as we see from Eq.~\eq{ii1i2} (see also the discussion in Section \ref{ica}). Under-determination is of course about theoretically {\it inequivalent} theories, and so looking for cases of under-determination goes in a direction opposite to most of our previous discussion so far, of establishing the conditions for theoretical {\it equivalence}. 

But under-determination also requires {\it empirical equivalence} (cf.~Section \ref{eronos}). Thus we have a triangle diagram analogous to the one in Figure \ref{Physeq}, where the bottom section of the diagram is only the observable part of the domain of application.

The idea is that {\it dual models are under-determined if they are empirically equivalent but theoretically inequivalent} (up to the caveat that we will discuss in the next Section). Thus external, rather than internal, interpretations are relevant. And to model theoretical inequivalence, we replace the lower triangle in Figure \ref{Physeq} by a square diagram that does not commute, as in Figure \ref{Physineq}. 
\begin{figure}
\begin{center}
\bea
\begin{array}{ccccc}&T&\\
~~~~~~~~~~~~~~~{\sm{$h_1$}}\swarrow\!\! &&\!\!\searrow{\sm{$h_2$}}~~~~~~~~~~~~~~~~
\\
~~~~~~~~~~~M_1\!\!\!\!\!\!\!\!&\xrightarrow{\makebox[.6cm]{$\sm{$d$}$}}&\!\!\!\!\!\!\!M_2~~~~~~~~~~\\
~~~~~~~~~~~~~{\sm{$i_1$}}\big\downarrow\!\!&&\!\!\big\downarrow {\sm{$i_2$}}~~~~~~~~~~~~~\\
~~~~~~~~~~~~~~~~~D_1&\not=&D_2~~~~~~~~~~~~~~~\end{array}\nonumber
\eea
\caption{\small A bare theory, $T$, and its dual models, $M_1$ and $M_2$. $d$ is the duality map from one model to the other, and $h_1$ and $h_2$ are representation maps for the models. The two models are mapped by their respective interpretation maps, $i_1$ and $i_2$, to the different domains of application, and so they are physically inequivalent.}
\label{Physineq}
\end{center}
\end{figure}

Looking back at the examples from Parts I and II, we do not find any such cases. The reason lies in the fact that under-determination requires duals that are empirically equivalent on their {\it external}, rather than internal, interpretations. And externally interpreted duals are in general {\it inequivalent}, as regards their observational statements, as well as their non-observational ones. Thus the correct diagram for externally interpretated duals is as in Figure \ref{Physineq}, for both the observational and the non-observational statements of the theory. And the diagram's lack of commutativity, at the bottom, for observational statements, blocks the under-determination argument.\footnote{This is of course to be expected: if two models are empirically equivalent, then they are not about completely different topics (as the external interpretations of several dualities appear to be: bosons versus fermions, Type IIA vs.~Type IIB strings, curved spaces vs.~gauge fields, etc.): there is already some agreement between them.}

So in order to allow for the empirical equivalence of external interpretations, so that the anti-realist can argue for a problem of under-determination as strongly as possible (indeed to allow that the argument can even get off the ground), we will be more liberal about the kinds of interpretations that we allow ourselves to use.\footnote{As we already noted in Section \ref{substd} (i), weaker notions of empirical equivalence, in the spirit of the one that we will introduce below, have been considered by Read and M\o ller-Nielsen (2020:~p.~263), Matsubara (2013:~pp.~477-479), Huggett (2017), and Huggett and W\"uthrich (2023:~Chapter 8).}

For example, in Kramers-Wannier duality, we will allow interpretation maps that exchange `high' and `low' temperature in an Ising model lattice, so that the duals come out as `both hot' or `both cold'. (Note that this is in general not even possible on an external interpretation, since a thermometer reading distinguishes hot from cold: but we can restrict ourselves to theories of the whole universe, where the measuring instruments are inside the world we are describing, as in the case of e.g.~T-duality discussed in Section \ref{SchemaT}. This then allows us to interpret a length measurement, that is normally seen as `large radius', as meaning `small radius'. Thus to allow our anti-realist opponent to make the strongest possible case, we here consider unextendable theories.)\footnote{We will argue below that extendable theories only give cases of transient underdetermination, and not of in-principle under-determination.}

The situation we require for an under-determination argument then looks like the one in Figure \ref{dtilde}. We have dual models with interpretation maps restricted to the observational statements of the models, $i^{\tn o}_1$ and $i^{\tn o}_2$ (these are the restrictions of the usual interpretation maps to the observable parts of models), and observable domains of application that are the ranges of these maps.

The key point about this square diagram is that it has an induced duality map, $\ti d$, that `co-varies' with the duality map $d$, and gives an isomorphism between the observable domains of application of the duals. This map is `induced' by the duality in the sense that it tracks terms from one model to the other using $d$, and makes their corresponding interpretations match across the observable domains of application (i.e.~$\ti d\,\circ\,i_1^{\tn o}=i_2^{\tn o}\,\circ\,d$). Of course, it is not in general guaranteed that such a map exists, since the interpretations could be very different (e.g.~the observable domains of application could happen to not be equinumerous). But to allow our anti-realist opponent to formulate their argument, we make the liberal assumption that $\ti d$ exists and that, just like the duality itself, it is an isomorphism. In effect, it interchanges `large radius' with `small radius', `electric' with `magnetic', etc. 

\begin{figure}
\begin{center}
\bea
\begin{array}{ccc}M_1&\overset{d}{\longrightarrow}&M_2\\
~~\Big\downarrow {\sm{$i^{\tn o}_1$}}&&~~\Big\downarrow {\sm{$i^{\tn o}_2$}}\\
O_1&\overset{\ti d}{\longrightarrow}&O_2
\end{array}\nonumber
\eea
\caption{\small The observable parts of the domains of application, $O_1$ and $O_2$, are isomorphic to each other. The isomorphism $\ti d$ between them is `co-varying' with the duality map, $d$.}
\label{dtilde}
\end{center}
\end{figure}

Depending on how we read the diagram in Figure \ref{dtilde}, we can make two different verdicts about empirical equivalence. The usual, i.e.~the right, reading for external interpretations, is that the two models in this diagram are in general empirically inequivalent: namely, the observable domains of application are not the same, i.e.~$O_1\not=O_2$.

But there is also a non-standard reading of the diagram, where we allow the induced duality map, $\ti d$, to enter into the definition of our interpretation maps. Thus define two {\bf non-standard interpretation maps} as follows:
\bea\label{perverse}
i^{\tn{o,p}}_1&:=&\ti d\,\circ\, i^{\tn o}_1:~~M_1\rightarrow O_2\nn
i^{\tn{o,p}}_2&:=&\ti d^{-1}\,\circ\, i^{\tn o}_2:~~M_2\rightarrow O_1\,,
\eea
where `p' is a mnemonic for `peculiar' or `perverse'. These maps ``interpret one model in terms of its dual's domain of application''.

These maps allow us to have any empirical interpretation we like. For if we take the (observable parts of) interpreted models to be $M_1$ with interpretation map $i^{\tn o}_1$, and $M_2$ with interpretation map $i^{\tn{o,p}}_2$, then these models have the same observable domain of application, $O_1$, and they are {\it empirically equivalent}. For both T-duals, the observed radius of space is {\it large}. And if we take our models to be $M_1$ with interpretation map $i^{\tn{o,p}}_1$, and $M_2$ with interpretation map $i^{\tn o}_2$, then both models have the observable domain of application $O_2$, and so the observed radius of space is always {\it small}. Thus we have succeeded in producing empirically equivalent models, and the anti-realist can now proceed with their argument.\footnote{At this point, and as we mentioned in Section \ref{comparison}, we see that, if one's notion of duality depends on a conception of `empirical equivalence', and if one is willing to allow isomorphic empirical domains to be empirically equivalent, this leads to different conclusions about under-determination. Van Fraassen (1980:~pp.~45-46) is one author who allows weakening `empirical equivalence' thus. By contrast, we argue that this can only be done by adopting non-standard interpretations (see also Butterfield (2021:~pp.~49, 55-56)).}\\
\\
But can we extend these interpretations, as is required for under-determination, so that we have {\it theoretically inequivalent} models? If we look at the examples in Parts I and II, this does {\it not} seem to be possible. For there is no remaining latitude in the interpretations: once we reinterpret the observational part of a model through a peculiar map, Eq.~\eq{perverse}, also the interpretation of the non-observational part has to be peculiar: else the interpretations of the observable and unobservable parts of the domain are inconsistent with one another (recall that we defined the interpretation maps in Figure \eq{dtilde} as {\it restrictions} of the full interpretations to the observational statements).

To give a simple explicit example of Kramers-Wannier duality (even if the Ising model is not unextendable), imagine that the thermometer (according to the standard $\b=1/k_{\tn B}T$ in Eq.~\eq{Zmod}) reads the temperature as `high', and we use $i_1^{\tn o}$ to interpret it (wisely!) as being {\it high}. Then (because of Eq.~\eq{Itemp}) the same thermometer will read the dual temperature, $\tilde\b=1/k_{\tn B}\ti T$, as `low': but, using $i_2^{\tn{o,p}}$, we interpret it (perversely!) as being {\it high}. The trouble is that this is incompatible with the non-observational terms of this dual model (where the spins of our model are unobservable to the unaided eye), because if $\ti T$ is numerically low but interpreted as high, then the dual Hamiltonian in Eq.~\eq{Savit2.9b} cannot be interpreted as the Hamiltonian of an Ising model! Thus the under-determination stumbles after all.\footnote{In this peculiar interpretation, for the Hamiltonian to be an Ising Hamiltonian, one needs to invert the temperature again, thus in effect reverting to the original Hamiltonian with temperature $T$ and spins $\{s_i\}_{i=1}^N$. In other words, the peculiar interpretation of the thermometer reading of $\ti T$, as being the same reading as that of $T$, is incompatible with the presence of the dual spins, $\{\s_i\}_{i=1}^N$, in the Hamiltonian. Thus we either write the Ising Hamiltonian in terms of $T$ and $\{s_i\}$, or in terms of $\ti T$ and $\{\s_i\}$, but we cannot mix the two sets of variables.}

Agreed: the under-determination argument gets into trouble because the observational terms of the model (namely, the dual temperature) appear in the model's dynamics mixed with non-observational terms: in this example, they both occur in the same term in the dual Hamiltonian, Eq.~\eq{Savit2.9b}. Thus if we use a peculiar interpretation for the observational terms, we must make it compatible with the interpretation of the non-observational terms, by {\it also} changing the interpretation of the latter. But if this is not possible (as we see that it is not possible in the dual Hamiltonian, Eq.~\eq{Savit2.9b}), then we get an incoherent theory: a theory without a well-defined physical semantics (or, in any case, it is not Ising model semantics). This `mixing' is more broadly seen as ruining under-determination arguments: they do not secure a clear enough separation of observational and non-observational sentences.\footnote{This is also one possible critique of the kinds of piecemeal changes, such as exchanging `electron' and `molecule', that Quine (1975:~p.~319) aims to do in his search for empirically equivalent alternatives. Permutations of words can of course be done, but the mathematical structure of the theories in Part II is such that any substantive changes of interpretation are disastrous. In other words, once a class of interpretations is fixed, the requirement of empirical equivalence turns out to be strong: it does not seem to admit alternatives. This critique echoes Lewis' (1984:~pp.~226-228; 1983:~pp.~346) refutation of Putnam's global descriptivism using causal descriptivism and natural properties.} 

The same applies to electric-magnetic duality. Assume, for the sake of the argument, a world such that charge is observable, and the field that mediates the force is unobservable (and the argument could be made with a different such division). Then, once we interpret the charge as `electric', we cannot interpret the vector field as `magnetic', because a magnetic vector field would interact with the electric charge in a way inconsistent with the Maxwell equations.\footnote{In AdS-CFT, once we have fixed our observations to be made e.g.~``on the boundary of a five-dimensional AdS spacetime'', we need to stick with the rest of the interpretation on the gravity side, and we cannot say that we have a four-dimensional non-abelian gauge field in our model: this would be inconsistent with the rest of the model.} 
The coherence between the interpretation and the model's structure requires that, once we fix the observational part of the interpretation `on one side', we fix the non-observational part `on the same side'. These difficulties are reminiscent of Quine's (1975:~p.~319) attempts to switch terms in his under-determination argument.

Thus we are back to the drawing board, looking---like Quine---for ways to tweak our non-observational terms to build coherent interpretations of dual models that are empirically equivalent, but theoretically inequivalent. On the examples from Part II, this does not seem to be possible: and so, we conclude that {\it there are no good examples of dualities that illustrate a case of under-determination} (where by `good', we mean that the examples satisfy Stanford's requirement in Section \ref{srr}). Dualities do not provide the anti-realist with ammunition. By itself, this suffices to argue that under-determination coming from dualities is {\it not a threat to scientific realism.}

\subsection{Cautious scientific realism and a structured view of theories}

What we mean by a `cautious' scientific realism (i.e.~that we must take into account the logico-semantic relations between theories and models that come with dualities) is best illustrated in contrast with a `flat' view, where such relations are not considered.

Thus, even though we have argued that there are no examples of dualities that illustrate in-principle under-determination (and indeed, somewhat against the spirit of our own endorsement of Stanford's quote in Section \ref{pud}), in this Section we temporarily assume, {\it for the sake of discussing a cautious scientific realism}, that the anti-realist can pick up the slack and present a genuine case of under-determination (perhaps there can be {\it sui generis} under-determination in a very special example, not generalizable to other dualities). From such an `analysis not substantiated by genuine examples' we will gain both further support for a structured view of theories (which the next Chapter will develop) and a view of cautious scientific realism that applies to both internal and external interpretations.

\begin{figure}
\begin{center}~~~~~~~~~
\includegraphics[height=5cm]{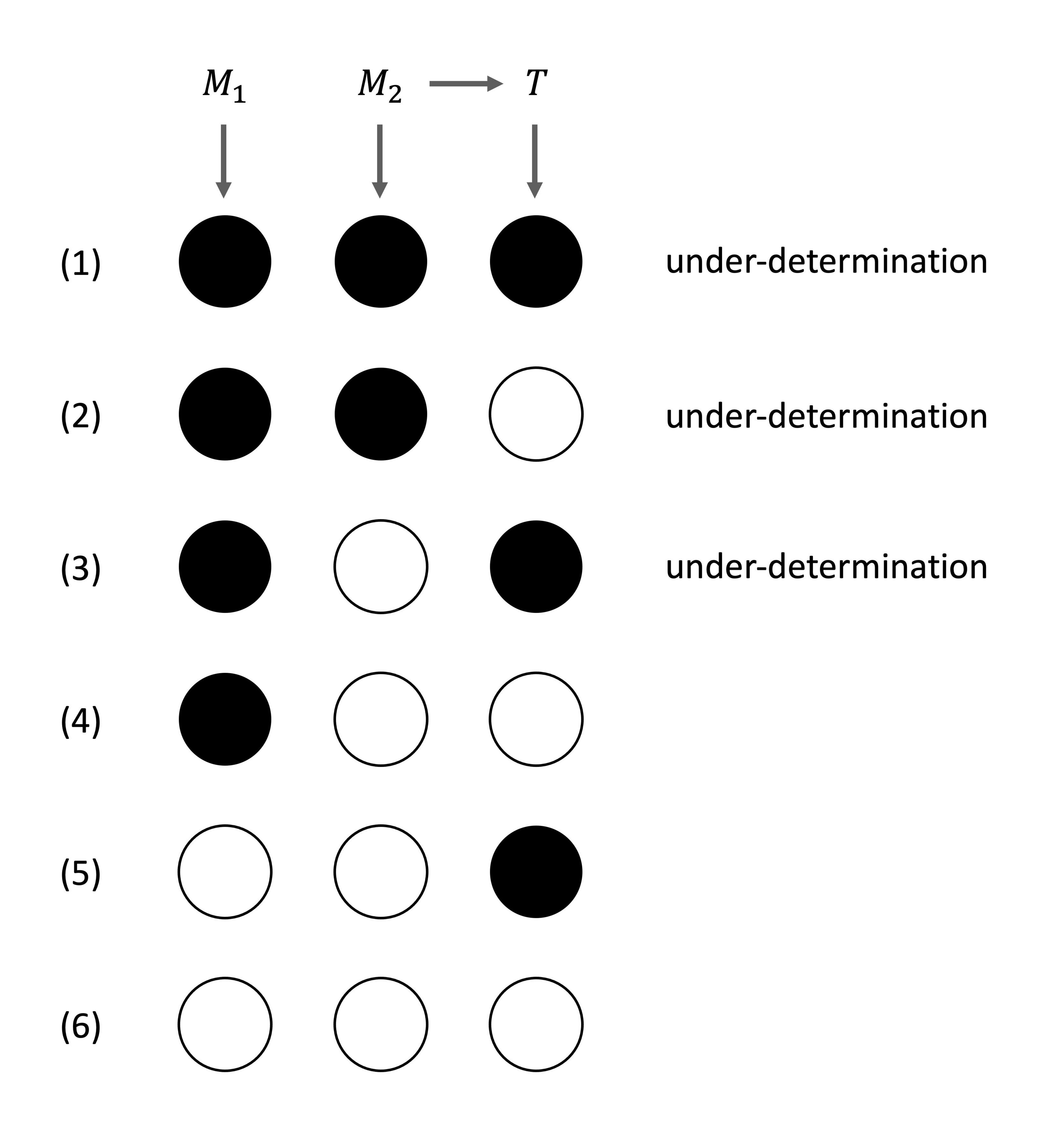}
\caption{\small Dual solutions and their common mathematical core. Such solutions are here interpreted as representing certain worlds (to which they are `isomorphic'). If a world is black, it is regarded as a legitimate candidate for being the actual world. If the world is white, it is not regarded as a legitimate candidate for being the actual world. Options (1)-(6) categorise different verdicts on which of these worlds are regarded as candidates for being the actual world. Based on Le Bihan and Read (2018:~p.~3).}
\label{UDworlds}
\end{center}
\end{figure}

Our analysis follows the discussion by Le Bihan and Read (2018:~p.~3), who have a helpful way to visualize the various options available, shown in Figure \ref{UDworlds}. $M_1$ and $M_2$ are solutions to the models that are related by the duality, and $T$ is the corresponding solution to their common core theory. (We here follow Le Bihan and Read's usage of taking the individual solutions of the theory as their models: see our discussion in `Model'-(i) of Section \ref{Ourthm}. However, this is of no consequence for our discussion.) The horizontal arrow at the top indicates that one way to construct this solution is by taking the common mathematical structure of the duals, as discussed in Section \ref{abstraction}. The black and white balloons are possible worlds described by the models, i.e.~the downward arrows are interpretation maps. The black worlds are legitimate candidates for being the actual world, while the white ones are not.

The number of distinct options is of course given by the elementary combinatorics of the possible black-and-white colourings of three worlds, of which two are symmetric, viz.~the pairs of worlds under $M_1$ and $M_2$. So there is one way to choose three black worlds, two distinct ways to choose two black worlds, two distinct ways to choose one black world, and one way to choose no black worlds. 

If a row has at least two black worlds, there are at least two models whose solutions are legitimate candidates for the actual world, and so the models are under-determined by the evidence. Thus three of the options in Figure \ref{UDworlds} are cases of under-determination.\footnote{Le Bihan and Read (2018:~pp.~6-7) discuss interesting interpretations of the other options, but we will not discuss them here, because they are not cases of under-determination.} 

Thus interpreted, it seems as if the existence of a common core theory does not only not help resolve the under-determination, but aggravates it. For of these three cases of under-determination, one stems from the addition of the common core theory in the third column (see the third row in Figure \ref{UDworlds}), and a second case, which was already under-determined before adding the third column (with two black balloons under the two models), becomes more under-determined after we add the common core theory (three black balloons in total: see the first row in Figure \ref{UDworlds}). Thus it appears that, despite the fact that the common core theory is supposed to ``express what is common to the two models'', in fact it makes matters worse, because it adds more options.

But we will argue that this appearance is misleading, and that `mere counting possibilities', without taking into account the logico-semantic relations between the common core theory and its models, gives incorrect results. Namely, the under-determination is a by-product of using a {\bf flat view} of theories and models that we argue is incorrect. For recall the entailments that we discovered in Eq.~\eq{matim}, i.e.~$M_1\,\vee\,M_2\vDash T$, which follow from the way the common core theory is related to its models: it is a more abstract theory that is true whenever either of the models is true (in other words, it is a theory whose truth conditions are weaker than those of the models). 
This eliminates the second and fourth row, because they are incompatible with the semantics of Eq.~\eq{matim}.

This leaves us with two possible cases of under-determination, namely (1) and (3). In case (1), all models are candidates for the actual world; in case (3), one of the models, and the common core, are candidates for the actual world.\footnote{Recall that we found no real examples of under-determination, since models that are theoretically inequivalent are also empirically inequivalent. We thus proceeded by allowing the anti-realist to suppose that an example, even though {\it sui generis}, might one day be found.}

We will now {\it not} argue that the under-determination is {\it eliminated} in the two remaining hypothetical cases, (1) and (3). 
Nor will we use specific arguments that e.g.~favour one model over another: which almost always allows us to undercut the under-determination.

Rather, we will argue that the kind of under-determination that we are left with is {\it benign}, and is not a threat to, a cautious scientific realism (see Section \ref{ncsr}). Thus in particular, we need to say more about cautious scientific realism: and the overall conclusion will be that {\it a cautious scientific realism takes the existence of inter-theoretic relations as part of the evidence used to select the theories that warrant belief.}

Note that although we can in principle, i.e.~given all the possible evidence, establish empirically that dual models are empirically equivalent and theoretically inequivalent, we cannot in the same way establish empirically which of the cases (1) and (3) in Figure \ref{UDworlds} reflects the fact that we cannot establish empirically whether either model, or both, describe the actual world, since that depends on unobservables. Thus we cannot know by empirical methods which case we are in. 

In such a situation, a cautious scientific realism notes that it is safe to commit only to the weaker, more general, model that is empirically equivalent to, and is fully compatible with, both duals, i.e.~the common core theory, $T$. For this theory is true in more cases than the dual models, since it is both true when the dual models are true, and in cases when the dual models are false (recall that, according to Eq.~\eq{ii1i2}, $T$ is true if either $M_1$ or $M_2$ are true).\footnote{Note that the aim here is not to find reasons to eliminate undesirable cases of under-determination, but rather to construct an epistemic argument that takes heed of the logico-semantic relations.} 

In other words, the logico-semantic relations between the common core theory and its duals, Eq.~\eq{ii1i2}, allow the cautious scientific realist to commit, not to either model, but to the logically weaker option available, $T$, that is {\it as empirically successful as} the dual models, and is isomorphic to them.\footnote{Recall that Section \ref{ica} recommended a common core theory that is isomorphic to its models as a preferred standard: which is also the case that Le Bihan and Read consider. (Furthermore, if $T$ is not isomorphic to its models, then there are significant differences that prompt a preference for either.) The condition of `empirical success' guarantees that the realist does not commit to true but uninteresting theories. For a formalization of the relation of abstraction between theories and models in a way that rules out true, but uninteresting, theories, see Fish (2023:~Ch.~5).} 
This theory does not ``stand next to'' the two duals, like a third {\it independent} model: rather, its logico-semantic dependence on (viz.~weakness compared to) the two models favours it.\footnote{In Section \ref{ica}, we noted that, given the interest in dualities, the isomorphism criterion that delivers the ``right'' level of abstraction, i.e.~$T\cong M_1\cong M_2$, is a preferred criterion. And the isomorphism $T\cong M_1\cong M_2$ is still the relevant case here, in our discussion of in-principle under-determination {\it for dual models}. For if we take the common core theory to be weaker than the dual models (and thus {\it not} isomorphic to them), so that it can represent new non-dual models, then those non-dual models are also legitimate candidates to describe the actual world, and any additional under-determination that may result from such a move in effect lies outside the discussion of a specific under-determination argument for dualities: ordinary under-determination (for non-dualities) is then what requires discussion. In other words, under-determination can be analysed in two steps: first, under-determination between duals, and then under-determination between duals and non-duals. Hence for our discussion of an under-determination argument for dualities, it suffices to take the common core theory to be isomorphic to its dual models.}

This idea is illustrated by T-duality (see Chapter \ref{String}), where the external interpretations of T-dual models ascribe different radii to target space. The common core of the models is a space with twice as many variables, i.e.~it comprises both the target space and the winding space, but it does not ascribe them independent determinate radii.

This resolves the under-determination in both cases (1) and (3) in Figure \ref{UDworlds}, because the semantic relations between the common core theory and its dual models secure that we may commit to only the common core theory. (The reason for the words `allow' and `may commit' is that, as always, choices based on semantic considerations can be over-ridden by other epistemic or metaphysical considerations. Thus, although a cautious choice is available to the scientific realist, there may be reasons for a different choice. One expects that these other reasons will be theory-dependent and not general.) 

Let us clarify this reply for case (3) in Figure \ref{UDworlds} in more detail, by answering a question that uses our earlier notation for models, $M=(m;\bar M)$ in Eq.~\eq{eqmodel}. The question is: while it is true that the first model, $M_1=(m_1;\bar M_1)$, and the common core theory are isomorphic, i.e.~$M_1\simeq T$: the model, $M_1$, differs from the common core theory, $T$, in that it has, in addition, the specific structure $\bar M_1$. And in case (3) it might just happen that the unobservable parts of $\bar M_1$ describe real features of the world, which by definition $T$ cannot describe, at least not using the additional structure $\bar M_1$. Thus there appears to be a downside to $T$'s being more abstract than its dual models: since it has less structure, it might not describe some relevant unobservable detail about the actual world. And so, $M_1$ gives a better, because more accurate, description of the unobservable parts of the world than $T$ does. (Recall that, in abstracting from the models to the common core theory, we gain in generality, but we lose in strength.)

Our reply is, first, to agree with the point of the question: it is indeed possible that $M_1$ describes the unobservable parts of the world in more detail than $T$. 

But is this under-determination a {\it threat to scientific realism}? No! First, remember that our realism is cautious. As a general strategy, it warrants the realist commitments that we make (so that we do not commit to a false model); but it does not guarantee that we always commit to the {\it best} (i.e.~the most approximately true) choice out of a set of correct and mutually compatible models.\footnote{As we emphasised above, this {\it general semantic strategy} aims to avoid errors in interpreting duals. As in any case of under-determination, there are usually theory-specific epistemic or metaphysical reasons that may prompt a different choice.}
In other words, the model that we pick in our epistemic situation, namely the common core theory $T$, is a safe choice.\footnote{Choosing $T$ as one's preferred theory of course requires that its ontology has been explicated, as in Read and M\o ller-Nielsen (2020:~p.~266) `motivationalism', which we endorsed in point (iii) of Section \ref{substd} and in Section \ref{nscte}.}
In our state of knowledge, we {\it do not make a mistake}, i.e.~are not in conflict with the evidence, or with logic or truth, by making this commitment. (The reason is the logico-semantic entailment Eq.~\eq{ii1i2}, illustrated for bosonization in Section \ref{lsrb}: our choice is {\it compatible}, also for unobservables, with other choices. We are justified in giving credence to the common core, but not to one of the models.) 

In these {\it hypothetical} examples, we may not be able to know about some aspects of the world. But remember, from the discussion in Section \ref{realism}, that the realist's claim is not that we can know {\it everything} about the world. Rather, scientific theories give us true and warranted (even if defeasible) knowledge about the world. In particular, this 
analysis overcomes the charge of the under-determination argument, described in Section \ref{pud}, that scientific realism makes us `accept the claims of {\it incompatible} theories, and that scientific theories `give us conflicting accounts of what the world is like'. Correctly taking into account the logico-semantic relations prevents this.

Second, even though the under-determination says that there may be a better description of the actual world---in the discussion above: $M_1$, or some other yet unknown dual model---the reasons why that may be a better choice are {\it sui generis}, case-specific, and not general. The difference between $M_1$ and $T$ is unobservable, and so no amount of empirical evidence will prompt a decision: but metaphysical or methodological arguments can help (see Sections \ref{realism} and \ref{semanticeq}). And such arguments will be specific not general, i.e.~they have to be worked out case by case, and weighed against the semantic argument given here. This is of course what we always do when we interpret scientific theories: no surprises here! (See e.g.~the hole argument, in Section \ref{holeA}).

Thus in the presence of the semantic relations that are implied by a duality, a cautious scientific realist is not saddled with an obligation to commit to dual models. Rather, in the various cases discussed, {\it the cautious scientific realist is always justified to commit only to the common core theory.} 
There is a safe and empirically adequate choice that the realist {\it can} make, and so there is no vicious under-determination. 

To summarize the argument: in {\it vicious} under-determination, one has a set of empirically equivalent and mutually incompatible models, and there is no warranted strategy to choose. The logico-semantic relations between duals and the common core replaces this by a set of empirically equivalent, possibly theoretically inequivalent models (in the case of external interpretations) but that have some common logical consequences that are true, or candidates for truth: namely, the common core theory. This common core theory itself has observational and non-observational sentences.

If this common core theory is rich enough that it satisfies whatever turn out to be the agreed requirements in philosophy of science, e.g.~it gives satisfactory explanations, as we have argued that common core theories can do, then the cautious scientific realist can happily live with this situation: the under-determination is {\it benign}. 

The more general lesson of this analysis is that `having a common core' can be more relevant for under-determination than the stronger `theoretical equivalence'. For, as we have seen, {\it theoretically inequivalent models can have a common core of statements that are candidates for truth}, in the sense secured by their logico-semantic entailments. And this lesson can, in principle, be applied more widely in discussions of under-determination: in particular, in historical cases of under-determination.\footnote{Extensional scientific realism is such a position: see Fish (2023) and De Haro (2020c).}

We conclude that dualities do not just bear on the individuation of theories that have theoretically {\it equivalent} models: it was sensible to expect that they did. Solving this problem is of course an important task. What is perhaps wholly {\it unexpected} is that, for theoretically {\it inequivalent} models, we can defeat the threat of under-determination by developing a structured view of theories, and a cautious realism that uses dualities to determine its warranted commitments. Indeed, dualities induce interesting logico-semantic relations between a theory and its models that bear on our ontological commitments: not just in cases of theoretical equivalence, but also in cases of theoretical inequivalence. 

\subsection{The common core ontology of duals}\label{cco}

Early work on dualities, discussed in Section \ref{comparison}, propounded that dualities undermine the usual types of scientific realism. Both Matsubara (2013:~p.~474) and Dawid (2013:~pp.~179-180) argue that, since dualities relate structurally similar, but ontologically very different, models, dualities favour structural realism over other types of realism. Dawid also has other reasons for this: he sees the notion of a physical object being eroded more broadly by modern fundamental physics. Thus in the light of dualities, he criticizes the following realist strategy:
\begin{quote}\small
[to] reduce the core characteristics of the ontological object in a way that jettisons those characteristics which can no longer be specified in a consistent way within the new theoretical framework ... Faced with duality, this strategy seems hopeless. It would amount to positing ontological objects whose defining characteristics specify neither the objects' dimensions nor their topological position ... The scientific realist would be left with a notion of ontological object that [is] virtually empty (Dawid (2013:~p.~182)).
\end{quote}
While there is some {\it prima facie} plausibility to these arguments, especially to the view that structural realism has some appeal in the face of dualities, we argue that it is no more than that: a `prima facie' appeal. For these arguments are not supported by our analysis in Part II, where the common core ontologies are rich and varied. 

We now give three arguments that both refute the idea that dualities favour structural realism, and support the idea that dualities go well with our cautious scientific realism. They are: (i) an analogy with the hole argument, (ii) an argument from our conception of theories, and (iii) an argument from the analysis of Part II.\\

(i)~~{\it The analogy with the hole argument}: Our advocacy of developing an ontology for the common core, i.e.~our endorsing motivationalism, is analogous to sophisticated substantivalism.\footnote{See our discussion of motivationalism and interpretationalism in Section \ref{contcons}. For sophisticated substantivalism, see Section \ref{responsesH}. Read (2016:~p.~224) has compared four possible responses to dualities in string theories with the responses to the hole argument. Two of them use {\it internal interpretations}, i.e.~one interprets duals as representing the same possible world. The other two use {\it external interpretations}, with different priorities (i.e.~not privileging either of the duals, or privileging one of them). We here focus on sophisticated substantivalism as the most relevant to motivationalism. Note that interpretationalism, because it infers that duals are theoretically equivalent without first requiring an explication of their common core ontology (i.e.~it regards the specific structure as not physically significant, so that the differences between duals are something like gauge redundancies), seems vulnerable to Dawid's critique that the remaining ontology, after one declares that the specific structure is `gauge', is `virtually empty'.} 
They are similar in that motivationalism requires an explication of the common core ontology: one changes one's (perhaps na\"ive) metaphysical view of the duals, and argues that they represent the same possible world. The common core ontology is not obtained just by declaring the specific structure to be not physically significant, but rather requires further analysis, and sometimes augmentation. This agrees with our cautious scientific realism, and it is not vulnerable to Dawid's critique of `virtual emptiness' (see points (ii) and (iii) below).\\

(ii)~~{\it The Schema's conception of theories and models}: First, our requirement, in Section \ref{isomdef}, that a duality has a common core that is itself a bare theory, i.e.~a triple of states, quantities, and dynamics, $T=\bra{\cal S},{\cal Q},{\cal D}\ket$, makes this theory capable of being assigned a referential semantics as a physical theory. Since not every isomorphism respects the structure of a bare theory, this is a strong requirement. The motivationalism discussed in (i) leads us to try and explicate the bare theory's ontology (as the examples in (iii) will illustrate). 

Second, our interpretative project of assigning a referential semantics (more specifically, an intensional semantics)
to bare theories and models explicitly goes beyond what a structural realist would say. As Sections \ref{refsr} and Chapter \ref{Theor} discussed, this bears on our agreement with the recent literature that the distinction between the syntactic and semantic conceptions of theories is not very firm or useful.\footnote{There are attempts at syntactic versions of structural realism: e.g.~Beni (2018) offers a syntactic informational structural realism. However, it is unclear whether, in the context of physical theories, such attempts remain true to structural realism's ambition of giving a minimal account of continuity through theory change.}
(We used the syntactic aspects explicitly in Section \ref{mtce}, where we gave a syntactic version of a duality: and in Section \ref{lsr}, where we discussed the logico-semantic relations between a theory and its models, which Section \ref{lsrr} used to give a common core for bosonization. Finally, Section \ref{theoreq} argued for a distinction between a model-theoretic semantics and a physical semantics, which was further developed in Chapter \ref{physeq}, and is incompatible with structural realism.)\\

(iii)~~{\it A rich common core ontology}: As our examples from Part II illustrate, the common core ontologies that one gets are rich, rather than empty (at least in the cases where the common core theory is, at least in part, known). In fact, they are often richer than the models themselves, because they include properties from both models.\footnote{Grimmer, Cinti and Jaksland (2024) claim that, in some cases, the common core may not be rich enough to support a realist commitment. This hypothesis was discussed earlier by Read and M\o ller-Nielsen (2020:~pp.~284-286). However, the claims by Grimmer et al.~seem to rely on a misconstrual of both the Schema and its examples. For example, they claim that the common core theory of position-momentum duality is a vector in Hilbert space. But this is incorrect, because a single vector is not a bare theory and, as Section \ref{pmd} showed, the common core of position-momentum duality is richer than a single vector. More generally, the Schema requires that the common core of duals is a triple, i.e.~a {\it bare theory} (see Section \ref{isomdef}). Thus, in this regard, there is nothing special about common core theories compared to other theories. Furthermore, our main text illustrates that none of the examples in this book lends credence to such claims. (See also our discussion in footnote \ref{richcc} in Chapter \ref{Thies} and the last paragraph of `About (i)' in Section \ref{lsr}.)}
We will now discuss several of these examples:\\
\\
(1)~~{\it Bosonization}: This is the most rigorous of the advanced examples that we discussed, i.e.~the one that is under the best control. Here, it is incorrect to argue that, because the {\it classical} bosonic and fermionic models have little in common (notably: the bosonic model contains no fermions, and the fermionic model contains no fundamental bosons!), the common core is virtually empty. For the common core theory is a quantum theory, and contains {\it both} bosonic and fermionic states, through the augmentation discussed in Eq.~\eq{expf} and at the end of Section \ref{bosoniz}: thus the structure of the quantum theory is not lost. (It is the classical underpinning of the theory that needs to be reinterpreted: see Section \ref{iob}. These two classical models are specific limits of the quantum theory, appropriate to specific states, and so it is unsurprising that {\it their} ontologies look very different: see Sections \ref{mvd} and \ref{emergence}.)\\
\\
(2)~~{\it Electric-magnetic dualities}: Also here we obtained (in Section \ref{MEMD}) a common core theory that is rich, and contains a Faraday-type tensor. In the more advanced examples, the invariant masses of BPS states depend on the electric-magnetic charges and the coupling constants, and the states can also be assigned other physical properties such as energy, angular momentum, spin, etc. 

Thus these are fairly familiar quantum field theory states, but without an assignment of `purely electric' or `purely magnetic' charge. Thus also here, it seems incorrect to argue that, because there is no distinction between `purely electric' and `purely magnetic' states, the ontology is virtually empty. Indeed, states that are normally assigned the properties `electric' and `magnetic' are {\it both} part of the spectrum (see Eq.~\eq{bps}). Instead, the correct conclusion seems to be that the ontology contains states with electric and magnetic properties, but which do not differ by their being assigned the properties `electric' or `magnetic'. (As we mentioned above, this move is analogous to sophisticated substantivalism: see Section \ref{responsesH}.) 

The clearest analogy, discussed in Section \ref{SchemaT}, is with special relativity, where there are no `purely electric' and `purely magnetic' forces, since they are related to each other by Lorentz transformations (and so being `purely electric' or `purely magnetic' depends on a choice of coordinates). In that case, one argues that underlying these coordinate-dependent phenomena there is an {\it electromagnetic field} that encompasses both, and captures what is real about them. Notice that this is not (only) a statement about the mathematical structure of the theory, but also about its interpretation using the electromagnetic field. 

On this view, the common core theory has a larger state-space than the state-space of duals with only electric or only magnetic states. `Purely electric' and `purely magnetic' properties express relational facts, relative to a particular classical limit or to a class of quantities that one is interested in. (Having said that, the relation betwen elementary and Noether charges, and between particle and soliton states, discussed in Chapters \ref{Advan} and \ref{EMYM}, does require further interpretative work. And here, one expects an analogy with bosonization.)\\
\\
(3)~~{\it AdS-CFT}: Here, we saw that the (partial) common core has a rich structure: namely, there is a set of states and quantities with specified values of masses and spins that can be assigned values for their energies, angular momenta etc. Also, the common core theory has a class of diffeomorphisms that it inherits from both duals: namely, conformal transformations. Thus a conclusion to the effect that the common core ontology is virtually empty seems unjustified. 

Having said that, there are two important points to make: (A) Section \ref{ggd} only described a small part of the renormalization group flow of a more general gauge-gravity duality: namely, the neighbourhood of the high-energy (or ultraviolet) conformal fixed point (see also Section \ref{3op}). (B) We do not have a fully non-perturbative common core theory for gauge-gravity dualities (not even at the conformal fixed point, let alone at other points), so that our programme of `explicating the ontology of the common core' is partial, and at present remains incomplete.\footnote{For a more detailed discussion of what is known about the common core ontology of AdS-CFT, see De Haro (2020a:~pp.~275-279) and De Haro, Teh and Butterfield (2016:~pp.~1395-1398).}
In other words: this is ongoing work in physics!\\
\\
(4)~~{\it String theory dualities}: As we discussed in Section \ref{SchemaT}, the interpretation of T-duality is similar to that of electric-magnetic duality. According to the common core, T-dual strings have both momentum and winding properties (as is e.g.~seen from the fact that the mass contains both kinds of excitations, see Eq.~\eq{M2}), but not as independent properties. The properties of classical space (such as having a large, or a small, radius) are only obtained by taking a semi-classical limit, where a given semi-classical world-sheet description is valid. Also S-duality, which relates e.g.~fundamental strings to Dirichlet strings (see Section \ref{S-d}), contains both kinds of states.\\
\\
Thus, in spite of the limitations of some of the more advanced dualities (especially (3) and (4)), for which there is no exact (i.e.~non-perturbative) formulation, and  which are therefore still conjectures: nevertheless, the evidence we have for them so far, as well as the evidence we have from the exact cases (1) and (2), does {\it not} vindicate a structural realist view, nor the view that the common core is virtually empty. In fact, common core theories are rich and sophisticated quantum (or classical) theories, which can be interpreted realistically. 

Chapter \ref{Heuri} will discuss the role of dualities in theory development and theory succession (and so, the possibilty that some of these common core theories are only particular limits of a deeper theory like M-theory), and also our `geometric view' of theories, which goes beyond the common core.

\section{Unextendability and epistemic warrant}\label{ejpe}

This Section takes up the important question of what warrants a verdict of theoretical equivalence, i.e.~given a pair of duals, what justifies using the diagram in Figure \ref{Physeq} in a domain like the actual empirical world, about which we have limited knowledge. 

Since epistemic warrant or justification involves issues of empirical evidence and experimentation, having such warrant is a matter of degree. Thus when using this phrase, we do not require {\it absolute certainty} about a verdict of physical equivalence or indefeasible knowledge about it (which realists like us never do: see the discussion in Section \ref{realism}). Instead, we require a condition for when we have sufficient warrant for such a verdict.

The question of warrant should be understood as follows. Given a pair of dual models, we ask whether a perspicuous interpretation gives a commuting diagram like the one in Figure \eq{Physeq}. Thus assuming that the model-theoretic semantics, i.e.~the top triangle in Figure \eq{Physeq}, is itself sound (which is a matter of mathematical physics), the question of epistemic warrant is about the bottom diagram in this Figure, i.e.~about the justification for the internal interpretation maps used. We ask: lacking perfect knowledge about the domain(s) of application, {\it what could possibly ``go wrong'' with interpreting duals as physically equivalent, i.e.~as having the same domain of application?} If we can find a satisfactory way that such interpretations ``cannot go wrong'', then this will give us a sufficient condition for judging duals to be physically equivalent. 

In the absence of perfect knowledge about the proper domain of application of two duals, one key worry is that more detailed knowledge could show that our best interpretations of two duals must {\it disagree}. Thus upon extending the models beyond their initial domain of application $D$, where they were physically equivalent, they might end up being physically inequivalent, i.e.~in need of external interpretations.\footnote{See for example Glymour's (1977:~p.~248) discussion of `gorce' and `morce' alternatives to Newton's gravitational force theory: `To test these hypotheses, the theory must be expanded still further, and in such a way as to make the universal force term [read instead: `gorce'] determinable'. Also in modern quantum field theory it is standard to regard the breakdown of field theory models at short distances as an indication of their being {\it effective field theories}, so that models that are indistinguishable above some short-distance cutoff may turn out to be inequivalent beyond the cutoff. Very different models may flow to the same infrared fixed point where they are indistinguishable, and so ascertaining their inequivalence requires a formulation that covers the whole domain. See De Haro (2020a:~pp.~273-274).}

While this is an unsurmountable problem if understood in terms of {\it certainty}, 
it is surmountable if understood in terms of warranted assertion or belief. Thus we will here give a sufficient condition that warrants a verdict of physical equivalence. It is a restriction of both the worlds and the models that we use; as follows. 

{\bf Unextendability} is the condition that the domain of application $D$ is already {\it as large and detailed as it can be}, so that the states of the models $M_1$ and $M_2$, under the given interpretations $i_1$ and $i_2$, already describe the whole world. In that case, there is no worry that an extension to some larger (including a more fine-grained) domain will reveal that the interpretations of the two models are different, because there is no such larger domain: the domain $D$ is already as large as it could be---it is already the relevant possible world (see our discussion in Section \ref{meshdi} (4)).

Contrast this conception of unextendability with {\it extendability}, which is illustrated by our standard examples of dual models that are physically inequivalent. Consider, for example, the duality between a hot and a cold Ising lattice (i.e.~Kramers-Wannier duality, in Section \ref{dualpf0}). The Ising lattice is extendable, because it is assumed to be at constant temperature, i.e.~at equilibrium with a heat bath. Thus the domain of application $D$ of the Ising model depends on a heat bath that is not itself part of $D$, and the coupling to it stands in the way of physical equivalence. (The model also admits other extensions, e.g.~by a coupling to an external magnetic field, as in Eq.~\eq{Isingb}, whose dynamics is not described by the Ising model.) The duality does {\it not} extend to the heat bath beyond $D$. Thus the extension of the interpretation maps reveals that the two temperatures (i.e.~the outputs of the interpretation maps) in this larger domain are different.

Also, recall our discussion of measurement in Section \ref{wde}, and how standards of measurement are often external to a theory. (For example, in the Ising model discussed above, a measurement of temperature uses a thermometer that is external to the Ising lattice.) 
The interaction between the system being studied and an external measurement apparatus is often represented through a coupling term. In particle-vortex duality in Section \ref{PVD}, whether a gauge field is {\it electric} or {\it magnetic} was decided from its coupling to an external current: a Noether-type coupling indicates an electric model, while a Hodge dual coupling indicates a magnetic model; and this recurred in several other examples.\footnote{See also the Seiberg-Witten theory in Section \ref{effD}, where the monopole fields are magnetic in $M$, but electric in $M'$. In these examples, whether a gauge field is electric or magnetic is {\it relative to a given standard}, given by an external current, or by the gauge group.}
The unextendability condition thus requires that such standards are {\it internal to the theory itself}, so that the relevant measurement couplings are {\it already} included in the theory and its dual models. 

Thus a first intuition behind the unextendability condition is that the theory in question, i.e.~its subject-matter, is a cosmology: a theory of the whole universe, including all of its forces and types of fields. Then the models already contain everything they should contain. 

A second intuition is that one has somehow a ``final'' theory, i.e.~the last item in a series of ever more encompassing and general models of a given type.\footnote{A strong version of this idea is advocated by Dawid (2013:~p.~131), who proposes that string theory has deep implications for scientific realism (which we discussed in Section \ref{comparison}), but who also (in part on the basis of non-empirical confirmation methods) defends realism about a final theory, understood as the only available theory that can make such a claim. We will compare unextendability and Dawid's idea of a final theory, at the end of this Section.}

Recall, from Section \ref{semanticeq}, that $W$ is the set of kinematically possible worlds---hence the word `relevant' in the unextendability condition above: $W$ contains the set of all and only worlds that are kinematically described by $T$. But also: $W$ contains as a subset the set of dynamically possible worlds, $W_{\sm{dyn}}$, i.e.~all the solutions of the theory's dynamics.\footnote{Thus, when we say that a theory $T$ `describes a set of possible worlds $W$', or that $D=W$, this implies that the dynamics ${\cal D}$ of $T$ (in the Schema's sense in Eq.~\eq{introtriple}) determines, from all the states ${\cal S}$, a subset of dynamical states that map onto the subset $W_{\tn{dyn}}\subseteq W$ (and likewise for quantities).} 

Since unextendability is a relation between theory and world, $T$'s formal properties are relevant in determining the set of possible worlds. In particular, {\it stipulated symmetries} (see Section \ref{salientstipul}) help in limiting this set (and, as we will see below, they sometimes help in determining it). For example, a theory with a stipulated Lorentz symmetry can only be extended with Lorentz-invariant states and quantities, and $W$ only contains worlds that satisfy Lorentz-invariance. (We will give more examples below, after we discuss a weakening of unextendability.)

There are two obvious ways to {\it weaken the condition of unextendability}, so that it can be used for theories with more limited domains of application. We will argue that, combined with the additional assumption of stipulated symmetries, these two weaker conditions can give sufficient warrant for verdicts of physical equivalence. 

First, we allow that $D$ is a proper subset of the relevant possible world, $D\subset W$, and that the bare theory and its domain of application can be extended in $W$, so long as the extension is such that the interpretation maps, $i_1$ and $i_2$, do not change significantly. For example, what were atoms in $D$ do not become collections of elementary particles in the extension.\footnote{Recently, Wallace (2022a, 2022b) has introduced a slighly different, but similar, notion of `subsystem recursivity', for making robust inferences on the basis of subsystems.}
In other words, the interpretation maps are {\it robust} against such extensions, so that they are already as general as they should be.\footnote{Recall, from Section \ref{wpd}, that Schr\"odinger (1926b:~pp.~58-59), in his discussion of the equivalence of matrix and wave mechanics, anticipated a version of the condition of unextendability. For him, two models that are mathematically equivalent can be said to be physically equivalent only if their domain of application, {\it even in future}, is the same.}

Second, even more weakly, we can allow that the interpretations may change (in the sense that some of the features of the referents in the domain of application may change), so long as they do so in the same way for both models. 

The robustness condition can often be secured by {\it stipulated symmetries}: which, as we discussed above, limit the types of formal extensions of a theory and of its models, as well as the types of domains that they describe.\footnote{In so far as this ontology regards both observable and unobservable entities, the discussion of unextendability and of the continuity of the ontology across extension of a domain presupposes scientific realism. Defending scientific realism in general also requires answering the pessimistic meta-induction argument. (The reply to this argument by one of us is an extensional scientific realism: see De Haro (2020c:~pp.~27-59) and Fish (2023).) Since the anti-realist like van Fraassen denies himself the epistemic warrant for his statements about unobservables, he cannot make this unextendability argument: which seems to us a drawback of epistemic anti-realism \`a la van Frassen that semantic anti-realism \`a la Reichenbach, despite its other flaws, did not have.} 
We now give examples of unextendable theories in this weaker sense.\footnote{For more examples of unextendable theories, and further details, see De Haro (2020a:~pp.~282; 2023:~Section 3.3). In the context of dualities, the condition of unextendability was foreshadowed by Huggett (2017:~p.~86).} 

One simple example of the second way is the position-momentum duality of quantum mechanics (see Section \ref{pmd}). Regardless of the precise form of the theory's Hamiltonian, we can always use either a position or a momentum representation, and the Fourier transformation secures that the two representations are dual to each other, and we may take the two versions to be physically equivalent. Thus quantum mechanics is an unextendable theory in the weaker sense, because the position and momentum representations can always be taken to be physically equivalent, even under the addition of new terms to the Hamiltonian (and so the dynamics will be different), so long as these terms preserve the linearity and unitarity of the theory; (and there will be corresponding changes in the domain of application). Quantum mechanics is unextendable in this weaker sense, because its linear and adjoint structure secure that unitary transformations (including the Fourier transformation) remain stipulated symmetries, quite independent of the specific dynamics that we consider.

While the unextendability of a theory requires that no states or quantities can be added that are candidates for interpretation in a given possible world $W$, the first weakening allows that states and quantities may be added, and that perhaps there will be changes in the dynamics. Thus also the domain of application $D$ may be extended, so long as this does not require substantial revisions of one's original domain $D$. 

By `substantial', we here mean within the admissible margins of error of one's theory. In other words, the weak unextendability argument requires that the theory's description of the domain $D$ remains valid within its envisaged validity. What is at stake here is not the theory's empirical adequacy or its giving a good description of the actual world, but rather whether a verdict of theoretical equivalence of duals is justified.\footnote{This distinguishes unextendability from some responses to the pessimistic meta-induction argument such as convergent realism, since they address different issues. (The pessimistic meta-induction, in short, argues that, since successful scientific theories previously believed to be true, were later found to be false, we have no reason to believe that our currently successful theories are true.) While convergent realism is a response to the challenge of historical theory change, especially in the face of new empirical evidence, unextendability is concerned with the domain of application of a theory's internal interpretation. It is thus not meant as a response to the pessimistic meta-induction.}
Thus what matters is whether the domain of application within the possible world of one's theory is robust against extensions, i.e.~whether one's internal interpretation remains adequate in the domain $D$, even after an extension to a larger domain. 

On the second weakening, modifications of the theory's original domain of application $D$ are allowed, so long as they are the same for both duals. This is what the linearity and unitarity of quantum mechanics secures. Although the Hamiltonian, and thus the states and quantities, may change by adding higher-order terms, these changes will not spoil the theory's internal interpretation: agreed, not in the sense that this interpretation does not change, but in the sense that, regardless of the detail of the domain of application, the Fourier transformation secures that the position and momentum representations give theoretically equivalent descriptions of the same domain of application.

In quantum field theories and string theories, symmetries such as supersymmetry, geometrical symmetries, holomorphy, etc.~are often sufficient to determine the basic features of a theory. For example, in Section \ref{N=4SYM}, we saw that the symmetries of the ${\cal N}=4$ super Yang-Mills theory suffice to determine its BPS spectrum. Also the action is determined by supersymmetry (up to $\mbox{SL}(2,\mathbb{Z})$ transformations of the coupling, which map the dual models: see Eq.~\eq{modular}). Thus this theory is unextendable in the sense that stipulating maximal supersymmetry gives us both the bare theory and its domain of application, and blocks the addition of other fields or couplings. 

The real world is not ${\cal N}=4$ supersymmetric. But, if Montonen-Olive duality is true (see Section \ref{M-O}), beings who inhabited a possible world with maximal supersymmetry would have epistemic warrant to take duals to be physically equivalent. Weak unextendability according to either the first or the second ways, together with the additional assumption of a stipulated symmetry, gives the epistemic warrant for a verdict of physical equivalence. Thus in such a world, our justification of physical equivalence would be {\it at least as strong} as our evidence for maximal symmetry.

Another putative example is string theory. Recall that the five superstring theories from Section \ref{DSTov} are distinguished, and determined, by the possible combinations of open and closed strings with supersymmetry. However, it is also believed that these five string theories are only perturbative (and thus extendable) models, related by non-perturbative dualities that point to the existence of M-theory. If the arguments for the existence and uniqueness of M-theory could be made rigorous (something that is currently out of reach!), then this might make M-theory into an unextendable theory.

In the context of string theory, unextendability is implied by Dawid's (2013:~p.~145) {\it structural uniqueness} thesis. This is the thesis that there are fundamental principles that any theory of quantum gravity is required to satisfy (e.g.~reproducing quantum field theory and general relativity in appropriate limits, gauge symmetry, and renormalizability), so that string theory is the only possible candidate. Structural uniqueness is stronger than unextendability because, while structural uniqueness can be taken to justify realism about string theory through its being the {\it final theory}, (weak) unextendability only requires that the theory cannot be extended (or, at least, that both its formalism and its ontology are robust against extensions). Thus unextendability admits that there can be alternatives.

\section{Conclusion}

This Chapter has developed two central epistemological topics that we delayed since Chapter \ref{Thies}: scientific realism, and the justification of verdicts of theoretical equivalence. Our realism is committed to a structured view of theories, that we introduced in the previous two Chapters for logico-semantic reasons, and that we will discuss in the next Chapter from the perspective of theory construction. 

After discussing realism and its rivals, we expounded the general idea of a {\it cautious scientific realism}, as a method to fix one's realist commitments, by: (i) {\it taking into account the logico-semantic relations} between a theory and its models; and (ii) securing commitments that do not lead to empirical error (logico-semantic considerations can of course be overruled by model-specific epistemic or metaphysical arguments).

Then we defended scientific realism against possible threats of in-principle under-determination. First, we noted that examples of dualities do not support the under-determination thesis, because whenever models are theoretically inequivalent, they are also empirically inequivalent. Strictly speaking, we do not find any genuine cases of under-determination.

But allowing that some {\it sui generis} example might one day be found, we saw that the relation of entailment between the common core theory and its dual models warrants a cautious scientific realist to commit to the common core: if the dual models contradict each other, there is in general no warrant to commit to one of them. 

The {\it compatibility} between the common core theory and its dual models suggests that, even if the models are mutually incompatible, we consider a structured view of theories. Namely, theories and models that are about the same topic can be theoretically {\it inequivalent and yet have a common core}. And allowing the structured view to inform the cautious realist means that there is warrant to commit to the common core theory, but in general not to one of the inequivalent duals. 

Thus we submit that, since theories and models are structured, and since there are theoretically {\it inequivalent theories with a common core}, the admission of empirically equivalent but theoretically inequivalent theories does not suffice to establish a case of vicious in-principle under-determination. Rather, the anti-realist who wishes to argue for under-determination must explore the inter-theoretic relations, and argue that the empirically equivalent models do {\it not} have a compatible common core. 

Gathering the threads from the examples in Part II, we discussed the common core ontology of duals. We argued, against what one might perhaps at first sight expect, that such ontologies are not empty but rather are rich, since they usually combine aspects of the ontologies of duals. Rather than simply deleting features from one's ontology that are not strictly common to duals, dualities require more sophisticated ways of thinking about such features.

Finally, we proposed unextendability as a sufficient condition for the epistemic justification of claims of theoretical equivalence. This condition admits a weakening, in which a theory is not strictly unextendable, but its extensions can be argued to be ``interpretation-preserving'', or at least ``duality-preserving''. A familiar example is quantum mechanics, where, regardless of the Hamiltonian, the interpretation of the variables as position or as momenta can be upheld. Weak unextendability is secured by the linearity and adjoint structure of the theory: and we argued that, also in other examples, unextendability can be secured by stipulated symmetries.

\chapter{The Geometric View of Theories}\label{Heuri} 
\markboth{\small{\textup{The Geometric View of Theories}}}{\textup{\small{The Geometric View of Theories}}}

As we explained in the preamble of Chapter \ref{Theor}: the practical functions of a duality ``look beyond'' the satisfaction of our Schema. This Chapter and the next take up four ways of ``looking beyond'': four practical functions of duality. 

The first (Section \ref{bcc}) is about guessing a theory beyond the `common core'---what we call the `heuristic function'. At present, this is the most important function of duality in the string theory programme, where a duality is seen as a clue towards a ``deeper theory'': we will dub it a `successor theory'. Section \ref{mvd} gives a specific proposal for a type of theory that goes beyond the common core: we dub it `the geometric view of theories'. On the geometric view, quasi-duals (our umbrella term, i.e.~more general than `effective duality': see Section \ref{featurerole}) are like open sets of a manifold, and quasi-duality relations are like transition functions on the overlaps between the open sets. Thus the heuristics of quasi-duals suggests that we move from thinking of a theory in terms of collections of models, to thinking of a theory as a manifold (or some other more general geometric structure).

The second way of ``looking beyond'' the Schema (Section \ref{emergence}) is emergence. `Emergence' is of course a prominent theme in philosophy of science. But in the last twenty years, it has also become a theme in string theory. For physicists discussing gauge-gravity duality often say that one of the dual models is emergent from the other. (Namely, the gravity theory is emergent from the gravity-free theory.) So we aim to bring the discussions in physics and philosophy into contact with each other. 

Section \ref{fundam} discusses a third topic: how dualities---especially dualities in string theories---bear on another notion that has been prominent in both physics and philosophy: {\it fundamentality}. The fourth (postponed until Chapter \ref{Understand}) is the role of duality in scientific understanding and explanation.

We will also discuss how the satisfaction of the Schema bears on each of the practical functions. This will illustrate how the Schema's logico-semantic analysis, and modifications and departures thereof, contributes to practical questions.

\section{Successor theories}\label{bcc}

One main heuristic function of a duality is to guess a theory ``beyond'' the bare theory: a theory of which the given duals are {\it not} isomorphic models, nor perhaps models (i.e.~precise representations) at all---they would only be some sort of approximation or idealization. This Section is devoted to this function.

\subsection{The practical function of duality: to find a successor theory}\label{2f}

Previous Chapters emphasized equivalences and inter-theoretic relations between already given models. Constructing a {\it common core} theory, $T$, out of a set of (dual) models meant finding an appropriate structure of which the models are representations, i.e.~it meant satisfying the Schema. 

But, as we saw in the examples from Part II, dualities play a more general role than this in theory development. For example, the 't Hooft-Mandelstam magnetic dual analogue of superconductivity, in Section \ref{cmds}, was essential in the application of the Higgs mechanism to the problem of confinement, i.e.~in seeing confined quarks, and the electric string between them, as the magnetic analogues (i.e.~quasi-duals) of an Abrikosov vortex. This idea led to a precise verification of monopole condensation as a mechanism for confinement, in the effective duality of the Seiberg-Witten theory (Section \ref{effD}). Dualities of course also played a major role in Polchinski's discovery of D-branes, in the M-theory conjecture (Section \ref{DSTov}), and in the AdS-CFT correspondence (Section \ref{ggd}). Finally, the counting of black hole microstates in the work of Strominger and Vafa made key use of the idea of duality (Section \ref{ocs}).

In most of these examples, the role of duality goes well beyond satisfying the Schema, since many of these examples are effective dualities, sometimes combined with analogies. Thus, to understand how dualities are used in physics, it is crucial to understand the heuristic function of dualities, i.e.~how dualities are used beyond the Schema.

The idea is that, while dualities can describe equivalent theories, and in particular the common core theory, they also help develop {\it new theories} that are not given by a common core. 

Thus while the {\bf theoretical function} of duality aims to: (1) prove a duality conjecture between already given models, and (2) develop the common core theory, $T$: the {\bf heuristic function} aims to find a {\bf successor theory}, $T_{\tn{S}}$, beyond the common core. This successor theory instantiates duality approximately (e.g.~in a limit, or for some ranges of parameters, or in some other approximation or idealization), and so there is almost always a {\it quasi-duality}:\footnote{`Quasi-duality' is our umbrella term, i.e.~more general than `effective duality': see Section \ref{featurerole}.} 
often an effective duality, i.e.~a map that is a duality only in an approximate sense, as an idealization, or in a limit (often at low energies). 

\subsection{Effective duality and the heuristic function}\label{effd}

The aim of the heuristic function is thus to find a {\it new theory} that is behind the duals. The novelty of this theory will be in the number and nature of the degrees of freedom, the dynamics, and the intepretation. 

This means that, unlike the theoretical function, where there are precise representation relations between a theory and its models, and the common core theory can in principle be found by a process of abstraction (Section \ref{abstraction}), and we can give a natural criterion for the desired logical strength of the common core (see Section \ref{ica}), in general none of this is strictly speaking true for the heuristic function. For there are no mechanical rules to find the successor theory, $T_{\tn S}$. Hence the word `heuristics', which Whewell (1847:~p.~480) aptly described as the `art of discovery'.\footnote{According to Whewell (1847:~p.~335), `speaking with strictness, an {\it Art of Discovery} is not possible ... we can give no Rules for the pursuit of truth which shall be universally and peremptorily applicable... Still, we trust it will be found that aids may be pointed out which are neither worthless nor uninstructive'. This theme, of `no strict rules, but useful aids or tools', will be important in the next Chapter, when we discuss scientific understanding.\label{whewell}} 

In our examples from Part II, we already distinguished between dualities and {\it quasi-dualities}. The first group comprises sine-Gordon-Thirring duality and bosonization, electric-magnetic duality, S-duality in ${\cal N}=4$ supersymmetric Yang-Mills theory, T-duality and S-duality in string theory, and (perhaps) AdS-CFT.\footnote{Among these examples, dualities in high-dimensional quantum field theory and string theory have only been {\it conjectured} to be dualities.\label{conj}} 

Among the quasi-dualities, including effective dualities, there are various generalizations or weakenings of the above examples: e.g.~bosonization in dimensions higher than two, S-duality in ${\cal N}=2$ SYM, more general versions of gauge-gravity duality, particle-vortex duality, and the duality between Type IIA string theory and eleven-dimensional supergravity. For example, bosonization in three and four dimensions is not an exact duality, but holds only in a limit, where the boson arises as the infrared (i.e.~low-energy) limit of a complicated fermionic system.\footnote{See Karch and Tong (2016), and Seiberg et al.~(2016).}

In these latter cases, there is {\it no duality}, i.e.~no isomorphism of model roots, although there is something similar to it: a quasi-duality. Thus in general there is also no common core theory $T$ (perhaps there is one in an approximation or limit). Instead, physicists often expect to find a successor theory, $T_{\tn S}$, often with a pair of successor models, $M_1^{\tn S}$ or $M_2^{\tn S}$, that are in general {\it not} duals, but quasi-duals. In other words, the contrast is between two triples, $(T, M_1,M_2)$ and $(T_{\tn S},M_1^{\tn S},M_2^{\tn S})$: the first triple instantiates a duality, while the second triple instantiates a quasi-duality. 

The heuristic role of the idea of duality is that it aids, and motivates, physicists in constructing this successor theory: and that if they do indeed find one, this will explain the existence of the duality or quasi-duality (see Section \ref{Mth}). 

\begin{figure}
\begin{center}
\includegraphics[height=2.5cm]{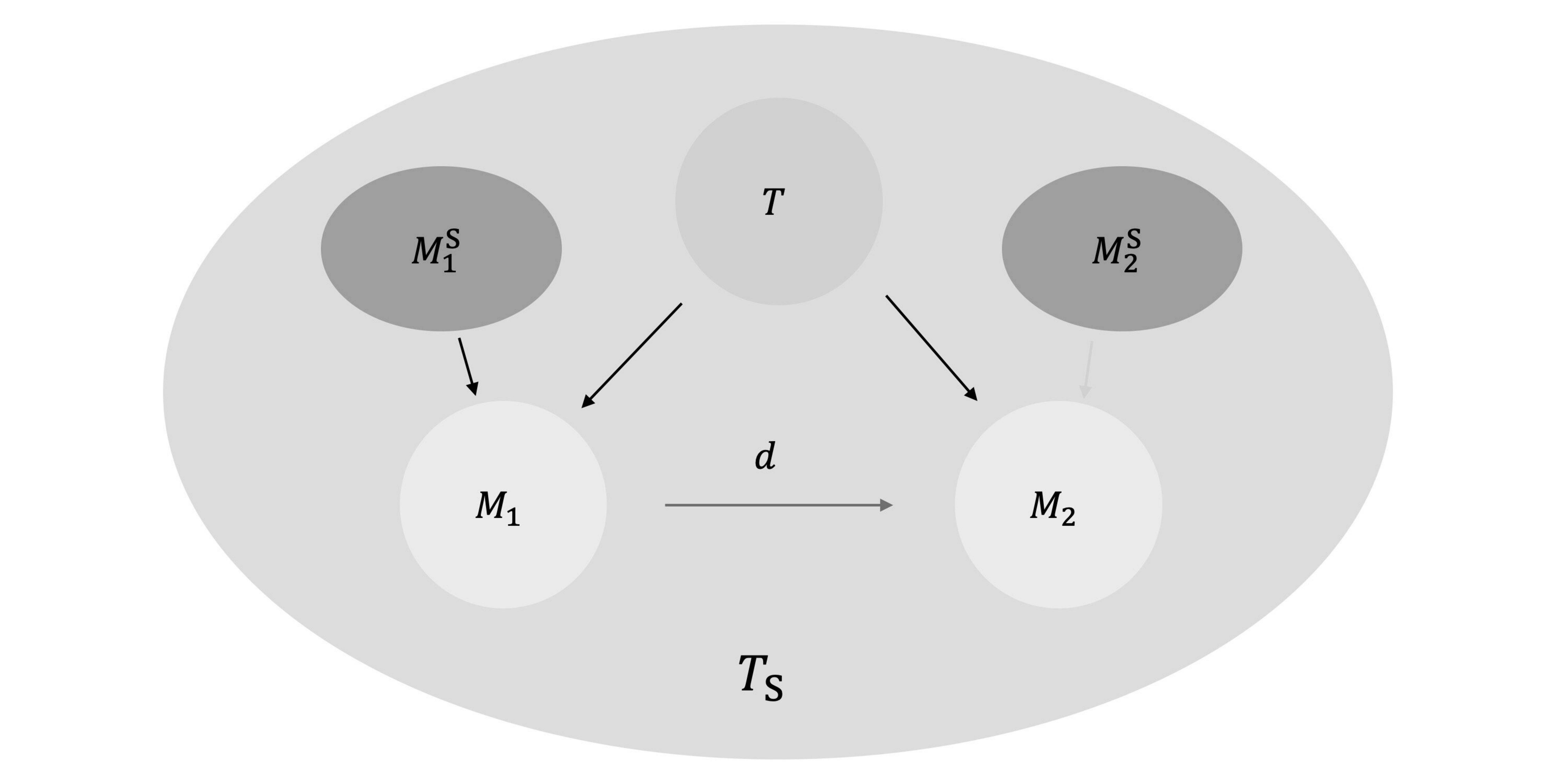}
\caption{\small Schematic indication of the relations between the successor theory, $T_{\tn S}$ and two effective duals, $M_1^{\tn S}$ and $M_2^{\tn S}$. The central triangle, i.e.~the triple $(T,M_1,M_2)$, is a common core theory and its dual models embedded into the successor theory. The dual models can be obtained from the effective duals (as particular limits, or approximations, of them).}
\label{TS}
\end{center}
\end{figure}

However, notice that the idea of a quasi-duality and a successor theory, $T_{\tn S}$, is not incompatible with the existence of a common core theory, $T$, since the above two triples can both exist---hence the above phrase, `in general'. The idea is illustrated in Figure \ref{TS}. Here, the successor theory, $T_{\tn S}$, encompasses quasi-dual models, $M_1^{\tn S}$ and $M_2^{\tn S}$, that in general do not instantiate a duality. But the figure also shows dual models, $M_1$ and $M_2$, and their common core theory, $T$, if these exist. Once the successor models are constructed, it may be possible to recover the dual models $M_1$ and $M_2$ from them (and-or from the successor theory) by some limit or approximation. (Although the figure locates $T,M_1$, and $M_2$ ``inside'' the successor theory, there is no implication that the models are strictly subsets of the theory: they might be limits, approximations to, or idealizations of, parts of the successor theory). 

Although one often thinks of the relations between the successor theory and the common core theory, and between the successor models and the dual models, in terms of limits, one should keep in mind that this relation need not be a limit in the mathematical sense: it can be taking some parameters to specific values, or a suitable approximation scheme, idealization, particularization, etc. 

With this notation, the main point is that {\it the heuristic function aims to construct the successor theory}, $T_{\tn S}\not=T$ (and $T$ does not always exist). 

\subsection{Duality and the M-theory programme}\label{Mth}

Quantum gravity and, more specifically, string theory, is the main place where finding a theory that unifies the duals is a programmatic goal.

However, there is no general agreement on the precise meaning of the word `unification' as used here. We should clarify whether, by `unifying the duals', we mean finding the common core theory, $T$, the successor theory, $T_{\tn S}$, or both. Indeed, the literature is often ambiguous about this question. And in view of the definitions of the previous Section, there are, given a {\it duality conjecture}, three positions that one can adopt.\footnote{We will not consider a sceptical position that is sceptical of, or not interested in finding, a theory ``behind the duals''.}\\
\\
(1)~~Duals (and-or quasi-duals) point only to the existence of a common core theory, $T$.\\
(2)~~Quasi-duals point only to the existence of a successor theory, $T_{\tn S}$.\footnote{Since duals by definition share a common core theory, this option is only available for quasi-dualities and not for dualities.}\\
(3)~~Duals (and-or quasi-duals) point to the existence of both a common core theory, $T$, and a successor theory behind it, $T_{\tn S}$.\\
\\
While all three positions are compatible with having (non-conclusive) evidence for a duality {\it conjecture}, one important reason for the contrast between (1) and (2) is the belief that there {\it is} a duality, as against the belief that there is {\it not} a duality, i.e.~that there is only a quasi-duality (another reason for the contrast is of course the assumption that there is only a quasi-duality, and that this only suggests a common core theory or a successor theory). Position (3) is the most inclusive position, i.e.~the one in Figure \ref{TS}. (From the above positions, position (1) takes the duality in question to only fulfil the theoretical function, while (2) and (3) (also) take it to fulfil the heuristic function).\footnote{Dawid (2006:~p.~319) emphasizes the different nature of the scientific advance produced by the discovery of a successor theory in string theory, from the process of replacement of one theory by another that is assumed in usual discussions of theory succession (and, in particular, in the context of transient under-determination). The process of finding a common core theory is `intra-theoretic', which as we have argued belongs to the theoretical function. On the other hand, finding an appropriate successor theory, $T_{\tn S}$, requires heuristics (including empirical input), i.e.~$T_{\tn S}$ is not simply determined by the duals (see Figure \ref{TS}). Nevertheless, the successor theory is usually obtained by studying effective duals, to which it is strongly analogous. The relation between the (quasi-) duals and the successor theory is one of (two-way) correspondence: see van Dongen et al.~(2020:~pp.~122-125) for the M-theory programme, and references therein for the generalized correspondence principle.} 

Although, as we mentioned, different authors take different positions,\footnote{For a survey of the literature, see De Haro (2019b:~pp.~5184-5186). Cf.~also Rickles (2013).} 
given the different kinds of dualities involved in the M-theory conjecture, our own preference, as a starting point for the investigation, is to take as our hypothesis the more inclusive position (3). This is because some of the string theory dualities are believed to both be precise dualities, at least for appropriate regimes of parameters (see footnote \ref{conj}), and suggest the existence of M-theory, i.e.~a successor theory beyond the duals. (This hypothesis also lends itself better to discussing all the aspects of successor theories: but nothing in what follows will depend on this choice.)

For example, T-duality between Type IIA and Type IIB string theories is widely believed to be a perturbative duality, order by order in $\alpha'$ (see the explanation following Eq.~\eq{WS2}; for T-duality in general, see Section \ref{T-d}). Likewise for the S-duality of Type IIB string theory (see Section \ref{S-d}). Also some of the basic AdS-CFT cases are believed, by many authors, to be dualities (see Section \ref{ggd}). 

On the other hand, M-theory is likely to be a successor theory. Also its interpretation seems very different. For while Type IIA string theory (which is one of the limiting models of this theory, valid at small string coupling, cf.~Eq.~\eq{R10}) is a theory of strings, and eleven-dimensional supergravity (another model of it, valid at strong coupling but low energies, i.e.~in the point-particle limit) is a theory of gravity with a four-form field strength and solitonic solutions, Chapter \ref{STII} discussed that the elementary objects of M-theory, away from these two limits, may yet be different. They may be elementary or solitonic membranes or fivebranes, and they may also be D0-branes. 

\section{The geometric view of theories}\label{mvd}

As we announced in Section \ref{refsr}-(i), and in the examples in Chapter \ref{Simple} and in Part II (especially in Section \ref{SWmanifold}), there is a more specific view on dualities and quasi-dualities, about the form of the successor theory, $T_{\tn S}$, and the kind of description that the various duals (if there is a non-trivial duality group, more than two) give of it. We will call this view {\bf the geometric view of theories}. 

This view motivates the depiction of M-theory in Figure \ref{Mthfig}, and the most detailed example that we have discussed is the Seiberg-Witten theory, in Section \ref{effD}. Below we will discuss Kramers-Wannier duality as another example.\footnote{Something like this view expressed in: Schwarz (1997:~p.~1) and Dijkgraaf (1997:~pp.~8-9).}
Since this view holds for both dualities and quasi-dualities, it is more general than our Schema: so that, for most dualities, the relations between a common core and its duals are special cases of the geometric view.

We will dub the theory that we get by ``putting together the quasi-duals'' a {\bf comprehensive theory}, $T_{\sm c}$. Then, at the end of this Section, we will argue that, in most cases, the comprehensive theory either is itself a successor theory, or that it points to one (especially, by being the low-energy theory of a successor theory, about which it provides information). Section \ref{basicM} first gives the geometric view, Section \ref{dep} illustrates it in examples from physics, and Section \ref{ctccsc} discusses the relation between the comprehensive theory, the common core, and successor theories.

\subsection{The basic ideas of the geometric view}\label{basicM}

In this Section, after expounding the main idea of the geometric view, we will discuss possible generalizations of the geometric view, and compare with the semantic conception of theories.

On the {\bf geometric view of theories}, a comprehensive theory (often, but not always, at low energies) can be described as a {\it geometric space}, in particular as a manifold, of dimension $n$, such that each dual model, $M_i$ (in particular its set of states ${\cal S}_i$, on which the quantities take values), is like an open set, such that the union of all the (quasi-)duals is the manifold i.e.~the comprehensive theory, $T_{\sm c}$ (in whatever regime this theory is defined: at low energies, or at any energy). 

\begin{figure}
\begin{center}
\includegraphics[height=5cm]{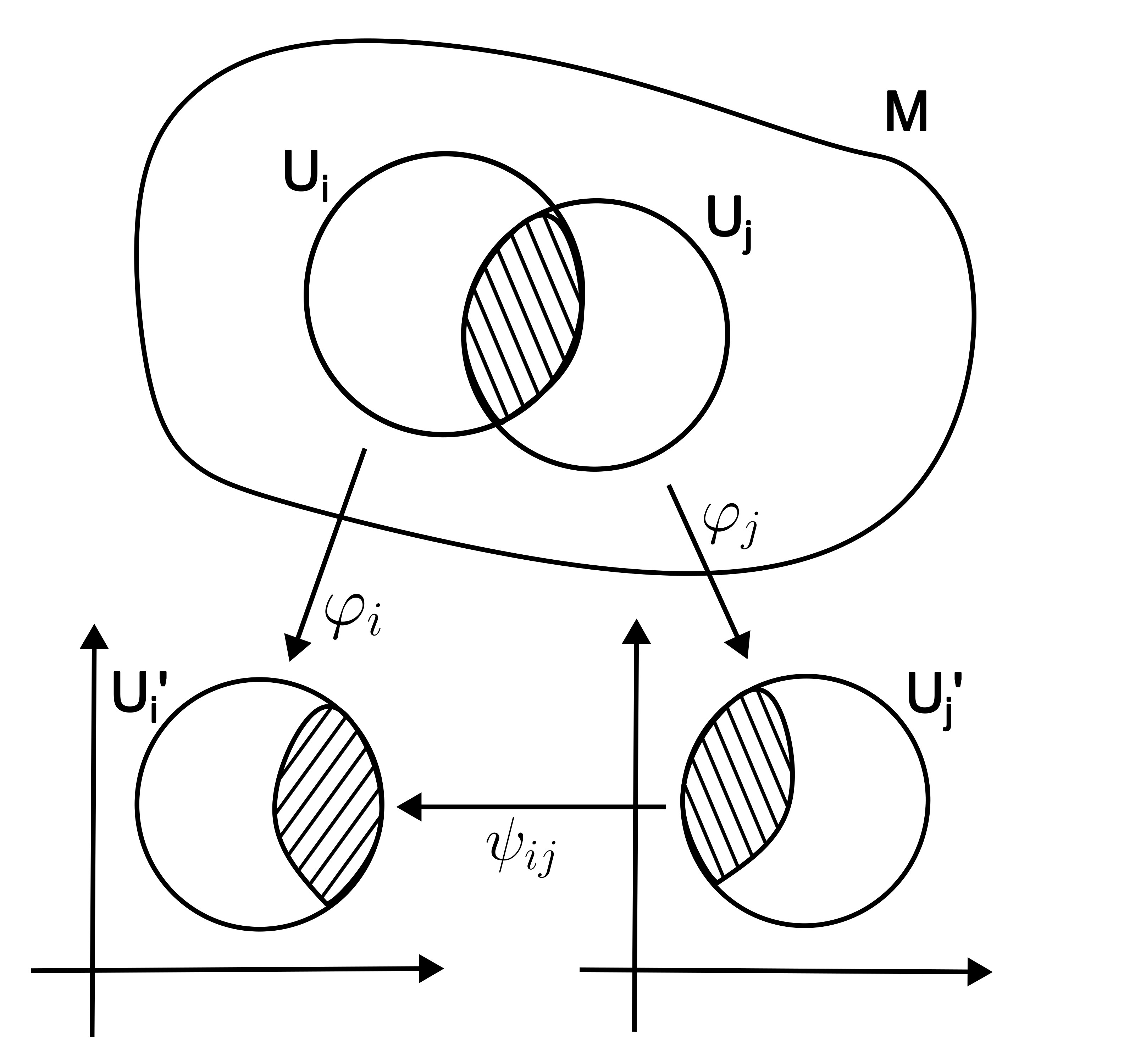}
\caption{\small The geometric view of theories. A chart $(M_i,\f_i)$ is a pair of a model and a modulus (usually the vacuum expectation value of a suitable operator), valid in a limited region of the manifold, i.e.~the comprehensive theory.}
\label{manifoldview}
\end{center}
\end{figure}

For each $i$ there is a homeomorphism, $\f_i$, from the open set ${\cal S}_i$, to an open subset of $\mathbb{R}^n$. Each of the $\f_i$'s is an appropriate (sub)set of physical quantities or variables (usually, operators) in the model. These quantities or variables form a distinguished set called {\bf moduli}. They are distinguished because, by analogy with order parameters in statistical mechanics, they distinguish the various phases of the theory, i.e.~equivalence classes of states with the same qualitative behaviour: see the discussion in Section \ref{SWmanifold}. Thus each model, with its moduli, $(M_i,\f_i)$, is like a chart. Where models overlap, a duality or a quasi-duality maps one set of coordinates into the other, and thus one model into the other: see Figure \ref{manifoldview}.

We have used the phrase `is like', rather than `is', because in general there is only an analogy between the comprehensive theory and a manifold. More precisely, the doctrine is that a comprehensive theory is often a differentiable manifold, and in all cases known to us, something like a manifold. Thus we first discuss the basic case of a manifold: at the end of this Section, we will discuss possible generalizations.

To make the state space into a manifold, three additional conditions are required: (a) The moduli $\f_i$ must be ``sufficiently varied'', i.e.~there are at least $n$ independent moduli on each open set, so that the Euclidean spaces have dimension $n$. (b) The union and intersection of two state spaces (with the quantities defined on them) must again be state spaces. (c) The quasi-duality map must be bijective. These conditions are indeed satisfied, in our two ``running examples'' in this Section: namely, the Seiberg-Witten theory and the Ising model: and, in what follows, we will take moduli to be a distinguished set of quantities that satisfy these conditions.\footnote{Thus the analogy with a manifold is strengthened by noting that, if the (quasi-)duality map is bijective, then it {\it is} a transition function. For, given two states of the two models, $s\in {\cal S}_i$, $s'\in {\cal S}_j$, related by $s'=d_{ij}(s)$ (where $d_{ij}$ is a (quasi-)duality map, $d_{ij}:{\cal S}_i\rightarrow {\cal S}_j$), we have, at the overlaps (by matching of quantities: see Section \ref{isomdef} (i)): $\f_i(s)=\f_j(s')$, and so (temporarily assuming that $\f_j$ is invertible, as in point (c) above): $d_{ij}=\f_j^{-1}\,\circ\,\f_i$, so that the (quasi-)duality map {\it is} indeed a transition function between two open sets.}

As we have here described them, the state spaces ${\cal S}_i$ of our models that, together, make up the moduli space, are abstract, and it is useful to describe them using coordinates: such coordinates are provided by the distinguished variables that are the moduli. In the examples from physics, the moduli space can sometimes only be described at low energies, or in the thermodynamic limit. 

For example, in the thermodynamic limit of the Ising model, the partition function and the free energy only depend on {\it macroscopic variables}, such as the reduced temperature $t$ and perhaps the external magnetic field: other quantities are also insensitive to the individual spin states, and their behaviour depends on a macroscopic distance on the lattice (cf.~Eq.~\eq{Mtilde}). To allow for such cases, the points on the moduli space may be labelled by variables, such as temperature, coupling and distance, that describe the macroscopic features of the system. The order parameters, such as the magnetisation and the dual magnetisation, i.e.~$M(t)$ and $\ti M(t)$, are then functions of these variables that indicate the macroscopic state (i.e.~the phase) of the system. Taking $\tau_{\tn{order}}$ to be the ordered or ferromagnetic phase (i.e.~the interval of the real variable $t$ for which the temperature is below the critical temperature), and $\tau_{\tn{disorder}}$ to be the disordered or paramagnetic phase (i.e.~the interval of the real variable $t$ for which the temperature is above the critical temperature), the moduli space of the Ising model from Section \ref{dualpf0} is the union of the two real-line segments: ${\cal M}_{\tn{Ising}}=\t_{\tn{order}}\cup\t_{\tn{disorder}}$. And these regions are coordinatized by the magnetisation (below the critical temperature) and by its dual (above the critical temperature). For more details, see Section \ref{dep}. 

That coordinates on the moduli space play a role as order parameters means that, in the low-energy limit, they indicate the {\it semi-classical state} that solves the theory's low-energy equations of motion, including the quantum corrections (namely, the order parameters minimize the quantum potential), i.e.~the theory's {\it vacua}. The semi-classical state thus depends, through its moduli, on the theory's parameters and boundary conditions, which are usually determined by the momentum and energy scales of the particles and values of background fields used in the laboratory, and by other parameters such as the temperature. (In string theory, the moduli also depend on the parameters of the internal dimensions on which the theory is compactified.)

As we discussed above, the topology of the manifold is given by the models' state-spaces and by the moduli, which are a subset of the physical quantities, i.e.~homeomorphisms between the state spaces and Euclidean spaces (see also the `two main premisses that we endorse', in Section \ref{illH}, about the quantities: namely, their (i) respecting the structure of the state-space, and (ii) being ``sufficiently varied''). 

The manifold usually comes endowed with differentiable, symplectic, metric, and perhaps other, structure.\footnote{For a formulation of classical mechanics in the Hamiltonian formalism, in terms of a symplectic manifold with an algebra of quantities on it, where the algebraic structure is given by the Poisson bracket, see Wald (1994:~pp.~11-14). For more details and examples, see Arnold (1989:~Chapter 8). In the Seiberg-Witten theory, the symplectic structure of the moduli space is given by the Seiberg-Witten differential: see Section \ref{effD}.} 
Depending on the formulation and the type of theory, this additional structure is induced from:

(A)~~For {\it quantum theories}: the (Wilsonian quantum effective) action, e.g.~Eq.~\eq{Seff} in the Seiberg-Witten theory, and its associated pre-potential: alternatively, by the generating functional of connected correlation functions (see Section \ref{matchpf} for gauge-gravity duality); 

(B)~~For {\it statistical mechanical theories}: the free energy, e.g.~Eq.~\eq{sinhb} for the Ising model: alternatively, the Hamiltonian and its associated Boltzmann weights and the partition function, i.e.~Eqs.~\eq{isingH} and \eq{ising0}.

If the theory is formulated such that the state space ${\cal S}$ is just the space of moduli (as in the Seiberg-Witten theory), then the manifold i.e.~the union of all $(M_i,\f_i)$, is the {\it moduli space} of the theory (see Sections \ref{swt} and \ref{SWmanifold}).\footnote{A distinction is sometimes made between the classical moduli space (whose coordinates are the set of fields that satisfy the equations of motion) and the quantum moduli space (whose coordinates are the set of fields that satisfy the quantum equations of motion, as given by the quantum (effective) action). Our discussion here focusses on the latter.}
The geometry of the moduli space (e.g.~its metric and complex structure) encodes the information about the {\it quantum states}.\footnote{In the Seiberg-Witten theory, these geometric quantities only contain part of the information such as the electric and magnetic charges. To encode more refined information such as the spin of the BPS states, other geometric quantities are required.}
However, this need not be the case in general. (For example, in T-duality i.e.~for a string theory on a circle, the moduli space depends on the radius of the circle and other parameters, but the whole state space also depends on the string modes in the nine non-compactified dimensions, which are not part of the moduli space.)

Curiel (2014:~p.~276) notes a familiar feature of state-spaces in classical mechanics: they `always have the structure of a ... differentiable manifold'. While our approach builds on this feature, it differs from it in two ways: first, recall, from Section \ref{ThisB}, that we lifted our usage of `theory' and `model' ``one level up'' (see the discussion of this point below). Second, our proposal purports to give an accurate description of systems other than deterministic classical systems, i.e.~quantum theories and statistical mechanical systems (see the cases (A) and (B) discussed above).\\
\\
{\bf Generalizations of the geometric view:} We expect that our presentation of physical theories that have (quasi-)duals as manifolds can, and perhaps even should, be generalized to more general geometrical structures, especially in the context of string theory. For in string theory, moduli spaces are often algebraic varieties (an algebraic variety already appears in the Ising model for complex values of the temperature: cf.~footnote \ref{algebraicV} below). 

Thus we propose the geometric view of theories in the same undogmatic spirit in which we introduced our Schema for dualities in Section \ref{giantS}. We admit that it may be improved, but we argue that it is appropriate for us to begin with the simplest possible description, rather than postulating more advanced mathematical structures without worked-out examples from science. We are indeed sanguine that the geometric view has the great advantage of concreteness and connection to scientific practice. For it is straightforward, and it is illustrated by examples from physics, even elementary ones, that we developed in Parts I and II, as well as by many other examples in quantum field theory and string theory.\footnote{For a discussion of moduli spaces in string theory, see Nelson (1987:~p.~343); for moduli spaces in topological quantum field theories, see Montano and Sonnenschein (1989:~pp.~350-352).}

Here, we also wish to echo Lehmkuhl's (2017:~pp.~1, 9-10) two complementary attitudes towards developing a `theory of theories':\footnote{Indeed, Lehmkuhl (2017:~pp.~2) has advocated to develop `a theory of spacetime theories', although in a sense different from ours.} 
(a) He stresses the difficulty of the task, and hence the importance of going `step by step' (`It seems a daunting and impossible task', p.~1). (b) This task, even though incomplete, is important (`we cannot afford not to take it on', p.~1), and small steps can be significant (pp.~9-10).

It is crucial to note that the geometric view is not an alternative to our description of theories in Section \ref{Ourthm} as structured triples of state-space, quantities, and dynamics, since it is {\it compatible} with it. Indeed, the geometric view provides a theory, defined as a triple, with additional structure, through its two main requirements: (a) for quasi-duals, the state-spaces and algebras of all the models taken together form a geometric space, often a differentiable manifold; (b) a subset of the physical quantities are moduli that describe some appropriate regime of parameters.\\
\\
{\bf Comparison with the semantic conception of theories.} Our view of theories as geometric objects, such as manifolds, vindicates the shift we made in Section \ref{ThisB}---as our comparison between the Schema and the semantic conception of theories in Section \ref{mtce} already did---of our usage of both `theory' and `model' ``one level up''. For our proposal is that a comprehensive theory of quasi-duals is a manifold with open sets that are models: and so, by shifting up our usage of both `theory' and `model', we avoid saying that `a theory is a manifold of which the open sets are other theories'.

Thus the geometric view of theories is a generalization of the semantic conception of theories, where the (quasi-)duals do not simply sit together as independent elements of a set, but form a geometric object such as a manifold, or an object close to it. This vindicates the structured view discussed in Section \ref{mtce}, as against the flat model-theoretic view (especially in Eq.~\eq{Mmodels}). The comprehensive theory, $T_{\sm c}$, is itself a manifold with a topology, and each dual describes an open subset.\footnote{Dawid (2017:~pp.~27-28) seems to have something like this view in mind when he writes that `[r]ather than talking about different theories, it seems more accurate to talk about different dual perspectives on one theory ... in the case of string dualities, taking all empirically equivalent theories into account is essential for acquiring an understanding of the full theory'. In turn, Dawid's recommendation to talk about `different dual perspectives on one theory' vindicates our lifting the use of `model' ``one level up'', and calling the comprehensive theory a `theory'. See also Rickles (2013:~p.~65).}

In other words, we can still maintain our notion of a `theory', as a triple of state-space, set of quantities, and dynamics (cf.~(3) of `Theory' of Section \ref{Ourthm}). But the state-space is topologically non-trivial, and its global structure is given by a distinguished subset of quantities. 

As we will further discuss in Section \ref{emergence}, the geometric view allows us to regard the Euclidean spaces $\mathbb{R}^n$, where the theory's quantities take values, as `model spaces' (see Lee (2018:~pp.~55, 143)), i.e.~as structures that give our theory, with its abstract sets of states and quantities, a model-theoretic semantics (in the sense of Section \ref{lsps}). The interpretation of these quantities, especially the moduli, in a domain of application, further endow the theory with a physical semantics.\\

Given the idea of a moduli space in statistical mechanics, quantum field theories and string theory, one gets the topological and geometric structure ``for free''. For as we have explained, the topological structure is given by the open sets i.e.~the state spaces of the models with quantities on them, with coordinatizations given by the (expectation) values of quantities, and transition functions that are quasi-duality relations. As we discussed in (A) above, for quantum field theories, the geometric structure, especially the metric, is given by the theory's Wilsonian effective action. 

\subsection{Illustrating the geometric view}\label{dep}

In this Section, we briefly discuss how two of our running examples instantiate the geometric view: namely, the Kramers-Wannier duality of the Ising model, and Seiberg-Witten theory. Of these two, the first is simpler and more familiar, but the second is closer to the Schema's formulation of theories, because expressed as a triple of states, quantities, and dynamics.\footnote{In the case of gauge-gravity dualities, the radial variable of the bulk space can be seen as a coordinate in moduli space, which is equipped with an anti-de Sitter metric. For an exact AdS-CFT duality, the common core theory has two different semi-classical limits, each of which is described by one dual, in an appropriate regime of parameters (see Section \ref{motgg}). For further discussion, see Section \ref{3op}.}\\
\\
{\it The thermodynamic limit of the Ising model.} As we discussed in Section \ref{dualpf0}, in the limit $N\rightarrow\infty$ the (simplest case of the) Ising model has two phases, with a phase transition at the critical temperature $T_{\sm c}$. Thus the relevant moduli space is a line, parametrized by the reduced temperature $t=(T-T_{\sm c})/T_{\sm c}$, with two (semi-)open segments.

(1)~~The {\it ordered or ferromagnetic phase} with broken macroscopic symmetry, where all the spins over a macroscopic region have the same orientation, below the critical temperature, i.e.~$T<T_{\sm c}$ or in the interval $t\in[-1,0)$ (since $T$ is bounded below by 0). We will denote this interval by $\t_{\tn{order}}$.

(2)~~The {\it disordered or paramagnetic phase} with macroscopic $\mathbb{Z}_2$ symmetry (or, for spins that can take any values, rotational symmetry), i.e.~there is no preferred direction, and $T>T_{\sm c}$ or $t\in(0,\infty)$. We will denote this interval by $\t_{\tn{disorder}}$.

Thus in the simplest, i.e.~square-lattice and isotropic, Ising model, the moduli space is one-dimensional and disconnected:\footnote{One can make the moduli space {\it connected} by considering the analytic continuation of the Ising model to complex values of the temperature, i.e.~by considering the two-dimensional space of complex values of $t$ (an algebraic variety): even though in two dimensions one can analytically continue the quantities, $t=0$ is still a singular point of that space (see Matveev and Shrock (1995a)). The convergence of the low- and high-temperature series expansions of the susceptibility are discussed at pp.~1552, 1567 and pp.~1576, respectively. There are striking and helpful analogies between the behaviour of the Ising model for complex values of $t$ and the Seiberg-Witten theory in the two-dimensional space of values of complex values of the Higgs scalar fields that, in our view, have not been sufficiently emphasised in the quantum field theory literature. Matveev and Shrock (1995b) is a generalization to other lattices: at p.~5251, it gives the equation for the algebraic curve in the isotropic case, which shows that the two-dimensional moduli space is topologically much more non-trivial than the one-dimensional case.\label{algebraicV}} 
it is the union of the two intervals, i.e.~the line segments ${\cal M}_{\tn{Ising}}=\t_{\tn{order}}\cup\t_{\tn{disorder}}$. The point $t=0$ is {\it not} part of this manifold, because it is a singular point where (in the thermodynamic limit that we are considering) physical quantities such as the specific heat are not well-defined.\footnote{We thank Silvester Borsboom for a discussion of this point.}

\begin{figure}
\begin{center}
\includegraphics[height=2cm]{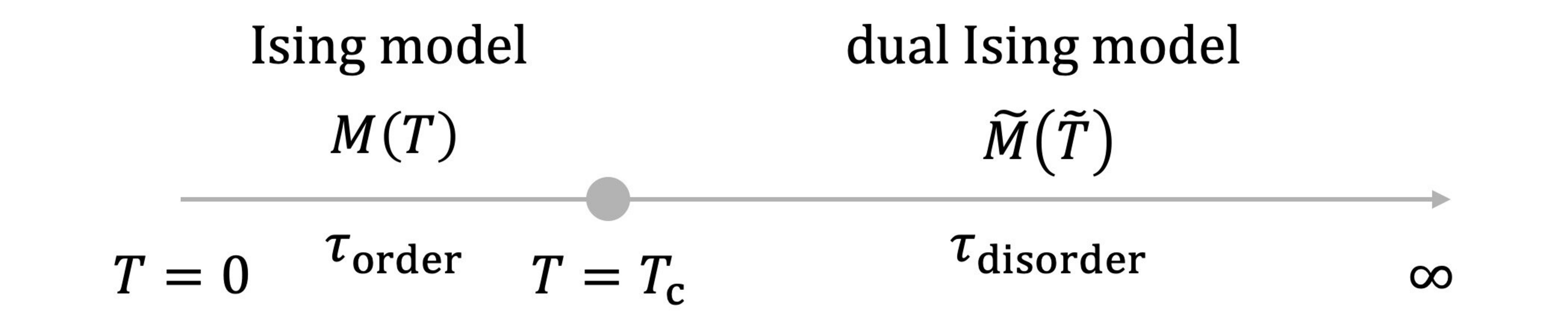}
\caption{\small The one-dimensional parameter space of the Ising model as a union of two (semi-) open intervals, and two quantities i.e.~homeomorphisms to $\mathbb{R}$ on each interval.}
\label{IsingManif}
\end{center}
\end{figure}

As Figure \ref{DisorderP} shows (see also Eqs.~\eq{Mt} and \eq{Mtilde}), the magnetisation $M(t)$ and dual magnetisation $\ti M(t)$ are order parameters that are non-zero only in their corresponding phases, i.e.~$M(t)\not=0$ iff $t\in\t_{\tn{order}}$, and $\ti M(t)\not=0$ iff $t\in\t_{\tn{disorder}}$. They are good order parameters because they satisfy the three conditions (a) to (c) from the previous Section: (a) the moduli space is one-dimensional, and so it is enough to consider, on each line segment, a single real quantity: $M:\t_{\tn{order}}\rightarrow\mathbb{R}$ in the ferromagnetic phase, and its dual, $\ti M:\t_{\tn{disorder}}\rightarrow\mathbb{R}$ in the paramagnetic phase: see Figure \ref{IsingManif}. (b) The union of the two intervals is indeed the whole state-space, and there is no intersection because the state-space is disconnected. (c) It follows from the expression for the magnetisation, Eq.~\eq{Mt}, and from the corresponding expression for the dual magnetisation (see the comment following Eq.~\eq{Mtilde}) that the magnetisation and its dual are bijective functions, and so that they are good order parameters.

Since Kramers-Wannier duality is a duality, one model is mapped into the other under the duality transformation. But this is independent of the manifold description, since {\it two} quantities are required to describe the whole moduli space, regardless of which model we use: in other words, the moduli space is topologically non-trivial.\\
\\
{\it The Seiberg-Witten theory.} Here, the expectation value of the (square of the) Higgs field is a gauge-invariant order parameter that gives a local coordinate on an open set of the moduli space (see Eq.~\eq{Higgsvev}). The moduli space is a two-dimensional space, namely the complex plane with three punctures. Thus two open sets (i.e.~two models) are required to cover the whole manifold, and each modulus is a map from an open set into $\mathbb{C}$. (As we discussed in Section \ref{effD}, we use, for reasons of mathematical convergence, three models rather than two.) The moduli space comes equipped with a metric (see Eq.~\eq{SWmetric}), and each of the three local models, $M_i$ ($i=1,2,3$), is characterized by a prepotential (see Eq.~\eq{Fsum}); and, as we discussed, the coefficients of this prepotential differ in each of the open sets, but the prepotential can be transformed from one region to another.\footnote{Thus the prepotential is not a function, but a multi-valued section, on the moduli space: see Lerche (1998:~pp.~176, 179).} 
The coordinate transformation from one open set to another is the quasi-duality map, i.e.~an $\mbox{SL}(2,\mathbb{Z})$ transformation, Eq.~\eq{dualHiggs}.

\subsection{Comparing comprehensive theories with common core and successor theories}\label{ctccsc}

In this Section, we first discuss the relation between a comprehensive and a common core theory, by considering dualities as ``minimal cases'' of the geometric view. Then we discuss the sense in which a comprehensive theory can be a successor theory.\footnote{This Section has benefitted from discussions with Nick Huggett and Richard Dawid.}\\
\\
{\bf Dualities as ``minimal cases'' of the geometric view}. Dualities are ``minimal cases'' in the sense that, since they are isomorphisms between whole models i.e.~between all the states and quantities, the transition functions are well-defined on the {\it whole open covering}: thus the open sets are isomorphic. Thus for dualities, each model covers the whole manifold that is the common core theory. However, unlike in cases where we begin with a moduli space with open sets whose union covers the whole manifold, for general duals it is not guaranteed that each dual is homeomorphic to $\mathbb{R}^n$, and so work is still required to find an open covering and to satisfy the conditions (a) to (c) at the beginning of Section \ref{basicM}: in particular, to find a set of moduli that is sufficiently varied. Thus some work still needs to be done, which we will not pursue here.

The point is that the topology of the manifold may be very involved for a theory at high energies, where a large number of open sets is required to cover the whole manifold. On the other hand, in the thermodynamic limit, or at low energies, we expect this number to decrease, because of the aggregation of microscopic states into classes, i.e.~macroscopic states or phases that are less discerning. So here, fewer open sets are required; viz.~two in both the Ising model and in the simplest Seiberg-Witten theory.

For example, in our example of elementary quantum mechanics from Section \ref{pmd}, the Hilbert space is infinite dimensional, and there may not be much gain in making the Hilbert space into a Hilbert manifold, i.e.~in constructing a manifold in which each point has a neighbourhood that is homeomorphic to the Hilbert space, rather than working with the Hilbert space itself.

We will take this point in our stride, i.e.~as part of what the previous Section called the `undogmatic spirit' in which we propose the geometric view of theories. Namely, for theories with a single, {\it infinite-dimensional, state-space}, there may be no practical gain in making the state-space into a manifold; and in that case we may work directly with the state-space and the algebra on it. For, as we said in the previous Section, the geometric view is not an alternative to our description of {\it theories} as triples, but rather specifies structure that characterises quasi-dualities and thereby generalises the Schema for dualities.

Having said that, we can go ahead and check, in our examples from Chapters \ref{Simple} and \ref{Advan}, the main point that we need for our comparison between dualities and the geometric view. Namely, we have a single open set for duals, so that the state-space is the whole manifold (or whatever structure comes to replace it in more general cases), and we can compare the differences between the comprehensive theory $T_{\sm c}$ and the common core theory $T$. We will argue that, in these cases, either $T=T_{\sm c}$ or $T\subset T_{\sm c}$, where in the latter case the comprehensive theory adds to the common core a notion of a (semi-)classical limit (there is also a special case $T_{\sm c}\subset T$, for reasons that we will explain).\footnote{Two comments about the way in which the relation $T=T_{\tn c}$ should be understood: (i) Given dual models, $T$ is defined up to isomorphism, and so all that we can require is that the two are isomorphic, i.e.~$T\simeq T_{\tn c}$. However, this will not play any role, and we can simply assume that $T$ and $T_{\tn c}$ are ``formulated in the same way''. (ii) In terms of their sets of states and quantities, the two theories are indeed the same. The equality sign requires that $T$ has also been equipped with the structure of a manifold, i.e.~that its duals are its open sets, the duality is the transition function, and that a distinguished (sub)set of quantities plays the role of coordinates. Since this is precisely the structure that a duality is defined to preserve (see Section \ref{isomdef}), it is natural to define the common core theory in this way. In other words, the common core theory $T$ stands for the bare theory together with its dual representations.\label{cciso}}

(i) In the case of {\it position-momentum duality} in elementary quantum mechanics from Section \ref{pmd}, already discussed above, the spaces $L^2(\mathbb{R}^3)$ of wave-functions in the position or the momentum representation are each a representation of the Hilbert space, with a ``choice of basis'' i.e.~a choice of a spectral family of projectors. Thus all the states and all the operators are in each of the models and in the common core theory, so that $T=T_{\sm c}$. 

(ii) For {\it bosonization}, as e.g.~formulated in Section \ref{lsrb}, the specific structure (i.e.~the two axioms $B6$ and $F6$) adds, to the common core, the notion of a classical limit (alternatively, a classical starting point for quantization that delivers the common core as a quantum theory). Thus, although these axioms do not introduce new states of quantities, they do introduce additional structure: namely, they take the form of {\it bridge laws} that allow a reduction of a specific classical limit to the quantum theory (`reduction' is here taken in the sense of Section \ref{eandr}, below). This means that $T\subset T_{\sm c}$, where: (a) the common core $T$ is the full quantum theory, without a notion of a classical limit, (b) the comprehensive theory $T_{\sm c}$, in addition, specifies two classical limits (so that the additional structure required includes e.g.~a notion of convergence). 

(iii) For {\it electric-magnetic duality}, discussed in Section \ref{EMduality}, the state-spaces of the models are again representations of the whole state-space, with the specific structure being a choice of electric and magnetic components of the state-space (alternatively, a choice of representation of the Lorentz group, a choice of Hodge star, or a choice of complex structure: cf.~Section \ref{FHodge}). These differences can be encoded in the homeomorphisms, i.e.~in how the state-space is mapped to $\mathbb{R}^n$. Thus this is also a case of $T=T_{\sm c}$. 

On the other hand, in the {\it Lagrangian formulation} of electric-magnetic duality from Section \ref{MEMD}, the common core is strictly larger than the comprehensive theory, i.e.~$T_{\sm c}\subset T$. This difference concerns only the set of kinematically possible states and not the set of dynamically possible states (which is indeed isomorphic to the set of dynamically possible states of the duals). For, as we discussed, the common core theory uses more variables than the two duals (namely, an additional field is introduced as a Lagrange multiplier into the set of kinematically possible states). 

This was to be expected, since, as we already discussed, the use of a Lagrange multiplier means that the set of kinematically possible states of the common core is strictly larger than that of its models, i.e.~the common core is not isomorphic to its dual models. As a consequence, the common core is also strictly larger than the comprehensive theory. Thus a comparison between the common core and the comprehensive theory presupposes that the common core has the `desired logical strength' discussed in Section \ref{ica}. We will assume this from now on.\\
\\
{\bf Comprehensive theories as successor theories.} We have so far remained neutral about whether a comprehensive theory is a successor theory (in the sense of Section \ref{bcc}). Here, we should distinguish between duals and quasi-duals.

For duals, as we discussed above, the comprehensive theory $T_{\sm c}$ is either equal to, or it contains, the common core theory $T$ (assuming that the latter has the `desired logical strength'). So we distinguish these two cases: (a) If $T=T_{\sm c}$, the comprehensive theory is not a successor theory of the common core, since it is equal to the common core. (b) If $T\subset T_{\sm c}$, then $T_{\sm c}$ says things that $T$ does not say, and it is a successor theory. However, since the extra things that $T_{\sm c}$ says might be consistent with $T$, not contradicting it, this would not necessarily be a successor in the sense of a `corrector', but in the sense of its saying more (e.g.~giving more detail: see the example (ii) above).

For quasi-duals, we argue that we can regard the comprehensive theory itself as a successor theory, in our sense from Section \ref{bcc}: namely, a theory ``beyond'' the common core (if there is a precise common core in some limit). Notice that the quasi-duals are not isomorphic models, or models i.e.~precise representations, of this successor theory, at all. And the comprehensive theory deserves the name `successor theory', relative to both the quasi-duals and the common core (if there is one in some limit), because it is more widely applicable, i.e.~applicable to more situations. For, without the global information, i.e.~without the information about the quasi-dualities, we do not know which quasi-dual we should use. Indeed, the global information is required to be able to use the theory for all the states in the state-space.

Thus there is one first straightforward sense in which the comprehensive theory `succeeds' the quasi-duals: viz.~absent the global information, the quasi-duals are regarded as disconnected `theories'. By contrast, with the global information supplied, one regards the duals as `local models', i.e.~open subsets, of the comprehensive theory: so that, as `theories', they are superseded. Thus the comprehensive theory succeeds the models in the literal sense that it takes over their role of theories, because it describes more cases.

Second, in some cases, the comprehensive theory may be not just a low-energy theory, or a theory defined in the thermodynamic limit. Indeed, by putting together information about all the states, one may be able to derive information about the high-energy limit. In the Seiberg-Witten theory, putting together the information about the states of the various duals one can reconstruct the BPS spectrum of ${\cal N}=2$ SYM theory, as in Figure \ref{N=4N=2spectrum}. 

\section{Emergence}\label{emergence}

`Emergence' is (like `equivalence', in Section \ref{realism}) a term of art. Indeed, how best to define it and how it relates to other notions in philosophy of science, such as reduction, is probably more disputed than are the corresponding questions about theoretical equivalence, not least because the latter is more often considered a formal notion. 

But on anyone's usage, there is a obvious contrast between emergence, and duality. Namely: emergence is obviously an asymmetric relation---if A is emergent from B then B is not emergent form A---while duality {\it is} a symmetric relation. Nevertheless, the notions used in our account of the interpretation of theories (especially Section \ref{itm}), and of duality (especially Section \ref{isomdef}), give us the wherewithal for an account of emergence (Section \ref{Sem}). One main initial step towards this is to take the items A and B related by emergence, to be theories (in the usual general sense, not our ``above dual models" sense) rather than entities. Thus we will talk of one theory being emergent from another. And we will adopt a widespread mnemonic labelling: we will label  the emergent theory the `top theory',  written $T_{\sm t}$, and the  theory from which it emerges the `bottom theory', written $T_{\sm b}$.\footnote{Another mnemonic is `tainted, tangible' and `better, basic', respectively.} 
Then we will discuss how the Schema's logico-semantic analysis bears on emergence (Sections \ref{eandr} and \ref{dande}), and the options available to combine emergence and duality---thus dissolving the contrast between these two notions (Section \ref{3op}). 

As a general conception of emergence, we will take the following:\footnote{See Butterfield (2011a:~p.~921), who also takes the relata of emergence to be theories, labelled $T_b$ and $T_t$. This account is echoed by recent discussions of emergence in the philosophy of physics literature, which takes emergence to be a ``delicate balance'' between dependence, or rootedness, and independence, or autonomy. See, for example, Crowther (2016:~pp.~42, 50), Humphreys (2016:~p.~26), Bedau (1997:~p.~375), Bedau and Humphreys (2008:~p.~1), and De Haro (2019a:~p.~7).}
\begin{quote}\small
I shall take emergence to mean: properties or behaviour of a system which are {\it novel} ... relative to some appropriate comparison class. Here `novel' means something like: `not definable from the comparison class', and maybe `showing features (maybe striking ones) absent from the comparison class'.\footnote{Butterfield also requires that emergence be {\it robust}: a feature that is indeed present in emergent behaviour, but which we will here not require for emergence. For a discussion, see De Haro (2019a:~pp.~10, 18).}
\end{quote}

There is a traditional distinction, between epistemic and ontological emergence, depending on whether `novelty' is construed as ontological (i.e.~in the world) or as epistemic (i.e.~in our theoretical representations of the world).\footnote{We follow here the account of the different ``dimensions'' of emergence by Guay and Sartenaer (2016:~p.~299).} 
We will notice that these two aspects, viz.~`the world' and `our theoretical representations of the world', align with the two aspects of scientific theories that we have distinguished in previous Chapters: namely, the domain of application, and the bare theory (with its interpretation map). Another relevant distinction often made in the literature is between weak and strong emergence, i.e.~`in-practice' vs.~`in-principle' emergence (seen as a contrast of degree within a given type, e.g.~epistemic vs.~ontological emergence). 

Our main topic in this Section is ontological emergence, in relation to dualities. We will postpone epistemic emergence until Section \ref{3op} (3). 

\subsection{The Schema's conception of ontological emergence}\label{Sem}

\begin{figure}
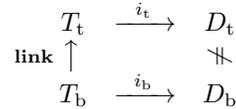

\begin{center}
\bea
\begin{array}{ccccc}
~~~~~~~~~~~~~~~~~T_{\sm t}
&\xrightarrow{\makebox[.6cm]{$\sm{$i_{\tn t}$}$}}
&D_{\sm t}~~~~~~~~~~~~~~~~\\
~~~~~~~~~~{\sm{${\bf link}$}}~\big\uparrow\!\!
&
&\rotatebox[origin=c]{-90}{$\not=$}~~~~~~~~~~~~~~~~\\
~~~~~~~~~~~~~~~~~T_{\sm b}&\xrightarrow{\makebox[.6cm]{$\sm{$i_{\tn b}$}$}}
&D_{\sm b}~~~~~~~~~~~~~~~~\end{array}\nonumber
\eea
\caption{\small Ontological emergence. The failure of interpretation and linkage to commute ($i_{\sm b}\not=i_{\sm t}\circ\,\mbox{link}$) gives different interpretations, with different domains of application, $D_{\sm b}\not=D_{\sm t}$.}
\label{Oemergence}
\end{center}
\end{figure}

There is a straightforward conception of {\bf ontological emergence} that the Schema suggests.\footnote{This conception of emergence was introduced in De Haro (2019a), which works out several examples.} 
This conception combines two distinctive features of emergence in Butterfield's account: novelty and an asymmetric linkage of levels or scales. As we mentioned above, in ontological emergence, novelty is construed as being `in the world'. Thus in the Schema's account of scientific theories, ontological novelty is construed as being in the domain of application, and not in the bare theory. In this Subsection, we will first present these two features; then we will illustrate them with the example of spontaneous magnetisation; and finally, we will return to our geometric view of theories.\\
\\
\begin{figure}
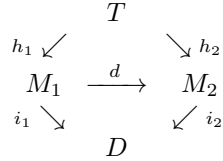

\begin{center}
\bea
\begin{array}{ccccc}&T&\\
~~~~~~~~~~~~~{\sm{$h_1$}}\swarrow\!\! &&\!\!\searrow{\sm{$h_2$}}~~~~~~~~~~~~~~
\\
~~~~~~~~~~~M_1\!\!\!\!\!\!\!\!&\xrightarrow{\makebox[.6cm]{$\sm{$d$}$}}&\!\!\!\!\!\!\!M_2~~~~~~~~~~\\
~~~~~~~~~~~~~{\sm{$i_1$}}\searrow\!\!&&\!\!\swarrow {\sm{$i_2$}}~~~~~~~~~~~~~\\
&D&\end{array}\nonumber
\eea
\caption{\small Interpretative commuting diagram (this is again Figure \ref{Physeq} from Chapter \ref{physeq}).}
\label{Physeq14}
\end{center}
\end{figure}
First, this conception of {\bf novelty} can be made precise by comparing two interpretative diagrams: the commuting one in Figure \ref{Physeq14} and the non-commuting one in Figure \ref{Oemergence}.\\
 
Note first that in the commuting diagram, there is no novelty in the domain of application (i.e.~the case that is of interest for theoretical equivalence), because the domains of the two models are the same. On the other hand, in the non-commuting diagram, the two domains are different (both in the sense that there is some difference between them, and in the sense that one domain of application is not a proper subdomain of the other). We have novelty in the domain of application, or ontological novelty.

More precisely, recall our endorsement of intensional semantics in Section \ref{itm}, where interpretations have an intensional and an extensional component (see Section \ref{dualint}). Thus the diagram in Figure \ref{Oemergence} has an intensional and an extensional variant. The condition of {\it ontological novelty}, that the linkage and the interpretation maps ``do not mesh'', i.e.~do not form a commuting diagram, should be understood as the condition that the linkage map does not commute with either the map that is the intension {\it or} with the map that is the extension.\footnote{In classical referential semantics, extensions are determined by intensions, circumstances of evaluation, and context of utterance. See our endorsement of this framework in Section \ref{itm}.}
In other words, the bottom and the top theories $T_{\sm b}$ and $T_{\sm t}$ differ in the (features of the) systems that they describe. Thus ontological emergence can obtain between theories that have the same extension but different intensions, and between theories that have both different extensions and intensions. In the latter case, one domain of application has {\it referential novelty} compared to the other. 

This condition of referential novelty can be understood as an {\bf idealization}, in the following sense. The top theory describes a possible world (in particular, the states describe a system or state of affairs) that is different from the possible world or system described by the bottom theory. Thus both the bottom and the top theory are, in this sense, idealizations.\footnote{For the motivation for taking ontological novelty (as against e.g.~irreducibility) as the mark of emergence, see De Haro (2019a:~p.~7). For further discussion of ontological emergence as novel reference, see (ibid, pp.~10-19). Different realist ontologies are discussed in (ibid, pp.~25-26). See also Norton's (2012:~p.~209) discussion of the difference between idealisations and non-idealising approximations, which is the mark of novel reference.}

This condition for ontological emergence should be understood as a {\it semantic} or interpretative, and not as an epistemic or metaphysical, condition for emergence (see the discussion of this distinction in Section \ref{realism}). Thus there are further questions, that we will postpone until Sections \ref{fundam} and Chapter \ref{Understand}, as to whether: (i) one is a realist about the bottom theory, the top theory, or both; (ii) whether the target system that one wishes to describe is more adequately, or more exactly, described by the bottom, or by the top, theory. (In answering these questions, the cautious scientific realist takes into account any logico-semantic constraints, of the type discussed in Section \ref{lsr}, that are relevant.)\footnote{These logico-semantic constraints are indeed the starting point of an extensional realist view: see De Haro (2020c) and Fish (2023).}\\ 

Second, ontological emergence requires the idea of {\bf asymmetric linkage} between the levels or scales, where `asymmetric' means the relation is both antisymmetric and irreflexive. This is an additional requirement, because ontological novelty does not give us an asymmetric relation: it is simply non-symmetric.\footnote{In the context of weak epistemic emergence (Section \ref{3op}) the asymmetry between the levels will be realized in a different way, and thus we will be able to weaken this condition on the linkage map.\label{weakEE}} 
An antisymmetric relation between the domains would of course be given if e.g.~one domain of application is microscopic and the other is macroscopic (as measured by e.g.~the typical size of objects or some other length scale), or one domain has a finite number of components, while the other has a much larger (perhaps infinite) number of components. And, according to our principle of univocality of the interpretation of physical theories (Section \ref{lessons}), these numerical and physical relations should have a formal correlate in the bare theory: e.g.~in the number of states or in the values of quantities. 

Thus the idea is to realize the linkage of the levels or scales, and its asymmetry, by replacing the duality map in Figure \ref{noncomm} by an appropriate asymmetric map between the `bottom' and `top' theories or models, i.e.~$T_{\sm b}$ and $T_{\sm t}$, respectively. We label this linkage map as follows:
\begin{center}
 ${\bf link}: T_{\sm b}\rightarrow T_{\sm t}$ is a {\it non-injective, partial map, that is a partial homomorphism}.\footnote{We only require that the map be {\it partial}, because in some examples, some states of the bottom theory do not approximate any states of the top theory.  For example, both loop quantum gravity and causal set theory (two research programmes in quantum gravity) postulate many discrete states in their bottom theory that do not approximate any  classical spacetime (so such states are in effect non-geometric). For more details on the linkage map, see De Haro (2019a:~p.~9). As for other maps between theories that we have discussed, like duality maps, $\mbox{\bf link}$ is a {\it pair} of maps, one on states and one on quantities.}
\end{center}
This map gives an ordering between the theories or models, and it is illustrated in Figure \ref{Oemergence}.

The map's being {\it non-injective} formalizes the notion of coarse-graining, as we go from the bottom to the top theory, i.e.~from a ``more detailed'' to a ``less detailed'' theory, where several ``microscopic'' states are mapped onto a single ``macroscopic'' state. For example, real-space renormalization of an Ising lattice defines block variables in the top theory, in terms of spin variables in the bottom theory, by a linkage map.\footnote{For a detailed account, see Butterfield and Bouatta (2011:~pp.~17-19). See also Fisher (1998:~p.~667) and Binney et al.~(1992:~pp.~115-120).}
(Section \ref{3op} will discuss linkage through renormalization of quantum field theories.)\footnote{See also Section \ref{effD} and Butterfield (2014:~pp.~25-28, 34-26) for a discussion of low-energy effective quantum field theories. The latter reference also discusses renormalization as a case of Nagelian reduction (see Section \ref{sse}, and the next Section).}

Recall, from the description of emergence in the preamble to Section \ref{emergence}, that novelty is defined relative to some appropriate comparison class. As we will discuss below, the {\bf comparison class} is usually a family of related theories, i.e.~a set $\{T_{\sm b}(x)\}_{x<x_0}$ of distinct theories of the same type, labelled by a parameter or set of parameters, $x$.\footnote{This then makes for {\it robust} emergence: for a discussion and details, see De Haro (2019a:~pp.~18-19).}
Given a comparison class thus presented, we can think of non-injectivity as ``losing information'' as follows.

We take an (in principle arbitrary) representative of the comparison class, i.e.~$T_{\sm b}(x)$, as the domain of our linkage map that is mapped onto the top theory. Thus presented, the comparison class is a sequence labelled by $x<x_0$, where $x_0$ labels the limit of the sequence, and the limit is in the top theory: $\lim_{x\rightarrow x_0}T_{\sm b}(x)\subseteq T_{\sm t}$. If the limit {\it is} the top theory, then we have a case of {\it reduction}, for which see the next Section; here, we allow that the top theory may have additional states or quantities that are not obtained by taking the limit. 

Thus the top theory is the codomain of the linkage map, through its containing the limit of the sequence. The top theory at $x=x_0$ no longer has the information about the particular representative of the sequence that was picked as bottom theory.\footnote{The (set of) parameter(s) $x$ can be discrete or continuous, and $x_0$ can be either finite or infinite. To avoid introducing additional notation, we are here making a simplification. Strictly speaking, the non-injectivity that we are now discussing is the injectivity of a more general map $\mbox{\bf link}'$, whose domain is a whole class of theories labelled by the parameter $x$, rather than a single bottom theory: hence our speaking of the `representative of the class'. But we will not need to explicitly distinguish between our official definition of $\mbox{\bf link}$ and its generalization $\mbox{\bf link}'$, because the latter is independent of the representative that we pick, and so we can think of it simply as a map with a single bottom theory on its domain, as we did in the main text. Also notice that, even the map from a single bottom theory on its domain, i.e.~the corresponding pair of maps on the states and quantities, will in most cases of interest be non-injective.}
Also, it has lost the information about the values of most of the quantities in the comparison class. Any other member of the comparison class that we pick as our bottom theory is mapped, by our linkage map, onto the same top theory. In general, taking the limit also loses other information, e.g.~it maps different states in the bottom theory onto the same state in the top theory.

As we discussed above, emergence takes place when, in the relevant approximation or limit, there is a difference in the domains of application, i.e.~a difference in the intensions and-or in the extensions. Novel reference, i.e.~an idealization, is there only in the limit: although it is almost always already visible {\it before} reaching the limit.\footnote{Butterfield (2011b:~pp.~1067, 1069) calls this behaviour `before the limit', that already resembles the emergent behaviour and thus indicates the onset of emergence, a `weaker, yet still vivid sense', of emergence.\label{JNBbeforelimit}}

The map being a {\it partial homomorphism}, i.e.~its preserving some, but perhaps not all, of the theory's defined structure, illustrates the idea of linkage not being `mere' coarse graining. That is: it is not just lumping together states that have the same structure or features (e.g.~those that are related by a symmetry), but lumping together states that do not have similar features or are not related by a symmetry.\footnote{Agreed, `partial homomorphism' (like `partial isomorphism') is a vague word, since we did not say {\it how much} structure must be preserved: so that novelty in the bare theory is a matter of degree, that depends on how much structure is preserved. If {\it all} structure is preserved, we have a case of mere coarse-graining, and no novelty in the bare theory (at least no novelty in the range of the map; if the map is not surjective, there will be novelty in the rest of its codomain).} 
More generally, the linkage map being a partial homomorphism means that it effaces relevant differences, and information is ``lost''. Note that, in mapping states thus, in a way that does not respect the structures (features), new features may appear.\footnote{The idea of `emergence as abstraction', i.e.~omission of detail, as giving novelty (in particular, giving novel explanations), has been discussed by Knox (2016:~57), Franklin and Knox (2018:~p.~68), and Franklin and Robertson (forthcoming). For us, abstraction is neither a necessary nor a sufficient condition for ontological emergence: see the discussion in De Haro (2019:~pp.~29-34).} 

Section \ref{3op} (3)-(i) will discuss another type of novelty that arises in the bare top theory, due to the fact that the range of the linkage map is not in general equal to its codomain.

This idea of `novelty in the top bare theory' is consistent with our principle
of univocality in the physical sciences, i.e.~that the novelty in the interpretation should have a correlate in the bare theory.\\
\\
{\bf Example: the emergence of spontaneous magnetisation.} We are going to argue that, in the limit of an infinite lattice (i.e.~$N\rightarrow\infty$), the phenomenon of {\it spontaneous magnetisation} emerges in the two-dimensional Ising model of Section \ref{dualpf0}, as we lower the temperature towards the critical value, $T=T_{\sm c}$ (see the end of Section \ref{dualpf}).

We take as our comparison class, i.e.~our (set of) bottom theories, the $N\rightarrow\infty$ limit of the ferromagnetic Ising model on a square lattice, defined by the Hamiltonian in Eq.~\eq{isingH}, and denoted: $\{\mbox{Ising}(t)\}$, for any values $t>0$, where we follow the convention in the literature of using the dimensionless {\it reduced temperature}, $t:=(T-T_{\tn c})/T_{\tn c}$.\footnote{The free energy of the Ising model in the limit $N\rightarrow\infty$ limit satisfies Eq.~\eq{sinhb}, but we will only require the value of the magnetisation, below.}
This comparison class is a {\it set} of theories, one for each value of the temperature. As we discussed above, we can pick any member of this class as our bottom theory, i.e.~$T_{\sm b}(t):=\mbox{Ising}(t)$ (the theory defined by Eq.~\eq{isingH}, in the limit $N\rightarrow\infty$), at some $t>0$.\footnote{This agrees with Butterfield's (2011a:~p.~921) requirement that emergence is {\it robust}: we always obtain the same top theory in the range of our linkage map, regardless of the value of the reduced temperature $t>0$ in the domain of our linkage map. In other words, $T_{\tn t}$ is independent of the differences between the members of our comparison class $\{T_{\tn b}(t)\}$.} 
Our top theory is the Ising model {\it at the critical temperature}, i.e.~$T_{\sm t}:=\lim_{t\rightarrow0}T_{\sm b}(t)=\lim_{t\rightarrow0}\mbox{Ising}(t)$.

\begin{figure}
\begin{center}
\includegraphics[height=2.8cm]{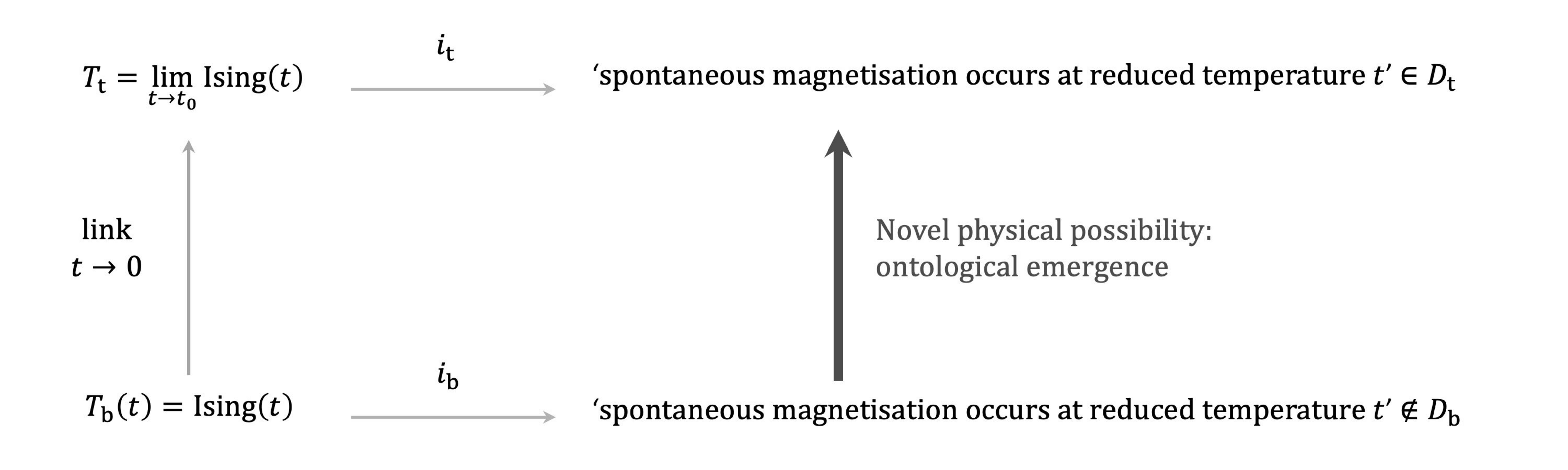}
\caption{\small Ontological emergence in the Ising model. In the bottom theory at high temperatures there is no spontaneous magnetisation. Decreasing the temperature, when we reach the critical temperature in the top theory, a non-zero spontaneous magnetisation appears at $t=0$.}
\label{IsingEm}
\end{center}
\end{figure}

The novel phenomenon or event in the domain of application of the top theory, relative to the bottom theory, is spontaneous magnetization: in the bottom theory, spontaneous magnetisation is {\it impossible}, because the energy of the spins is too high. This impossibility is reflected in the absence of the following event in the domain of application:
\bea
\mbox{`spontaneous magnetisation occurs at reduced temperature $t$'}\notin D_{\sm b}\,.\label{occursnot}
\eea
This event {\it is} possible, and does occur, in the top theory:\footnote{A subtlety: although, as we see from Eq.~\eq{Mt}, the {\it value} of the spontaneous magnetization is zero at $t=0$, it {\it starts to rise} at $t=0$ (its first derivative is discontinuous), so that, as a function in the neighbourhood of $t=0$, it is non-trivial: in that sense, we say that spontaneous magnetisation {\it occurs} at $t=0$. (In real systems, there is, in addition to the temperature, the phenomenon of {\it magnetic hysteresis}, which smoothens the dependence of the magnetisation of the ferromagnet on the external magnetic field.)}
\bea
\mbox{`spontaneous magnetisation occurs at reduced temperature $t$'}\in D_{\sm t}\,.\label{occurs}
\eea
Thus there is a {\it new physical possibility} in the top theory, relative to the bottom theory, and a physical event in the top theory's domain of application that is novel relative to the bottom theory: thus we have ontological emergence. This is summarized in Figure \ref{IsingEm}. 

To see that spontaneous magnetisation is possible below the critical temperature, but not above, we recall the form of $M(t)$, defined as the average of the individual magnetisations, in Eq.~\eq{Mt}.\footnote{As we discussed in Section \ref{dualpf}, the (sharp) phase transition is an idealisation introduced by taking the limit of an infinite lattice. However, this is unimportant for us, since what matters is that the magnetisation appears {\it as we take the temperature to the critical value}. Also, note that infinitely many degrees of freedom are not necessary to get ontological emergence. Since lattices are always finite, a more realistic treatment of our example would begin with the bottom theory defined at finite $N$, and take the linkage map to consist of two limits, $t\rightarrow0$ and $N\rightarrow\infty$. We have here followed the former (because simpler) presentation.}

The diagram in Figure \ref{IsingEm} also illustrates the non-commutativity of the linkage and interpretation maps, as in Figure \ref{Oemergence}. Namely, starting from the interpretation map $i_{\sm b}$ of the bottom theory, the domain of application cannot have the event stated in Eq.~\eq{occurs}, i.e.~the occurrence of spontaneous magnetisation. This is because this event is not in $D_{\sm b}$, regardless of the value of $t$, i.e.~also not if we take $t$ to negative values (see Eq.~\eq{occursnot}). Thus the domain of application of the top theory is not the domain of application for the bottom theory at some value $t'$ different from $t$, i.e.~$\nexists t,t': D_{\sm t}(t')=D_{\sm b}(t)$. Rather, the domain of application of the top theory contains the {\it novel event} in Eq.~\eq{occurs}. The occurrence of this event cannot be derived from the events in the domain of application of the bottom theory: instead, to ``find'' it, we need to return to the bare theory and interpret $T_{\sm t}$ using its own interpretation map, $i_{\sm t}$. Thus $i_{\sm b}\not=i_{\sm t}\,\circ\,\mbox{link}$.\\
\\
{\bf Emergence and the geometric view of theories.} We have here described ontological novelty in terms of physical properties, physical possibilities, etc.~in the domains of application. This agrees with the `two interpretative steps' that we distinguished in Section \ref{lsps}: namely, the model-theoretic semantics, and the physical semantics.

Thus it is useful to return to the model-theoretic semantics, and try to sketch a more precise characterisation of ontological emergence between phases, as in the Ising model, so as to relate emergence to our geometric view of theories. For the non-commutativity of the diagram in Figure \ref{IsingEm} can be traced back to the fact that the state-space of the Ising model is a disconnected line with two open segments (see Figure \ref{IsingManif}). Namely, the magnetisation $M:{\cal M}_{\tn{Ising}}\rightarrow\mathbb{R}$ and its dual, $\ti M:{\cal M}_{\tn{Ising}}\rightarrow\mathbb{R}$, each provide a one-dimensional model, i.e.~a model-theoretic semantics, for respectively the top and the bottom theories. In the bottom theory, the event `spontaneous magnetisation does not occur' is the physical interpretation of the zero value of the magnetisation above the critical temperature. Alternatively, since the order parameters of the two phases have disjoint support, it is the physical interpretation of the non-zero value of the {\it dual} magnetisation. And the event `spontaneous magnetisation occurs' is the physical interpretation of the non-zero expectation value of the magnetisation above the critical temperature. 

Thus the difference between the bottom and top domains in Figure \ref{IsingEm}, i.e.~the fact that this diagram does not commute, relies on the {\it discontinuity} of the underlying model-theoretic semantics with linkage. Namely: going from bottom to top, we need to change our quantity or order parameter: which in this case we can do by using a duality transformation. In other words, ontological emergence and, in particular, the non-commutativity of the diagram, is here intimately related to the fact that the manifold of the Ising model is topologically non-trivial: namely, that it is disconnected. This suggests a close relationship between ontological emergence and the topology of the manifold for the theories involved. (Similar remarks apply to the Seiberg-Witten theory, which we will discuss in the next Section.)\footnote{Although we will not further pursue the relation betwen ontological emergence and topology, we wish to point out that there is a mathematical notion, namely the Lusternik-Schnirelmann category, that might do some work for ontological emergence. The {\it Lusternik-Schnirelmann category} of a topological space ${\cal M}$ is the least natural number $n$ such that there is a covering of ${\cal M}$ by $n+1$ open sets, each of which can be contracted to a point in the space ${\cal M}$ (see Cornea et al.~(2003:~p.~1)). Thus the Ising model and the Seiberg-Witten theory both have the minimum non-trivial Lusternik-Schnirelmann category, viz.~$n=1$. Spheres $S^m$ also have $n=1$, while tori $T^m$ have $n=m$. Note that $n+1$ is also a lower bound for the minimal number of `critical points' of a function $\f:{\cal M}\rightarrow\mathbb{R}$ (see Cornea et al.~(2003:~pp.~7-8)).}

\subsection{Reduction combined with emergence}\label{eandr}

Recall the idea of theoretical reduction: namely, of reducing one theory to another by encompassing it (cf.~Section \ref{sse} (B)). Thus the idea is to reduce $T_{\sm t}$ to $T_{\sm b}$ by encompassing $T_{\sm t}$ into $T_{\sm b}$. Although reduction and emergence have traditionally been seen as opposed, we will now argue that we see them as compatible.\footnote{This view is in e.g.~Butterfield (2011a,b), Guay and Sartenaer (2016), De Haro (2019a).}

First, we ask how our linkage map can be adapted to give {\bf Nagelian reduction}. This is easily seen from our discussion, in Section \ref{lsr}, of a structure-preserving map interpreted as a relation of entailment in the opposite direction.\footnote{The traditional account of reduction, which we here endorse, is: Nagel (1961:~pp.~351-363; 1979), Hempel (1966: Ch.~8), Schaffner (2012). For defences of this account, see Dizadji-Bahmani et al.~(2010:~pp.~403-410), Butterfield (2011a:~Section 3) and Butterfield and Gomes (2023:~Sections 2-4).}  

A theory $T_{\sm t}$ can be (Nagel) reduced to $T_{\sm b}$ iff $T_{\sm t}$ can be deduced from $T_{\sm b}$, using appropriate bridge laws or bridge principles (which are hypotheses relating the terms of the two theories, and may include specific or initial conditions).\footnote{Note that Nagel himself took `reduction as deduction'  to include `reduction as approximate deduction', as do his defenders in the previous footnote. That is: rather than deriving $T_{\sm t}$ from $T_{\sm b}$, one derives a close cousin of it (what Schaffner (1967:~p.~144) dubs a `corrected' top theory), $T_{\sm t}^*$, which provides more accurate experimentally verifiable predictions than $T_{\sm t}$, and is strongly analogous to it. See also Nickles (1973:~p.~185, 197-201), who also considers limits, and Sklar (1967:~p.~111).\label{Nagelc}} 
Nagel's (1961:~pp.~339-340, 345-354) notion of reduction of course assumes a syntactic formulation of a theory, and not a set-theoretic one like ours.\footnote{More precisely, Nagel (1961:~p.~354) discusses a `condition of connectability' or linkage between the terms of the two theories, and a `condition of [logical] derivability'. As Butterfield and Isham (1999:~p.~8) notice, these two conditions together amount to requiring that $T_{\tn t}$ is a {\it definitional extension} of $T_{\tn b}$ in exactly the sense we discussed in Sections \ref{synsemeq} and \ref{sse}). The point is that the bridge laws (hypotheses) that enable the deduction of $T_{\tn t}$ can be definitions in the sense of  those Sections. Also, these authors give an early discussion, and defence, of Nagelian reduction applied to quantum gravity (pp.~6-12).}

We will implement the Nagelian idea of reduction, i.e.~that ``the top bare theory has no more content than the bottom bare theory'', with the following {\it additional} requirement (recall that the linkage map is non-injective and partly structure-preserving):\footnote{There is here an analogy with forgetful functors in category theory. Recall that categorical equivalence is the existence of a functor that is full, faithful and essentially surjective, so that it ``forgets nothing''. By contrast, a functor is forgetful if it lacks at least one of these three properties. Our condition of surjectivity for a linkage map to be reductive is analogous to the condition that a functor is essentially surjective (i.e.~surjective up to isomorphism on the objects). Weatherall (2016b:~p.~1043) takes categorical equivalence as a criterion of equivalence, and uses forgetful functors as tools to decide whether a formulation of a theory has excess structure.} 
\begin{center}
the linkage map ${\bf link}: T_{\sm b}\rightarrow T_{\sm t}$ is {\it surjective}.
\end{center}
Thus every element in the top theory has a pre-image in the bottom theory, and so all states and quantities in the top theory have correlates in the bottom theory (Nagel, 1961:~pp.~364-365). This is consistent with Nagel's account, where the linkage map can be given as a collection of bridge laws that link the terms of the two theories. Thus the linkage map may specify the conditions under which reduction obtains (e.g.~for a large value of a parameter like $N$, or a small value of a parameter like $\hbar$, etc.).\footnote{If, as in Section \ref{lsr}, there is a syntactic formulation that relates these two theories, then the direction of the deduction is the direction of the abstraction, i.e.~$T_{\tn b}\vDash T_{\tn t}$: the more abstract theory is the one that ``says less'', and thus is logically weaker. That is also how we think of the top theory, which we obtain by ``lumping together'' states from the bottom theory, and thereby erasing information, so that the truth-values of the sentences of $T_{\tn t}$ are determined by those of $T_{\tn b}$, but not the other way around. For a recent discussion of these roles, and of the more general notion of correspondence in quantum gravity, see Crowther (2018:~pp.~77-80) and van Dongen et al.~(2020:~pp.~122-125). Dizadji-Bahmani et al.~(2010:~pp.~397-399) discusses the `generalized Nagel-Schaffner model' of reduction in some detail and Hartmann (2002:~pp.~85-89) discusses various relevant types of correspondence relations.\label{linguisticA}}
This then models a very common case of ontological emergence in physics, i.e.~reductive emergence, as a linkage map that is surjective.

In the notation from the previous Section, when the comparison class is labelled by a parameter $x$, the top theory reduces to the bottom theory if the top theory {\it is} the limit of the sequence: $T_{\sm t}=\lim_{x\rightarrow x_0}T_{\sm b}(x)$. The Ising model was our first example of a linkage map that is surjective in this sense. Another example is as follows:\\
\\
{\it Example: massive vs.~massless particles.} Take as our bottom theory $T_{\sm b}(m)$ a theory of a massive relativistic particle of mass $m$, and as our top theory $T_{\sm t}$ a theory of a massless relativistic particle, so that the linkage map `loses the information about the mass'. As De Haro (2019a:~pp.~40-41) showed, this is a case of {\it ontological emergence}. For a `massless particle' is {\it not} a `particle of mass $m=0$', since e.g.~they have different symmetries, and also there are physical possibilities in $D_{\sm t}$ that are impossibilities in $D_{\sm b}$. For example, `any two particles sent from the same point, and in the same direction, at a non-zero relative time interval, will never collide' is true, and is a real possibility, in $D_{\sm t}$; while it is false in $D_{\sm b}$, because in $D_{\sm b}$, particles can have any speed, and thus can overtake each other, while in $D_{\sm t}$ their speed is always $c$. Thus in particular, $D_{\sm t}$ admits a temporal ordering of particles on a given trajectory in space, that $D_{\sm b}$ does not admit. More generally, the causal structures of the two domains are very different. This illustrates our statement, in Section \ref{Sem}, that erasing structure, or ``losing information'', may lead to new features and physical possibilities. 

The linkage map is surjective, i.e.~a case of reduction, because the equations of motion of the massless particle can be derived from the equations of motion of the massive particle. In other words, while the domains of application are different, $D_{\sm t}\not=D_{\sm b}(m=0)$, we do have reduction of the bare theories, i.e.~$T_{\sm t}=\lim_{m\rightarrow0}T_{\sm b}(m)$.

With these construals---of emergence as non-injectivity and reduction as surjectivity---we turn to arguing that these two notions are compatible:\\
\\
{\bf Reduction as formal, and as respecting extensions but not intensions.} The reason why, in our account, reduction is compatible with ontological emergence, is that {\it reduction is a formal i.e.~non-interpretative notion, while ontological emergence is interpretative}. A similar point is made by Howard (2007:~p.~143), who distinguishes reduction as a logical relationship between syntactically formulated theories, from ontological relations (like supervenience) between theories understood semantically.\footnote{Also Nickles (1973:~pp.~184-185) distinguishes Nagelian reduction, whose function is primarily logical and explanatory, from a second type of reduction, whose function is heuristic and justificatory (and which, like Nagelian reduction, is also {\it not} ontological reduction but, unlike Nagelian reduction, is not explanatory). For our own reasons for keeping the notion of `reduction' formal i.e.~non-interpretative, see De Haro (2019a:~p.~19).}

Our statement that `reduction is formal i.e.~non-interpretative' requires unpacking. For even if one's scheme for making deductive inferences is very formal, {\it the meanings of the reduced statements need to be respected}, if there is to be reduction.\footnote{See Benacerraf (1965:~pp.~53, 56, 58), and especially the discussion in Butterfield and Gomes (2023:~Section 2.2.1), who call this the problem of `Faithlessness'.} 
For example, there may be bridge laws defining numerals `1', `2', etc.~in terms of set theory that enable a deduction of arithmetical sentences like `1 + 1 = 2'---but only as thus interpreted, i.e.~interpreted as in set theory. Which we might say is not what we originally meant by our arithmetical statements!

More precisely, and without entering into technical details here: {\it Nagelian reduction only respects extensions, and not intensions}. This is because bridge laws do not in general respect the intensions, while (if successful, under specific physical conditions) they respect the extensions. Thus they only respect part of the meanings.\footnote{Nagel was aware that his bridge laws only respected extensions, but not intensions. Witness his reply to Feyerabend in Nagel (1979:~p.~914). But this did not prevent him from using the word `reduction'. This is precisely our situation here: where, in cases of ontological emergence, the bottom and top theories may have the same extensions, but they must have different intensions, and thus different meanings.}
This is compatible with emergence, because novelty in the intensions suffices to have emergence (see Section \ref{Sem}). 

This agrees with the usage of `reduction' in physics, which does not fully respect meanings either. For example, `temperature' has a different meaning in thermodynamics and in statistical mechanics, even if a bridge law relates the corresponding expressions. Under some circumstances, the extensions of these expressions may be the same, but their intensions are different: {\it bridge laws do not in general fully respect meanings}. Nor is the meaning of `massless particle' the same as that of `particle of mass $m=0$'. And yet scientists use the word `reduction' in just these sorts of examples.

Also recall our distinction between the `model-theoretic' and `physical semantics' (see the discussion of this distinction in Section \ref{lsps}, and also the contrast (A) `formal' vs.~(B) `interpretative'). Although reduction does not fully respect physical meanings, it may respect meanings in the sense of a model-theoretic semantics (but it will not always do, especially in the presence of limits): namely, meanings in a formal structure that are not (by themselves) items in the world. After all, our bare theories presented as triples are semantic, yet formal i.e.~uninterpreted. Thus they are not fully physical meanings, in the sense of interpretation that we have discussed in this book, but only necessary conditions for such meanings. For example, the statement `spontaneous magnetisation occurs at reduced temperature $t$' (see Eq.~\eq{occursnot}) is not a model-theoretic statement, but a physical one. 

We see it as a virtue of Nagelian reduction that it is weak in this sense, of not requiring ontological reduction (i.e.~subsumption of both the extensions and the intensions of the top theory into those of the bottom theory), and only respecting the extensions.\\
\\
{\it Example: emergence in Seiberg-Witten quasi-duality.} A clear example of emergence and reduction is the Seiberg-Witten theory in Section \ref{effD}, where the low-energy models emerge from the underlying ${\cal N}=2$ SU(2) SYM theory in various regions of the moduli space, i.e.~for various expectation values of the Higgs field. 

For example, consider the linkage map between $T$ and $M$ (where $M$ is the low-energy ${\cal N}=2$ U(1) SYM). This map is {\it non-injective}, because the fields in the Wilsonian low-energy effective action in $M$, Eq.~\eq{Seff}, arise by coarse-graining of the high-energy fields under renormalization, in the low-energy limit of $T$. Thus there are many field configurations at high energies that give the same low-energy field, and the map is non-injective. The map is a {\it partial homomorphism}, because it preserves some of $T$'s structure: e.g.~the ${\cal N}=2$ supersymmetry is preserved, and the gauge symmetry is spontaneously broken from SU(2) to U(1).\footnote{See our discussion of spontaneous symmetry breaking in Chapter \ref{EMDuality}.} The map is {\it surjective}, and so reductive, because all of the states of the models are mapped from the BPS states in the supersymmetry algebra of $T$ (and also the quantities of the models are derived from $T$ through the Wilsonian effective action). 

And there is {\it novel low-energy behaviour}, because the low-energy fields have classical properties under the Higgs mechanism, such as the classical value of the Higgs field and the factorization of the correlation functions, that a quantum theory does not have. Also, near a singularity, the novelty is more striking. For there, magnetic monopoles, which were massive {\it solitons} far away from the singularity, become (after including in the linkage map a duality transformation) light {\it particles}, and we get a model $M'$ (i.e.~${\cal N}=2$ SQCD) where the monopoles are electrically, rather than magnetically, charged (i.e.~they couple to abelian gauge fields through the usual electric current coupling, rather than through a dual current).\footnote{For a detailed discussion of emergence in the Seiberg-Witten theory, see Vergouwen and De Haro (2024).}

\subsection{Emergence and duality}\label{dande}

Over the past two decades, emergence has been an important motivation for the studies of dualities in string theory, and in particular gauge-gravity dualities. The promise is that spacetime and gravity could emerge from a more fundamental structure that is either not spatio-temporal, or based on a non-gravitational spacetime of a different type.\footnote{The idea of emergence also crops up in other contexts in string theory, such as the derivation of the Einstein field equations from the conformal invariance of the world-sheet theory, which is often seen as a case of emergence of gravity. See Huggett and Vistarini (2015:~p.~1163). An early philosophical discussion of the emergence of space and time in quantum gravity is Butterfield and Isham (1999).}
 
The obvious example is of course AdS-CFT, where the bulk model is dual to a quantum field theory model in one dimension less. Furthermore, the quantum field theory is generally better understood than the string theory, which has led some to conclude that the bulk physics `emerges' from the boundary. For example, Horowitz and Polchinski (2009:~p.~178) write about the emergence of the bulk from the boundary model as follows:

\begin{quote}\small
AdS/CFT duality is an example of emergent gravity, emergent spacetime, and emergent general coordinate invariance. But it is also an example of emergent strings! We should note that the terms `gauge/gravity duality' and `gauge/string duality' are often used ... to reflect these emergent properties. \end{quote}

Having discussed the Schema's construals of duality and of (ontological) emergence, there is good reason to be cautious about such enthusiastic pronouncements. For duality requires isomorphism of models, while emergence requires a non-injective map: and these requirements are prima facie in {\it tension} with each other. Thus we can now be more precise about the conditions under which we can have duality and emergence between a pair of models, and the kinds of emergence (if any) that we can have. To see whether they are compatible, we will distinguish the interpretative and formal i.e.~non-interpretative aspects of the emergence and duality maps, and compare them for compatibility or incompatibility. 

We first discuss the interpretative aspects of both emergence and duality (and we consider two interpretative options, namely that a duality is a case of theoretical equivalence or a case of theoretical inequivalence). Thus we first compare the non-commuting diagram for ontological emergence in Figure \ref{Oemergence} with the commuting (bottom) triangle diagram for theoretical equivalence in Figure \ref{Physeq14}. We see that duals that are theoretically equivalent exclude emergence, because theoretical equivalence requires that the domains of application are the same, while emergence implies that they are different. And thus, there can be ontological emergence of duals only if they are theoretically {\it inequivalent}. This is because Figure \ref{Physineq} {\it is} compatible with Figure \ref{Oemergence}. (Note that physicists tend to interpret duals of fundamental theories, such as string theory, as theoretically equivalent: and so, this analysis implies that Horowitz and Polchinski's talk of `emergence' cannot be understood as ontological emergence in our sense.)

While the interpretative aspects of the maps do allow that emergence can be compatible with duality if the latter relates theoretically inequivalent models, we will conclude from the formal i.e.~non-interpretative aspect that duality and emergence are incompatible. The reason for this is a map cannot be simultaneously a duality (i.e.~an isomorphism) and a linkage map (i.e.~a non-injective, partial homomorphism), and so if duality and linkage are to coexist, they must be different maps. 

But even if we use different maps, the two main cases of ontological emergence from Section \ref{eandr} exclude duality. We discuss this first for reductive emergence, and then for emergence without reduction (irreducible emergence).

First, duality is incompatible with reductive emergence because duality requires equinumerous models, while reductive emergence requires a non-injective map that is also surjective, and so the models have different equinumerosity.\footnote{While this argument strictly holds for maps that relate sets with a finite number of elements, it is, on physical grounds, unlikely that a salient sense of emergence-cum-duality can be realized in theories with infinite numbers of elements. For consider regularizing a theory with an infinite number of degrees of freedom by e.g.~putting it on a lattice, so that the lattice theory has a finite number. Since emergence is a meso- and macroscopic phenomenon, it should surely be independent of the details of the short-distance behaviour, i.e.~of the lattice spacing. Namely, one expects to have emergence already `before' taking the limit (see Butterfield (2011b:~p.~1078) and footnote \ref{JNBbeforelimit}). Thus, also in the familiar theories with infinite numbers of degrees of freedom, like the ones discussed in Part II, duality and emergence of this type are incompatible.} 

Second, the other main case of emergence is irreducible emergence, where the linkage map is not surjective, and it is required that no surjective map between the models exists (on pain of one model being reducible to the other by a different map). But duality is a surjective, i.e.~reductive, map between two models: and so, duality also excludes irreducible emergence (we will discuss irreducible emergence in Section \ref{3op} (3) (i)).\footnote{While the idea of reducibility is that the linkage map is surjective, the idea of irreducibility is that the linkage map is not surjective and there exists no surjective linkage map with the same domain and range (thus the models are not only not reduced, but {\it cannot} be reduced, i.e.~there is a {\it lack of derivability}). The intermediate case, of a linkage map that is not surjective but where a surjective map exists, is much less interesting, because it means that the model can be reduced, although the emergence map does not exhibit this reduction. Hence the importance, on physical grounds, of {\it aligning} the emergence and reduction (i.e.~duality) relations, i.e.~of making a judicious choice of a single inter-theoretic map that encompasses both reduction and emergence. Thus De Haro (2019a:~p.~8) only considered reductive emergence.\label{surjective}} 

Thus the prospects for duality and ontological emergence are slim, because the options left are not physically salient or interesting, and are surely not along the lines of the quote above.\footnote{Thus the only remaining logico-semantic option is to combine duality between theoretically {\it inequivalent} models with emergence that is in principle reducible, i.e.~an emergence map that does not align with the reduction map: see footnote \ref{surjective}. As we discussed, this case is not physically salient, because the emergence map only tells us how {\it part} of the top theory (namely, the image of the emergence map) is linked to the bottom theory, while another reduction maps the {\it whole} top theory back to the bottom theory. And thus the emergence map seems artificially restricted: additional motivation is required to consider this to be interesting emergence.}

But not all is lost. For, while we have argued that there are no interesting cases of duality-cum-ontological-emergence, in the next Section we will review three additional ways to have emergence and duality, including epistemic emergence.

\subsection{Three options for emergence and duality}\label{3op}

\begin{figure}
\begin{center}
\includegraphics[height=3cm]{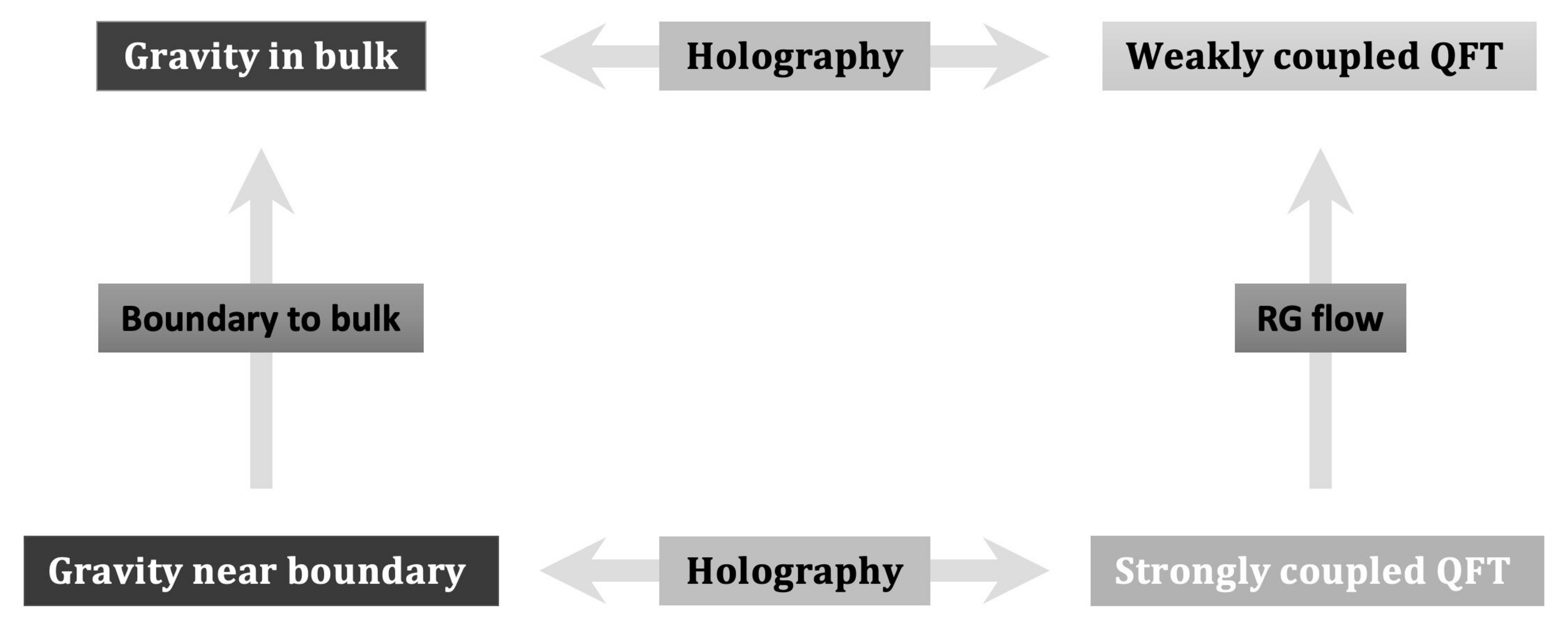}
\caption{\small Emergence in AdS-CFT according to option (2). The holographic duality is in the horizontal direction, and the linkage map is given by the renormalization group flow (or boundary-to-bulk map) in the vertical direction. The horizontal duality map extends to other points along the vertical direction, so that for each value of the renormalization group parameter the left and right models are conjectured to be duals.}
\label{EmAdS}
\end{center}
\end{figure}

There are three basic options to combine duality with emergence. These three options are independent, but not mutually exclusive. The first option is to have emergence ``in a different direction'' than the ``direction'' of duality: so that there are four models instead of two. The second weakens the idea of duality; and the third weakens the idea of emergence so as to have epistemic emergence. We discuss them here in turn, with emphasis on the first and third options:\\
\\
\begin{figure}
\begin{center}
\bea
\begin{array}{ccccc}
~~~~~~~~~~~~~~~~~M_{\sm t}^{\tn L}
&\xrightarrow{\makebox[.6cm]{$\sm{\it d}_{\tn t}$}}
&M_{\sm t}^{\tn R}~~~~~~~~~~~~~~~~\\
~~~~~~~~{\sm{${\bf link}_{\tn L}$}}~\big\uparrow\!\!
&
&\big\uparrow~{\sm{${\bf link}_{\tn R}$}}~~~~~~~~~~\\
~~~~~~~~~~~~~~~~~M_{\sm b}^{\tn L}&\xrightarrow{\makebox[.6cm]{$\sm{\it d}_{\tn b}$}}
&M_{\sm b}^{\tn R}~~~~~~~~~~~~~~~~\end{array}\nonumber
\eea
\caption{\small Option (1), emergence beside duality. The two duality maps go horizontally from left to right, while the two linkage maps go upwards.}
\label{ebd}
\end{center}
\end{figure}
{\bf (1)~~Emergence beside duality:} The idea here is that the duality and emergence maps lie ``along different axes'', i.e.~that we have a square with four models rather than two, pairwise related by duality in the horizontal direction, and by linkage in the vertical direction (see Figure \ref{ebd}). One major example of this option is AdS-CFT duality, as shown in Figure \ref{EmAdS}, which is an adaptation of Figure \ref{ebd} to this particular example. As we discussed in Section \ref{ggd}, changing the radial parameter in AdS is dual to changing the RG flow parameter in the quantum field theory, which is the vertical direction. The horizontal direction is the holographic duality map, evaluated at each scale of the parameter.\footnote{This analysis of the relation between duality and emergence was first given in Dieks et al.~(2015:~p.~213). See also van Dongen et al.~(2020:~p.~118).}

More specifically, the bottom-left corner is the (super)gravity model in the AdS space, which gives a good approximation to the physics in the region near the boundary, i.e.~the bulk long-distance physics (the supergravity model thus gives a good approximation of the string theory in the region near the boundary). This is dual to the bottom-right corner, where the quantum field theory model at high energies is strongly coupled. This connection between long-distance physics in the bulk, and high-energy physics on the boundary, is called the {\bf IR/UV connection}. 

The linkage map on the right represents the renormalization group flow in the quantum field theory, going from the strongly coupled fixed point to the weakly coupled one, i.e.~from the ultraviolet at high energies to the infrared at low energies. At these fixed points, the field theory is a conformal field theory. Following the renormalization group flow, from the strongly coupled ultraviolet regime towards the infrared in the top-right corner, the model becomes weakly coupled. It is dual to the top-left corner, where the model's cutoff is in the deep bulk of the AdS, i.e.~we have a short-distance gravitational theory, which in general requires us to use the full string theory (i.e.~in general, the supergravity approximation is not valid here). On the left, the linkage map (which is dual to the renormalization group map) is the boundary-to-bulk map, and goes in the same direction.\footnote{Recent work on quantum information models for gauge-gravity quasi-dualities generalize this picture by allowing that the boundary-to-bulk map can ``go in both directions'', i.e.~it need not always follow the RG flow, but one can consider the reverse of the coarse-graining procedure. See e.g.~Gharibian et al.~(2014:~pp.~227-228). See also Chen et al.~(2022), Jahn and Eisert (2021) and Harlow (2018).}

There is an intuitive analogy that can help us understand, from the bulk perspective, the {\it direction} of the flow, which is the direction of the emergence map.\footnote{The analogy used here is between the geodesic equation for a massive point particle and the Klein-Gordon equation for a massive scalar field (which is dual to a source in the quantum field theory with a first derivative whose sign gives the running of the coupling, i.e.~the RG equation). For the correspondence between the bulk equations of motion and the RG equation, see Balasubramanian and Kraus (1999:~pp.~3607).}
An object shrinks in size as it is brought from the bulk out to the boundary, so that paths that come close to the boundary are in general longer than paths that go through the bulk. For example, a geodesic between two points near the boundary is a chord that recedes deep into the bulk, rather than a line that connects the two points along a path that stays close to the boundary. Thus the proper length of a path between two points is in general smaller if it goes through the centre of the space and stays away from the boundary. In other words, massive objects are pulled away from the boundary towards the bulk, and so objects that are in free fall can go from the boundary to the bulk, but not from the bulk to the boundary.

In Figure \ref{EmAdS}, each label stands for the model that is applicable, in each of the four corners, in the corresponding regime of parameters.\footnote{As we discussed above, the linkage map between the bottom and the top left models is labelled by the radial cutoff. Thus we have a one-parameter family of gravity models. Although a string theory with a particular value of the cutoff underlies each vertical model, the label `Gravity' on the left of Figure \ref{EmAdS} indicates that, for the given value of the cutoff, we use the gravity theory that gives an appropriate approximate description of the string theory for that value of the cutoff. For sufficiently favourable dynamical set-ups, a (super)gravity approximation may exist at a given point in the vertical direction. But in general such approximations only exist at the bottom, i.e.~near the boundary where the curvature is small, and not at the top, where the curvature corrections may be large.}
As we go upwards on the left, we integrate out the outer regions of the AdS space. Assuming that the quantum field theory has a fixed point,\footnote{Such solutions have been constructed explicitly in Freedman et al.~(1999:~pp.~376-381), who deformed the ${\cal N}=4$ SYM theory by a mass term, to get an ${\cal N}=1$ gauge theory. In practice, the behaviour found in the QFT is more complicated than having two isolated fixed points. For example, Freedman et al.~(1999:~p.~364) found {\it lines} of fixed points, both in the ultraviolet and in the infrared, where the line is defined by the expectation value of an operator in the CFT that does not change the flow. Also, the 't Hooft coupling is large both in the UV and in the IR, so that the Einstein field equations can be used reliably on both ends.} 
the RG equation is dual to Einstein's field equation for the corresponding regions. Thus, in the bottom left corner, Einstein's equations are only valid in the very small neighbourhood of the boundary (and the rest of the space needs to be described using string theory); while in the top left corner Einstein's equations hold near the interior region, and we are left with a much smaller region that can only be described using the full string theory. In other words, a semi-classical spacetime emerges as we integrate out more and more high-energy modes: Einstein's field equations themselves are the effective equations describing this process of integrating out.

The direction of the arrow in Figure \ref{EmAdS} does not imply that the (short-distance) bulk gravity emerges from the (long-distance) gravity near the boundary. Rather, the arrow indicates that, as we integrate out the spatial modes of the fields (namely, as we move the cutoff {\it from the boundary into the bulk}), the long-distance modes appear in the effective action as classical supergravity fields that provide a boundary action (or a boundary state) for the other bulk modes. (Using the IR/UV connection, we can think of this as the spatial analogue of integrating out high-energy modes in quantum field theory, where the integrated out modes appear as semi-classical corrections in the Wilsonian effective action of the other modes.\footnote{For a discussion of the Wilsonian effective action in quantum field theory, see Section \ref{effD}.}) 
The effective boundary action for the integrated out modes is then conjectured to be dual, by the AdS-CFT correspondence, to the UV boundary action.\footnote{For a detailed discussion of this point, including examples of such flows, see Faulkner et al.~(2011:~pp.~2-6).}\\
\\
{\bf (2)~~Effective duality:} If the models are quasi-duals, more specifically if they are only approximate duals in a regime of energies and-or parameters, then there is of course no isomorphism, and so the main obstruction against emergence is lifted. Namely, there need not be a bijective map between effective duals, and there can be a non-injective emergence map between them---recall our discussion of equinumerosity in Section \ref{dande}. Thus we can combine {\it effective} dualities with ontological emergence. For example, the above quotation by Horowitz and Polchinski could be interpreted as making this claim. Other examples include more general gauge-gravity dualities,\footnote{For an example of a holographic scenario in which Newton's equations emerge from an underlying quantum theory, see Dieks et al.~(2015:~p.~210) and De Haro (2017a:~p.~120). Also, recent work on realizing analogues of holographic dualities in quantum information theory, and in particular on reconstructing spacetime geometry using a quantum error-correction code, go in this direction. For recent reviews, see Chen et al.~(2022), Jahn and Eisert (2021) and Harlow (2018).} 
and bosonization in higher dimensions, which are only effective dualities.\footnote{For bosonization in three dimensions, see Seiberg et al.~(2016:~p.~404) and Karch and Tong (2016:~p.~2).}\\
\\
{\bf (3)~~Epistemic emergence:} While our discussion in this Section has focussed on ontological emergence, where the domains of application of the putative emergents are different, there is another interesting case, where: (i) the domains of application are the same (or, in a logically weaker sense, the domain of application of the top theory is a proper subset of that of the bottom theory), but (ii) there are differences in our theories' {\it descriptions} of it. Thus there is emergence in the bare theories or models, but it does not correspond to novelty in the domain of application.\footnote{Since the novelty is now not in the domain of application, but there are differences in the theory's description of it, we can weaken the non-injectivity requirement of the linkage map, which formalized the idea of mapping several microscopic states onto a single macroscopic state (which is the formal correlate of ontological emergence, i.e.~of having fewer states or fewer components). This is not a necessary requirement for epistemic emergence, because other differences between the bottom and top bare theories can also give significant epistemic differences (including e.g.~differences of calculability in practice), without a correlate in the world.\label{Winj}} 
This is a very interesting case, because it is ubiquitous in the physical sciences.\footnote{We endorse Humphreys' (2016:~pp.~54-55) critique of what he calls the `rarity heuristic', i.e.~the idea that for an account of emergence to be correct, it must make emergence a rare property. For a discussion of the right balance between the ubiquitousness of the examples of emergence and the desirability to regiment the notion, so as to reduce the number of putative cases of emergence, see De Haro (2019a:~pp.~19-23).}

Since the novelty of {\it epistemic} emergence is not in the domain of application but in the descriptions given by the two theories, one useful way to analyse epistemic emergence is to distinguish various kinds of linkage maps. Below we discuss two main cases: the first is {\it irreducibility}, and uses a strengthening of the linkage map that we discussed in the previous Sections. The second is {\it weak epistemic emergence}, and uses a weakening of our usage, so far, of the linkage map (our examples will focus on this case). The two cases are as follows:\footnote{There is a third case, of partial homomorphism: if the linkage map is not a full, but a partial, homomorphism, then some of the model's structure is not preserved, and so there is the possibility of structural novelty of the top theory, relative to the bottom theory. Cases of `mere coarse graining', where we use less fine-grained, aggregate variables, are often of this type. Fodor (1974:~pp.~102-111) gives a general syntactic analysis of such cases, where the map does not preserve natural kinds (i.e.~it does not map natural kinds to natural kinds), and it does not preserve lawhood. Sober (1999:~p.~553) is a critique of a special case of this argument, where the theories in question are formalized in first-order logic.}\\

(i)~~{\it Irreducibility, or lack of derivability}:\footnote{In the terminology of Guay and Sartenaer (2016:~p.~299), this is `in principle', or strong, epistemic emergence.} 
As we discussed in Section \ref{dande}, irreducibility means that there is no reductive (linkage) map reducing the top theory to the bottom theory, i.e.~no surjective map of the right kind. This lack of formal reduction is an interesting case of epistemic emergence, because it gives a clear sense of the epistemic novelty of the top bare model: there are formal aspects about the states or quantities that cannot be reduced to the states or quantities of the bottom bare model. The novelty is here epistemic, because recall that ontological novelty is a matter of the domain of application, not of the theories: thus we are ``an interpretation map away'' from ontological novelty.

Examples are quasi-dualities and similar mathematical or logical equivalence relations that are not defined over the whole domain or co-domain. That is: either (a) they are partial functions, so that the pre-image of the map is a proper subset of the domain, or (b) they are not surjective, i.e.~the image is a proper subset of the codomain. Such a quasi-duality can have an internal interpretation, so that the domain of application is the same for both models (this contrasts with case (2), which also involved quasi-dualities, but with interpretation maps that do not match).\footnote{As we noted in footnote \ref{Winj}, the non-injectivity requirement can be weakened for epistemic emergence, thus allowing injective linkage maps as we will do in case (ii) below. In case (i), even with an injective linkage map, the emergence is strongly epistemic because the top theory cannot be reduced to the bottom theory.} 
The lack of derivability is because the quasi-duality does not map the whole domain and-or co-domain. A paradigmatic example, from Section \ref{eil}, is the relationship between modal logic and classical logic. In that example, there is an appropriate bisimulation that preserves the meanings under the translation, but the range of the bisimulation is limited to a {\it fragment} of classical logic.

An example in physics is the relation between Hamiltonian and Lagrangian mechanics,\footnote{Agreed, one might object to calling all of the classical (Lagrangian or Hamiltonian) mechanics, a `theory'. But note that even for the Lagrangian and Hamiltonian of a single point particle, where the dimension of the state space is fixed, the question of hyperregularity already arises (see the discussion in point (2) of Section \ref{Ourthm}). However, the point here is not about what is the best notion of `theory' or `model', but rather about the epistemic emergence being irreducible outside of the hyperregular case.} 
which is an isomorphism, but only for hyperregular Lagrangians and Hamiltonians, i.e.~the case in which the Legendre transformation (and its inverse) is a diffeomorphism.\footnote{This is the condition that the Legendre transformation and its inverse are invertible and differentiable, which is the case if the Hessian is non-degenerate, i.e.~if its determinant is non-zero. For the proof of equivalence, see Marsden and Ratiu (1998:~p.~188) and Abraham and Marsden (1978:~p.~222).}\\

(ii)~~{\it Weak epistemic emergence:} There is a weaker sense of epistemic emergence `in practice, but not in principle', e.g.~because of the computational complexity of the duality map. We can get this type of emergence by weakening the requirement of non-injectivity of the linkage map, thus allowing it to be injective (even on the whole domain and range of the duality).\footnote{As we anticipated in footnote \ref{weakEE}, we may here allow cases of injective linkage maps (including dualities), since the `in practice'-character of weak epistemic emergence entails that the novelty is a practical one, introduced by the difficulty of evaluating the linkage map, and not by the mathematical properties that define it.} 
There is weak epistemic emergence when it is in practice very difficult to find the linkage between the top and the bottom theories, e.g.~because it is difficult to calculate the outputs of the duality map, or because the map itself is difficult to construct. Even though there is in principle no emergence, since the two models are duals and thus formally equivalent, there {\it is} in practice, i.e.~relative to the computational or other epistemic resources available.

This notion of weak epistemic emergence, as `lack of derivability in practice', is closely related to Bedau's (1997:~p.~378) weak emergence, where a macrostate is weakly emergent iff it can be derived from the microsates and microdynamics, but only by simulation. More specifically, although the macrostate can in principle be derived from the microstates,\footnote{Bedau (1997:~p.~393) acknowledges that the behaviour can be {\it predicted}, even if only by simulation: and since a simulation counts as a {\it derivation}, the behaviour follows deterministically. Thus any two simulations that use the same microdynamics, and the same external conditions, predict the same behaviour.} 
this is algorithmically difficult.

The notion of weak epistemic emergence emphasizes how dualities can render dual models more tractable, through their mapping between models of different tractability. For, as we discussed in Section \ref{featurerole}, `hard-easy' is one of the scientifically important features that dualities often have: and, as the examples in Part II illustrated, this feature is often reflected in the inversion of the coupling parameter (see Eqs.~\eq{Diracq}, \eq{Itemp}, \eq{betag}, \eq{Tduality} and \eq{Sdual}). A calculation that is hard in one regime of parameters, is rendered easy in the dual regime of the parameters. Even in cases where the domain of application does not change, the duality is useful and the emergence shows surprising novel behaviour. 

This also emphasizes the importance of what physicists call `finding the right variables', i.e.~variables that are well-adapted to a regime of parameters, so that calculations can be readily done. Finding a duality is often a matter of finding the right set of variables that turns a difficult problem into a tractable one (as in e.g.~solving a problem by doing a Fourier transformation). 

Bosonization in two dimensions is an example, where the relation between the scalar field and the fermion is highly non-linear: the fermionic field is dual to a `vertex operator' for the bosonic field (see Eq.~\ref{expf}). Recent work has used measures of quantum complexity for the bosonic and fermionic models, which are then compared across the duality.\footnote{The quantum complexity of a (final) state $\psi_{\tn f}$, relative to a given initial state $\psi_{\tn i}$, is the minimum number of gates (i.e.~elementary unitary operators) required to produce the final state from the initial state. For quantum field theories with continuous variables, Ge and Policastro (2019:~p.~5) (see also Hackl and Myers, 2018:~pp.~4-5) adopt as measures of (path) length the Fubini-Study metric and the measure by Nielsen and collaborators (both of which are geodesic distances: directly on the Hilbert space, and on the manifold of the group of operators, respectively). (See Dowling and Nielsen (2007) and Nielsen and Chuang (2010).)
The set of unitary operators used is in principle the set of all the unitary opertors available in the theory, although in practice this is restricted to an algebraically closed tractable subset, for example the set of gates that are quadratic in the fields. `Universal' here means that every unitary operation can be approximated to within any desired degree of accuracy by some combination of gates from the finite universal gate set. For a discussion, and examples of such finite universal gate sets, see Watrous (2008:~pp.~8-9). In the case of bosonization, the set of all quadratic bosonic gates is of course not the same as the set of all quadratic fermionic gates, and so Ge and Policastro (2019:~pp.~8-9) use a set of operators that they claim are well-adapted to the fermionic and bosonic models whose complexity they calculate.} 
For example, Ge and Policastro (2019:~p.~5) report `striking differences between the bosonic and the fermionic result. The bosonic complexity for these states is cutoff-independent, and smoothly dependent on the parameters corresponding to the Fourier modes, whereas for fermions it grows like [the logarithm of the cut-off], and it is much less regular'.\footnote{For related work that defines measures of quantum complexity of the boundary quantum field theory by using bulk quantities in gauge-gravity dualities, see Chen et al.~(2022), Jahn and Eisert (2021), Harlow (2018), Brown et al.~(2016:~p.~1), and Susskind (2016:~p.~24).}

Although defining quantitative measures of epistemic emergence along these lines is well beyond the aim of this book, it is clear that there is a vast and rich field here that is bound to give rise to interesting to new conceptual questions.\footnote{There is a growing body of literature on defining measures of emergence and distances between theories (although the various approaches tend to be disconnected, and ontological vs.~epistemic emergence are not always clearly distinguished). For a discussion of (mostly syntax-based, i.e.~regardless of meaning) measures of emergence, see Gershenson et al.~(2012) and Crutchfield (1994). For model theory-based and geometric metrics, such as the `conceptual distance' between theories, see Friend et al.~(2020), Khaled and Sz\'ekely (2020), Balasubramanian (1997), and Balasubramanian et al.~(2015). For an overview of the use of geometric methods in information theory, see Amari and Nagaoka (2000).}

\section{What is fundamental?}\label{fundam}

In both physics and philosophy, there is a central question about what is fundamental and what is derivative. And in both fields, this question is understood as being about what has priority, what is more basic, independent, etc. In philosophy, fundamentality underpins the epistemic and semantic aspects of reduction and emergence that we discussed in previous Sections: it aims to give an account of what makes some entities, facts or theories more basic or fundamental than others, and how they relate to each other in terms of dependence and explanation. This Section explores these questions, first in general (Section \ref{fundapp}) and then in combination with dualities (Section \ref{fundaD}) and emergence (Section \ref{fundaE}).

\subsection{`Fundamental' in physics and philosophy}\label{fundapp}

Fundamentality, i.e.~the view that some theories, entities, behaviours, etc.~are more fundamental than others, appears to require a stratification of the world into levels or layers.\footnote{See Schaffer (2003:~p.~498) and Kim (2002:~p.~4).} 
This stratification is often formalized by introducing an asymmetric relation between the theories, entities, etc.~that is transitive and irreflexive, i.e.~a strict partial ordering.\footnote{See Tahko (2018:~Section 1) and Kim (2002:~p.~9).}
However, philosophers' views about what fundamentality entails differ widely. We will here discuss four views.\footnote{For another discussion of various views on fundamentality, see Glick et al.~(2020:~p.~2).} 

(i)~~One widely held view is that of {\it mereological priority}, in other words, that {\it parthood} is the fundamental relation that does the work of distinguishing levels: the parts out of which something is made are more fundamental than the whole.\footnote{Incidentally, the opposite view also occurs, i.e.~holism, the view that the whole is more fundamental than its parts. For a clear formulation, see Humphreys (2016:~pp.~35-36).} 
This view of course motivates some of the programmes of reduction, both in the physical sciences and in philosophy.\footnote{See for example Weinberg (1993:~p.~55) and Oppenheim and Putnam (1958:~p.~9). For a critique of this picture, see Kim (2002:~p.~16-17) and Wimsatt (1997:~p.~382).} 

(ii)~~Fundamentality in contemporary physics, rather than being given in terms of parthood, is often defined in terms of appropriately chosen {\it physical scales}, especially distance or energy, as they appear in quantum (field) theories. As we have already discussed (see Section \ref{effD}, and also gauge-gravity duality in Section \ref{3op} (1)), renormalization group procedures in effective field theories, either in momentum or in real space, in effect assume this kind of hierarchy, where the energy range (and-or length range) is divided into scales, each of which exhibits a certain kind of physics, with the high-energy scales being more fundamental because they cover all of the relevant physics, and the low-energy scales being less fundamental (hence the word `effective'), because the corresponding description is approximate. Furthermore, the definition of the low-energy physics depends on the high-energy physics (which is effectively incorporated into the renormalization of the low-energy parameters), and usually not the other way around.\footnote{It is of course possible to, in an epistemic sense, invert the scales by relativizing to a certain epistemic situation. Thus, relative to a set of quantities (observables) valid at low energies, low-energy physics may be taken to be more fundamental than high-energy physics, which is irrelevant to the given epistemic situation.} 
Thus one can, in some sense, regard this kind of fundamentality as the formal correlate of mereology, where each successively higher energy scale shows more of the ``internal structure'' of matter.\footnote{This is Weinberg's (1993:~Ch.~III, 2008:~p.~30; 2001:~Ch.~10) `reductive' view, famously contested by Anderson (1972:~p.~393).}

But this sense of `fundamentality' in physics is not restricted to specific quantum field theory methods such as the renormalization group. In general, there is a sense in which a more fundamental theory has a more fundamental {\it set of (coupling) parameters}---often called {\it fundamental constants}. Part of the early success of quantum theory lay in its ability to make accurate predictions for experimentally determined parameters that appear in macroscopic theories (for example, the fine-structure constant $\a=e^2/2\e_0hc$, the Rydberg constant $R=m_e e^4/8\e_0^2h^3c$, the Stefan-Boltzmann constant $\s=\pi^2k^4_{\tn B}/60\hbar^3c^2$, etc.). The heuristic is that if we can predict such parameters from a microscopic theory, we have reached a more fundamental successor theory. In this sense, our current best theories contain a number of parameters ($G$, $c$, $\hbar$, etc.) whose values are determined experimentally, and are not predicted from other theories: hence the honorific `fundamental constants'.\footnote{In string theory, although Newton's constant can be derived in terms of the parameters of the string (see Section \ref{IIAsugra}), their measured values can still not be predicted.}
(The currently fundamental parameters can of course be expected to be predicted by a yet more fundamental theory.)

Finally, there is the notion of fundamentality as {\it dependence}, which again has two variants, one of which is logical and linguistic, and the other is ontological: 

(iii)~~In fundamentality as {\it definitional dependence}, A is prior to B iff the correct definition of B mentions A, while the definition of A is independent of B.\footnote{We take the `correct definition' to be the best or agreed definition, given a body of evidence.}
As we will discuss in Section \ref{fundaE}, Nagelian reduction establishes this type of fundamentality for linguistically formulated theories, because it reduces one theory to another by logically deriving the reduced theory from the reducing theory. Thus the reducing theory is, in this formal sense, more fundamental.

(iv)~~In its ontological variant, fundamentality is often defined as ontological dependence or {\it grounding}, in the broad sense that if A grounds B then B exists in virtue of A, or A accounts for or provides a metaphysical explanation of B.\footnote{For a discussion of various notions of grounding, see Bliss and Trogdon (2021:~Section 1.1), McKenzie (2022:~p.~9) and Peramatzis (2011:~pp.~6, 12-13, 23). Schaffer (2009:~pp.~354-355) contrasts Quine's {\it flat} ontology with the {\it ordered} ontologies that are generated by grounding relations. See also Wilson (2014:~p.~536).} 
Our discussion will not focus on this case.

\subsection{Priority and duality}\label{fundaD}

In view of the Schema's connection to various inter-theoretic relations that we discussed in previous Sections, it is natural to ask whether some of these enable a conclusion about priority along the lines of some of the four notions of priority discussed above. We will first ask this question for dualities and theoretical equivalence, and then (in the next Subsection) for emergence.

Do {\it dualities} give us reason to think that one model is more fundamental than another? Note that dualities, as formal equivalence relations, are symmetric and so do not establish a partial ordering in the strict sense---and so the general answer is {\it no}. However, {\it effective} dualities do give a strict partial ordering, and also interpretative considerations may give a strict partial ordering. 

Interpretation first: we distinguish cases of theoretical equivalence and inequivalence, i.e.~internal and external interpretations. Duals that are theoretically equivalent do not have a relation of priority, since theoretical equivalence is a criterion for two models to be formulations of one and the {\it same} theory---again, there is no strict partial ordering available. 

But in cases of theoretical inequivalence, one dual could be more fundamental than the other in a substantive sense. Thus we should ask: what could lead us to give priority, in some objective sense, to one dual over the other? If we had sufficient evidence that one model is superior to the other, in the sense that it gives a better description or a better explanation of the intended domain of application (either of its empirical or of its theoretical aspects), then one might take that to be a good reason to regard it as more fundamental. Thus for example, if it turned out that our universe is better described by the gravity side of a gauge-gravity duality than by the gauge side (e.g.~if a version of a gauge-gravity duality were true for a spacetime with a positive cosmological constant),\footnote{See Strominger (2001), Witten (2001), and Maldacena (2003a, 2003b). For an overview of the various proposals, see De Haro, Mayerson and Butterfield (2016:~pp.~1407-1410).} 
then one might take that to be a good reason to give it ontological priority. Thus the same reasons that may prompt the use of external interpretations of duals, may also suggest the priority of one dual over the other.

This kind of priority is clearest for {\it effective dualities}, where one side of the duality describes the real world (e.g.~a strongly coupled condensed matter system, such as a superconductor: see Section \ref{themesgg}) and the other side is an effective, auxiliary system that we use to carry out some otherwise intractable calculations (a four-dimensional black hole).\footnote{See also Horowitz (2011:~pp.~318-323). For a general introduction, see Zaanen et al.~(2015).} 
The real world is described by the three-dimensional quantum theory of the superconductor, which in this case is the fundamental theory; the gravity side is a mere calculational device. (For more details about effective dualities, in combination with emergence, see below).

In these examples, there is a clear ontological priority of one dual over the other (see case (4) of Figure \ref{UDworlds}, where in this case there is no relevant common core theory). The side of the duality that describes our world is ontologically more fundamental, while the other side is dependent on it, since it is a mere calculational device.

\subsection{Priority and emergence}\label{fundaE}

Does the appearance of ontological emergence, with its antisymmetry in the linkage relation between the bottom and top theories (see Section \ref{Sem}), give us reason to regard one theory as more fundamental than the other? We now argue that, in general, the answer to this question is `No', but that there is a special sort of case---a distinguished kind of linkage---where the `Yes' answer is admissible. Various examples will then illustrate this idea.

The general argument first: although ontological emergence is antisymmetric and thus gives a partial ordering between theories, this partial ordering does not support, i.e.~does not imply, a notion of {\it priority}, in any of the four senses of priority that we discussed in Section \ref{fundapp} (i.e.~mereology, scales and fundamental constants, definitional dependence or ontological dependence). Namely, the conception of ontological emergence is distinct from the notions of priority that we have discussed. One may of course find specific examples where the specific emergent behaviour falls under one of these types of priority, and in that sense one theory is more fundamental than the another: but these will be theory-specific reasons, rather than following from a purported general connection between ontological emergence and fundamentality. Or perhaps one can find a fifth sense of fundamentality, according to which ontological emergence is a fundamentality relation: but to avoid circularity, this should be supported by an independent argument.

We argue that it is incorrect to say that a top (emergent) theory is always less fundamental than the bottom theory from which it emerges, because an emergent theory has novel aspects relative to the bottom theory. As we will see in our examples below, whether a relation of fundamentality obtains depends on the target system described  by the two theories.

But if we consider a logically stronger linkage relation, then there {\it is} a sense in which the top theory is less fundamental than the bottom theory. The strengthening that we will add (the `special sort of case' above) is to take the linkage map to be {\it reductive}. Thus, as in Section \ref{eandr}, we consider `emergence and reduction'.\footnote{This case gives priority in the sense of definitional dependence. Other strengthenings can give the other senses of priority, but these will be theory-specific rather than general.}

Recall that we formalized reduction by requiring that the linkage map is surjective. If there is a syntactic construal of the theory that instantiates the entailment relation $T_{\sm b}\vDash T_{\sm t}$ (see footnote \ref{linguisticA} and Section \ref{lsr}), then the bare theory at the top can be {\it deduced} from the bare theory at the bottom---which is Nagelian reduction. Thus there is a definitional dependence between the two theories, in the following formal i.e.~non-interpretative sense: even though it may well be possible to formulate $T_{\sm t}$ without reference to $T_{\sm b}$ (e.g.~because the top theory has an independent set of axioms), the entailment relation from the bottom to the top bare theory means that the bottom theory is logically stronger, more detailed or specific, than the top bare theory. This greater specificity or detail is priority as definitional dependence: the bottom bare theory provides an appropriate theoretical underpinning of the axioms of the top bare theory, in that they are given more specificity, and perhaps they are unified, simplified or clarified, by the bottom bare theory. Thus there is a sense in which a `correct definition' of the top bare theory (see the sense of `correct definition' at the end of Section \ref{fundapp}: here, `correct' can be taken in the sense of `in view of the evidence that the bottom theory is the more accurate theory') requires understanding it through its linkage to the bottom theory.\footnote{That the bottom bare theory gives a fuller, more correct, definition of the top bare theory does not follow from the fact that it implies the top bare theory as a matter of logic, but requires an additional ontological condition: namely, that the target system of the two theories is the same, or at least that they are about the same topic, so that the bottom theory indeed gives more detail than the top theory about the same target system or topic, i.e.~it can say everything that the top theory says about the target system or topic, and more. The other epistemic qualifications, i.e.~that the bottom bare theory gives a more unified or simplified description of the same topic than the top bare theory does, requires saying more about how theories are formulated and understood by epistemic communities of scientists. We will expound our views on this topic in the next Chapter.}

Our first two examples illustrate combining priority with ontological emergence. Then we give an example of priority with weak ontological emergence.\\
\\
{\it Example: thermodynamics and statistical mechanics.} A good illustration of priority combined with ontological emergence that makes no reference to dualities is the relation between thermodynamics and statistical mechanics.\footnote{For a nuanced early philosophical analysis of this question, see Nagel (1949:~pp.~294-296, 303-307; 1961:~pp.~340-345; 362-365). For a critique of the idea that full reduction succeeds, see Howard (2007:~pp.~145-146). In particular, Howard argues that the ergodic hypothesis in general cannot be derived from statistical mechanics. For a philosophical discussion of the relations between thermodynamics and statistical mechanics, see Myrvold (2022).} 
Thermodynamics is an abstract (in the sense of Section \ref{abstraction}), high-level, top theory that applies to many systems, as long as they have quantities that correspond to temperature, energy, heat, entropy, etc.~that satisfy the laws of thermodynamics. Statistical mechanics is a more detailed theory, about systems of many (usually point) particles.\footnote{Thus we only in part agree with Dieks and de Regt (1998:~p.~58) that `[d]eeper theories, valid on deeper levels of reality, are more general; they have a wider field of application than the older theories and cover more phenomena. They are therefore better theories in a quite simple way: they better fulfil the central aim of science to develop theories that accord with as many phenomena as possible in the most accurate possible way'. While we agree that deeper theories are more accurate (we have here spoken about `specificity' and `detail'), usually for a greater variety of systems, we deny that this is always related to their being more general---since the notions of `accuracy' and `generality' are quite independent of each other. Deeper theories can be more general in the sense that they describe systems under a wider class of physical conditions (i.e.~their domain of application is larger than that of shallower theories), and in particular various kinds of linkage maps can apply to them: thus, as we have argued, they strictly speaking `say more' than the top theories to which they are linked. But saying more does not always mean being more general: top theories, by saying less, can often describe a wider class of phenomena, albeit in less detail. For example, the laws of thermodynamics (as a bare theory) apply equally to physical systems that statistical mechanics would differentiate, as well as to systems to which statistical mechanics does not apply in a direct way (for example, the communication of information, \`a la Shannon: notice that we are here considering {\it bare} theories). Another example is of course the universality of infrared fixed points under the renormalization group flow: many different bottom bare theories that describe systems of particles of various masses flow, under the renormalization group, towards the same infrared system. The detailed information (e.g.~about the masses of particles) is thereby lost, since the top bare theory only describes the massless sector of {\it all} the bottom bare theories that flow to it. It is its being less detailed that allows it to be more general---hence the label {\it universality} that is used for this kind of property of a top theory.} 
And there is a clear sense in which e.g.~the second law can be derived, under suitable assumptions, from considerations of the probability of states of a statistical mechanical ensemble, while the opposite is of course not true. Furthermore, some facts about thermodynamical systems that are not well explained, or not at all explained, by thermodynamics (such as the van der Waals equation, which corrects the Boyle-Charles ideal gas law at high densities, where the corrections depend on parameters that depend on the substance considered, and are determined experimentally), have an explanation in statistical mechanics, as resulting from the effects of the interactions between the molecules and their finite sizes (and so, in statistical mechanics one can calculate these parameters, by considering a non-idealized gas, and taking into account the non-zero size of the molecules). In other words, the word `correct' in the conception of definitional dependence (Section \ref{fundapp}) should not only be taken in a mathematical sense; correctness here means that epistemic or empirical arguments may give a {\it better} definition of the definitionally dependent term or axiom.\\
\\
{\it Example: gravity and string theory.} An example of emergence and priority, as in reduction, for {\it effective dualities} is the emergence of a theory of gravity as the effective dual of a quantum field theory. Although Einstein's theory of general relativity can be defined without reference to quantum field theory, it is likely that the quantum corrections can only be calculated by using the linkage map from the microscopic theory, so that the microscopic theory is required to define the top theory with quantum corrections. 

This is also the case for the quantum corrections to supergravity theory provided by string theory. The precise coefficients of these corrections can be constrained by, but cannot be calculated in, supergravity, while they can be calculated in the underlying string theory.\footnote{For example, Green (1999) shows that supersymmetry constrains the coefficients of the effective action of Type IIB supergravity to be modular forms with certain weights (depending on the dimension of the terms they multiply), as a function of the complexified Type IIB string coupling, under the $\mbox{SL}(2,\mathbb{Z})$ symmetry of type IIB supergravity (see Section \ref{S-d}). But this only constrains the {\it form} of these coefficients, i.e.~their functional dependence on the complexified coupling. To calculate the coefficient of each of these terms, supersymmetry and supegravity do not suffice, and a (microscopic) M-theory or string theory calculation is required. See also Tseytlin (2000) and D'Hoker, Freedman et al.~(1999).}

Thus we argue that, in so far as string- and supergravity theories can be understood as emerging from M-theory, M-theory is, in this sense, more fundamental. This is because, in the view from Section \ref{DSTov}, string- and supergravity theories are different limits of M-theory: and yet their ontologies (of quantum strings and classical point particles and classical fields, respectively) are different from the ontology of M-theory (which is still unknown, although various proposals exist: see Section \ref{11DM}). The bottom and top theories are related by linkage: see Eq.~\eq{R10}, where the radius of the eleventh dimension, related to the string coupling, is the parameter that we called $x$ in Section \ref{emergence}). Thus if string- and supergravity theories can be reduced to M-theory, then M-theory is more fundamental in the sense of scales {\it and} definitional dependence (Section \ref{fundapp}): namely, in the light of the M-theory developments, a correct future definition of these theories (specifically, the non-perturbative aspects of string theory) will require mentioning M-theory. 

This logical dependence also correlates with the inter-dependence of the various {\it scales}, as illustrated by the relation between the radius of the eleventh dimension and the string coupling, Eq.~\eq{R10} (and the relations between the other coupling parameters, in Eqs.~\eq{GNewton}-\eq{D0mass}, i.e.~Newton's constant, the Planck length, the string constant) and other physical quantities, such as the masses of states. For the linkage map links: (i) the eleven-dimensional spacetime, to (ii) the effectively ten-dimensional spacetime that is probed by the string. And so, this is also a case of fundamentality relative to physical scales, as given by each theory's fundamental constants and other parameters.\\
\\
{\it Example of priority with weak epistemic emergence: bosonization.} Section \ref{3op} discussed weak epistemic emergence as another interesting sort of emergent behaviour, i.e.~emergence in practice, but not in principle. Bosonization (and its generalization, sine-Gordon-Thirring duality) was an important example of this, and in view of the duality relations between the bosons and the fermions (see Eq.~\ref{expf}), a natural question is whether this can be seen as a case of fundamentality in the sense of mereological priority. 

There is a prima facie plausible line of thought, that a boson is made out of two fermions, because a bosonic operator is quadratic in the fermions, and so that the fermions are mereologically prior to the bosons: and the other way around, that a fermion is a coherent state of bosons, and in that sense the latter are mereologically prior. In other words, the prima facie plausible line of thought is that {\it what we regard as elementary or as composite is model-relative}.

We basically agree with this line of thought, provided we interpret it as an epistemic statement, i.e.~not as a statement about the nature of bosons or fermions, but about how our {\it models} describe them, in various regimes of validity.\footnote{The reason to be cautious about an ontological interpretation in terms of mereology is that, as before, if we interpret the duals as theoretically equivalent, then there is no strict partial ordering. Priority is only possible if duals are taken to be theoretically inequivalent, or if one considers an effective duality: for example, in bosonization in three or four spacetime dimensions. See also the discussion of Castellani (2017) in Section \ref{contcons}. One way that this model-relative priority could be in the world is if the kinds of situations, in which e.g.~the boson is elementary and the fermion is composite, are the only ones that occur in a given domain, while the opposite situation never occurs. But this in effect amounts to the physical inequivalence of the duals.} 
The idea of weak epistemic emergence is of course that, given the values of the physical parameters (e.g.~for low values of the energy as defined in one of the two models), it is easier to work with one model than with the other: the bosonic model, say, so that we consider the boson as elementary and the fermion as composite (and likewise for other soliton-particle dualities). And so, in that epistemic situation, the bosonic model and its interpretation have priority, but in an epistemic sense, i.e.~as approximations introduced by the model itself, and not in the sense of ontological priority.\footnote{For the new options and relations between priority and emergence that this introduces, in particular the notion of fundamentality at a given scale, and the independence of fundamentality and emergence, see Castellani and De Haro (2020).} 

This is seen from the additional axioms, in Section \ref{lsrr}, that we add to link the common core theory to a specific classical model. That this is model-relative is clear from the fact that, when we go to a different regime of parameters (e.g.~to high energies) the original model loses applicability and practical predictive power. In that case, a dual description emerges which is now readily applicable, and in this regime of parameters it is more convenient to regard the fermion as elementary, and the boson as composite. However, if one takes these two models to be theoretically equivalent, then these descriptions in terms of `elementary' or `composite' fields are part of the theory's descriptive apparatus: they do not have a correlate in the common core of the two duals. Thus they are aids that are suitably added to the models for certain tasks, like the calculation of expectation values of operators using Feynman diagrams. 

This discussion carries over to the particle-soliton dualities that we discussed in Part II, especially regarding the spectrum of BPS states of ${\cal N}=4$ SYM theory, and D-branes in string theory (for example, the exchange of the fundamental string and the D1-string). Also in those cases, the applicability of `particle' and `soliton' is model-relative and a matter of convenience of the description. But this should not be taken as an ontological statement, but rather as an epistemic one. For on an internal interpretation, there are no purely `elementary' or purely `solitonic' objects, and so neither of the two is more fundamental.

This type of priority has been dubbed by Castellani (2017:~p.~108) {\it epistemic, representational} priority (see also Section \ref{comparison}). It resonates with Coleman (1985:~p.~252), who famously declared that:
\begin{quote}\small
There is no way of deciding whether the fermion is fundamental and the boson a bound state or the boson is fundamental and the fermion a quantum lump. One is the natural way of putting things if one is describing the massive Thirring model and the other is the natural way of putting things if one is describing the sine-Gordon equation, but these two theories define identical physics. Which you choose to use is purely a matter of taste.
\end{quote}

\section{Conclusion}

This Chapter has explored the compatibility of duality with three practical functions that the literature associates with dualities, namely for finding: a successor theory, emergence, and fundamentality. Each of these notions is either {\it incompatible} with duality, or it is {\it not logically implied}, or even suggested, by duality. Rather, making them compatible requires philosophical work. This is not surprising, given that the preambles of Chapters \ref{Theor} and \ref{Heuri} already announced that the practical functions would `look beyond' satisfying the Schema (i.e.~beyond the idea of models that instantiate a precise duality).

But this tension between duality, successor theory, emergence, and fundamentality is interesting, because it runs against the gist of several suggestions in the physics literature, some of which we have discussed. For clarifying the language and the compatibility between concepts is one of the jobs of philosophy. Thus this Chapter has emphasised that, while the notion of duality is prima facie in tension with these other three ideas, the notion of an {\it effective duality}, which is a quasi-duality rather than a duality, is compatible with all of them, and does naturally suggest them. Thus while the literature often associates these practical functions with the idea of duality, it is {\it effective dualities} and quasi-dualities that it usually has in mind when the practical functions are at play.

Alternatively, there are several ways in which duality can be made compatible with these notions: so that the Schema gives an interesting classification of some salient cases.

M-theory is the main example of the search for a successor theory, with three ways in which it can be understood: searching for a common core $T$, a successor theory $T_{\tn S}$, or {\it both} a common core and a successor theory. An influential idea that bears on the conception of a (successor) theory is what we have called the {\it geometric view} of theories, where a theory is a set of models with topological and geometrical structures on it. The various regions of the manifold are parametrized by {\it moduli} $\f_i$, i.e.~operators whose expectation values function as (order) parameters that distinguish phases of the theory, i.e.~its models. 

The Schema suggests a notion of {\it ontological emergence} that is a conjunction of two conditions (see Section \ref{emergence}, and Figure \ref{Oemergence}): (i) an asymmetric linkage map (of levels, scales or layers) between bottom and top bare theories, and (ii) ontological novelty, formalized as a non-commuting diagram between the linkage and interpretation maps with domains of application as their codomains.

{\it Reduction} is formalized as a linkage map that is surjective, so that all the states and all the quantities of the top bare model correlate with those of the bottom bare model. This notion of reduction meshes not only with the Nagelian tradition of reduction, but also with current physical practice, which almost always takes reduction to be a formal relation between theories, i.e.~a relation in mathematical physics (so that the possibility of deduction, regardless of ontology, usually suffices for physicists to talk about reduction).

These conceptions explain how emergence can be compatible with reduction: namely, emergence is an interpretative notion (i.e.~it has both (i) formal i.e.~non-interpretative, and (ii) ontological components), while reduction is formal i.e.~non-interpretative: it is a notion in the realm of mathematical physics. 

We argued that duality is incompatible with irreducible emergence (because `irreducible' means that the top model cannot be reduced to the bottom model, i.e.~there is no reducing linkage map).\footnote{An argument to this effect was first given in Dieks et al.~(2015).} 
The reason is that duality requires that the models are equinumerous, while irreducibility requires a linkage map between models that is not one-to-one. 

There are three natural ways to combine duality and emergence, all of which are illustrated by examples in physics: (1) emergence beside duality (e.g.~as in AdS-CFT, where a new dimension on the AdS side represents the renormalization group flow in the CFT); (2) effective duality; and (3) epistemic emergence.

Section \ref{fundam} discussed four types of fundamentality (mereological, priority based on physical scales or parameters, definitional dependence, and ontological dependence), none of which are in general exemplified by dualities. But we discussed two ways in which duality {\it can} lead to fundamentality: namely, when the same reasons that may lead us to adopt an external interpretation also lead us to regard one model as more fundamental than its dual; and again for effective dualities, where one ``dual'' is more fundamental than the other.

Finally, about the possibility of combining fundamentality and emergence (regardless of dualities): although emergence does not by itself give a relation of fundamentality, there is fundamentality in cases of reductive emergence (e.g.~in the relation between thermodynamics and statistical mechanics, or in renormalization group approximations).

\chapter{Understanding and Explanation}\label{Understand}
\markboth{\small{\textup{Understanding, Explanation and Spacetime}}}{\textup{\small{Understanding, Explanation and Spacetime}}}

The previous Chapter discussed the heuristic roles that duality plays in three types of inter-theoretic relations: (i) the relation between a successor theory and its set of duals, (ii) relations of emergence, and (iii) relations of fundamentality or priority. These functions go beyond the satisfaction of the Schema, but as we also saw, the Schema gives an interesting classification of some salient cases.

This Chapter discusses the fourth role of dualities: namely, providing scientific understanding and explanations. We will focus on dualities in string theory and quantum gravity, which, as we saw in Chapters \ref{String} to \ref{HABHM}, have sparked much of the recent interest in dualities.

First, Section \ref{understanding} first discusses explanation and understanding in philosophy of science. While the former is a familiar (well-trodden!) topic, the surge of interest in the latter is much more recent. Section \ref{pragmatic} introduces {\it pragmatic} theories of understanding, where `pragmatic' here concerns `use'. In particular, we discuss de Regt's (2017) theory of understanding, which Section \ref{UEwST} uses to answer the question of how theories of quantum gravity (which often purport to have no spacetime at the fundamental level) can be intelligible, i.e.~how they can give us understanding. Section \ref{DandU} expounds how the Schema's tripartite classification of the interpretation of dualities (in terms of physical inequivalence of duals, physical equivalence of duals, and a successor theory beyond the duals) bears on scientific understanding. 

In the first three Sections, our use of `models' will be the usual one in the philosophy of science: namely, we will follow either the model-theoretic, or the semantic conception, of `model'. We will revert to the Schema's use of `model' in Section \ref{DandU}, which is specifically about dualities.

\section{Understanding and explanation in philosophy of science}\label{understanding}

Among the logical empiricists, an influential view, espoused by for example Hempel (1965), was that {\it understanding} is a subjective activity that is of interest to psychology, but not to the philosophy of science.\footnote{Recently, Trout (2002:~p.~213; see also 2005, 2007) has discussed a psychological variant of understanding that he dubs the `sense of understanding' (e.g.~as in a {\it Eureka!}-type of experience) and that he argues is the result of well-known cognitive biases, such as overconfidence and hindsight, and therefore not in any sense a cue to genuine understanding nor an indicator of truth. However, none of the recent accounts that we are aware of construe understanding along the lines of Trout's `sense of understanding'.} 
For example, in a reply to Scriven, he writes that `such expressions as `realm of understanding' and `comprehensible' do not belong to the vocabulary of logic, for they refer to the psychological or pragmatic aspects of explanation' (Hempel, 1965:~p.~413).\footnote{Not all logical empiricists downplayed the place of understanding in the philosophy of science. Notably, the last Section of Carnap's (1939) {\it Foundations of Logic and Mathematics} bears the title {\it ``Understanding'' in Physics}: and the scare quotes warn us that this vague and ambiguous word requires philosophical analysis (i.e.~it requires an explication: see footnote \ref{explication} in Chapter \ref{Intro}). Thus Carnap distinguishes two senses of `understanding' of an expression, a sentence, or a theory: first, the `capability of its use for the description of known facts or the prediction of new facts'. This conception of understanding as `the ability to use' he admits as {\it legitimate}, while he rejects the idea of an ``intuitive'' understanding of e.g.~the axioms of a theory, or of physical magnitudes and quantities like the electric field and the wave-function. Also in his (1947), the notion of `understanding' plays an important role: it often means something like `knowing the truth conditions', and also something like `the ability to use'.}

More moderately, the view was that {\it explanation is understanding enough}, because the notion of explanation already does all the work that we wish it to do, and so there is no need for a separate philosophical theory of understanding\footnote{A sophisticated version of this view is held by Khalifa (2012:~p.~17, 2017:~p.~14). In extreme brevity: understanding is a form of knowledge, reducible to explanation and cognate notions; and this is compatible with his pluralism about explanation. Thus although Khalifa's views about understanding are close to those of the logical empiricists, he has, by contrast, developed them in to a theory of understanding.} 
(even though Hempel (1965:~p.~337) admits that explanation `enables us to {\it understand why} the phenomenon occurred', thus in effect conceding that understanding {\it why} the phenomenon occurred is a valid topic).\footnote{The topic of {\it explanation} was initially also regarded with suspicion by the logical positivists. For since Aristotle, explaning a phenomenon meant `citing the causes' for that phenomenon: for example, Moravcsik (1974:~p.~3) argues that Aristotle's theory of {\it aitiai} is a theory about the structure of {\it explanations}. And the post-Kantian and post-Hegelian German idealist tradition explicitly sought the metaphysical causes behind the phenomena (for a discussion, see Carnap, 1966:~p.~12, and Salmon, 1989:~p.~4). Thus whether one considers {\it aitiai} or causes: explanations smack of metaphysics, which early logical positivism wished to eliminate from the realm of scholarship. Also Duhem (1991 [1914]:~p.~10) argued that including explanations among the aims of science is giving hostages to fortune: it subordinates science to metaphysics, thus threatening the autonomy of science. Nevertheless, the ever-open-minded Carnap rejected Duhem's specific claim about explanation, in so far as he admits explanations that involve empirical laws, as did Hempel.} 

In Hempel's (1965:~pp.~337-338) highly influential `deductive-nomological' model of explanation,\footnote{More generally, this is also called the `covering-law account of explanation' which, besides its deductive-nomological variant, also comes with a statistical variant, in two versions: deductive-statistical and inductive-statistical explanation (Hempel, 1965:~pp.~380-382).} 
an {\bf explanation} is an objective relation between a theory and a phenomenon, independent of the agent's subjectivity: namely, an explanation is a {\it deductive argument} in which the occurrence of the phenomenon to be explained (dubbed the `explanandum') is deduced from appropriate premisses (the `explanans'): namely, general laws of nature, and particular circumstances (i.e.~appropriate temporal or spatial conditions).\footnote{The deductive-nomological model was subjected to numerous counter-examples and criticisms (for a discussion of the argument and an overview of the debates, see Salmon, 1989; Woodward, 2021; and Curd and Cover, 1998: Chapter 6). Thus, according to Friedman (1974:~p.~9), Hempel himself would admit that the model `provides at best necessary conditions for the explanation of particular events'. Alternative models were developed that emphasized different aspects, some of them better applicable outside the physical sciences: such as statistical relevance, causal explanations, mechanisms, unification, various sorts of manipulation and intervention, etc. Yet other theories emphasize the `pragmatic' aspects of explanation, i.e.~its making reference to contextual factors, the intentions of agents, and the scientific community. According to van Fraassen (1980:~pp.~141-157), an explanation is an answer to a {\it why?}-question, where what counts as an adequate explanation crucially depends on the context (e.g.~on the interests of the person who asks the question). Indeed, according to van Fraassen (1977:~pp.~149-150), `science contains no explanations ... Explanation is indeed a virtue; but still, less a virtue than an anthropocentric pleasure'. According to Achinstein, an explanation is a certain kind of linguistic performance, an illocutionary act, i.e.~something that people do in certain contexts, and with certain intentions. Both van Fraassen and Achinstein emphasize the continuity between explanations in science and other kinds of explanations. For an exposition and critique, see Salmon (1989:~pp.~135-149).} Thus both views (Hempel's eliminativist view in his reply to Scriven, and the more moderate reductivist view) agree that understanding is not a topic of primary interest for the philosophy of science. On the former view, understanding is {\it pragmatic}, here used in the sense of `subjective' (not our sense!), i.e.~having to do with scientists' personal interests and preferences, rather than with science as the endeavour to acquire objective and universal knowledge about the world.\footnote{The theme of the unity of science, in its several variants, was an important motivation for the work of the logical empiricists, most notably Oppenheim and Putnam (1958:~pp.~3-5), and Carnap (1995 [1935]:~p.~32). A modern development of this view is in Kitcher (1981:~p.~507). For a general introduction, see Cat (2017:~Section 1.4).} 
In other words, by pursuing an investigation of understanding, we may gain knowledge about people's cognitive preferences and interests; but such an investigation is not a reliable way of gaining knowledge about the topic that those preferences and interests are about.\footnote{This view is developed in detail by Trout (2002, 2005, 2007).}

Bas van Fraassen is a prominent contemporary philosopher who, for different reasons, holds something in the neighbourhood of Hempel's line. We say `in the neighbourhood of', because van Fraassen does not relegate the pragmatics of explanation and understanding to the realm of psychology: in fact, pragmatics is an important element of his philosophy of science (and he devotes to it an entire chapter of {\it The Scientific Image}).

But, for van Fraassen, explanation and understanding are part of the pragmatic dimension of science, which he {\it contrasts} with its epistemic dimension. More precisely, explanation is part of the reasons that we may have to {\it accept} a theory as useful for particular purposes (e.g.~to make predictions, and to answer questions), but it does not add anything to our beliefs about the relation between the theory and the world. Thus to explain is not an epistemic aim of science: it is part of the pragmatic dimension of theory acceptance.

Friedman (1974:~pp.~7-8) long ago, and more recently Woodward and Ross (2021:~Section 6.1), highlighted an equivocation in Hempel's use of the word `pragmatic' (which we also recognize in van Fraassen). Indeed, we can (indeed, should!) distinguish, `pragmatic' as in `subjective', from `pragmatic' in the broadly Wittgensteinian sense of `use(able) for certain aims'. For a number of recent authors have argued that this latter sense, i.e.~pragmatic as in `use(able)', is compatible with an objective (suitably inter-subjective) notion of understanding.\footnote{We take `objectivity' here in the sense of Friedman (1974 p.~14): `what counts as an explanation should not depend on the idiosyncracies and changing tastes of scientists and historical periods. It should not depend on such nonrational factors as which phenomena one happens to find somehow more natural, intelligible, or self-explanatory than others'. Indeed, the objectivity that is relevant to scientific understanding has nothing to do with idiosyncracies and tastes, or with non-rational factors. Also Schutz and Lambert (1994:~p.~66) write that `scientific understanding is an intersubjective (objective) notion and independent of the psychological features of given persons'.\label{objectivity}} 
Thus they have taken understanding to be an epistemic aim of science, that is achieved by giving appropriate explanations.\footnote{Dieks and de Regt (1998:~p.~51; 2005:~p.~142) distinguish the aim of understanding from the aim of prediction thus: an oracle that makes perfectly true predictions gives no understanding, since it offers no insight into how the predictions are made. We discuss how understanding differs from explanation in Section \ref{pragmatic}.} 

We will endorse this consensus.\footnote{Authors who endorse this consensus include Friedman (1974:~pp.~18-19), Salmon (1978:~p.~684), Kitcher (1981:~pp.~509, 529; 1989:~p.~419), Lipton (2004:~p.~30), de Regt and Dieks (2005:~p.~150), Grimm (2010:~p.~337), and Reutlinger et al.~(2018:~p.~1081). Khalifa (2017:~pp.~16-20) discusses `the received view of understanding', and the main objections that have been leveled against it.\label{UthroughEx}}
Thus, for example, Salmon (1978:~p.~684) writes:\footnote{There is a pinch of irony in Salmon's mention of `high-sounding goals'. Since Salmon himself endorses the idea that science aims at understanding, and that this is achieved through explanations, we take the reason for the irony to be that he is critical of what he dubs ``the received view'' of explanation, i.e.~Hempel's deductive-nomological model. Instead, he defends a causal-mechanical account of explanation.} 
\begin{quote}\small
Science, the majority say, has at least two principal aims---prediction (construed broadly enough to include inference from the observed to the unobserved, regardless of temporal relations) and explanation. The first of these provides knowledge of {\it what} happens; the second is supposed to furnish knowledge of {\it why} things happen as they do ... It is now fashionable to say that science aims not merely at describing the world---it also provides {\it understanding}, {\it comprehension}, and {\it enlightenment}. Science presumably accomplishes such high-sounding goals by supplying scientific explanations.
\end{quote}

But a number of questions arise: (i) What is understanding, and how does it differ from explanation? (ii) What makes understanding a pragmatic notion?

\section{Pragmatic theories of understanding}\label{pragmatic}

Section \ref{dRU} first answers the two questions above by discussing three general characteristics that are common to various theories of scientific understanding. In particular, we focus on de Regt's account, which Section \ref{UEwST} will use to clarify how understanding without postulating a fundamental spacetime can be had, and how dualities give scientific understanding. To that end, Section \ref{UaS} first identifies four points where the Schema meshes with de Regt's account. 

\subsection{De Regt and the pragmatic theories of understanding}\label{dRU}

Reutlinger et al.~(2018:~p.~1081) note that the account of understanding by de Regt and Dieks (2005) `is, in several respects, a typical account of scientific understanding'. And then they go on to give three {\it common requirements} of these accounts of understanding, which we will here characterize as follows:\footnote{Some authors who endorse the general idea of `understanding through explanations' are listed in footnote \ref{UthroughEx}. Some recent authors for whom understanding involves nothing beyond knowledge include Lipton (2004:~p.~30) and Khalifa (2012:~p.~17; 2017:~p.~14).}\\
\\
{\bf (1)~~Explanation:} A requirement for the scientific understanding of a phenomenon is that there is a {\it scientific explanation} of this phenomenon. While different philosophers favour different accounts of explanation, de Regt in particular endorses a weakening of the deductive-nomological model of explanation. Namely, an explanation is an appropriate argument that links the occurrence of a phenomenon (the explanandum) to appropriate background knowledge (i.e.~a theory, and other assumptions). 

There are two reasons why this is said to be a {\it weakening} of the deductive-nomological model of explanation. First, while in the deductive-nomological model of explanation, the explanandum is logically deduced from the explanans, de Regt follows the literature of models in allowing {\it arguments}, rather than formal deduction, to give the right relation between the theory and the occurrence and phenomena.\footnote{This first point is hardly a weakening of the received view of explanation, since the logical empiricists themselves did not always see `deduction' strictly as `derivation in a formal logic'. For, depending on the particular science involved, they allowed several degrees of formalization and even informality, as is clear from e.g.~Oppenheim and Putnam (1958:~pp.~21-23) and Nagel (1961:~pp.~349, 366). Thus formalization functioned as a {\it regulative ideal}, rather than as a condition that can in practice always be required.}
Second, de Regt follows the literature in allowing {\it models} to take on the function of bridge laws in enabling the argument to succeed. For models mediate between theory and phenomena, thus allowing interpretations of abstract theory in (usually idealized) models that resemble those phenomena. \\
\\
{\bf (2)~~Requirements of adequacy:} The theory is required to satisfy epistemic criteria of adequacy, which are usually linked to the theoretical virtues of the theories in question, or epistemic values.\footnote{See De Regt (2009:~p.~592). For a general discussion of values in science, see Kuhn (1977:~pp.~321-322).} 
De Regt singles out two that he argues are widely supported: the theory should be internally consistent, and it should be empirically well-confirmed.\footnote{These two criteria can already be found as part of Heisenberg's (1927:~p.~478) re-definition of `understanding' ({\it Anschaulichkeit}), in his 1927 uncertainty paper. See also Section \ref{wpd}.}
For a scientific realist, these features will be grounds for inferring that the theory is true or approximately true.\\
\\
{\bf (3)~~Useability:} It should be possible to use the theory to construct adequate explanations (Reutlinger et al.~(2018:~p.~1082) call this `epistemic accessibility').\footnote{Their formulation of the condition is that `If an individual scientist understands phenomenon P, then he or she has epistemic access to an explanation of P'.} De Regt (2017:~p.~40) requires that a theory be intelligible, where {\bf intelligibility} is `the value that scientists attribute to the cluster of qualities of a theory (in one or more of its representations) that facilitate the use of the theory'. Thus for example, the pronouncements of an oracle that makes perfect predictions are not an intelligible theory, because one does not know how the oracle arrives at those predictions: understanding requires a degree of knowledge of how the theory works. It requires that scientists can see what is ``under the hood''. Thus understanding `why $P$ occurs' requires that appropriate short-cuts and-or evaluation strategies are available for how the theory works and how it arrives at certain answers. 

De Regt analyses the role of the theory's useability in both justifying a given explanation and in formulating new explanations. Both require a theory that is readily useable by the scientific community, and de Regt provides a criterion for a theory to be intelligible and thus useable (see below).\\

Conditions (1) and (2) give an account of scientific explanation that is broadly Hempelian, and which we thus endorse: these two conditions regard an explanation as a specific kind of knowledge provision, i.e.~justified and true belief. Condition (1), that there is an explanation of the phenomenon, is the existence of a certain type of argument for some state of affairs in the world, i.e.~it is a belief regarding the relation between the theory and the world. Condition (2) requires that, for this belief to give knowledge, it must be adequate in an appropriate sense: it must be true (in any case about observables, and in a realist sense also about unobservables), and it must be justified by the empirical evidence and by its internal consistency, predictive power, etc.

Condition (3) is a rather different condition that distinguishes genuine scientific understanding from ``mere'' explanations (which need to fulfill ``only'' (1) and (2)). Namely, it requires that the explanation be epistemically available, or useable, by the scientific community. De Regt emphasizes the {\it pragmatic} nature of the useability of a theory, which in particular involves skills and judgment. In other words, this condition requires that it is possible for scientists to {\it do} something (namely, to use the theory), rather than requiring a belief. Thus de Regt's account, and others like it, goes against the idea that understanding can be reduced to a specific kind of knowledge, which was implicit in many traditional approaches to the philosophy of science and epistemology. For example, Lipton (2004:~p.~30) writes: `Understanding is not some sort of super-knowledge, but simply more knowledge: knowledge of causes.' The difference, then, with these older ideas about understanding lies in including condition (3).

To emphasize the objectivity of this condition, de Regt and Dieks (2005:~p.~139) distinguish three levels of analysis of the scientific community: `the macro-level of science as a whole; the meso-level of scientific communities; and the micro-level of individual scientists ... The three-level distinction reconciles the existence of universal aims of science with the existence of variation in the precise specification and/or application of these general aims'. Thus they regard understanding as a macro-level aim of science, with shared standards of intelligibility that are not universally determined for all of science, but admit variation across different scientific communities, i.e.~at the meso-level of a discipline, while still being objective, relative to the level of progress of a given discipline.\footnote{For the meaning of `objectivity', see footnote \ref{objectivity}.} 
Thus, although the strength of the standards of intelligibility can vary with contextual factors, the criterion of intelligibility itself does not change, since it is a macro-aim. 

Reutlinger et al.~(2018:~p.~1082) emphasize the fact that the various available accounts of understanding differ in how they fill in the details about the conditions (1) to (3), with (3) being the most significant one.\\

De Regt's (2017:~p.~92) {\bf criterion for understanding phenomena} is his actual definition of scientific understanding, which encompasses all three aspects above, and is as follows (`intelligibility' is here defined as in requirement (3) above): 
\begin{quote}\small
A phenomenon $P$ is understood scientifically if and only if there is an explanation of $P$ that is based on an intelligible theory $T$ and conforms to the basic epistemic values of empirical adequacy and internal consistency.
\end{quote}

De Regt also has a {\bf criterion for the intelligibility of theories}, which does not aim to give an independent definition of intelligibility, but rather is provided as an objective test of whether the theory is in fact intelligible: 
\begin{quote}\small
A scientific theory $T$ (in one or more of its representations) is intelligible for scientists (in context $C$) if they can recognize qualitatively characteristic consequences of $T$ without performing exact calculations
\end{quote}
(or, alternatively, if they can recognize these consequences without fully explicit theoretical argumentation). 

This criterion is attuned to the idea, in de Regt's definition of intelligibility, that a theory has `a cluster of qualities' that facilitate its use in recognizing qualitatively its characteristic consequences. And de Regt emphasizes the development of appropriate {\it tools} that thus facilitate the use of the theory, or tools for understanding. If such tools can be developed within a theory, then the theory is (rendered) intelligible. Thus much of the discussion has focussed on which tools are suited for which theories. We will take this up in Section \ref{UQG} for quantum gravity, after we discuss how this theory of understanding fits with our Schema.

\subsection{Understanding and the Schema}\label{UaS}

There are four aspects of the above description of understanding, and in particular of de Regt's theory, that mesh with the Schema's treatment of theories and models, in Chapters \ref{Thies} and \ref{Theor}.

First, de Regt (2017:~pp.~97-98) emphasizes that the use of theories is pervasive in science, and that purported examples of ``models constructed without theory'' either do involve theory after all, or indeed fail to provide understanding. In this vein, he notes that the kind of theory-independence of models that has been defended by e.g.~Su\'arez and Cartwright (2008:~p.~66) and Morrison (2007:~p.~198) is less radical than it appears, and that the main target of their critique is not the use of theory per se, but rather the stronger view that models that explain specific phenomena {\it follow deductively} from theories. This meshes with our own defence of `theory', in Section \ref{Ourthm}, as a useful and important notion in all of science.\footnote{See also our discussion of French's `eliminativism' about theories, in item (1) of `Theory' in Section \ref{Ourthm}.} 
But also, models {\it do not in general follow} from theories, but the other way around: theories are abstractions that are often {\it obtained from} models (see Eq.~\eq{matim}). Thus the theoretical function of duality aims to construct a theory from the dual models, and the heuristic function aims to construct a successor theory.

Second, dualities motivated our specific usage of `theory' and `model', with theories being more general and models being more specific formulations, or versions, or realizations of a theory. Thus we lifted the use of `model' ``one level up'', which marks the Schema's flexibility in what it calls `theory' as against `model'. 

Section \ref{lsps} distinguished between model-theoretic semantics and physical semantics---both of which are used by the Schema. The model-theoretic semantics is a formal view of theories and models that allows us to discuss much important work in mathematical physics, and it was crucial to the logico-semantic discussions of Chapter \ref{Theor}. On the other hand, the physical semantics is not necessarily formal(ized). This distinction leads in to the third and fourth aspects where de Regt's treatment of scientific understanding meshes with the Schema.

Third, because the physical semantics is not necessarily formal(ized), it often plays the role of `models' as mediators between theory and phenomena, that we discussed in Section \ref{thmscph}. That `mediation' is what we have called an interpretation {\it map} between a bare theory and its domain of application. Thus a `model' (not only in the mathematical and logical sense, but also in the iconic or representational sense used in physics, e.g.~Bohr's model of the atom, or the liquid-drop model of the atomic nucleus) can be used to link a theory to the empirical world and thereby make predictions. And such a model can be used to give a map from theory to world, in the sense of Section \ref{itm}.\footnote{For further discussion of this point, see De Haro and De Regt (2018:~pp.~633-638).}

Fourth, while the relation between a bare theory and the model-theoretic semantics is, almost always, one of representation in the mathematical sense, the relation between a bare theory and its physical semantics is in general {\it not} one of mathematical representation or of deduction (the constraints were discussed in Sections \ref{lsps}, \ref{lessons}, and in the previous two Chapters). And it is the physical semantics that is most relevant for the questions of explanation and understanding. And this of course agrees with de Regt's views on the relation between theory and models as not always being one of deduction.

\section{Understanding and explanation---even without spacetime?}\label{UEwST}

This Section takes up the question of the intelligibility of theories of quantum gravity, which purportedly have no spacetime at the fundamental level. Section \ref{UQG} discusses this question in general. Then the main point of Sections \ref{visualizing} and \ref{3T} is to show {\it how} understanding can be had in theories without a spacetime: namely, through the development of specific tools for understanding. Thus these Sections discuss four tools that are available to make such theories intelligible, including in cases of dualities. The first of these tools (Section \ref{visualizing}) involves visualization in space(time), but the other three (Section \ref{3T}) do not require visualization. 

\subsection{Understanding in quantum gravity}\label{UQG}

At the end of Section \ref{dRU}, we discussed the idea of {\it tools for understanding}, which make theories intelligible, i.e.~the existence of such tools is a sufficient condition for theories to be intelligible. Thus if a theory admits the development of tools that facilitate its use, according to de Regt's criterion for the intelligibility of theories, the theory is intelligible, and can provide understanding of phenomena. 

In this Section, we focus on theories of quantum gravity, and the claim, by both philosophers and physicists, that at least some of these theories fundamentally lack a space-time framework. For example, Huggett and W\"uthrich (2013:~p.~277) analyse the problem of {\it empirical incoherence}, i.e.~the idea that:\footnote{The idea of empirical (in)coherence is described by Barrett (1999:~pp.~116-117) thus: `We can only accept a physical theory on empirical grounds if we can explain, if it were in fact true, one might have empirical justification for accepting it ... If a theory fails to be empirically coherent, it might be true, but if true, then one could never have empirical reason for accepting it as true.'} 
\begin{quote}\small
the truth of the theory undermines our empirical justification for believing it to be true. Thus, goes the worry, if a theory rejects the fundamental existence of spacetime, it is threatened with empirical incoherence because it entails that there are, fundamentally, no local beables situated in spacetime; but since any observations are of local beables, does not it then follow that none of our supposed observations are anything of the kind? The only escape would be if spacetime were in some way derived or ... `emergent' from the theory.
\end{quote}
But if there is a problem of empirical incoherence, then {\it a fortiori} there is a more general problem of understanding---as their paper also suggests (pp.~276-277, 284).

They go on to classify several approaches to quantum gravity (including lattice approaches, loop quantum gravity, string dualities, and non-commutative geometry) that, in one way or another, deviate from the normal relativistic spacetime description, and thus are in principle liable to the accusation of empirical incoherence. 

While the {\it emergence} of spacetime is one plausible answer to the question about empirical incoherence,\footnote{To substantiate this answer, Huggett and W\"uthrich also discuss what are appropriate standards of what they call `physical salience', and argue that these standards should not be copied from our past theories, but that a theory of quantum gravity will have its own standards of physical salience, and that these standards will be vindicated by the empirical success of the theory. This discussion is very close to our own discussion of the requirements of adequacy, in Section \ref{dRU}.} this cannot be an answer to the question about the {\it scientific understanding} of theories that do not have a fundamental spacetime. Thus, although such a theory may provide an explanation of spacetime that is adequate according to the criteria in Section \ref{dRU} (and this adequacy is not undermined by the absence of local beables at the fundamental level), understanding requires asking how such a theory can be {\it used} to give explanations. More precisely, the question is what conceptual tools are available that render the theory intelligible.

There are three reasons why this question is important. First, understanding requires more than an explanation. For, while the emergence of spacetime explains how a theory of quantum gravity is empirically coherent, the question of understanding is not just about the spacetime of our (direct) empirical experience. Understanding quantum gravity requires tools that enable us to work with these theories: not just in the regime where spacetime emerges, but also at the fundamental level (i.e.~at the Planck scale). It would indeed be disappointing if theories of quantum gravity were black boxes that can be understood only at the level of the emergent spacetime, but not at the fundamental level. Theory development obviously requires that physicists can use the theory well beyond the observable regime. And, to that end, appropriate tools are required.

The second reason is the salience that physicists in general ascribe to spacetime models. For example, Feynman diagrams and Penrose diagrams are prevalent not just in the pedagogy of quantum field theory, but also in scientific research.\footnote{For an example in the context of dualities, see Section \ref{visualizing}. For a general discussion of the role of Feynman diagrams in scientific understanding, see de Regt (2014:~pp.~389-392), who in part draws from Kaiser (2005).} 
Thus if a theory has no spacetime at the fundamental level, and if tools for visualization in spacetime at that level cannot be developed, then the question of how such a theory can be understood becomes urgent.\footnote{For example, the Kantian challenge was posed by Butterfield and Isham (1999:~p.~25, our emphasis): `we should take note of the Kantian position that it is not merely very difficult to dispense with space and time in thought about the empirical world...: it is {\it downright impossible}'. They, in effect, go on to argue that it would be acceptable if the spacetime of our experience can emerge from the more fundamental non-spatiotemporal structure, but that this non-spatiotemporal structure remains outside Kant's realm of appearance. See also their (2004:~pp.~44-45).} 
This is surely behind Maudlin's (2007) worry about theories that lack fundamental local beables. It is not just a demand for an adequate explanation, but also for a minimal standard of intelligibility.\footnote{In De Haro and De Regt (2020:~p.~3125), we argued that `understanding-as-knowledge' accounts, i.e.~accounts where understanding is reduced to explanation, fail to make sense of the controversy between Maudlin and Earman over a `frozen' spacetime, in which no physical quantities change over time: while an account of scientific understanding that includes condition (3) can make sense of it. We also argued that the disagreements between Maudlin and Earman can be traced back to different expectations about---different standards of---intelligibility, i.e.~ultimately different ideas about what scientific understanding entails.} 

The third reason to focus on conceptual tools for understanding is that there are at present no direct data about quantum gravity phenomena (there are models, empirical and theoretical constraints, and some indirect testing).\footnote{For a philosophical discussion of the consequences of this lack of data for philosophical research, see Butterfield and Isham (2004:~pp.~36-41), and also Section \ref{friendf}. For an overview of string phenomenology in particle physics, see Ib\'a\~nez and Uranga (2012). For applications of holographic duality in condensed matter physics, see Zaanen et al.~(2015).} 
Given this lack of empirical data, it is crucial to have a clear sense of what other tools we have for making these theories not just abstract recipes for calculation, but useable theories---and this also benefits theory development. Thus in the face of the current lack of empirical evidence, it is crucial to have a clear view of the tools that physicists use for understanding.\footnote{For recent philosophical work on analogue experiments and simulations, and the question whether these can provide confirmation of Hawking radiation, and more generally cast light on black holes, see Dardashti et al.~(2017:~p.~56, 2019:~p.~1), Crowther et al.~(2021:~pp.~3703-3704) and Field (2022).}

There is an interesting analogy between the intelligibility of (quantum) gravity theories that lack appropriate notions of spacetime (and some of the debates that have taken place around this analogy), and early debates over the {\it Anschaulichkeit} of quantum mechanics.\footnote{For a discussion, see de Regt (2017:~p.~230-236) and De Haro and De Regt (2020:~pp.~3127-3129).} 
As we discussed in Section \ref{adcom}, Schr\"odinger (1926a:~p.~27) regarded having a spacetime framework as an essential quality of a theory: `we cannot really alter our manner of thinking in space and time, and what we cannot comprehend within it we cannot understand at all'. The lack of a spacetime picture in Heisenberg's matrix mechanics motivated him to develop his own wave mechanics, which, as it happened, was more succesful than matrix mechanics (Beller, 1999:~pp.~36-38).

But visualizability, or having a spacetime framework, is not essential for scientific understanding. Heisenberg (1927:~p.~478) proposed a modification of the notion of {\it Anschaulichkeit}, along the lines of de Regt's {\it criterion for the intelligibility of theories} from Section \ref{dRU}:

\begin{quote}\small
We believe [ourselves] to understand a theory intuitively [{\it anschaulich zu verstehen}], if in all simple cases we can qualitatively imagine the theory's experimental consequences and if we have simultaneosuly realized that the application of the theory excludes internal contradictions.
\end{quote}

In short, visualization is an important tool in physics, and it is not immediately clear that visualization is available for theories of quantum gravity that claim that their fundamental structures are not spatio-temporal. 
The next Section analyses how visualization can be used as a tool that renders the intelligibility of theories of quantum gravity.

\subsection{Visualizing quantum gravity}\label{visualizing}

The previous Subsection discussed how some theories of quantum gravity claim that spacetime is fundamentally absent, so that it is not immediately clear whether these theories can be visualized. Thus the situation is similar to the early days of quantum theory. In quantum gravity, either the spacetime is fuzzy, or spacetime has been replaced by some other fundamental quantum object altogether. We will argue that, in such cases, visualization is an important tool for understanding, because it aids the heuristic view of the theory (see the examples below). 

\begin{figure}
\begin{center}
\includegraphics[height=5cm]{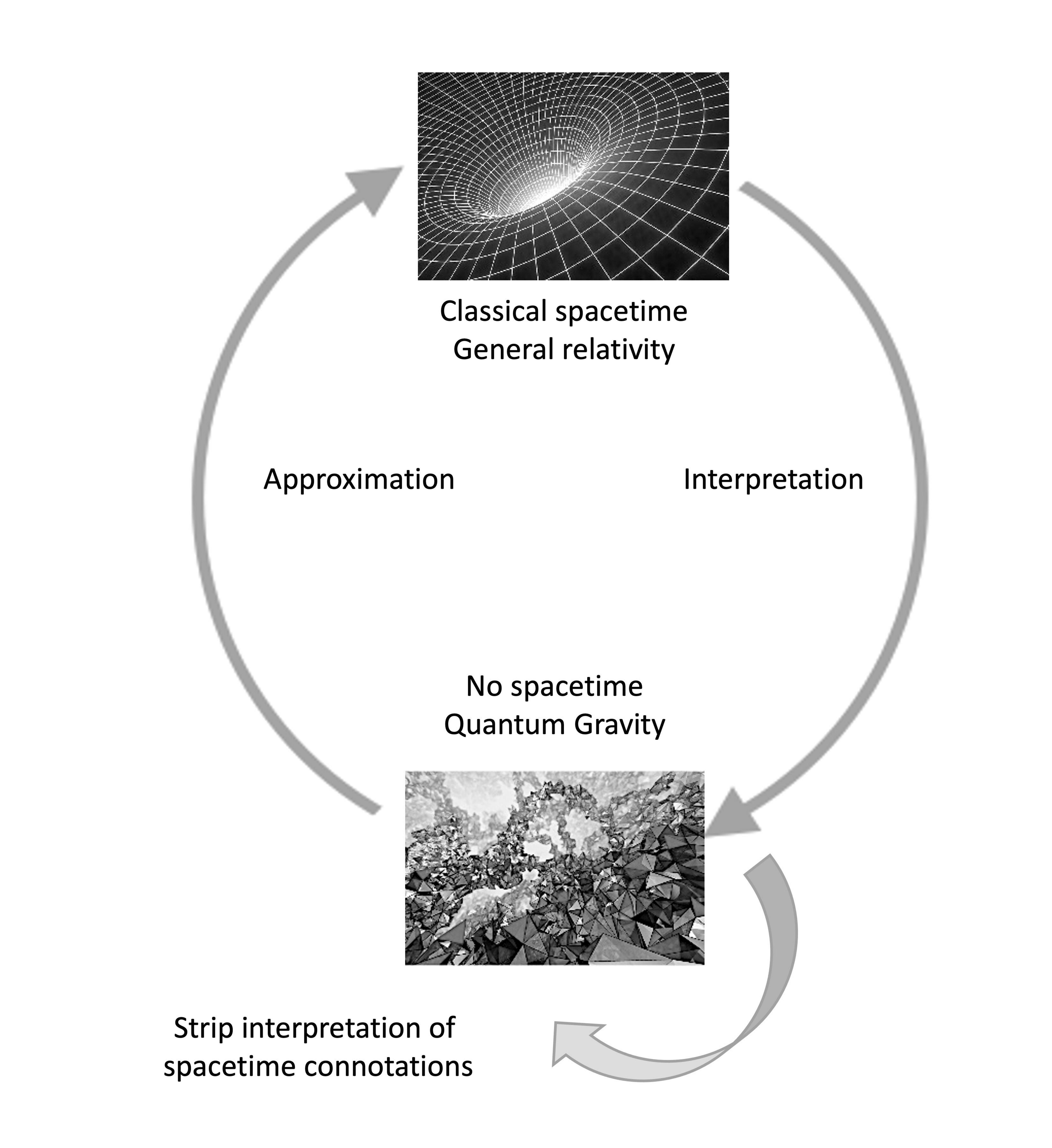}
\caption{\small Circle of approximation and interpretation, i.e.~iterations that are used in constructing and interpreting a theory of quantum gravity `with no spacetime'.}
\label{hcircle}
\end{center}
\end{figure}

There are two main strategies to visualize a theory that lacks a spacetime. 

(i)~~One often-used strategy is to derive a relativistic spacetime through a suitable approximation. For any theory of quantum gravity must of course reproduce the spacetime of general relativity in an appropriate approximation or limit.

Such visualization leads to understanding of quantum gravity through an interpretation of the terms of the theory, which then facilitates its use. In particular, we can develop a circle of approximation and interpretation, as in Figure \ref{hcircle}. We enter the circle at any point---at the bottom, say, where we have constructed a theory of quantum gravity that has no spacetime, and which we wish to interpret. If, through an appropriate approximation, we are able to (re-)derive a relativistic spacetime of general relativity, then we can use this approximation to interpret some of the terms of the bottom theory, by pairing the images of the approximation map with its pre-images. Then we can go on to interpret other terms in the bottom theory (also using the tools discussed in the next Section). We can also strip the interpretation of the bottom theory of (some of) its spacetime connotations, to get an interpretation that is more suited to a non-spatiotemporal theory. Two comments about Figure \ref{hcircle} are in order.

(a)~~Notice that the upward arrow of approximation, to get a relativistic spacetime, is in some cases a Nagelian reduction, i.e.~an `approximate deduction', or deduction of a {\it corrected top theory}, $T_{\sm t}^*$, rather than the original theory, $T_{\sm t}:=\mbox{`classical general relativity'}$ (see the discussion of Nagelian reduction in Section \ref{eandr}, in particular footnote \ref{Nagelc}). For example, if $T_{\sm b}=\mbox{`string/M-theory'}$, then $T_{\sm t}^*=\mbox{`supergravity as effective field theory'}$, i.e.~Eq.~\eq{S10}. The arrows shown in Figure \ref{hcircle} are bridge laws that connect the top theory and the bottom theory, similarly to our added axioms and signature elements in Section \ref{theoreq}.

(b)~~Although one might prefer to think of the bottom and top theories as `formulated and interpreted only once', the {\it iterative} procedure shown in Figure \ref{hcircle} reflects better the actual process of theory construction, which is our focus in this Chapter. The upward arrows indicate that one does not get finished theories in one step of approximation and interpretation. Rather, one gets fragments that one can improve and enlarge with each iteration, until (if the programme is successful) one gets finished theories. 

In the step of `stripping the interpretation of its spacetime connotations', one develops a more accurate theory of quantum gravity, e.g.~by distinguishing the fundamental from the non-fundamental degrees of freedom---something that Section \ref{sssm} discussed {\it in extenso} for M-theory. For dualities, `stripping the interpretation of its spacetime connotations' includes the process of getting an internal interpretation from (one, or more, historically given) external interpretations, along the lines discussed in Sections \ref{abstraction} and \ref{lsr}. For example, we discussed in Section \ref{SchemaT} that, for T-duality, the internal interpretation strips states of their (external) interpretations as `pure momentum' or `pure winding'. Likewise, in AdS-CFT, the radial direction of the AdS spacetime is stripped of its interpretation as the `radial direction in an AdS spacetime'.

So far, so clear. And this procedure brings out one defect of the use of spacetime visualization for interpretation: namely, spacetime visualization is not enough to interpret the bottom theory (and so, an interpretation that only uses visualization may in fact be inadequate). For the intelligibility of the theory requires further tools for understanding that we will discuss in the next Section.

(ii)~~A second strategy for visualizing a theory that lacks a spacetime is to derive other effective spacetimes (relativistic or not), which are {\it not} interpreted as spacetimes of experience, but nevertheless can be used for visualization and to render the theory intelligible: usually, by suggesting mathematical reformulations of the theory that aid qualitative reasoning and allow performing difficult calculations. Thus scientists can recognize qualitatively characteristic consequences of the theory without performing exact calculations---thus fulfilling de Regt's criterion for the intelligibility of theories. 

\begin{figure}
\begin{center}
\includegraphics[height=2cm]{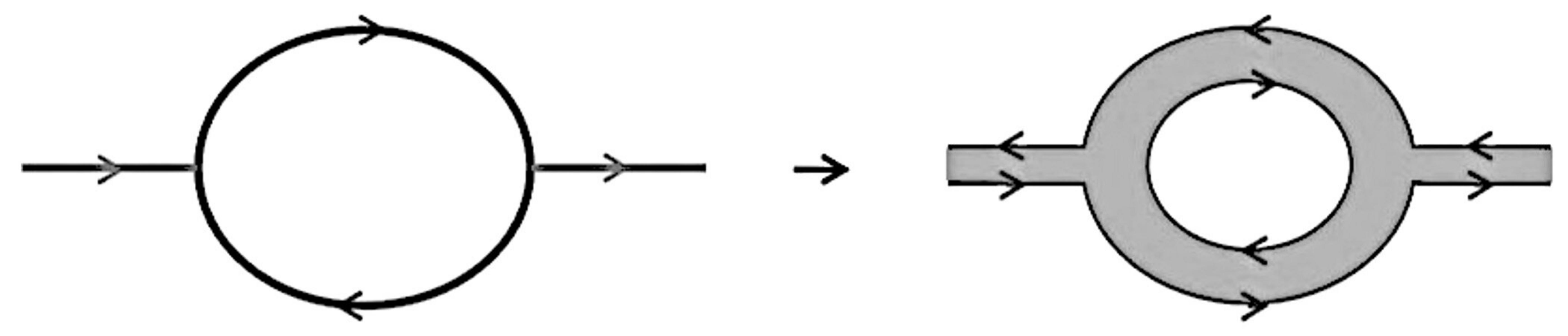}
\caption{\small 't Hooft's thickening of Feynman diagrams for point-particles, into string diagrams.}
\label{tHplanar}
\end{center}
\end{figure}

An example of this is the reformulation of Yang-Mills theory at large $N$ that enabled the idea of gauge-gravity duality from Section \ref{ggd}.\footnote{This is an example where one spacetime (the flat spacetime of Yang-Mills theory) is transmuted into another spacetime (that of the string). Several detailed examples where a spacetime emerges from a non-spatiotemporal situation are discussed in De Haro and De Regt (2018:~pp.~654-668).} 
This is a modern realization (with supersymmetry added) of 't Hooft's (1974a) idea of relating the theory of gluons to a theory of strings (i.e.~a field theory in $(1+1)$ dimensions) by taking the {\bf large $N$ limit} (see Section \ref{ggd}). First, he took the colour charge, $N$, of the gluons to be large, and the interaction strength $g$ small. Then he thickened the Feynman diagrams of the point particles to reflect the fact that the properties of gluons, at large $N$, go over into string properties. By this thickening of the diagram as in Figure \ref{tHplanar}, he was able to reinterpret a theory of point particles as a theory of strings. In the limit $N\rightarrow\infty$, only diagrams that can be drawn on the plane (so-called `planar diagrams') give a non-zero contribution.\footnote{Bouatta and Butterfield (2015:~p.~73-80) discuss the emergence of the planar properties of both QCD and super-Yang Mills theory, as well as some of their special properties, in the large-$N$ limit. At (p.~67-69) they also discuss the mathematical rigour of these theories, and conclude that they do exist.}

But---and this is a key step---he was able to use this new interpretation in terms of strings, to characterize the behaviour of the theory also {\it away} from the $N\rightarrow\infty$ limit. Namely, the original Feynman diagram expansion of the quantities of the theory could be reorganized as a topological expansion, of the type that one gets in string theory.

\begin{figure}
\begin{center}
\includegraphics[height=1.5cm]{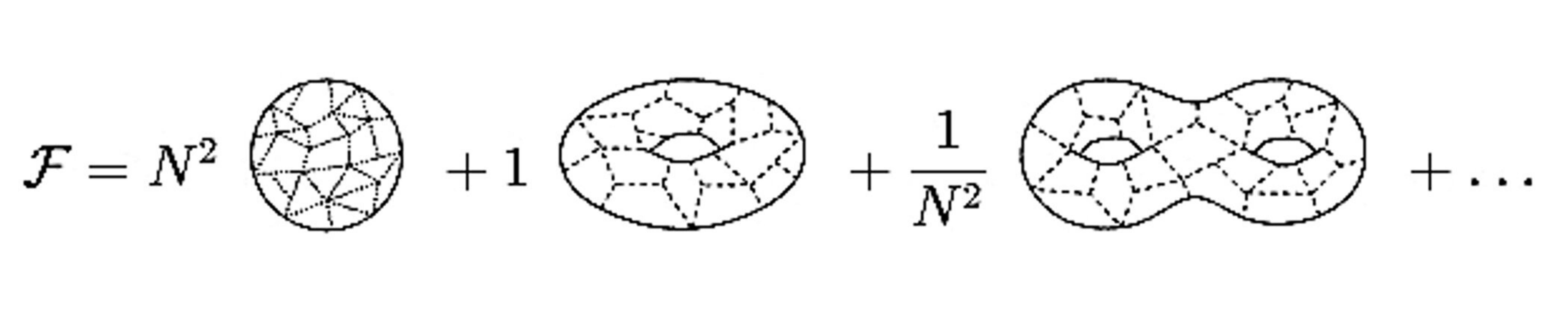}
\caption{\small In the large $N$ limit, the diagrams can be rearranged as a sum over topologies, with each topology multiplied by a particular power of $N$. The leading contribution has the topology of the two-sphere, the next one is a two-torus, etc.}
\label{tHlarge}
\end{center}
\end{figure}

In mored detail: an infinite number of Feynman diagrams contribute to an expression with a fixed power of $N$. In Figure \ref{tHlarge}, the leading contribution is of order $N^2$, and the corresponding diagrams (in the double-line notation) have the topology of the sphere. The next contribution is of order $1$, and the corresponding diagrams have the topology of a two-torus, etc. Thus for each power of $N$, an infinite number of Feynman diagrams contribute, and 't Hooft's clever diagrammatic notation, and the mathematical expansion that it represents, gives a very efficient reorganization of the Feynman diagrams. Namely, it picks out the ones that contribute at each order in $N$. In this way, he could use topological reasoning in string theory to make predictions about Yang-Mills theory. This illustrates de Regt's criterion of intelligibility of theories (in Section \ref{dRU}).\\

Although visualization, following either of these strategies (i) and (ii) (through the emergence of a relativistic spacetime, or through an effective spacetime that is not necessarily realistic), is very useful and makes theories intelligible, visualization is of course neither a necessary condition for scientific understanding (see de Regt, 2001:~p.~243), nor the only tool that physicists have to understand quantum gravity. And thus it is crucial that we discuss other tools that are available to render quantum gravity theories intelligible.

\subsection{Three tools for understanding}\label{3T}

We will here discuss three other conceptual tools that are used to develop understanding in modern physics, and more specifically in theories of quantum gravity that do not have a spacetime at the fundamental level:

(Approximations): This is the use of approximative relations. For example, one uses a classical limit (like the large $N$ limit: cf.~Section \ref{visualizing}) to understand a quantum theory. Since approximation often allows visualization, as in the case of the planar diagrams of the previous Section, approximations are often combined with visualization.

The approximated top theory can be both an {\it idealization}, in the sense that it describes another system whose properties are similar to the properties of the target system, and provides an inexact description of it (see Norton (2012:~p.~207)): and a {\it non-idealizing approximation} that does not describe another system, but gives an approximate (inexact) description of the target system.\footnote{Our use of `idealization' and `appoximation' here follows De Haro (2019a:~p.~12), itself modelled on, but slightly differing from, Norton (2012:~p.~207).}

Like in the case of visualizing theories of quantum gravity that we discussed in Section \ref{visualizing}, an approximation of a top theory $T_{\sm t}$ by a bottom theory $T_{\sm b}$ is usually a case of Nagel-Schaffner {\it deduction of a corrected theory}, $T_{\sm t}^*$, that approximates the top theory, $T_{\sm t}$, or is analogous to it.\footnote{See Schaffner (1967:~p.~114; 2012:~p.~540). For more on the principle of correspondence, especially between quantum and classical theories, see Radder (1991:~p.~199) and Post (1971:~p.~228).} 

(Similar): Similarities in the use of concepts between theories, either formal or interpretative, are also useful conceptual tools. For example, the formal similarity of charge conservation in the Maxwell theory and probability conservation in quantum mechanics was of instrumental value in developing the Born interpretation of probabilities in quantum mechanics.\footnote{These similarities give {\it analogies} between theories: see for example Hesse (2000:~p.~299).}

(Internal): Internal criteria can be used to interpret the concepts or terms in a bare theory or model. As for (Similar), internal criteria can be formal or interpretative similarities. 

On an internal interpretation, dualities are examples of (Internal). For example, in T-duality, once one has worked out, using the duality, that the radial direction is `large', and that one's coordinate $p$ is `momentum', then by interpretative consistency the conjugate coordinate $w$ must be `winding' (see Section \ref{SchemaT}). Likewise for other dualities, where, having fixed the interpretation of some quantity or state, one can fix the interpretation of the other quantities or states by using the duality map.

Dualities contribute to all three tools, and they usually combine them. Thus (Approximations) are often used in quantum gravity dualities, as we already saw in the case of the 't Hooft planar diagram expansion---the supersymmetric version of this being the large $N$ limit of Yang-Mills theory in four dimensions, which is dual to string theory in a space with a negative cosmological constant. Since most dualities relate a strong-coupling to a weak-coupling regime, a strong-coupling situation can be simulated and understood by a duality transformation on a weak-coupling situation, suitably approximated. (See for example the discussion of bosonization and sine-Gordon-Thirring model duality, in Section \ref{3op}).

For example, an effective duality between a five-dimensional black hole in AdS spacetime and a four-dimensional high-temperature superconducting material (a `strange metal') gives the only known explanation for the momentum-dependence of the self-energy of the electron in the strange metal. Thus a system that is intractable (because strongly coupled) in solid state physics becomes tractable after mapping it to the linearized Einstein-Maxwell equations. These equations give quantitative and qualitative arguments for the momentum-dependence of the self-energy in a way that (at present) cannot be done using solid state physics. Thus the presence of a duality has advantages for both explanation and understanding. \\

The previous examples involved not only (Approximations), but also the fact that the duals are formally (Similar), as in an effective duality. Besides these formal similarities, interpretative similarities are also often crucial: for example, 't Hooft and Mandelstam's realization that monopoles can give a magnetic analogue of the condensation of Cooper pairs, whereby monopole pairs condense in the vacuum, to give a mechanism of electric confinement, analogous to the Meissner effect (Section \ref{cmds}).

Holographic dualities are also examples of how similarities are used for understanding. In the large $N$ limit, an amplitude is reorganized as a topological expansion in the genus of the string world-sheet, so that the calculation is conveniently reformulated in terms of strings (\`a la 't Hooft, see Sections \ref{visualizing} and \ref{ggd}). Likewise, particle vs.~soliton dualities, where a disordered phase is reformulated using a magnetic order parameter, are also examples of (Similar).

The tool (Internal) is related, but it is in the context of a single theory: especially, formulating its common core (see the discussion below). Furthermore, (Internal) is used in cases where dualities serve as formal tools for theory construction within a given theory, at different values of the parameters. Thus duality was instrumental in Kramer and Wannier's derivation of the critical point of the phase transition of the two-dimensional Ising model (see Section \ref{dualpf0}). 

(Internal) is also used to construct a duality map between Noether currents and topological currents within a single theory, and between the solutions that carry the corresponding charges (see Section \ref{eagf}). The (Internal) comparison then enables the construction of the currents, and of the corresponding solutions, according to the duality conjecture. Likewise for Montonen-Olive duality, where the duality conjecture helps find the states with the correct (i.e.~duality-related) charges (Section \ref{M-O}).

The construction of the common core by abstraction, discussed in Section \ref{abstraction}, is also an example of a formal use of (Internal), where we use the duality map to construct a formal common core theory. 

Mirror symmetry, which unexpectedly connects manifolds of different topology (i.e.~Calabi-Yau manifolds with a K\"ahler form vs.~those with a holomorphic three-form on them)\footnote{See Witten (1991) and Strominger, Yau, and Zaslow (1996).}
also gave physicists and mathematicians an important formal and interpretative tool to study the topology of manifolds.

The above examples illustrate how dualities figure as parts of scientific explanations, and are routinely used in constructing novel explanations, thus providing understanding. In the next Section, we will make a distinction between the three interpretative cases of dualities that we have discussed in previous Chapters, to see how each of them contributes to understanding. (In the next Section, we will label these three interpetetative cases by (i) to (iii). And even though each of them can be linked to the above tools, we will set this aside).

\section{Dualities, explanation, and understanding}\label{DandU}

Our discussion in the previous Section used the examples from Part II to illustrate how dualities contribute to the three tools for understanding that we have identified. 

But the specifics of the Schema's analysis of dualities, especially in Chapters \ref{physeq} and \ref{Heuri}, contributes to the intelligibilty of dual models and to scientific understanding in a more direct way. For each of the interpretations in the tripartite classification of the interpretation of dualities (i.e.~interpreting duals as physically inequivalent, as physically equivalent, and as having a successor theory) contributes to understanding in its own way. We next briefly turn to each of these three cases, thereby focussing on the interpretative not formal issues:\\
\\
{\bf(i)~~Physical inequivalence:} There are two obvious ways in which the existence of a duality between physically inequivalent models helps understanding, especially if the duals are very disparate, so that the duality is surprising. Usually, we begin with two models that seem unrelated, and then it is surprising to discover that there is a duality between them.

First, discovering a duality between physically inequivalent models is discovering an interesting relation between models with different ontologies. Thus if our interest is in model $M_1$, the discovery of a duality with $M_2$ helps understanding because $M_2$'s ontology allows us to reason through problems in a way that is often not possible or feasible with $M_1$. In other words, the two ontologies, together with the translation map between them that is induced by the duality map,\footnote{This translation map is induced from the physical inequivalence of duals in Figure \ref{Physineq}. The translation is imperfect, because the interpretation maps $i_1$ and $i_2$ are in general not invertible. Figure \ref{dtilde} shows a case where the observable part of the domains of application are isomorphic, and in this case there is a perfect translation between the observable subdomains.} 
enables arguments that are unavailable with a single model and so helps with solving problems. To use a Kuhnian phrase, the duality map allows a change of perspective that helps us solve problems and further develop or articulate our models. Some examples are among the ones we already mentioned: particle-soliton dualities, and elementary vs.~solitonic D-branes (Appendix 8.A.2 and Section \ref{sssm}), are all used to find new solutions that would be very difficult to find without a duality.

Second, duals that are physically inequivalent often propose {\it different mechanisms} that again help us find solutions to problems, thereby gaining understanding of specific phenomena. A major example of this is the idea of an (effective) magnetic dual of the Ginzburg-Landau model of superconductivity (see Chapter \ref{EMDuality}). This prompted 't Hooft and Mandelstam to suggest that monopole condensation might be the mechanism behind confinement: namely, the quasi-dual of Cooper pair condensation. And as we discussed in Chapter \ref{EMYM}, this effect has been confirmed in the Seiberg-Witten theory and in lattice models.\\
\\
{\bf(ii)~~Physical equivalence:} In this case, dual models are ``mere'' reformulations of a single theory. There are two interesting parts of the Schema that contribute to scientific understanding in this case.

First, in analogy with the case of physically inequivalent models, the {\it specific structure} of physically equivalent models also helps us to construct explanations. Specifically, recall from Section \ref{abstraction} that internal interpretations are often constructed by abstraction from external intepretations, which one can then use to gain understanding of the duals (these external interpretations can be either true or false: for the logico-semantic relations with the common core theory, see Section \ref{lsr}). Thus there is a good use in the {\it theoretical functions} of dualities for understanding (see especially Chapter \ref{physeq}).

For example, T-duality---itself a case of expected physical equivalence---has been used to gain understanding of the early phase of the universe as described by string theory, including near the singularity, by using the models' external interpretations.\footnote{See Section \ref{T-d}, Brandenberger and Vafa (1989) and Huggett (2017).}

All the other string dualities (string-fivebrane duality, string-supermembrane duality, gauge-gravity duality, open-closed string duality), in so far as they are cases of physical equivalence, also contribute to understanding in this way. This is manifest in AdS-CFT, which has been employed to study several open problems on both sides of the duality, using both the formal aspects of the duality and the (external) interpretation of the dual models.

Second, the {\it common core theory} itself also contributes to intelligibility. For it gives a more concise formulation of the models, including their core semantics. And this often gives not only a concise formulation of the theory, but also a very general interpretation of it that helps further theory development. For example, T-duality relates IIA and IIB string theories, thereby in some sense capturing the ``formal essence'' of these models, i.e.~their common core. Likewise, the common core of AdS$_5$-CFT$_4$ is often seen as giving a very useful perspective on Type IIB string theory based on the common core between the two models.\\
\\
{\bf(iii)~~Successor theories:} Here, an important example is the M-theory programme, where various string theories are regarded as different limits of M-theory. Furthermore, we discussed the {\it geometric view of dualities}, illustrated in detail by Seiberg-Witten duality (Section \ref{effD}), where all the models are needed for a full formulation of the theory. Finding the complete form of the Seiberg-Witten theory (specifically, evaluating the prepotential, Eq.~\eq{Fsum}, in the three relevant regions of the manifold) requires knowledge of the various patches, and of the (duality) transformation rules. This then provides understanding of the role of the singularities---which it rendered explanatory, rather than problematic---and led to the confirmation of the 't Hooft-Mandelstam mechanism of monopole condensation as an explanation of quark confinement.

In sum, the logico-semantic analysis of dualities is very relevant to the specifics of scientific understanding, since the basic cases considered (of physical inequivalence, physical equivalence, and successor theories) provide understanding each in their own way. 

\section{Conclusion}

In this Chapter, we have endorsed Hempel's traditional view of explanation and a view of scientific understanding that is pragmatic, where `pragmatic' is here related to `use', and so does not have the connotation of `subjective'. The view is that understanding a phenomenon, $P$, using some theory, $T$, consists in $T$'s being useable to construct an adequate {\it explanation} of $P$. A condition for $T$ to be thus useable, according to de Regt (2017), is that $T$ is intelligible: namely, that $T$ has a cluster of qualities that allow scientists to use $T$. 

Visualization is one of the qualities of an intelligible theory, but it is not necessary for intelligibility: and, as we have argued, it is by itself not a sufficient condition either, since it requires other tools. 

In theories of quantum gravity that do not have a spacetime at the fundamental level, physicists use approximations, similarities and internal comparisons to develop understanding, and dualities contribute to each of these three tools. 

We also discussed that several cases of the Schema (in particular, physical equivalence and inequivalence), and the notion of a successor theory that goes beyond the Schema, each contribute to the intelligibility of theories in their own way---as we saw in examples taken from Part II. This understanding is sometimes provided by the discovery of a new ontology or of alternative mechanisms, by the specific structure of a model, or by the common core of the dual models. 

Thus dualities contribute in a variety of ways to the construction of explanations and to scientific understanding---which justifies (explains!) the emphasis that physicists have given to them in the past few decades.

\chapter{Conclusion}\label{Concl}
\markboth{\small{\textup{Conclusion}}}{\textup{\small{Conclusion}}}

In this book, we have explored the philosophy and physics of duality, a phenomenon in physics that relates apparently (very) different models. We close with a discussion of the Schema's contribution to questions in the philosophy of science, especially our view of scientific theories, and the relation between physics and philosophy.

\section{What is a scientific theory?}\label{Sec16.1}

Dualities bear on inter-theoretic relations: most of all, our Schema for them bears on the question {\it What is a scientific theory?} According to the Schema, a scientific theory is a bare theory with an interpretation, understood as a referential semantics that assigns both intensions and extensions. 

Duality gives a criterion of formal equivalence of bare theories and models: namely, dual models are isomorphic representations of a single bare theory. The models are {\it theoretically equivalent} just in case they are duals, and their interpretation maps (for both intensions and extensions) commute with the duality map, so that they have the same domain of application. This gives an {\it internal interpretation} of the dual models and, consequently, of the theory.

Our notion of a bare theory is in terms of the semantic conception of theories, i.e.~a bare theory is a triple of states, quantities, and dynamics. However, we saw in Section \ref{theoreq} that, since a duality is an isomorphism of models, a theory and its (dual) models can also be formulated syntactically, in terms of the theory's sentences, and especially its axioms. Our examples vindicated the claim that, as a formal i.e.~non-interpretative criterion of equivalence, duality has the right logical strength. 

A useful way to understand the distinction between a model root and the specific structure, in some of the examples, is in terms of the additional or specific axioms that distinguish each dual from the common core. For quantum theories, these specific axioms relate the theory to one of its classical limits. On an external interpretation, these classical limits are {\it complementaries} (in Bohr's sense), rather than duals. An external interpretation also interprets the specific structure, i.e.~the specific axioms of the dual models. The complementarity lies in the mutual exclusivity of the external interpretations of these specific axioms, i.e.~their theoretical inequivalence. 

An internal interpretation is silent about these specific axioms, and only interprets the common core: so that the models are compatible with each other and with the theory. 

In some of our examples, the common core has additional variables that make the duality map local in the fields. In other examples, the model roots require augmentation for them to be duals. In the examples of duals with an internal interpretation, the common core, even if it is isomorphic to the dual models, is a more abstract theory than the two duals. 

Regarding the {\it epistemology} of dualities, our conclusions were rather sober. Justifying the interpretations of duals requires taking into account the logico-semantic relations between theories and models. Thus the interpretation of duals requires an analysis of these logico-semantic relations. 

We did not find evidence that dualities favour specific types of scientific realism like structural realism, or that they undermine scientific realism. We argued against the claim that dualities create a problem of in-principle under-determination. Our logico-semantic analysis prompted the conclusion that the under-determination argument as usually stated is too strong and is not a threat to a cautious scientific realism: formulating a correct argument requires first exploring the space of theories. Otherwise, the argument must in general be stated in terms of the weaker notion of {\it compatibility}, rather than in terms of {\it theoretical equivalence}. 

\section{The geometric view of theories}

Quasi-dualities suggest a different relation between a successor theory and its models to the one that follows from dualities: we have called it the {\it geometric view} of theories, and it gives a useful schematization of the low-energy phases of quantum field theories, and of string theory itself. 

On this view, a model describes a region (an open set) of the moduli space of a successor theory, and together the quasi-duals cover the whole moduli space. In the overlaps between two regions, a quasi-duality map is like a coordinate transformation from one open set to another. The various regions correspond to different semi-classical limits, that are in general mutually incompatible. The various regions can be interpreted as {\it phases} of the successor theory, where the system has the same qualitative behaviour. These phases are characterised by the expectation value of a field (or set of fields) that becomes classical (i.e.~the field's correlation functions factorize) near one of the field's extrema. More precisely, a singularity is a point where the expansions of the values of the quantities of some model, in its semi-classical variable, do not converge. Note that having {\it quasi-duals}, as against arbitrary models, is required to secure an appropriate change of coordinates on the overlaps.

The relation between a {\it common core theory} and its models is the special case, in which (at any energy) the whole manifold can be covered by a single open set, and dual models are like isomorphic systems of coordinates that cover the whole manifold. 

Even though the geometric view gives a description of e.g.~the Seiberg-Witten theory valid only at low energies, in this limit the moduli space and its properties emerge as robust concepts. Thus the low-energy description favours not a flat, but rather a {\it structured view of theories}.\\

Three other {\it inter-theoretic relations} illustrate the practical functions of dualities: namely, the heuristic function (i.e.~finding a successor theory), emergence, and fundamentality. Like our conclusions about epistemology in Section \ref{Sec16.1}, our conclusions about the relation between dualities and other inter-theoretic relations are sober. Dualities are by themselves not indicative of emergence or fundamentality, nor are they---except in the significant case of AdS-CFT, where an extra dimension appears---compatible with anything other than an epistemic conception of emergence. 

On the other hand, for the same reasons that {\it effective dualities} suggest the existence of a successor theory (in the sense just described, i.e.~a theory whose low-energy limit moduli space is covered by quasi-duals), they also suggest both emergence and fundamentality. For quasi-duals often represent semi-classical limits of a successor theory, and so one can speak of the emergence of various states and quantities in the low-energy limit, and also of the relative emergence of one model relative to another, through a change of phase. In this case, the successor theory (if it exists) is more fundamental than its models, which give various approximative descriptions of it. 

Our distinction between a bare theory and its interpretation clarifies the {\it compatibility between reduction and emergence}, which the literature has in our view not satisfactorily established. We took reduction as a formal i.e.~non-interpretative notion, as in the Nagelian logico-semantic tradition (and also as it is usually described in physics). And we took emergence as a notion with both a formal and an interpretative component. The reason these notions are compatible is that reduction, being a formal i.e.~non-interpretative notion, cannot forbid novelty in the interpretation: but the novelty requirement of ontological emergence is interpretative. (If one's preferred notion of reduction is interpretative, then reduction and emergence are incompatible. But we argued that the Nagelian notion is correct.) 

\section{Epilogue: physics and philosophy}

Through the lens of dualities, it is clear how, over the past fifty years, high-energy physics and quantum gravity have been rooted in other branches of physics: quantum mechanics, general relativity, statistical physics, and condensed matter physics. The Chapters in Part II attest to how, from phase transitions to quark confinement, from quantum mechanics to non-perturbative quantum field theories, and from string theory to black hole microstates, many important developments in physics stem from the feasibility of identifying, developing, and then exploiting, dualities. This alone makes the study of dualities expedient and timely for both physicists and aspiring philosophers of physics. 

In relating disparate areas of physics, and disparate ideas in these areas, dualities are technical and heuristic tools that suggest new perspectives, give us understanding, and point to new physical mechanisms and explanations. They make tantalizing physical predictions that cannot easily be made by other means: such as the Dirac quantisation condition, the critical point of statistical mechanical systems, the appearance of disordered phases, and the condensation of pairs of solitons to a new phase of emergent topological order. Dualities in string theory allow the calculation of fundamental parameters, such as Newton's constant and the string length, in a dual or a quasi-dual model. Furthermore, duality arguments are key to insights into the degrees of freedom of successor theories such as M-theory.

This book is of course not intended to be the last word on dualities. Indeed, it might have even opened more questions that it answered. Here follows our attempt to list some of these open questions.

As to physics, the main open questions concern: first, proving the extant duality conjectures; second, using dualities and quasi-dualities to make progress in fundamental physics, especially quantum field theory and string theory. And third, experimental progress will be required to further develop dualities and to get guidance into quantum gravity.

As to philosophy, the main open questions concern, first, for the profound dualities we nowadays see in current fundamental physics, especially string theory, isolating more definitively, and in more detail, what is the bare theory, the common core. For example, a good number of dualities exchange Noether and topological charges: but not all dualities do (notably, holographic dualities do not appear to do this). Also, there is much more to be understood about the examples of dualities from Part II: for example, about the analogies between condensed matter theory, quantum field theory and string theory, about the cosmological applications of dualities, and about the application of quantum information theory methods to dualities. Finally, we have not touched upon dualities in ${\cal N}=1$ supersymmetric gauge theories, which exhibit rich behaviour. 

Second, further philosophical work is required to provide more detailed ontologies for the common core theories, so as to articulate detailed internal interpretations of duals.

Our study, in Part II, of solitons and non-perturbative aspects of quantum field theory, has revealed fascinating behaviours, such as the condensation of solitons (analogous to the condensation of electrons in Cooper pairs) and the changes of phase that this can trigger; the remarkable duality between electric and magnetic charges in quantum field theories, and between theories in different spacetime dimensions in string theory. As we have stressed, this suggests that solitons and their vacua are important parts of the ontologies of quantum field theories. 

Third, we have only outlined a {\it conception of quasi-dualities}, and the associated geometric view of theories, both of which require further work. 

Thus further detailed formal and interpretative studies, for both duals and quasi-duals, should bear on the main philosophical questions that have engaged us in Part III: especially, theoretical equivalence and the individuation of scientific theories, the interpretative correlates of logico-semantic relations between theories, and the inter-theoretic relations of emergence and reduction.

Our aim has not been to strictly {\it integrate} physics and philosophy. For this might, in some ways, be artificial. After all, physics and philosophy are different fields, each with its own methods and preoccupations. We have attempted to put them next to each other: even more, to interlace them, so that one informs the other, and each can bear on the other. This is one method in philosophy of physics that we believe works, and is relevant to current work in physics. 

\chapter*{References}
\addcontentsline{toc}{chapter}{References}
\markboth{\small{\textup{References}}}{\textup{\small{References}}}

Abazov, V.~M., Abbott, B., Acharya, B.~S., Adams, M., Adams, T., Agnew, J.~P., ... and Fuess, S.~(2021). `Odderon Exchange from Elastic Scattering Differences between $p p$ and $p \bar{p}$ Data at $1.96$ TeV and from $p p$ Forward Scattering Measurements'. {\it Physical Review Letters}, 127 (6), 062003, pp.~1-10.

\ \\Abraham, R.~and Marsden, J.~E.~(1978). {\it Foundations of Mechanics}. Second Edition, 1987. Addison-Wesley.

\ \\Abrikosov, A.~A.~(1957). `On the Magnetic Properties of Superconductors of the Second Group'. {\it Soviet Physics JETP}, 5 (6), pp.~1174-1182.

\ \\Abrikosov, A.~A.~(2004). `Nobel Lecture: Type-II Superconductors and the Vortex Lattice'. {\it Reviews of Modern Physics}, 76 (3), pp.~975-979.

\ \\Aharonov, Y.~and Bohm, D.~(1959). `Significance of Electromagnetic Potentials in the Quantum Theory'. {\it The Physical Review,} 115 (3), pp.~485-491.

\ \\Aharony, O., Gubser, S.S., Maldacena, J.M., Ooguri, H.,~and~Oz, Y.~(2000). `Large $N$ Field Theories, String Theory and Gravity'. {\it Physics Reports}, 323, pp.~183-386.

\ \\Ahlfors, L.~V.~(1979). {\it Complex Analysis}. New York: McGraw-Hill.

\ \\Aitchison, I.~J.~R.~and Hey, A.~J.~G.~(2013). {\it Gauge Theories in Particle Physics}. Volume 1. London: Taylor and Francis.

\ \\Aitchison, I.~J.~R.~and Hey, A.~J.~G.~(2013). {\it Gauge Theories in Particle Physics}. Volume 2. London: Taylor and Francis.

\ \\Alim, M.~(2012). `Lectures on Mirror Symmetry and Topological String Theory'. arXiv:1207.0496.

\ \\Alty, L.~J.~(1994). `Kleinian Signature Change'. {\it Classical and Quantum Gravity}, 11, pp.~2523-2536.

\ \\Alvarez, E., Alvarez-Gaume, L.~and Lozano, Y.~(1995). `An Introduction to T-Duality in String Theory'. {\it Nuclear Physics} B, 41, pp.~1-20.

\ \\\'Alvarez-Gaum\'e, L.~and Hassan, S.~F.~(1997). `Introduction to S-Duality in ${\cal N}=2$ Supersymmetric Gauge Theories'. {\it Fortschritte der Physik}, 45 (3-4), pp.~159-236.

\ \\\'Alvarez-Gaum\'e, L.~and Zamora, F.~(1998). `Duality in Quantum Field Theory (and String Theory)'. {\it AIP Conference Proceedings}, 423, pp.~46-83.

\ \\Amari, S.~and Nagaoka, H. (2000). `Methods of Information Geometry', in {\it Translations of Mathematical Monographs}, volume 191. American Mathematical Society. Oxford University Press.

\ \\Ammon, M.~and Erdmenger, J.~(2015). {\it Gauge/Gravity Duality}. Cambridge: Cambridge University Press.

\ \\Anderson, P.~W.~(1958). `Coherent Excited States in the Theory of Superconductivity: Gauge Invariance and the Meissner Effect'. {\it Physical Review}, 110 (4), pp.~827-835.

\ \\Anderson, P.~W.~(1963). `Plasmons, Gauge Invariance, and Mass'. {\it Physical Review}, 130 (1), pp.~439-442.

\ \\Anderson, P.~W.~(1972). `More Is Different'. {\it Science}, 177 (4047), pp.~393-396.

\ \\Andr\'eka, H., N\'emeti, I.~and van Benthem, J.~(1998). `Modal Languages and Bounded Fragments of Predicate Logic'. {\it Journal of Philosophical Logic}, 27, pp.~217-274.

\ \\Andr\'eka, H., Madar\'asz, J.~X., N\'emeti, I.~(2008). `Defining New Universes in Many-Sorted Logic'. Unpublished, \\https://citeseerx.ist.psu.edu/viewdoc/download?doi=10.1.1.184.1313\&rep=rep1\&type=pdf.

\ \\Anninos, D., Ng, G.~S.~and Strominger, A.~(2011). `Asymptotic Symmetries and Charges in De Sitter Space'. {\it Classical and Quantum Gravity}, 28, 175019. 

\ \\Arafune, J., Freund, P.~G.~O.~and Goebel, C.~J.~(1975). `Topology of Higgs Fields'. {\it Journal of Mathematical Physics}, 16, pp.~433-437.

\ \\Arageorgis, A.~(1995). {\it Fields, Particles, and Curvature: Foundations and Philosophical Aspects of Quantum Field Theory in Curved Spacetime}. PhD dissertation, UMI 9614172, University of Pittsburgh.

\ \\Ariew, R.~(Ed.)~(2000). {\it G.~W.~Leibniz and Samuel Clarke. Correspondence}. Indianapolis: Hackett Publishing.

\ \\Armoni, A.~(2023). `S-Dual of Maxwell--Chern-Simons Theory'. {\it Physical Review Letters}, 130, 141601.

\ \\Arnold, V.~I.~(1989). {\it Mathematical Methods of Classical Mechanics}. New York: Springer.

\ \\Arnowitt, R., Deser, S., and Misner, C.~W.~(1959). `Dynamical structure and definition of energy in general relativity'. {\it Physical Review}, 116 (5), p.~1322.

\ \\Arnowitt, R.~L., Deser, S., and Misner, C.~W.~(2008). `Republication of: The dynamics of general relativity'. {\it General Relativity and Gravitation}, 40 (9), pp.~1997-2027.

\ \\Ashtekar, A.~and Hansen, R.~O.~(1978). `A unified treatment of null and spatial infinity in general relativity. I. Universal structure, asymptotic symmetries, and conserved quantities at spatial infinity'. {\it Journal of Mathematical Physics} 19, pp.~1542-1566.

\ \\Atiyah, M.~F.~(1988). `Topological Quantum Field Theories'. {\it Publications math\'ematiques de l'I.H.\'E.S}, 68, pp.~175-186.

\ \\Atland, A.~and Simons, B.~(2010). {\it Condensed Matter Field Theory}. Cambridge: Cambridge University Press.

\ \\Auslander, L.~and MacKenzie, R.~(1963). {\it Introduction to Differentiable Manifolds}. Dover: McGraw-Hill, reprint 2007.

\ \\Avis, S.~J., Isham, C.~J.~and Storey, D.~(1978). `Quantum Field Theory in anti-de Sitter Space-Time'. {\it Physical Review} D, 18 (10), pp.~3565-3576.

\ \\Bain, J.~(2020). `Spacetime as a Quantum Error-Correcting Code?' {\it Studies in History and Philosophy of Modern Physics}, 71, pp.~26-36.

\ \\Bain, J.~(2021). `The RT Formula and its Discontents: Spacetime and Entanglement'. {\it Synthese}, 198 (12), pp.~11833-11860.

\ \\Bais, F.~A.~(1978). `Charge-Monopole Duality in Spontaneously Broken Gauge Theories'. {\it Physical Review} D, 18 (4), pp.~1206-1210.

\ \\Balasubramanian, V.~(1997). `Statistical Inference, Occam's Razor, and Statistical Mechanics on the Space of Probability Distributions'. {\it Neural Computation}, 9, pp.~349-368.

\ \\Balasubramanian, V., de Boer and Minic, D.~(2001). `Mass, Entropy, and Holography in Asymptotically de Sitter spaces'. {\it Physical Review}, D, 65, 123508, 1-15.

\ \\Balasubramanian, V., Heckman, J.~J.~and Maloney, A.~(2015). `Relative Entropy and Proximity of Quantum Field Theories'. {\it Journal of High-Energy Physics}, 104, pp.~1-13.

\ \\Balasubramanian, V.~and Kraus, P.~(1999). `Spacetime and the Holographic Renormalization Group'. {\it Physical Review Letters}, 83 (18), pp.~3605-3608.

\ \\Bali, G.~S.~(2000). `The Mechanism of Quark Confinement'. In: {\it Quark Confinement and the Hadron Spectrum} III, Isgur, N.~(Ed.), pp.~17-36. Singapore New Jersey: World Scientific.

\ \\Baltag, A.~and Smets, S.~(Eds.) (2014). {\it Johan van Benthem on Logic and Information Dynamics}. Cham: Springer.

\ \\Ba\~nados, M., Teitelboim, C.~and Zanelli, J.~(1992). `Black Hole in Three-Dimensional Spacetime'. {\it Physical Review Letters}, 69 (13), pp.~1849-1851.

\ \\Banks, T., Fischler, W., Shenker, S. H.~and Susskind, L.~(1997). `M- Theory as a Matrix Model: A Conjecture'. {\it Physical Review} D, 55 (8), pp.~5112-5128.

\ \\Banks, T., Myerson, R.~and Kogut, J.~(1977). `Phase Transitions in Abelian Lattice Gauge Theories'. {\it Nuclear Physics} B, pp.~493-510.

\ \\Barbour, J.~B.(1982). `Relational Concepts of Space and Time'. {\it The British Journal for the Philosophy of Science}, 33 (3), pp.~251-274.

\ \\Barbour, J.~B.~(1999). `The Development of Machian Themes in the Twentieth Century'. In: Butterfield (1999), pp.~83-109.

\ \\Barbour, J.~B.~and Bertotti, B.~(1982). `Mach's Principle and the Structure of Dynamical Theories'. {\it Proceedings of the Royal Society of London}, A, 382 (1783), pp.~295-306.

\ \\Bardeen, J.~M., Carter, B.,~and Hawking, S.~W.~(1973). `The Four Laws of Black Hole Mechanics'. {\it Communications in Mathematical Physics}, 31, pp.~161-170.

\ \\Bardeen, J., Cooper, L.~N., and Schrieffer, J.~R.~(1957). `Theory of Superconductivity'. {\it Physical Review}, 108 (5), pp.~1175-1205.

\ \\Bargmann, V.~and Wigner, E.~P.~(1948). `Group Theoretical Discussions of Relativistic Wave Equations'. {\it Proceedings of the National Academy of Sciences}, 34 (5), pp.~211-223.

\ \\Barrett, J.~A.~(1999). {\it The Quantum Mechanics of Minds and Worlds}. New York: Oxford University Press.

\ \\Barrett, T.~W.~(2015). `On the Structure of Classical Mechanics'. {\it British Journal for the Philosophy of Science}, 66, pp.~801-828.

\ \\Barrett, T.~W.~(2019). `Equivalent and Inequivalent Formulations of Classical Mechanics'. {\it British Journal for the Philosophy of Science}, 70, pp.~1167-1199.

\ \\Barrett, T.~W.~(2020). `Structure and Equivalence'. {\it Philosophy of Science}, 87, pp.~1184-1196.

\ \\Barrett, T.~W.~and Halvorson, H.~(2016a). `Glymour and Quine on theoretical equivalence'. {\it Journal of Philosophical Logic}, 45 (5), pp.~467-483.

\ \\Barrett, T.~W.~and Halvorson, H.~(2016b). `Morita Equivalence'. {\it The Review of Symbolic Logic}, 9 (3), pp.~556-582.

\ \\Bars, I.~(2001). `Survey of Two-Time Physics'. {\it Classical and Quantum Gravity}, 18, pp.~3113-3130.

\ \\Bars, I., Deliduman, C.~and Andreev, O.~(1998). `Gauged Duality, Conformal Symmetry, and Spacetime with Two Times'. {\it Physical Review} D, 58, 066004, pp.~1-10.

\ \\Bartels, A.~(1996). `Modern Essentialism and the Problem of Individuation of Spacetime Points'. {\it Erkenntnis}, 45 (1), pp.~25-43.

\ \\Barwise, J.~and Moss, L.~(1996). {\it Vicious Circles}. Stanford, CA: Center for the Study of Language and Information.

\ \\Batterman, R.~W.~(2011). `Emergence, Singularities, and Symmetry Breaking'. {\it Foundations of Physics}, 41, pp.~1031-1050.

\ \\Baxter, R.~J.~(1982). {\it Exactly Solved Models in Statistical Mechanics}. London: Academic Press.

\ \\Becker, K., Becker, M.~and Schwarz, J.~(2007). {\it String Theory and M-Theory}. Cambridge: Cambridge University Press.

\ \\Bedau, M.~A.~(1997). `Weak Emergence'. {\it Philosophical Perspectives}, 11, pp.~375-399.

\ \\Bedau, M.~A.~and Humphreys, P.~(2008). `Emergence: Contemporary Readings in Philosophy and Science.' Cambridge, MA: The MIT Press.

\ \\Beekman, A.~J., Rademaker, L.~and van Wezel, J.~(2019). `An Introduction to Spontaneous Symmetry Breaking'. {\it Sci|Post Physics Lecture Notes}, 11, pp.~1-140.

\ \\Beisert, N., Ahn, C., Alday, L.~F., Bajnok, Z., Drummond, J.~M., Freyhult, L.,... and Zoubos, K.~(2012). `Review of AdS/CFT Integrability: An Overview'. {\it Letters in Mathematical Physics}, 99 (1), pp.~3-32.

\ \\Bekenstein, J.~D.~(1972). `Black Holes and the Second Law'. {\it Lettere Al Nuovo Cimento}, 4 (15), pp.~737-740.

\ \\Bekenstein, J.~D.~(1973). `Black Holes and Entropy'. {\it Physical Review} D, 7 (8), pp.~2333-2346.

\ \\Bekenstein, J.~D.~(1974). `Generalized Second Law of Thermodynamics in Black-Hole Physics'. {\it Physical Review} D, 9 (12), pp.~3292-3300.

\ \\Beller, M.~(1999). {\it Quantum Dialogue}. Chicago: The University of Chicago Press.

\ \\Belot, G.~(1998). `Understanding Electromagnetism'. {\it The British Journal for the Philosophy of Science}, 49, pp.~531-555.

\ \\Belot, G.~(2018). `Fifty Million Elvis Fans Can't be Wrong'. {\it Nous}, 52 (4), pp.~946-981.

\ \\Belot, G., Earman, J.~and Ruetsche, L.~(1999). `The Hawking Information Loss Paradox: The Anatomy of a Controversy'. {\it The British Journal for the Philosophy of Science}, 50, pp.~189-229.

\ \\Belot, G~and Earman, J.~(2001). `Pre-Socratic Quantum Gravity'. In: Callender and Huggett (2001).

\ \\Benacerraf, P.~(1965). `What Numbers Could not Be'. {\it The Philosophical Review}, 74 (1), pp.~47-73.

\ \\Beni, M.~D.~(2018). `Syntactical Informational Structural Realism'. {\it Minds and Machines}, 28, pp.~623-643. 

\ \\Berezin, F.~A.~(1966). {\it The Method of Second Quantization}. London: Academic Press.

\ \\Berezinskii, V.~L.~(1971). `Destruction of Long-Range Order in One-Dimensional and Two-Dimensional Systems Having a Continuous Symmetry Group. I. Classical Systems'. {\it Soviet Physics JETP}, 32 (3), pp.~493-500.

\ \\Berezinskii, V.~L.~(1972). `Destruction of Long-Range Order in One-Dimensional and Two-Dimensional Systems Having a Continuous Symmetry Group. II. Quantum Systems'. {\it Soviet Physics JETP}, 34 (3), pp.~610-616.

\ \\Berghofer, P., Fran\'cois, J., Friederich, S., Gomes, H., Hetzroni, G., Maas, A.~and Sondenheimer, R.~(2023). {\it Gauge Symmetries, Symmetry Breaking, and Gauge-Invariant Approaches}. Cambridge: Cambridge University Press.

\ \\Bergshoeff, E., Hull, C.~and Ort\'in, T.~(1995). `Duality in the Type-II Superstring Effective Action'. {\it Nuclear Physics} B, 451, pp.~547-575.

\ \\Bergshoeff, E., Sezgin, E.~and Townsend, P.~K.~(1987). `Supermembranes and Eleven-Dimensional Supergravity'. {\it Physics Letters} B, 189 (1-2), pp.~75-78.

\ \\Berman, D.~S.~(2007). `M-Theory Branes and Their Interactions'. {\it Physics Report}, 456, pp.~89-126.

\ \\Berman, D.~S.~and Thompson, D.~C.~(2015). `Duality Symmetric String and M-Theory'. {\it Physics Reports}, 566, pp.~1-60.

\ \\Beth, E.~W.~(1960). `Semantics of Physical Theories'. {\it Synthese}, 12 (2/3), pp.~172-175.

\ \\Beth, E.~W.~(1969). `Semantic Entailment and Formal Derivability'. In: Hintikka, J.~(Ed.), {\it The Philosophy of Mathematics}, Oxford: Oxford University Press.

\ \\Bilal, A.~(1997). `Duality in ${\cal N}=2$ Susy SU(2) Yang-Mills Theory: A Pedagogical Introduction to the Work of Seiberg and Witten'. {\it Les rencontres physiciens-math\'ematiciens de Strasbourg}, RCP25, 48, 87-120.

\ \\Binney, J.~J., Dowrick, N.~J., Fisher, A.~J.~and Newman, M.~E.~J.~(1992). {\it The Theory of Critical Phenomena}. Oxford: Oxford University Press.

\ \\Biquard, O.~(2005). {\it AdS-CFT Correspondence: Einstein Metrics and Their Conformal Boundaries}. European Mathematical Society Publishing House.

\ \\Birkhoff, G.~and Von Neumann, J.~(1936). `The Logic of Quantum Mechanics'. {\it Annals of Mathematics}, 37 (4), pp.~823-843.

\ \\Blencowe, M.~P.~and Duff, M.~J.~(1988). `Supermembranes and the Signature of Spacetime'. {\it Nuclear Physics}, B310, pp.~387-404.

\ \\Bliss, R.~and Trogdon, K.~(2021). `Metaphysical Grounding'. {\it Stanford Encyclopedia of Philosophy}. https://plato.stanford.edu/entries/grounding.

\ \\Bogomol'nyi, E.~B.~(1976). `The Stability of Classical Solutions'. {\it Soviet Journal of Nuclear Physics}, 24 (4), pp.~449-454. Reprinted in: {\it Solitons and Particles}, Rebbi, C.~and Soliani, G.~(Eds.), pp.~389-394. Singapore: World Scientific.

\ \\Bohr, N.~(1928). `The Quantum Postulate and the Recent Development of Atomic Theory'. {\it Nature}, Supplement, pp.~580-590.

\ \\Bohr, N.~(1985). {\it Niels Bohr Collected Works}. Volume 6. Kalckar, J.~(Ed.). Amsterdam: North-Holland.

\ \\Bohr, N.~(1996). {\it Niels Bohr Collected Works}. Volume 7. Kalckar, J.~(Ed.). Amsterdam: North-Holland.

\ \\Bondi, H., van der Burg, M.~G.~J., Metzner, A.~W.~K.~(1962). `Gravitational Waves in General Relativity. VII. Waves from Axi-Symmetric Isolated Systems', {\it Proceedings of the Royal Society of London. Series A, Mathematical and Physical Sciences}, 269 (1336), pp.~21-52.

\ \\Born, M., Heisenberg, W.~and Jordan, P.~(1925). `On Quantum Mechanics II'. In: van der Waerden (1967), pp.~321-385.

\ \\Born, M.~and P.~Jordan~(1925). `Zur Quantenmechanik'. {\it Zeitschrift f\"ur Physik}, 34, pp.~858-888. Translated in van der Waerden (1967), pp.~277-306. Page numbers are as in van der Waerden.

\ \\Borrelli, A.~(2015). `The Story of the Higgs Boson: the Origin of Mass in Early Particle Physics'. {\it The European Physical Journal} H, 50, pp.~1-50.

\ \\Bouatta, N.~and Butterfield, J.~(2015). `On Emergence in Gauge Theories at the 't Hooft Limit'. {\it European Journal for Philosophy of Science}, 5, pp.~55-87.

\ \\Brading, K.~and Brown, H.~R.~(2004). `Are Gauge Symmetry Transformations Observable?' {\it The British Journal for the Philosophy of Science,} 55, pp.~645-665.

\ \\Brading, K.~and Castellani, E.~(2003). {\it Symmetries in Physics: Philosophical Reflections}. Cambridge: Cambridge University Press.

\ \\Brading, K, Castellani, E.~and Teh, N.~(2017). `Symmetry and Symmetry Breaking', {\it Stanford Encyclopedia of Philosophy}. https://plato.stanford.edu/entries/symmetry-breaking.

\ \\Bradley, C.~and Weatherall, J.~O.~(2022). `Mathematical Responses to the Hole Argument: Then and Now'. {\it Philosophy of Science}, 89 (5), pp.~1223-1232.

\ \\Brandenberger, R.~and Vafa, C.~(1989). `Superstrings in the Early Universe'. {\it Nuclear Physics} B, 316 (2), pp.~391-410.

\ \\Breckenridge, J. C., Myers, R. C., Peet, A. W.~and Vafa, C.~(1997). `D-branes and Spinning Black Holes. {\it Physics Letters} B, 391(1-2), pp.~93-98.

\ \\Brighouse, C.~(1994). `Spacetime and Holes'. {\it PSA: Proceedings of the Biennial Meeting of the Philosophy of Science Association}, 1994 (1), pp.~117-125. 

\ \\Brink, L.~and Schwarz, J.~H.~(1977). `Supersymmetric Yang-Mills Theories'. {\it Nuclear Physics} B, 121, pp.~77-92.

\ \\Brown, A.~R., Roberts, D.~A., Susskind, L., Swingle, B., and Zhao, Y.~(2016). `Holographic complexity equals bulk action?' {\it Physical Review Letters}, 116(19), 191301.

\ \\Brown, J.~D.~and Henneaux, M.~(1986). `Central Charges in the Canonical Realization of Asymptotic Symmetries: An Example from Three Dimensional Gravity'. {\it Communications in Mathematical Physics}, 104, pp.~207-226.

\ \\Brown, J.D., York, Jr., J.W.~(1993). `Quasilocal energy and conserved charges derived from the gravitational action', {\it Physical Review D}, 47, pp.~1407-1419.

\ \\Brush, S.~G.~(1967). `History of the Lenz-Ising Model'. {\it Reviews of Modern Physics}, 39 (4), pp.~883-893.

\ \\Bub, J.~(1982). `Quantum Logic, Conditional Probability, and Inference'. {\it Philosophy of Science}, 49 (3), pp.~402-421.

\ \\Bub, J.~and Clifton, R.~(1996). `A Uniqueness Theorem for `No Collapse' Interpretations of Quantum Mechanics'. {\it Studies in History and Philosophy of Modern Physics}, 27 (2), pp.~181-219.

\ \\Burgess, J.~P.~(2016). {\it Rigor and Structure}. New York: Oxford University Press.

\ \\Butterfield, J.~N.~(1989). `The Hole Truth'. {\it The British Journal for the Philosophy of Science}, 40, pp.~1-28.

\ \\Butterfield, J.~N.~(1999). {\it The Arguments of Time}. Oxford: Oxford University Press. 

\ \\Butterfield, J.~(2007). `On Symplectic Reduction in Classical Mechanics'. In: {\it Philosophy of Physics}, Butterfield, J.~and Earman, J.~(Eds.), pp.~1-131. Amsterdam: Elsevier.

\ \\Butterfield, J.~N.~(2011a). `Emergence, Reduction and Supervenience: A Varied Landscape'. {\it Foundations of Physics}, 41, pp.~920-959.

\ \\Butterfield, J.~N.~(2011b). `Less is Different: emergence and Reduction Reconciled'. {\it Foundations of Physics}, 41, pp.~1065-1135.

\ \\Butterfield, J.~N.~(2014). `Reduction, Emergence, and Renormalization'. {\it The Journal of Philosophy}, 111 (1), pp.~5-49.

\ \\Butterfield, J.~N~(2021). `On Dualities and Equivalences Between Physical Theories'. In: {\it Philosophy Beyond Spacetime}, Huggett, N., Le Bihan B.~and W\"uthrich, C.~(Eds.), pp.~41-77. Oxford: Oxford University Press.

\ \\Butterfield, J.~N.~and Bouatta, N.~(2011). `Emergence and Reduction Combined in Phase Transitions'. arXiv preprint arXiv:1104.1371.

\ \\Butterfield, J.~N.~and Bouatta, N.~(2015). `Renormalization for Philosophers'. In: {\it Metaphysics in Contemporary Physics}. Bigaj, T.~and W\"uthrich, C.~(Eds.), pp.~437-485. Leiden: Brill.

\ \\Butterfield, J.~N.~and Gomes, H.~(2023). `Functionalism as a species of reduction'. In: {\it Current Debates in Philosophy of Science}, Soto, C.~(Ed.), pp.~123-200. Cham: Springer.

\ \\Butterfield, J.~N.~and Isham, C.~(1999). `On the Emergence of Time in Quantum Gravity'. In: Butterfield (1999), pp.~111-168. 

\ \\Butterfield, J.~N.~and Isham, C.~(2001). `Spacetime and the Philosophical Challenge of Quantum Gravity'. In: {\it Physics Meets Philosophy at the Planck Scale}, Callender, C.~and Huggett, N.~(Eds.), pp.~33-89. Cambridge: Cambridge University Press.

\ \\Button, T.~and Walsh, S.~(2018). {\it Philosophy and Model Theory}. Oxford: Oxford University Press.

\ \\Callan, C.~G, Dashen, R.~F.~and Gross, D.J.~(1976). `The Structure of the Gauge Theory Vacuum'. {\it Physics Letters} 63 B (3), pp.~334-340.

\ \\Callender, C.~and Cohen, J.~(2006). `There Is No Special Problem About Scientific Representation'. {\it Theoria} 55, pp.~7-25.

\ \\Callender, C.~and N.~Huggett (Eds.) (2001). {\it Physics Meets Philosophy at the Planck Scale}. Cambridge: Cambridge University Press.

\ \\Camilleri, K.~(2006). `Heisenberg and the Wave-Particle Duality'. {\it Studies in History and Philosophy of Modern Physics}, 37, pp.~298-315.

\ \\Camilleri, K.~(2009). {\it Heisenberg and the Interpretation of Quantum Mechanics}. Cambridge: Cambridge University Press.

\ \\Cao, T.~Y.~(1997). {\it Conceptual Developments of 20th Century Field Theories}. Second Edition, 2019. Cambridge: Cambridge University Press.

\ \\Cardy, J.~L.~(1982). `Duality and the $\th$ Parameter in Abelian Lattice Models'. {\it Nuclear Physics B}, 205 [FS5], pp.~17-26.

\ \\Cardy, J.~L.~and Rabinovici, E.~(1982). `Phase Structure of $\mathbb{Z}_p$ Models in the Presence of a $\th$ Parameter'. {\it Nuclear Physics} B 205 [FS5], pp.~1-16.

\ \\Carnap, R.~(1939). `Foundations of Logic and Mathematics'. In: {\it International Encyclopedia of Unified Science. Foundations of the Unity of Science} (volumes I-II). Neurath, O., Carnap, R.~and Morris, C.~(Eds.). Chicago: The University of Chicago Press.

\ \\Carnap, R.~(1945). `The Two Concepts of Probability: The Problem of Probability'. {\it Philosophy and Phenomenological Research}, 5 (4), pp.~513-532.

\ \\Carnap, R.~(1947). {\it Meaning and Necessity}, Chicago: University of Chicago Press.

\ \\Carnap, R.~(1950). {\it Logical Foundations of Probability}. Second Edition, 1962. The University of Chicago Press.

\ \\Carnap, R.~(1966). {\it Philosophical Foundations of Physics. An Introduction to the Philosophy of Science}. Edited by Martin Gardner. New York and London: Basic Books, Inc.

\ \\Carnap, R.~(1995) [1934]. {\it The Unity of Science}. Bristol: Thoemmes Press, reprint of the 1934 Edition.

\ \\Castellani, E.~(2017). `Duality and `particle' democracy'. {\it Studies in History and Philosophy of Modern Physics}, 59, pp.~100-108.

\ \\Castellani, E.~and De Haro, S.~(2020). `Duality, Fundamentality, and Emergence'. In: Glick et al.~(2020), pp.~195-216. 

\ \\Castellani, E.~and Rickles, D.~(2017). `Introduction to special issue on dualities', {\it Studies in History and Philosophy of Modern Physics}, forthcoming. doi.org/10.1016/j.shpsb.2016.10.004.

\ \\Cat, J.~(2017). `The Unity of Science'. {\it Stanford Encyclopedia of Philosophy}. https://plato.stanford.edu/entries/scientific-unity.

\ \\Caulton, A.~(2015). `The role of symmetry in the interpretation of physical theories'. {\it Studies in History and Philosophy of Modern Physics}, 52, pp.~153-162.

\ \\Chambers, R.~G.~(1960). `Shift of an Electron Interference Pattern by Enclosed Magnetic Field'. {\it Physical Review Letters,} 5 (1), pp.~3-5.

\ \\Chen, B., Czech, B.~and Wang, Z.-Z.~(2022). `Quantum Information in Holographic Duality'. {\it Reports on Progress in Physics}, 85, 046001, pp.~1-59.

\ \\Christensen, D.~(1983). `Glymour on Evidential Relevance'. {\it Philosophy of Science}, 50 (3), pp.~471-481.

\ \\Cinti, E., Corti, A.~and Sanchioni, M.~(2022). `On Entanglement as a Relation. {\it European Journal for Philosophy of Science}, 12 (1), 10, pp.~1-29.

\ \\Coffey, K.~(2014). `Theoretical Equivalence as Interpretative Equivalence'. {\it The British Journal for the Philosophy of Science}, 65, pp.~821-844.

\ \\Coleman, S.~(1975). `Quantum sine-Gordon Equation as the Massive Thirring Model'. {\it Physical Review} D, 11 (8), pp.~2088-2097.

\ \\Coleman, S.~(1985). {\it Aspects of Symmetry}. Cambridge: Cambridge University Press.

\ \\Coleman, S.~and Mandula, J.~(1967). `All Possible Symmetries of the S-Matrix'. {\it Physical Review}, 159 (5), pp.~1251-1256.

\ \\Coleman, S., Neveu, A.~and Sommerfield, C.~(1977). `Can One Dent a Dyon?' {\it Physical Review} D, 15 (2), pp.~544-545.

\ \\Cooper, L.~N.~(1956). `Bound Electron Pairs in a Degenerate Fermi Gas'. {\it Physical Review,} 104 (4), pp.~1189-1190.

\ \\Corfield, D.~(2017). `Duality as a Category-Theoretic Concept'. {\it Studies in History and Philosophy of Modern Physics}, 59, pp.~55-61.

\ \\Cornea, O., Lupton, G., Oprea, J.~and Tanr\'e, D.~(2003). {\it Lusternik-Schnirelmann Category}, Providence, RI: American Mathematical Society.

\ \\Cowling, S.~(2022). `Haecceitism'. {\it Stanford Encyclopedia of Philosophy}.

\ \\Crede, V.~and Meyer, C.~A.~(2009). `The Experimental Status of Glueballs'. {\it Progress in Particle and Nuclear Physics}, 63, pp.~74-116.

\ \\Cremmer, E., B.~Julia and J.~Scherk (1978). `Supergravity Theory in Eleven Dimensions'. {\it Physics Letters B}, 74 (4), pp.~409-412.

\ \\Creutz, M.~(1980). `Asymptotic-Freedom Scales'. {\it Physical Review Letters}, 45 (5), pp.~313-316.

\ \\Creutz, M.~(1983). {\it Quarks, Gluons, and Lattices}. Cambridge: Cambridge University Press.

\ \\Crowther, K.~(2016). {\it Effective Spacetime. Understanding Emergence in Effective Field Theory and Quantum Gravity}. Springer.

\ \\Crowther, K.~(2018). `Inter-Theory Relations in Quantum Gravity: Correspondence, Reduction, and Emergence'. {\it Studies in History and Philosophy of Modern Physics}, 63, pp.~74-85.

\ \\Crowther, K.~(2021). `Defining a Crisis: the Roles of Principles in the Search for a Theory of Quantum Gravity'. {\it Synthese}, 198 (Supplement 14), pp.~3489-3516.

\ \\Crowther, K.~and De Haro, S.~(2022). `Four Attitudes Towards Singularities in the Search for a Theory of Quantum Gravity'. In: {\it The Foundations of Spacetime Physics. Philosophical Perspectives}, Vassallo, A.~(Ed.), pp.~223-250. London and New York: Routledge.

\ \\Crowther, K., Linnemann, N.~S.~and W\"uthrich, C.~(2021). `What we cannot learn from analogue experiments'. {\it Synthese}, 198 (Suppl.~16), S3701-S3726.

\ \\Crutchfield, J.~P.~(1994). `The Calculi of Emergence: Computation, Dynamics and Induction'. {\it Physica}, D 75, pp.~11-54.

\ \\Csikor, F., Fodor, Z.~and Heitger, J.~(1999). `End Point of the Hot Electroweak Phase Transition'. {\it Physical Review Letters}, 82 (1), pp.~21-24.

\ \\Cuffaro, M.~E.~and Hartmann, S.~(2023). `The Open Systems View and the Everett Interpretation'. {\it Quantum Reports}, 5 (2), pp.~418-425.

\ \\Curd, M.~and Cover, J.~A.~(1998). {\it Philosophy of Science}. New York: W.~W.~Norton.

\ \\Curiel, E.~(2014). `Classical Mechanics Is Lagrangian; It Is Not Hamiltonian'. {\it The British Journal for the Philosophy of Science}, 65, pp.~269-321.

\ \\Cushing, J.~T.~(1994). {\it Quantum Mechanics. Historical Contingency and the Copenhagen Hegemony}. Chicago: The University of Chicago Press.

\ \\Cyrot, M.~(1973). `Ginzburg-Landau Theory for Superconductors'. {\it Preports on Progress in Physics}, 36, pp.~103-158.

\ \\Dabholkar, A., Gibbons, G., Harvey, J.~A.~and Ruiz Ruiz, F.~(1990). `Superstrings and Solitons'. {\it Nuclear Physics} B, 340, pp.~33-55.

\ \\Da Costa, N.~C.~A.~and French, S.~(1990). `The Model-Theoretic Approach in the Philosophy of Science'. {\it Philosophy of Science}, 57 (2), pp.~248-265.

\ \\Dai, J., Leigh, R.~G.~and Polchinski, J.~(1989). `New Connections Between String Theories. {\it Modern Physics Letters} A, 4 (21), pp.~2073-2083.

\ \\D'Alessandro, A., D'Elia, M.~and Shuryak, E.~V.~(2010). `Thermal Monopole Condensation and Confinement in Finite Temperature Yang-Mills Theories'. {\it Physical Review} D, 81, 094501, pp.~1-13.

\ \\Dardashti, R., Hartmann, S., Th\'ebault, K.~and Winsberg, E.~(2019). `Hawking radiation and analogue experiments: A Bayesian analysis'. {\it Studies in History and Philosophy of Modern Physics}, 67, pp.~1-11.

\ \\Dardashti, R., Th\'ebault, P.~Y.~and Winsberg, E.~(2017). `Confirmation via Analogue Simulation: What Dumb Holes Could Tell Us about Gravity'. {\it The British Journal for the Philosophy of Science}, 68, pp.~55-89.

\ \\Darrigol, O.~(1986). `The Origin of Quantized Matter Waves'. {\it Historical Studies in the Physical and Biological Sciences}, 16 (2), pp.~197-253.

\ \\Dasgupta, A., Nicolai, H.~and Plefka, J.~(2002). `An Introduction to the Quantum Supermembrane'. {\it Gravitation and Cosmology}, 8, p.~1.

\ \\Dawid, R.~(2006). `Underdetermination and Theory Succession from the Perspective of String Theory'. {\it Philosophy of Science}, 73 (3), pp.~298-322.

\ \\Dawid, R~(2007). `Scientific Realism in the Age of String Theory'. {\it Physics and Philosophy}, 11, pp.~1-35.

\ \\Dawid, R.~(2013). {\it String Theory and the Scientific Method}. Cambridge: Cambridge University Press.

\ \\Dawid, R.~(2017). `String Dualities and Empirical Equivalence'. {\it Studies in History and Philosophy of Modern Physics}, 59, pp.~21-29.

\ \\De Bouvere, K.~(1965). `Synonymous Theories'. In: Addison, J.~W., Henkin, L.~and Tarski, A.~(Eds.), {\it Studies in Logic and the Foundations of Mathematics}, pp.~402-406. Amsterdam: North-Holland.

\ \\de Broglie, L.~(1930). {\it An Introduction to the Study of Wave Mechanics}. Methuen.

\ \\de Broglie, L.~(1960). {\it Non-Linear Wave Mechanics}. Amsterdam: Elsevier.

\ \\DeGrand, T.~A.~and Toussaint, D.~(1980). `Topological Excitations and Monte Carlo Simulation of Abelian Gauge Theory'. {\it Physical Review} D, 22 (10), pp.~2478-2489.

\ \\De Haro, S.~(2017a). `Dualities and emergent gravity: Gauge/gravity duality'. {\it Studies in History and Philosophy of Modern Physics}, 59, pp.~109-125. 

\ \\De Haro, S.~(2017b). `The Invisibility of Diffeomorphisms'. {\it Foundations of Physics} 47 (11), pp.~1464-1497.

\ \\De Haro, S.~(2019a). `Towards a Theory of Emergence for the Physical Sciences'. {\it European Journal for Philosophy of Science}, 9 (38), pp.~1-52.

\ \\De Haro, S.~(2019b). `The Heuristic Function of Duality'. {\it Synthese}, 196, pp.~5169-5203.

\ \\De Haro, S.~(2020a). `Spacetime and Physical Equivalence'. In: {\it Beyond Spacetime. The Foundations of Quantum Gravity}, Huggett, N., Matsubara, K.~and W\"uthrich, C.~(Eds.), pp.~257-283. Cambridge: Cambridge University Press. 

\ \\De Haro, S.~(2020b). `On Empirical Equivalence and Duality'. In: {\it One Hundred Years of Gauge Theory}, De Bianchi, S.~and Kiefer, C. (Eds.), pp.~91-106. Cham: Springer. 

\ \\De Haro, S.~(2020c). {\it On Inter-Theoretic Relations and Scientific Realism}. PhD Thesis, University of Cambridge. https://www.repository.cam.ac.uk/handle/1810/307749.

\ \\De Haro, S.~(2021). `Theoretical Equivalence and Duality'. {\it Synthese}, 198:~pp.~5139-5177.

\ \\De Haro, S.~(2022). `Noether's Theorems and Energy in General Relativity'. In: {\it The Philosophy and Physics of Noether's Theorems}, J.~Read and N.~Teh (Eds.), pp.~197-256. Cambridge: Cambridge University Press. 

\ \\De Haro, S.~(2023). `The Empirical Under-Determination Argument Against Scientific Realism for Dual Theories'. {\it Erkenntnis}, 88, pp.~117-145.

\ \\De Haro, S.~and Butterfield, J.~N.~(2018). `A Schema for Duality, Illustrated by Bosonization'. In: {\it Foundations of Mathematics and Physics one Century after Hilbert}. Kouneiher, J.~(Ed.), pp.~305-376. Cham: Springer.

\ \\De Haro, S.~and Butterfield, J.~N.~(2021). `On Symmetry and Duality'. {\it Synthese}, 198, pp.~2973-3013.

\ \\De Haro, S.~and De Regt, H.~W.~(2018). `Interpreting Theories without a Spacetime'. {\it European Journal for Philosophy of Science}, 8, pp.~631-670.

\ \\De Haro, S.~and De Regt, H.~W.~(2020). `A Precipice Below Which Lies Absurdity? Theories without a Spacetime and Scientific Understanding'. {\it Synthese}, 197, pp.~3121-3149.

\ \\De Haro, S., D.~Dieks, G.~'t Hooft and E.~Verlinde (2013). `Forty Years of String Theory Reflecting on the Foundations'. {\it Foundations of Physics}, 43, pp.~1-7.

\ \\De Haro, S., Mayerson, D., Butterfield, J.N. (2016). `Conceptual Aspects of Gauge/Gravity Duality'. {\it Foundations of Physics}, 46 (11), pp.~1381-1425. 

\ \\De Haro, S., Teh, N., Butterfield, J.N.~(2016). `On the Relation between Dualities and Gauge Symmetries'. {\it Philosophy of Science}, 83 (5), pp.~1059-1069. 

\ \\De Haro, S., Teh, N., Butterfield, J.N.~(2017). `Comparing dualities and gauge symmetries'. {\it Studies in History and Philosophy of Modern Physics}, 59, pp.~68-80.

\ \\De Haro, S., van Dongen, J., Visser, M.~and Butterfield, J.~(2020). `Conceptual Analysis of Black Hole Entropy in String Theory'. {\it Studies in History and Philosophy of Modern Physics}, 69, pp.~82-111.

\ \\Del Giudice, E., Di Vecchia, P.~and Fubini, S.~(1972). `General Properties of the Dual Resonance Model'. {\it Annals of Physics}, 70, pp.~378-398.

\ \\Dell'Antonio, G.~F., Frishman, Y.~and Zwanziger, D.~(1972). `Thirring Model in Terms of Currents: Solution and Light-Cone Expansions'. {\it Physical Review} D, 6 (4), pp.~988-1007.

\ \\De Regt, H.~W.~(2001). `Spacetime Visualisation and the Intelligibility of Physical Theories'. {\it Studies in History and Philosophy of Modern Physics}, 32 (2), pp.~243-265.

\ \\De Regt, H.~W.~(2009). `The Epistemic Value of Understanding'. {\it Philosophy of Science}, 76 (5), pp.~585-597.

\ \\De Regt, H.~W.~(2014). `Visualization as a Tool for Understanding'. {\it Perspectives on Science}, 22 (3), pp.~377-396.

\ \\De Regt, H.~W.~(2017). {\it Understanding Scientific Understanding}. New York: Oxford University Press.

\ \\De Regt, H.~W.~(2020). `Understanding, Values, and the Aims of Science'. {\it Philosophy of Science}, 87, pp.~921-932.

\ \\De Regt, H.~W.~and Dieks, D.~(2005). `A Contextual Approach to Scientific Understanding'. {\it Synthese}, 144, pp.~137-170.

\ \\Dewar, N.~(2015). ``Symmetries and the Philosophy of Language'. {\it Studies in History and Philosophy of Modern Physics}, 52, pp.~317-327.

\ \\Dewar, N.~(2019). `Sophistication about Symmetries'. {\it The British Journal for the Philosophy of Science}, 70 (2), pp.~485-521.

\ \\Dewar, N.~(2021). `There Are No Such Things as Theories, by Steven French'. {\it Mind}, 3, pp.~1-11.

\ \\De Wit, B., Hoppe, J.~and Nicolai, H.~(1988). `On the Quantum Mechanics of Supermembranes'. {\it Nuclear Physics} B, 305, pp.~545-581.

\ \\De Wit, B., L\"uscher, M.~and Nicolai, H.~(1989). `The Supermembrane is Unstable'. {\it Nuclear Physics} B, 320, pp.~135-159.

\ \\De Wit, B.~and Smith, J.~(1986). {\it Field Theory in Particle Physics}. Amsterdam: North-Holland.

\ \\D'Hoker, E., Freedman, D.~Z., Mathur, S.~D., Matusis, A.~and Rastelli, L.~(1999). `Graviton Exchange and Complete Four-Point Functions in the AdS/CFT Correspondence'. {\it Nuclear Physics} B, 562 (1-2), pp.~353-394.

\ \\D'Hoker, E.~and D.H.~Phong (1999). `Lectures on Supersymmetric Yang-Mills Theory and Integrable Systems'. arXiv:hep-th/991227.

\ \\D'Hoker, E.~and D.~H.~Phong (2002). `Two-Loop Superstrings. I. Main Formulas'. {\it Physics Letters B}, 529, pp.~241-255.

\ \\Dieks, D.~and de Regt, H.~W.~(1998). `Reduction and Understanding'. {\it Foundations of Physics}, 1, pp.~45-59.

\ \\Dieks, D., van Dongen, J.~and De Haro, S.~(2015). `Emergence in Holographic Scenarios for Gravity'. {\it Studies in History and Philosophy of Modern Physics}, 52, pp.~203-216.

\ \\Di Francesco, P., Mathieu, P.~and S\'en\'echal, D.~(1997). {\it Conformal Field Theory}. New York: Springer.

\ \\Dijkgraaf, R.~(1997). `Les Houches Lectures on Fields, Strings and Duality'. arXiv: hep-th/9703136.

\ \\Dijkgraaf, R.~and Vafa, C.~(2002). `On Geometry and Matrix Models'. {\it Nuclear Physics} B, 644, pp.~21-39.

\ \\Dingle, R.~B.~(1973). {\it Asymptotic Expansions: Their Derivation and Interpretation}. London: Academic Press.

\ \\Dirac, P.~A.~M.~(1925). `The Fundamental Equations of Quantum Mechanics'. In: van der Warden (1967), pp.~307-320.

\ \\Dirac, P.~A.~M.~(1927a). `The Physical Interpretation of the Quantum Dynamics. {\it Proceedings of the Royal Society of London} A 113:~pp.~621-641.

\ \\Dirac, P.~A.~M.~(1927b). `The Quantum Theory of the Emission and Absorption of Radiation'. {\it Proceedings of the Royal Society of London} A114:~pp.~243-265.

\ \\Dirac, P.~A.~M.~(1931). `Quantised Singularities in the Electromagnetic Field'. {\it Proceedings of the Royal Society of London. Series A}, 133 (821), pp.~60-72.

\ \\Dirac, P.~A.~M.~(1948). `The Theory of Magnetic Poles'. {\it Physical Review}, 74 (7), pp.~817-830.

\ \\Dirac, P.~A.~M.~(1958). {\it The Principles of Quantum Mechanics}. Oxford: Oxford University Press.

\ \\Dizadji-Bahmani, F., Frigg, R.~and Hartmann, S.~(2010). `Who's afraid of Nagelian reduction?' {\it Erkenntnis}, 73 (3), pp.~393-412.

\ \\Dolen, R., Horn, D.~and Schmid, C.~(1968). `Finite-Energy Sum Rules and Their Application to $\pi N$ Charge Exchange'. {\it Physical Review}, 166 (5), pp.~1768-1781.

\ \\Dorato, M.~(2000). `Substantivalism, Relationism, and Structural Spacetime Realism'. {\it Foundations of Physics}, 30 (10), pp.~1605-1628.

\ \\Dorey, N., Fraser, C., Hollowood, T.~J.~and Kneipp, M.~A.~C.~(1996). `S-Duality in ${\cal N}=4$ Supersymmetric Gauge Theories with Arbitrary Gauge Group'. {\it Physics Letters} B, 383, pp.~422-428.

\ \\Dorling, J.~(1970). `The Dimensionality of Time'. {\it American Journal of Physics}, 38, pp.~539-540.

\ \\Dougherty, J.~(2021). `The Substantial Role of Weyl Symmetry in Deriving General Relativity from String Theory'. {\it Philosophy of Science}, 88, pp.~1149-1160.

\ \\Dougherty, J.~and Callender, C.~(2016). `Black-Hole Thermodynamics: More Than an Analogy?' Forthcoming in: {\it A Guide to the Philosophy of Cosmology}, Ijjas, A.~and Loewer, B.~(Eds.). Oxford: Oxford University Press.

\ \\Dowling, M.~R.~and Nielsen, M.~A.~(2007). `The Geometry of Quantum Computation'. arXiv:quant-ph/0701004.

\ \\Duff, M.~J., Howe, P.~S., Inami, T.~and Stelle, K.~S.~(1987). `Superstrings in $D=10$ from Supermembranes in $D=11$'. {\it Physics Letters} B, 191 (70-4), pp.~205-209.

\ \\Duff, M.~J.~(1996). `Supermembranes'. arXiv:hep-th/9611203.

\ \\Duff, M.~J., Gibbons, G.~W.~and Townsend, P.~K.~(1994). `Macroscopic Superstrings as Interpolating Solitons'. {\it Physics Letters} B, 332, pp.~321-328.

\ \\Duff, M.~J, Khuri, R.~R.~and Lu, J.~X.~(1995). `String Solitons'. {\it Physics Reports}, 259, pp.~213-326.

\ \\Duff, M.~J.~and Lu, J.~X.~(1993). `Black and Super $p$-Branes in Diverse Dimensions'. {\it Nuclear Physics} B, 416, pp.~301-334.

\ \\Duff, M.~J.~and Stelle, K.~S.~(1991). `Multi-Membrane Solutions of $D=11$ Supergravity'. {\it Physics Letters} B, 253 (1-2), pp.~113-118.

\ \\Duhem, P.~(1991) [1914]. {\it The Aim and Structure of Physical Structure}. Princeton: Princeton University Press.

\ \\Duncan, A.~(2012). {\it The Conceptual Framework of Quantum Field Theory}. Oxford: Oxford University Press.

\ \\Duncan, A.~and Janssen, M.~(2008). `Pascual Jordan's Resolution of the Conundrum of Wave-Particle Duality'. {\it Studies in History and Philosophy of Modern Physics}, pp.~634-666.

\ \\Duncan, A.~and Janssen, M.~(2023). {\it Constructing Quantum Mechanics}. Volume 1: The Scaffold 1900-1923. Oxford: Oxford University Press.

\ \\Earman, J.~(1989). {\it World Enough and Space-Time. Absolute versus Relational Theories of Space and Time.} Cambridge, MA: The MIT Press.

\ \\Earman, J.~(1993). `Underdetermination Realism Reason'. {\it Midwest Studies in Philosophy}, XVIII, pp.~19-38.

\ \\Earman, J.~(2003). `Rough Guide to Spontaneous Symmetry Breaking'. In: Brading and Castellani (2003), pp.~335-346.

\ \\Earman, J.~(2004). `Curie's Principle and Spontaneous Symmetry Breaking'. {\it International Studies in the Philosophy of Science}, 18 (2-3), pp.~173-198.

\ \\Earman, J.~(2004). `Laws, Symmetry, and Symmetry Breaking: Invariance, Conservation Principles, and Objectivity'. {\it Philosophy of Science}, 71, pp.~1227-1241.

\ \\Earman, J.~(2011). `The Unruh Effect for Philosophers'. {\it Studies in History and Philosophy of Modern Physics}, 42, pp.~81-97.

\ \\Earman, J.~and Norton, J.~(1987). `What Price Spacetime Substantivalism? The Hole Story'. {\it The British Journal for the Philosophy of Science}, 38, pp.~515-525.

\ \\Earman, J.~and Valente, G.~(2014). `Relativistic Causality in Algebraic Quantum Field Theory'. {\it International Studies in the Philosophy of Science}, 28 (1), pp.~1-48.

\ \\Edwards, S.~F.~and Anderson, P.~W.~(1975). `Theory of Spin Glasses'. {\it Journal of Physics F: Metal Physics}, 5, pp.~965-974.

\ \\Einstein, A.~(1905). `On the Electrodynamics of Moving Bodies'. {\it Annalen der Physik}, 17 (10), pp.~891-921.

\ \\Einstein, A.~(1914a). `Covariance Properties of the Field Equations of the Theory of Gravitation Based on the General Theory of Relativity'. In: {\it The Collected Papers of Albert Einstein,} Vol.~6, Doc.~2, pp.~6-16. Princeton: Princeton University Press.

\ \\Einstein, A.~(1914b). `The Formal Foundation of the General Theory of Relativity'. In: {\it The Collected Papers of Albert Einstein,} Vol.~6, Doc.~9, pp.~30-84. Princeton: Princeton University Press.

\ \\Einstein, A.~and Grossman, M.~(1913). `Entwurf einer Verallgemeinerten Relativit\"tstheorie und einer Theorie der Gravitation'. Teubner, Leipzig. In: {\it The Collected Papers of Albert Einstein,} Vol.~4, Doc.~13, pp.~302-343. Princeton: Princeton University Press.

\ \\El Skaf, R.~and Palacios, P.~(2022). `What Can We Learn (and not learn) from Thought Experiments in Black Hole Thermodynamics?' {\it Synthese} 200, 434, pp.~1-27.

\ \\Englert, F.~and Brout, R.~(1964). `Broken Symmetry and the Mass of Gauge Vector Mesons'. {\it Physical Review Letters}, 13 (9), pp.~321-323.

\ \\Englert, F.~and Windey, P.~(1976). `Quantization Condition for 't Hooft Monopoles in Compact Simple Lie Groups'. {\it Physical Review} D, 14 (10), pp.~2728-2731.

\ \\Esfeld, M.~and Lam, V.~(2008). `Moderate Structural Realism about Space-Time'. {\it Synthese}, 160, pp.~27-46.

\ \\Falguera, J.~L., Mart\'inez-Vidal, C.~and Rosen, G.~(2021). `Abstract Objects'. {\it Stanford Encyclopedia of Philosophy}. https://plato.stanford.edu/entries/abstract-objects.

\ \\Faulkner, T., Liu, H.~and Rangamani, M.~(2011). `Integrating Out Geometry: Holographic Wilsonian RG and the Membrane Paradigm'. {\it Journal of High-Energy Physics}, 8, 51, pp.~1-38.

\ \\Faye, J.~(2019). `Copenhagen Interpretation of Quantum Mechanics'. {\it Stanford Encyclopedia of Philosophy}.

\ \\Fefferman, C.~and Graham, C.~R.~(1985). `Conformal Invariants'. {\it Soci\'et\'e Math\'ematique de France}, Ast\'erisque, S131, pp.~95-116.

\ \\Fefferman, C.~and Graham, C.~R.~(2012). {\it The Ambient Metric}. Princeton: Princeton University Press.

\ \\Feintzeig, B. (2018). `Toward an Understanding of Parochial Observables'. {\it The British Journal for the Philosophy of Science}, 69, pp.~161-191.

\ \\Ferrara, S.~(1987). {\it Supersymmetry}, Volume 1. Amsterdam: North Holland.

\ \\Ferrara, S., Zumino, B.~and Wess, J.~(1974). `Supergauge Multiplets and Superfields'. {\it Physics Letters} B, 51, pp.~239-241. Reprinted in: Ferrara (1987), pp.~41-43.

\ \\Feynman, R.~P.~(1955). `Application of Quantum Mechanics to Liquid Helium'. In: Gorter, C.~J.~(Ed.), {\it Progress in Low Temperature Physics I}, pp.~17-53. Amsterdam: North-Holland.

\ \\Field, G.~(2022). `Putting Theory in Its Place: The Relationship between Universality Arguments and Empirical Constraints'. {\it The British Journal for the Philosophy of Science}, forthcoming. 

\ \\Figueroa-O'Farrill, J.~M.~(1998). `Electromagnetic Duality for Children'. Online: https://www.maths.ed.ac.uk/~jmf/Teaching/Lectures/EDC.pdf.

\ \\Fine, A.~(1984). `The Natural Ontological Attitude'. In: {\it Scientific Realism}, J.~Leplin (Ed.), pp.~83-107. Berkeley: University of California Press.

\ \\Fine, A.~(1986). `Unnatural Attitudes: Realist and Instrumentalist Attachments to Science'. {\it Mind}, 95 (378), pp.~149-179.

\ \\Fish, A.~(2023). `Extensional Realism: Interesting and Uninteresting Truths'. Master of Logic Thesis Series, University of Amsterdam.\\ https://eprints.illc.uva.nl/id/eprint/2237.

\ \\Fisher, M.~E.~(1967). `The Theory of Equilibrium of Critical Phenomena'. {\it Reports on Progress in Physics}, 30, pp.~615-730.

\ \\Fisher, M.~E.~(1998). `Renormalization Group Theory: Its Basis and Formulation in Statistical Physics'. {\it Reviews of Modern Physics}, 70 (2), pp.~653-681.

\ \\Fisher, M.~E.~and Ferdinand, A.~E.~(1967). `Interfactial, Boundary, and Size Effects at Critical Points'. {\it Physical Review Letters}, 19 (4), pp.~169-172.

\ \\Fletcher, S.~C.~(2016). `Similarity, Topology, and Physical Significance in Relativity Theory'. {\it The British Journal for the Philosophy of Science}, 67, pp.~365-389.

\ \\Fodor, J.~A.~(1974). `Special Sciences (or: the Disunity of Science as a Working Hypothesis'. {\it Synthese}, 28, pp.~97-115.

\ \\Fradkin, E.~(2017). `Disorder Operators and Their Descendants'. {\it Journal of Statistical Physics}, 167, pp.~427-461.

\ \\Fodor, Z.~(2000). `Electroweak Phase Transitions'. {\it Nuclear Physics} B, 83-84, pp.~121-125.

\ \\Fradkin, E.~and Shenker, S.~H.~(1979). `Phase Diagrams of Lattice Gauge Theories with Higgs Fields'. {\it Physical Review} D, 19 (12), pp.~3682-3697.

\ \\Franklin, A.~and Knox, E.~(2018). `Emergence without Limits: The Case of Phonons'. {\it Studies in History and Philosophy of Modern Physics}, 64, pp.~68-78.

\ \\Franklin, A.~and Robertson, K.~(forthcoming). `Emerging into the Rainforest: Emergence and Special Science Ontology', 2021. http://philsci-archive.pitt.edu/19912.

\ \\Fraser, D.~(2009). `Quantum Field Theory: Underdetermination, Inconsistency, and Idealization'. {\it Philosophy of Science}, 76, pp.~536-567.

\ \\Fraser, D.~(2011). `How to Take Particle Physics Seriously: A Further Defence of Axiomatic Quantum Field Theory'. {\it Studies in History and Philosophy of Modern Physics}, 42, pp.~126-135.

\ \\Fraser, D.~(2017). `Formal and physical equivalence in two cases in contemporary quantum physics'. {\it Studies in History and Philosophy of Modern Physics}, 59, 30-43. 

\ \\Fraser, D.~and Koberinski, A.~(2016). `The Higgs Mechanism and Superconductivity: A Case Study of Formal Analogies'. {\it Studies in History and Philosophy of Modern Physics}, 55, pp.~72-91.

\ \\Freedman, D.~Z., Gubser, S.~S., Pilch, K.,~and Warner, N.~P.~(1999). `Renormalization Group Flows from Holography---Supersymmetry and a c-Theorem'. arXiv preprint hep-th/9904017.

\ \\Frege, G.~(1892) [1948]. `\"Uber Sinn und Bedeutung'. {\it Zeitschrift f\"ur Philosophie und philosophische Kritik}, 100, pp.~25-50. Translated into English as `Sense and Reference', in: {\it The Philosophical Review}, 1948, 57 (3), pp.~209-230.\\

\ \\Frege, G.~(1956). {\it The Foundations of Arithmetic}. New York: Harper and Brothers.

\ \\French, S.~(2020). {\it There Are No Such Things as Theories}. Oxford: Oxford University Press.

\ \\Friederich, S.~(2013). `Gauge Symmetry Breaking in Gauge Theories---in Search of Clarification'. {\it European Journal for Philosophy of Science}, 3, pp.~157-182.

\ \\Friederich, S.~(2015). `Symmetry, Empirical Equivalence, and Identity'. {\it The British Journal for the Philosophy of Science}, 66, pp.~537-559.

\ \\Friedman, M.~(1974). `Explanation and Scientific Understanding'. {\it The Journal of Philosophy}, 71 (1), pp.~5-19.

\ \\Friend, M., Khaled, M., Lefever, K.~and Sz\'ekely, G.~(2020). `Distances Between Formal Theories'. {\it The Review of Symbolic Logic}, 13 (3), pp.~633-654.

\ \\Frigg, R.~(2023). {\it Models and Theories}. London: Routledge.

\ \\Frigg, R.~and Votsis, I.~(2011). `Everything you always wanted to know about structural realism but were afraid to ask'. {\it European Journal for Philosophy of Science}, 1, pp.~227-276.

\ \\Frishman, Y.~and Sonnenschein, J.~(2010). {\it Non-Perturbative Field Theory}. Cambridge: Cambridge University Press.

\ \\Gaiotto, D.~and Witten, E.~(2022). `Gauge Theory and the Analytic Form of the Geometric Langlands Program'. {\it Annales Henri Poincar\'e}, pp.~1-115. Cham: Springer.

\ \\Garson, J.~(2018). `Modal Logic'. {\it Stanford Encyclopedia of Philosophy}.\\ https://plato.stanford.edu/entries/logic-modal.

\ \\Ge, D.~and and Policastro, G.~(2019). `Circuit Complexity and 2D Bosonisation'. {\it Journal of High Energy Physics}, 10, pp.~1-30.

\ \\Georgi, H.~(1999). {\it Lie Algebras in Particle Physics.} Second Edition. Westview Press.

\ \\Georgi, H.~and Glashow, S.~L.~(1972). `Unified Weak and Electromagnetic Interactions without Neutral Currents'. {\it Physical Review Letters}, 28 (22), pp.~1494-1497.

\ \\Georgi, H.~and Glashow, S.~L.~(1974). `Unity of All Elementary-Particle Forces'. {\it Physical Review Letters}, 32 (8), pp.~438-441.

\ \\Geroch, R.~(1969). `Limits of Spacetimes'. {\it Communications in Mathematical Physics}, 13, pp.~180-193.

\ \\Geroch, R.~(1972). `Structure of the Gravitational Field at Spatial Infinity', {\it Journal of Mathematical Physics}, 13, pp.~956-968.

\ \\Geroch, R.~(1978). {\it General Relativity from A to B}. Chicago: The University of Chicago Press.

\ \\Gershenson, C.~and Fern\'andez, N.~(2012). `Complexity and Information: Measuring Emergence, Self-Organization, and Homeostasis at Multiple Scales'. {\it Complexity}, 18 (2), pp.~29-44.

\ \\Gharibian, S., Huang, Y., Landau, Z.~and Shin, S.~W.~(2014). `Quantum Hamiltonian Complexity'. {\it Foundations and Trends in Theoretical Computer Science}, 10 (3), pp.~159-282.

\ \\Giamarchi, T.~(2003). {\it Quantum Physics in One Dimension}. Oxford: Oxford University Press.

\ \\Gibbons, G.~W.~(1994). `Changes of Topology and Changes of Signature'. {\it International Journal of Modern Physics} D, 3 (1), pp.~61-70.

\ \\Gibbons, G.~W.~(2012). `The Emergent Nature of Time and the Complex Numbers in Quantum Cosmology'. In: Mersini-Houghton, L.~and Vaas, R., {\it The Arrows of Time. A Debate in Cosmology}, pp.~109-148. Heidelberg: Springer.

\ \\Gibbons, G.~W.~and Hartle, J.~B.~(1990). `Real Tunneling Geometries and the Large-Scale Topology of the Universe'. {\it Physical Review} D, 42 (8), pp.~2458-2468.

\ \\Giddings, S.~B.~(2015). `Hilbert Space Structure in Quantum Gravity: An Algebraic Perspective'. {\it Journal of High-Energy Physics}, 99, pp.~1-20.

\ \\Gilbert, W.~(1964). `Broken Symmetries and Massless Particles'. {\it Physical Review Letters}, 12 (25), pp.~713-714.

\ \\Gilmore, R.~(2008). {\it Lie Groups, Physics, and Geometry}. Cambridge: Cambridge University Press.

\ \\Ginsparg, P.~(1988). {\it Applied Conformal Field Theory}. arXiv:hep-th/9108028.

\ \\Ginzburg, V.~L., and Landau, L.~D.~(1950). `On the Theory of Superconductivity'. {\it Soviet Physics JETP}, 20, 1064. In: ter Haar (Ed.), {\it Collected Papers of L.~D.~Landau,} pp.~546-568. New York and London: Pergamon Press, 1965.

\ \\Girardello, L., Giveon, A., Porrati, M.~and Zaffaroni, A.~(1994). `Non-Abelian Strong-Weak Coupling Duality in (String-Derived) ${\cal N}=4$ Supersymmetric Yang-Mills Theories'. {\it Physics Letters} B, 334 (3-4), pp.~331-338.

\ \\Girardello, L., Giveon, A., Porrati, M.~and Zaffaroni, A.~(1995). `S-Duality in ${\cal N}=4$ Yang-Mills Theories with General Gauge Groups'. {\it Nuclear Physics} B, 448, pp.~127-165.

\ \\Glashow, S.~L.~(1959), `The Renormalizability of Vector Meson Interactions'. {\it Nuclear Physics}, 10, pp.~107-117.

\ \\Glashow, S.~L.~(1961). `Partial-Symmetries of Weak Interactions'. {\it Nuclear Physics}, 22, pp.~579-588.

\ \\Glick, D., Darby, G.~and Marmodoro, A.~(2020). {\it The Foundation of Reality}. Oxford: Oxford University Press.

\ \\Glymour, C.~(1970). `Theoretical Realism and Theoretical Equivalence'. {\it PSA Proceedings of the Biennial Meeting of the Philosophy of Science Association}, pp.~275-288.

\ \\Glymour, C.~(1977). `The epistemology of geometry'. {\it No$\hat u$s}, pp.~227-251.

\ \\Glymour, C.~(1980). {\it Theory and Evidence}. Princeton: Princeton University Press.

\ \\Glymour, C.~(1983). `Revisions of Bootstrap Testing'. {\it Philosophy of Science}, 50 (4), pp.~626-629.

\ \\Glymour, C.~(2013). `Theoretical Equivalence and the Semantic View of Theories'. {\it Philosophy of Science}, 80, pp.~286-297.

\ \\Goddard, P., Nuyts, J.~and Olive, D.~(1977). `Gauge Theories and Magnetic Charge'. {\it Nuclear Physics} B, 125, pp.~1-28.

\ \\Goddard, P.~and Olive, D.~I.~(1978). `Magnetic Monopoles in Gauge Field Theories. {\it Reports on Progress in Physics}, 41, pp.~1357-1437.

\ \\Goldstone, J.~(1961). `Field Theories with ``Superconductor'' Solutions'. {\it Il Nuovo Cimento}, 19 (1), pp.~154-164.

\ \\Goldstone, J., Salam, A.~and Weinberg, S.~(1962). `Broken Symmetries'. {\it Physical Review}, 127 (3), p.~965-970.

\ \\Gomes, H.~(2021). `Holism as the Empirical Significance of Symmetries'. {\it European Journal for Philosophy of Science}, 11 (87), pp.~1-41.

\ \\Gomes, H.~and Butterfield, J.~(2023). `The Hole Argument and Beyond: Part I: The Story so Far'. {\it Journal of Physics: Conference Series}, 2533, 012002, pp.~1-27.

\ \\G\"opfert, M.~and Mack, G.~(1982). `Proof of Confinement of Static Quarks in 3-Dimensional U(1) Lattice Gauge Theory for all Values of the Coupling Constant'. {\it Communications in Mathematical Physics}, 82, pp.~545-606.

\ \\Gor'kov, L.~P.~(1958). `On the Energy Spectrum of Superconductors'. {\it Soviet Physics JETP}, 34 (7), 3, pp.~505-508.

\ \\Gor'kov, L.~P.~(1959). `Microscopic Derivation of the Ginzburg-Landau Equations in the Theory of Superconductivity'. {\it Soviet Physics JETP}, 36 (9), 6, pp.~1364-1367.

\ \\Goto, T.~(1971). `Relativistic Quantum Mechanics of One-Dimensional Mechanical Continuum and Subsidiary Condition of Dual Resonance Model'. {\it Progress of Theoretical Physics}, 46 (5), pp.~1560-1569.

\ \\Gra\"adel, E.~and Otto, M.~(2014). `The Freedoms of (Guarded) Bisimulation'. In: Baltag and Smets (Eds.) (2014), pp.~3-31.

\ \\Greaves, H.~and Wallace, D.~(2014). `Empirical Consequences of Symmetries'. {\it The British Journal for the Philosophy of Science,} 65, pp.~59-89. 

\ \\Green, M.~B.~(1999). `Interconnections between Type II superstrings, M-theory and ${\cal N}= 4$ Yang-Mills'. In: {\it Quantum Aspects of Gauge Theories, Supersymmetry and Unification}, Ceresole, A., Kounnas, C., L\"ust, D.~and Theisen, S. (Eds.), pp.~22-96. Berlin, Heidelberg: Springer.

\ \\Green, M.~B., Schwarz, J.~H.~and Witten, E.~(1987). {\it Superstring Theory}. Volume 1: Introduction. Cambridge: Cambridge University Press.

\ \\Greensite, J.~(2020). {\it An Introduction to the Confinement Problem}. Heidelberg: Springer. Second Edition.

\ \\Greiner, W.~and Reinhardt, J.~(1996). {\it Field Quantization}. Berlin: Springer.

\ \\Griffiths, D.~J.~(1999). {\it Introduction to Electrodynamics}. Third Edition. New Jersey: Prentice Hall.

\ \\Grimm, S.~R.~(2010).`The Goal of Explanation'. {\it Studies in History and Philosophy of Science}, 41, pp.~337-344.

\ \\Grimmer, D., Cinti, E.~and Jaksland, R.~(2024). `Duality, Underdetermination, and the Uncommon Common Core'. 

\ \\Gryb, S., Palacios, P.~and Th\'ebault, P.~Y.~(2021). `On the Universality of Hawking Radiation'. {\it The British Journal for the Philosophy of Science}, 72 (3), pp.~809-837.

\ \\Guay, A., Sartenaer, O.~(2016). `A new look at emergence. Or when after is different'. {\it European Journal for Philosophy of Science}, 6 (2), pp.~297-322.

\ \\Gubser, S.~S., Klebanov, I.~R.~amd Polyakov, A.~M. (1998). `Gauge Theory Correlators from Non-critical String Theory'. {\it Physics Letters} B, 428 (1-2), pp.~105-114.

\ \\Gukov, S.~and Witten, E.~(2006). `Gauge Theory, Ramification, and the Geometric Langlands Program'. {\it Current Developments in Mathematics}, 1, pp.~35-180.

\ \\Guralnik, G.~S., Hagen, C.~R.~and Kibble, T.~W.~B.~(1964). `Global Conservation Laws and Massless Particles'. {\it Physical Review Letters}, 13 (20), pp.~585-587.

\ \\Gustafson, S.~and Sigal, I.~(2020). {\it Mathematical Concepts of Quantum Mechanics}, Third Edition. Cham: Springer.

\ \\Haag, R., Lopuszanski, J.~T.~and Sohnius, M.~(1975). `All Possible Generators of Supersymmetries of the S-Matrix'. {\it Nuclear Physics} B, 88, pp.~257-274.

\ \\Hackl, L., and Myers, R.~C.~(2018). `Circuit complexity for free fermions'. {\it Journal of High Energy Physics}, 7, pp.~1-72.

\ \\Halvorson, H.~(2007). `Algebraic Quantum Field Theory'. In: {\it Handbook of the Philosophy of Science}, Butterfield, J.~and Earman, J.~(Eds.). Amsterdam: Elsevier.

\ \\Halvorson, H.~(2012). `What Scientific Theories Could Not Be'. {\it Philosophy of Science}, 79, pp.~183-206.

\ \\Halvorson, H.~(2013). `The Semantic View, If Plausible, Is Syntactic'. {\it Philosophy of Science}, 80, pp.~475-478.

\ \\Halvorson, H.~(2021). `Steven French: {\it There Are No Such Things as Theories}'. {\it Journal for General Philosophy of Science}, 52, pp.~609-612.

\ \\Halvorson, H.~and Clifton, R.~(1999). `Maximal Beable Subalgebras of Quantum Mechanical Observables'. {\it International Journal of Theoretical Physics}, 38 (10), pp.~2441-2484.

\ \\Halvorson, H.~and Clifton, R.~(2001). `No Place for Particles in Relativistic Quantum Theories?' {\it Philosophy of Science}, 69, pp.~1-28.

\ \\Halvorson, H.~and Clifton, R.~(2002). `Reconsidering Bohr's Reply to EPR'. In: {\it Non-locality and Modality}, Placek, T.~and Butterfield, J.~(Eds.), pp.~3-18.

\ \\Halvorson, H.~and Manchak, J.~B.~(2022). `Closing the Hole Argument'. {\it The British Society for the Philosophy of Science}, in press.

\ \\Halvorson, H.~and Tsementzis, D.~(2017). `Categories of Scientific Theories'. In: Landry, E.~(Ed.), {\it Categories for the Working Philosopher}, pp.~402-429. Oxford: Oxford University Press.

\ \\Hamilton, M.~J.~D.~(2017). {\it Mathematical Gauge Theory}. Cham: Springer.

\ \\Han, M.~and Hung, L.-Y.~(2017). `Loop Quantum Gravity, Exact Holographic Mapping, and Holographic Entanglement Entropy'. {\it Physical Review} D, 95, 024011, pp.~1-24.

\ \\Harlow, D.~(2018). `TASI Lectures on the Emergence of Bulk Physics in AdS/CFT'. {\it Proceedings of Science}, pp.~1-52.

\ \\Hartmann, H.~(1999). `Models and Stories in Hadron Physics'. In: {\it Models as Mediators}, Morgan, M.~and Morrison, M.~(Eds.), pp.~326-346. Cambridge: Cambridge University Press.

\ \\Hartmann, S.~(2001). `Effective Field Theories, Reductionism and Scientific Explanation'. {\it Studies in History and Philosophy of Modern Physics}, 32 (2), pp.~267-304.

\ \\Hartmann, S.~(2002). `On Correspondence'. {\it Studies in History and Philosophy of Modern Physics}, 33, pp.~79-94.

\ \\Harvey, J.~A.~(1996). `Magnetic Monopoles, Duality, and Supersymmetry'. arXiv: hep-th/9603086.

\ \\Hawking, S.~W.~(1974). `Black Hole Explosions?' {\it Nature}, 248, pp.~30-31.

\ \\Hawking, S.~W.~(1975). `Particle Creation by Black Holes'. {\it Communications in Mathematical Physics}, 43, pp.~199-220.

\ \\Hawking, S.~W.~(1976). `Breakdown of Predictability in Gravitational Collapse'. {\it Physical Review} D, 14 (10), pp.~2460-2473.

\ \\Hawking, S.~W.~and Ellis, G.~F.~(1973). {\it The Large Scale Structure of Space-Time}. Cambridge: Cambridge University Press.

\ \\Hayward, S.~A.~(1992). `Signature Change in General Relativity'. {\it Classical and Quantum Gravity}, 9, pp.~1851-1862.

\ \\Healey, R.~(2007). {\it Gauging What's Real}. Oxford: Oxford University Press.

\ \\Healey, R.~(2009). `Perfect Symmetries'. {\it The British Journal for the Philosophy of Science,} 60, pp.~697-720.

\ \\Heckman, J.~J., Joyce, A., Sakstein, J.~and Trodden, M.~(2022). `Exploring $2+2$ Answers to $3+1$ Questions'. arXiv:2208.02267, pp.~1-22. https://arxiv.org/abs/2208.02267.

\ \\Hegerfeldt, G.~C.~(1998a). `Causality, Particle Localization and Positivity of the Energy'. In: {\it Irreversibility and Causality}, Bo\"ohm, A.~et al.~(Eds.), pp.~238-245. New York: Springer.

\ \\Hegerfeldt, G.~C.~(1998b). `Instantaneous Spreading and Einstein Causality in Quantum Theory'. {\it Annals of Physics}, 7 (7-8), pp.~716-725.

\ \\Heim, I.~and Kratzer, A.~(1998). {\it Semantics in Generative Grammar}. Malden and Oxford: Blackwell.

\ \\Heisenberg, W.~(1925a). `Zur Quantentheoie der Multiplettstruktur und der anomalen Zeemaneffekte'. {\it Zeitschrift f\"ur Physik}, pp.~841-860. In: Heisenberg (1985), pp.~307-325. Berlin: Springer.

\ \\Heisenberg, W.~(1925b). `\"Uber quantentheoretische Umdeutung kinematischer und mechanischer Beziehungen'. {\it Zeitschrift f\"ur Physics}, 33, pp.~879-893. In: Heisenberg (1985), pp.~382-396. Berlin: Springer.

\ \\Heisenberg, W.~(1926). `Quantenmechanik'. {\it Die Naturwissenschaften} 14, pp.~989-994. In: Heisenberg (1984), pp.~52-57. Berlin: Springer.

\ \\Heisenberg, W.~(1927). `\"Uber den anschaulischen Inhalt der quantentheoretischen Kinematik und Mechanik'. In: Heisenberg (1985), pp.~478-504. English translation: `The Actual Content of Quantum Theoretical Kinematics and Mechanics', NASA Technical Memorandum, 1.15: 77379.

\ \\Heisenberg, W.~(1929). `Die Entwicklung der Quantentheorie 1918-1928'. {\it Die Naturwissenschaften}, 17, pp.~490-496. In: Heisenberg (1984), pp.~109-115.

\ \\Heisenberg, W.~(1930). {\it The Physical Principle of the Quantum Theory}. Chicago: The University of Chicago Press. In: Heisenberg (1984), p.~117-166.

\ \\Heisenberg, W.~(1984). {\it Collected Works of Werner Heisenberg}, B, Blum, W., D\"urr, H.-P.~and Rechenberg, H.~(Eds.). Berlin: Springer.

\ \\Heisenberg, W.~(1985). {\it Collected Works of Werner Heisenberg}, A-I, Blum, W., D\"urr, H.-P.~and Rechenberg, H.~(Eds.). Berlin: Springer.

\ \\Held, C.~(1994). `The Meaning of Complementarity'. {\it Studies in History and Philosophy of Science}, 25 (6), pp.~871-893.

\ \\Hempel, C.~G.~(1965). {\it Scientific Explanation. Essays in the Philosophy of Science.} New York: The Free Press.

\ \\Hempel, C.~(1966). {\it Philosophy of Natural Science}. New Jersey: Prentice Hall.

\ \\Hesse, M.~(2000). `Models and Analogies'. In: {\it A Companion to the Philosophy of Science}, Newton-Smith, W.~H.~(Ed.), pp.~299-307. Blackwell Publishers.

\ \\Higgs, P.~W.~(1964). `Broken Symmetries and the Masses of Gauge Bosons'. {\it Physical Review Letters}, 13 (16), pp.~508-509.

\ \\Higgs, P.~W.~(1966). `Spontaneous Symmetry Breakdown without Massless Bosons'. {\it Physical Review}, 145 (4), pp.~1156-1163.

\ \\Hilgevoord, J.~and Uffink, J.~(2016). `The Uncertainty Principle'. {\it Stanford Encyclopedia of Philosophy}.

\ \\Hodges, W.~(1993). {\it Model Theory}. Cambridge: Cambridge University Press.

\ \\Hodges, W.~(1997). {\it A Shorter Model Theory}. Cambridge: Cambridge University Press.

\ \\Hoefer, C.~(1996). `The Metaphysics of Space-Time Substantivalism. {\it The Journal of Philosophy}, 93 (1), pp.~5-27.

\ \\Hollands, S.~and Wald, R.~M.~(2010). `Axiomatic Quantum Field Theory in Curved Spacetime'. {\it Communications in Mathematical Physics}, 293, pp.~85-125.

\ \\Hollowood, T.~J.~and Kumar, S.~P.~(2015). `Partition Function of ${\cal N}=2^*$ on a Large Four-Sphere'. {\it Journal of High-Energy Physics}, 12, 16, pp.~1-41.

\ \\Horowitz, G.~T.~(2011). `Introduction to Holographic Superconductors'. In: {\it From Gravity to Themral Gauge Theories: The AdS/CFT Correspondence}, Papantonopoulos, E.~(Ed.), pp.~313-347. Heidelberg: Springer.

\ \\Horowitz, G.~T.~and Polchinski, J.~(2009). `Gauge/gravity Duality'. In: {\it Approaches to Quantum Gravity}, Oriti, D.~(Ed.). Cambridge: Cambridge University Press.

\ \\Horowitz, G.~T.~and Polchinski, J.~(2009). `Gauge/gravity Duatlity'. In: {\it Approaches to Quantum Gravity}, Oriti, D.~(Ed.), pp.~169-186. Cambridge: Cambridge University Press.

\ \\Horowitz, G.~T.~and Strominger, A.~(1991). `Black Strings and $p$-Branes. {\it Nuclear Physics} B, 360, pp.~197-209.

\ \\Howard, D.~(2007). `Reduction and Emergence in the Physical Sciences: Some Lessons from the Particle Physics-Condensed Matter Physics Debate'. In: Murphy, N.~and Stoeger, W.~R.~(Eds.), {\it Evolution and Emergence}. New York: Oxford University Press.

\ \\Huang, P., Joglekar, A., Li, B.~and Wagner, C.~E.~M.~(2016). `Probing the Electroweak Phase Transition at the LHC'. {\it Physical Review} D, 93, 055049, pp.~1-21.

\ \\Hudetz, L.~(2019). `Definable Categorical Equivalence'. PhilSci: http://philsci-archive.pitt.edu/14297.

\ \\Huggett, N.~(1999). {\it Space from Zeno to Einstein. Classic Readings with a Contemporary Commentary}. Cambridge, MA: The MIT Press.

\ \\Huggett, N.~(2017), `Target space $\neq$ space'. {\it Studies in History and Philosophy of Modern Physics}, 59, pp.~81-88.

\ \\Huggett, N.~and Vistarini, T.~(2015). `Deriving General Relativity from String Theory'. {\it Philosophy of Science}, 82, pp.~1163-1174.

\ \\Huggett, N.~and W\"uthrich, C.~(2013). `Emergent Spacetime and Empirical (In)coherence'. {\it Studies in History and Philosophy of Modern Physics}, 44, pp.~276-285.

\ \\Huggett, N.~and W\"uthrich, C.~(2023). {\it Out of Nowhere: The Emergence of Spacetime in Quantum Theories of Gravity}. Forthcoming, 
Oxford: Oxford University Press.

\ \\Hughes, G.~E.~and Cresswell, M.~J.~(1968). {\it An Introduction to Modal Logic}. London: Methuen.

\ \\Hughes, J., Liu, J.~and Polchinski, J.~(1986). `Supermembranes'. {\it Physics Letters} B, 180 (4), pp.~370-374.

\ \\Hull, C.~M.~and Townsend, P.~K.~(1995). `Unity of Superstring Dualities'. {\it Nuclear Physics} B, 438, pp.~109-137.

\ \\Humphreys, J.~E.~(1972). {\it Introduction to Lie Algebras and Representation Theory}. New York: Springer.

\ \\Humphreys, P.~(2016). {\it Emergence. A Philosophical Accounts}. New York and Oxford: Oxford University Press.

\ \\Ib\'a\~nez, L.~E.~and Uranga, A.~M.~(2012). {\it String Theory and Particle Physics. An Introduction to String Phenomenology}. Cambridge: Cambridge University Press.

\ \\Imbimbo, C., Schwimmer, A., Theisen, S.~and Yankielowicz, S.~(2000). `Diffeomorphisms and Holographic Anomalies. {\it Classical and Quantum Gravity}, 17, pp.~1129-1138.

\ \\Intrilligator, K.~and Seiberg, N.~(1996). `Lectures on Supersymmetric Gauge Theories and Electric-Magnetic Duality'. {\it Nuclear Physics} B, 45 (b-c), pp.~1-28.

\ \\Jackiw, R.~and Rebbi, C.~(1976). `Vacuum Periodicity in a Yang-Mills Quantum Theory'. {\it Physical Review Letters}, 37 (3), pp.~172-175.

\ \\Jackson, J.~D.~(1962). {\it Classical Electrodynamics}. New York: John Wiley and Sons.

\ \\Jaeger, G.~(1998). `The Ehrenfest Classification of Phase Transitions: Introduction and Evolution'. {\it Archive for History of Exact Sciences,} 53, pp.~51-81.

\ \\Jafferis, D., Zlokapa, A., Lykken, J. D., Kolchmeyer, D. K., Davis, S. I., Lauk, N., ... and Spiropulu, M.~(2022). `Traversable Wormhole Dynamics on a Quantum Processor. {\it Nature}, 612 (7938), pp.~51-55.

\ \\Jahn, A.~and Eisert, J.~(2021). `Holographic Tensor Network Models and Quantum Error Correction: A Topical Review'. {\it Quantum Science and Technology}, 6, 033002, pp.~1-30.

\ \\Jaksland, R.~(2021). `Entanglement as the World-Making Relation: Distance from Entanglement'. {\it Synthese}, 198 (10), pp.~9661-9693.

\ \\Jammer, M.~(1989). {\it The Conceptual Development of Quantum Mechanics}. Tomash Publishers.

\ \\Janssen, M.~and Renn, J.~(2022). {\it How Einstein Found His Field Equations}. Cham: Springer.

\ \\Jauch, J-M.~(1968). {\it Foundations of Quantum Mechanics}. Reading, MA: Addison-Wesley.

\ \\Jeevanjee, N.~(2011). {\it An Introduction to Tensors and Group Theory for Physicists.} New York: Springer.

\ \\Johnson, C.~V.~(2003). {\it D-Branes}. Cambridge: Cambridge University Press.

\ \\Jordan, P.~(1927). `Zur Quantenmechanik der Gasentartung'. {\it Zeitschrift f"ur Physik}, 44 (6-7), pp.~473-480.

\ \\Jordan, P.~and Klein, O.~(1927). `Zum Mehrk\"orperproblem der Quantentheorie'. {\it Zeitschrift f\"ur Physik}, XLV, pp.~751-765.

\ \\Jordan, P.~and Wigner, E.~(1918). `\"Uber das Paulische \"Aquivalenzverbot'. {\it Zeitschrift f\"ur Physik}, 47 (42), pp.~631-651.

\ \\Jordan, T.~(1969). {\it Linear Operators for Quantum Mechanics}. Wiley 1969 (and Dover 2006).

\ \\Kac, V.~(1990). {\it Infinite-Dimensional Lie Algebras}. Cambridge: Cambridge University Press.

\ \\Kac, V.~(1997). {\it Vertex Algebras for Beginners}. American Mathematical Society. 

\ \\Kadanoff, L.~P.~(2000). {\it Statistical Physics. Statics, Dynamics, and Renormalization.} Singapore: World Scientific.

\ \\Kadanoff, L.~P.~and Ceva, H.~(1971). `Determination of an Operator Algebra for the Two-Dimensional Ising Model'. {\it Physical Review} B, 3 (11), pp.~3918-3939.

\ \\Kaiser, D.~(2005). {\it Drawing Theories Apart. The Dispersion of Feynman Diagrams in Postwar Physics}. Chicago and London: The University of Chicago Press.

\ \\Kapustin, A.~(2008). `Gauge Theory, Mirror Symmetry, and the Geometric Langlands Program'. In: {\it Homological Mirror Symmetry: New Developments and Perspectives}, pp.~1-22). Berlin, Heidelberg: Springer.

\ \\Kapustin, A.~and Witten, E.~(2006). `Electric-Magnetic Duality and the Geometric Langlands Program'. arXiv preprint hep-th/0604151.

\ \\Karaca, K.~(2013). `The Construction of the Higgs Mechanism and the Emergence of the Electroweak Theory'. {\it Studies in History and Philosophy of Modern Physics}, 44, pp.~1-16.

\ \\Karch, A.~and Tong, D.~(2016). `Particle-Vortex Duality from 3d Bosonization'. {\it Physical Review X}, 6, 031043, pp.~1-11.

\ \\Khaled, M.~and Sz\'ekely, G.~(2020). `Algebras of Concepts and Their Networks'. In: Allahviranloo, T., Salahshour, S. and Arica, N.~(Eds.), {\it Progress in Intelligent Decision Science}. Cham: Springer.

\ \\Khalifa, K.~(2012). `Inaugurating Understanding or Repackaging Explanation?' {\it Philosophy of Science}, 79, pp.~15-37.

\ \\Khalifa, K.~(2017). {\it Understanding, Explanation, and Scientific Knowledge}. Cambridge: Cambridge University Press.

\ \\Kim, J.~(2002). `The Layered Model: Metaphysical Considerations'. {\it Philosophical Explorations}, 5 (1), pp.~2-20.

\ \\Kitcher, P.~(1981). `Explanatory Unification'. {\it Philosophy of Science}, 48, pp.~507-531.

\ \\Kleiner, W.~H., Roth, L.~M., and Autler, S.~H.~(1964). `Bulk Solution of Ginzburg-Landau Equations for Type II Superconductors: Upper Critical Field Region'. {\it Physical Review}, 133 (5A), pp.~1226-1227.

\ \\Knox, E.~(2016). `Abstraction and its Limits: Finding Space For Novel Explanation'. {\it Nous}, 50 (1), pp.~41-60.

\ \\Kogut, J.~B.~(1979). `An Introduction to Lattice Gauge Theory and Spin Systems'. {\it Reviews of Modern Physics}, 51 (4), pp.~659-713.

\ \\Kock, J.~(2003). {\it Frobenius Algebras and 2D Topological Quantum Field Theories}. Cambridge: Cambridge University Press.

\ \\Komar, A.~and Salam, A.~(1960). `Renormalization Problem for Vector Meson Theories'. {\it Nuclear Physics}, 21, pp.~624-230.

\ \\Kosmann-Schwarzbach, Y.~(2011). {\it The Noether Theorems}. New York: Springer.

\ \\Kosso, P.~(2000). `The Empirical Status of Symmetries in Physics'. {\it The British Journal for the Philosophy of Science,} 51 (1), pp.~81-98.

\ \\Kosterlitz, J.~M.~and Thouless, D.~J.~(1972). `Long Range Order and Metastability in Two Dimensional Solids and Superfluids'. {\it Journal of Physics C: Solid State Physics}, 5 L, pp.124-126.

\ \\Kosterlitz, J.~M.~and Thouless, D.~J.~(1973). `Ordering, Metastability and Phase Transitions in Two-Dimensional Systems'. {\it Journal of Physics C: Solid State Physics}, 6, pp.~1181-1203.

\ \\Kraan, T.~C.~and van Baal, P.~(1998). `Periodic Instantons with Non-Trivial Holonomy'. {\it Nuclear Physics} B, 533 (1-3), pp.~627-659.

\ \\Kramers, H.~A.~and Wannier, G.~H.~(1941). `Statistics of the Two-Dimensional Ferromagnet. Part I'. {\it Physical Review}, 60, pp.~252-262.

\ \\Kripke, S.~A.~(1972). {\it Naming and Necessity}. Oxford: Basil Blackwell.

\ \\Kuhlmann, M.~(2010). {\it The Ultimate Constituents of the Material World}. Frankfurt: Ontos Verlag.

\ \\Kuhn, T.~S.~(1977). `Objectivity, Value Judgment, and Theory Choice'. In: {\it The Essential Tension}, pp.~320-339.

\ \\Kuhn, T.~S.~(1978). {\it Black-Body Theory and the Quantum Discontinuity, 1894-1912}. Chicago: The University of Chicago Press.

\ \\Ladyman, J.~(1998). `What is Structural Realism?' {\it Studies in History and Philosophy of Science}, 29 (3), pp.~409-424.

\ \\Ladyman, J.~(2014). `Structural Realism'. {\it Stanford Encyclopedia of Philosophy}, https://plato.stanford.edu/entries/structural-realism.

\ \\Ladyman, J., Ross, D., Spurrett, D.~and Collier, J.~(2007). {\it Every Thing Must Go: Metaphysics Naturalized}. Oxford: Oxford University Press.

\ \\Landau, L.~D.~(1936). `The Theory of Phase Transitions'. {\it Nature,} 138, pp.~840-841.

\ \\Landau, L.~D.~(1937). `On the Theory of Phase Transitions. Part I'. {\it JETP,} 7, p.~1. In: ter Haar (Ed.), {\it Collected Papers of L.~D.~Landau,} pp.~193-209. New York and London: Pergamon Press, 1965.

\ \\Landau, L.~D.~and Lifshitz, E.~M.~(1980). {\it Statistical Physics}. Course of Theoretical Physics, Volume 5, Part 1. Oxford: Pergamon Press. Third Edition.

\ \\Landsman, K.~(2017). {\it Foundations of Quantum Theory}. Cham: Springer.

\ \\Landsman, K.~(2022). `Reopening the Hole Argument'. arXiv:2206.04943.

\ \\Laudan, L.~(1990). `Demystifying Underdetermination'. {\it Minnesota Studies in the Philosophy of Science}, 14, pp.~267-297.

\ \\Laudan, L.~and Leplin, J.~(1991). `Empirical Equivalence and Underdetermination'. {\it The Journal of Philosophy}, 88 (9), pp.~449-472.

\ \\Lavis, D.~A.~and Bell, G.~M.~(1999). {\it Statistical Mechanics of Lattice Systems}. Berlin: Springer.

\ \\Le Bihan, B.~and Read, J.~(2018). `Duality and Ontology'. Published online in {\it Philosophy Compass}.

\ \\Lee, J.~M.~(2018). {\it Introduction to Riemannian Manifolds}. Cham: Springer.

\ \\Lee, K.~and Lu, C.~(1998). `SU(2) Calorons and Magnetic Monopoles'. {\it Physical Review} D, 58, 025011, pp.~1-7.

\ \\Leeds, S.~(1995). `Holes and Determinism: Another Look'. {\it Philosophy of Science}, 62 (3), pp.~425-437.

\ \\Lehmkuhl, D.~(2017). `Introduction: Towards a Theory of Spacetime Theories'. In: {\it Towards a Theory of Spacetime Theories}, Lehmkuhl, D., Schiemann, G.~and Scholz, E.~(Eds.). New York: Springer.

\ \\Lenzen, V.~F.~(1955). `Procedures of Empirical Science'. In: {\it International Encyclopedia of Unified Science}, Neurath, O., Bohr, N., Dewey, J., Russell, B., Carnap, R., and Morris, C. W. (Eds.). Volume I, pp.~280-339.

\ \\Lerche, W.~(1998). `Introduction to Seiberg-Witten Theory and its Stringy Origin'. {\it AIP Conference Proceedings} 419, pp.~171-217.

\ \\Lewis, D.~K.~(1968). `Counterpart Theory and Quantified Modal Logic'. {\it The Journal of Philosophy}, 65 (5), pp.~113-126.

\ \\Lewis, D.~K.~(1969). {\it Convention: A Philosophical Study}. Cambridge, MA: Harvard University Press.

\ \\Lewis, D.~K.~(1970). `General Semantics'. {\it Synthese} 22, pp.~18-67.

\ \\Lewis, D.~K.~(1973). `Counterfactuals and Comparative Possibility'. {\it Journal of Philosophical Logic}, 2, pp.~418-446.

\ \\Lewis, D.~K.~(1975). `Languages and Language'. In: {\it Language, Mind, and Knowledge}, Minnesota Studies in the Philosophy of Science, VII, Gunderson, K.~(Ed.), pp.~3-35. Minnesota: University of Minnesota Press.

\ \\Lewis, D.~K.~(1980). `Index, Context, and Content'. Reprinted in: {\it Papers in Philosophical Logic}, 1998, pp.~21-44. Cambridge: Cambridge University Press.

\ \\Lewis, D.~K.~(1983). `New Work for a Theory of Universals'. {\it Australasian Journal of Philosophy}, 61 (4), pp.~343-377.

\ \\Lewis, D.~K.~(1984). `Putnam's Paradox'. {\it Australasian Journal of Philosophy}, 62 (3), pp.~221-236.

\ \\Lewis, D.~K.~(1986). {\it On the Plurality of Worlds}. Oxford: Oxford University Press.

\ \\Lipton, P.~(2004). {\it Inference to the Best Explanation}. London and New York: Routledge.

\ \\London, F.~(1948). `On the Problem of the Molecular Theory of Superconductivity'. {\it Physical Review}, 74 (5), pp.~562-573.

\ \\London, F.~(1950). {\it Superfluids. Volume I: Macroscopic Theory of Superconducvity.} New York: Dover Publications, Inc. Second Revised Edition, 1961.

\ \\London, F.~and London, H.~(1935). `The Electromagnetic Equations of the Supraconductor'. {\it Proceedings of the Royal Society. Series A-Mathematical and Physical Sciences}, 149 (866), pp.~71-88.

\ \\Lowe, E.~J.~(2002). {\it A Survey of Metaphysics}. Oxford: Oxford University Press.

\ \\L\"ust, D.~and Theisen, S.~(1989). {\it Lectures on String Theory}. Berlin: Springer.

\ \\Lutz, S.~(2013). `On a Straw Man in the Philosophy of Science: A Defense of the Received View'. {\it HOPOS: The Journal of the International Society for the History of Philosophy of Science}, 2, pp.~77-120.

\ \\Lutz, S.~(2017). `What Was the Syntax-Semantics Debate in the Philosophy of Science About?' {\it Philosophy and Phenomenological Research}, XCV (2), pp.~319-352.

\ \\Lyre, H.~(2008). `Does the Higgs Mechanism Exist?' {\it International Studies in the Philosophy of Science}, 22 (2), pp.~119-133.

\ \\Lyre, H.~(2011). `Is Structural Underdetermination Possible?' {\it Synthese}, 180, pp.~235-247.

\ \\Maggiore, M.~(2005). {\it A Modern Introduction to Quantum Field Theory}. Oxford: Oxford University Press.

\ \\Magnin, P.~and Kahn, R.~(2017). {\it Superconductivity. An Introduction.} Grenoble: Springer.

\ \\Maldacena, J.M.~(1996). {\it Black Holes in String Theory}. Ph.D.~Thesis, Princeton University. arXiv:hep-th/9607235.

\ \\Maldacena, J.~(1997). `The large-$N$ Limit of Superconformal Field Theories and Supergravity'. {\it International Journal of Theoretical Physics}, 38 (4), 1999, pp.~1113-1133.

\ \\Malament, D.~B.~(1996). `In Defense of Dogma: Why There Cannot Be A Relativistic Quantum Mechanics of (Localizable) Particles. In: {\it Perspectives on Quantum Reality}, Clifton, R.~(Ed.), pp.~1-10. Dordrecht: Springer.

\ \\Malament, D.~B.~(2012). {\it Topics in the Foundations of General Relativity and Newtonian Gravitation Theory}. Chicago: The University of Chicago Press.

\ \\Maldacena, J.~M.~(2003a). `Non-Gaussian Features of Primordian Fluctuations in Single Field Inflationary Models'. {\it Journal of High-Energy Physics}, 13, pp.~1-31.

\ \\Maldacena, J.~M.~(2003b). `TASI Lectures 2003 on AdS/CFT'. In: {\it Progress in String Theory}. Proceedings, Summer School, TASI 2003, Boulder, USA, June 2–27, Hackensack. World Scientific, USA (2003). arXiv:hep-th/0309246.

\ \\Maldacena, J., Sheikh-Jabbari, M.~M.~and Van Raamsdonk, M.~(2003). `Transverse Fivebranes in Matrix Theory'. {\it Journal of High-Energy Physics}, 1, 38, pp.~1-21.

\ \\Mandelstam, S.~(1975a). `Vortices and Quark Confinement in Non-Abelian Gauge Theories'. {\it Physics Letters}, 53 B (5), pp.~476-478.

\ \\Mandelstam, S.~(1975b). `Soliton Operators for the Quantized sine-Gordon Equation'. {\it Physical Review} D, 11 (10), pp.~3026-3030.

\ \\Mandelstam, S.~(1976). `Vortices and Quark Confinement in Non-Abelian Gauge Theories. II'. In: Gervais, J.~L.~and Neveu, A., `Extended Systems in Field Theory', {\it Physics Reports}, 23 (3), pp.~245-249.

\ \\Manton, N.~and Sutcliffe, P.~(2004). {\it Topological Solitons}. Cambridge: Cambridge University Press.

\ \\Marsden, J.~E.~and Ratiu, T.~S.~(1998). {\it Introduction to Mechanics and Symmetry}. Second Edition, 2010. New York: Springer.

\ \\Martens, N.~C.~M.~and Read, J.~(2021). `Sophistry about Symmetries?' {\it Synthese}, 199, pp.~315-344.

\ \\Matsubara, K.~(2013). `Realism, underdetermination and string theory dualities'. {\it Synthese}, 190 (3), pp.~471-489.

\ \\Matveev, V.~and Shrock, R.~(1995). `Complex-Temperature Singularities of the Susceptibility in the $D=2$ Ising Model. I. Square Lattice'. {\it Journal of Physics A: Mathematical and General}, 28, pp.~1557-1583.

\ \\Matveev, V.~and Shrock, R.~(1995). `Complex-Temperature Properties of the Ising Model on 2D Heteropolygonal Lattices'. {\it Journal of Physics A: Mathematical and General}, 28, pp.~5235-5256.

\ \\Maudlin,T.~(2002). `Thoroughly muddled McTaggart: Or, how to abuse gauge freedom to create metaphysical monostrosities'. {\it Philosopher's Imprint}, 2(4), pp.~1-23.

\ \\Maudlin, T.~(2007). `Completeness, Supervenience and Ontology'. {\it Journal of Physics A: Mathematical and Theoretical}, 40, pp.~3151-3171.

\ \\McKenzie, K.~(2022). {\it Fundamentality and Grounding}. Cambridge: Cambridge University Press.

\ \\Mehra, J.~and Rechenberg, H.~(2000). {\it The Historical Development of Quantum Theory}. Volume 6, The Completion of Quantum Mechanics 1926-1941, Part I. New York: Springer.

\ \\Meissner, W.~and R.~Ochsenfeld~(1933). `Ein neuer effekt bei eintritt der supra\-leit\-f\"ahigkeit.' {\it Naturwissenschaften} 21 (44), pp.~787-788.

\ \\Menon, T.~(2021). `Taking up superspace'. In: {\it Philosophy Beyond Spacetime: Implications from Quantum Gravity}, p.~103.

\ \\Menon, T.~and Read, J.~(2023). `Some remarks on recent mathematical-{\it cum}-formalist responses to the Hole Argument'. In preparation.


\ \\Milnor, J.~W.~(1965). {\it Topology from the Differentiable Viewpoint}. Charlottesville: The University Press of Virginia.

\ \\M\o ller-Nielsen, T.~(2017). `Invariance, Interpretation, and Motivation'. {\it Philosophy of Science}, 84 (5), pp.~1253-1264.

\ \\Montague, R.~(1973). `The Proper Treatment of Quantification in Ordinary English'. In: {\it Approaches to Natural Language: Proceedings of the 1970 Stanford Workshop on Grammar and Semantics}, Hintikka, K.~J.~J., Moravcsik, J.~M.~E.~and Suppes, P., pp.~221-242. Dordrecht: Springer.

\ \\Montano, D.~and Sonnenschein, J.~(1989). `The Topology of Moduli Space and Quantum Field Theory'. {\it Nuclear Physics} B, 324, pp.~348-370.

\ \\Montonen, C.~and Olive, D.~(1977). `Magnetic Monopoles as Gauge Particles?' {\it Physics Letters}, B, 72 (1), pp.~117-120.

\ \\Moravcsik, J.~M.~E.~(1974). `Aristotle on Adequate Explanations'. {\it Synthese}, 28 (1), pp.~3-17.

\ \\Morgan, M.~and Morrison, M.~(1999). {\it Models as Mediators: Perspectives on Natural and Social Science.} Cambridge: Cambridge University Press.

\ \\Morrison, M.~(2007). `Where Have All the Theories Gone?' {\it Philosophy of Science}, 74, pp.~195-228.

\ \\Morrison, M.~(2015). {\it Reconstructing reality: Models, mathematics, and simulations.} New York: Oxford University Press.

\ \\M\"ossner, N.~(2018). {\it Visual Representations in Science: Concept and Epistemology}. London: Routledge.

\ \\Muller, F.~A.~(1997a). `The Equivalence Myth of Quantum Mechanics---Part I'. {\it Studies in History and Philosophy of Modern Physics}, 28 (1), pp.~35-61.

\ \\Muller, F.~A.~(1997b). `The Equivalence Myth of Quantum Mechanics---Part II'. {\it Studies in History and Philosophy of Modern Physics}, 28 (2), pp.~219-247.\\

\ \\Muller, F.~A.~(2014). `Circumveiloped by Obscuritads. The nature of interpretation in quantum mechanics, hermeneutic circles and physical reality, with cameos of James Joyce and Jacques Derrida'. ArXiv: 1406.6284 [physics.hist-ph].

\ \\Mundy, B.~(1992). `Space-time and isomorphism'. In: {\it PSA: Proceedings of the Biennial Meeting of the Philosophy of Science Association}, 1992 (1), pp.~515-527. Cambridge: Cambridge University Press.

\ \\Murdoch, D.~R.~(1987). {\it Niels Bohr's Philosophy of Physics}. Cambridge: Cambridge University Press.

\ \\Murgueitio Ram\'irez, S.~(2021). `A puzzle concerning local symmetries and their empirical significance'. {\it The British Journal for the Philosophy of Science}, 73, pp.~1021-1044.

\ \\Murgueitio Ram\'irez, S.~and Teh, N.~J.~(2021). `Abandoning Galileo's Ship: The quest for non-relational empirical significance'. {\it The British Journal for the Philosophy of Science.} https://doi.org/10.1086/718494.

\ \\Musgrave, A.~(1985). `Realism Versus Constructive Empiricism'. In: {\it Images of Science}, Churchland, P.~M.~and Hooker, C.~A.~(Eds.), pp.~197-221. Chicago: University of Chicago Press.

\ \\Muthukrishnan, S.~(2022). `Unpacking Black Hole Complementarity'. arXiv:2211.15650.

\ \\Myrvold, W.~C.~(2022). {\it Philosophical Issues in Thermal Physics}. Oxford Encyclopaedia of Physics.

\ \\Naber, G.~L.~(2011). {\it Topology, Geometry and Gauge Fields}. New York: Springer.

\ \\Nagel, E.~(1949). `The Meaning of Reduction in the Natural Sciences'. In: {\it Philosophy of Science}, Danto, A., Morgenbesser, S.~(Eds.), pp.~288-312. Cleveland: World Publishing Company.

\ \\Nagel, E.~(1961). {\it The Structure of Science. Problems in the Logic of Scientific Explanation}. New York: Harcourt, Brace and World.

\ \\Nagel, E.~(1979). `Issues in the Logic of Reductive Explanations'. In: {\it Teleology Revisited}, pp.~95-113. New York: Columbia University Press. Reprinted in: M.~Curd~and J.~A.~Cover, {\it Philosophy of Science}, New York London: W.~W.~Norton, pp.~905-921.

\ \\Nahm, W.~(1978). `Supersymmetries and their Representations'. {\it Nuclear Physics} B, 135, pp.~149-166.

\ \\Nakahara, M.~(2003). {\it Geometry, Topology and Physics}. Bristol: Institute of Physics.

\ \\Nambu, Y.~(1960). `Quasi-Particles and Gauge Invariance in the Theory of Superconductivity'. {\it Physical Review,} 117 (3), pp.~648-663.

\ \\Nambu, Y.~(1970). `Duality and Hadrodynamics'. In: {\it Broken Symmetry}, Eguchi, T.~and Nishijima, K.~(Eds.), 1995, pp.~280-301. Singapore: World Scientific.

\ \\Nambu, Y.~(1974). `Strings, Monopoles, and Gauge Fields'. {\it Physical Review D}, 10 (12), pp.~4262-4268.

\ \\Nambu, Y.~and Jona-Lasinio, G.~(1961). `Dynamical Model of Elementary Particles Based on an Analogy with Superconductivity. I'. {\it Physical Review}, 122 (1), pp.~345-358.

\ \\Nelson, P.~(1987). `Lectures on Strings and Moduli Space'. {\it Physics Reports}, 149 (6), pp.~337-375.

\ \\Nepomechie, R.~I.~(1985). `Magnetic Monopoles from Antisymmetric Tensor Gauge Fields'. {\it Physical Review} D, 31 (8), pp.~1921-1924.

\ \\Newman, E.~T.~and Penrose, R.~(1966). `Note on the Bondi-Metzner-Sachs Group', {\it Journal of Mathematical Physics}, 7, pp.~863-870.

\ \\Newman, M.~(2012). `An Inferential Model of Scientific Understanding'. {\it International Studies in the Philosophy of Science}, 26 (1), pp.~1-26.

\ \\Nickles, T.~(1973). `Two Concepts of Intertheoretic Reduction'. {\it The Journal of Philosophy}, 70 (7), pp.~181-201.

\ \\Nicolai, H. ~and Helling, R.~(1999). `Supermembranes and M(atrix) Theory'. In: {\it Non-Perturbative Aspects of Strings, Branes and Supersymmetry}, Duff, M.~et al., pp.~29-74. Singapore: World Scientific.

\ \\Nielsen, H.~B.~and Olesen, P.~(1973). `Vortex-Line Models for Dual Strings'. {\it Nuclear Physics} B, 61, pp.~45-61.

\ \\Nielsen, M.~A.~and Chuang, I.~L.~(2010). {\it Quantum Computation and Quantum Information}. Cambridge: Cambridge University Press.

\ \\North, J.~(2021). {\it Physics, Structure, and Reality}. Oxford: Oxford University Press. 

\ \\Norton, J.~D.~(2012). `Approximation and Idealization: Why the Difference Matters'. {\it Philosophy of Science}, 79, pp.~207-232.

\ \\Norton, J.~D.~(2003). `Must evidence underdetermine theory?' http://philsci-archive.pitt.edu/1257

\ \\Ochs, W.~(2013). `The Status of Glueballs'. {\it Journal of Physics G: Nuclear and Particle Physics}, 40, 043001, pp.~1-68.

\ \\Olive, D.~I.~(1995). `Exact Electromagnetic Duality'. arXiv:hep-th/9508089.

\ \\Olive, D.~I.~(1997). `Introduction to Electromagnetic Duality'. {\it Nuclear Physics} B (Proceedings Supplement), 58, pp.~43-55.

\ \\Onsager, L.~(1944). `Crystal Statistics. I. A Two-Dimensional Model with an Order-Disorder Transition'. {\it Physical Review}, 65, (3-4), pp.~117-149.

\ \\Onsager, L.~(1949). `Statistical Hydrodynamics'. {\it Il Nuovo Cimento (1943-1954)}, 6 (2), pp.~249-287.

\ \\Ooguri, H.~and Vafa, C.~(1991). `Geometry of ${\cal N}=2$ Strings'. {\it Nuclear Physics} B, 361, pp.~469-518.

\ \\Oppenheim, P.~and Putnam, H.~(1958). `Unity of Science as a Working Hypothesis'. In: {\it Concepts, Theories, and the Mind-Body Problem}, Feigl, H., Scriven, M.~and Maxwell, G.~(Eds.). Minnesota: University of Minnesota Press.

\ \\O'Raifeartaigh, L.~(1965). `Lorentz Invariance and Internal Symmetry'. {\it Physical Review}, 139 (4b), pp.~1052-1062.

\ \\Osborn, H.~(1979). `Topological Charges for ${\cal N}=4$ Supersymmetric Gauge Theories and Monopoles of Spin 1'. {\it Physics Letters} B, 83 (3-4), pp.~321-326.

\ \\Osborne, T.~J.~(2012). `Hamiltonian Complexity'. {\it Reports on Progress in Physics}, 74, 022001, pp.~1-10.

\ \\Osterwalder, K.~and Seiler, E.~(1978). `Gauge Field Theories on a Lattice'. {\it Annals of Physics}, 110, pp.~440-471.

\ \\Palti, E.~(2019). `The Swampland: Introduction and Review'. {\it Fortschritte der Physik}, 67, 1900037, pp.~1-95.

\ \\Paulsen, W.~(2014). {\it Asymptotic Analysis and Perturbation Theory}. CRC Press.

\ \\Peet, A.W.~(2001). `TASI Lectures on Black Holes in String Theory', pp.~353-433. In: {\it Strings, branes and gravity. Proceedings TASI '99}, Harvey, J.A., Kachru, S.~and Silverstein, E.~(eds.). New Jersey: World Scientific.

\ \\Penrose, R.~(1963). `Asymptotic Properties of Fields and Space-Times', {\it Physical Review Letters}, 10 (2), pp.~66-68.

\ \\Penrose, R.~(1964). `Conformal treatment of infinity', In: {\it Relativity, groups and topology}, pp.~565-584. DeWitt, B.~and DeWitt, C.~(Eds). New York and London: Gordon and Breach. Republished in: {\it General Relativity and Gravitation} (2011) 43, pp.~901-922.

\ \\Penrose, R.~(1967). `Twistor Algebra'. {\it Journal of Mathematical Physics}, 8 (2), pp.~345-366.

\ \\Penrose, R.~(1968). `Twistor Quantisation and Curved Space-Time'. {\it International Journal of Theoretical Physics}, 1 (1), pp.~61-99.

\ \\Penrose,~R. (1982). `Quasi-Local Mass and Angular Momentum in General Relativity'. {\it Proceedings of the Royal Society of London}, A, 381, pp.~53-63.

\ \\Penrose, R.~(1988). `Aspects of Quasi-Local Angular Momentum'. In: {\it Mathematics and General Relativity}, Isenberg, J.~A.~(Ed.), pp.~1-8. Providence, RI: American Mathematical Society.

\ \\Penrose, R.~and Rindler, W.~(1986). {\it Spinors and Space-Time}. Volume 2. Cambridge: Cambridge University Press.

\ \\Peramatzis, M.~(2011). {\it Priority in Aristotle's Metaphysics}. Oxford: Oxford University Press.

\ \\Peskin, M.~E.~(1978). `Mandelstam-'t Hooft Duality in Abelian Lattice Models'. {\it Annals of Physics,} 113, pp.~122-152.

\ \\Peskin, M.~E.~and Schroeder, D.~V.~(1995). {\it An Introduction to Quantum Field Theory}. Reading, Massachusetts: Addison-Wesley.

\ \\Pestun, V.~(2012). `Localization of Gauge Theory on a Four-Sphere and Supersymmetric Wilson Loops'. {\it Communications in Mathematical Physics}, 313, pp.~71-129.

\ \\Pippard, A.~B.~(1953). `An Experimental and Theoretical Study of the Relation between Magnetic Field and Current in a Superconductor'. {\it Proceedings of the Royal Society of London, Series A,} 216 (1127), pp.~547-568.

\ \\Polchinski, J.~(1995). `Dirichlet Branes and Ramond-Ramond Charges'. {\it Physical Review Letters}, 75 (26), pp.~4724-4727.

\ \\Polchinski, J.~(1998a). {\it String Theory}. Volume I. Cambridge: Cambridge University Press.

\ \\Polchinski, J.~(1998b). {\it String Theory}. Volume II. Cambridge: Cambridge University Press.

\ \\Polchinski, J.~(1995). `Dirichlet Branes and Ramond-Ramond Charges'. {\it Physical Review Letters} B, 75 (26), pp.~4724-4727.

\ \\Polchinski, J.~(2017). `Dualities of Fields and Strings'. {\it Studies in History and Philosophy of Modern Physics}, 59, pp.~6-20.

\ \\Polyakov, A.~M.~(1974). `Particle Spectrum in Quantum Field Theory'. {\it JETP Letters}, 20 (6), pp.~194-195.

\ \\Polyakov, A.~M.~(1975). `Compact Gauge Theories and the Infrared Catastrophe'. {\it Physics Letters}, 59 B (1), pp.~82-84.

\ \\Polyakov, A.~M.~(1977). `Quark Confinement and Topology of Gauge Theories', {\it Nuclear Physics} B, 120, pp.~429-458.

\ \\Pooley, O.~(2001). `Relationism Rehabiliated? II: Relativity'. Preprint, philsci-arthive.pitt.edu.

\ \\Pooley, O.~(2006). `Points, Particles, and Structural Realism'. In: Rickles, D., French, S.~and Saatsi, J.~(Eds.), {\it The structural foundations of quantum gravity}, pp.~83-120. Oxford: Oxford University Press.

\ \\Pooley, O.~(2013). `Substantivalist and Relationist Approaches to Spacetime'. In: {\it The Oxford Handbook of Philosophy of Physics}, Batterman, R., pp.~522-586. Oxford: Oxford University Press.

\ \\Pooley, O.~(2017). `Background Independence, Diffeomorphism Invariance and the Meaning of Coordinates. In: {\it Towards a Theory of Spacetime Theories,} Lehmkuhl, D., Schiemann, G., Scholz, E. (Eds.), pp.~105-143. New York: Birkh\"auser.

\ \\Pooley, O.~(2022). `The Hole Argument'. In: {\it Routledge Companion Philosophy of Physics}, Knox, E.~and Wilson, A., pp.~145-159. New York: Routledge.

\ \\Pooley, O.~and Brown, H.~R.~(2002). `Relationism Rehabilitated? I: Classical Mechanics'. {\it The British Journal for the Philosophy of Science}, 53 (2), pp.~183-204.

\ \\Pooley, O.~and Read, J.~A.~M.~(2021). `On the Mathematics and Metaphysics of the Hole Argument'. arXiv:2110.04036.

\ \\Post, H.~R.~(1971). `Correspondence, Invariance and Heuristics'. {\it Studies in History and Philosophy of Science}, 2 (3), pp.~213-255.

\ \\Prasad, M.~K.~and Sommerfield, C.~M.~(1975). `Exact Solution for the 't Hooft Monopole and the Julia-Zee Dyon'. {\it Physical Review Letters}, 35 (12), pp.~760-762.

\ \\Preskill, J.~(1984). `Magnetic Monopoles'. {\it Annual Review of Nuclear and Particle Science,} 84, pp.~461-530.

\ \\Prugovecki, E.~(1981). {\it Quantum Mechanics in Hilbert Space}. New York: Academic Press (and Dover 2006).

\ \\Putnam, H.~(1962). `What Theories Are Not'. {\it Logic, Methodology, and Philosophy of Science: Proceedings of the 1960 International Congress}, In: Nagel, E., Suppes, P.,~and Tarski, A. (Eds.), pp.~240-251. Stanford: Stanford University Press.

\ \\Putnam, H.~(1978). `Realism and Reason'. In: {\it Meaning and the Moral Sciences}, pp.~123-140. New York: Routledge.

\ \\Putnam, H.~(1996). `Equivalence'. In: {\it Realism and Reason, Philosophical Papers}, volume 3, pp.~26-45. Cambridge: Cambridge University Press.

\ \\Qi, X.-L.~(2013). `Exact Holographic Mapping and Emergent Space-Time Geometry'. arXiv preprint arXiv:1309.6282.

\ \\Quine, W.~V.~O.~(1960). {\it Word and Object}. New Edition, 2013, Cambridge, MA: The MIT Press.

\ \\Quine, W.~V.~O.~(1968). `Ontological relativity', {\it Journal of Philosophy} 65, pp.~185-212.

\ \\Quine, W.~V.~O.~(1975). `On Empirically Equivalent Systems of the World'. {\it Erkenntnis}, 9 (3), pp.~313-328.

\ \\Radder, H.~(1984). {\it The Material Realization of Science}. Revised Edition, with a New Postscript (2012). Dordrecht: Springer.

\ \\Radder, H.~(1991). `Heuristics and the Generalized Correspondence Principle'. {\it The British Journal for the Philosophy of Science}, 42, pp.~195-226.

\ \\Rajaraman, R.~(1975). `Some Non-Perturbative Semi-Classical Methods in Quantum Field Theory'. {\it Physics Reports}, 21 (5), pp.~227-313.

\ \\Ramond, P.~(2016). `Hommage \`a Nambu'. In Memorial Volume For Y. Nambu (pp.~105-124).
arXiv:1603.07795.

\ \\Ramsey-Musolf, M.~J.~(2020). `The Electroweak Phase Transition: A Collider Target'. {\it Journal of High-Energy Physics}, 9, 179, pp.~1-37.

\ \\Read, J.~(2016). `The Interpretation of String-Theoretic Dualities'. {\it Foundations of Physics}, 46, pp.~209-235.

\ \\Read, J.~and M\o ller-Nielsen, T.~(2020). `Motivating Dualities'. {\it Synthese}, 197, pp.~263-291.

\ \\Read, J.~and Teh, N.~J.~(2022). {\it The Philosophy and Physics of Noether's Theorems}. Cambridge: Cambridge University Press.

\ \\Redei, M.~(1998). {\it Quantum Logic in Algebraic Approach}. Dordrecht: Springer.

\ \\Reichenbach, H.~(1938). {\it Experience and Prediction}. Chicago: University of Chicago Press.

\ \\Reutlinger, A., Hangleiter, D.~and Hartmann, S.~(2018). `Understanding (with) Toy Models'. {\it British Journal for the Philosophy of Science}, 69, pp.~1069-1099.

\ \\Rickles, D.~(2008). `Quantum Gravity: A Primer for Philosophers'. In: {\it The Ashgate Companion to Contemporary Philosophy of Physics}, Rickles, D.~(Ed.), pp.~262-364. Aldershot: Ashgate.

\ \\Rickles, D.~(2011). `A Philosopher looks at String Dualities'. {\it Studies in History and Philosophy of Modern Physics}, 42 pp.~54-67.

\ \\Rickles, D.~(2013). `AdS/CFT duality and the emergence of spacetime'. {\it Studies in History and Philosophy of Modern Physics}, 44, pp.~312-320.

\ \\Rickles, D.~(2014). {\it A Brief History of String Theory. From Dual Models to M-Theory}. Heidelberg: Springer.

\ \\Rickles, D.~(2017). `Dual theories: `same but different' or different but same'?' {\it Studies in History and Philosophy of Modern Physics}, 59, 62-67.

\ \\Rivat, S.~(2014). `On the Heuristics of the Higgs Mechanism'. {\it Journal of General Philosophy of Science}, 45, pp.~351-367.

\ \\Roberts, B.~W.~and Weatherall, J.~O.~(2020). `New Perspectives on the Hole Argument'. {\it Foundations of Physics}, 50, pp.~217-227.

\ \\Rothe, H.~J.~(2005). {\it Lattice Gauge Theories}. New Jersey: World Scientific.

\ \\Ruetsche, L.~(2002). `Interpreting Quantum Field Theory'. {\it Philosophy of Science}, 69, pp.~348-378.

\ \\Ruetsche, L.~(2011). {\it Interpreting Quantum Theories}. New York: Oxford University Press.

\ \\Ruetsche, L.~and Earman, J.~(2011). `Interpreting Probabilities in Quantum Field Theory and Quantum Statistical Mechanics'. In: {\it Probabilities in Physics}, C.~Beisbart and Hartmann (Eds.), pp.~263-290. Oxford: Oxford University Press.

\ \\Russo, J.~G.~(1997). `Supermembrane Dynamics from Multiple Interacting Strings'. {\it Nuclear Physics} B, 492, pp.~205-222.

\ \\Russo, J.~G.~and Tseytlin, A.~A.~(1997). `Waves, Boosted Branes and BPS States in M-Theory'. {\it Nuclear Physics} B, 490, pp.~121-144.

\ \\Rynasiewicz, R.~(1994). `The lessons of the hole argument'. {\it The British Journal for the Philosophy of Science}, 45 (2), pp.~407-436.

\ \\Rynasiewicz, R.~(1996). `Is there a Syntactic Solution to the Hole Problem?' {\it Philosophy of Science}, 63, pp.~S55-S62.

\ \\Sachs, R.~(1961). `Gravitational Waves in General Relativity. VI. The Outgoing Radiation Condition', {\it Proceedings of the Royal Society of London. Series A, Mathematical and Physical Sciences}, 264 (1318), pp.~309-338.

\ \\Sachs, R.~(1962). `Asymptotic Symmetries in Gravitational Theory', {\it Physical Review}, 128 (6), pp.~2851-2864.

\ \\S\'a de Melo, C.~A.~R., Randeria, M.~and Engelbrecht, J.~R.~(1993). `Crossover from BCS to Bose Superconductivity: Transition Temperature and Time-Dependent Ginzburg-Landau Theory'. {\it Physical Review Letters}, 71 (19), pp.~3202-3205.

\ \\Saint-James, D., Sarma, G., Thomas, E.~J.~(1969). {\it Type II Superconductivity.} Oxford: Pergamon Press.

\ \\Sakharov, A.~D.~(1991) [1984]. `Cosmological Transitions with Changes in the Signature of the Metric'. {\it Soviet Physics Uspekhi}, 1991, 34 (5), pp.~409-413 [Soviet Physics JETP (1984) 60, pp.~214-218].

\ \\Salam, A.~(1962). `Renormalizability of Gauge Theories'. {\it Physical Review}, 127 (1), pp.~331-334.

\ \\Salam, A.~and Strathdee, J.~(1974). `Super-Gauge Transformations'. {\it Nuclear Physics} B, 76, pp.~477-482. Reprinted in: Ferrara (1987), pp.~35-40.

\ \\Salam, A.~and Ward, J.~C.~(1959). `Weak and Electromagnetic Interactions'. {\it Il Nuevo Cimento}, XI, 4, pp.~568-577.

\ \\Salmon, W.~C.~(1978). `Why Ask, ``Why?''? An Inquiry Conerning Scientific Explanation'. {\it Proceedings and Addresses of the American Philosophical Association}, 51 (6), pp.~683-705.

\ \\Salmon, W.~C.~(1989). {\it Four Decades of Scientific Explanation}. University of Pittsburgh Press.

\ \\Savit, R.~(1978). `Vortices and the Low-Temperature Structure of the $x$-$y$ Model'. {\it Physical Review B}, 17 (3), pp.~1340-1350.

\ \\Savit, R.~(1980). `Duality in Field Theory and Statistical Systems'. {\it Reviews of Modern Physics}, 52 (2), Part I, pp.~453-487.

\ \\Schaffer, J.~(2003). `Is There a Fundamental Level?' {\it No$\hat u$s}, pp.~498-517.

\ \\Schaffer, J.~(2009). `On What Grounds What'. In: {\it Metametaphysics. New Essays on the Foundations of Ontology}. Chalmers, D.~J., Manley, D.~and Wasserman, R.~(Eds.). New York: Oxford University Press.

\ \\Schaffner, K.~F.~(1967). `Approaches to Reduction'. {\it Philosophy of Science}, 34 (2), pp.~137-147.

\ \\Schaffner, K.~F.~(2012). `Ernest Nagel and reduction'. {\it The Journal of Philosophy}, 109 (8/9), pp.~534-565.

\ \\Scheibe, E.~(1973). {\it The Logical Analysis of Quantum Mechanics}. Oxford: Pergamon Press.

\ \\Schlesinger, K.~G.~(2010). `A Physics Perspective on Geometric Langlands Duality'. In: {\it Affine Flag Manifolds and Principal Bundles}, Schmitt, A. (Ed.). Basel: Springer.

\ \\Schr\"odinger, E.~(1926a). `Quantisation as a Problem of Proper Values. Part II.' In: {\it Collected Papers on Wave Mechanics}, pp.~13-40.

\ \\Schr\"odinger, E.~(1926b). `On the Relation between the Quantum Mechanics of Heisenberg, Born, and Jordan, and that of Schr\"odinger'. In: {\it Collected Papers on Wave Mechanics}, pp.~45-61.

\ \\Schroer, B.~and Truong, T.~(1977). `Equivalence of the sine-Gordon and Thirring Models and Cumulative Mass Effects'. {\it Physical Review} D, 15 (6), pp.~1684-1693.

\ \\Schutz, G.~and Lambert, K.~(1994). `Outline of a Theory of Understanding'. {\it Synthese}, 101, pp.~65-120.

\ \\Schwarz, J.~H.~(1996). `Superstring Dualities. {\it Nuclear Physics} B, 49, pp.~183-190.

\ \\Schwarz, J.~H.~(1997). `Lectures on Superstring and M Theory Dualities'. {\it Nuclear Physics} B, 55, pp.~1-32.

\ \\Schweber, S.~S.~(1961). {\it An Introduction to Relativistic Quantum Field Theory}. Evanston, IL: Row.

\ \\Schweber, S.~S.~(1994). {\it QED and the Men Who Made It}. Princeton: Princeton University Press.

\ \\Schwinger, J.~(1962). `Gauge Invariance and Mass'. {\it Physical Review}, 125 (1), pp.~397-398.

\ \\Sebens, C.~T.~(2022). `The Fundamentality of Fields'. {\it Synthese}, 200, 380, pp.~1-28.

\ \\Seiberg, N.~and Witten, E.~(1994a). `Electric-Magnetic Duality, Monopole Condensation, and Confinement in ${\cal N}=2$ Supersymmetric Yang-Mills Theory'. {\it Nuclear Physics} B, 426, pp.~19-52.

\ \\Seiberg, N.~and Witten, E.~(1994b). `Monopoles, Duality and Chiral Symmetry Breaking in ${\cal N}=2$ Supersymmetric QCD'. {\it Nuclear Physics} B, 431, pp.~484-550.

\ \\Seiberg, N., Senthil, T., Wang, C., and Witten, E.~(2016). `A Duality Web in 2+1 Dimensions and Condensed Matter Physics'. {\it Annals of Physics}, 374, pp.~395-433. 

\ \\Sen, A.~(1994). `Strong-Weak Coupling Duality in Four-Dimensional String Theory'. {\it International Journal of Modern Physics} A, 9 (21), pp.~3707-3750.

\ \\Sen, A.~(1999). `Duality Symmetries in String Theory'. {\it Current Science}, 77 (12), pp.~1635-1645.

\ \\Sen, A.~(2015). `Ultraviolet and Infrared Divergences in Superstring Theory'. arXiv:1512.00026. https://arxiv.org/abs/1512.00026

\ \\Shapere, A.~and Wilczek, F.~(1989). `Self-Dual Models with Theta Terms'. {\it Nuclear Physics} B 320, pp.~669-695.

\ \\Shizuya, K.~(1977). `Renormalization of Two-Dimensional Massive Yang-Mills Theory and Non-Renormalizability of Its Four-Dimensional Version'. {\it Nuclear Physics}, B 121, pp.~125-140.

\ \\Shuryak, E.~(2021). {\it Nonperturbative Topological Phenomena in QCD and Related Theories}. Cham: Springer.

\ \\Silberstein, L.~(1907a). `Elektromagnetische Grundgleichungen in bivektorieller Behandlung'. {\it Annalen der Physik}, 327 (3), pp.~579-586.

\ \\Skenderis, K.~(2001). `Asymptotically anti-de Sitter Spacetimes and their Stress-Energy Tensor'. {\it International Journal of Modern Physics} A, 16 (5), pp.~740-749.

\ \\Skenderis, K.~(2002). `Lecture Notes on Holographic Renormalization'. {\it Classical and Quantum Gravity}, 19, pp.~5849-5876.

\ \\Skenderis, K.~and van Rees, B.~C.~(2008). `Real-Time Gauge-Gravity Duality'. {\it Physical Review Letters}, 101, 081601, pp.~1-4.

\ \\Skenderis, K.~and van Rees, B.~C.~(2009). `Real-Time Gauge-Gravity Duality: Prescription, Renormalization and Examples'. {\it Journal of High-Energy Physics}, 5, 85, pp.~1-70.

\ \\Sklar, L.~(1967). `Types of Inter-Theoretic Reduction'. {\it The British Journal for the Philosophy of Science}, 18, pp.~109-124.

\ \\Sklar, L.~(1975). `Methodological Conservatism'. {\it The Philosophical Review}, 84 (3), pp.~374-400.

\ \\Sklar, L.~(1982). `Saving the Noumena'. {\it Philosophical Topics}, 13 (1), pp.~89-110.

\ \\Smeenk, C.~(2006). `The Elusive Higgs Mechanism'. {\it Philosophy of Science}, 73, pp.~487-499.

\ \\Smolin, L.~(2006). {\it The Trouble with Physics}. Boston: Houghton Mifflin.

\ \\Sober, E.~(1999). `The Multiple Realizability Argument against Reductionism'. {\it Philosophy of Science}, 66 (4), pp.~542-564.

\ \\Sohnius, M.~F.~(1985). `Introducing Supersymmetry'. {\it Physics Reports}, 128 (2-3), pp.~39-204.\\
\\
Speaks, J. (2019). `Theories of Meaning'. {\it Stanford Encyclopedia of Philosophy}. https://plato.stanford.edu/entries/meaning.

\ \\Stachel, J.~(1993). `The Meaning of General Covariance'. In: {\it Philosophical Problems of the Internal and
External Worlds: Essays on the Philosophy of Adolf Grunbaum}, Earman, J., Janis, A., Massey, G.~J.~and Rescher, N.~(Eds.), pp.~129-60. Pittsburgh: University of Pittsburgh Press.

\ \\Stanford, P.~K.~(2006). {\it Exceeding Our Grasp}. New York: Oxford University Press.

\ \\Stuewer, R.~H.~(1975). {\it The Compton Effect}. New York: Neale Watson Academic Publications.

\ \\St\"oltzner, M.~(2012). `Constraining the Higgs Mechanism: Ontological Worries and the Prospects for an Algebraic Curve'. {\it Philosophy of Science}, 79, pp.~930-941.

\ \\Strominger, A.~(2001). `The dS/CFT Correspondence'. {\it Journal of High-Energy Physics}, 10, 34, pp.~1-18.

\ \\Strominger, A.~and Vafa, C.~(1996). `Microscopic Origin of the Bekenstein-Hawking Entropy'. {\it Physics Letters} B, 379, pp.~99-104.

\ \\Strominger, A., Yau, S.~T.~and Zaslow, E.~(1996). `Mirror Symmetry is T-duality'. {\it Nuclear Physics} B, 479 (1-2), pp.~243-259.

\ \\Struyve, W.~(2011). `Gauge Invariant Accounts of the Higgs Mechanism'. {\it Studies in History and Philosophy of Modern Physics}, 42, pp.~226-236.

\ \\Su\'arez, M.~and Cartwright, N.~(2008). `Theories: Tools versus Models'. {\it Studies in History and Philosophy of Modern Physics}, 39, pp.~62-81.

\ \\Suppe, F.~(1974). `The Search for Philosophic Understanding of Scientific Theories'. In: Suppe, F.~(Ed.), {\it The Structure of Scientific Theories}, pp.~3-232.

\ \\Suppe, F.~(2000). `Understanding Scientific Theories: An Assessment of Developments, 1969-1998'. {\it Philosophy of Science}, 67, pp.~102-115.

\ \\Suppes, P.~(1962). `Models of Data'. In: Nagel, E., Suppes, P.~and Tarski, A.~(Eds.), {\it Logic, Methodology and Philosophy of Science. Proceedings of the 1960 International Congress}, pp.~252-261. Stanford, CA: Stanford University Press.

\ \\Susskind, L.~(2015). `The World as a Hologram'. {\it Journal of Mathematical Physics}, 36, pp.~6377-6396.

\ \\Susskind, L.~(2016). `Computational complexity and black hole horizons'. {\it Fortschritte der Physik}, 64.1, pp.~24-43.

\ \\Szabados, L.~B.~(2009). `Quasi-Local Energy-Momentum and Angular Momentum in General Relativity'. {\it Living Reviews in Relativity}, 12 (4), pp.~1-163.

\ \\Tahko, T.~E.~(2018). `Fundamentality'. {\it Stanford Encylopedia of Philosophy}. https://plato.stanford.edu/entries/fundamentality.

\ \\Takhtajan, L.~(2008). {\it Quantum Mechanics for Mathematicians}. American Mathematical Society.

\ \\Tarski, A.~(1983) [1931]. `Some Methodological Investigations on the Definability of Concepts'. In: {\it Logic, Semantics, Metamathematics. Papers from 1923 to 1938}, pp.~296-319. Indianapolis: Hackett Publishing Company, Second Edition.

\ \\Teh, N.~J.~(2013). `Holography and Emergence'. {\it Studies in History and Philosophy of Modern Physics}, 44, pp.~300-311.

\ \\Teh, N.~J.~(2016). `Galileo's Gauge: Understanding the Empirical Significance of Gauge Symmetry'. {\it Philosophy of Science,} 83, pp.~93-118.

\ \\Teitelboim, C.~(1986). `Monopoles of Higher Rank'. {\it Physics Letters}, 167 B (1), pp.~69-72.

\ \\Teller, P.~(1995). {\it An Interpretive Introduction to Quantum Field Theory}. Princeton: Princeton University Press.

\ \\ten Cate, B.~and Kolaitis, P.~G.~(2014). `Schema Mappings: A Case of Logical Dynamics in Database Theory'. In: Baltag and Smets (Eds.) (2014), pp.~67-100.

\ \\Thompson, C.~J.~(1972). {\it Mathematical Statistical Mechanics}. Princeton: Princeton University Press.

\ \\'t Hooft, G.~(1974a). `A Planar Diagram Theory for Strong Interactions'. {\it Nuclear Physics} B, 72, pp.~461-473.

\ \\'t Hooft, G.~(1974b). `Magnetic Monopoles in Unified Gauge Theories'. {\it Nuclear Physics} B, 79, pp.~276-284.

\ \\'t Hooft, G.~(1975). `Gauge Fields with Unified Weak, Electromagnetic, and Strong Interactions'. {\it International Physics Series}, pp.~1225-1249.

\ \\'t Hooft, G.~(1979). `A Property of Electric and Magnetic Flux in Non-Abelian Gauge Theories'. {\it Nuclear Physics} B, 153, pp.~141-160.

\ \\'t Hooft, G.~(1993). `Dimensional Reduction in Quantum Gravity'. In: {\it Salamfestschrift: A Collection of Talks}, 1994, 4 (A), 1-13. Ellis, A.J.~and Randjbar-Daemi, S.~(Eds.). World Scientific. arXiv preprint gr-qc/9310026.

\ \\'t Hooft, G.~(2007). `The Conceptual Basis of Quantum Field Theory'. In: {\it Handbook of the Philosophy of Science. Philosophy of Physics}, Butterfield, J.~and Earman, J.~(Eds.). Amsterdam: Elsevier.

\ \\Tinkham, M.~(1996). {\it Introduction to Superconductivity}. New York: McGraw-Hill.

\ \\Townsend, P.~K.~(1995). `The Eleven-Dimensional Supermembrane Revisited'. {\it Physics Letters} B, 350, pp.~184-188.

\ \\Trout, J.~D.~(2002). `Scientific Explanation and the Sense of Understanding'. {\it Philosophy of Science}, 69, pp.~212-233.

\ \\Trout, J.~D.~(2005). `Paying the Price for a Theory of Explanation: De Regt's Discussion of Trout'. {\it Philosophy of Science}, 72, pp.~198-208.

\ \\Trout, J.~D.~(2007). `The Psychology of Scientific Explanation'. {\it Philosophy Compass}, 2/3, pp.~564-591.

\ \\Tseytlin, A.~A.~(2000). `$R^4$ Terms in 11 Dimensions and Conformal Anomaly of $(2,0)$ Theory'. {\it Nuclear Physics} B, 584 (1-2), pp.~233-250.

\ \\Turaev, V.~G.~(1992). `Shadow Links and Face Models of Statistical Mechanics'. {\it Journal of Differential Geometry}, 36, pp.~35-74.

\ \\Turaev, V.~G.~(2010). {\it Quantum Invariants of Knots and 3-Manifolds}. Berlin: De Gruyter.

\ \\Vafa, C.~and Witten, E.~(1994). `A Strong Coupling Test of S-Duality'. {\it Nuclear Physics} B, 431, pp.~3-77.

\ \\van Benthem, J.~(1976). {\it Modal Correspondence Theory}. PhD thesis, University of Amsterdam.

\ \\van Benthem, J.~(1985). {\it Modal Logic and Classical Logic}. Napels: Bibliopolis.

\ \\van Benthem, J.~(2001). `Correspondence Theory'. In: Gabbay, D.~M.~and Guenthner, F.~(Eds.), {\it Handbook of Philosophical Logic}, Second Edition, Volume 3, pp.~325-408. Springer.

\ \\van Benthem, J.~(2010a). `Logic between Expressivity and Complexity'. In: Giesl, J.~and H\"ahnle, R.~(Eds.), {\it International Joint Conference on Automated Reasoning}, pp.~122-126. Berlin, Heidelberg: Springer.

\ \\van Benthem, J.~(2010b). {\it Modal Logic for Open Minds}. Stanford, CA: CSLI Publications.

\ \\van der Waerden, B.~L.~(1967). {\it Sources of Quantum Mechanics}. New York: Dover.

\ \\van Dongen, J.~and De Haro, S.~(2004). `On Black Hole Complementarity'. {\it Studies in History and Philosophy of Modern Physics}, 35 (3), pp.~509-525.

\ \\van Dongen, J., De Haro, S., Visser, M.~and Butterfield, J.~(2020). `Emergence and Correspondence for String Theory Black Holes'. {\it Studies in History and Philosophy of Modern Physics}, 69, pp.~112-127.

\ \\van Fraassen, B.~C.~(1970). `On the Extension of Beth's Semantics of Physical Theories'. {\it Philosophy of Science}, 37 (3), pp.~325-339.

\ \\van Fraassen, B.~C.~(1974). `The Formal Representation of Physical Quantities'. {\it Boston Studis in the Philosophy of Science}, XIII, pp.~196-209.

\ \\van Fraassen, B.~C.~(1977). `The Pragmatics of Explanation'. {\it American Philosophical Quarterly}, 14 (2), pp.~143-150.

\ \\van Fraassen, B.~C.~(1980). {\it The Scientific Image.} Oxford: Clarendon Press.

\ \\van Fraassen, B.~C.~(1989). {\it Laws and Symmetry}. Oxford: Oxford University Press.

\ \\van Fraassen, B.~C.~(2014). `One or Two Gentle Remarks about Hans Halvorson's Critique of the Semantic View'. {\it Philosophy of Science}, 81, pp.~276-283.

\ \\van Riet, T.~and Zoccarato, G.~(2023). `Beginners Lectures on Flux Compactifications and Related Swampland Topics'. arXiv preprint arXiv:2305.01722.

\ \\Vassallo, A.~and Esfeld, M.~(2016). `Leibnizian Relationism for General Relativistic Physics'. {\it Studies in History and Philosophy of Modern Physics}, 55, pp.~101-107.

\ \\Venema, Y.~(2014). `Expressiveness Modulo Bisimilarity: A Coalgebraic Perspective'. In: Baltag and Smets (Eds.) (2014), pp.~33-65.

\ \\Vergouwen, S.~(2022). {\it Emergent Solitons and the Philosophy of Non-Perturbative Quantum Field Theory}. MSc thesis, Utrecht University.\\ https://studenttheses.uu.nl/handle/20.500.12932/42660

\ \\Vergouwen, S.~and De Haro, S.~(2024). `Supersymmetry in the Seiberg-Witten Theory: A Window into Quantum Field Theory'. {\it Synthese}, forthcoming. 

\ \\Verlinde, E.~(1995). `Global Aspects of Electric-Magnetic Duality'. {\it Nuclear Physics} B, 455, pp.~211-225.

\ \\Vonk, M.~(2005). `A Mini-Course on Topological strings'. arXiv preprint hep-th/0504147.

\ \\Von Neumann, J.~(1932) [1955]. {\it Mathematical Foundations of Quantum Mechanics}. Princeton: Princeton University Press.

\ \\Wald, R.~M.~(1994). {\it Quanfum Field Theory in Curved Spacetime and Black Hole Thermodynamics}. Chicago: The University of Chicago Press.

\ \\Wallace, D.~(2011). `Taking Particle Physics Seriously: A Critique of the Algebraic Approach to Quantum Field Theory'. {\it Studies in History and Philosophy of Modern Physics}, 42, pp.~116-125.

\ \\Wallace, D.~(2018). `The Case for Black Hole Thermodynamics Part I: Phenomenological Thermodynamics'. {\it Studies in History and Philosophy of Modern Physics}, 64, pp.~52-67.

\ \\Wallace, D.~(2019). `The Case for Black Hole Thermodynamics Part II: Statistical Mechanics'. {\it Studies in History and Philosophy of Modern Physics}, 66, pp.~103-117.

\ \\Wallace, D.~(2022a). `Isolated Systems and their Symmetries, Part I: General Framework and Particle-Mechanics Examples'. {\it Studies in History and Philosophy of Science}, 92, pp.~239-248.

\ \\Wallace, D.~(2022b). `Isolated Systems and their Symmetries, Part II: Local and Global Symmetries of Field Theories'. {\it Studies in History and Philosophy of Science}, 92, pp.~249-259.

\ \\Wallace, D.~(2022c). `Stating Structural Realism: Mathematics-First Approaches to Physics and Metaphysics'. {\it Philosophical Perspectives}, 36, pp.~345-378.

\ \\Wang, M.~T.~(2015). `Four Lectures on Quasi-Local Mass'. arXiv preprint arXiv:1510.02931.

\ \\Watrous, J.~(2008). `Quantum Computational Complexity'. arXiv preprint arXiv:0804.3401.

\ \\Weatherall, J.~O.~(2016a). `Are Newtonian gravitation and geometrized Newtonian gravitation theoretically equivalent?' {\it Erkenntnis}, 81 (5), pp.~1073-1091.

\ \\Weatherall, J.~O.~(2016b). `Understanding Gauge'. {\it Philosophy of Science}, 83, pp.~1039-1049.

\ \\Weatherall, J.~O.~(2018). `Regarding the `Hole Argument'{}'. {\it The British Journal for the Philosophy of Science}, 69, pp.~329-350.

\ \\Weatherall, J.~O.~(2020). `Equivalence and Duality in Electromagnetism'. {\it Philosophy of Science}, 87 (5), pp.~1172-1183.

\ \\Weinberg, E.~J.~(1980). `Fundamental Monopoles and Multimonopole Solutions for Arbitrary Simple Gauge Groups'. {\it Nuclear Physics} B, 167, pp.~500-524.

\ \\Weinberg, E.~J.~(1984). `Monopoles and Grand Unification'. In: {\it Monopole '83,} Stone, J.~L. (Ed.), pp.~1-16. New York: Plenum Press.

\ \\Weinberg, E.~J.~(2012). {\it Classical Solutions in Quantum Field Theory}. Cambridge: Cambridge University Press.

\ \\Weinberg, S.~(1967). `A Model of Leptons'. {\it Physical Review Letters}, 19 (21), pp.~1264-1266.

\ \\Weinberg, S.~(1993). {\it Dreams of a Final Theory. The Scientist's Search for the Ultimate Laws of Nature.} New York: Vintage.

\ \\Weinberg, S.~(1995). {\it The Quantum Theory of Fields}. Volume I. Cambridge: Cambridge University Press.

\ \\Weinberg, S.~(2001). {\it Facing Up}. Cambridge, MA: Harvard University Press.

\ \\Weinberg, S.~(2008). `From BCS to the LCH'. {\it AAPPS Bulletin}, 18 (2), pp.~30-35.

\ \\Wess, J.~and Bagger, J.~(1992). {\it Supersymmetry and Supergravity}. Princeton: Princeton University Press.

\ \\Wess, J.~and Zumino, B.~(1974a). `A Lagrangian Model Invariant Under Supergauge Transformations'. {\it Physics Letters} B, 49 (1), pp.~52-54.

\ \\Wess, J.~and Zumino, B.~(1974b). `Supergauge Transformations in Four Dimensions'. {\it Nuclear Physics} B, 70, 39-50.

\ \\Wess, J.~and Zumino, B.~(1974c). `Supergauge Invariant Extension of Quantum Electrodynamics'. {\it Nuclear Physics} B, 78, pp.~1-13.

\ \\Weyl, H.~(1934). `Mind and Nature'. In: {\it Hermann Weyl, Mind and Nature, Selected Writings on Philosophy, Mathematics, and Physics}, Pesic, P.~(Ed.), 2009, pp.~95-96. Princeton: Princeton University Press.

\ \\Whewell, W.~(1876). `Letter to Prof.~J.~D.~Forbes (1860)'. In: {\it William Whewell, D.~D., Master of Trinity College, Cambridge: an account of his literary and scientific correspondence}, vol.~2, Todhunter, I. London: Macmillan 1876.

\ \\Whewell, W.~(1847) [1967]. {\it The Philosophy of the Inductive Sciences}, Volume 2. New York and London: Johnson.

\ \\Wigner, E.~P.~(1931). {\it Gruppentheorie und ihre Anwendung auf die Quantenmechanik der Atomspektren}. Braunschweig: Fried.~Vieweg und Sohn. The page numbers refer to the English translation: {\it Group Theory}, 1959, New York: Academic Press.

\ \\Wigner, E.~P.~(1939). `On Unitary Representations of the Inhomogeneous Lorentz Group'. {\it Nuclear Physics B}, 6, 1989, pp.~9-64.

\ \\Wilkenfeld, D.~A.~(2013). `Understanding as Representation Manipulability'. {\it Synthese}, 190, pp.~997-1016.

\ \\Williams, P.~(2019). `Scientific Realism Made Effective'. {\it The British Journal for the Philosophy of Science}, 70, pp.~209-237.

\ \\Wilson, K.~G.~(1974). `Confinement of Quarks'. {\it Physical Review} D, 10 (8), pp.~2445-2459.

\ \\Wilson, J.~M.~(2014). `No Work for a Theory of Grounding'. {\it Inquiry}, 57 (5-6), pp.~535-579.

\ \\Wilson, K.~G.~(1974). `Confinement of Quarks'. {\it Physical Review D}, 10 (8), pp.~2445-2459.

\ \\Wimsatt, W.~C.~(1997). `Aggregativity: Reductive Heuristics for Finding Emergence'. {\it Philosophy of Science}, 64, Supplement, pp.~S372-S384.

\ \\Witten, E.~(1979). `Dyons of Charge $e\th/2\pi$'. {\it Physics Letters} 86 B (3), pp.~283-287.

\ \\Witten, E.~(1980). `The $1/N$ expansion in atomic and particle physics.' {\it Recent Developments in Gauge Theories}, pp.~403-419. Springer US. 

\ \\Witten, E.~(1984). `Nonabelian Bosonization in Two Dimensions'. {\it Communications in Mathematical Physics} 92, 455-472.

\ \\Witten, E.~(1987). `Elliptic genera and quantum field theory', {\it Communications in Mathematical Physics}, 109, pp.~525-536.

\ \\Witten, E.~(1988a). `Topological Quantum Field Theory'. {\it Communications in Mathematical Physics}, 117, pp.~353-386.

\ \\Witten, E.~(1988b). `Topological Sigma Models. {\it Communications in Mathematical Physics}, 118 (3), pp.~411-449.

\ \\Witten, E.~(1988c). `2+ 1 Dimensional Gravity as an Exactly Soluble System'. {\it Nuclear Physics B}, 311 (1), pp.~46-78.

\ \\Witten, E.(1989). `Quantum Field Theory and the Jones Polynomial'. {\it Communications in Mathematical Physics}, 121 (3), pp.~351-399.

\ \\Witten, E.~(1991). `String Theory and Black Holes'. {\it Physical Review} D, 44 (2), pp.~314-324.

\ \\Witten, E.~(1991). `Mirror Manifolds and Topological Field Theory'. arXiv preprint hep-th/9112056.

\ \\Witten, E.~(1994). `Perturbative Gauge Theory as a String Theory in Twistor Space'. {\it Communications in Mathematical Physics}, 252, pp.~189-258.

\ \\Witten, E.~(1995a). `String Theory Dynamics in Various Dimensions'. {\it Nuclear Physics} B, 443, pp.~85-126.

\ \\Witten, E.~(1995b). `On S-duality in Abelian Gauge Theory'. {\it Selecta Mathematica, New Series}, 1 (2), pp.~383-410.

\ \\Witten, E.~(1995c). `Some Comments on String Dynamics'. In: {\it Strings '95}, Bars, I., Bouwknegt, P., Minahan, J., Nemeschansky, D., Pilch, K., Saleur, H.~and Warner, N.~(Eds.), pp.~501-523. Singapore: World Scientific.

\ \\Witten, E.~(1996). `Five-Branes and M-Theory on an Orbifold'. {\it Nuclear Physics} B, 463, pp.~383-397.

\ \\Witten, E.~(1998). `Anti-de Sitter Space and Holography'. arXiv: hep-th/9802150.

\ \\Witten, E.~(2001). `Quantum Gravity in de Sitter space'. arXiv: hep-th/0106109. 

\ \\Witten, E.~(2004). `Perturbative Gauge Theory as a String Theory in Twistor Space'. {\it Communications in Mathematical Physics}, 252, pp.~189-258.

\ \\Witten, E.~and Olive, D.~(1978). `Supersymmetry Algebras that Include Topological Charges'. {\it Physics Letters} B, 78, pp.~97-101.

\ \\Woodward, J.~and Ross, L.~(2021). `Scientific Explanation'. {\it Stanford Encyclopedia of Philosophy}. https://plato.stanford.edu/entries/scientific-explanation. 

\ \\Worrall, J.~(1989). `Structural Realism: The Best of both Worlds?' {\it Dialectica}, 43 (1‐2), 99-124.

\ \\Wu, T.~T.~and Yang, C.~N.~(1975). `Concept of Non-Integrable Phase Factors and Global Formulation of Gauge Fields'. {\it Physical Review D,} 12 (12), pp.~3845-3857.

\ \\Yang, C.~N.~(1952). `The Spontaneous Magnetization of a Two-Dimensional Ising Model'. {\it Physical Review}, 85 (5), pp.~808-816.

\ \\Yang, C.~N.~(1970). `Charge Quantization, Compactness of the Gauge Group, and Flux Quantization'. {\it Physical Review D}, 1 (8), p.~2360.

\ \\Zaanen, J., Sun, Y.-W., Liu, Y.~and Schalm, K.~(2015). {\it Holographic Duality in Condensed Matter Physics}. Cambridge: Cambridge University Press.

\ \\Zee, A.~(2003). {\it Quantum Field Theory in a Nutshell}. Princeton and Oxford: Princeton University Press.

\ \\Zinn-Justin, J.~(2002). {\it Quantum Field Theory and Critical Phenomena}. Oxford: Oxford University Press.

\ \\Zwiebach, B.~(2009). {\it A First Course in String Theory}. Cambridge: Cambridge University Press.

\end{document}